\documentclass{emulateapj}
\usepackage{apjfonts}
\usepackage{lscape}
\usepackage{amsmath}

\newcommand{\Halpha}{H{$\alpha$}}

\newcommand{\LB}{\ensuremath{L_{\mathrm{B}}}}

\newcommand{\Lsol}{\ensuremath{L_{\mathrm{\odot}}}}

\newcommand{\LX}{\ensuremath{L_{\mathrm{X}}}}

\newcommand{\NH}{\ensuremath{N_{\mathrm{H}}}}

\newcommand{\arcs}{\ensuremath{^{\prime\prime}}}

\newcommand{\pcmsq}{\ensuremath{\cm^{-2}}}

\newcommand{\erg}{\ensuremath{\mbox{erg}}}

\newcommand{\cm}{\ensuremath{\mbox{cm}}}

\newcommand{\nm}{\ensuremath{\mbox{\nm}}}

\newcommand{\ps}{\ensuremath{\s^{-1}}}

\newcommand{\s}{\ensuremath{\mbox{s}}}

\newcommand{\ergps}{\ensuremath{\erg~\ps}}

\newcommand{\etal}{{\it et al.\thinspace}}

\newcommand{\CHANDRA}{\emph{Chandra}}

\newcommand{\chisq}{\ensuremath{\chi^2}}
\newcommand{\rchisq}{\ensuremath{\chi^2_\nu}}



\begin{document}

\title{Deep \CHANDRA\ Monitoring Observations of NGC 4278: Catalog of 
 Source Properties  \\}

\author{N. J. Brassington, G. Fabbiano, D.-W. Kim, A. Zezas}
\affil{Harvard-Smithsonian Center for Astrophysics, 60 Garden
Street, Cambridge, MA 02138}
\email{nbrassington@head.cfa.harvard.edu}
\author{S. Zepf, A. Kundu}
\affil{Department of Physics and Astronomy, Michigan State University, East Lansing, MI 48824-2320}
\author{L. Angelini}
\affil{Laboratory for X-Ray Astrophysics, NASA Goddard Space Flight Center, Greenbelt, MD 20771}
\author{R. L. Davies}
\affil{Sub-Department of Astrophysics, University of Oxford, Oxford
OX1 3RH, UK}
\author{J. Gallagher}
\affil{Department of Astronomy, University of Wisconsin, Madison, WI 53706-1582}
\author{V. Kalogera, T. Fragos}
\affil{Department of Physics and Astronomy, Northwestern University, Evanston, IL 60208}
\author{A. R. King}
\affil{Theoretical Astrophysics Group, University of Leicester, Leicester 
LE1 7RH, UK}
\author{S. Pellegrini}
\affil{Dipartimento di Astronomia, Universita di Bologna, Via Ranzani 1, 40127 Bologna, Italy}
\author{G. Trinchieri}
\affil{INAF-Osservatorio Astronomico di Brera, Via Brera 28, 20121 Milan, Italy}
\author{}
\affil{}
\author{}
\affil{}

\shorttitle{Catalog of the X-ray sources in NGC 4278}
\shortauthors{Brassington \etal}
\bigskip

\begin{abstract}
 
We present the properties of the discrete X-ray sources detected in
our monitoring program of the globular cluster (GC) rich elliptical galaxy, NGC
4278, observed with {\em Chandra} ACIS-S in six separate pointings,
resulting in a co-added exposure of 458-ks. From this deep
observation, 236 sources have been detected
within the region overlapped by all observations, 180 of which lie within the
$D_{25}$ ellipse of the galaxy. These 236 sources range in \LX\
from 3.5$\times 10^{36}$ \ergps\ (with 3$\sigma$ upper limit $\le$1$\times 10^{37}$ \ergps) to $\sim2~\times 10^{40}$\ergps, including the central nuclear source which has been classified as a LINER. 
From optical data, 39 X-ray sources have been determined to be
coincident with a globular cluster, these sources tend to have high X-ray luminosity, with
ten of these sources exhibiting \LX$> 1\times10^{38}$\ergps. From X-ray source
photometry, it has been determined that the majority of the 236 point sources that
have well constrained colors, have values that are consistent with
typical LMXB spectra, with 29 of these sources expected to be background objects from the $logN - logS$ relation.
There are 103 sources in this population that exhibit long-term
variability, indicating that they are accreting compact objects. 3 of
these sources have been identified as transient candidates, with a further 3
possible transients. Spectral variations have also been identified in
the majority of the source population, where a diverse range of
variability has been identified, indicating that there are many
different source classes located within this galaxy.

\end{abstract}

\section{Introduction}

NGC 4278 is the second of two nearby, isolated, normal elliptical galaxies for which we were awarded deep monitoring observations with \CHANDRA\ ACIS-S (Weisskopf \etal\ 2000). The purpose of these observations was to study the low-mass X-ray binary (LMXB) population down to limiting luminosities of a few (0.3$-$8.0 keV) $10^{36}$ \ergps, well in the range of those LMXBs in the Galaxy and M31. 
LMXBs are the only direct fossil evidence of
the formation and evolution of binary stars in the old  stellar
populations of early-type galaxies. First discovered in the Milky Way
(see Giacconi 1974), these binaries are composed of a compact
accretor, neutron star or black hole, and a late-type stellar
donor. The origin and evolution of Galactic LMXBs has been the subject
of much discussion, centered on two main evolution paths (see Grindlay
1984; review by Verbunt \& Van den Heuvel 1995): the evolution of
primordial binary systems in the stellar field, or formation and
evolution in Globular Cluster. 

With the advent of \CHANDRA, many LMXB
populations have been discovered in early-type galaxies (see review
Fabbiano 2006), and the same evolutionary themes (field or GC
formation and evolution) have again surfaced, supported and stimulated
by a considerably larger and growing body of data. With our unique observations of NGC 3379 (Brassington \etal\ 2008) and NGC 4278 (this paper), we can now take a fresh look at these issues.
Moreover, our observations have a time sampling that permits
variability studies and the identification of X-ray
transients. Such data will extend the limited sample of existing multi-epoch
observations (albeit with higher limiting luminosities; e.g. Irwin 2006; Sivakoff \etal\ 2008), where a variety of different variability behaviours of LMXBs have already been observed.
Both multi-epoch observations and low luminosity thresholds are
important aspects of the observational characteristics of 
Galactic LMXBs and are needed for constraining the evolution of these
populations (e.g., Piro \& Bildsten 2002, Bildsten \& Deloye
2004).

The first galaxy that was observed in this program was NGC 3379, in the nearby poor group Leo (D=10.6~Mpc; Tonry \etal\ 2001). This system was chosen for this study because it is a relatively isolated unperturbed `typical' elliptical galaxy, with an old stellar population (age of 9.3~Gyr, Terlevich \& Forbes 2002) and a poor globular cluster system ($S_  {GC}=1.3 \pm 0.7$, Harris 1991; where $S_{GC}$ = No.GC$\times 10^{(0.4(M_V+15))}$). From the 324-ks \CHANDRA\ observation, reported in detail in Brassington \etal\ (2008), 132 X-ray point sources were detected, 98 of which were found to lie within the $D_{25}$ ellipse of the galaxy. Through {\em HST} observations of the galaxy the globular cluster system was identified in the central region and ten GC-LMXBs detected. These GC-LMXB associations were found to have high luminosities, with three sources exhibiting \LX$\ge 1\times10^{38}$\ergps. Also from studying this population it was determined that 48\% of the detected LMXBs were found to show some type of long-term variability, with 5 of these sources being classified as transients and a further 3 of these as possible transients.

The characteristics of NGC 4278 are similar to those of NGC 3379 in all but the GC population, where NGC 4278 has a rich globular cluster system ($S_{GC}=6.1 \pm 1.5$, Harris 1991), making it an ideal galaxy to explore the evolution of LMXB from primordial field binaries in a rich environment. Like NGC 3379, NGC 4278 is a nearby (D=16.1~Mpc; Tonry \etal\ 2001) relatively isolated, unperturbed `typical' elliptical galaxy with an old stellar population (age of 10.7~Gyr, Terlevich \& Forbes 2002).

Observationally, NGC 4278 is an ideal target for LMXB population
studies, because of its proximity, resulting in a resolution of
$\sim36$~pc with {\em Chandra}, and its lack of a prominent hot
gaseous halo. These characteristics optimize the
detection of fainter LMXBs, and minimize source confusion; because of
its angular diameter ($D_{25} = 4.0$ arcmin, RC3), NGC 4278 is entirely
contained in the ACIS-S3 CCD chip, and is not affected by the
degradation of the {\it Chandra} PSF at large radii. 

Here we publish the catalog of LMXBs with their properties
resulting from the entire observational campaign of NGC 4278 (five
observations between March 2006 and February 2007, for a total of $\sim430$~ks), which has been recently completed, and includes the first 37~ks observation taken in
2005, from the {\it Chandra} archive.
In addition to this catalog paper, further highlights from the X-ray binary population of NGC 4278 will be presented in forthcoming papers. In the first (Kim \etal\ 2009, in prep) we compare the X-ray luminosity function of GC and field LMXBs detected in both NGC 4278 and NGC 3379.  This will be followed by a paper presenting the properties of the transient population of both galaxies (Brassington \etal\ 2009, in prep). 
Further papers will be presented on: the properties of the nuclear source and the diffuse emission of NGC 4278, as well
as the nonuniform distribution of the LMXBs within the galaxy and the intensity and spectral variability of the luminous X-ray binary population.

This paper is organized as follows: \S 2. details the
observational program and describes the data analysis methods and
results, including pipeline processing of the data, source detection,
astrometry and matching of sources from the different observations,
X-ray photometry and overall population results, variability analysis
and optical counterpart matching (GC and background objects); \S
3. is the source catalog, including the results from the individual
observations and the co-added data; \S 4 presents the discussion
of the properties of the sources catalog; \S 5 summarizes the conclusions of this work.

\section{Observations and Data Analysis}
\label{sec:process}

The six separate \CHANDRA\ observations of NGC 4278 have been carried
out over a two year baseline, with the first of these,
a 37 ks pointing, being performed in February 2005. This observation has been followed by five
deeper pointings, all carried out between March
2006 and February 2007, resulting in a total exposure time of 471-ks.
 
The initial data processing to correct for the motion of the
spacecraft and apply instrument calibration was carried out with the
Standard Data Processing (SDP) at the {\em Chandra} X-ray Center
(CXC). The data products were then analysed using the CXC CIAO
software suite (v3.4)\footnote{http://asc.harvard.edu/ciao} and
HEASOFT (v5.3.1). The data were reprocessed, screened for bad pixels,
and time filtered to remove periods of high background.
Following the methods of Kim \etal\ (2004a), time filtering was done
by making a background light curve and then excluding those time
intervals beyond a 3$\sigma$  fluctuation above the mean background
count rate, where the mean rate was determined iteratively after
excluding the high background intervals. This resulted in a total
corrected exposure time of 458-ks, the log of these exposures is
presented in Table \ref{tab:log}.

From the six individual data sets, a combined observation was produced, using the script {\em merge\_all }\footnote{http://asc.harvard.edu/ciao/ahelp/merge\_all.html}, where the reprocessed level 2 event files from each observation were reprojected to a given RA and Dec, and then combined, and a combined exposure map was also created. Each of the individual observations used in the {\em merge\_all } script were first corrected for relative astrometry before being reprocessed and reprojected. These corrections were done by using the CIAO tool {\em reproject\_aspect}\footnote{http://asc.harvard.edu/ciao/threads/reproject\_aspect/} to update the aspect solution files by comparing the source list from an individual observation to the source list from a single reference observation (in this case the longest cleaned exposure obs-7078). These corrected aspect solution files were then used in all subsequent analysis. 

\begin{figure}
\begin{centering}
  \begin{minipage}{0.72\linewidth}
  \vspace{0.7cm}
\includegraphics[width=\linewidth]{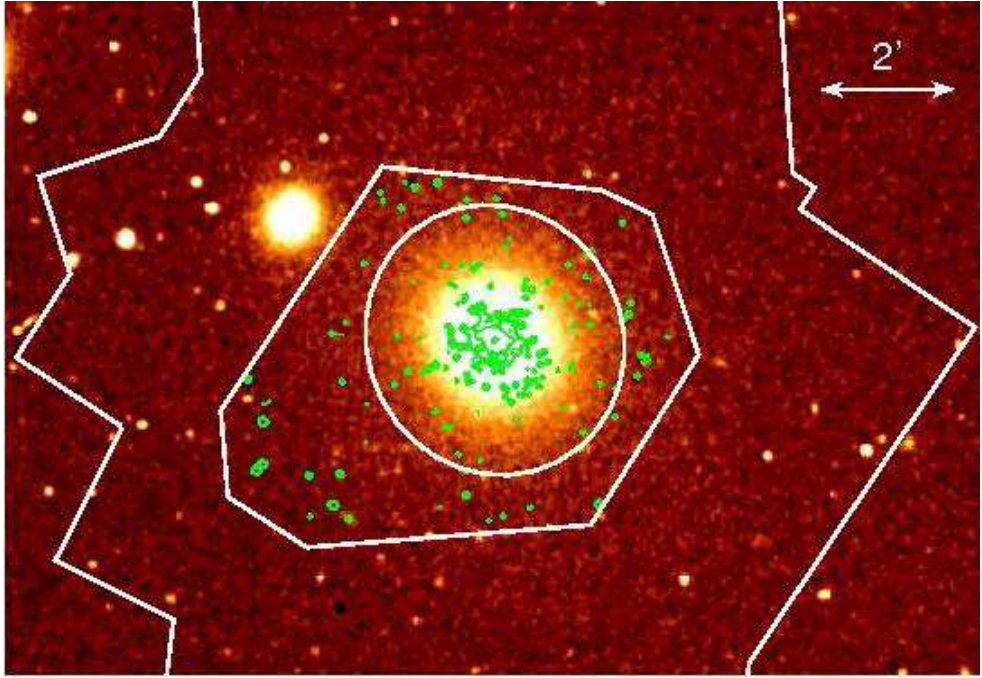}
\end{minipage}
\vspace{1cm}

  \begin{minipage}{0.72\linewidth}

\includegraphics[width=\linewidth]{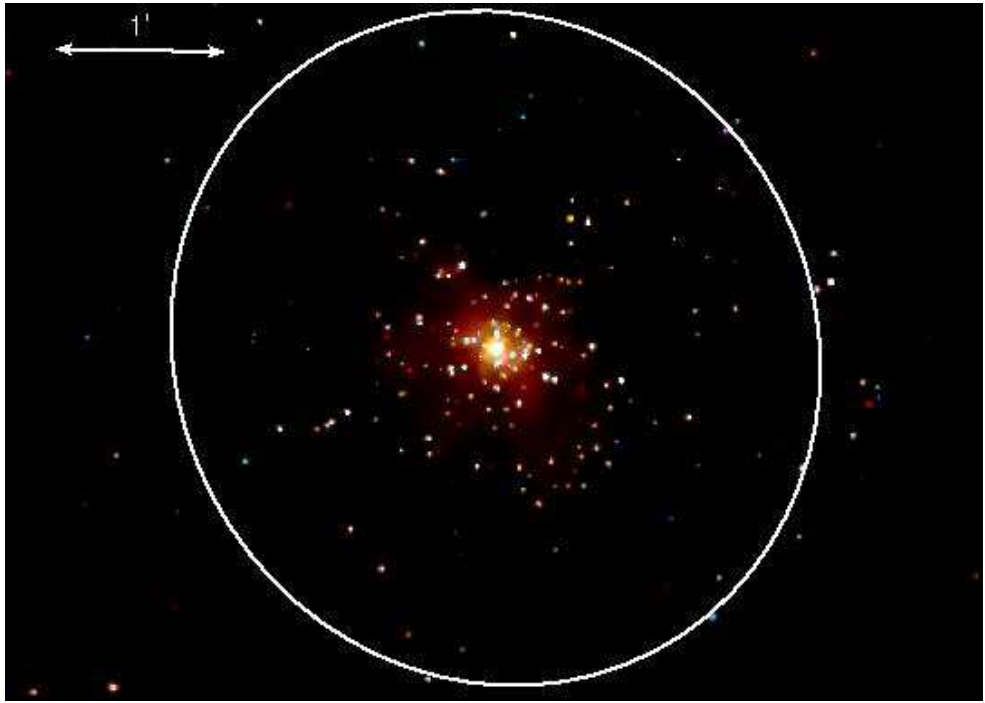}

\end{minipage}
\caption{Top: An optical image of NGC 4278, with the outline of the total area covered by the ACIS-S3 chips, the region overlapped by all six
of the pointings and the $D_{25}$ ellipse of this galaxy shown in white. Also overlaid within the overlap region are the full band adaptively
smoothed, X-ray contours. Bottom: A `true color' image of the galaxy, where red
corresponds to 0.3$-$0.9 keV, green to 0.9$-$2.5 keV and blue to
2.5$-$8.0 keV. The $D_{25}$ ellipse of this galaxy is also shown.}\label{fig:image}
\end{centering}
\end{figure}

From the co-added dataset of these six observations, a
0.3$-$8.0\,keV (from here on referred to as `full band') {\em Chandra}
image was created and adaptively smoothed
using the CIAO task {\em csmooth}. This uses a smoothing kernel to
preserve an approximately constant signal to noise ratio across the
image, which was constrained to be between 2.4$\sigma$ and 4$\sigma$. In Figure
\ref{fig:image}, both the optical image, with the
full band X-ray contours overlaid (top), and the `true color' image
of the galaxy system (bottom) are shown. The `true color' image was
created by combining three separate smoothed, and exposure corrected,
images in three energy bands; 0.3$-$0.9 keV, 0.9$-$2.5 keV and
2.5$-$8.0 keV, using the same smoothing scale for each image. These
energy bands correspond to red, green and blue respectively. 

\subsection{Source Detection and Count Extraction Regions}
\label{sec:src_props}

Discrete X-ray sources were searched for over each observation (the
six single observations and the co-added observation) using the
CIAO tool {\em wavdetect}, where the full
band, with a significance threshold parameter of 1$\times10^{-6}$,
corresponding to roughly one spurious source over one CCD, was searched
over. This CIAO tool searches for localized
enhancements of the X-ray emission, and does not set any apriori
thresholds on the S/N of each source (in contrast to sliding cell
algorithms: Freeman \etal\ 2002). In Kim \etal\ (2004a) simulations were carried out to investigate the number of false detections compared to the expected $\sim$1 false source per image provided by the threshold significance of 1$\times10^{-6}$. These simulations and results are detailed in \S4 of the paper, where they find that the performance of {\em wavdetect} is as expected, resulting in $\sim$1 spurious source per image. These simulations cover the values of background ($\sim$0.28 counts/pixel) of our co-added observation. Further to this, these simulations were compared to \CHANDRA\ observations with relatively long exposures ($\sim$100 ks), where 0.3 spurious sources per exposure were detected, fully consistent with the simulation results. A similar approach was also used by Kenter \etal\ (2005).

Following the results of these simulations it is clear that, when setting a detection threshold of 1$\times10^{-6}$ in {\em wavdetect}, only one of the formally identified sources is expected to be a false detection per image and so this prescription has been followed here. 
We reiterate that using this method does not set any apriori thresholds on the S/N of each source, it is therefore possible to include sources that have a high detection significance but at the same time a low flux significance, or S/N, therefore resulting in sources with poorly constrained flux.

When running {\em wavdetect} a range of 1, 2, 4, 8,
16 and 32 pixel wavelet scales were selected (where pixel width
is 0.49\arcs), with all other parameters set at the default
values. Exposure maps were created for the S3 chip from each
observation, at 1.5 keV. The {\em wavdetect} tool was used in
preference to other source detection software, as this detection
package can be used within the low counts regime, as it does not
require a minimum number of background counts per pixel for the
accurate computation of source detection thresholds. Further to this,
{\em wavdetect} also performs better in confused regions, which is the
case in the nuclear region of elliptical galaxies (Freeman \etal\ 2002).

Once the X-ray sources had been detected, and their position had been
determined by {\em wavdetect}, counts were extracted from a circular
region, centered on the {\em wavdetect} position, with background counts determined
locally, in an annulus surrounding the source, following the
prescription of Kim \etal\ (2004a). The extraction radius 
for each source was chosen to be the 95\% encircled energy radius at
1.5 keV (which varies as a function of the off-axis angle\footnote{see
http://cxc.harvard.edu/cal/Hrma/psf/index.html}), with a
minimum of 3\arcs\ near the aim point. Similarly, background counts for
each source were estimated from a concentric annulus, with inner and
outer radii of two and five times the source radius respectively. 

When nearby sources were found within the background region, they were
excluded before measuring the background counts. Net count rates were
then calculated with the effective exposure (including vignetting) for
both the source and background regions. Errors on counts were derived
following Gehrels (1986). For cases where sources have fewer than 4 counts, the Gehrels
approximation begins to differ to Poissonian errors. However,
these error values are still accurate to 1\%, and, if anything,
provide a more conservative estimate as Gehrels approximation
does not account for the smaller error value at the lower limit.
When the source extraction regions of nearby
regions were found to overlap, to avoid an overestimate of their
source count rates, counts were calculated from a pie-sector, excluding
the nearby source region, and then rescaled, based on the area ratio
of the chosen pie to the full circular region. Once the correction
factor was determined, it was applied to correct the counts in all
energy bands. For a small number of sources that overlapped with
nearby sources in a more complex way (e.g. overlapped with more than 2
sources), instead of correcting the aperture photometry, the source
cell determined by {\em wavdetect} was used to extract the source
counts in each energy band.

From these source counts, fluxes and luminosities were calculated in
the 0.3$-$8.0 keV band, with an energy conversion factor (ECF)
corresponding to an assumed power law spectral shape, with $\Gamma$ =
1.7 and Galactic \NH\footnote{\NH=1.76$\times10^{20}$\pcmsq\ (from
COLDEN: http://cxc.harvard.edu/toolkit/\\colden.jsp).} (see Figure
\ref{fig:cc_pop} for a justification of this assumption). The ECF was
calculated with the {\em arf} (auxiliary 
response file) and the {\em rmf} (redistribution matrix file) generated for
each source in each observation. For each source, the spatial and temporal quantum
efficiency variations\footnote{See http://cxc.harvard.edu/cal/Acis/Cal\_prods/qeDeg/
 for the low energy QE degradation.} were accounted for by calculating
the ECF in each observation and then taking an exposure-weighted mean
ECF. The ECF over the 0.3$-$8.0~keV band varied by only $\sim$2\% between
all six of the observations
taken between 2005 and 2007\footnote{http://asc.harvard.edu/ciao/why/acisqedeg.html}. This procedure was applied to each single observation
and to the total co-added exposure. 

In the instances where {\em wavdetect} did not formally identify a
source in a single observation, source counts have been extracted from a
circle with a 95\% encircled energy radius, centered on the position
from the co-added observation (or in cases where the source was not
formally detected in the co-added observation, the source position from
the single observation was used). The definition of background regions and the treatment
of overlapping sources are outlined above. From these extracted source counts, a Bayesian
approach, developed by Park \etal\ (2006), has been used to provide
68\% source intensity upper confidence bounds on the full band counts. These values have
then been used to calculate upper limits on the flux and luminosity of
these sources.

\subsection{Source Correlation}
\label{sec:astrometry}

\begin{figure}
\begin{centering}

   \begin{minipage}{0.48\linewidth}

  \includegraphics[width=\linewidth]{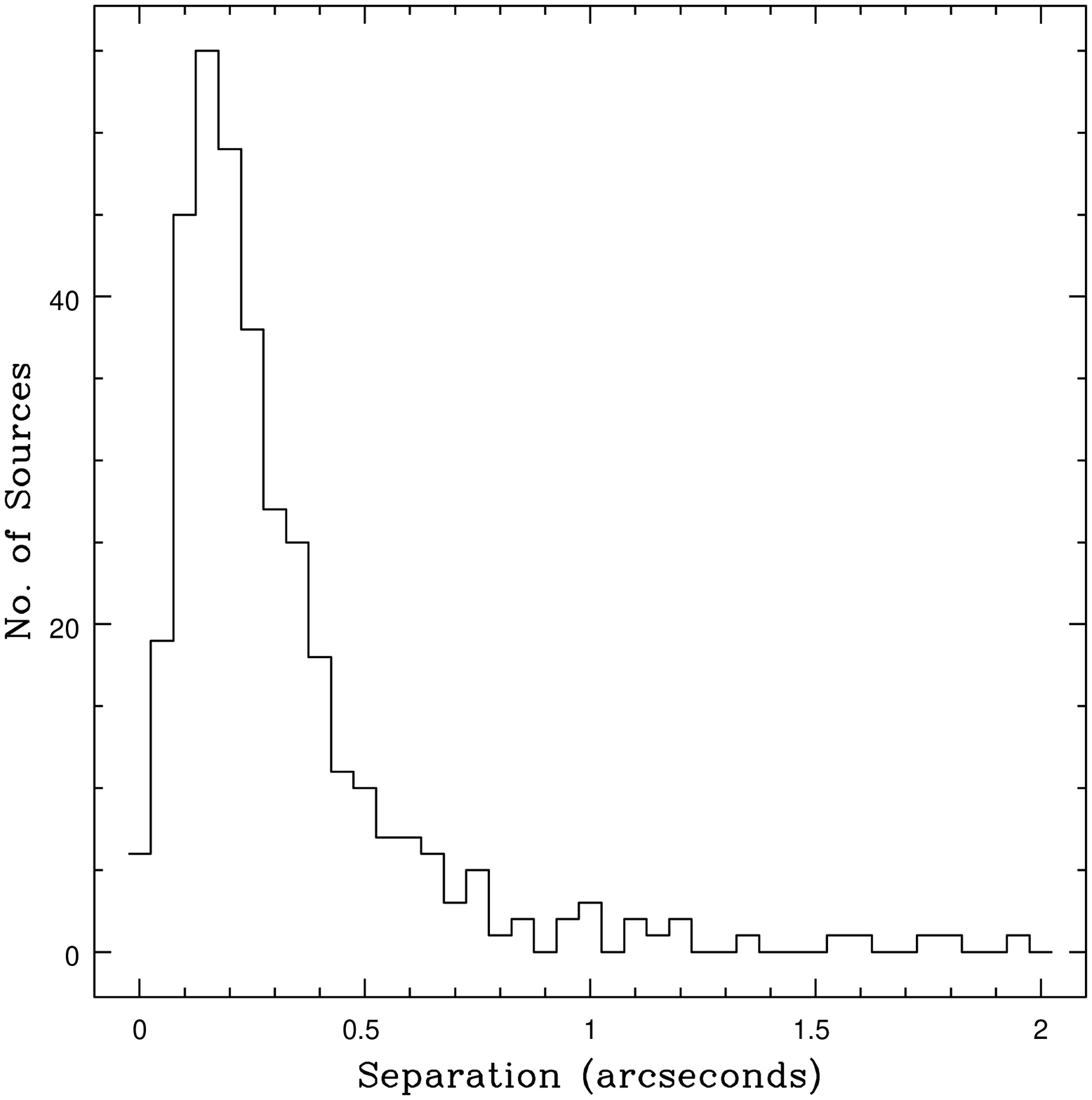}
 
  \end{minipage}
  \begin{minipage}{0.48\linewidth}

  \includegraphics[width=\linewidth]{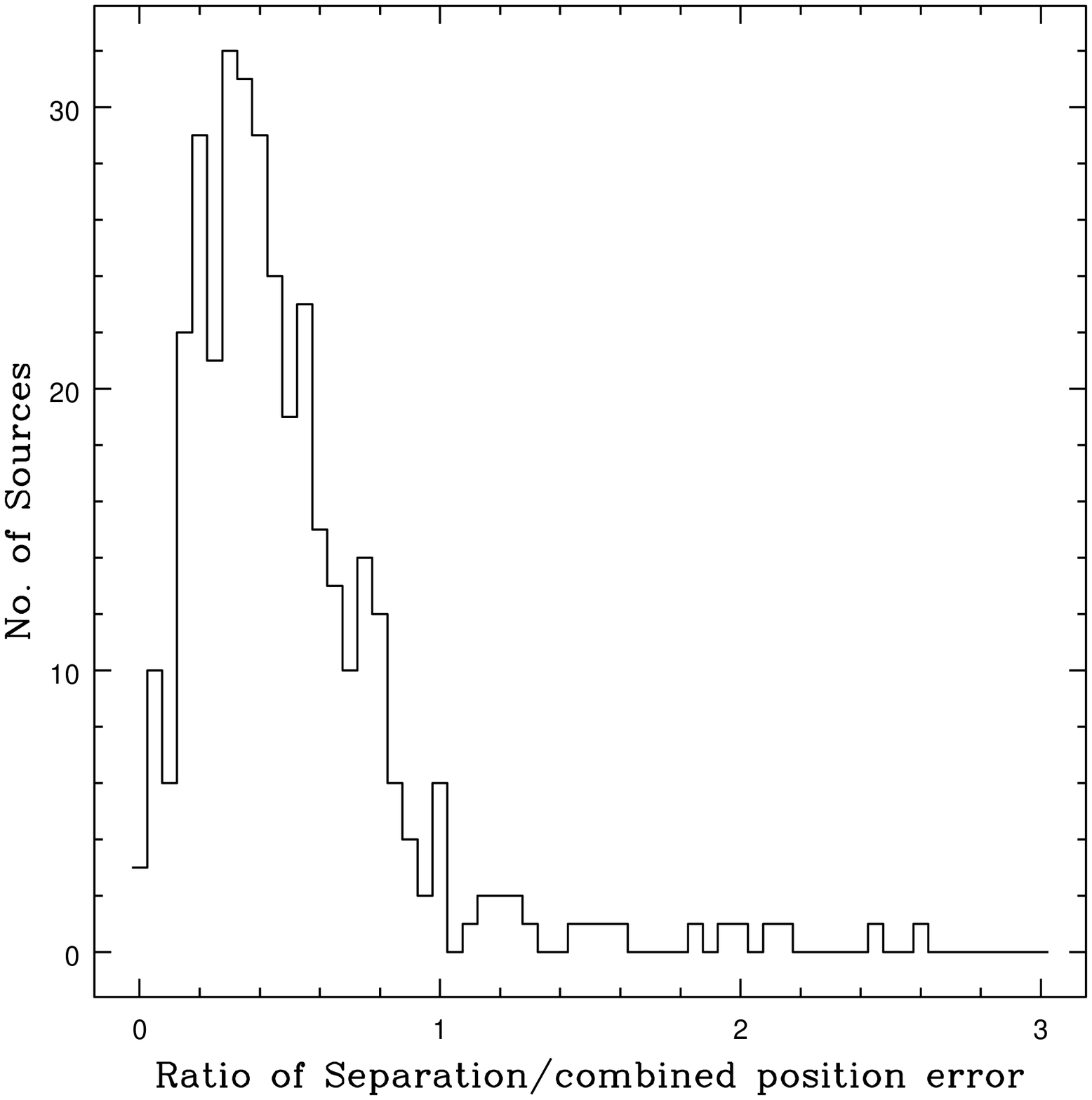}
	\end{minipage}
  \caption{Left: Histogram of the separation between sources detected in the
co-added observation and sources detected in single
observations. Right: Histogram of the ratio of separation between sources
detected in the co-added observation and sources detected in single
observations, divided by the combined position uncertainty of these sources.}
  \label{fig:sep_histo}

\end{centering}
\end{figure}

From the co-added observation only, 271 sources were detected by {\em
wavdetect}. From this list, sources external to the  overlapping area covered by
the S3 chip in
all six individual observations were excluded, reducing this total
number to 220 point sources. Using this source list from the co-added
observation, sources detected
in the individual observations were matched with this list, where source correlations were
searched for up to a separation of 2\arcs. In the cases where multiple
matches were detected for a source, the closer correlation was
selected. From these matches, a histogram of source separations, shown
in the left panel of Figure \ref{fig:sep_histo}, was produced. In this figure it
is clear that the peak separation between sources lies $\sim$0.15\arcs,
with the number of correlated sources dropping at $\sim$1.0\arcs, and this
is therefore the value we set for maximum separation when cross-correlating sources.

Once a cut of 1.0\arcs\ had been applied to the cross-matched source
list, the remaining unmatched sources, detected in the separate
pointings only, were investigated individually, resulting in further
potential matches being established. These potential matches correspond to sources with
fewer counts, and hence a greater positional uncertainty, leading to
larger values of separation. 
These potential source correlations were then further investigated by
calculating the ratio of
the source separation and the combined position uncertainty. Where the
position uncertainty at the 95\% confidence level has been defined by
Kim \etal\ (2007a), as:
\begin{equation}
\label{equ:pu}
\mathrm{log PU}= 
\begin{cases} 0.1145\times \mathrm{OAA}- 0.4958 \times \mathrm{log C} +0.1932, 
\\
\qquad 0.0000 < \mathrm{log C} \le 2.1393,
\\
0.0968\times \mathrm{OAA} -0.2064 \times \mathrm{log C} -0.4260,
\\ 
\qquad 2.1393 < \mathrm{log C} \le 3.3000
\end{cases}
\end{equation}
where the position uncertainty, $PU$, is in arcseconds, and the off
axis angle, $OAA$, is in arcminutes. Source counts, $C$, are as
extracted by {\em wavdetect}. Using this ratio of source
separation and position uncertainty allows low \LX\ source
correlations to be identified. Often these sources, particularly at
greater off axis angles, cannot be matched by source separation cuts
alone, due to the increasing PSF spread out and asymmetry at larger
$OAA$\footnote{See \S 5 in Kim \etal\ (2004a) for a full
discussion}. Therefore, by using this source separation - $PU$ ratio, the greater
position uncertainties in these weak sources can be accounted for, resulting in
smaller ratios, and thereby identifying correlations that would
otherwise be missed with source separation cuts alone.

In the right panel of Figure \ref{fig:sep_histo}, a histogram of the ratio of separation and the
combined position uncertainty is shown, where sources with a ratio of
greater than 1 were investigated individually. In all but one
instance it was found that these higher ratio sources lie in the
central region of the galaxy, where both source confusion is likely
and diffuse gas may be present. This
emission results in higher background fluctuations, which can lead to
the $PU$ of these sources to be underestimated, therefore resulting in
a falsely high ratio value. The source
that was detected outside the central region is too faint ({\em net
counts} $< \sim$ 100 counts) to allow its radial profiles to be
compared with a corresponding model PSF profile, generated  for
the source position using the CIAO tool {\em mkpsf}, and has
been flagged as a possible double sources.

\begin{figure}
\begin{centering}
  
  \includegraphics[angle=-90,width=\linewidth]{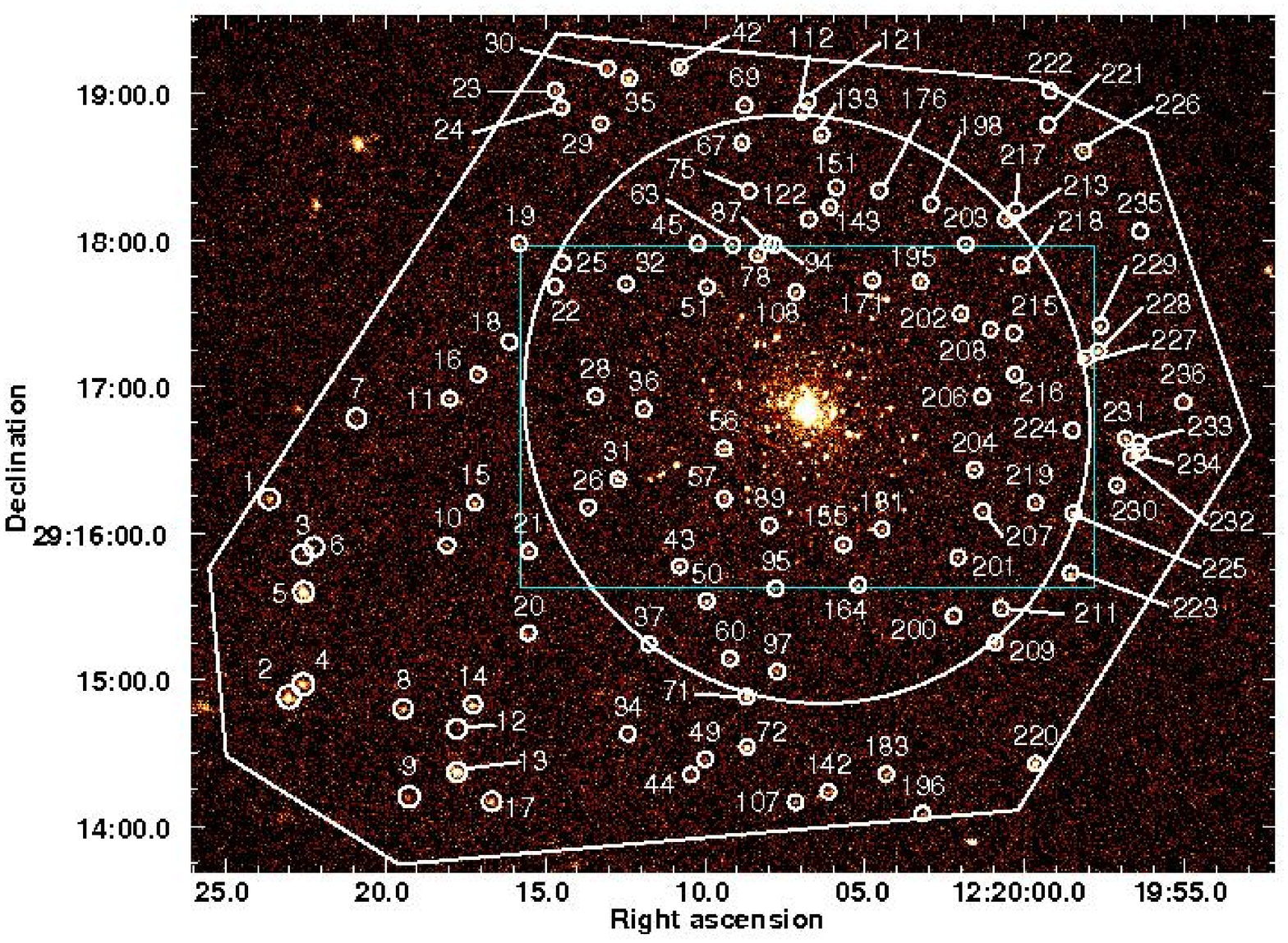}
  \caption{The main image presents a full band, raw (unsmoothed, with no exposure
correction) image from the co-added
observation of NGC 4278, with the $D_{25}$ ellipse and the region
overlapped by all six observations overlaid. Source region
numbering corresponds to the 
naming convention in Table \ref{tab:Mainprops} and regions represent
the 95\% encircled energy radius at 1.5 keV. The box in the central region
indicates the area shown in the next image, 
the central region of the galaxy, with sources labeled with the same convention
as in the main image. The box shown here encloses the
nuclear region of the galaxy, where there is a dense population of
sources. This is presented in the final image, where these
individual sources can be more clearly seen.}
  \label{fig:rgb}
 \end{centering}
\end{figure}

After confirming that these additional sources with separation $\ge$1\arcsec, were correlated with sources detected in the co-added observation, the remaining list of sources detected in individual observation was reduced to sixteen. All of these sixteen sources were determined to be well separated from the sources detected in the co-added observation, and were therefore included in the final list of detected sources, increasing the total number to 236. These 236 sources are presented in Figure \ref{fig:rgb}, where the
unsmoothed full band image from the co-added dataset, with regions
overlaid in white, is shown.

\subsection{Hardness Ratios and X-ray colors}
\label{sec:HR}

Within NGC 4278, the range of net counts for the pointlike sources in
the co-added observation is
$\sim~7-48772$ (with signal-to-noise ratio (S/N) values ranging from
1.1 to 218.8), corresponding to 0.3$-$8.0 keV luminosities of 3.5
$\times 10^{36}$ \ergps (3$\sigma$ upper limit $\le$1$\times 10^{37}$ \ergps) $-~2~\times 10^{40}$\ergps, when using the
energy conversion factor described in \S
\ref{sec:src_props}. Most of these sources are too faint for detailed
spectral analysis, therefore their hardness ratio and X-ray colors
were calculated in order to characterize their spectral
properties. The X-ray hardness ratio is defined as
HR~$=$~(Hc$-$Sc)$/$(Hc+Sc), where Sc and Hc are the net counts in the
0.5$-$2.0 keV and 2.0$-$8.0 keV band respectively. Following the
prescription of Kim \etal\ (2004b), the X-ray colors are defined as
C21~=~log(S$_1$/S$_2$) and C32~=~log(S$_2$/H), where S$_1$, S$_2$ and
H are the net counts respectively in the energy bands of 0.3$-$0.9
keV, 0.9$-$2.5 keV and 2.5$-$8.0 keV (energy bands and definitions are
summarized in Table \ref{tab:bands}). These counts were corrected for
the spatial and temporal QE variation, referring them all to the aim point of the first, recent
observing epoch (Feb. 2006, Table \ref{tab:log}), and for the effect
of the Galactic absorption, using \NH=1.76$\times10^{20}$\pcmsq\ (from
COLDEN: http://cxc.harvard.edu/toolkit/colden.jsp). 

By definition, as the X-ray spectra become
harder, the HR increases and the X-ray colors decrease. For faint
sources with a small number of counts, the formal calculation of the
HR and colors often results in unreliable errors, because of negative
net counts in one band and an asymmetric Poisson
distribution. Therefore a Bayesian approach has been applied to derive the
uncertainties associated with the HR and colors. This model was developed by
Park \etal\ (2006) and calculates values using a method based on the
Bayesian estimation of the `real' source intensity, which takes into
account the Poisson nature of the probability distribution of the
source and background counts, as well as the effective area at the
position of the source (van Dyk \etal\ 2001), resulting in HR and color values that are
more accurate than the classical method, especially in the
small-number-of-counts regime (less than 10 counts), where the Poisson
distributions become distinctly asymmetric.

\subsection{Source Variability}
\label{sec:var}

Due to the monitoring approach that has been used when observing NGC
4278, both long-term and short-term variations have been
able to be searched for in the galaxy's LMXB population.
Long-term variability was defined by the chi-squared test, where 
a straight line model was fitted to the luminosities derived for each
individual observation, with errors based on the Gehrels approximation
(Gehrels 1986). For
the cases where sources only had upper limit values of \LX, the
associated error was defined to be the standard deviation of 
the upper limit from the mode value attained from the Bayesian
estimates method, resulting in a conservatively large error, due to
the nature of the Poissonian statistics. From these best fit models,
sources were determined to be variable if \rchisq$>$1.2, and those with
fits with \rchisq$<$1.2 were defined as non-variable sources. For
sources that were only detected in the co-added observation,
long-term variability was not searched for. This long-term behaviour
will be further investigated in a forthcoming paper, where full Poissonian
error treatment will be applied to sources with very low observed counts.

In addition to the chi-squared test variability criterion, transient candidates (TC), sources that either appear or disappear, or are only detected for a limited amount
of `contiguous' time during the observations, were searched
for. Typically, sources are defined to be TCs if the ratio between the
`on-state', the peak \LX\ luminosity, and the `off-state', the lower
\LX\ luminosity or non-detection upper limit, is greater than a certain value (usually between 5$-$10; e.g. Williams \etal\ 2008). However, such a criterion can overestimate the number of transient candidates, when the `on-state' X-ray luminosity is poorly constrained. To address this, the Bayesian model developed by Park \etal\ (2006) was used to derive the uncertainties associated with the ratio between `on-state' and `off-state'. In this model, source and background counts from both the peak \LX\ luminosity and the non-detection observations were used to estimate the ratio, where the differences in both the exposure and ECF values were also accounted for. From this Bayesian approach a value of peak \LX/non-detection upper limit was calculated, along with a lower bound value of this ratio. This lower bound value was then used to determine the transient nature of the source, where a ratio of greater than 10 indicated a TC and sources with a ratio between 5 and 10 were labeled as possible transient candidates (PTC). This transient behavior was only searched for in sources that were only detected for a limited amount of `contiguous' time during the observations and were determined to be variable using the chi-squared test.

Further to these four long-term variability classifications, the
variation of the source luminosity between each observation was also
investigated, by comparing the significance (in $\sigma$) of the change in
luminosity between exposures, where the significance has been estimated by:
\begin{equation}
\label{equ:sig}
signif = \frac{|\ensuremath{L_{\mathrm{X1}}}-\ensuremath{L_{\mathrm{X2}}}|}{\sqrt{(\sigma_1^2 + \sigma_2^2)}},
\end{equation}
where $\sigma_n$ is the error value of the luminosity from that
individual observation, based on the Gehrels approximation, or, where upper
limits have been used, the standard deviation of the estimated luminosity. 

Short-term variations in each source were investigated when $net$ 
$counts > 20$ in a single observation. In these instances, the
variability was identified by using the Kolmogorov-Smirnov test
(K-S test), where sources with variability values $>$90\% confidence were
labeled as possible variable sources and sources with values $>$99\% confidence were
defined as variable sources. This short-term variability was also
quantified by using the Bayesian blocks method (BB) (Scargle 1998;
Scargle \etal\ 2009, in prep). This method searches for abrupt changes
in the source intensity during an observation, and therefore is very
efficient for detecting bursts or state changes. Because it is based
on the Poisson likelihood it can be used on the unbinned lightcurves
of sources with very few counts. The implementation of the method used
in this analysis is the same as in the ChaMP pipeline (see
\S3.3.2 in Kim \etal\ 2004a). This assumes a
prior of $\gamma=4.0$ which {\em roughly} translates to a significance
level of $\sim99$\% for each detected block (however see Scargle \etal\ (2009, in prep)\footnote{see also
http://space.mit.edu/CXC/analysis/SITAR/functions.html}, for a caveat
on this interpretation of the  value of the prior). 

\subsection{Radial Profile}
\label{sec:prof}

\begin{figure}

\begin{centering}
\includegraphics[width=\linewidth]{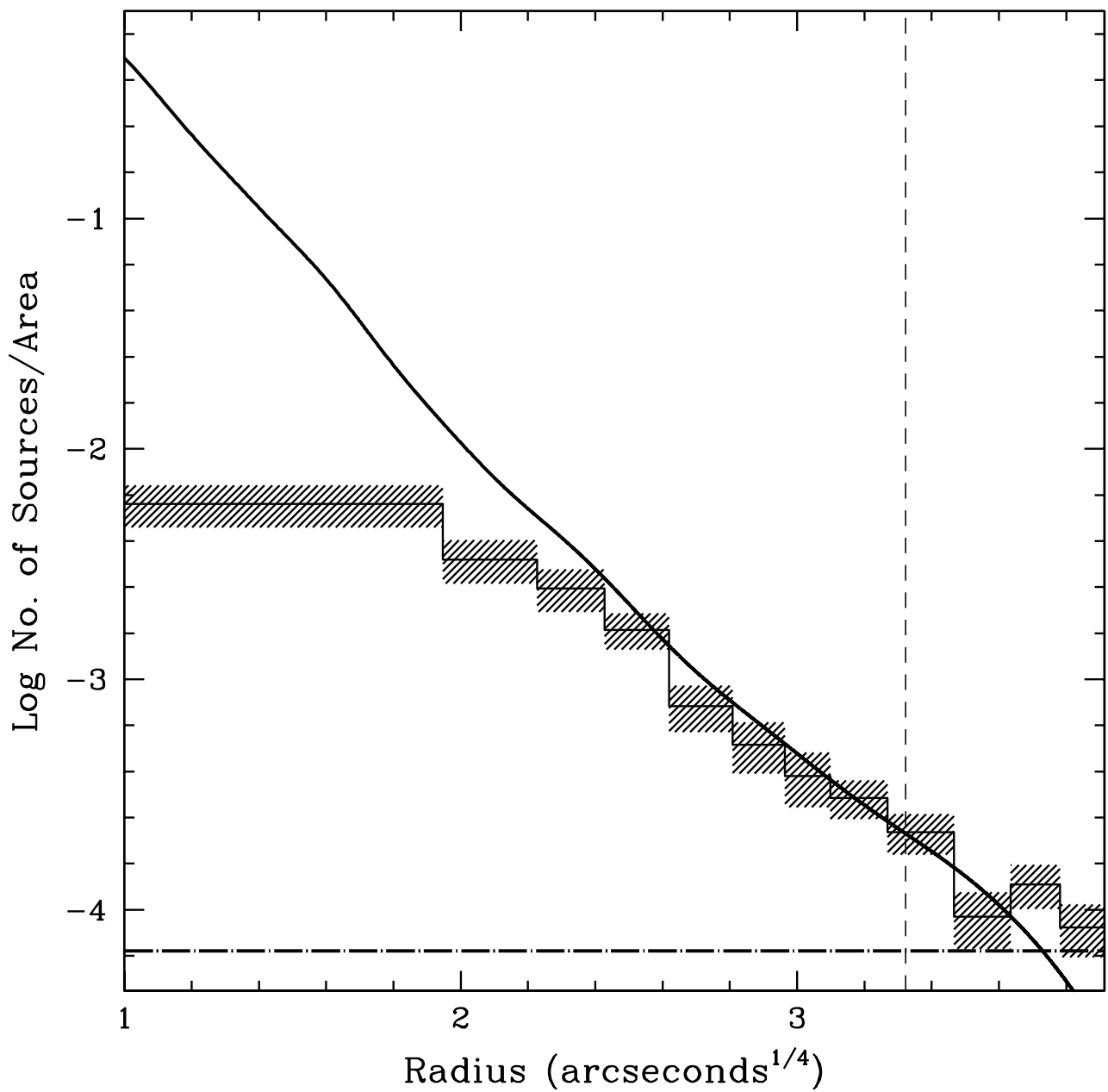}

\caption{X-ray source density profile compared to the optical
profile. The histogram indicates the X-ray data and the thick black
line is the I-band surface brightness best fit of Cappellari \etal\
(2006). The vertical dashed line is the $D_{25}$ ellipse and the
horizontal dot-dashed line indicates the expected number of background
sources. The central 1\arcsec\ has been excluded in this plot due to the excess optical light contribution from the LINER at the center of this galaxy.}\label{fig:profile}
\end{centering}
\end{figure}

From the complete source list from the co-added observation a radial
distribution of LMXBs has been created, using annuli centered on the
nucleus of the galaxy (source 117). This profile has been compared to a
multi-Gaussian expansion model of the I-band optical data (Cappellari
\etal\ 2006), which is assumed to follow the stellar mass of the
galaxy (Gilfanov 2004). This X-ray source density profile is presented
in Figure \ref{fig:profile}, where the optical profile has been normalized
to the X-ray data by way of a \chisq\ fit. The central 1\arcsec\ profile of the galaxy has not been plotted due to the excess of optical light, arising from the LINER that lies at the center of NGC 4278. Also indicated in this figure is the $D_{25}$ ellipse and the number of background sources,
which has been estimated from the hard-band $ChaMP+CDF$ $logN - logS$
relation (Kim \etal\ 2007b), where $\sim$29 sources are expected to be
objects not associated with  NGC 4278. 
From this figure it can be seen that the X-ray profile follows the
optical surface density profile at larger radii, with the flattening
in the central region (r$\le$15\arcs) a consequence of source confusion. This
indicates that the number and spatial distribution of LMXBs follows
that of their parent population, the old stellar population.

\subsection{Optical Counterparts}
\label{sec:opt}

The globular cluster system of NGC 4278, observed with WFPC2, on-board
{\em HST}, is reported in Kundu \& Whitmore (2001), where images in
both the {\em V} and {\em I} bands have been analyzed. 
In addition to this GC system identified in the {\em HST} data,
background objects have also been classified (Kundu, A. 2007, private
communication). These have been identified as objects that were well
resolved in the {\em HST} images and were clearly more extended than any
known globular cluster. Further to this, these background objects often had other features,
such as visible disks, indicative of a galaxy rather than a globular
cluster. 

Right ascension and declination corrections have been applied to the
astrometry of the {\em HST} data, relative to the co-added {\em Chandra} observation. This was done by comparing the positions of all GC-LMXB correlations with separations $\le$1\arcs\ and with net counts$\ge$50, resulting in 37 matches. From these matches relative offsets of $\Delta$RA=0.48\arcs\ and $\Delta$Dec=0.22\arcs\ were identified and removed from the {\em HST} data. 

\begin{figure}
\begin{centering}
   \begin{minipage}{0.48\linewidth}
  \includegraphics[width=\linewidth]{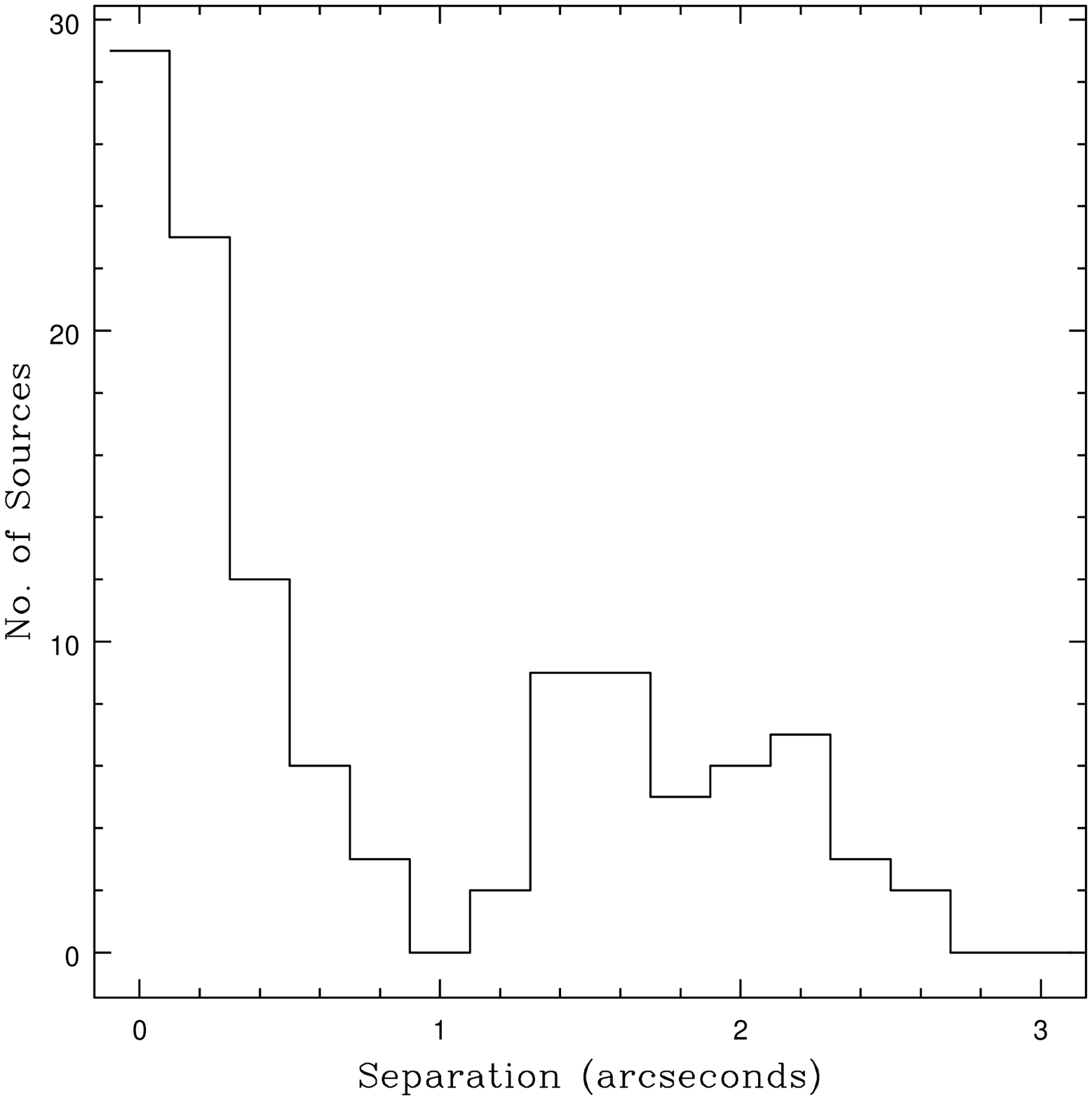}

  \end{minipage}
  \begin{minipage}{0.48\linewidth}

  \includegraphics[width=\linewidth]{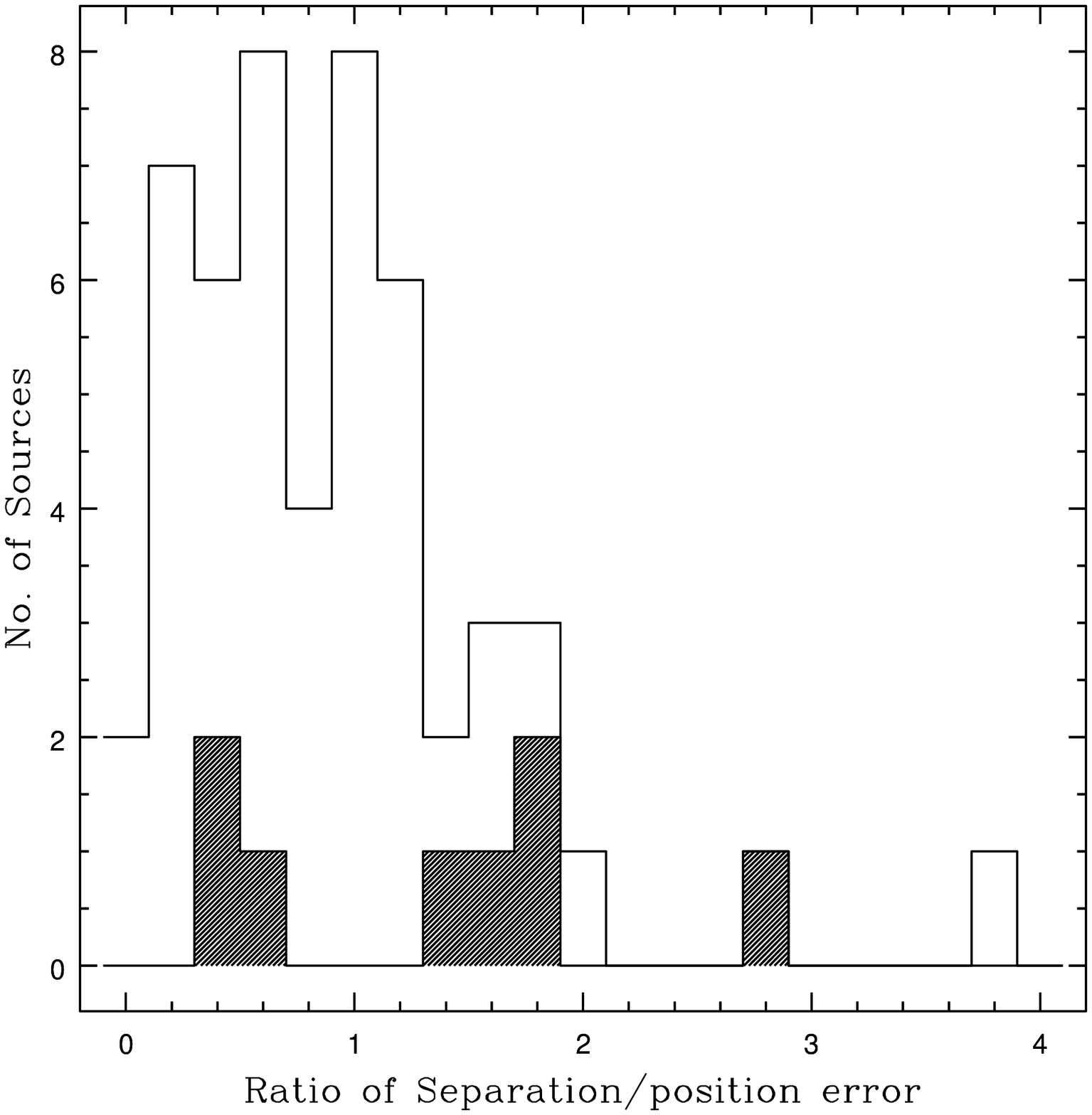}
	\end{minipage}

  \caption{Left:A histogram of the separation between the co-added X-ray source
position and the optical counterpart. Right: Histogram of the ratio of separation divided by the
position uncertainty from the X-ray point source for all optical-X-ray
correlations with separations smaller than 1\arcs. Shaded regions
indicate correlations with optical objects that have been classified
as background sources (details of this classification are given in the
text).}
  \label{fig:pser_GChisto}

 \end{centering}
\end{figure}

After correcting the astrometry of these optical data, correlations up
to an offset of 3\arcs\ with the X-ray sources, were searched
for. When multiple matches were found, the closer matching object was
selected. In the left panel of Figure \ref{fig:pser_GChisto}, a histogram of these matches
is shown, where it can be seen that the number of source correlations decreases at $\sim$1\arcs\ before increasing at greater separations. Because of this clear distinction, the source radius was set to 1\arcs\ and sources between 1\arcs\ and 3\arcs\ were
defined as `excluded matches'. This cut off radius value was then tested by
comparing these correlations with the ratio of the separation divided
by the combined position
uncertainty from the co-added X-ray point sources (the definition of
this is given in equation \ref{equ:pu}) and the
uncertainty in the astrometry in the optical data, which has been
conservatively set at 0.2\arcs.

These ratios are shown in the right panel in Figure \ref{fig:pser_GChisto}, where a
histogram of all optical-X-ray correlations is presented, with a
shaded histogram of the background correlations only, overlaid. From
this figure, it is shown that the majority of the confirmed GC correlated sources have a separation-position uncertainty ratio of less than 1.4. All of the sources with a ratio of greater than 1.4 were visually inspected and only two sources are found to have small separations $<$0.5\arcs. The remaining sources all have separations $>0.6$\arcs\ and have been identified as excluded matches. From this further analysis the separation value cut off was redefined to
be 0.6\arcs. This results in 45 X-ray-optical correlations, 6 of which
have been classified as background objects, which is consistent with the number of sources that are expected to be background objects from the $logN - logS$ relation (5). This leaves 39 GC-X-ray source that are classified as correlations. The optical properties of these GC-LMXB sources, and the `excluded matches' are shown in Tables \ref{tab:GCprops_corr} and
\ref{tab:GCprops_near} respectively (Full descriptions of these
tables are given in \S \ref{sec:atlas}). 

In order to estimate the chance coincidence probability of the sources
within the {\em HST} FOV, the same method as in Zezas \etal\ (2002) was
followed, where the positions of the globular clusters were randomized
by adding a random shift between 0.6\arcs and 30\arcs, and for
each new fake dataset the cross-correlation was performed using the
same search radius as for the observed list of globular clusters. The
limits of the shifts were chosen so that the new positions did not fall
within the search radius and that they follow the general spatial
distribution of the globular clusters. 500 such simulations were
performed, resulting in 2.05$\pm$1.43 associations expected by
chance. If the cross-correlation radius is increased to 1\arcs, the
chance associations rises to 4.96$\pm$2.10. Increasing this radius to
3\arcs\ results in 35.85$\pm$6.00 associations expected by chance, which is slightly higher than
the 31 `excluded matches' that have been found within this radius.

\begin{figure}
\begin{centering}
   \begin{minipage}{0.65\linewidth}
\includegraphics[width=\linewidth]{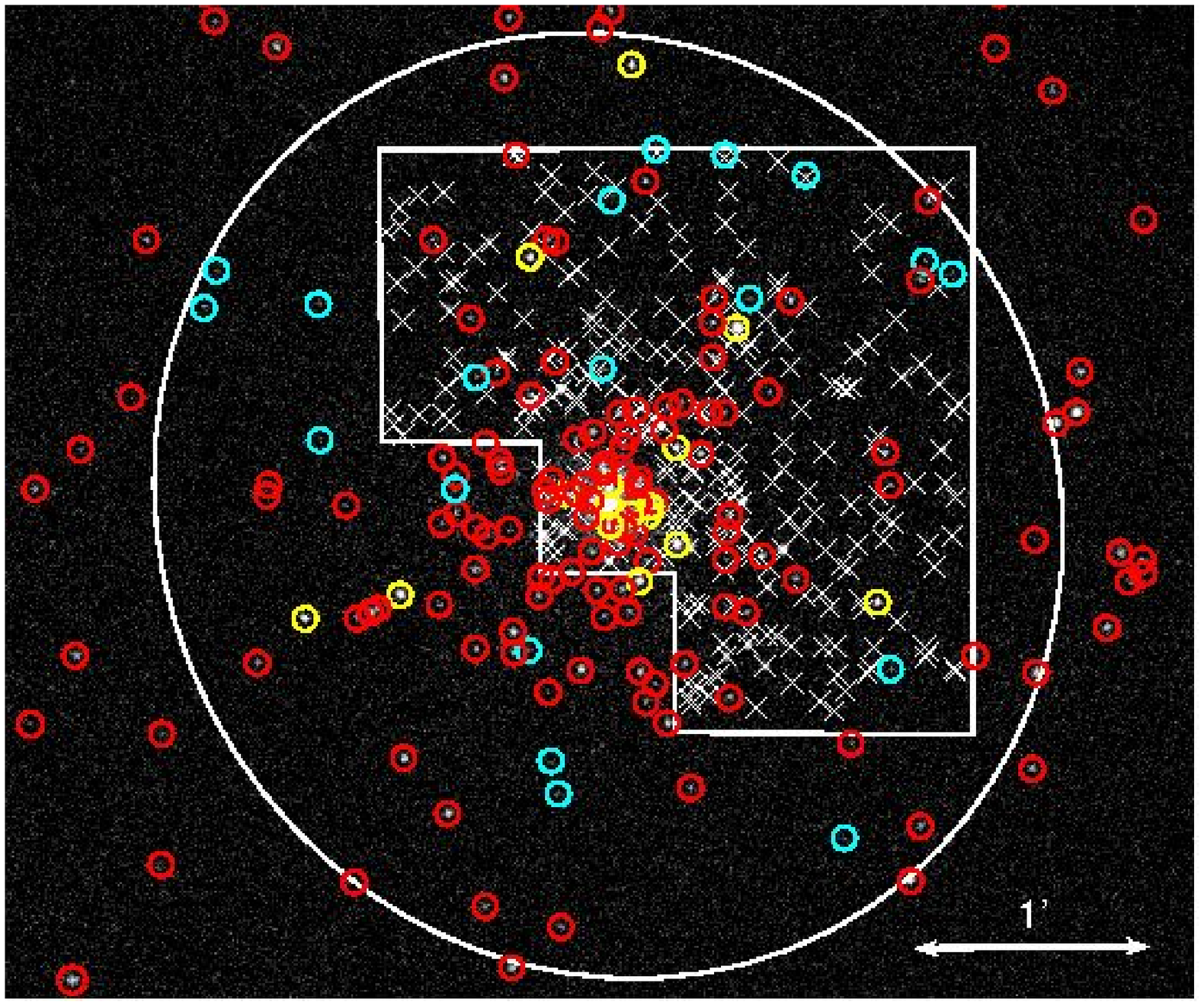}

\end{minipage}

  \begin{minipage}{0.65\linewidth}

  \includegraphics[width=\linewidth]{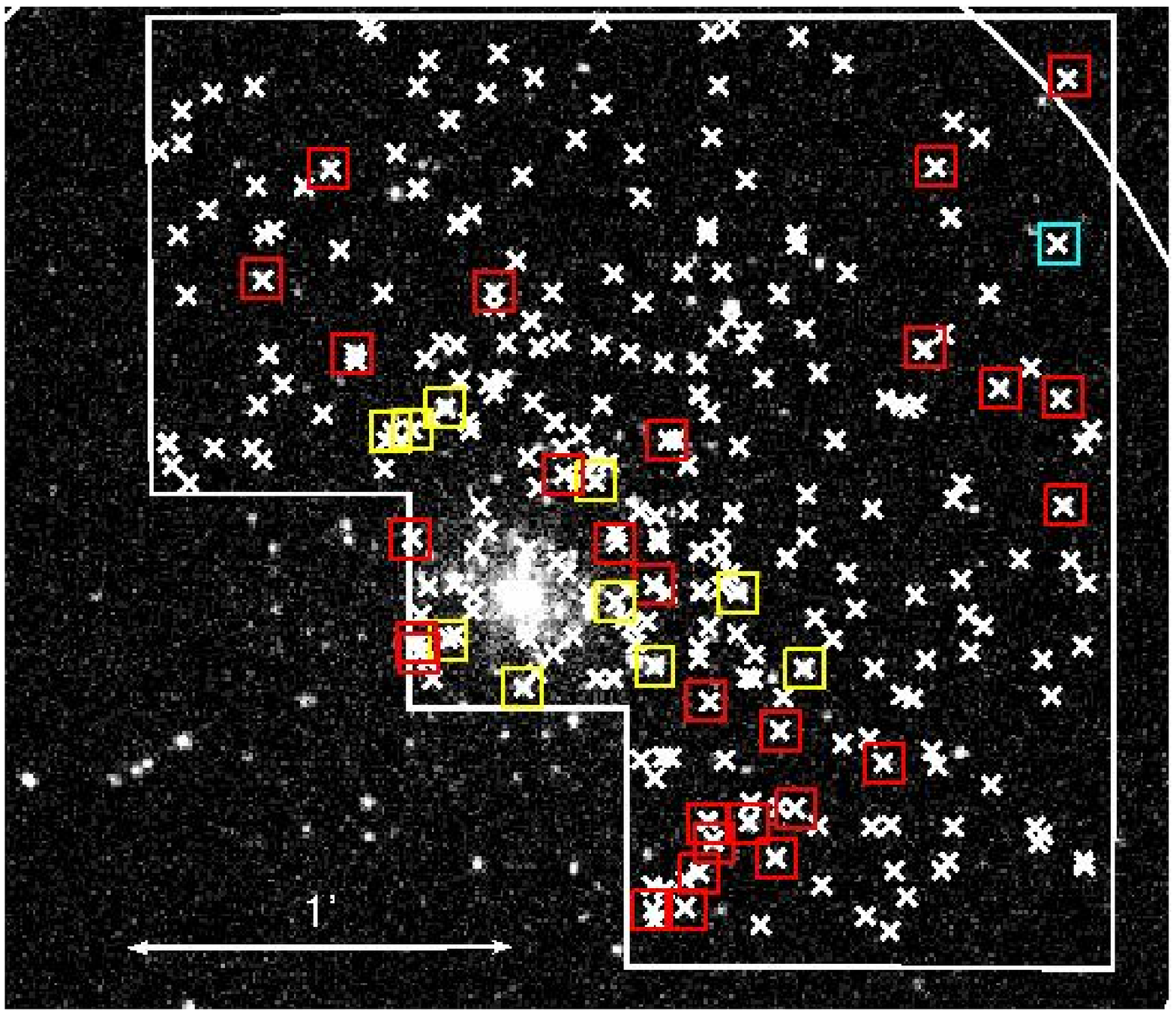}
	\end{minipage}
\caption{Full band X-ray images from the co-added observation of NGC 4278. Both images show
confirmed GCs from the {\em HST} observation indicated by white
`X' marks. The top image shows all X-ray sources {\em not} correlated with a GC and the bottom image indicates all correlated X-ray sources. Region colors indicate the 0.3$-$8.0 keV luminosity of the source from the co-added observation;
yellow regions indicate \LX$\ge$1$\times
10^{38}$\ergps, red regions have
1$\times 10^{38}\ge$\LX$\ge$1$\times 10^{37}$\ergps, and cyan regions
show sources with \LX$\le$1$\times 10^{37}$\ergps. Also shown in white are the
$D_{25}$ ellipse and the {\em HST} FOV. }\label{fig:gccorr}
\end{centering}
\end{figure}

In Figure \ref{fig:gccorr} the confirmed GC and X-ray sources, as well associated correlations, are indicated. In the top image circular regions indicating the X-ray sources with no GC counterpart are overlaid on a full band X-ray image, where the confirmed GCs are indicated by white 'X' marks. Also in this image the $D_{25}$ ellipse and {\em HST} field of view are also shown. In the bottom image a full band X-ray image covering the {\em HST} FOV is presented, where the correlated X-ray sources are indicated by box regions and the white 'X' marks indicate the confirmed GCs. In both images X-ray luminosities are indicated by color, where sources with \LX$\ge$1$\times
10^{38}$\ergps\ are shown in yellow, sources with 1$\times
10^{38}\ge$\LX$\ge$1$\times 10^{37}$\ergps\ are shown in red and
sources with \LX$\le$1$\times 10^{37}$\ergps\ are indicated in cyan. 

\section{Source Catalog and Variability Atlas}
\label{sec:atlas}

Table \ref{tab:Mainprops} presents the properties of the master list
of the 236 X-ray sources detected within NGC 4278, from the co-added
observation of 458-ks. This table has been divided into two sections, where the first part presents all sources with S/N$>$3 in at least one observation, and the second part lists all sources with S./N$<$3. In this table column (1) gives the source number used
through out this series of papers, column (2) gives the IAU name
(following the convention ``CXOU
Jhhmmss.s$+/-$ddmmss''), columns (3) and (4) give the R.A. and
Dec. of the source aperture, columns (5) and (6) give the
radius and the position uncertainty ({\em PU}) of the source
(both in arcseconds), column (7) gives the S/N,
column (8) gives the log value of the co-added luminosity in the 0.3$-$8.0 keV
energy band (for sources with S/N$<$3, 3$\sigma$ upper limit values are also presented in brackets). For sources detected in a single observation only, 1$\sigma$ upper limit from the co-added observation are shown, with 3$\sigma$ upper limit values from the detected observation presented in brackets.
Column (9) provides information about the long-term variability of the
source,
indicating if the source is non-variable (N), variable (V), a
transient candidate (TC) or a possible transient (PTC). In all other
cases the source was only detected in the combined observation,
providing insufficient information to investigate long-term
variability. In columns (10)
and (11) the short-term variability of the source is indicated from both
Bayesian block analysis (BB) and the Kolmogorov-Smirnov test (K-S),
where `V' indicates that the source is variable in at least one
observation and `N' indicates that is has been found to be non-variable
in all six observations. In the K-S column, sources have  also been
labeled as possible variable sources (P) (see \S \ref{sec:var}
for further information). In all other cases there were insufficient counts
to investigate the short-term variability. In column
(12) the optical associations with the X-ray source are indicated, where `GC'
indicates that the associated optical sources has been confirmed as a
globular cluster, and `BG' indicates that the sources has been
classified as a background object. `corr' denotes matches that have
been defined as correlations, and `exmt' denotes the `excluded matches', between
0.8\arcs\ and 3\arcs\ in separation. Sources with a `none' label were
inside the field of view of the {\em HST} observation, but have no
optical counterpart. All other sources were external to the {\em HST}
FOV. Column (13) gives the distance
from the galactic center (in arcseconds), where values in bold type face
indicate sources that lie within the $D_{25}$ ellipse. Column (14) provides
source flag information, indicating sources that have been detected in
a single observation only (X), overlapping sources (O1 for single
overlaps and O2 for more complicated cases) and possible double
sources (double?).

In this table, the 236 sources presented are the complete list
detected by {\em wavdetect}, for which we estimate that $\sim$1 source
is a spurious detection (see \S \ref{sec:src_props}). Since this
catalog of X-ray sources is intended to 
be as complete a study as possible, all detected sources are
included in the complete list, although for sources with S/N$<$3 source parameters such as flux, hardness ratio and color values are not as well constrained as sources with higher flux significance. We have therefore separated the table into two sections, where the first part presents sources with S/N $>$3 in at least one observation and well constrained properties, and the second part lists the sources with low S/N values.

Table \ref{tab:counts_main} presents the detailed source parameters
from the co-added observation; column (1) gives
the source number, columns (2)$-$(8) give the net counts, in each of
the 7 energy bands (see Table \ref{tab:bands} for definitions of these
bands), column (9) indicates the hardness ratio, columns (10) and (11)
show the color-color values and column (12) gives the log value of the
luminosity in the 0.3$-$8.0 keV energy band. Where sources were not
detected in the co-added observation, upper limits for net broad band
counts and \LX\ are given. 

Tables \ref{tab:counts_ob1}$-$\ref{tab:counts_ob6} present the source parameters,
measured for each observation, where columns (1)$-$(11)
provide the same information presented in Table \ref{tab:counts_main},
but further provided in this table is source variability information,
where columns (12)$-$(14) present results of Bayesian block analysis (BB), the
Kolmogorov-Smirnov test (K-S) and the significance of the change in
\LX\ between the previous observation and the current observation
respectively. Column (15) indicates the log value of the
luminosity in the 0.3$-$8.0 keV energy band.

Table \ref{tab:GCprops_corr} presents the optical properties of the
counterparts found from the
optical data of NGC 4278, where 39 GCs and 6 background objects have
been found to be coincident with X-ray sources. Table
\ref{tab:GCprops_near} summarizes the results for the `excluded matches'
sources. In both tables
column (1) gives the X-ray source number, column (2) the {\em V} band
magnitude, column (3) the {\em I} band magnitude, column (4) {\em
V$-$I} colors, column (5) the separation between the X-ray source and the
GC and column (6) the ratio between separation and the combined position
error. The horizontal line in both tables separates the confirmed GCs (top
section of table) from the background objects (bottom section of table).

\begin{figure}
  \begin{minipage}{0.485\linewidth}
  \centering
  
    \includegraphics[width=\linewidth]{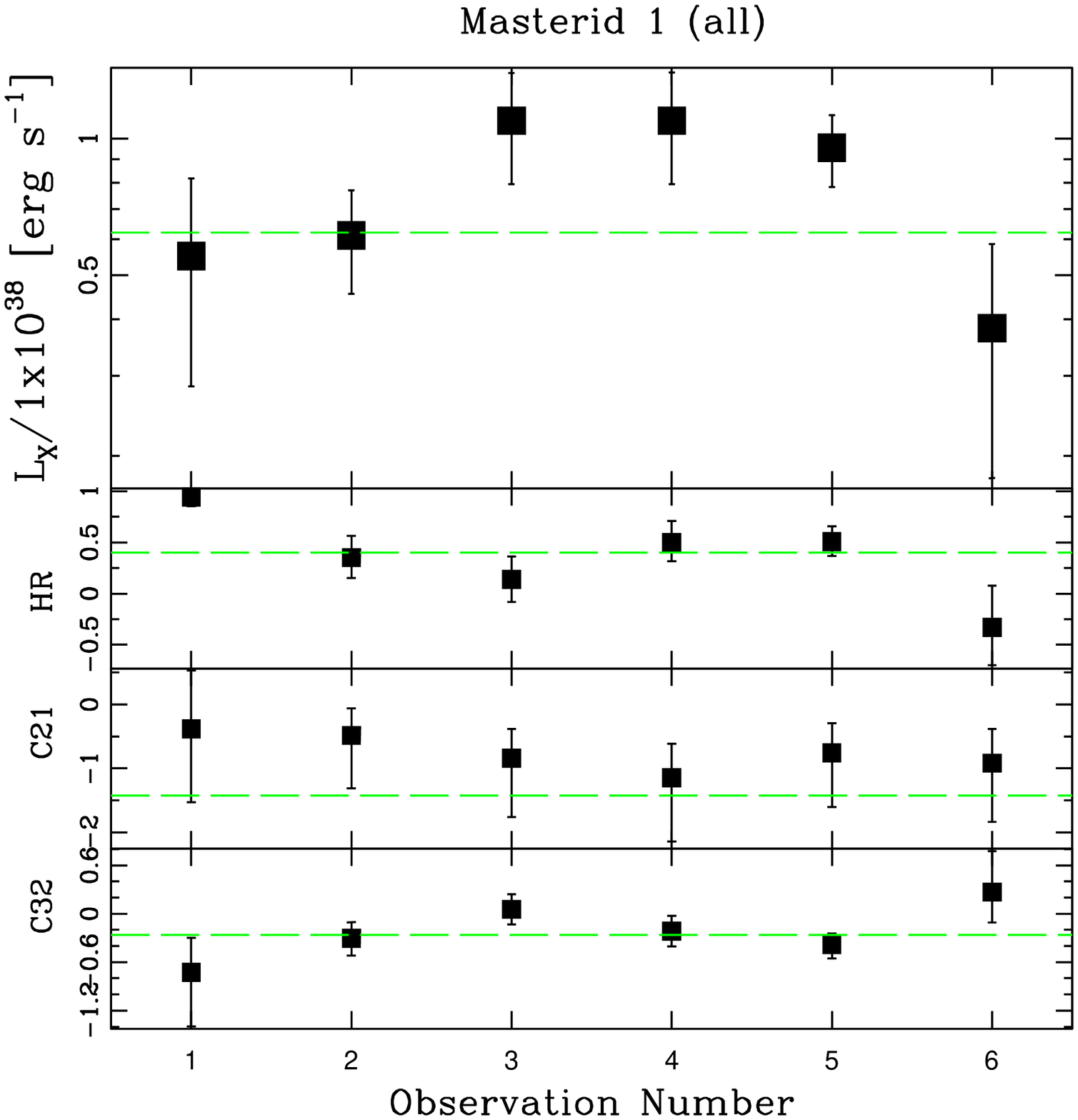}

  \end{minipage}\hspace{0.02\linewidth}
  \begin{minipage}{0.485\linewidth}
  \centering

    \includegraphics[width=\linewidth]{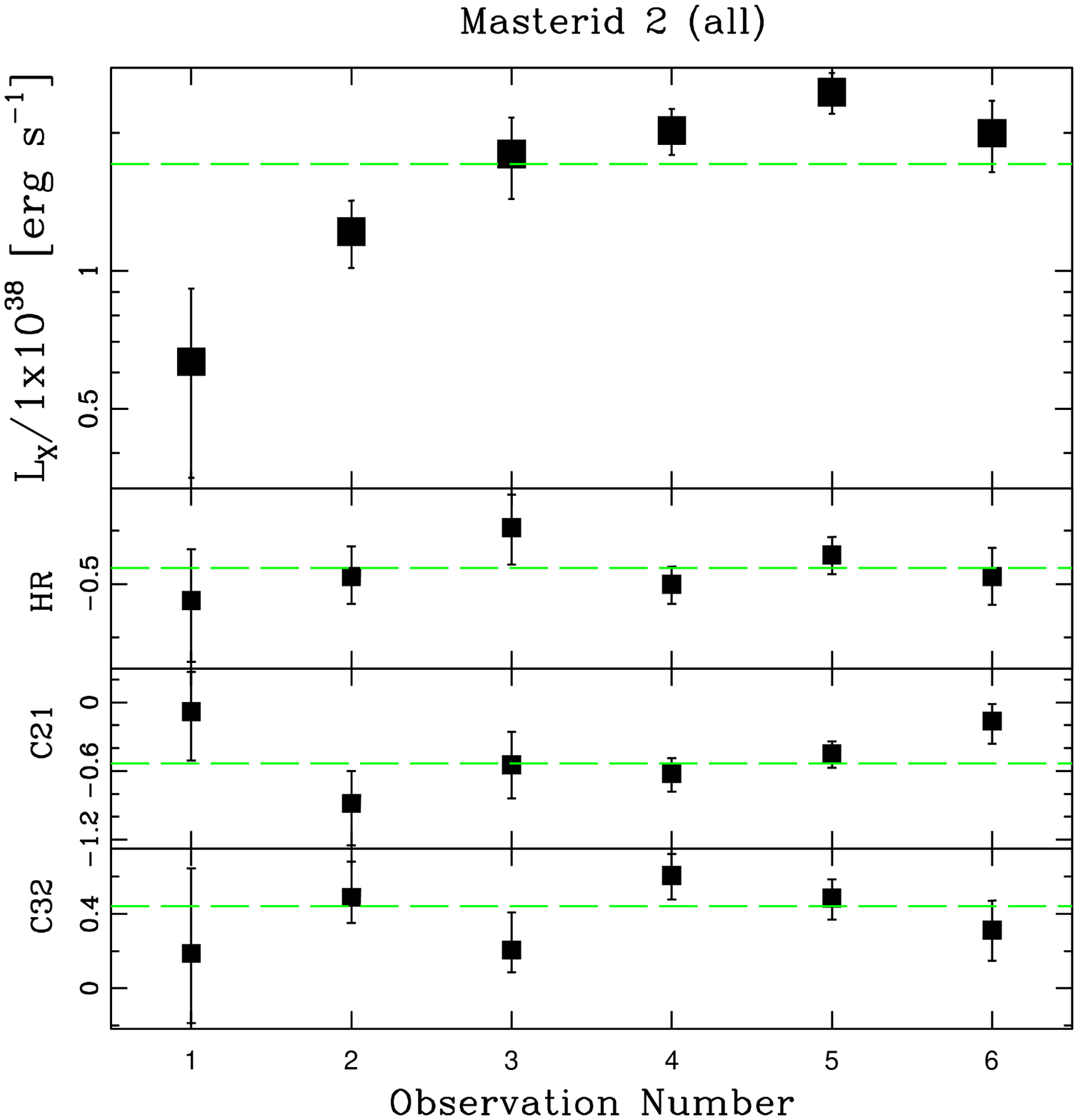}

\end{minipage}

\begin{minipage}{0.485\linewidth}
  \centering

    \includegraphics[width=\linewidth]{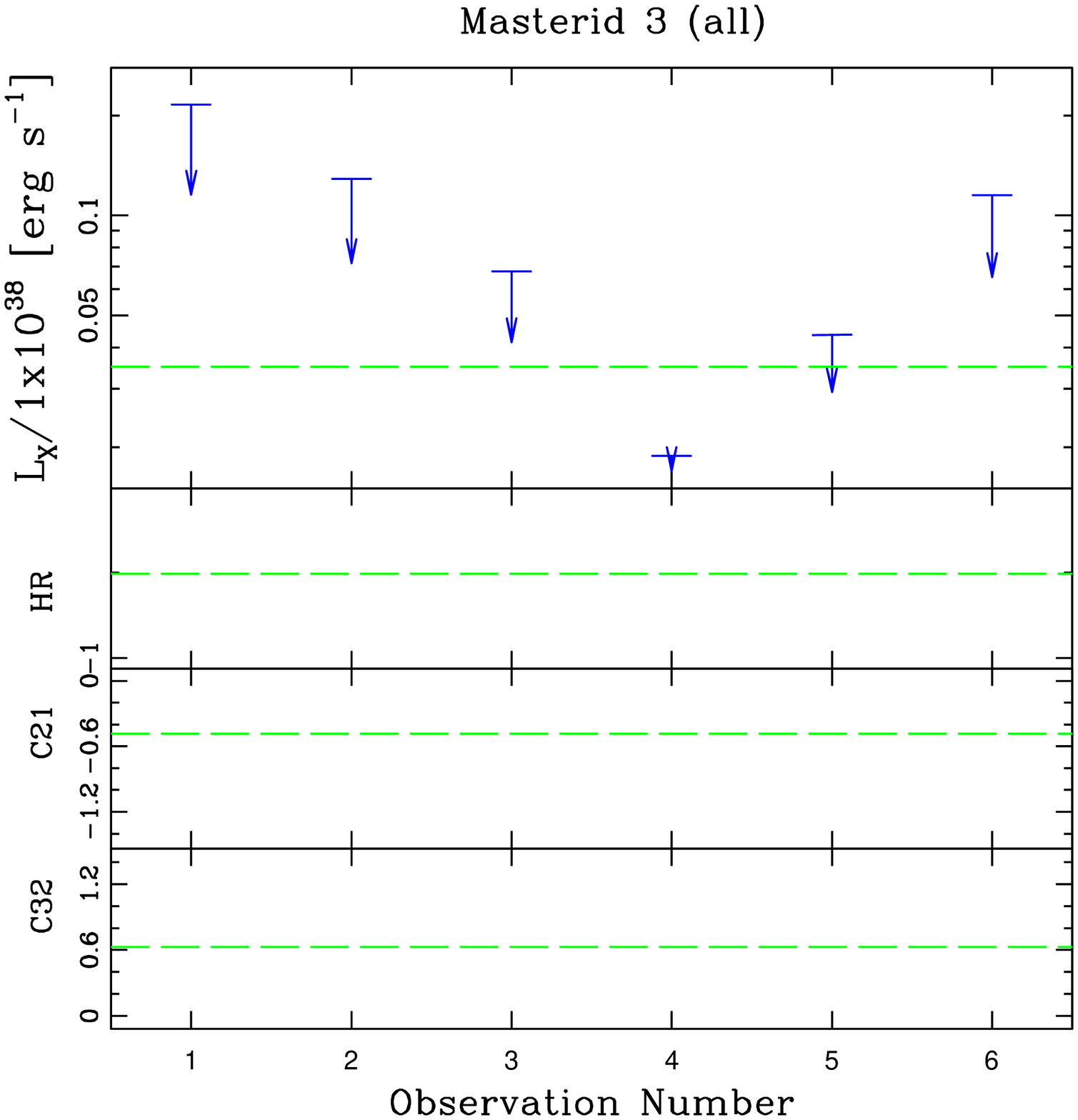}

 \end{minipage}\hspace{0.02\linewidth}
\begin{minipage}{0.485\linewidth}
  \centering
  
    \includegraphics[width=\linewidth]{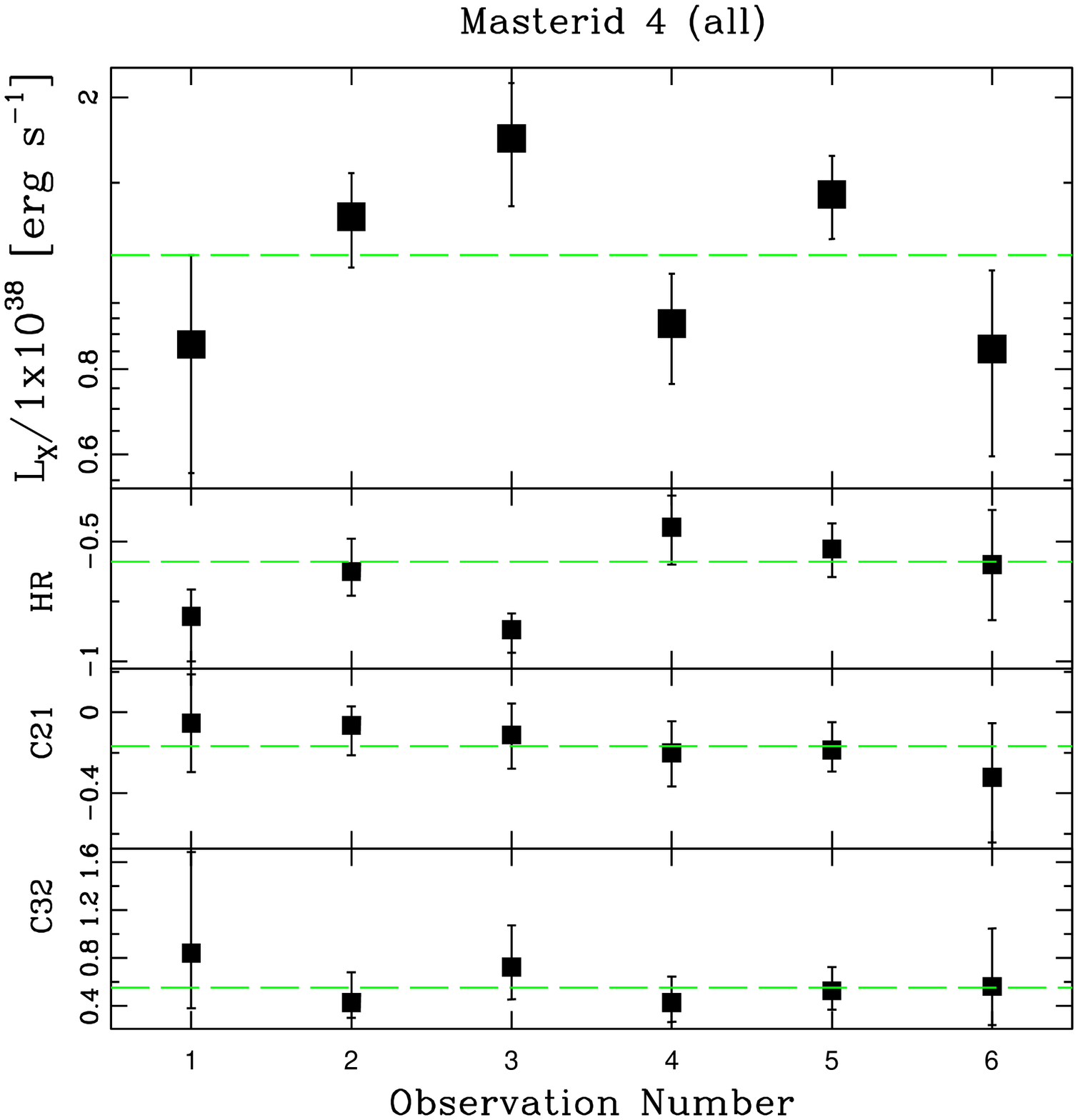}

  \end{minipage}\hspace{0.02\linewidth}

\caption{Plots of the the 236 detected sources, summarizing the
variations in properties of each source between each observation. In
the top panel the long-term light curves are shown. In the second
panel down, the hardness ratios are indicated. These are defined to be
 HR = Hc$-$Sc/Hc+Sc, where Hc is the number of counts in the
hard band (2.0$-$8.0 keV) and Sc is the number of counts in the soft
band (0.5$-$2.0 keV). These bands have been selected to show
comparisons with HR ratios derived in the literature. In the third and
fourth panels the color ratios; C21 and C32, are
plotted, where C21=logS2+logS1 and C32=$-$logH+logS2. For
the color ratios the bandwidths are defined to be S1=0.3$-$0.9
keV, S2=0.9$-$2.5 keV and H=2.5$-$8.0 keV. 
In cases where a source was not detected in a single observation, an
upper limit of the X-ray luminosity is indicated, this is the 90\%
completeness limit for that observation. In all panels the green horizontal
line indicates the value derived from the co-added observation. The blue
horizontal line indicates the 90\% completness limit for instances
where no source was detected in the co-added observation.}
\label{fig:all_LC}

\end{figure}

Figure \ref{fig:all_LC} presents the intensity and spectral
variability of each of the 236 X-ray sources, over all six 
pointings, where the
temporal properties of each point source are shown in four separate
panels. In the top panel the long-term light curve of each source is
presented, with errorbars indicating the 1$\sigma$ uncertainty in the
intensity of the source, with upper limit values provided for sources
that were not detected in a single observation. The second panel shows
the hardness ratio variation of each source, and panels three and four, show
 the temporal properties of C21 and C32 respectively. In all four
panels, the co-added values are also indicated, by a horizontal
dashed green line. In instances where the source was not detected in
the co-added observation, a blue line indicates the upper limit of the
source luminosity.

\begin{figure}
  \begin{minipage}{0.32\linewidth}
  \centering
  
    \includegraphics[width=\linewidth]{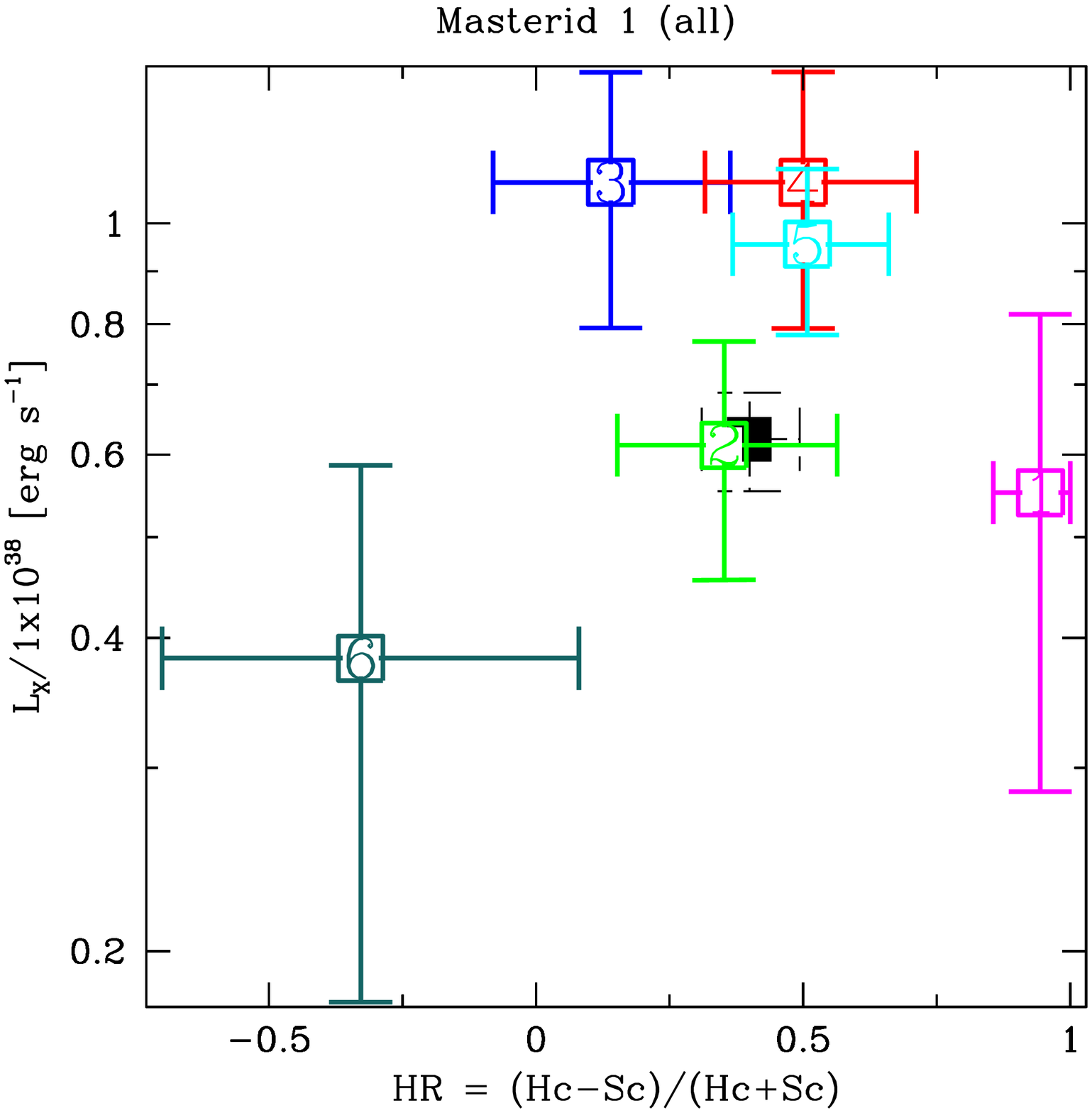}

  \end{minipage}
  \begin{minipage}{0.32\linewidth}
  \centering

    \includegraphics[width=\linewidth]{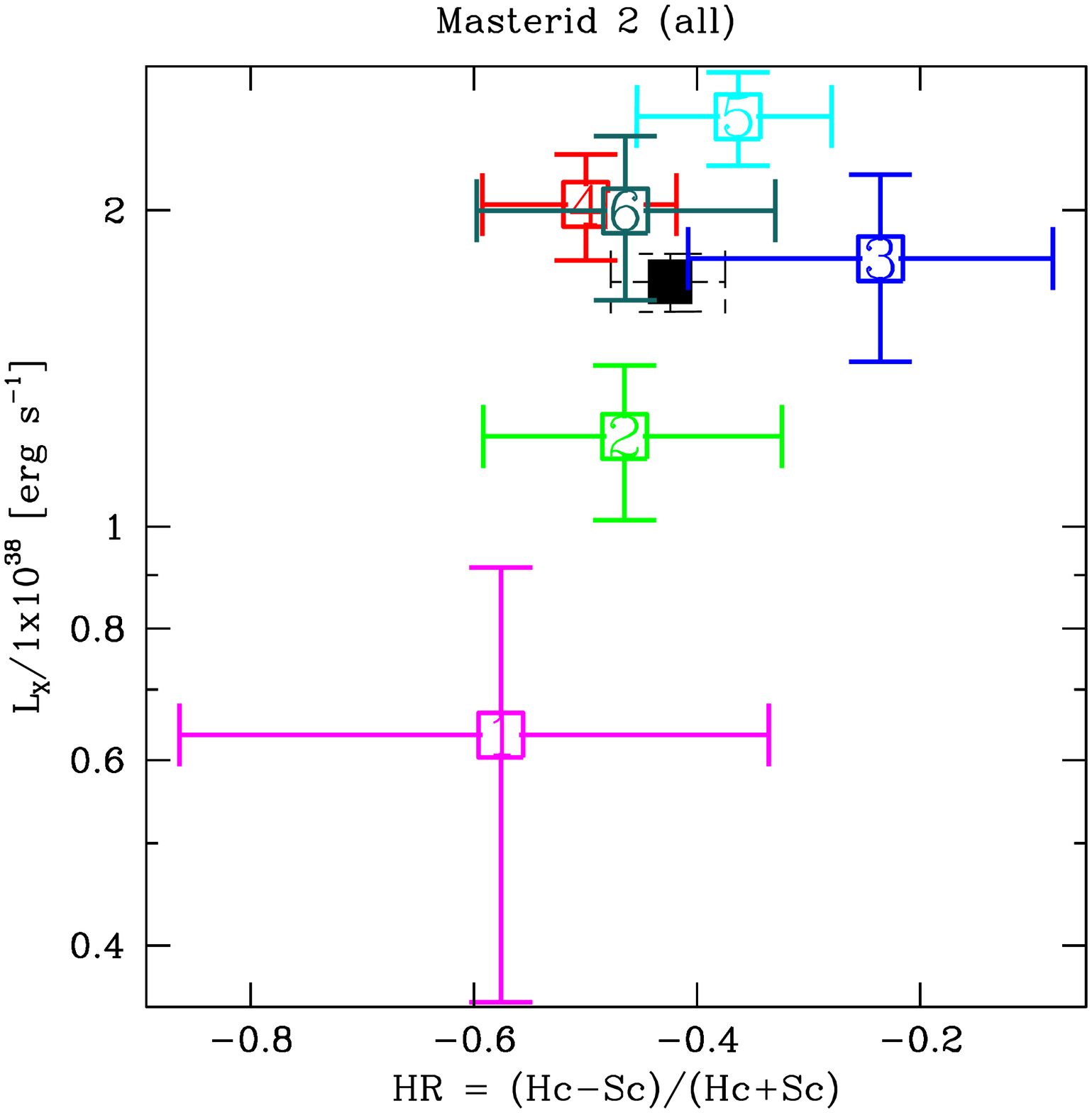}

\end{minipage}
\begin{minipage}{0.32\linewidth}
  \centering

    \includegraphics[width=\linewidth]{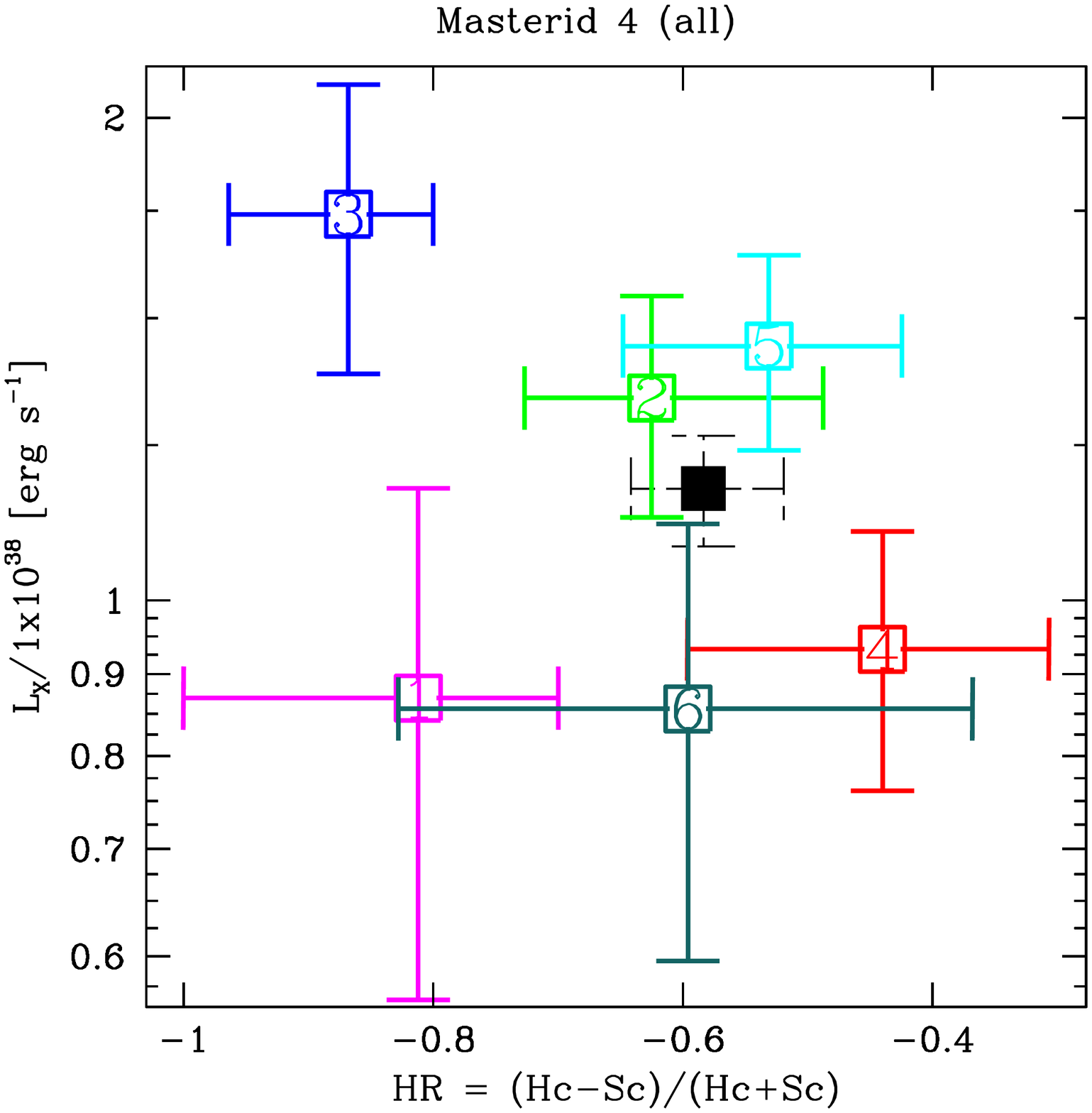}

 \end{minipage}

\begin{minipage}{0.32\linewidth}
  \centering
  
    \includegraphics[width=\linewidth]{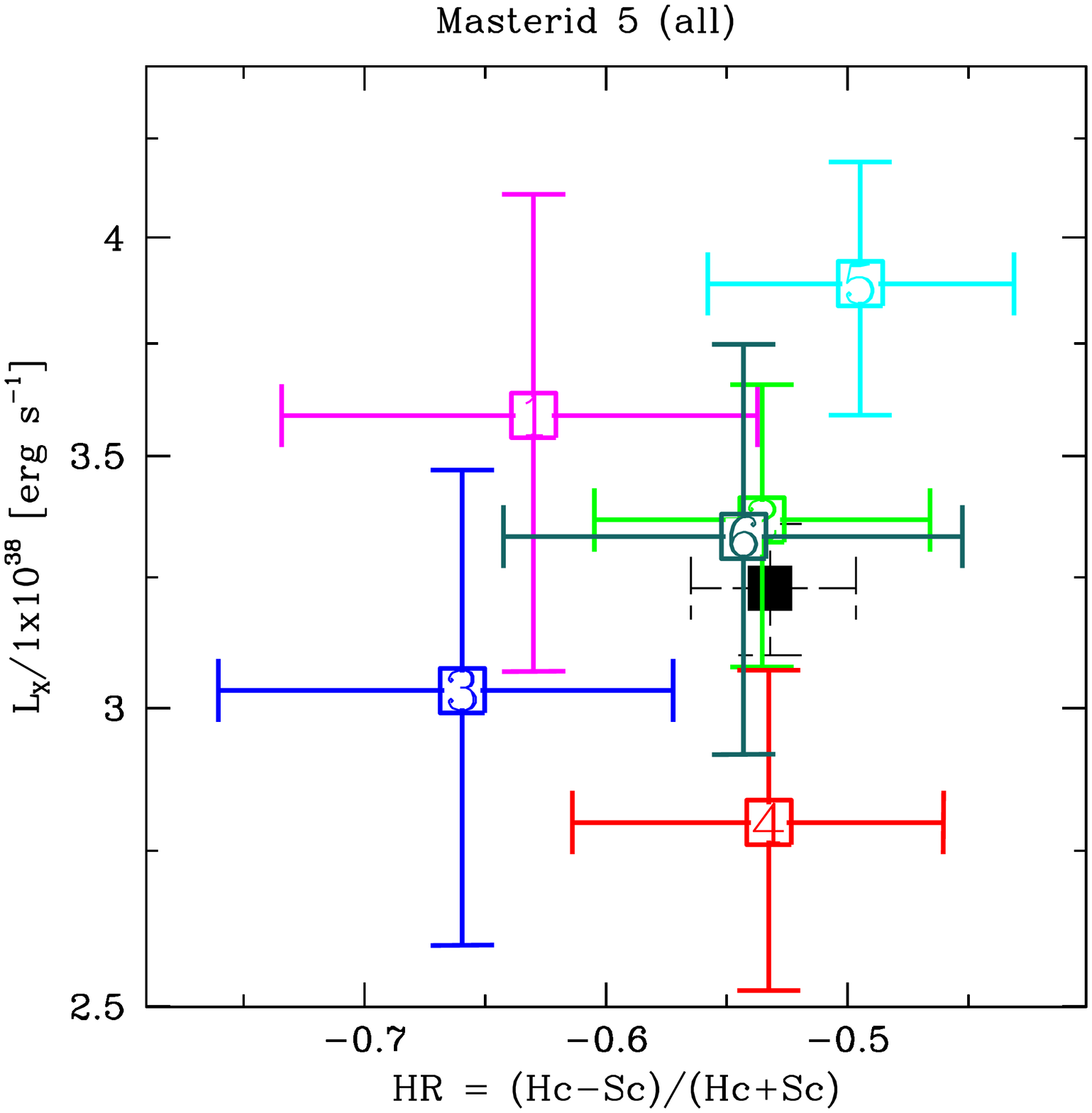}

  \end{minipage}
  \begin{minipage}{0.32\linewidth}
  \centering

    \includegraphics[width=\linewidth]{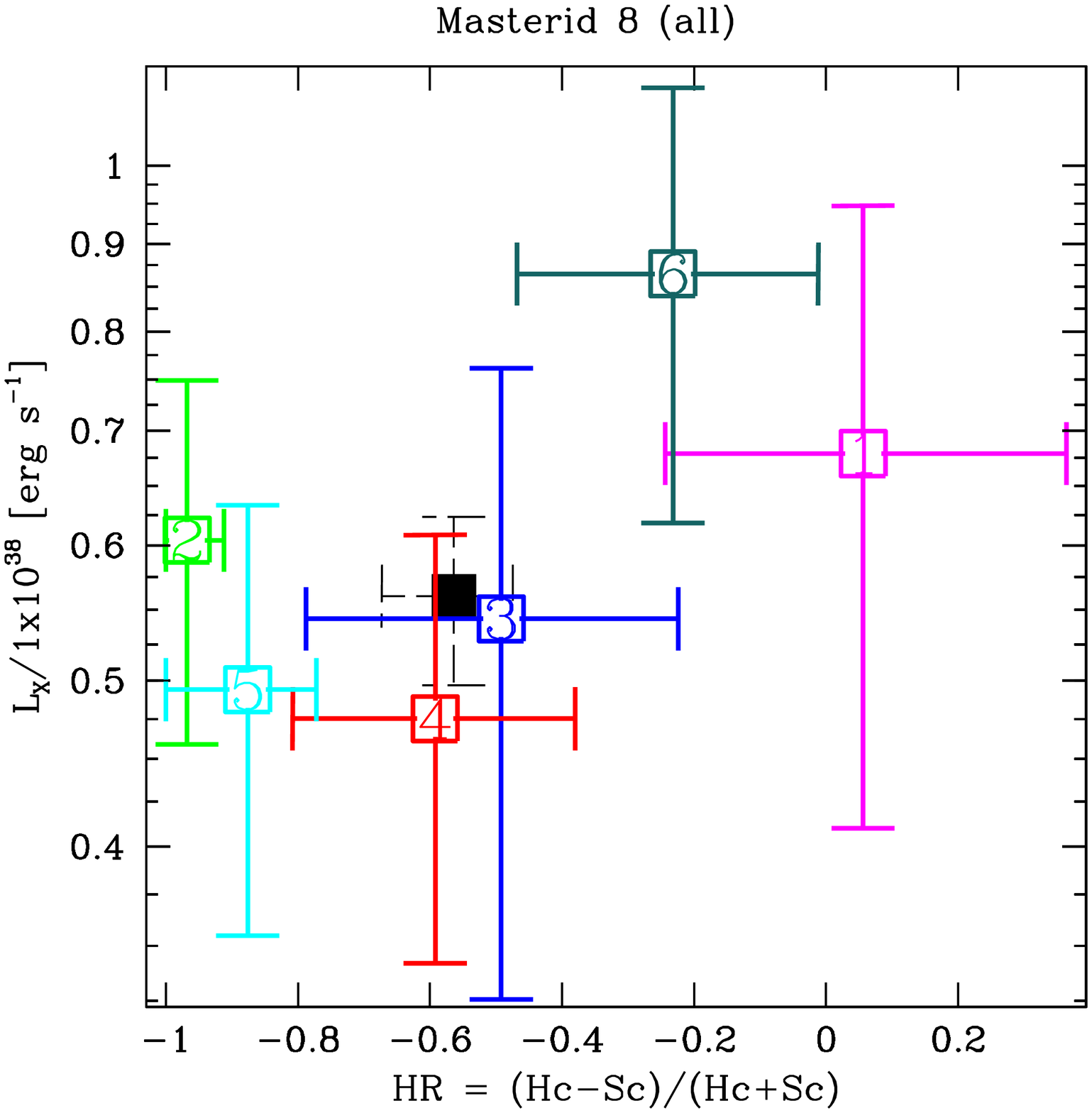}

\end{minipage}
\begin{minipage}{0.32\linewidth}
  \centering

    \includegraphics[width=\linewidth]{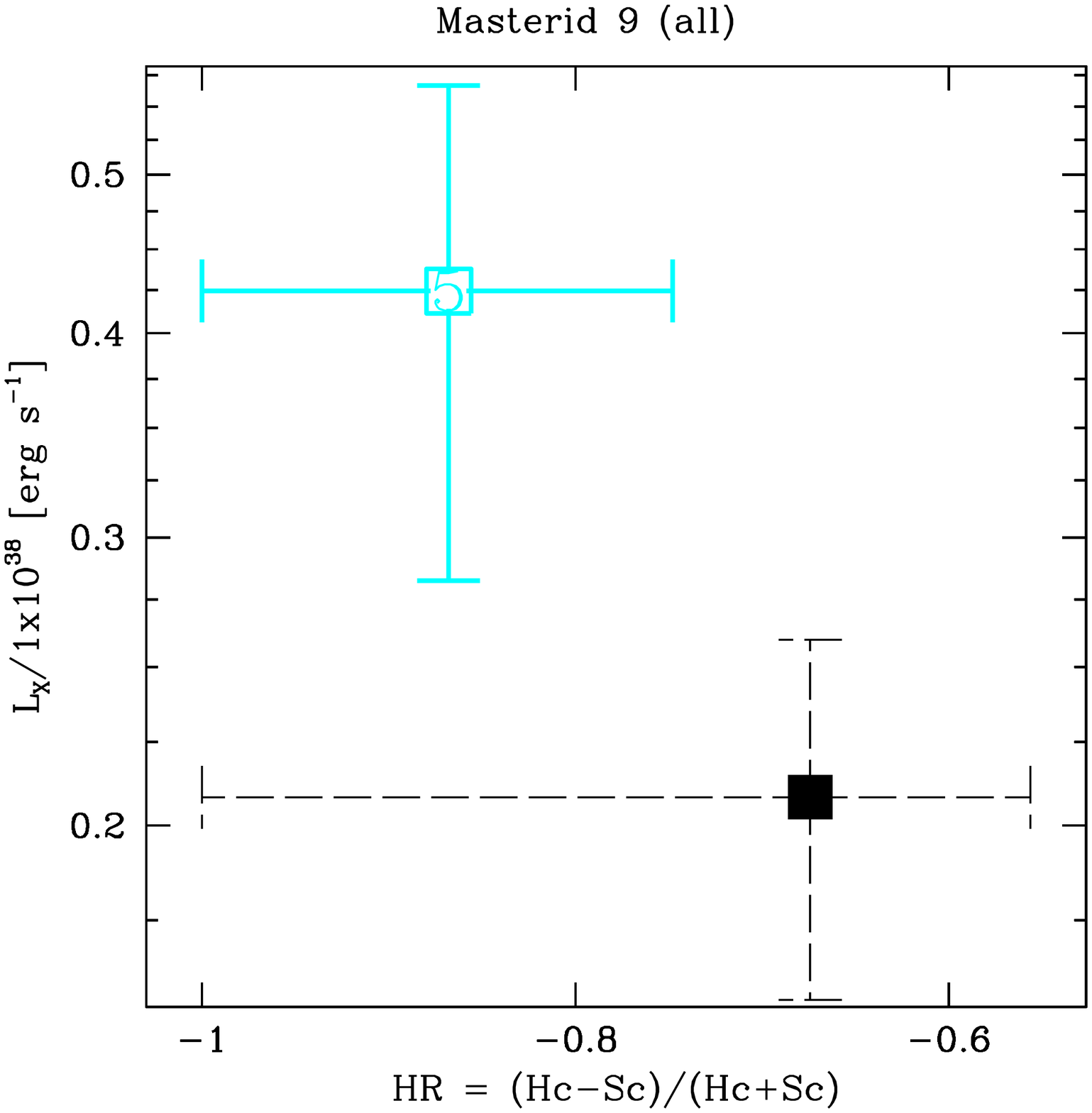}

 \end{minipage}

  \begin{minipage}{0.32\linewidth}
  \centering
  
    \includegraphics[width=\linewidth]{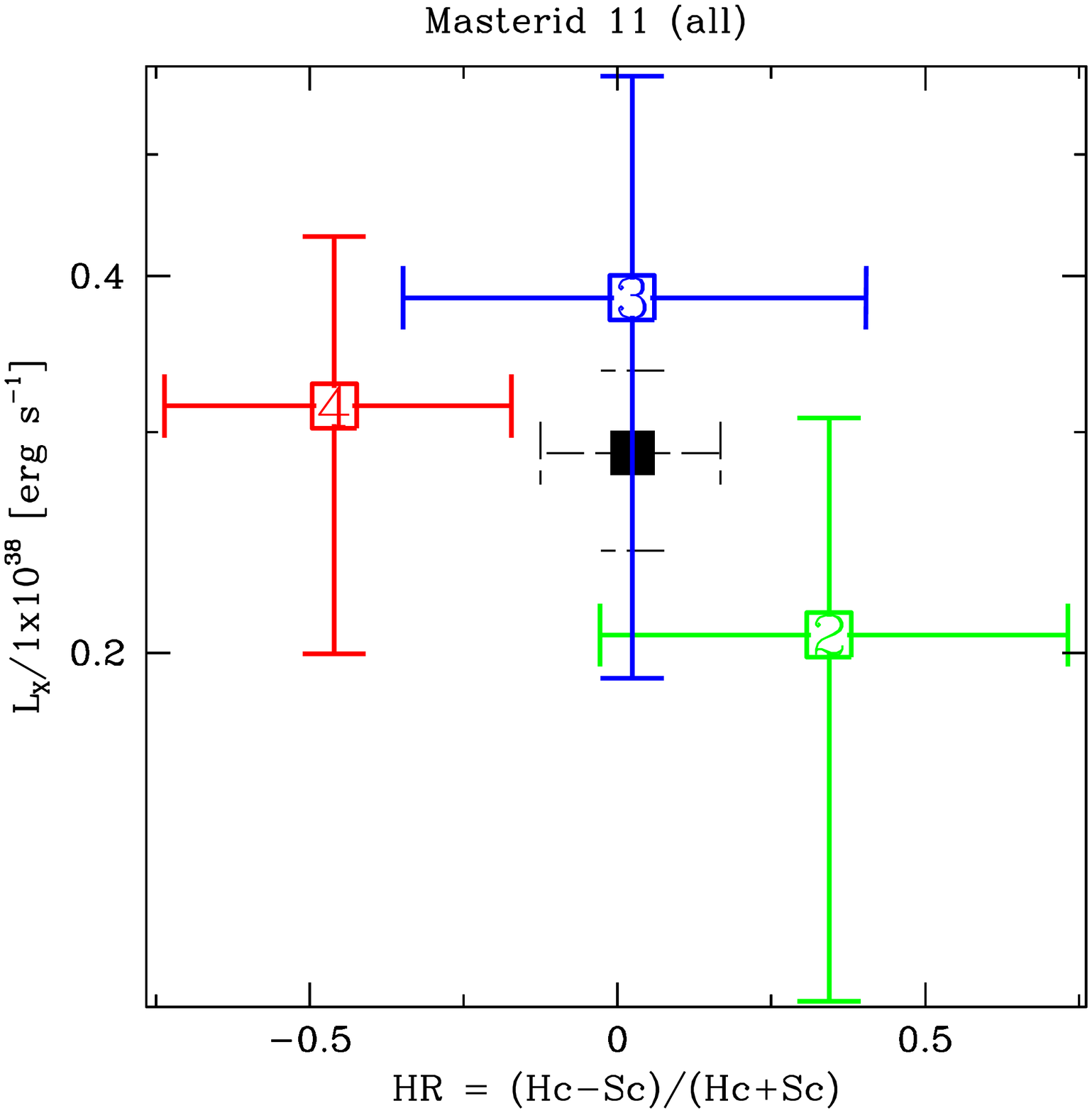}

  \end{minipage}
  \begin{minipage}{0.32\linewidth}
  \centering

    \includegraphics[width=\linewidth]{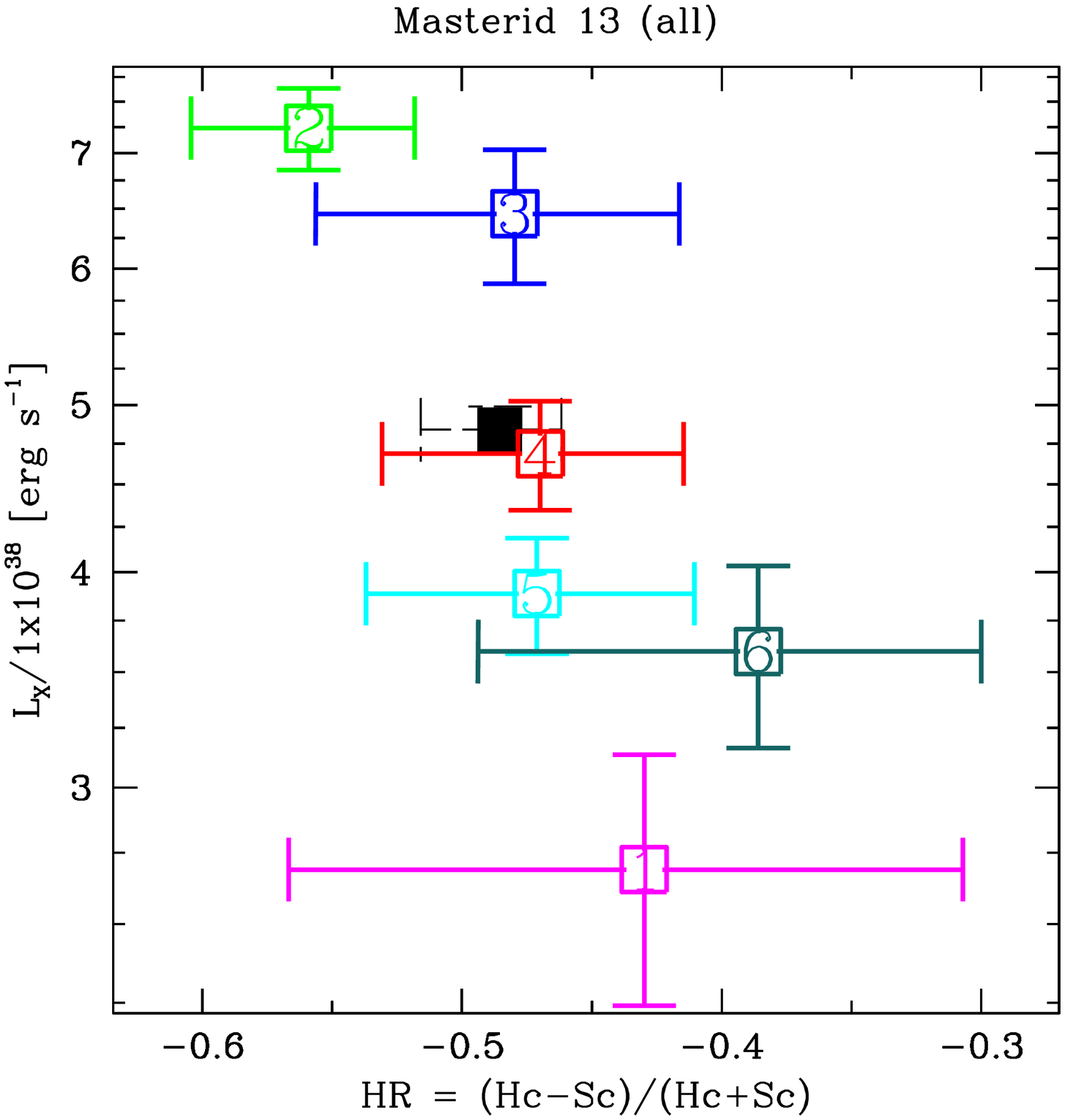}

\end{minipage}
\begin{minipage}{0.32\linewidth}
  \centering

    \includegraphics[width=\linewidth]{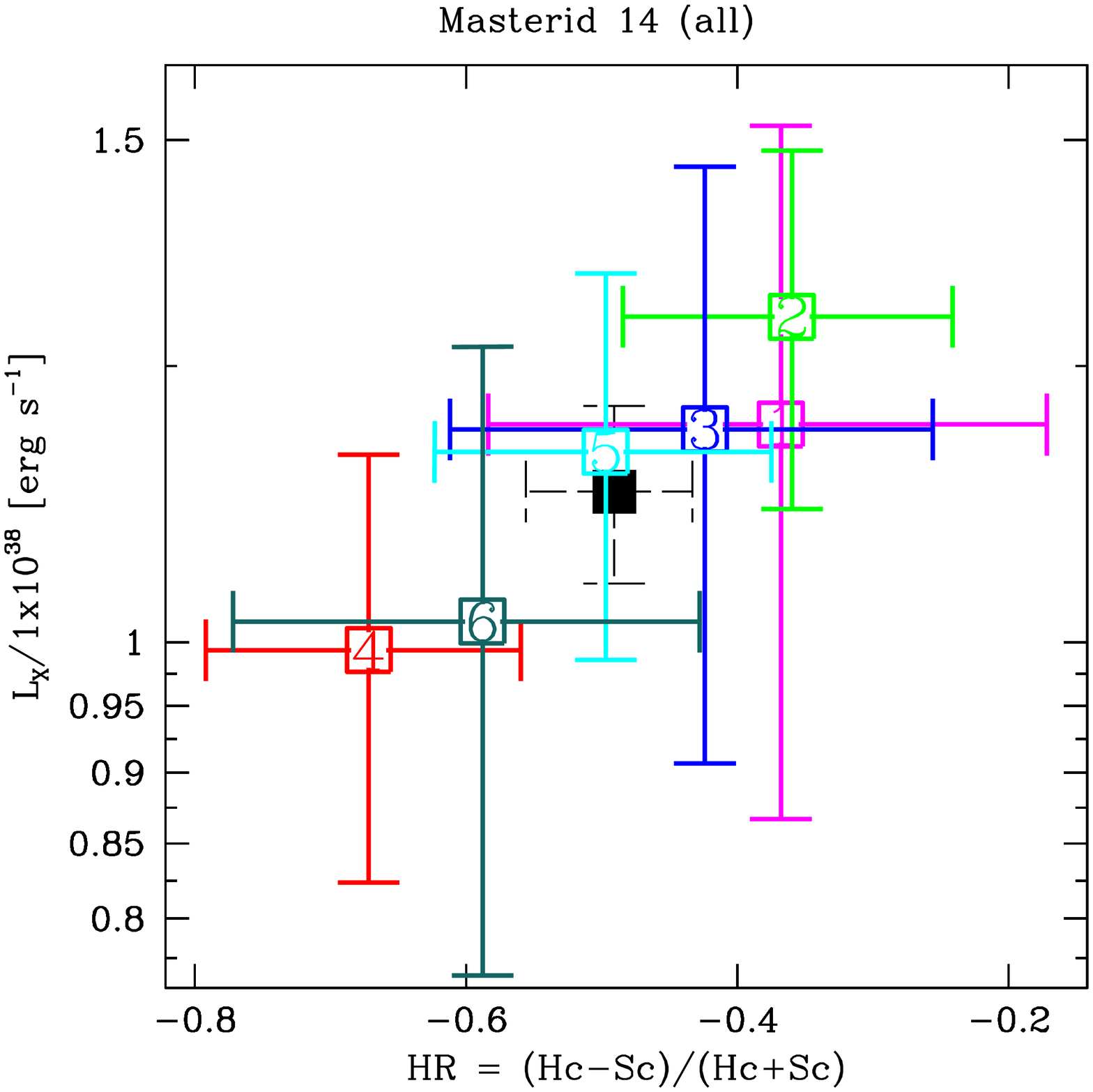}

 \end{minipage}

\caption{HR$-\ensuremath{L_{\mathrm{X}}}$ plots for each
source detected in more that one individual observation, with each observation plotted in a different color; observation 1
is magenta, observation 2 is green, observation 3 blue, observation 4
red, observation 5 cyan and observation 6 is dark green. The HR and
$\ensuremath{L_{\mathrm{X}}}$ values for the co-added observation are
also shown, plotted in black. The hardness ratios are defined to be
HR = Hc$-$Sc/Hc+Sc, where Hc is the number of counts in the
hard band (2.0$-$8.0 keV) and Sc is the number of counts in the soft
band (0.5$-$2.0 keV).  }
\label{fig:lxhrindiv}
\end{figure}

\begin{figure}
  \begin{minipage}{0.32\linewidth}
  \centering
  
    \includegraphics[width=\linewidth]{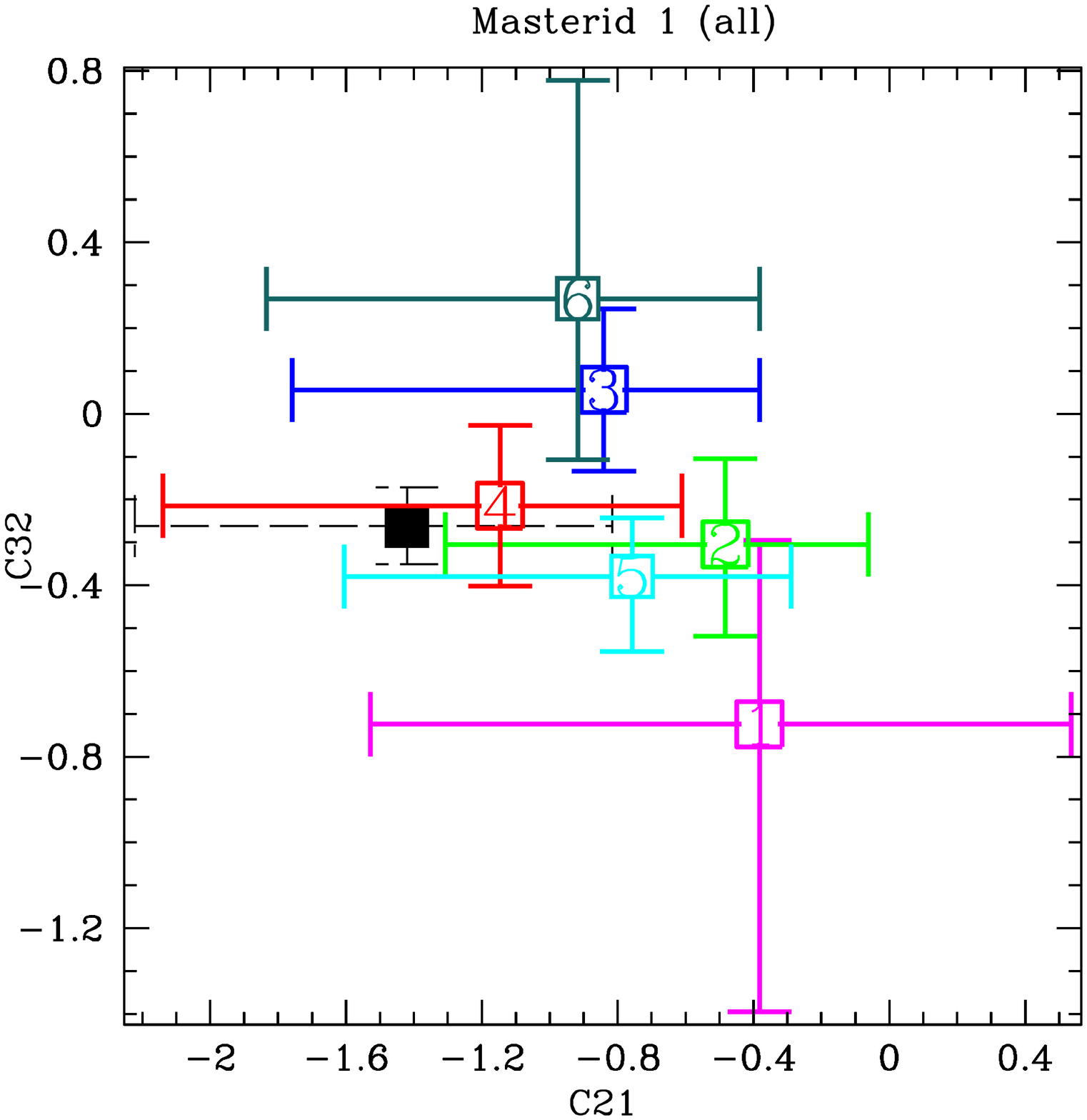}
 
  \end{minipage}
  \begin{minipage}{0.32\linewidth}
  \centering

    \includegraphics[width=\linewidth]{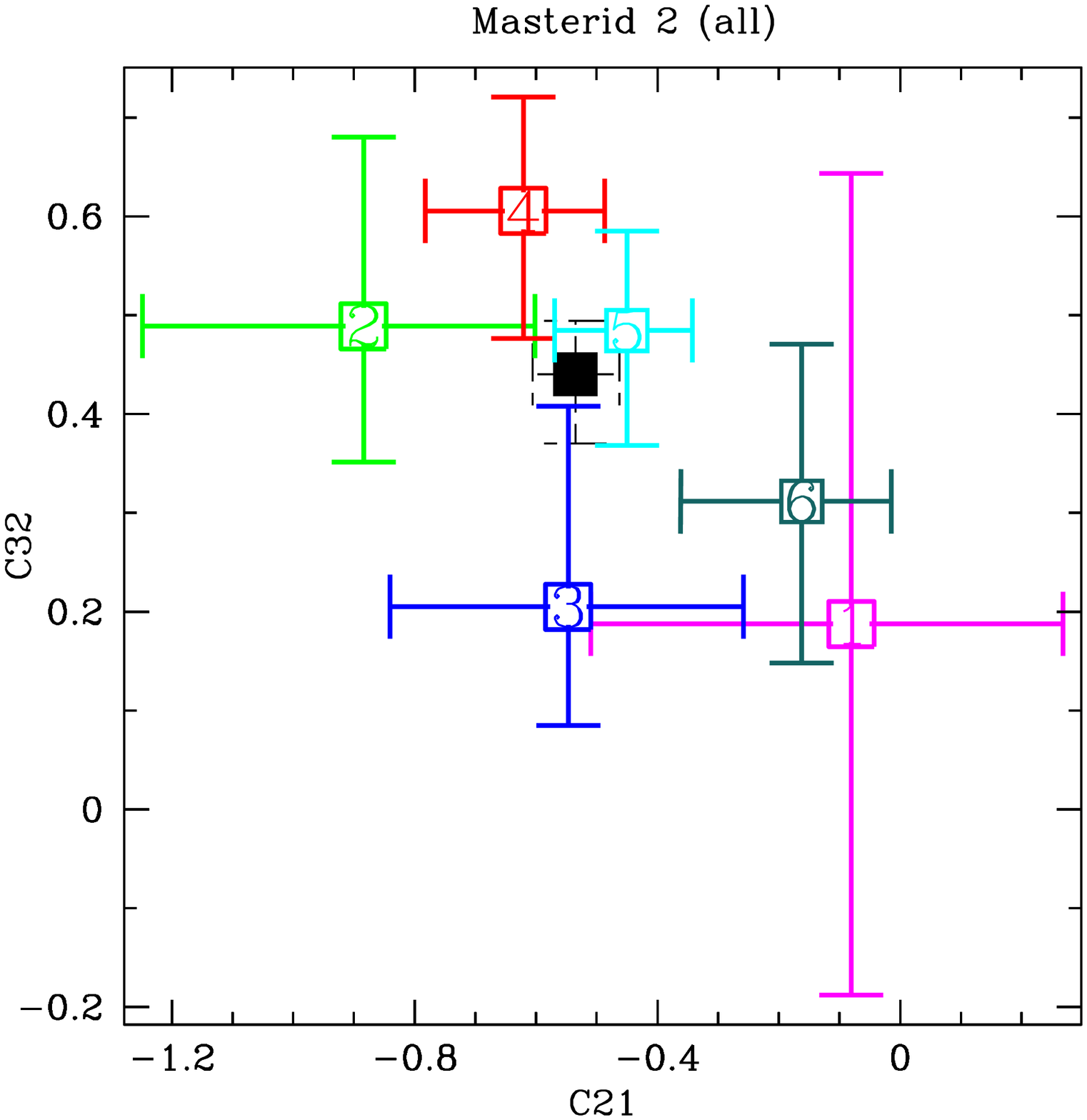}

\end{minipage}
\begin{minipage}{0.32\linewidth}
  \centering

    \includegraphics[width=\linewidth]{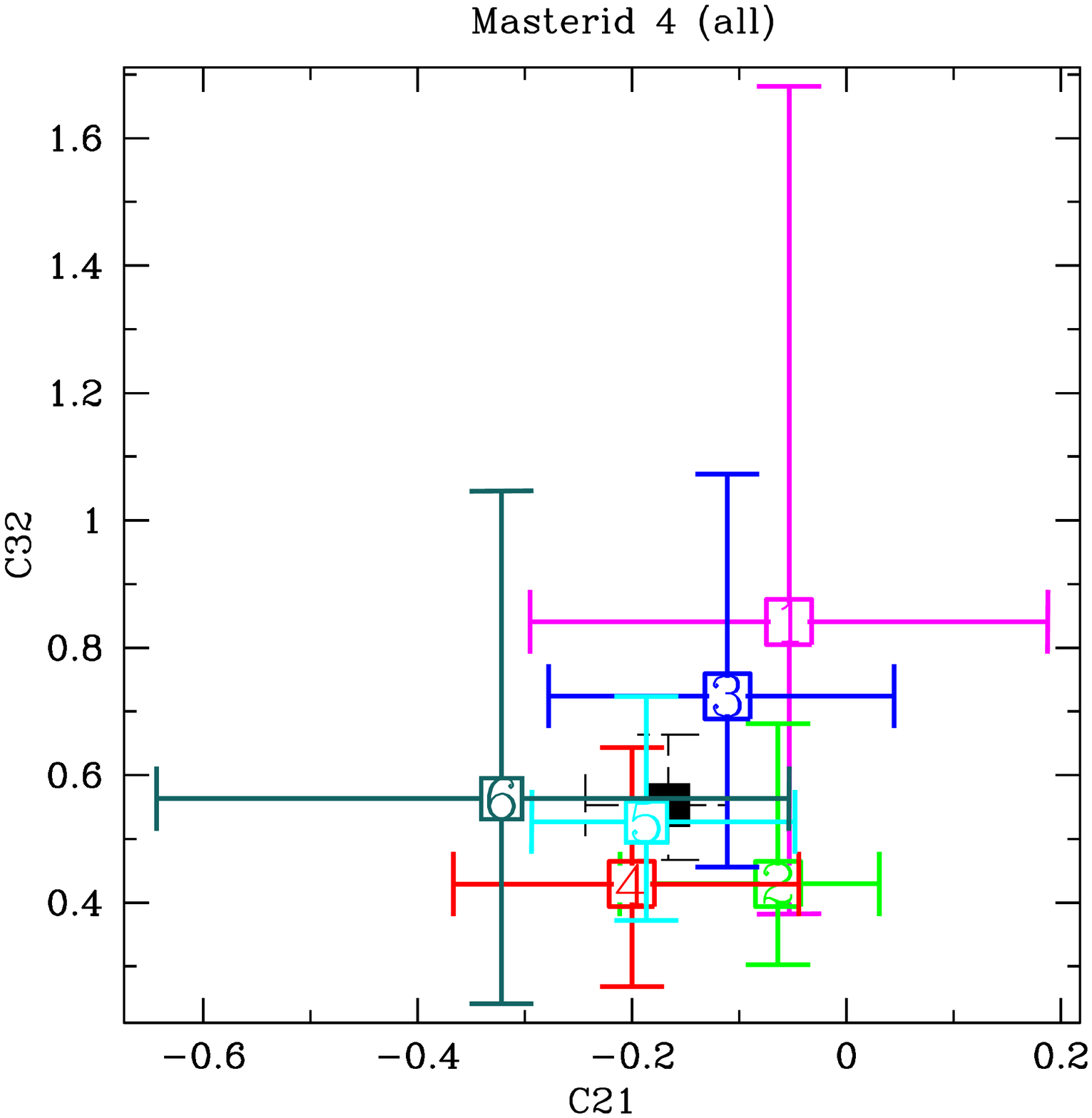}

 \end{minipage}

\begin{minipage}{0.32\linewidth}
  \centering
  
    \includegraphics[width=\linewidth]{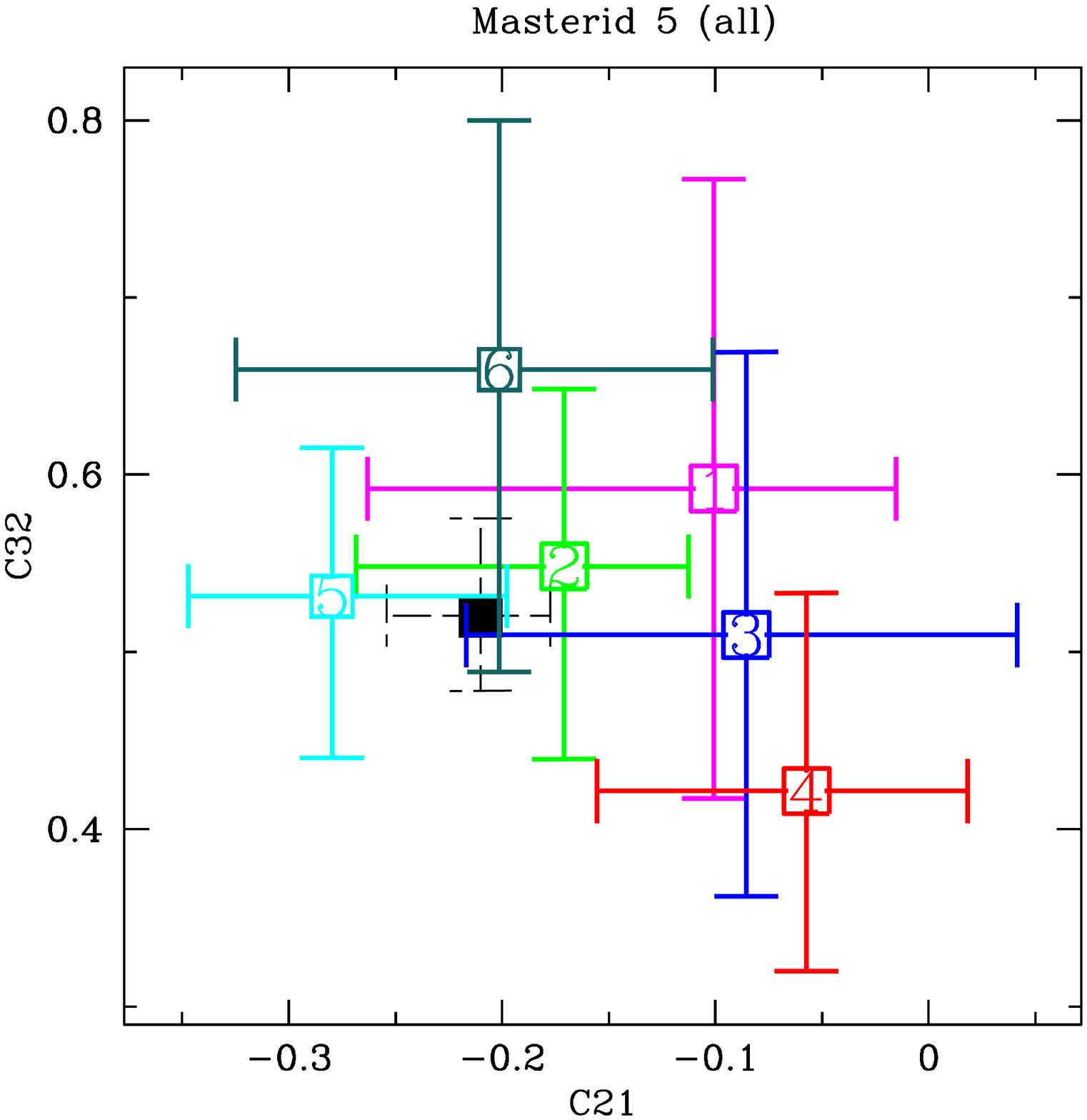}
  
  \end{minipage}
  \begin{minipage}{0.32\linewidth}
  \centering

    \includegraphics[width=\linewidth]{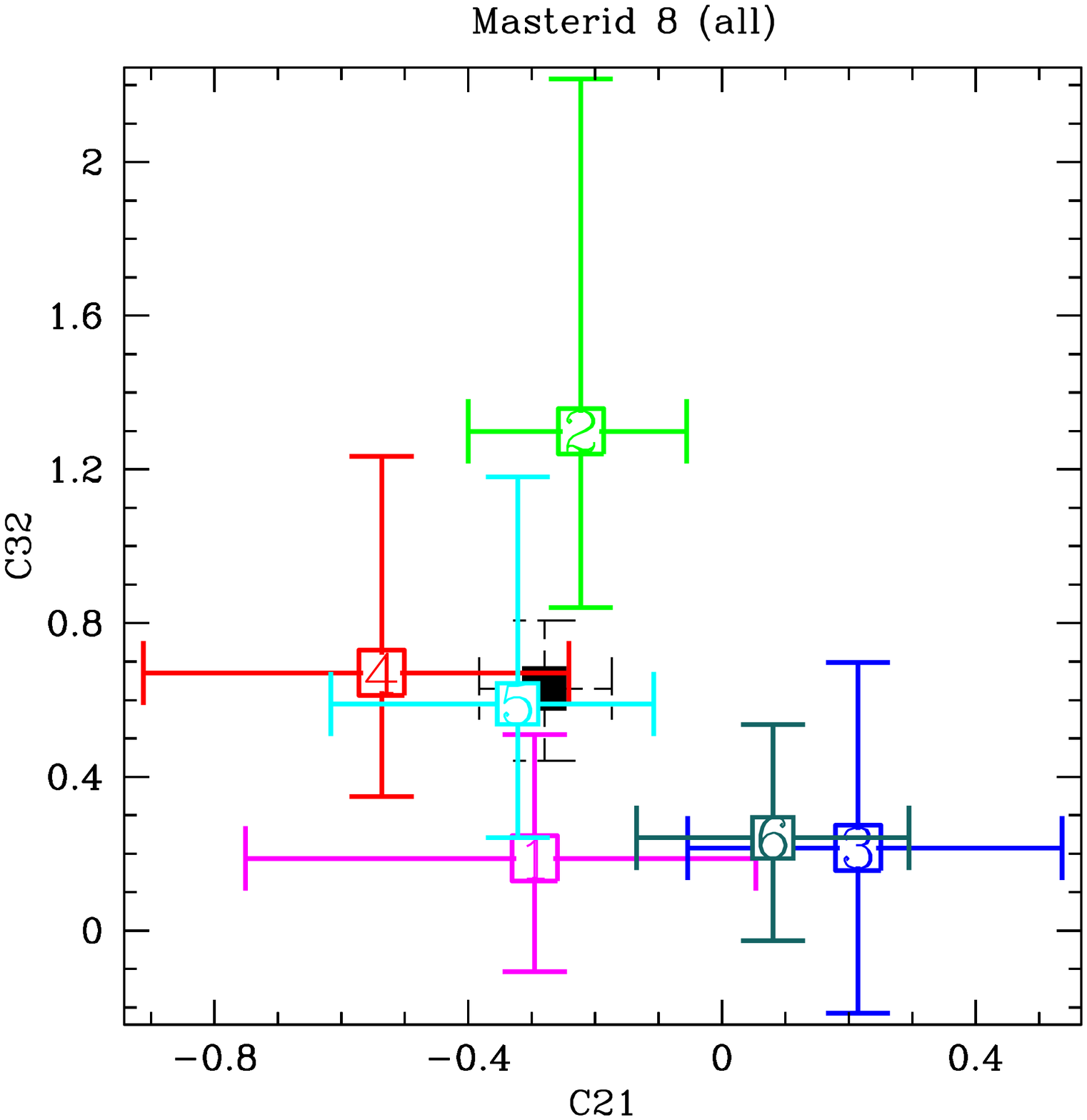}

\end{minipage}
\begin{minipage}{0.32\linewidth}
  \centering

    \includegraphics[width=\linewidth]{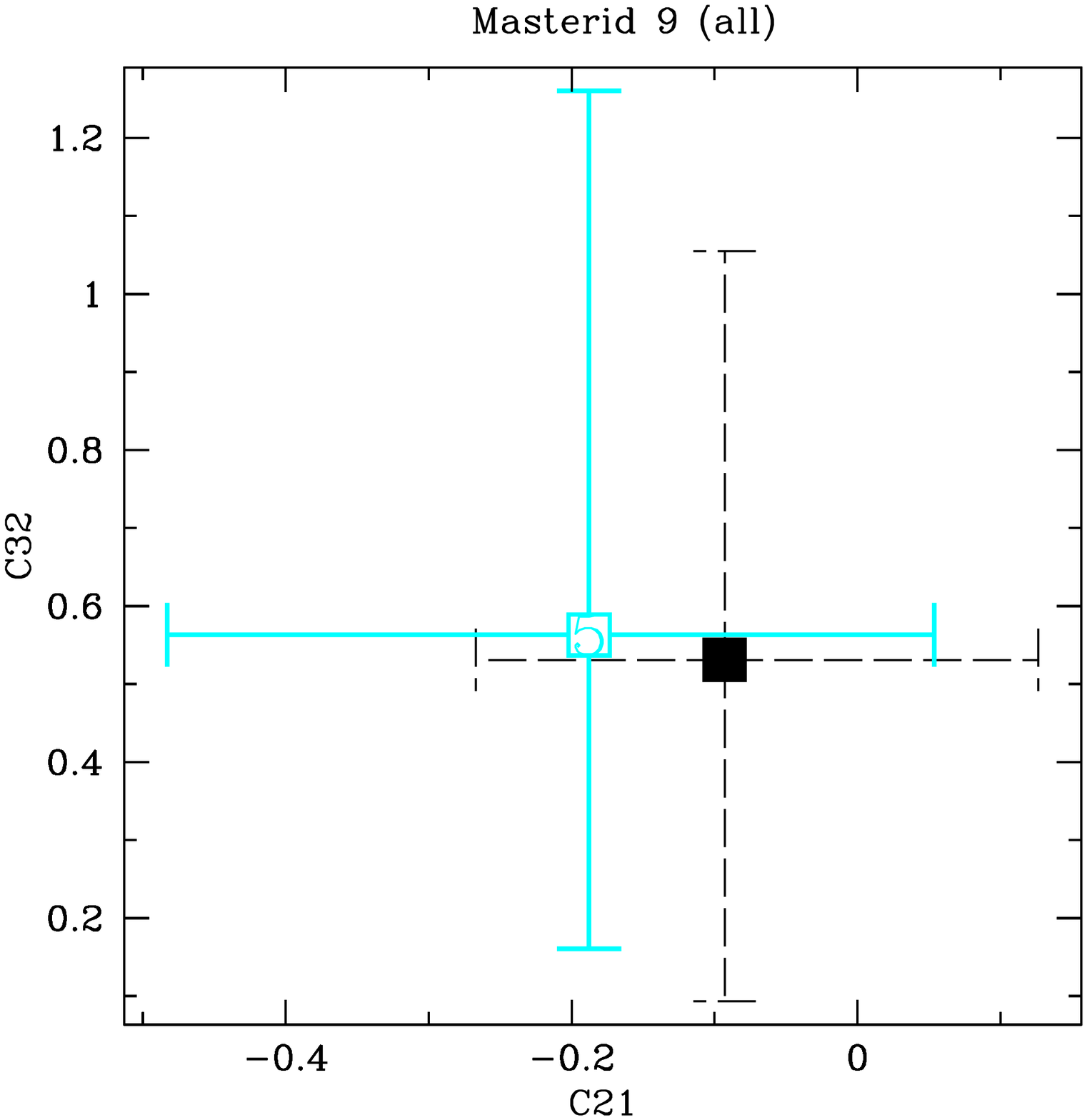}

 \end{minipage}

  \begin{minipage}{0.32\linewidth}
  \centering
  
    \includegraphics[width=\linewidth]{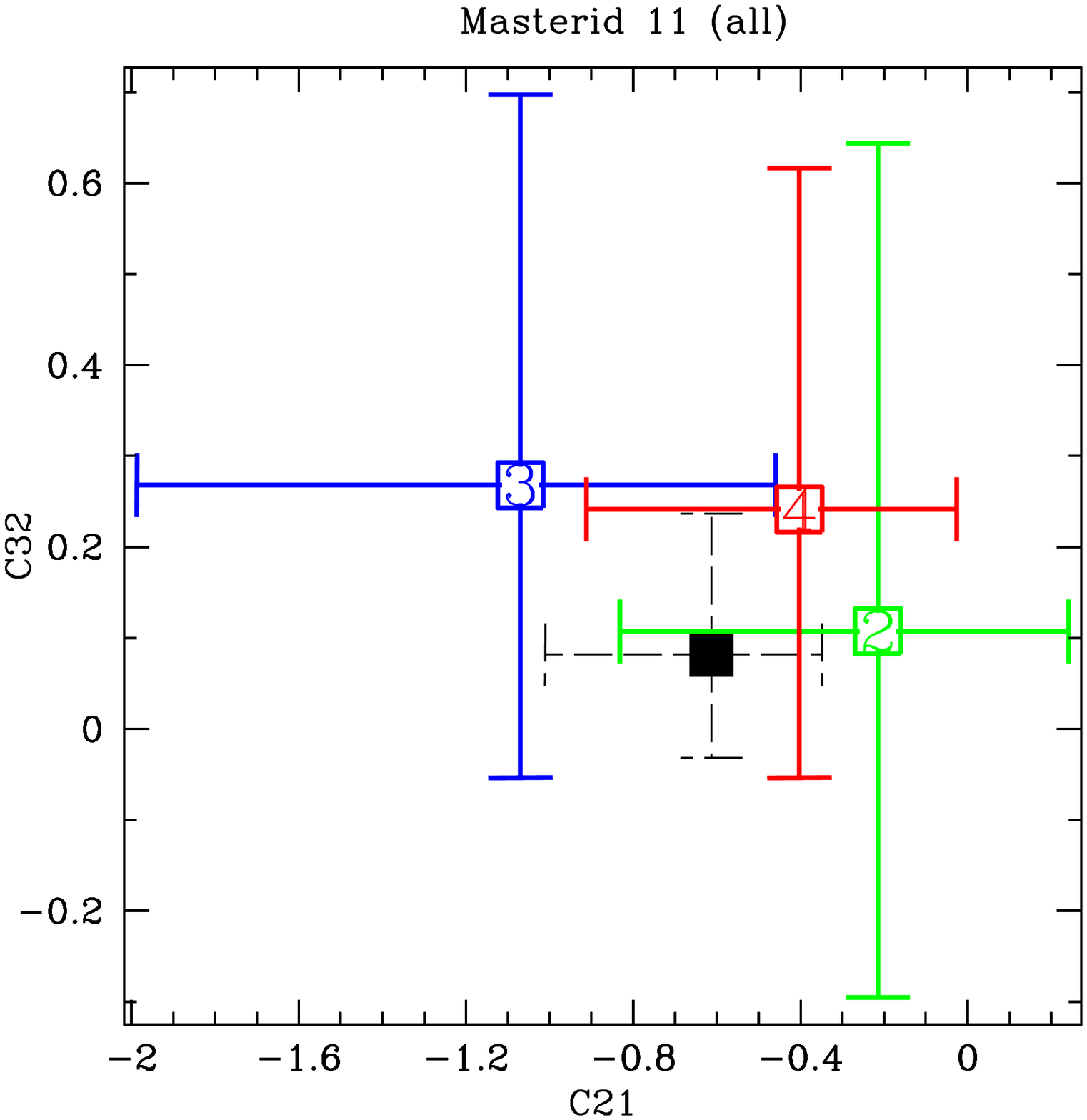}
  
  \end{minipage}
  \begin{minipage}{0.32\linewidth}
  \centering

    \includegraphics[width=\linewidth]{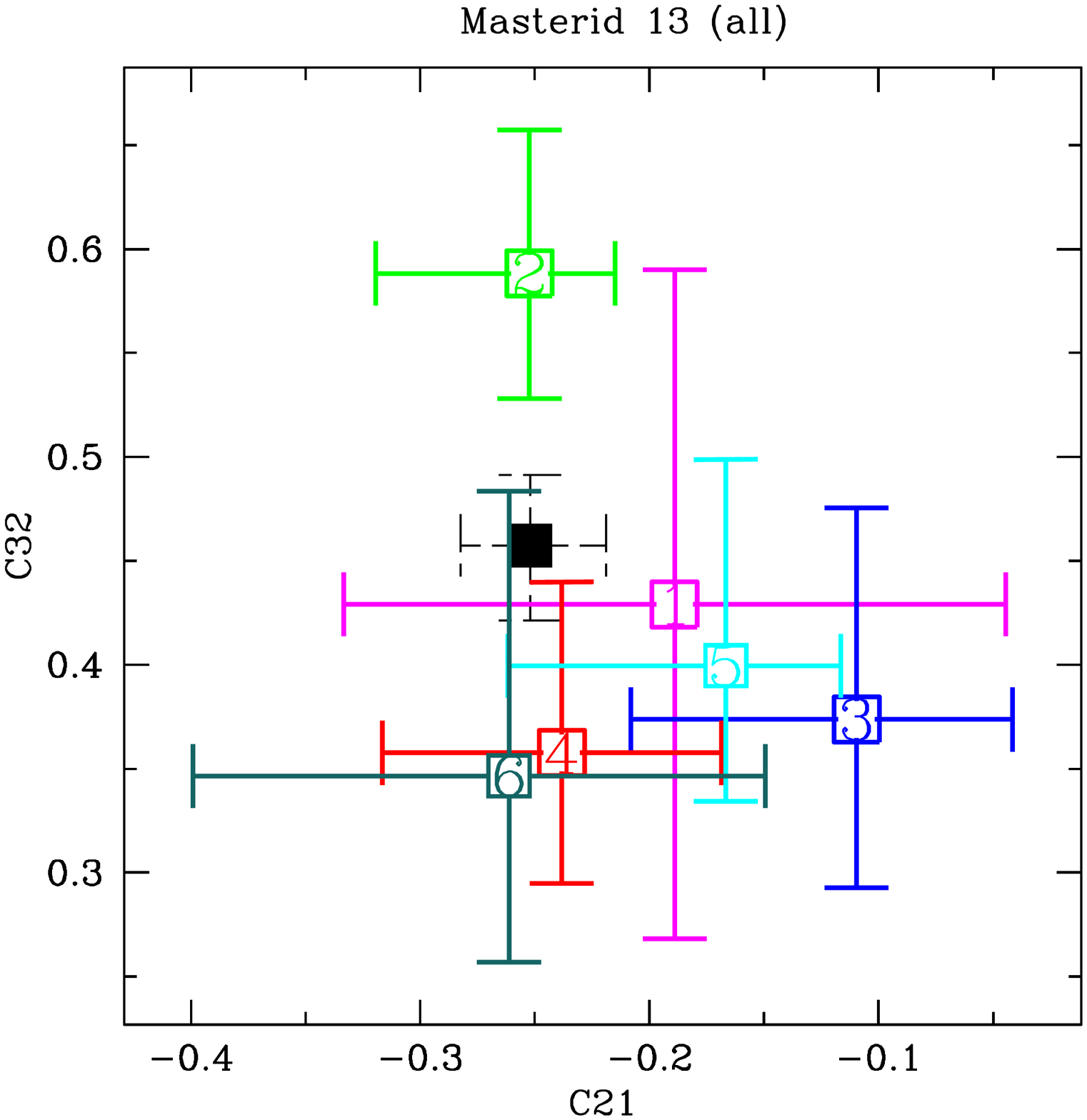}

\end{minipage}
\begin{minipage}{0.32\linewidth}
  \centering

    \includegraphics[width=\linewidth]{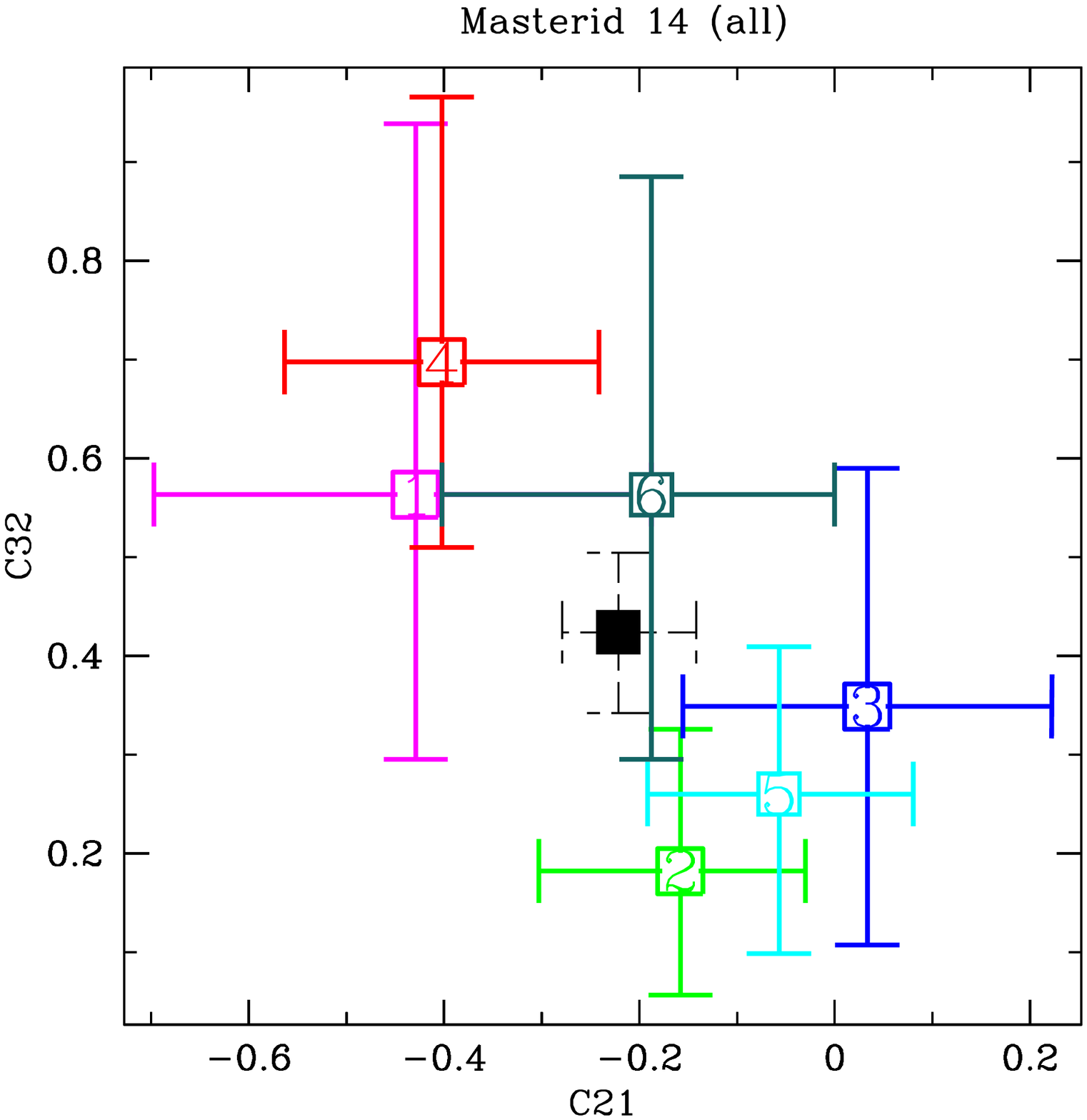}

 \end{minipage}

	\caption{Color-color plots for each
source that has been detected in more than one pointing. Each individual observation is plotted in a different color; observation 1
is magenta, observation 2 is green, observation 3 blue, observation 4
red, observation 5 is cyan and observation 6 is dark green. The co-added observation is also
plotted in black. The color ratios; C21 and C32, are
plotted, where C21=logS2+logS1 and C32=$-$logH+logS2. For
the color ratios the bandwidths are defined to be S1=0.3$-$0.9
keV, S2=0.9$-$2.5 keV and H=2.5$-$8.0 keV.  }
\label{fig:CCindiv}
\end{figure}

Figure \ref{fig:lxhrindiv} presents the \LX-HR plots for sources with
measured hardness ratios in at least two observations. Each point
shows the X-ray luminosity and hardness ratio value of a source during
each pointing, as well as the values derived from the co-added
observation. Each point is labeled and color coded, where
magenta, green, blue, red, cyan and dark green indicate observations 1$-$6 respectively, and
black represents the co-added observation value. Similarly, Figure
\ref{fig:CCindiv} presents the color-color values 
for sources with measured color-color values in at least two
observations, where again individual observations are labeled and
color coded (following the same color scheme as in Figure
\ref{fig:lxhrindiv}), with the co-added observation indicated in
black. 

\section{Discussion}

\subsection{X-ray Source Population}

In the previous sections the data analysis methods, used to determine
the properties of the X-ray binary population of NGC 4278, have been
presented. From the six individual \CHANDRA\ pointings, taken
between February 2005 and February 2007, a co-added observation,
totaling an exposure time of 458-ks, has been produced. From this deep
observation of the galaxy, 236 X-ray point sources have been detected
in the region overlapped by all of the individual pointings, with 180 of these
sources residing within the $D_{25}$ ellipse of the system. These 236
sources are presented in Figure \ref{fig:rgb} where a raw, full band
image from the co-added observation, with the overlap region and
$D_{25}$ ellipse overlaid, is presented in the main image, with source regions
also indicated. The two following images present the central region of the
galaxy, where the dense population of sources can be
more clearly seen.
Of these 236 sources, based on the hard-band $ChaMP+CDF$ $logN - logS$
relation (Kim \etal\ 2007b), $\sim$29 sources detected in the 
co-added observation are expected to be objects not associated with
NGC 4278. Within the $D_{25}$ ellipse of the galaxy it is expected
that $\sim$12 of these sources are background objects. In Figure
\ref{fig:profile} the number of expected background objects is
indicated in the X-ray source number density profile of the galaxy and appears consistent with the flattening of the profile at larger radii.

The total number of LMXBs residing in NGC 4278 is much greater than that of NGC 3379 (notwithstanding the 90\% completeness limit of \LX $\sim5\times 10^{36}$ \ergps\ compared to \LX $\sim1\times 10^{37}$ \ergps\ in NGC 4278), where only 98 sources were detected within the $D_{25}$ ellipse of the galaxy, compared to 180 in the $D_{25}$ ellipse of NGC 4278. This increase in the total LMXB population in NGC 4278 follows the known correlation of increasing X-ray luminosity arising from LMXBs (normalized by \LB) with increasing globular cluster specific frequency ($S_{GC}$) (e.g Kim \etal\ 2006, Irwin 2005). Such a relationship confirms the importance of LMXB formation in GCs, and through the deep observations that have presented here and in Brassington \etal\ (2008) this correlation will be further investigated in a forthcoming paper, where the hypothesis that {\em all} LMXBs were formed in GCs will be tested.

The X-ray luminosity of the sources detected within NGC 4278
ranges from 3.5$\times~10^{36}$~\ergps\ (with 3$\sigma$ upper limit $\le$1$\times 10^{37}$ \ergps) up to $\sim2\times10^{40}$\ergps, where the brightest source, 117, lies at the nucleus of the galaxy and has been classed as a LINER with \Halpha\ emission (Ho \etal\ 1997). The full X-ray data analysis of this source from these six observations will be reported in a forthcoming paper. Apart from this central source, no LMXBs exceeding $1\times10^{39}$\ergps\ have been detected in this galaxy, unlike NGC 3379, where a ULX source with a co-added X-ray luminosity of $\sim2\times10^{39}$\ergps\ was discovered. This lack of ULXs in NGC 4278 is unsurprising as it is rare to find such a high luminosity object (external to the nuclear region) in E and S0 galaxies (Irwin, Bregman \& Athey 2004).

\begin{figure}
\begin{centering}
  \begin{minipage}{0.68\linewidth}
  \centering
  
    \includegraphics[width=\linewidth]{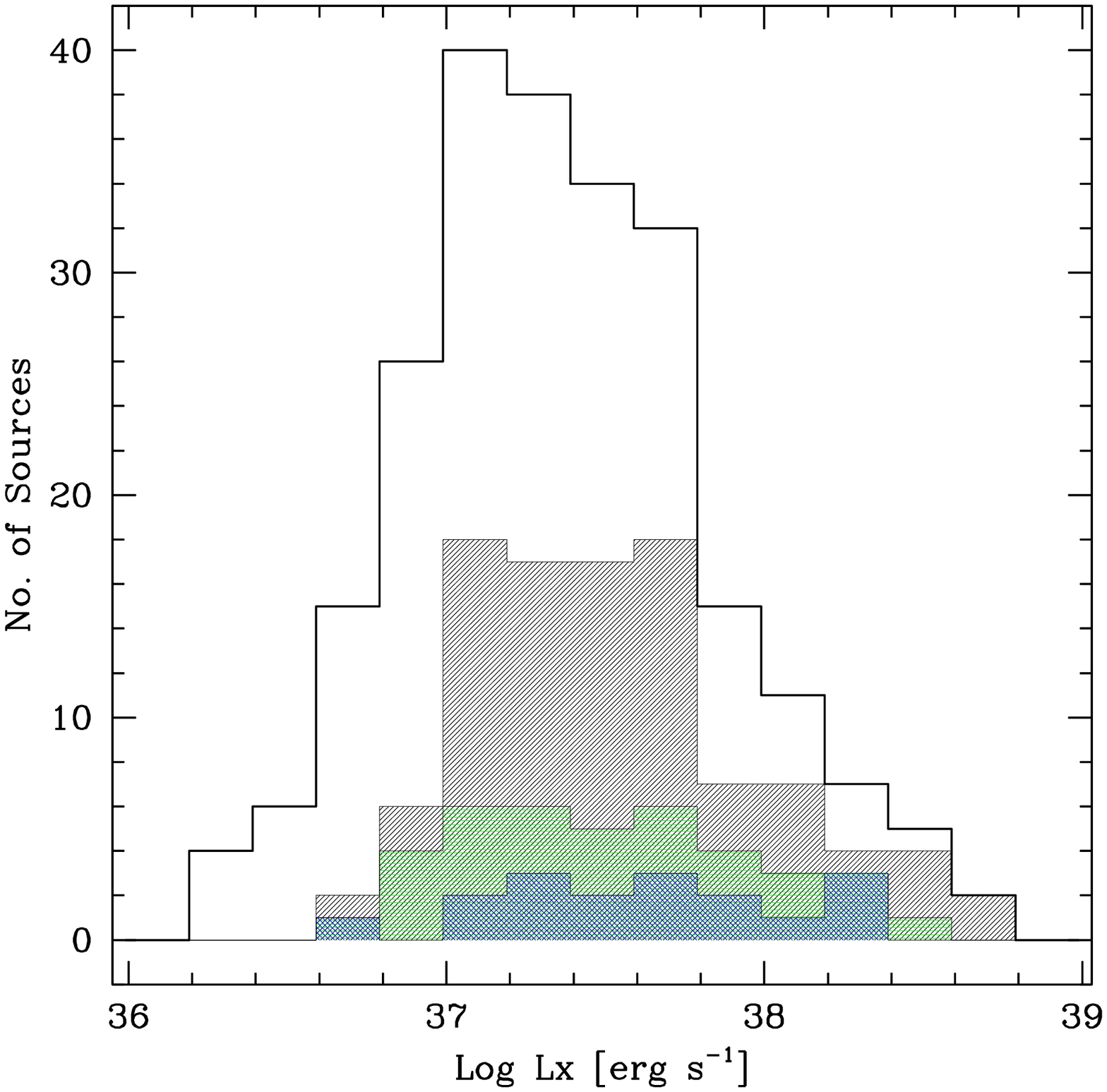}
  
  \end{minipage}
 
\begin{minipage}{0.68\linewidth}
  \centering

    \includegraphics[width=\linewidth]{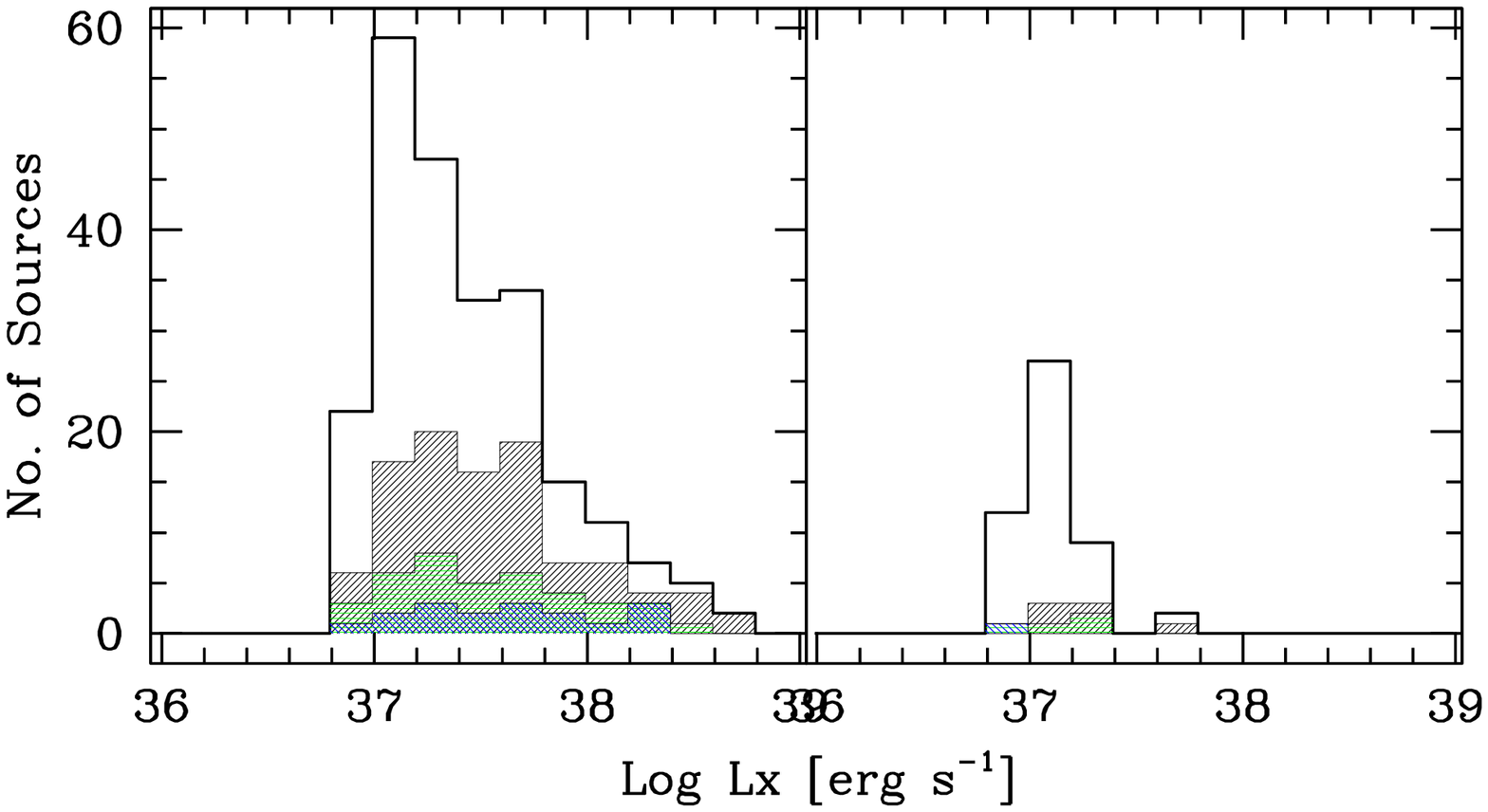}

 \end{minipage}

\caption{The top figure presents the \LX\ distribution of the 236 sources detected within the
overlapping region, covered by all six \CHANDRA\ pointings. The unshaded histogram
indicates all detected sources. The lightly shaded (gray)
region shows all variable sources (including both transient classes). The darker (green) histogram indicates sources
associated with a GC and the darkest (blue) histogram shows varying sources
that have a confirmed GC counterpart. The bottom left image indicates the same 236 sources, but for those with S/N$<$3, 3$\sigma$ upper limit values have been used in place of \LX. The bottom right image presents these upper limit values only. The shading for these two figures are the same as 
described for the main histogram.}\label{fig:lxhist}
\end{centering}
\end{figure}

The \LX\ distribution of  all of the detected X-ray sources within NGC 4278 is shown in Figure \ref{fig:lxhist}, where the GC associations are also indicated. In this figure the main histogram presents the calculated \LX\ values from all sources (with 1$\sigma$ upper limits from the co-added observation provided for sources only detected in a single observation), except for the central bright source 117. The bottom left histogram presents these same sources, but for those with S/N$<$3, 3$\sigma$ upper limits are shown, these upper limit values are then presented separately in the bottom right histogram.
From the main figure it can be seen that the majority of
sources detected from this observation lie in the luminosity range of
$1\times10^{37}$\ergps$-6\times10^{37}$\ergps, with a mode luminosity
of $\sim1.5\times10^{37}$\ergps\ and with source
incompleteness beginning to affect the source distribution \LX$\le1\times10^{37}$\ergps. From the histogram including 3$\sigma$ upper limit values, this mode value remains the same, as does source incompleteness.

In the forthcoming
paper Kim \etal\ (2009, in prep), the X-ray luminosity function (XLF)
of NGC 4278 will be investigated, and a correction to allow for
source incompleteness will be applied. Some preliminary results, investigating
the XLF of NGC 4278, alongside NGC 3379, have been reported in Kim
\etal\ (2006), where sources, down to a 90\% completeness limit of
\LX$\sim$3$\times10^{37}$\ergps, from the first two
observations, have been detected. From the even greater sensitivity
afforded to us by combining the six separate pointings, we can
investigate the XLF close to the X-ray luminosity range of normal Galactic
LMXBs. Previously, this has only been possible for the nearby radio galaxy
Centaurus A (NGC 5128), where the XLF has been measured down to
$\sim2\times10^{36}$\ergps\ (Kraft \etal\ 2001; Voss \& Gilfanov
2006). With our greater sensitivity we can compare our results to
these studies, allowing us to investigate the shape of the low
luminosity LMXB XLF, although it should be noted that both NGC 3379 and NGC 4278 are much more `normal' galaxy than Centaurus A.

In addition to the X-ray point sources that have been presented in
this catalog, the optical sources within NGC 4278 have also been
identified. These were detected in a WFPC2 {\em HST} observation,
where 266 confirmed globular clusters have been identified. From these
266 sources, 39 GC-LMXB, with separations $<$ 0.6\arcsec, have been
detected. From Figure \ref{fig:lxhist} it can bee seen that these 39 GC-LMXB sources appear to predominantly lie at the high X-ray luminosity end of this distribution. The luminosity function of these GC-LMXB sources from both galaxies and its implications for the
understanding of LMXB evolution will also be presented in Kim \etal\ (2009, in prep), where the X-ray luminosity functions of both GC and field LMXBs will be compared.

\subsection{X-ray Colors}
\label{sec:xcol}

\begin{figure}

\begin{centering}
  \begin{minipage}{0.58\linewidth}
  \vspace{-0.5cm}
\includegraphics[width=\linewidth]{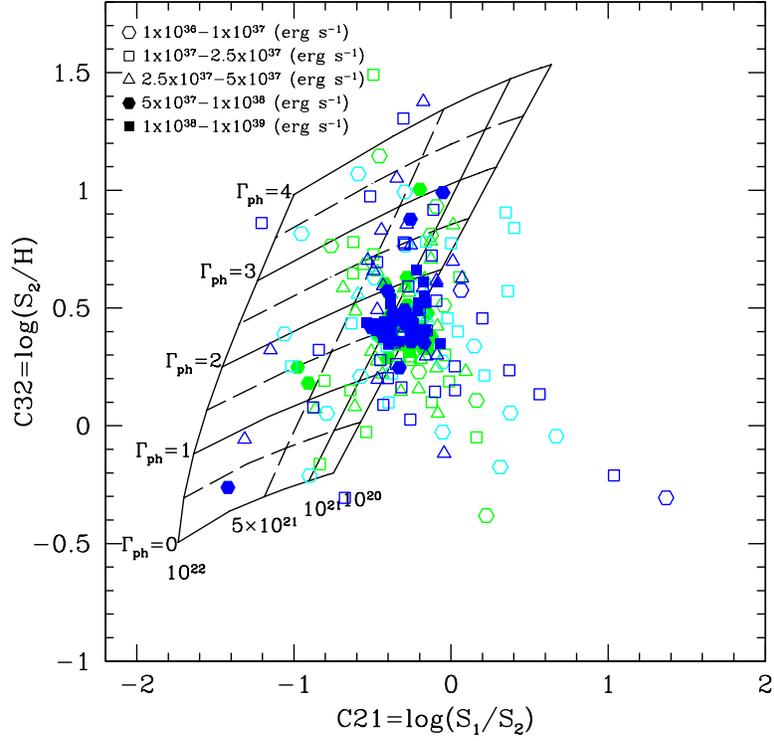}

\end{minipage}

  \begin{minipage}{0.58\linewidth}

\includegraphics[width=\linewidth]{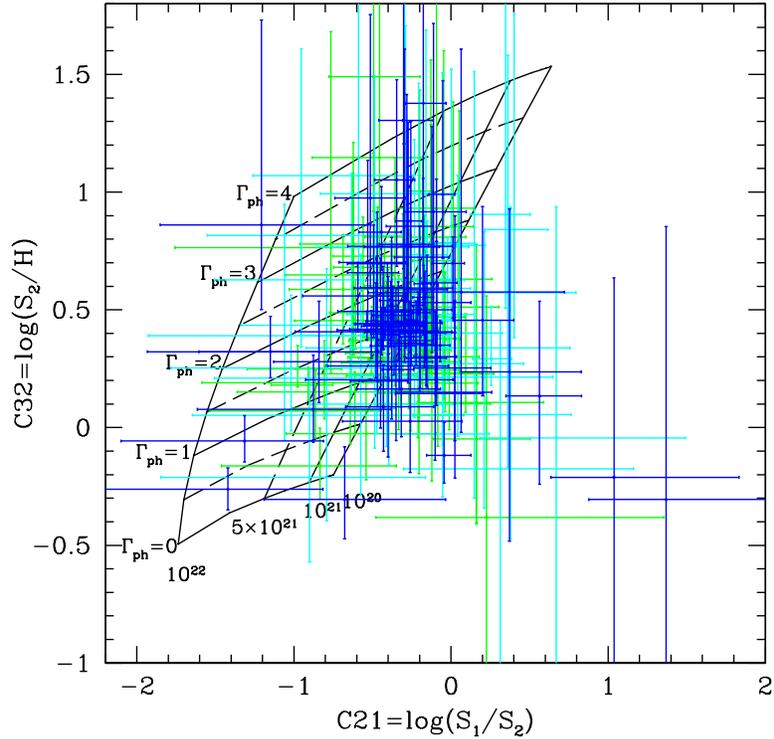}

\end{minipage}
\caption{The color-color diagram of the X-ray point sources detected
in the co-added observation. In the top panel color-color values are
plotted, with the sources divided into
luminosity bins, with symbols of each bin indicated by the labeling in
the panel. Variability is also indicated, where variable sources are
shown in blue, non-variable source are indicated in green and source that have insufficient counts to identify variability are shown in cyan. In the
lower panel the error values for each of the sources are presented. In
both of the panels, the grid indicates the predicted locations of the sources at
redshift $z$=0 with various photon indices (0$\le\Gamma_{ph}\le4$,
from top to bottom.) and absorption column densities (10$^{20}\le
$\NH\ $\le10^{22}$ \pcmsq, from right to left).}
  \label{fig:cc_pop}
\end{centering}
\end{figure}

\begin{figure}

\begin{centering}
  \begin{minipage}{0.58\linewidth}
  \vspace{-0.5cm}
\includegraphics[width=\linewidth]{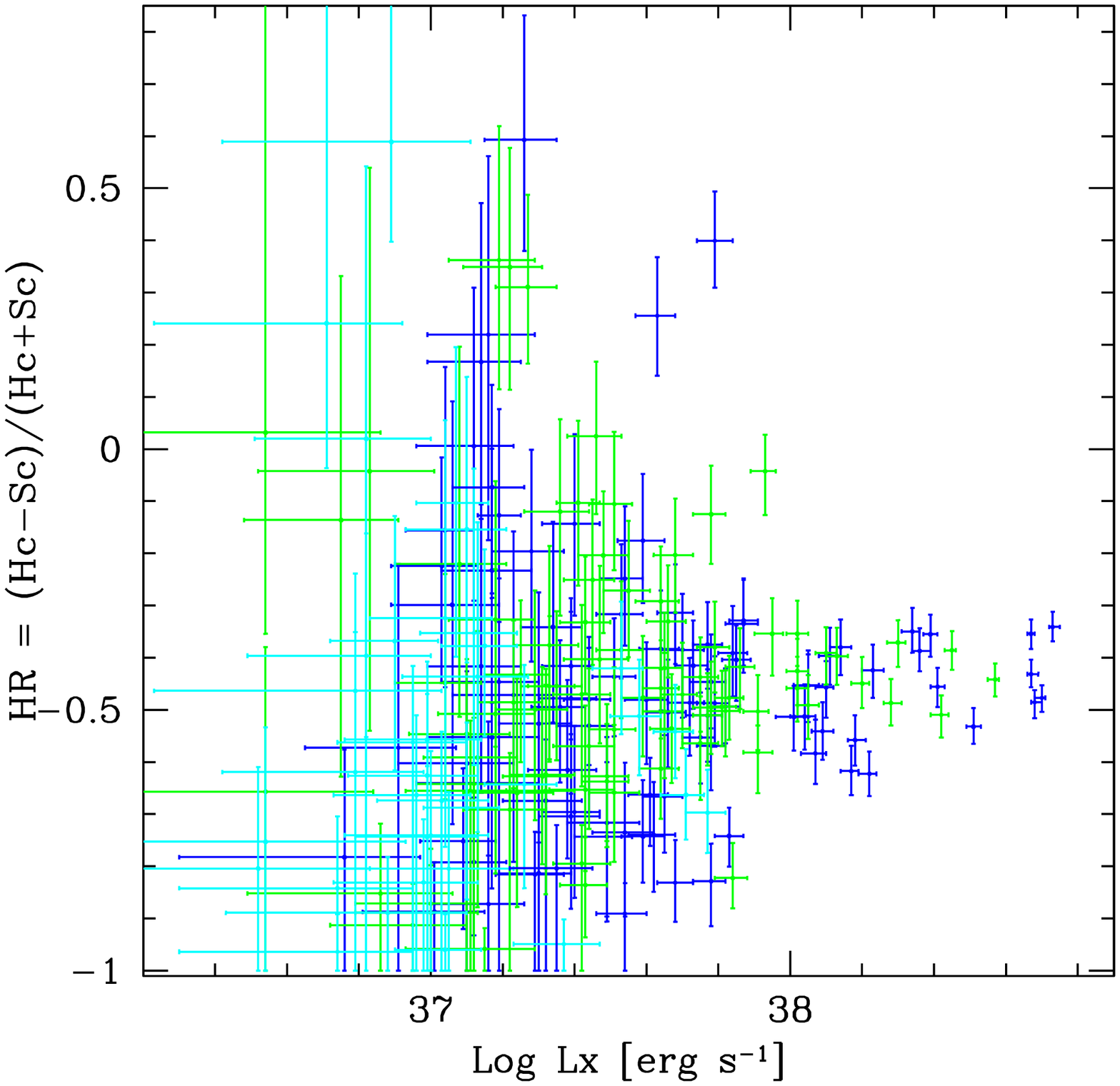}

\end{minipage}

  \begin{minipage}{0.58\linewidth}
\hspace{0.05cm}
\includegraphics[width=\linewidth]{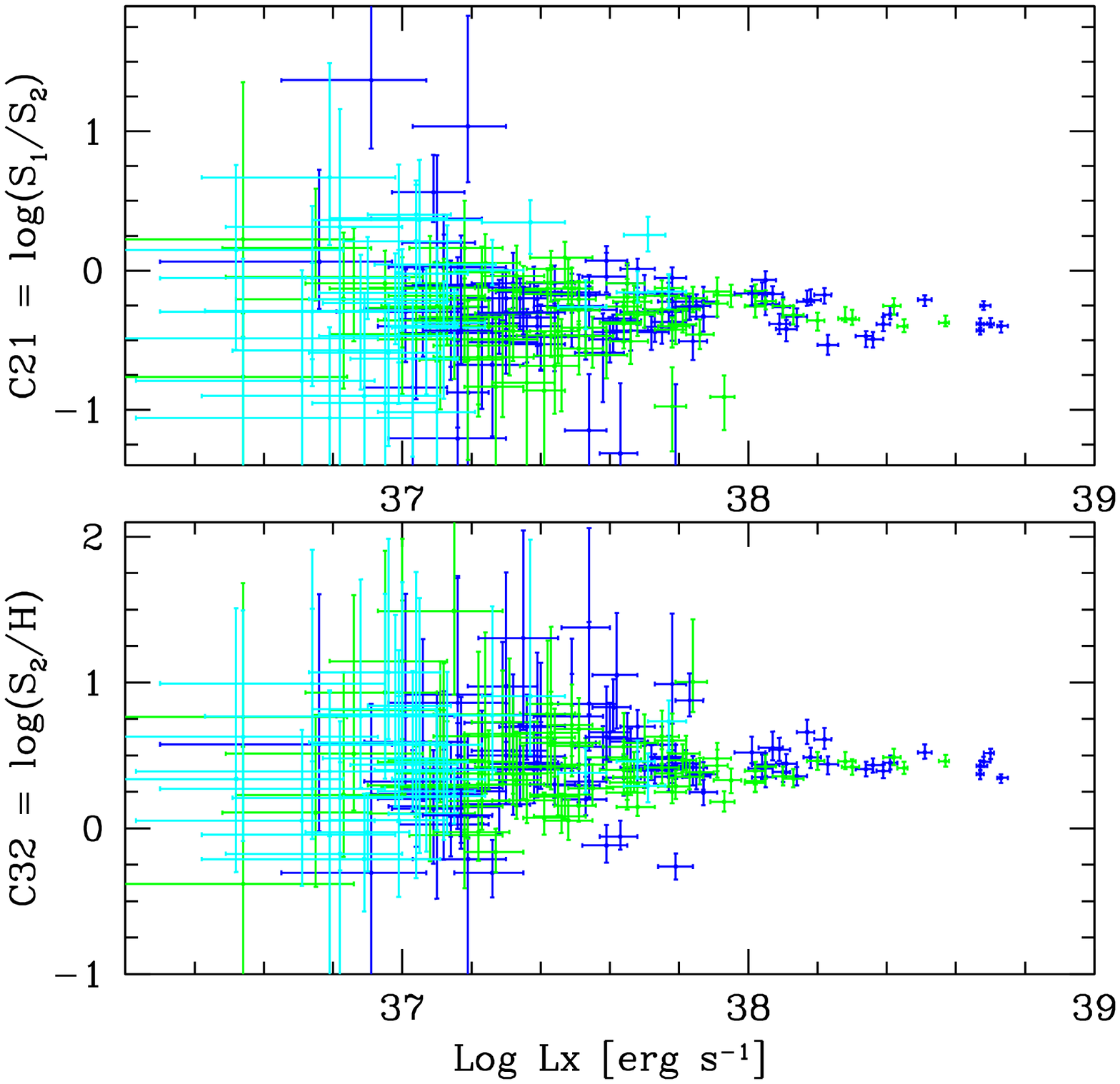}

\end{minipage}
\caption{The top panel presents the \LX-HR diagram of the X-ray point sources detected
in the co-added observation. The second panel shows the \LX-C21 plot
for this population and the bottom panel shows the \LX-C32 values. In
all three panels the variable sources are plotted in blue, non-variable sources in green and the sources without determined variability are plotted in cyan.}
  \label{fig:lxhr_pop}
\end{centering}
\end{figure}

In Figure \ref{fig:cc_pop} the LMXB population color-color diagram, based on the
photometry of the co-added observation, is presented. In the top panel
color-color values are plotted, with the sources divided into
luminosity bins, with symbols of each bin indicated by the labeling in
the panel. In the bottom panel, the errorbars for each of these points
are plotted. Also in this figure source variability is indicated,
where variable sources are plotted in blue, non-variable sources
are shown in green and sources with undetermined variability are indicated in cyan. Additionally, in both of the panels a grid has
been overlaid to indicate the predicted locations of the sources at 
redshift $z$=0 for different spectra, described by a power law with
various photon indices (0$\le\Gamma_{ph}\le4$, 
from top to bottom.) and absorption column densities (10$^{20}\le
$\NH\ $\le10^{22}$ \pcmsq, from right to left).
In Figure \ref{fig:lxhr_pop} the \LX-HR, \LX-C21 and \LX-C32
population plots are presented, where variability is again indicated
by color, with variable sources shown in blue, non-variable
sources are plotted in green and sources with undetermined variability are shown in cyan.

From the color-color diagram, presented in Figure \ref{fig:cc_pop}, it
can be seen that most of the well defined colors lie within 
the area of a typical LMXB spectrum of $\Gamma=1.5-2.0$, with no
intrinsic absorption (e.g. Irwin, Athey \& Bregman 2003; Fabbiano
2006). However, there also appears to be a population of sources that
have much harder spectra, again with either no intrinsic absorption,
or sources with a possible soft excess, albeit
with colors that are not as well defined. The \LX-HR distribution of this subpopulation was explored to identify sources with higher hardness ratios than one would expect from LMXBs but it was found that the sources that were identified in the color-color diagram did not exhibit harder HR values (see the top panel of Figure \ref{fig:lxhr_pop}). Because of this, while it is possible that these sources could be absorbed background AGN, the poorly constrained colors of this subpopulation alone do not provide adequate information to confirm this interpretation. The overall color-color and \LX-HR distributions of the LMXB population within NGC 4278 is very similar to that of NGC 3379, where again, most of the sources with well constrained colors were determined to lie within the area of a typical LMXB spectrum. Also in NGC 3379, like NGC 4278, a subpopulation of sources with harder spectra with either no intrinsic absorption, or a soft excess, was identified. However, unlike the subpopulation of NGC 4278, these sources were determined to show not only color-color values indicating their harder spectra, but also exhibited higher HR values and for this reason were flagged as possible absorbed background AGN.

\subsection{Source Variability}

A characteristic of compact accretion sources such as LMXBs is
variability, and, as a result of the monitoring nature of the
observing campaign, we have been able to search for this variability,
in both the long-term regime, and also over short-term baselines,
where changes over hours and days have been identified. One of the
specific aims of our monitoring campaign has been to identify transient
candidate sources as it has been suggested that field LMXBs are expected
to be transients (Piro \& Bildsten 2002; King 2002) and low luminosity
ultracompact binaries in GCs are also expected to be transient in
nature (Bildsten \& Deloye 2004). In the forthcoming paper Brassington
\etal\ (2009, in prep) we
investigate the subpopulation of transient candidates that has been discovered
in both NGC 4278 and NGC 3379.

Our data represent the most complete variability study for an
extragalactic LMXB population (see Fabbiano 2006; Xu \etal\ 2005),
investigating both long and short term behaviour. In the case of the
long-term variability, sources have been separated into four different
classifications; non-variable and variable sources, and also transient
candidates (TC)
or possible transients candidates (PTC). These two latter definitions have
been applied to sources that either appear or disappear, or are only
detected for a limited amount of `contiguous' time during the
observations, with a lower bound ratio of greater than 10 between the
`on-state' and the `off-state', for TCs, or a lower bound ratio
between 5 and 10 for the PTCs (see \S\ref{sec:var} for a full discussion of this definition). 

The 13 sources that were investigated for transient behavior are presented in Table \ref{tab:trans}, where both the ratio and lower bound ratio, calculated from Bayesian modeling, are presented, along with each source's variability classification. From this table is can be seen that many of these sources appear to be PTCs and TCs from their ratio alone, but when allowing for the uncertainties from their source and background counts, they can only be classified as variable sources. Including the uncertainties when determining TCs is particularly important when dealing with sources with low S/N values, as is the case for a number of sources in this catalog.

Out of the 236
sources, 97, 41\% of the sources within NGC 4278, have been defined as
variable sources. A further 3 sources are TCs, and 3 are
PTCs, with 97 sources found to be
non-varying in intensity over the six observations. The remaining 36
sources have insufficient data to investigate their long-term variability. 
These levels of long-term variability observed in NGC 4278 are very similar to those of NGC 3379, where 42\% of the sources were determined to be variable, with a further 5 sources identified as TCs and 3 as PTCs. This similar number of TCs/PTCs in both galaxies is maybe somewhat surprising, given the larger number of LMXBs detected within NGC 4278. However, from theory it has been suggested that transient sources will predominantly be field LMXBs (Piro \& Bildsten 2002) and in fact, in NGC 4278 five of the six TC/PTC sources have been determined to be field sources and in NGC 3379 three of the four TC/PTCs that we have optical coverage for are also field sources (the remaining 4 TC/PTCs are external to the {\em HST} FOV). Both NGC 3379 and NGC 4278 were selected for this study as they have very similar \LB\ values (1.35$\times10^{10}$\Lsol, 1.63$\times10^{10}$\Lsol), and by comparing the number of field TCs/PTCs in each galaxy to their field population we find similar values of 11\% and 9\% respectively, compared to 8\% and 3\% when including the whole X-ray source population within the $D_{25}$ ellipse. These values are consistent with the suggestion that transient sources will be predominantly found in the field, and this result will be further explored in the forthcoming paper Brassington \etal\ (2009, in prep).

The long and short-term variability of the X-ray sources with NGC 4278, are summarized
in Table \ref{tab:varsum}, where these two variability parameters have
been cross-correlated, to indicate the number of sources exhibiting
both long and short-term variations, although,
the majority of these sources do not have sufficient counts in each
observation to determine their short-term variability. The numbers
within this table indicate the number of sources from the whole observation
and the numbers in brackets represent the sources within the $D_{25}$
ellipse. 
From this table it can be seen that, for the sources with
a defined short-term variability measure, both long-term variable and non-variable
sources have a variety of short-term behavior. For the transient candidates no observations had sufficient counts with which to investigate their short-term variability. Also, as an additional point,
all of the TCs, and PTCs found within NGC 4278 reside well within
the $D_{25}$ ellipse of the galaxy, indicating that they are likely
LMXBs associated with NGC 4278.

In addition to the \LX\ variability, spectral variations have also
been investigated. These are presented in Figures \ref{fig:lxhrindiv}
and \ref{fig:CCindiv}, where \LX-HR and color-color plots for each
source are shown. From these figures it is clear that the majority of
sources within NGC 4278 exhibit some sort of variability, with a variety of different
spectral variations shown within this population. 
This spectral
variability is similar to behaviour
discussed in McClintock \& Remillard (2006), with a 
significant number of sources emitting softer spectra as \LX\
increases (e.g sources 76 and 147). It has been suggested that this increasing softness represents a thermal state, where the flux is dominated by the heat radiation from the inner accretion disk. Conversely, sources exhibiting spectral hardening with increasing \LX\ are also present within the galaxy (e.g sources 14 and 119). This harder state has been explained as a steep power law state, where a highly energized corona is present (e.g. Feng \& Kaaret 2006). In addition to these spectral behaviors, sources that show little to no spectral variation with increasing luminosity (e.g. sources 40 and 169), and sources that show no discernible pattern at all (e.g. sources 118
and 158) have also been detected. A more detailed discussion of the spectral variability of all of these X-ray sources presented in this catalog will be the subject of a
forthcoming paper.

\section{Conclusions}

We have presented a source catalog and variability atlas resulting from
our monitoring deep observations of the nearby elliptical NGC 4278
with \CHANDRA\ ACIS-S. Our results can be summarized as follows:

\begin{itemize}

\item{236 X-ray point sources have been detected within NGC 4278,
ranging in luminosity from 3.5$\times~10^{36}$~\ergps\ (with 3$\sigma$ upper limit $\le$1$\times 10^{37}$ \ergps) to $\sim2~\times 10^{40}$\ergps, with 180 of these sources residing within the $D_{25}$
ellipse of the galaxy.}

\item{The nuclear source in this galaxy has been classified as a LINER with \Halpha\ emission (Ho \etal\ 1997) and has a luminosity of $\sim2\times10^{40}$\ergps\ from the co-added observation.}

\item{39 globular clusters have been identified to be coincident with
X-ray sources, all of which lie within the $D_{25}$ ellipse of
the galaxy. These GC-LMXB associations tend to have high X-ray
luminosities, with ten of these sources exhibiting \LX$>~1\times10^{38}$\ergps.}

\item{From source photometry, it has been determined that the majority
of source with well constrained colors have values that are consistent
with a typical LMXB spectrum of $\Gamma=1.5-2.0$, with no intrinsic absorption.}

\item{103 sources, 44\% of the X-ray source population, have been
found to exhibit some type of long-term variability, which clearly
identifies them as accreting compact objects. 3 of these variable
sources have been identified as transient candidates, with a further 3
identified as possible transients.}

\item{Spectral variability analysis has revealed that the sources within NGC
4278 exhibit a range of variability patterns, where both
high/soft$-$low/hard and low/soft$-$high/hard spectral transitions
have been observed, as well as sources that vary in luminosity, but
exhibit no spectral variation, indicating that there are many different
source classes within this galaxy.}

\end{itemize}

Following this catalog paper we will be presenting highlights from the X-ray binary population of this galaxy in forthcoming papers. The first will compare the X-ray luminosity function of GC and field LMXBs detected in both NGC 4278 and NGC 3379 (Kim \etal\ 2009, in prep). This will be followed by a paper presenting the properties of the transient population of both galaxies (Brassington \etal\ 2009, in prep). 
In addition to these we will also present papers on: the properties of the nuclear source and the diffuse emission of NGC 4278, as well
as the nonuniform distribution of the sources within the galaxy and the intensity and spectral variability of the luminous X-ray binary population.

\acknowledgments

 We thank the CXC DS and SDS teams for their efforts in reducing the data and 
developing the software used for the reduction (SDP) and analysis
(CIAO). We would also like to thank the referee, Jimmy Irwin, for his helpful
comments which have improved this paper. 
This work was supported by {\em Chandra} G0 grant G06-7079A
(PI:Fabbiano) and subcontract G06-7079B (PI:Kalogera). We acknowledge
partial support from NASA contract NAS8-39073(CXC). A. Zezas
acknowledges support from NASA LTSA grant NAG5-13056. S. Pellegrini
acknowledges partial financial support from the Italian Space Agency
ASI (Agenzia Spaziale Italiana) through grant ASI-INAF I/023/05/0.


{}

\clearpage

\LongTables
\begin{deluxetable}{ccccc}
\tabletypesize{\scriptsize}
\tablecolumns{5}
\tablewidth{0pt}
\tablecaption{Observation Log\label{tab:log}}
\tablehead{\colhead{Obs. Num.} & \colhead{OBSID} & \colhead{Date} &
\colhead{Exposure (sec)} & \colhead{Cleaned Exposure (sec)}  }
\startdata
1  & 4741 & 2005-02-03  &    37462.0  &  37264.5 \\
2  & 7077 & 2006-03-16  &   110303.8  & 107736.7 \\
3  & 7078 & 2006-07-25  &    51433.2  &  48076.2 \\
4  & 7079 & 2006-10-24  &   105071.7  & 102504.6 \\
5  & 7081 & 2007-02-20  &   110724.0  & 107564.5 \\
6  & 7080 & 2007-04-20  &    55824.8  &  54837.5 \\

Total & - & -           &   470819.5  & 457984.0 \\
\enddata
\tablecomments{The pointing OBSID 7181 was taken before OBSID 7080, so to maintain the time sequence of the exposures these observation numbers have been labeled as above in this paper.}
\end{deluxetable}


\begin{deluxetable}{lc}
\tabletypesize{\scriptsize}
\tablecolumns{2}
\tablewidth{0pt}
\tablecaption{Definition of Energy Bands and X-ray Colors \label{tab:bands}}
\tablehead{\colhead{Band} & \colhead{Definition} }
\startdata
Broad (B)	& 0.3$-$8 keV \\
Soft (S)	& 0.3$-$2.5 keV \\
Hard (H)	& 2.5$-$8 keV \\
Soft 1 (S$_1$)	& 0.3$-$0.9 keV \\
Soft 2 (S$_2$)	& 0.9$-$2.5 keV \\
\hline
Conventional Broad (Bc)	& 0.5$-$8 keV \\
Conventional Soft (Sc) & 0.5$-$2 keV \\
Conventional Hard (Hc) & 2$-$8 keV \\
\hline
Hardness Ratio $HR$	& (Hc$-$Sc)/(Hc+Sc) \\
X-ray Color C21 & $-log($S$_2)+log($S$_1)=log($S$_1/$S$_2$) \\
X-ray Color C32 & $-log($H$)+log($S$_2)=log($S$_2/$H) \\
\enddata
\end{deluxetable}

\clearpage
\begin{landscape}
\begin{deluxetable}{lcccccc@{}c@{}cccccc}
\tabletypesize{\scriptsize}
\tablecolumns{14}
\tablewidth{0pt}
\tablecaption{Master Source List \label{tab:Mainprops} }
\tablehead{  \colhead{Masterid}   & \colhead{CXOU Name} &
\colhead{RA}  & \colhead{Dec}  & \colhead{Radius} &
\colhead{PU} & \colhead{S/N}  & \colhead{Log \LX (0.3$-$8.0 keV)} &
\multicolumn{3}{c}{Variability} & \colhead{Opt Corr}  & \colhead{DG} &
\colhead{Flag} \\
 \colhead{}   &
\colhead{}  & \colhead{(J2000)} & \colhead{(J2000)}  & \colhead{(\arcs)} &
\colhead{(\arcs)} & \colhead{}  & \colhead{(\ergps)} &
\colhead{LT} & \colhead{BB} & \colhead{K-S} & \colhead{}  & \colhead{(\arcs)} &
\colhead{} \\
 \colhead{(1)}   &
\colhead{(2)}  & \colhead{(3)} & \colhead{(4)}  & \colhead{(5)} &
\colhead{(6)} & \colhead{(7)}  & \colhead{(8)} &
\colhead{(9)} & \colhead{(10)}  & \colhead{(11)} &
\colhead{(12)} & \colhead{(13)}  & \colhead{(14)} }
\startdata
1 & J122023.6+291614 & 12:20:23.640 & +29:16:13.92 & 4.13 & 0.39 & 9.2 & 37.8  & V & - & - & - & 223.1 & - 	 \\
2 & J122023.0+291453 & 12:20:23.021 & +29:14:52.76 & 4.57 & 0.34 & 15.8 & 38.2 & V & - & P & - & 242.6 & O1 	 \\
4 & J122022.6+291458 & 12:20:22.570 & +29:14:58.08 & 4.36 & 0.36 & 12.6 & 38.1 & V & - & P & - & 234.8 & O1 	 \\
5 & J122022.6+291536 & 12:20:22.558 & +29:15:35.89 & 4.09 & 0.27 & 24.9 & 38.5 & V & N & V & - & 219.1 & - 	 \\
8 & J122019.4+291448 & 12:20:19.435 & +29:14:47.98 & 3.87 & 0.45 & 8.8 & 37.8  & N & - & - & - & 205.7 & - 	 \\
9 & J122019.3+291412 & 12:20:19.262 & +29:14:12.11 & 4.17 & 0.87 & 4.0 & 37.3  & V & - & - & - & 227.3 & - 	 \\
11 & J122018.0+291655 & 12:20:17.988 & +29:16:54.97 & 3.00 & 0.45 & 6.1 & 37.5 & N & - & - & - & 146.1 & - 	 \\
13 & J122017.8+291422 & 12:20:17.753 & +29:14:22.22 & 3.84 & 0.24 & 31.6 & 38.7 & V & N & N & - & 206.2 & - 	 \\
\enddata
\end{deluxetable}
\clearpage
\end{landscape}

\clearpage
\begin{landscape}
\begin{deluxetable}{l@{}cccccccccc@{}c@{}}
\tabletypesize{\scriptsize}
\tablecolumns{12}
\tablewidth{0pt}
\tablecaption{Source counts, hardness ratios and color-color values:
Coadded Observation \label{tab:counts_main} }
\tablehead{  \colhead{MID}   & \multicolumn{7}{c}{Net Counts} &
\colhead{HR} & \colhead{C21}  & \colhead{C32} &
\colhead{Log \LX} \\
\colhead{}   & \colhead{B-band} &
\colhead{S1-band}  & \colhead{S2-band}  & \colhead{S-band} &
\colhead{H-band} & \colhead{Sc-band}  & \colhead{Hc-band} &
\colhead{} & \colhead{}  & \colhead{} &
\colhead{(0.3$-$8.0 keV)} \\
 \colhead{(1)}   &
\colhead{(2)}  & \colhead{(3)} & \colhead{(4)}  & \colhead{(5)} &
\colhead{(6)} & \colhead{(7)}  & \colhead{(8)} &
\colhead{(9)} & \colhead{(10)}  & \colhead{(11)} &
\colhead{(12)} }
\startdata
1 &139.8$\pm$15.2 &-2.3$\pm$4.7 &48.0$\pm$8.9 &45.6$\pm$9.7 &94.2$\pm$12.2 &39.1$\pm$8.4 &103.4$\pm$12.7 &0.40	$_ { -0.09}^ { +0.09}$	&	-1.42$_ { -0.80}^ { +0.60}$	&	-0.26$_ { -0.09}^ { +0.09}$	&	37.8\\																	
2 &386.2$\pm$24.4 &57.5$\pm$9.6 &237.4$\pm$17.0 &294.9$\pm$19.1 &91.8$\pm$12.4 &258.6$\pm$17.7 &117.5$\pm$13.7 &-0.42	$_ { -0.06}^ { +0.04}$	&	-0.54$_ { -0.07}^ { +0.08}$	&	0.44$_ { -0.07}^ { +0.05}$	&	38.2\\																	
3 &7.8$\pm$11.2 &-0.8$\pm$5.0 &9.5$\pm$6.1 &8.7$\pm$7.4 &-1.6$\pm$6.5 &10.3$\pm$6.5 &-2.3$\pm$6.7 &-0.75	$_ { -0.25}^ { +0.22}$	&	-0.49$_ { -1.02}^ { +0.57}$	&	0.63$_ { -0.72}^ { +0.86}$	&	36.5\\																	
4 &266.1$\pm$21.1 &78.8$\pm$10.7 &145.3$\pm$13.8 &224.1$\pm$17.0 &42.2$\pm$9.8 &195.7$\pm$15.7 &58.1$\pm$10.9 &-0.58	$_ { -0.06}^ { +0.06}$	&	-0.17$_ { -0.07}^ { +0.06}$	&	0.55$_ { -0.08}^ { +0.11}$	&	38.1\\																	
5 &725.7$\pm$29.1 &202.2$\pm$15.8 &398.3$\pm$21.3 &600.5$\pm$26.1 &125.2$\pm$13.6 &508.6$\pm$24.0 &174.7$\pm$15.5 &-0.53	$_ { -0.03}^ { +0.03}$	&	-0.21$_ { -0.04}^ { +0.03}$	&	0.52$_ { -0.04}^ { +0.06}$	&	38.5\\																	
6 &17.5$\pm$11.7 &-6.7$\pm$4.1 &9.1$\pm$5.9 &2.5$\pm$6.7 &14.4$\pm$7.6 &4.7$\pm$5.8 &17.9$\pm$8.1 &0.59	$_ { -0.19}^ { +0.41}$	&	-0.90$_ { -0.95}^ { +0.74}$	&	-0.21$_ { -0.36}^ { +0.45}$	&	36.9\\																	
7 &$\le$10.8 &- &- &- &- &- &- &-		&	-	&	-	&	$\le$36.7\\																	
8 &128.6$\pm$14.5 &32.9$\pm$7.7 &76.6$\pm$10.5 &109.6$\pm$12.6 &19.0$\pm$7.9 &93.8$\pm$11.4 &30.1$\pm$8.9 &-0.56	$_ { -0.11}^ { +0.09}$	&	-0.28$_ { -0.10}^ { +0.11}$	&	0.63$_ { -0.19}^ { +0.18}$	&	37.7\\
\enddata
\tablecomments{Col. (1): Master ID, cols. (2)$-$(8): net counts, in each of
the 7 energy bands (see Table \ref{tab:bands} for definitions of these
bands), col. (9): hardness ratio, cols. (10) and (11) color values,
col. (12): log \LX\ (0.3$-$8.0 keV). Upper limit \LX\ values are at
the 68\% confidence level. }
\end{deluxetable}
\clearpage
\end{landscape}

\clearpage
\begin{landscape}
\begin{deluxetable}{l@{}cccccc@{}cccc@{}c@{}c@{}c@{}c@{}}
\tabletypesize{\scriptsize}
\tablecolumns{15}
\tablewidth{0pt}
\tablecaption{Source counts, hardness ratios, color-color values and variability:
Observation 1 \label{tab:counts_ob1} }
\tablehead{  \colhead{Masterid}   & \multicolumn{7}{c}{Net Counts}
& \colhead{HR} & \colhead{C21}  & \colhead{C32} &
  \multicolumn{3}{c}{Variability} & \colhead{Log \LX } \\
\colhead{}   & \colhead{B-band} &
\colhead{S1-band}  & \colhead{S2-band}  & \colhead{S-band} &
\colhead{H-band} & \colhead{Sc-band}  & \colhead{Hc-band} &
\colhead{} & \colhead{}  & \colhead{} &
\colhead{BB} & \colhead{k-S} & \colhead{signif.} &\colhead{(0.3$-$8.0 keV)} \\
 \colhead{(1)}   &
\colhead{(2)}  & \colhead{(3)} & \colhead{(4)}  & \colhead{(5)} &
\colhead{(6)} & \colhead{(7)}  & \colhead{(8)} &
\colhead{(9)} & \colhead{(10)}  & \colhead{(11)} &
\colhead{(12)} & \colhead{(13)}  & \colhead{(14)}& \colhead{(15)} }
\startdata
1 &9.8$\pm$4.7 &-1.0$\pm$1.9 &1.1$\pm$2.7 &0.1$\pm$2.7 &9.7$\pm$4.4 &0.0$\pm$2.3 &10.5$\pm$4.6 &0.94	$_ { -0.08}^ { +0.06}$	&	-0.38$_ { -1.15}^ { +0.91}$	&	-0.72$_ { -0.67}^ { +0.43}$	&	-	&	-	&	-	&	37.7\\														
2 &11.3$\pm$5.0 &3.0$\pm$3.2 &5.1$\pm$3.6 &8.0$\pm$4.3 &3.3$\pm$3.4 &7.9$\pm$4.1 &3.0$\pm$3.4 &-0.58	$_ { -0.28}^ { +0.24}$	&	-0.08$_ { -0.43}^ { +0.35}$	&	0.19$_ { -0.38}^ { +0.45}$	&	-	&	-	&	-	&	37.8\\														
3 &$\le$3.9 &- &- &- &- &- &- &-		&	-	&	-	&	-	&	-	&	-	&	$\le$37.3\\														
4 &15.5$\pm$5.5 &6.2$\pm$3.8 &9.1$\pm$4.3 &15.3$\pm$5.2 &0.3$\pm$2.7 &12.9$\pm$4.8 &2.1$\pm$3.2 &-0.81	$_ { -0.19}^ { +0.11}$	&	-0.05$_ { -0.24}^ { +0.24}$	&	0.84$_ { -0.46}^ { +0.84}$	&	-	&	-	&	-	&	37.9\\														
5 &64.3$\pm$9.3 &19.8$\pm$5.7 &35.3$\pm$7.1 &55.1$\pm$8.6 &9.1$\pm$4.4 &49.9$\pm$8.2 &13.0$\pm$5.0 &-0.63	$_ { -0.10}^ { +0.09}$	&	-0.10$_ { -0.16}^ { +0.08}$	&	0.59$_ { -0.17}^ { +0.18}$	&	N	&	N	&	-	&	38.6\\														
6 &$\le$2.0 &- &- &- &- &- &- &-		&	-	&	-	&	-	&	-	&	-	&	$\le$37.0\\														
7 &$\le$0.6 &- &- &- &- &- &- &-		&	-	&	-	&	-	&	-	&	-	&	$\le$36.5\\														
8 &12.6$\pm$5.0 &2.2$\pm$2.9 &6.3$\pm$3.8 &8.6$\pm$4.3 &4.0$\pm$3.4 &5.3$\pm$3.6 &6.8$\pm$4.0 &0.06	$_ { -0.30}^ { +0.30}$	&	-0.29$_ { -0.46}^ { +0.34}$	&	0.19$_ { -0.30}^ { +0.32}$	&	-	&	-	&	-	&	37.8\\
\enddata
\tablecomments{Col. (1): Master ID, cols. (2)$-$(8): net counts, in each of
the 7 energy bands (see Table \ref{tab:bands} for definitions of these
bands), col. (9): hardness ratio, cols. (10) and (11) color values, errors are given as 1$\sigma$,
cols. (12) and (13): short-term variability, where (BB) indicate
Bayesian block analysis and (K-S) indicates the Kolmogorov-Smirnov test, in both columns symbols
indicate - (N) non-variable in all observations, (V) variable in at
least one observation, (P) possible variability in at least one
observation, col. (14): the significance of the change in
\LX\ between the previous observation and the current observation
respectively (equation \ref{equ:sig}), col. (15): log \LX\ (0.3$-$8.0
keV).  Upper limit values of net B and \LX\ are at
the 68\% confidence level.}
\end{deluxetable}
\clearpage
\end{landscape}

\clearpage
\pagestyle{empty}
\clearpage
\begin{landscape}
\begin{deluxetable}{l@{}cccccc@{}cccc@{}c@{}c@{}c@{}c@{}}
\tabletypesize{\scriptsize}
\tablecolumns{15}
\tablewidth{0pt}
\tablecaption{Source counts, hardness ratios, color-color values and variability:
Observation 2 \label{tab:counts_ob2} }
\tablehead{  \colhead{Masterid}   & \multicolumn{7}{c}{Net Counts}
& \colhead{HR} & \colhead{C21}  & \colhead{C32} &
  \multicolumn{3}{c}{Variability} & \colhead{Log \LX } \\
\colhead{}   & \colhead{B-band} &
\colhead{S1-band}  & \colhead{S2-band}  & \colhead{S-band} &
\colhead{H-band} & \colhead{Sc-band}  & \colhead{Hc-band} &
\colhead{} & \colhead{}  & \colhead{} &
\colhead{BB} & \colhead{k-S} & \colhead{signif.} &\colhead{(0.3$-$8.0 keV)} \\
 \colhead{(1)}   &
\colhead{(2)}  & \colhead{(3)} & \colhead{(4)}  & \colhead{(5)} &
\colhead{(6)} & \colhead{(7)}  & \colhead{(8)} &
\colhead{(9)} & \colhead{(10)}  & \colhead{(11)} &
\colhead{(12)} & \colhead{(13)}  & \colhead{(14)}& \colhead{(15)} }
\startdata
1 &30.6$\pm$7.9 &1.3$\pm$3.4 &9.1$\pm$4.6 &10.4$\pm$5.2 &20.2$\pm$6.5 &9.0$\pm$4.6 &21.2$\pm$6.6 &0.35	$_ { -0.20}^ { +0.21}$	&	-0.48$_ { -0.83}^ { +0.42}$	&	-0.30$_ { -0.22}^ { +0.19}$	&	N	&	N	&	0.2	&	37.8\\														
2 &64.6$\pm$10.8 &4.3$\pm$4.0 &45.5$\pm$8.1 &49.8$\pm$8.6 &14.9$\pm$6.0 &42.9$\pm$7.9 &18.0$\pm$6.4 &-0.47	$_ { -0.12}^ { +0.15}$	&	-0.88$_ { -0.37}^ { +0.28}$	&	0.49$_ { -0.14}^ { +0.19}$	&	N	&	P	&	1.7	&	38.1\\														
3 &$\le$6.6 &- &- &- &- &- &- &-		&	-	&	-	&	-	&	-	&	0.0	&	$\le$37.1\\														
4 &71.0$\pm$11.2 &23.3$\pm$6.2 &35.4$\pm$7.3 &58.7$\pm$9.1 &12.1$\pm$5.7 &52.3$\pm$8.5 &14.2$\pm$6.0 &-0.62	$_ { -0.11}^ { +0.13}$	&	-0.06$_ { -0.15}^ { +0.09}$	&	0.43$_ { -0.13}^ { +0.25}$	&	N	&	P	&	1.3	&	38.1\\														
5 &170.3$\pm$14.7 &48.7$\pm$8.3 &93.7$\pm$10.9 &142.3$\pm$13.2 &28.0$\pm$7.1 &120.1$\pm$12.2 &41.2$\pm$8.1 &-0.54	$_ { -0.06}^ { +0.07}$	&	-0.17$_ { -0.10}^ { +0.06}$	&	0.55$_ { -0.11}^ { +0.10}$	&	N	&	N	&	0.4	&	38.5\\														
6 &$\le$1.8 &- &- &- &- &- &- &-		&	-	&	-	&	-	&	-	&	0.0	&	$\le$36.5\\														
7 &$\le$4.4 &- &- &- &- &- &- &-		&	-	&	-	&	-	&	-	&	0.0	&	$\le$36.9\\														
8 &32.2$\pm$7.7 &11.0$\pm$4.7 &23.0$\pm$6.2 &34.0$\pm$7.3 &-1.8$\pm$3.4 &28.9$\pm$6.7 &0.4$\pm$4.0 &-0.97	$_ { -0.03}^ { +0.06}$	&	-0.22$_ { -0.18}^ { +0.16}$	&	1.30$_ { -0.46}^ { +0.92}$	&	N	&	N	&	0.2	&	37.8\\
\enddata
\tablecomments{Col. (1): Master ID, cols. (2)$-$(8): net counts, in each of
the 7 energy bands (see Table \ref{tab:bands} for definitions of these
bands), col. (9): hardness ratio, cols. (10) and (11) color values, errors are given as 1$\sigma$,
cols. (12) and (13): short-term variability, where (BB) indicate
Bayesian block analysis and (K-S) indicates the Kolmogorov-Smirnov test, in both columns symbols
indicate - (N) non-variable in all observations, (V) variable in at
least one observation, (P) possible variability in at least one
observation, col. (14): the significance of the change in
\LX\ between the previous observation and the current observation
respectively (equation \ref{equ:sig}), col. (15): log \LX\ (0.3$-$8.0
keV).  Upper limit values of net B and \LX\ are at
the 68\% confidence level.}
\end{deluxetable}
\clearpage
\end{landscape}

\clearpage
\begin{landscape}
\begin{deluxetable}{l@{}cccccc@{}cccc@{}c@{}c@{}c@{}c@{}}
\tabletypesize{\scriptsize}
\tablecolumns{15}
\tablewidth{0pt}
\tablecaption{Source counts, hardness ratios, color-color values and variability:
Observation 3 \label{tab:counts_ob3} }
\tablehead{  \colhead{Masterid}   & \multicolumn{7}{c}{Net Counts}
& \colhead{HR} & \colhead{C21}  & \colhead{C32} &
  \multicolumn{3}{c}{Variability} & \colhead{Log \LX } \\
\colhead{}   & \colhead{B-band} &
\colhead{S1-band}  & \colhead{S2-band}  & \colhead{S-band} &
\colhead{H-band} & \colhead{Sc-band}  & \colhead{Hc-band} &
\colhead{} & \colhead{}  & \colhead{} &
\colhead{BB} & \colhead{k-S} & \colhead{signif.} &\colhead{(0.3$-$8.0 keV)} \\
 \colhead{(1)}   &
\colhead{(2)}  & \colhead{(3)} & \colhead{(4)}  & \colhead{(5)} &
\colhead{(6)} & \colhead{(7)}  & \colhead{(8)} &
\colhead{(9)} & \colhead{(10)}  & \colhead{(11)} &
\colhead{(12)} & \colhead{(13)}  & \colhead{(14)}& \colhead{(15)} }
\startdata
1 &22.5$\pm$6.2 &0.1$\pm$2.3 &11.1$\pm$4.6 &11.2$\pm$4.7 &11.3$\pm$4.7 &8.3$\pm$4.1 &14.8$\pm$5.2 &0.14	$_ { -0.22}^ { +0.22}$	&	-0.84$_ { -0.92}^ { +0.46}$	&	0.06$_ { -0.19}^ { +0.18}$	&	N	&	N	&	1.4	&	38.0\\														
2 &37.1$\pm$7.5 &3.3$\pm$3.2 &20.1$\pm$5.7 &23.4$\pm$6.1 &13.7$\pm$5.1 &19.9$\pm$5.7 &16.4$\pm$5.4 &-0.24	$_ { -0.17}^ { +0.16}$	&	-0.55$_ { -0.29}^ { +0.29}$	&	0.21$_ { -0.12}^ { +0.20}$	&	N	&	N	&	1.4	&	38.3\\														
3 &$\le$1.4 &- &- &- &- &- &- &-		&	-	&	-	&	-	&	-	&	0.0	&	$\le$36.8\\														
4 &36.6$\pm$7.5 &10.5$\pm$4.4 &21.9$\pm$5.9 &32.4$\pm$6.9 &4.2$\pm$3.8 &32.1$\pm$6.8 &3.7$\pm$3.8 &-0.87	$_ { -0.09}^ { +0.07}$	&	-0.11$_ { -0.17}^ { +0.15}$	&	0.72$_ { -0.26}^ { +0.35}$	&	N	&	N	&	1.0	&	38.2\\														
5 &64.5$\pm$9.3 &17.5$\pm$5.3 &34.9$\pm$7.1 &52.4$\pm$8.4 &12.1$\pm$4.8 &47.9$\pm$8.1 &12.7$\pm$5.0 &-0.66	$_ { -0.10}^ { +0.09}$	&	-0.09$_ { -0.13}^ { +0.13}$	&	0.51$_ { -0.15}^ { +0.16}$	&	N	&	N	&	0.6	&	38.5\\														
6 &$\le$2.1 &- &- &- &- &- &- &-		&	-	&	-	&	-	&	-	&	0.0	&	$\le$37.0\\														
7 &$\le$1.3 &- &- &- &- &- &- &-		&	-	&	-	&	-	&	-	&	0.0	&	$\le$36.8\\														
8 &12.4$\pm$5.0 &5.5$\pm$3.6 &4.3$\pm$3.4 &9.8$\pm$4.4 &2.6$\pm$3.2 &8.0$\pm$4.1 &3.5$\pm$3.4 &-0.49	$_ { -0.30}^ { +0.27}$	&	0.21$_ { -0.26}^ { +0.33}$	&	0.21$_ { -0.42}^ { +0.49}$	&	-	&	-	&	0.2	&	37.7\\
\enddata
\tablecomments{Col. (1): Master ID, cols. (2)$-$(8): net counts, in each of
the 7 energy bands (see Table \ref{tab:bands} for definitions of these
bands), col. (9): hardness ratio, cols. (10) and (11) color values, errors are given as 1$\sigma$,
cols. (12) and (13): short-term variability, where (BB) indicate
Bayesian block analysis and (K-S) indicates the Kolmogorov-Smirnov test, in both columns symbols
indicate - (N) non-variable in all observations, (V) variable in at
least one observation, (P) possible variability in at least one
observation, col. (14): the significance of the change in
\LX\ between the previous observation and the current observation
respectively (equation \ref{equ:sig}), col. (15): log \LX\ (0.3$-$8.0
keV).  Upper limit values of net B and \LX\ are at
the 68\% confidence level.}
\end{deluxetable}
\clearpage
\end{landscape}

\clearpage
\begin{landscape}
\begin{deluxetable}{l@{}cccccc@{}cccc@{}c@{}c@{}c@{}c@{}}
\tabletypesize{\scriptsize}
\tablecolumns{15}
\tablewidth{0pt}
\tablecaption{Source counts, hardness ratios, color-color values and variability:
Observation 4 \label{tab:counts_ob4} }
\tablehead{  \colhead{Masterid}   & \multicolumn{7}{c}{Net Counts}
& \colhead{HR} & \colhead{C21}  & \colhead{C32} &
  \multicolumn{3}{c}{Variability} & \colhead{Log \LX } \\
\colhead{}   & \colhead{B-band} &
\colhead{S1-band}  & \colhead{S2-band}  & \colhead{S-band} &
\colhead{H-band} & \colhead{Sc-band}  & \colhead{Hc-band} &
\colhead{} & \colhead{}  & \colhead{} &
\colhead{BB} & \colhead{k-S} & \colhead{signif.} &\colhead{(0.3$-$8.0 keV)} \\
 \colhead{(1)}   &
\colhead{(2)}  & \colhead{(3)} & \colhead{(4)}  & \colhead{(5)} &
\colhead{(6)} & \colhead{(7)}  & \colhead{(8)} &
\colhead{(9)} & \colhead{(10)}  & \colhead{(11)} &
\colhead{(12)} & \colhead{(13)}  & \colhead{(14)}& \colhead{(15)} }
\startdata
1 &22.1$\pm$6.1 &-0.8$\pm$1.9 &8.4$\pm$4.1 &7.6$\pm$4.1 &14.5$\pm$5.1 &5.2$\pm$3.6 &17.4$\pm$5.4 &0.50	$_ { -0.18}^ { +0.21}$	&	-1.15$_ { -0.99}^ { +0.54}$	&	-0.21$_ { -0.19}^ { +0.18}$	&	N	&	N	&	0.0	&	38.0\\														
2 &97.9$\pm$11.3 &11.4$\pm$4.7 &68.0$\pm$9.4 &79.4$\pm$10.2 &18.4$\pm$5.8 &69.9$\pm$9.5 &27.8$\pm$6.7 &-0.50	$_ { -0.09}^ { +0.08}$	&	-0.62$_ { -0.16}^ { +0.13}$	&	0.61$_ { -0.13}^ { +0.11}$	&	N	&	N	&	0.5	&	38.3\\														
3 &$\le$0.9 &- &- &- &- &- &- &-		&	-	&	-	&	-	&	-	&	0.0	&	$\le$36.3\\														
4 &45.5$\pm$8.4 &11.4$\pm$4.7 &24.8$\pm$6.3 &36.1$\pm$7.4 &9.4$\pm$4.7 &29.8$\pm$6.7 &13.8$\pm$5.3 &-0.44	$_ { -0.16}^ { +0.13}$	&	-0.20$_ { -0.17}^ { +0.16}$	&	0.43$_ { -0.16}^ { +0.21}$	&	N	&	N	&	2.0	&	38.0\\														
5 &131.6$\pm$12.8 &42.5$\pm$7.7 &63.8$\pm$9.2 &106.3$\pm$11.5 &25.4$\pm$6.5 &88.4$\pm$10.6 &31.0$\pm$7.0 &-0.53	$_ { -0.08}^ { +0.07}$	&	-0.06$_ { -0.10}^ { +0.08}$	&	0.42$_ { -0.10}^ { +0.11}$	&	N	&	Y	&	0.5	&	38.4\\														
6 &$\le$10.4 &- &- &- &- &- &- &-		&	-	&	-	&	-	&	-	&	0.8	&	$\le$37.3\\														
7 &$\le$2.6 &- &- &- &- &- &- &-		&	-	&	-	&	-	&	-	&	0.0	&	$\le$36.7\\														
8 &23.8$\pm$6.7 &3.4$\pm$3.4 &17.3$\pm$5.4 &20.7$\pm$6.0 &3.1$\pm$3.8 &17.1$\pm$5.4 &5.7$\pm$4.3 &-0.59	$_ { -0.22}^ { +0.21}$	&	-0.54$_ { -0.37}^ { +0.30}$	&	0.67$_ { -0.32}^ { +0.56}$	&	N	&	N	&	0.3	&	37.7\\
\enddata
\tablecomments{Col. (1): Master ID, cols. (2)$-$(8): net counts, in each of
the 7 energy bands (see Table \ref{tab:bands} for definitions of these
bands), col. (9): hardness ratio, cols. (10) and (11) color values, errors are given as 1$\sigma$,
cols. (12) and (13): short-term variability, where (BB) indicate
Bayesian block analysis and (K-S) indicates the Kolmogorov-Smirnov test, in both columns symbols
indicate - (N) non-variable in all observations, (V) variable in at
least one observation, (P) possible variability in at least one
observation, col. (14): the significance of the change in
\LX\ between the previous observation and the current observation
respectively (equation \ref{equ:sig}), col. (15): log \LX\ (0.3$-$8.0
keV).  Upper limit values of net B and \LX\ are at
the 68\% confidence level.}
\end{deluxetable}
\clearpage
\end{landscape}

\clearpage
\begin{landscape}
\begin{deluxetable}{l@{}cccccc@{}cccc@{}c@{}c@{}c@{}c@{}}
\tabletypesize{\scriptsize}
\tablecolumns{15}
\tablewidth{0pt}
\tablecaption{Source counts, hardness ratios, color-color values and variability:
Observation 5 \label{tab:counts_ob5} }
\tablehead{  \colhead{Masterid}   & \multicolumn{7}{c}{Net Counts}
& \colhead{HR} & \colhead{C21}  & \colhead{C32} &
  \multicolumn{3}{c}{Variability} & \colhead{Log \LX } \\
\colhead{}   & \colhead{B-band} &
\colhead{S1-band}  & \colhead{S2-band}  & \colhead{S-band} &
\colhead{H-band} & \colhead{Sc-band}  & \colhead{Hc-band} &
\colhead{} & \colhead{}  & \colhead{} &
\colhead{BB} & \colhead{k-S} & \colhead{signif.} &\colhead{(0.3$-$8.0 keV)} \\
 \colhead{(1)}   &
\colhead{(2)}  & \colhead{(3)} & \colhead{(4)}  & \colhead{(5)} &
\colhead{(6)} & \colhead{(7)}  & \colhead{(8)} &
\colhead{(9)} & \colhead{(10)}  & \colhead{(11)} &
\colhead{(12)} & \colhead{(13)}  & \colhead{(14)}& \colhead{(15)} }
\startdata
1 &49.4$\pm$9.0 &0.4$\pm$3.2 &13.1$\pm$5.1 &13.6$\pm$5.6 &35.8$\pm$7.5 &10.6$\pm$4.8 &37.1$\pm$7.7 &0.51	$_ { -0.14}^ { +0.15}$	&	-0.76$_ { -0.85}^ { +0.47}$	&	-0.38$_ { -0.17}^ { +0.14}$	&	N	&	N	&	0.4	&	38.0\\														
2 &129.3$\pm$13.2 &22.1$\pm$6.3 &79.4$\pm$10.2 &101.5$\pm$11.5 &27.8$\pm$7.1 &83.1$\pm$10.4 &44.1$\pm$8.4 &-0.36	$_ { -0.09}^ { +0.08}$	&	-0.45$_ { -0.12}^ { +0.11}$	&	0.48$_ { -0.11}^ { +0.11}$	&	N	&	N	&	1.3	&	38.4\\														
3 &$\le$2.3 &- &- &- &- &- &- &-		&	-	&	-	&	-	&	-	&	0.0	&	$\le$36.6\\														
4 &76.2$\pm$10.6 &22.0$\pm$6.2 &41.7$\pm$7.8 &63.7$\pm$9.5 &12.6$\pm$5.6 &53.7$\pm$8.7 &18.9$\pm$6.3 &-0.53	$_ { -0.12}^ { +0.11}$	&	-0.19$_ { -0.10}^ { +0.14}$	&	0.53$_ { -0.16}^ { +0.19}$	&	N	&	N	&	1.9	&	38.2\\														
5 &204.7$\pm$15.8 &49.8$\pm$8.4 &118.2$\pm$12.0 &168.0$\pm$14.3 &36.7$\pm$7.6 &140.7$\pm$13.1 &54.1$\pm$8.9 &-0.49	$_ { -0.07}^ { +0.06}$	&	-0.28$_ { -0.07}^ { +0.08}$	&	0.53$_ { -0.09}^ { +0.09}$	&	N	&	N	&	2.7	&	38.6\\														
6 &$\le$1.8 &- &- &- &- &- &- &-		&	-	&	-	&	-	&	-	&	1.2	&	$\le$36.5\\														
7 &$\le$2.1 &- &- &- &- &- &- &-		&	-	&	-	&	-	&	-	&	0.0	&	$\le$36.6\\														
8 &26.4$\pm$7.4 &6.0$\pm$4.1 &16.9$\pm$5.6 &22.9$\pm$6.5 &3.5$\pm$4.4 &23.5$\pm$6.3 &2.8$\pm$4.4 &-0.88	$_ { -0.12}^ { +0.11}$	&	-0.32$_ { -0.30}^ { +0.21}$	&	0.59$_ { -0.35}^ { +0.59}$	&	N	&	N	&	0.1	&	37.7\\
\enddata
\tablecomments{Col. (1): Master ID, cols. (2)$-$(8): net counts, in each of
the 7 energy bands (see Table \ref{tab:bands} for definitions of these
bands), col. (9): hardness ratio, cols. (10) and (11) color values, errors are given as 1$\sigma$,
cols. (12) and (13): short-term variability, where (BB) indicate
Bayesian block analysis and (K-S) indicates the Kolmogorov-Smirnov test, in both columns symbols
indicate - (N) non-variable in all observations, (V) variable in at
least one observation, (P) possible variability in at least one
observation, col. (14): the significance of the change in
\LX\ between the previous observation and the current observation
respectively (equation \ref{equ:sig}), col. (15): log \LX\ (0.3$-$8.0
keV).  Upper limit values of net B and \LX\ are at
the 68\% confidence level.}
\end{deluxetable}
\clearpage
\end{landscape}

\clearpage
\begin{landscape}
\begin{deluxetable}{l@{}cccccc@{}cccc@{}c@{}c@{}c@{}c@{}}
\tabletypesize{\scriptsize}
\tablecolumns{15}
\tablewidth{0pt}
\tablecaption{Source counts, hardness ratios, color-color values and variability:
Observation 6 \label{tab:counts_ob6} }
\tablehead{  \colhead{Masterid}   & \multicolumn{7}{c}{Net Counts}
& \colhead{HR} & \colhead{C21}  & \colhead{C32} &
  \multicolumn{3}{c}{Variability} & \colhead{Log \LX } \\
\colhead{}   & \colhead{B-band} &
\colhead{S1-band}  & \colhead{S2-band}  & \colhead{S-band} &
\colhead{H-band} & \colhead{Sc-band}  & \colhead{Hc-band} &
\colhead{} & \colhead{}  & \colhead{} &
\colhead{BB} & \colhead{k-S} & \colhead{signif.} &\colhead{(0.3$-$8.0 keV)} \\
 \colhead{(1)}   &
\colhead{(2)}  & \colhead{(3)} & \colhead{(4)}  & \colhead{(5)} &
\colhead{(6)} & \colhead{(7)}  & \colhead{(8)} &
\colhead{(9)} & \colhead{(10)}  & \colhead{(11)} &
\colhead{(12)} & \colhead{(13)}  & \colhead{(14)}& \colhead{(15)} }
\startdata
1 &10.1$\pm$5.4 &-1.0$\pm$2.3 &7.3$\pm$4.1 &6.4$\pm$4.3 &3.7$\pm$4.0 &5.9$\pm$4.0 &4.2$\pm$4.1 &-0.33	$_ { -0.37}^ { +0.41}$	&	-0.92$_ { -0.91}^ { +0.54}$	&	0.27$_ { -0.38}^ { +0.51}$	&	-	&	-	&	2.1	&	37.6\\														
2 &47.4$\pm$8.4 &9.8$\pm$4.4 &24.2$\pm$6.2 &34.0$\pm$7.1 &13.3$\pm$5.2 &31.0$\pm$6.8 &14.9$\pm$5.4 &-0.46	$_ { -0.14}^ { +0.13}$	&	-0.16$_ { -0.20}^ { +0.15}$	&	0.31$_ { -0.16}^ { +0.16}$	&	N	&	N	&	1.1	&	38.3\\														
3 &$\le$3.0 &- &- &- &- &- &- &-		&	-	&	-	&	-	&	-	&	0.0	&	$\le$37.1\\														
4 &20.7$\pm$6.3 &3.8$\pm$3.4 &13.5$\pm$5.0 &17.3$\pm$5.6 &3.4$\pm$3.8 &13.2$\pm$5.0 &4.9$\pm$4.1 &-0.60	$_ { -0.23}^ { +0.23}$	&	-0.32$_ { -0.32}^ { +0.27}$	&	0.56$_ { -0.32}^ { +0.49}$	&	N	&	N	&	1.8	&	37.9\\														
5 &85.8$\pm$10.7 &22.6$\pm$6.0 &51.1$\pm$8.3 &73.7$\pm$9.8 &12.1$\pm$5.1 &58.5$\pm$8.9 &20.9$\pm$6.1 &-0.54	$_ { -0.10}^ { +0.09}$	&	-0.20$_ { -0.12}^ { +0.10}$	&	0.66$_ { -0.17}^ { +0.14}$	&	N	&	N	&	1.1	&	38.5\\														
6 &$\le$4.7 &- &- &- &- &- &- &-		&	-	&	-	&	-	&	-	&	0.0	&	$\le$37.3\\														
7 &6.8$\pm$4.6 &2.1$\pm$2.9 &2.6$\pm$3.2 &4.7$\pm$3.8 &2.1$\pm$3.4 &1.6$\pm$2.9 &3.7$\pm$3.8 &0.34	$_ { -0.28}^ { +0.39}$	&	0.03$_ { -0.62}^ { +0.59}$	&	0.08$_ { -0.67}^ { +0.75}$	&	-	&	-	&	1.4	&	37.4\\														
8 &21.7$\pm$6.2 &7.4$\pm$4.0 &8.9$\pm$4.3 &16.3$\pm$5.3 &5.5$\pm$4.0 &11.8$\pm$4.7 &9.2$\pm$4.6 &-0.23	$_ { -0.24}^ { +0.22}$	&	0.08$_ { -0.21}^ { +0.21}$	&	0.24$_ { -0.27}^ { +0.30}$	&	N	&	N	&	1.3	&	37.9\\
\enddata
\tablecomments{Col. (1): Master ID, cols. (2)$-$(8): net counts, in each of
the 7 energy bands (see Table \ref{tab:bands} for definitions of these
bands), col. (9): hardness ratio, cols. (10) and (11) color values, errors are given as 1$\sigma$,
cols. (12) and (13): short-term variability, where (BB) indicate
Bayesian block analysis and (K-S) indicates the Kolmogorov-Smirnov test, in both columns symbols
indicate - (N) non-variable in all observations, (V) variable in at
least one observation, (P) possible variability in at least one
observation, col. (14): the significance of the change in
\LX\ between the previous observation and the current observation
respectively (equation \ref{equ:sig}), col. (15): log \LX\ (0.3$-$8.0
keV).  Upper limit values of net B and \LX\ are at
the 68\% confidence level.}
\end{deluxetable}
\clearpage
\end{landscape}

%
\makeatletter
\def\jnl@aj{AJ}
\ifx\revtex@jnl\jnl@aj\let\tablebreak=\nl\fi
\makeatother
\begin{deluxetable}{lccccc}
\tabletypesize{\scriptsize}
\tablecolumns{6}
\tablewidth{0pt}
\tablecaption{Properties of optical sources that are correlated with
an X-ray point source\label{tab:GCprops_corr} }
\tablehead{  \colhead{Masterid}   &
\colhead{V}  & \colhead{I} &
\colhead{V$-$I} & \colhead{GC separation} &
\colhead{Ratio} \\
 \colhead{}   &
\colhead{}  & \colhead{}  & \colhead{} &
 \colhead{(arcsec)}  & \colhead{} \\
\colhead{(1)}   &
\colhead{(2)}  & \colhead{(3)}  & \colhead{(4)} &
\colhead{(5)} & \colhead{(6)}  }

\startdata
51 & 21.21$\pm$0.01 & 20.11$\pm$0.01 & 1.11$\pm$0.02 & 0.40 & 1.02 \\
63 & 22.31$\pm$0.02 & 21.17$\pm$0.03 & 1.14$\pm$0.03 & 0.48 & 1.73 \\
68 & 20.25$\pm$0.01 & 19.21$\pm$0.01 & 1.04$\pm$0.01 & 0.38 & 1.45 \\
76 & 22.75$\pm$0.03 & 21.66$\pm$0.05 & 1.09$\pm$0.06 & 0.26 & 1.01 \\
81 & 22.10$\pm$0.02 & 20.94$\pm$0.04 & 1.16$\pm$0.04 & 0.23 & 0.70 \\
83 & 21.63$\pm$0.01 & 20.63$\pm$0.02 & 1.00$\pm$0.02 & 0.19 & 0.76 \\
84 & 22.14$\pm$0.02 & 20.91$\pm$0.03 & 1.23$\pm$0.04 & 0.19 & 0.72 \\
85 & 22.29$\pm$0.02 & 21.39$\pm$0.05 & 0.90$\pm$0.05 & 0.33 & 1.22 \\
\enddata
\tablecomments{The sources in the top section of the table, denoted by
the horizontal line, have been
confirmed as globular clusters, while those in the bottom section of
the table have all been identified as background objects.}

\end{deluxetable}

\clearpage

\makeatletter
\def\jnl@aj{AJ}
\ifx\revtex@jnl\jnl@aj\let\tablebreak=\nl\fi
\makeatother
\begin{deluxetable}{lccccc}
\tabletypesize{\scriptsize}
\tablecolumns{6}
\tablewidth{0pt}
\tablecaption{Properties of optical sources that have been classified
as `excluded matches'; sources detected between 0.6\arcs\ and 3\arcs\
of an X-ray point source\label{tab:GCprops_near}  }
\tablehead{  \colhead{Masterid}   &
\colhead{V}  & \colhead{I} &
\colhead{V$-$I} & \colhead{GC separation} &
\colhead{Ratio} \\
 \colhead{}   &
\colhead{}  & \colhead{}  & \colhead{} &
 \colhead{(arcsec)}  & \colhead{} \\
\colhead{(1)}   &
\colhead{(2)}  & \colhead{(3)}  & \colhead{(4)} &
\colhead{(5)} & \colhead{(6)} }

\startdata
75 & 23.21$\pm$0.04 & 22.09$\pm$0.06 & 1.12$\pm$0.07 & 0.66 & 1.13 \\
88 & 23.35$\pm$0.06 & 22.22$\pm$0.10 & 1.13$\pm$0.12 & 0.66 & 1.99 \\
93 & 22.76$\pm$0.03 & 21.70$\pm$0.04 & 1.07$\pm$0.05 & 1.61 & 2.49 \\
99 & 22.11$\pm$0.03 & 20.78$\pm$0.03 & 1.33$\pm$0.04 & 0.98 & 3.91 \\
104 & 22.56$\pm$0.04 & 21.52$\pm$0.07 & 1.04$\pm$0.08 & 2.26 & 8.64 \\
105 & 22.97$\pm$0.07 & 21.76$\pm$0.08 & 1.21$\pm$0.10 & 2.23 & 5.57 \\
106 & 22.16$\pm$0.17 & 21.09$\pm$0.23 & 1.07$\pm$0.29 & 2.27 & 8.92 \\
113 & 23.96$\pm$0.08 & 22.73$\pm$0.11 & 1.24$\pm$0.14 & 1.68 & 3.42 \\
\enddata
\tablecomments{The sources in the top section of the table, denoted by
the horizontal line, have been
confirmed as globular clusters, while those in the bottom section of
the table have all been identified as background objects.}

\end{deluxetable}

\clearpage 


\begin{deluxetable}{cccc}
\tabletypesize{\scriptsize}
\tablecolumns{4}
\tablewidth{0pt}
\tablecaption{Ratio Values of Potential Transient Candidates \label{tab:trans}}
\tablehead{\colhead{Masterid} & \colhead{Mode Ratio} &\colhead{Lower Bound Ratio}& \colhead{Variability}}
\startdata
27 & 5.0 & 3.6 & V \\
30 & 3.2 & 2.2 & V \\
32 & 4.5 & 3.7 & V \\
59 & 4.8 & 3.5 & V \\
83 & 57.8 & 35.5 & TC \\
91 & 9.0 & 8.3 & PTC \\
92 & 5.8 & 4.3 & V \\
113 & 3.5 & 2.4 & V \\
139 & 26.1 & 18.9 & TC \\
152 & 5.3 & 4.2 & V \\
156 & 6.2 & 5.5 & PTC \\
159 & 13.6 & 5.8 & PTC \\
204 & 871.7 & 202.9 & TC \\
\enddata
\tablecomments{The 13 sources that were identified as potential transients candidates. Mode ratios and lower bound ratios were derived from Bayesian modeling, see \S\ref{sec:var} for more details. Sources were determined to be transient candidates (TC) if the lower bound $>$10, possible transient candidates (PTC) if the lower bound $>$5 and variable (V), if lower than 5.}
\end{deluxetable}

\begin{deluxetable}{ccccccc}
\tabletypesize{\scriptsize}
\tablecolumns{6}
\tablewidth{0pt}
\tablecaption{Summary of Long-Term and Short-Term Source Variability \label{tab:varsum}}
\tablehead{\colhead{Short-Term} & \multicolumn{5}{c}{Long-Term
Variability} &\colhead{Total} \\
\colhead{Variability} & \colhead{N} & \colhead{V} & \colhead{TC} &
\colhead{PTC} &\colhead{U} &\colhead{}}
\startdata
N & 7 {\bf (7)} & 11  {\bf (9)} & 0  {\bf (0)} & 0 {\bf (0)} & 0 {\bf (0)} & 18 {\bf (16)} \\ 
V & 3 {\bf (2)} & 6  {\bf (5)} & 0  {\bf (0)} & 0 {\bf (0)} & 0 {\bf (0)} & 9 {\bf (7)} \\
P & 11 {\bf (7)} & 14  {\bf (10)} & 0  {\bf (0)} & 0 {\bf (0)} & 0 {\bf (0)}& 25 {\bf (17)} \\ 
U & 76 {\bf (54)} & 66  {\bf (52)} & 3  {\bf (3)} & 3 {\bf (3)} & 36 {\bf (28)} & 184 {\bf (140)} \\
\hline
Total & 97 {\bf (70)} & 97 {\bf (76)} & 3  {\bf (3)} & 3 {\bf (3)} & 36 {\bf (28)} & 236 {\bf (180)} \\
\enddata
\tablecomments{The long-term variability definitions: N-non-variable,
V-variable, TC-transient candidate, PTC-possible transient
candidate and U unable to determine variability as all individual
observations were upper limits. Short-term variability definitions: N-non-variable in all
six observation, V-variable in a least one observation, P-possible
variability in at least one observation (see \S \ref{sec:var} for
full definition), U too few counts in all six observations to
determine variability. Bold values in brackets indicate the number of
sources within the $D_{25}$ ellipse.}

\end{deluxetable}

\clearpage

%
\begin{landscape}
\begin{deluxetable}{lccccccccccccccc}
\tabletypesize{\scriptsize}
\tablecolumns{16}
\tablewidth{0pt}
\tablecaption{Raw source and background counts from the co-added observation \label{tab:raw} }
\tablehead{ \colhead{Masterid}  & \colhead{Net B Counts} &
\multicolumn{14}{c}{Raw Counts} \\
\colhead{}   & \colhead{} &
\multicolumn{2}{c}{B-band}
& \multicolumn{2}{c}{S1-band} & \multicolumn{2}{c}{S2-band} &
\multicolumn{2}{c}{H-band} & \multicolumn{2}{c}{Bc-band} & \multicolumn{2}{c}{Sc-band} & 
\multicolumn{2}{c}{Hc-band} \\
 \colhead{}   &
\colhead{}  & \colhead{(Src )} & \colhead{(Bkg )}  &
\colhead{(Src )} &
\colhead{(Bkg )} & \colhead{(Src )}  & \colhead{(Bkg )} &
\colhead{(Src )} & \colhead{(Bkg )}  & \colhead{(Src )} &
\colhead{(Bkg )} & \colhead{(Src )}  & \colhead{(Bkg )}  & \colhead{(Src )} &
\colhead{(Bkg )} \\
 \colhead{(1)}   &
\colhead{(2)}  & \colhead{(3)} & \colhead{(4)}  & \colhead{(5)} &
\colhead{(6)} & \colhead{(7)}  & \colhead{(8)} &
\colhead{(9)} & \colhead{(10)}  & \colhead{(11)} &
\colhead{(12)} & \colhead{(13)}  & \colhead{(14)}& \colhead{(15)} & \colhead{(16)} }
\startdata
1 & 139.8 & 198.0 & 58.2 & 13.0 & 15.3 & 61.0 & 13.0 & 124.0 & 29.8 & 191.0 & 48.5 & 54.0 & 14.9 & 137.0 & 33.6 \\
2 & 386.2 & 458.0 & 71.3 & 74.0 & 16.5 & 254.0 & 16.6 & 130.0 & 38.2 & 438.0 & 61.9 & 278.0 & 19.4 & 160.0 & 42.5 \\
3 & 7.8 & 69.0 & 61.9 & 15.0 & 15.8 & 25.0 & 15.5 & 29.0 & 30.6 & 61.0 & 53.1 & 29.0 & 18.7 & 32.0 & 34.3 \\
4 & 266.1 & 333.0 & 66.6 & 94.0 & 15.2 & 162.0 & 16.7 & 77.0 & 34.8 & 312.0 & 58.1 & 215.0 & 19.3 & 97.0 & 38.9 \\
5 & 725.7 & 787.0 & 61.3 & 217.0 & 14.8 & 413.0 & 14.7 & 157.0 & 31.8 & 737.0 & 53.6 & 527.0 & 18.4 & 210.0 & 35.3 \\
6 & 17.5 & 75.0 & 58.1 & 9.0 & 15.7 & 23.0 & 13.9 & 43.0 & 28.6 & 72.0 & 49.4 & 22.0 & 17.3 & 50.0 & 32.1 \\
7 & - & - & - & - & - & - & - & - & - & - & - & - & - & - & - \\
8 & 128.6 & 180.0 & 51.4 & 44.0 & 11.1 & 89.0 & 12.4 & 47.0 & 28.0 & 169.0 & 45.0 & 108.0 & 14.2 & 61.0 & 30.9 \\
\enddata
\tablecomments{Col. (1): Master ID, cols. (2): net broad-band
counts. Cols. (3), (5), (7), (9), (11), (13) and (15): raw source counts in each of
the 7 energy bands (see Table \ref{tab:bands} for definitions of these
bands), cols. (4), (6), (8), (10), (12), (14) and (16): background
counts in each of the 7 energy bands.
In some instances background counts are very low. For these sources
standard aperture photometry results in negative net counts, so,
instead, the source cell determined by {\em wavdetect} has been used. This
results in a large area ratio between the background and source
regions and therefore a low background count value is derived.}
\end{deluxetable}
\clearpage
\end{landscape}

\clearpage

\clearpage

\begin{figure}
\begin{centering}

  \includegraphics[angle=90,width=0.8\linewidth]{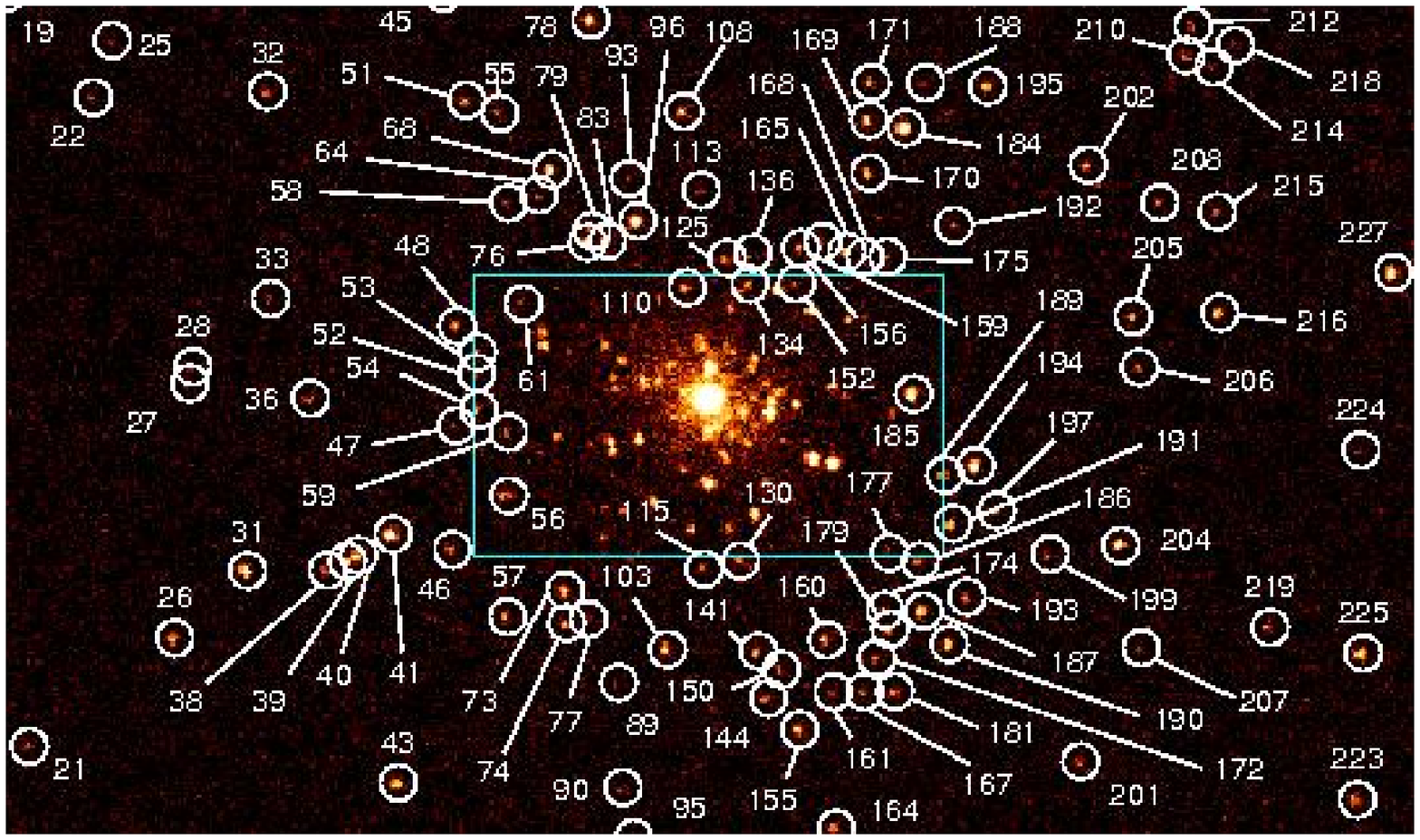}

\end{centering}
\end{figure}

  \begin{figure}
\begin{centering}

  \includegraphics[angle=90,width=0.8\linewidth]{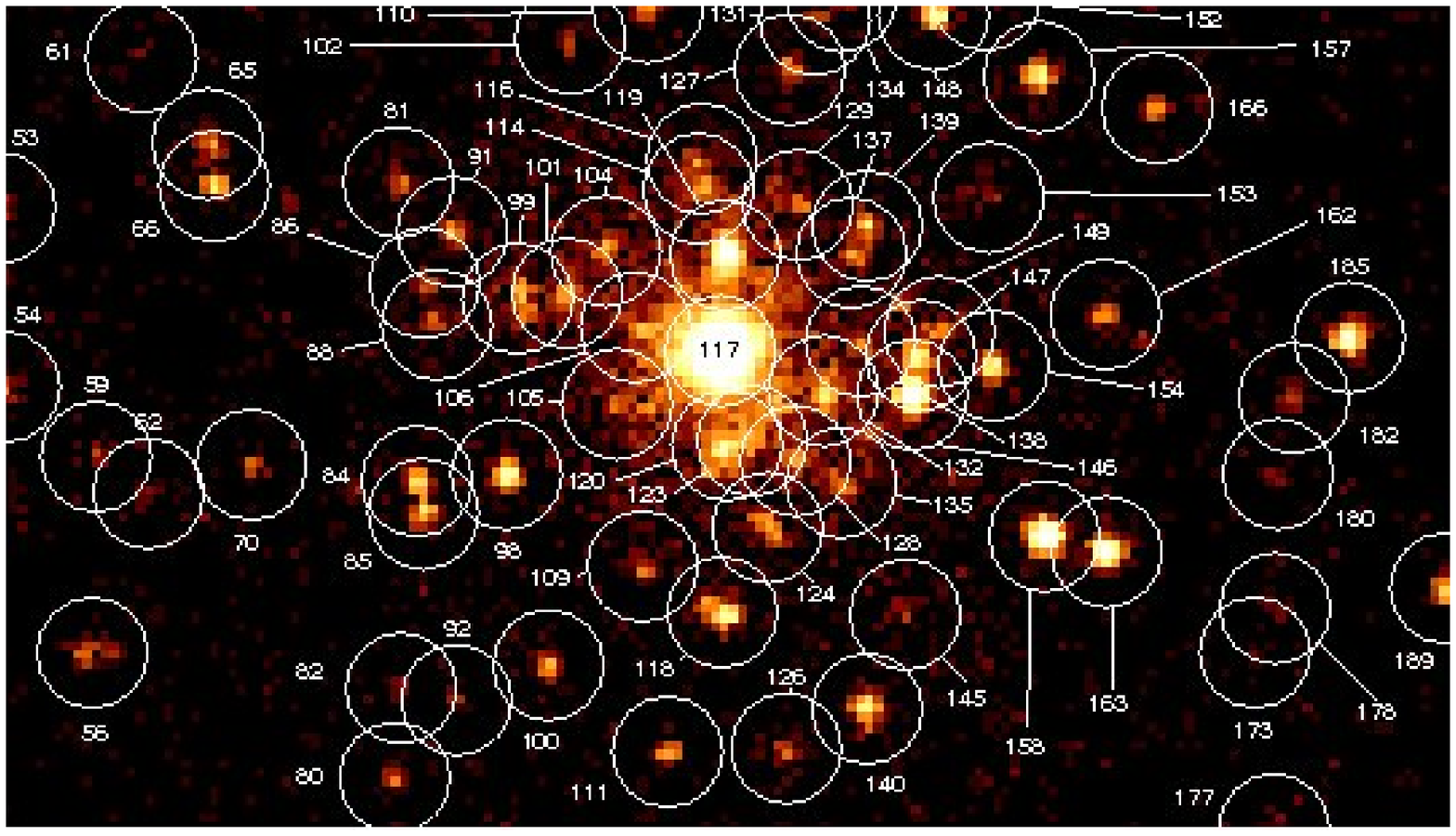}

\end{centering}
\end{figure}
\clearpage

\begin{figure}

  \begin{minipage}{0.485\linewidth}
  \centering

    \includegraphics[width=\linewidth]{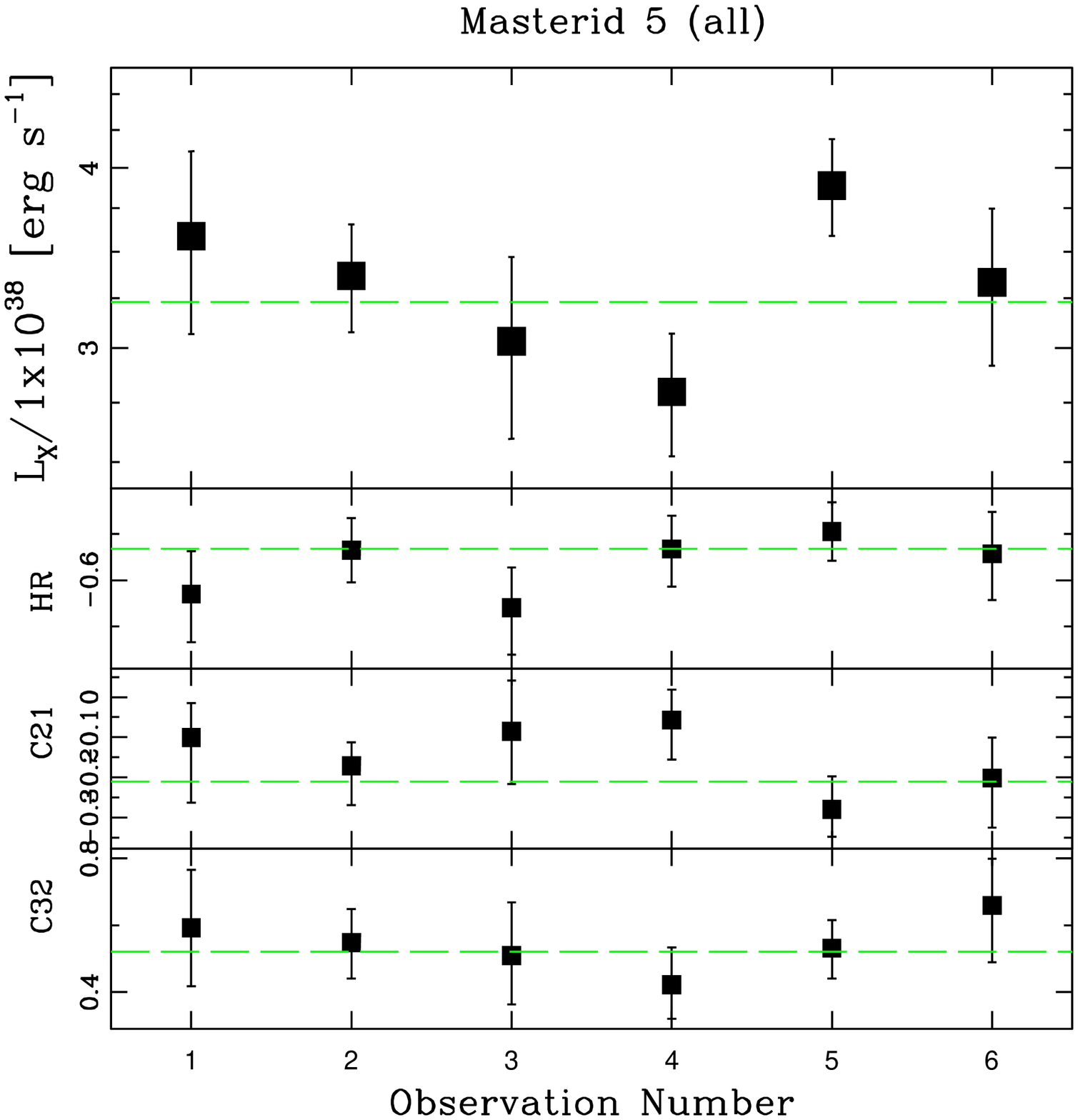}

\end{minipage}\hspace{0.02\linewidth}
\begin{minipage}{0.485\linewidth}
  \centering

    \includegraphics[width=\linewidth]{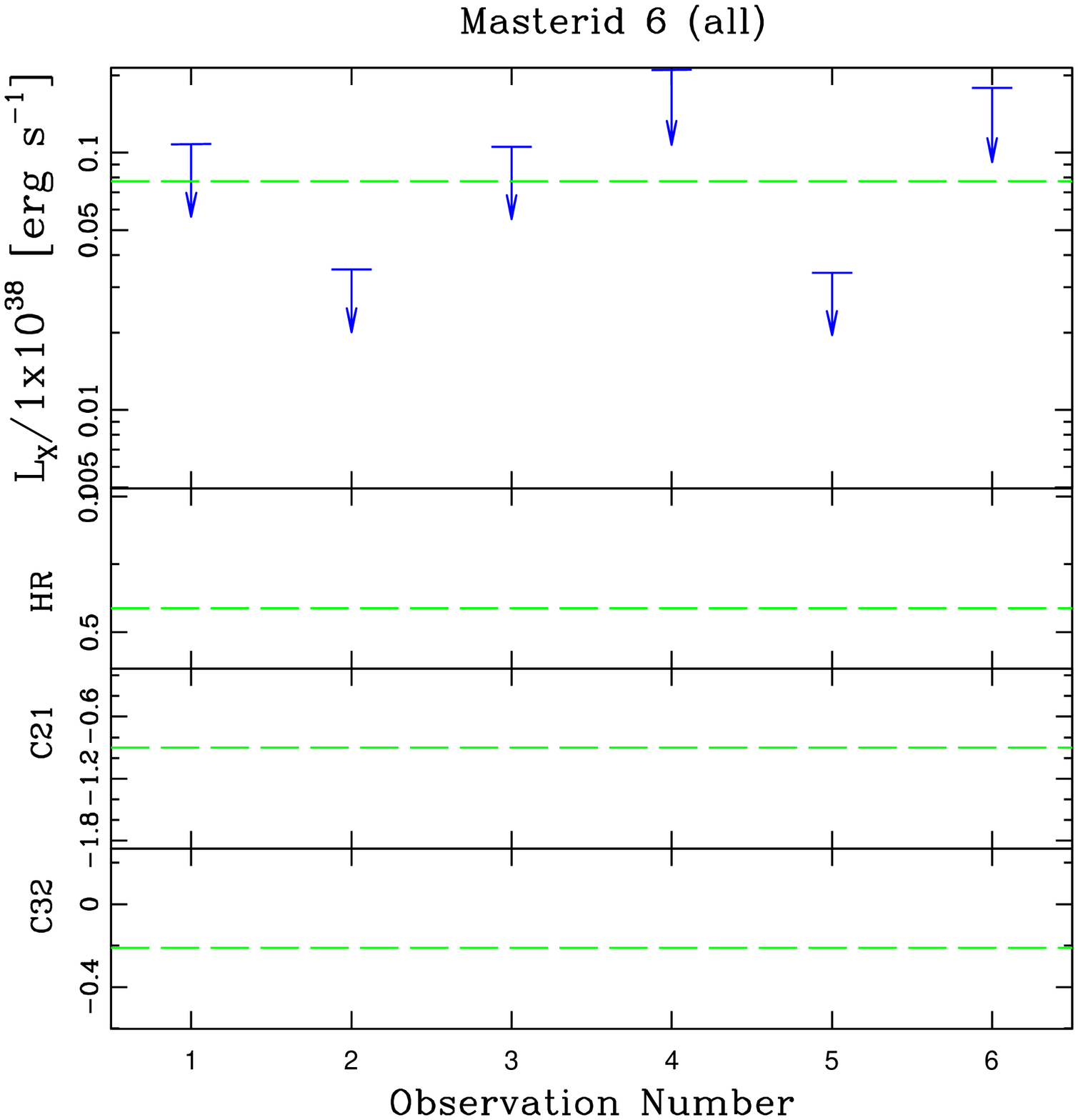}

 \end{minipage}\hspace{0.02\linewidth}

  \begin{minipage}{0.485\linewidth}
  \centering
  
    \includegraphics[width=\linewidth]{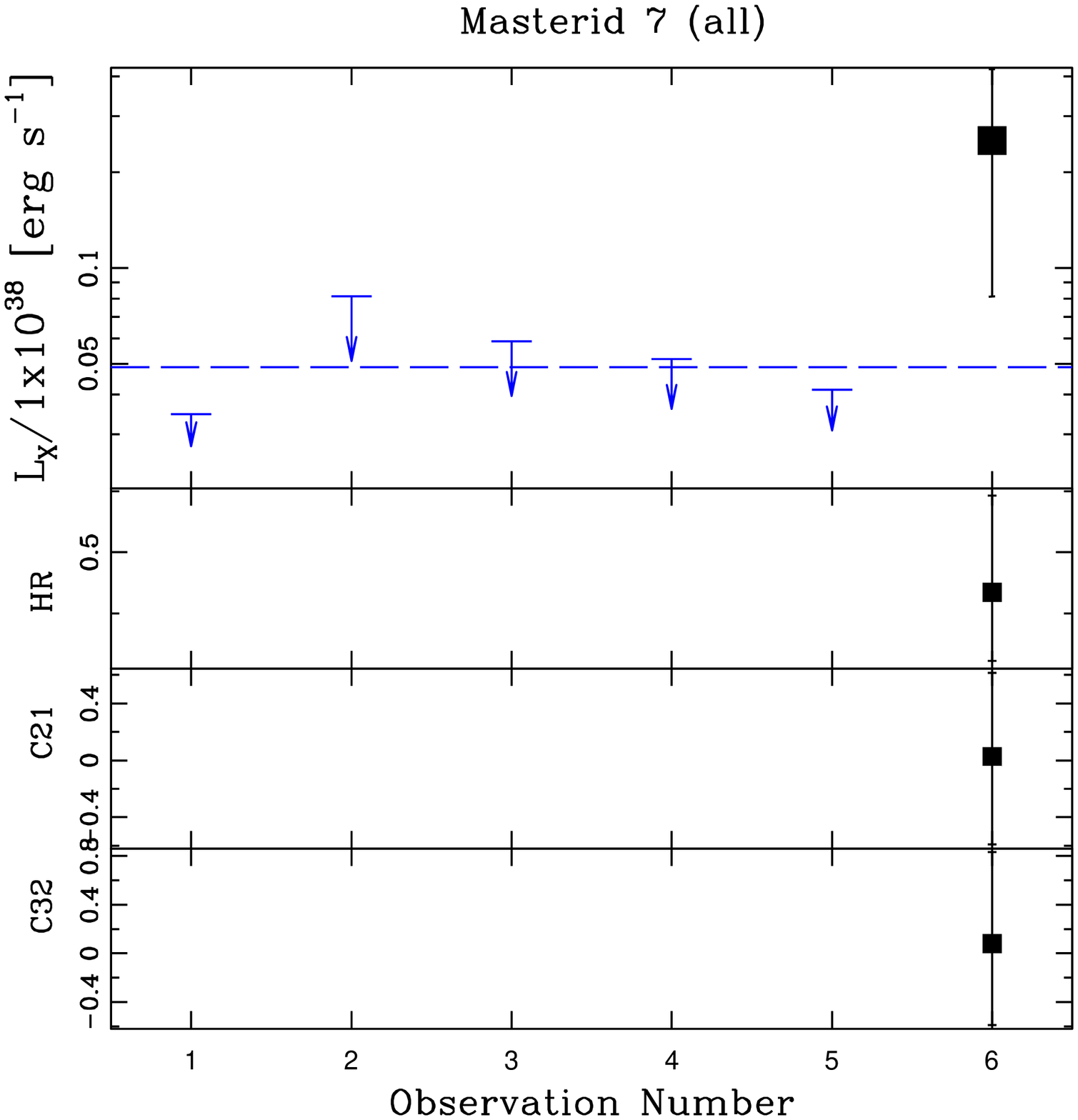}

  \end{minipage}\hspace{0.02\linewidth}
  \begin{minipage}{0.485\linewidth}
  \centering

    \includegraphics[width=\linewidth]{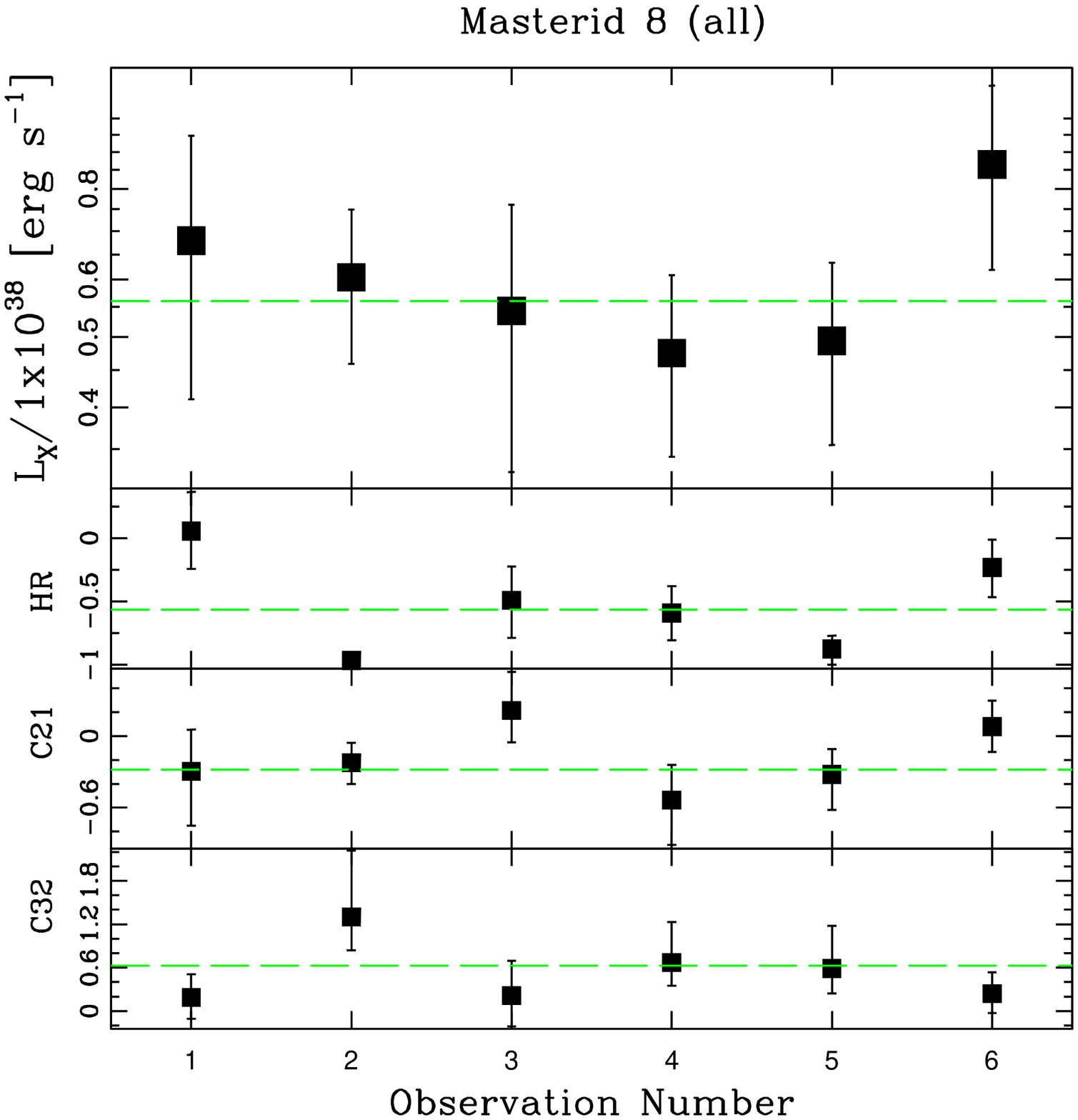}

\end{minipage}
\begin{minipage}{0.485\linewidth}
  \centering

    \includegraphics[width=\linewidth]{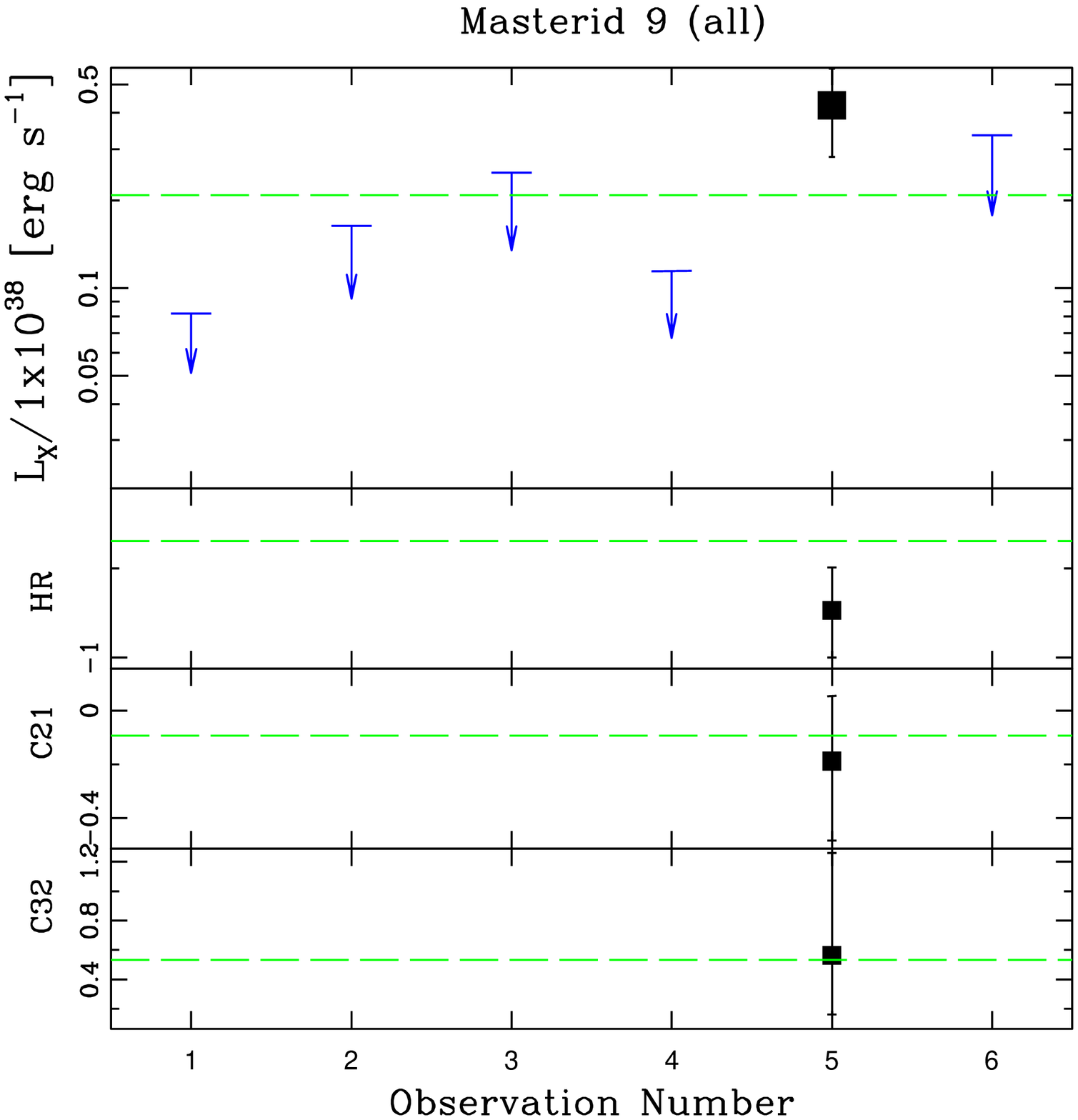}

 \end{minipage}\hspace{0.02\linewidth}
\begin{minipage}{0.485\linewidth}
  \centering
  
    \includegraphics[width=\linewidth]{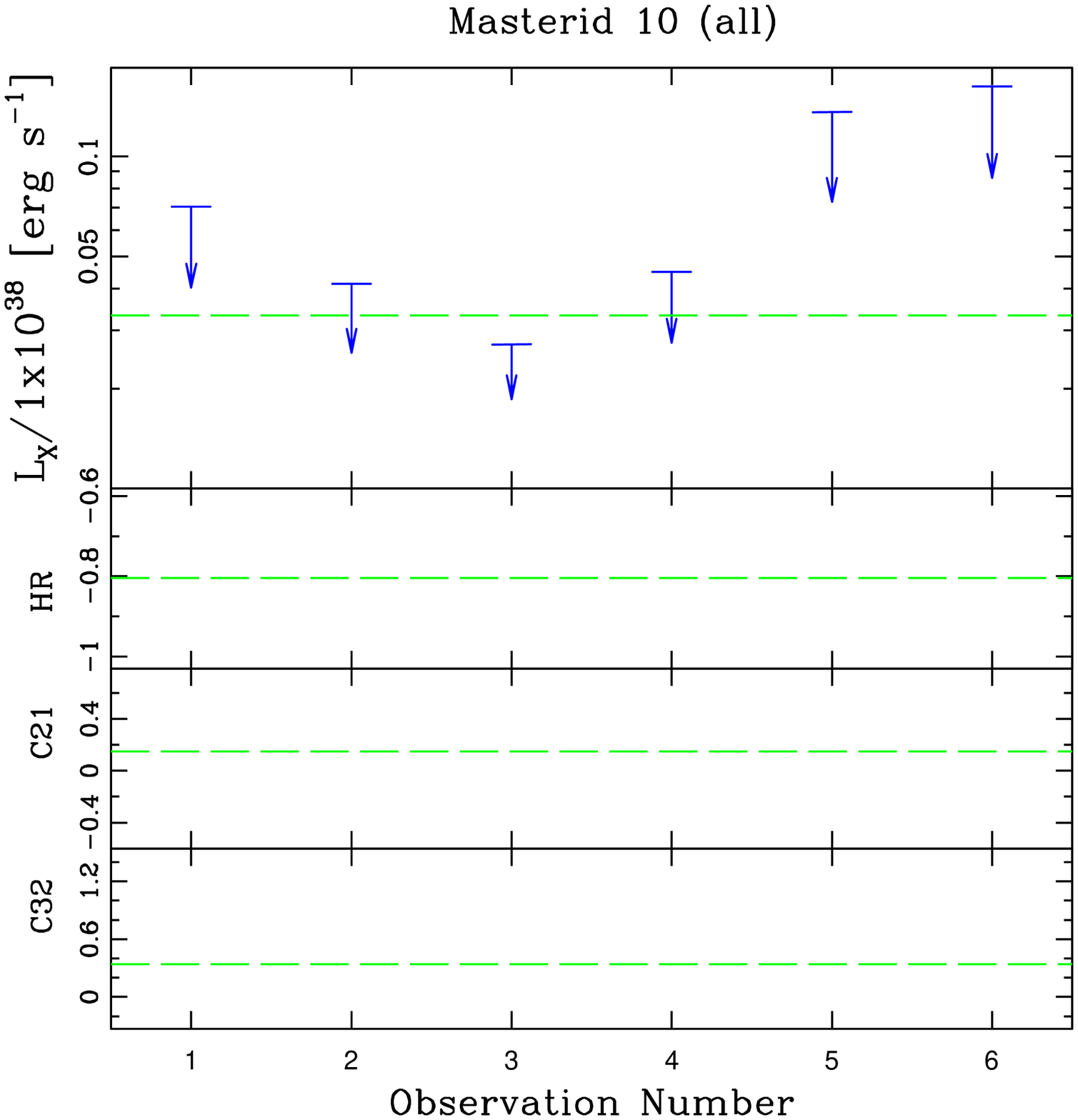}

  \end{minipage}\hspace{0.02\linewidth}

\end{figure}

\begin{figure}

  \begin{minipage}{0.485\linewidth}
  \centering

    \includegraphics[width=\linewidth]{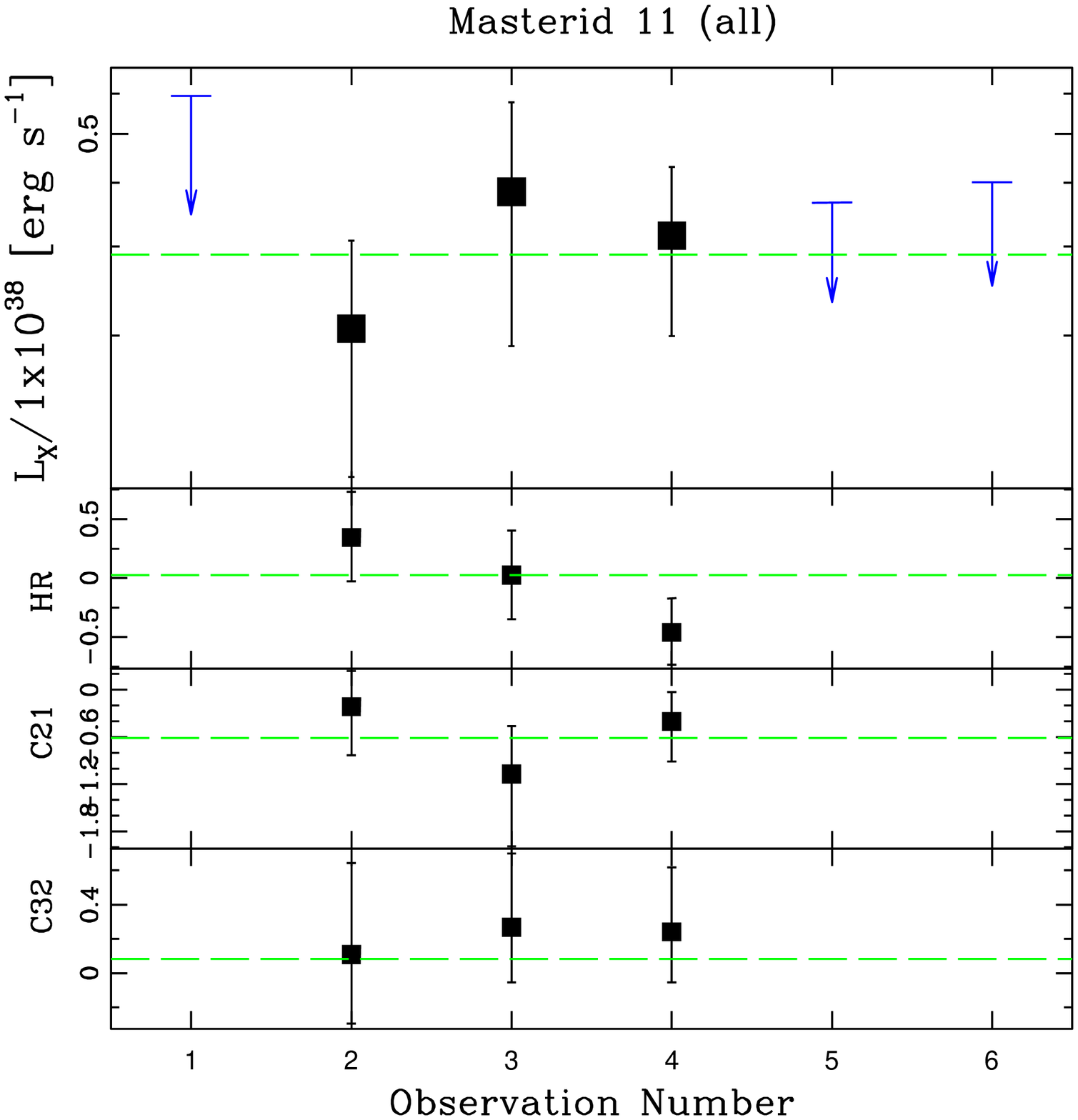}

\end{minipage}\hspace{0.02\linewidth}
\begin{minipage}{0.485\linewidth}
  \centering

    \includegraphics[width=\linewidth]{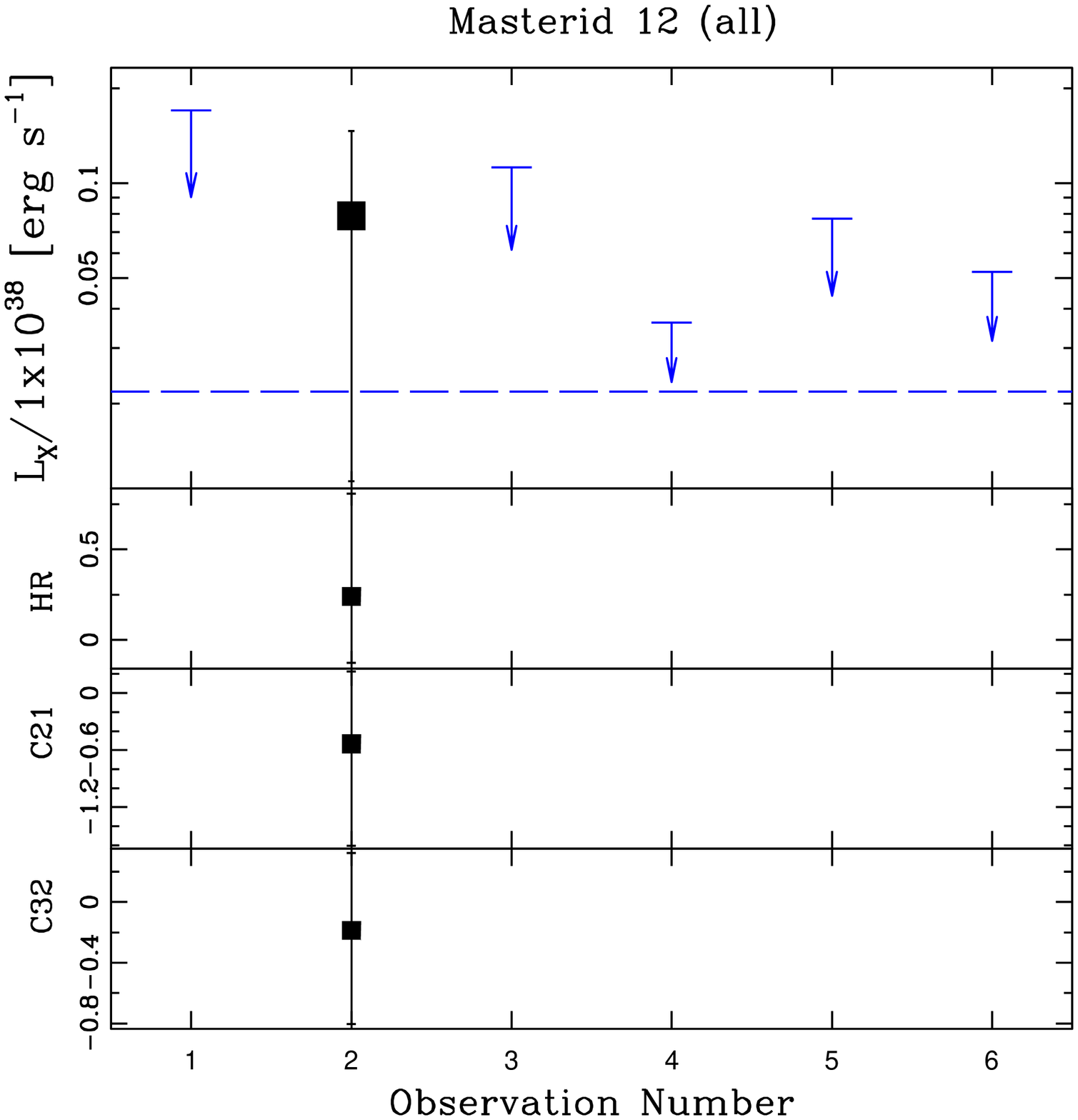}

 \end{minipage}\hspace{0.02\linewidth}

  \begin{minipage}{0.485\linewidth}
  \centering
  
    \includegraphics[width=\linewidth]{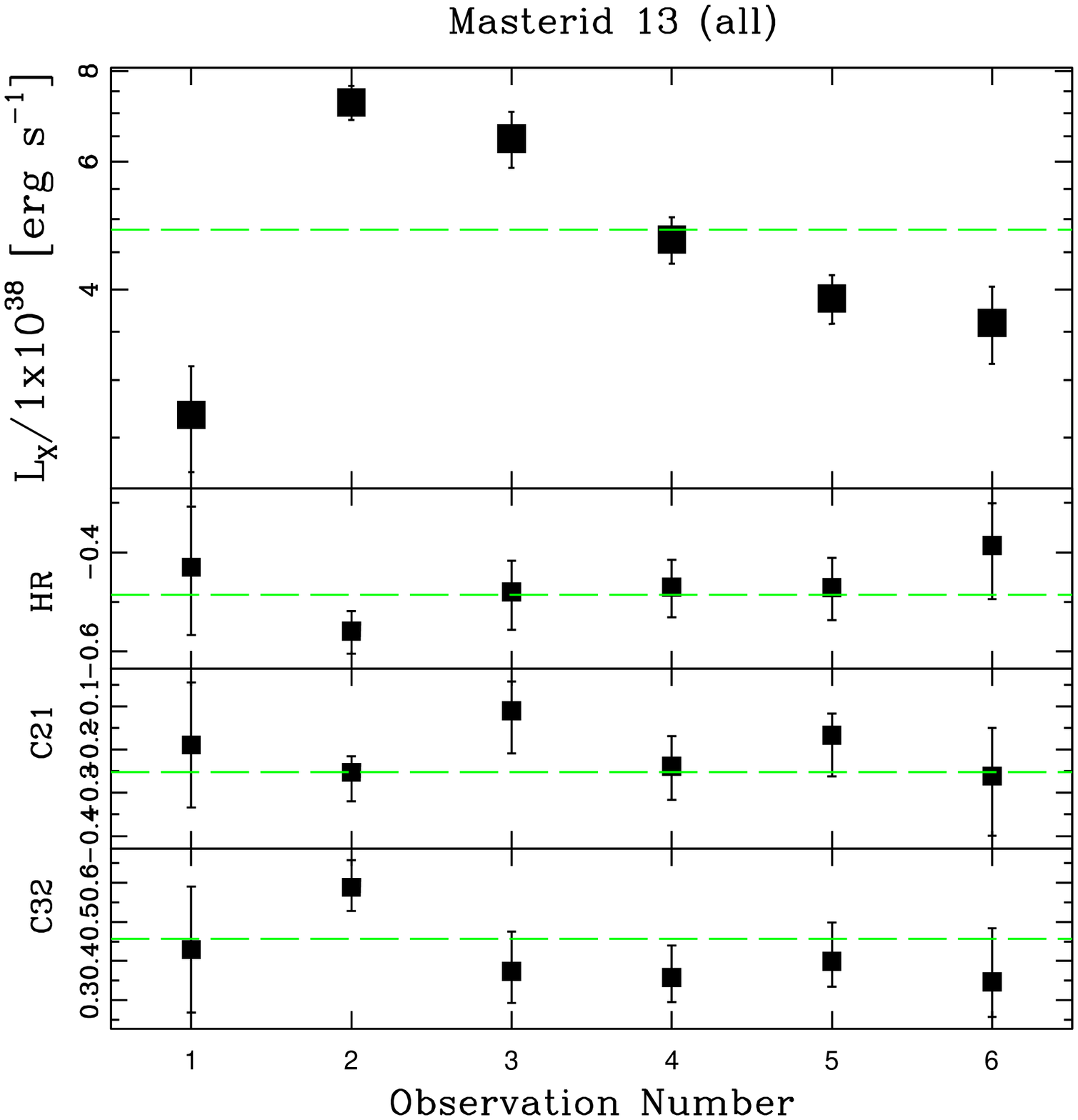}

  \end{minipage}\hspace{0.02\linewidth}
  \begin{minipage}{0.485\linewidth}
  \centering

    \includegraphics[width=\linewidth]{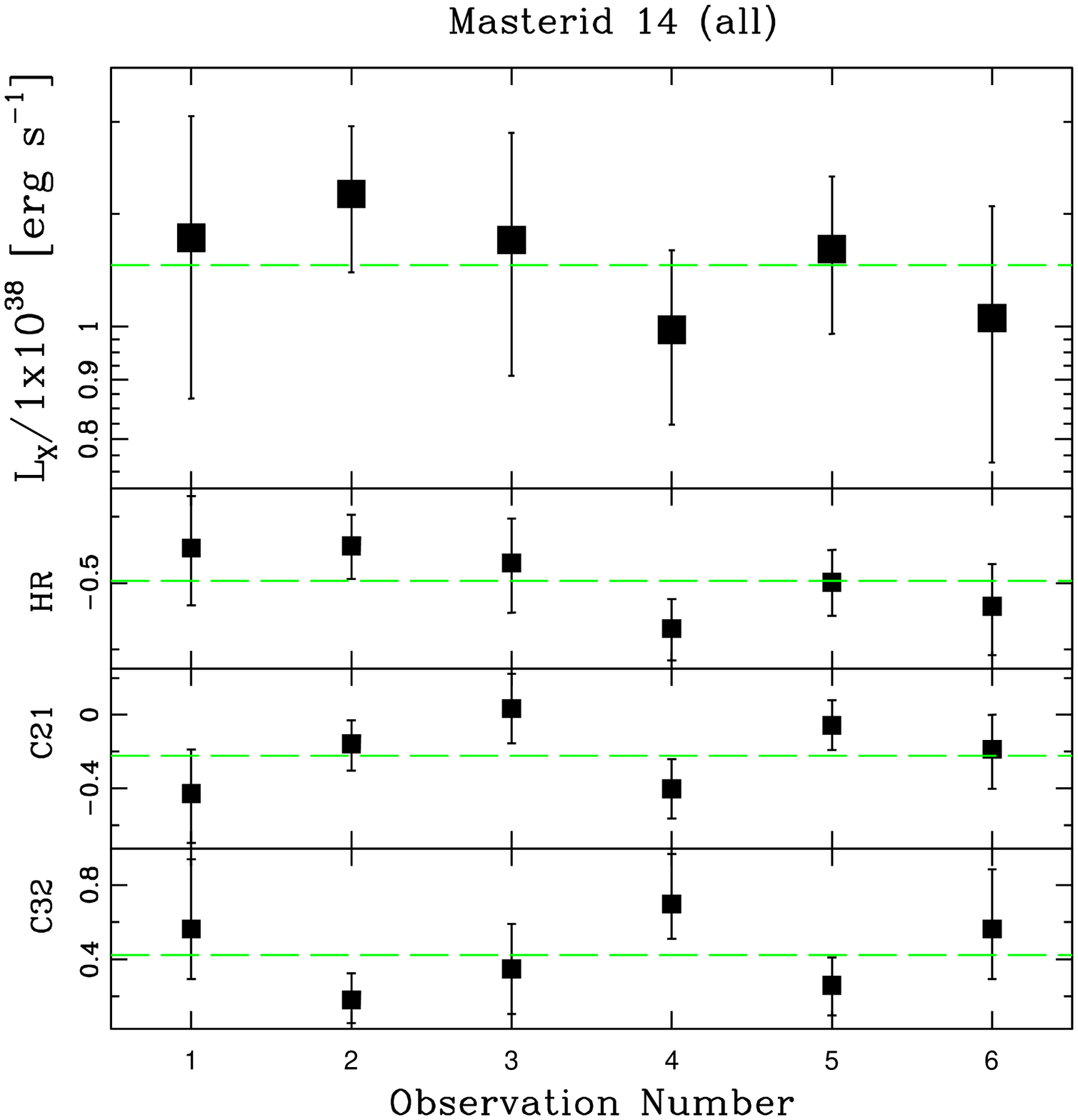}

\end{minipage}\hspace{0.02\linewidth}

\begin{minipage}{0.485\linewidth}
  \centering

    \includegraphics[width=\linewidth]{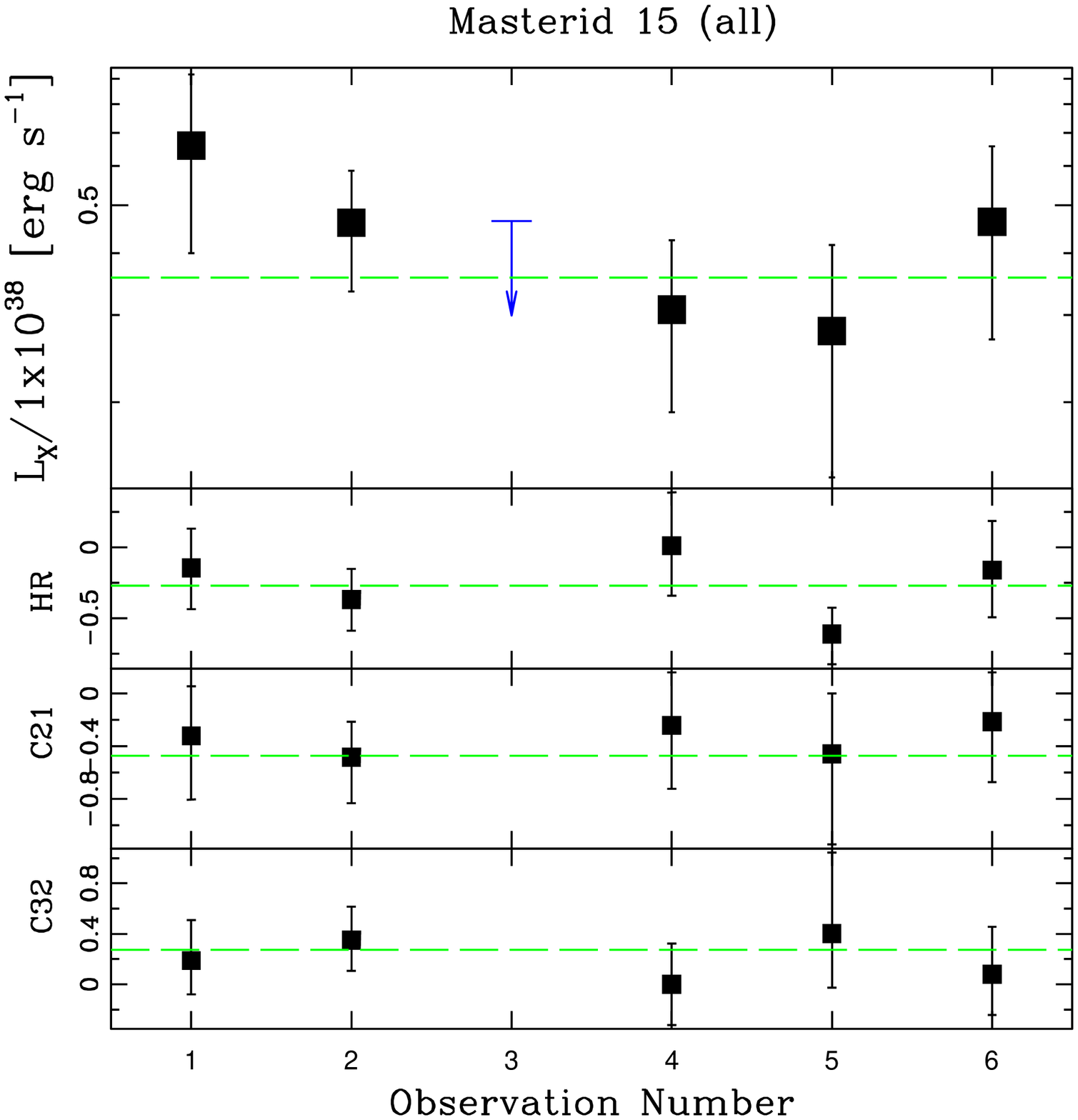}

 \end{minipage}\hspace{0.02\linewidth}
\begin{minipage}{0.485\linewidth}
  \centering
  
    \includegraphics[width=\linewidth]{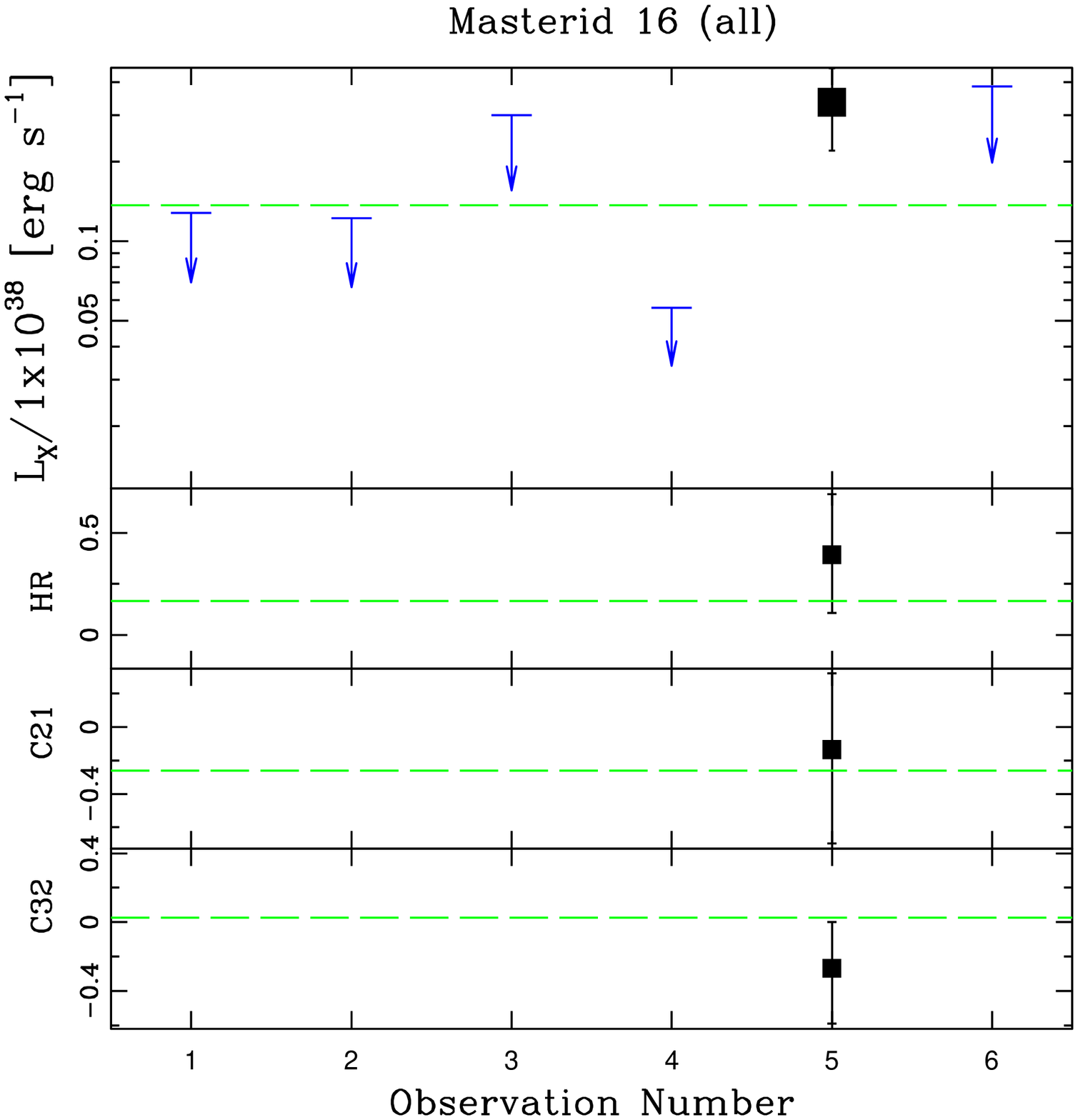}

  \end{minipage}\hspace{0.02\linewidth}

\end{figure}

\begin{figure}

  \begin{minipage}{0.485\linewidth}
  \centering

    \includegraphics[width=\linewidth]{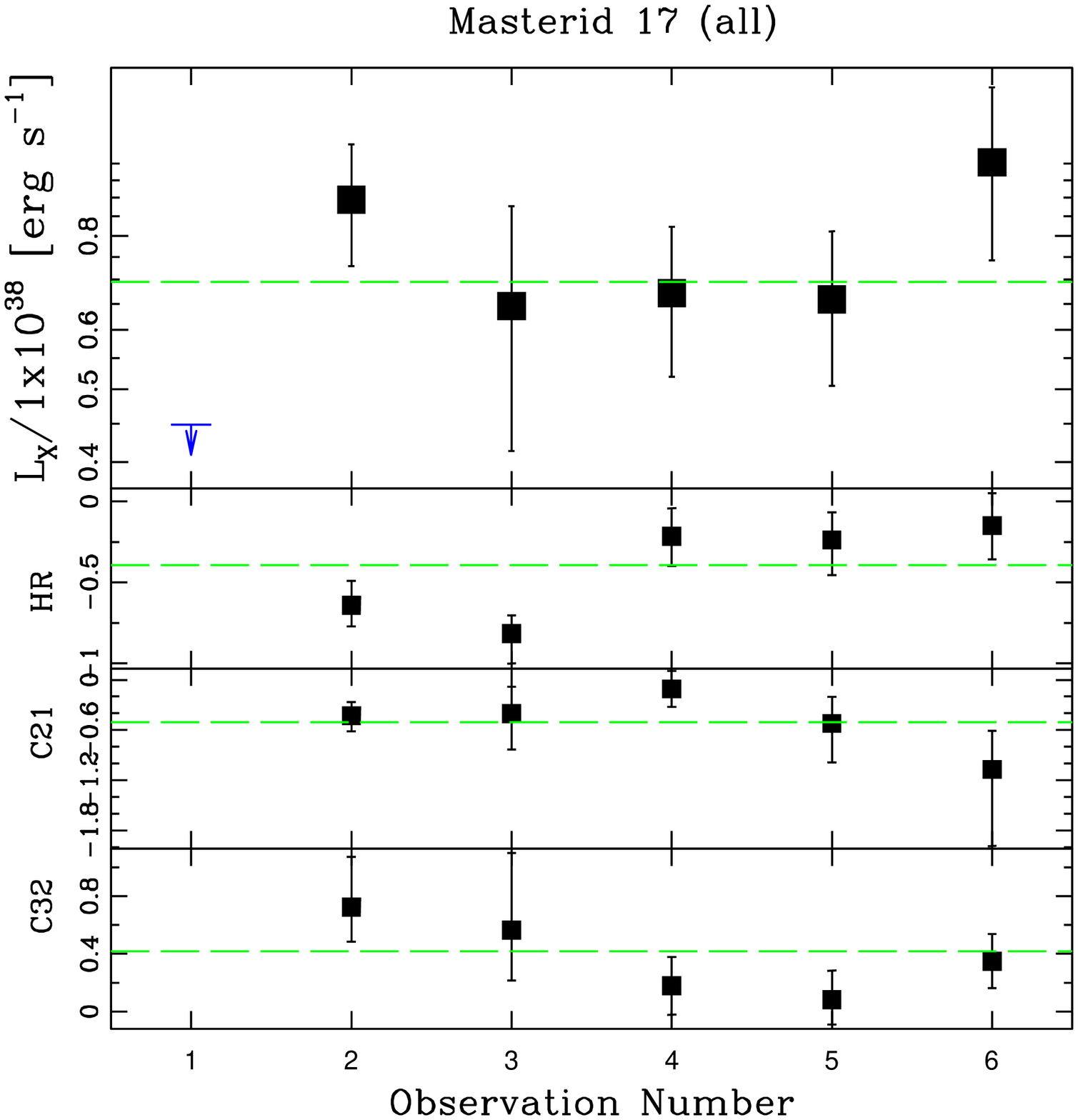}

\end{minipage}\hspace{0.02\linewidth}
\begin{minipage}{0.485\linewidth}
  \centering

    \includegraphics[width=\linewidth]{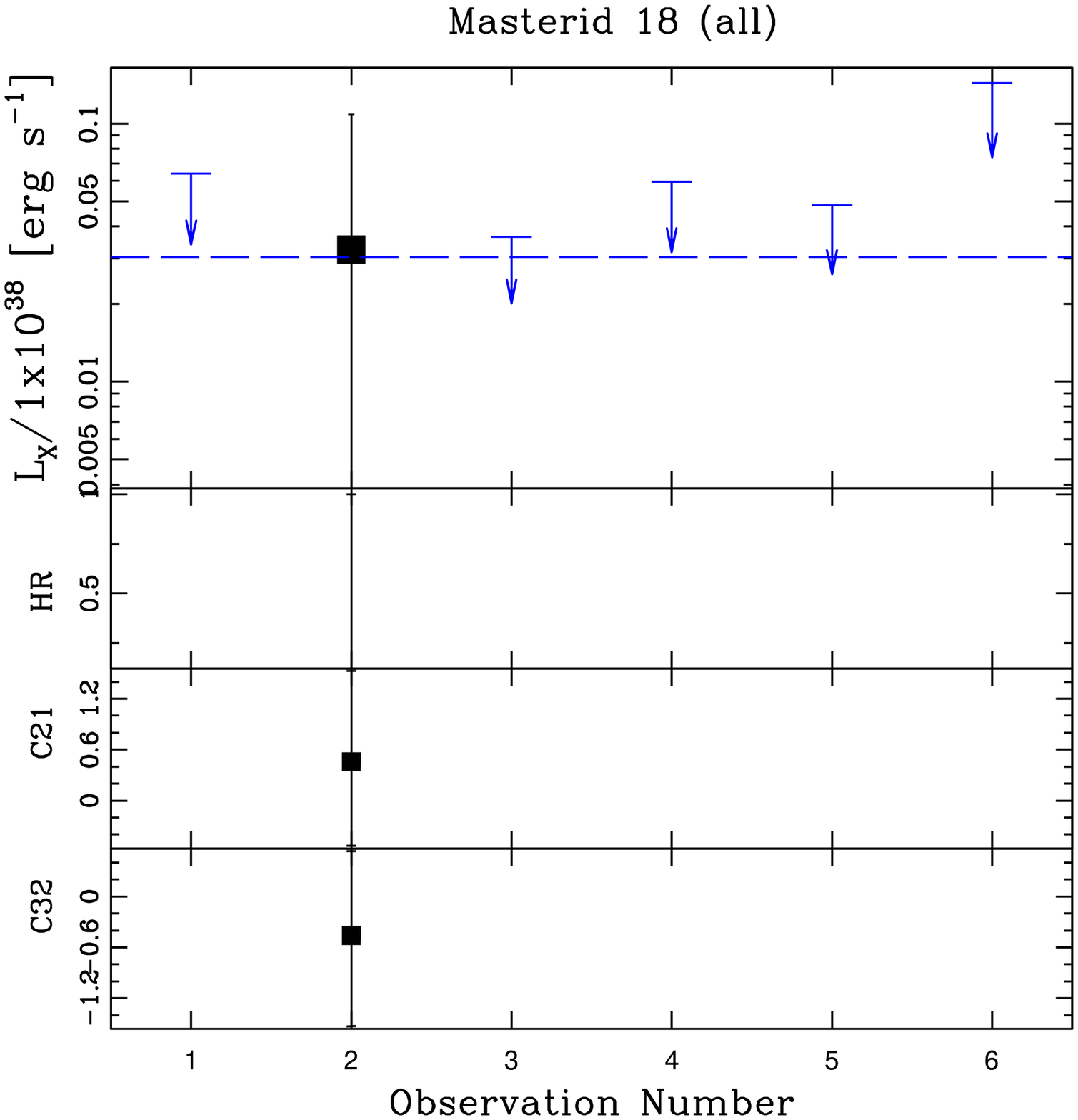}

 \end{minipage}\hspace{0.02\linewidth}
  
  \begin{minipage}{0.485\linewidth}
  \centering
  
    \includegraphics[width=\linewidth]{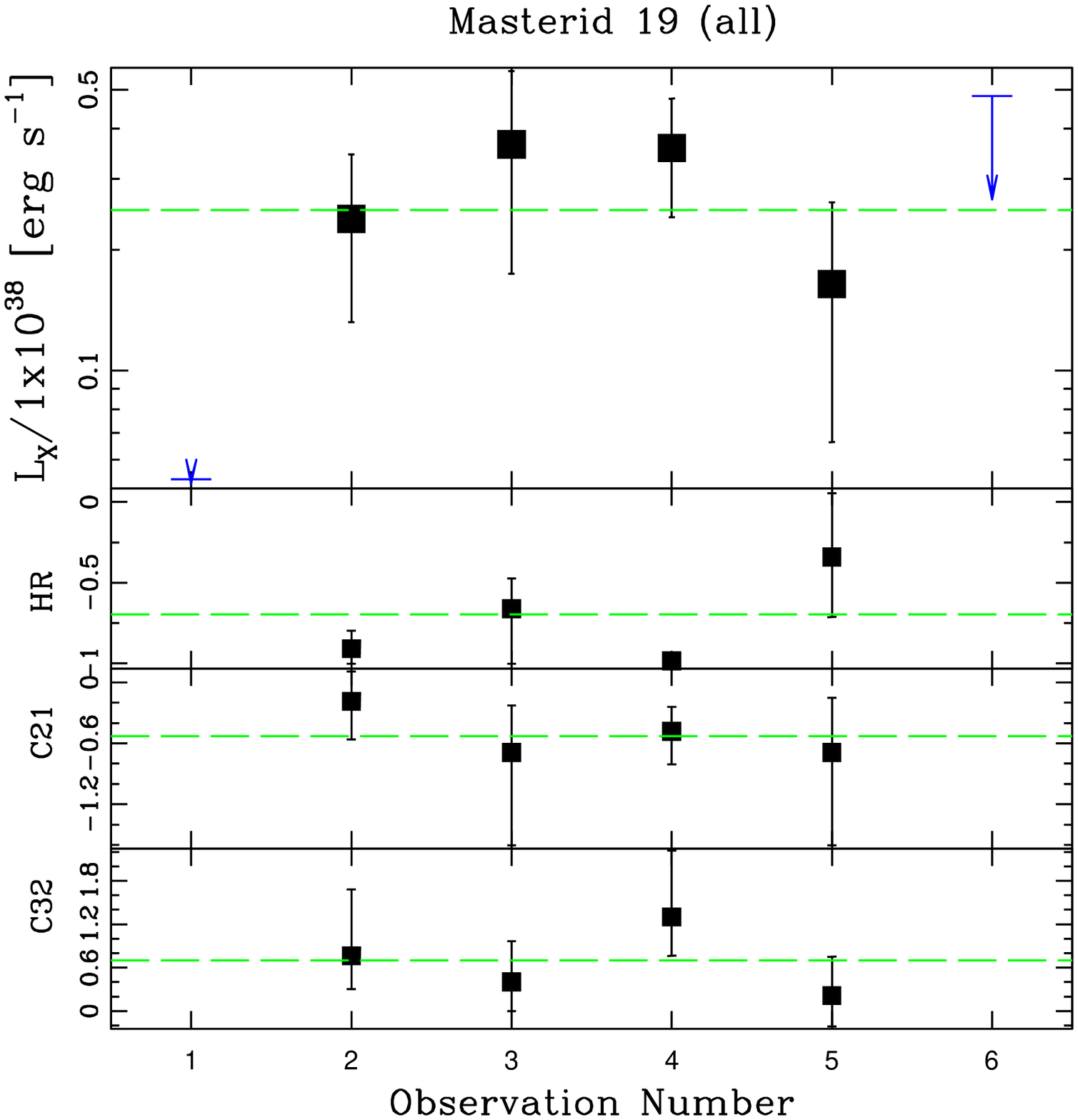}

  \end{minipage}\hspace{0.02\linewidth}
  \begin{minipage}{0.485\linewidth}
  \centering

    \includegraphics[width=\linewidth]{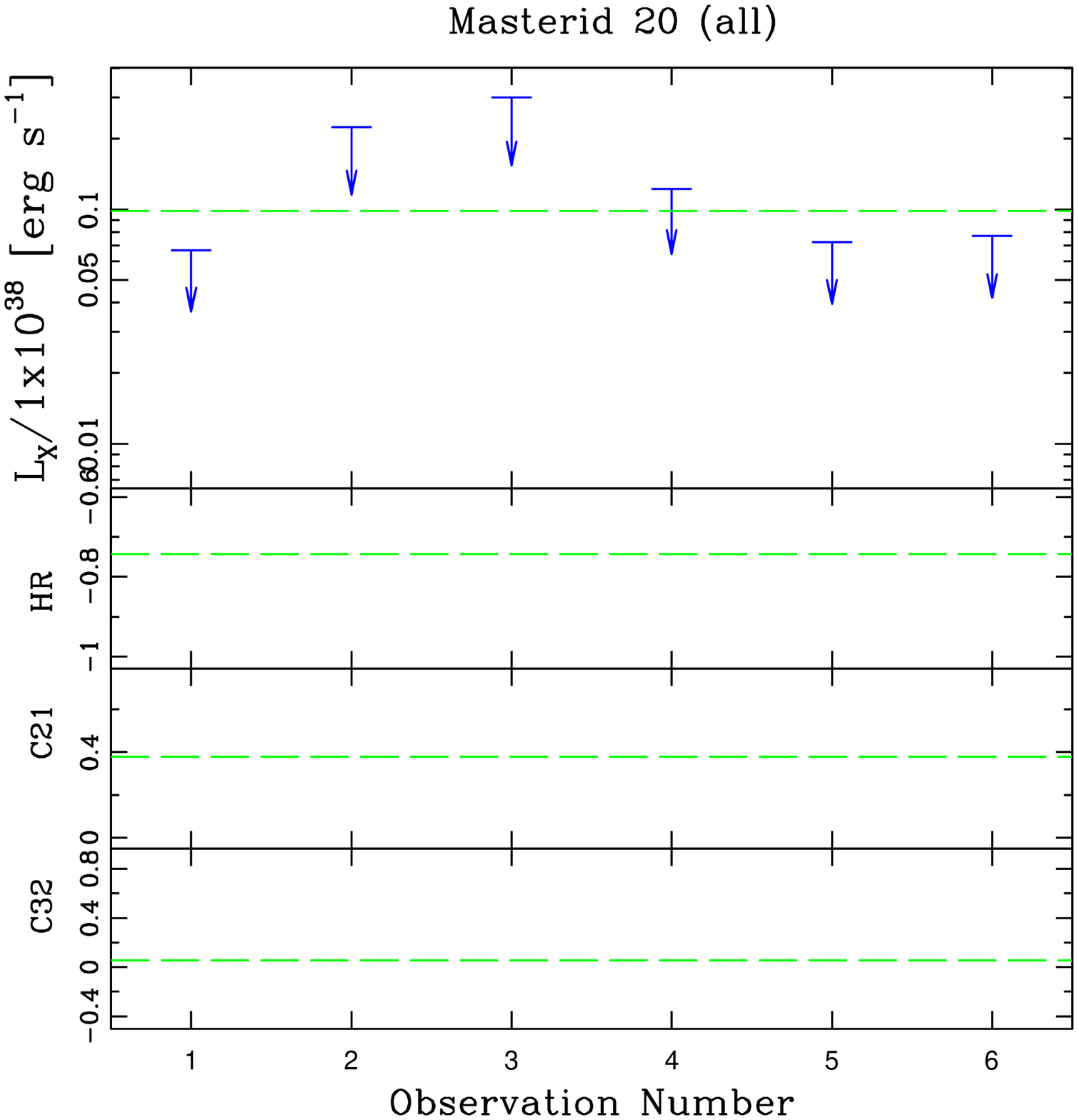}

\end{minipage}\hspace{0.02\linewidth}

\begin{minipage}{0.485\linewidth}
  \centering

    \includegraphics[width=\linewidth]{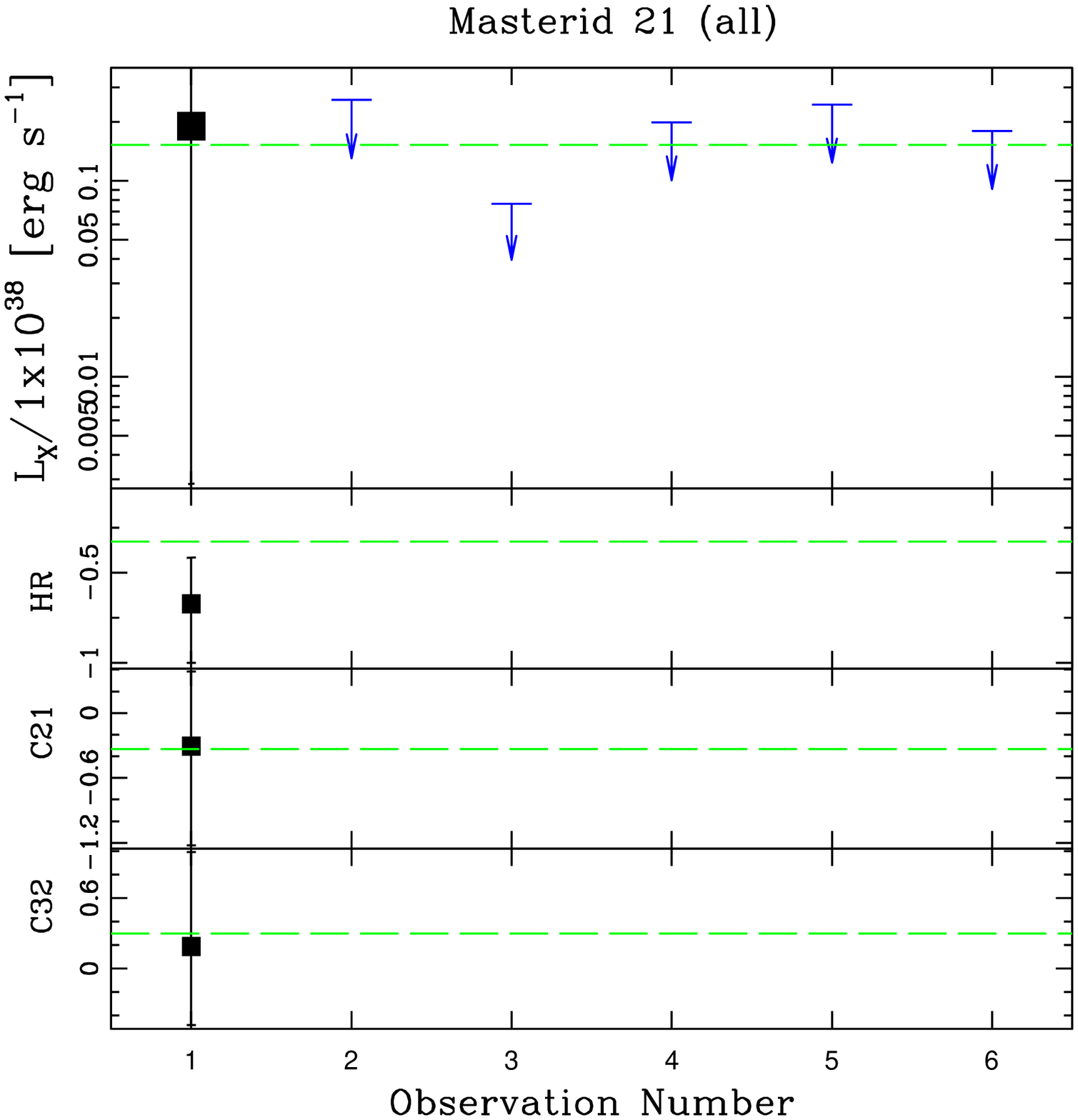}

 \end{minipage}\hspace{0.02\linewidth}
\begin{minipage}{0.485\linewidth}
  \centering
  
    \includegraphics[width=\linewidth]{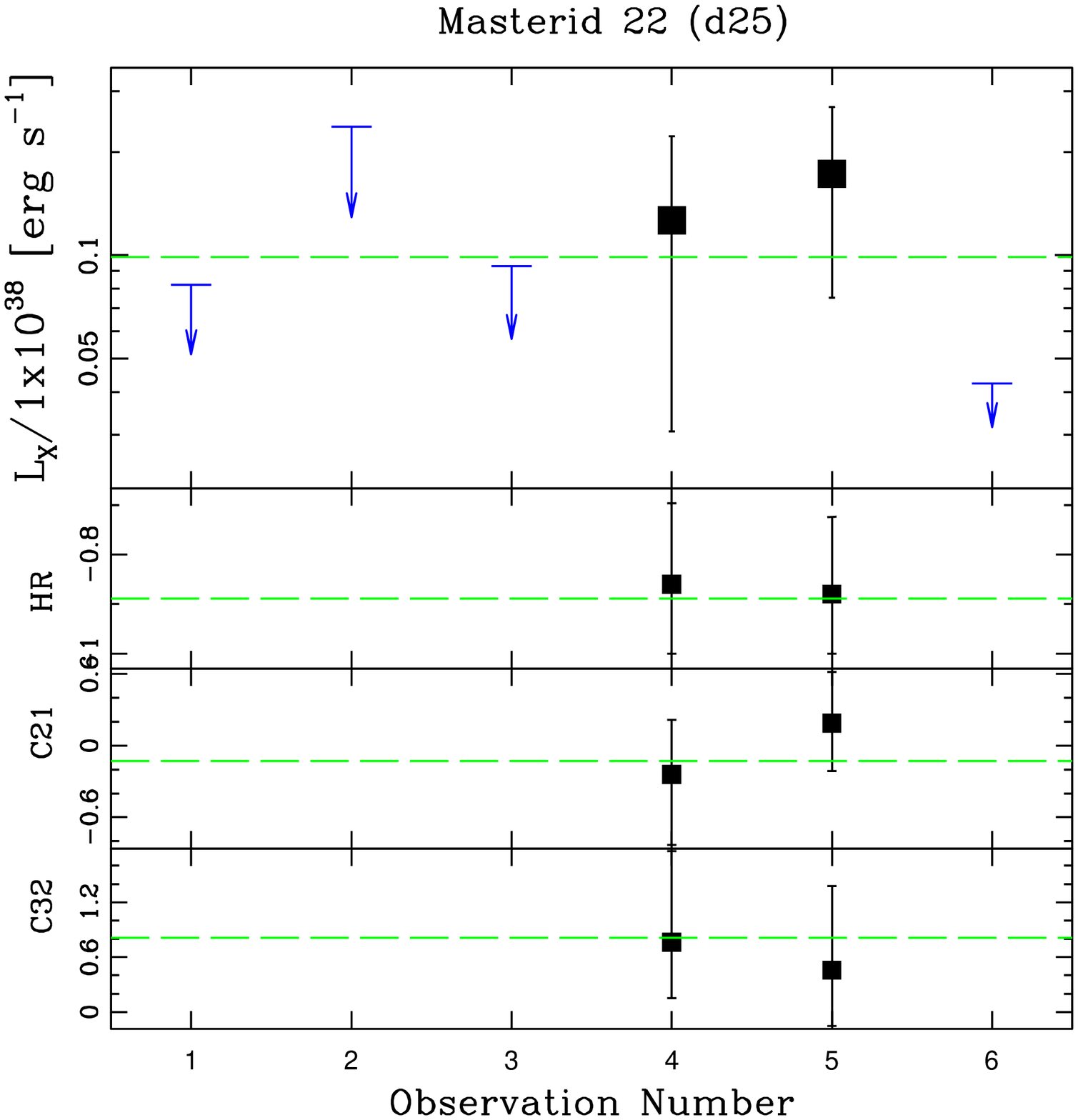}

  \end{minipage}\hspace{0.02\linewidth}

\end{figure}

\begin{figure}

  \begin{minipage}{0.485\linewidth}
  \centering

    \includegraphics[width=\linewidth]{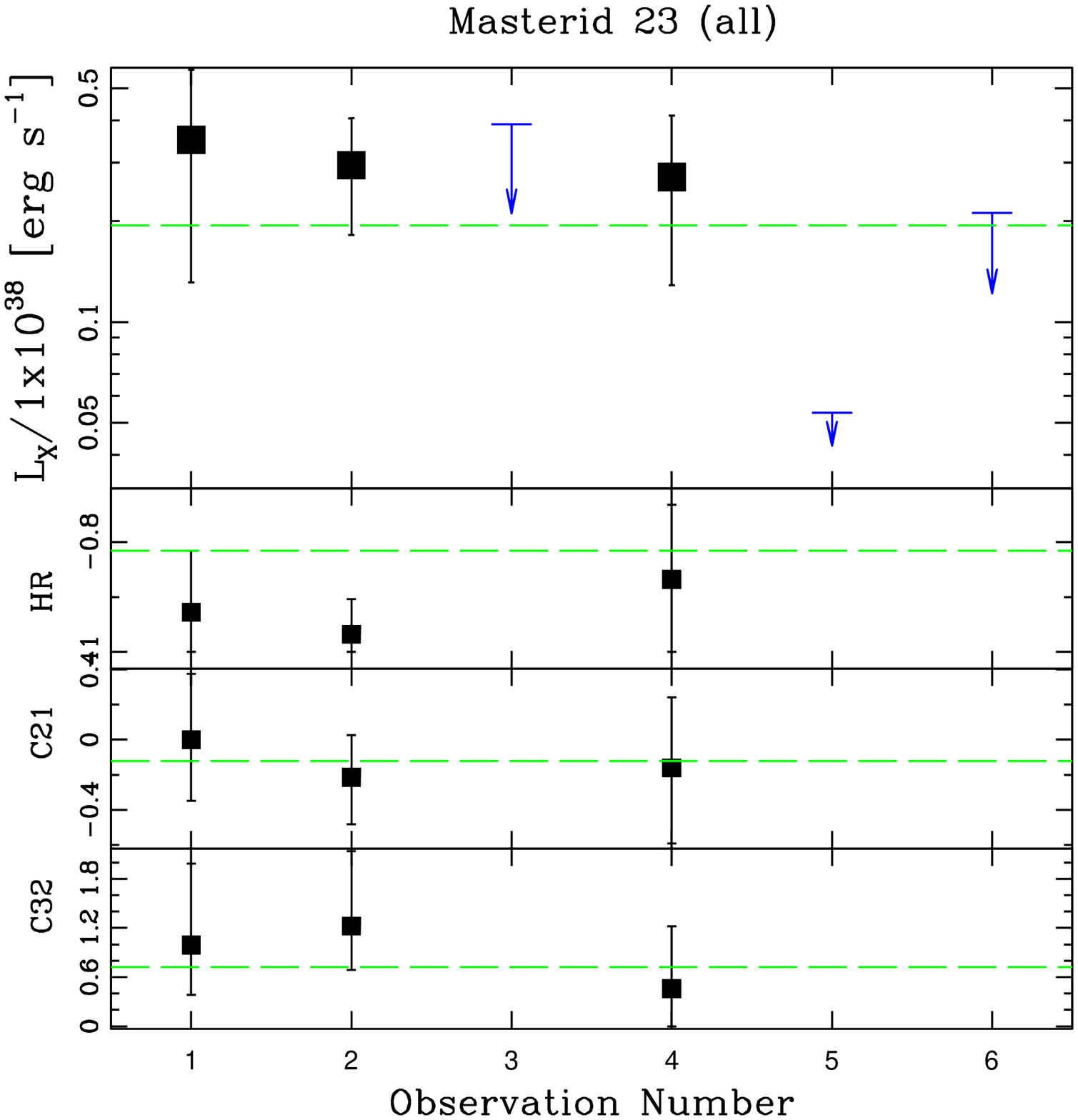}

\end{minipage}\hspace{0.02\linewidth}
\begin{minipage}{0.485\linewidth}
  \centering

    \includegraphics[width=\linewidth]{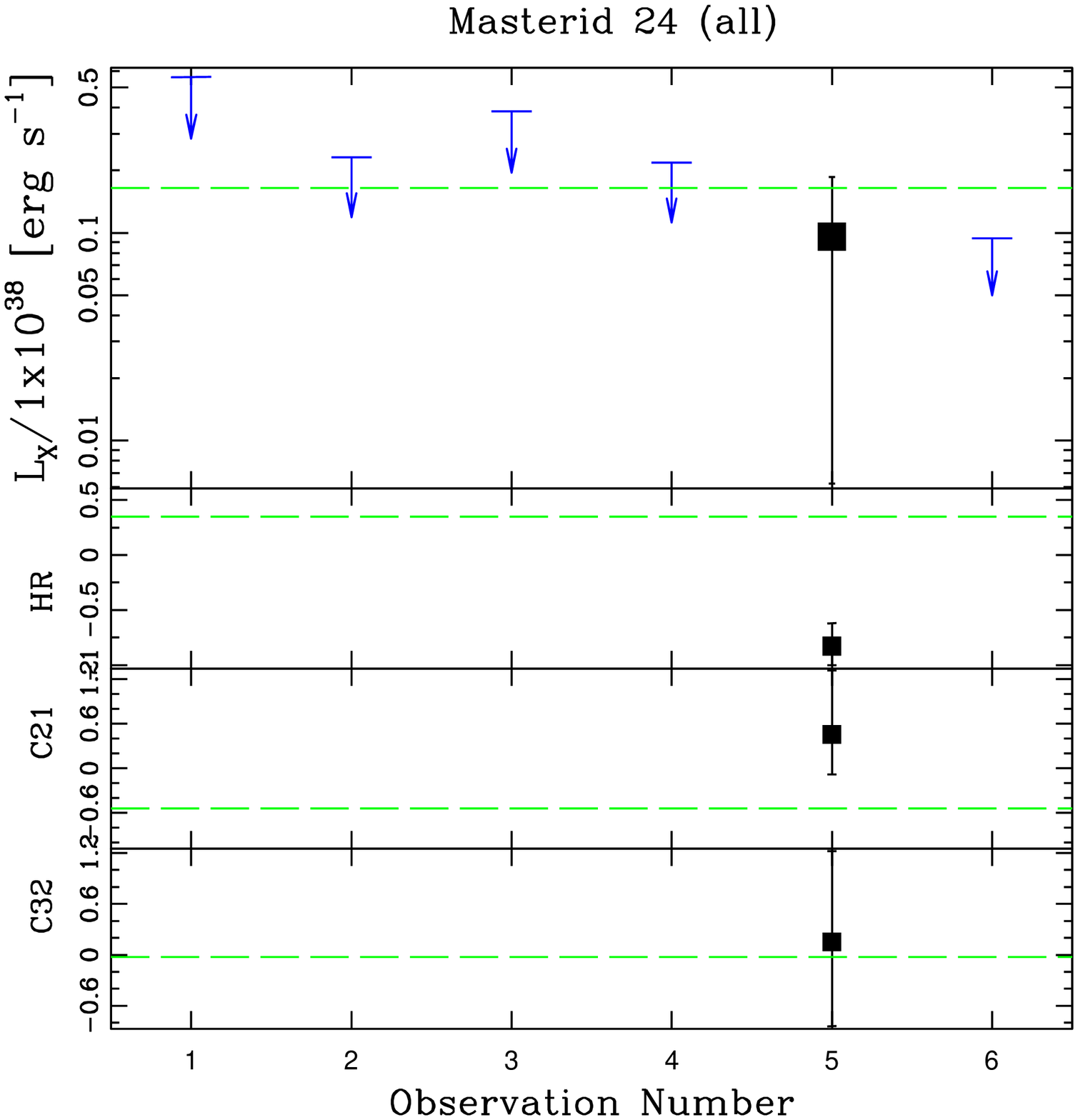}

 \end{minipage}\hspace{0.02\linewidth}

  \begin{minipage}{0.485\linewidth}
  \centering
  
    \includegraphics[width=\linewidth]{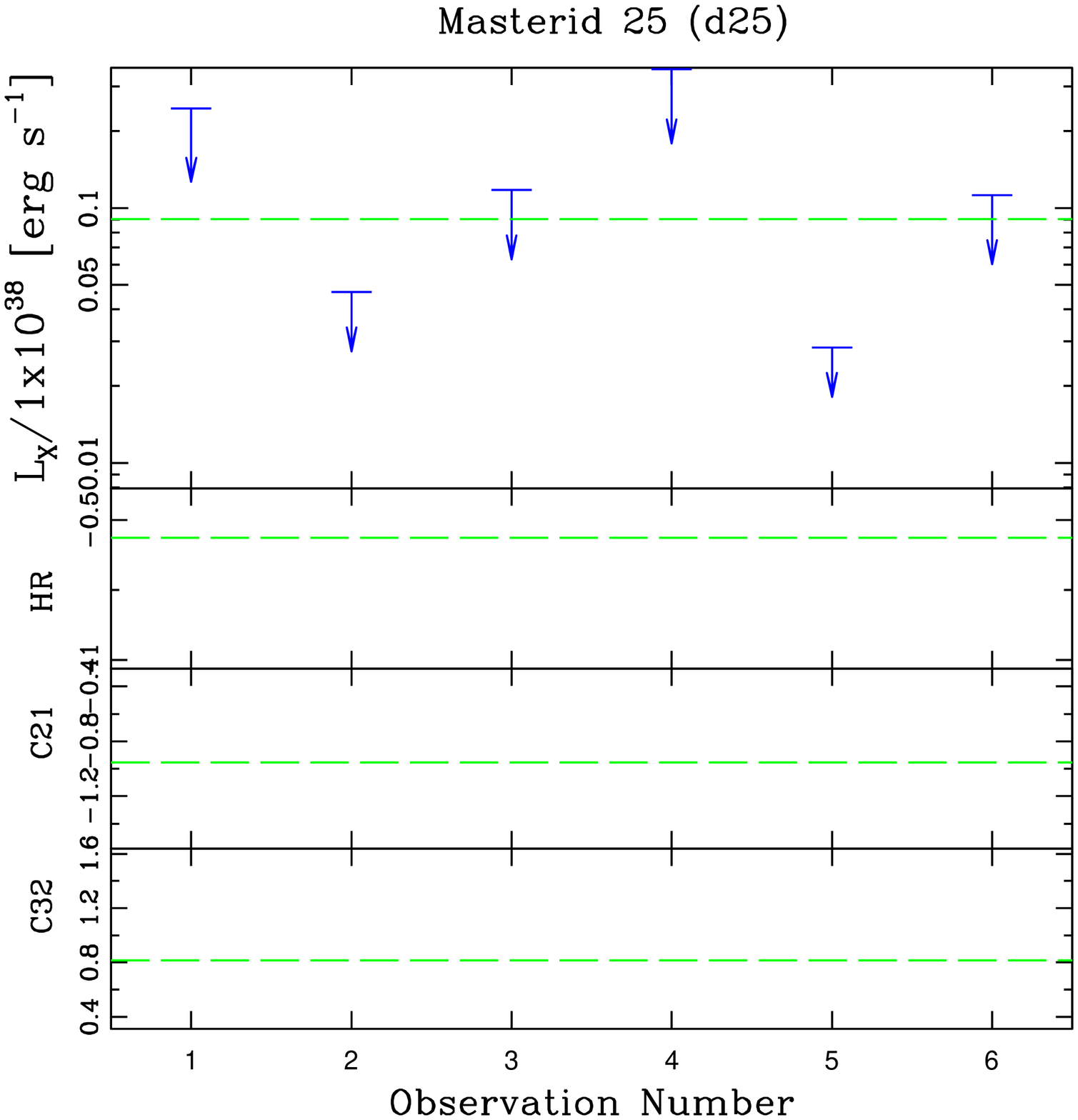}

  \end{minipage}\hspace{0.02\linewidth}
  \begin{minipage}{0.485\linewidth}
  \centering

    \includegraphics[width=\linewidth]{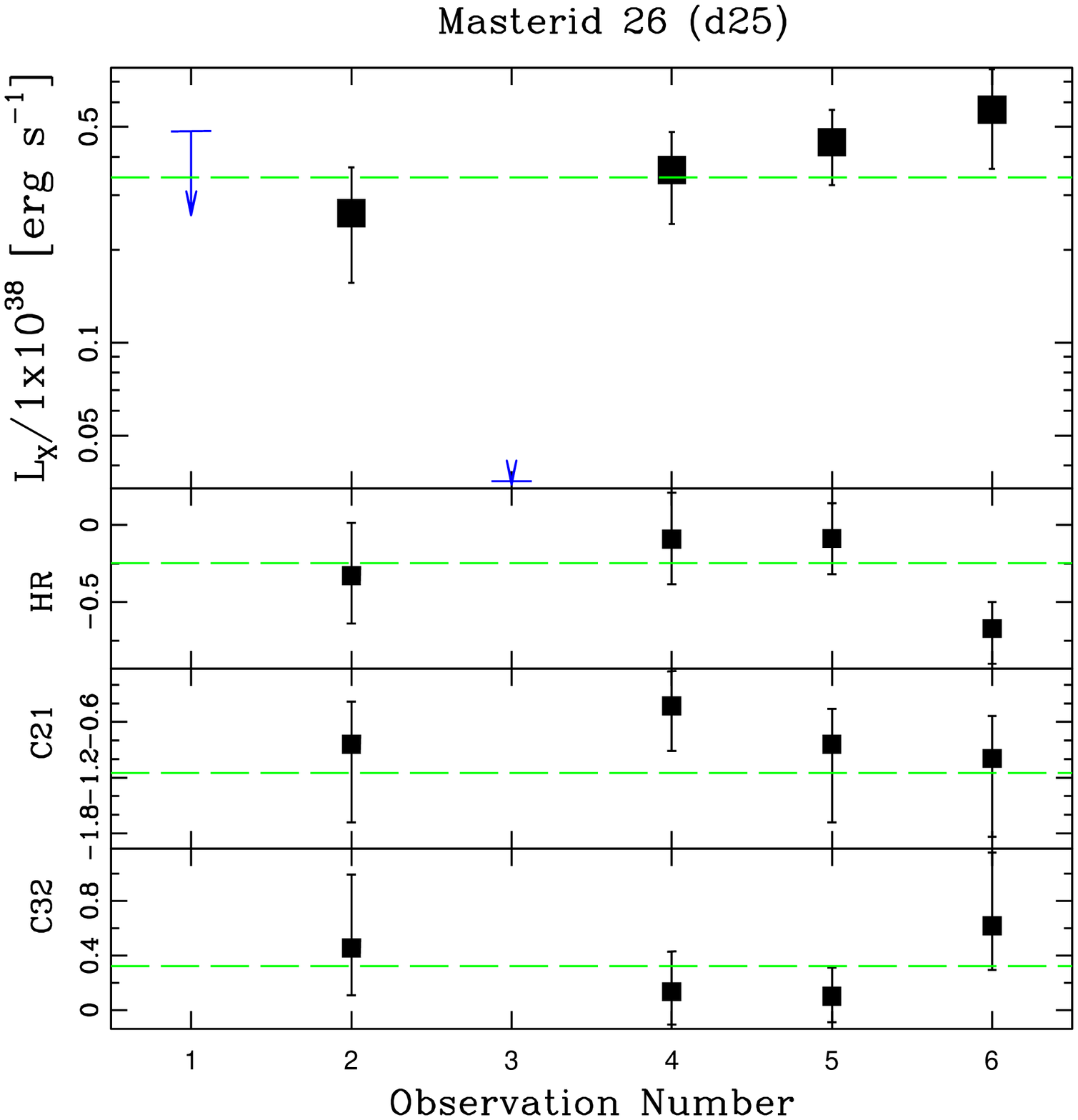}

\end{minipage}\hspace{0.02\linewidth}

\begin{minipage}{0.485\linewidth}
  \centering

    \includegraphics[width=\linewidth]{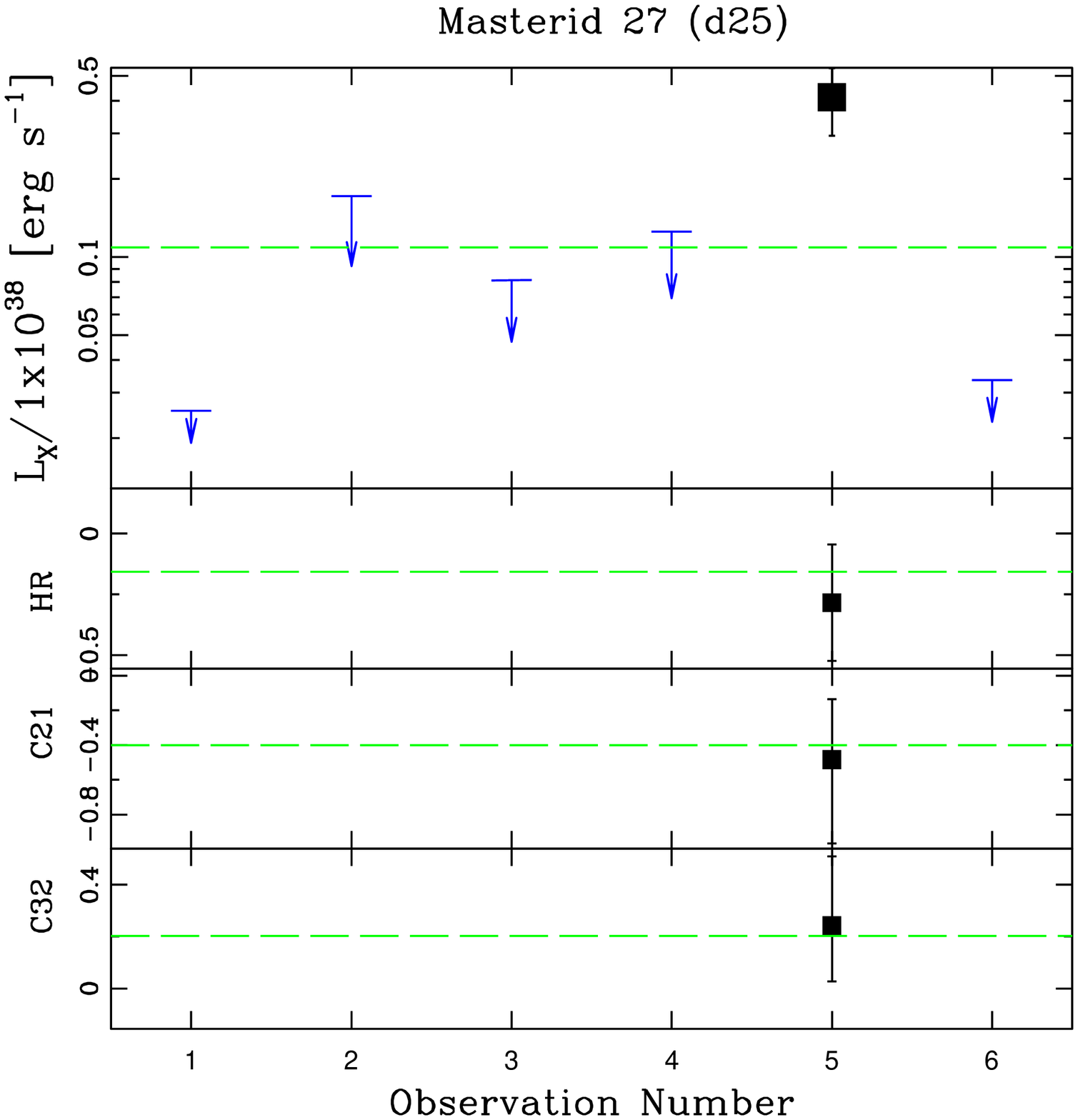}

 \end{minipage}\hspace{0.02\linewidth}
\begin{minipage}{0.485\linewidth}
  \centering
  
    \includegraphics[width=\linewidth]{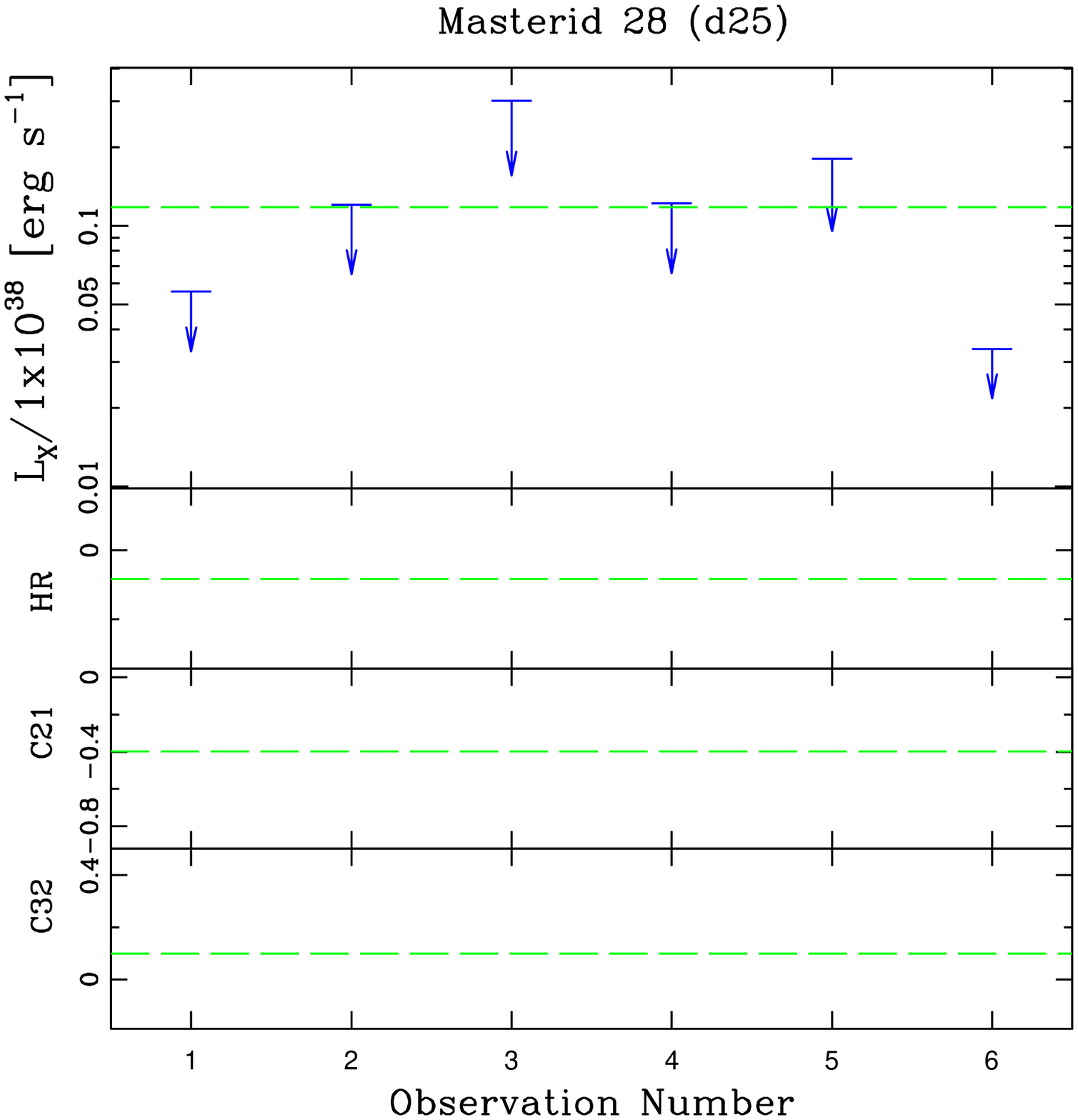}

  \end{minipage}\hspace{0.02\linewidth}
  
\end{figure}

\clearpage

\begin{figure}

  \begin{minipage}{0.485\linewidth}
  \centering

    \includegraphics[width=\linewidth]{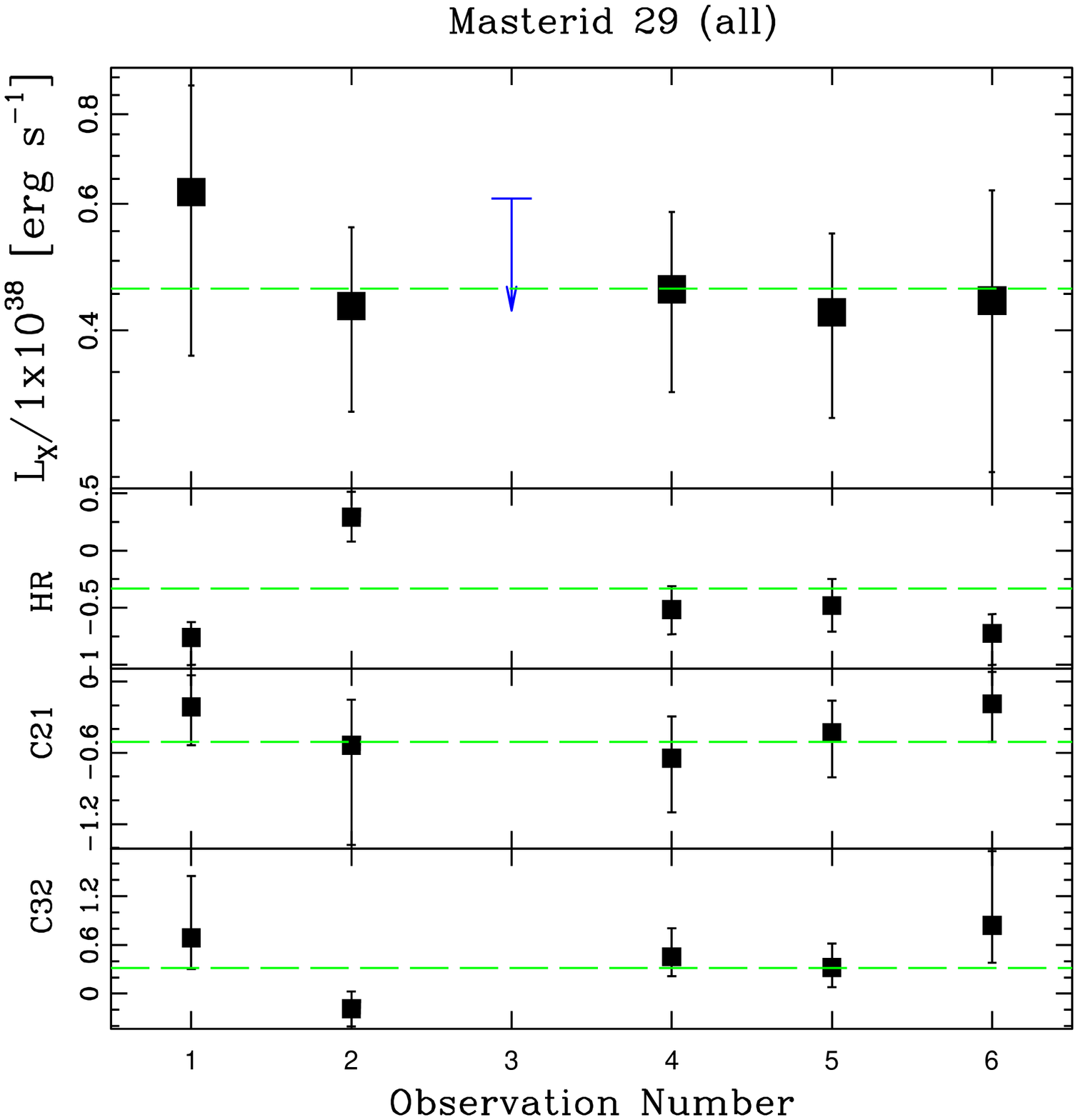}

\end{minipage}\hspace{0.02\linewidth}
\begin{minipage}{0.485\linewidth}
  \centering

    \includegraphics[width=\linewidth]{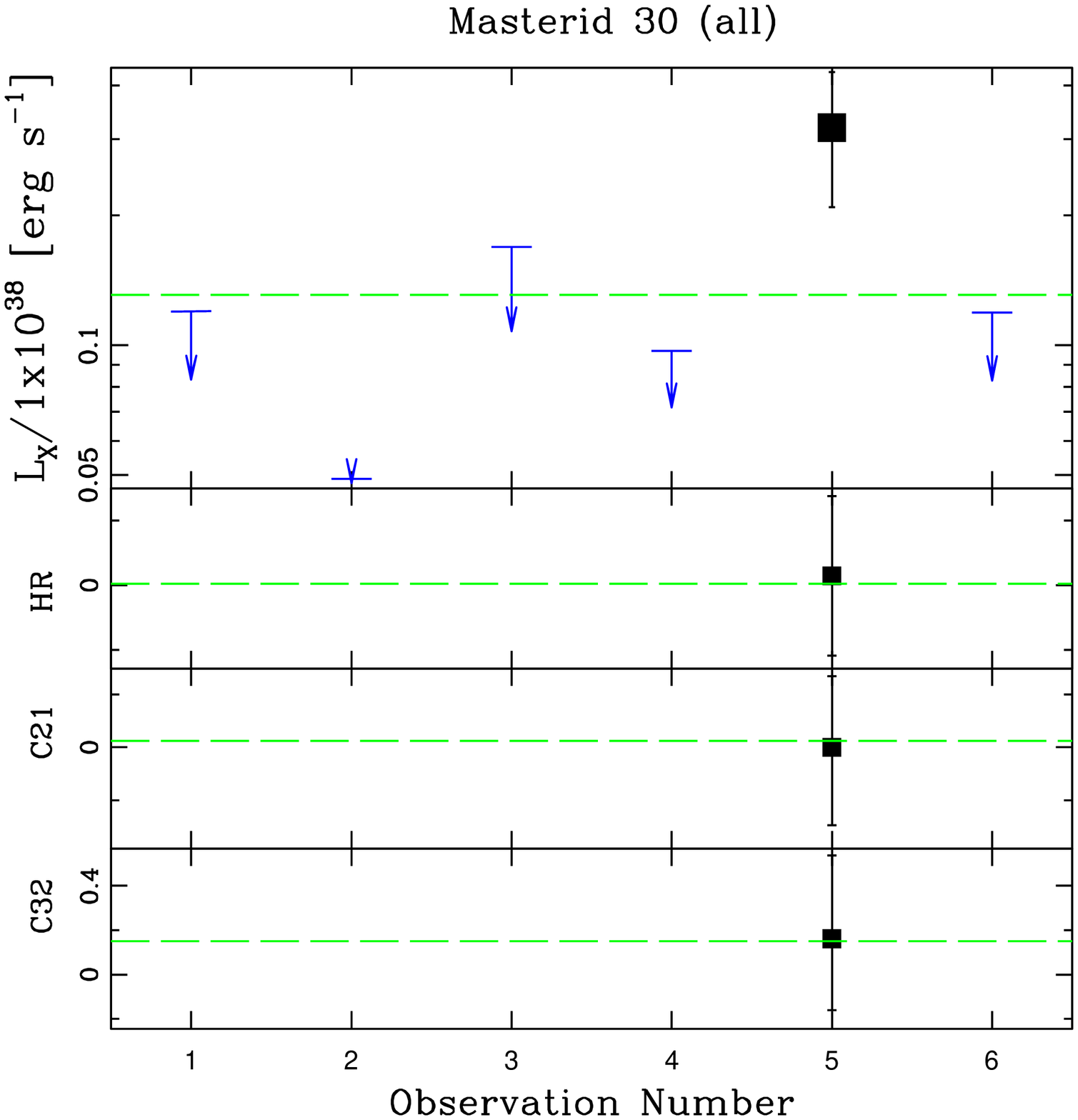}

 \end{minipage}\hspace{0.02\linewidth}

  \begin{minipage}{0.485\linewidth}
  \centering
  
    \includegraphics[width=\linewidth]{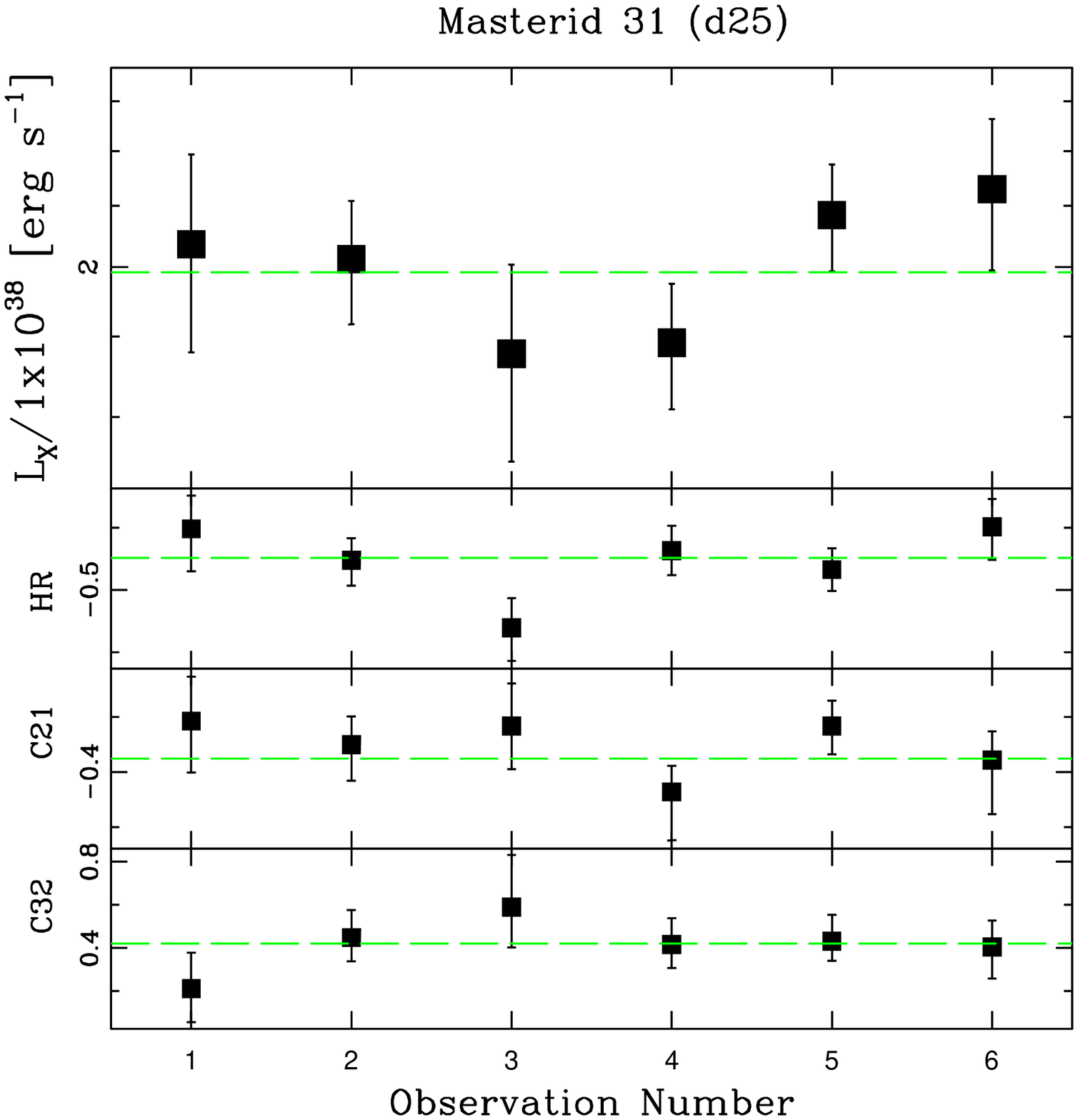}

  \end{minipage}\hspace{0.02\linewidth}
  \begin{minipage}{0.485\linewidth}
  \centering

    \includegraphics[width=\linewidth]{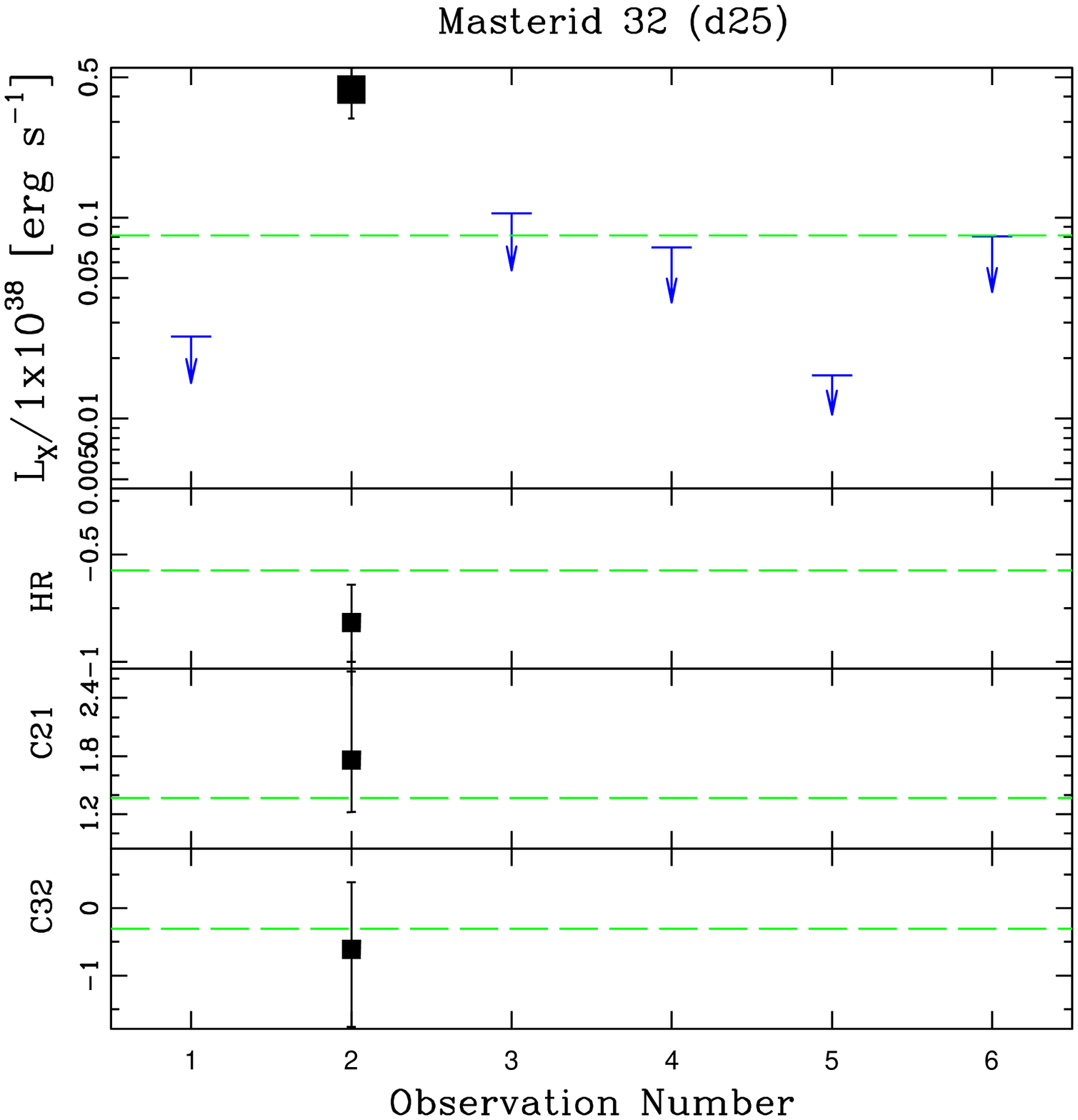}

\end{minipage}\hspace{0.02\linewidth}

\begin{minipage}{0.485\linewidth}
  \centering

    \includegraphics[width=\linewidth]{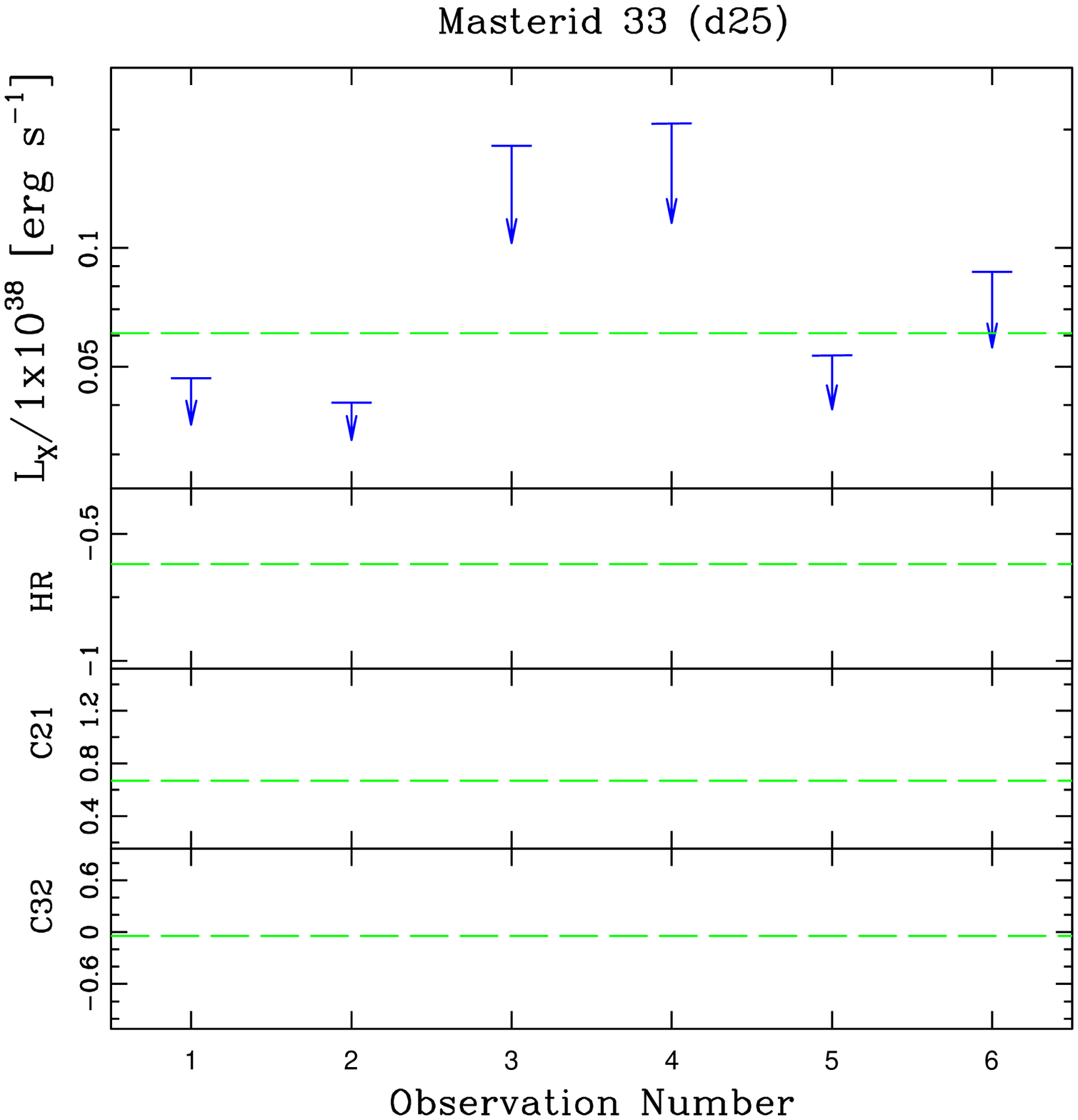}

 \end{minipage}\hspace{0.02\linewidth}
\begin{minipage}{0.485\linewidth}
  \centering
  
    \includegraphics[width=\linewidth]{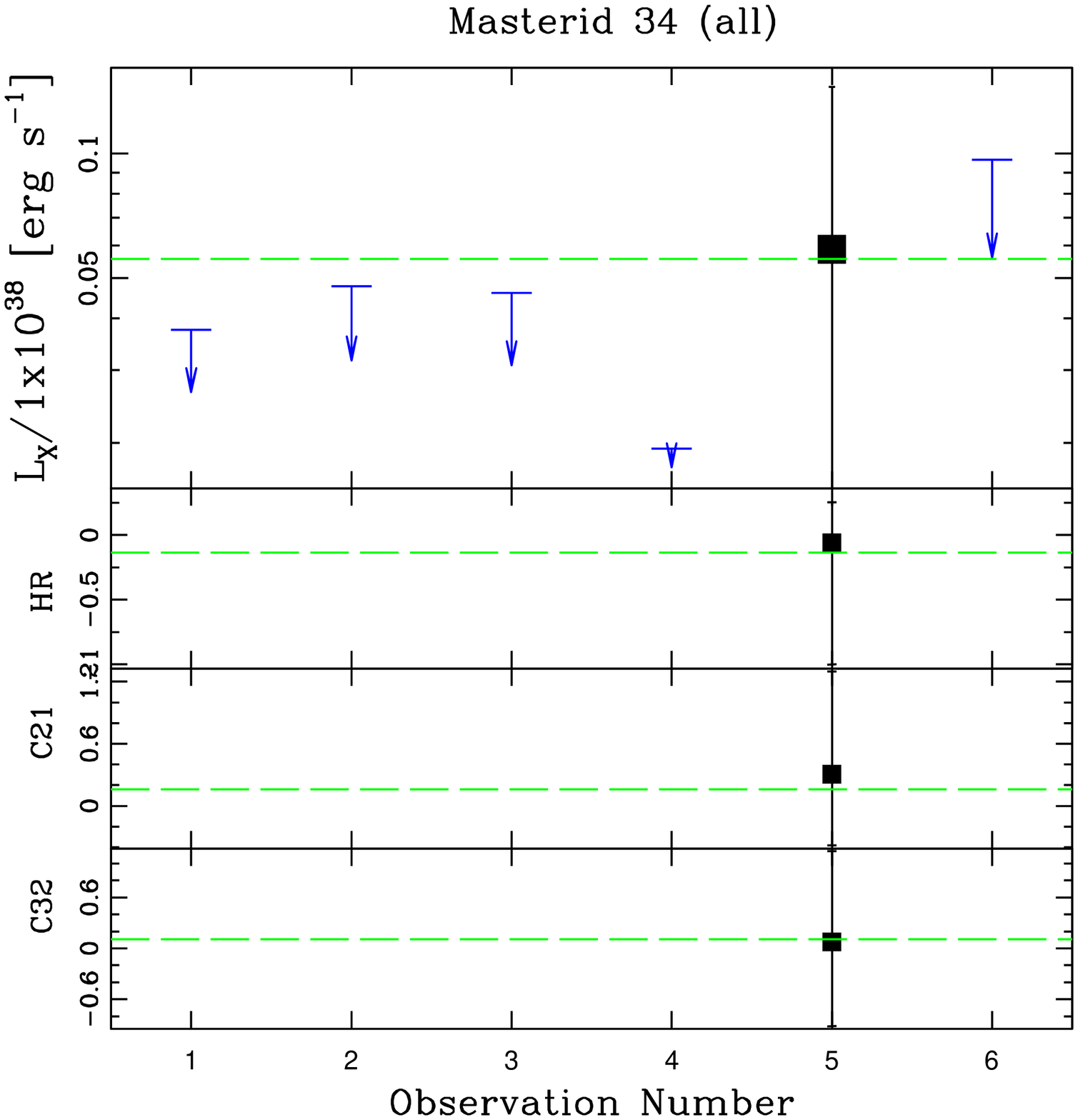}

  \end{minipage}\hspace{0.02\linewidth}

\end{figure}

\begin{figure}

  \begin{minipage}{0.485\linewidth}
  \centering

    \includegraphics[width=\linewidth]{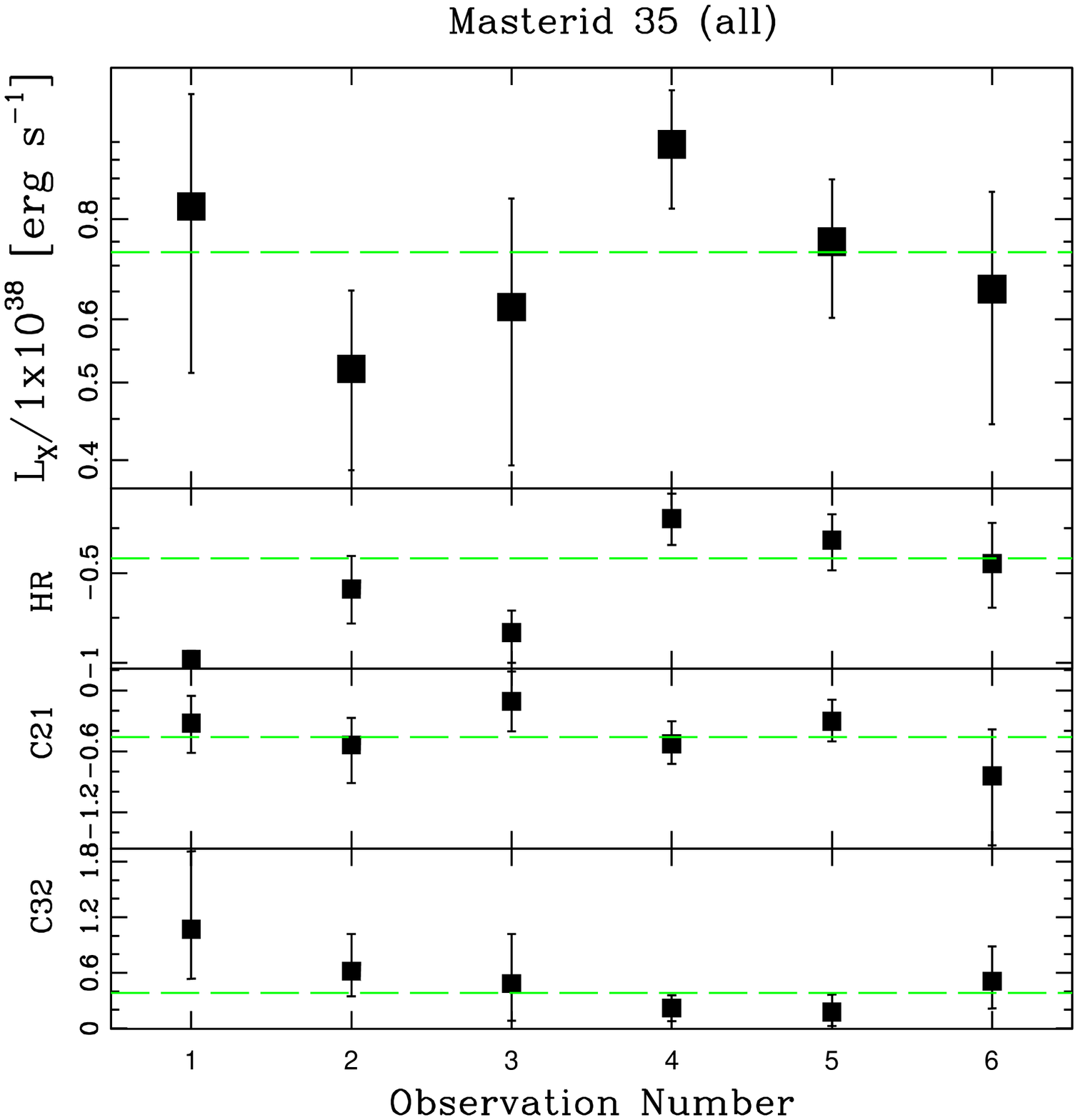}

\end{minipage}\hspace{0.02\linewidth}
\begin{minipage}{0.485\linewidth}
  \centering

    \includegraphics[width=\linewidth]{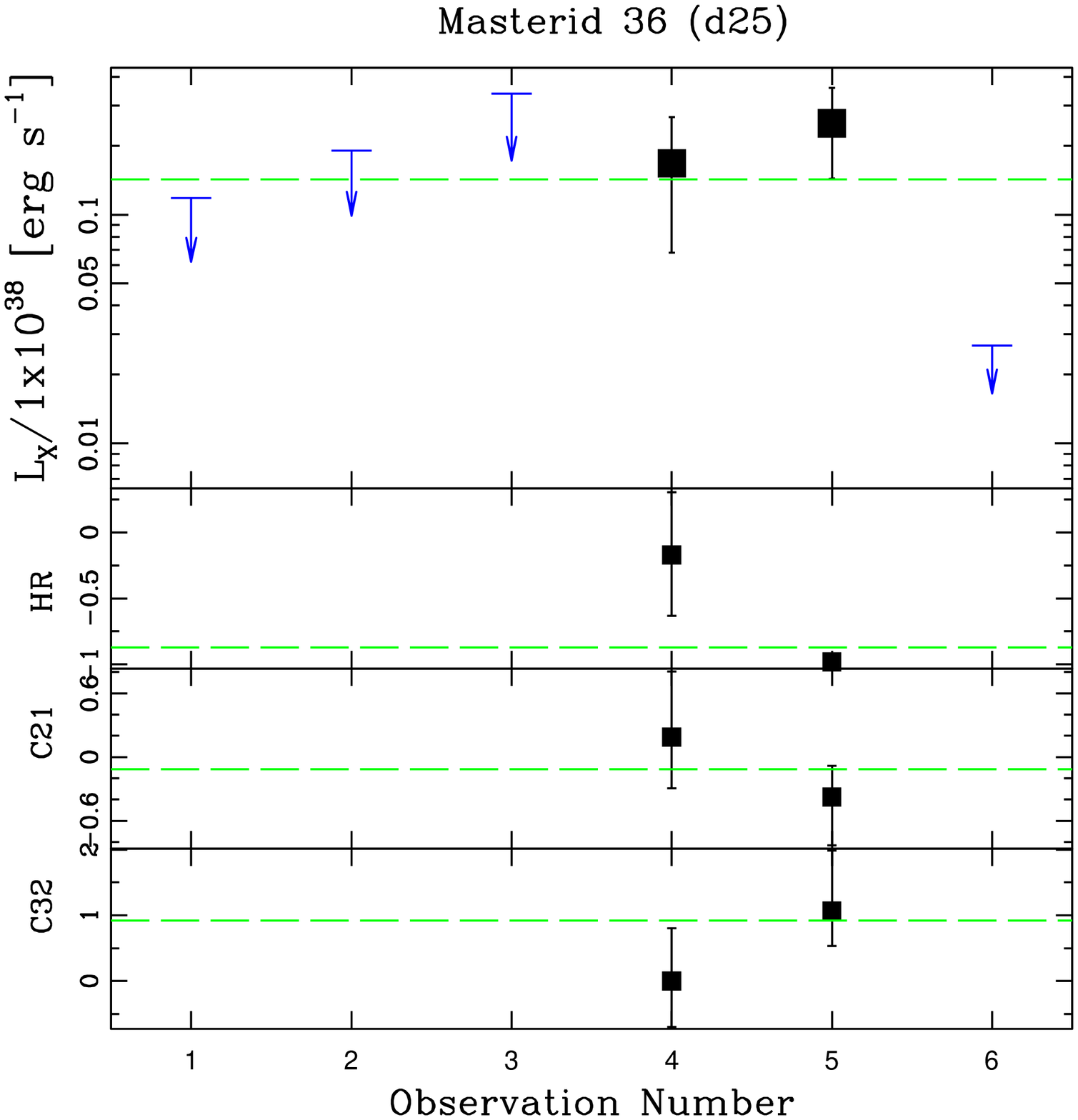}

 \end{minipage}\hspace{0.02\linewidth}

  \begin{minipage}{0.485\linewidth}
  \centering
  
    \includegraphics[width=\linewidth]{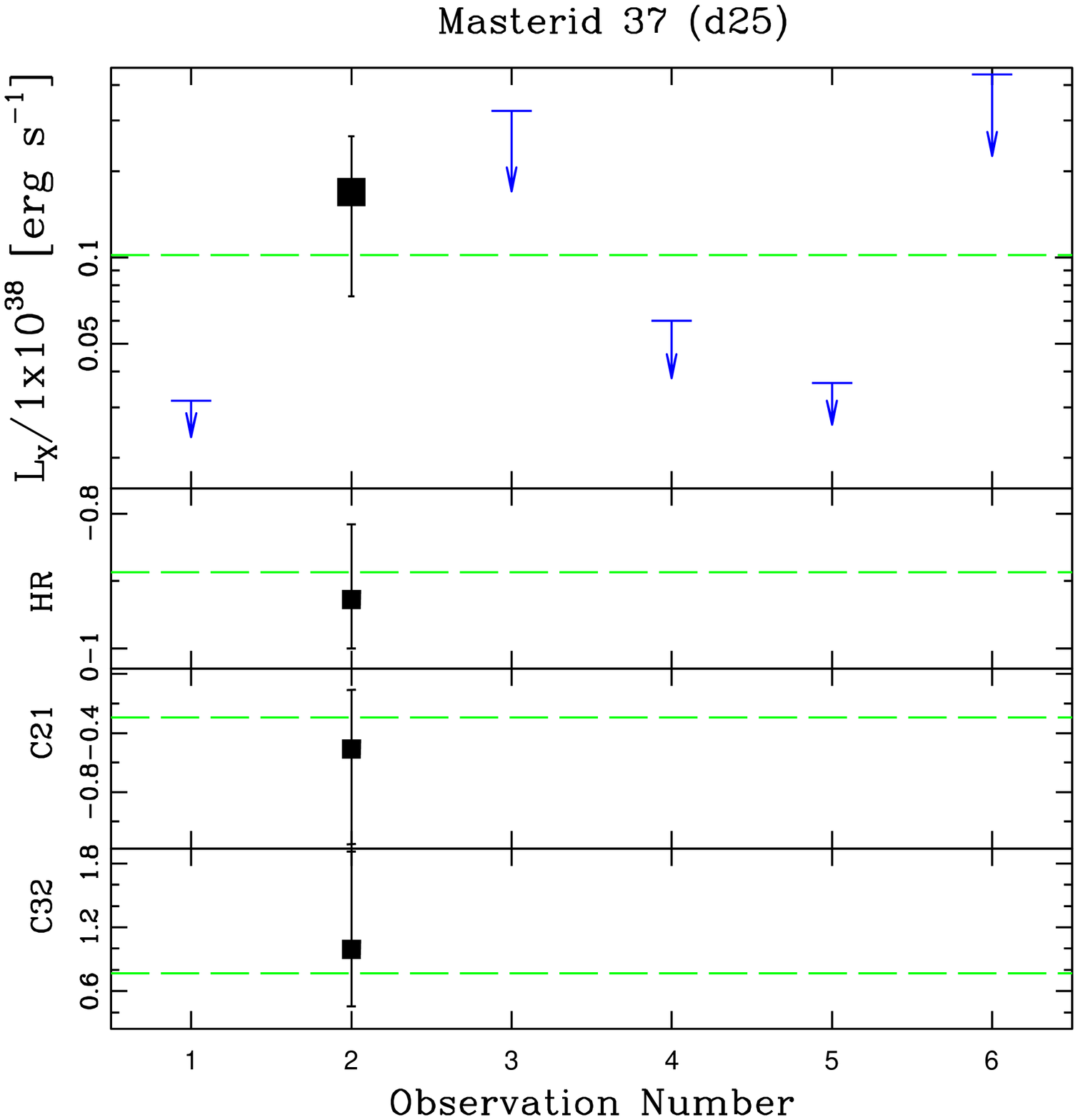}

  \end{minipage}\hspace{0.02\linewidth}
  \begin{minipage}{0.485\linewidth}
  \centering

    \includegraphics[width=\linewidth]{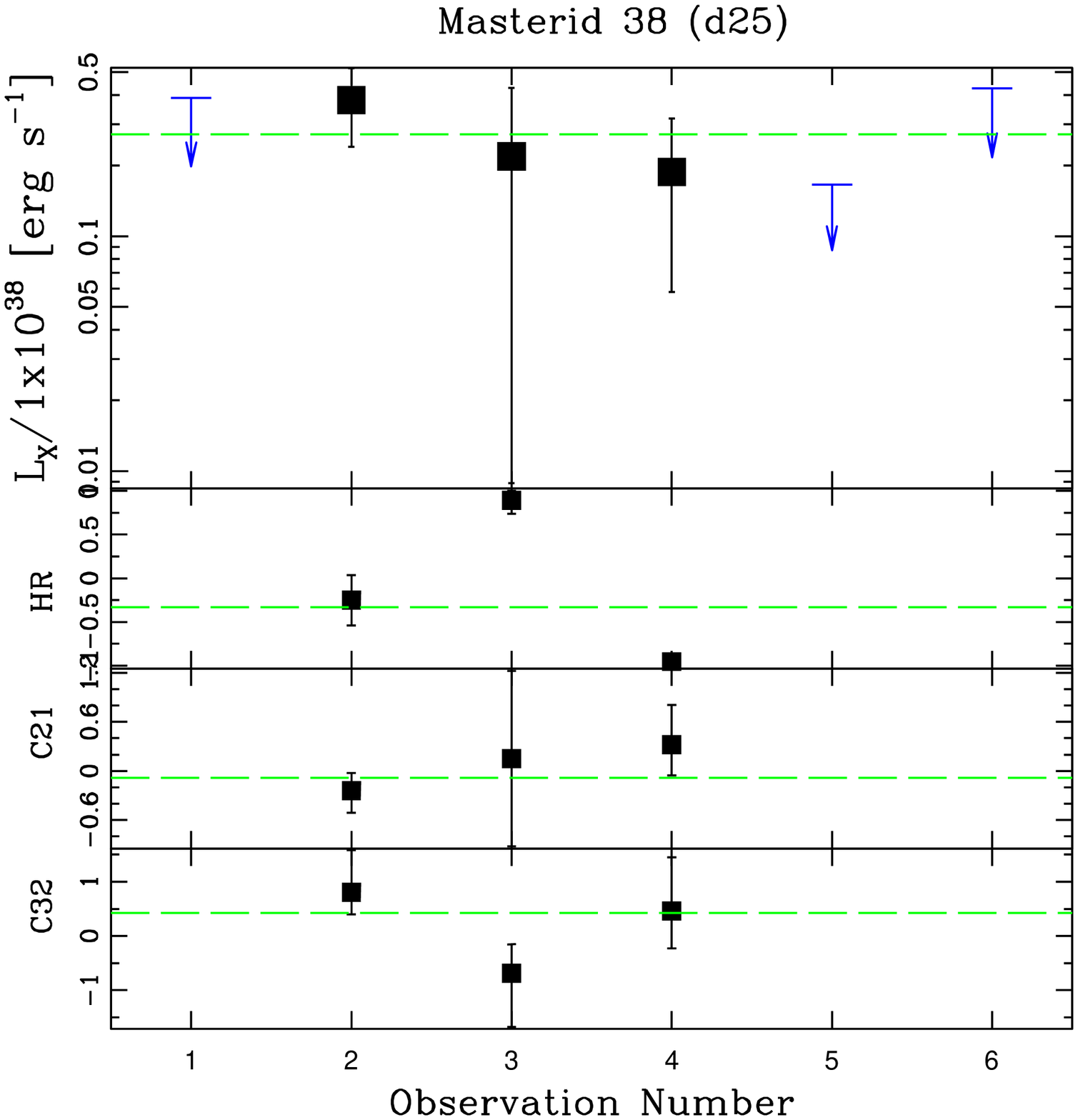}

\end{minipage}\hspace{0.02\linewidth}

\begin{minipage}{0.485\linewidth}
  \centering

    \includegraphics[width=\linewidth]{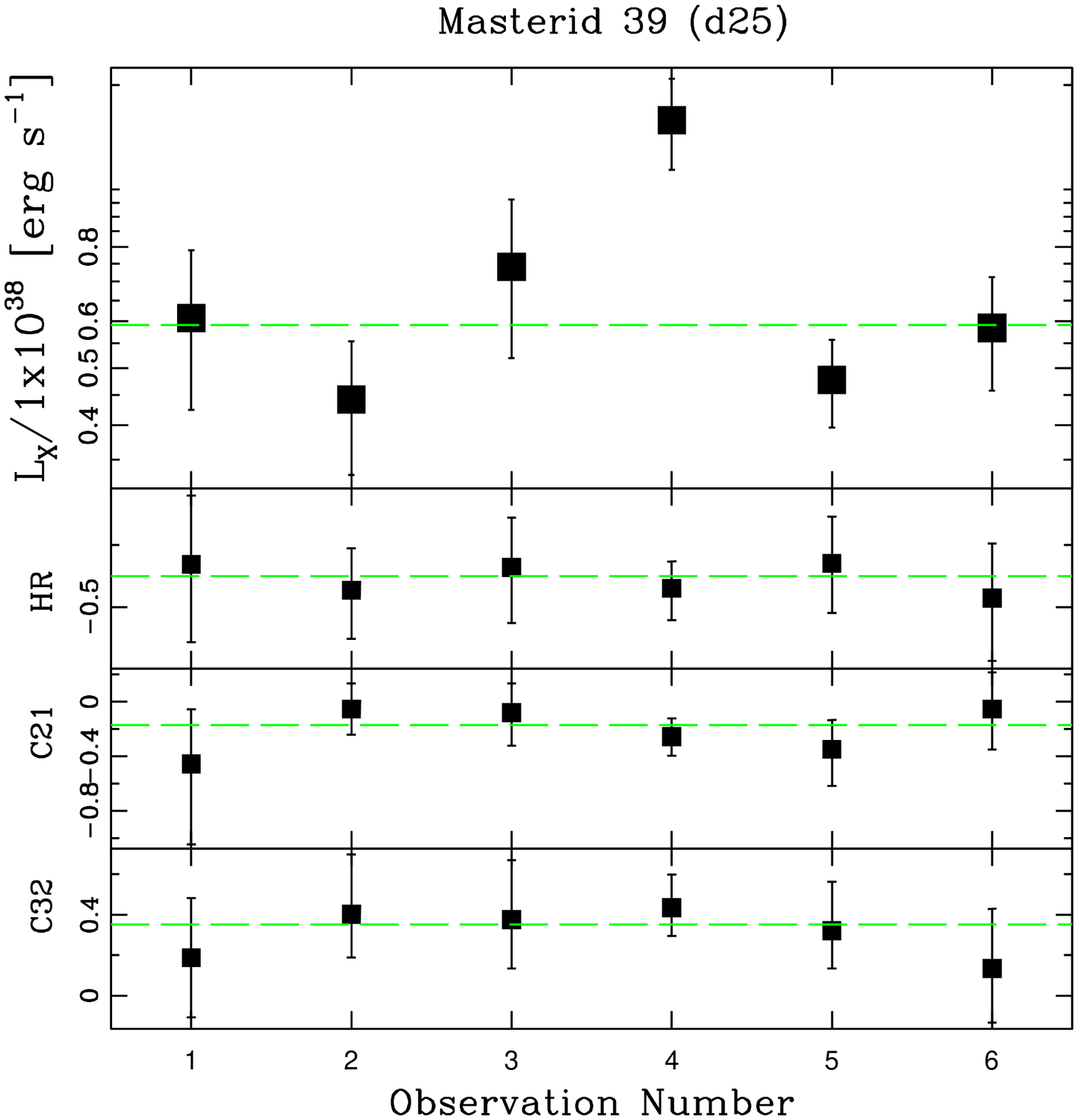}

 \end{minipage}\hspace{0.02\linewidth}
\begin{minipage}{0.485\linewidth}
  \centering
  
    \includegraphics[width=\linewidth]{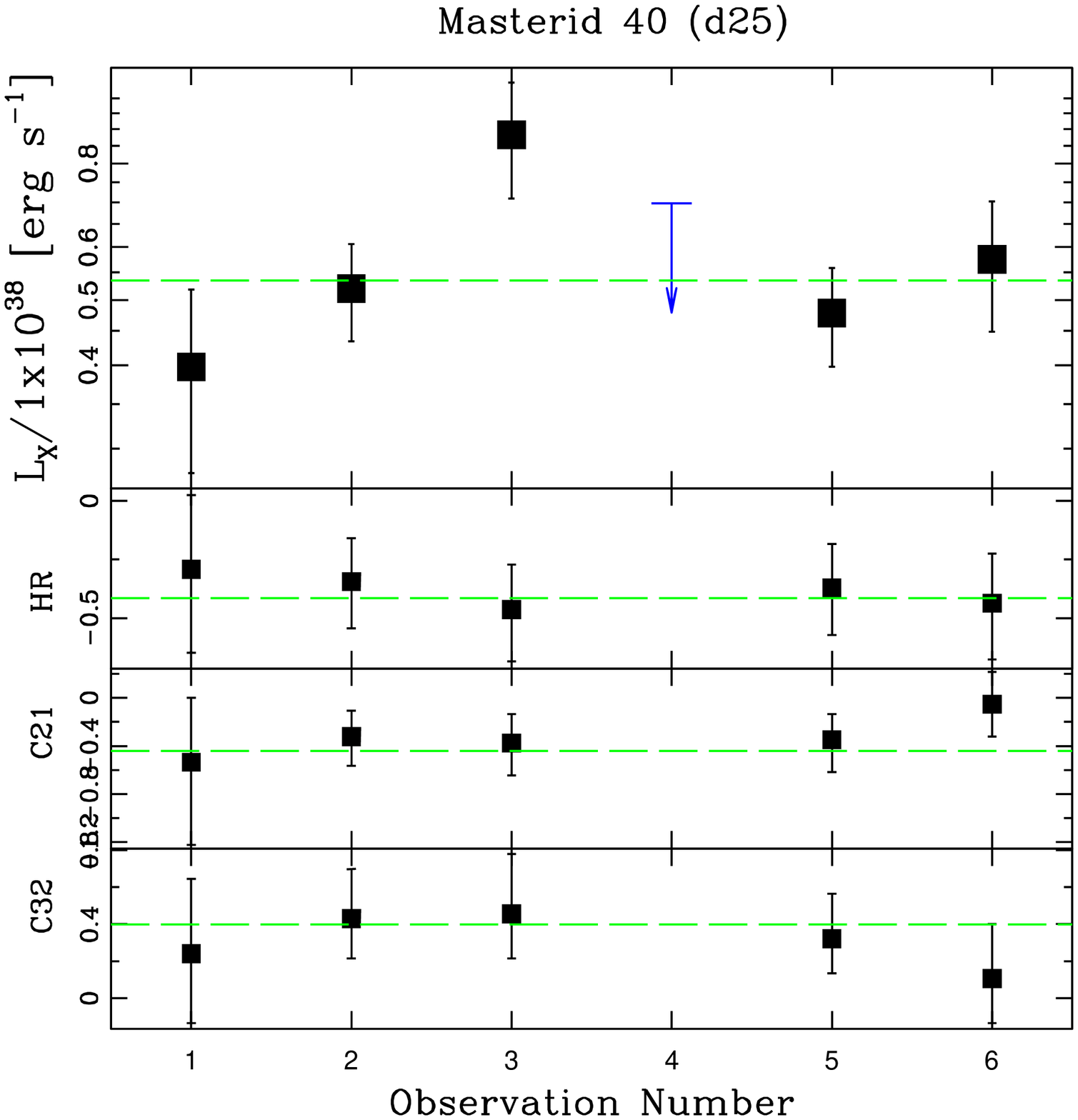}

  \end{minipage}\hspace{0.02\linewidth}

\end{figure}

\begin{figure}

  \begin{minipage}{0.485\linewidth}
  \centering

    \includegraphics[width=\linewidth]{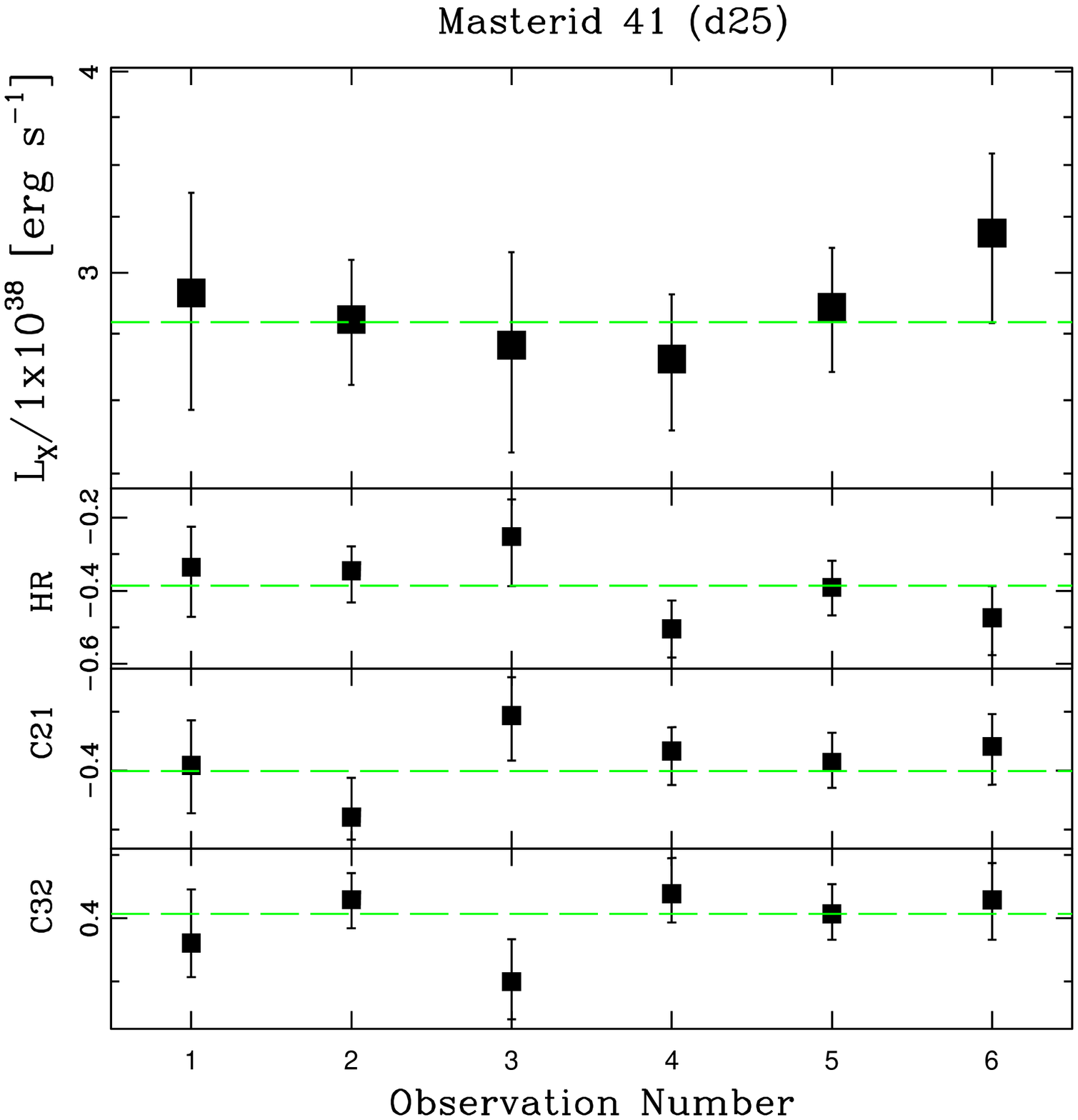}

\end{minipage}\hspace{0.02\linewidth}
\begin{minipage}{0.485\linewidth}
  \centering

    \includegraphics[width=\linewidth]{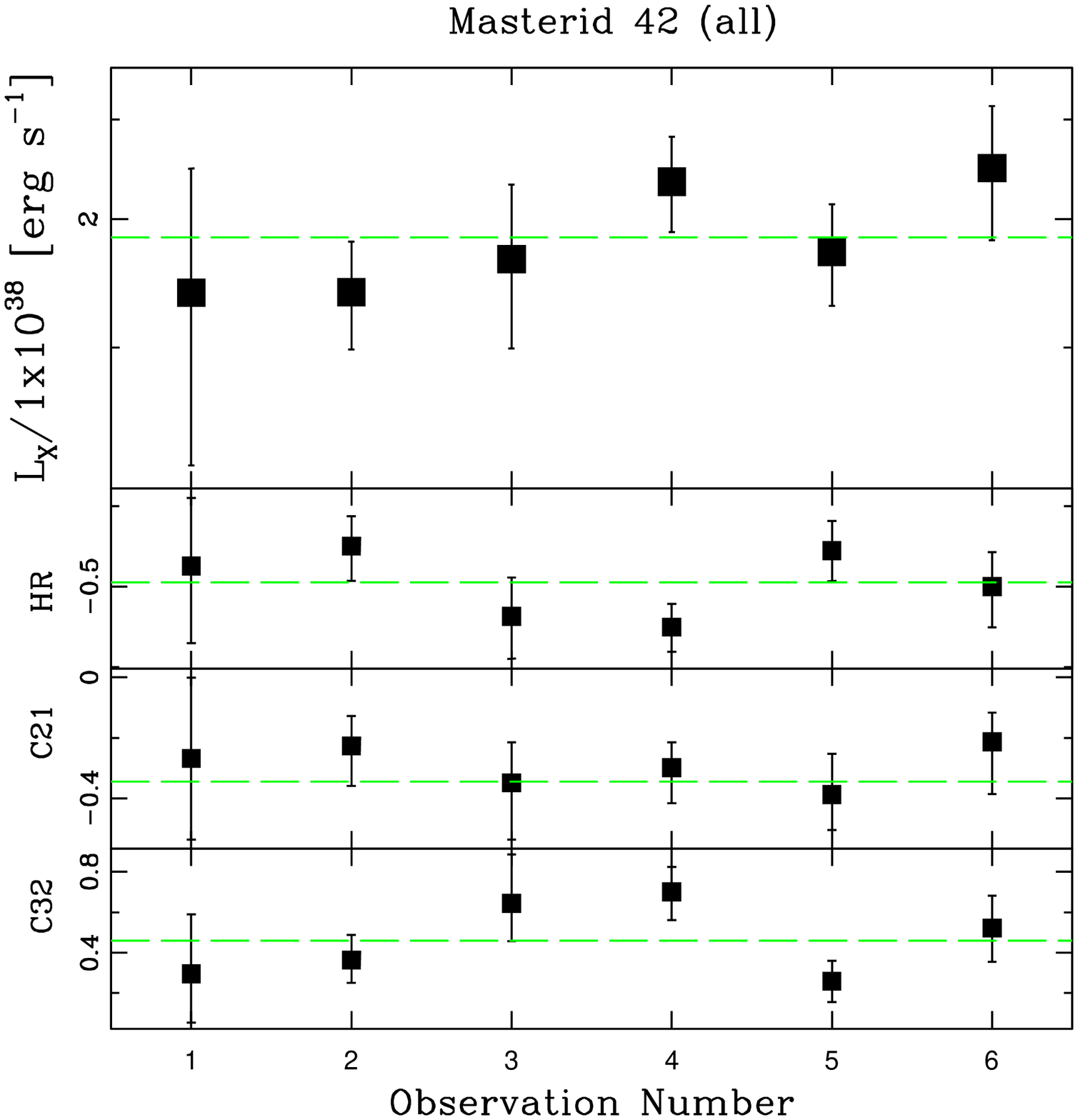}

 \end{minipage}\hspace{0.02\linewidth}

  \begin{minipage}{0.485\linewidth}
  \centering
  
    \includegraphics[width=\linewidth]{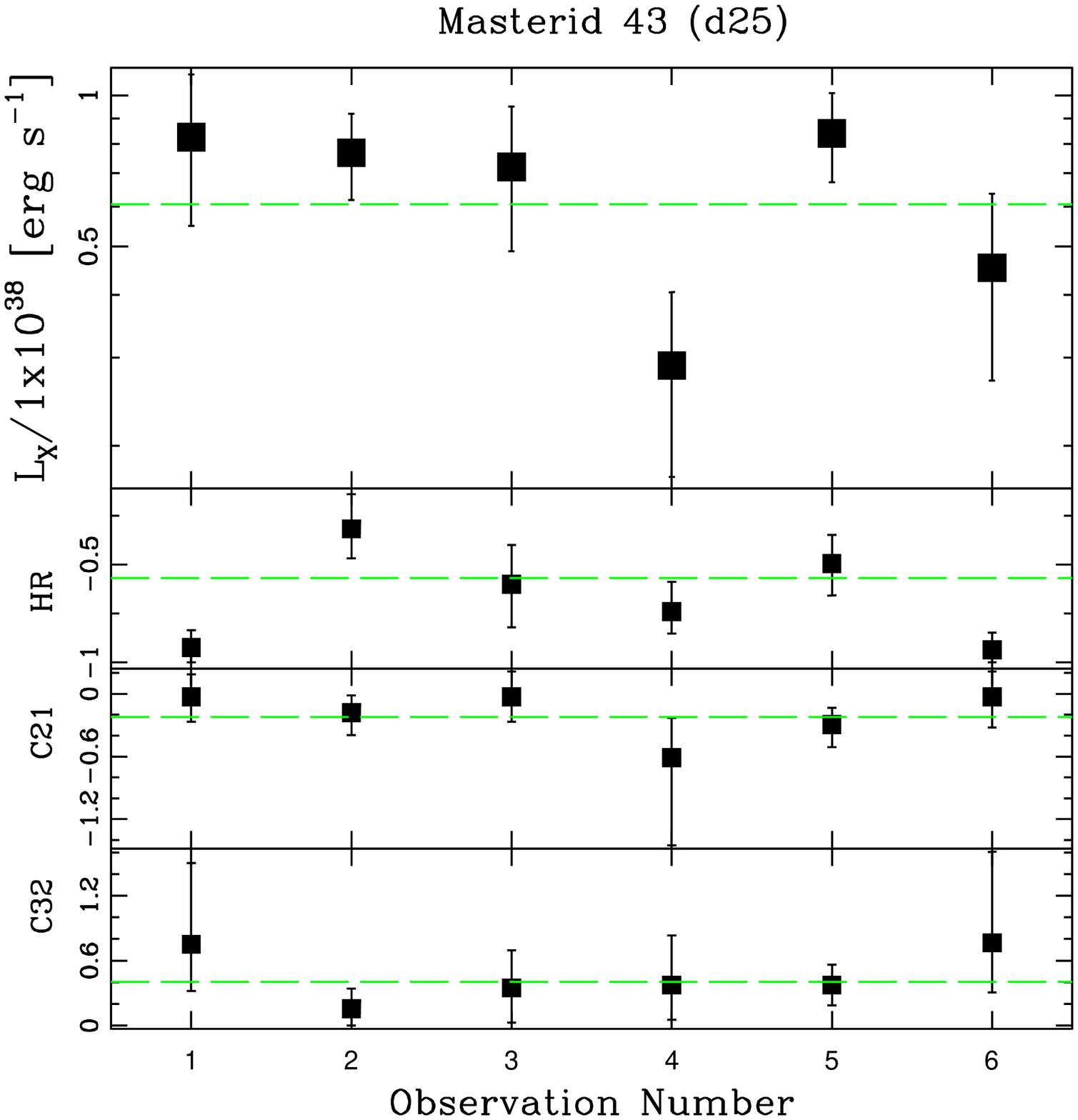}

  \end{minipage}\hspace{0.02\linewidth}
  \begin{minipage}{0.485\linewidth}
  \centering

    \includegraphics[width=\linewidth]{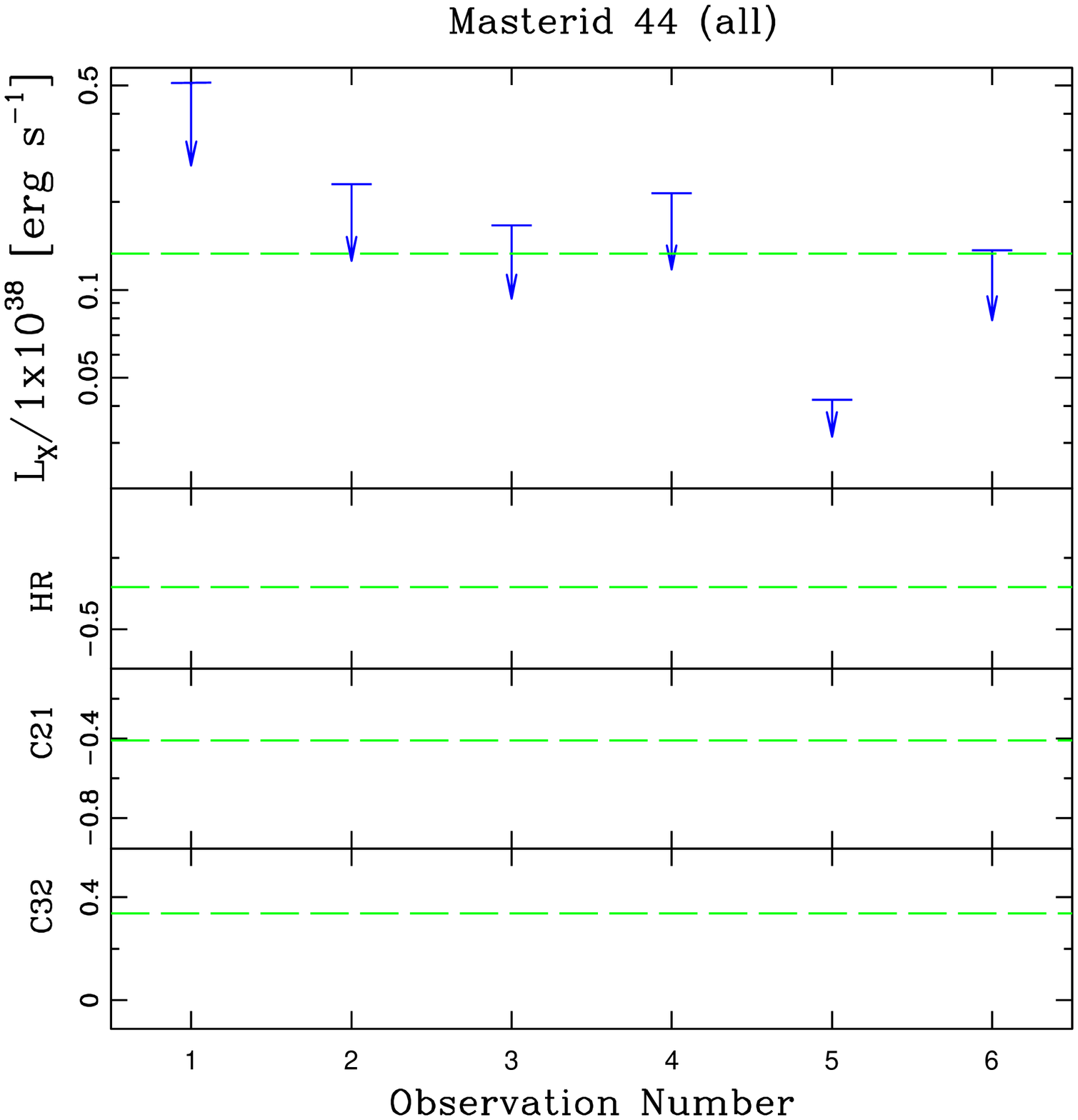}

\end{minipage}\hspace{0.02\linewidth}

\begin{minipage}{0.485\linewidth}
  \centering

    \includegraphics[width=\linewidth]{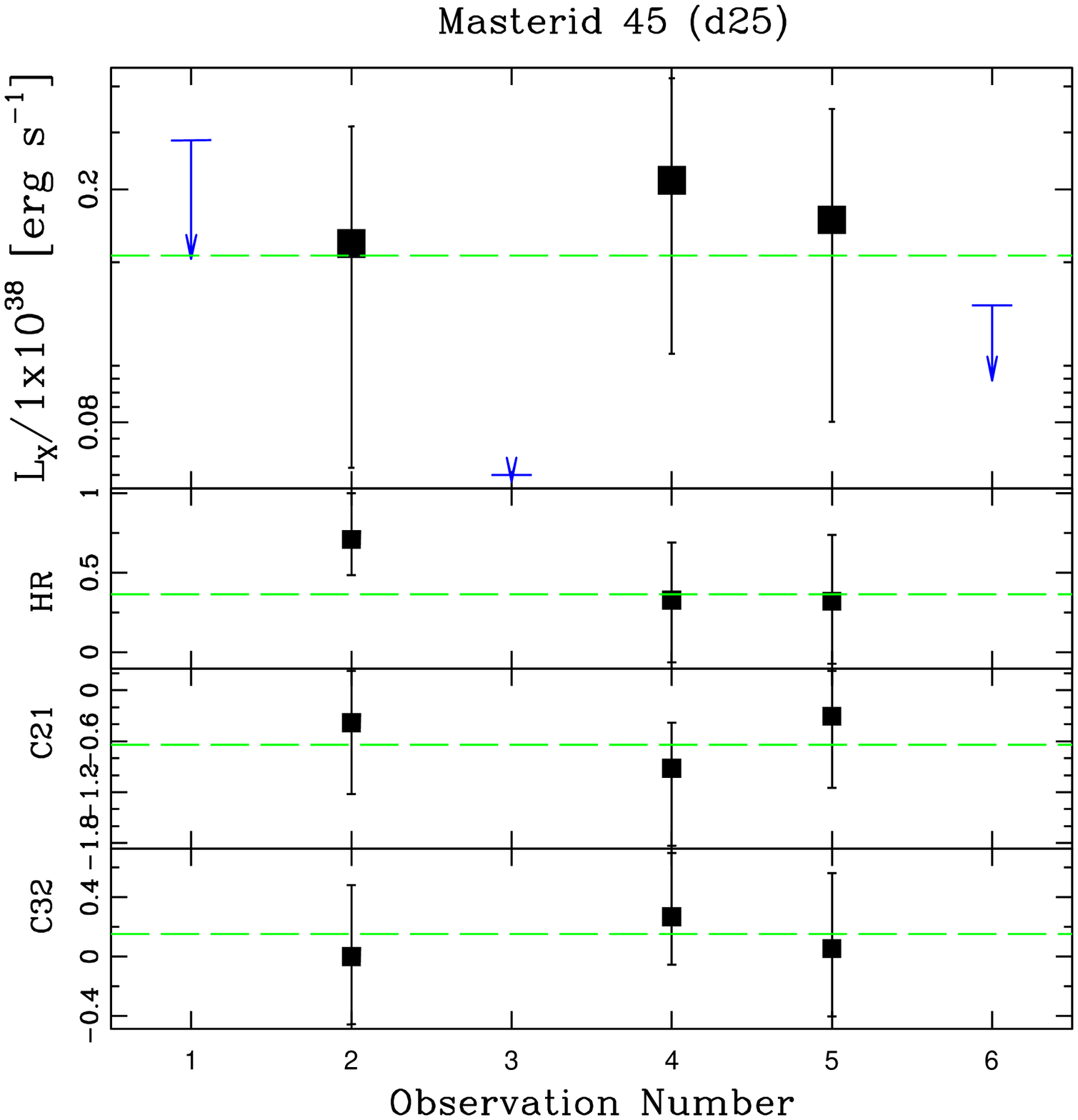}

 \end{minipage}\hspace{0.02\linewidth}
\begin{minipage}{0.485\linewidth}
  \centering
  
    \includegraphics[width=\linewidth]{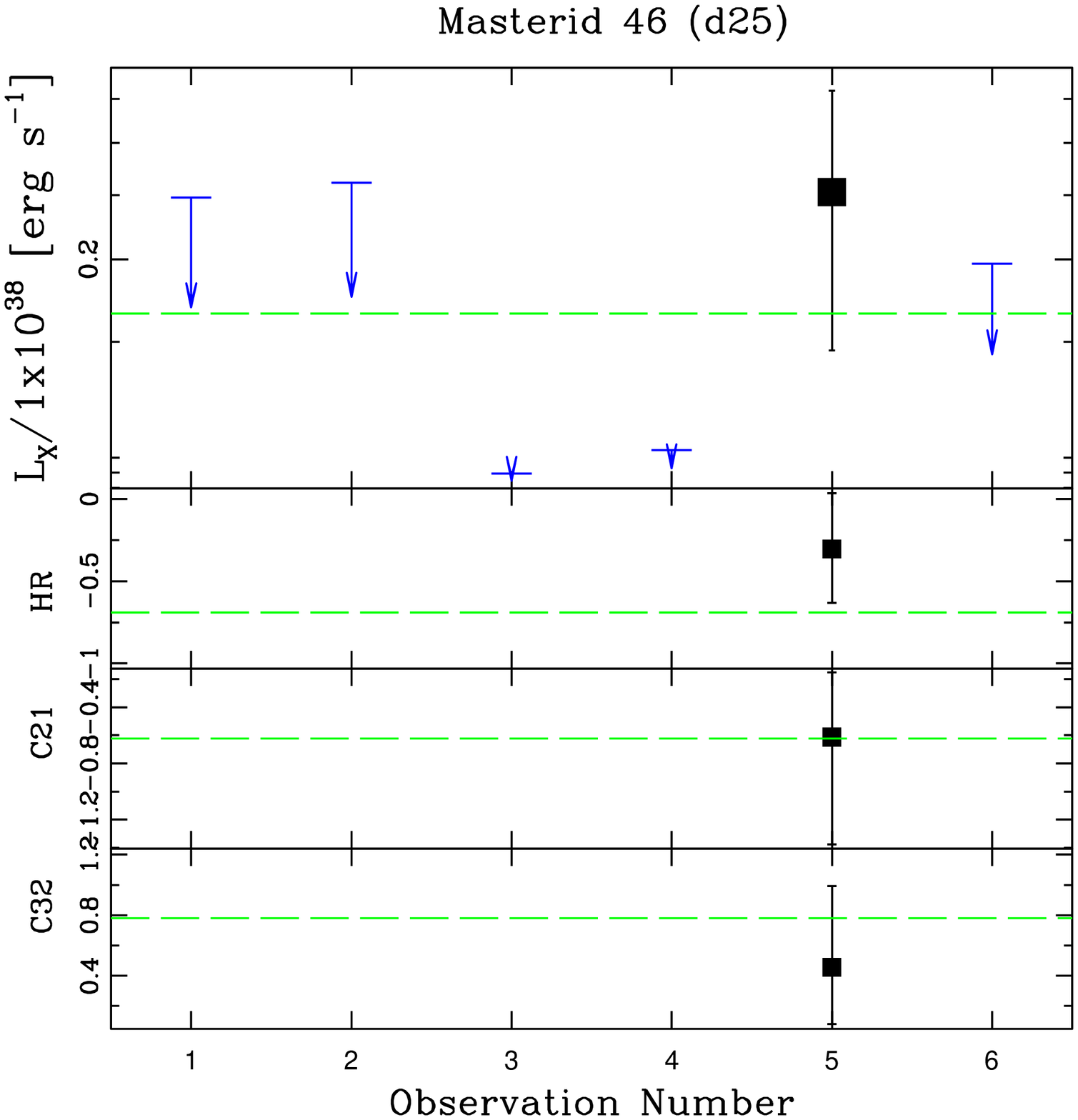}

  \end{minipage}\hspace{0.02\linewidth}
  
\end{figure}

\begin{figure}

  \begin{minipage}{0.485\linewidth}
  \centering

    \includegraphics[width=\linewidth]{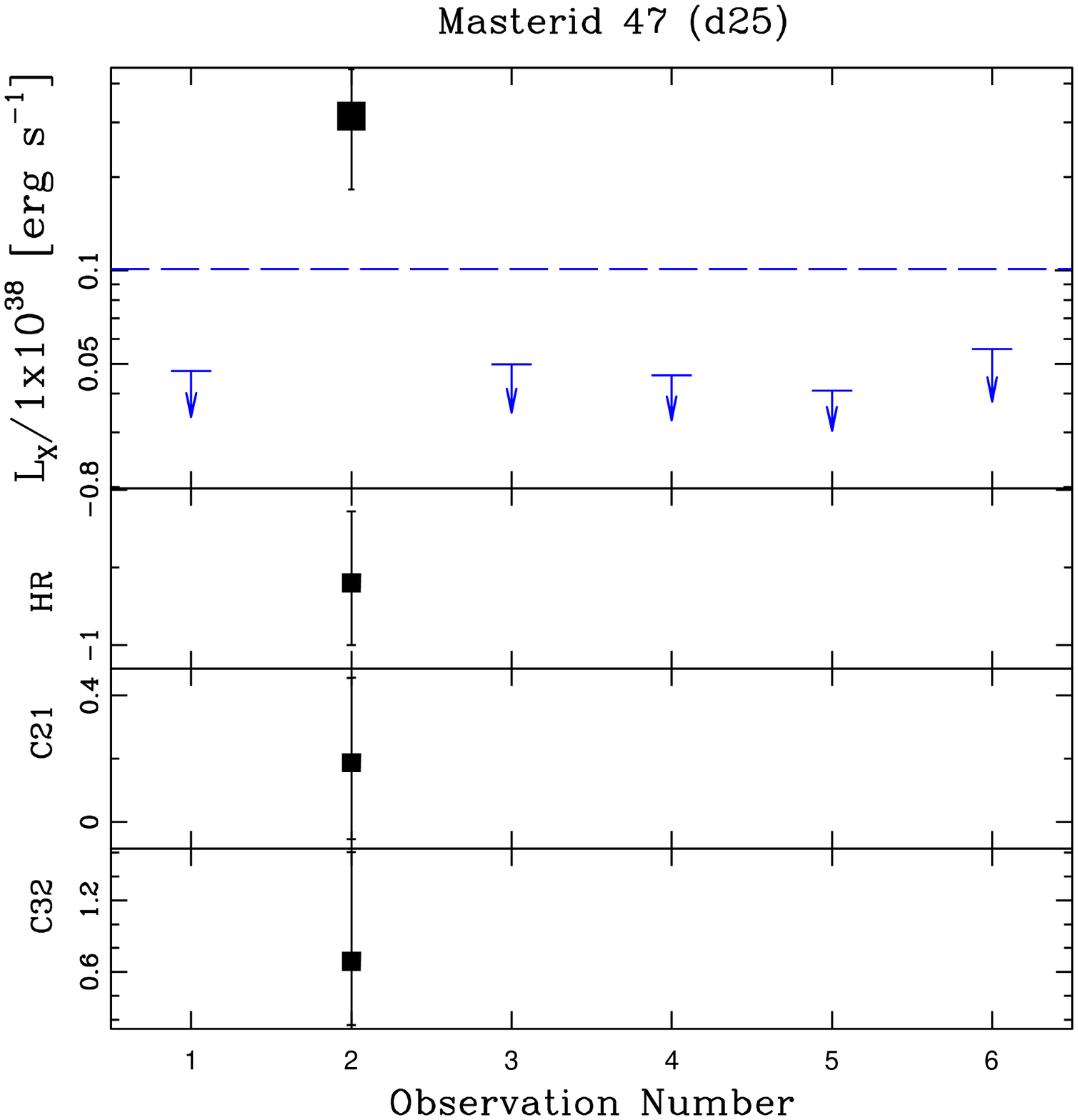}

\end{minipage}\hspace{0.02\linewidth}
\begin{minipage}{0.485\linewidth}
  \centering

    \includegraphics[width=\linewidth]{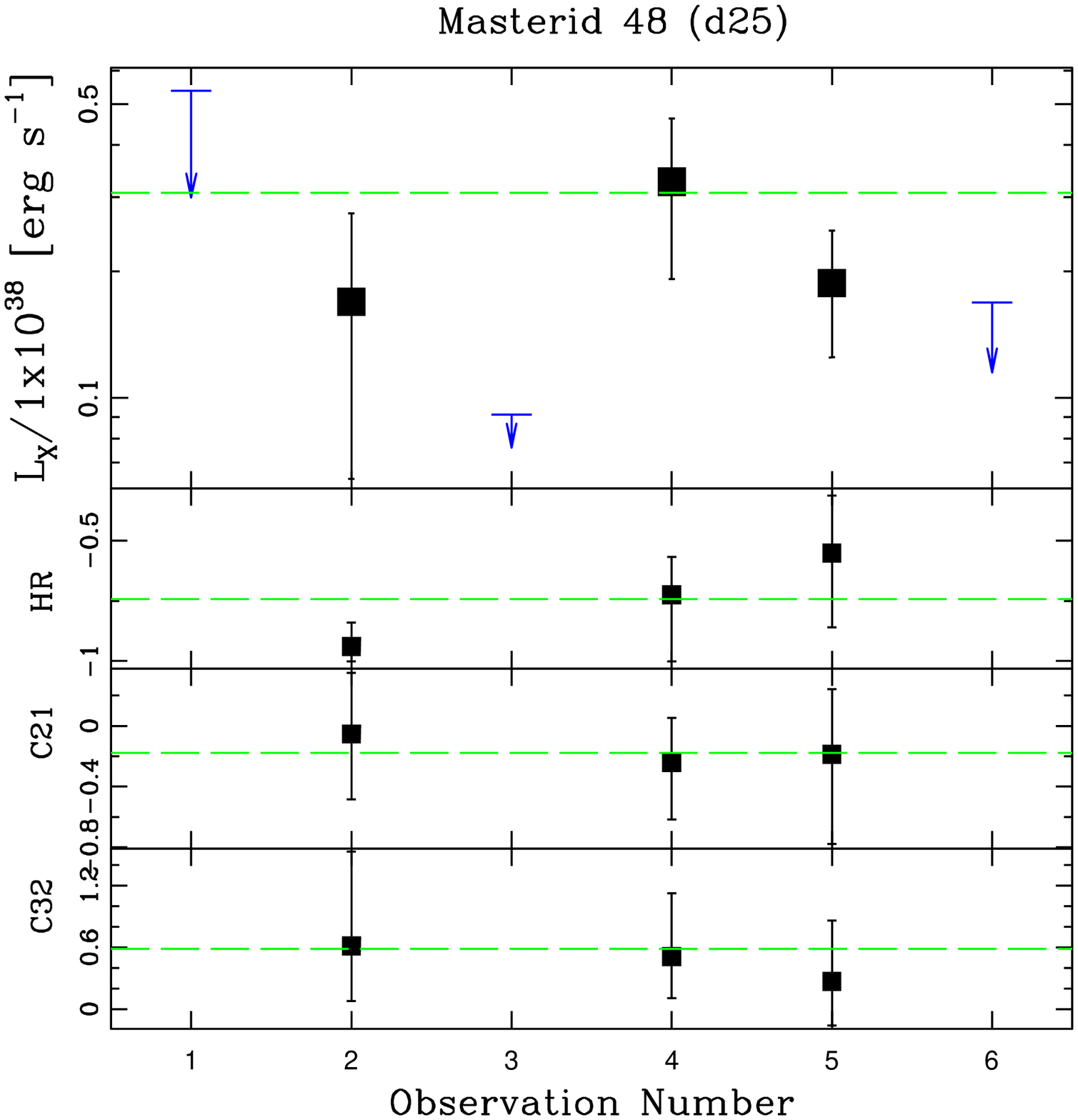}

 \end{minipage}\hspace{0.02\linewidth}

  \begin{minipage}{0.485\linewidth}
  \centering
  
    \includegraphics[width=\linewidth]{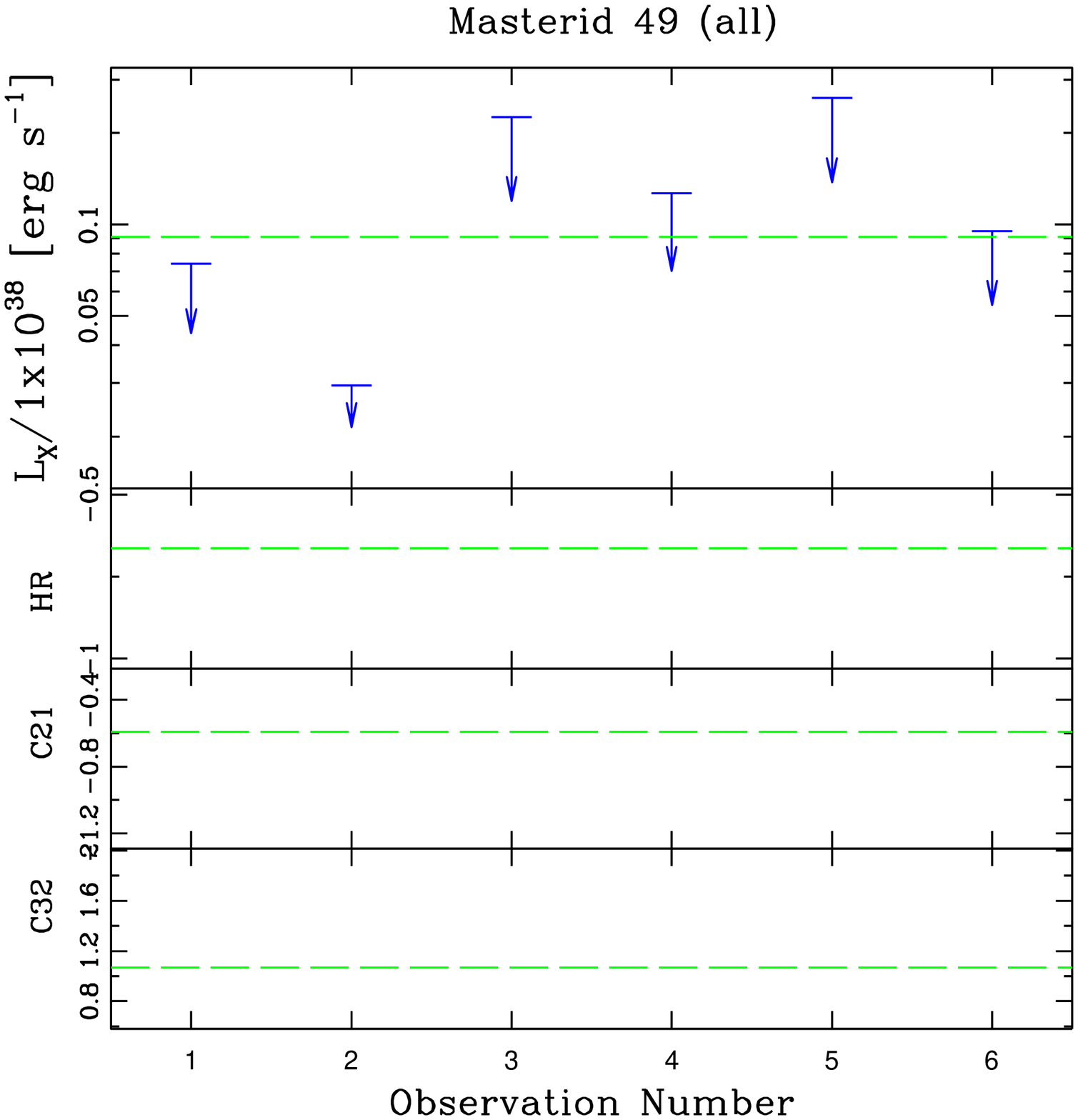}

  \end{minipage}\hspace{0.02\linewidth}
  \begin{minipage}{0.485\linewidth}
  \centering

    \includegraphics[width=\linewidth]{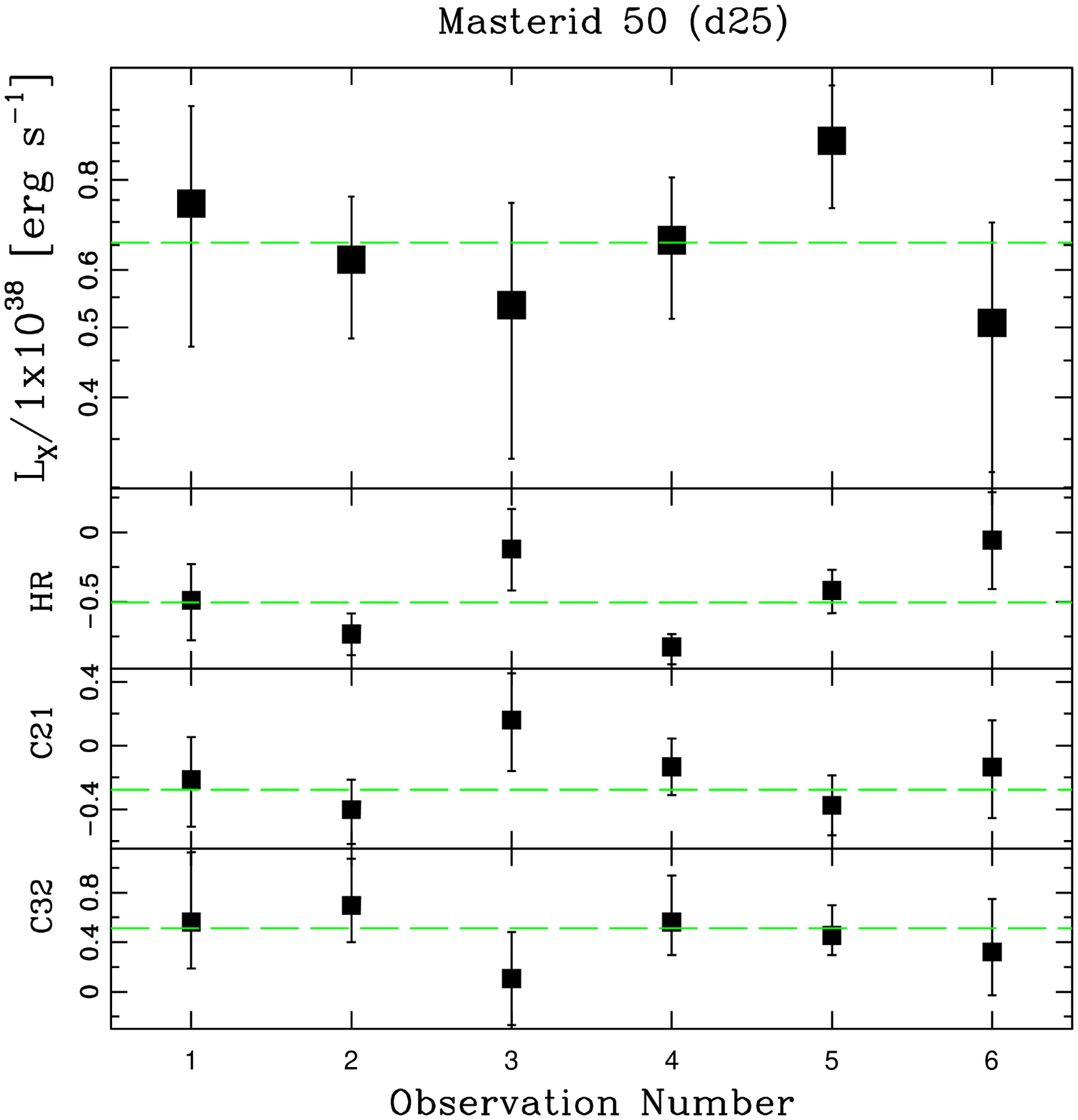}

\end{minipage}\hspace{0.02\linewidth}

\begin{minipage}{0.485\linewidth}
  \centering

    \includegraphics[width=\linewidth]{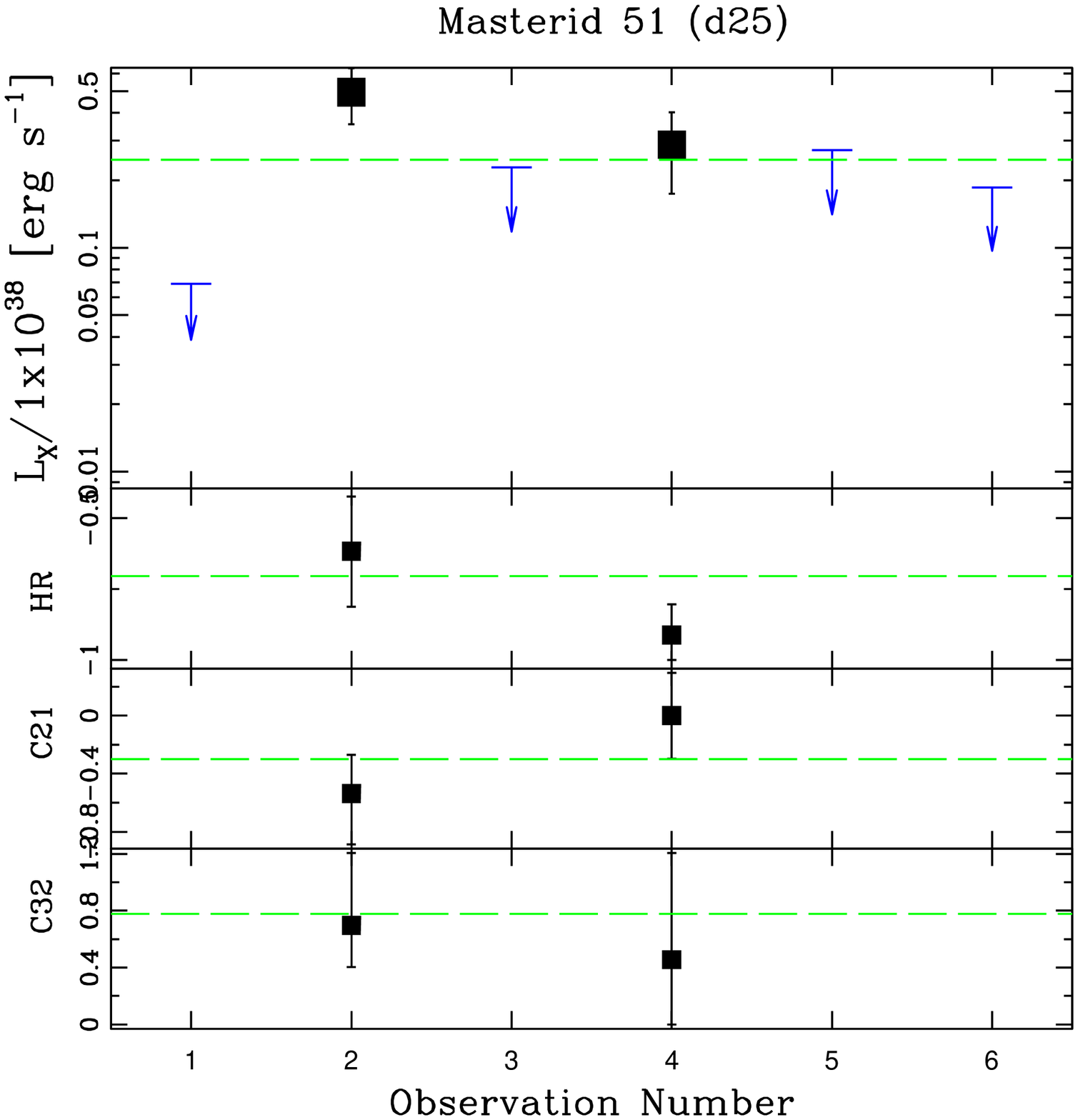}

 \end{minipage}\hspace{0.02\linewidth}
\begin{minipage}{0.485\linewidth}
  \centering
  
    \includegraphics[width=\linewidth]{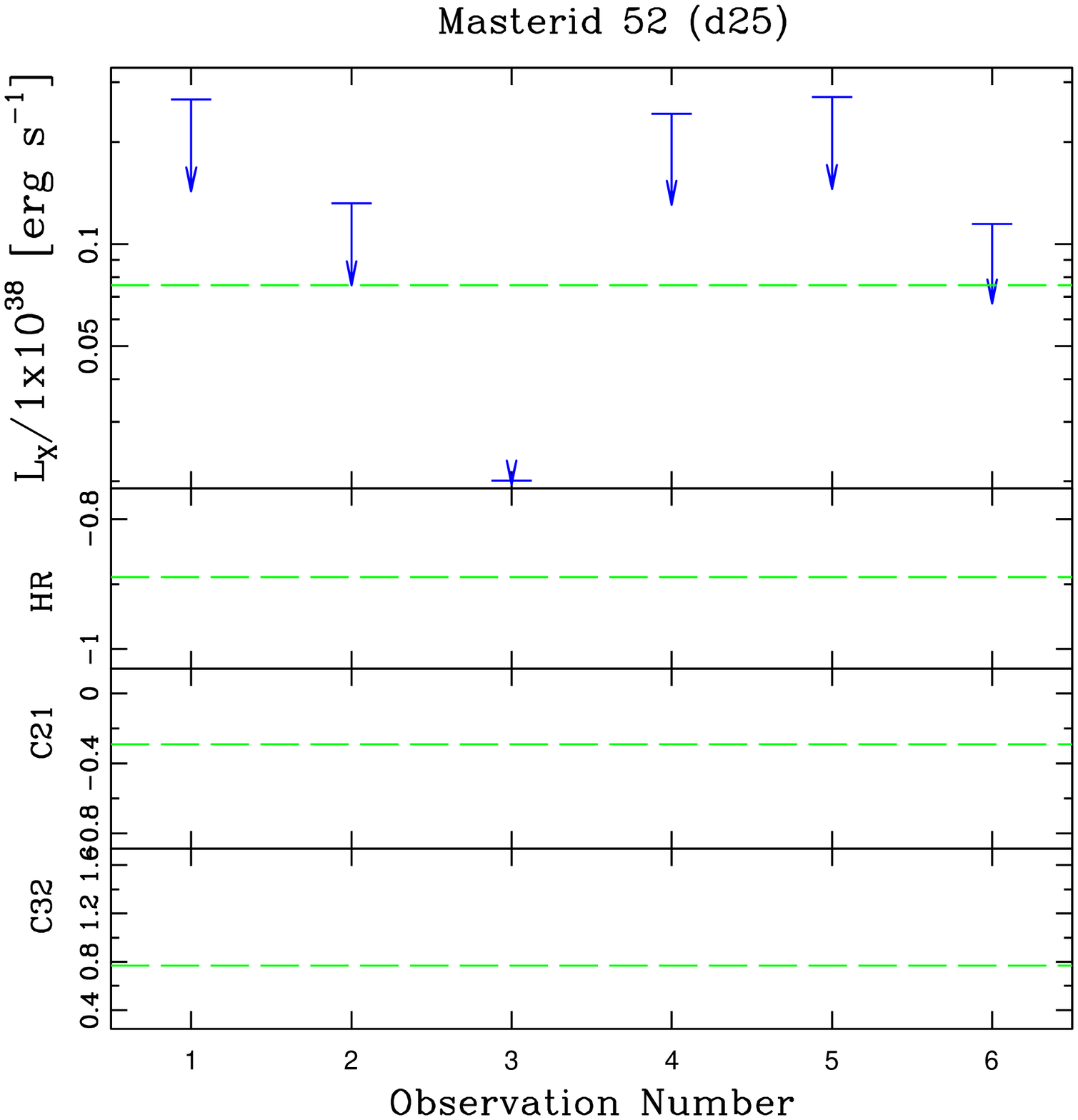}

  \end{minipage}\hspace{0.02\linewidth}

\end{figure}

\begin{figure}

  \begin{minipage}{0.485\linewidth}
  \centering

    \includegraphics[width=\linewidth]{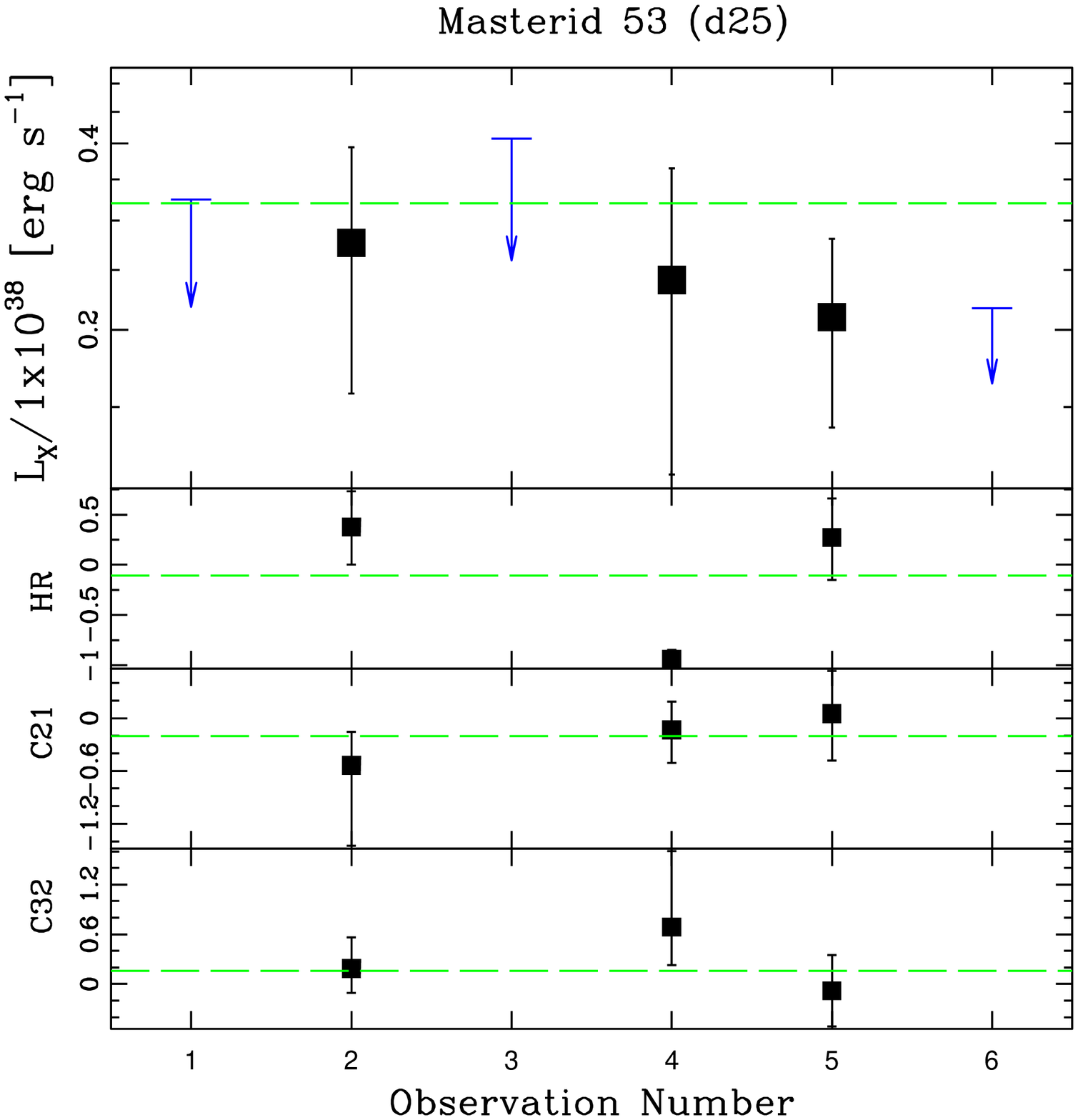}

\end{minipage}\hspace{0.02\linewidth}
\begin{minipage}{0.485\linewidth}
  \centering

    \includegraphics[width=\linewidth]{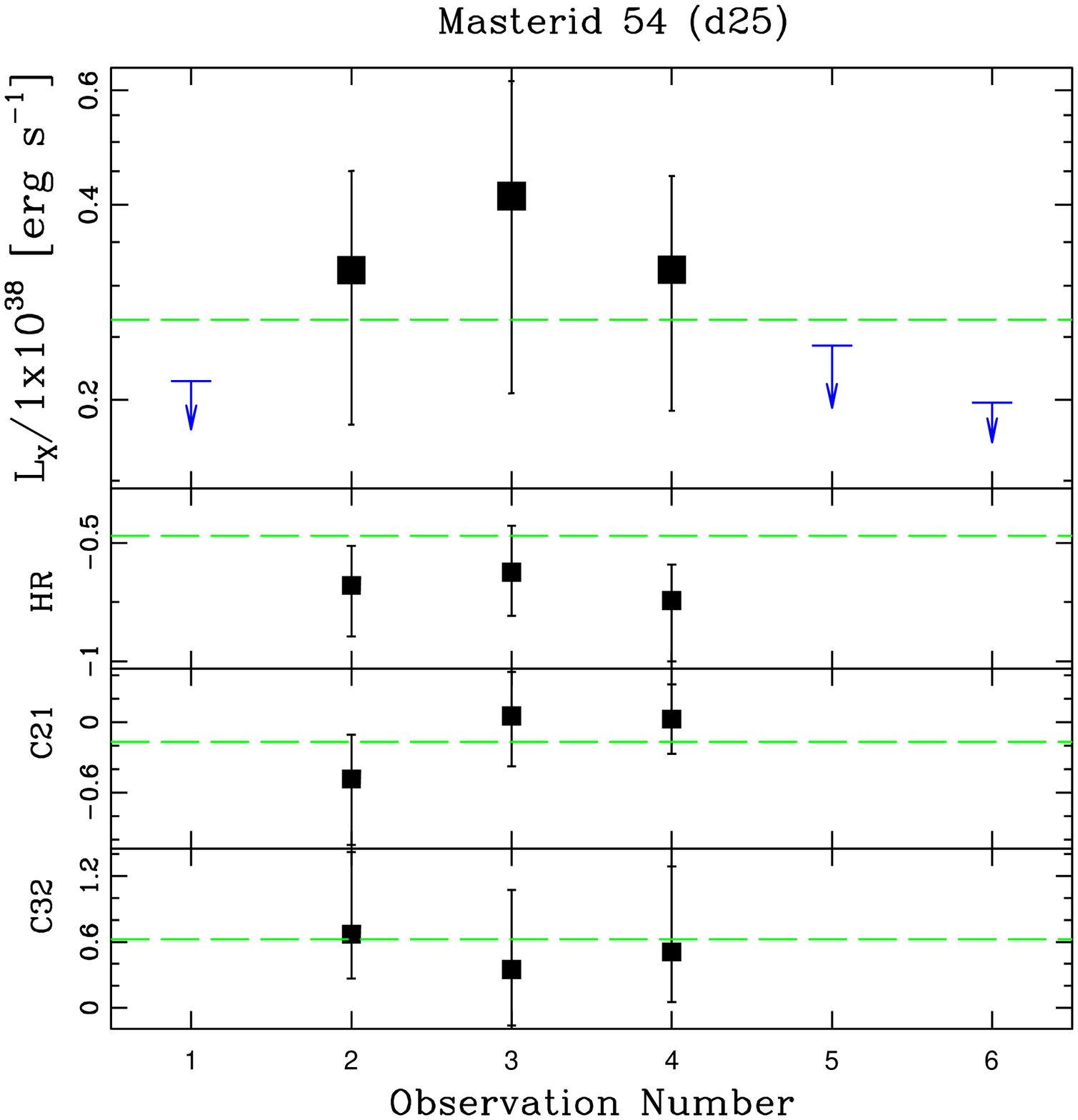}

 \end{minipage}\hspace{0.02\linewidth}
  
  \begin{minipage}{0.485\linewidth}
  \centering
  
    \includegraphics[width=\linewidth]{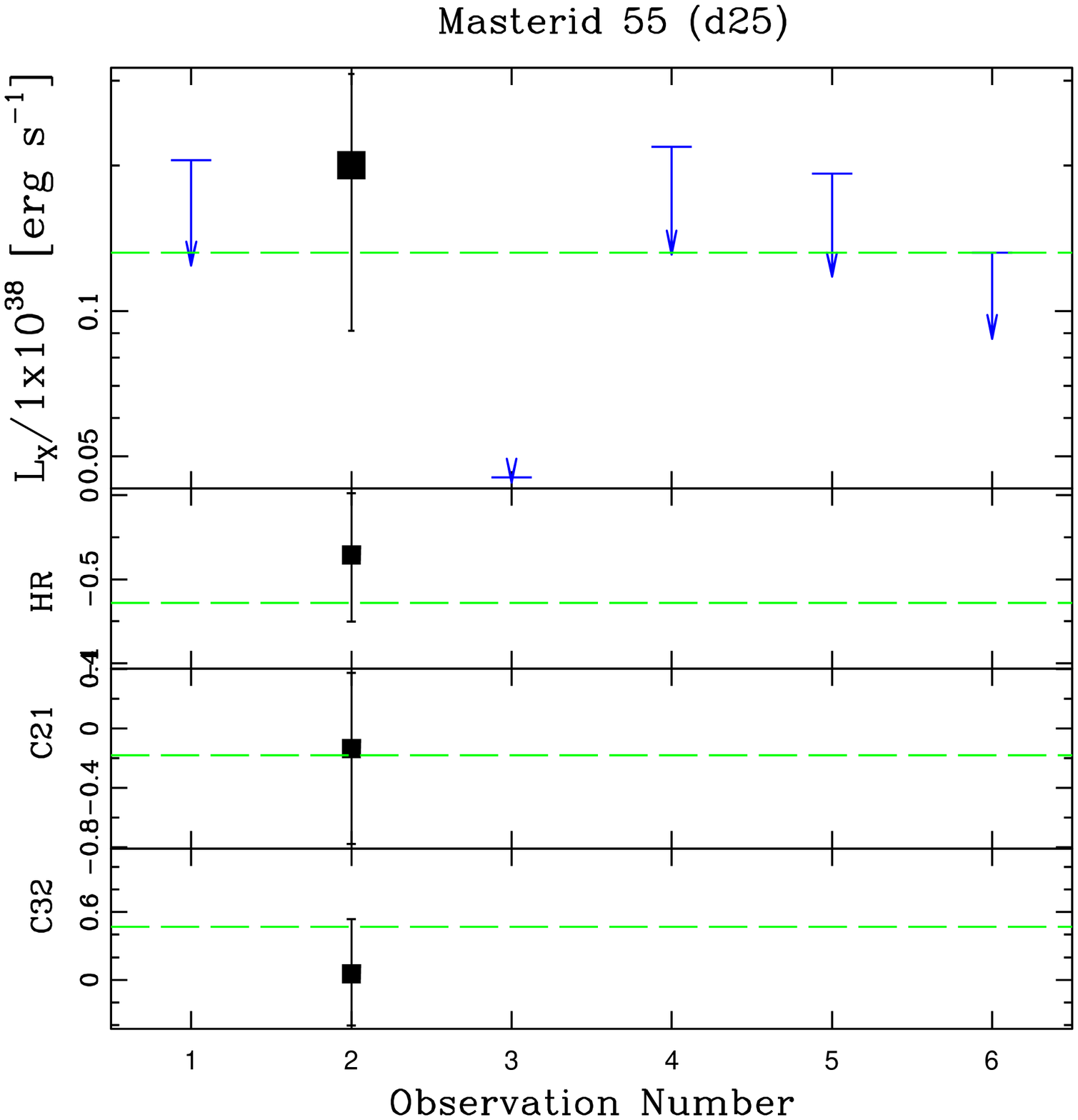}

  \end{minipage}\hspace{0.02\linewidth}
  \begin{minipage}{0.485\linewidth}
  \centering

    \includegraphics[width=\linewidth]{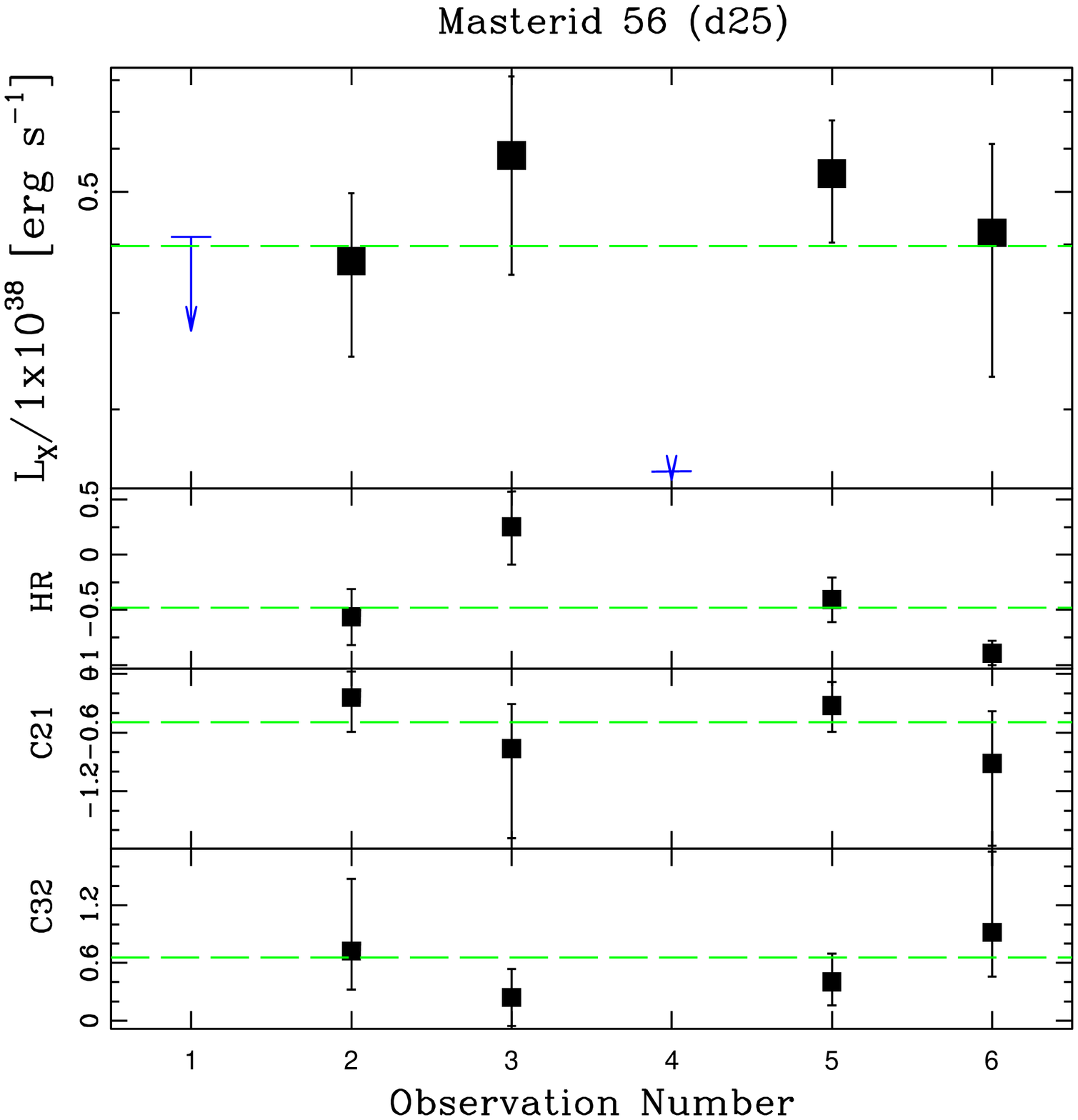}

\end{minipage}\hspace{0.02\linewidth}

\begin{minipage}{0.485\linewidth}
  \centering

    \includegraphics[width=\linewidth]{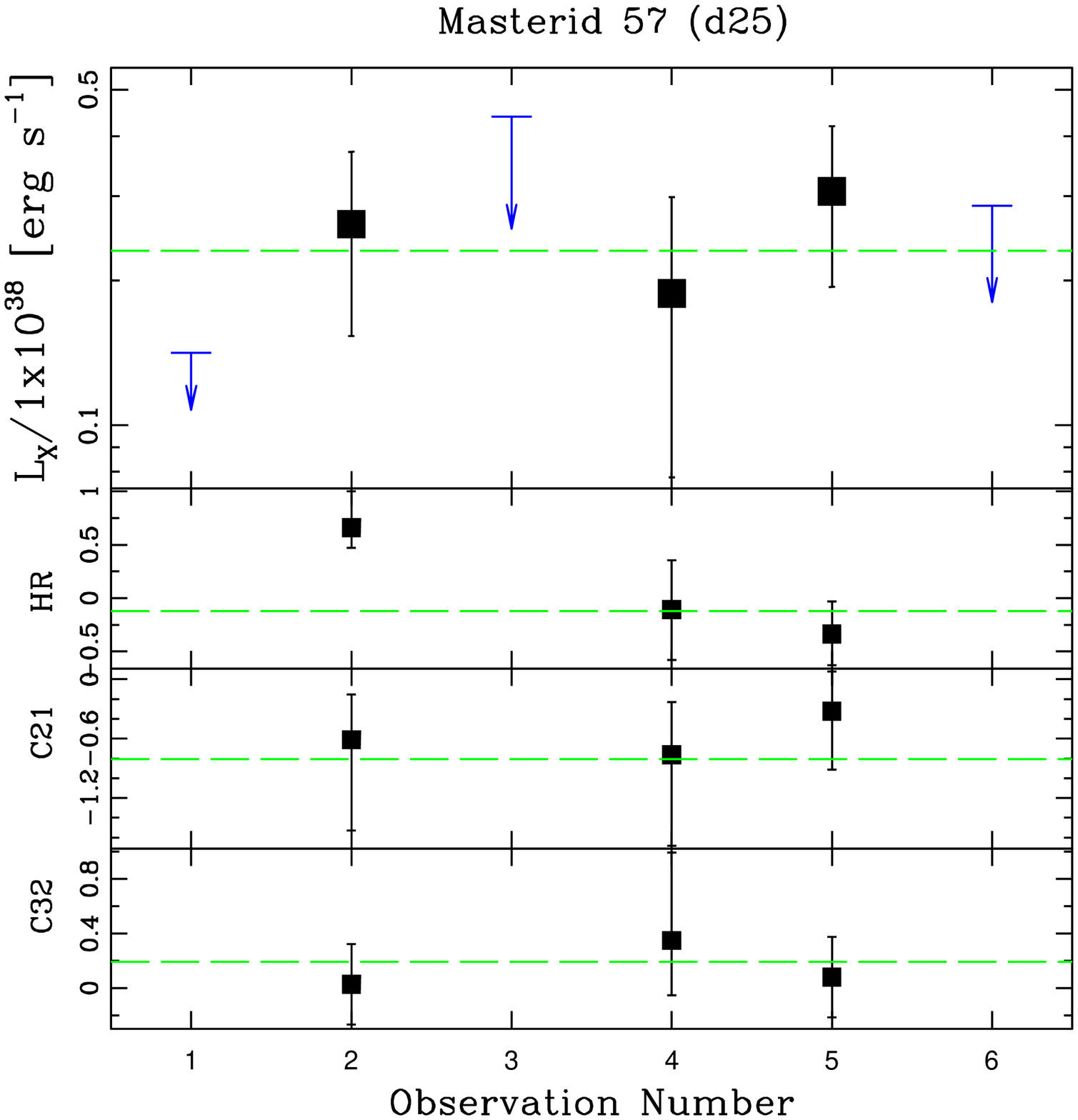}

 \end{minipage}\hspace{0.02\linewidth}
\begin{minipage}{0.485\linewidth}
  \centering
  
    \includegraphics[width=\linewidth]{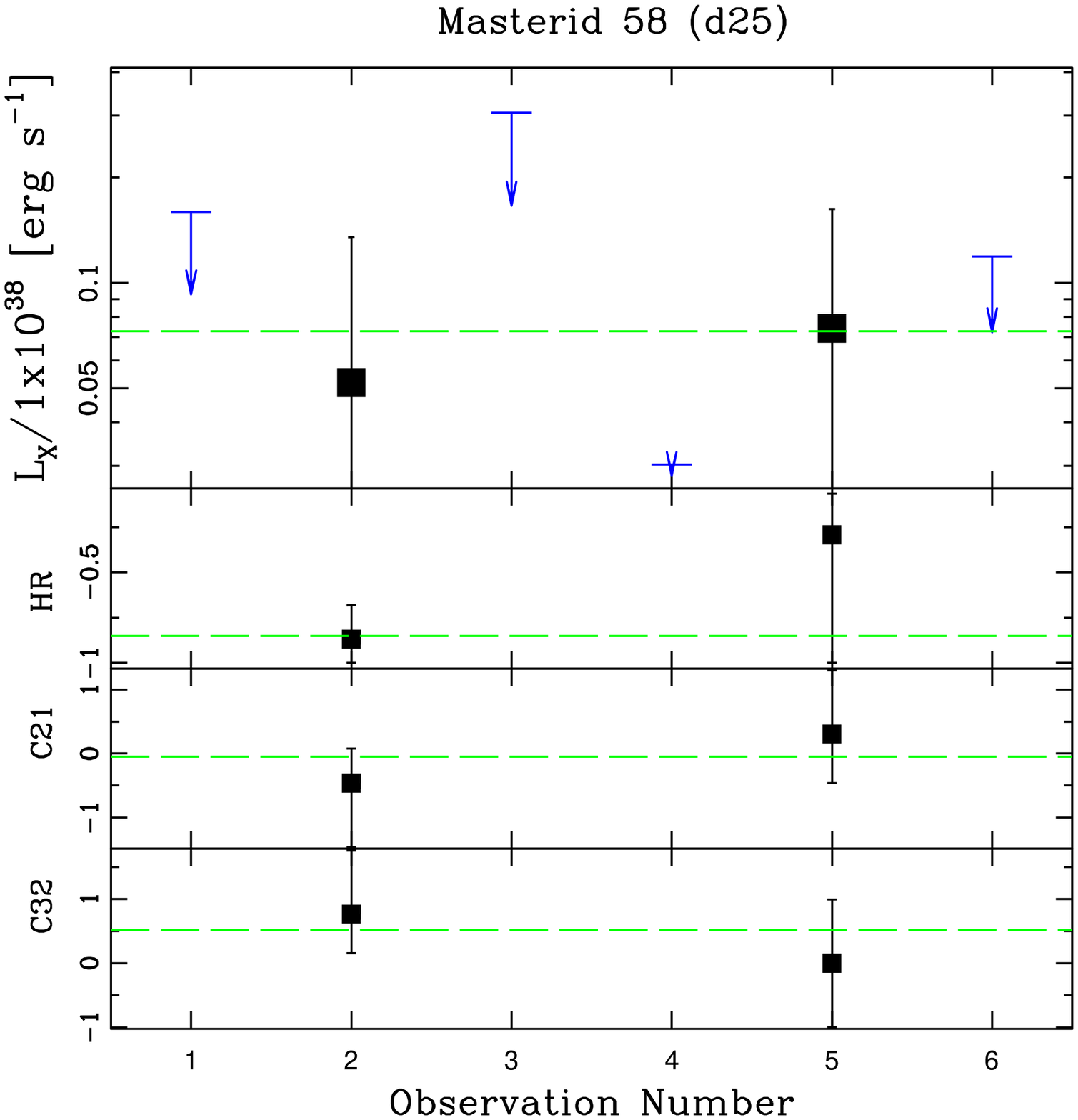}

  \end{minipage}\hspace{0.02\linewidth}

\end{figure}

\begin{figure}

  \begin{minipage}{0.485\linewidth}
  \centering

    \includegraphics[width=\linewidth]{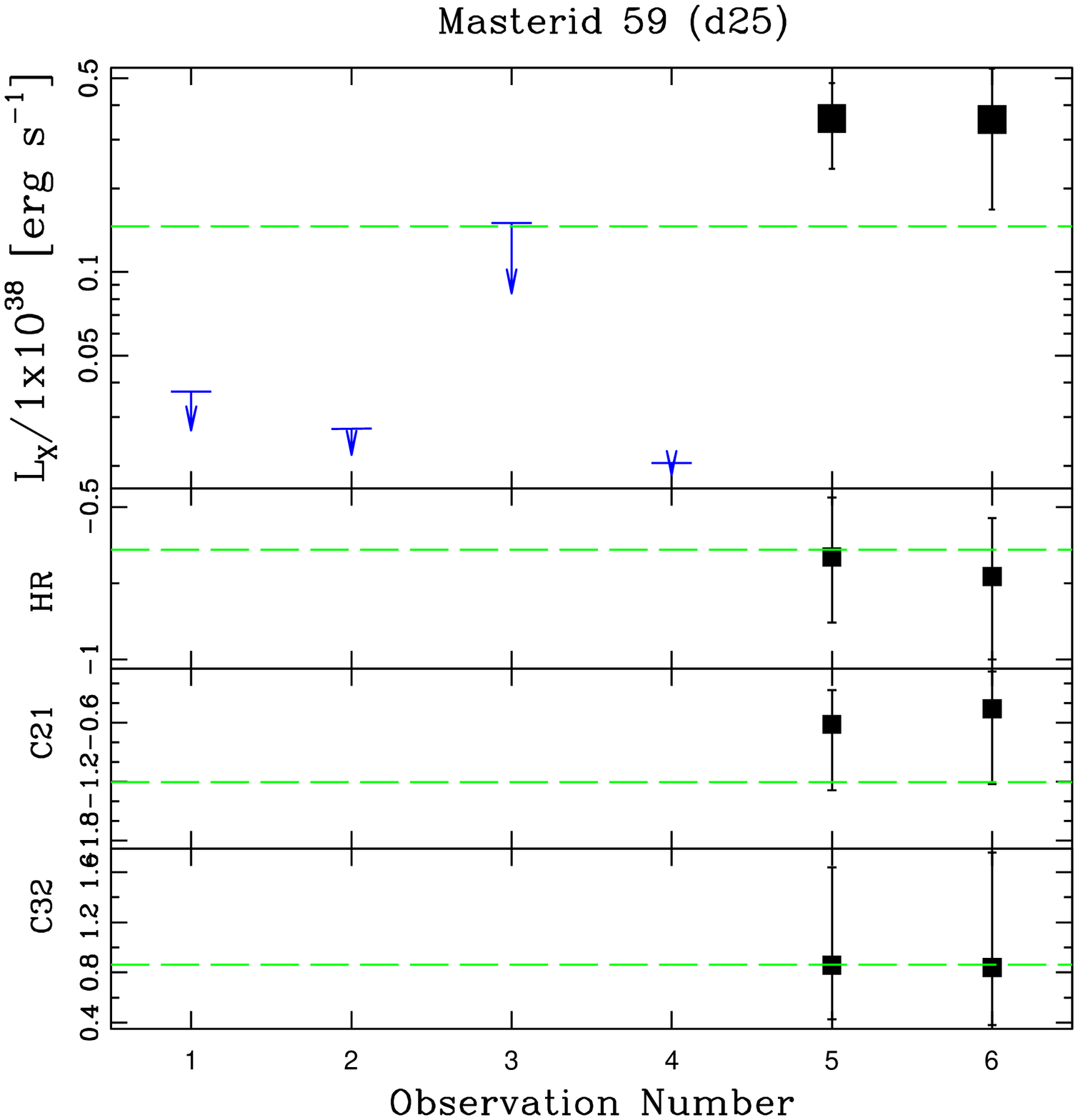}

\end{minipage}\hspace{0.02\linewidth}
\begin{minipage}{0.485\linewidth}
  \centering

    \includegraphics[width=\linewidth]{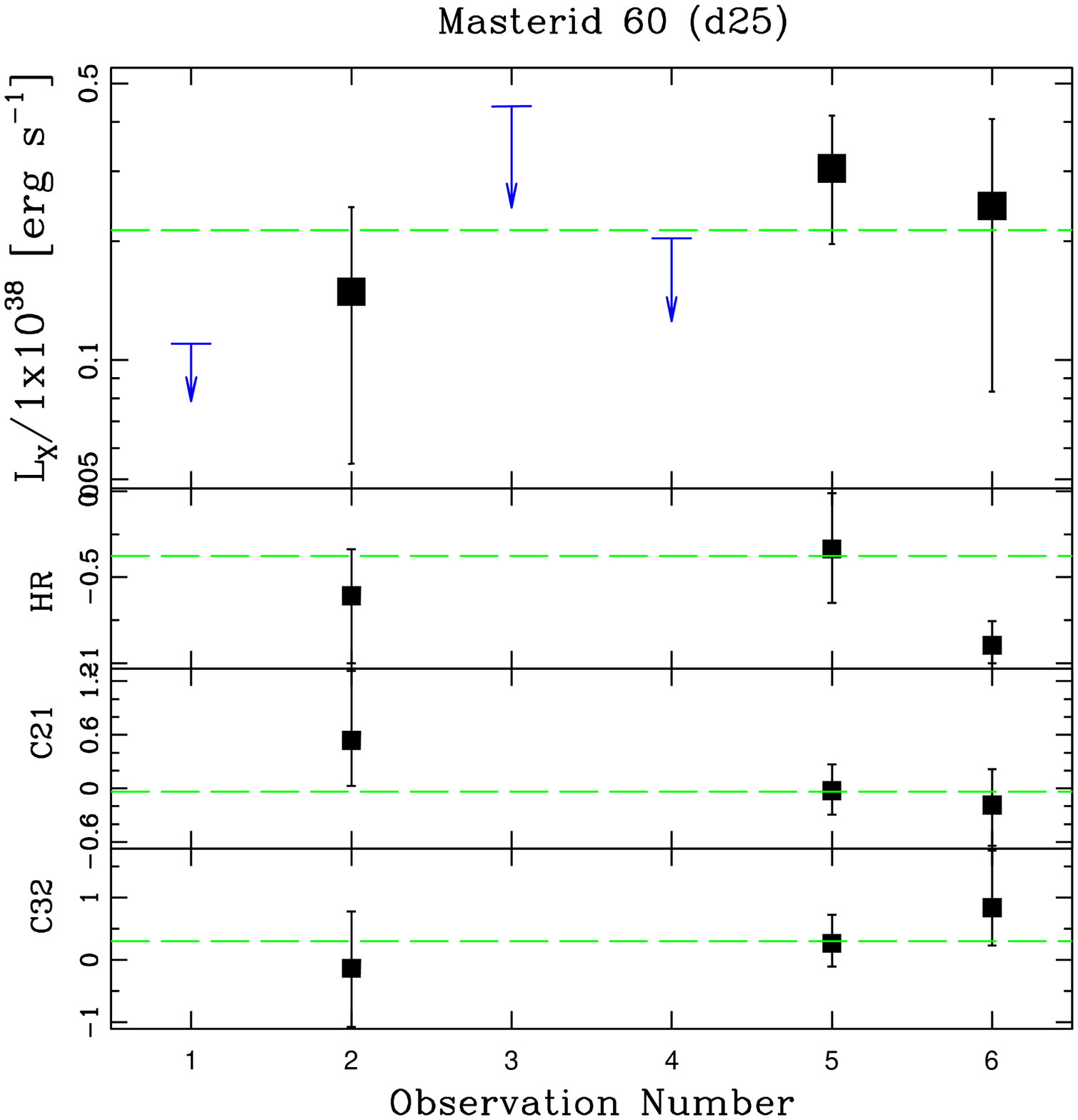}

 \end{minipage}\hspace{0.02\linewidth}

  \begin{minipage}{0.485\linewidth}
  \centering
  
    \includegraphics[width=\linewidth]{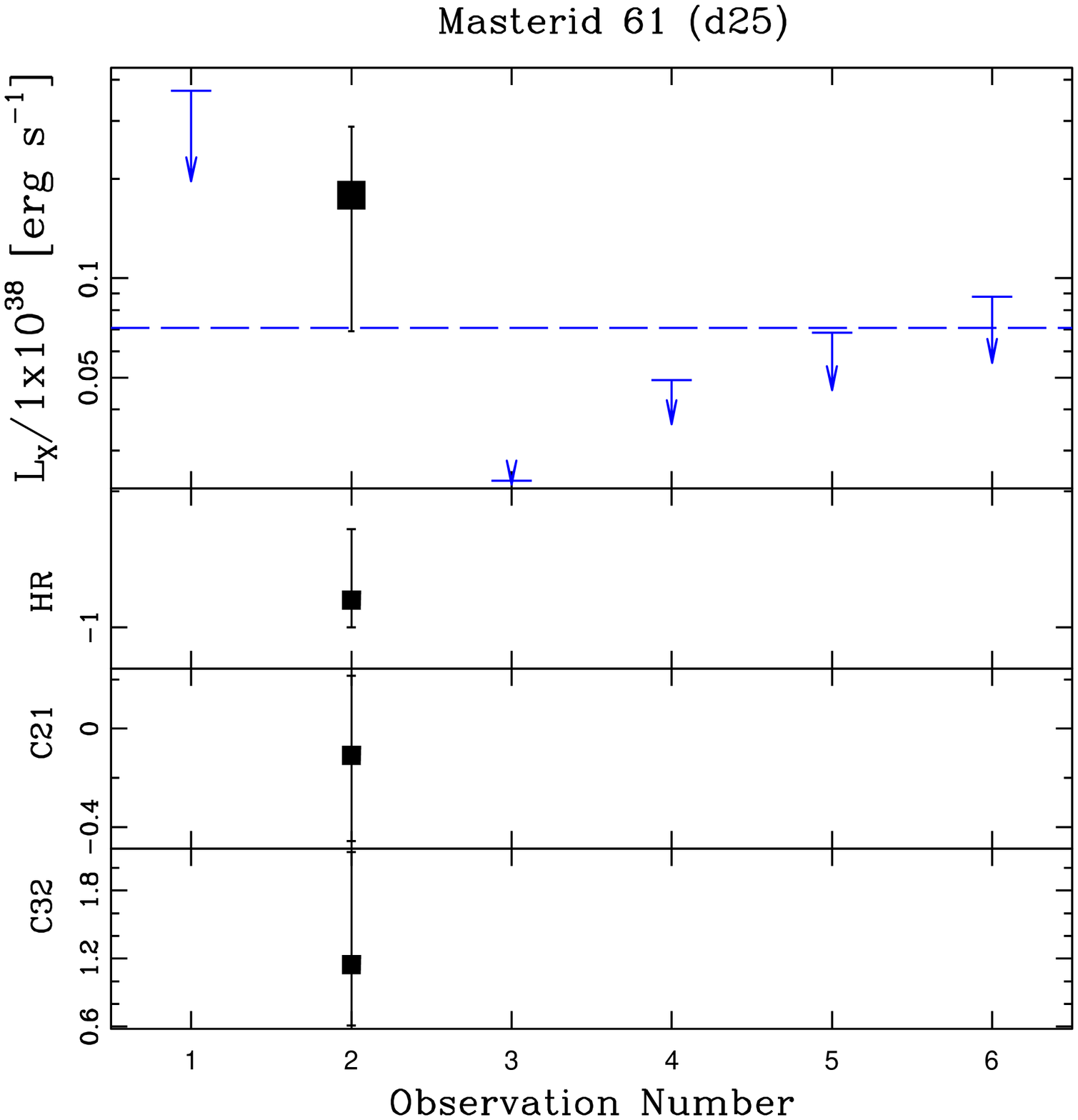}

  \end{minipage}\hspace{0.02\linewidth}
  \begin{minipage}{0.485\linewidth}
  \centering

    \includegraphics[width=\linewidth]{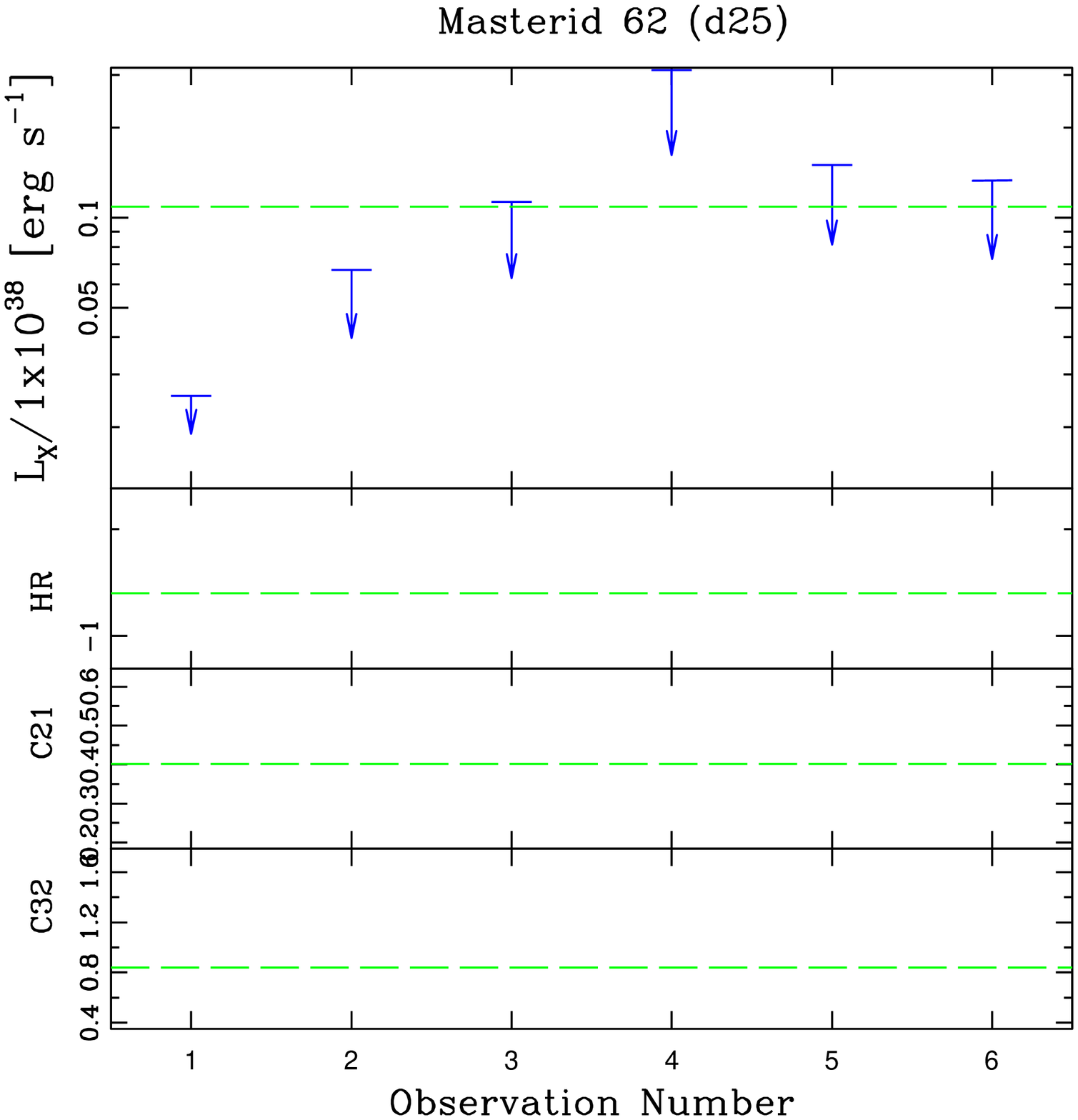}

\end{minipage}\hspace{0.02\linewidth}

\begin{minipage}{0.485\linewidth}
  \centering

    \includegraphics[width=\linewidth]{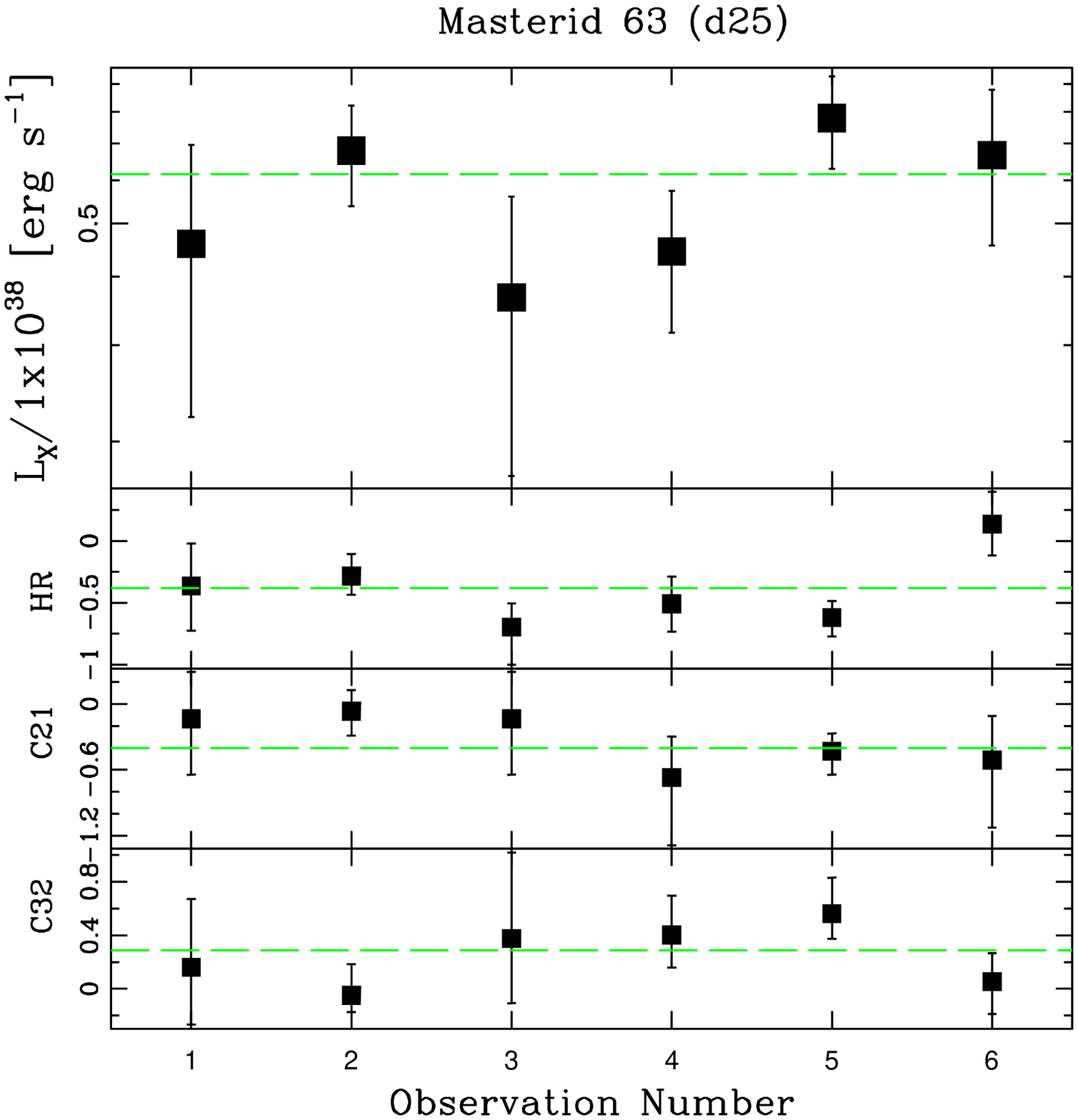}

\end{minipage}\hspace{0.02\linewidth}
\begin{minipage}{0.485\linewidth}
  \centering
  
    \includegraphics[width=\linewidth]{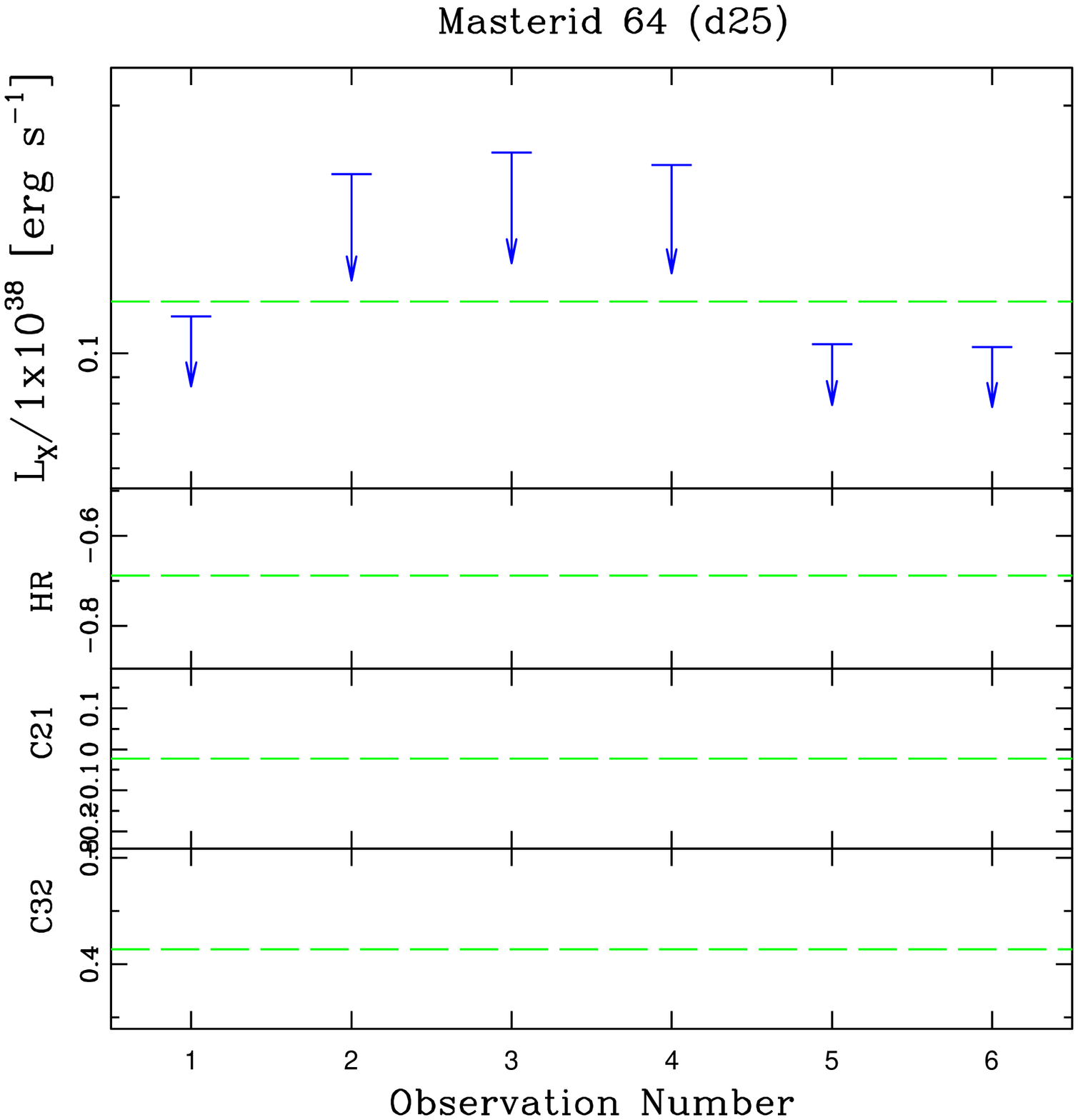}

  \end{minipage}\hspace{0.02\linewidth}

\end{figure}

\begin{figure}
  \begin{minipage}{0.485\linewidth}
  \centering

    \includegraphics[width=\linewidth]{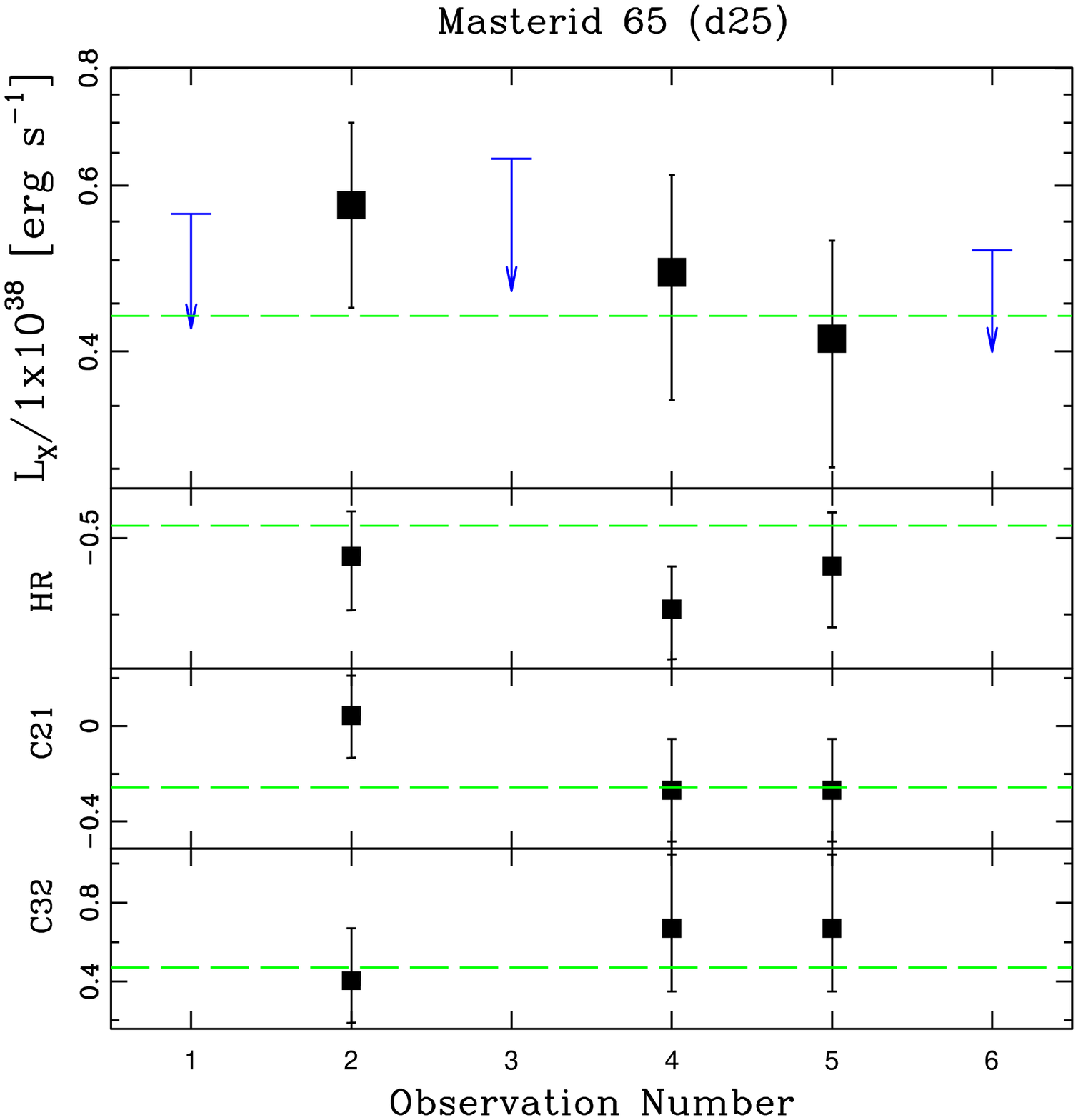}

\end{minipage}\hspace{0.02\linewidth}
\begin{minipage}{0.485\linewidth}
  \centering

    \includegraphics[width=\linewidth]{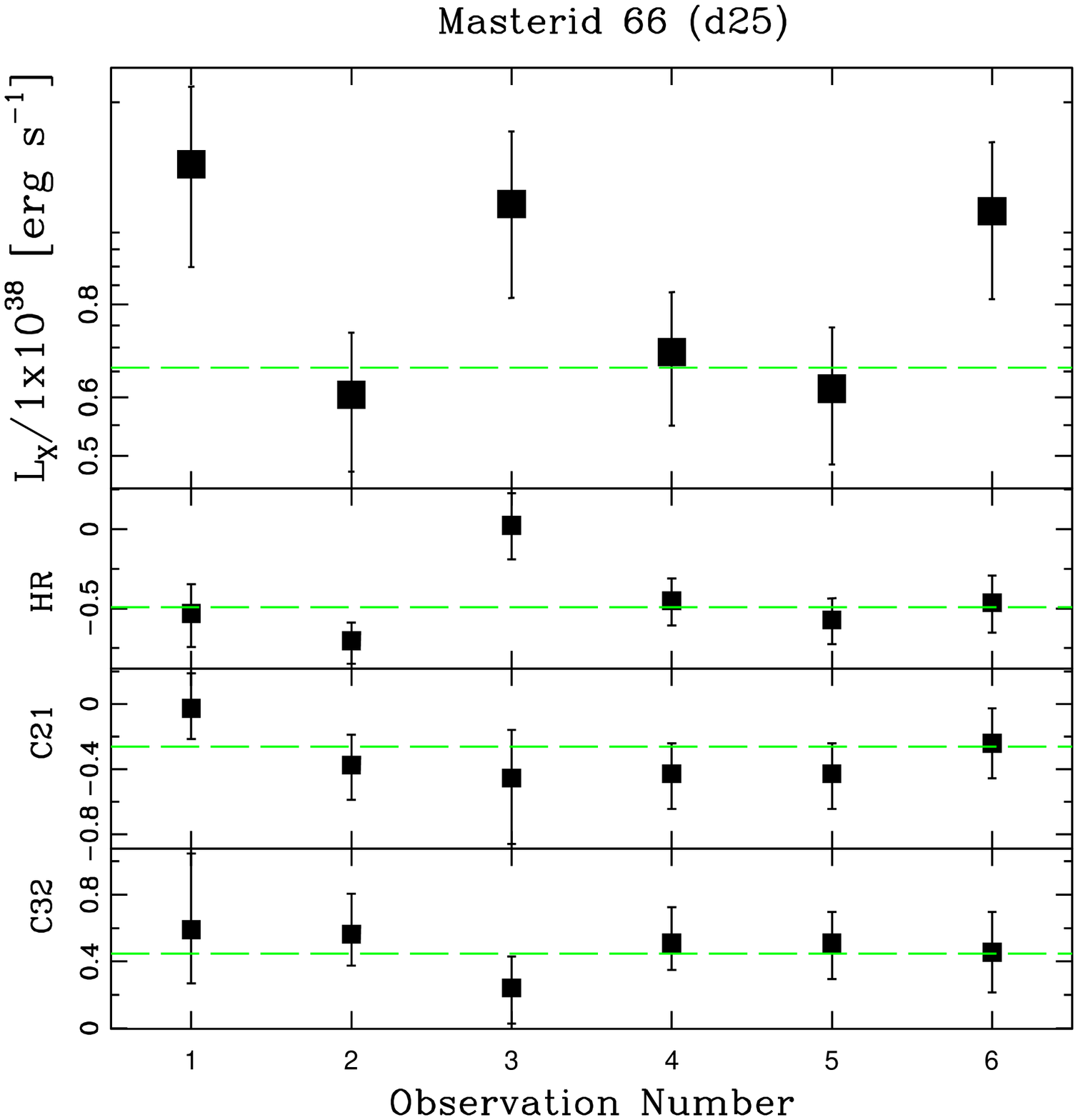}

\end{minipage}\hspace{0.02\linewidth}

  \begin{minipage}{0.485\linewidth}
  \centering
  
    \includegraphics[width=\linewidth]{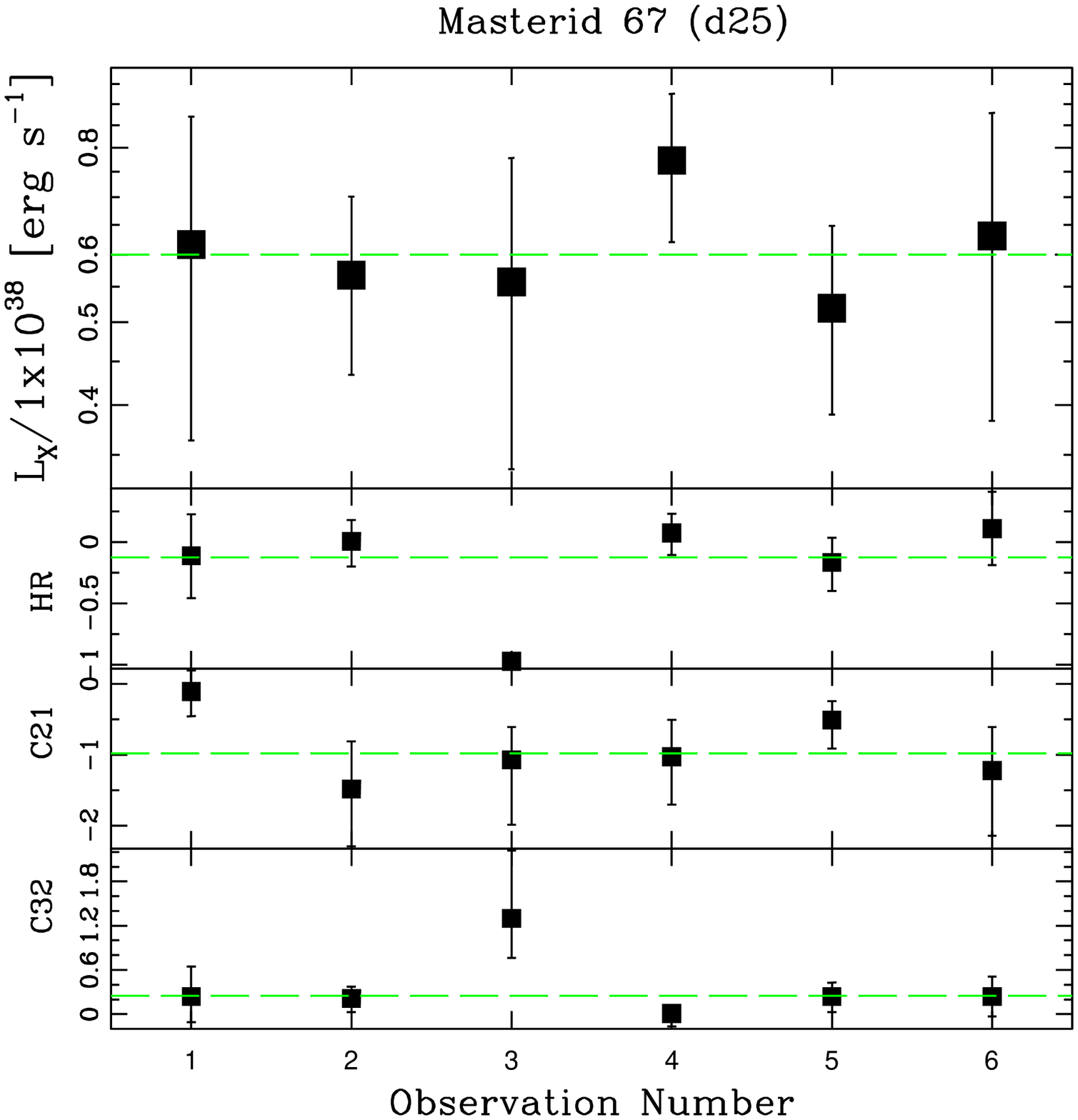}

  \end{minipage}\hspace{0.02\linewidth}
  \begin{minipage}{0.485\linewidth}
  \centering

    \includegraphics[width=\linewidth]{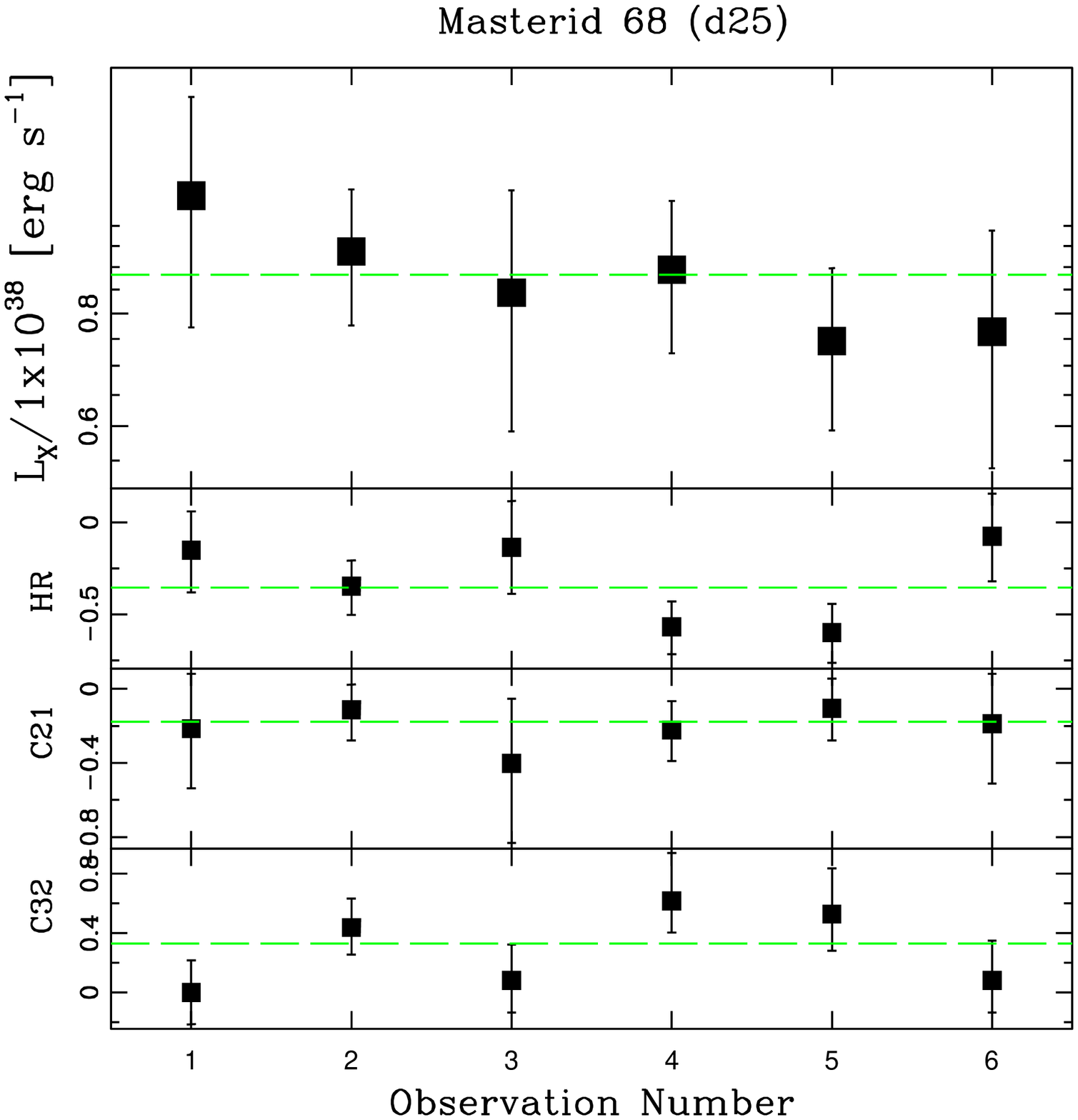}

\end{minipage}\hspace{0.02\linewidth}

\begin{minipage}{0.485\linewidth}
  \centering

    \includegraphics[width=\linewidth]{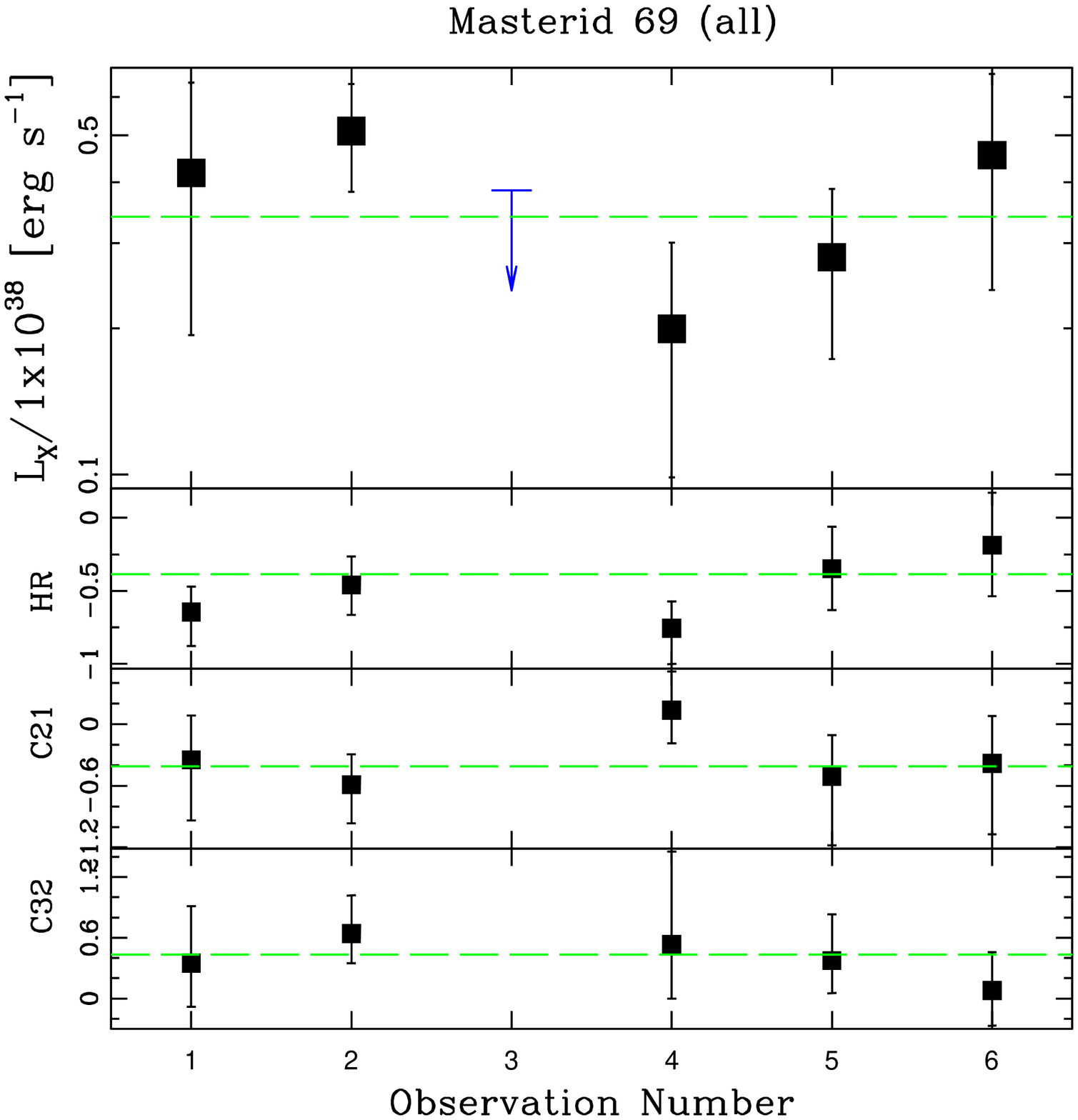}

 \end{minipage}\hspace{0.02\linewidth}
\begin{minipage}{0.485\linewidth}
  \centering
  
    \includegraphics[width=\linewidth]{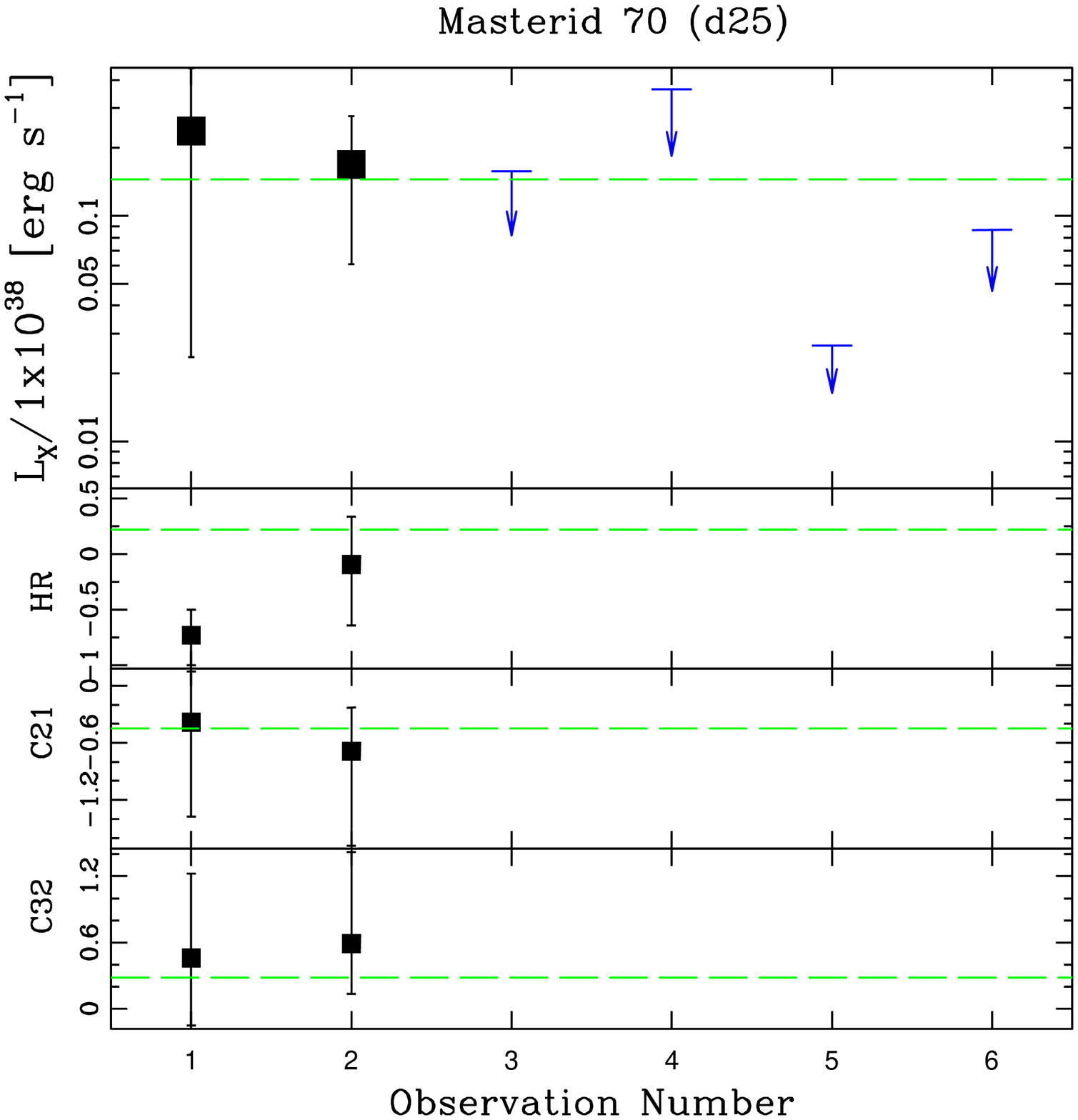}

  \end{minipage}\hspace{0.02\linewidth}
\end{figure}

\begin{figure}
  \begin{minipage}{0.485\linewidth}
  \centering

    \includegraphics[width=\linewidth]{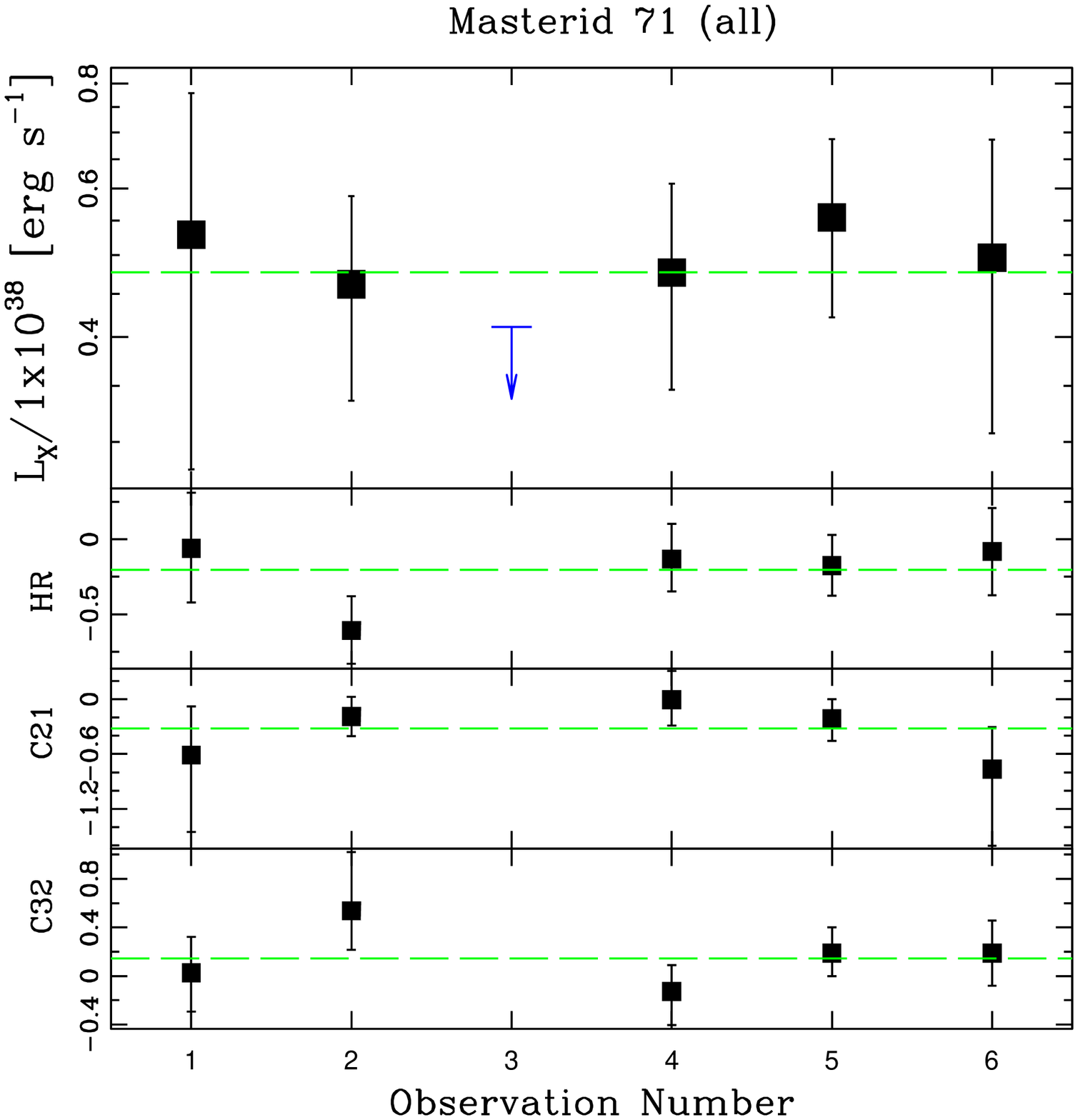}

\end{minipage}\hspace{0.02\linewidth}
\begin{minipage}{0.485\linewidth}
  \centering

    \includegraphics[width=\linewidth]{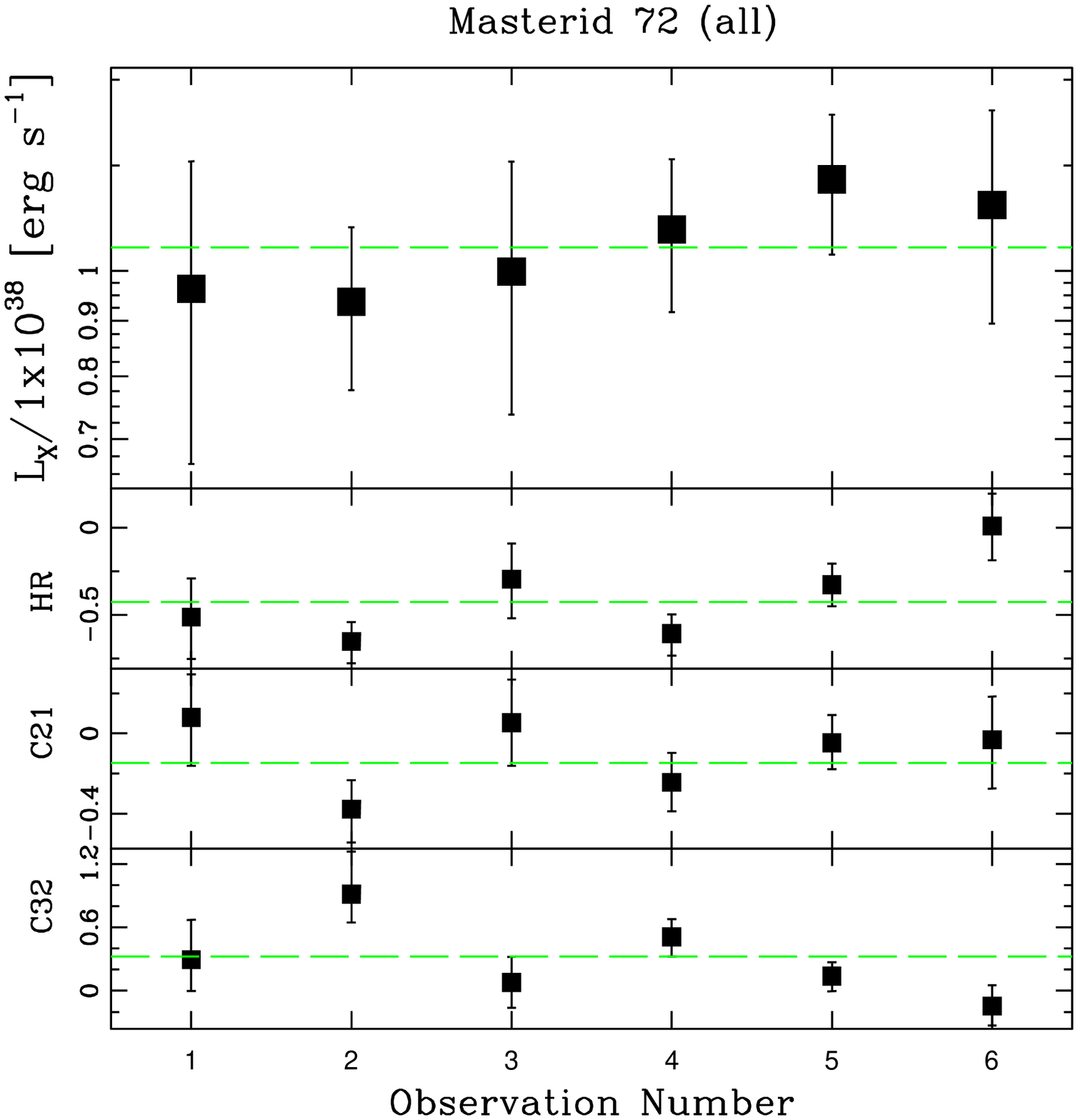}

 \end{minipage}\hspace{0.02\linewidth}

  \begin{minipage}{0.485\linewidth}
  \centering
  
    \includegraphics[width=\linewidth]{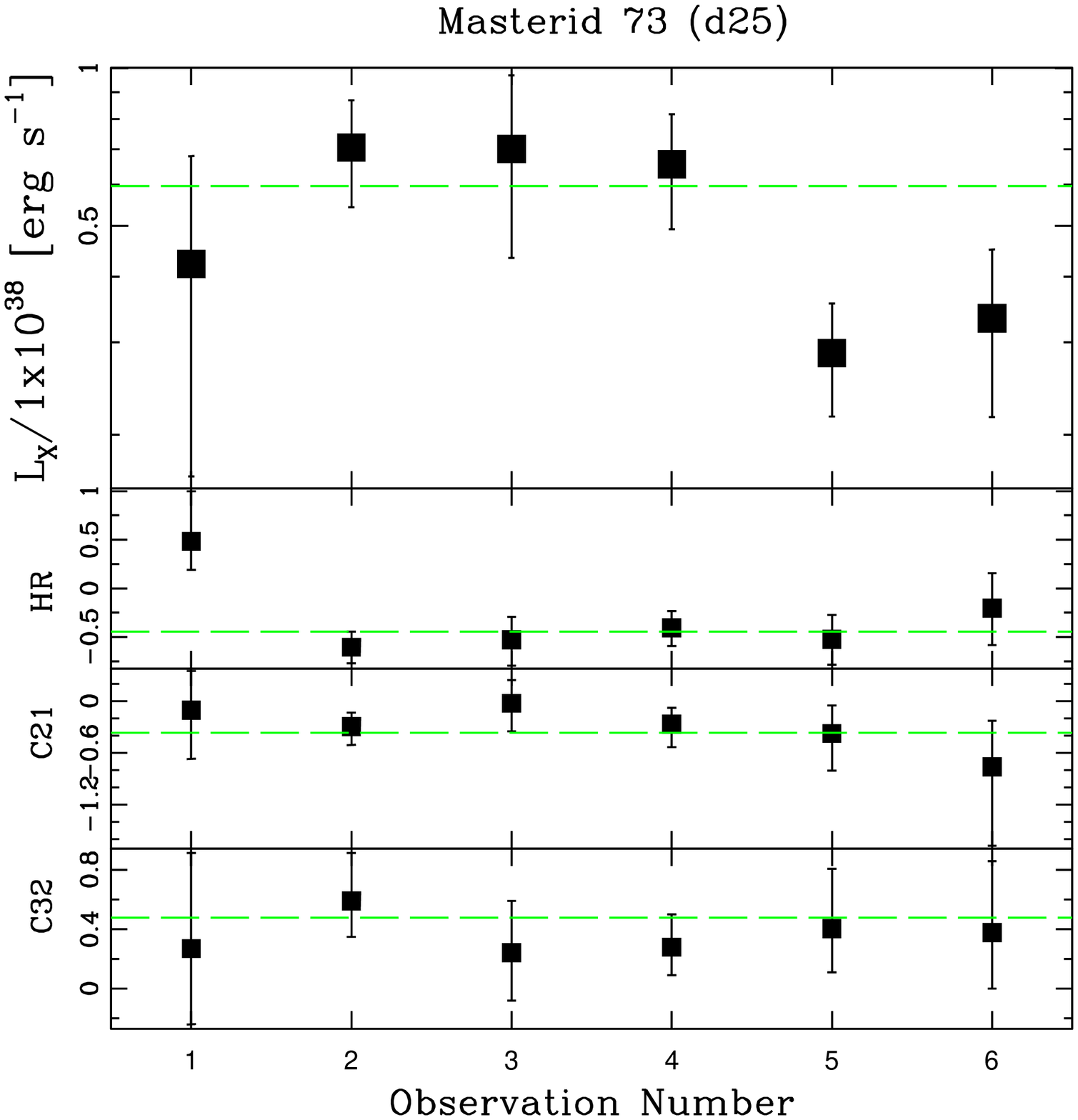}

  \end{minipage}\hspace{0.02\linewidth}
  \begin{minipage}{0.485\linewidth}
  \centering

    \includegraphics[width=\linewidth]{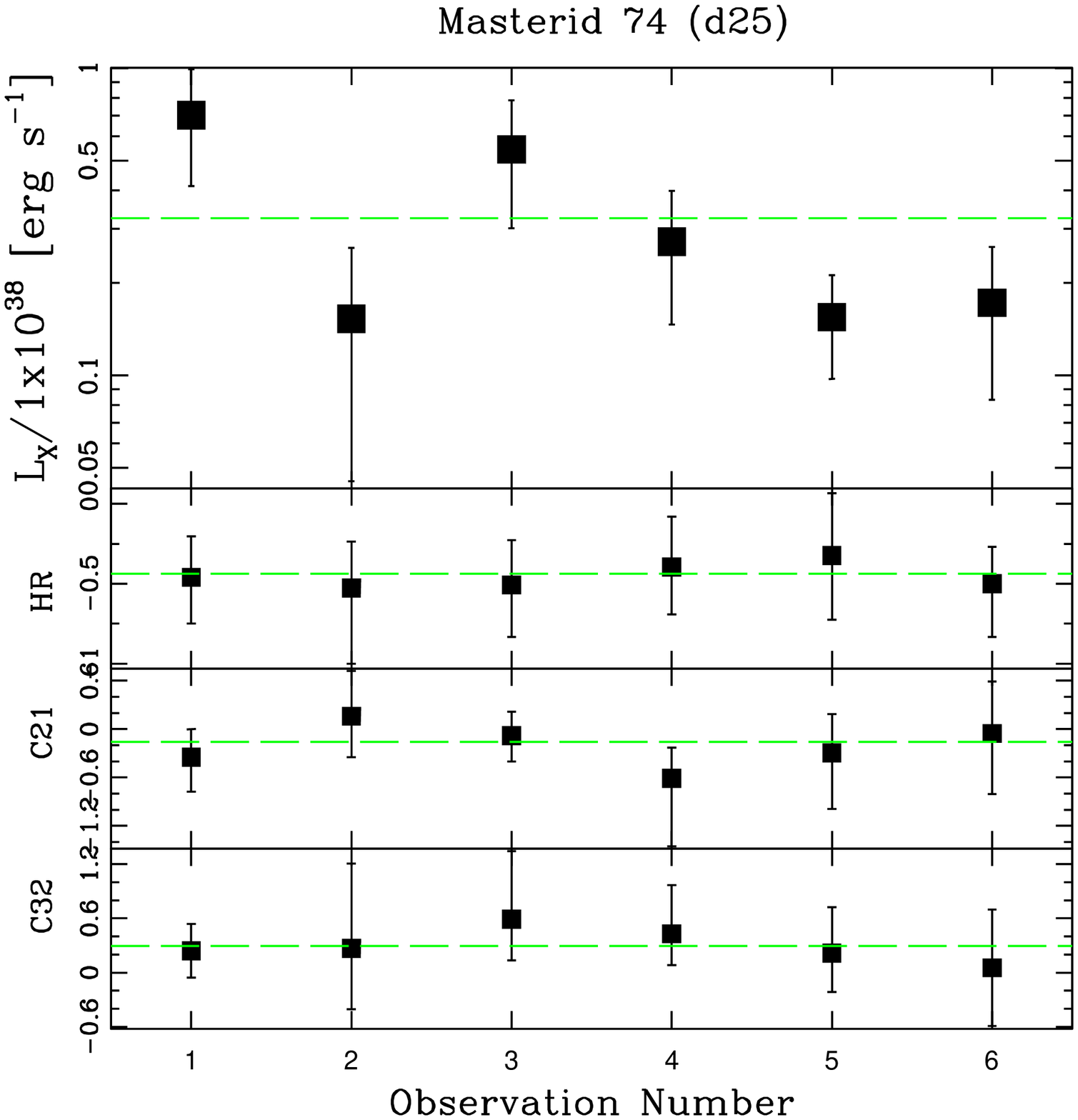}

\end{minipage}\hspace{0.02\linewidth}

\begin{minipage}{0.485\linewidth}
  \centering

    \includegraphics[width=\linewidth]{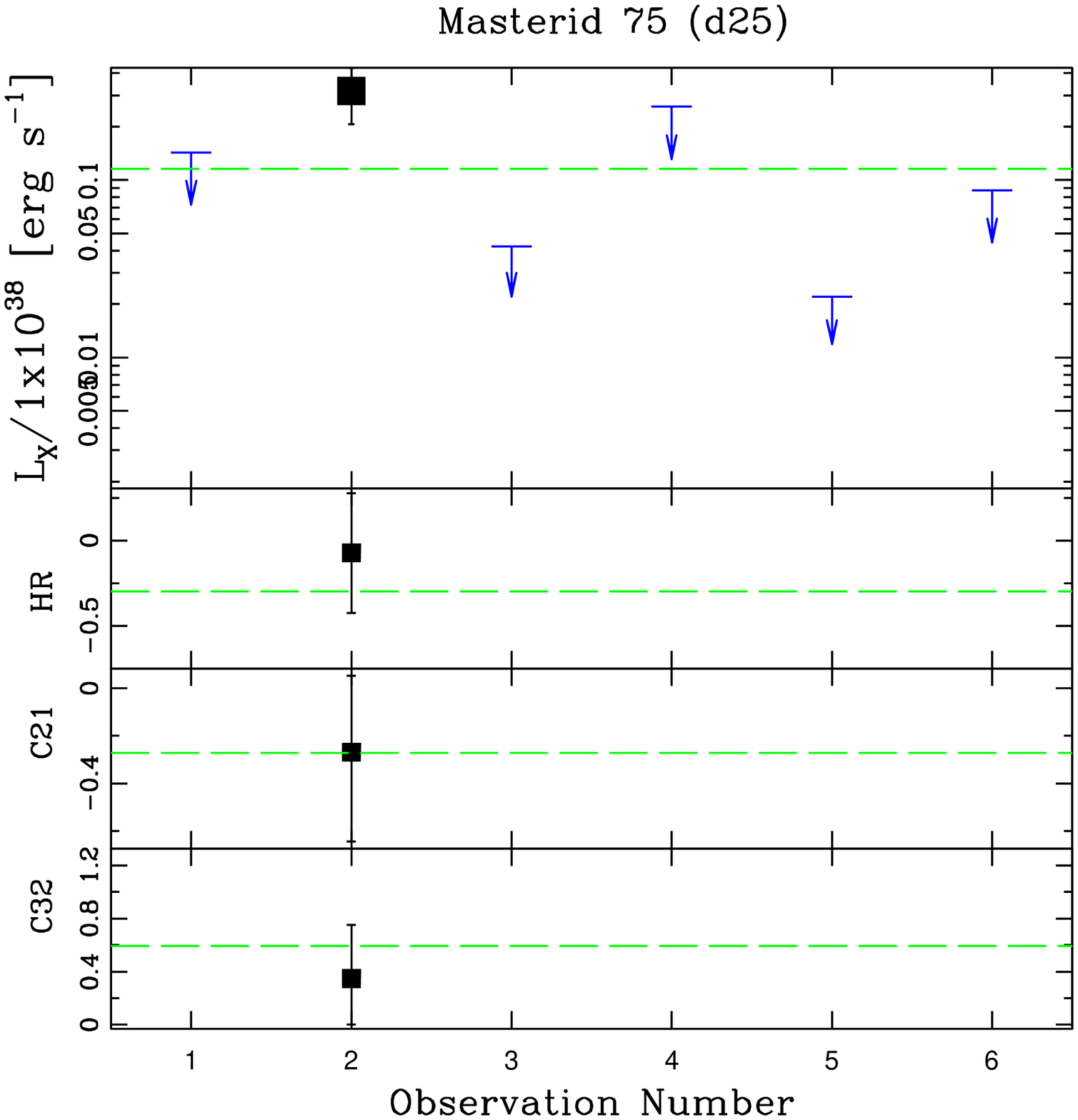}

\end{minipage}\hspace{0.02\linewidth}
\begin{minipage}{0.485\linewidth}
  \centering
  
    \includegraphics[width=\linewidth]{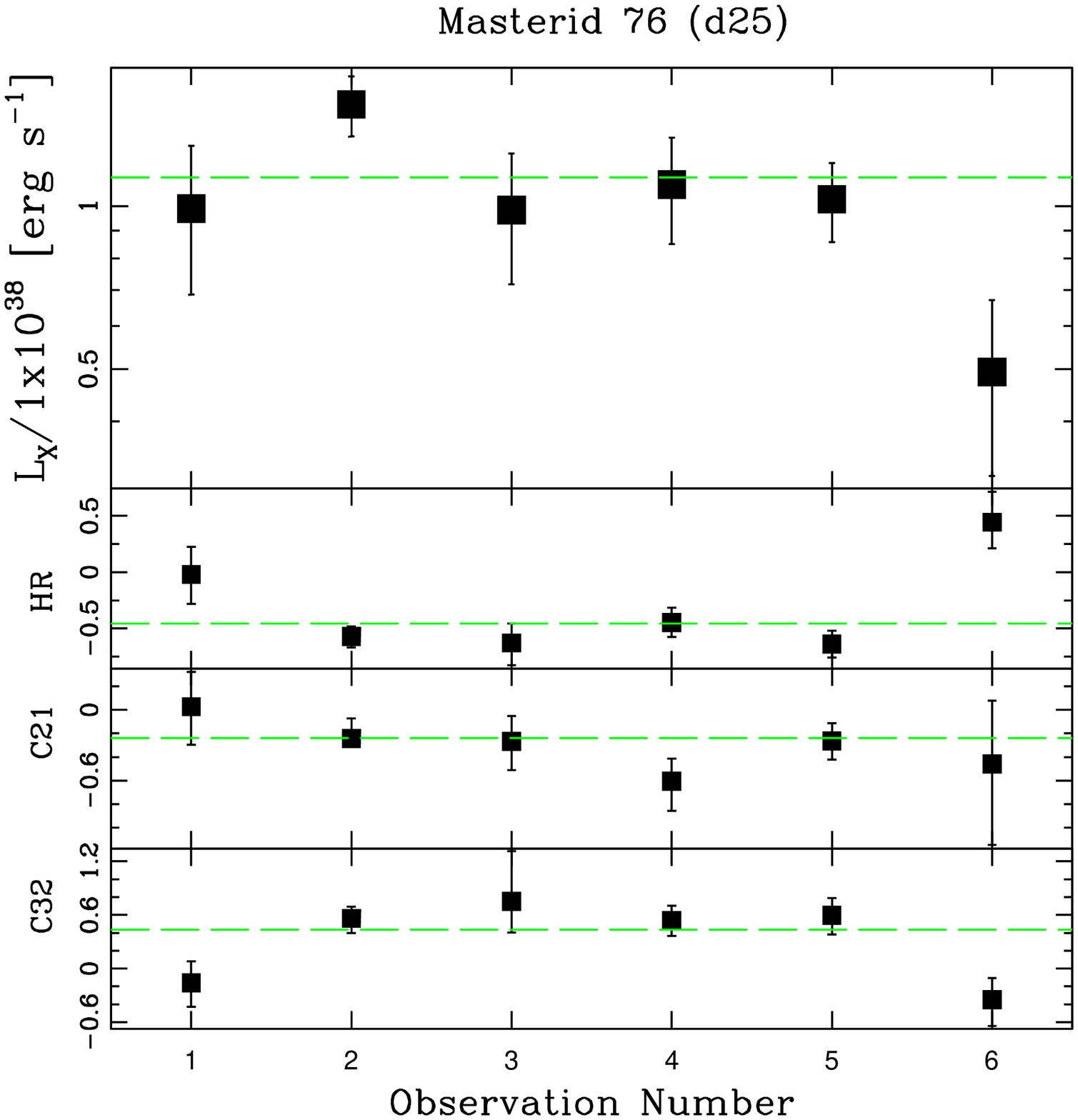}

  \end{minipage}\hspace{0.02\linewidth}
\end{figure}

\clearpage

\begin{figure}
  \begin{minipage}{0.485\linewidth}
  \centering

    \includegraphics[width=\linewidth]{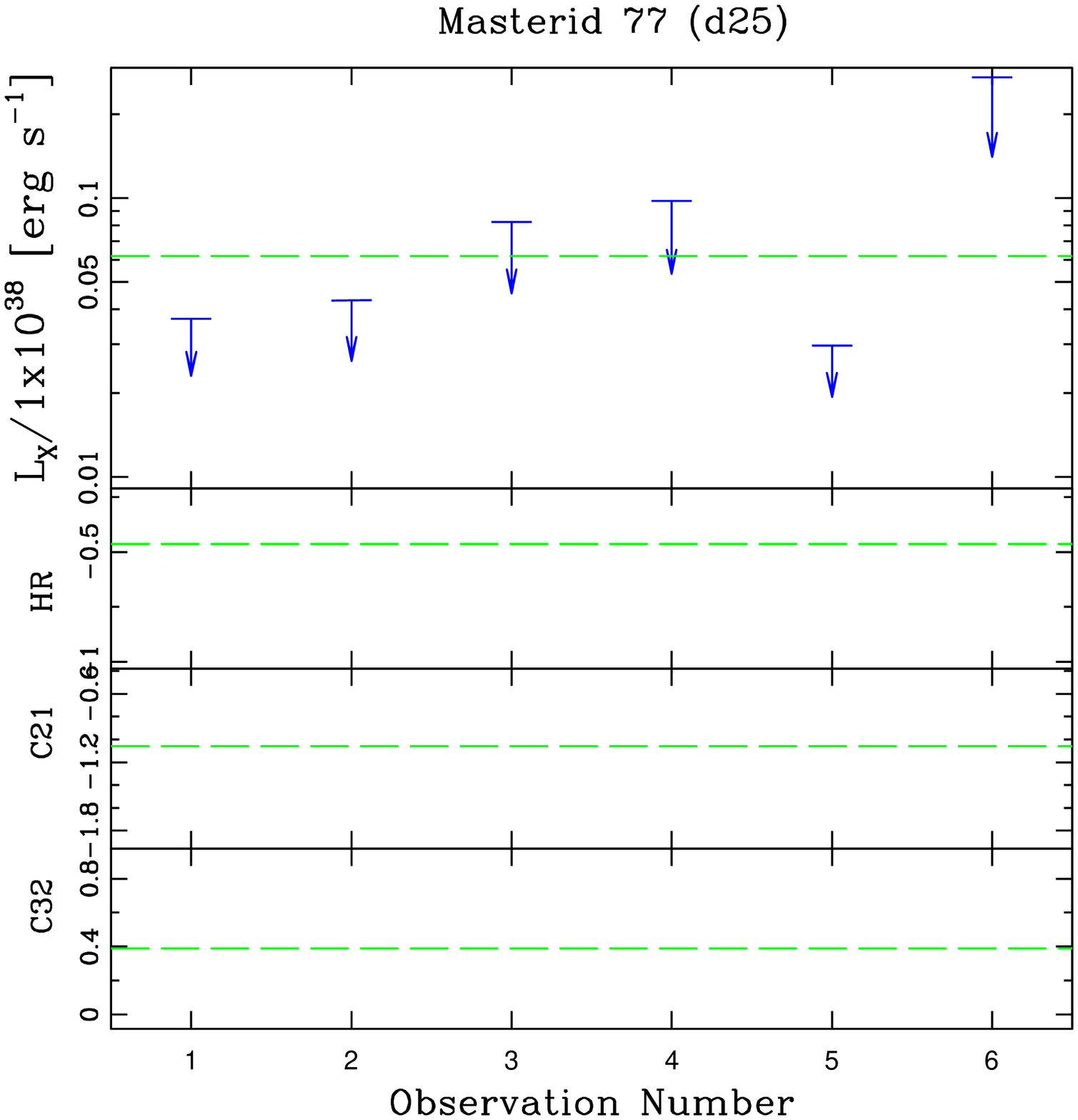}

\end{minipage}\hspace{0.02\linewidth}
\begin{minipage}{0.485\linewidth}
  \centering

    \includegraphics[width=\linewidth]{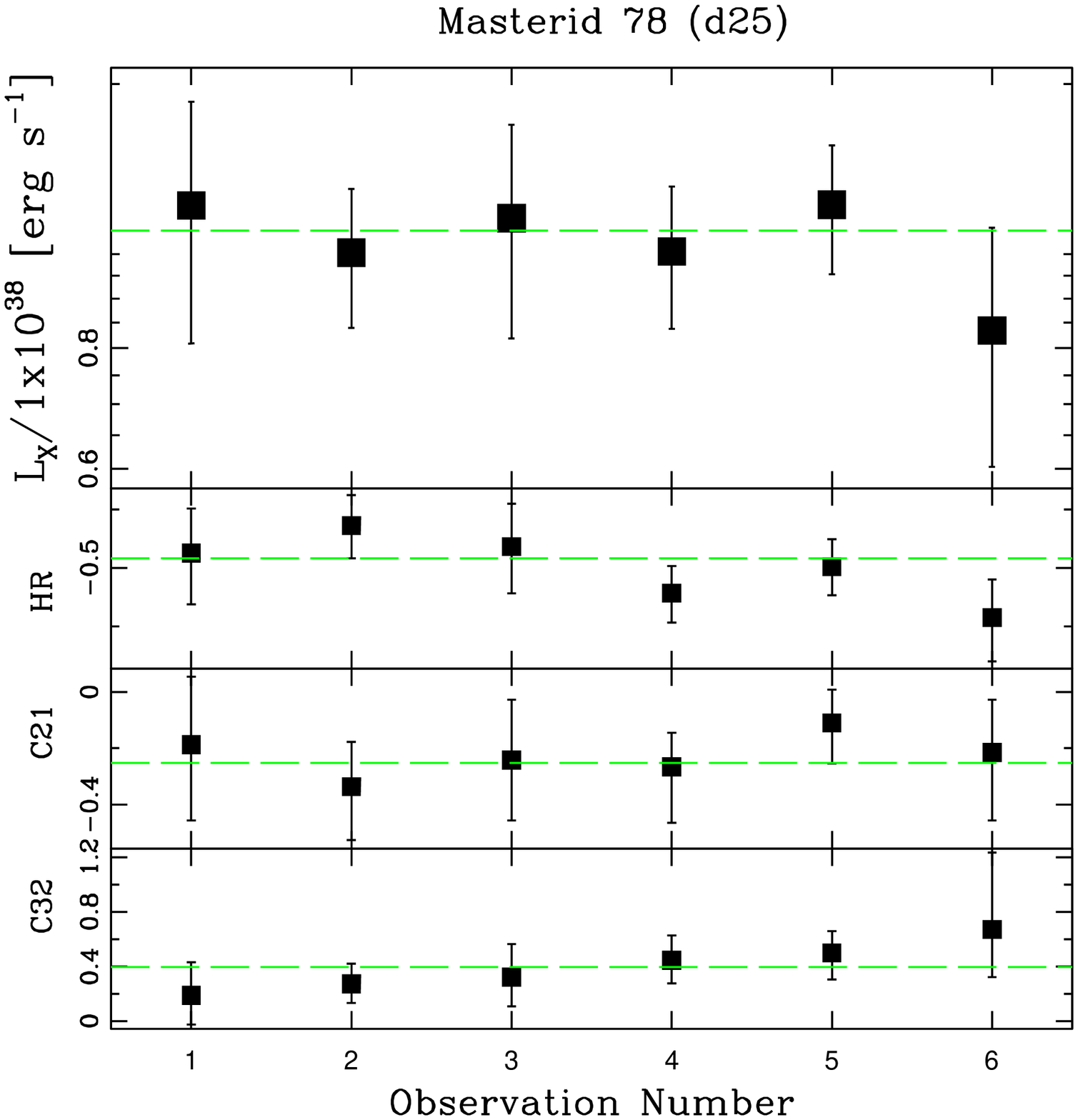}

\end{minipage}\hspace{0.02\linewidth}

  \begin{minipage}{0.485\linewidth}
  \centering
  
    \includegraphics[width=\linewidth]{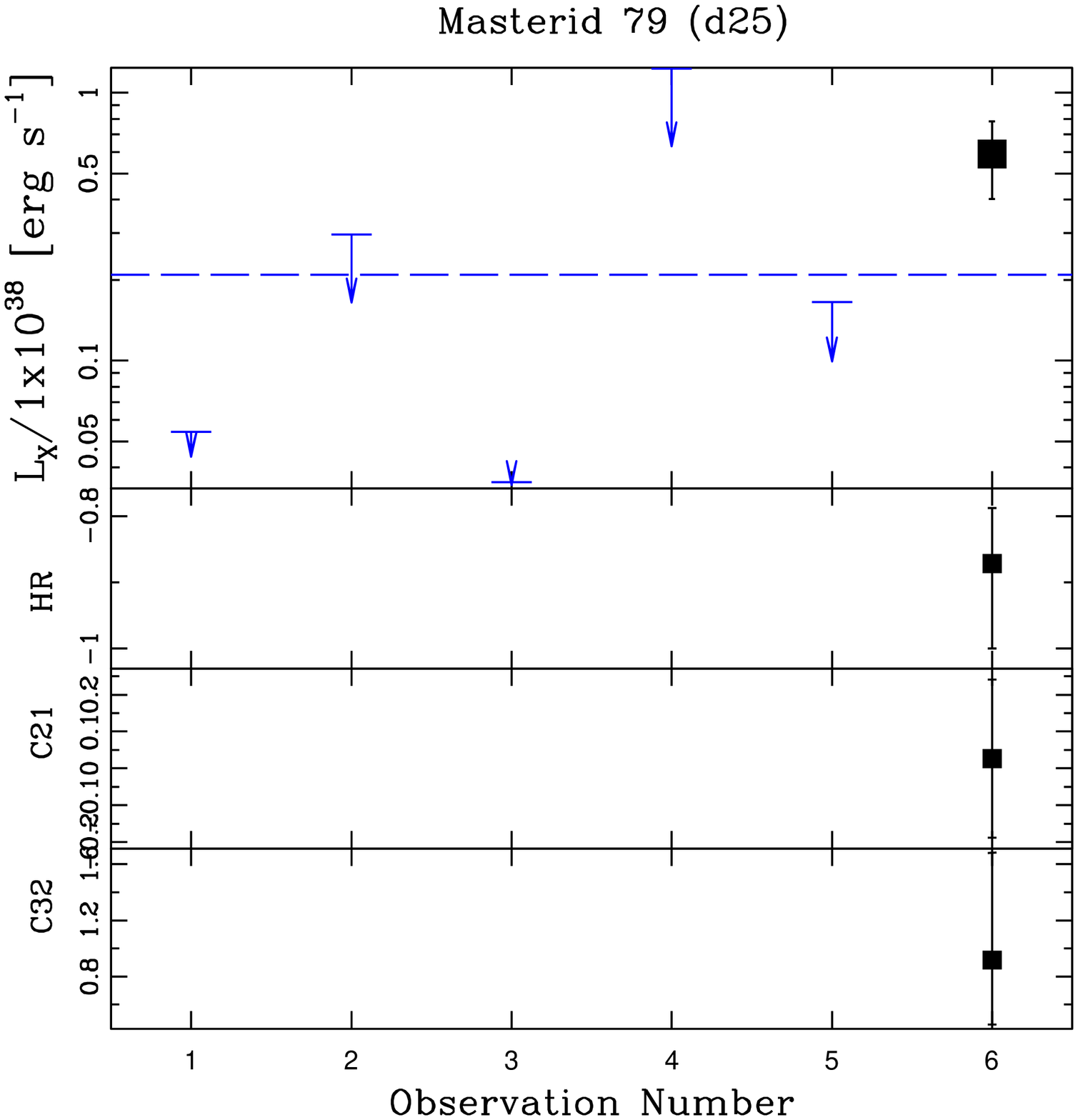}

  \end{minipage}\hspace{0.02\linewidth}
  \begin{minipage}{0.485\linewidth}
  \centering

    \includegraphics[width=\linewidth]{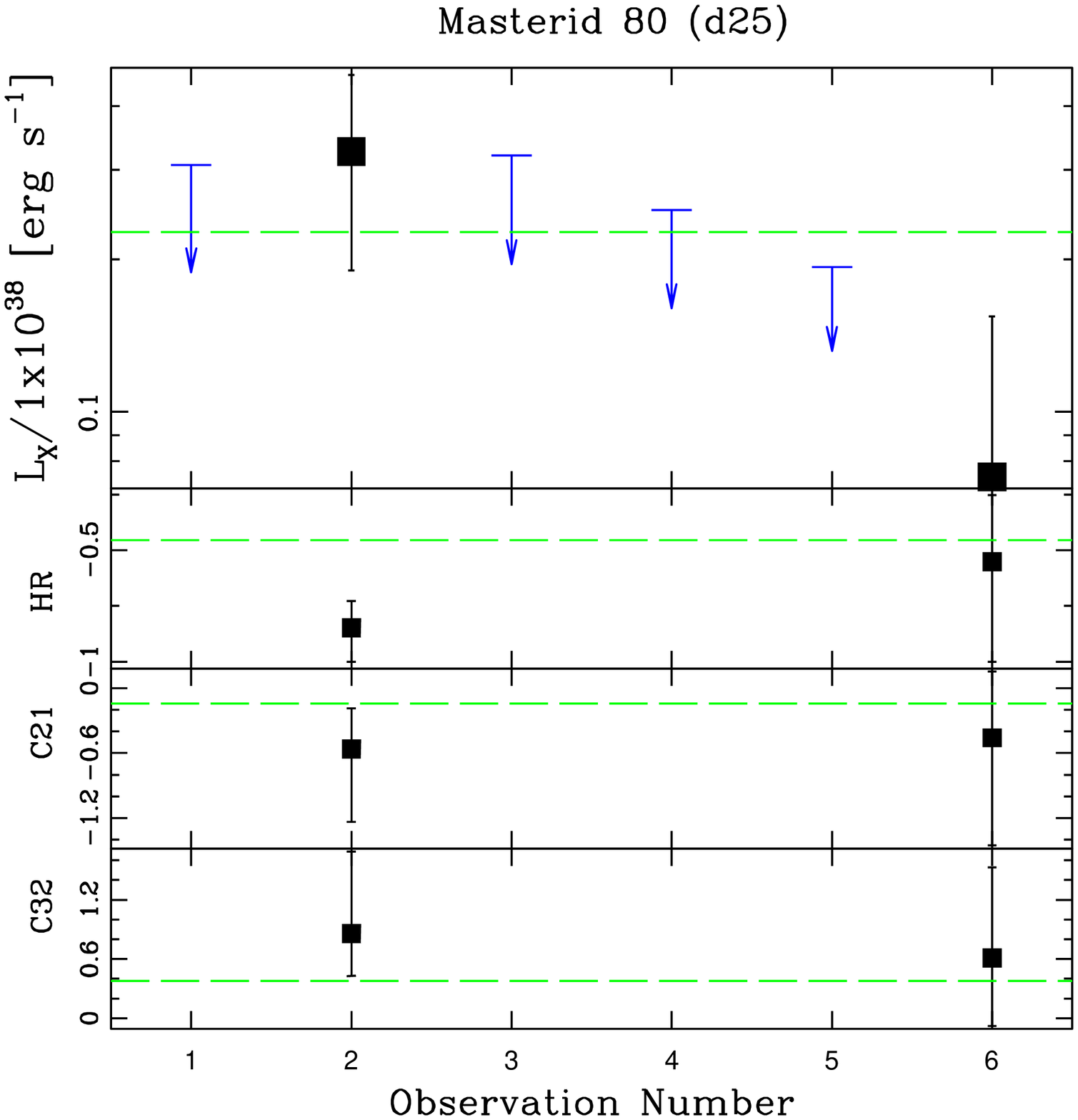}

\end{minipage}\hspace{0.02\linewidth}

\begin{minipage}{0.485\linewidth}
  \centering

    \includegraphics[width=\linewidth]{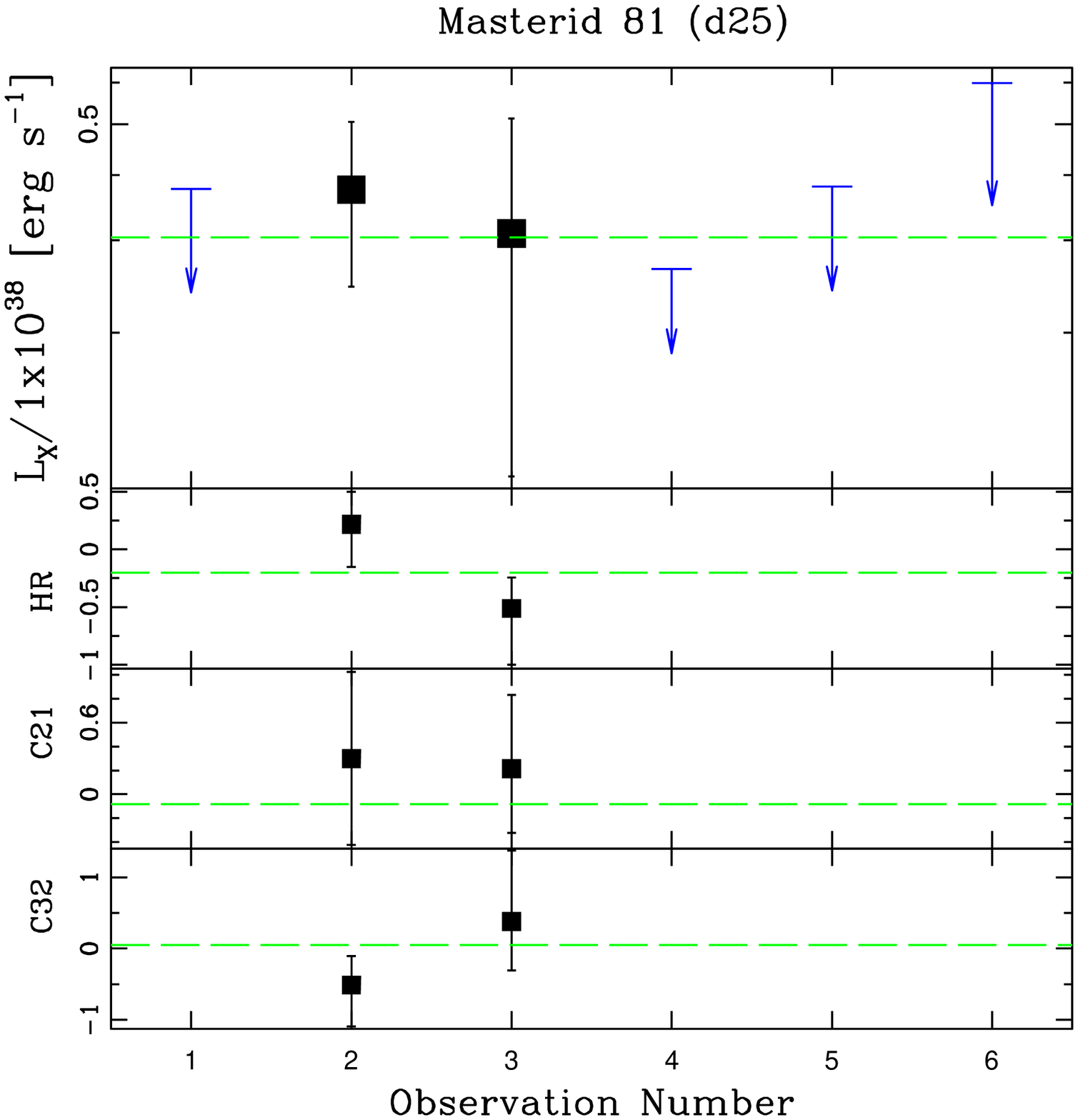}

 \end{minipage}\hspace{0.02\linewidth}
\begin{minipage}{0.485\linewidth}
  \centering
  
    \includegraphics[width=\linewidth]{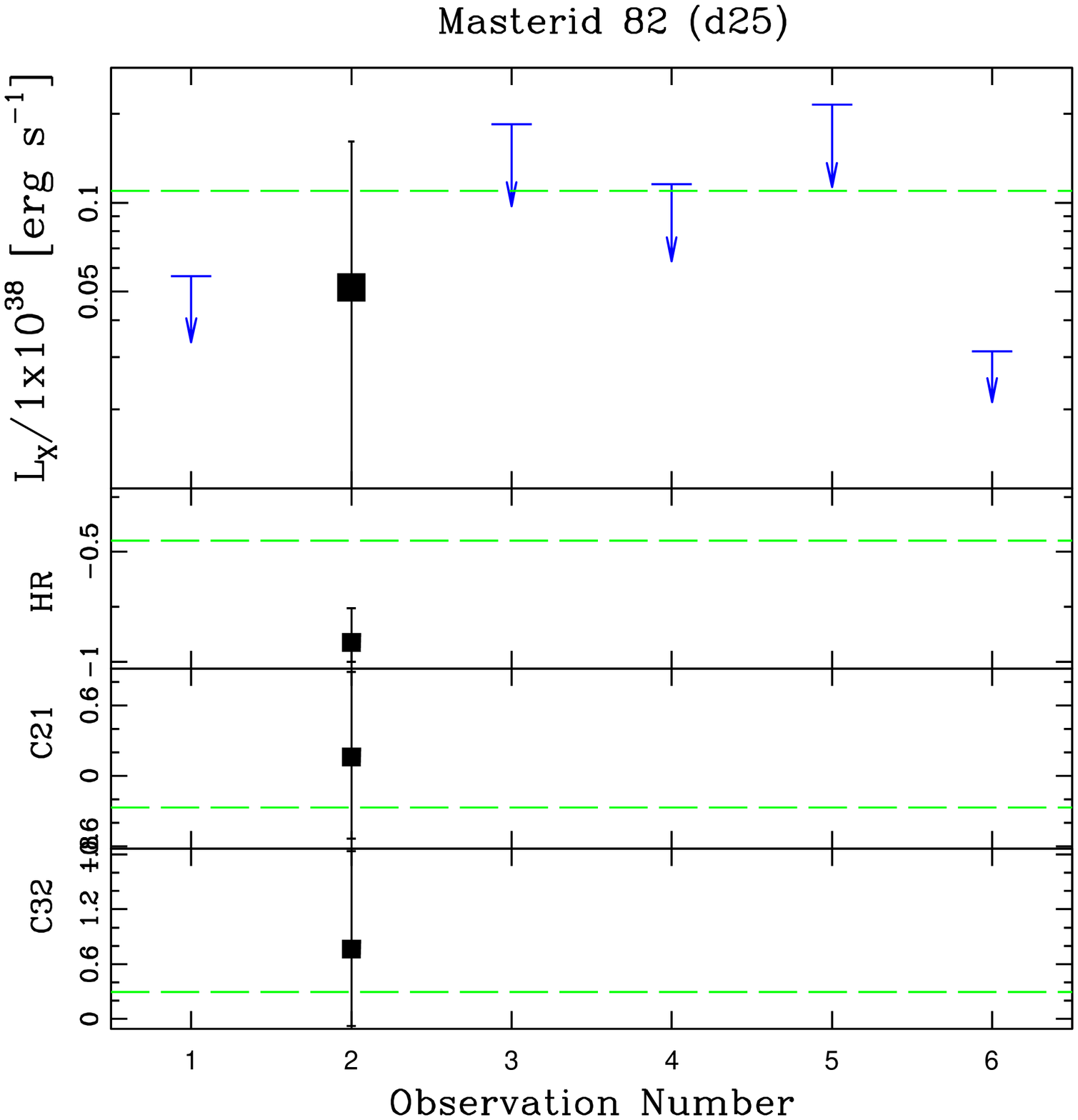}

  \end{minipage}\hspace{0.02\linewidth}
\end{figure}

\begin{figure}
  \begin{minipage}{0.485\linewidth}
  \centering

    \includegraphics[width=\linewidth]{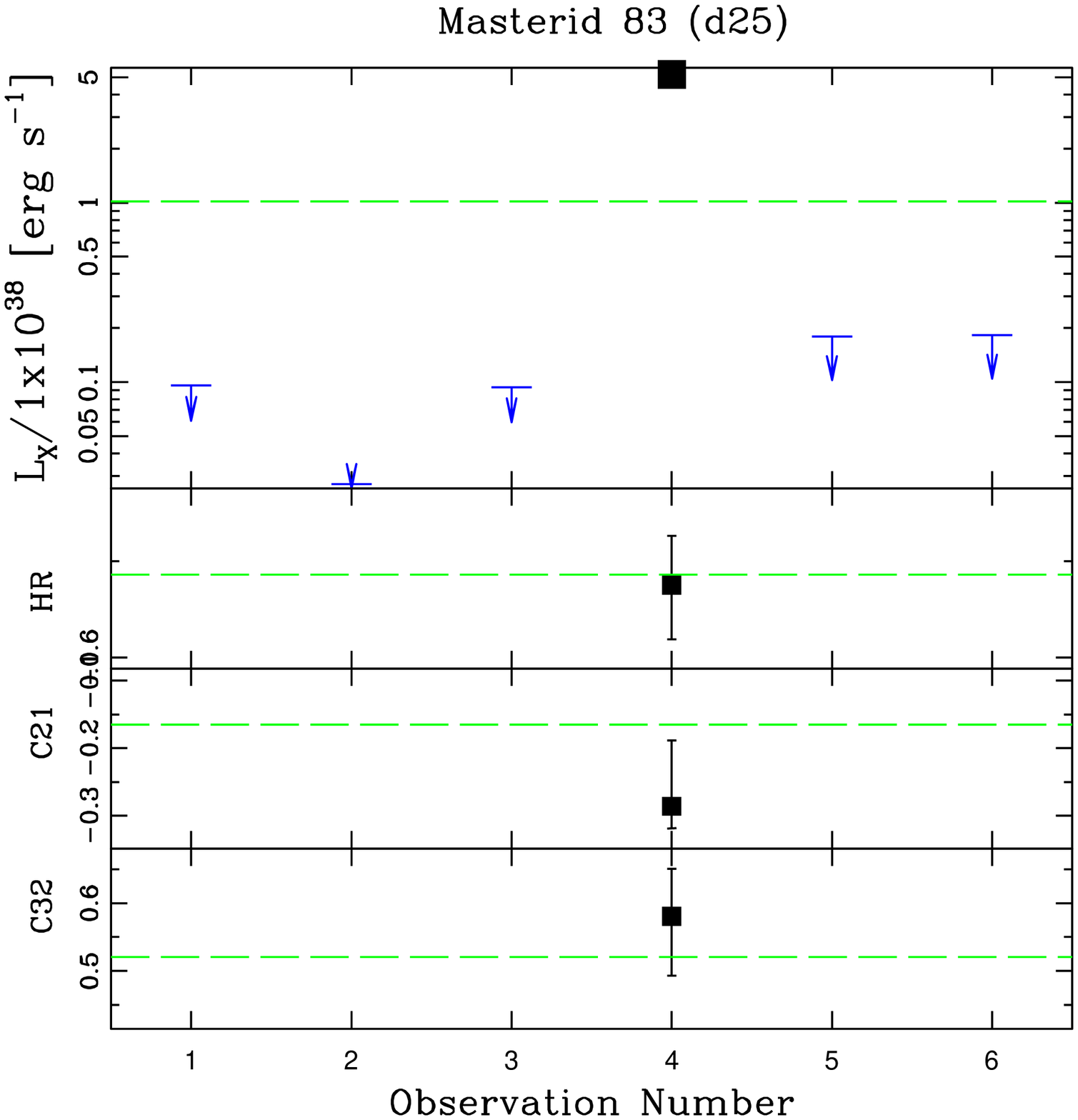}

\end{minipage}\hspace{0.02\linewidth}
\begin{minipage}{0.485\linewidth}
  \centering

    \includegraphics[width=\linewidth]{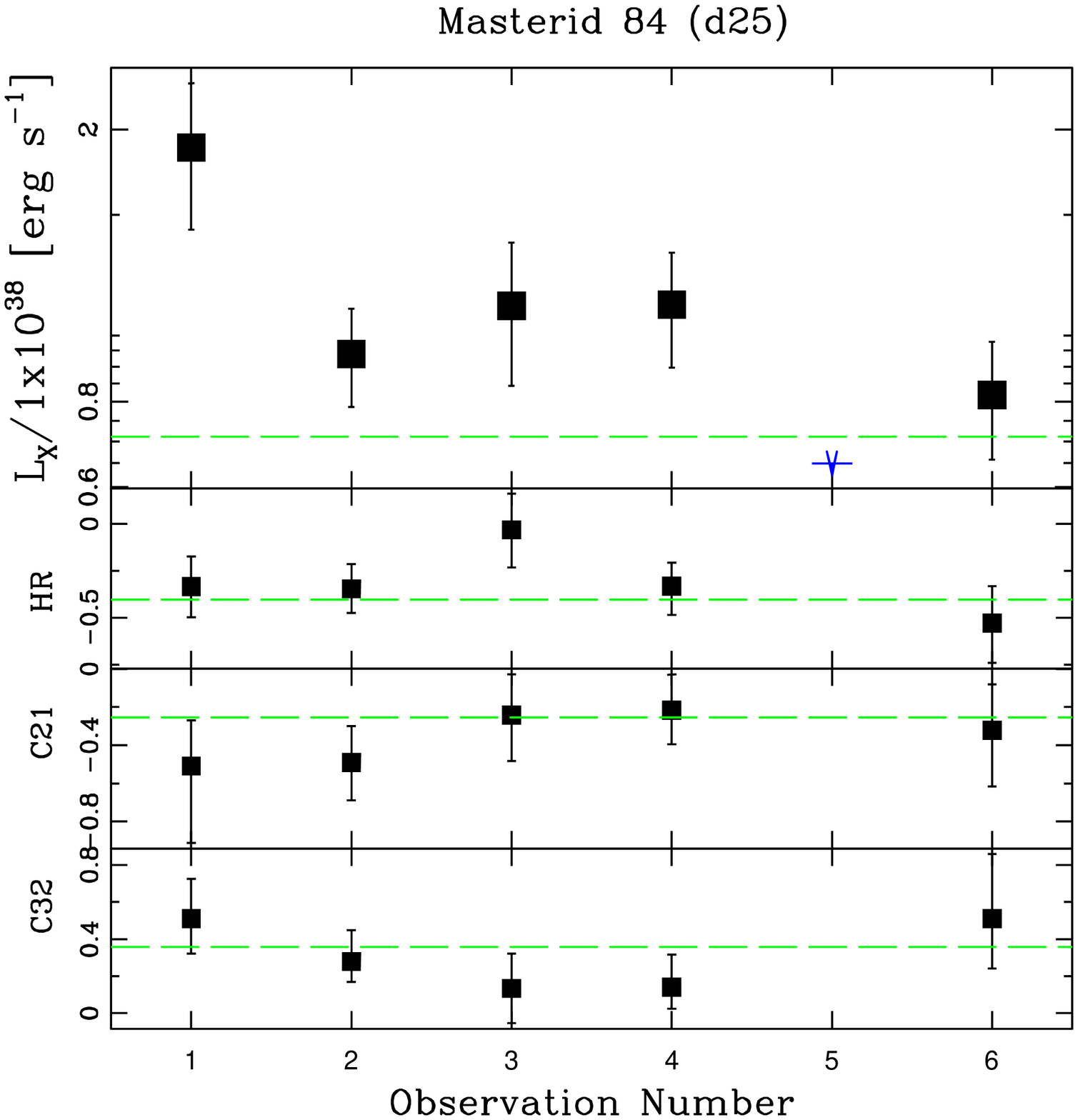}

 \end{minipage}\hspace{0.02\linewidth}

  \begin{minipage}{0.485\linewidth}
  \centering
  
    \includegraphics[width=\linewidth]{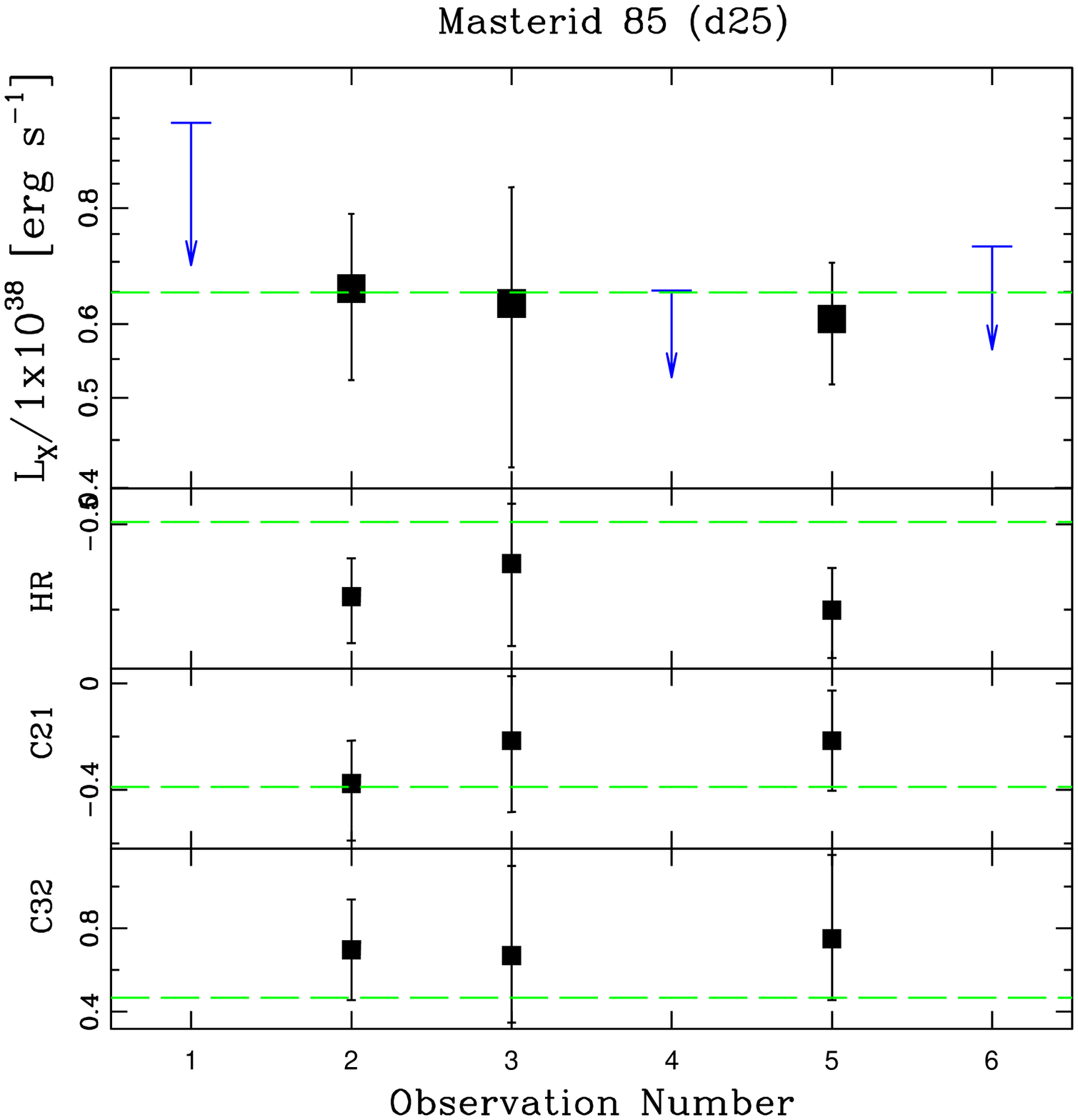}

  \end{minipage}\hspace{0.02\linewidth}
  \begin{minipage}{0.485\linewidth}
  \centering

    \includegraphics[width=\linewidth]{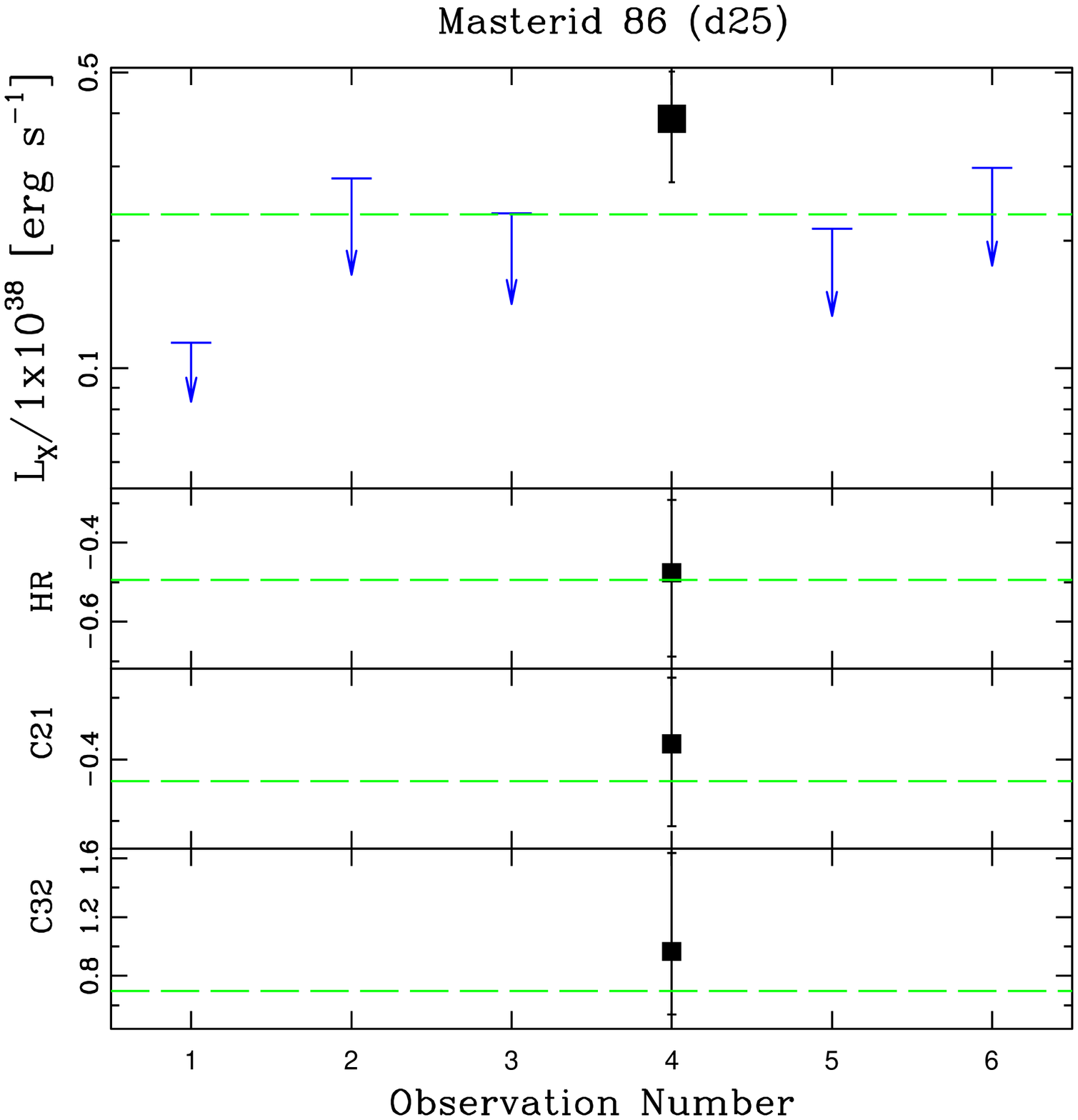}

\end{minipage}\hspace{0.02\linewidth}

\begin{minipage}{0.485\linewidth}
  \centering

    \includegraphics[width=\linewidth]{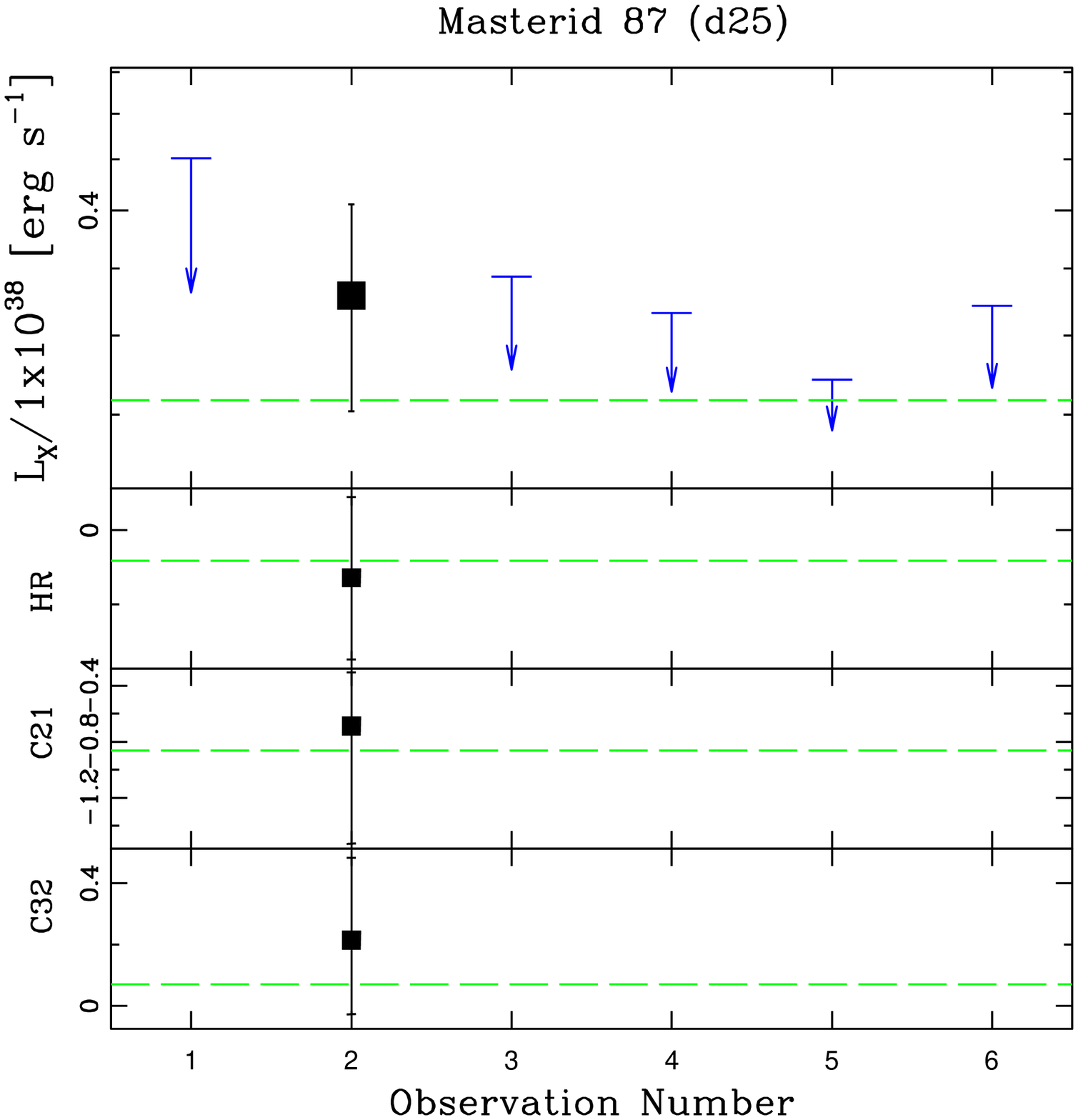}

\end{minipage}\hspace{0.02\linewidth}
\begin{minipage}{0.485\linewidth}
  \centering
  
    \includegraphics[width=\linewidth]{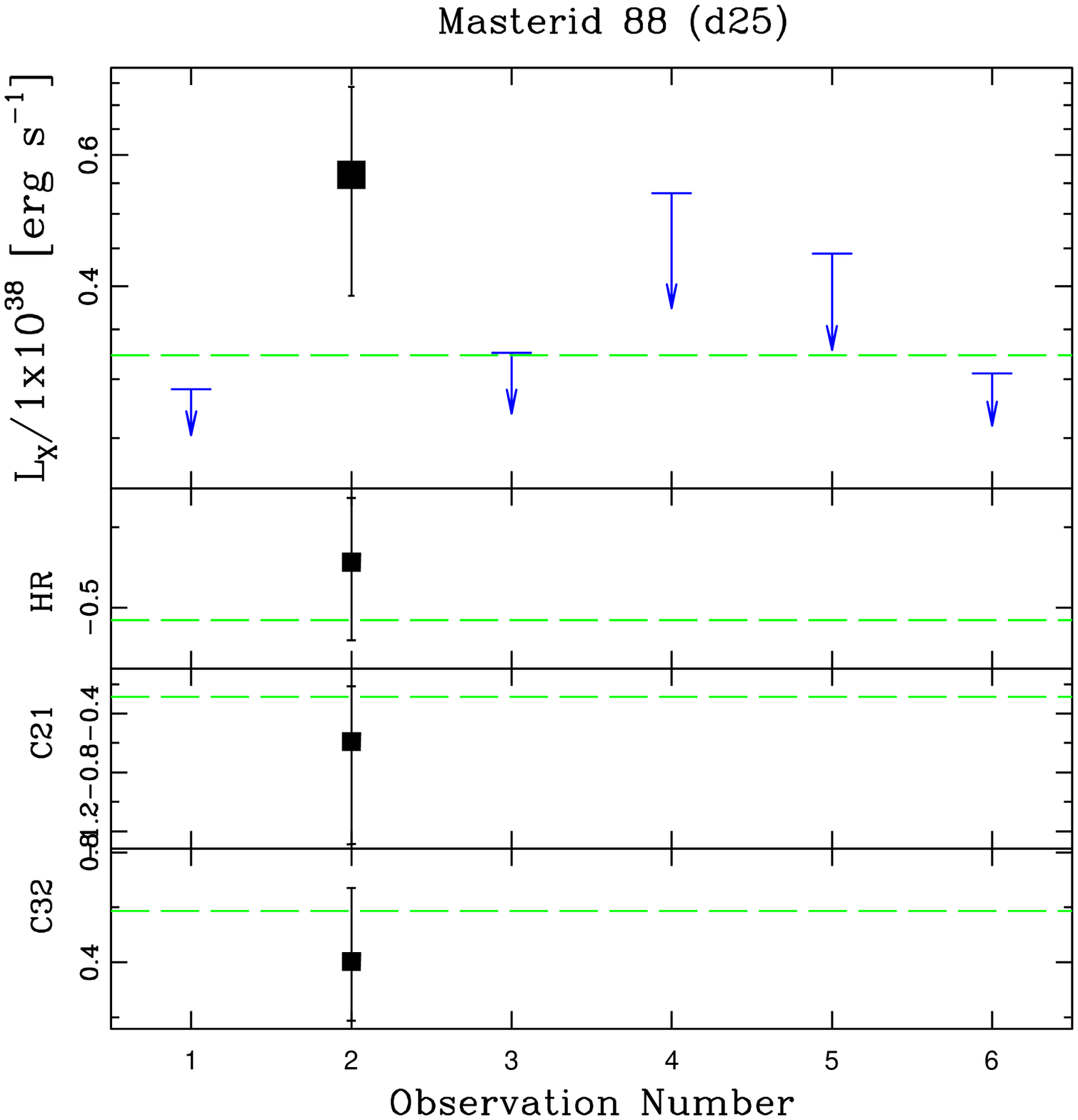}

  \end{minipage}\hspace{0.02\linewidth}

\end{figure}

\begin{figure}
  \begin{minipage}{0.485\linewidth}
  \centering

    \includegraphics[width=\linewidth]{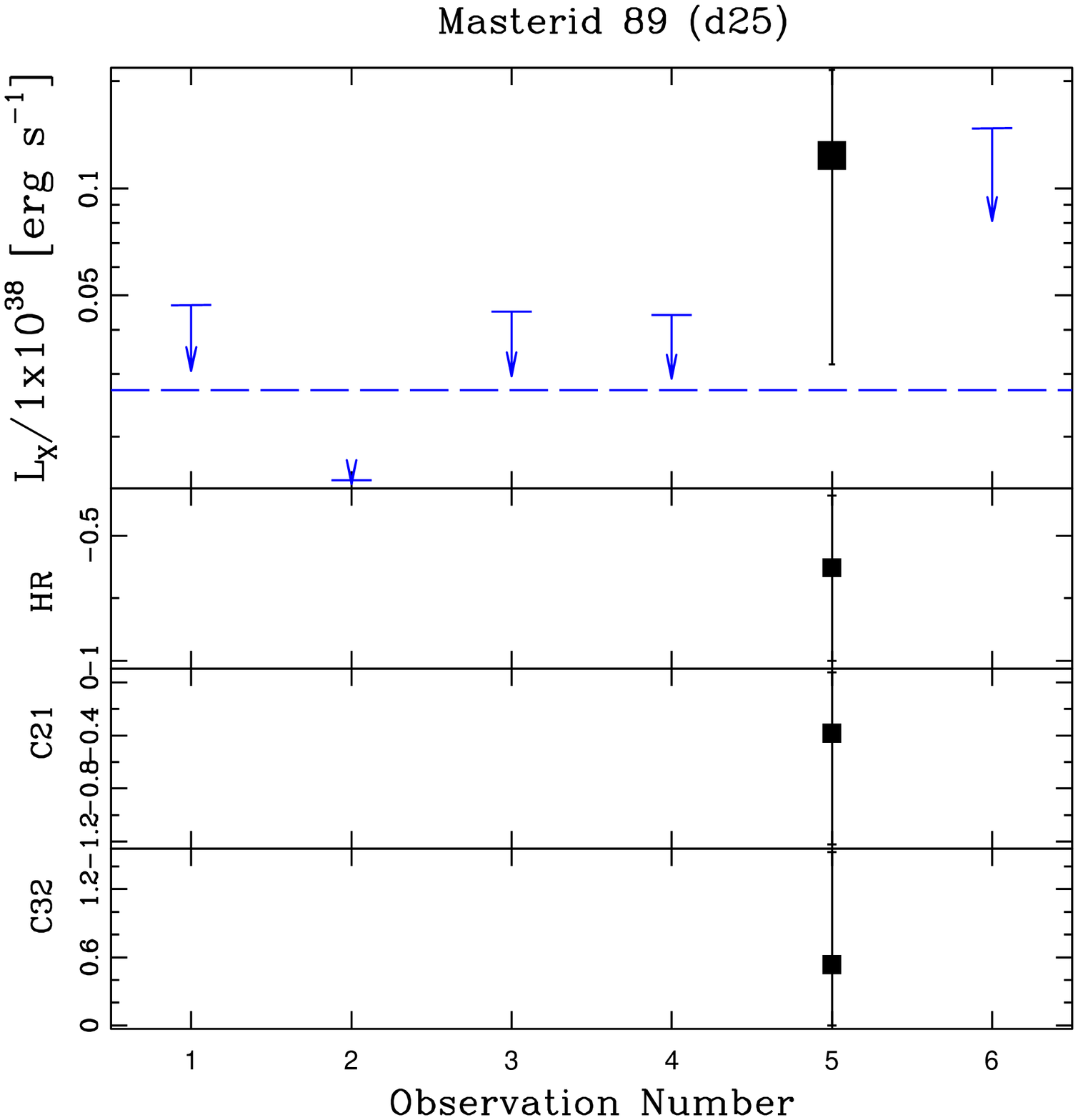}

\end{minipage}\hspace{0.02\linewidth}
\begin{minipage}{0.485\linewidth}
  \centering

    \includegraphics[width=\linewidth]{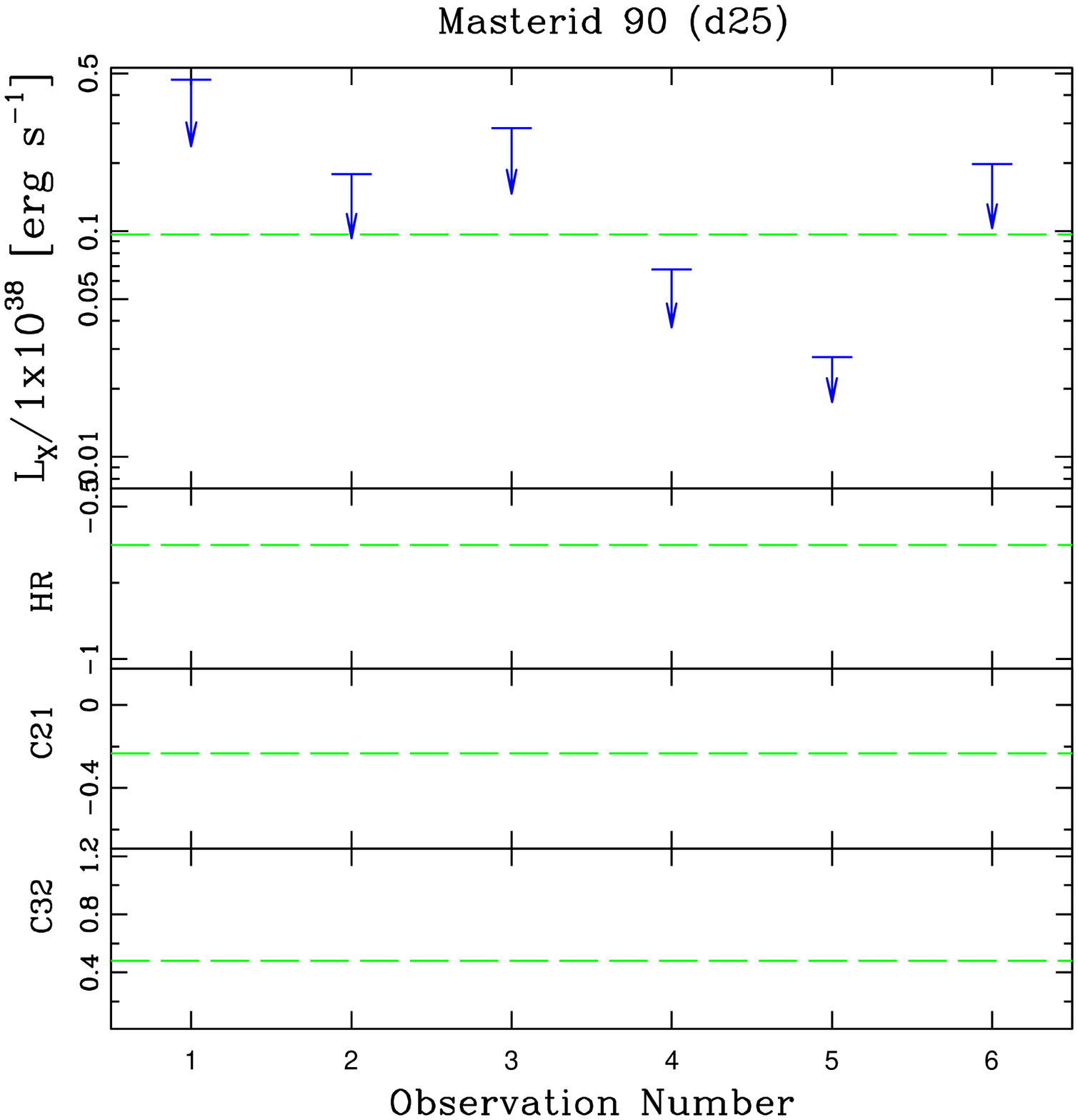}

\end{minipage}\hspace{0.02\linewidth}

  \begin{minipage}{0.485\linewidth}
  \centering
  
    \includegraphics[width=\linewidth]{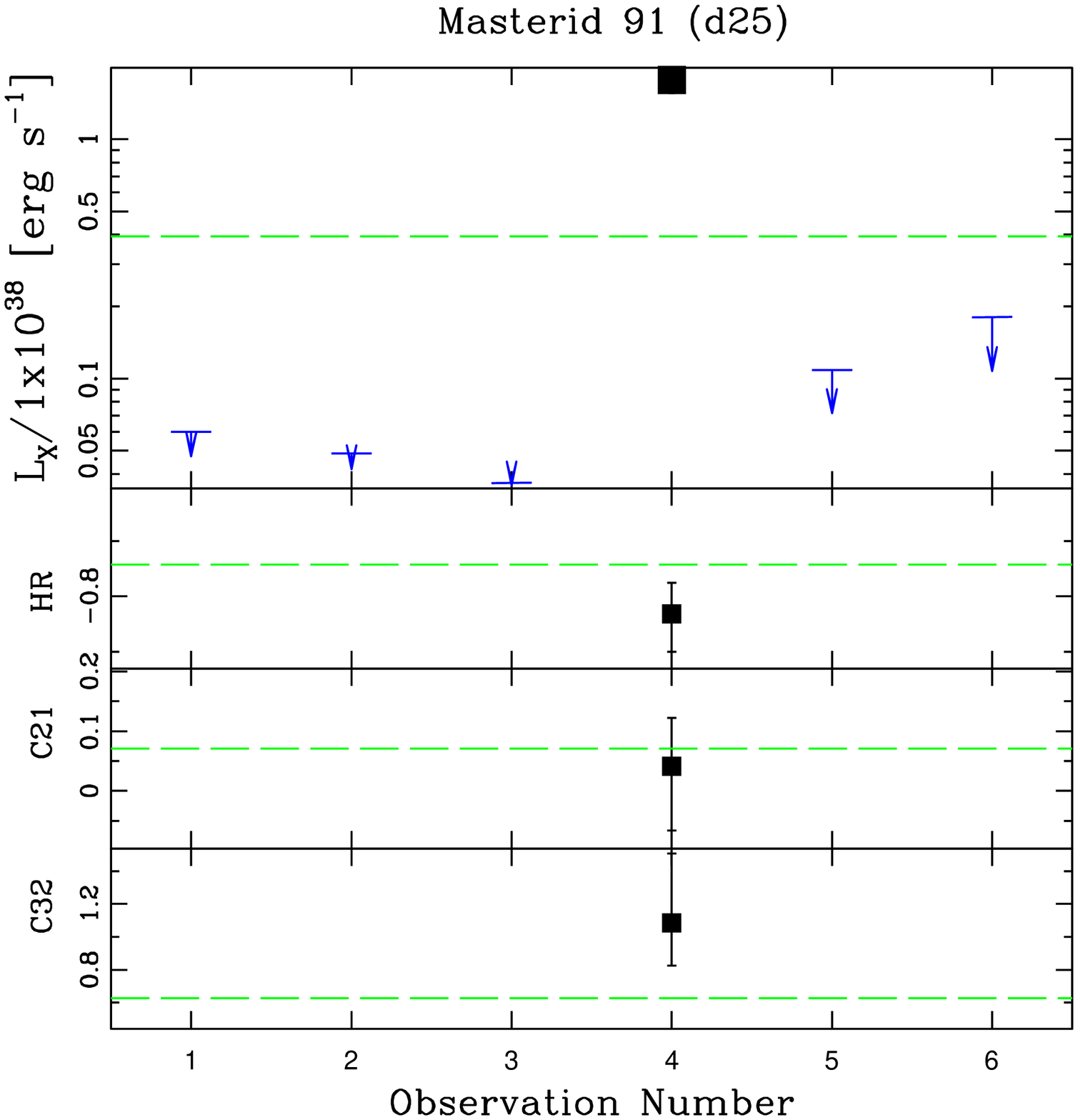}

  \end{minipage}\hspace{0.02\linewidth}
  \begin{minipage}{0.485\linewidth}
  \centering

    \includegraphics[width=\linewidth]{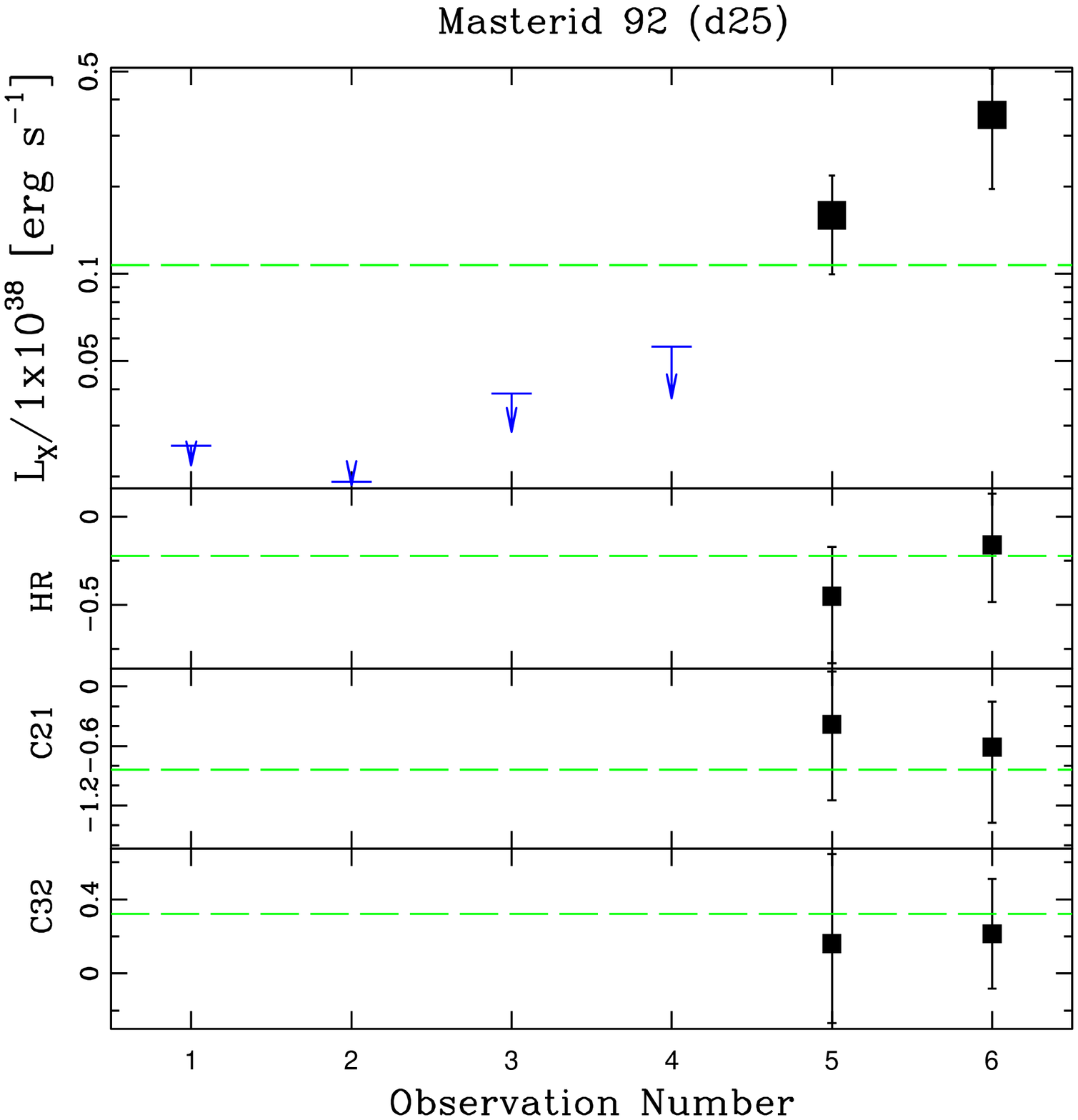}

\end{minipage}\hspace{0.02\linewidth}

\begin{minipage}{0.485\linewidth}
  \centering

    \includegraphics[width=\linewidth]{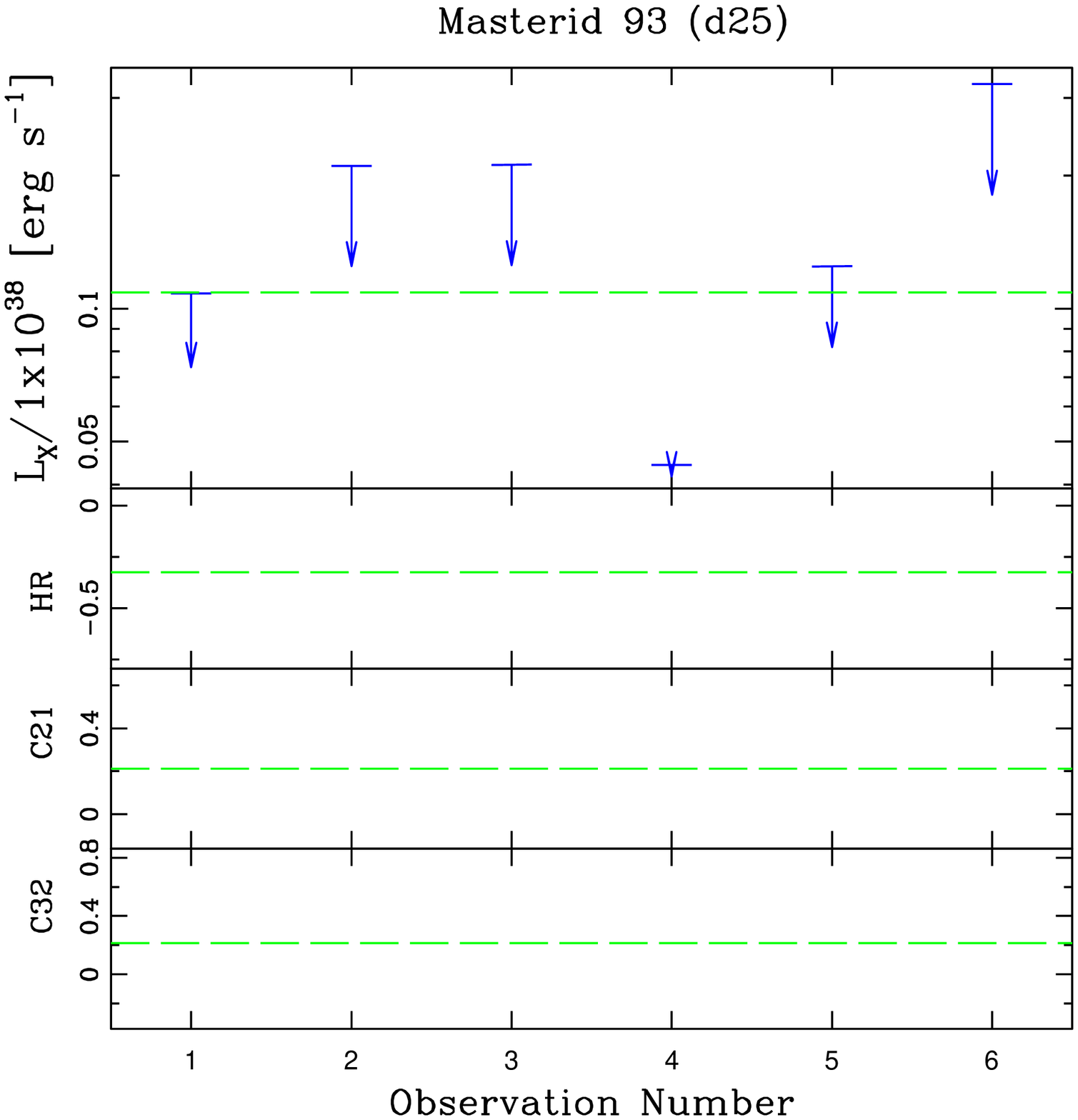}

 \end{minipage}\hspace{0.02\linewidth}
\begin{minipage}{0.485\linewidth}
  \centering
  
    \includegraphics[width=\linewidth]{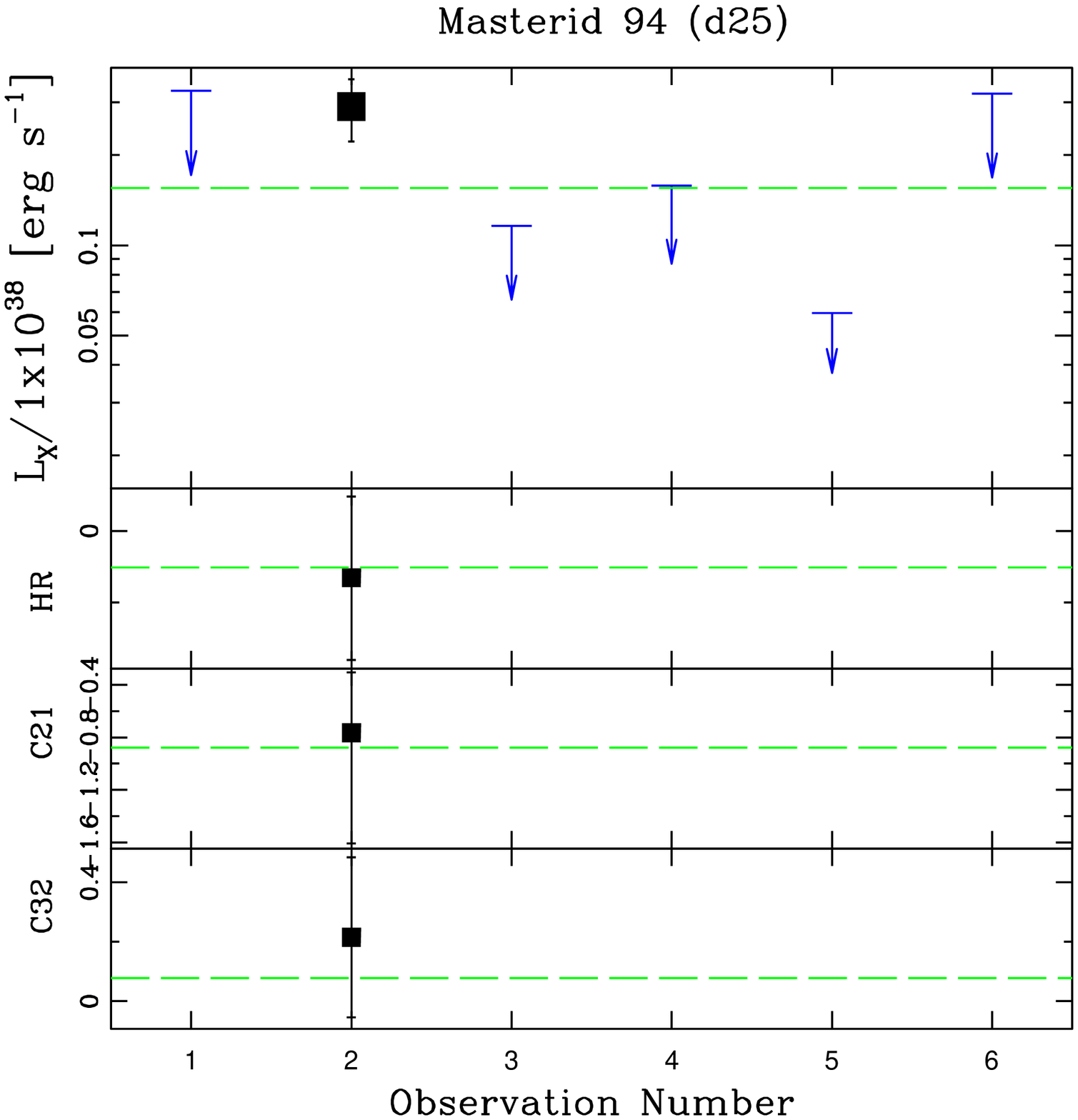}

  \end{minipage}\hspace{0.02\linewidth}
\end{figure}

\begin{figure}
  \begin{minipage}{0.485\linewidth}
  \centering

    \includegraphics[width=\linewidth]{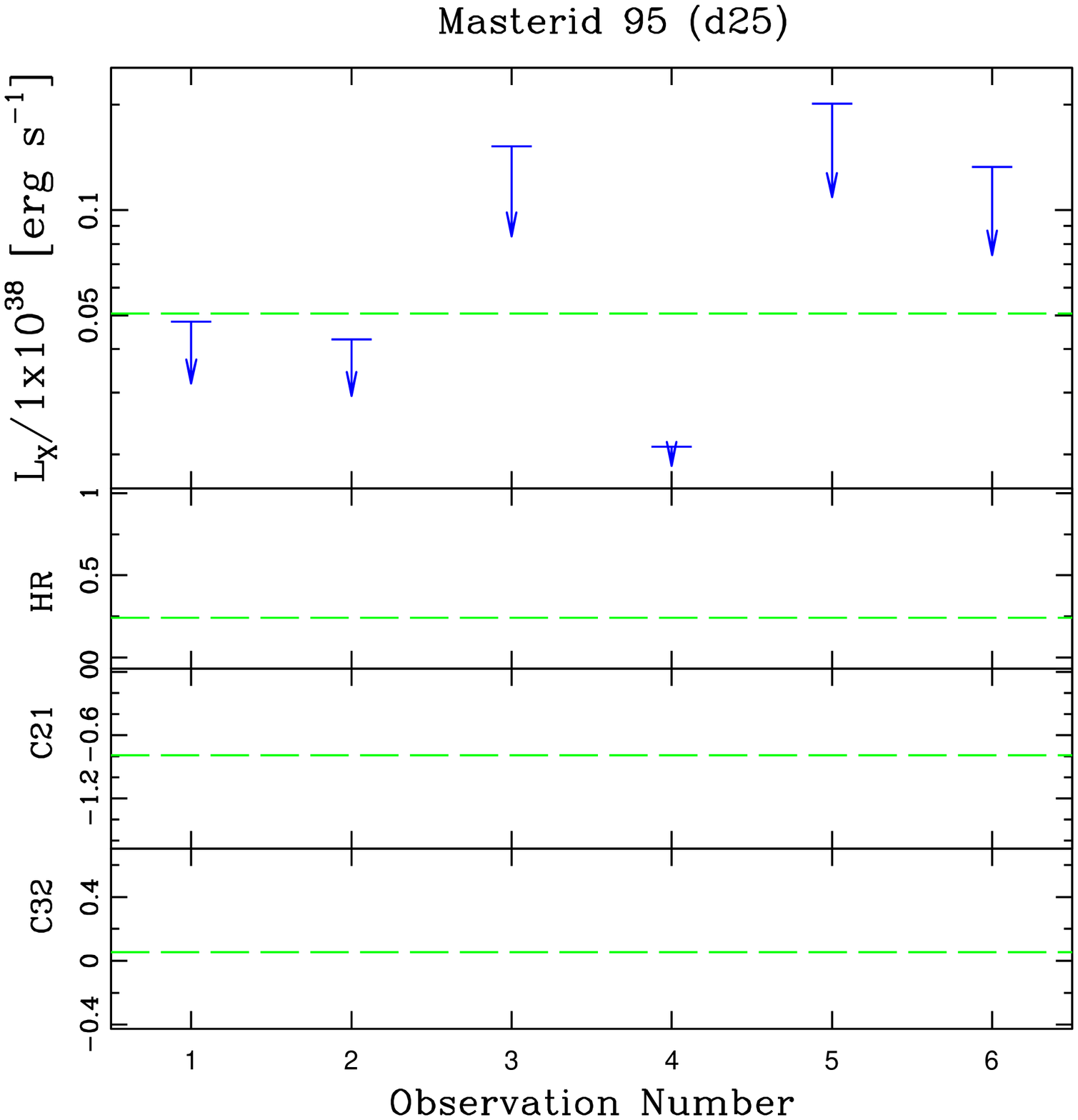}

\end{minipage}\hspace{0.02\linewidth}
\begin{minipage}{0.485\linewidth}
  \centering

    \includegraphics[width=\linewidth]{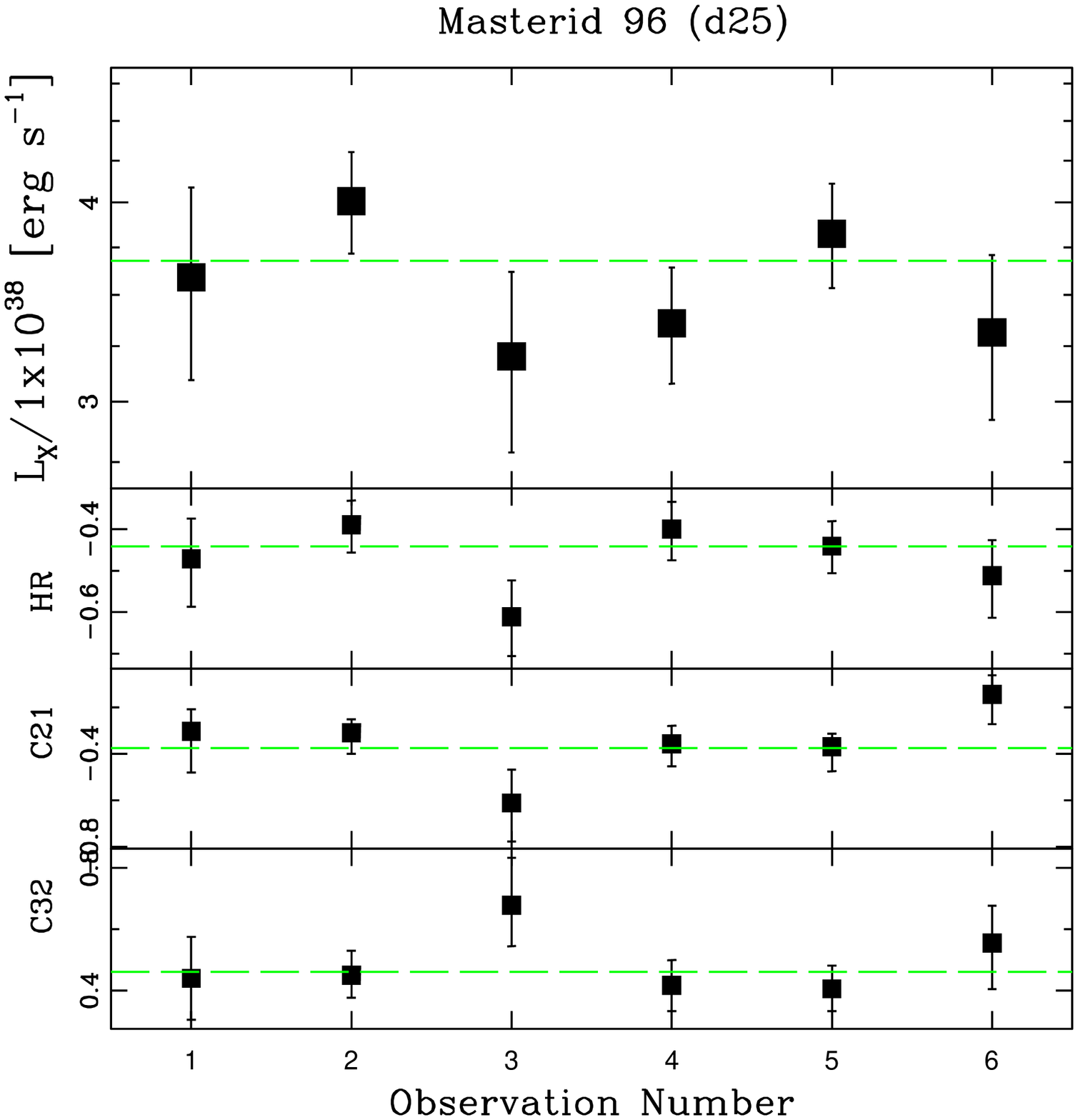}

 \end{minipage}\hspace{0.02\linewidth}

  \begin{minipage}{0.485\linewidth}
  \centering
  
    \includegraphics[width=\linewidth]{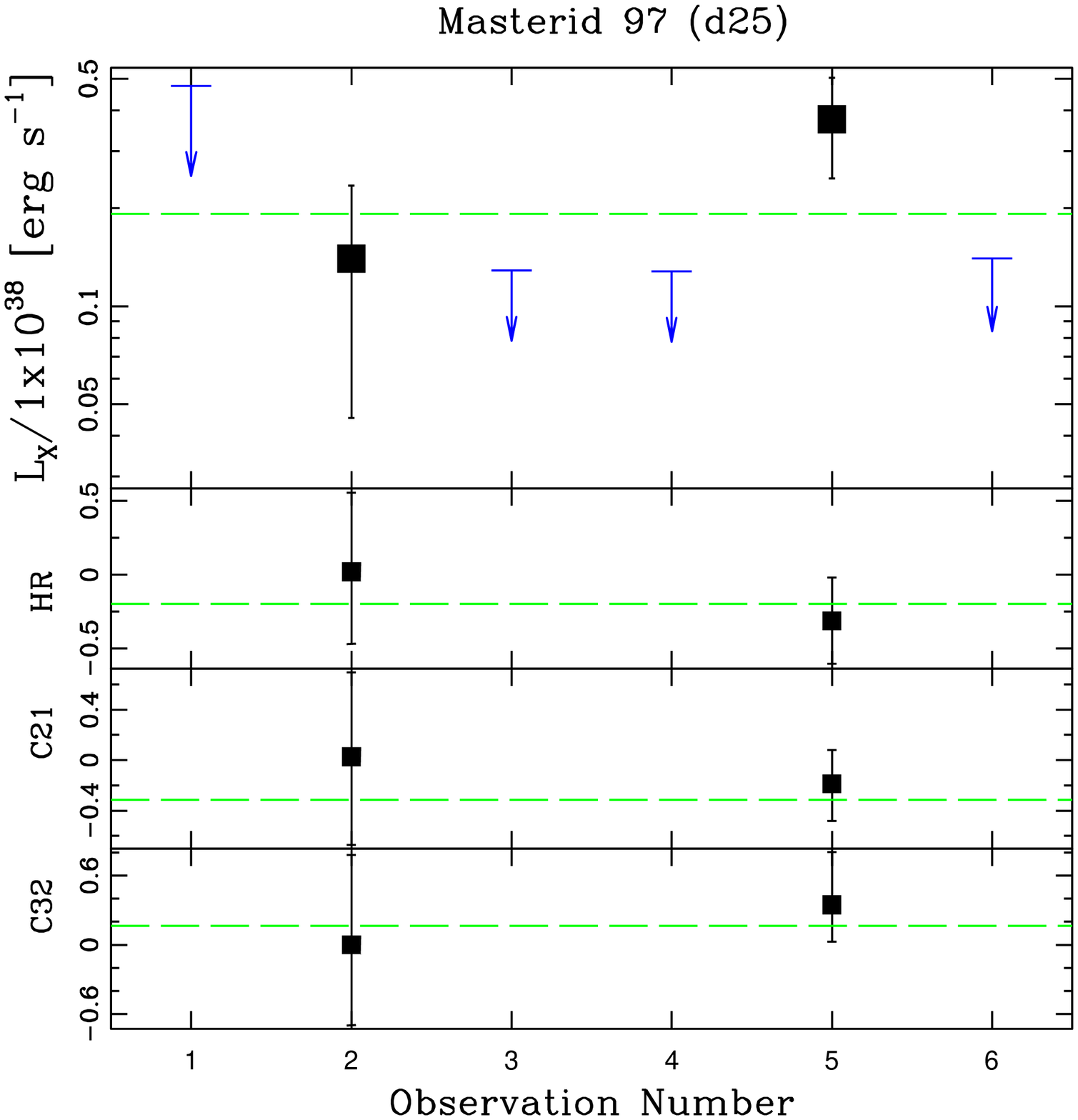}

  \end{minipage}\hspace{0.02\linewidth}
  \begin{minipage}{0.485\linewidth}
  \centering

    \includegraphics[width=\linewidth]{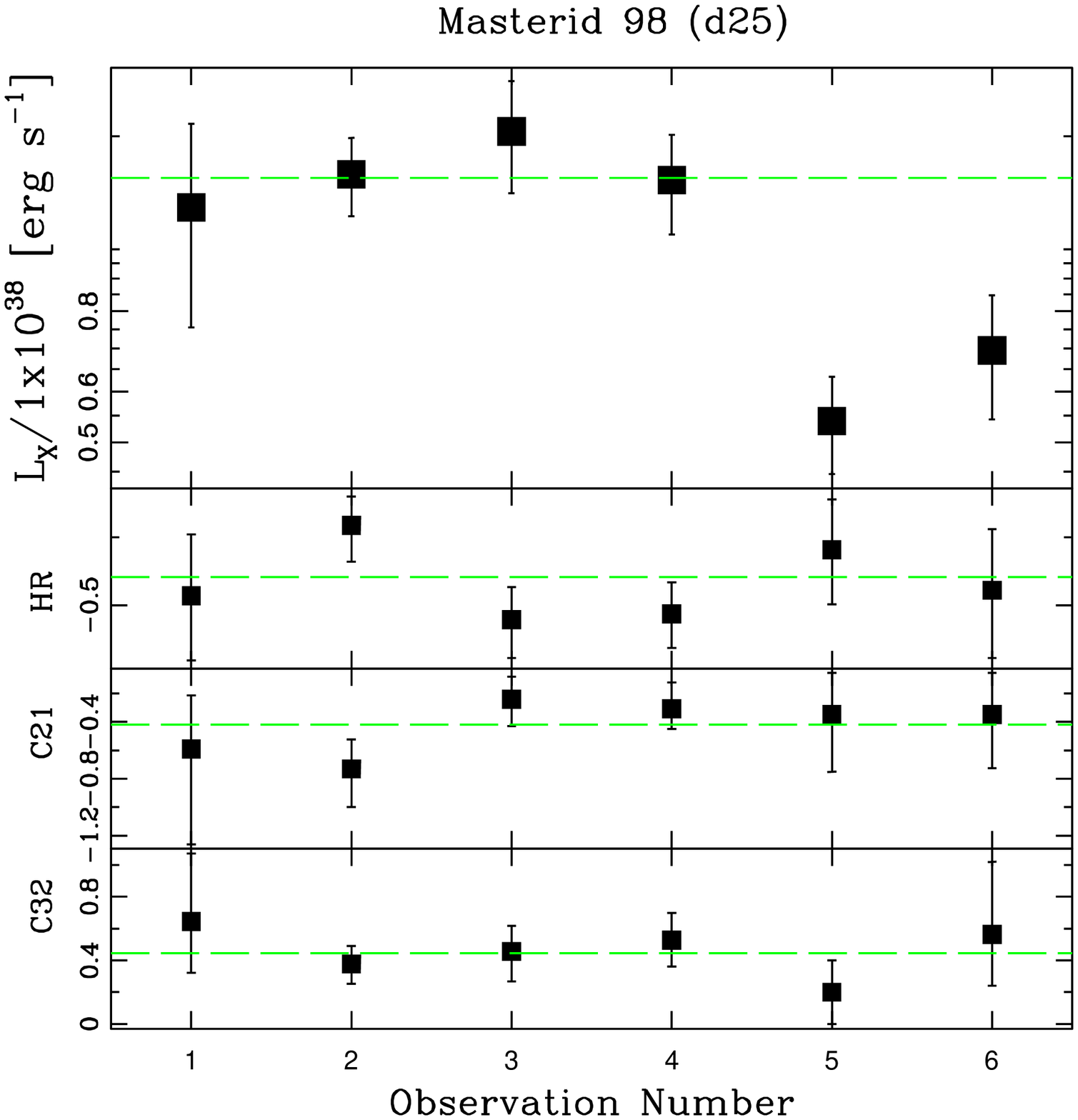}

\end{minipage}\hspace{0.02\linewidth}

\begin{minipage}{0.485\linewidth}
  \centering

    \includegraphics[width=\linewidth]{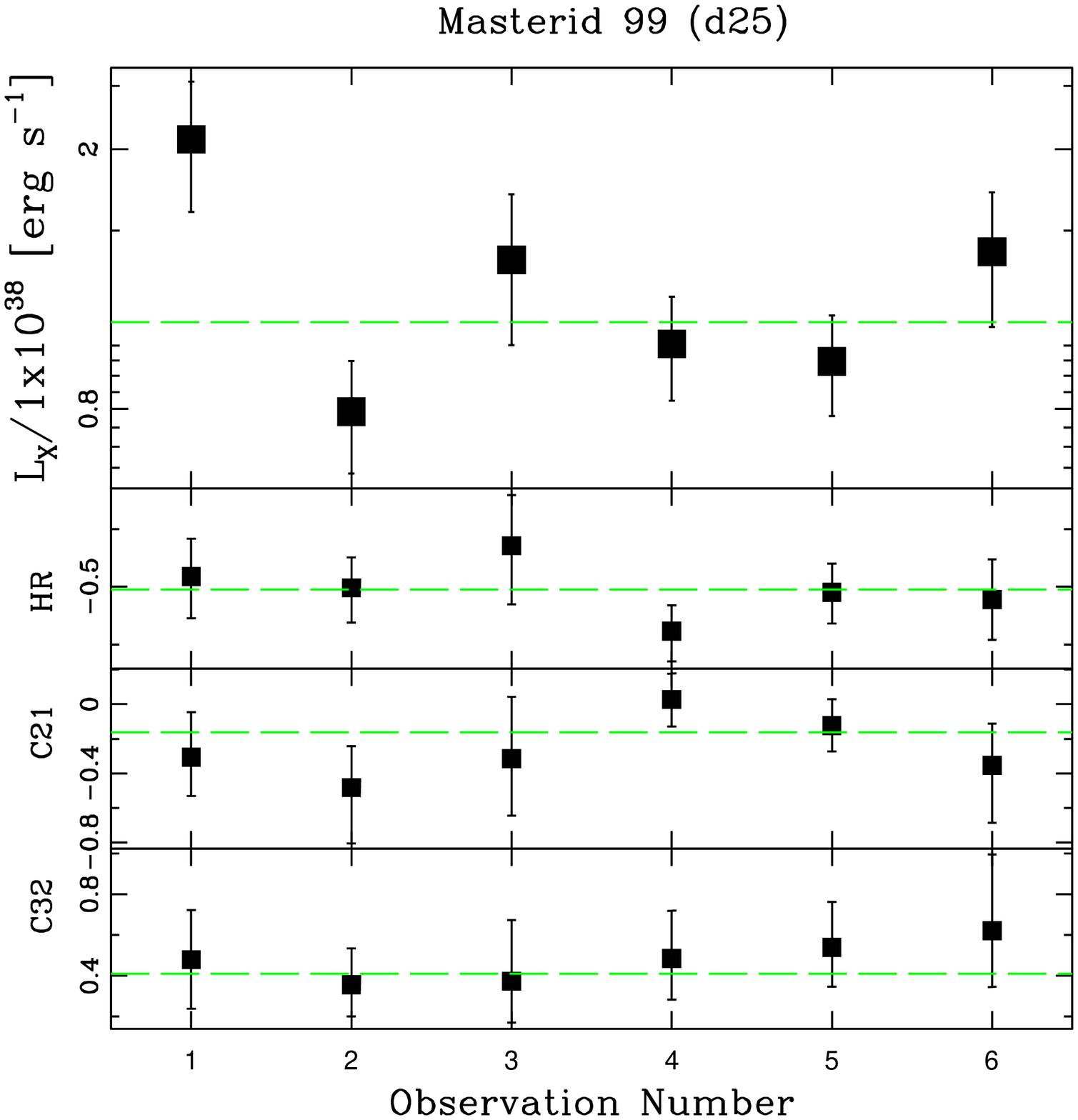}

\end{minipage}\hspace{0.02\linewidth}
\begin{minipage}{0.485\linewidth}
  \centering
  
    \includegraphics[width=\linewidth]{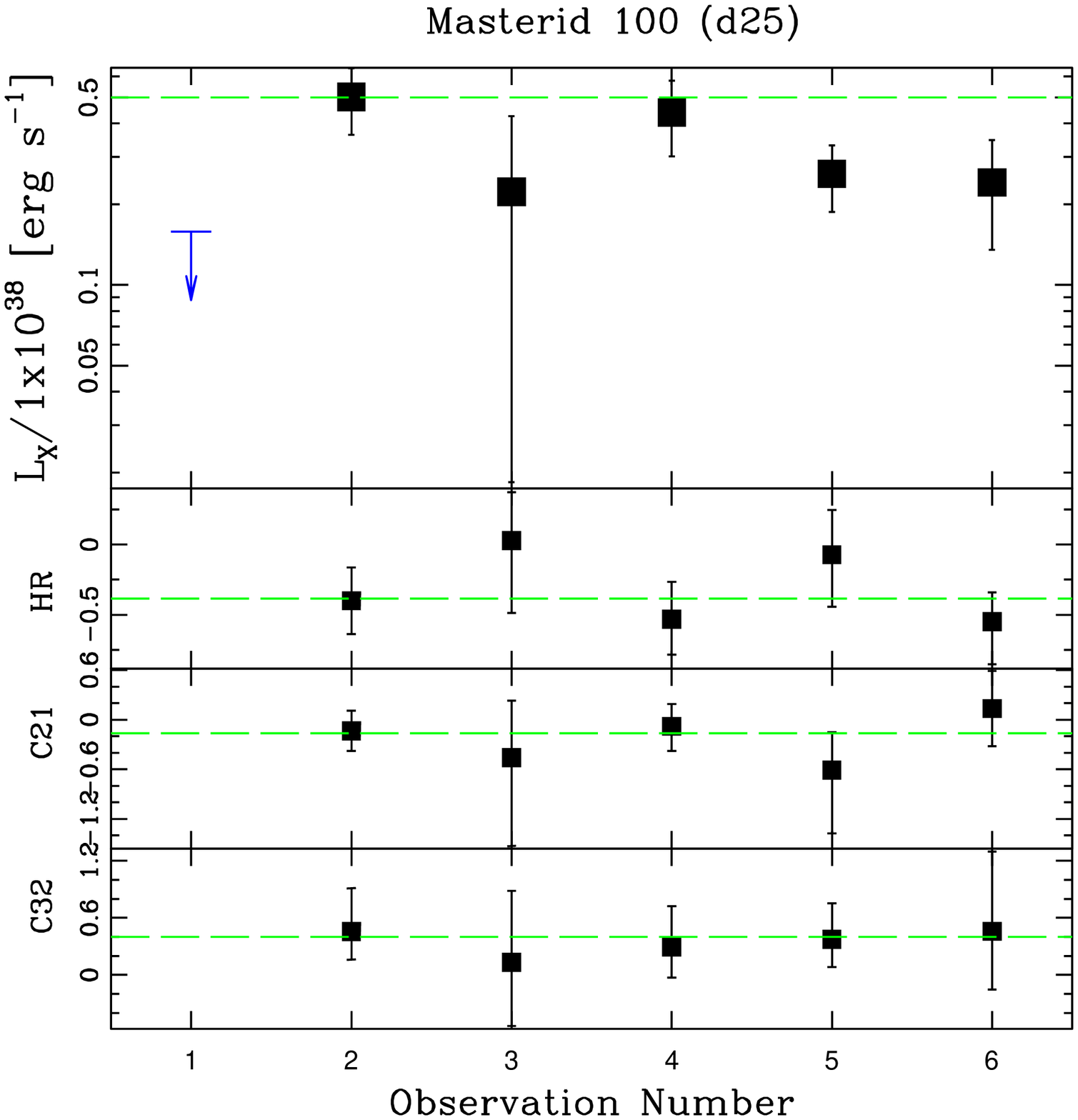}

  \end{minipage}\hspace{0.02\linewidth}
\end{figure}

\clearpage

\begin{figure}
  \begin{minipage}{0.485\linewidth}
  \centering

    \includegraphics[width=\linewidth]{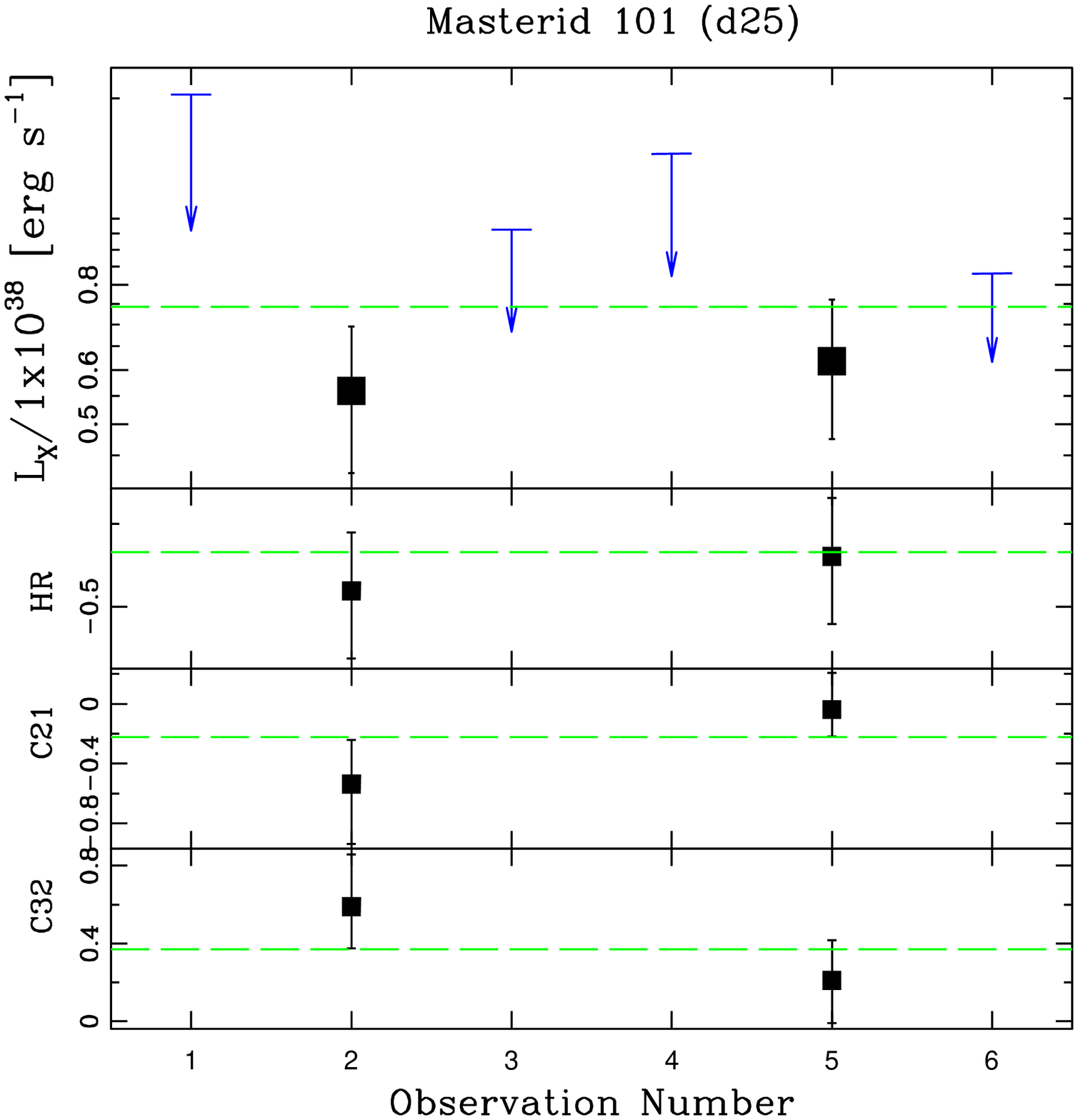}

\end{minipage}\hspace{0.02\linewidth}
\begin{minipage}{0.485\linewidth}
  \centering

    \includegraphics[width=\linewidth]{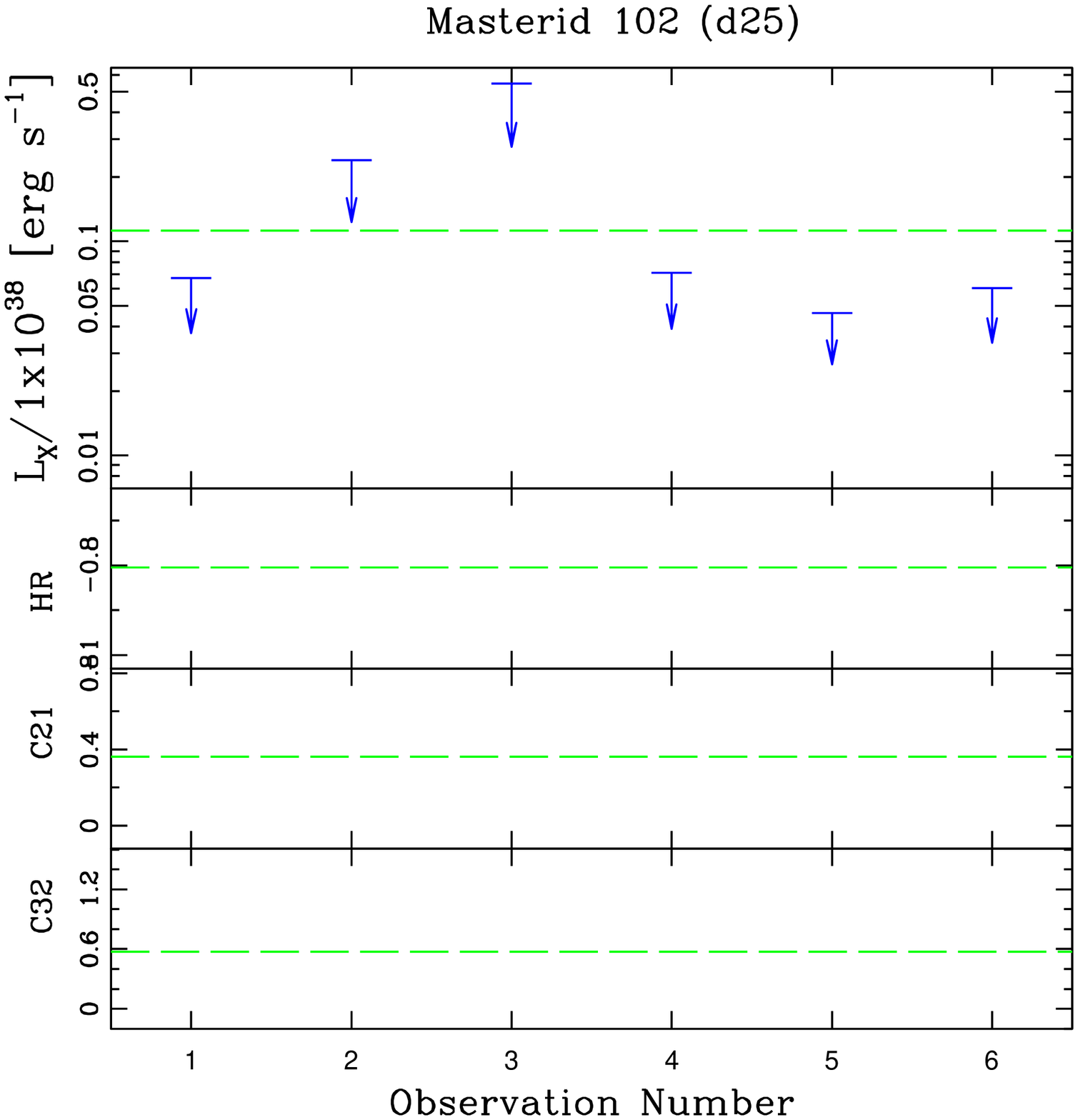}

\end{minipage}\hspace{0.02\linewidth}

  \begin{minipage}{0.485\linewidth}
  \centering
  
    \includegraphics[width=\linewidth]{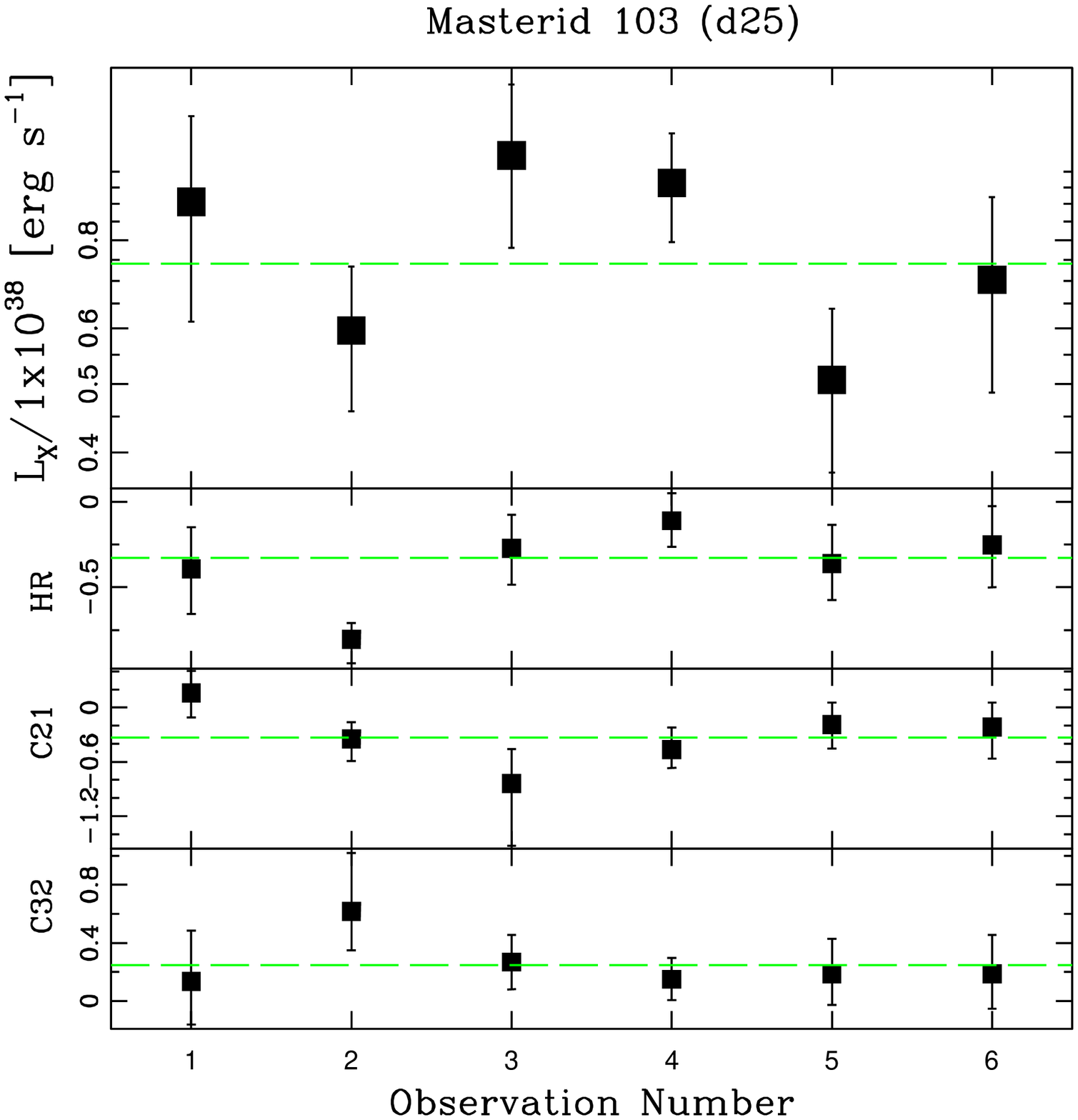}

  \end{minipage}\hspace{0.02\linewidth}
  \begin{minipage}{0.485\linewidth}
  \centering

    \includegraphics[width=\linewidth]{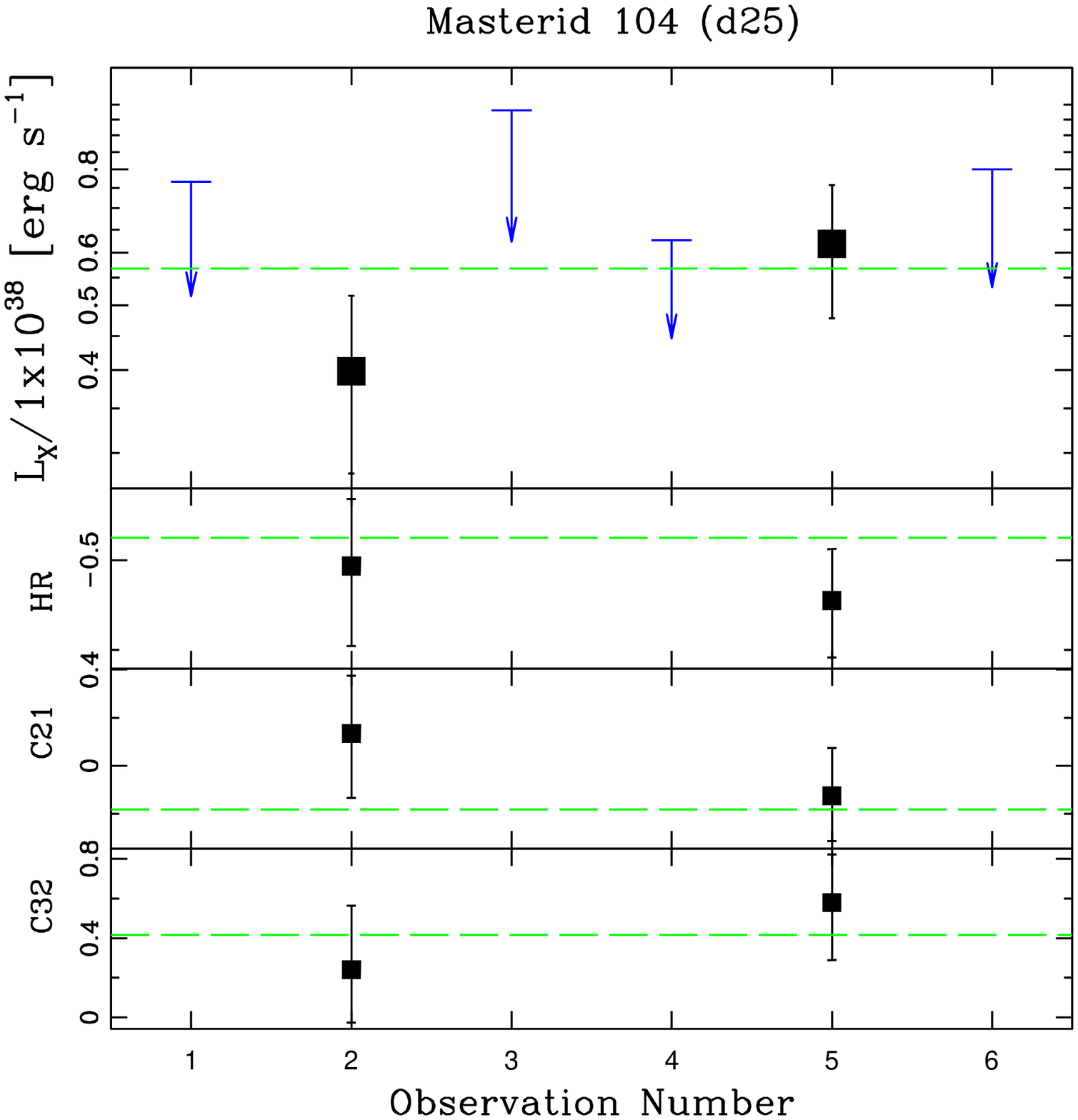}

\end{minipage}\hspace{0.02\linewidth}

\begin{minipage}{0.485\linewidth}
  \centering

    \includegraphics[width=\linewidth]{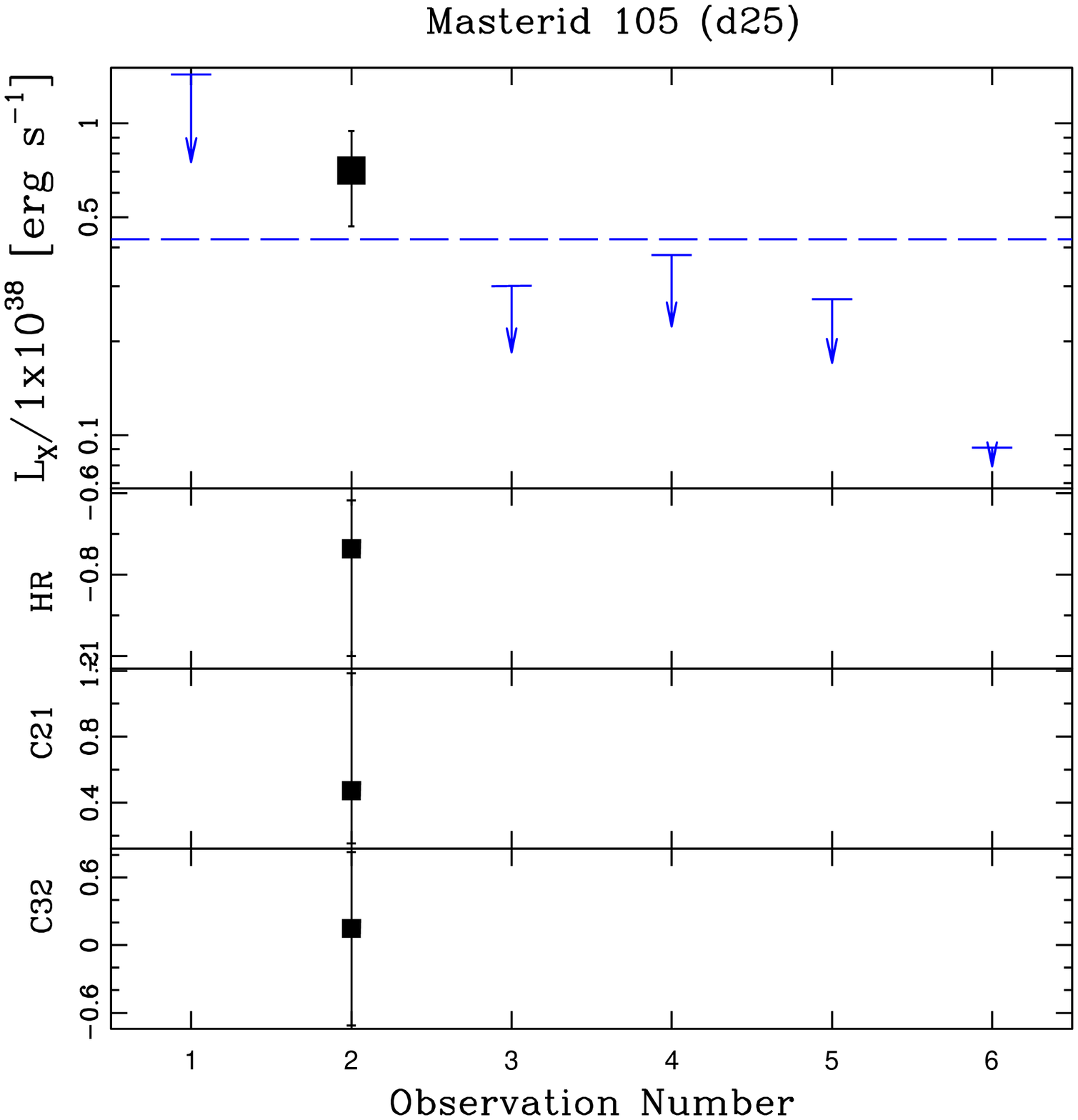}

 \end{minipage}\hspace{0.02\linewidth}
\begin{minipage}{0.485\linewidth}
  \centering
  
    \includegraphics[width=\linewidth]{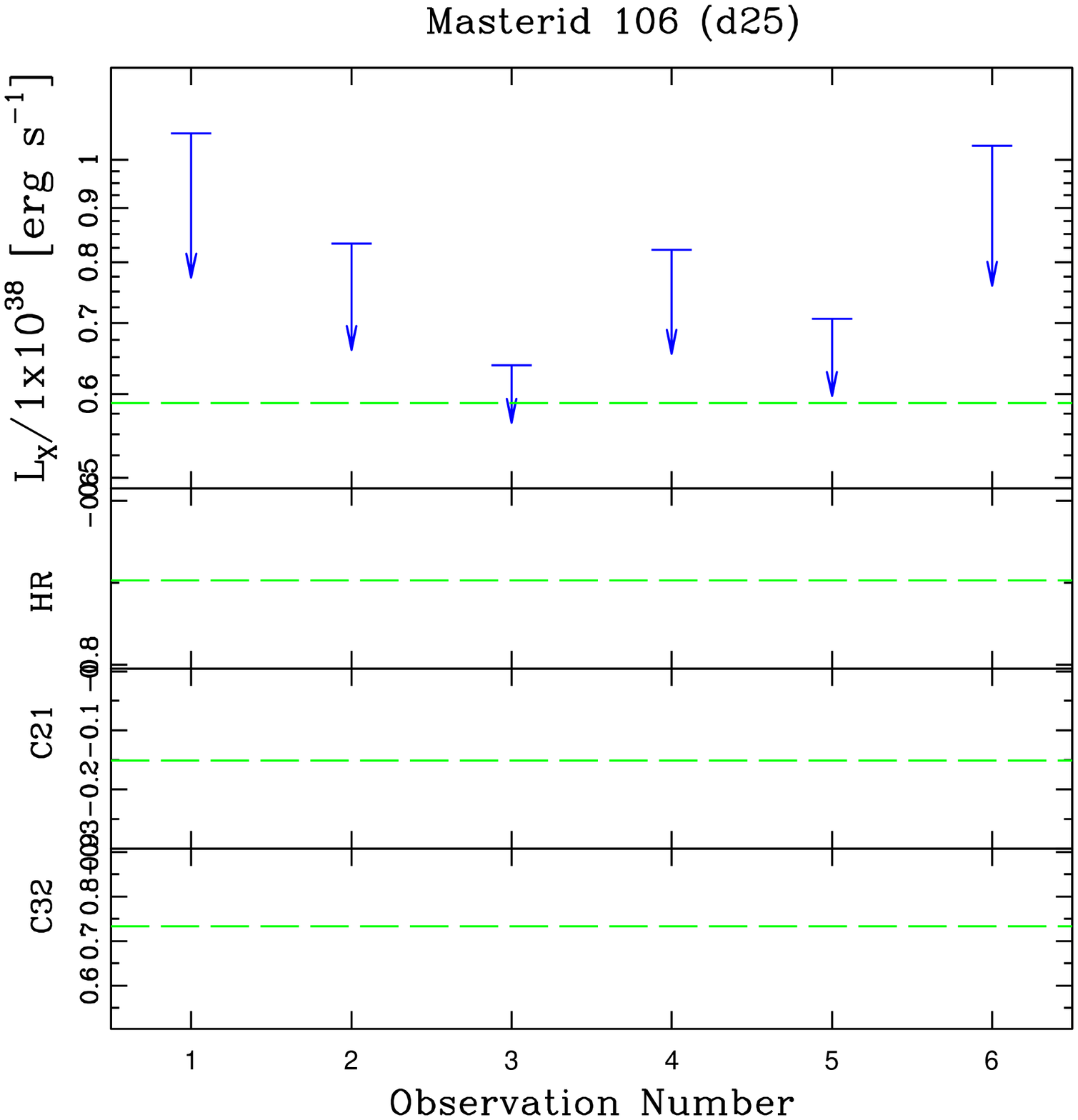}

  \end{minipage}\hspace{0.02\linewidth}
\end{figure}

\begin{figure}
  \begin{minipage}{0.485\linewidth}
  \centering

    \includegraphics[width=\linewidth]{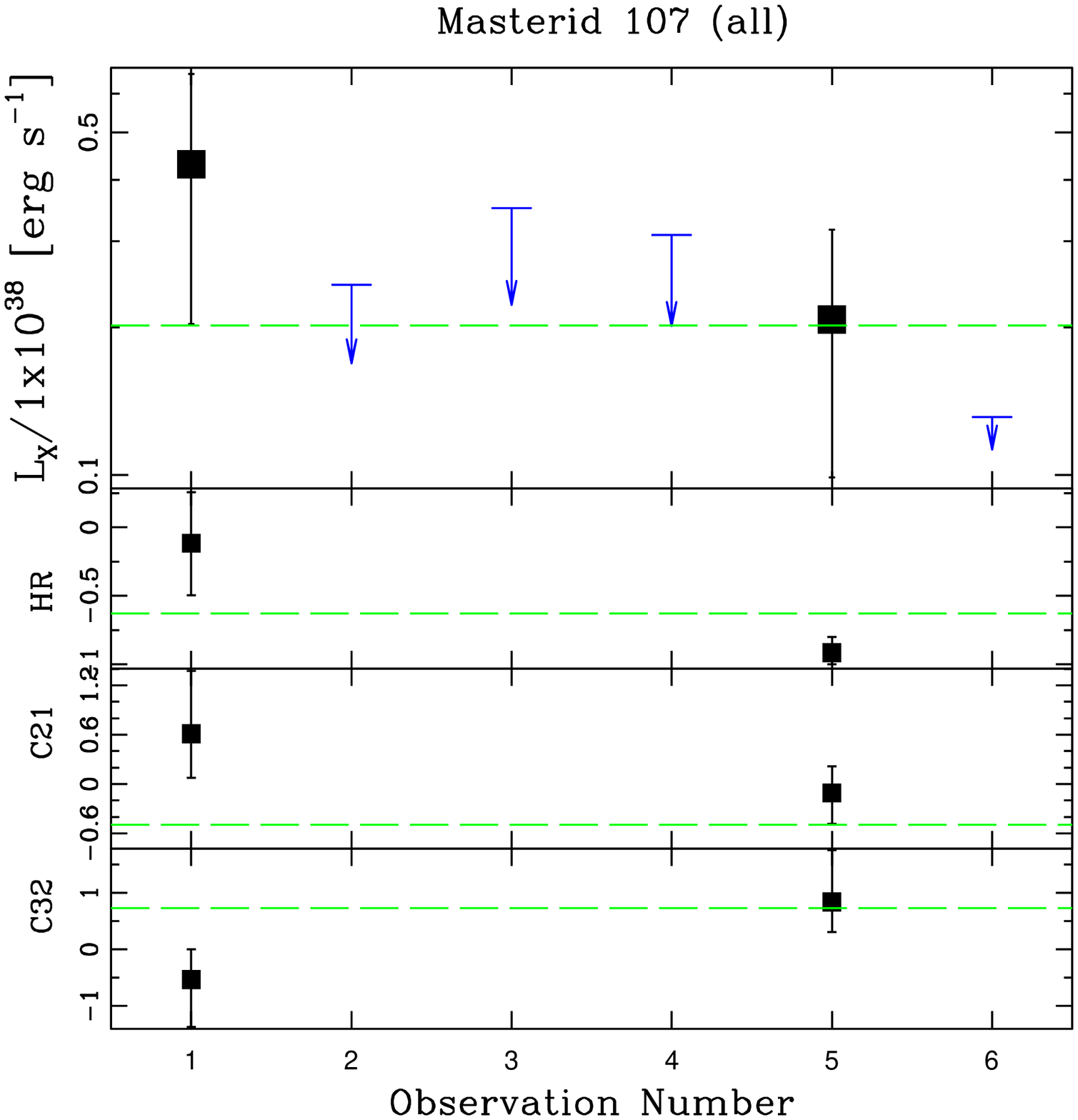}

\end{minipage}\hspace{0.02\linewidth}
\begin{minipage}{0.485\linewidth}
  \centering

    \includegraphics[width=\linewidth]{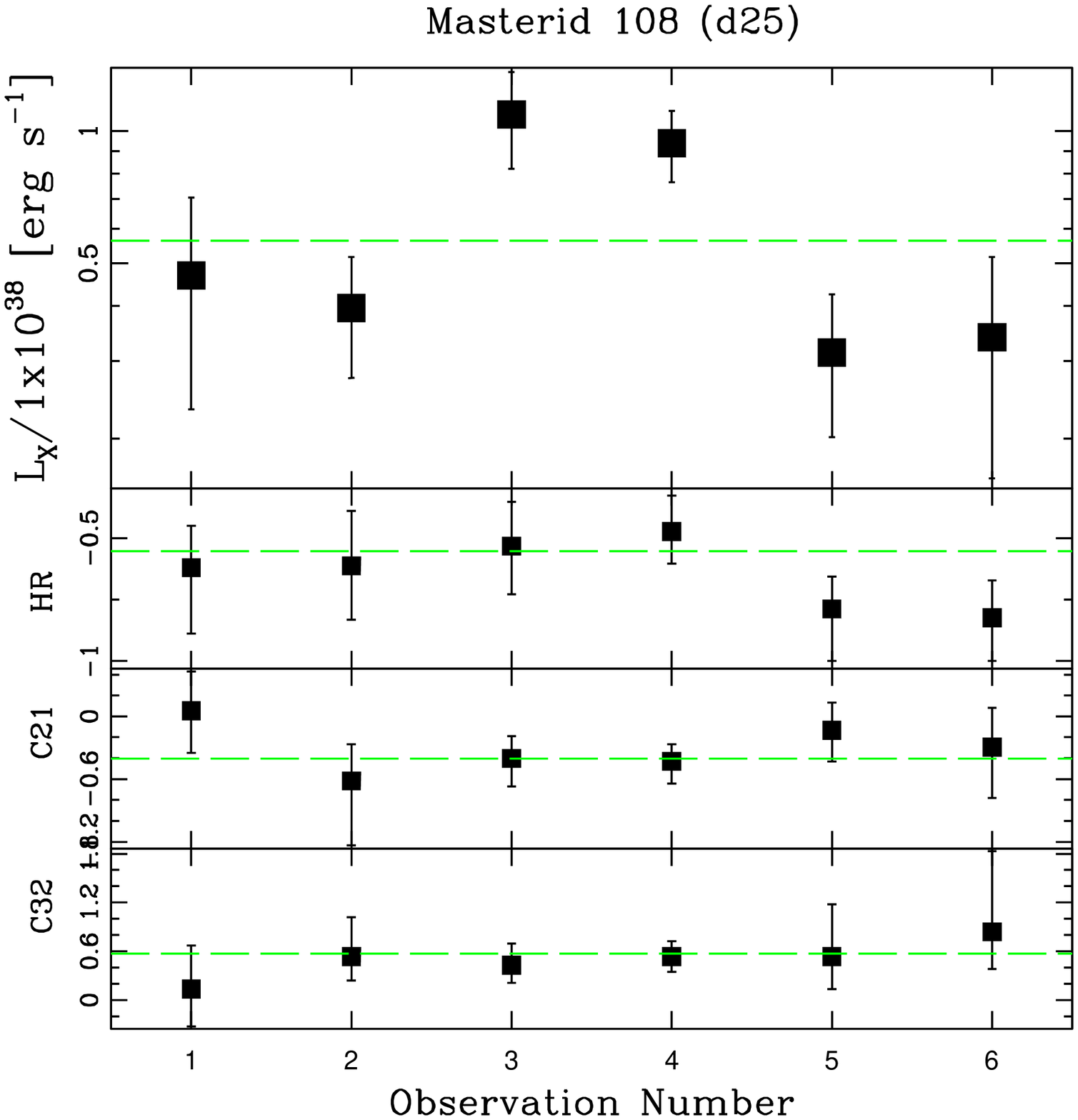}

 \end{minipage}\hspace{0.02\linewidth}

  \begin{minipage}{0.485\linewidth}
  \centering
  
    \includegraphics[width=\linewidth]{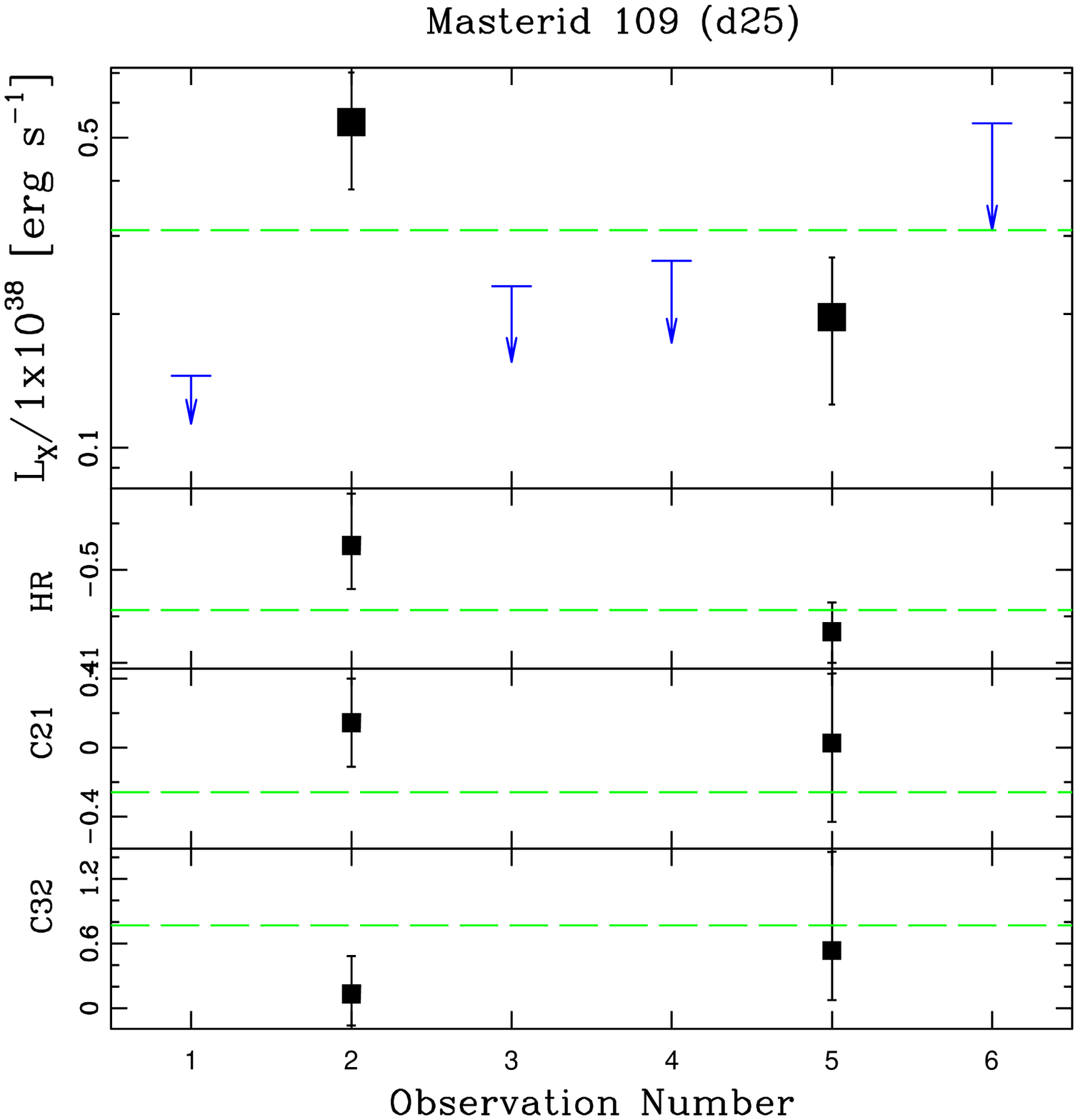}

  \end{minipage}\hspace{0.02\linewidth}
  \begin{minipage}{0.485\linewidth}
  \centering

    \includegraphics[width=\linewidth]{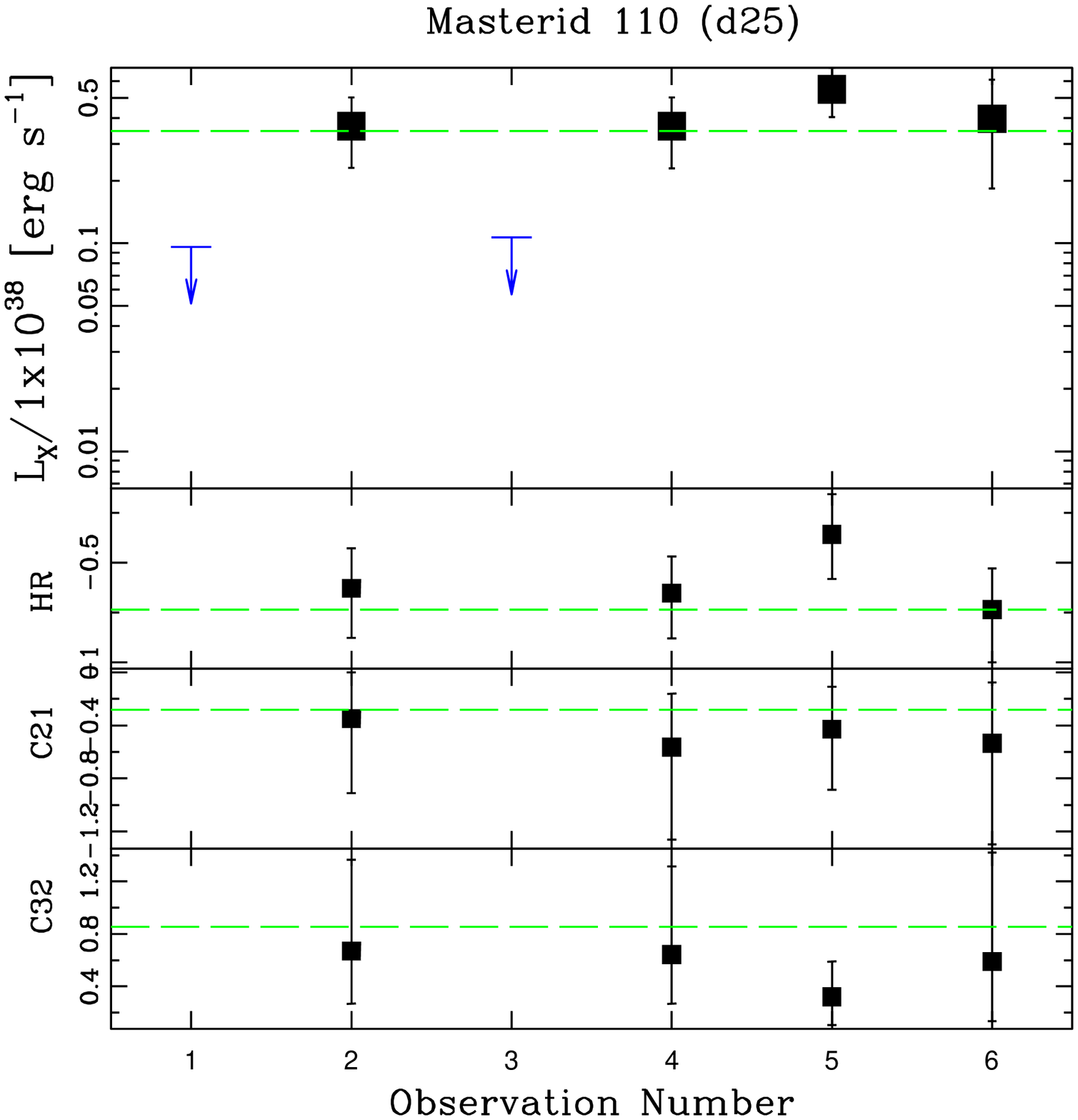}

\end{minipage}\hspace{0.02\linewidth}

\begin{minipage}{0.485\linewidth}
  \centering

    \includegraphics[width=\linewidth]{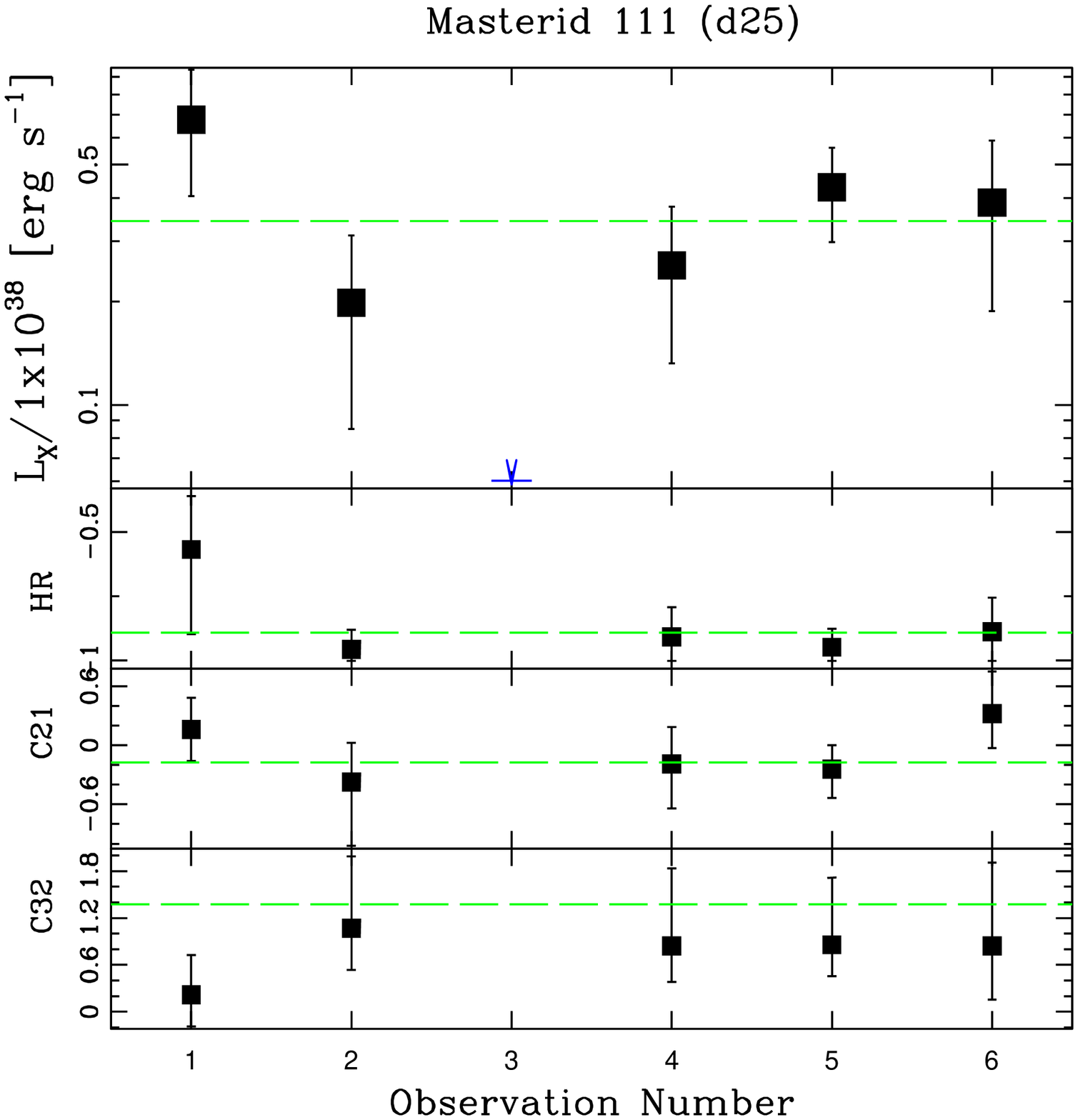}

\end{minipage}\hspace{0.02\linewidth}
\begin{minipage}{0.485\linewidth}
  \centering
  
    \includegraphics[width=\linewidth]{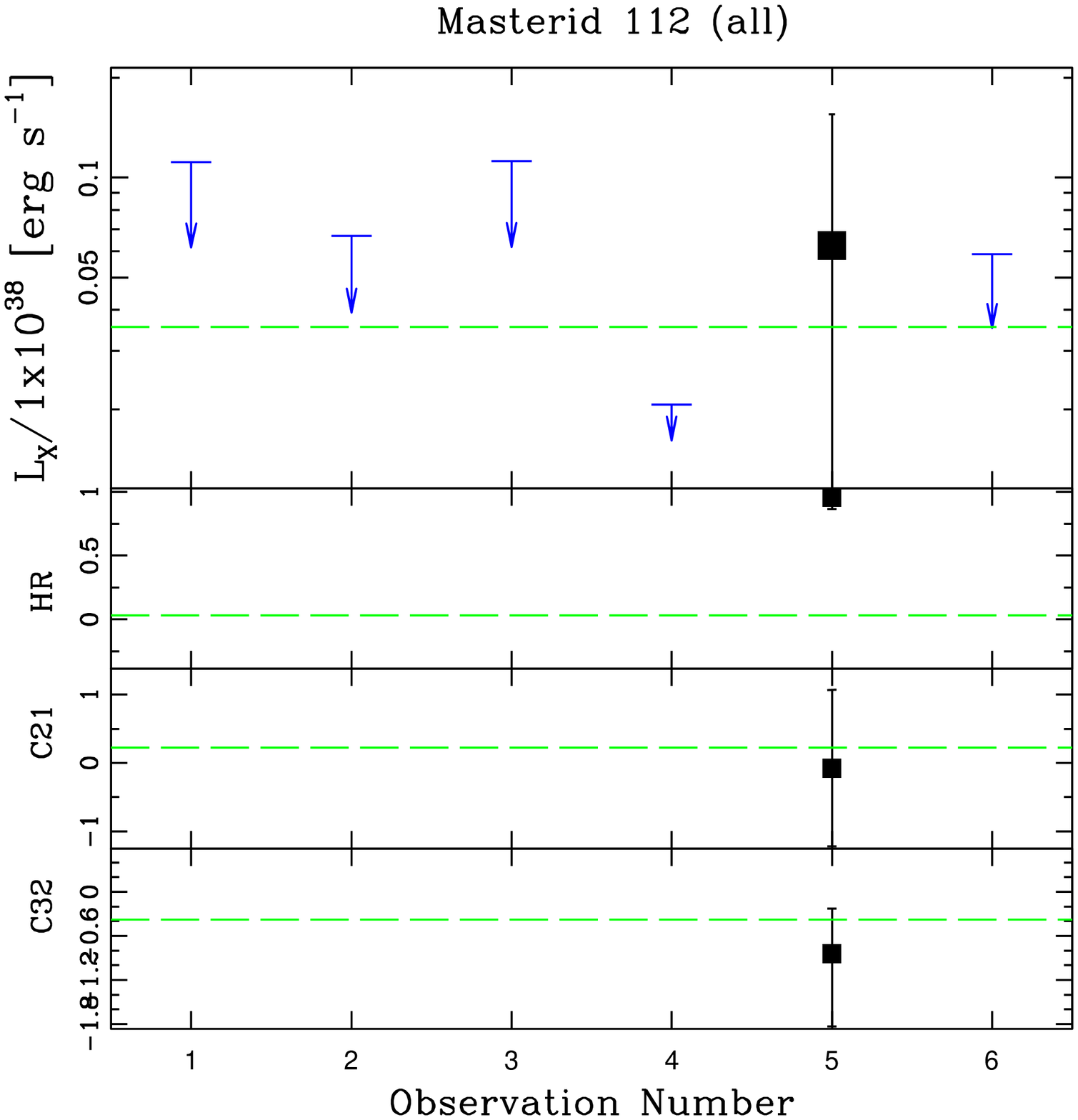}

  \end{minipage}\hspace{0.02\linewidth}

\end{figure}

\begin{figure}
  \begin{minipage}{0.485\linewidth}
  \centering

    \includegraphics[width=\linewidth]{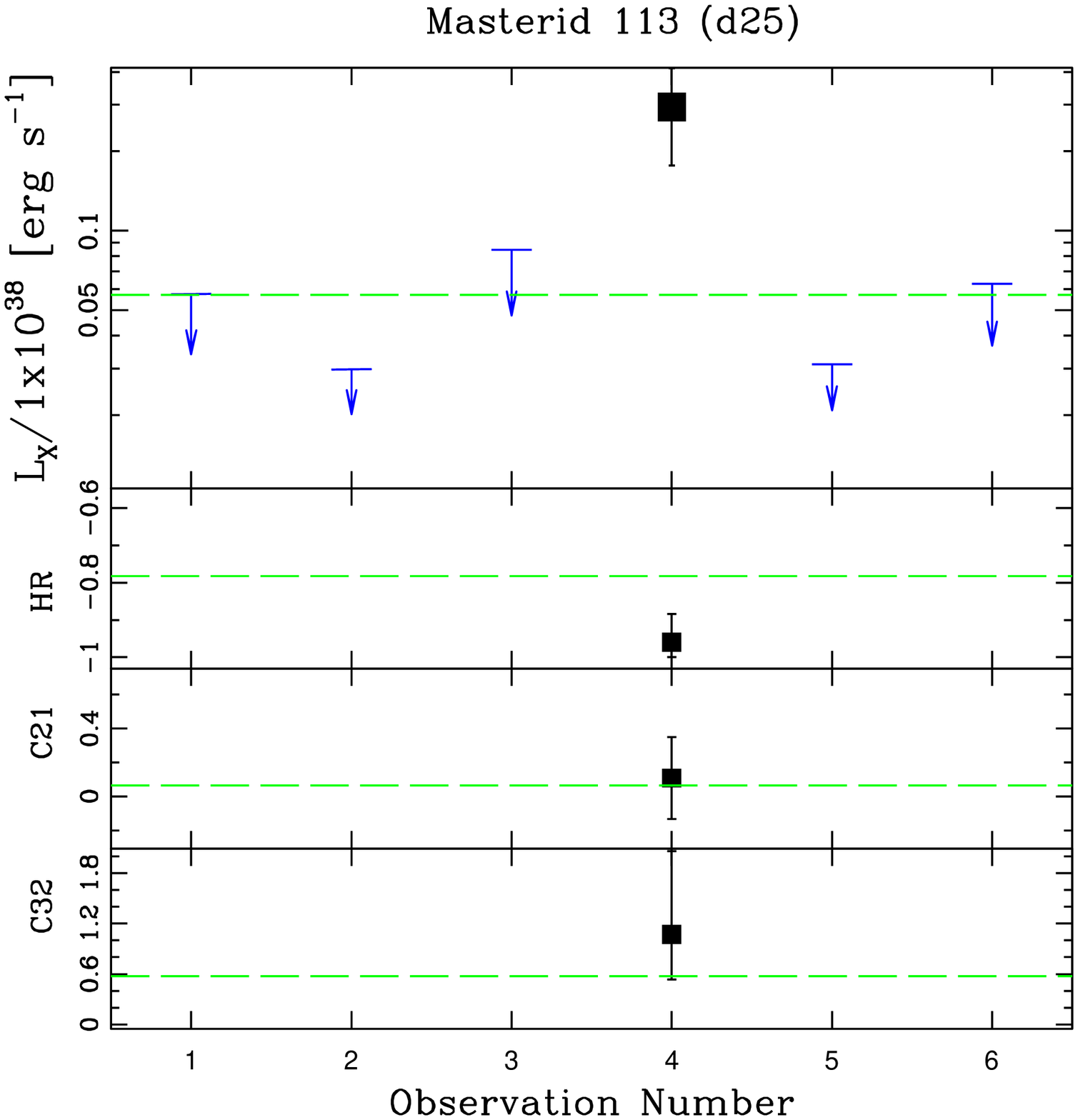}

\end{minipage}\hspace{0.02\linewidth}
\begin{minipage}{0.485\linewidth}
  \centering

    \includegraphics[width=\linewidth]{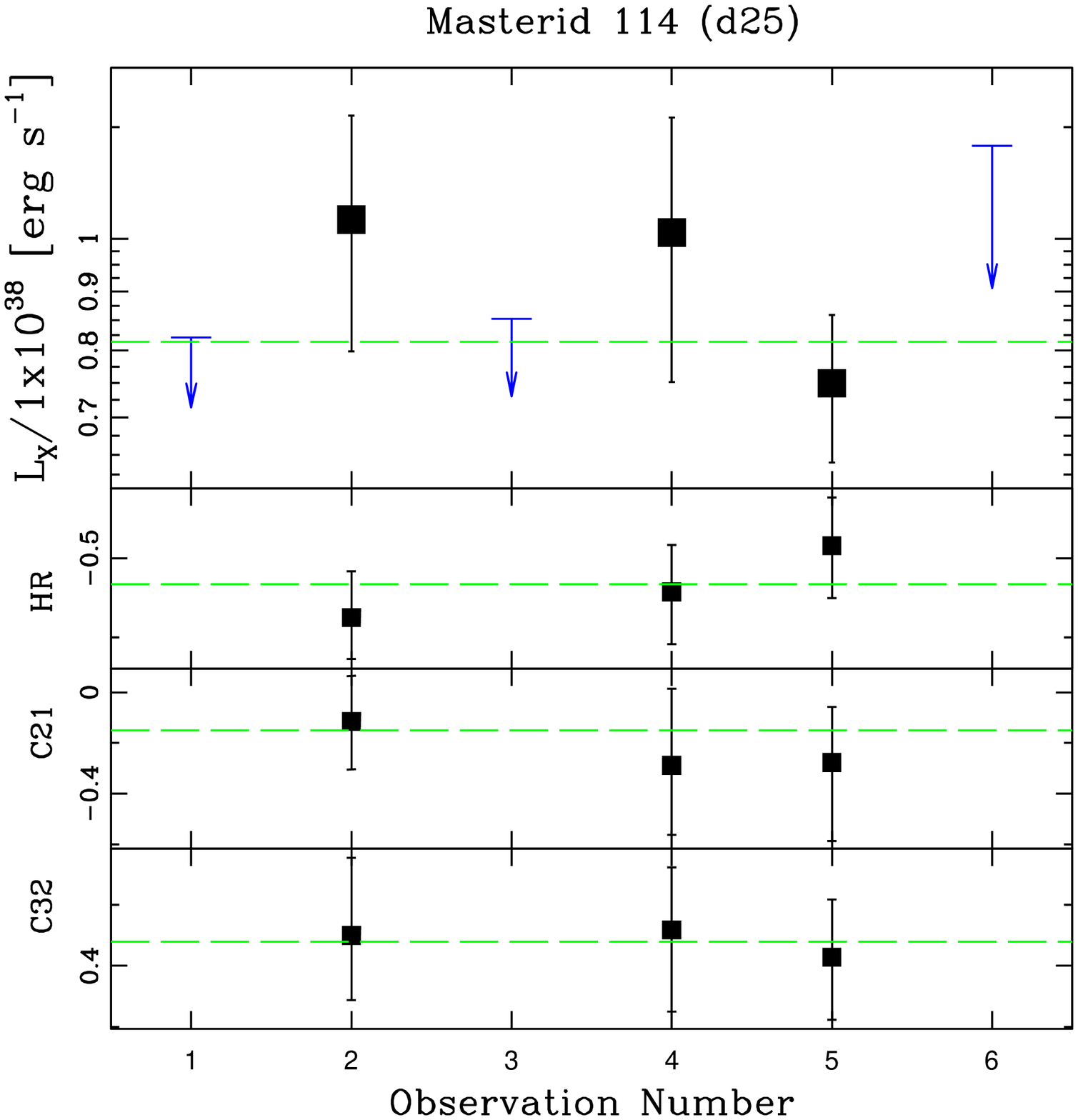}

\end{minipage}\hspace{0.02\linewidth}

  \begin{minipage}{0.485\linewidth}
  \centering
  
    \includegraphics[width=\linewidth]{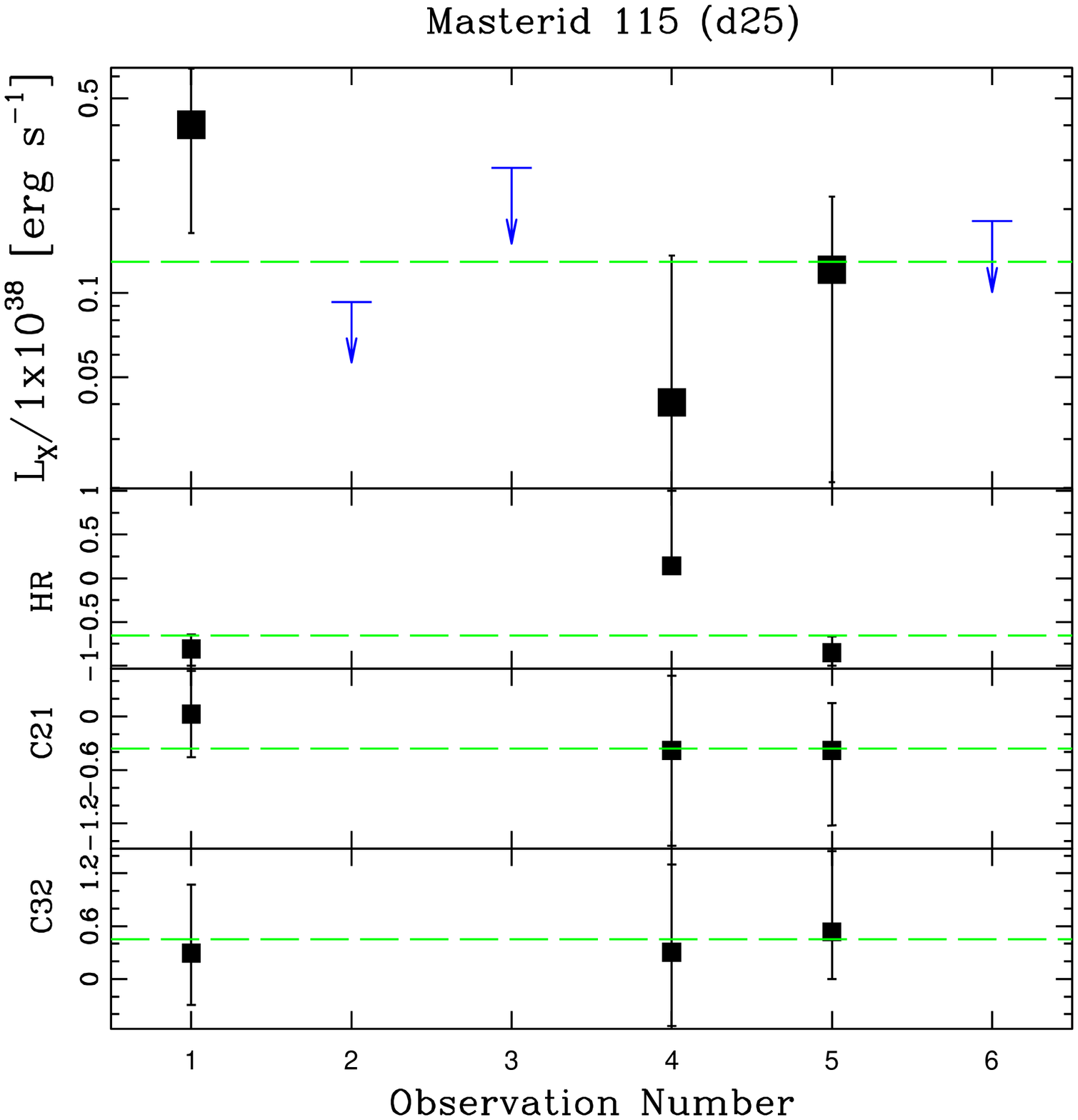}

  \end{minipage}\hspace{0.02\linewidth}
  \begin{minipage}{0.485\linewidth}
  \centering

    \includegraphics[width=\linewidth]{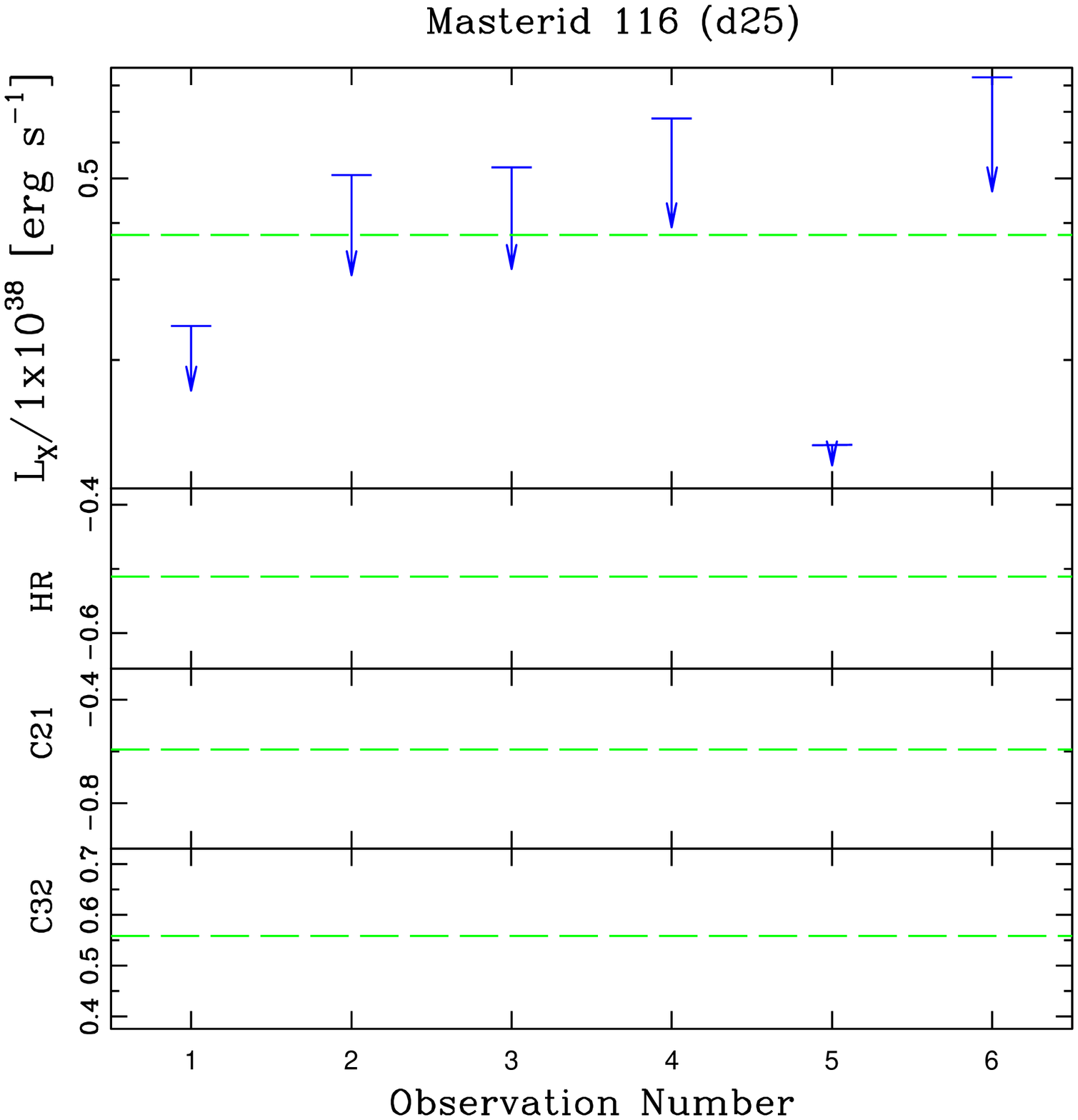}

\end{minipage}

\begin{minipage}{0.485\linewidth}
  \centering

    \includegraphics[width=\linewidth]{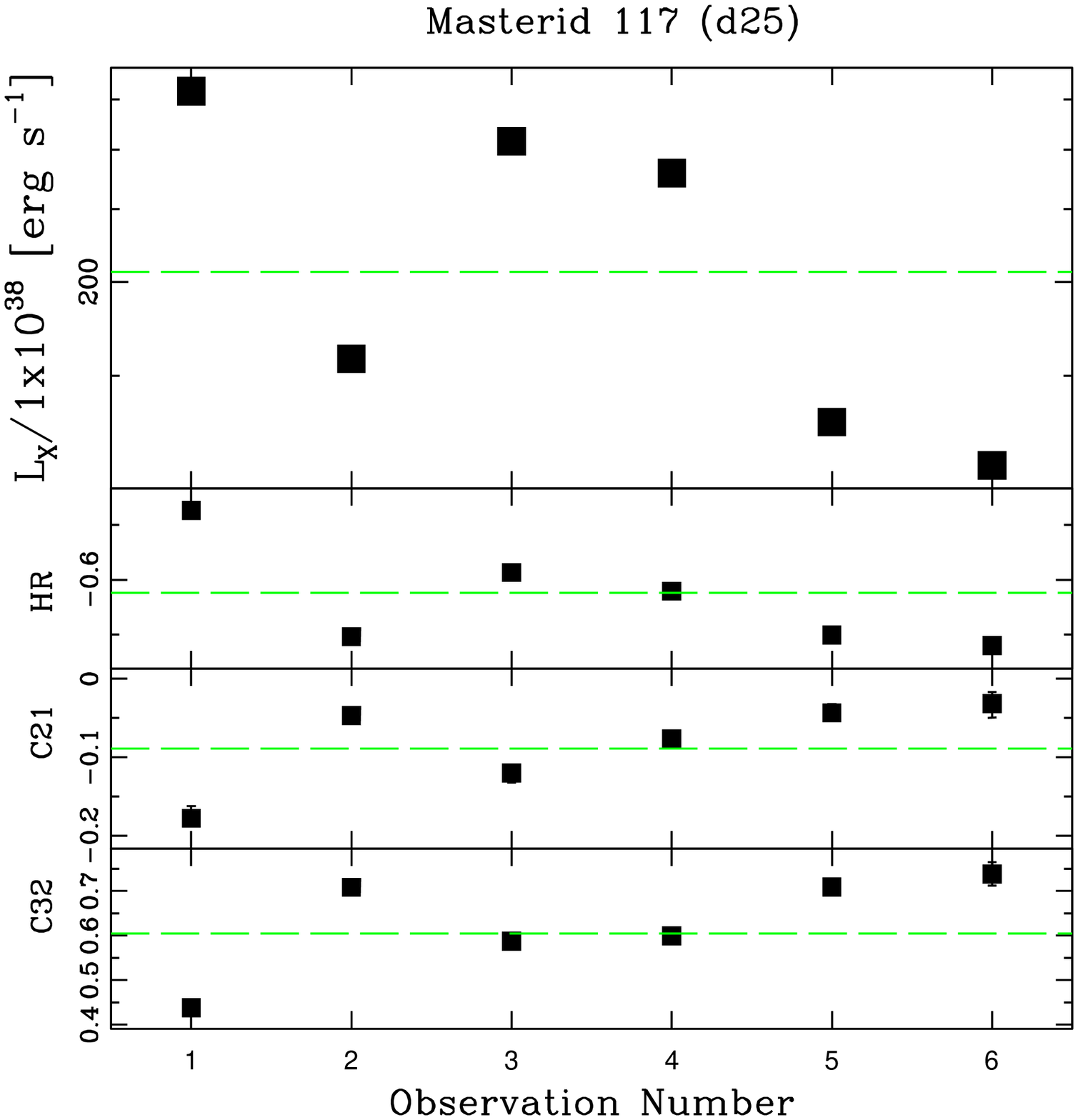}

 \end{minipage}\hspace{0.02\linewidth}
\begin{minipage}{0.485\linewidth}
  \centering
  
    \includegraphics[width=\linewidth]{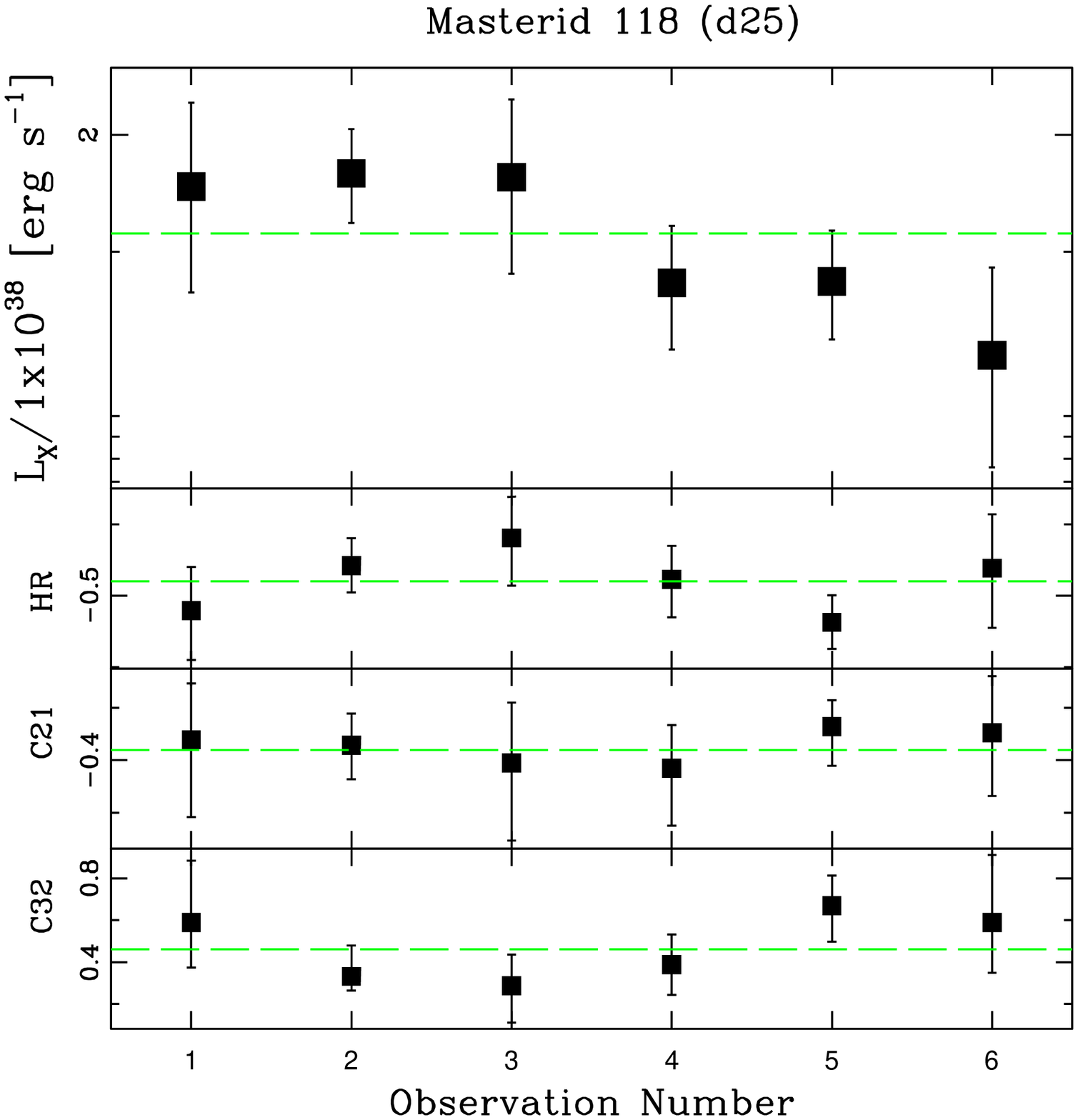}

  \end{minipage}\hspace{0.02\linewidth}
\end{figure}

\clearpage

\begin{figure}
  \begin{minipage}{0.485\linewidth}
  \centering

    \includegraphics[width=\linewidth]{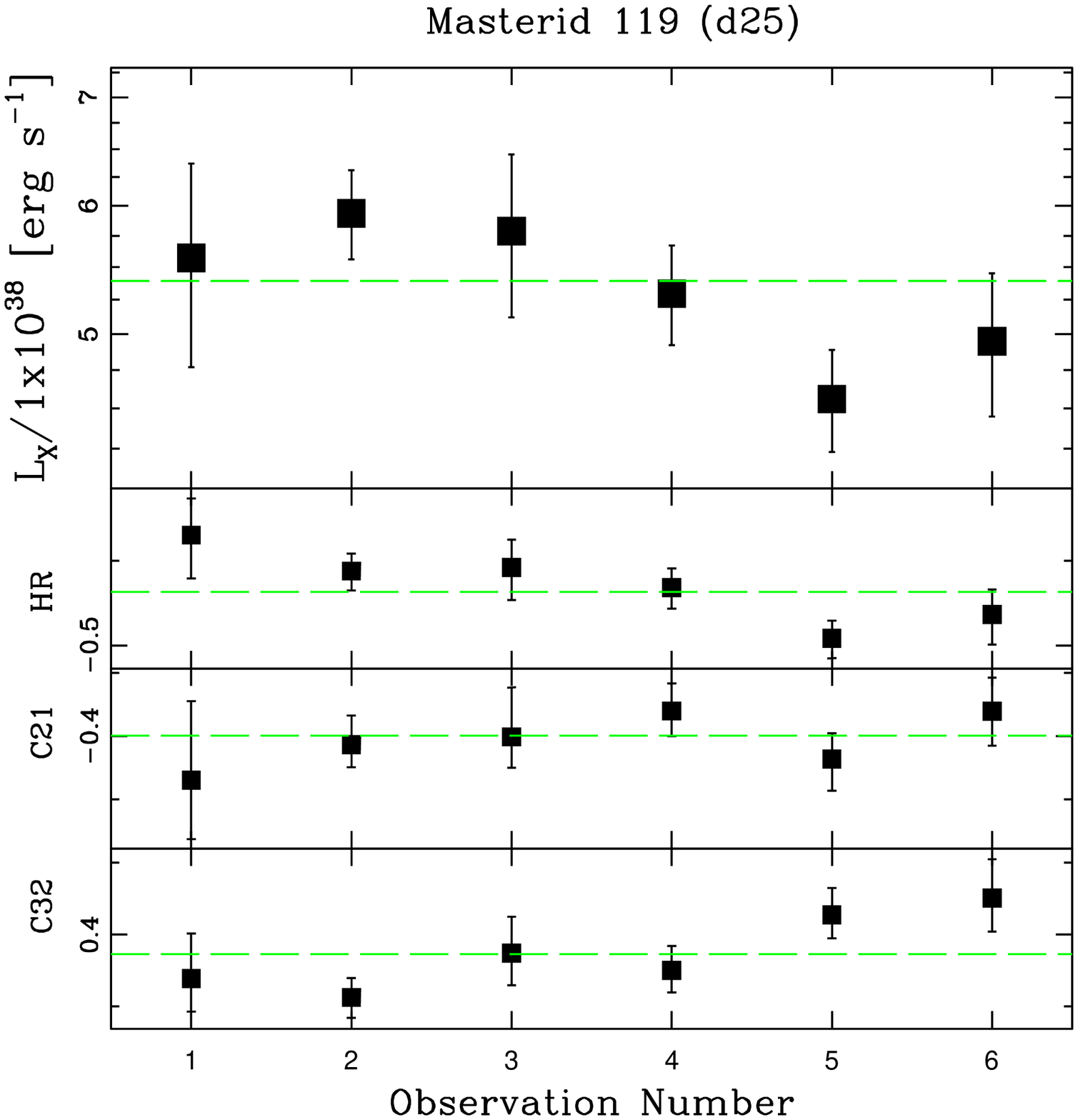}

\end{minipage}\hspace{0.02\linewidth}
\begin{minipage}{0.485\linewidth}
  \centering

    \includegraphics[width=\linewidth]{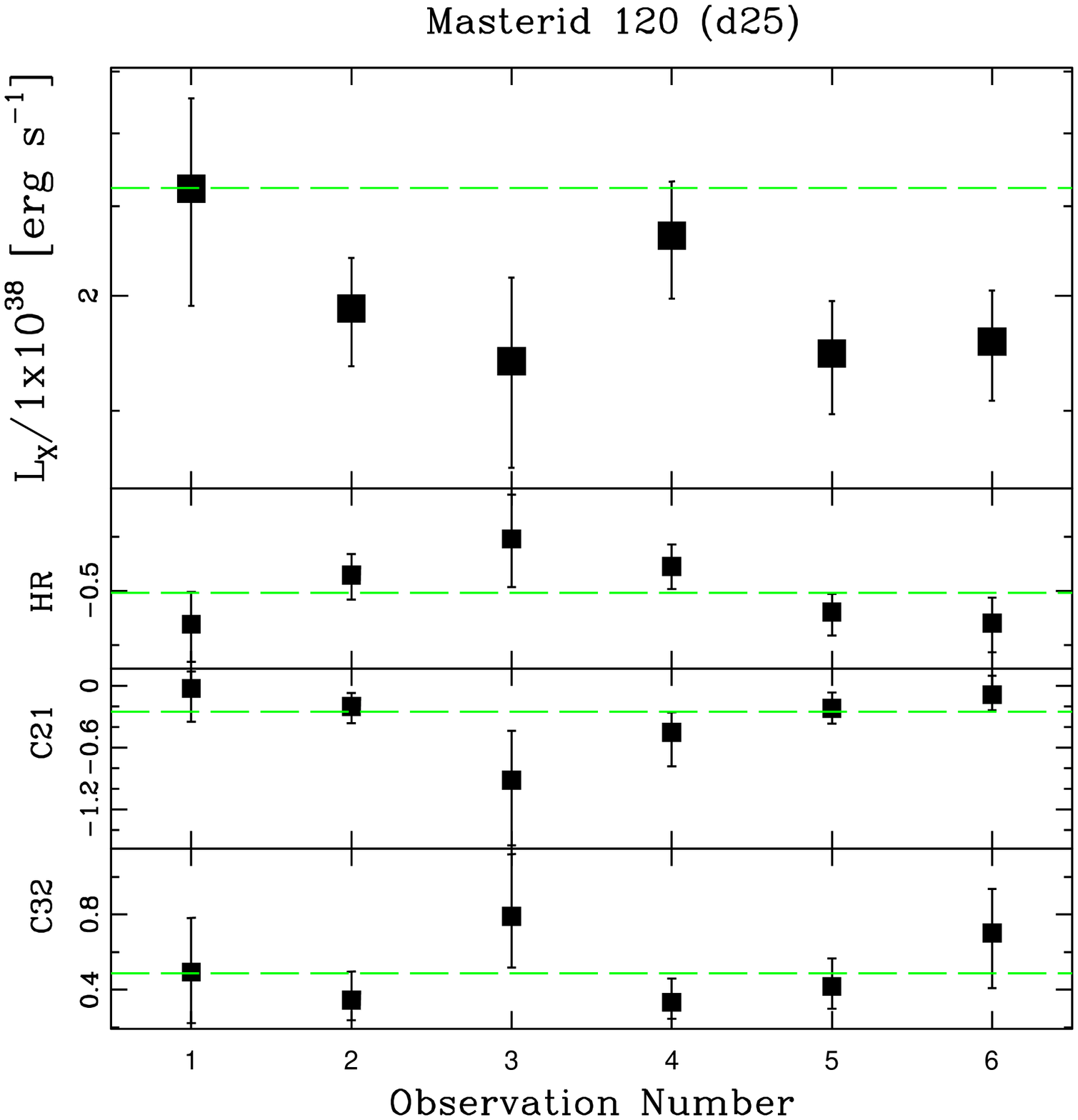}

 \end{minipage}\hspace{0.02\linewidth}

  \begin{minipage}{0.485\linewidth}
  \centering
  
    \includegraphics[width=\linewidth]{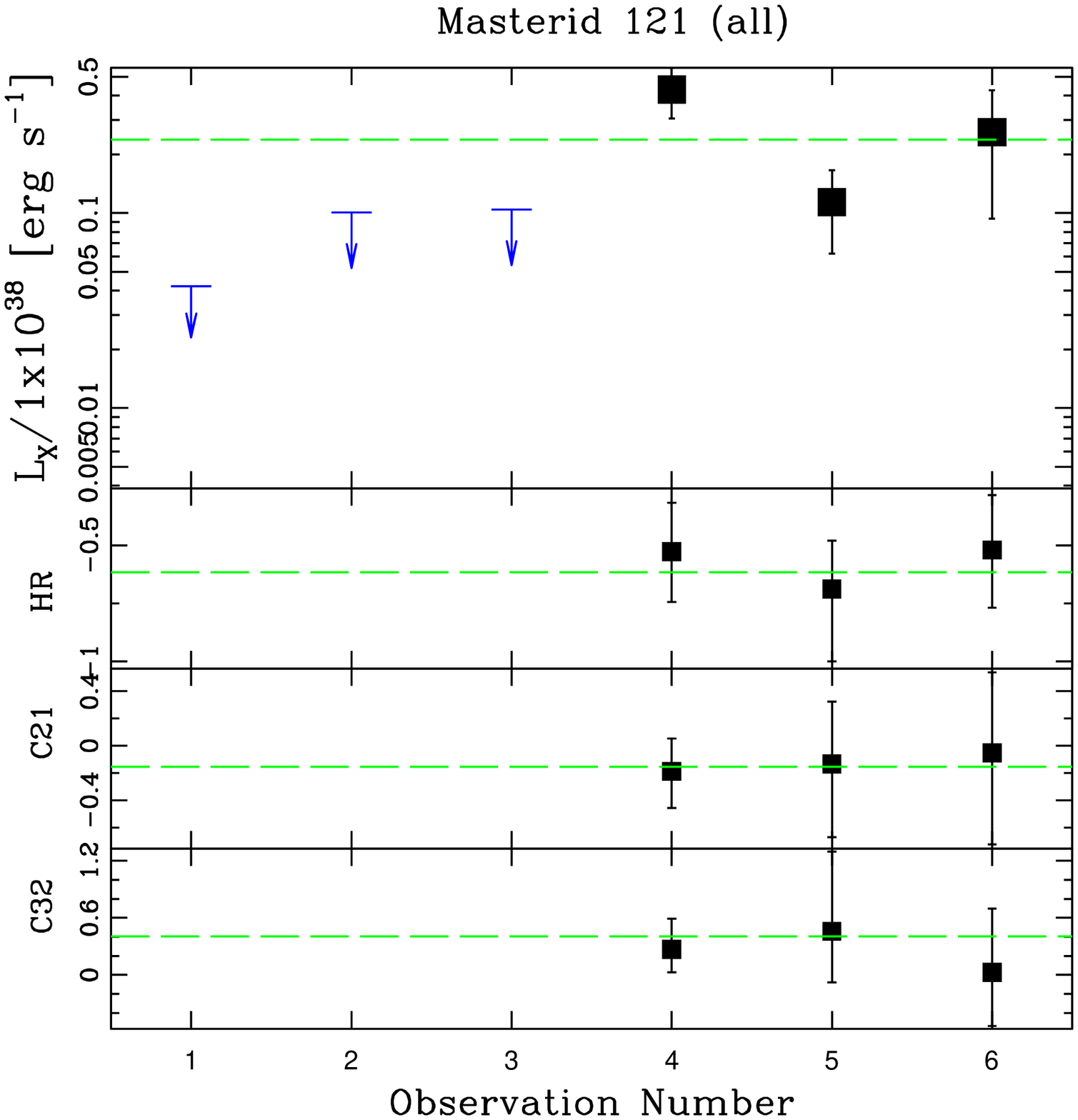}

  \end{minipage}\hspace{0.02\linewidth}
  \begin{minipage}{0.485\linewidth}
  \centering

    \includegraphics[width=\linewidth]{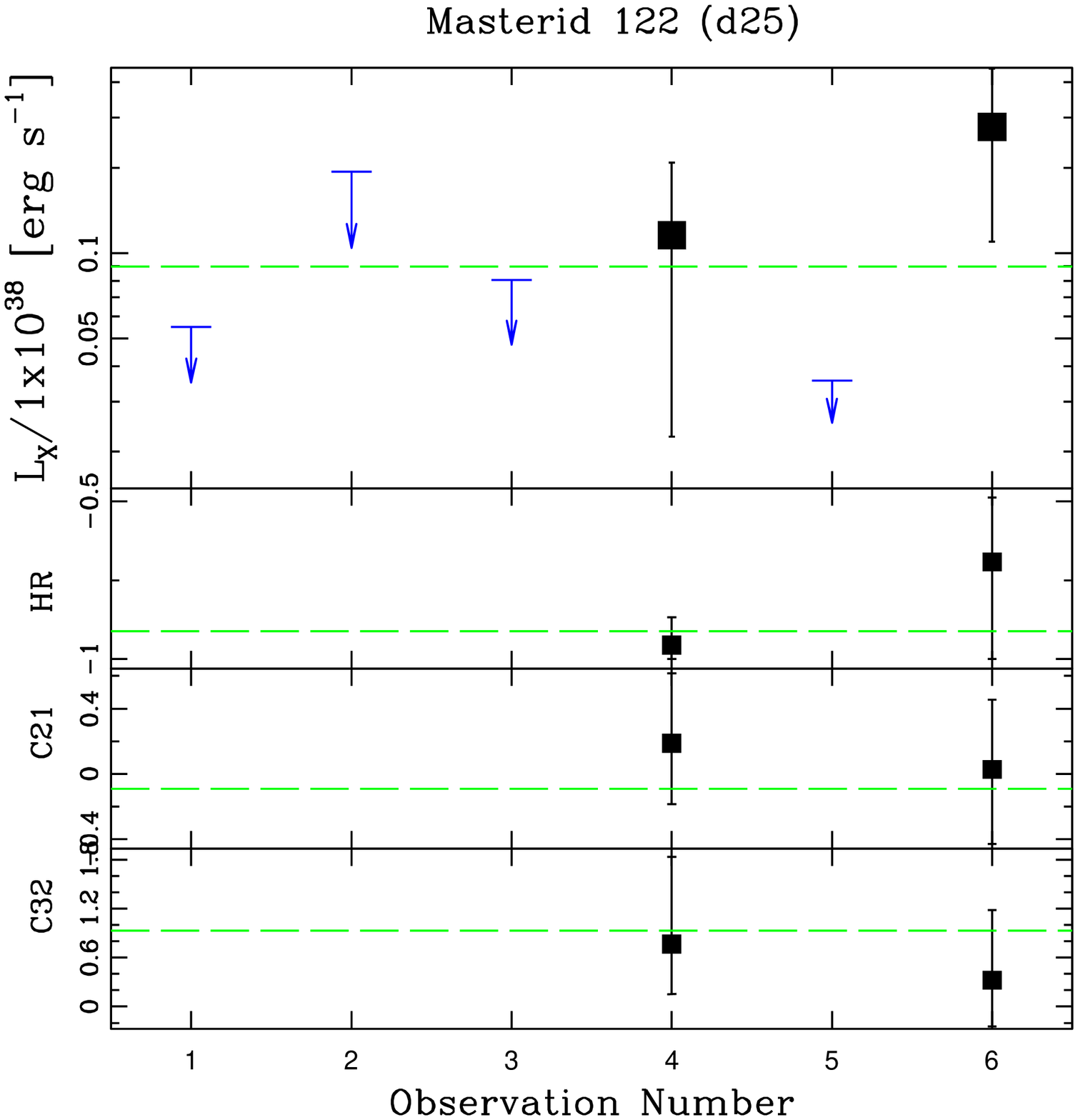}

\end{minipage}\hspace{0.02\linewidth}

\begin{minipage}{0.485\linewidth}
  \centering

    \includegraphics[width=\linewidth]{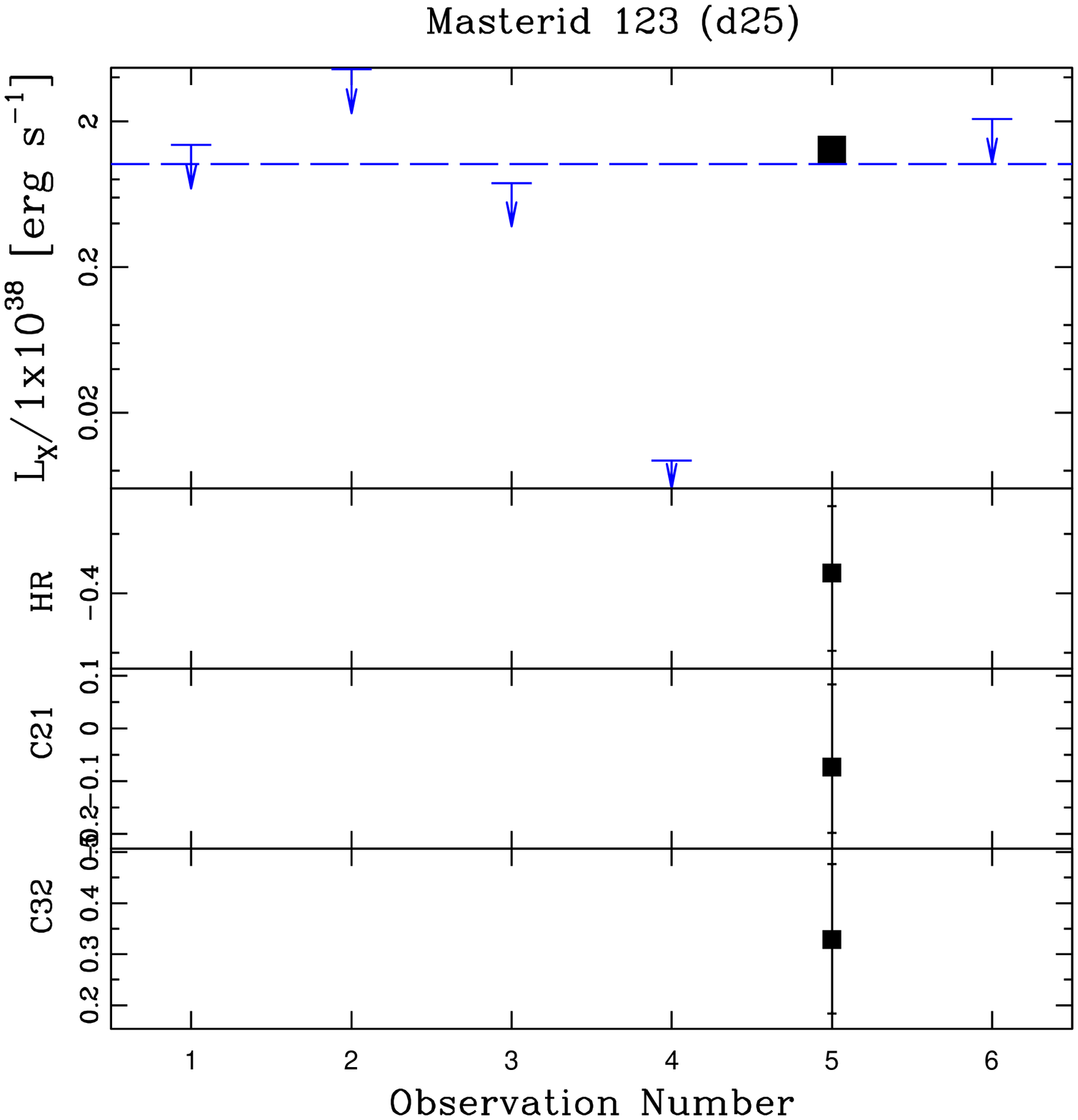}

\end{minipage}\hspace{0.02\linewidth}
\begin{minipage}{0.485\linewidth}
  \centering
  
    \includegraphics[width=\linewidth]{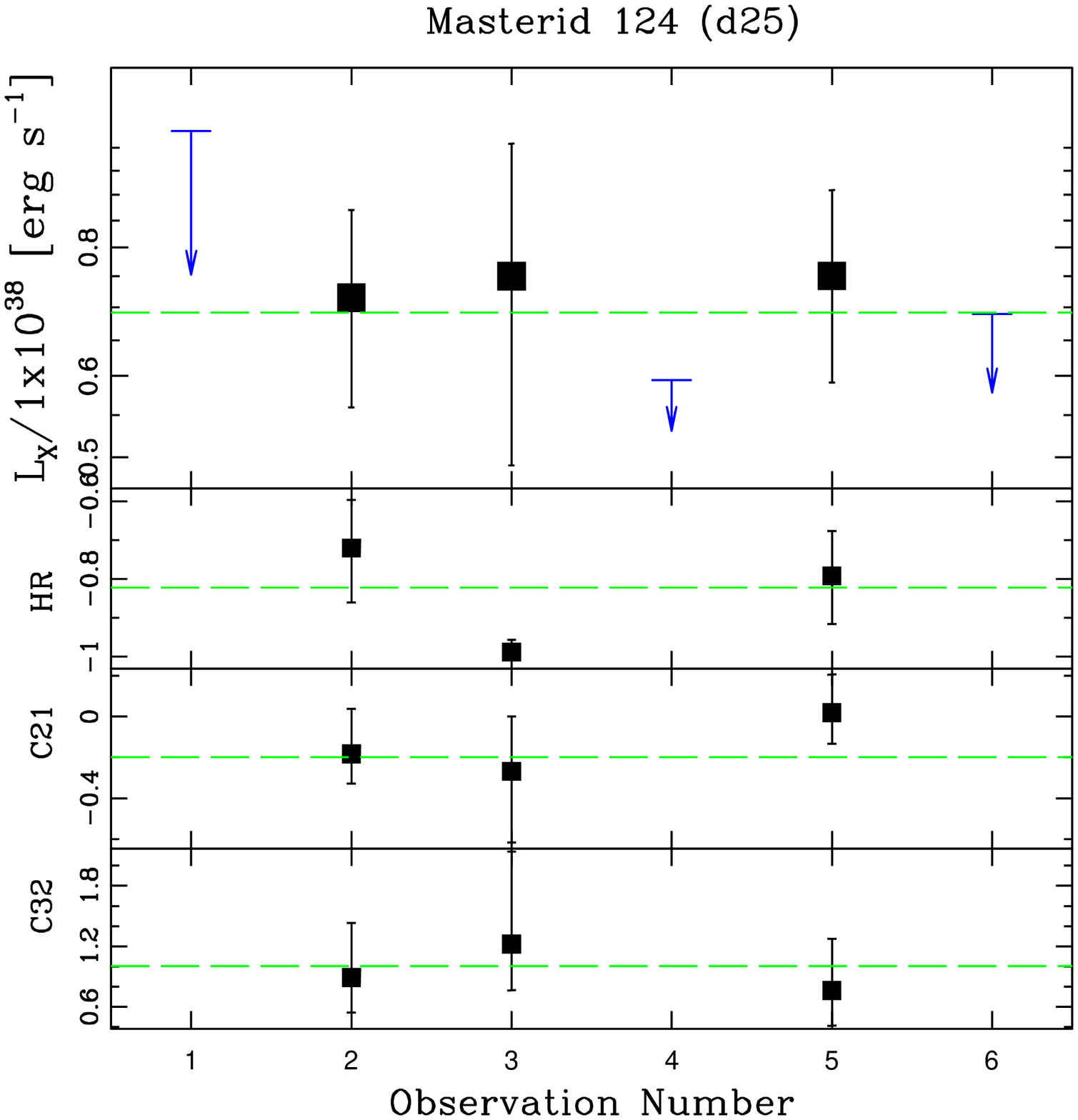}

  \end{minipage}\hspace{0.02\linewidth}

\end{figure}

\begin{figure}
  \begin{minipage}{0.485\linewidth}
  \centering

    \includegraphics[width=\linewidth]{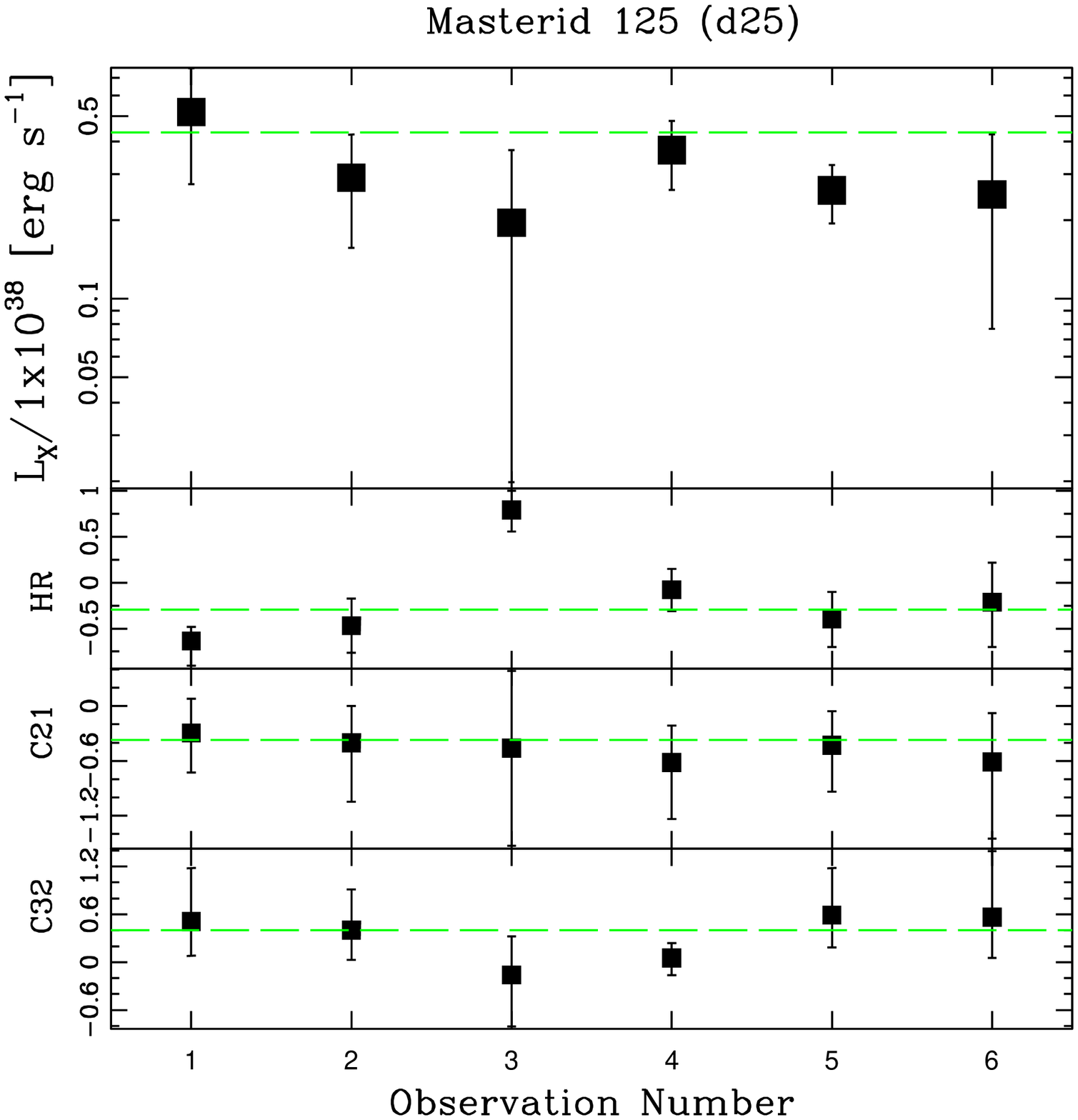}

\end{minipage}\hspace{0.02\linewidth}
\begin{minipage}{0.485\linewidth}
  \centering

    \includegraphics[width=\linewidth]{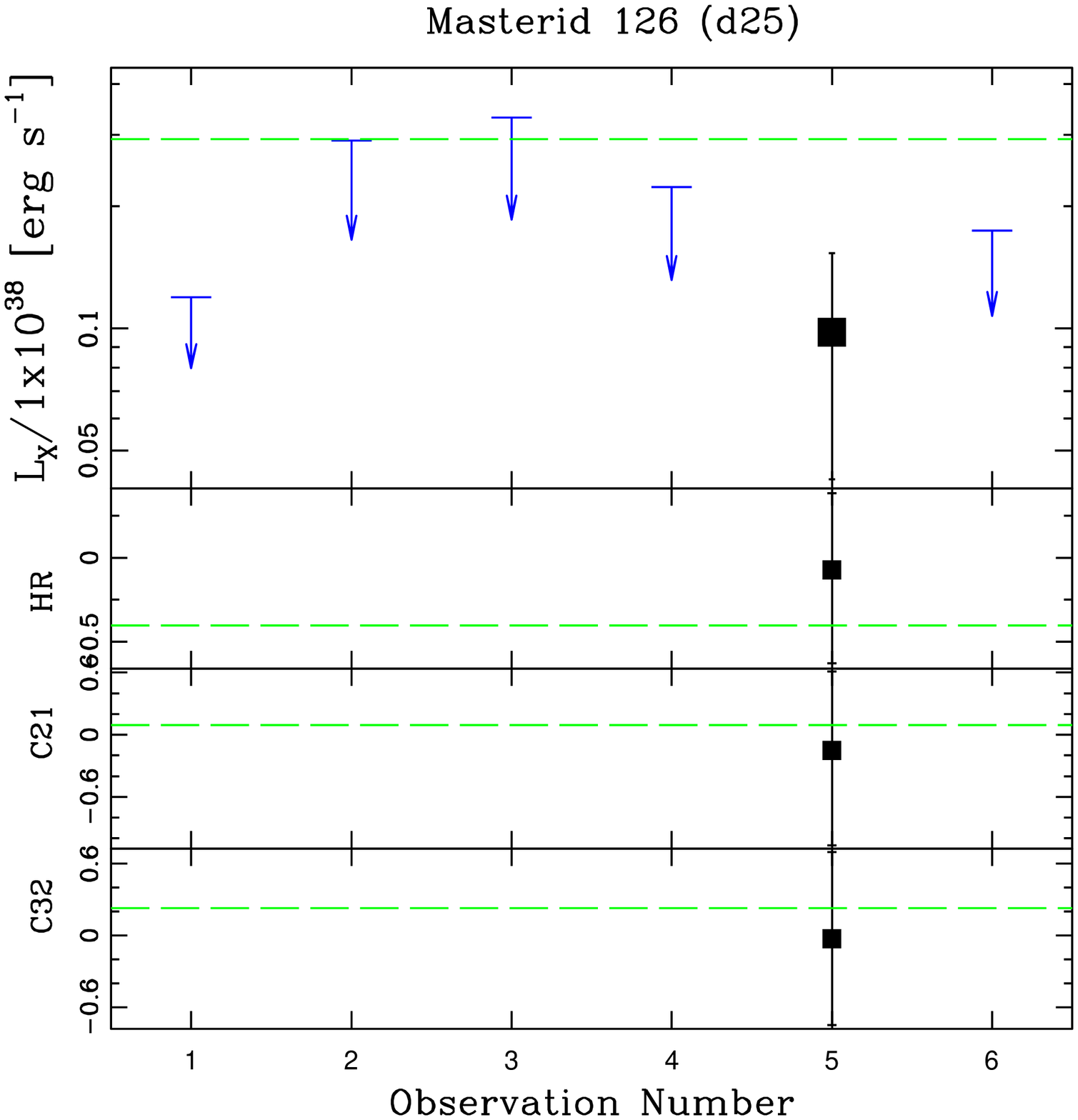}

\end{minipage}\hspace{0.02\linewidth}

  \begin{minipage}{0.485\linewidth}
  \centering
  
    \includegraphics[width=\linewidth]{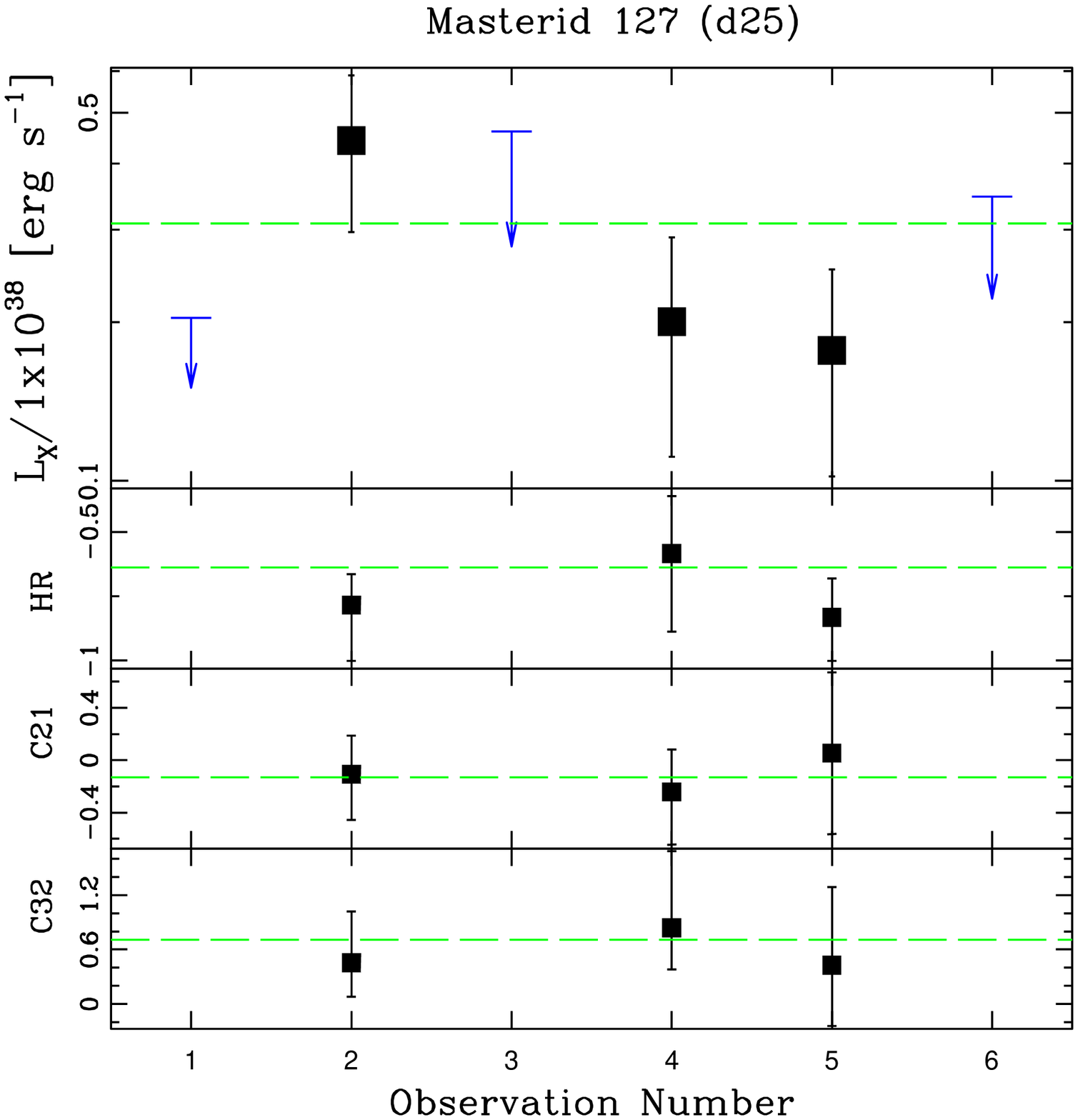}

  \end{minipage}\hspace{0.02\linewidth}
  \begin{minipage}{0.485\linewidth}
  \centering

    \includegraphics[width=\linewidth]{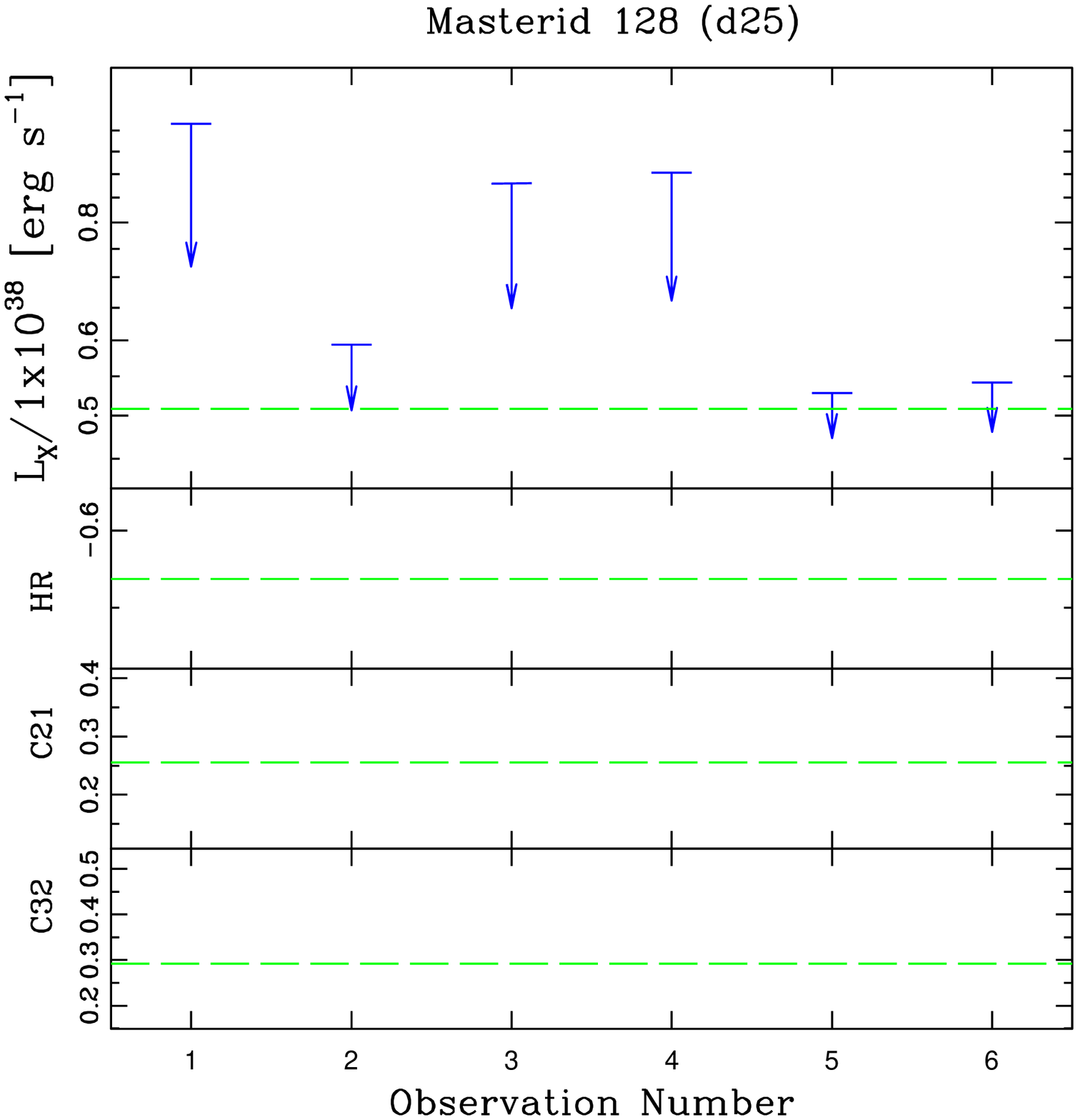}

\end{minipage}\hspace{0.02\linewidth}

\begin{minipage}{0.485\linewidth}
  \centering

    \includegraphics[width=\linewidth]{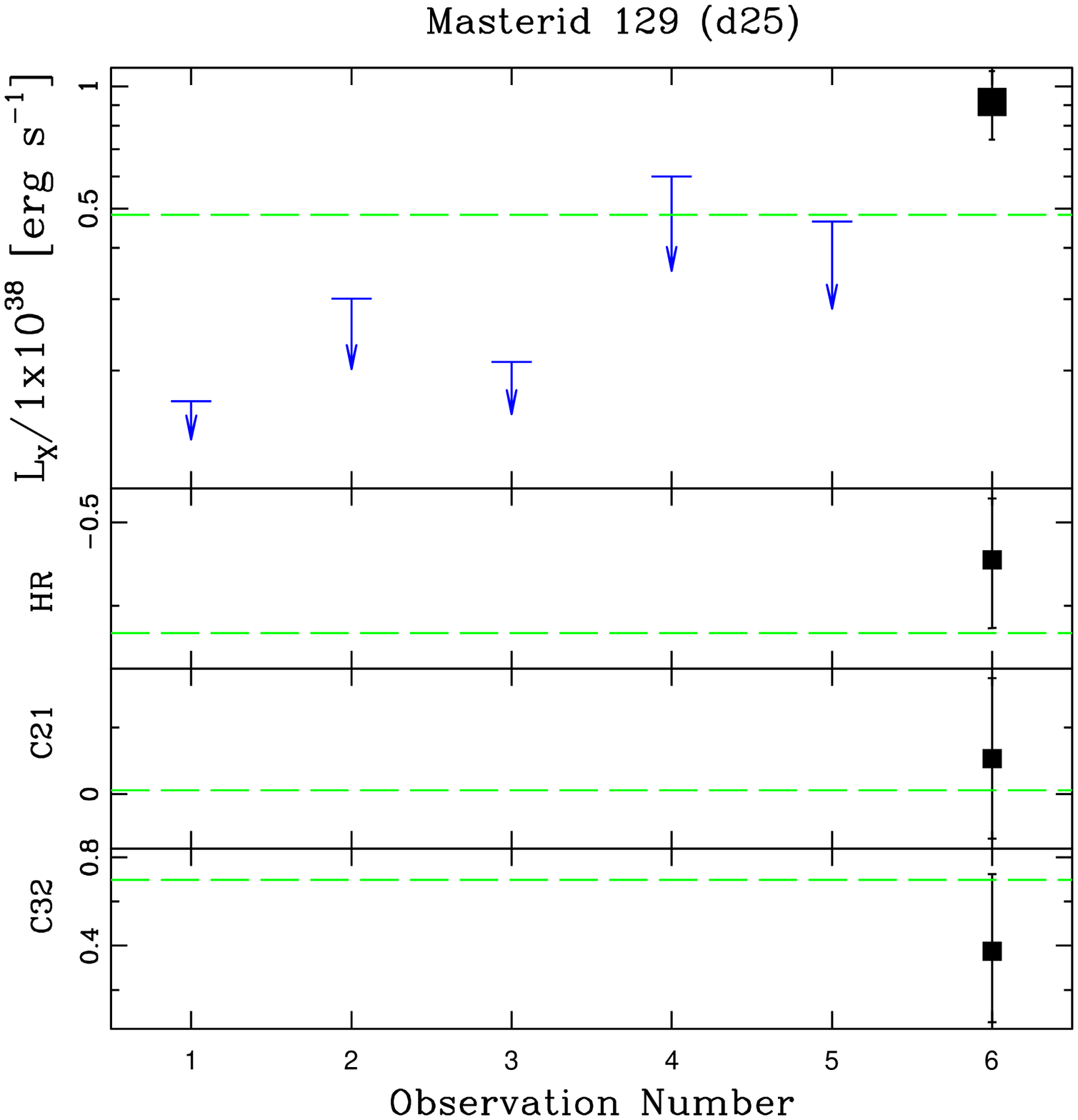}

 \end{minipage}\hspace{0.02\linewidth}
\begin{minipage}{0.485\linewidth}
  \centering
  
    \includegraphics[width=\linewidth]{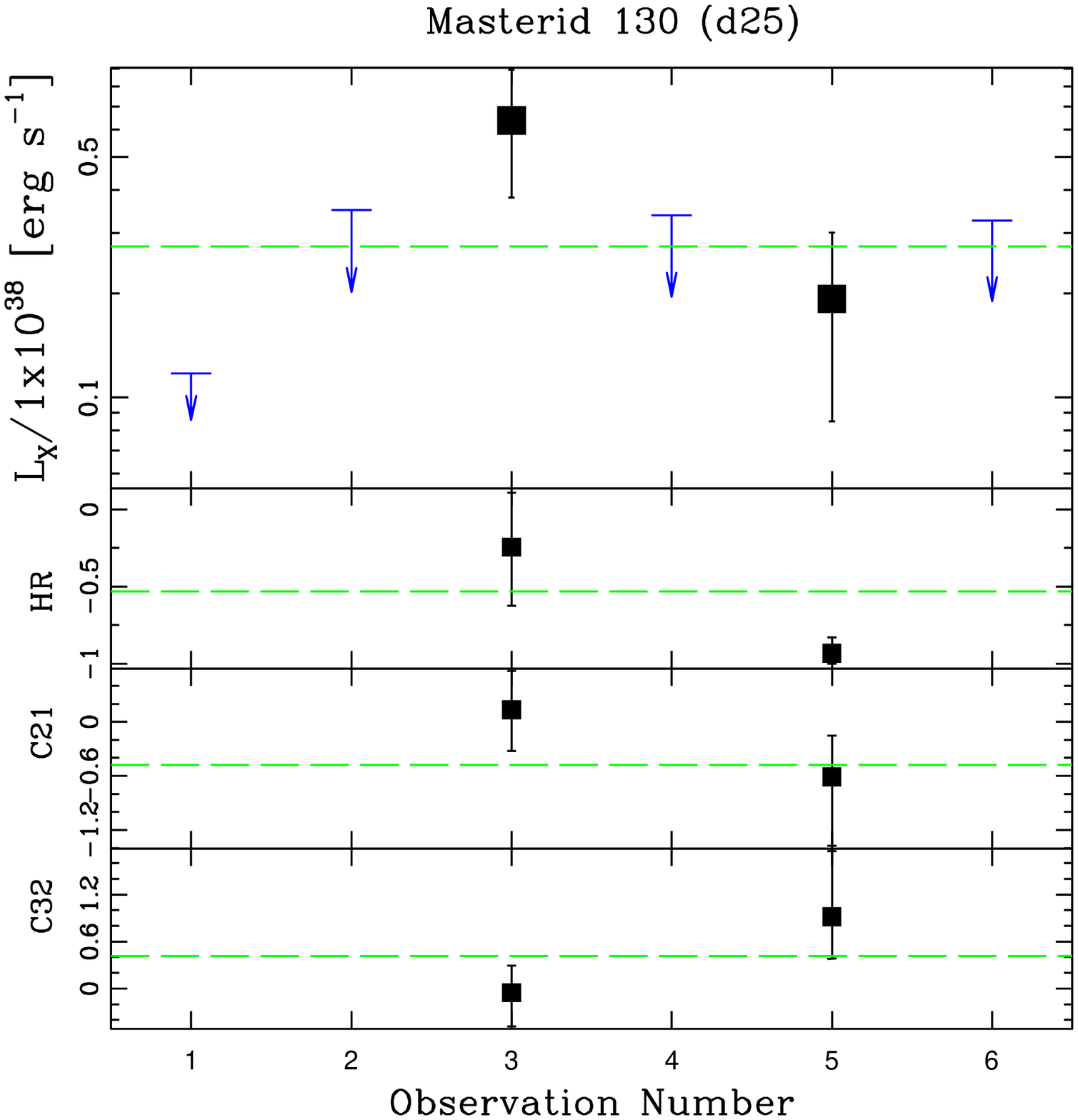}

  \end{minipage}\hspace{0.02\linewidth}
\end{figure}

\begin{figure}
  \begin{minipage}{0.485\linewidth}
  \centering

    \includegraphics[width=\linewidth]{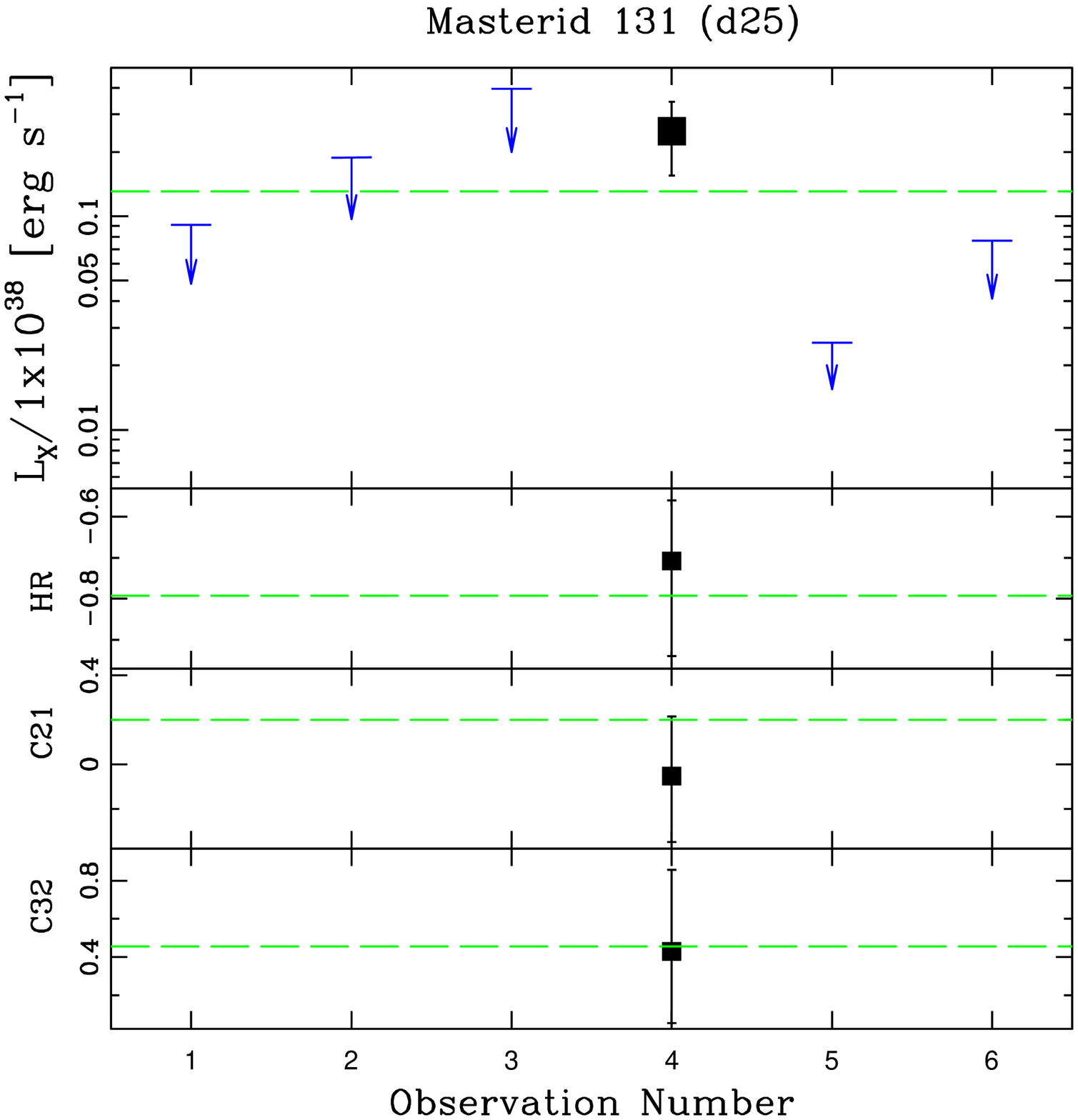}

\end{minipage}\hspace{0.02\linewidth}
\begin{minipage}{0.485\linewidth}
  \centering

    \includegraphics[width=\linewidth]{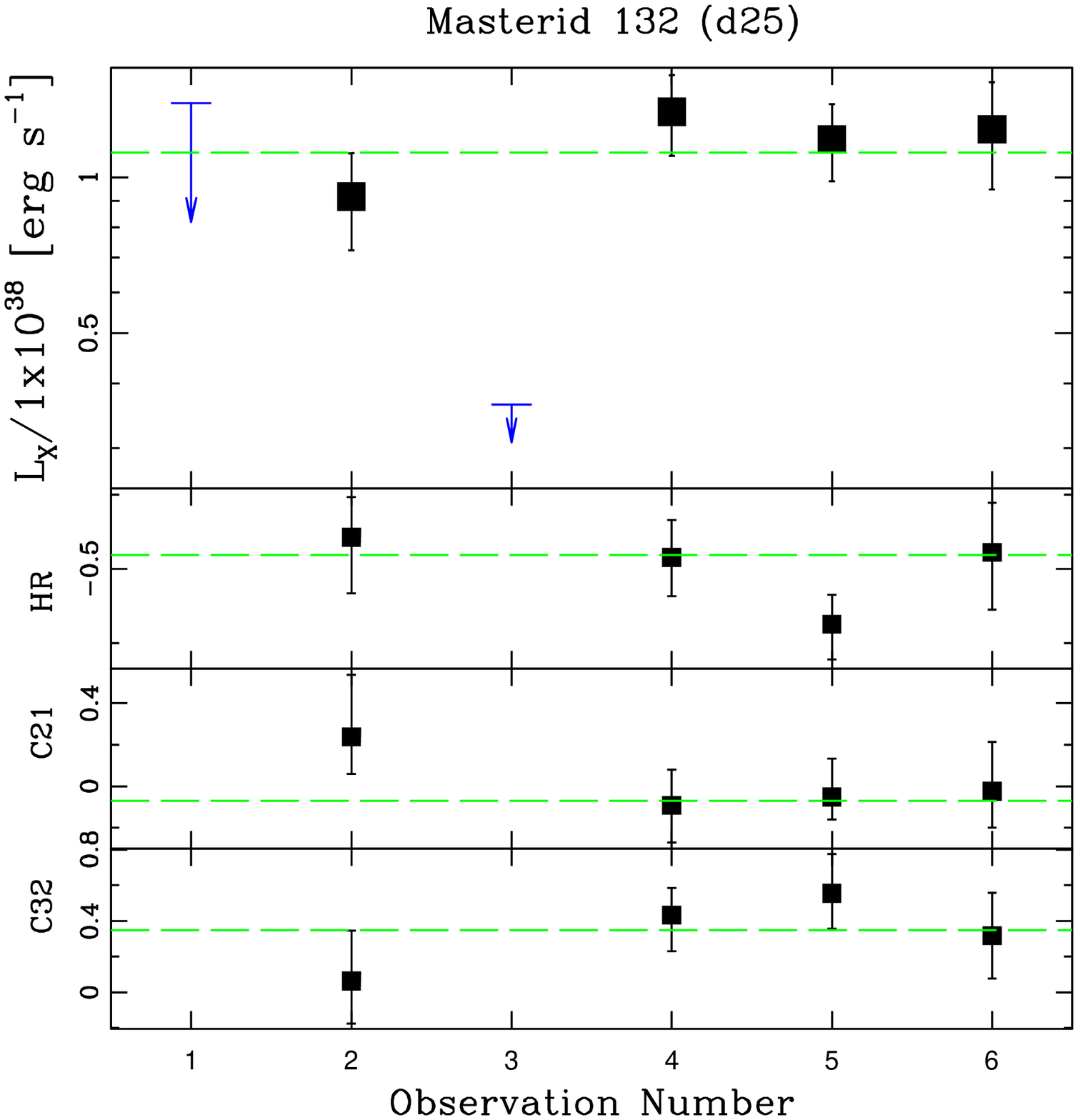}

\end{minipage}\hspace{0.02\linewidth}

  \begin{minipage}{0.485\linewidth}
  \centering
  
    \includegraphics[width=\linewidth]{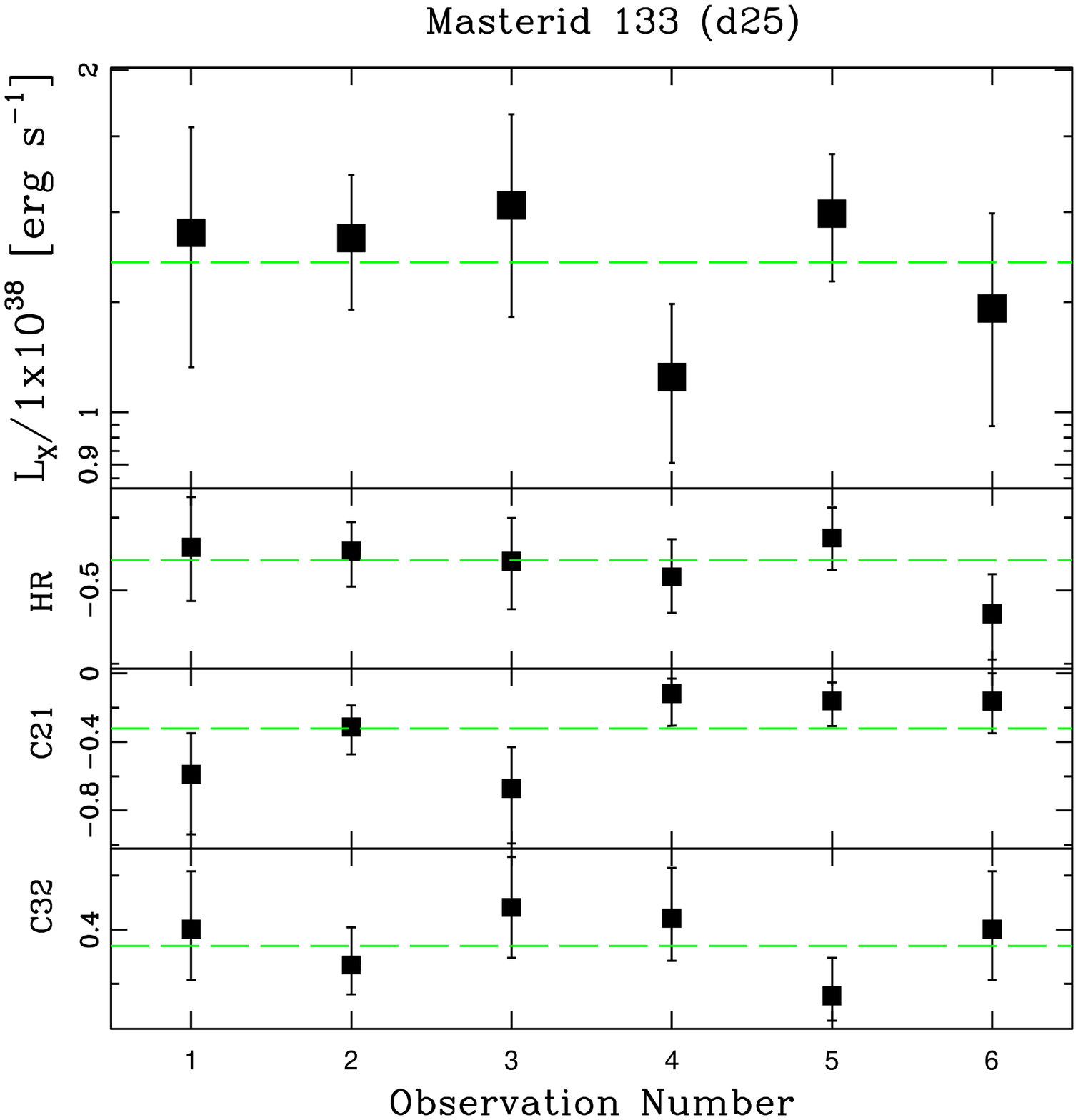}

  \end{minipage}\hspace{0.02\linewidth}
  \begin{minipage}{0.485\linewidth}
  \centering

    \includegraphics[width=\linewidth]{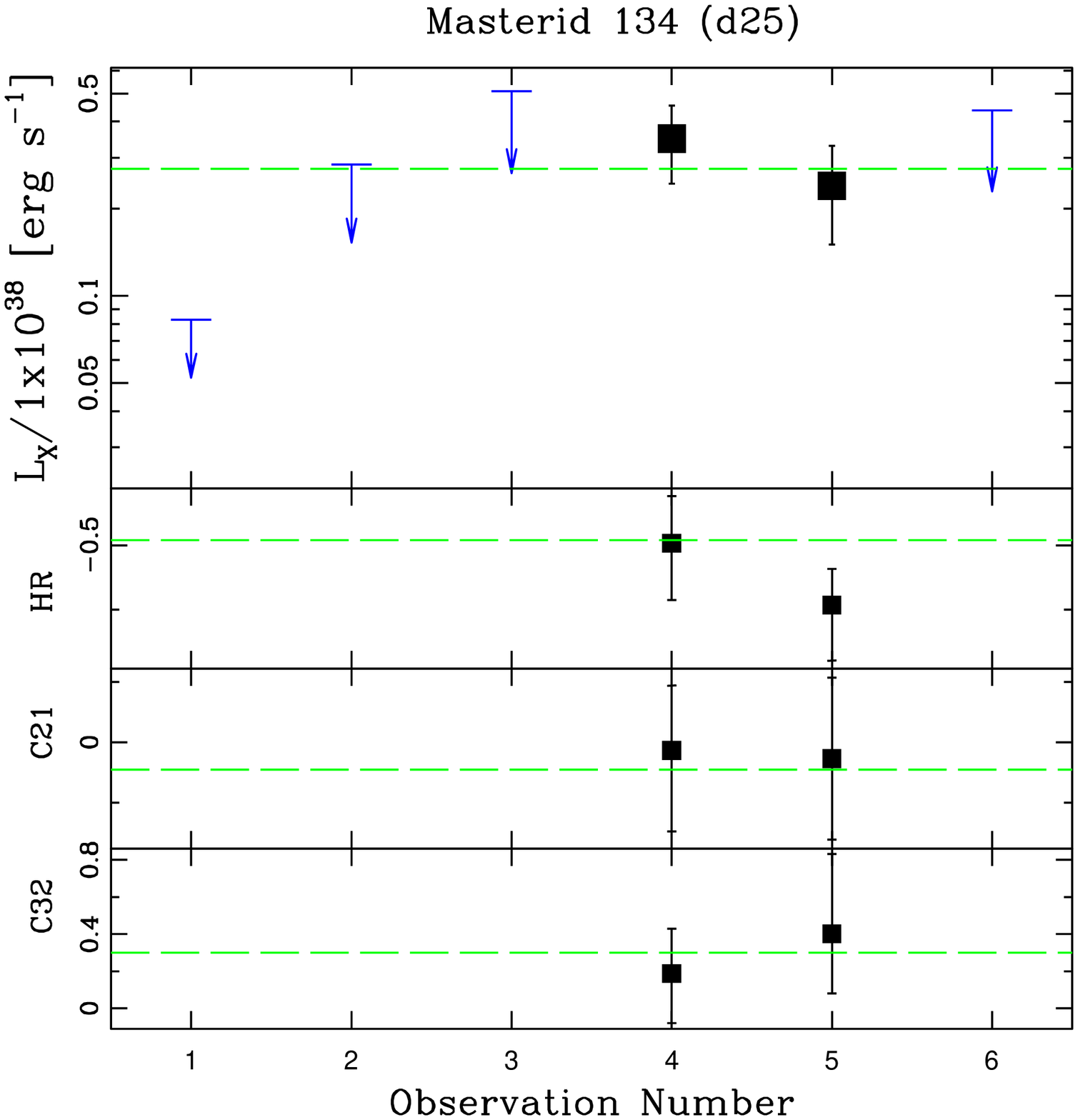}

\end{minipage}\hspace{0.02\linewidth}

\begin{minipage}{0.485\linewidth}
  \centering

    \includegraphics[width=\linewidth]{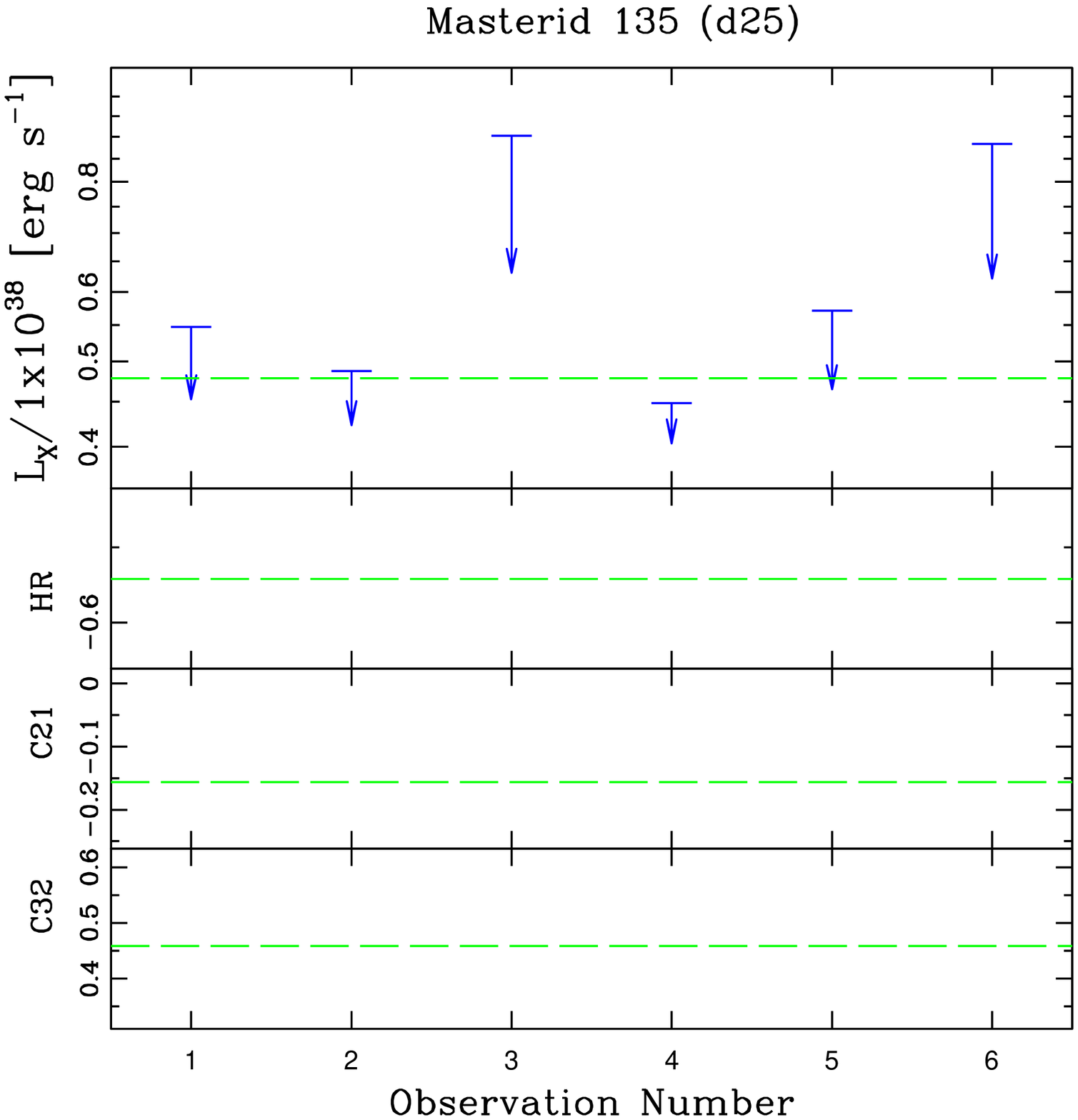}

 \end{minipage}\hspace{0.02\linewidth}
\begin{minipage}{0.485\linewidth}
  \centering
  
    \includegraphics[width=\linewidth]{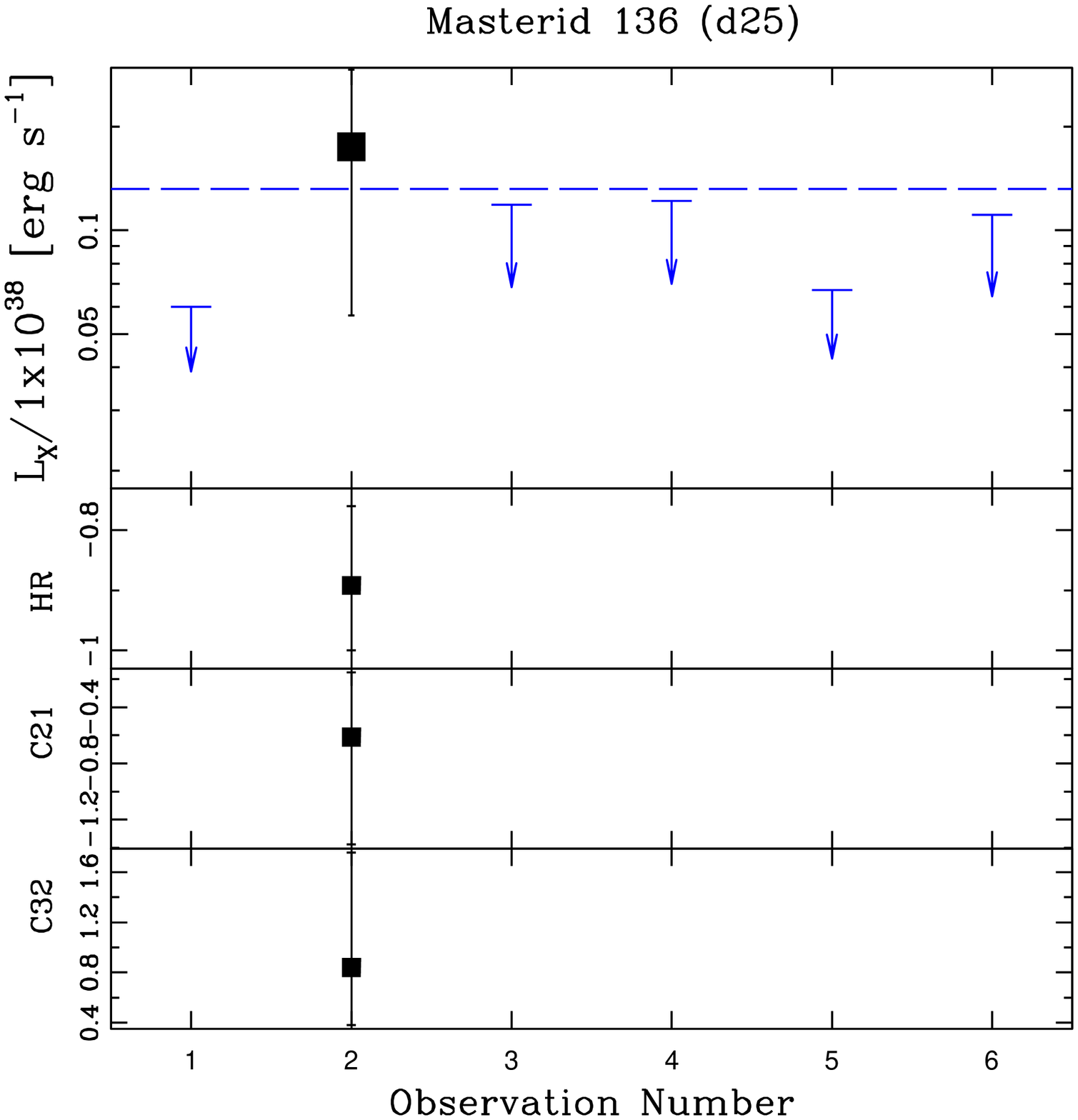}

  \end{minipage}\hspace{0.02\linewidth}
\end{figure}

\begin{figure}
  \begin{minipage}{0.485\linewidth}
  \centering

    \includegraphics[width=\linewidth]{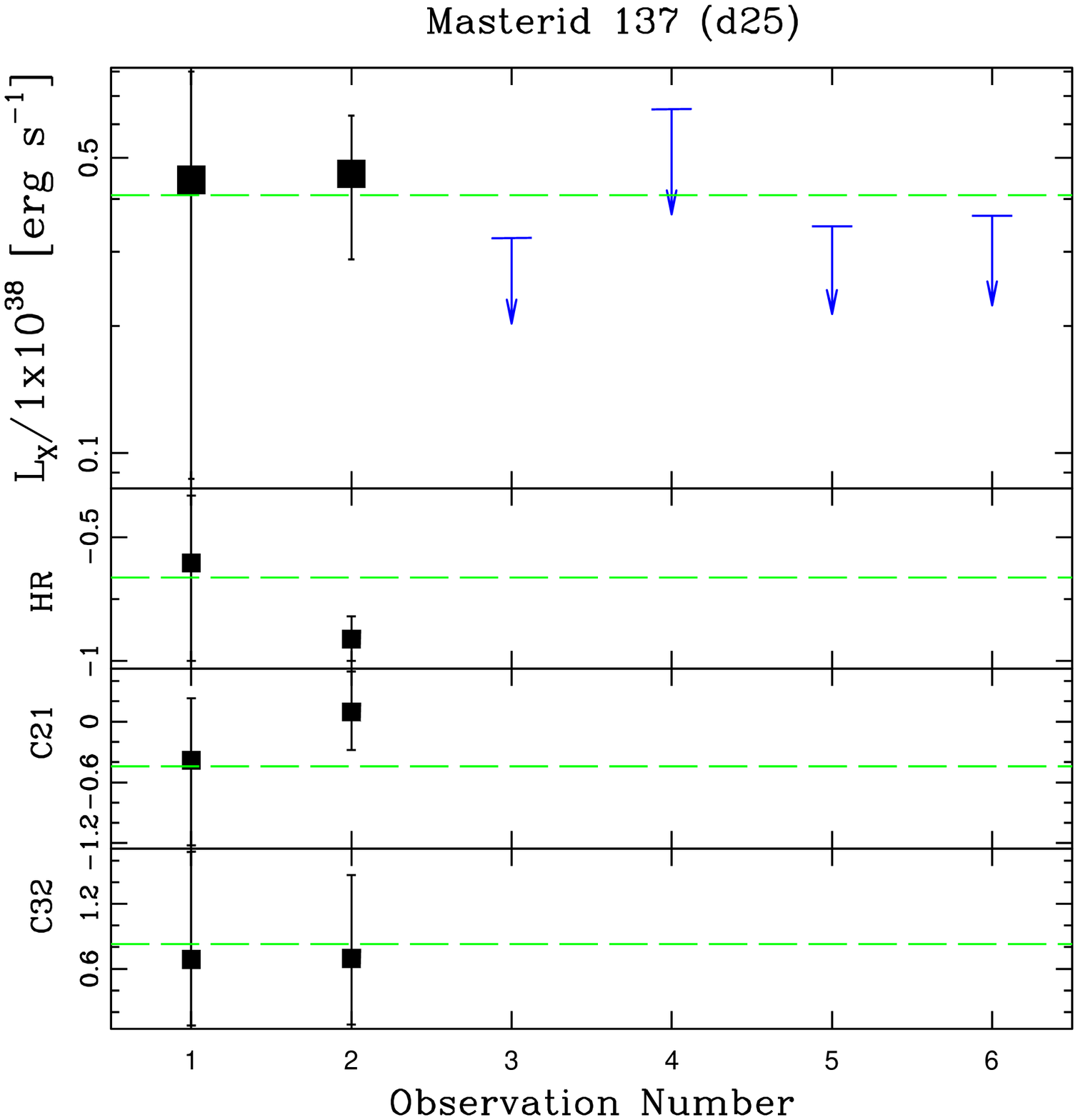}

\end{minipage}\hspace{0.02\linewidth}
\begin{minipage}{0.485\linewidth}
  \centering

    \includegraphics[width=\linewidth]{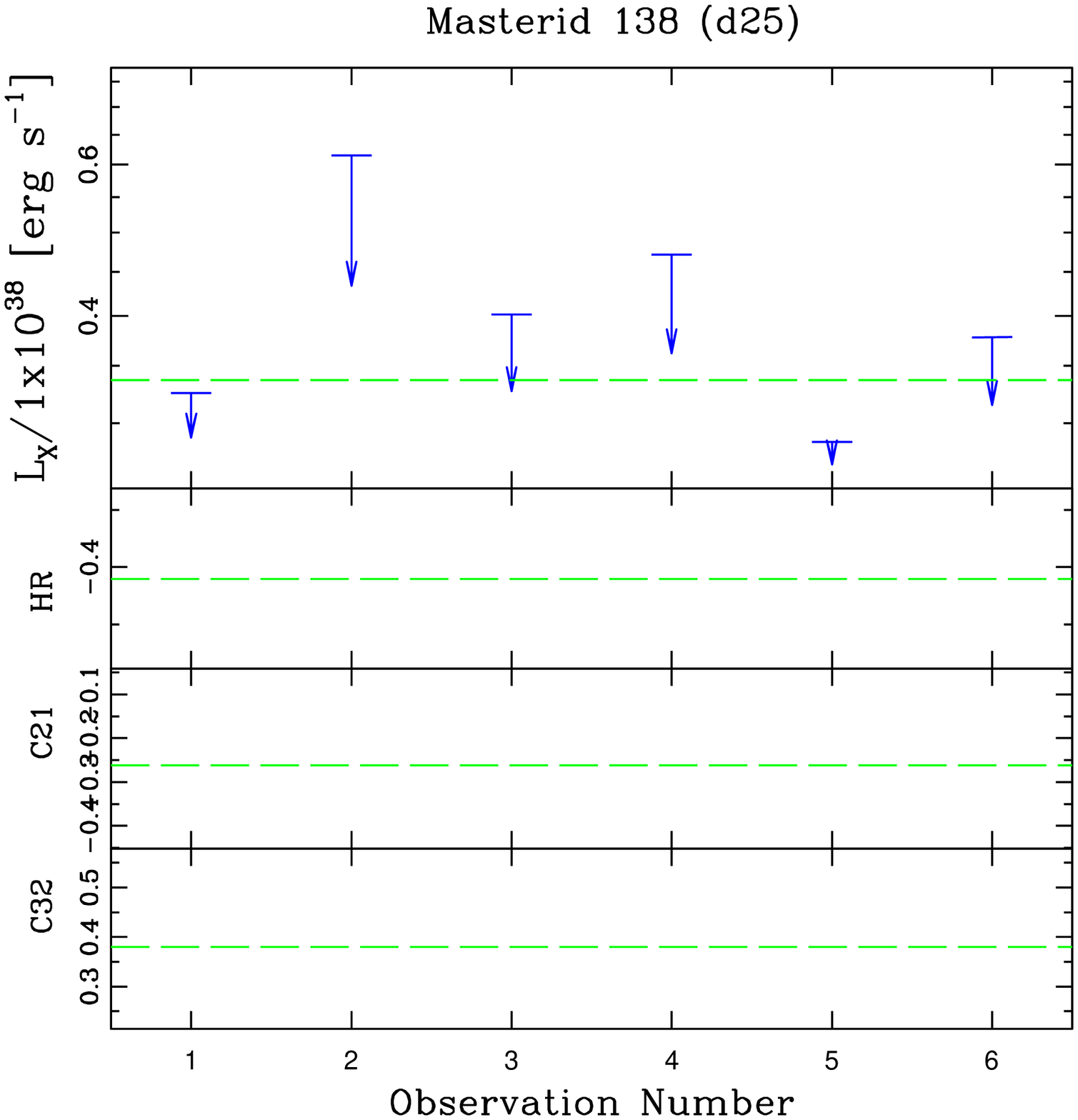}

\end{minipage}\hspace{0.02\linewidth}

  \begin{minipage}{0.485\linewidth}
  \centering
  
    \includegraphics[width=\linewidth]{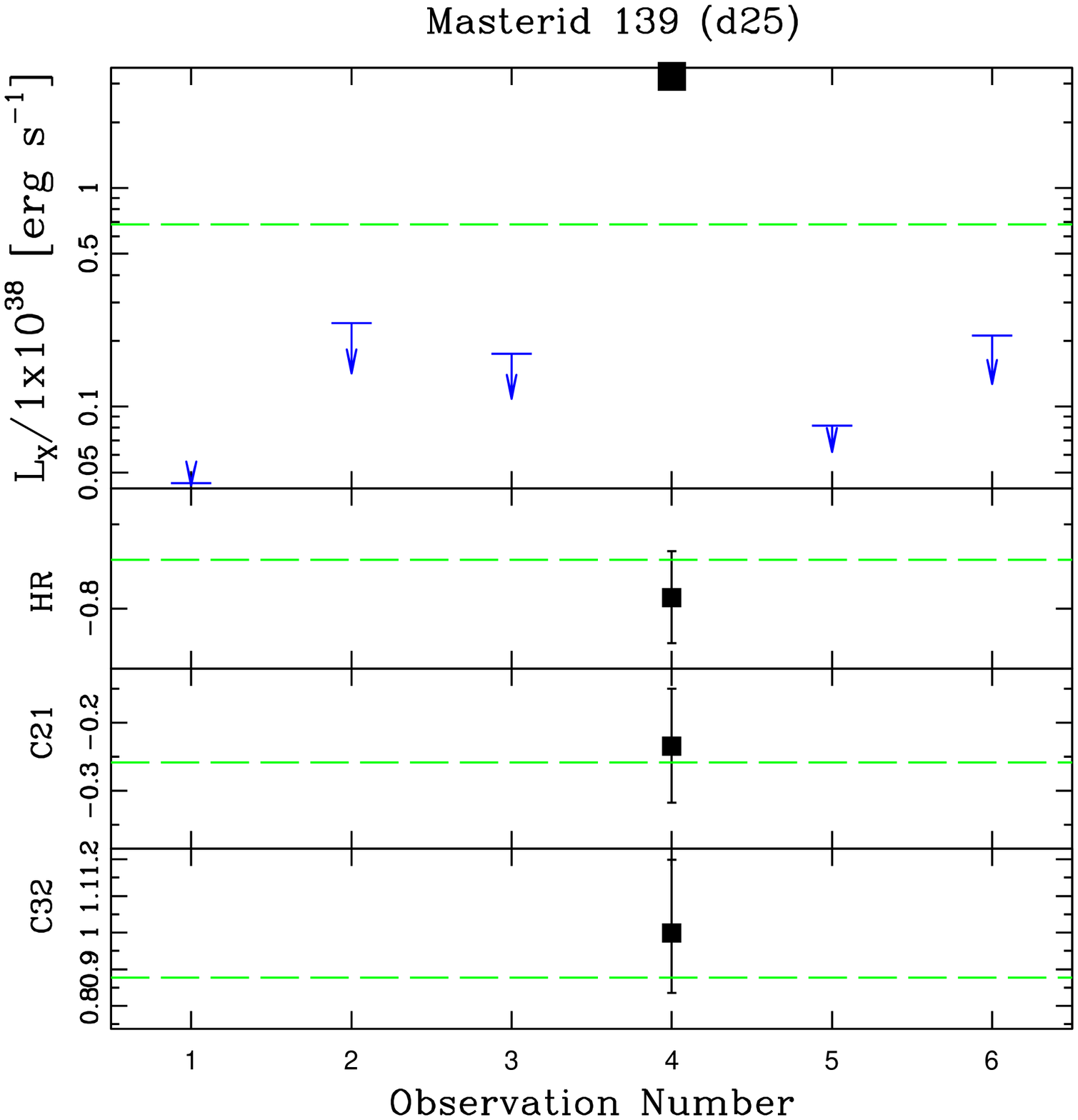}

  \end{minipage}\hspace{0.02\linewidth}
  \begin{minipage}{0.485\linewidth}
  \centering

    \includegraphics[width=\linewidth]{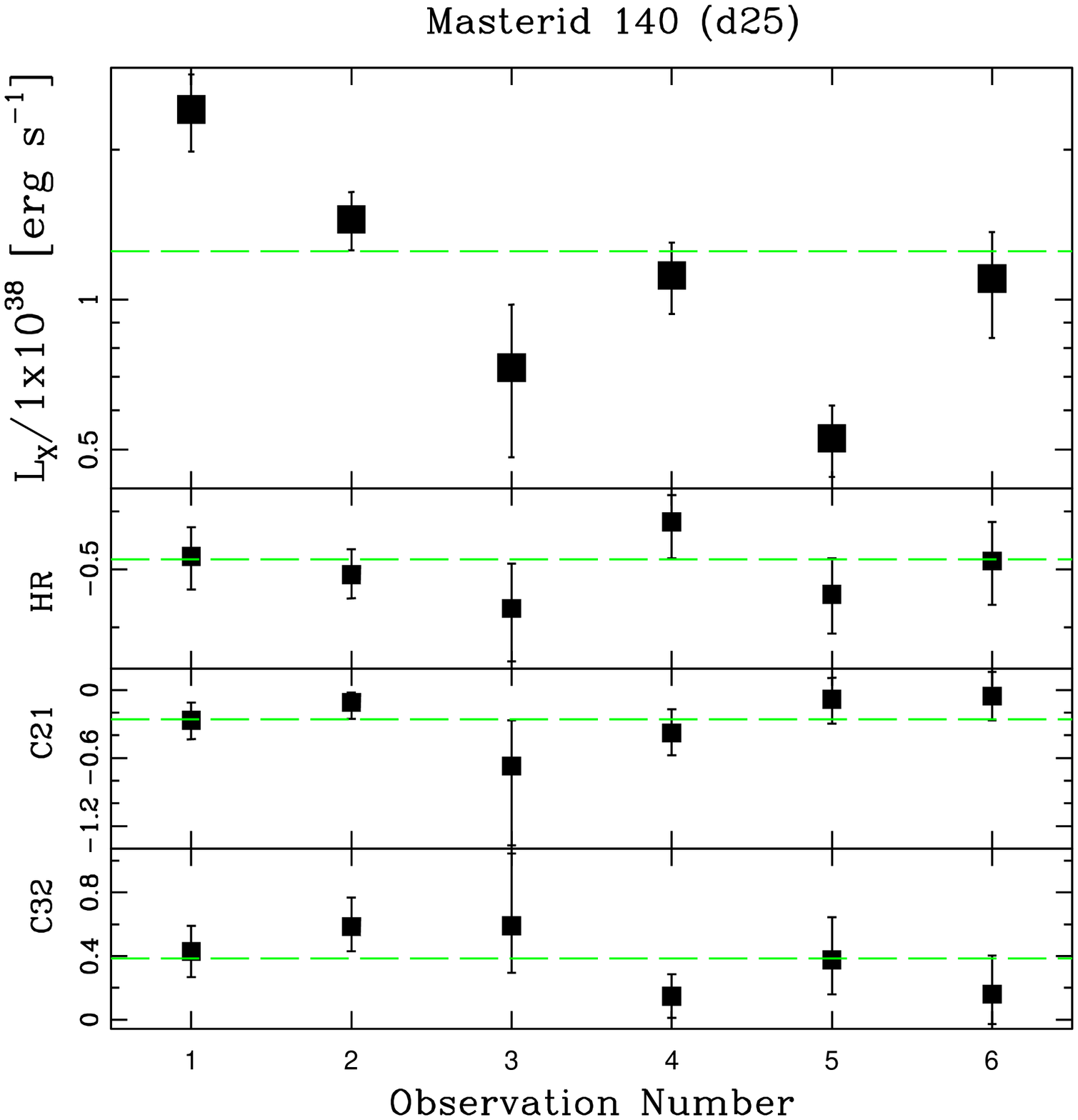}

\end{minipage}\hspace{0.02\linewidth}

\begin{minipage}{0.485\linewidth}
  \centering

    \includegraphics[width=\linewidth]{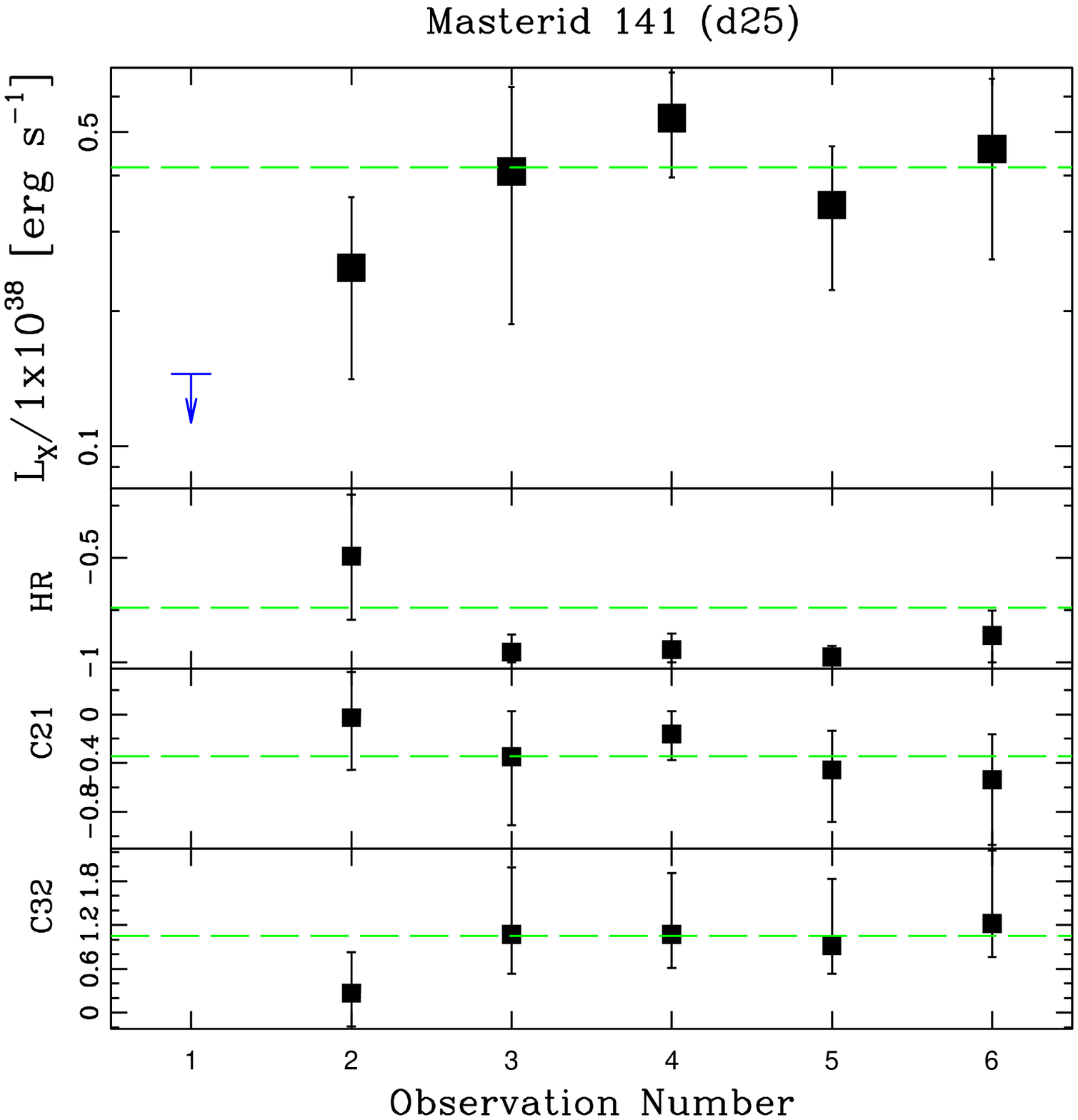}

 \end{minipage}\hspace{0.02\linewidth}
\begin{minipage}{0.485\linewidth}
  \centering
  
    \includegraphics[width=\linewidth]{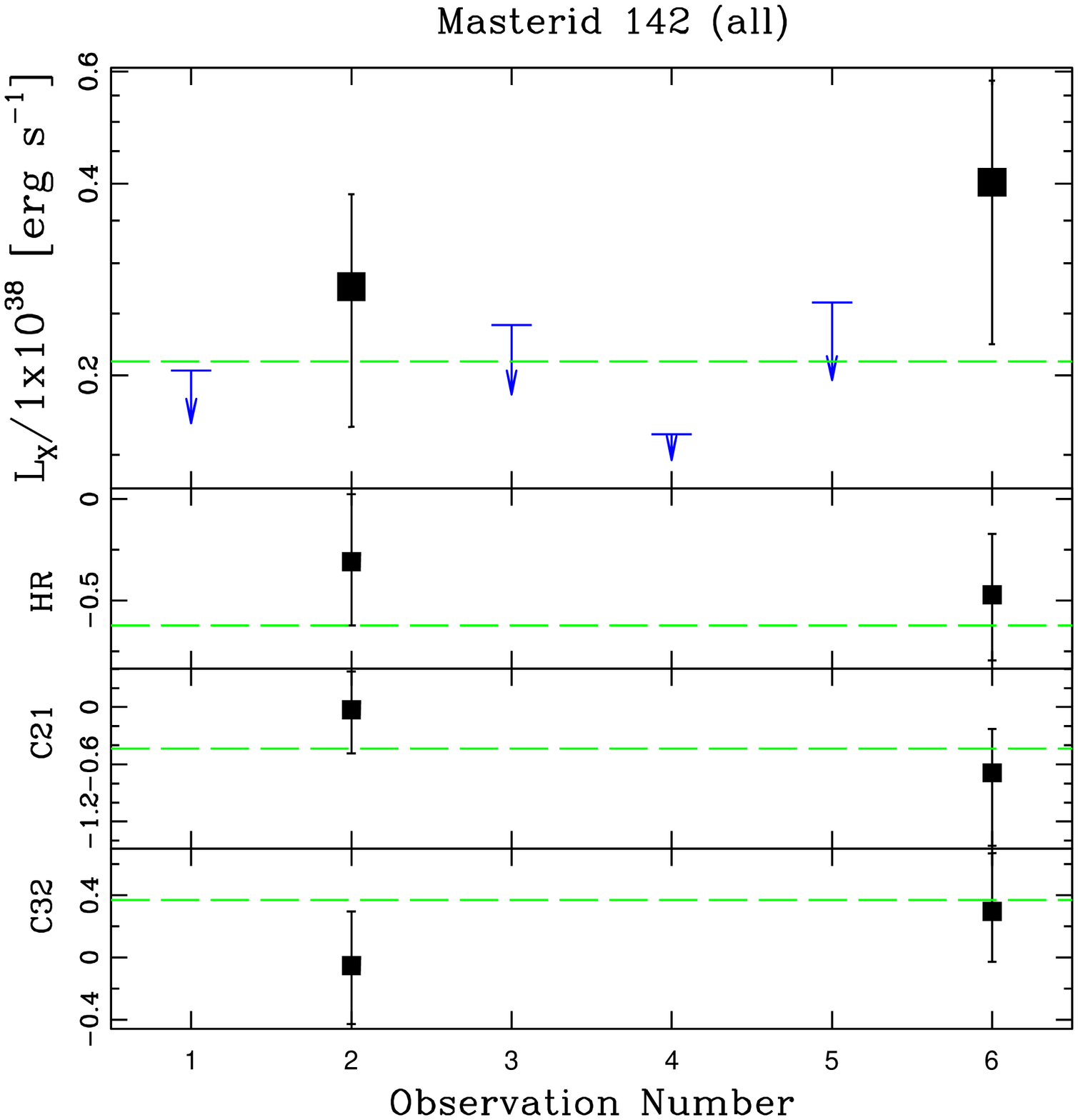}

  \end{minipage}\hspace{0.02\linewidth}
\end{figure}

\begin{figure}

  \begin{minipage}{0.485\linewidth}
  \centering

    \includegraphics[width=\linewidth]{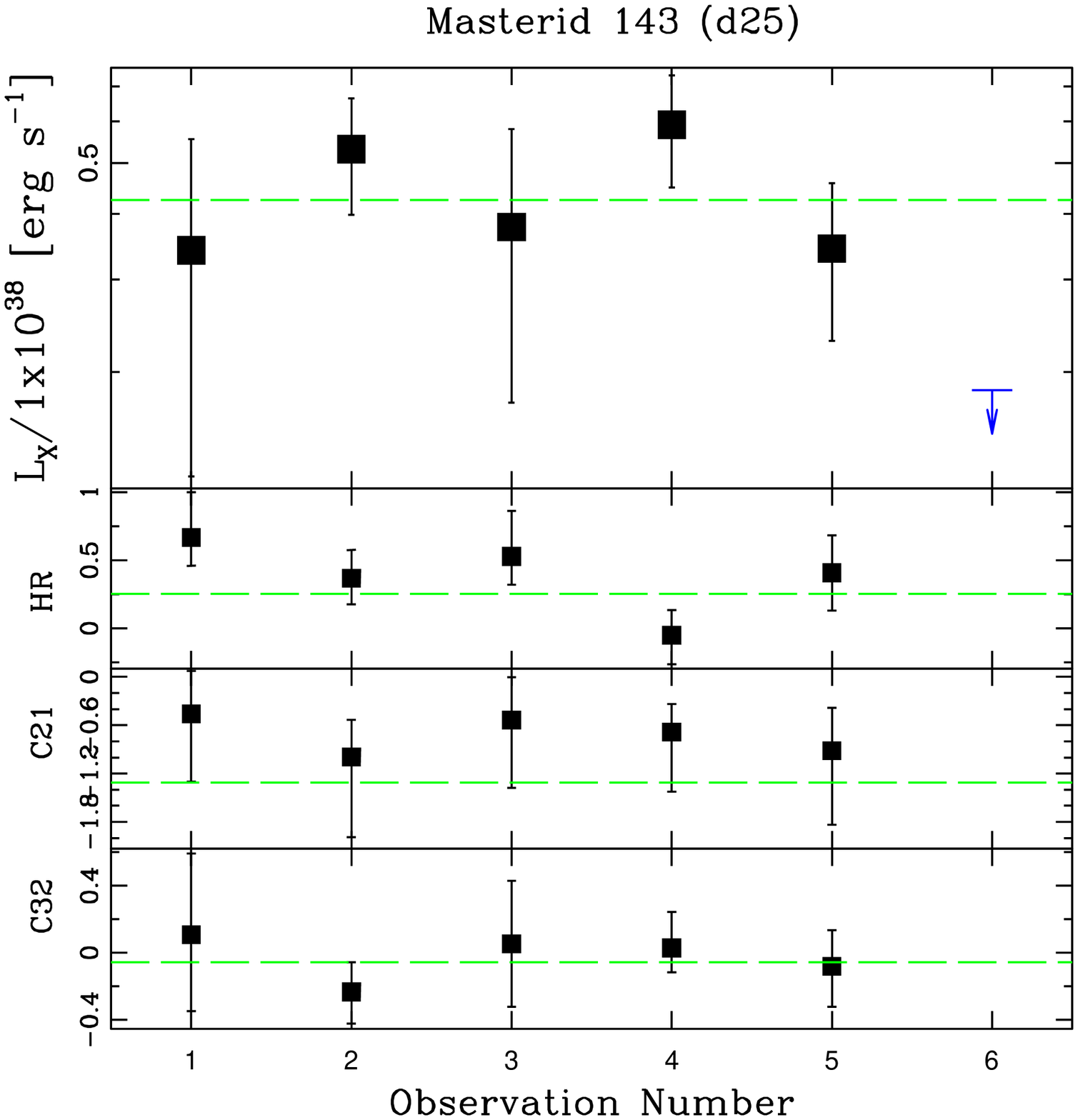}

\end{minipage}\hspace{0.02\linewidth}
\begin{minipage}{0.485\linewidth}
  \centering

    \includegraphics[width=\linewidth]{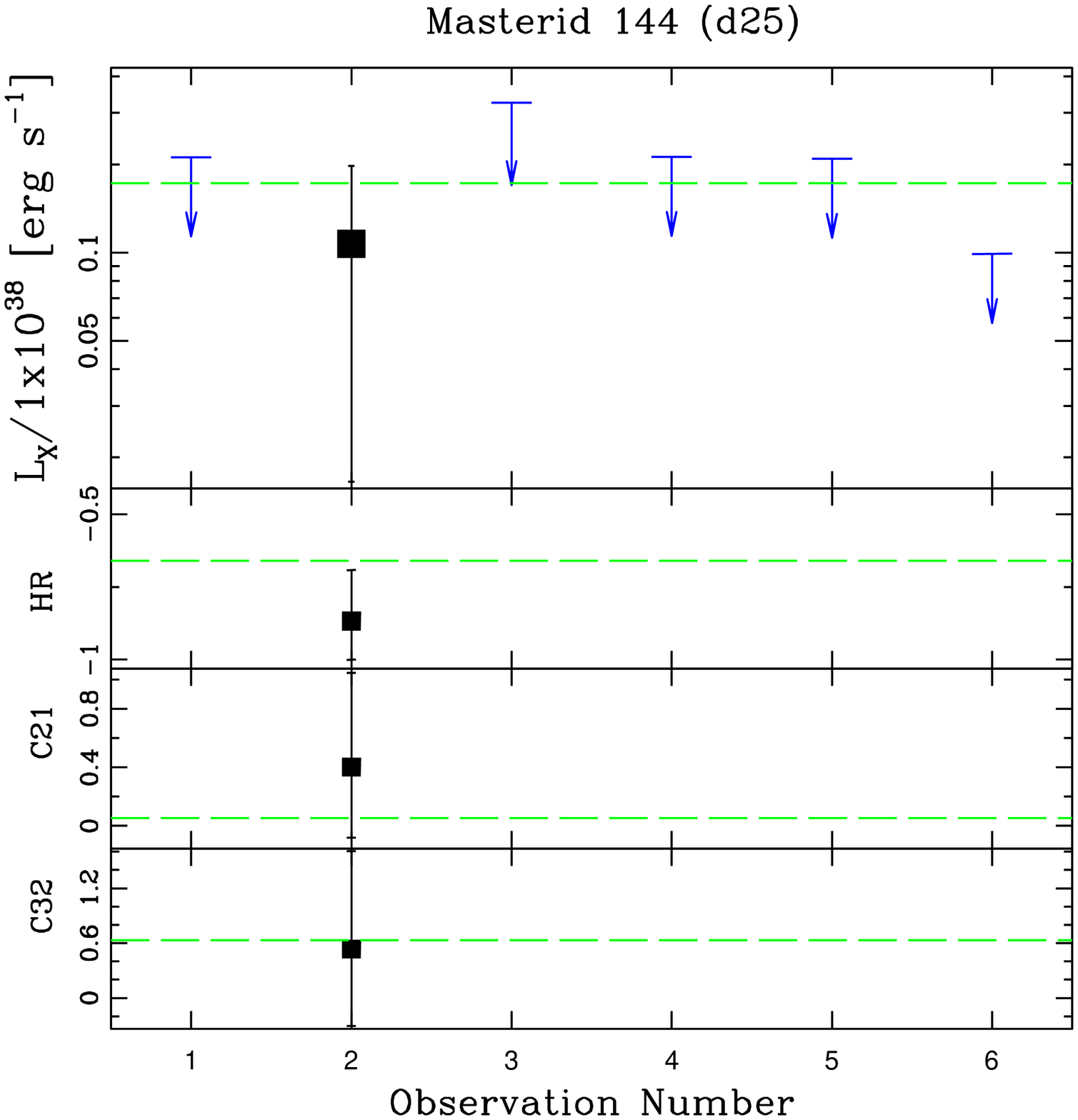}

 \end{minipage}\hspace{0.02\linewidth}

  \begin{minipage}{0.485\linewidth}
  \centering
  
    \includegraphics[width=\linewidth]{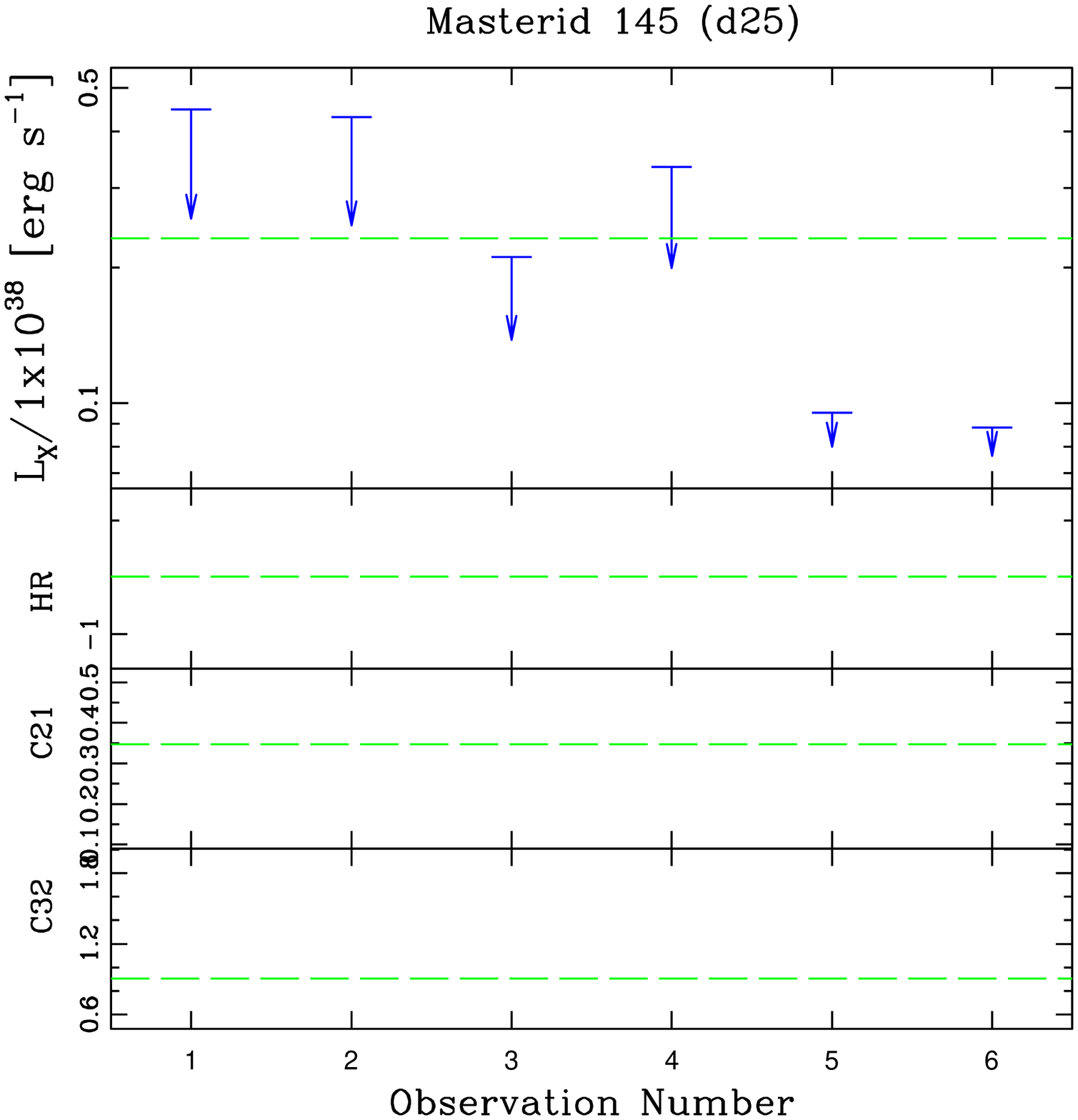}

  \end{minipage}\hspace{0.02\linewidth}
  \begin{minipage}{0.485\linewidth}
  \centering

    \includegraphics[width=\linewidth]{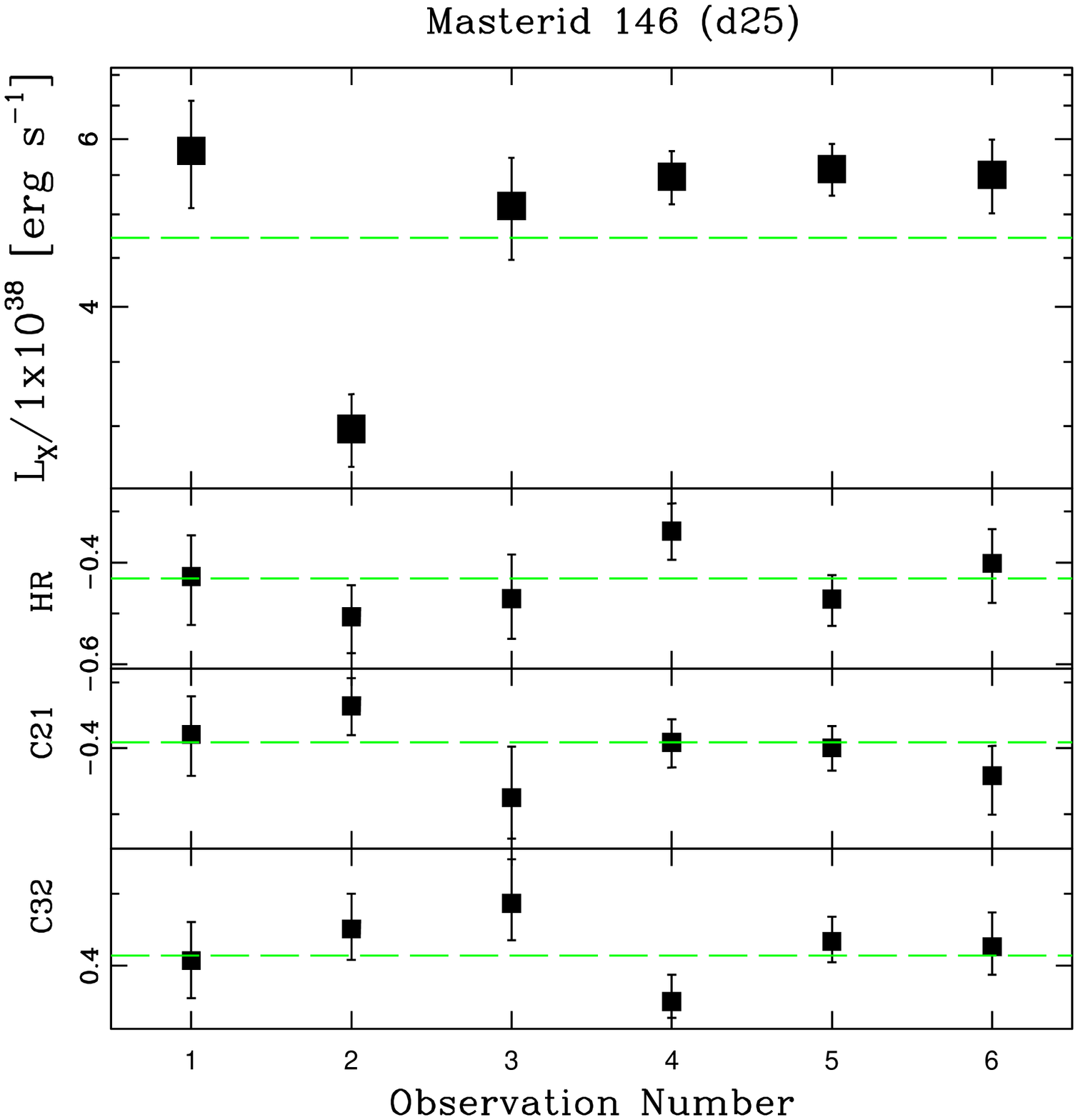}

\end{minipage}\hspace{0.02\linewidth}

\begin{minipage}{0.485\linewidth}
  \centering

    \includegraphics[width=\linewidth]{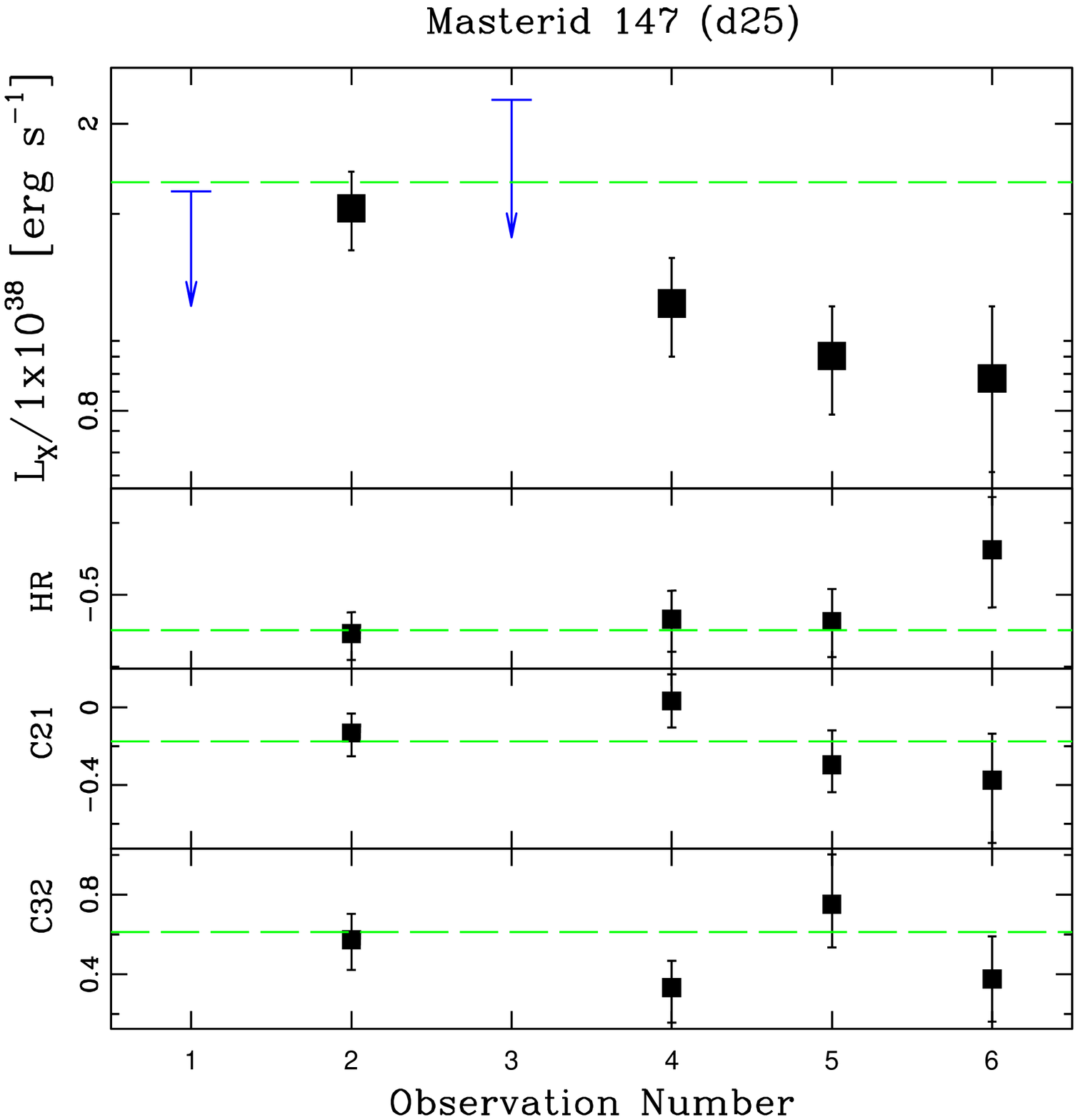}

 \end{minipage}\hspace{0.02\linewidth}
\begin{minipage}{0.485\linewidth}
  \centering
  
    \includegraphics[width=\linewidth]{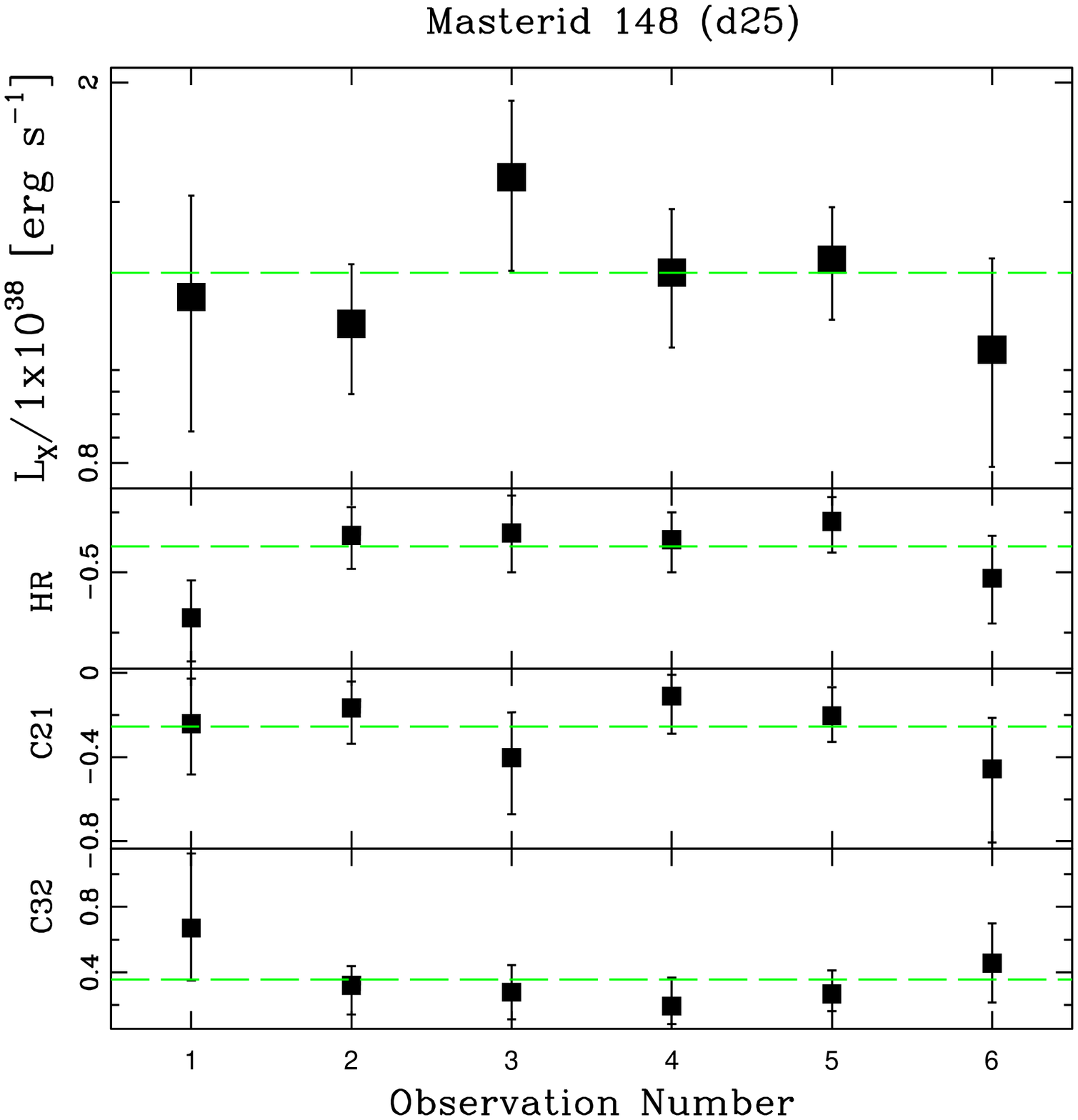}

  \end{minipage}\hspace{0.02\linewidth}
  
\end{figure}

\begin{figure}

  \begin{minipage}{0.485\linewidth}
  \centering

    \includegraphics[width=\linewidth]{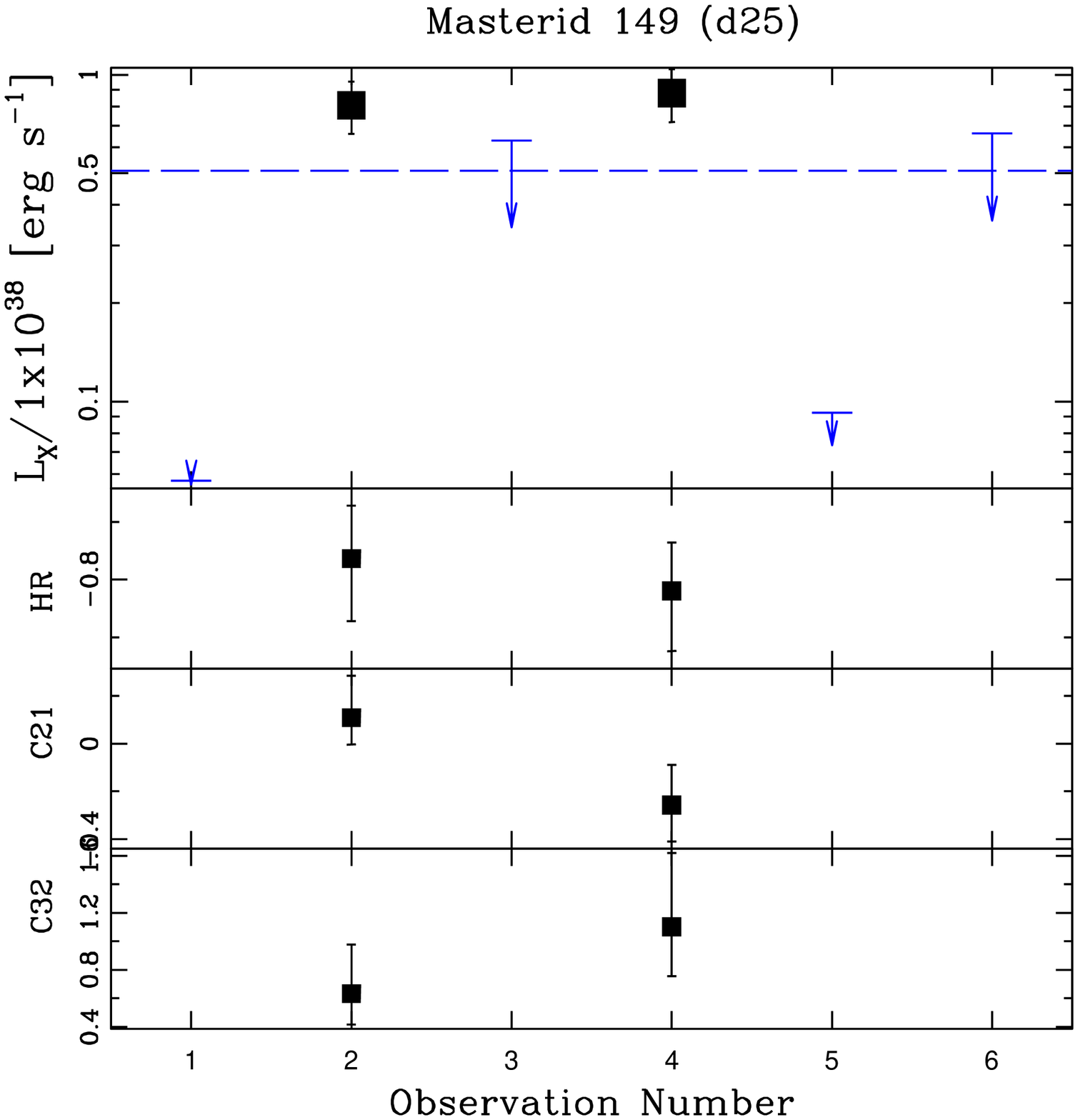}

\end{minipage}\hspace{0.02\linewidth}
\begin{minipage}{0.485\linewidth}
  \centering

    \includegraphics[width=\linewidth]{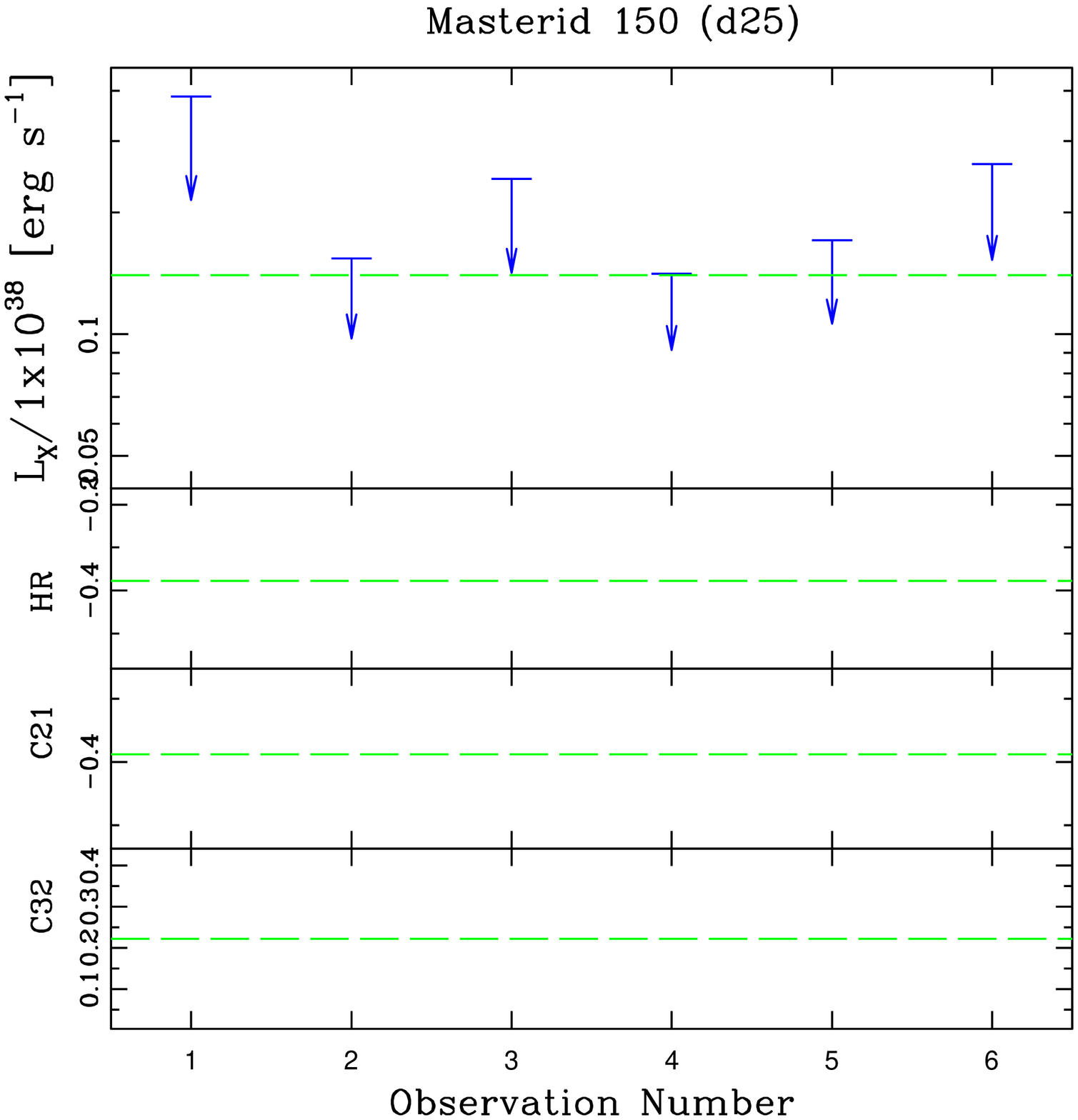}

 \end{minipage}\hspace{0.02\linewidth}

  \begin{minipage}{0.485\linewidth}
  \centering
  
    \includegraphics[width=\linewidth]{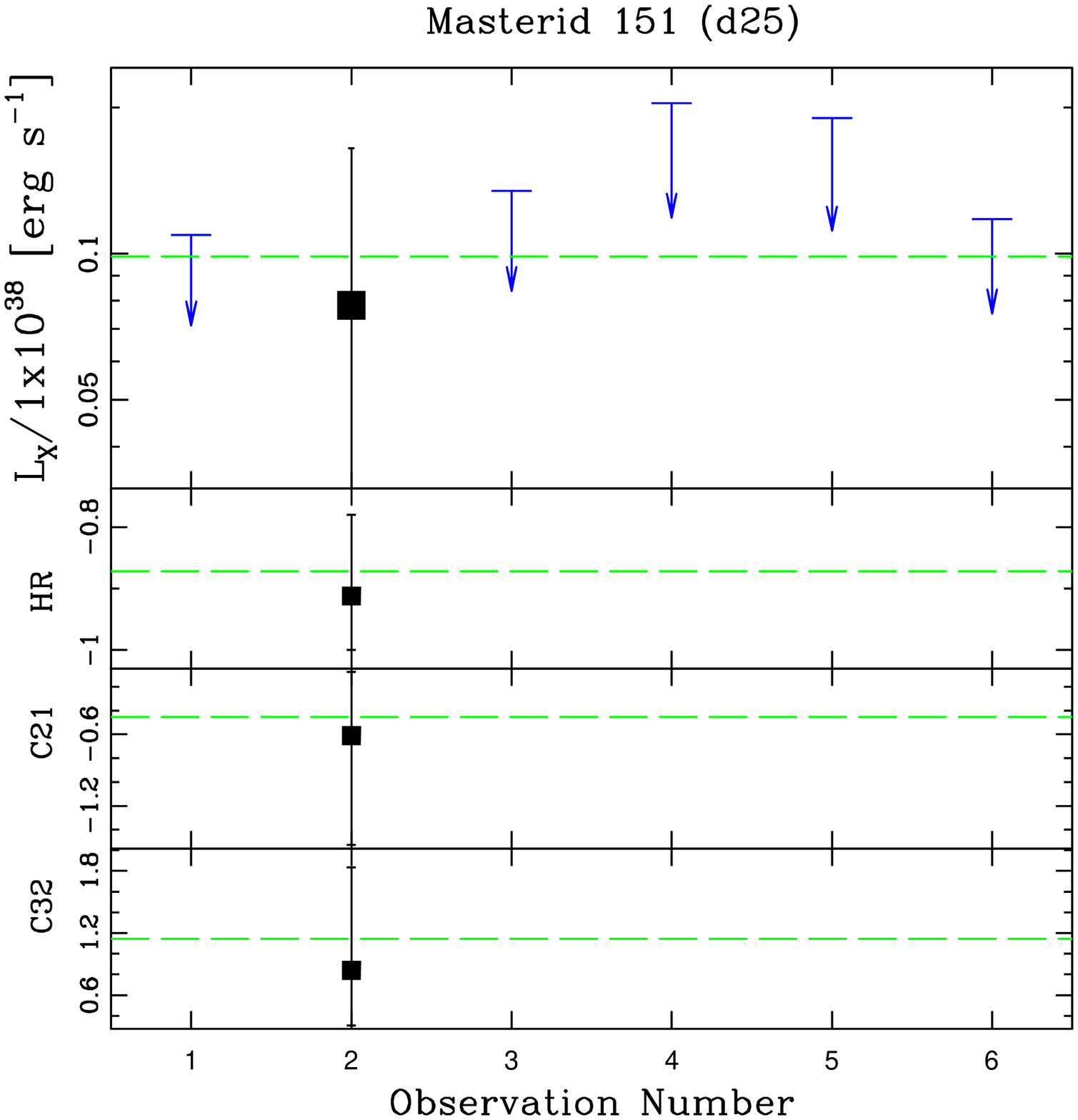}

  \end{minipage}\hspace{0.02\linewidth}
  \begin{minipage}{0.485\linewidth}
  \centering

    \includegraphics[width=\linewidth]{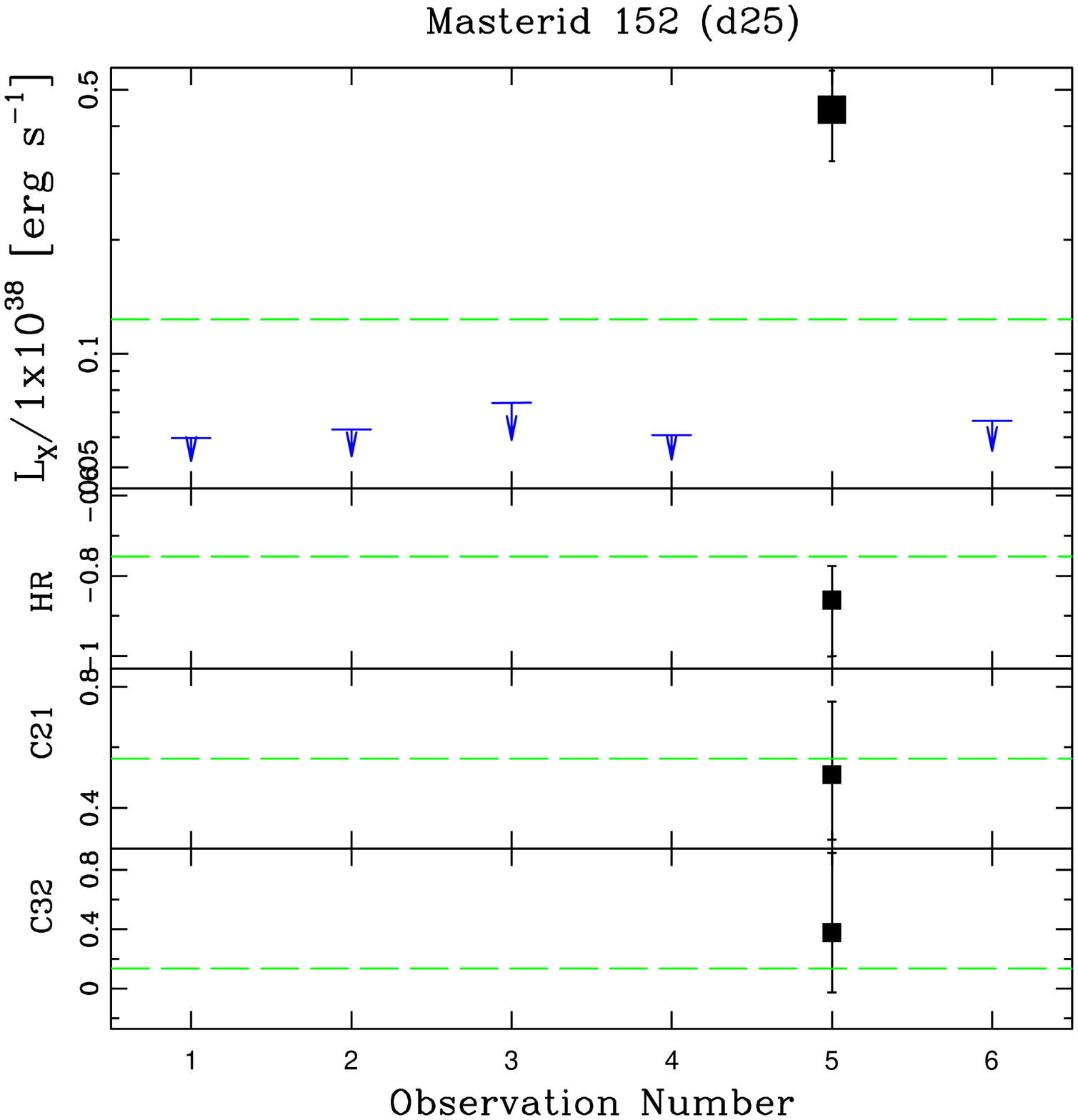}

\end{minipage}\hspace{0.02\linewidth}

\begin{minipage}{0.485\linewidth}
  \centering

    \includegraphics[width=\linewidth]{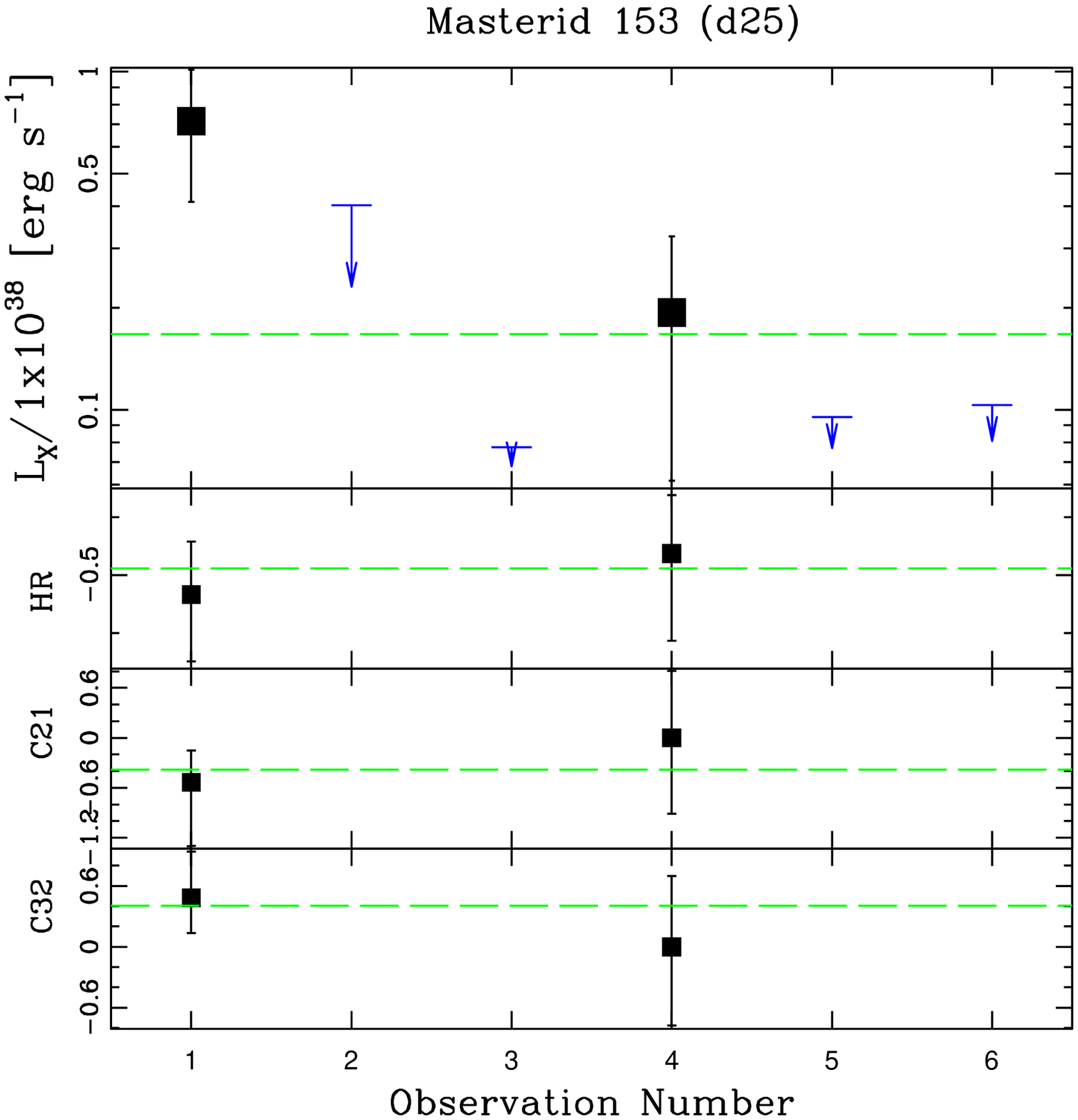}

 \end{minipage}\hspace{0.02\linewidth}
\begin{minipage}{0.485\linewidth}
  \centering
  
    \includegraphics[width=\linewidth]{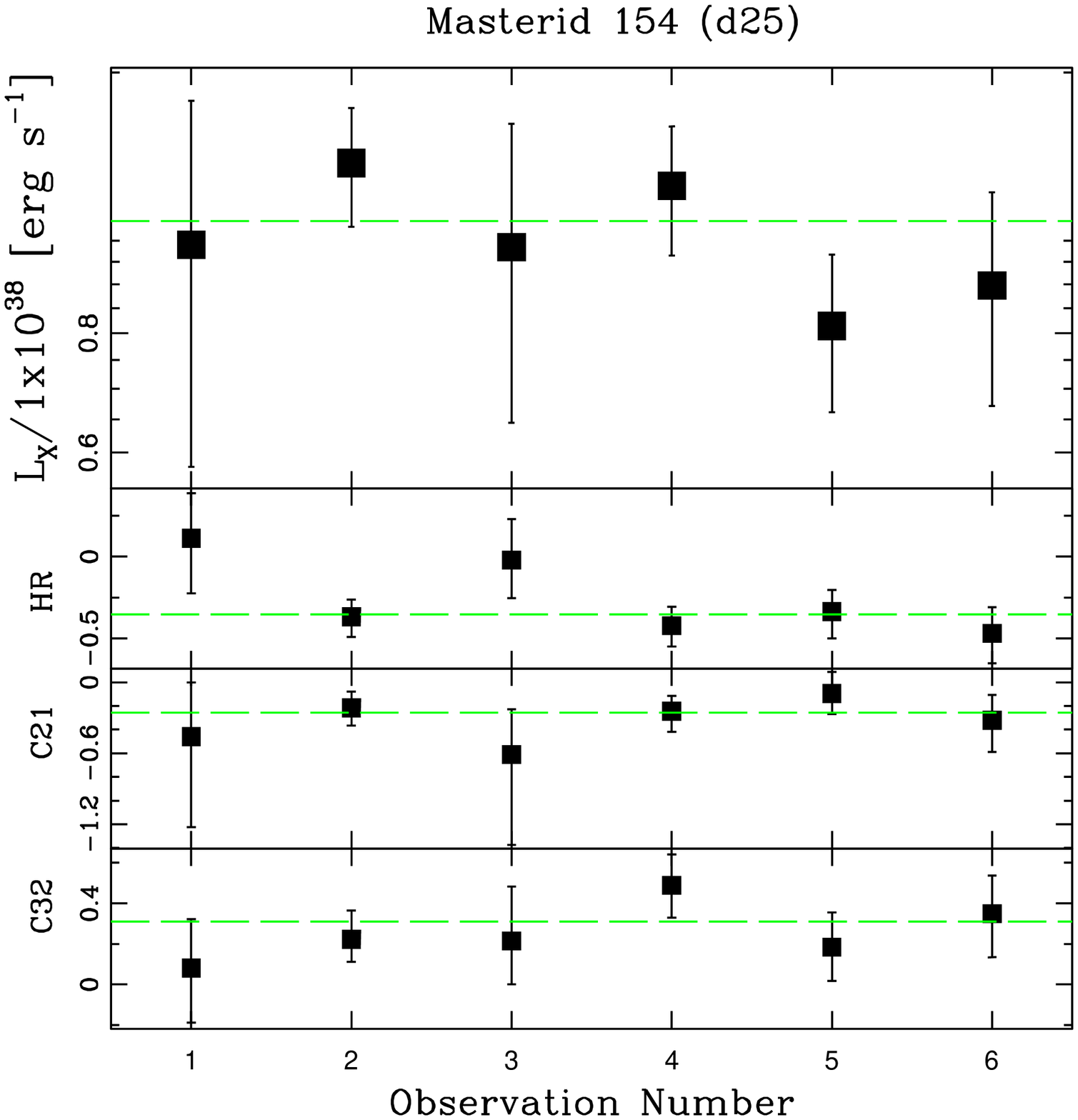}

  \end{minipage}\hspace{0.02\linewidth}
  
\end{figure}

\begin{figure}

  \begin{minipage}{0.485\linewidth}
  \centering

    \includegraphics[width=\linewidth]{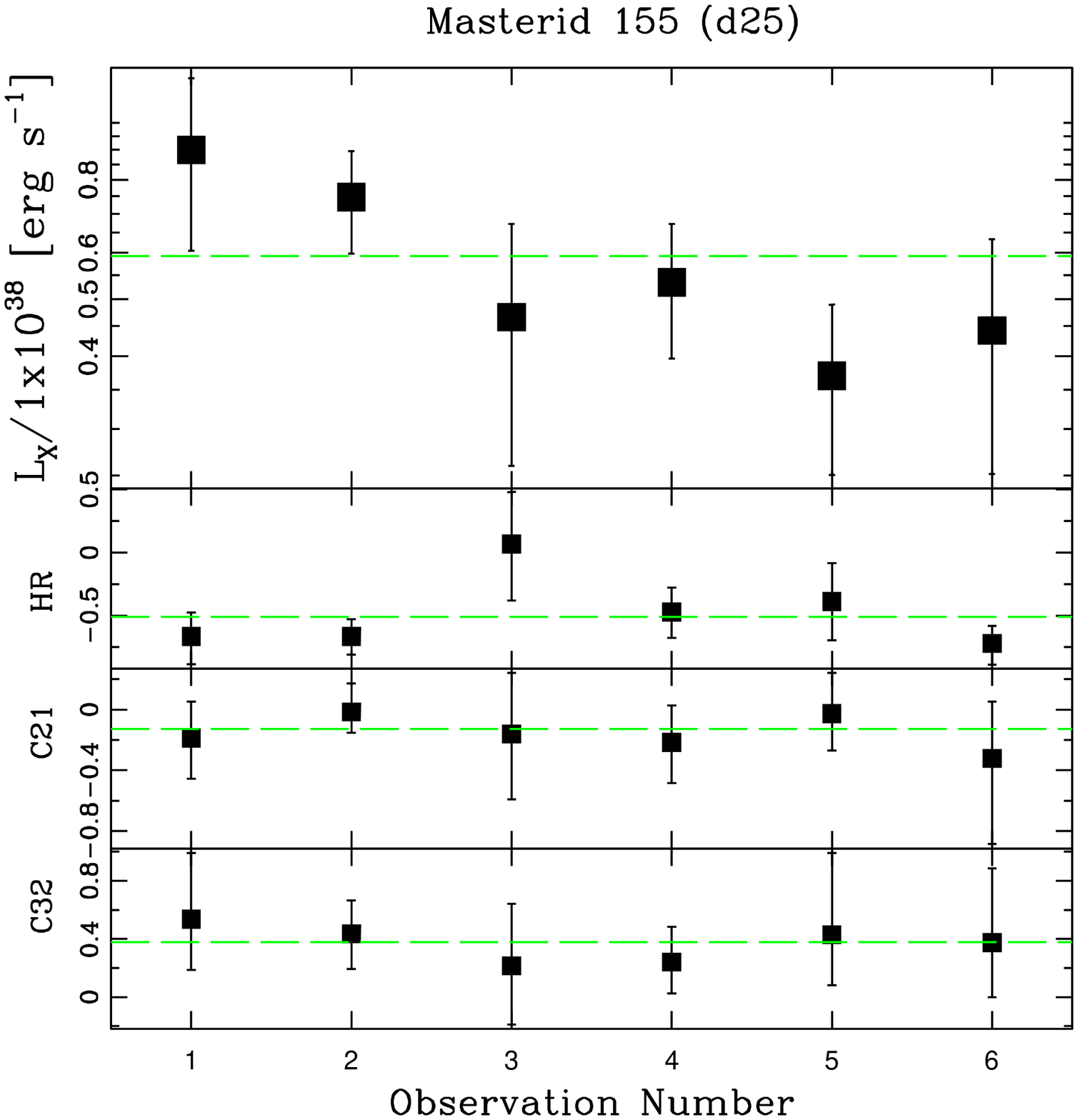}

\end{minipage}\hspace{0.02\linewidth}
\begin{minipage}{0.485\linewidth}
  \centering

    \includegraphics[width=\linewidth]{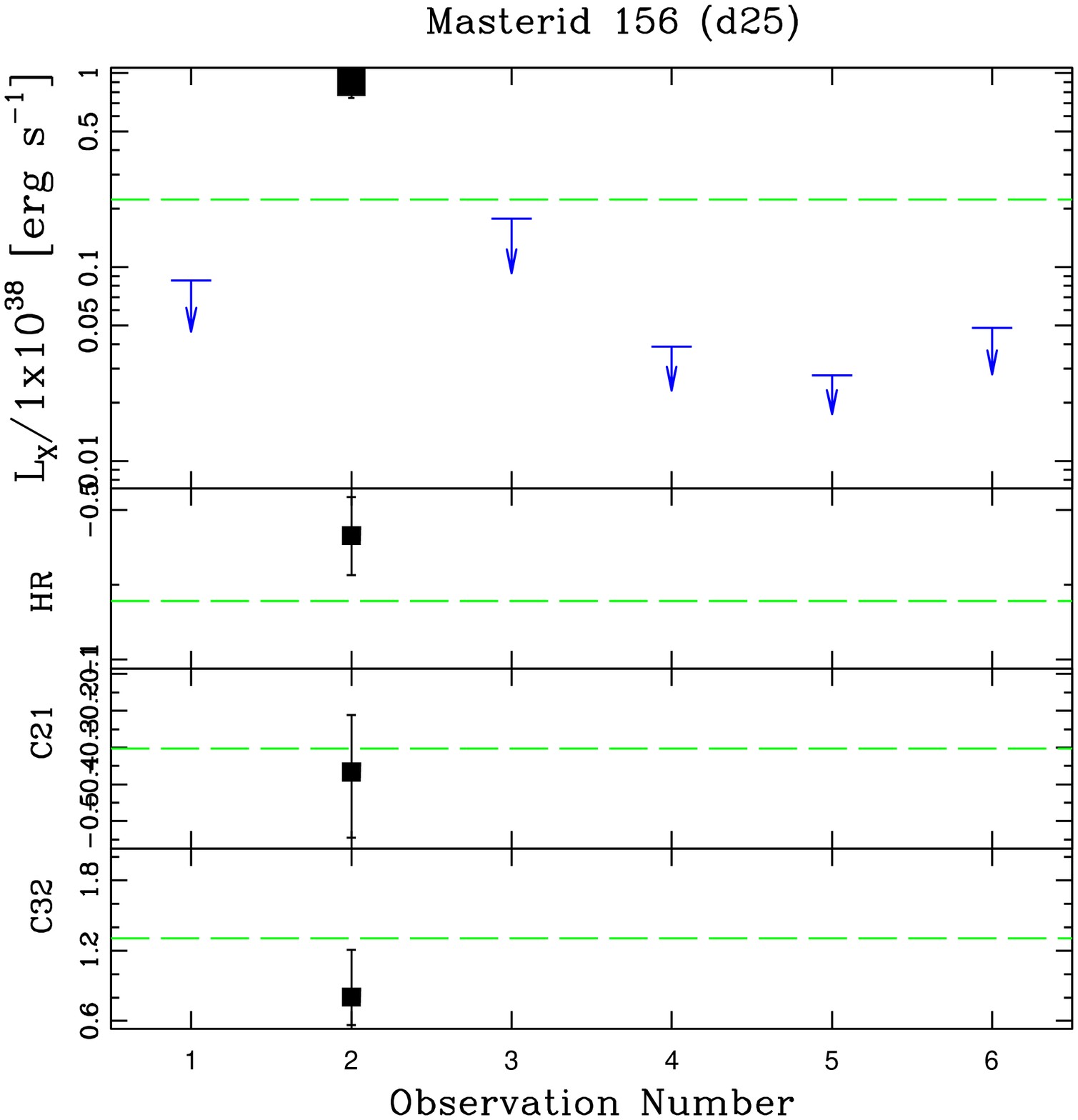}

 \end{minipage}\hspace{0.02\linewidth}

  \begin{minipage}{0.485\linewidth}
  \centering
  
    \includegraphics[width=\linewidth]{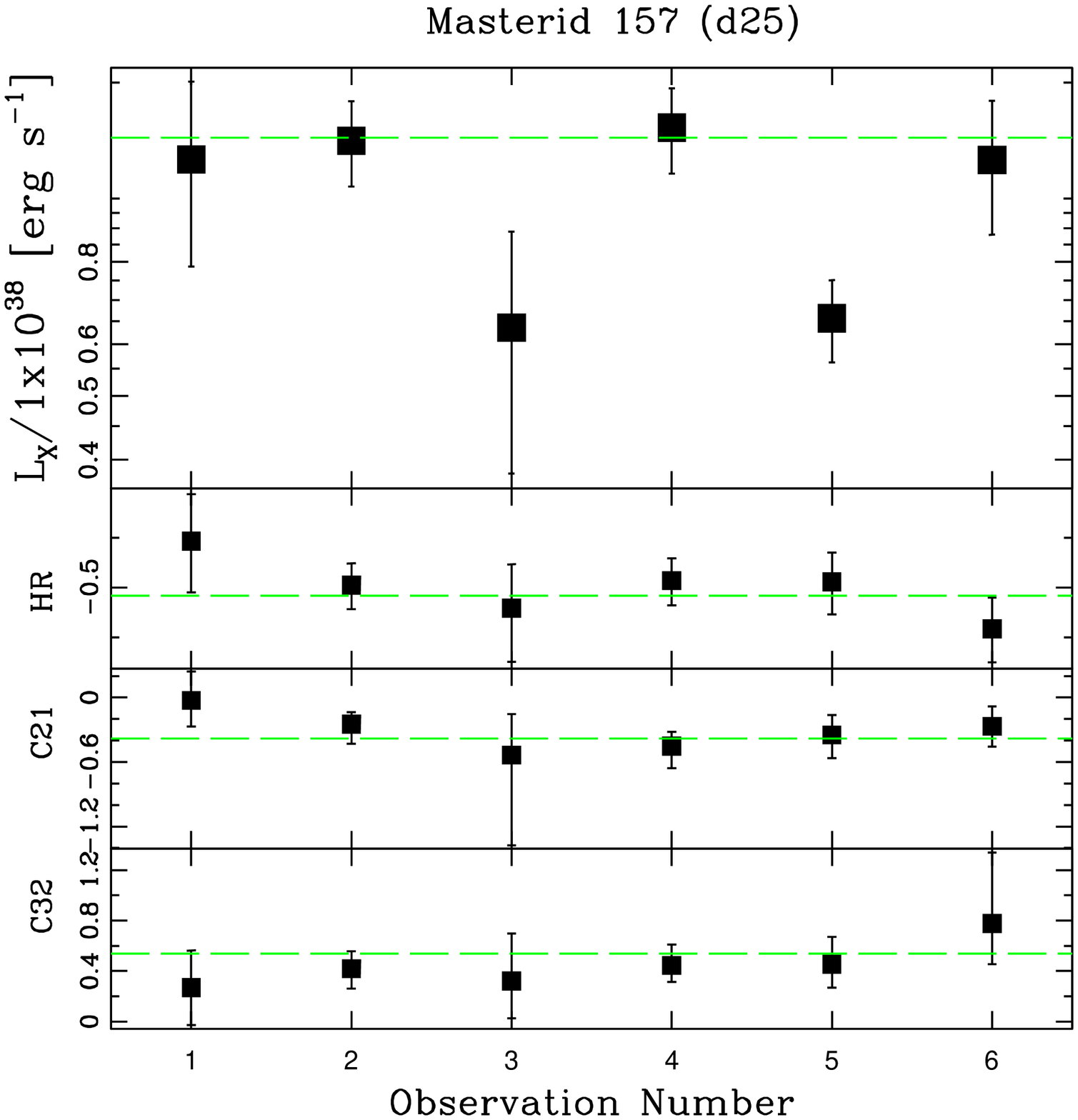}

  \end{minipage}\hspace{0.02\linewidth}
  \begin{minipage}{0.485\linewidth}
  \centering

    \includegraphics[width=\linewidth]{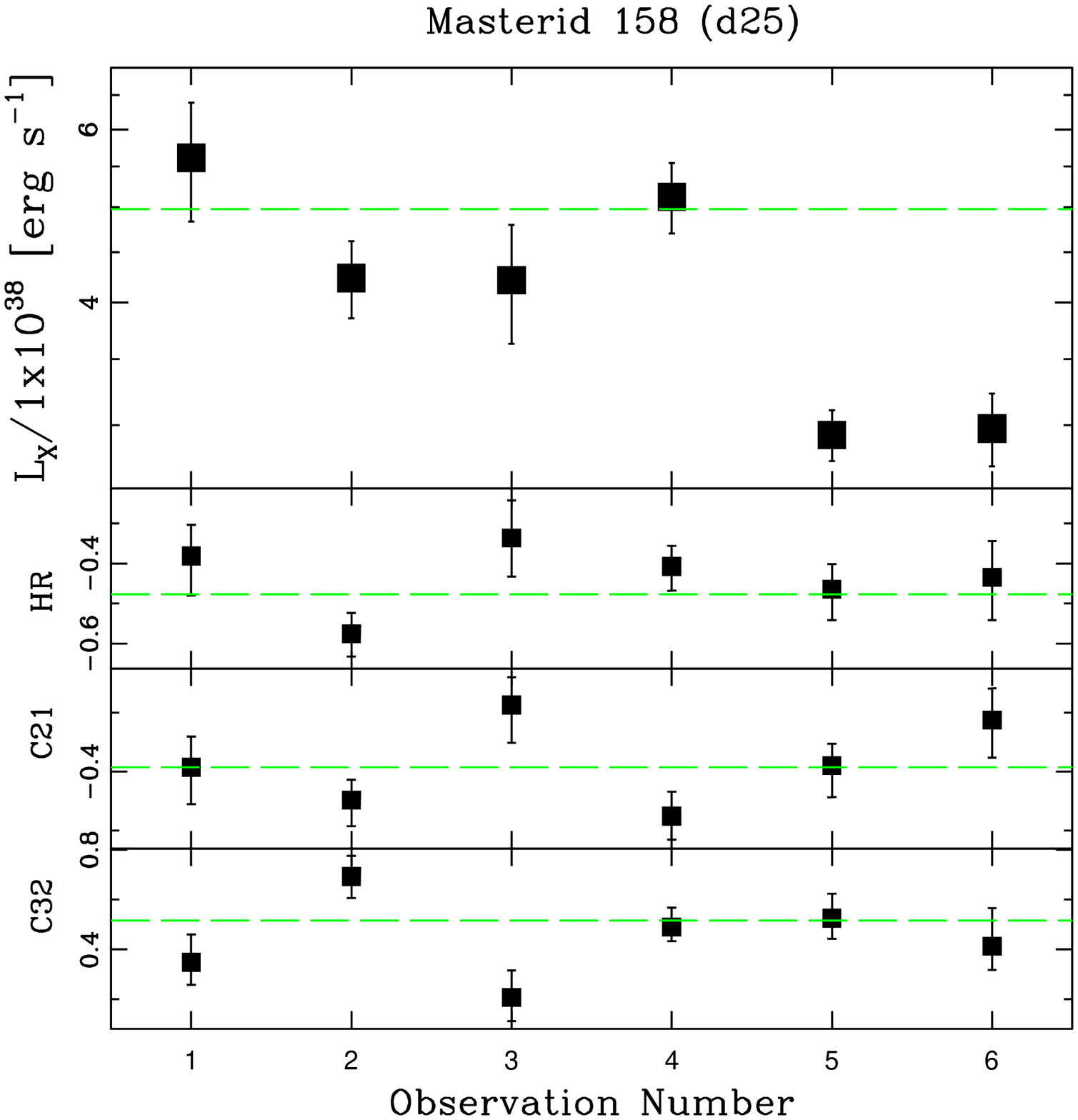}

\end{minipage}\hspace{0.02\linewidth}

\begin{minipage}{0.485\linewidth}
  \centering

    \includegraphics[width=\linewidth]{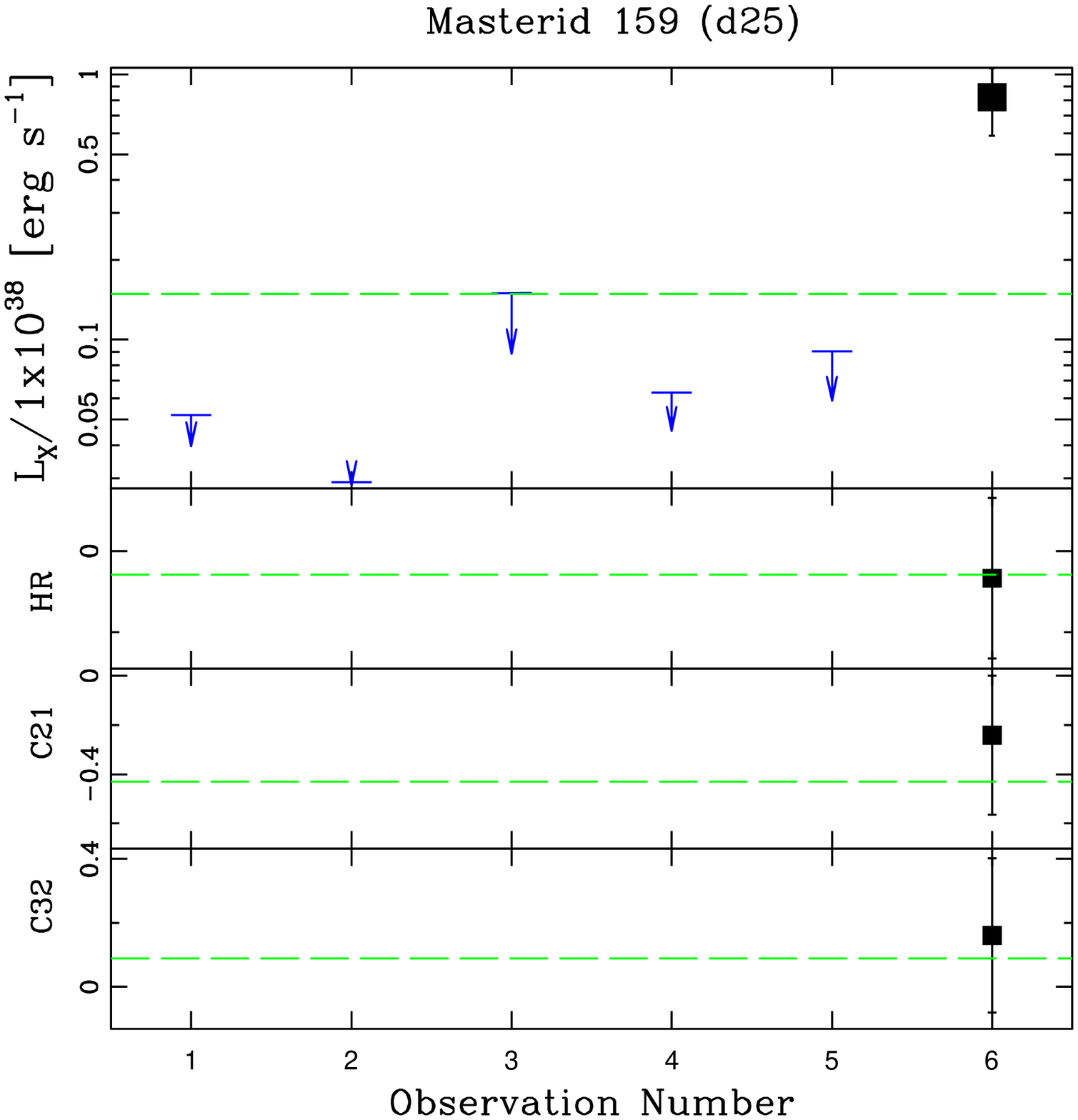}

 \end{minipage}\hspace{0.02\linewidth}
\begin{minipage}{0.485\linewidth}
  \centering
  
    \includegraphics[width=\linewidth]{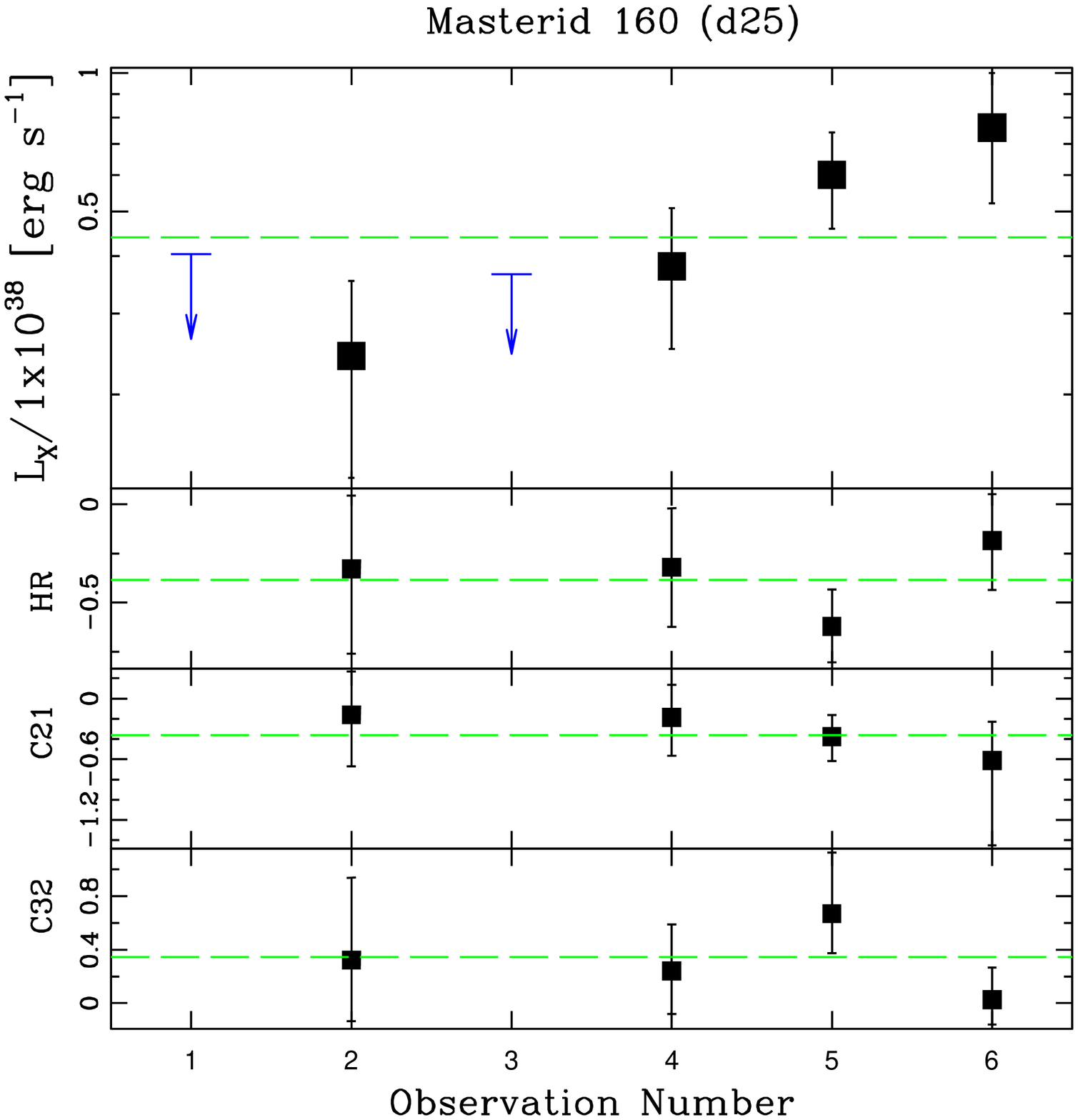}

  \end{minipage}\hspace{0.02\linewidth}
  
\end{figure}

\begin{figure}

  \begin{minipage}{0.485\linewidth}
  \centering

    \includegraphics[width=\linewidth]{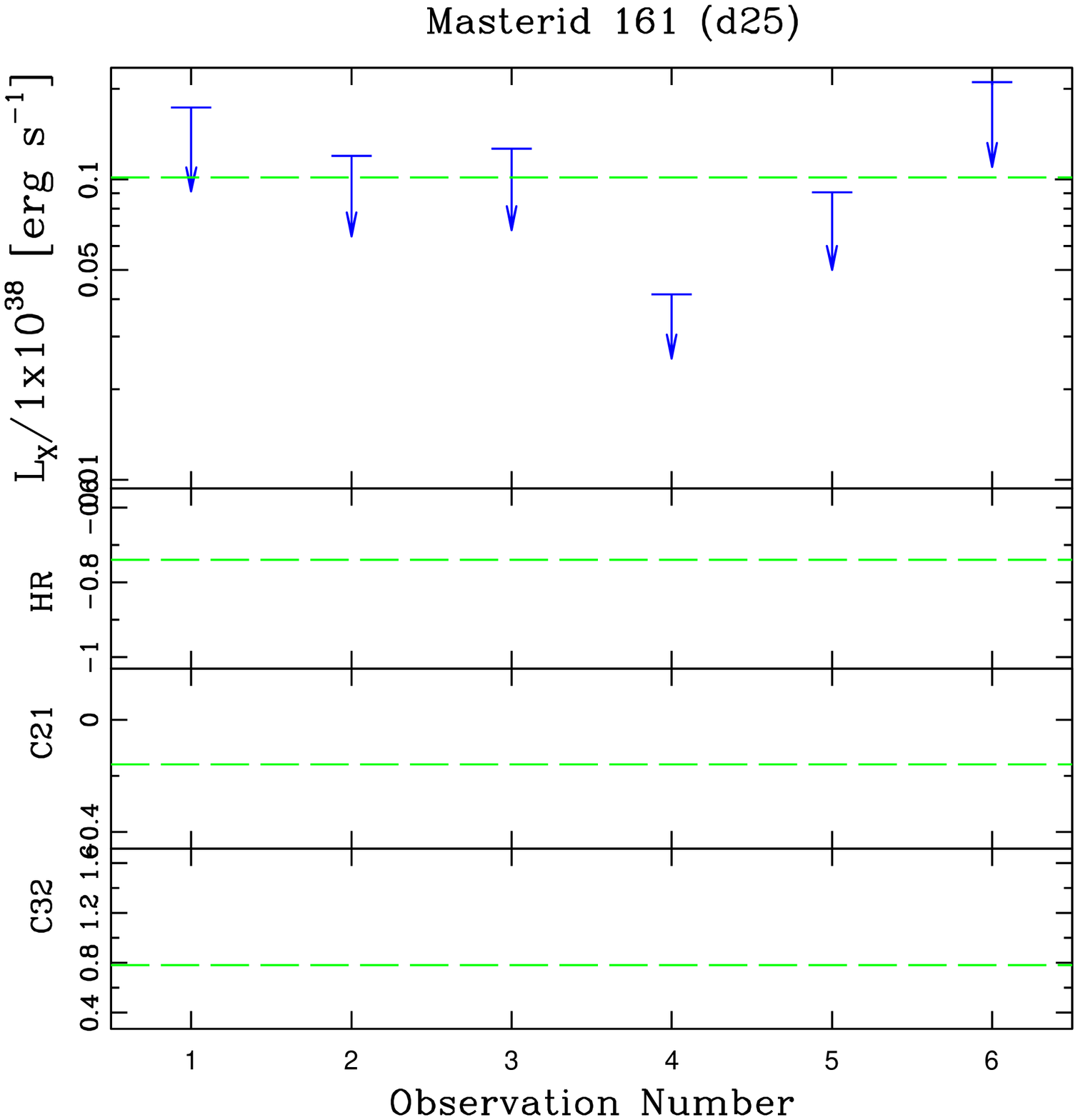}

\end{minipage}\hspace{0.02\linewidth}
\begin{minipage}{0.485\linewidth}
  \centering

    \includegraphics[width=\linewidth]{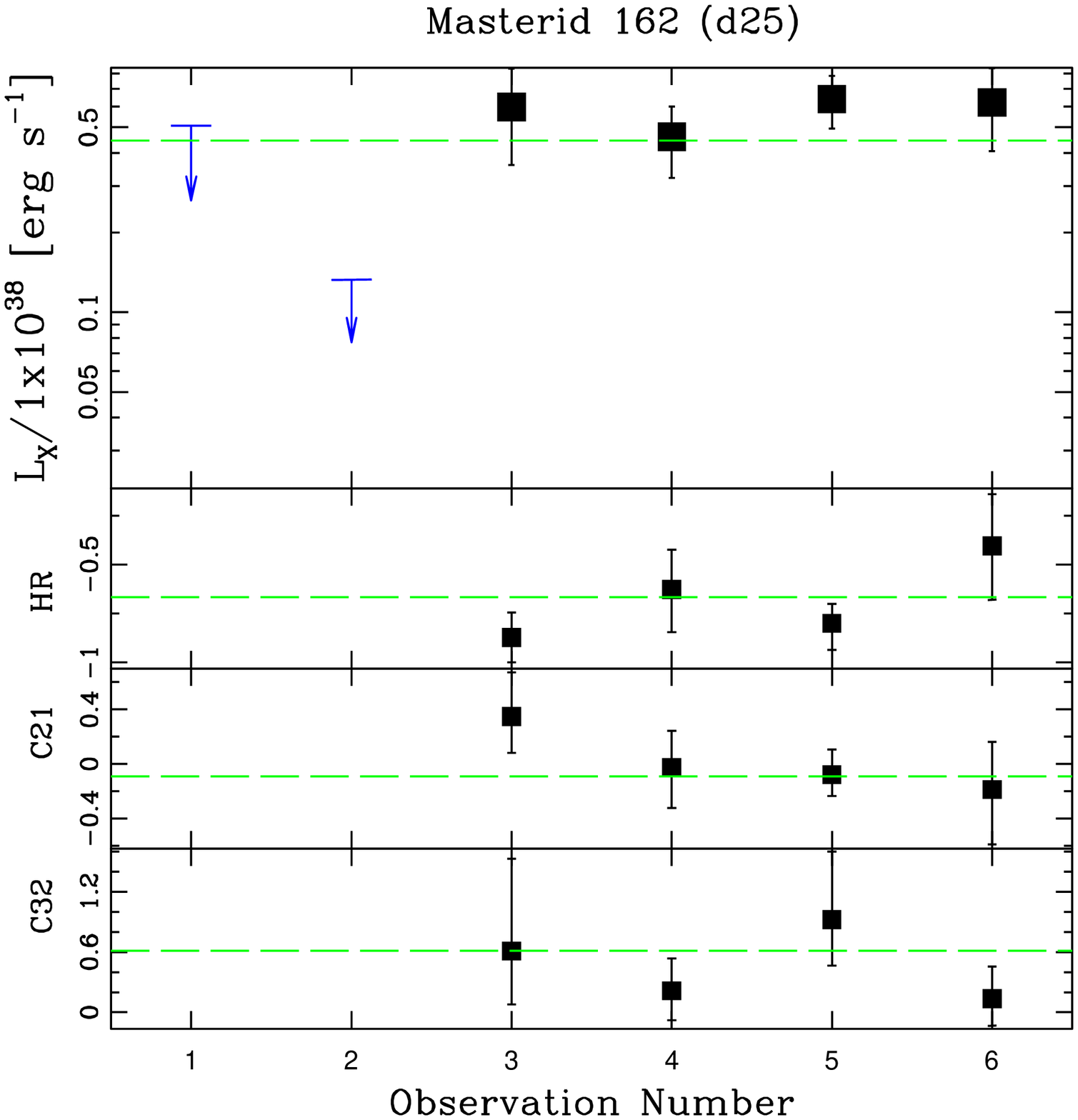}

 \end{minipage}\hspace{0.02\linewidth}

  \begin{minipage}{0.485\linewidth}
  \centering
  
    \includegraphics[width=\linewidth]{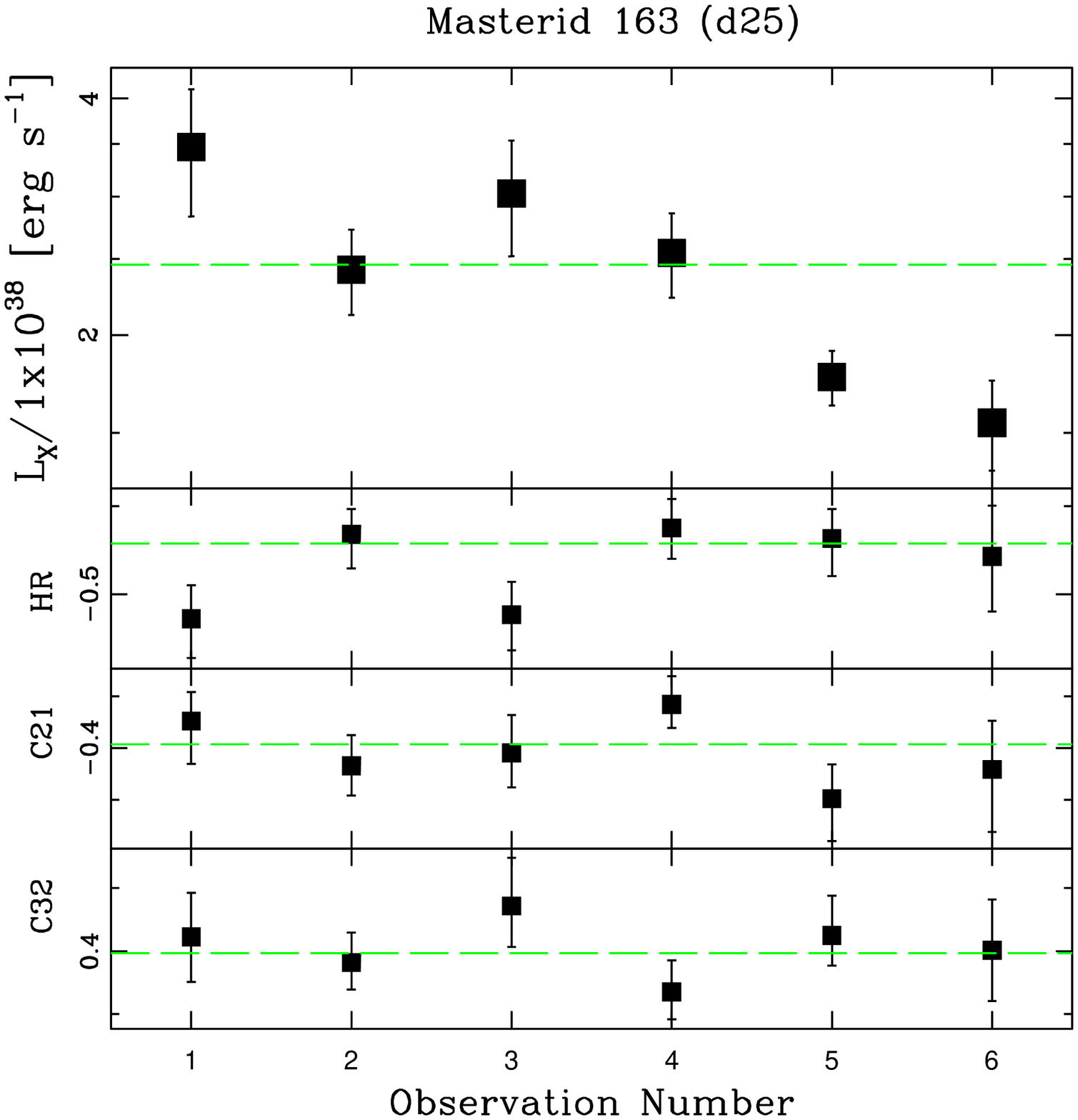}

  \end{minipage}\hspace{0.02\linewidth}
  \begin{minipage}{0.485\linewidth}
  \centering

    \includegraphics[width=\linewidth]{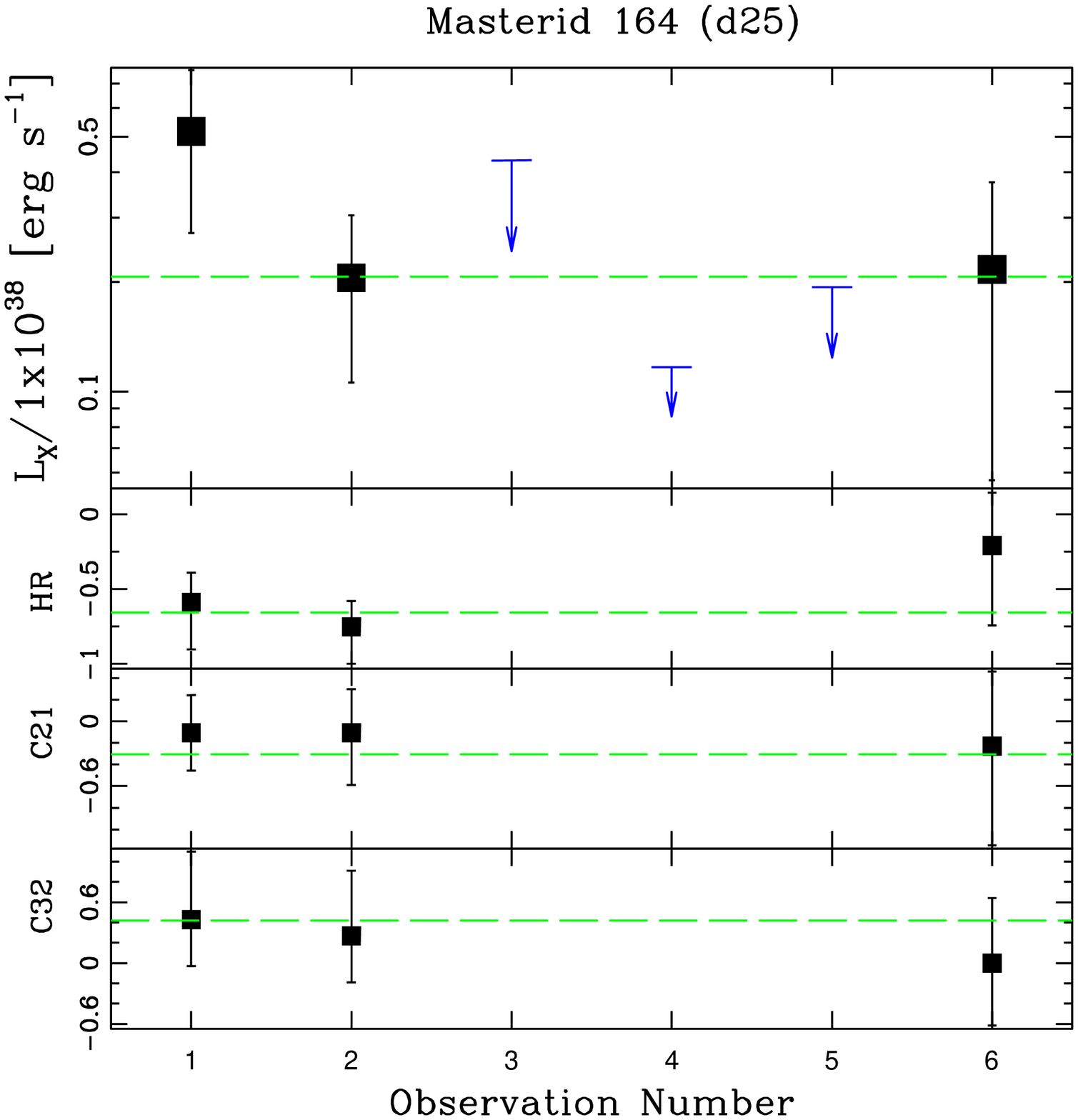}

\end{minipage}\hspace{0.02\linewidth}

\begin{minipage}{0.485\linewidth}
  \centering

    \includegraphics[width=\linewidth]{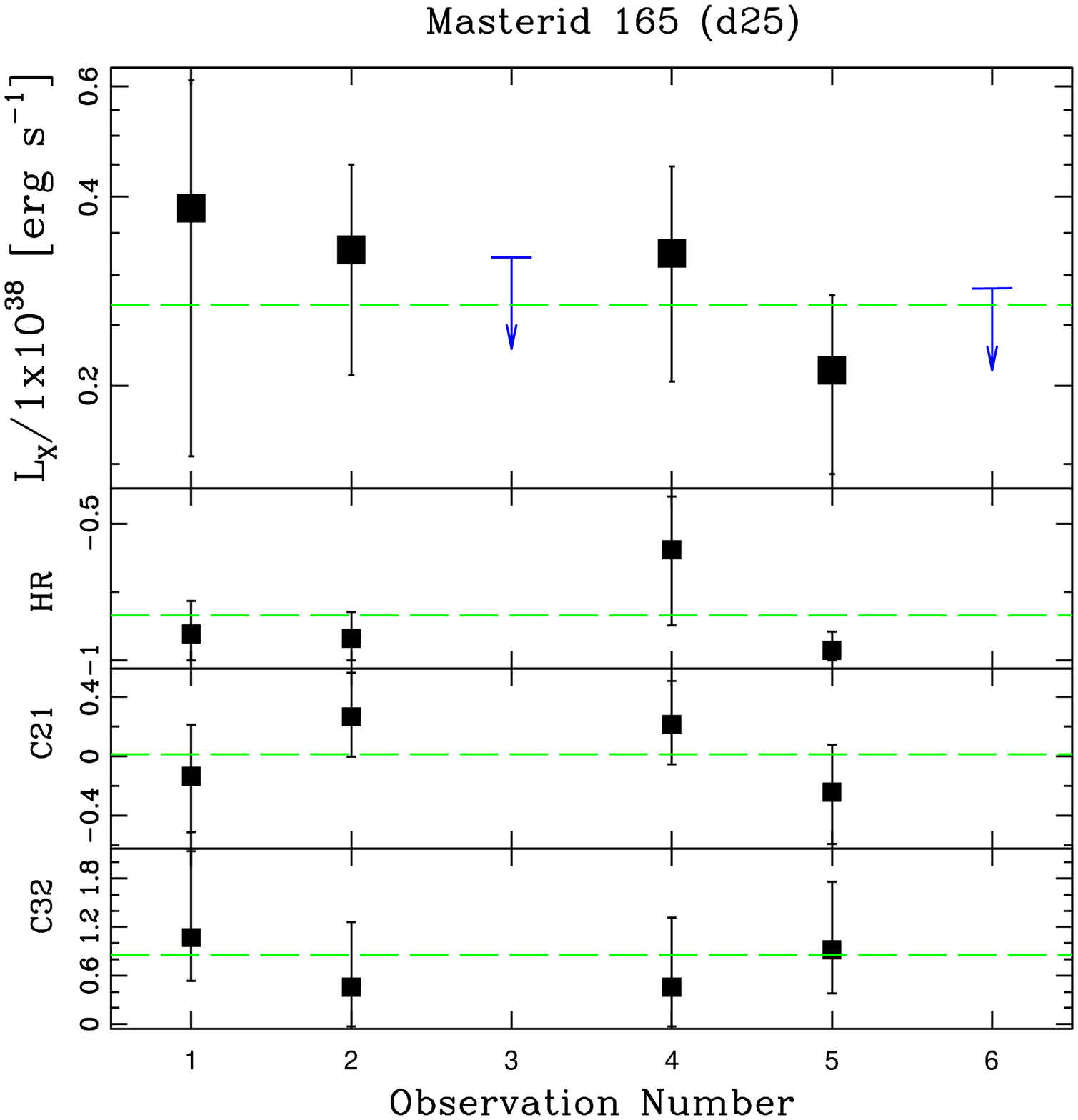}

 \end{minipage}\hspace{0.02\linewidth}
\begin{minipage}{0.485\linewidth}
  \centering
  
    \includegraphics[width=\linewidth]{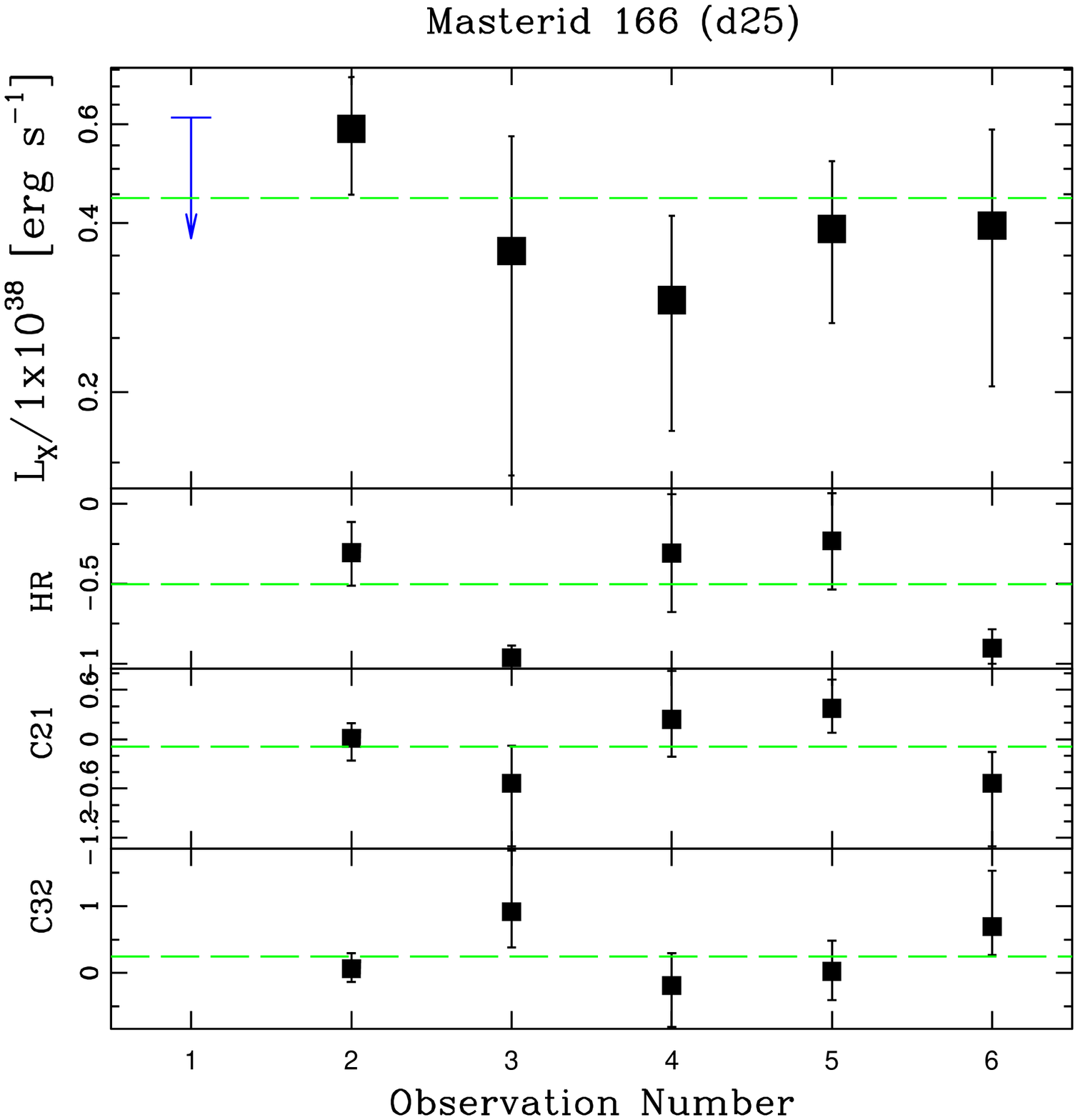}

  \end{minipage}\hspace{0.02\linewidth}
  
\end{figure}

\begin{figure}

  \begin{minipage}{0.485\linewidth}
  \centering

    \includegraphics[width=\linewidth]{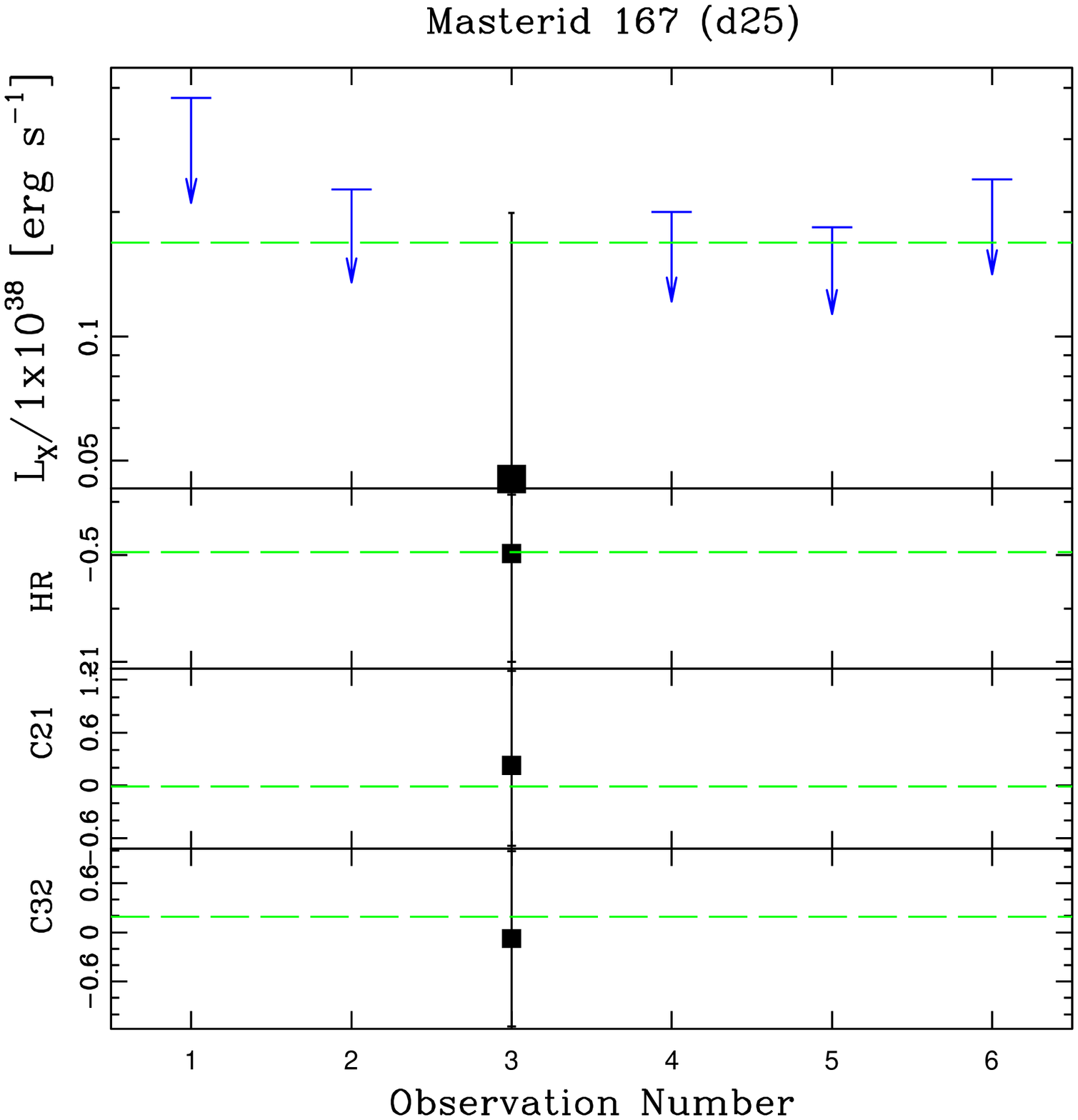}

\end{minipage}\hspace{0.02\linewidth}
\begin{minipage}{0.485\linewidth}
  \centering

    \includegraphics[width=\linewidth]{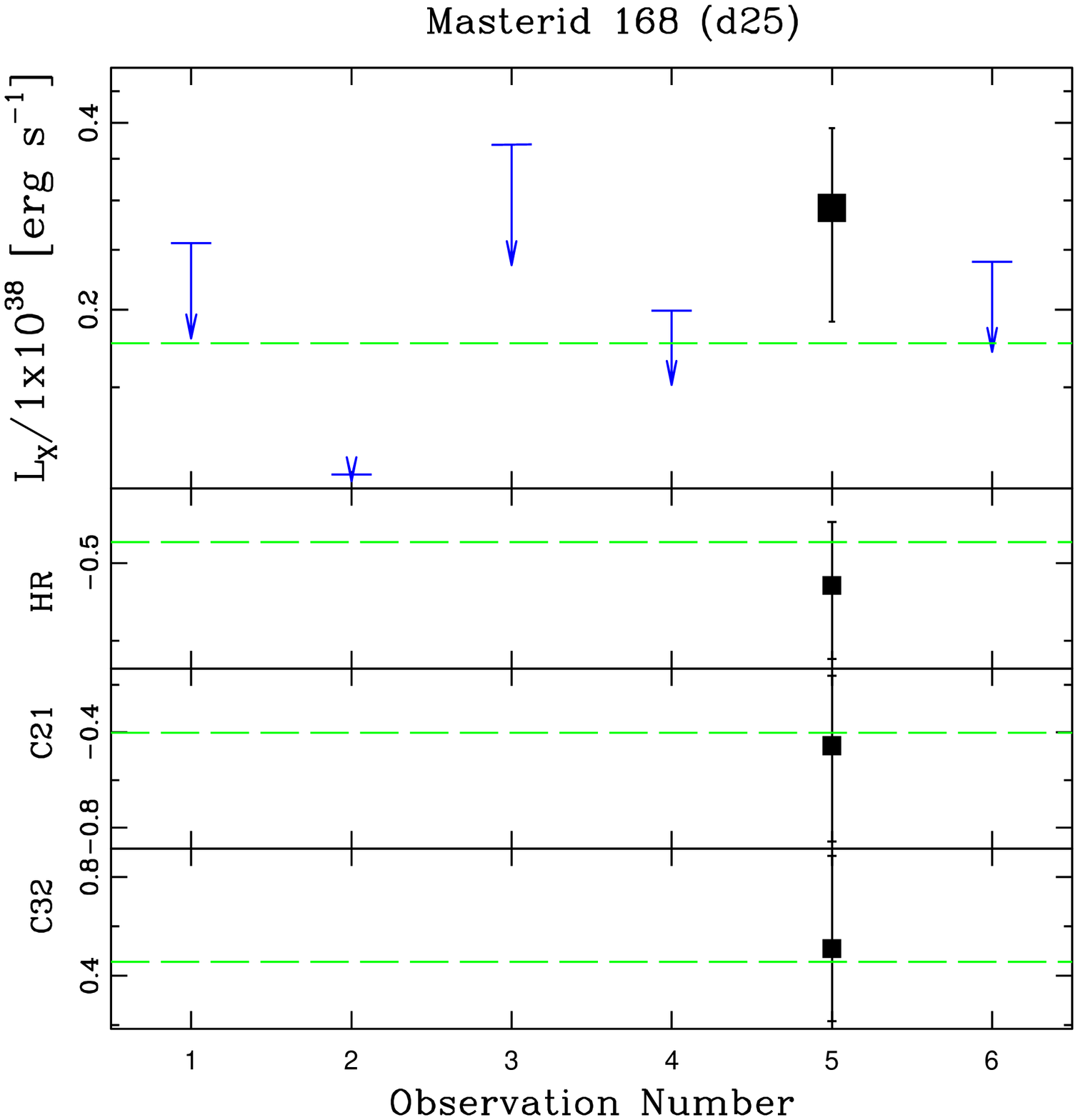}

 \end{minipage}\hspace{0.02\linewidth}

  \begin{minipage}{0.485\linewidth}
  \centering
  
    \includegraphics[width=\linewidth]{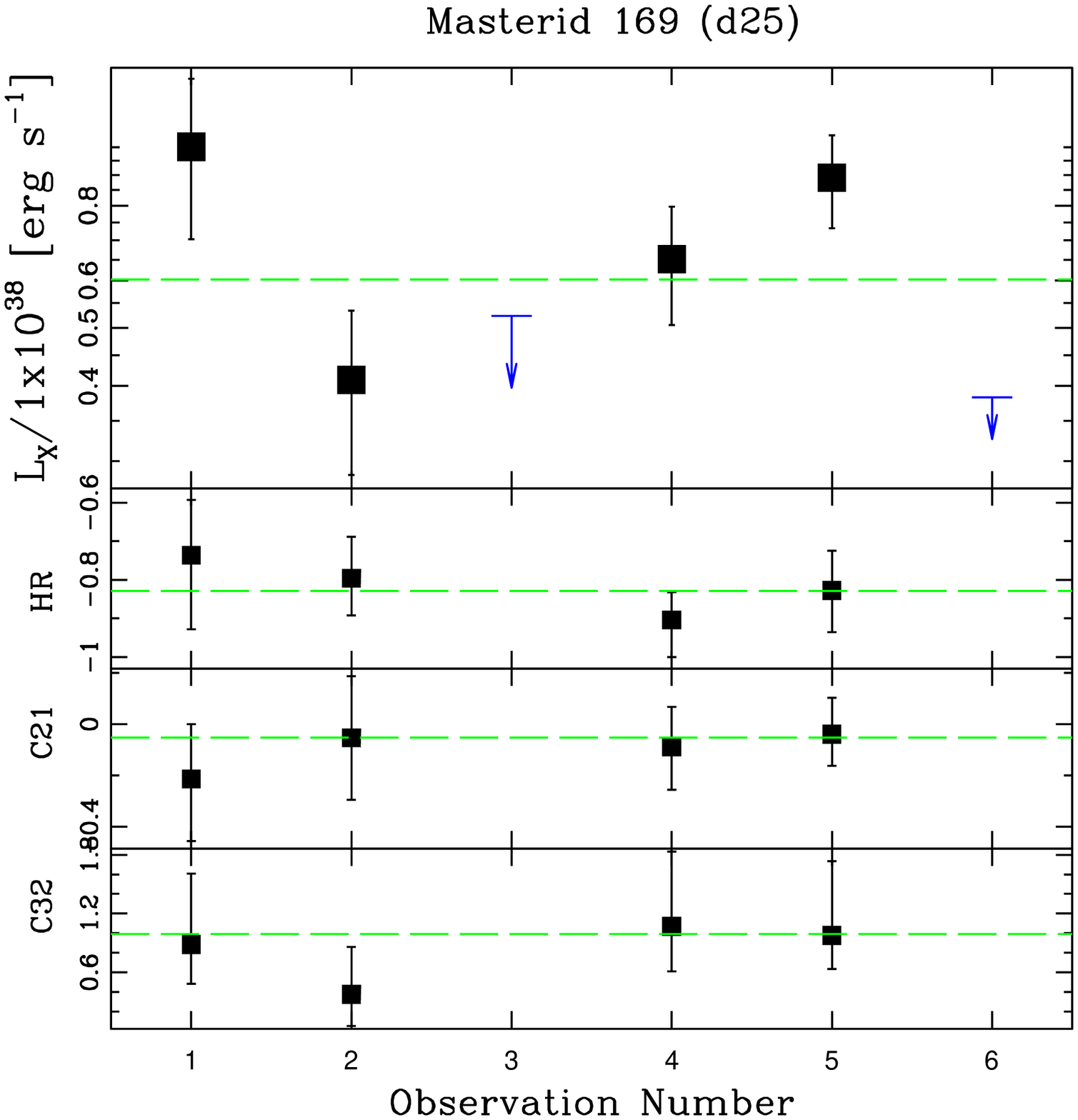}

  \end{minipage}\hspace{0.02\linewidth}
  \begin{minipage}{0.485\linewidth}
  \centering

    \includegraphics[width=\linewidth]{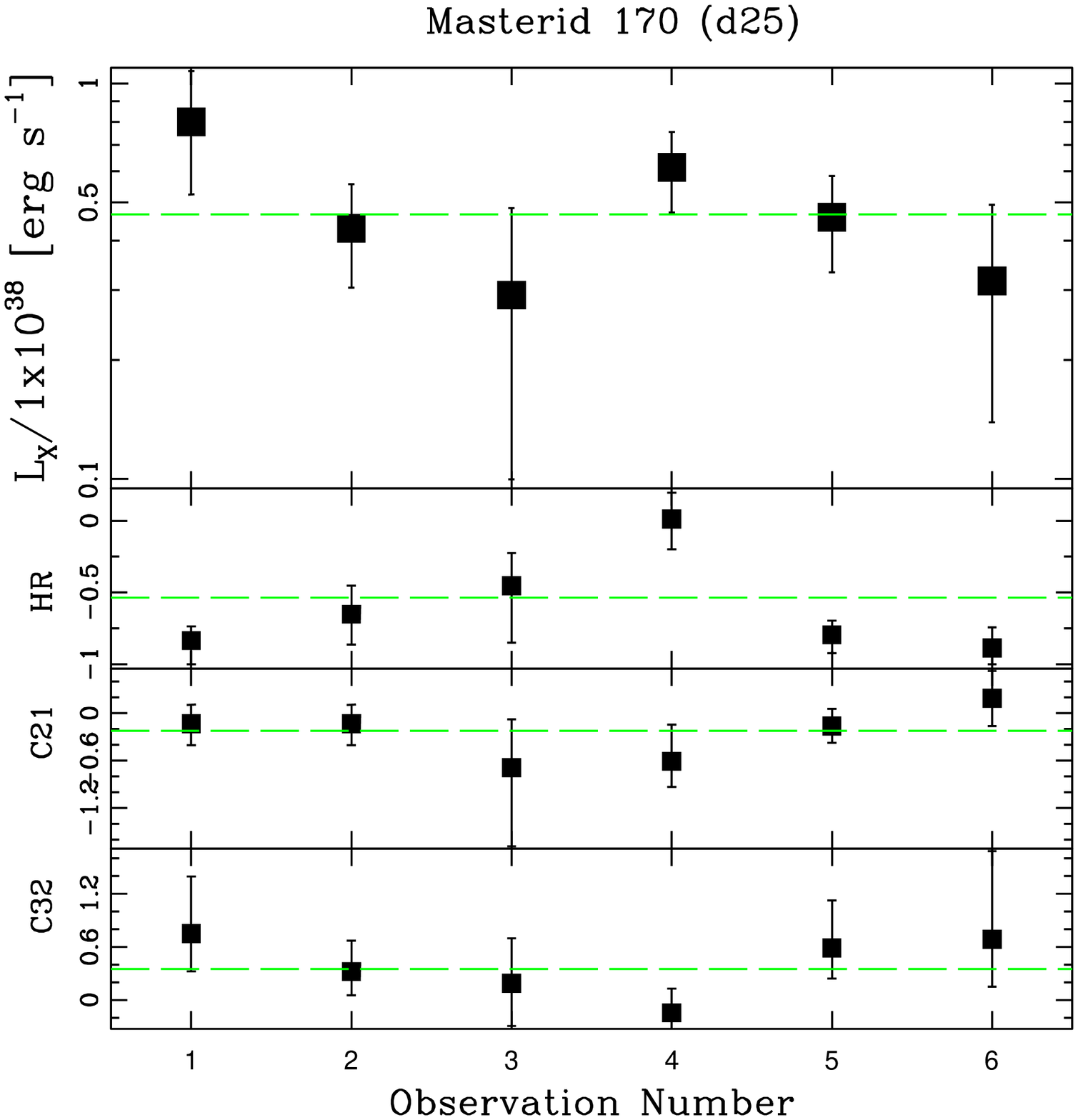}

\end{minipage}\hspace{0.02\linewidth}

\begin{minipage}{0.485\linewidth}
  \centering

    \includegraphics[width=\linewidth]{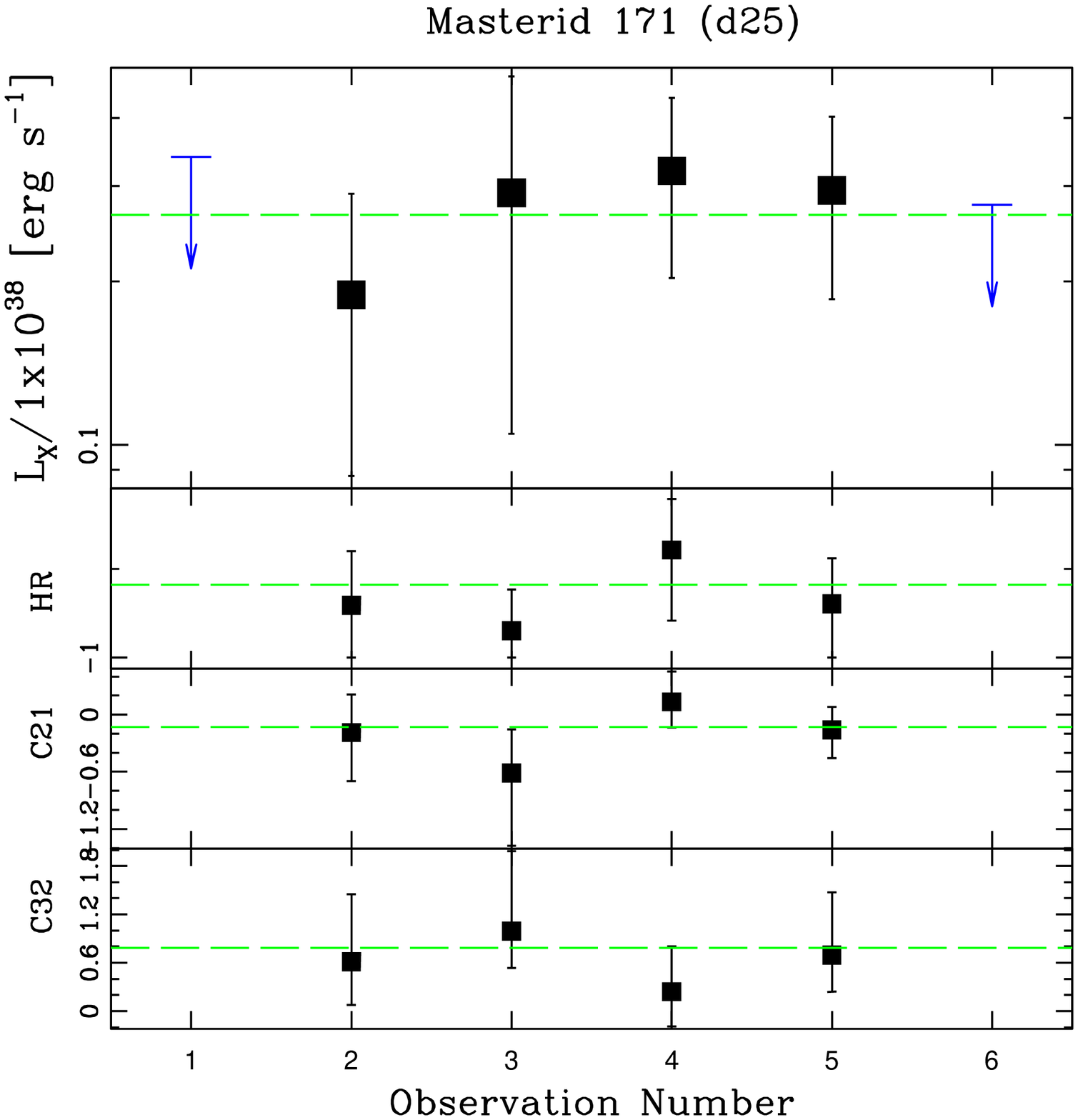}

 \end{minipage}\hspace{0.02\linewidth}
\begin{minipage}{0.485\linewidth}
  \centering
  
    \includegraphics[width=\linewidth]{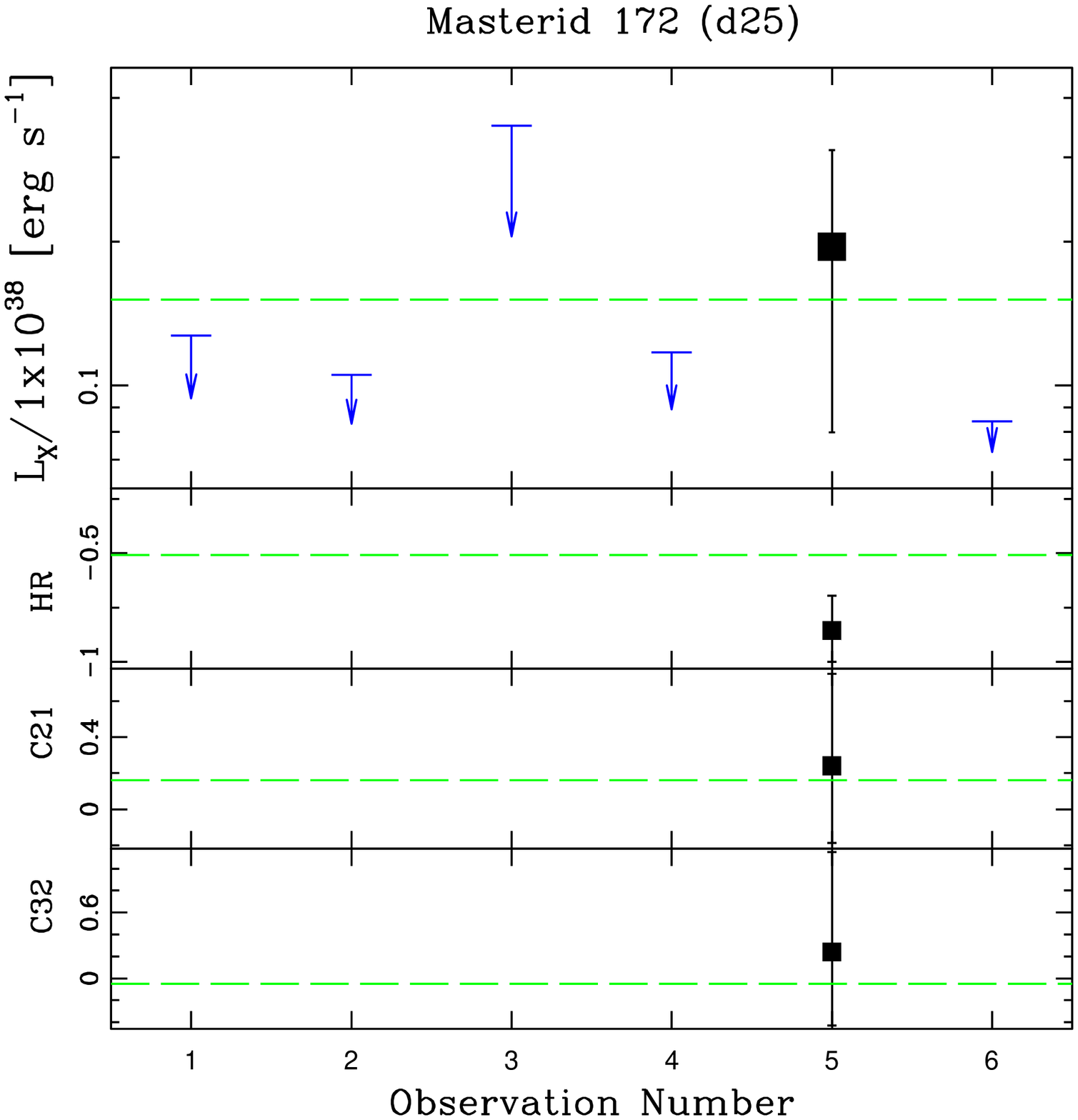}

  \end{minipage}\hspace{0.02\linewidth}
  
\end{figure}

\begin{figure}

  \begin{minipage}{0.485\linewidth}
  \centering

    \includegraphics[width=\linewidth]{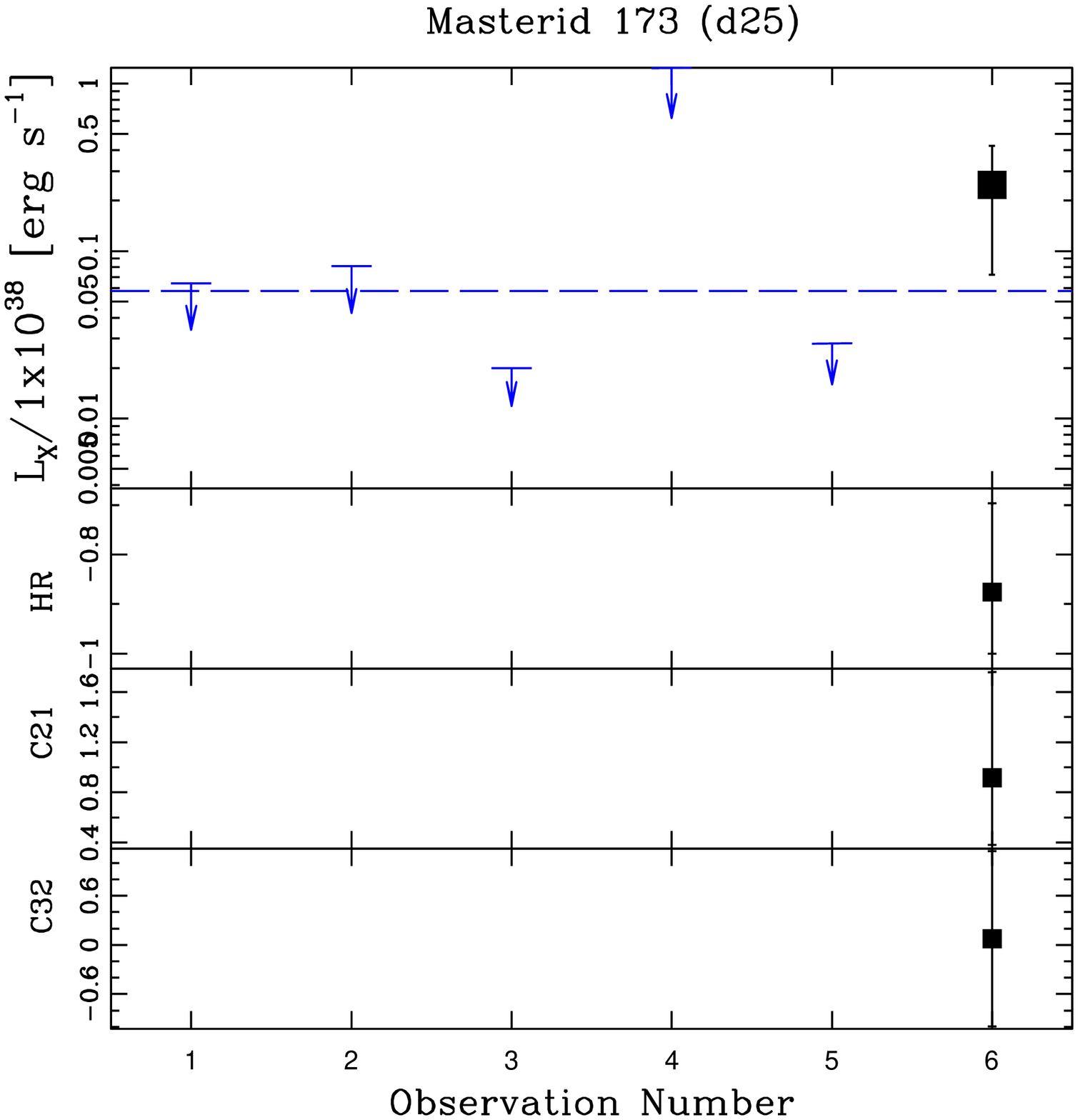}

\end{minipage}\hspace{0.02\linewidth}
\begin{minipage}{0.485\linewidth}
  \centering

    \includegraphics[width=\linewidth]{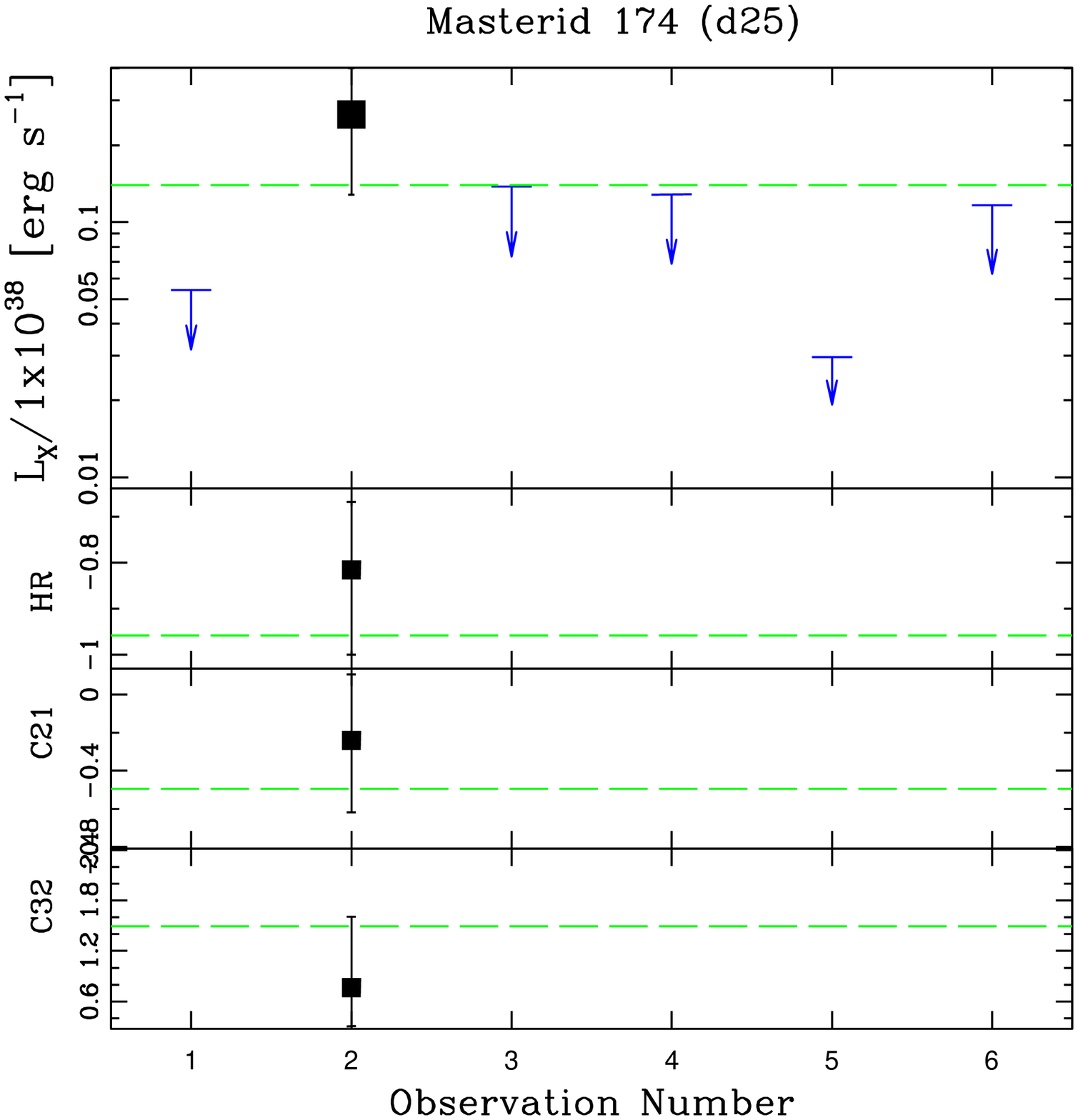}

 \end{minipage}\hspace{0.02\linewidth}

  \begin{minipage}{0.485\linewidth}
  \centering
  
    \includegraphics[width=\linewidth]{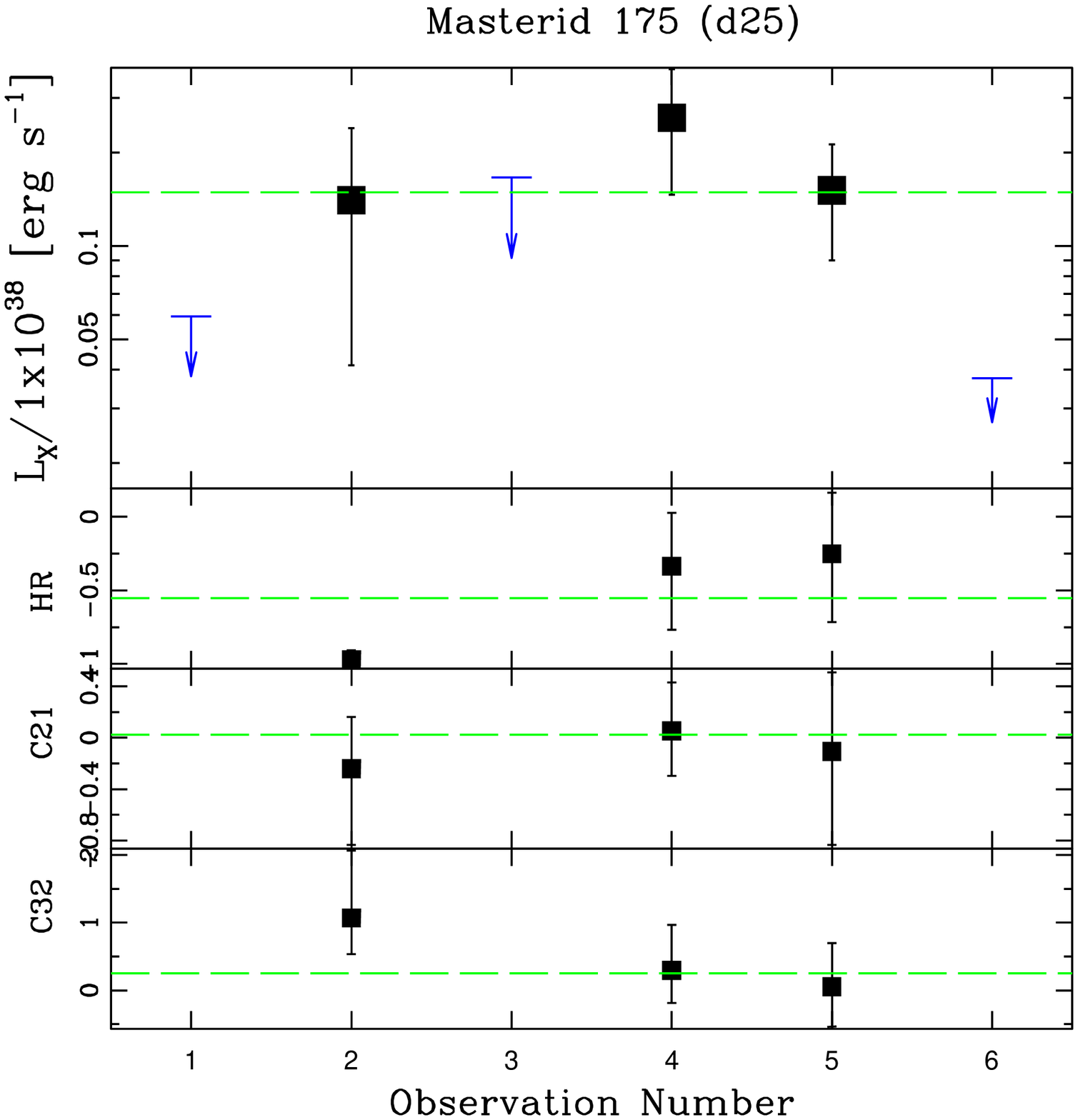}

  \end{minipage}\hspace{0.02\linewidth}
  \begin{minipage}{0.485\linewidth}
  \centering

    \includegraphics[width=\linewidth]{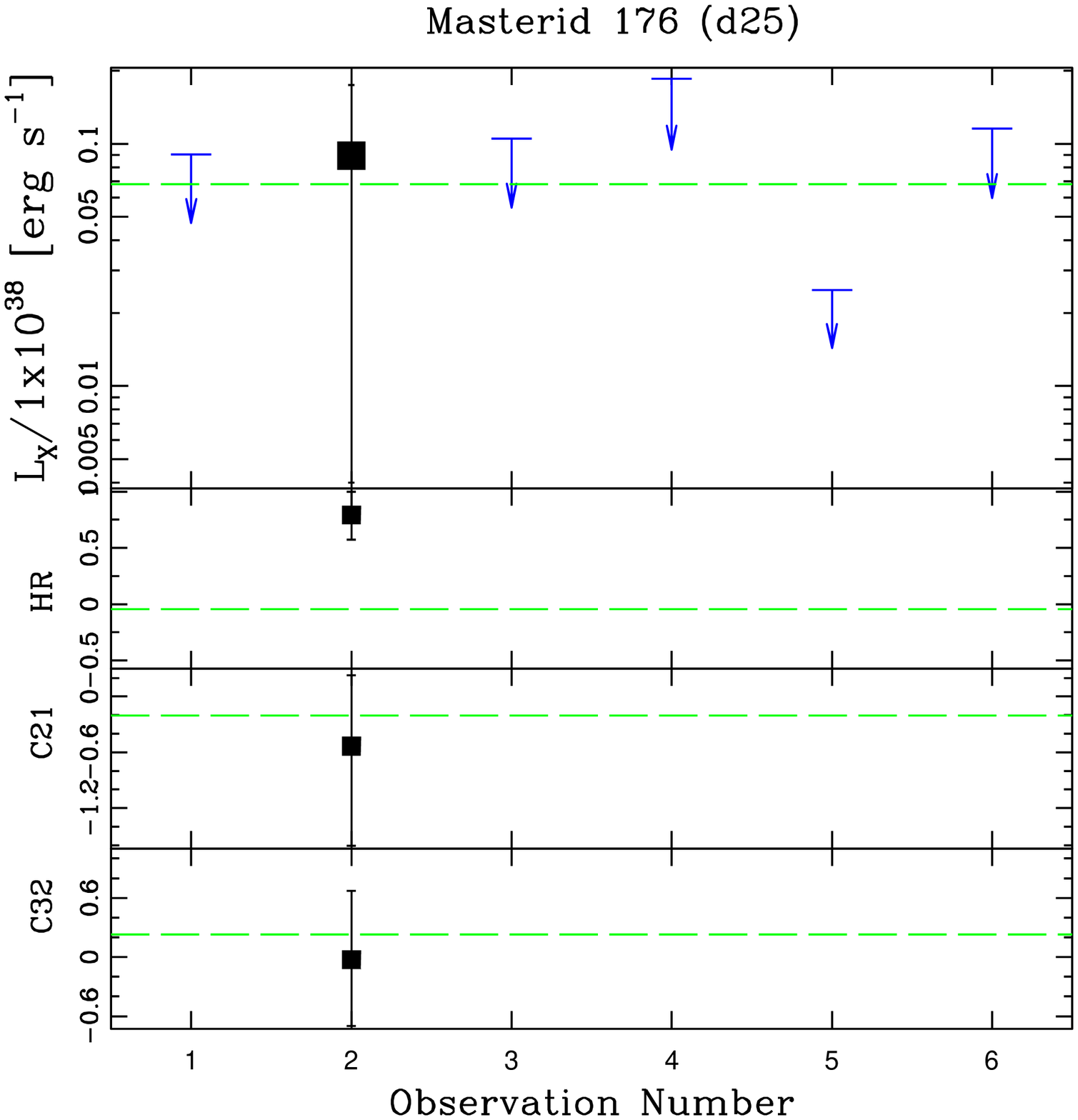}

\end{minipage}\hspace{0.02\linewidth}

\begin{minipage}{0.485\linewidth}
  \centering

    \includegraphics[width=\linewidth]{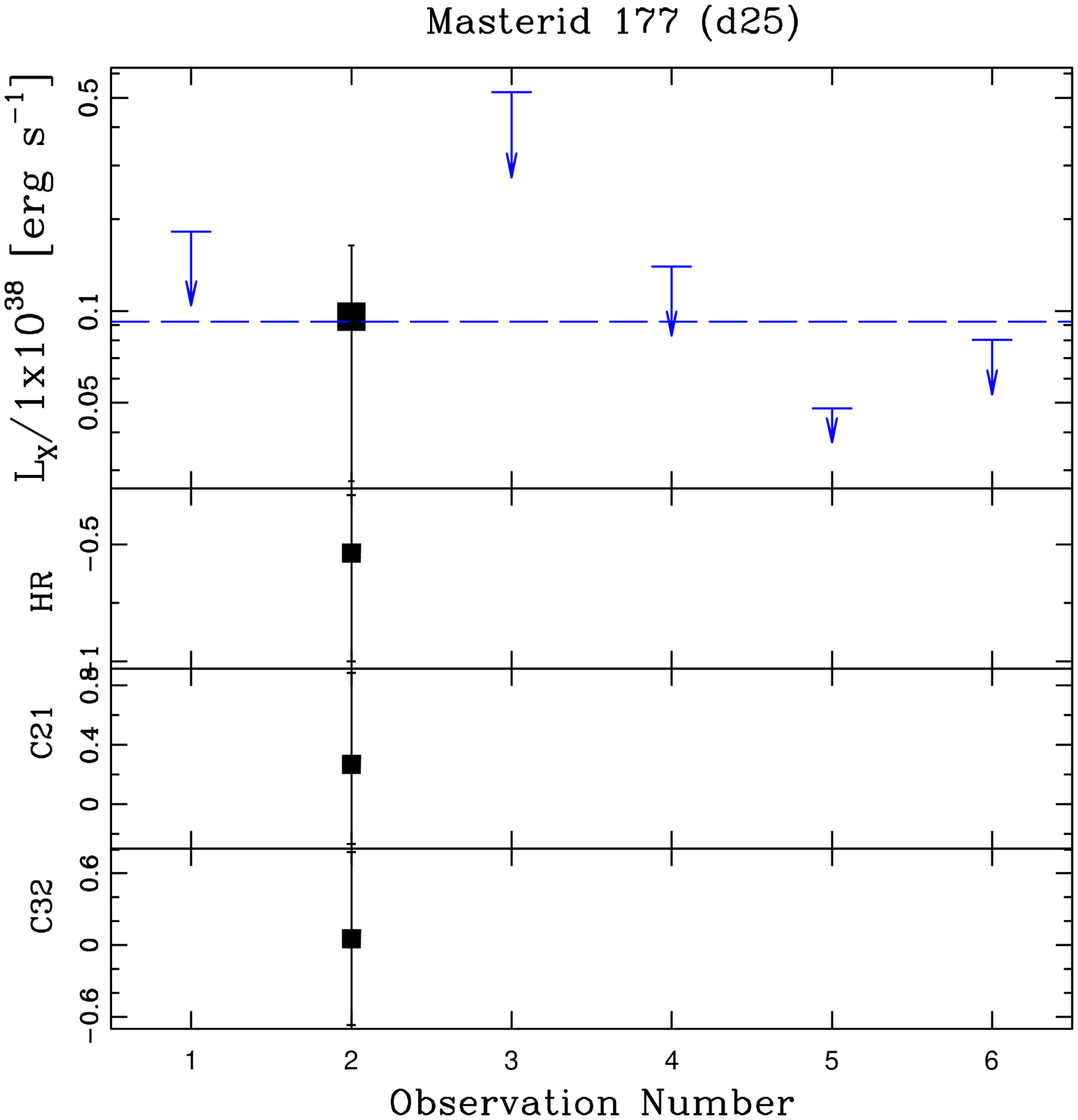}

 \end{minipage}\hspace{0.02\linewidth}
\begin{minipage}{0.485\linewidth}
  \centering
  
    \includegraphics[width=\linewidth]{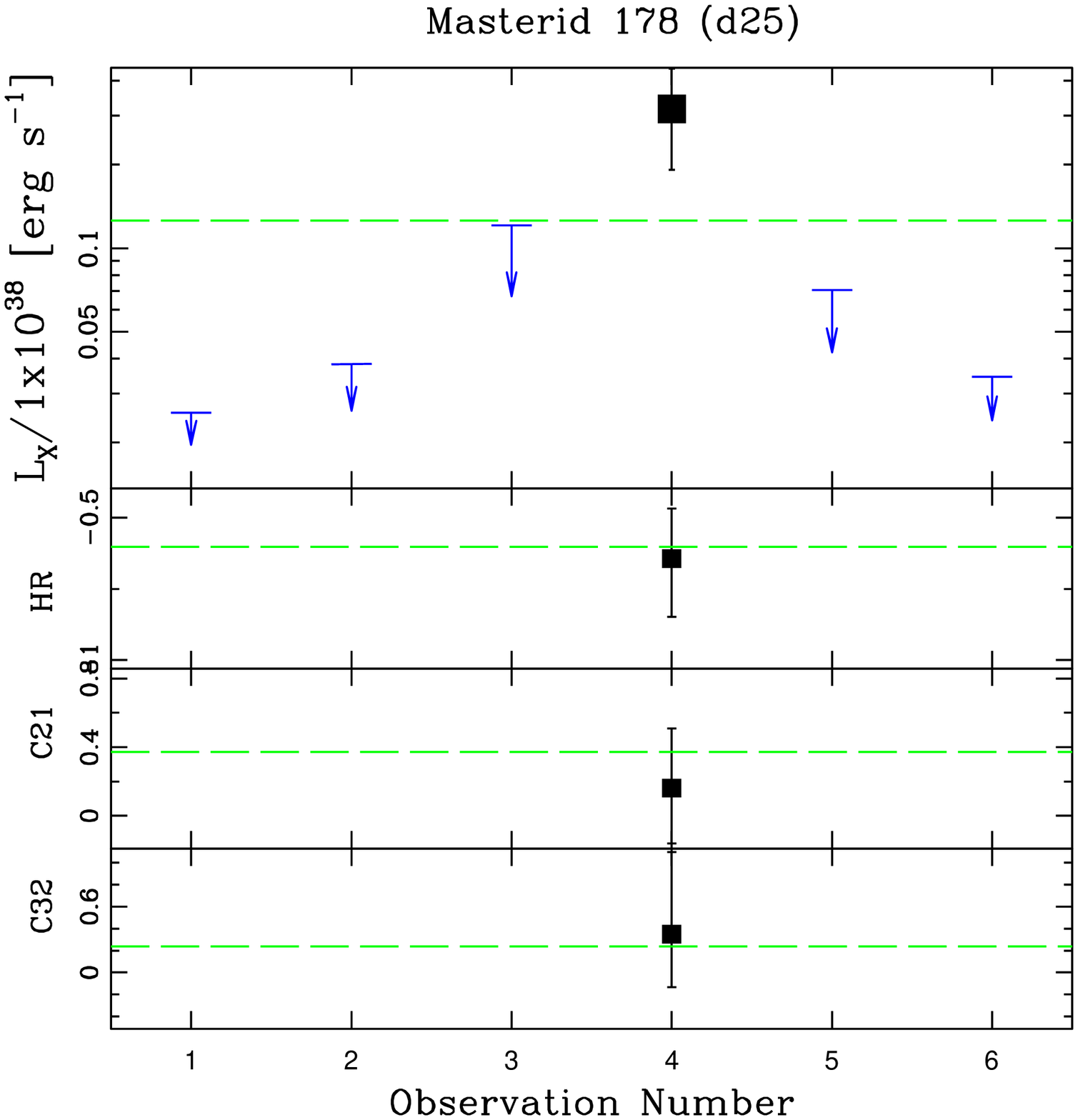}

  \end{minipage}\hspace{0.02\linewidth}
  
\end{figure}

\begin{figure}

  \begin{minipage}{0.485\linewidth}
  \centering

    \includegraphics[width=\linewidth]{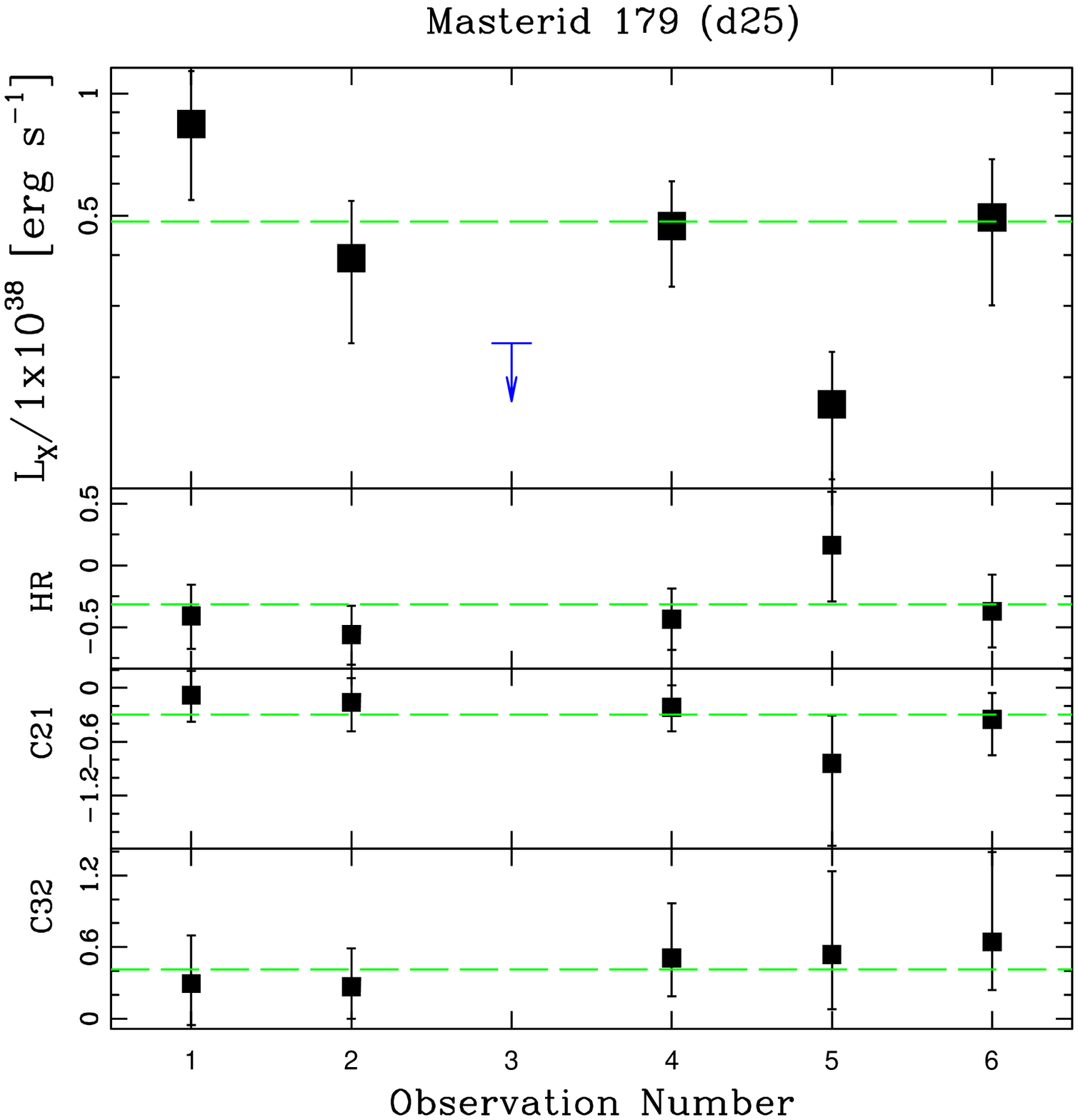}

\end{minipage}\hspace{0.02\linewidth}
\begin{minipage}{0.485\linewidth}
  \centering

    \includegraphics[width=\linewidth]{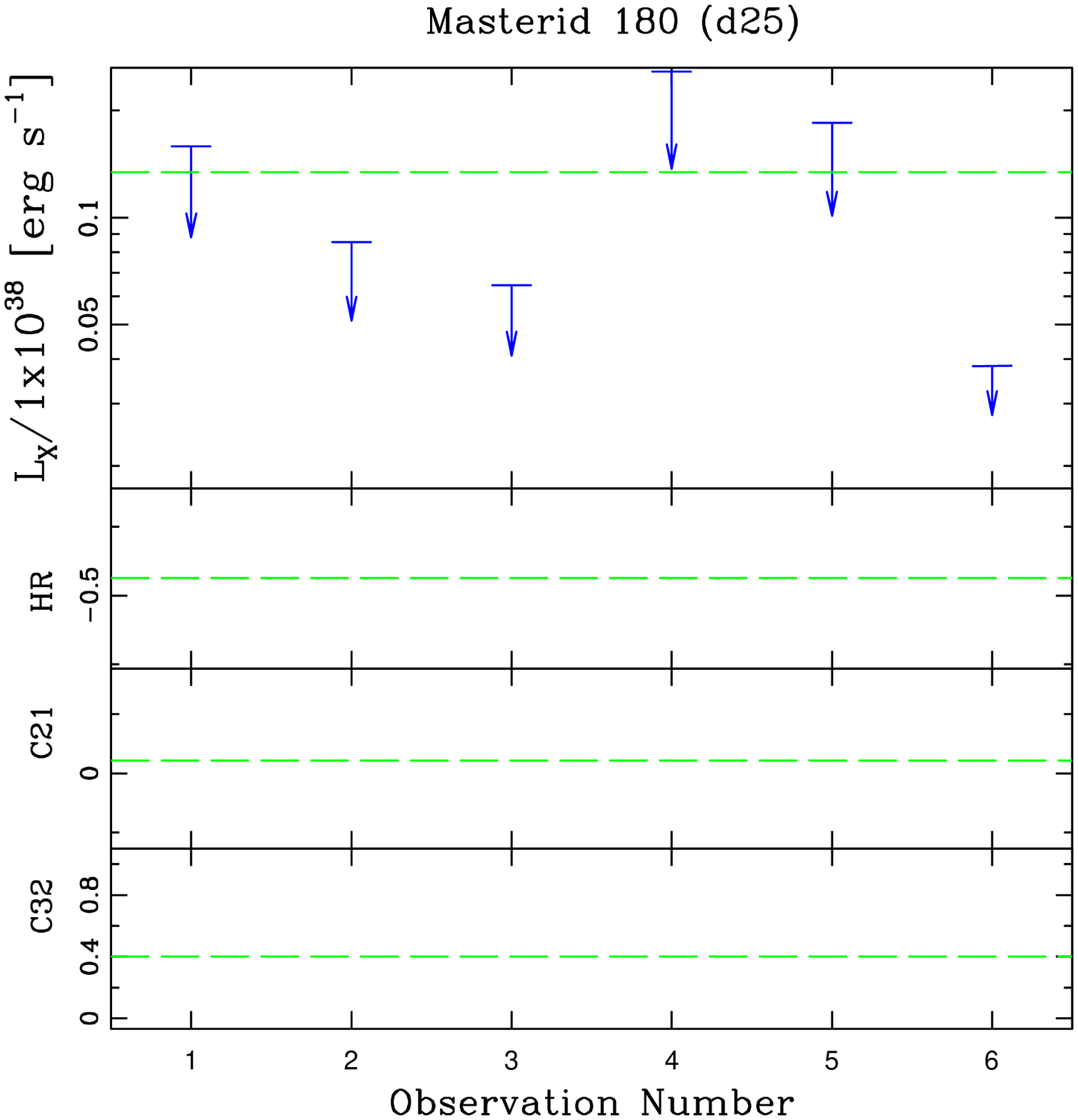}

 \end{minipage}\hspace{0.02\linewidth}

  \begin{minipage}{0.485\linewidth}
  \centering
  
    \includegraphics[width=\linewidth]{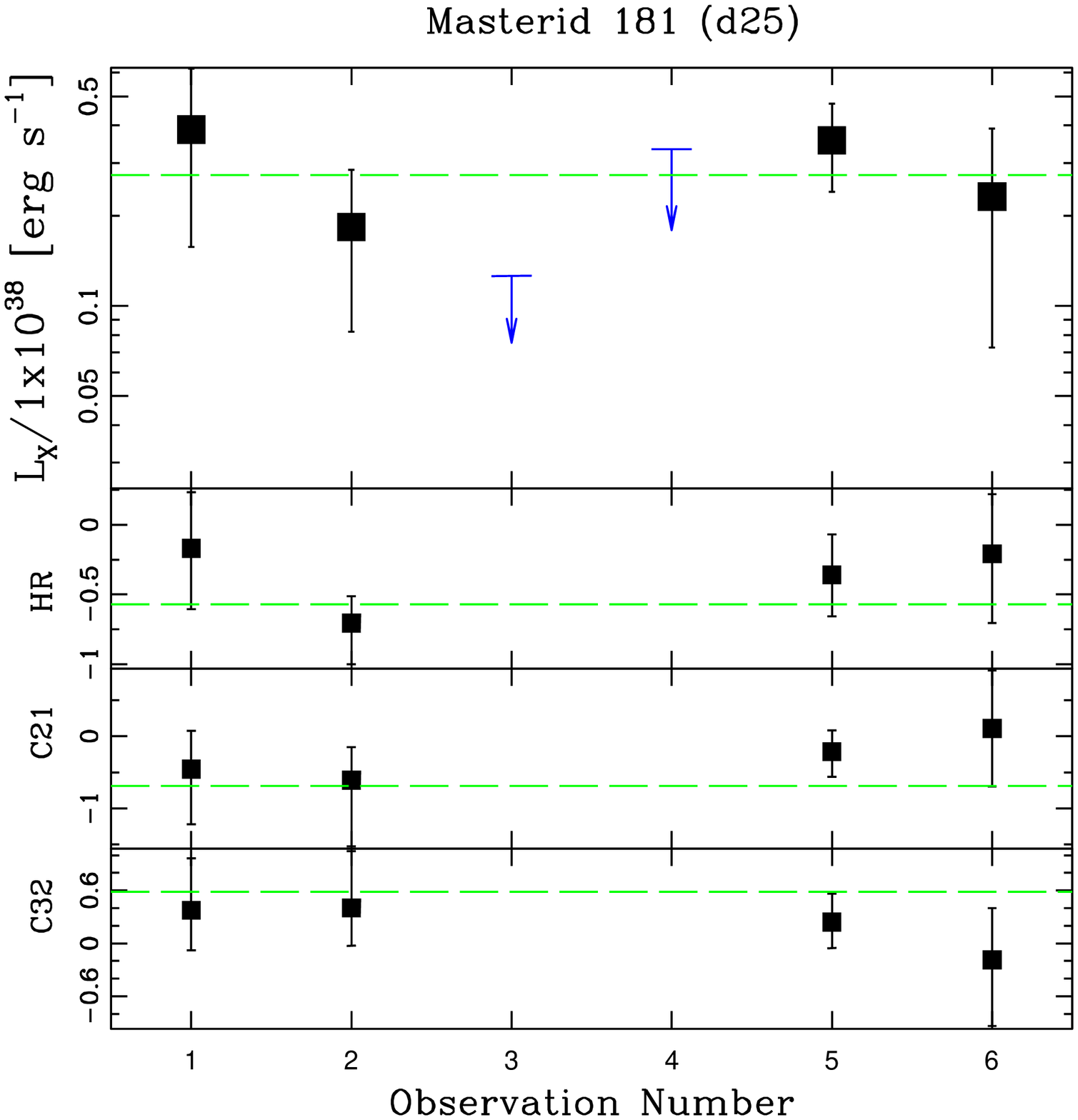}

  \end{minipage}\hspace{0.02\linewidth}
  \begin{minipage}{0.485\linewidth}
  \centering

    \includegraphics[width=\linewidth]{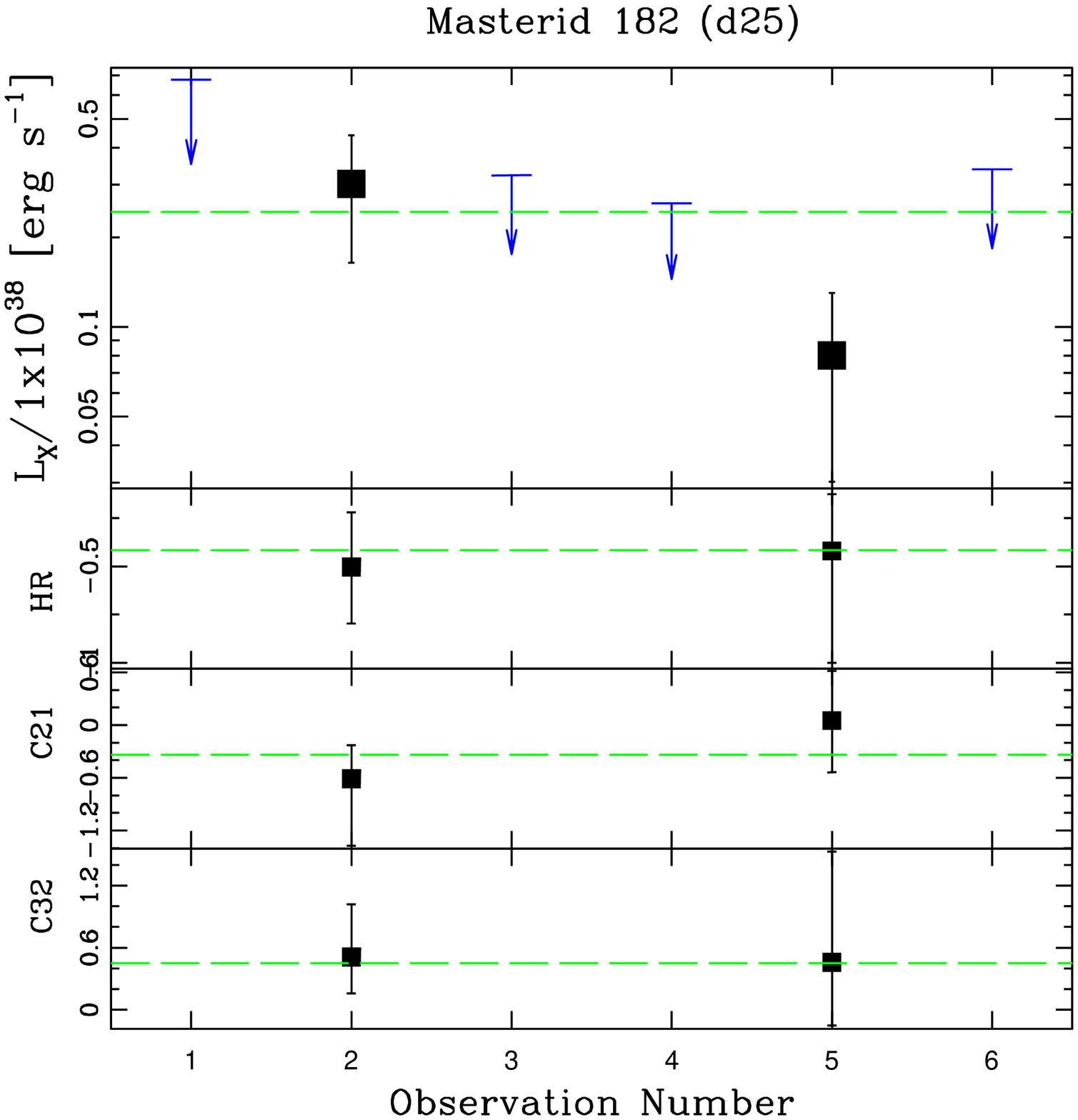}

\end{minipage}\hspace{0.02\linewidth}

\begin{minipage}{0.485\linewidth}
  \centering

    \includegraphics[width=\linewidth]{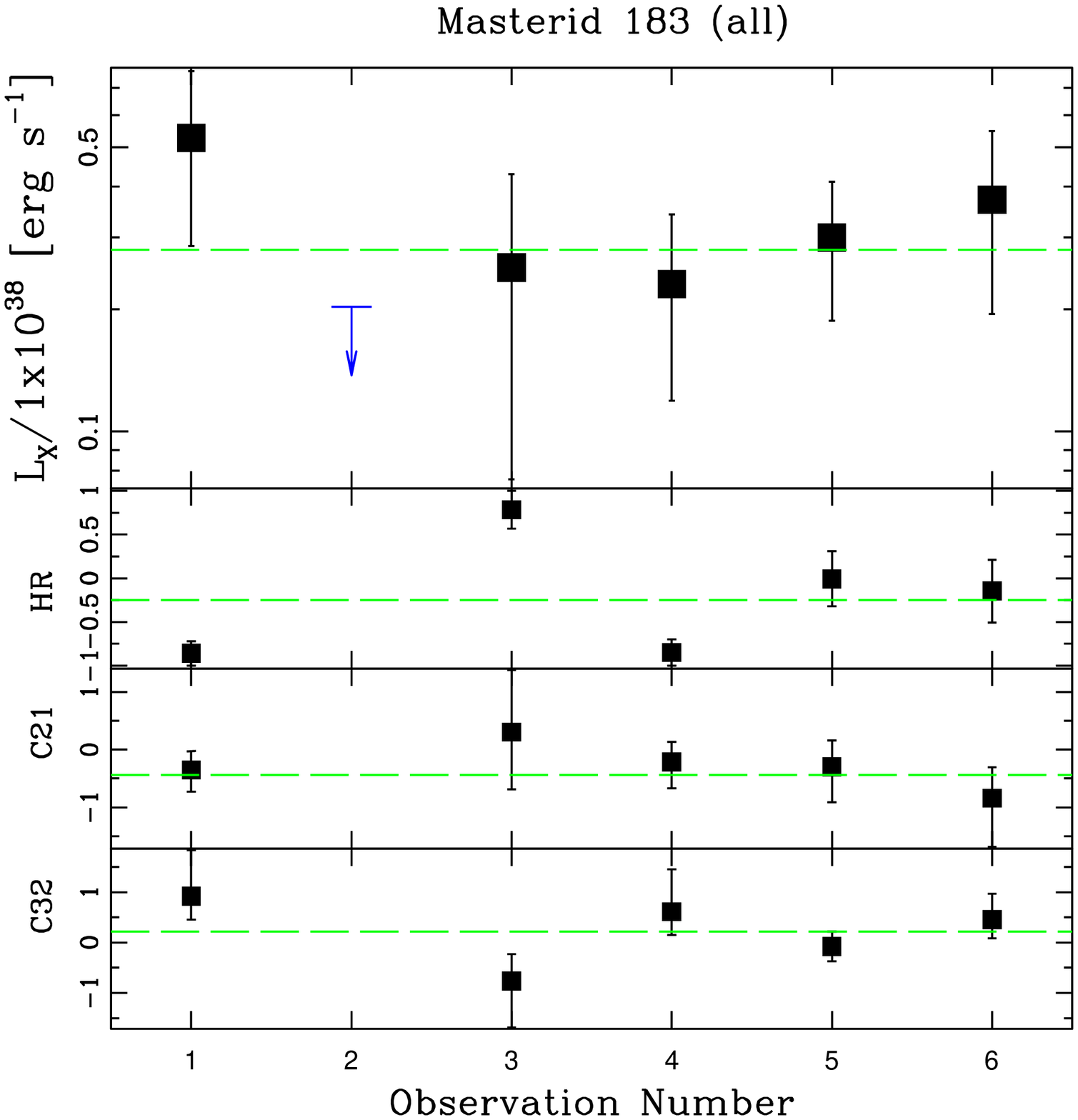}

 \end{minipage}\hspace{0.02\linewidth}
\begin{minipage}{0.485\linewidth}
  \centering
  
    \includegraphics[width=\linewidth]{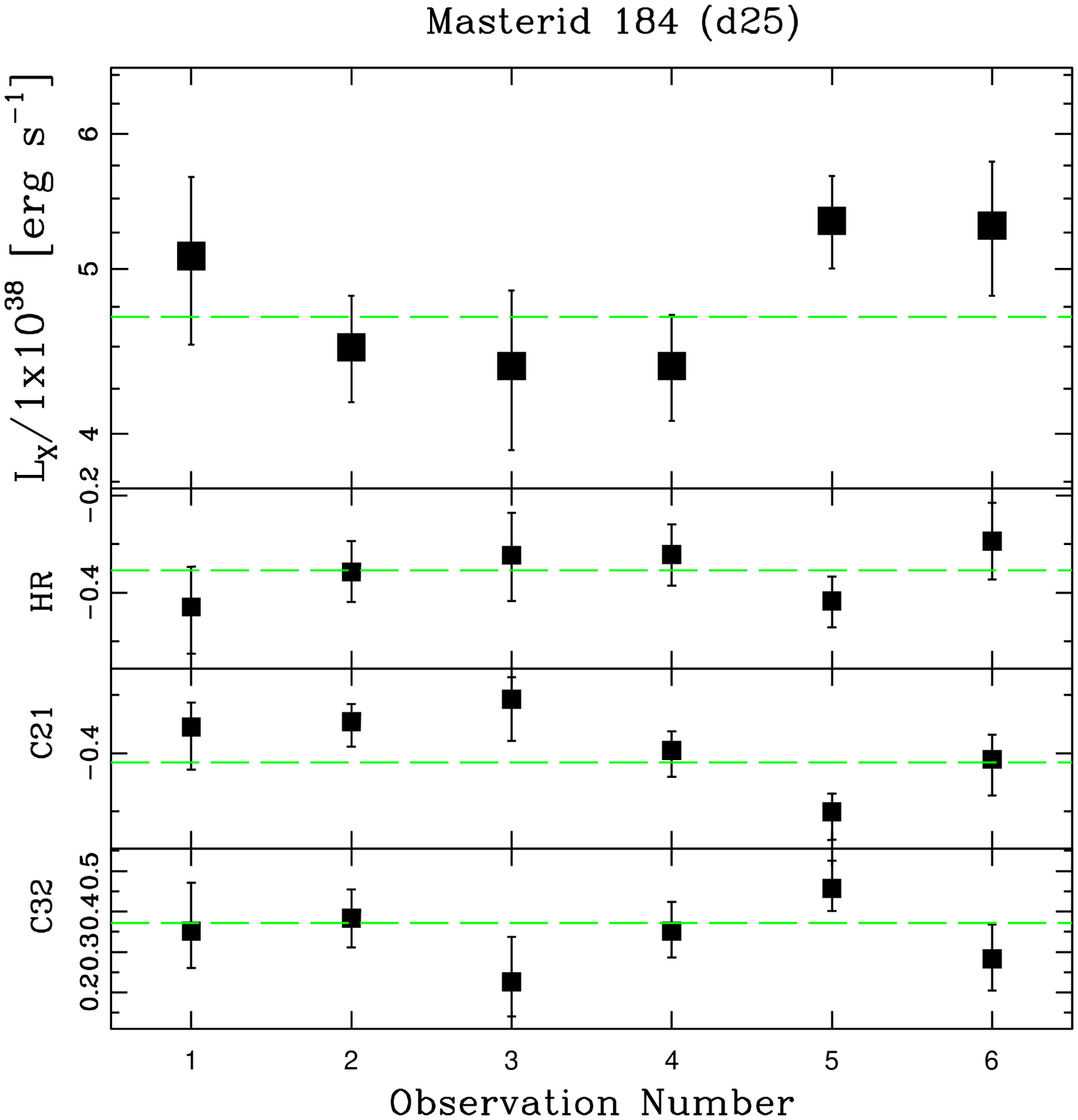}

  \end{minipage}\hspace{0.02\linewidth}
  
\end{figure}

\begin{figure}

  \begin{minipage}{0.485\linewidth}
  \centering

    \includegraphics[width=\linewidth]{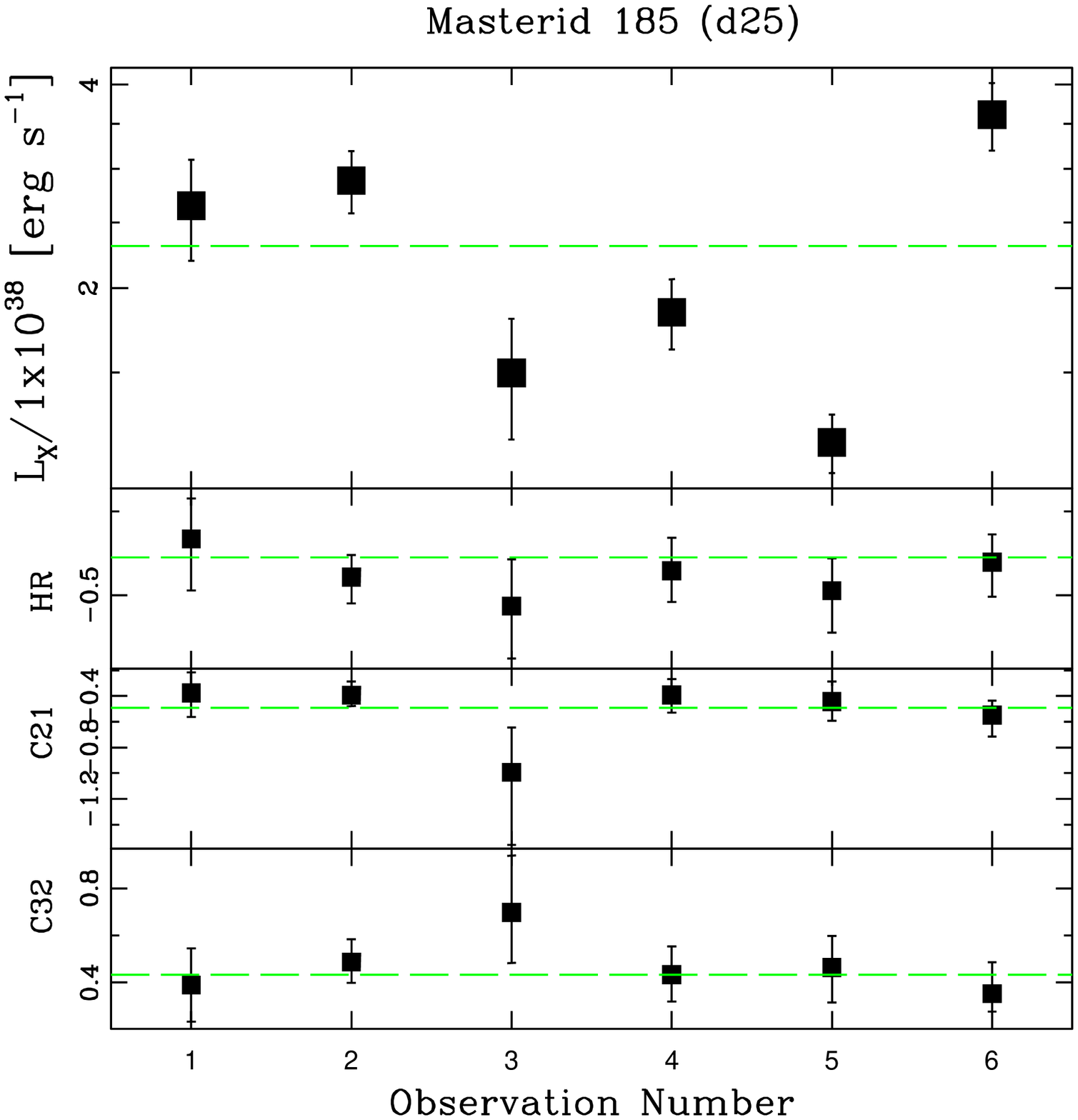}

\end{minipage}\hspace{0.02\linewidth}
\begin{minipage}{0.485\linewidth}
  \centering

    \includegraphics[width=\linewidth]{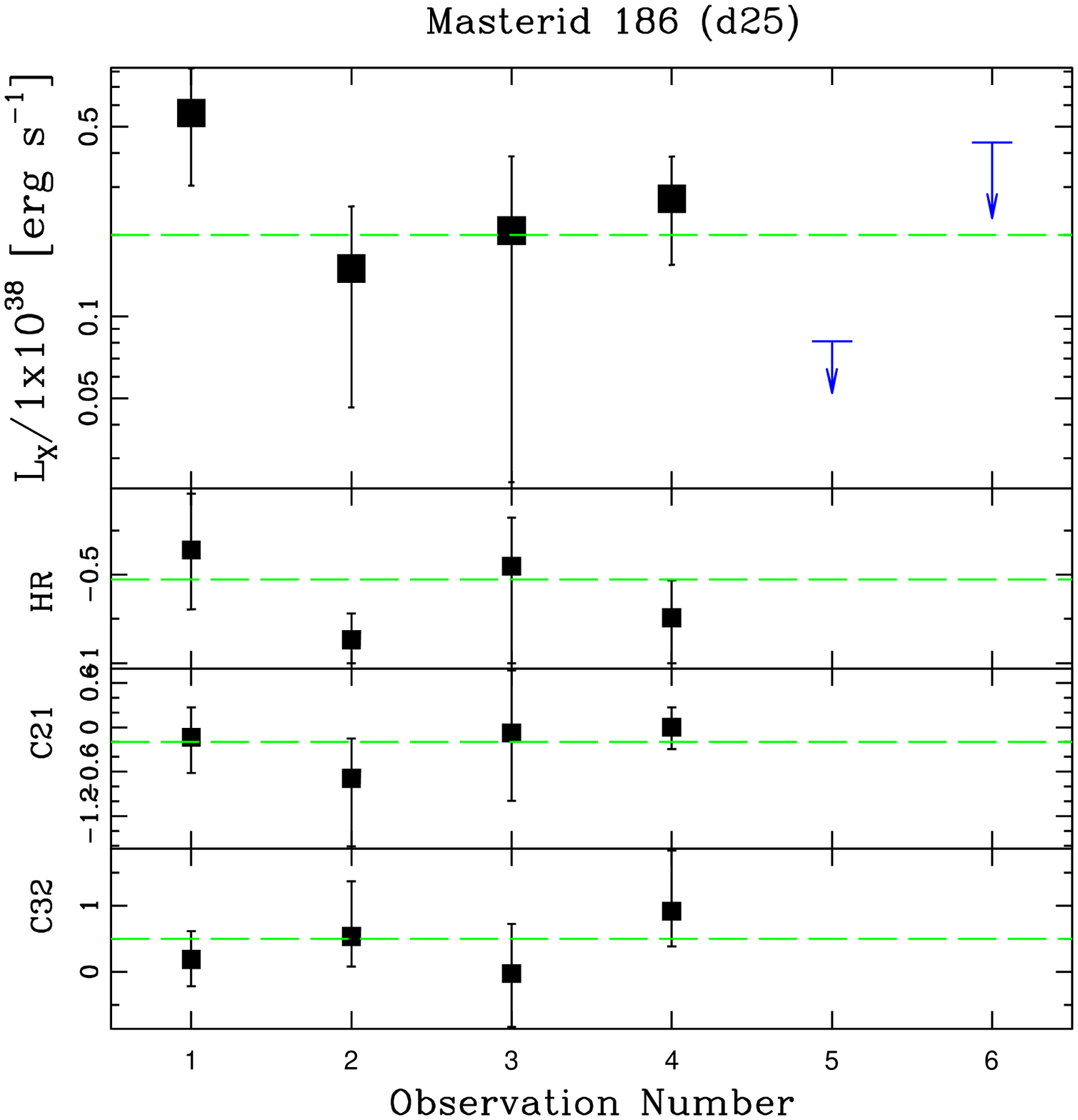}

 \end{minipage}\hspace{0.02\linewidth}

  \begin{minipage}{0.485\linewidth}
  \centering
  
    \includegraphics[width=\linewidth]{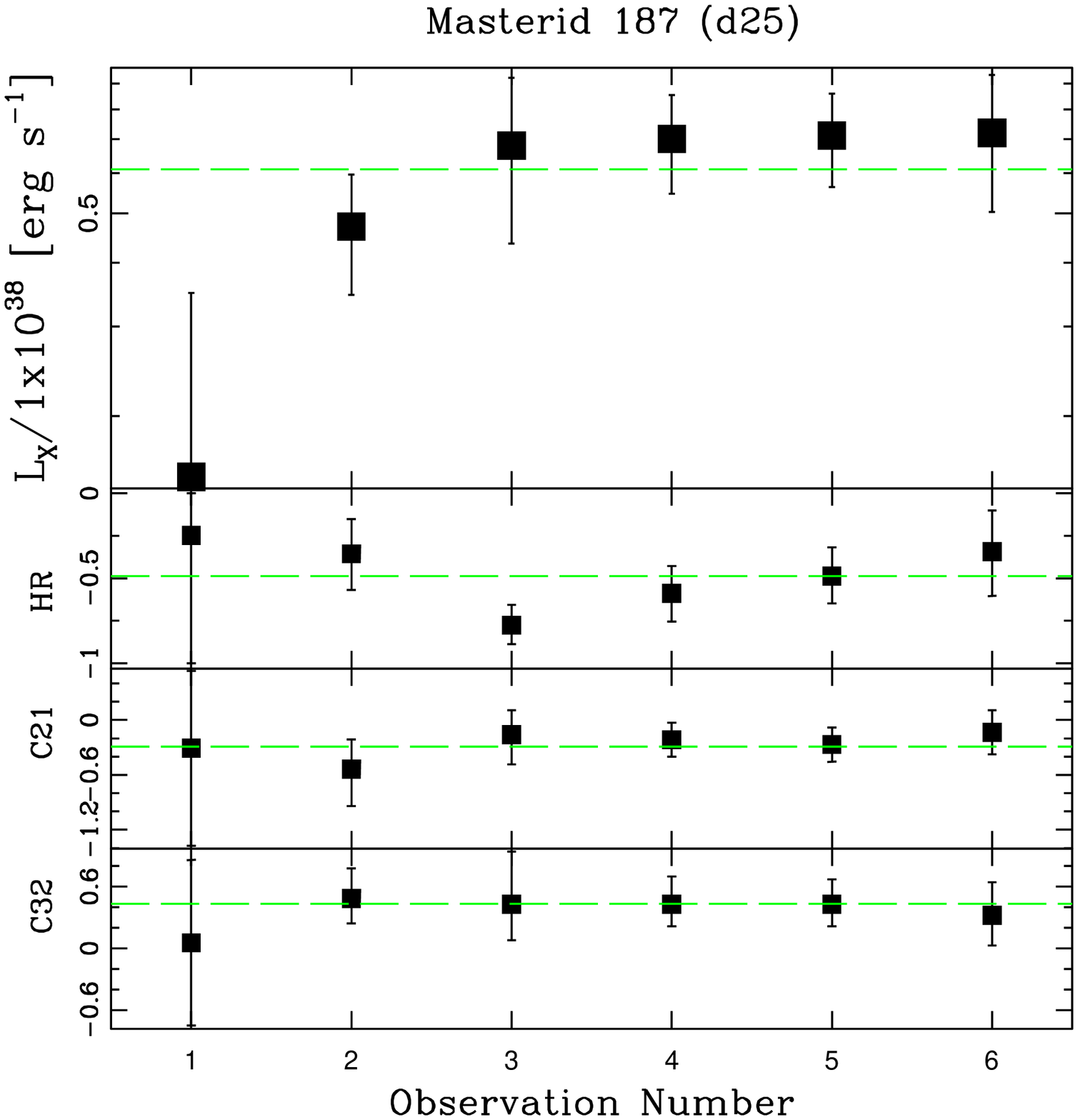}

  \end{minipage}\hspace{0.02\linewidth}
  \begin{minipage}{0.485\linewidth}
  \centering

    \includegraphics[width=\linewidth]{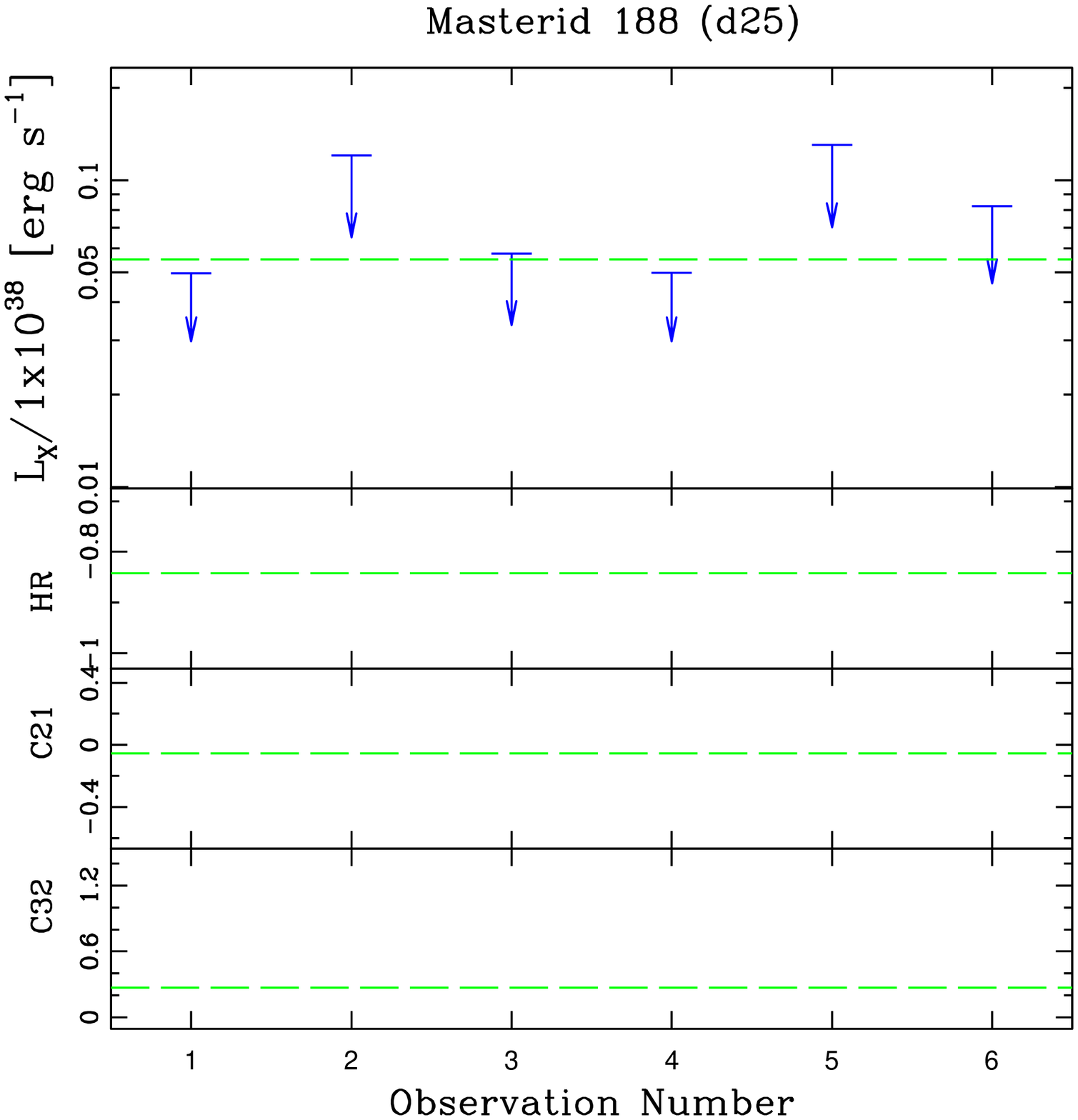}

\end{minipage}\hspace{0.02\linewidth}

\begin{minipage}{0.485\linewidth}
  \centering

    \includegraphics[width=\linewidth]{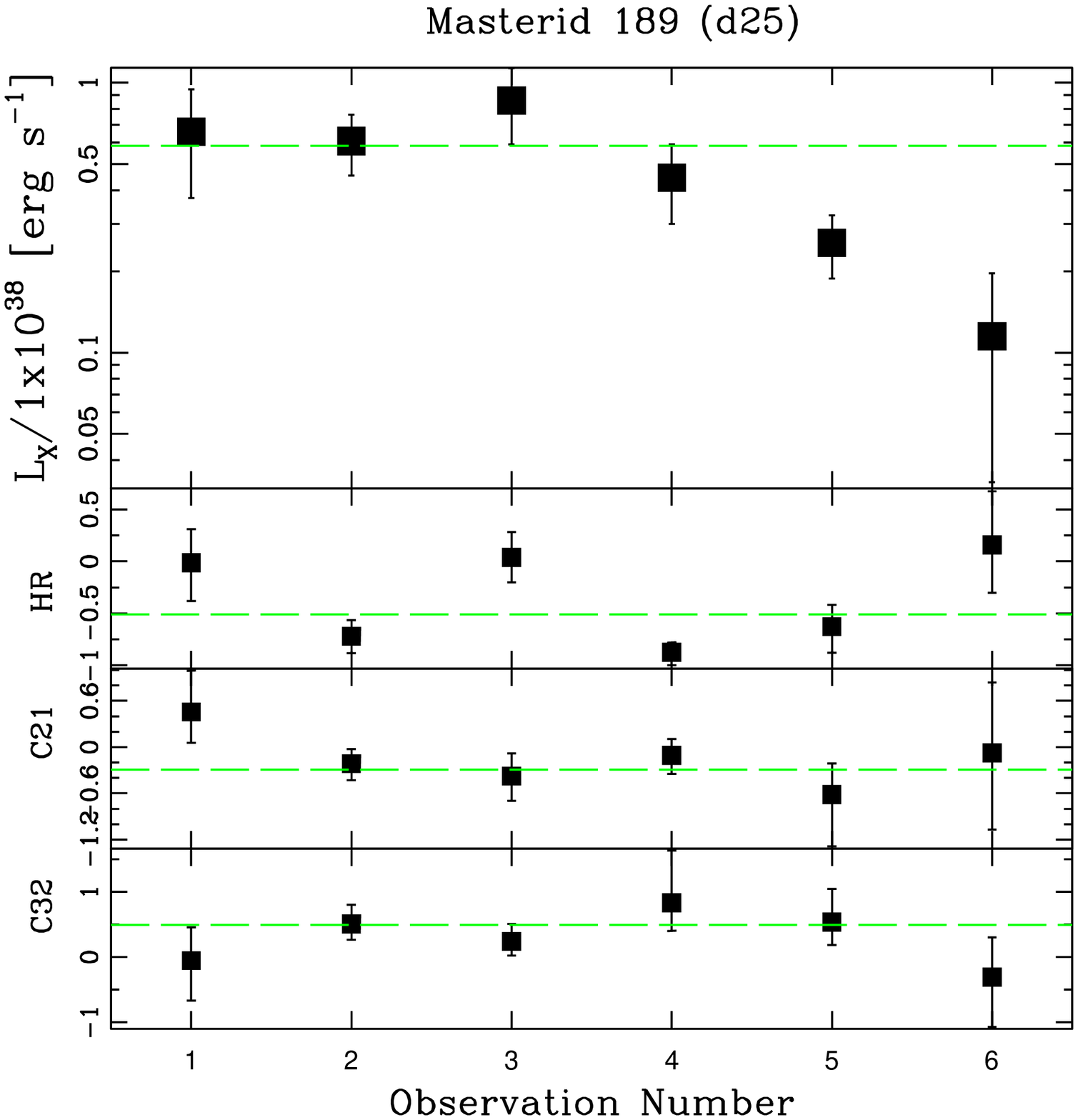}

 \end{minipage}\hspace{0.02\linewidth}
\begin{minipage}{0.485\linewidth}
  \centering
  
    \includegraphics[width=\linewidth]{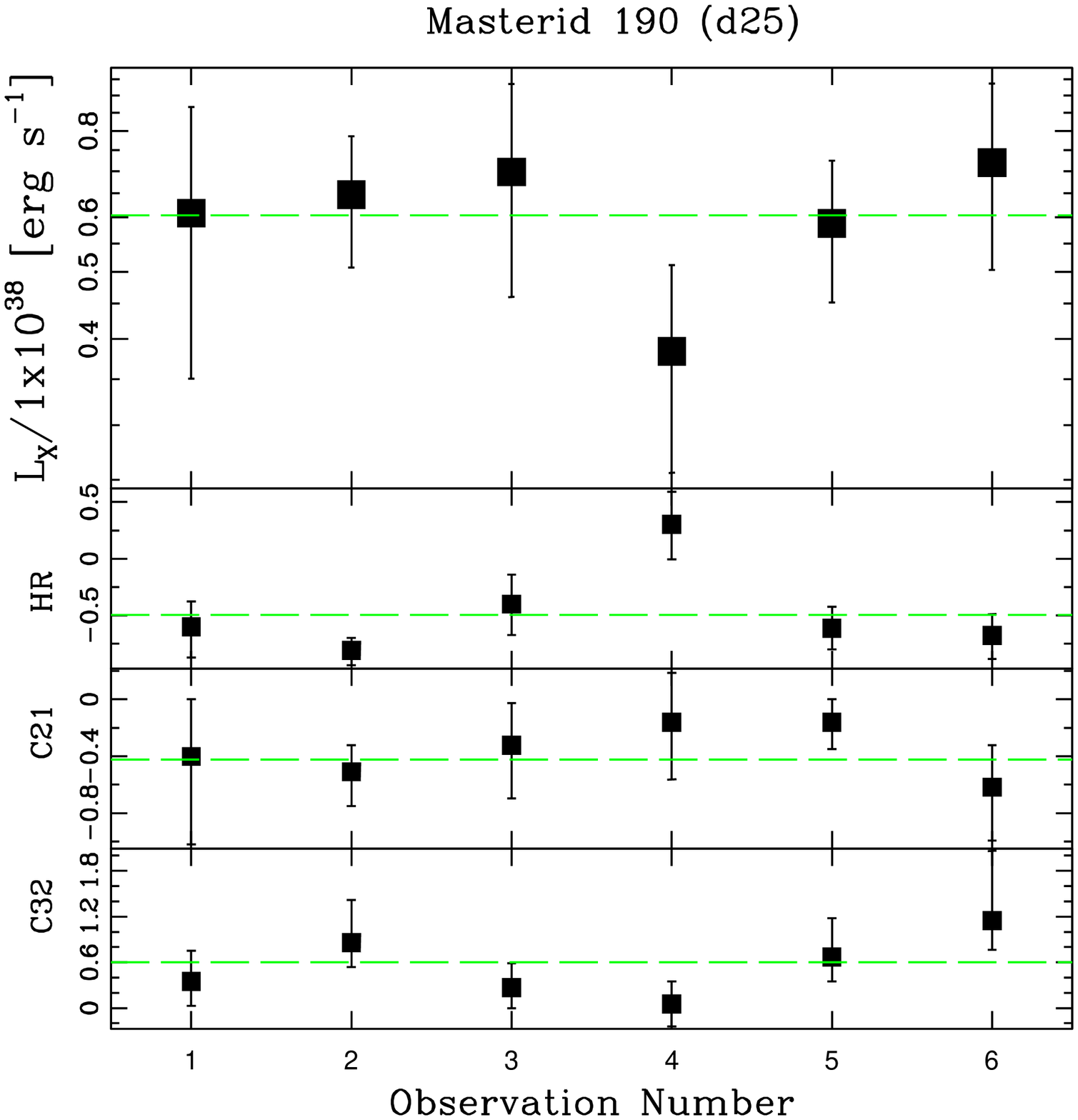}

  \end{minipage}\hspace{0.02\linewidth}
  
\end{figure}

\clearpage

\begin{figure}

  \begin{minipage}{0.485\linewidth}
  \centering

    \includegraphics[width=\linewidth]{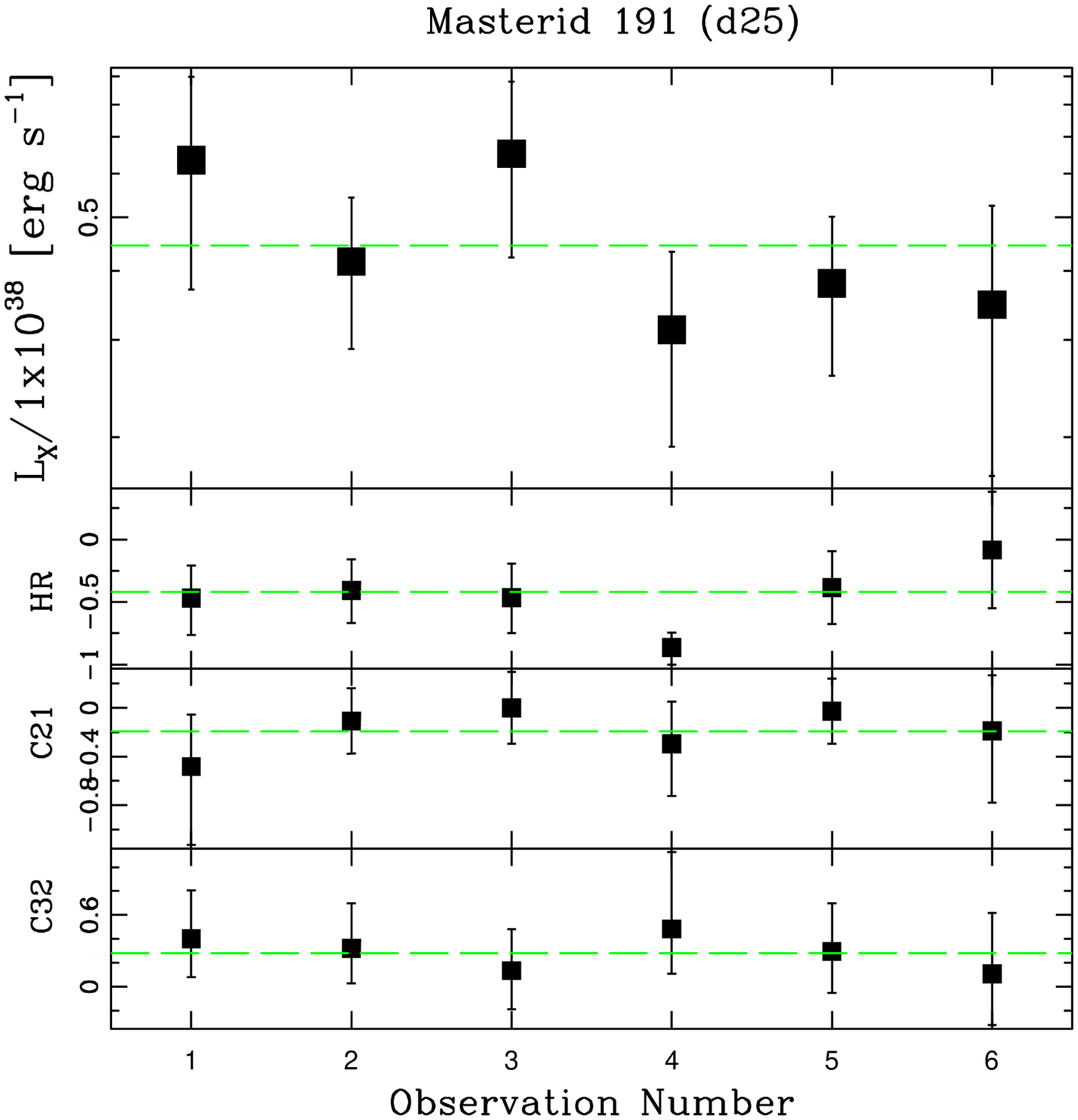}

\end{minipage}\hspace{0.02\linewidth}
\begin{minipage}{0.485\linewidth}
  \centering

    \includegraphics[width=\linewidth]{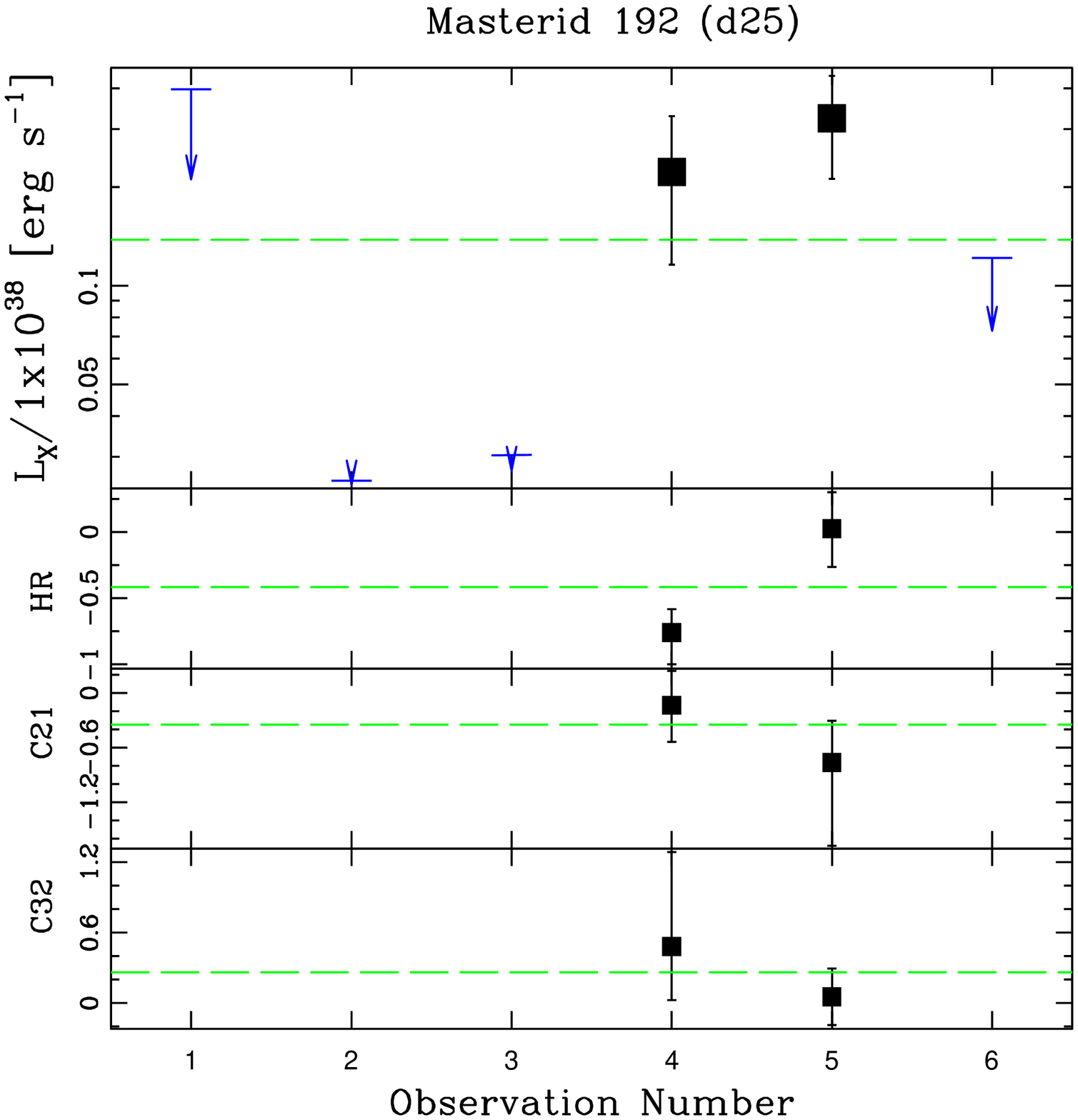}

 \end{minipage}\hspace{0.02\linewidth}

  \begin{minipage}{0.485\linewidth}
  \centering
  
    \includegraphics[width=\linewidth]{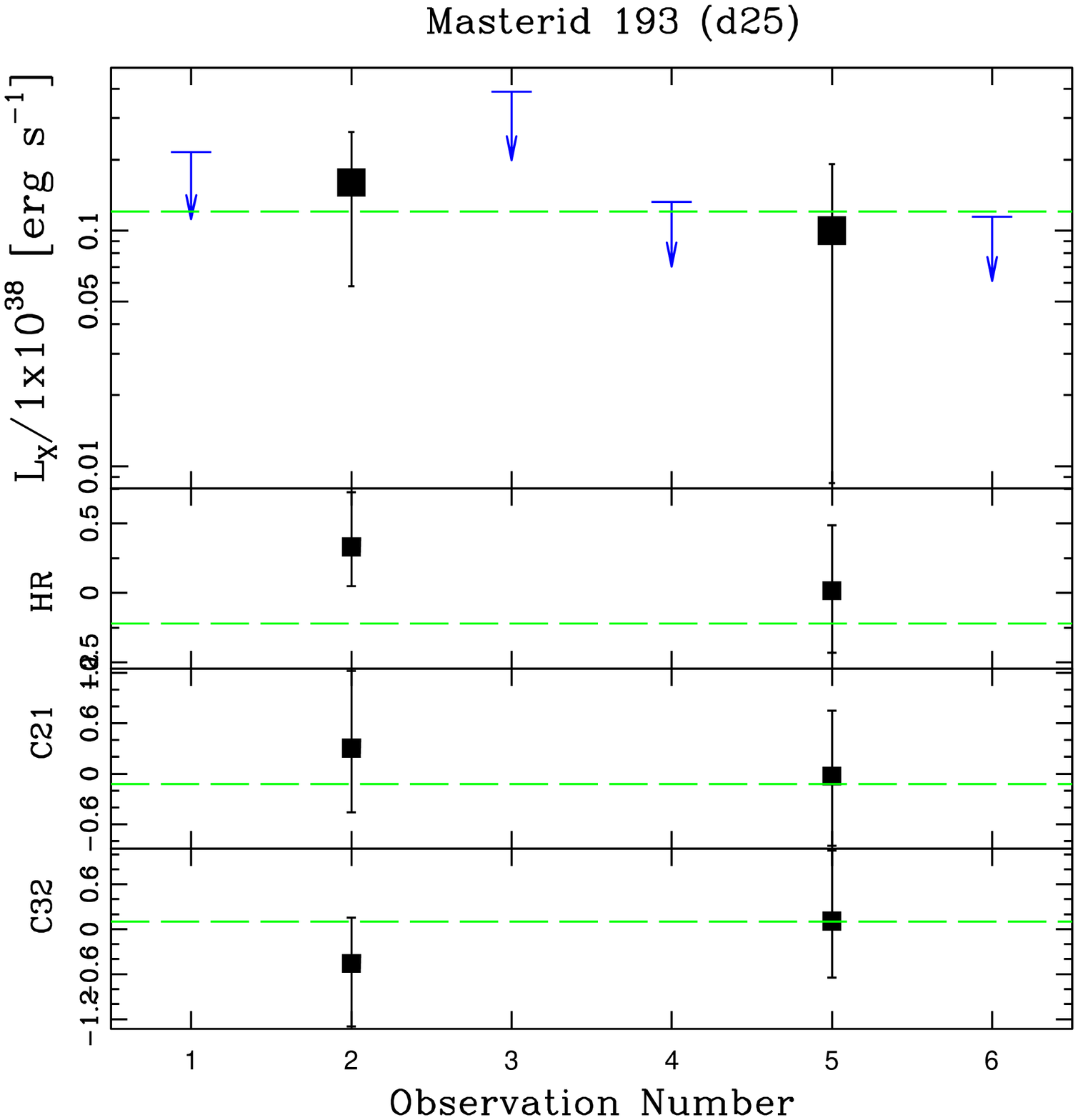}

  \end{minipage}\hspace{0.02\linewidth}
  \begin{minipage}{0.485\linewidth}
  \centering

    \includegraphics[width=\linewidth]{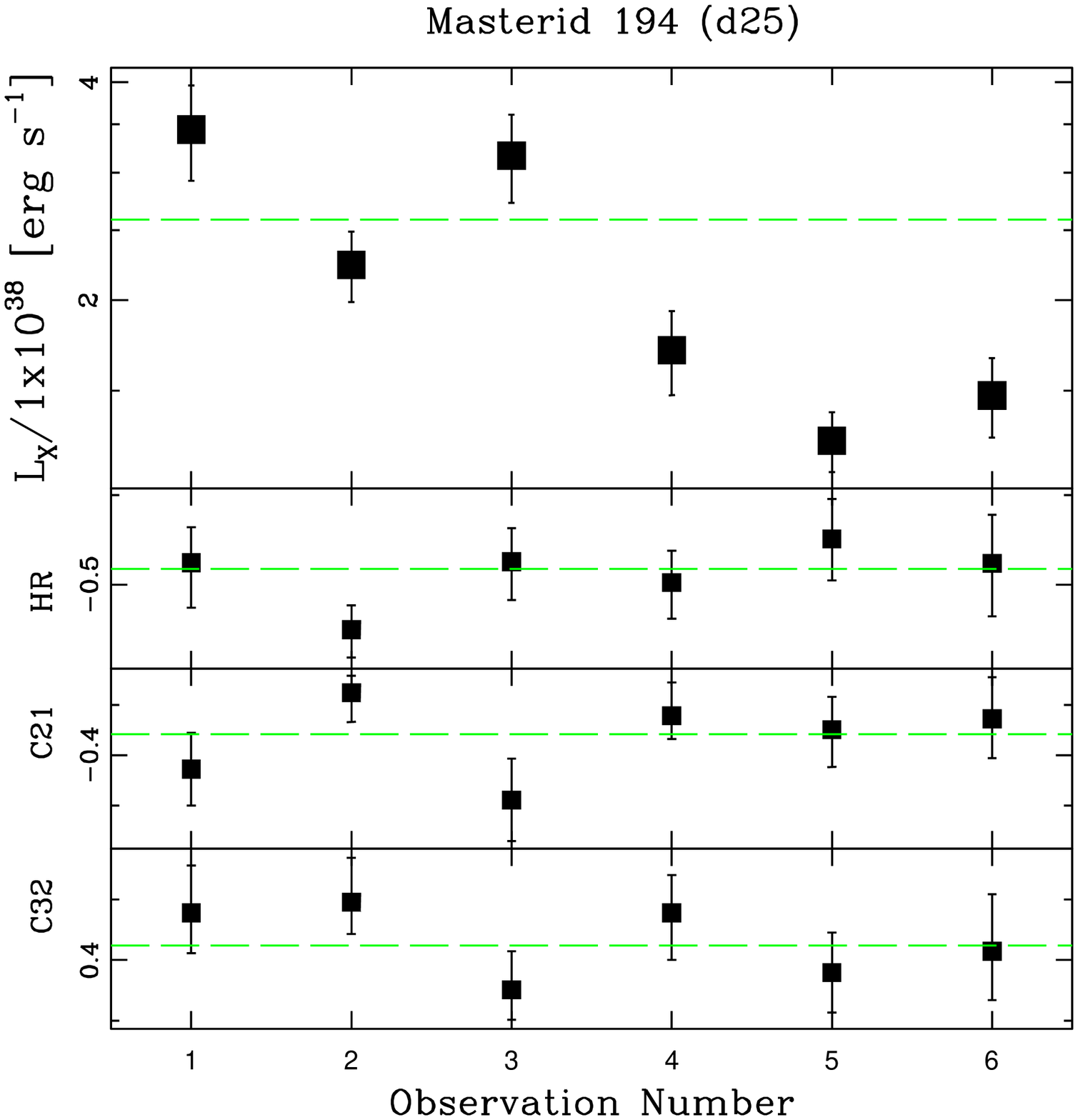}

\end{minipage}\hspace{0.02\linewidth}

\begin{minipage}{0.485\linewidth}
  \centering

    \includegraphics[width=\linewidth]{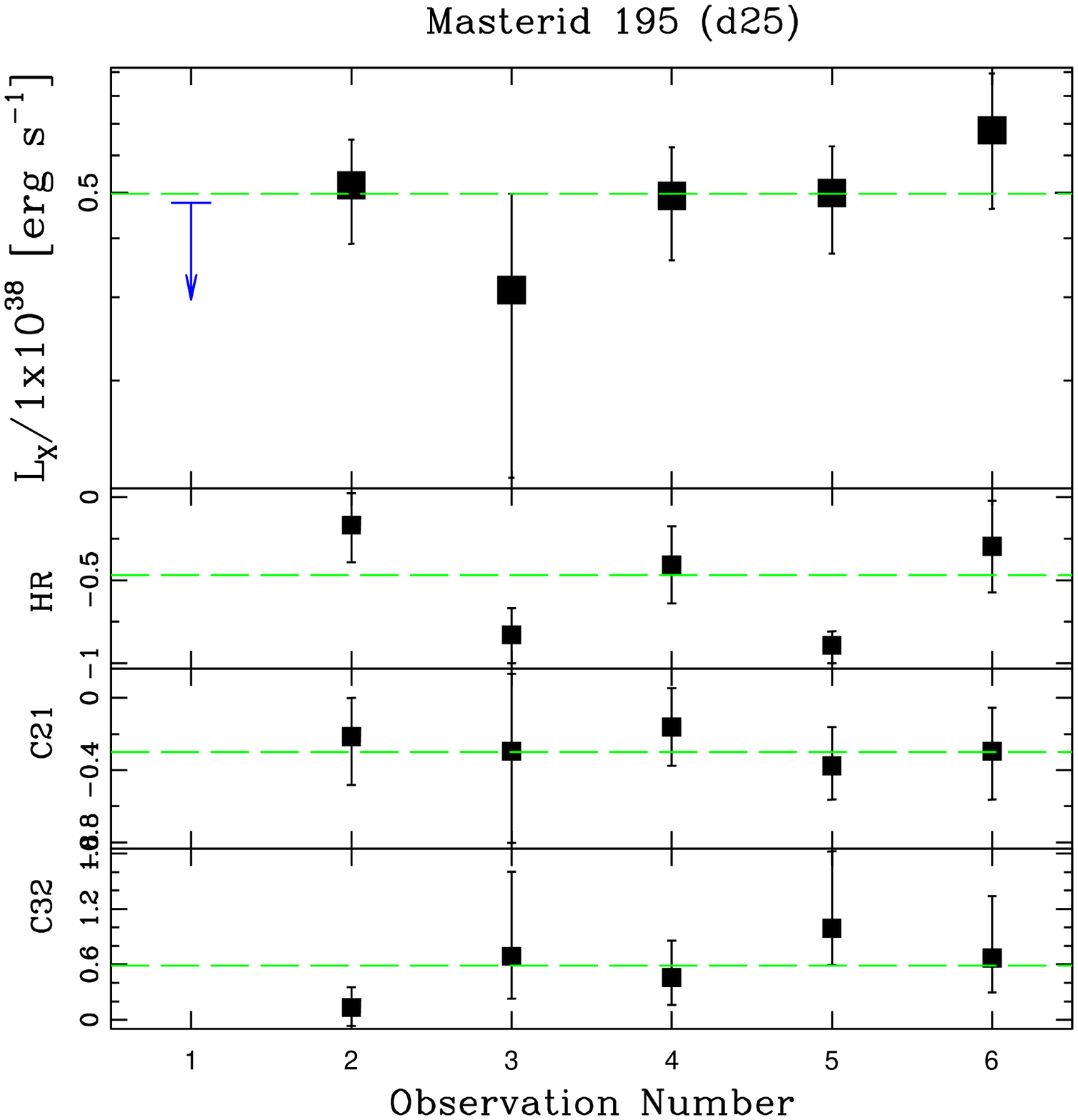}

 \end{minipage}\hspace{0.02\linewidth}
\begin{minipage}{0.485\linewidth}
  \centering
  
    \includegraphics[width=\linewidth]{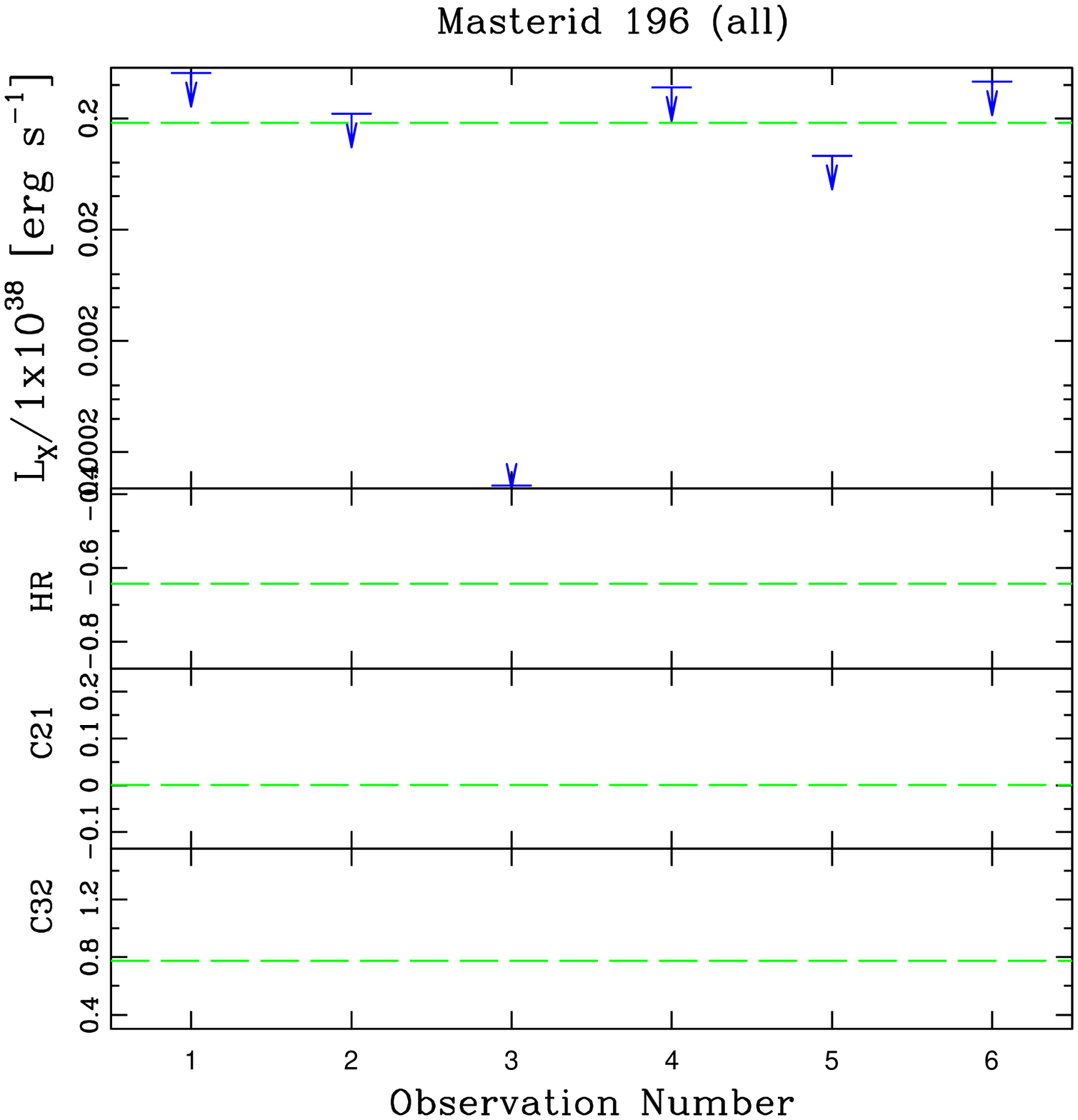}

  \end{minipage}\hspace{0.02\linewidth}
  
\end{figure}

\begin{figure}

  \begin{minipage}{0.485\linewidth}
  \centering

    \includegraphics[width=\linewidth]{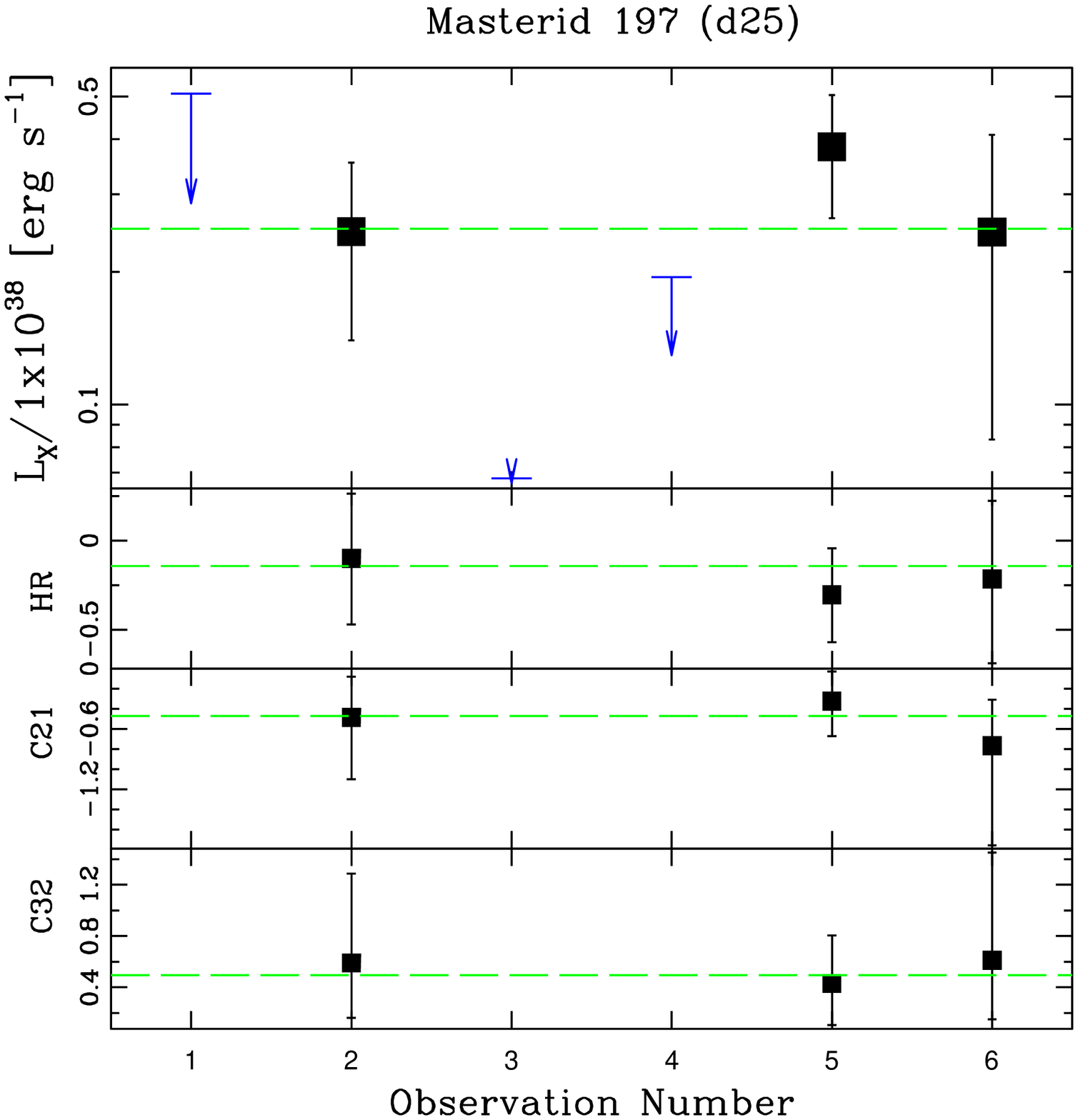}

\end{minipage}\hspace{0.02\linewidth}
\begin{minipage}{0.485\linewidth}
  \centering

    \includegraphics[width=\linewidth]{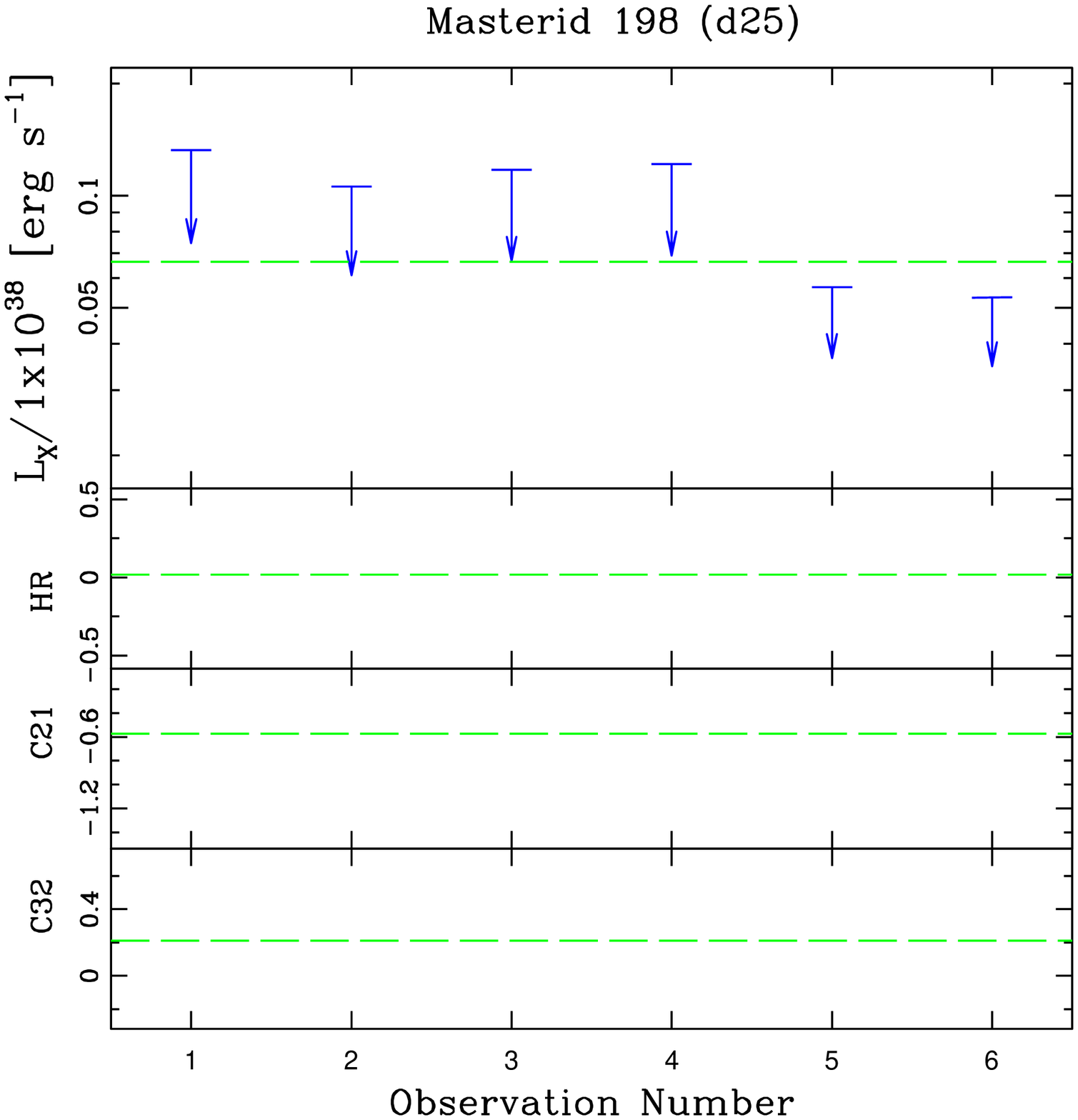}

 \end{minipage}\hspace{0.02\linewidth}

  \begin{minipage}{0.485\linewidth}
  \centering
  
    \includegraphics[width=\linewidth]{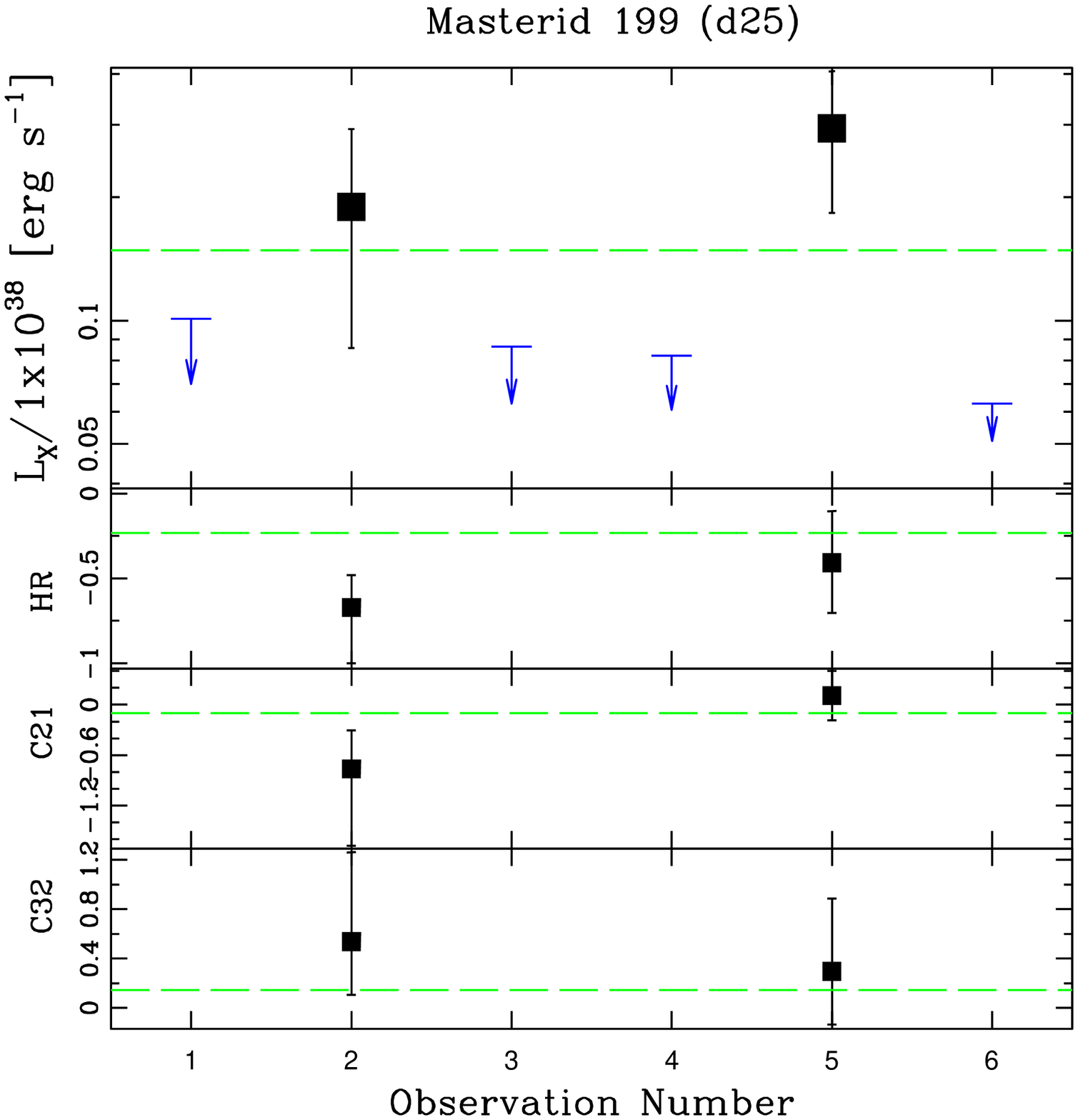}

  \end{minipage}\hspace{0.02\linewidth}
  \begin{minipage}{0.485\linewidth}
  \centering

    \includegraphics[width=\linewidth]{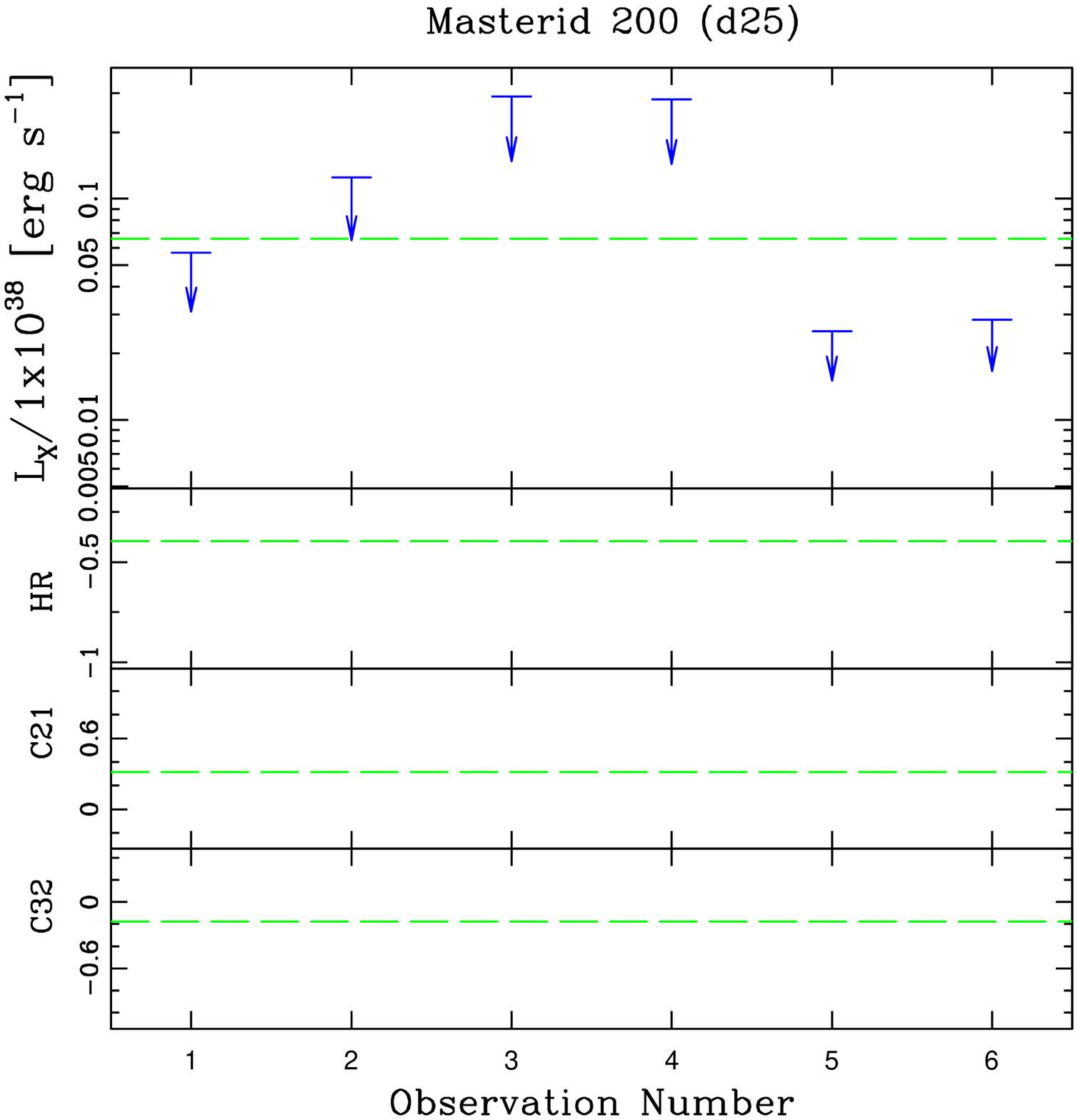}

\end{minipage}\hspace{0.02\linewidth}

\begin{minipage}{0.485\linewidth}
  \centering

    \includegraphics[width=\linewidth]{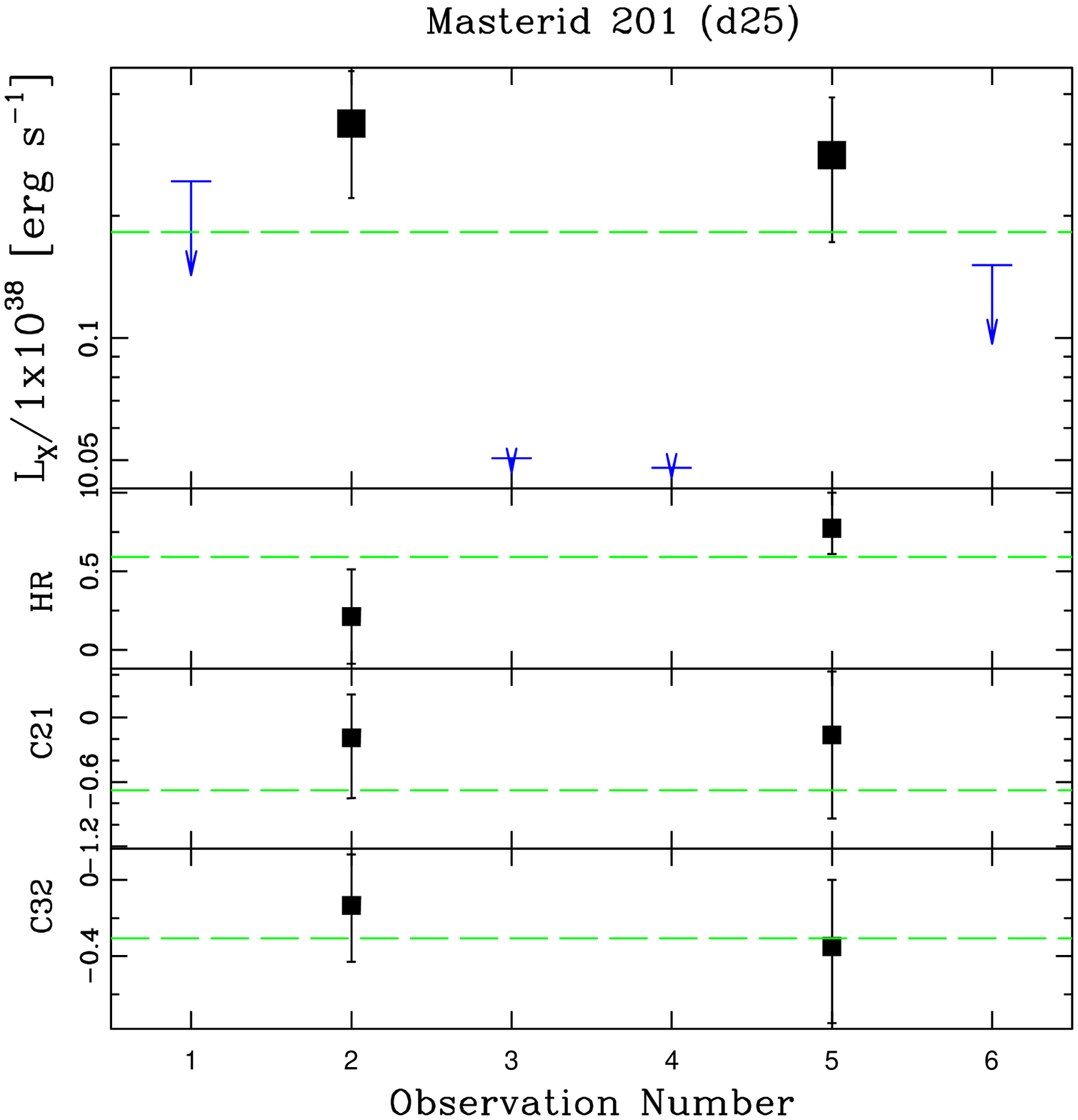}

 \end{minipage}\hspace{0.02\linewidth}
\begin{minipage}{0.485\linewidth}
  \centering
  
    \includegraphics[width=\linewidth]{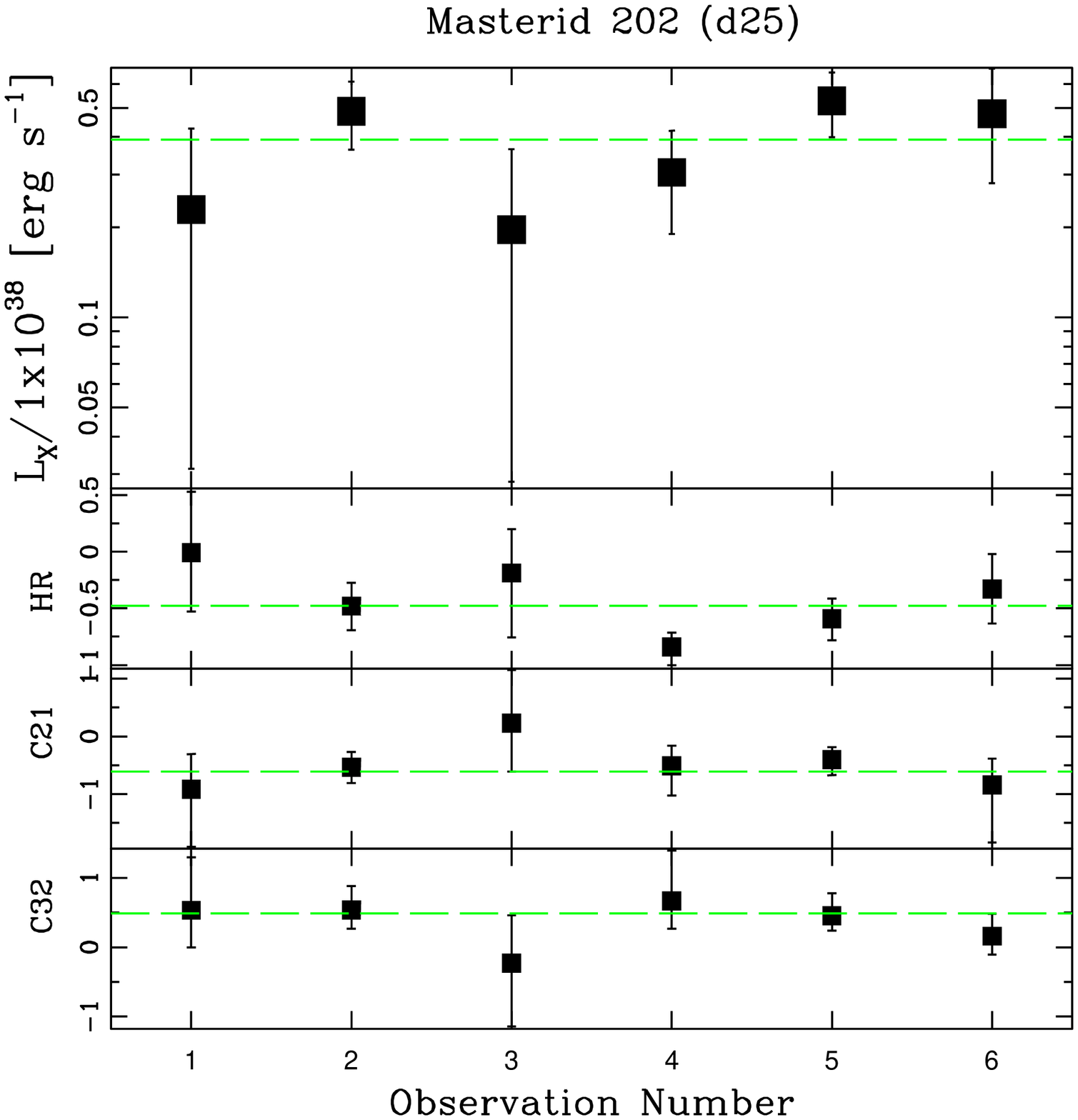}

  \end{minipage}\hspace{0.02\linewidth}
  
\end{figure}

\begin{figure}

  \begin{minipage}{0.485\linewidth}
  \centering

    \includegraphics[width=\linewidth]{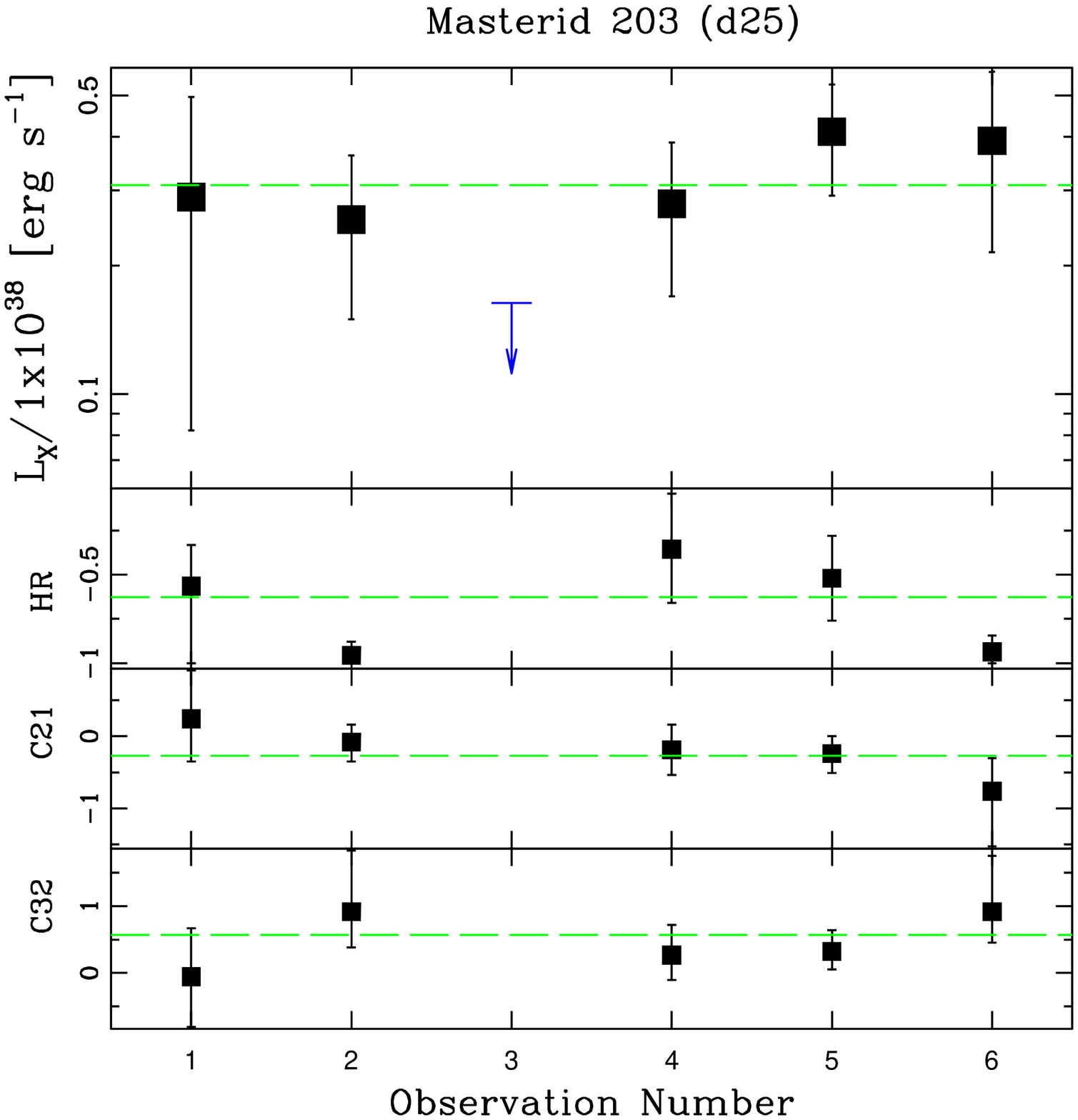}

\end{minipage}\hspace{0.02\linewidth}
\begin{minipage}{0.485\linewidth}
  \centering

    \includegraphics[width=\linewidth]{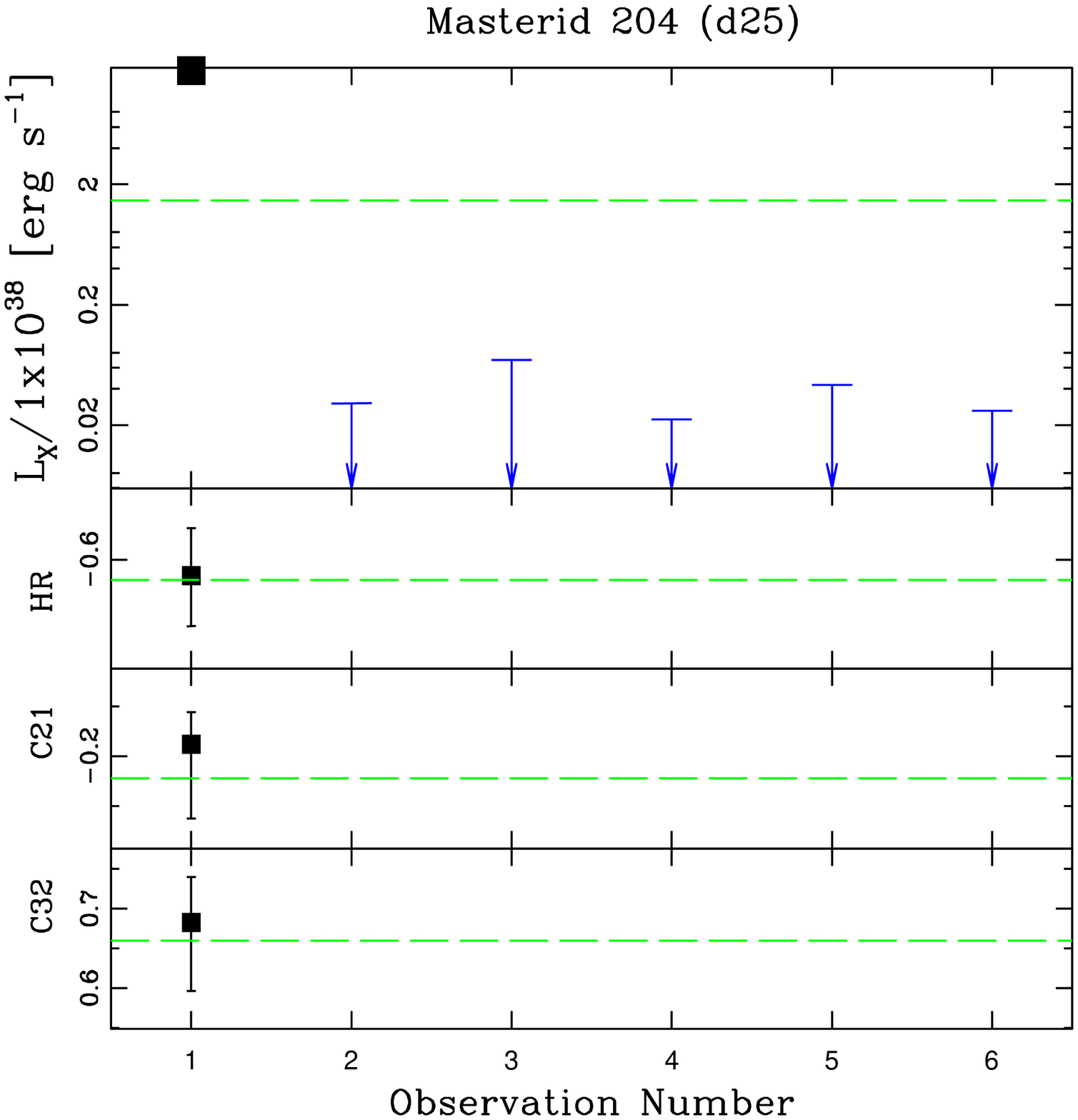}

 \end{minipage}\hspace{0.02\linewidth}

  \begin{minipage}{0.485\linewidth}
  \centering
  
    \includegraphics[width=\linewidth]{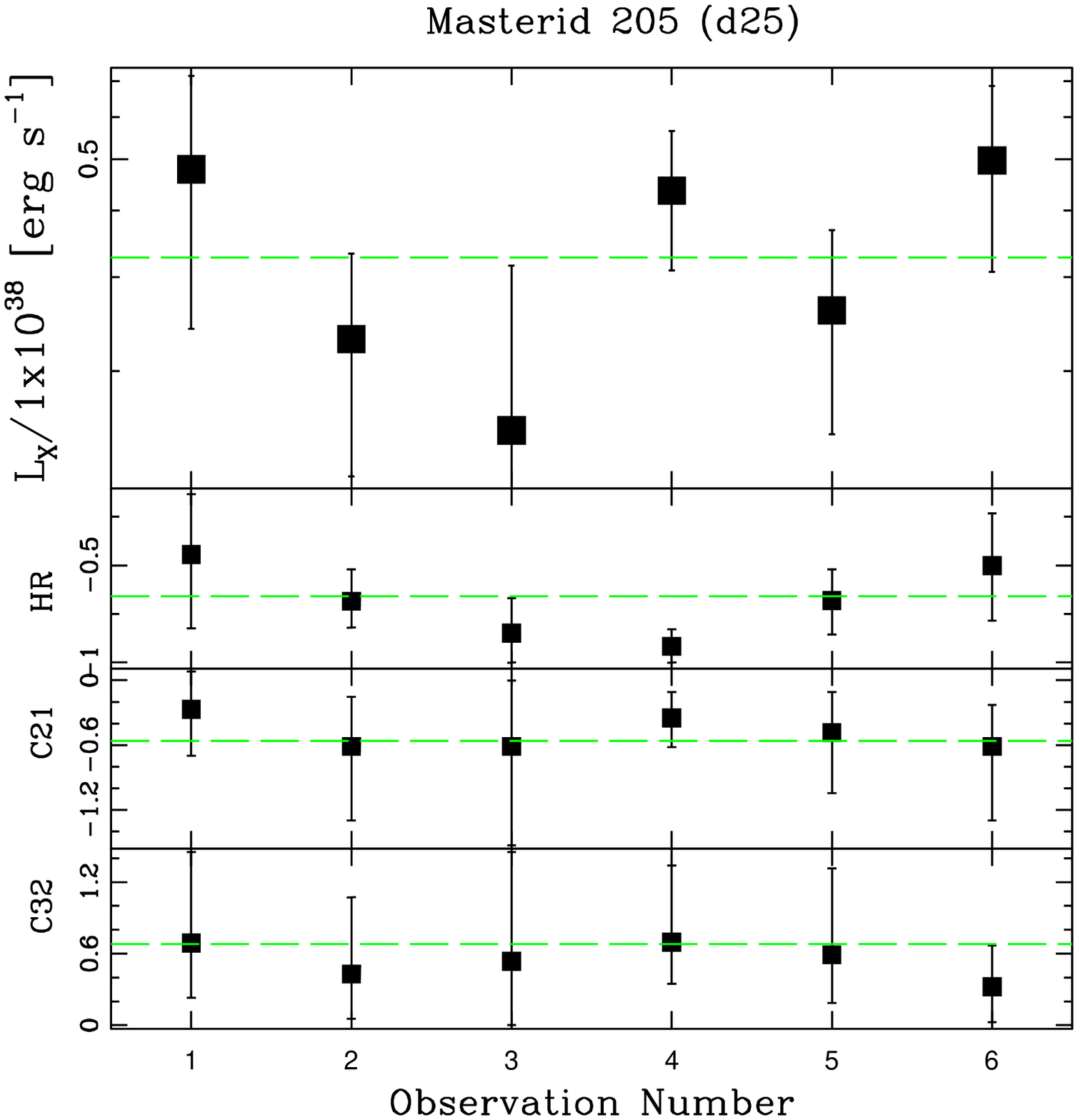}

  \end{minipage}\hspace{0.02\linewidth}
  \begin{minipage}{0.485\linewidth}
  \centering

    \includegraphics[width=\linewidth]{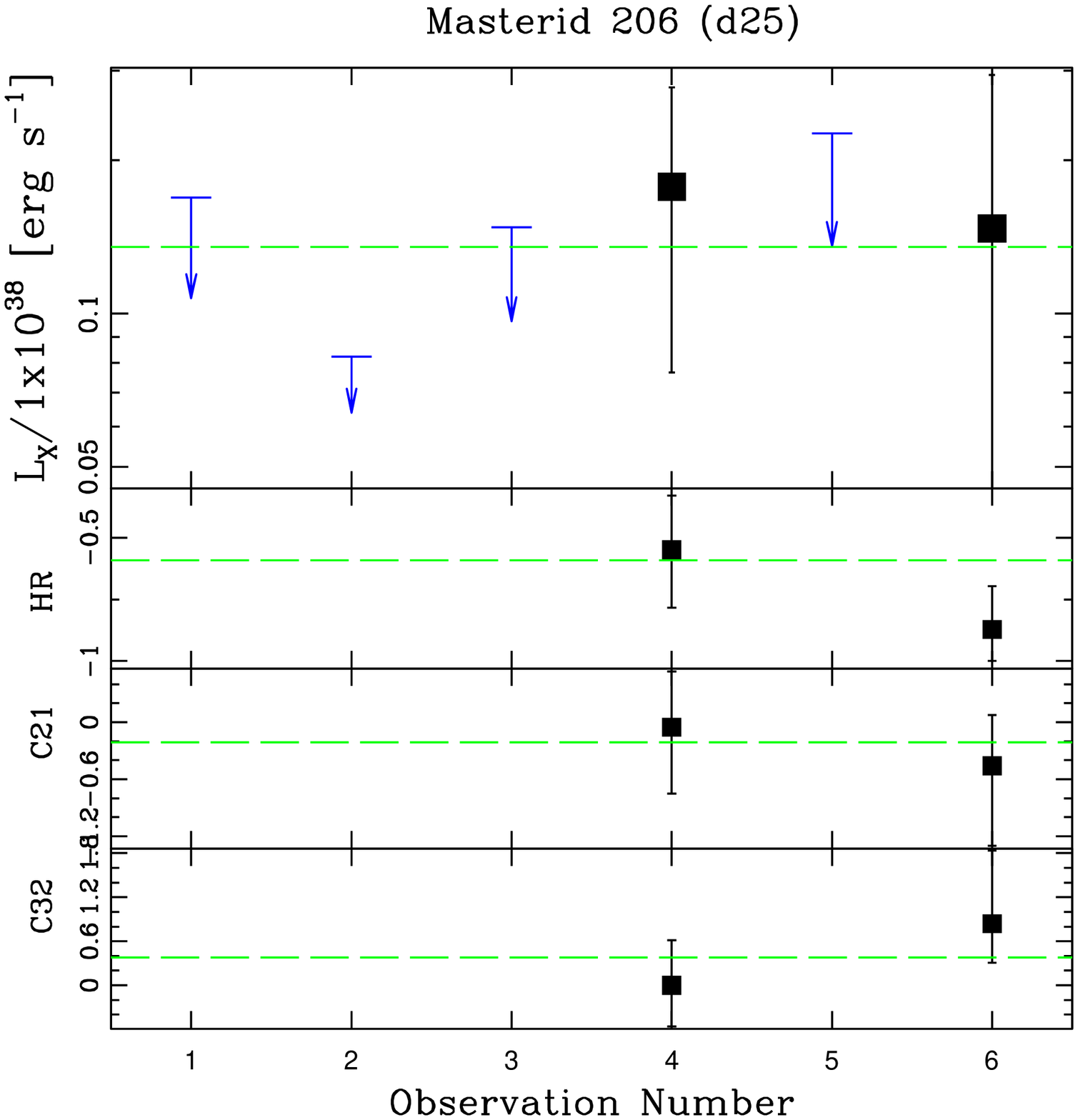}

\end{minipage}\hspace{0.02\linewidth}

\begin{minipage}{0.485\linewidth}
  \centering

    \includegraphics[width=\linewidth]{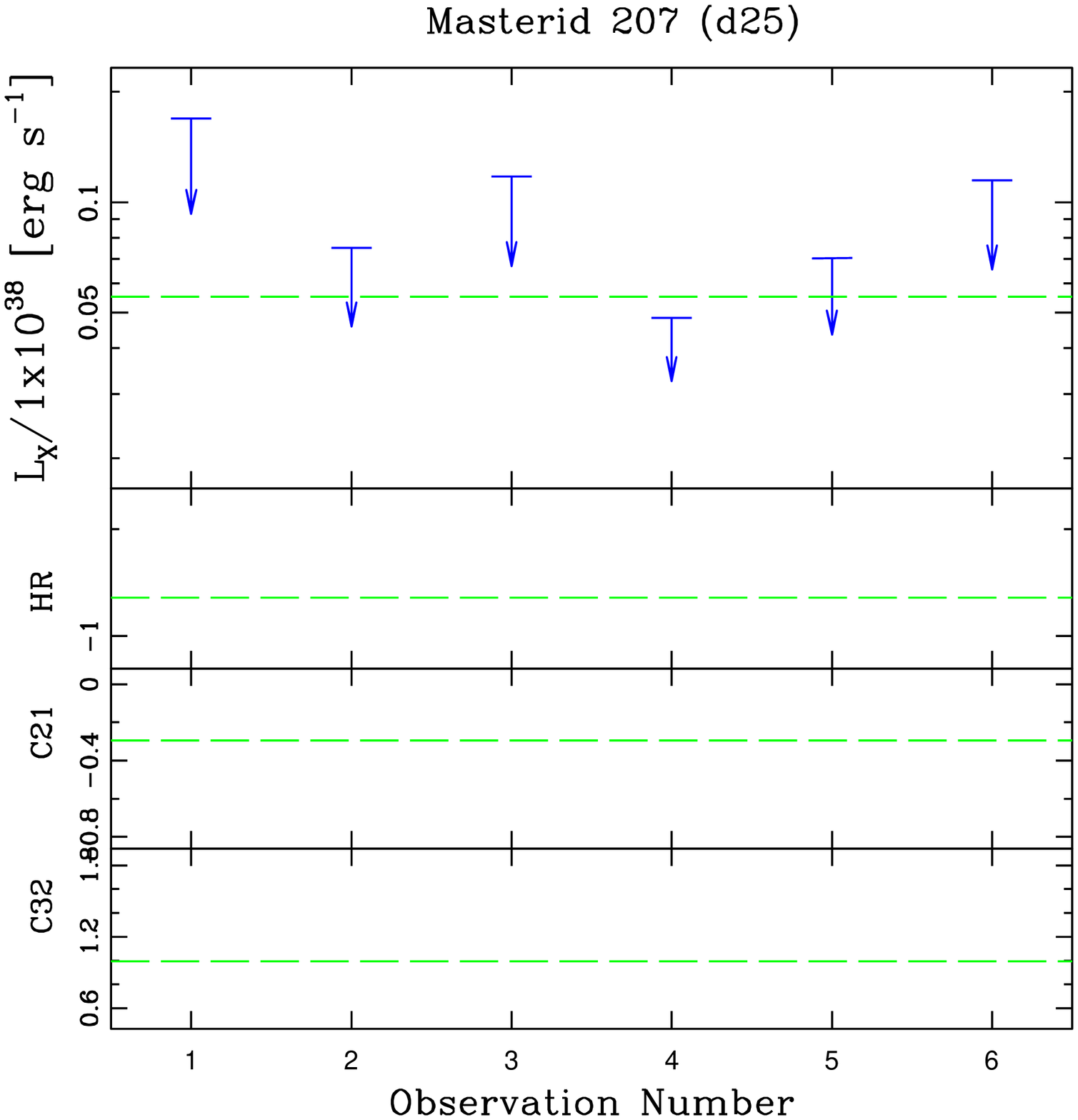}

 \end{minipage}\hspace{0.02\linewidth}
\begin{minipage}{0.485\linewidth}
  \centering
  
    \includegraphics[width=\linewidth]{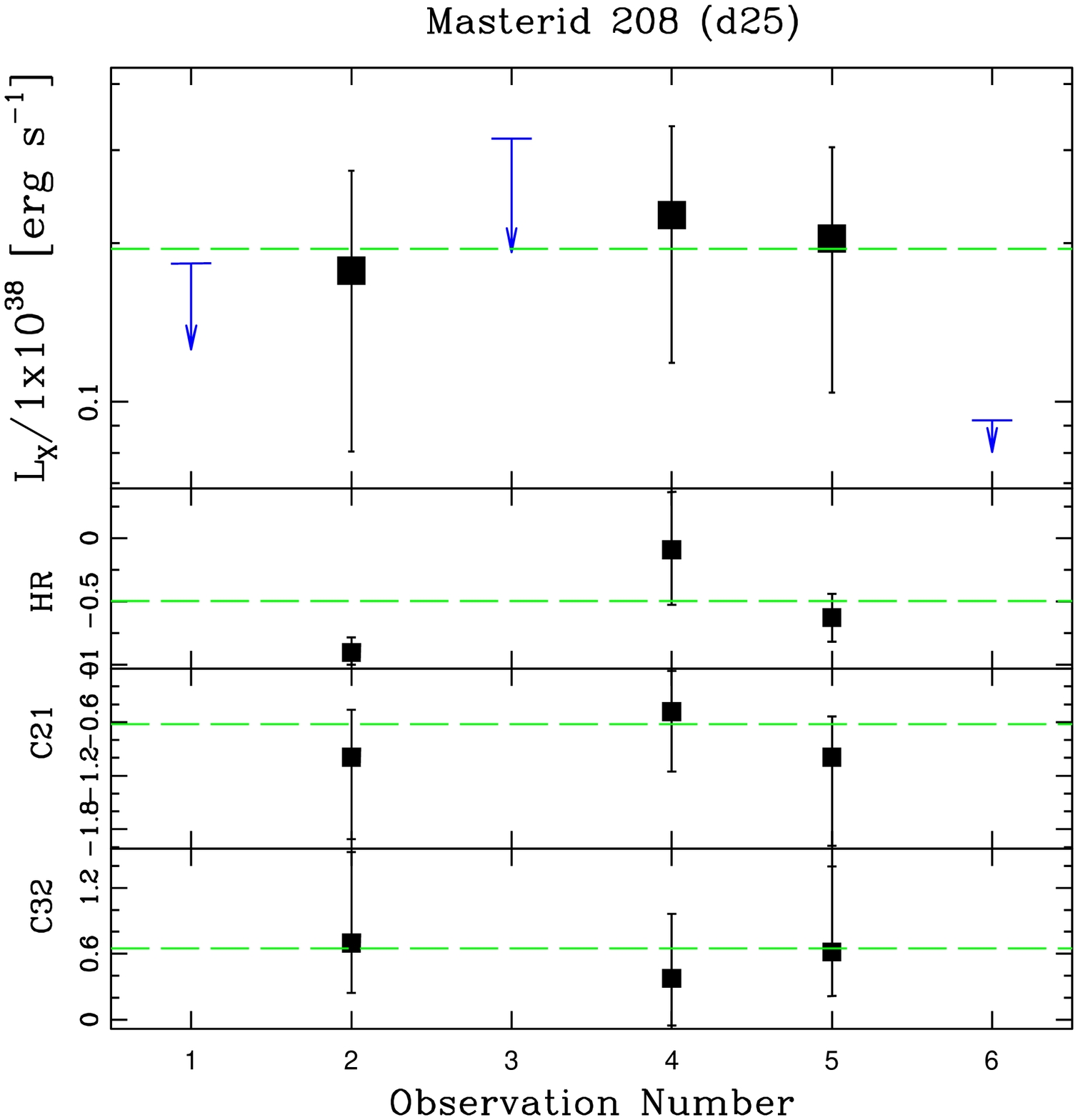}

  \end{minipage}\hspace{0.02\linewidth}
  
\end{figure}

\begin{figure}

  \begin{minipage}{0.485\linewidth}
  \centering

    \includegraphics[width=\linewidth]{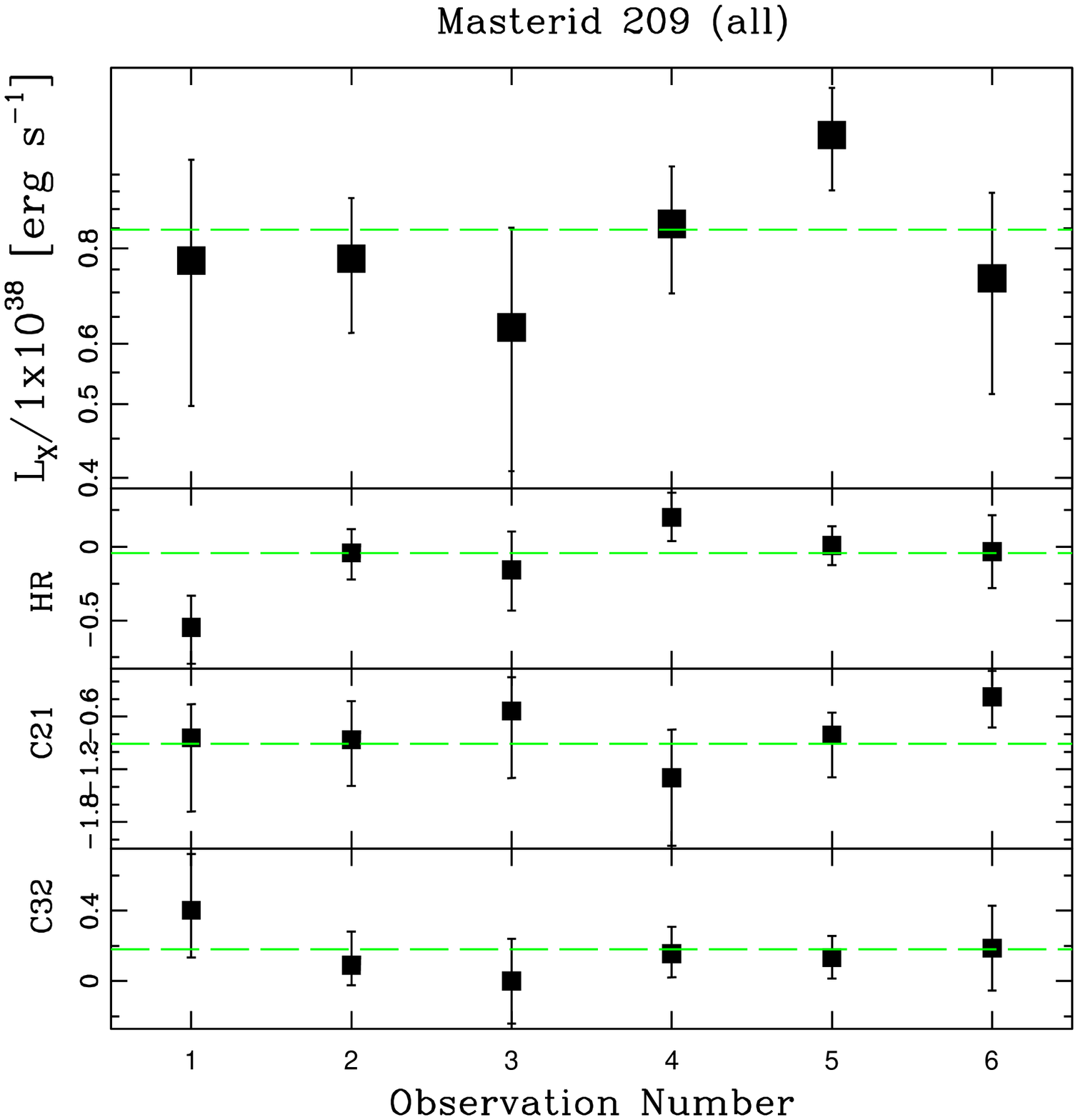}

\end{minipage}\hspace{0.02\linewidth}
\begin{minipage}{0.485\linewidth}
  \centering

    \includegraphics[width=\linewidth]{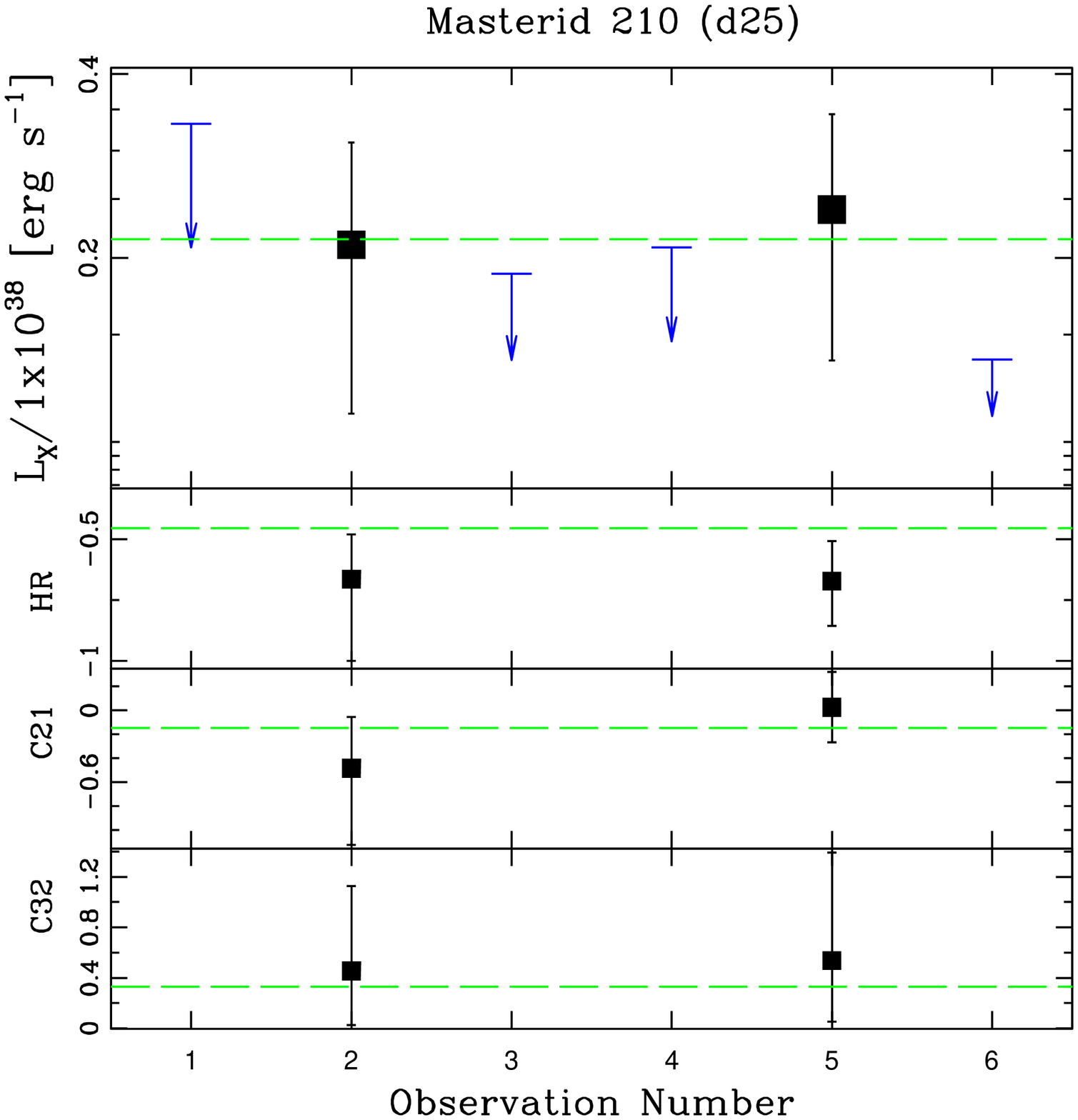}

 \end{minipage}\hspace{0.02\linewidth}

  \begin{minipage}{0.485\linewidth}
  \centering
  
    \includegraphics[width=\linewidth]{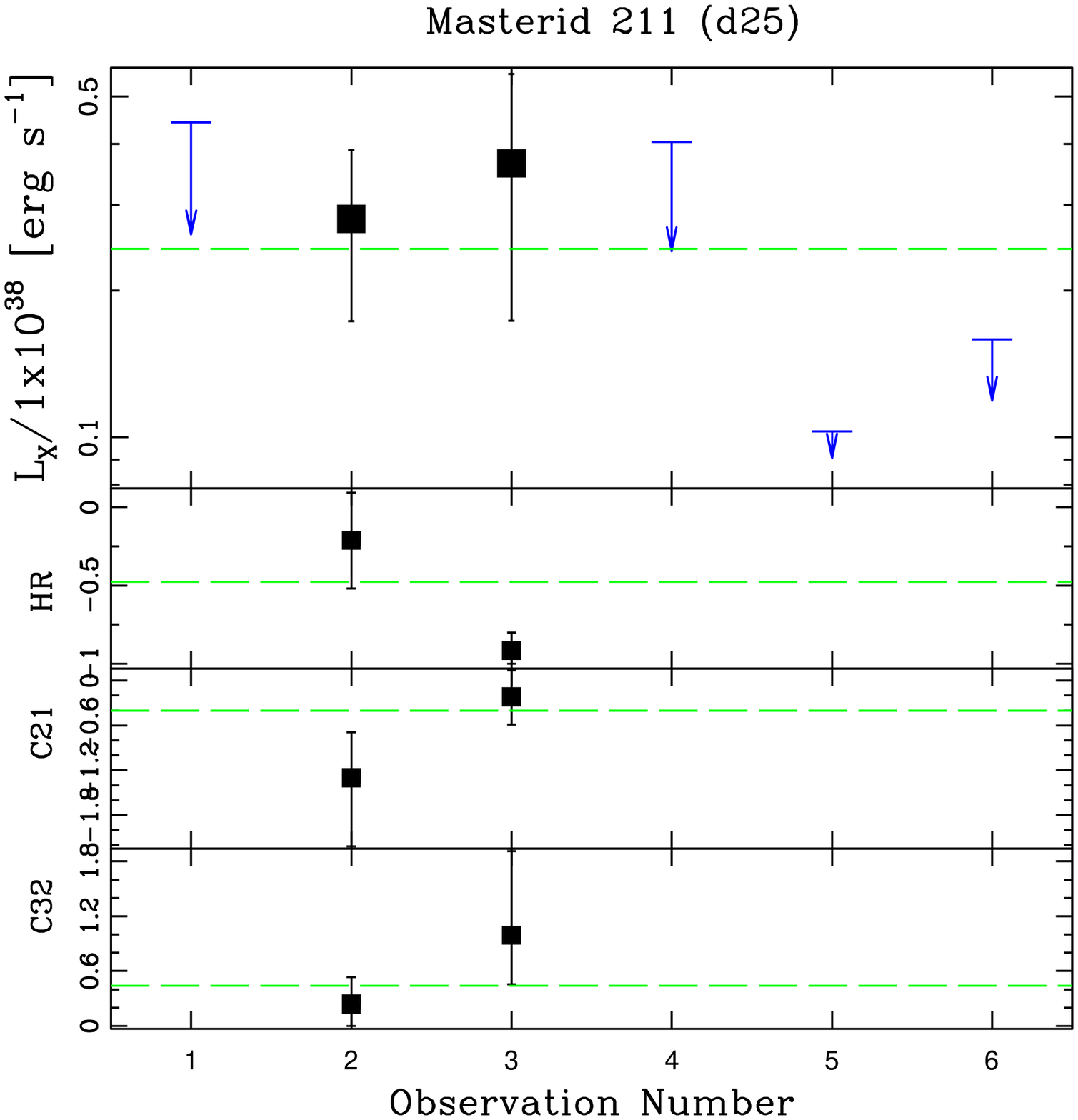}

  \end{minipage}\hspace{0.02\linewidth}
  \begin{minipage}{0.485\linewidth}
  \centering

    \includegraphics[width=\linewidth]{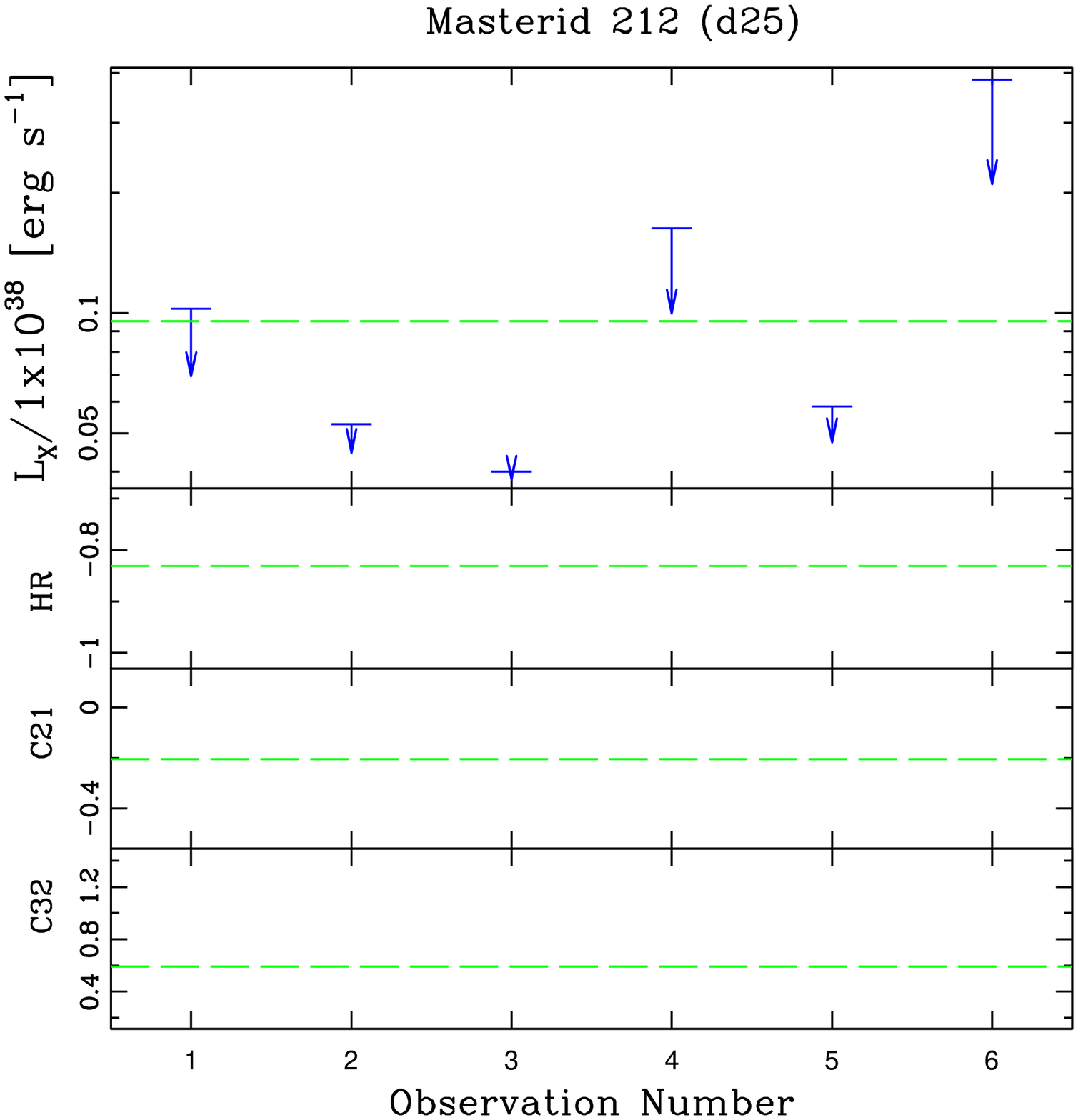}

\end{minipage}\hspace{0.02\linewidth}

\begin{minipage}{0.485\linewidth}
  \centering

    \includegraphics[width=\linewidth]{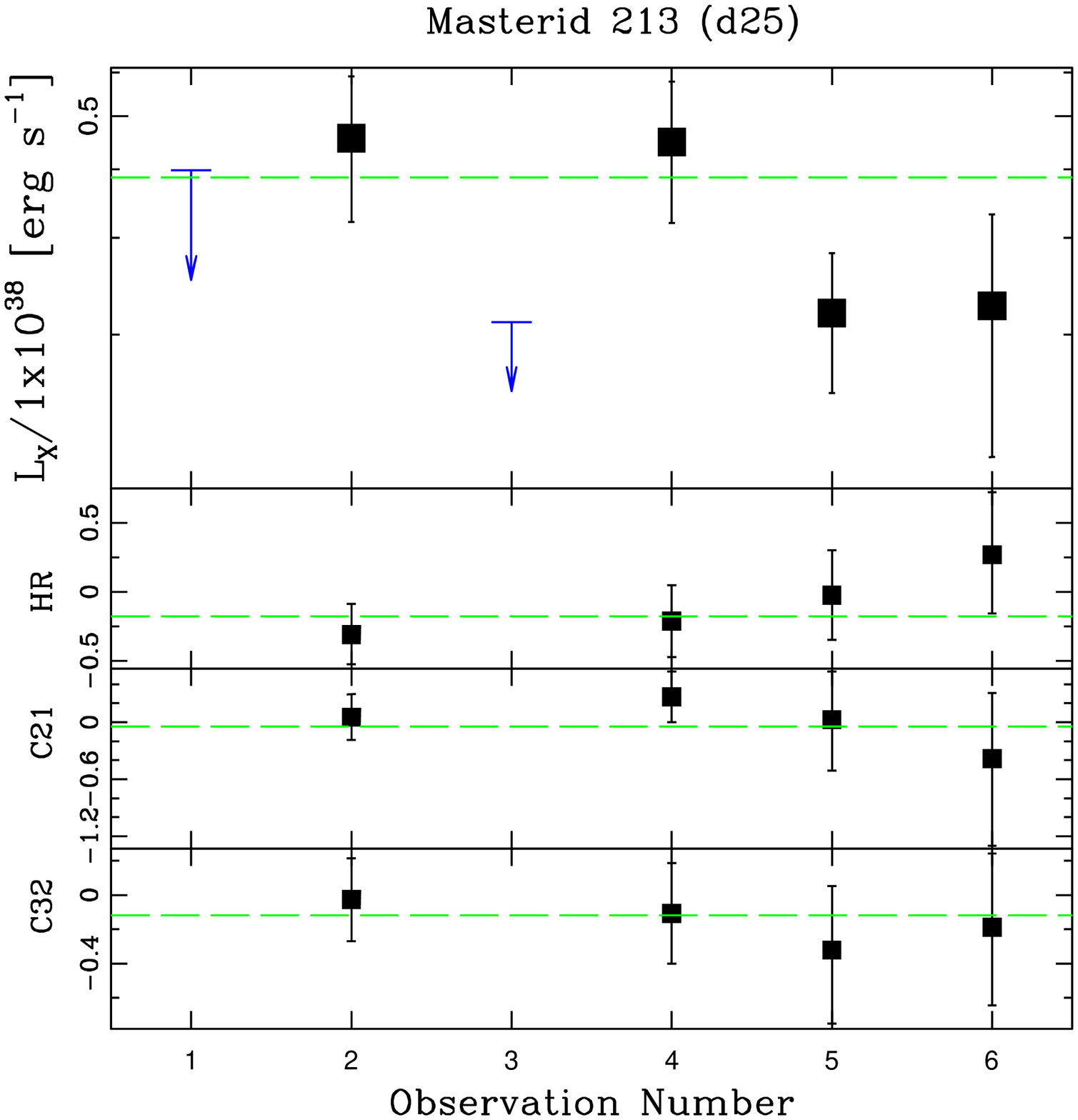}

 \end{minipage}\hspace{0.02\linewidth}
\begin{minipage}{0.485\linewidth}
  \centering
  
    \includegraphics[width=\linewidth]{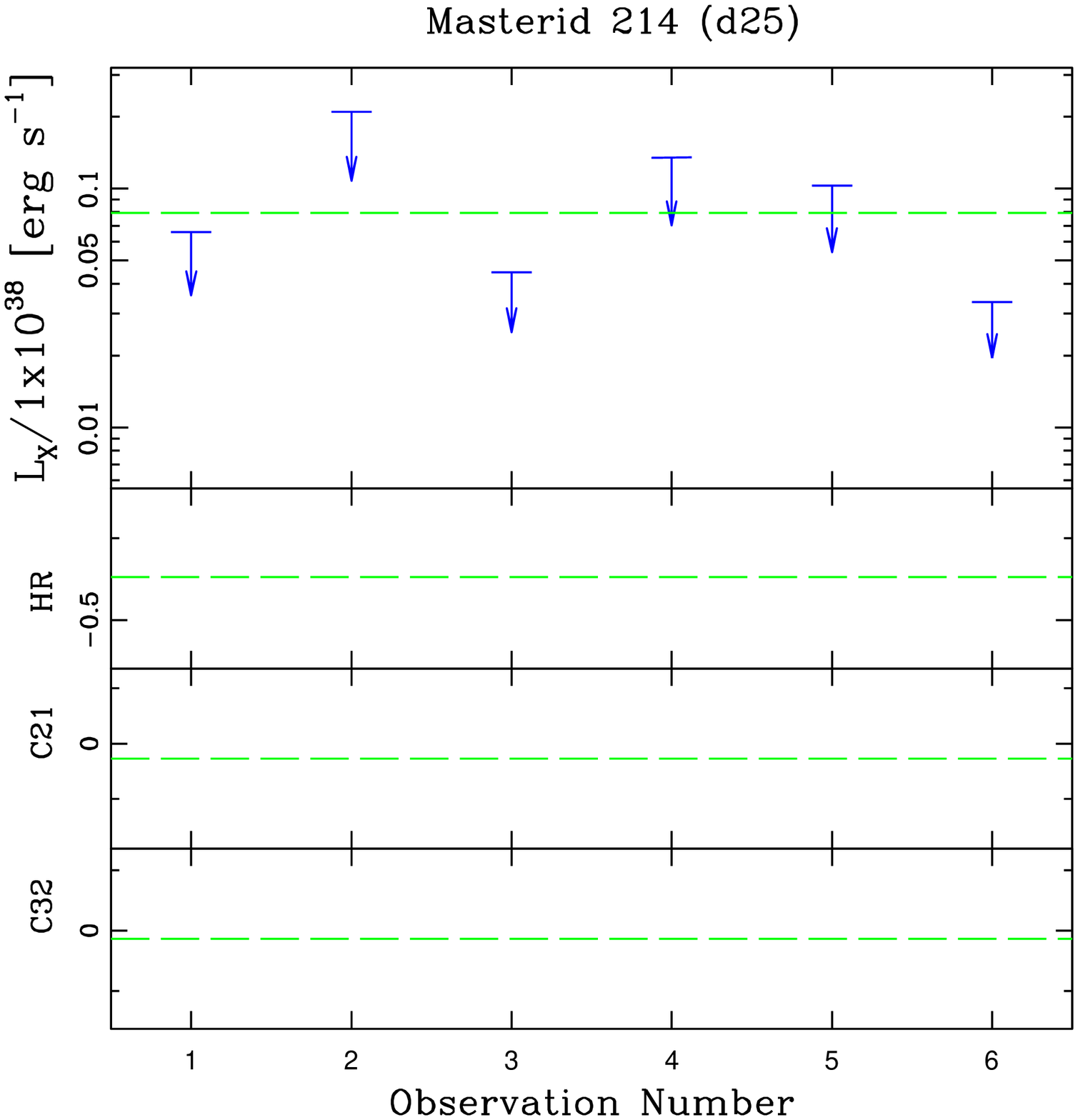}

  \end{minipage}\hspace{0.02\linewidth}
  
\end{figure}

\begin{figure}

  \begin{minipage}{0.485\linewidth}
  \centering

    \includegraphics[width=\linewidth]{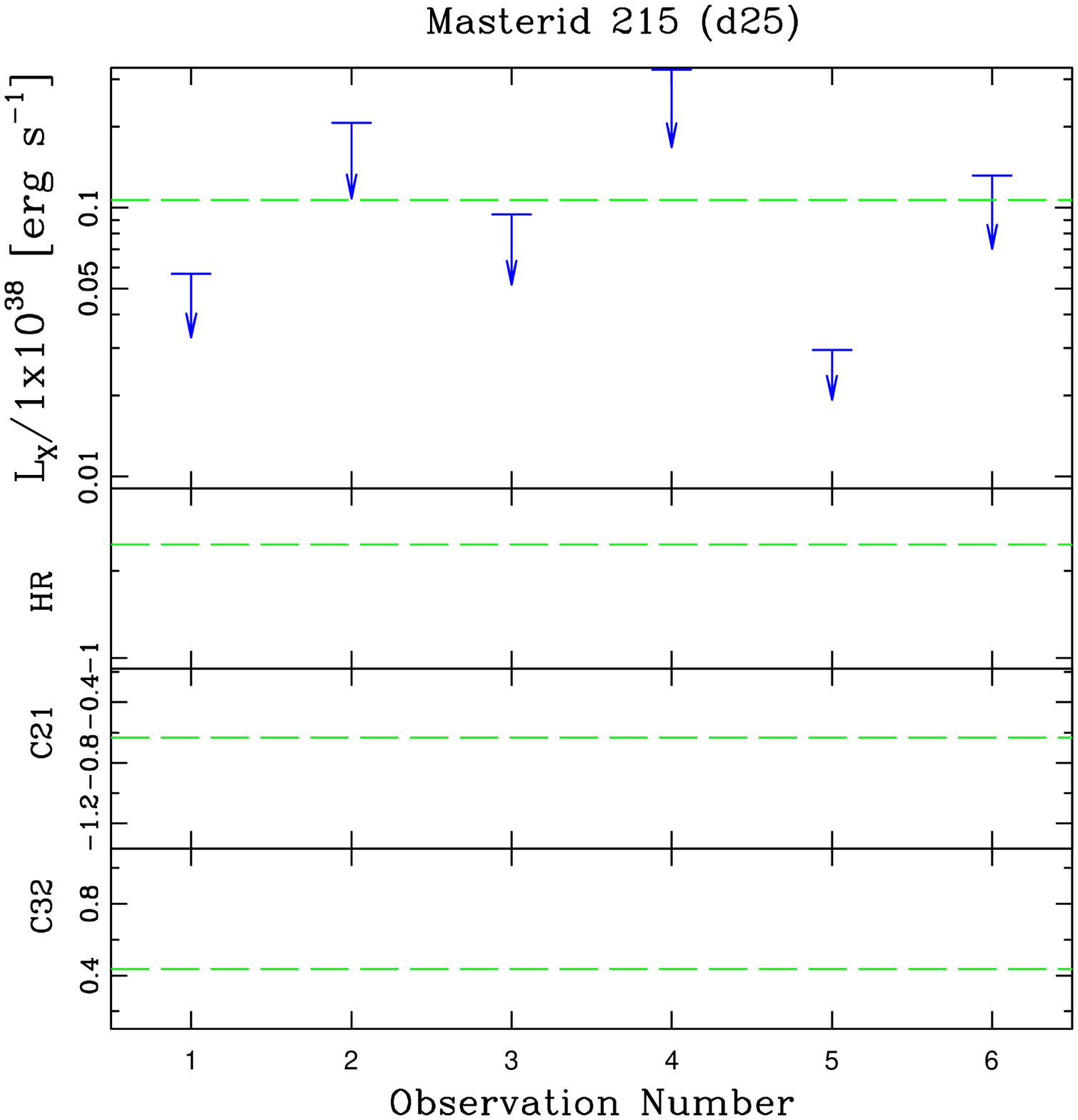}

\end{minipage}\hspace{0.02\linewidth}
\begin{minipage}{0.485\linewidth}
  \centering

    \includegraphics[width=\linewidth]{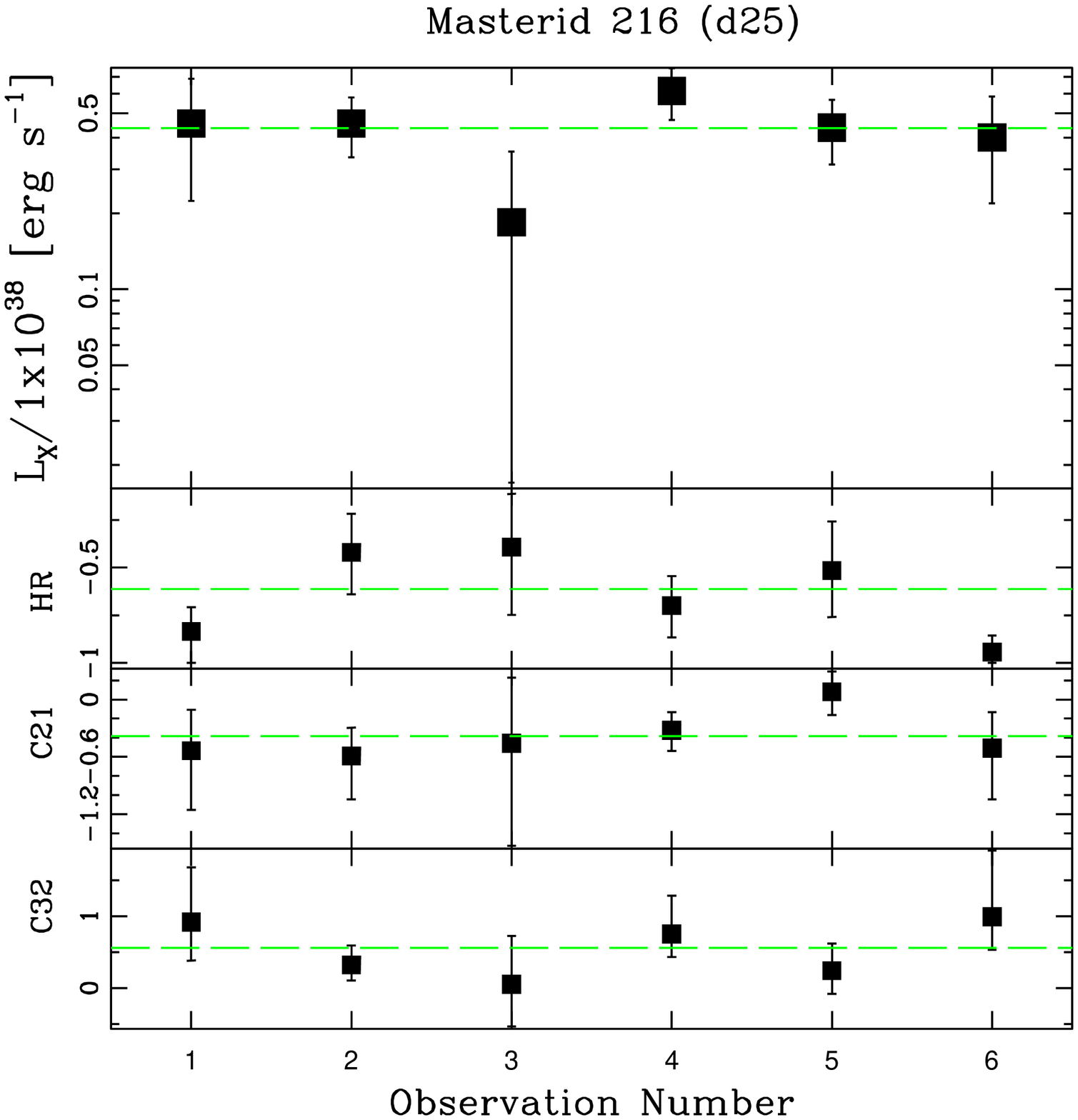}

 \end{minipage}\hspace{0.02\linewidth}

  \begin{minipage}{0.485\linewidth}
  \centering
  
    \includegraphics[width=\linewidth]{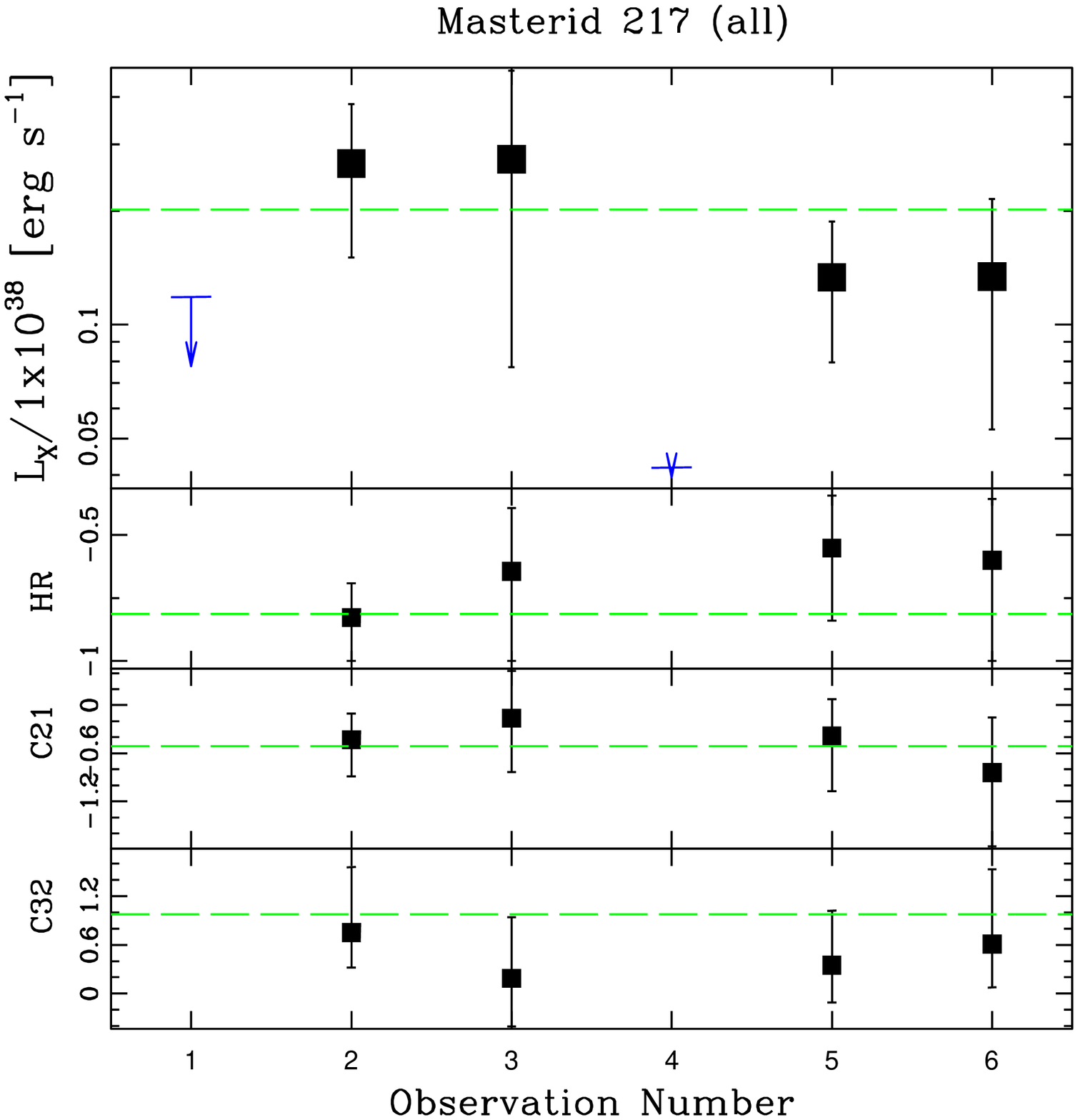}

  \end{minipage}\hspace{0.02\linewidth}
  \begin{minipage}{0.485\linewidth}
  \centering

    \includegraphics[width=\linewidth]{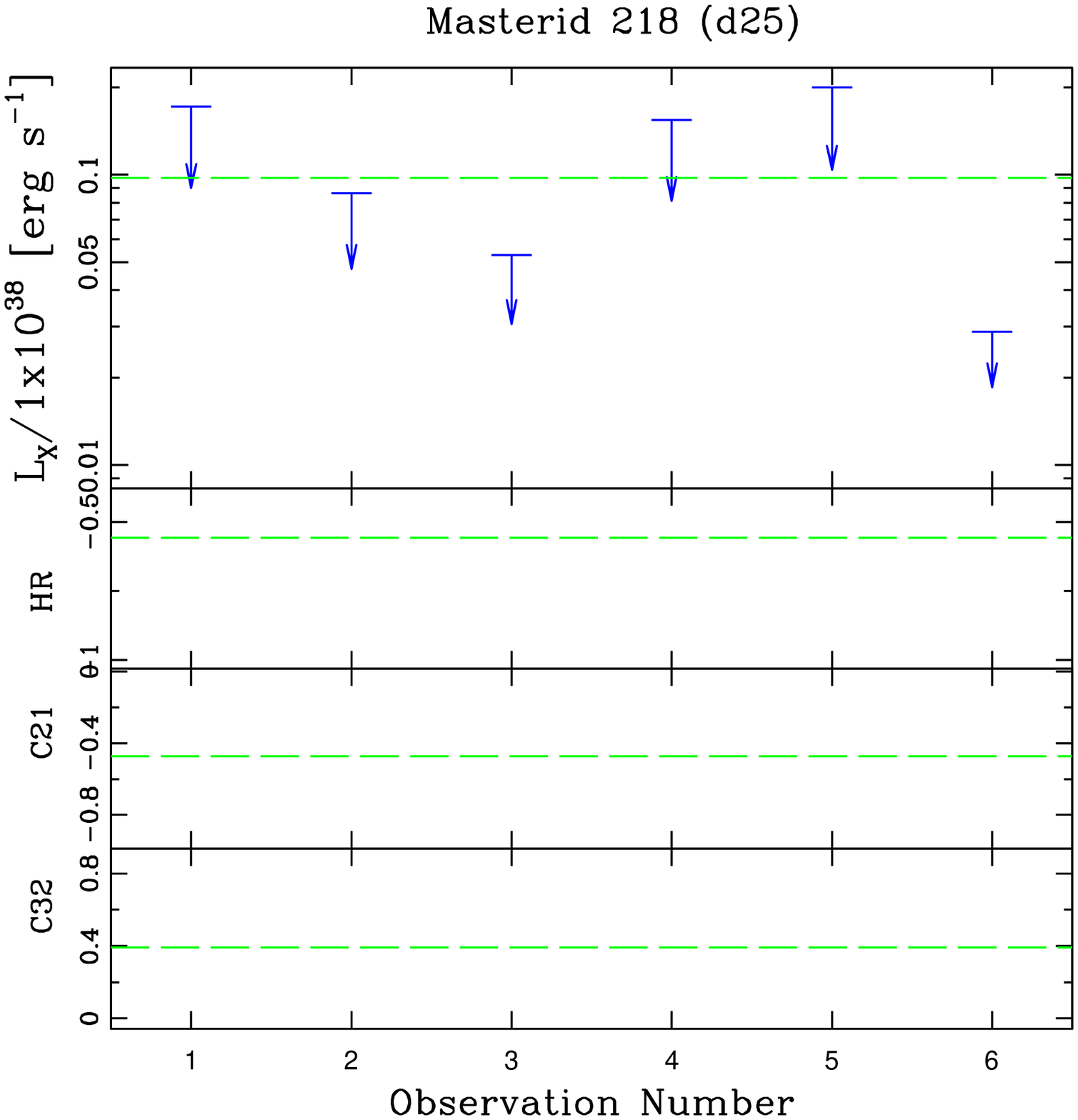}

\end{minipage}\hspace{0.02\linewidth}

\begin{minipage}{0.485\linewidth}
  \centering

    \includegraphics[width=\linewidth]{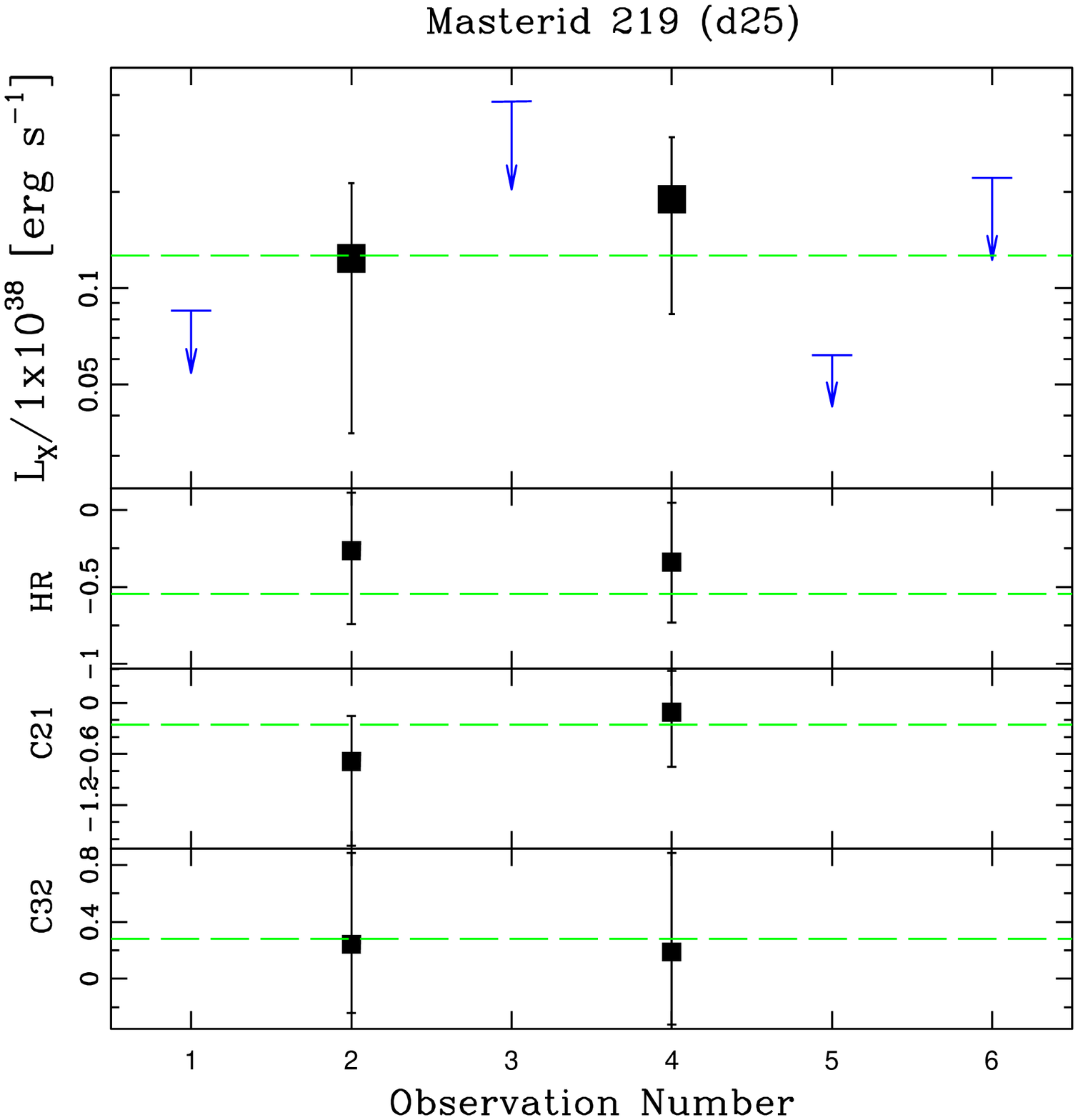}

 \end{minipage}\hspace{0.02\linewidth}
\begin{minipage}{0.485\linewidth}
  \centering
  
    \includegraphics[width=\linewidth]{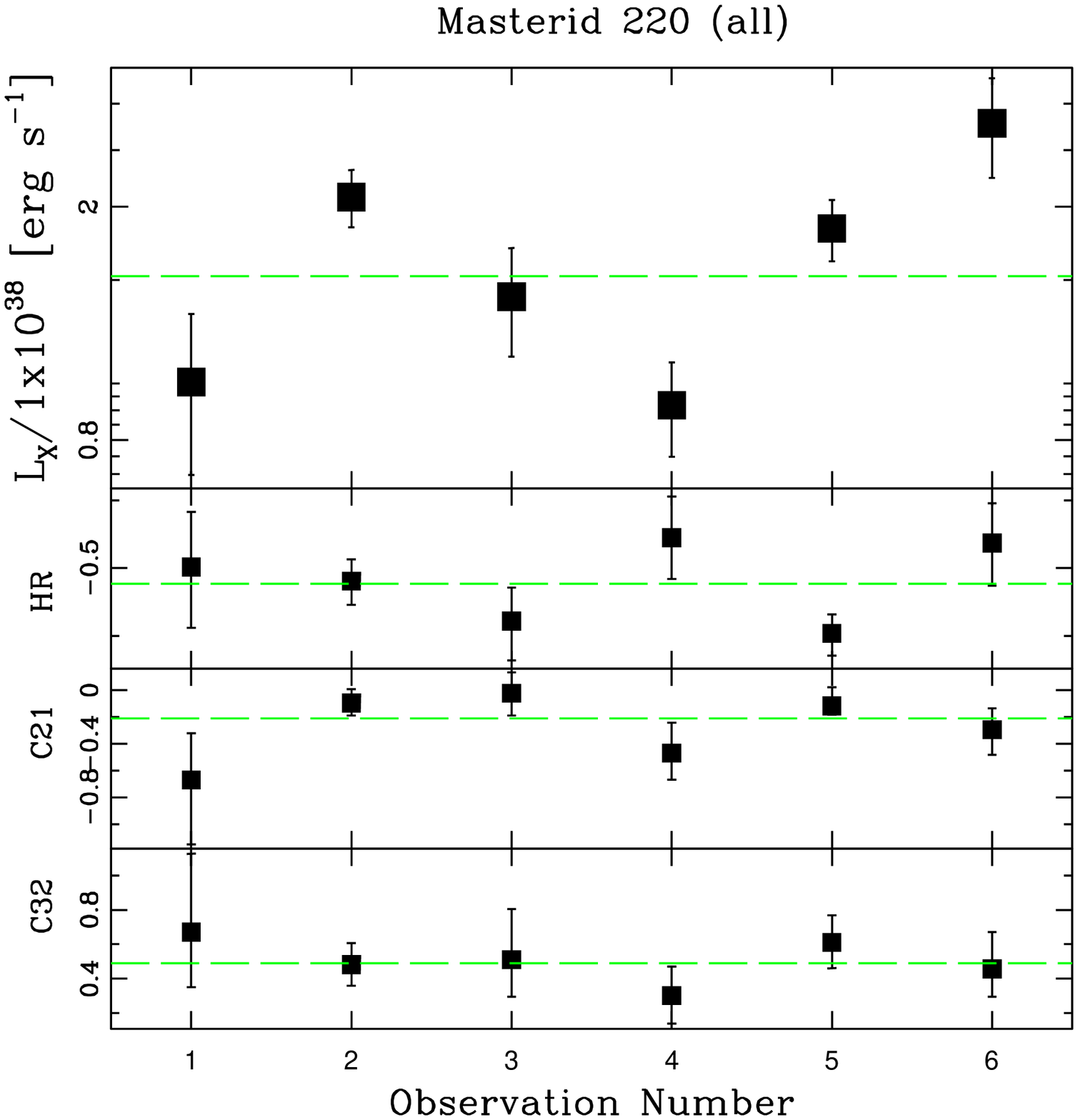}

  \end{minipage}\hspace{0.02\linewidth}
  
\end{figure}

\begin{figure}

  \begin{minipage}{0.485\linewidth}
  \centering

    \includegraphics[width=\linewidth]{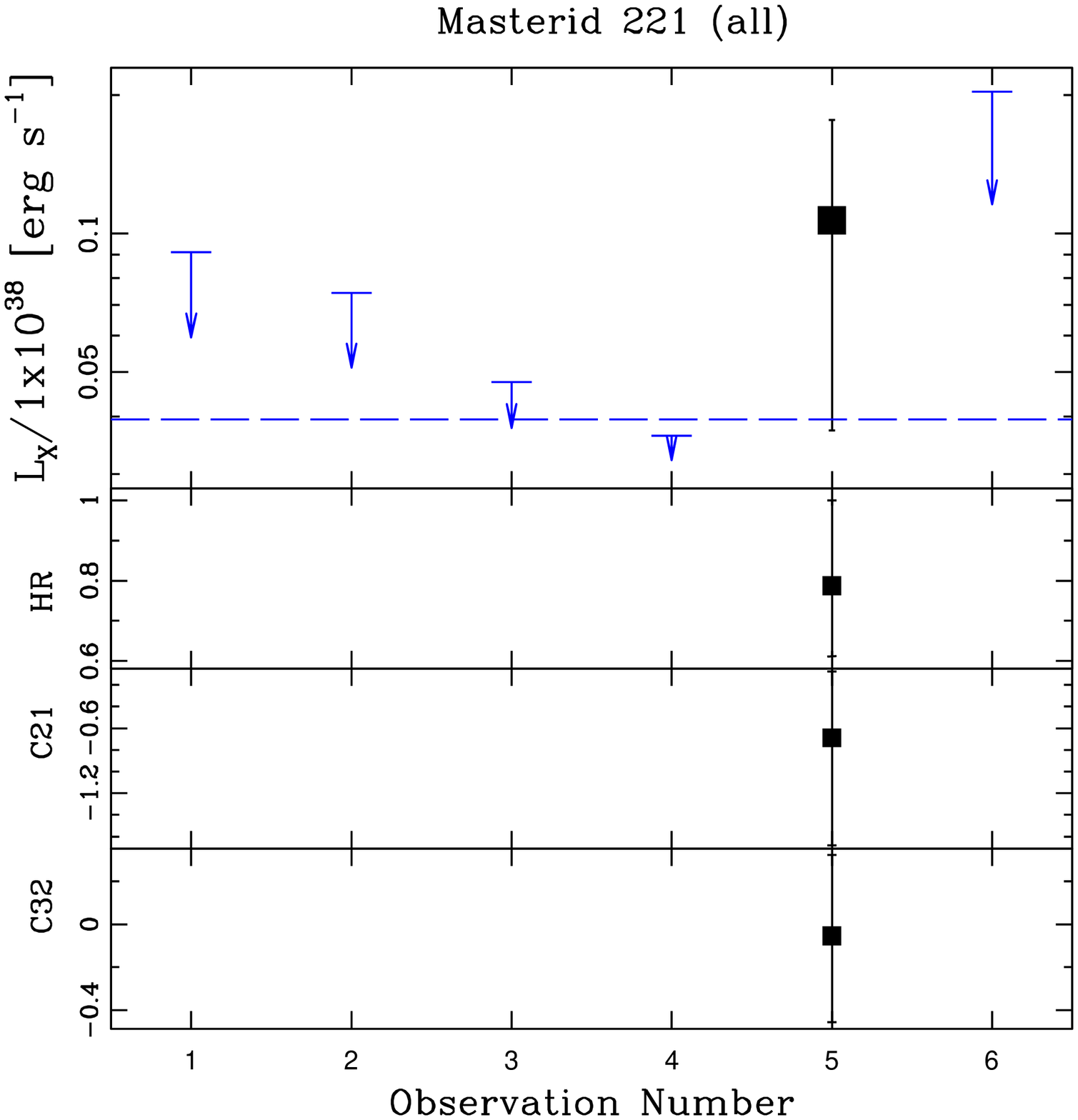}

\end{minipage}\hspace{0.02\linewidth}
\begin{minipage}{0.485\linewidth}
  \centering

    \includegraphics[width=\linewidth]{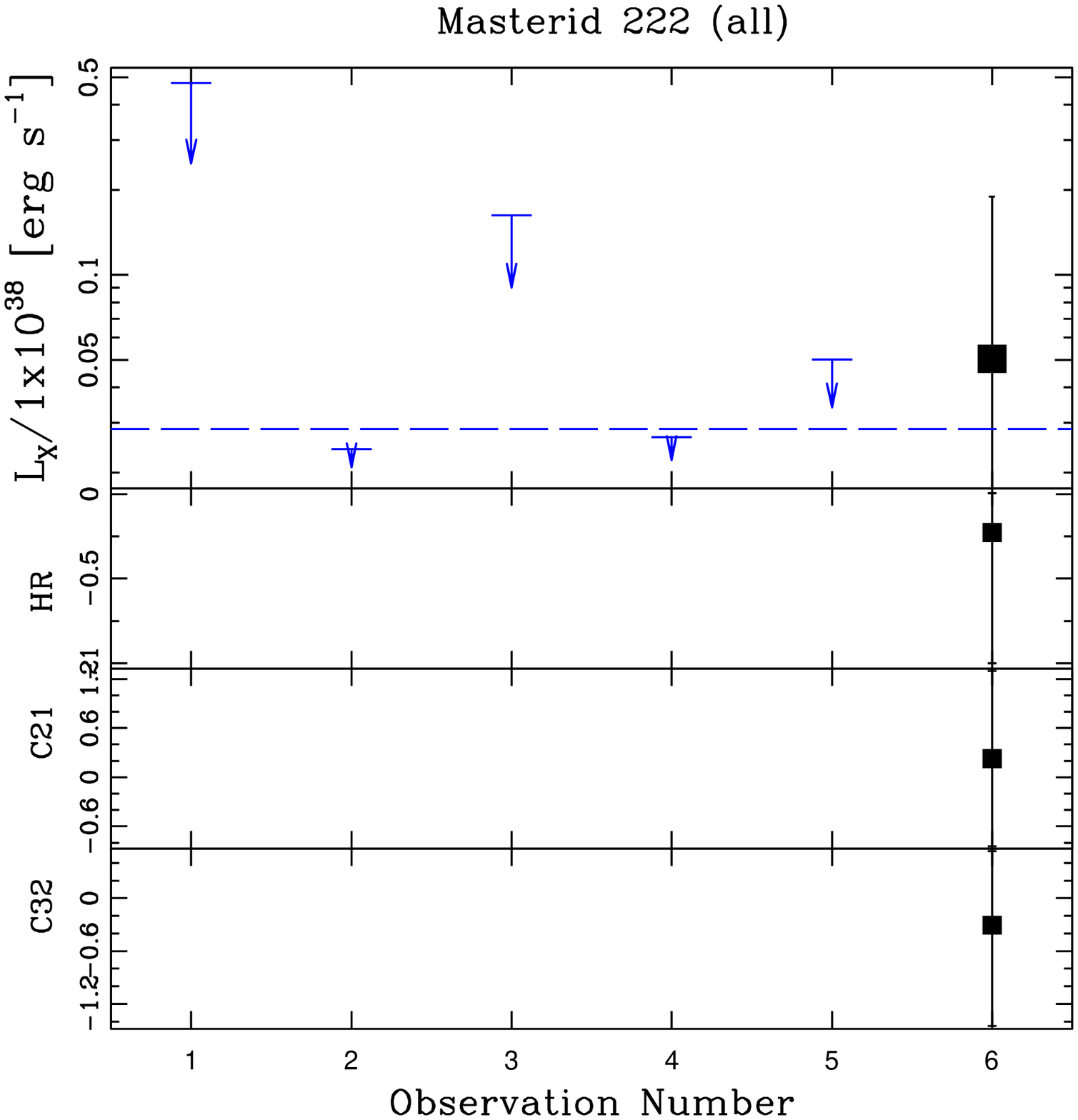}

 \end{minipage}\hspace{0.02\linewidth}

  \begin{minipage}{0.485\linewidth}
  \centering
  
    \includegraphics[width=\linewidth]{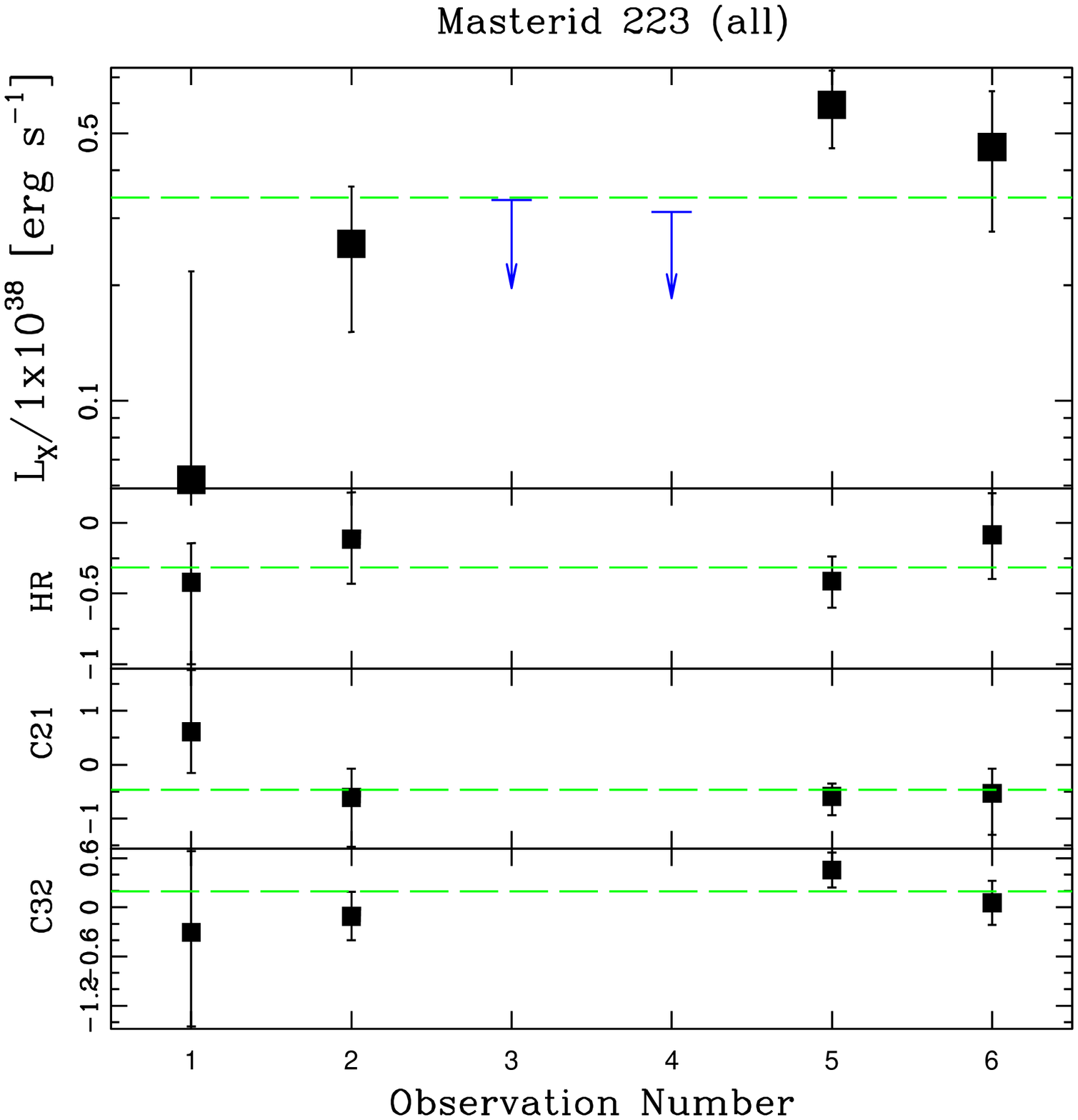}

  \end{minipage}\hspace{0.02\linewidth}
  \begin{minipage}{0.485\linewidth}
  \centering

    \includegraphics[width=\linewidth]{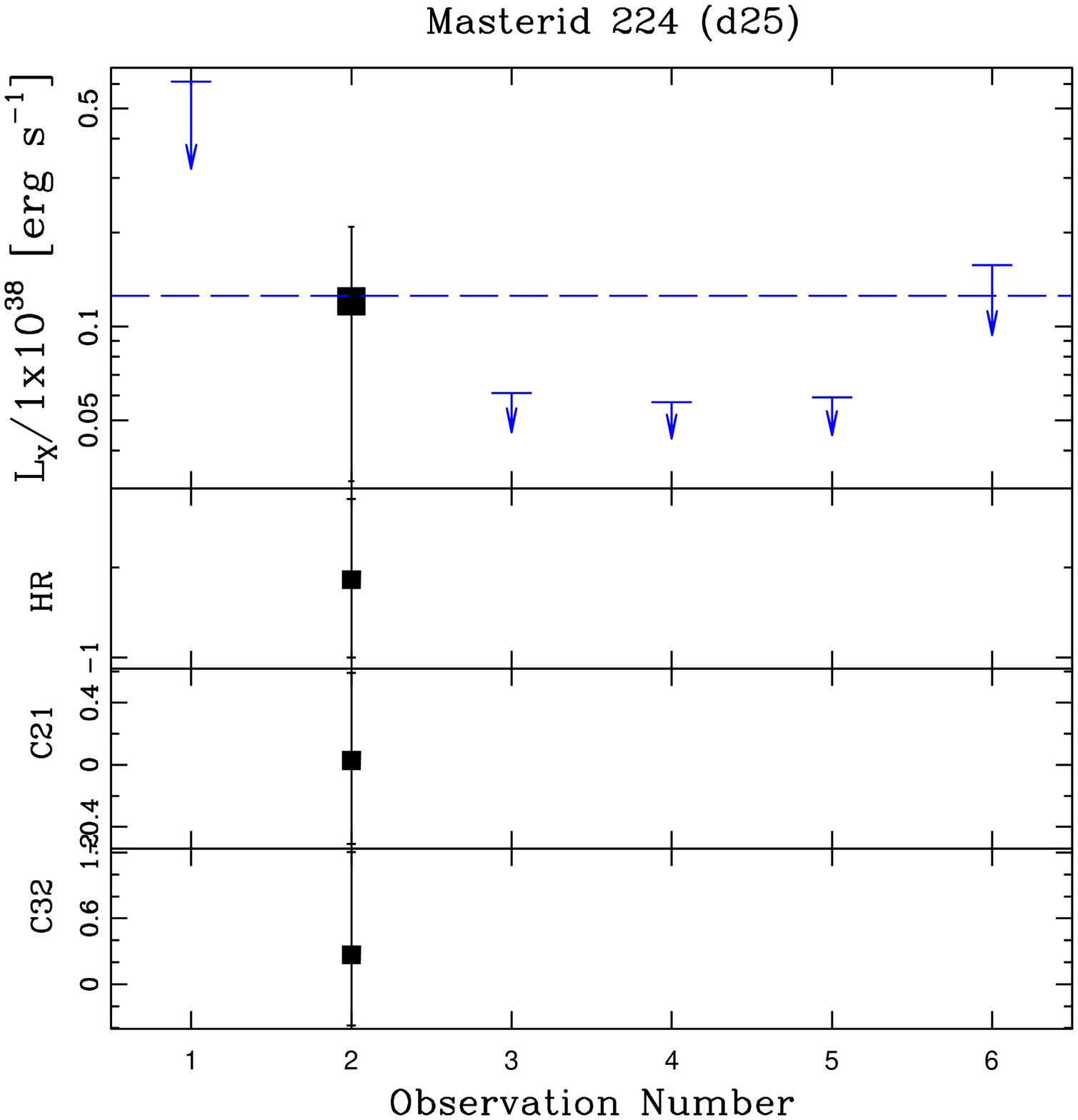}

\end{minipage}\hspace{0.02\linewidth}

\begin{minipage}{0.485\linewidth}
  \centering

    \includegraphics[width=\linewidth]{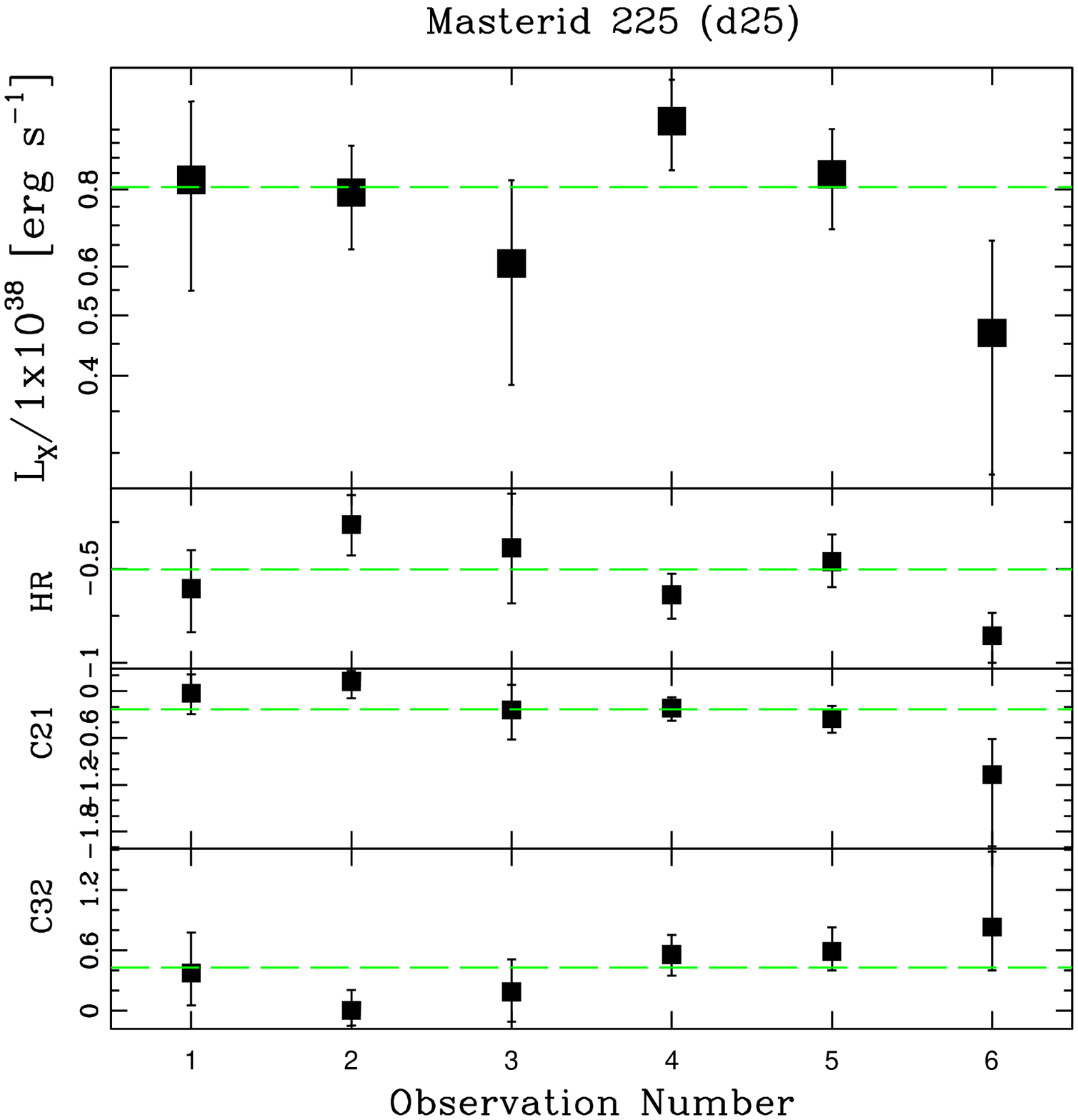}

 \end{minipage}\hspace{0.02\linewidth}
\begin{minipage}{0.485\linewidth}
  \centering
  
    \includegraphics[width=\linewidth]{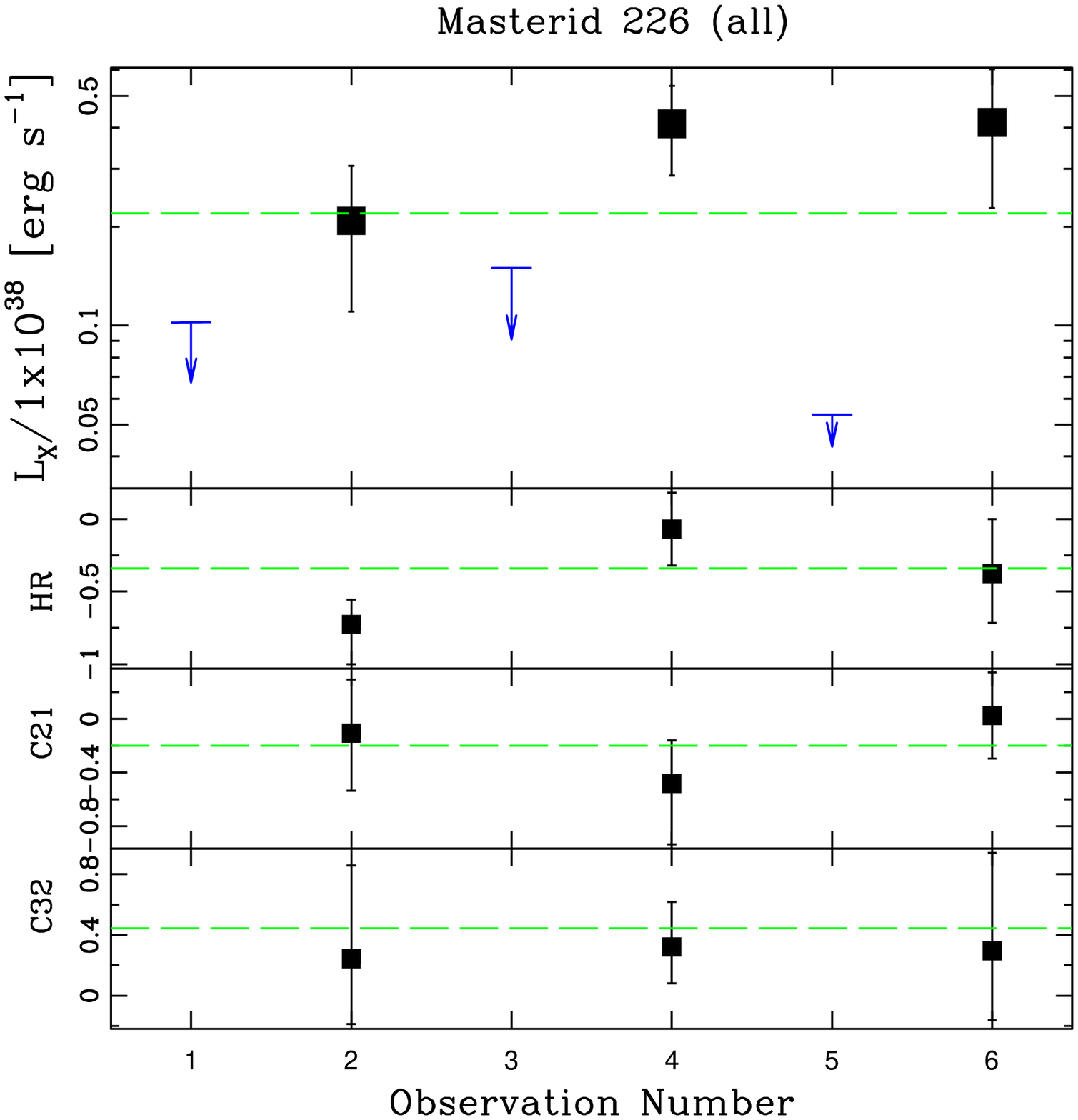}

  \end{minipage}\hspace{0.02\linewidth}
  
\end{figure}

\begin{figure}

  \begin{minipage}{0.485\linewidth}
  \centering

    \includegraphics[width=\linewidth]{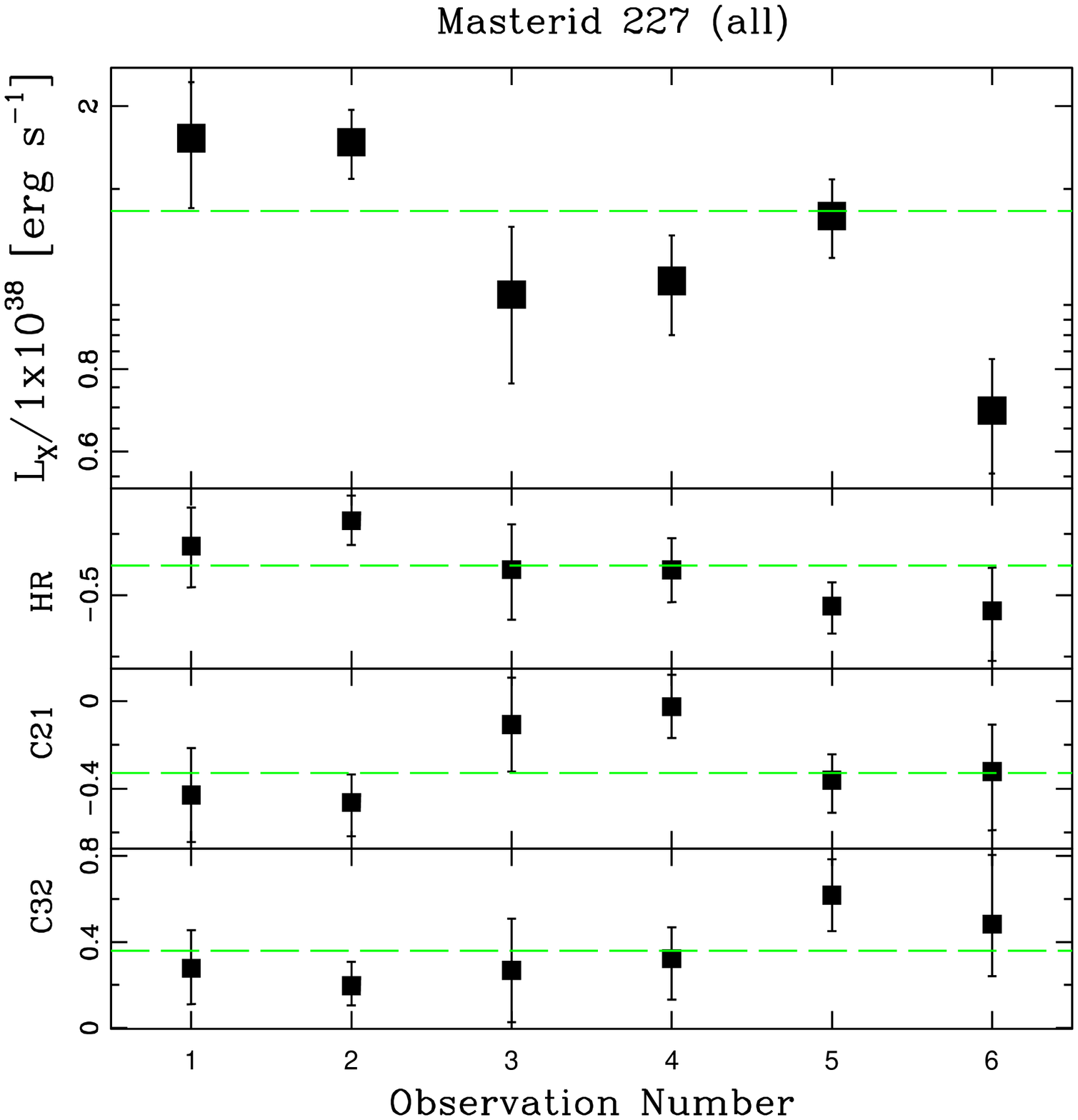}

\end{minipage}\hspace{0.02\linewidth}
\begin{minipage}{0.485\linewidth}
  \centering

    \includegraphics[width=\linewidth]{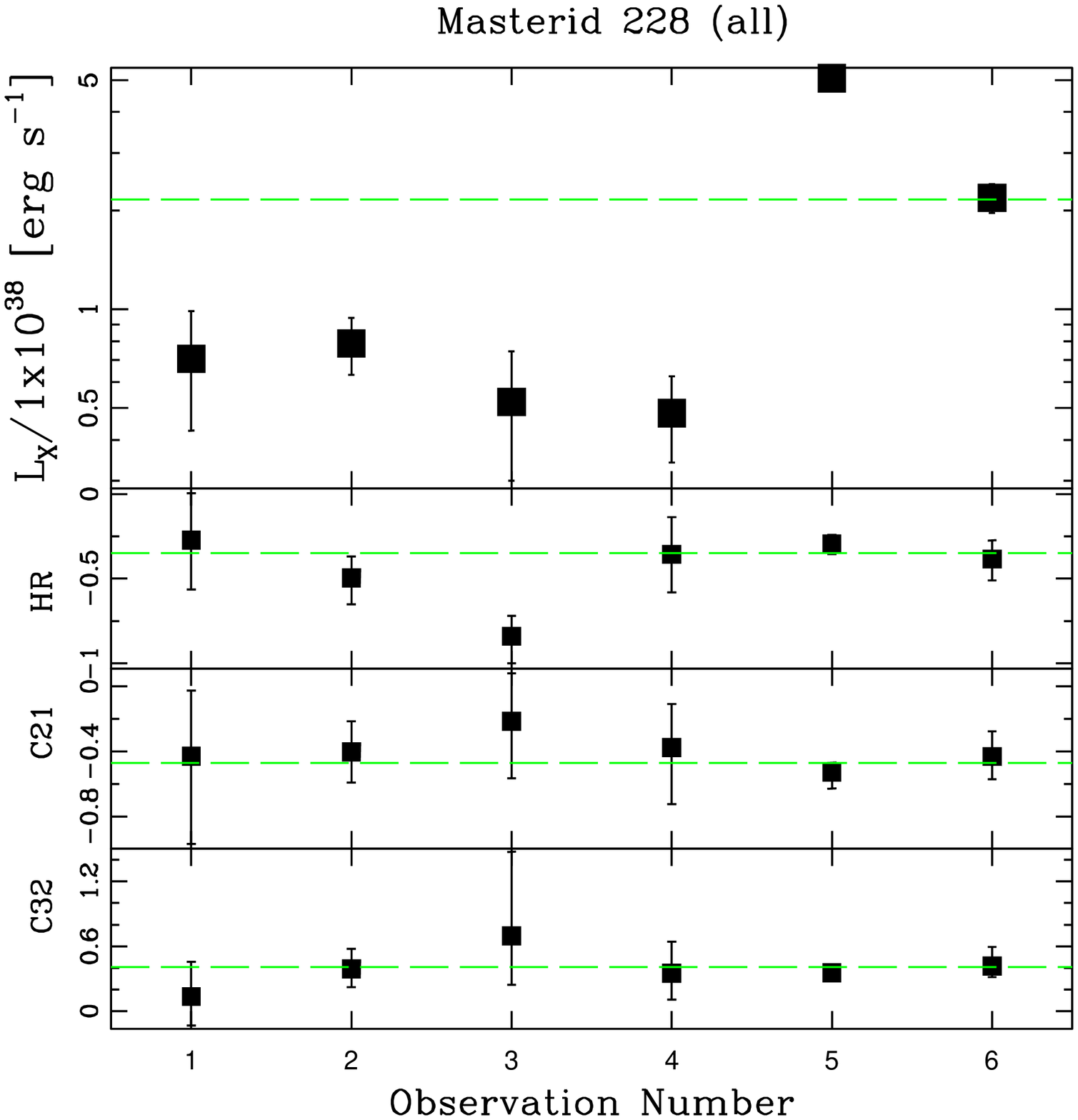}

 \end{minipage}\hspace{0.02\linewidth}

  \begin{minipage}{0.485\linewidth}
  \centering
  
    \includegraphics[width=\linewidth]{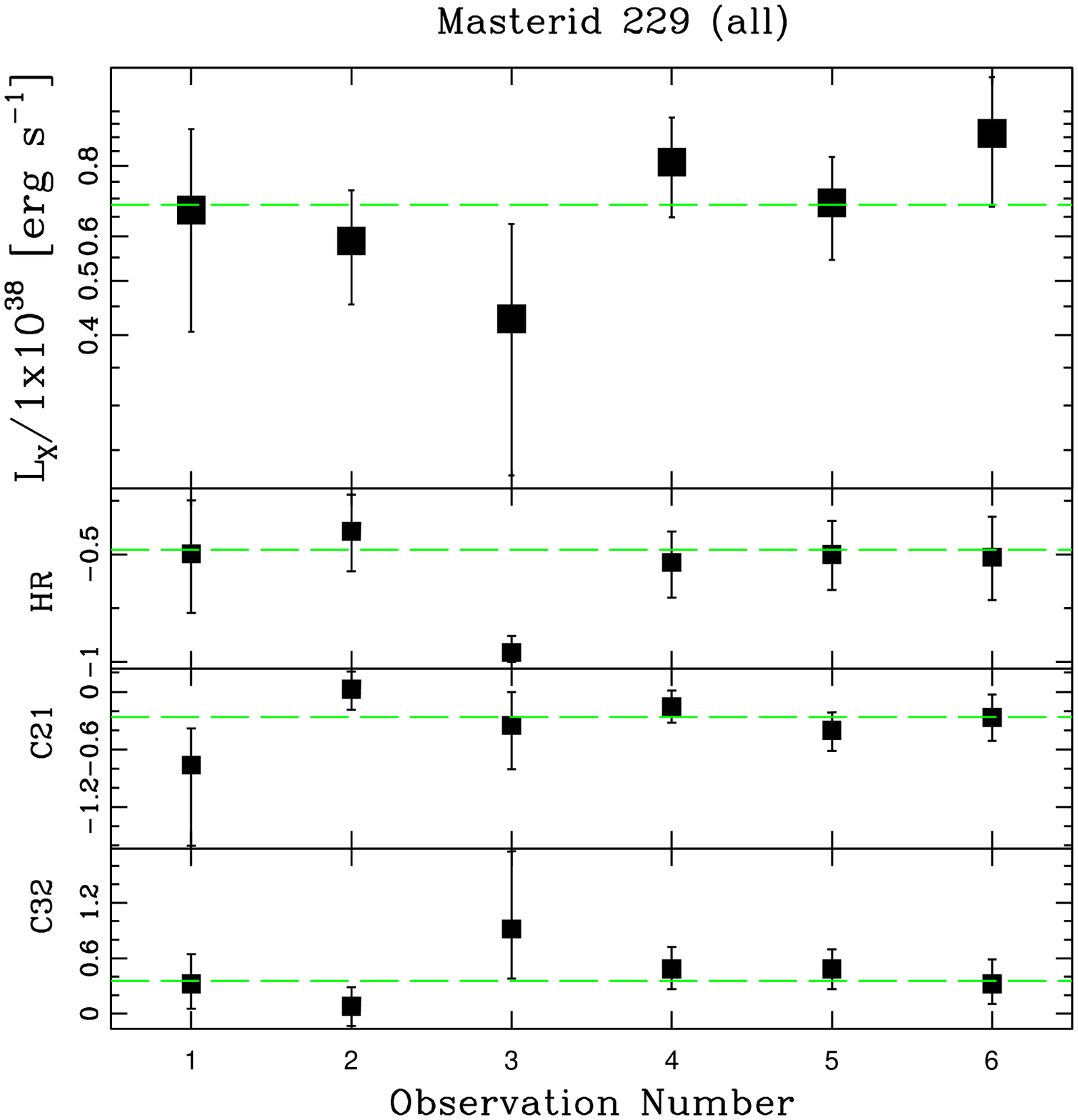}

  \end{minipage}\hspace{0.02\linewidth}
  \begin{minipage}{0.485\linewidth}
  \centering

    \includegraphics[width=\linewidth]{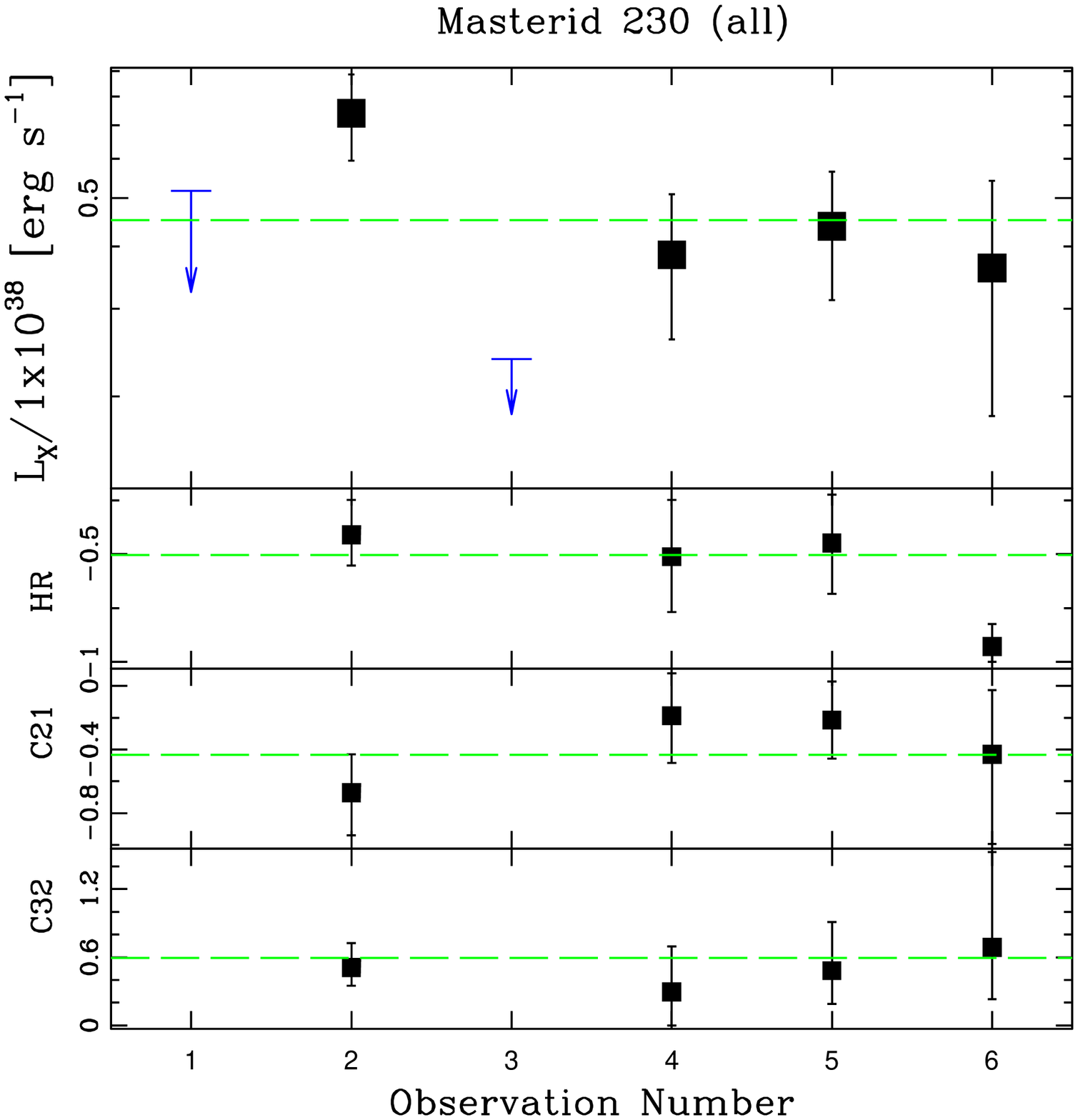}

\end{minipage}\hspace{0.02\linewidth}

\begin{minipage}{0.485\linewidth}
  \centering

    \includegraphics[width=\linewidth]{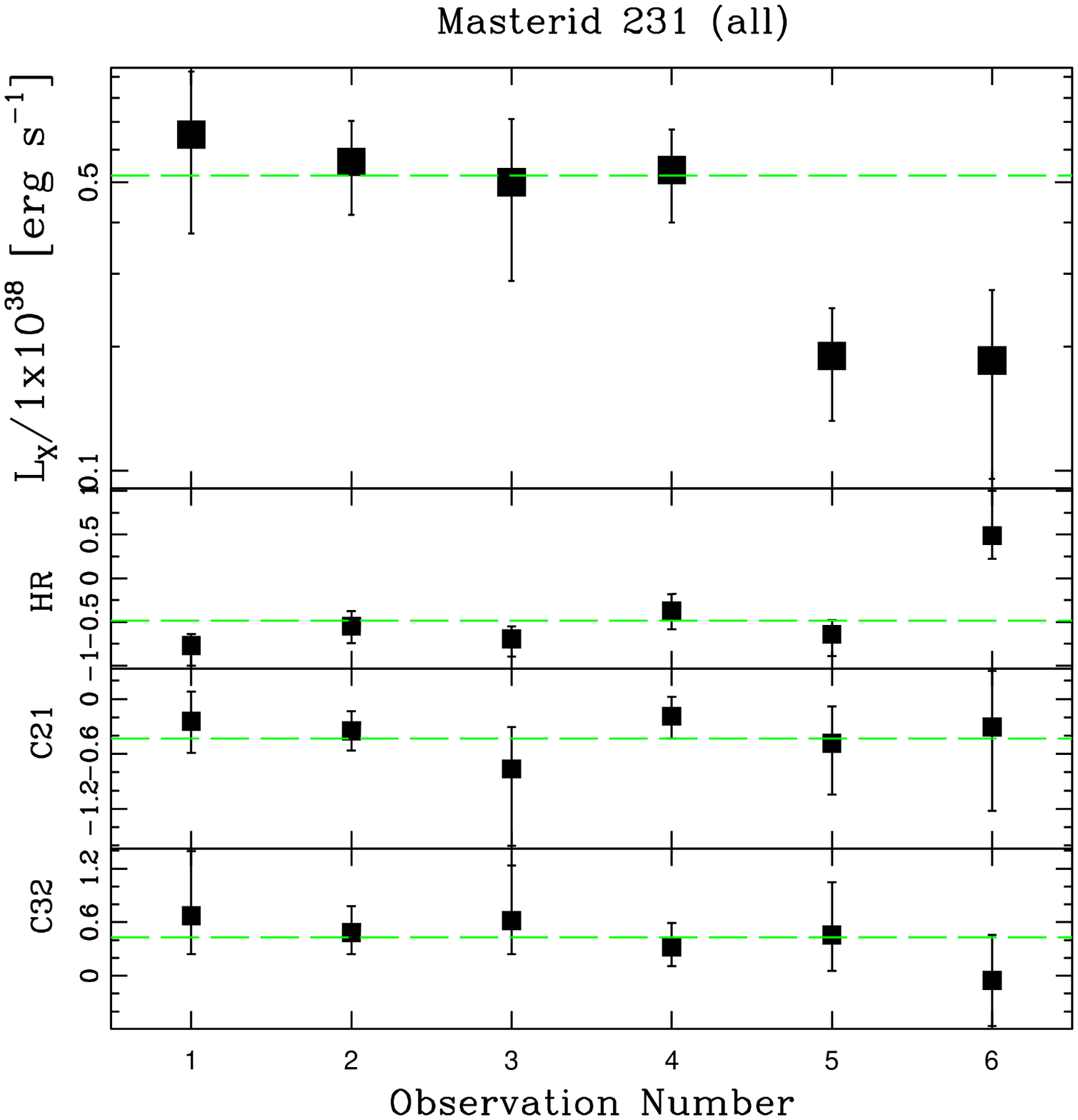}

 \end{minipage}\hspace{0.02\linewidth}
\begin{minipage}{0.485\linewidth}
  \centering
  
    \includegraphics[width=\linewidth]{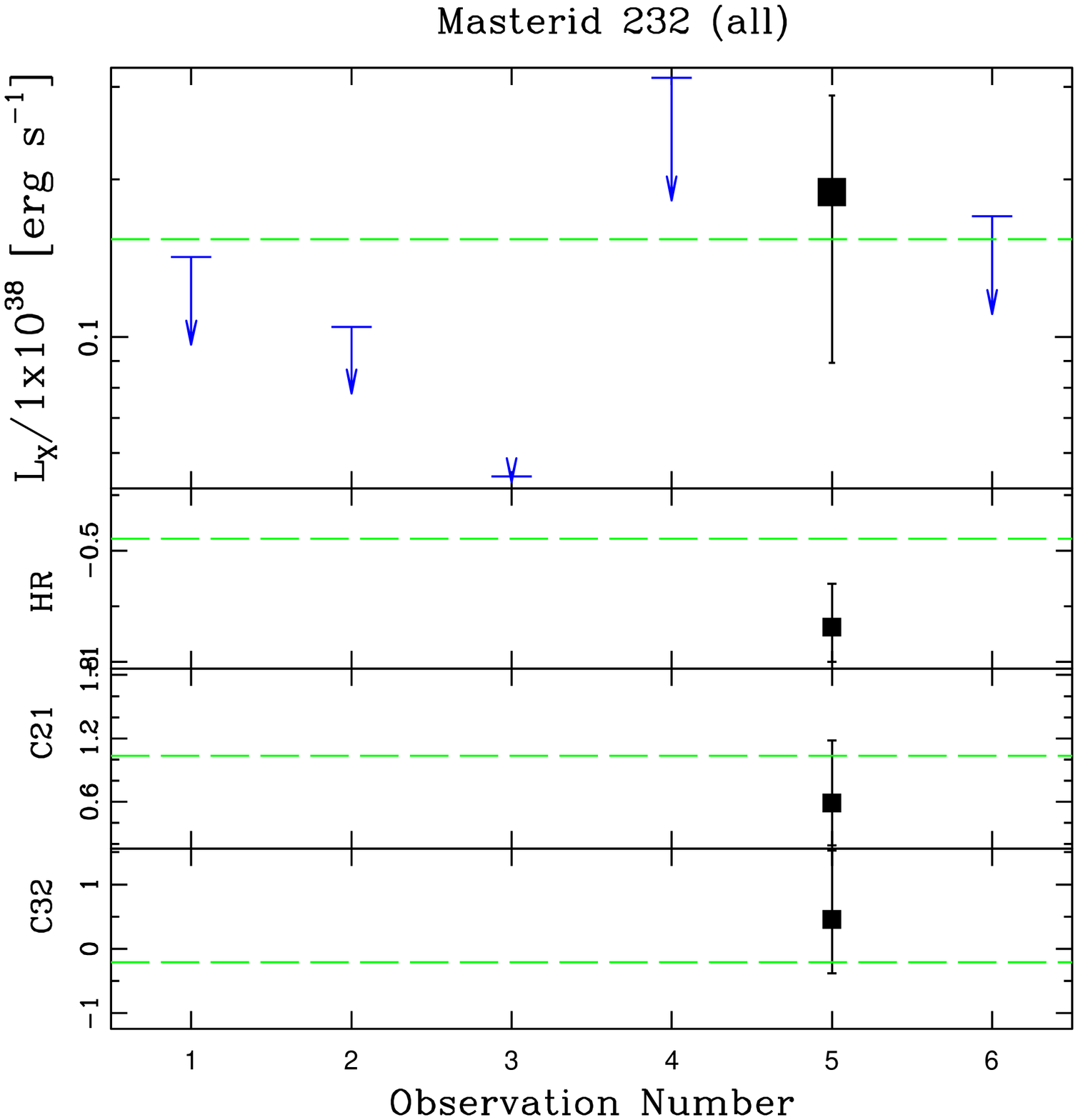}

  \end{minipage}\hspace{0.02\linewidth}
  
\end{figure}

\begin{figure}

  \begin{minipage}{0.485\linewidth}
  \centering

    \includegraphics[width=\linewidth]{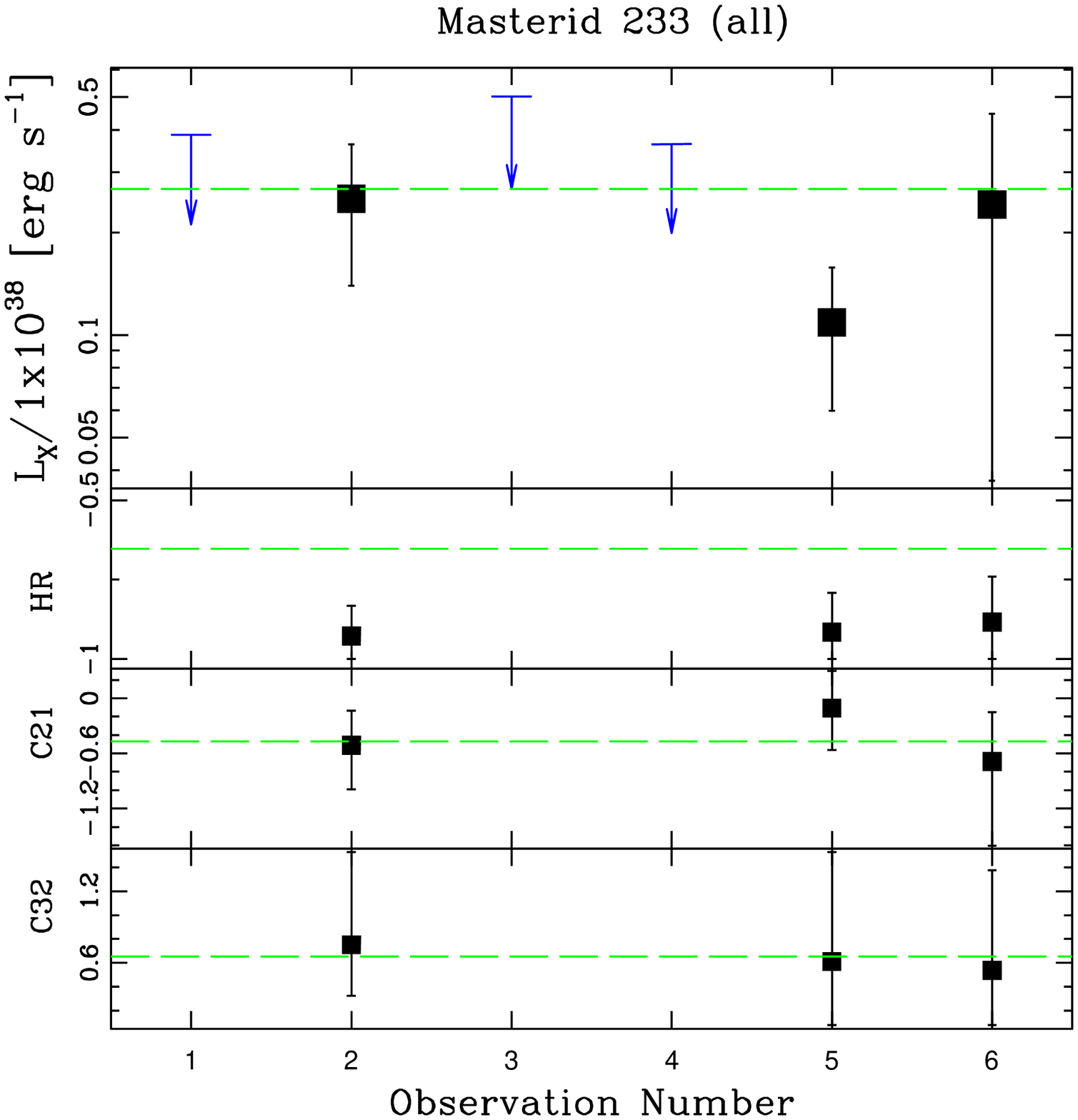}

\end{minipage}\hspace{0.02\linewidth}
\begin{minipage}{0.485\linewidth}
  \centering

    \includegraphics[width=\linewidth]{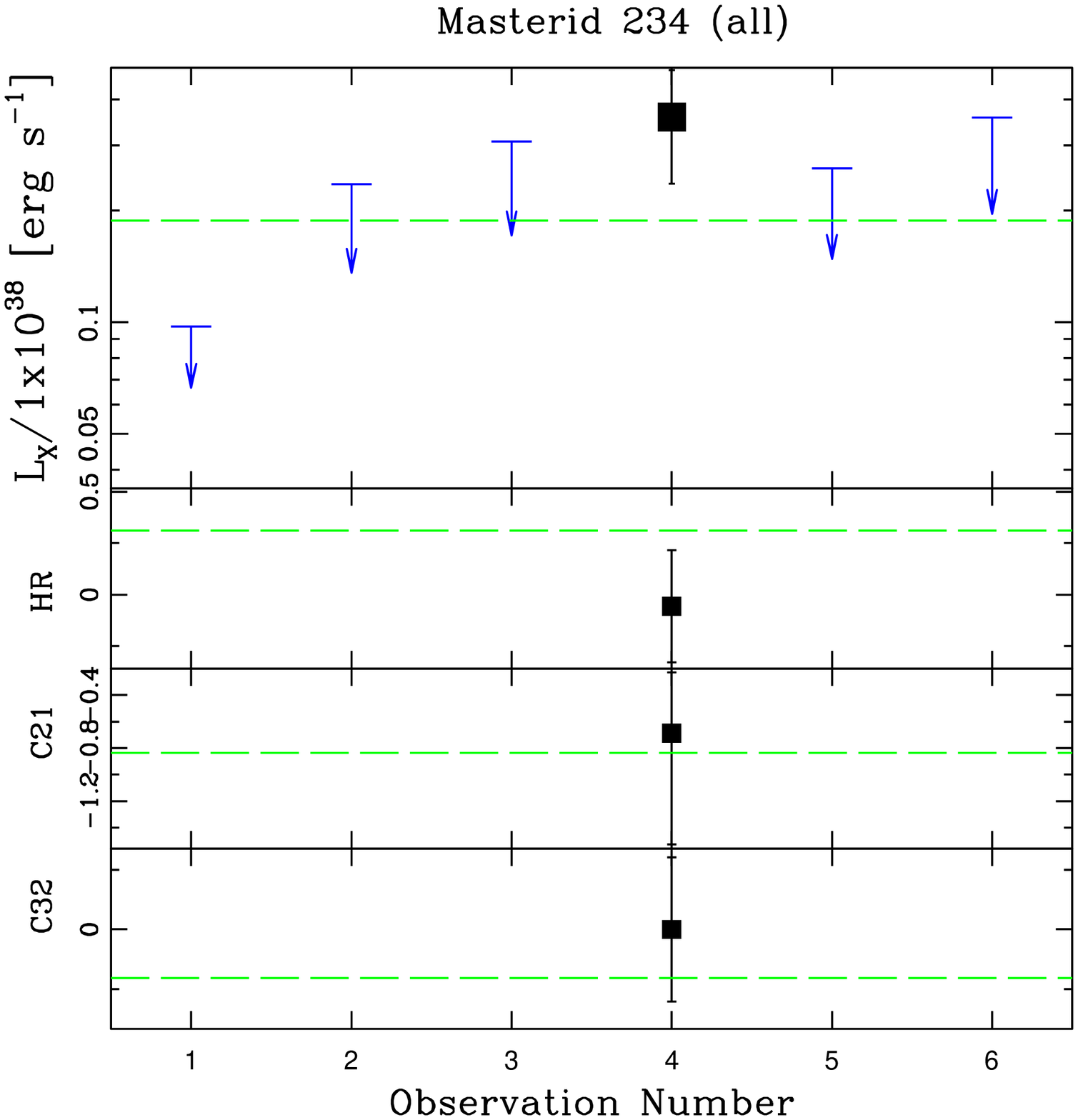}

 \end{minipage}\hspace{0.02\linewidth}

  \begin{minipage}{0.485\linewidth}
  \centering
  
    \includegraphics[width=\linewidth]{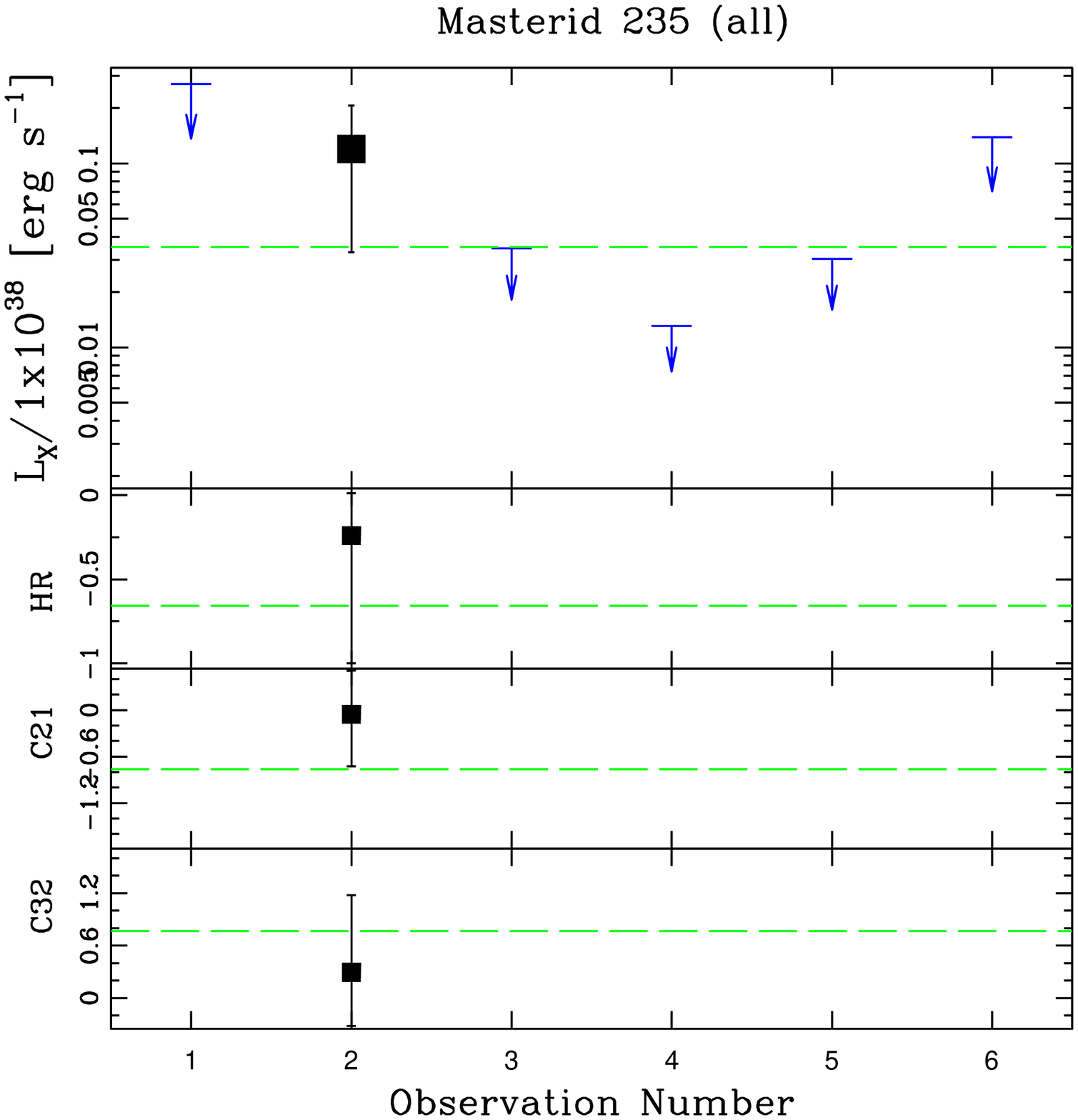}

  \end{minipage}\hspace{0.02\linewidth}
  \begin{minipage}{0.485\linewidth}
  \centering

    \includegraphics[width=\linewidth]{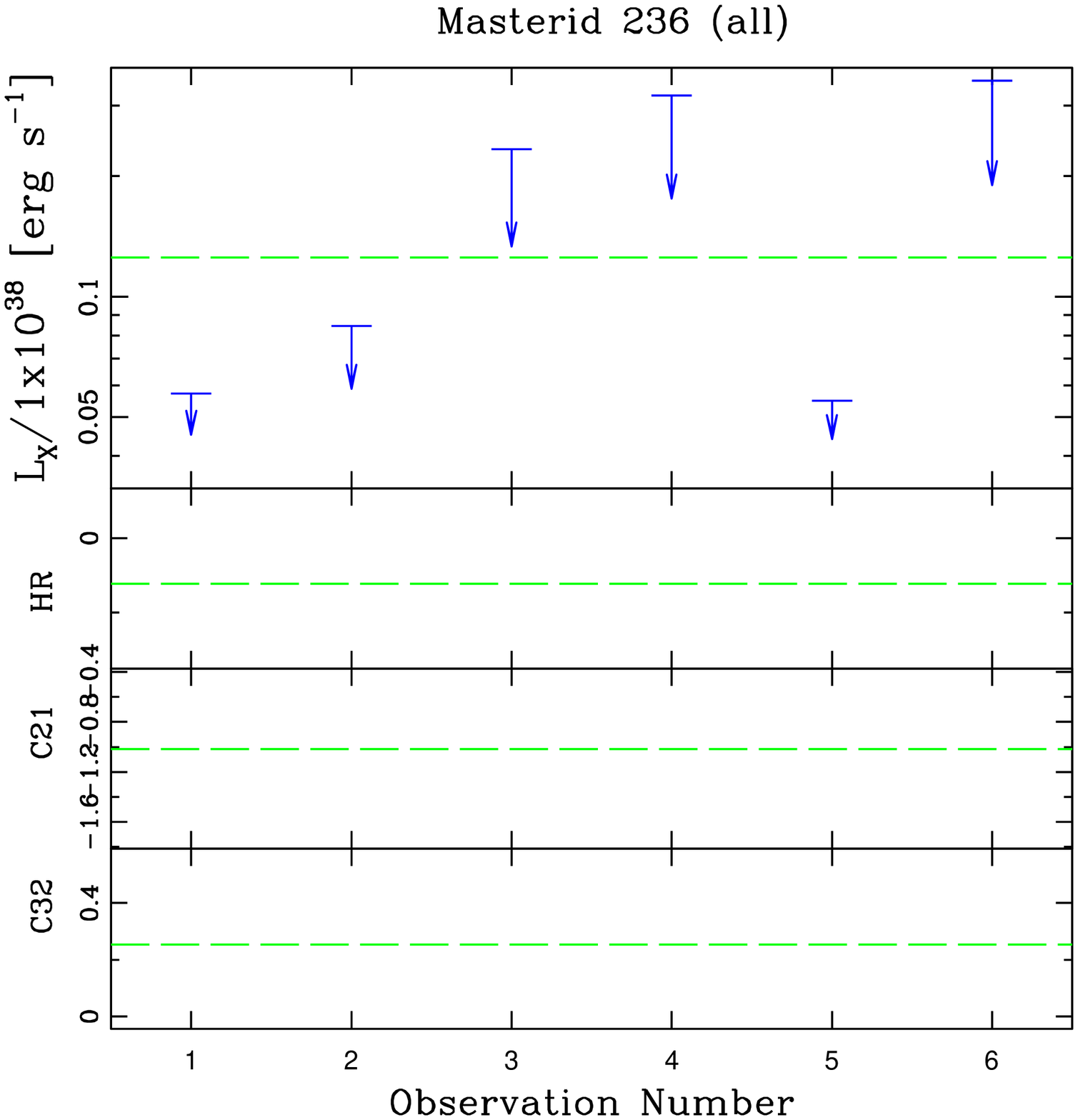}

\end{minipage}\hspace{0.02\linewidth}
 
\end{figure}

\clearpage

\begin{figure}
  \begin{minipage}{0.32\linewidth}
  \centering
  
    \includegraphics[width=\linewidth]{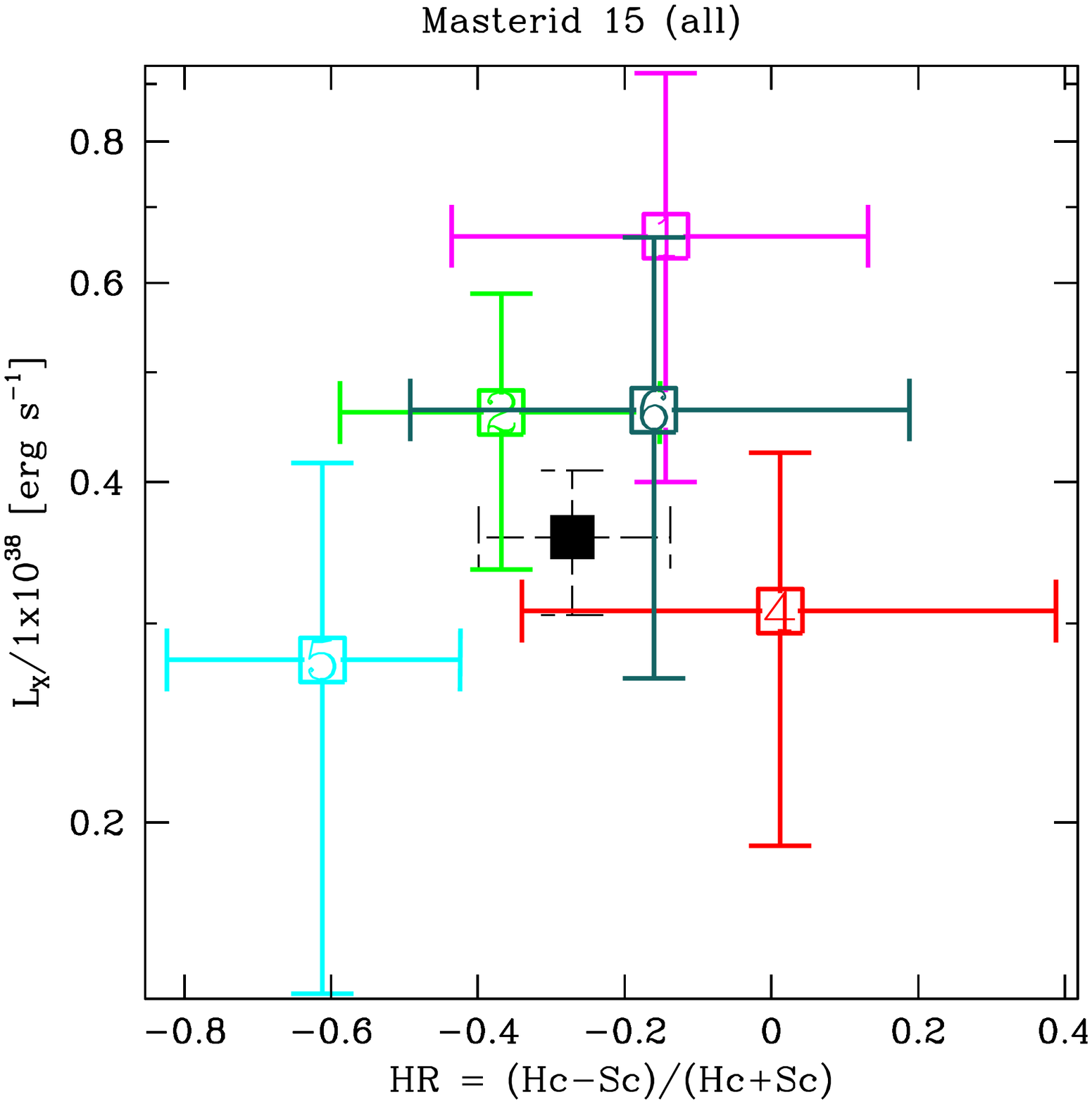}

  \end{minipage}
  \begin{minipage}{0.32\linewidth}
  \centering

    \includegraphics[width=\linewidth]{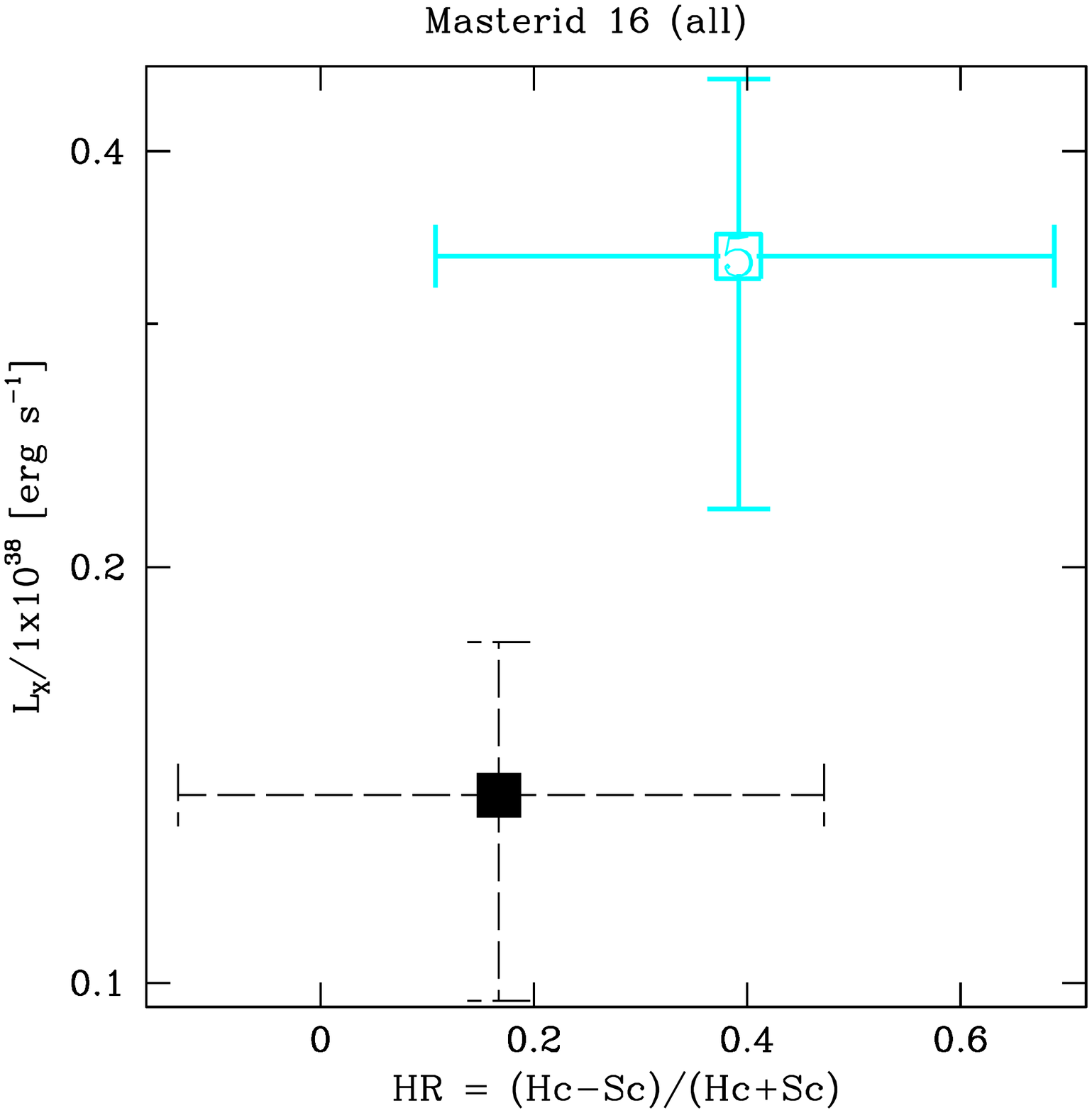}

\end{minipage}
\begin{minipage}{0.32\linewidth}
  \centering

    \includegraphics[width=\linewidth]{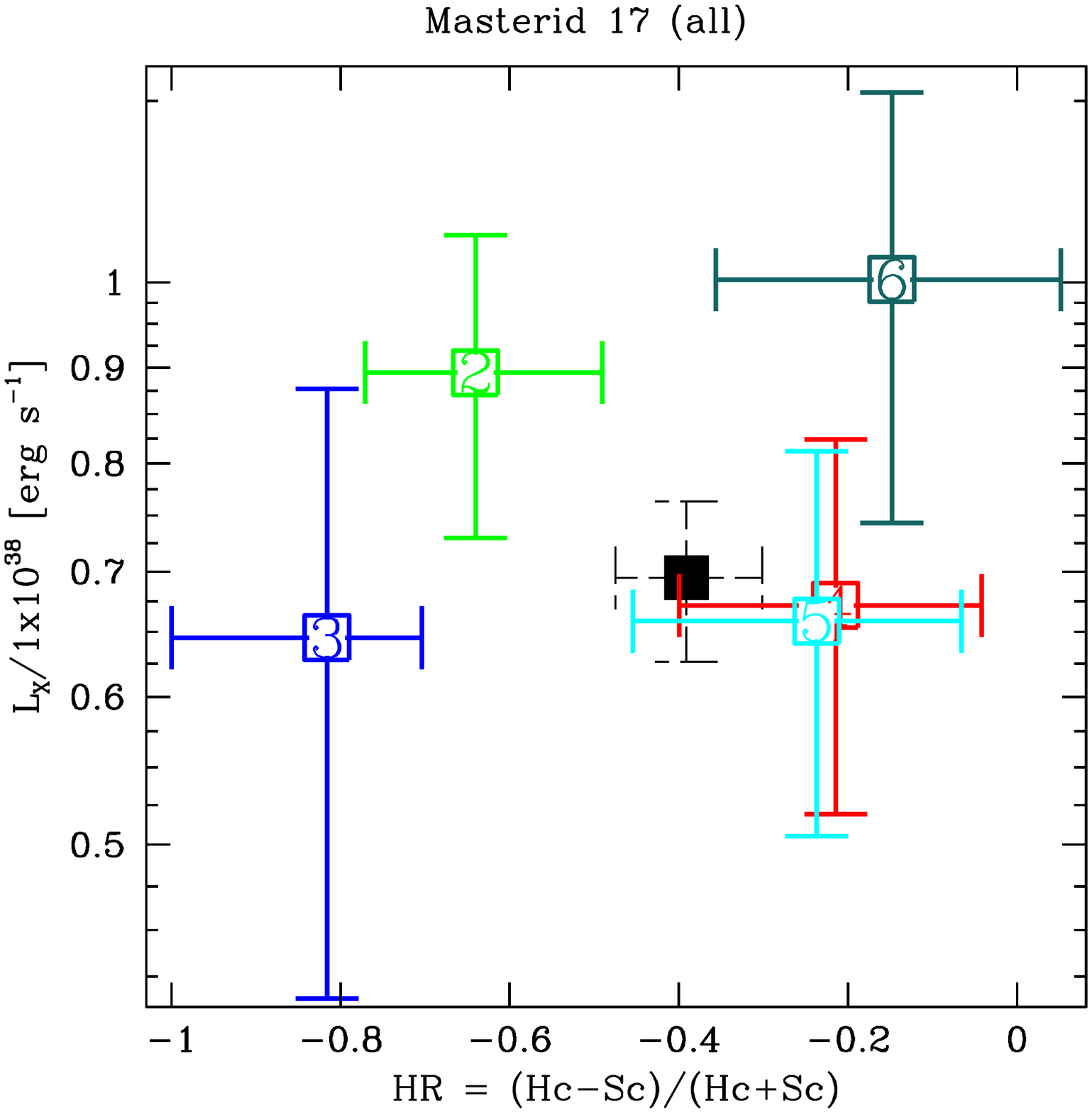}

 \end{minipage}

\begin{minipage}{0.32\linewidth}
  \centering
  
    \includegraphics[width=\linewidth]{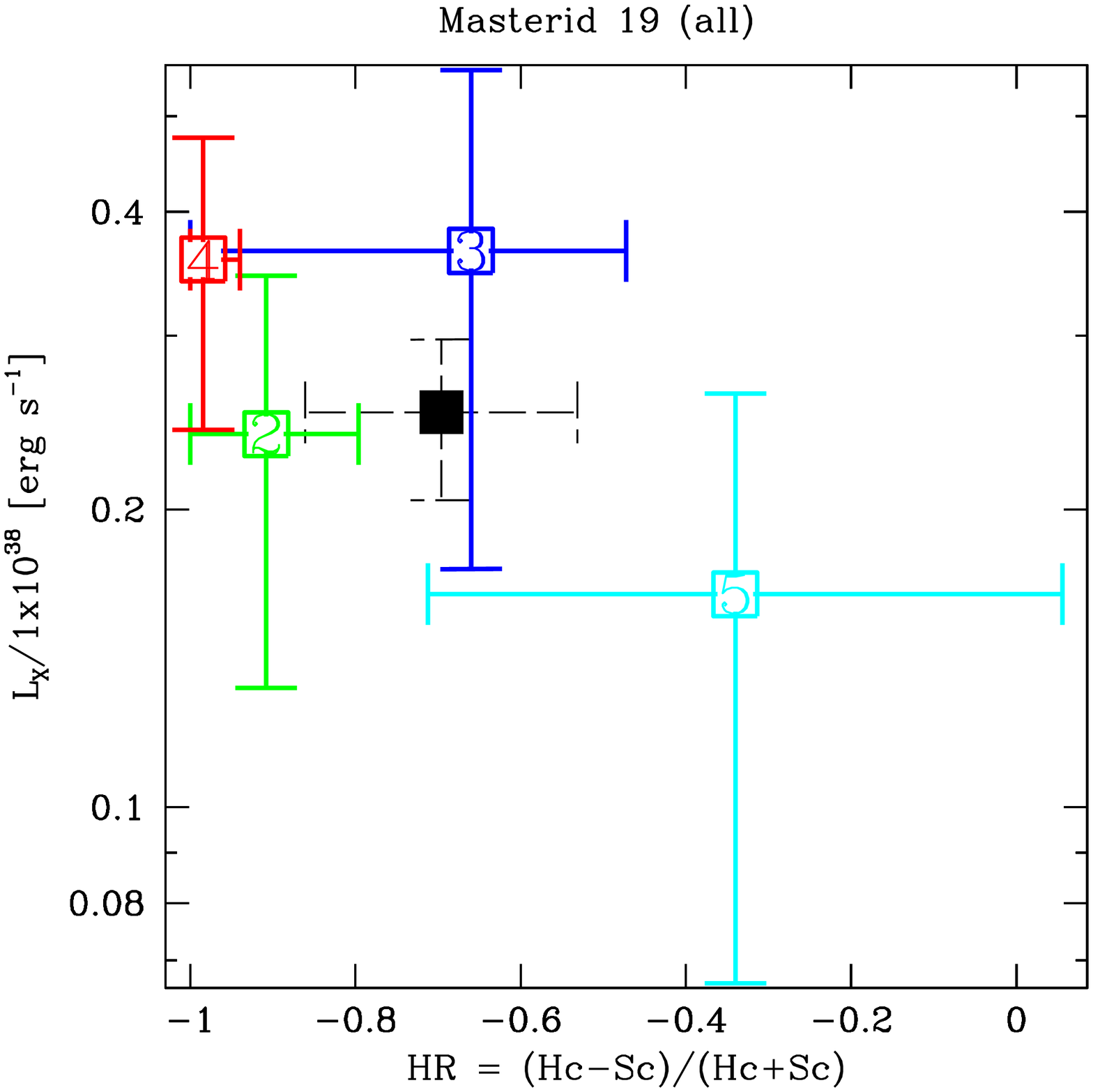}

  \end{minipage}
  \begin{minipage}{0.32\linewidth}
  \centering

    \includegraphics[width=\linewidth]{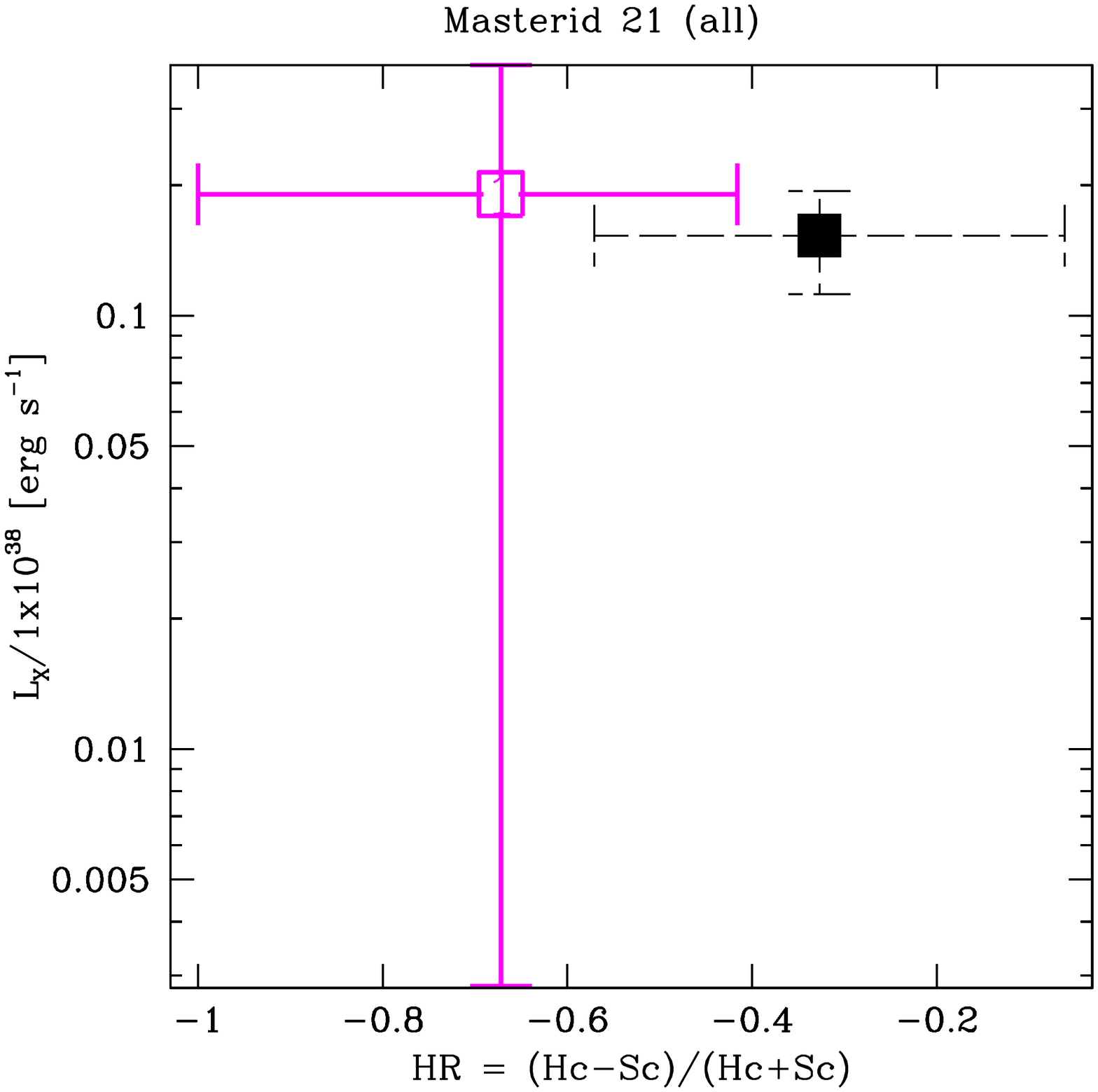}

\end{minipage}
\begin{minipage}{0.32\linewidth}
  \centering

    \includegraphics[width=\linewidth]{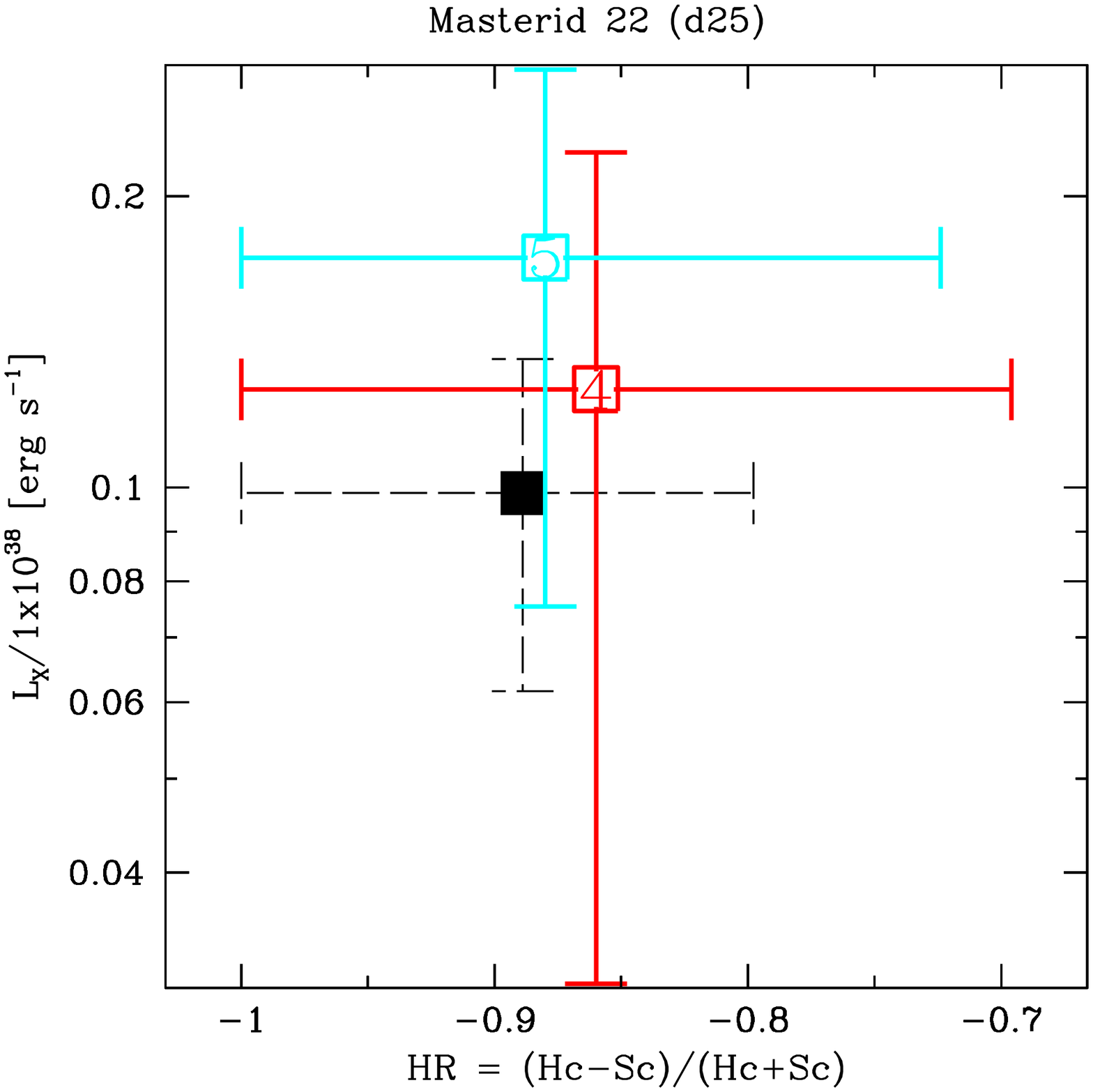}

 \end{minipage}

  \begin{minipage}{0.32\linewidth}
  \centering
  
    \includegraphics[width=\linewidth]{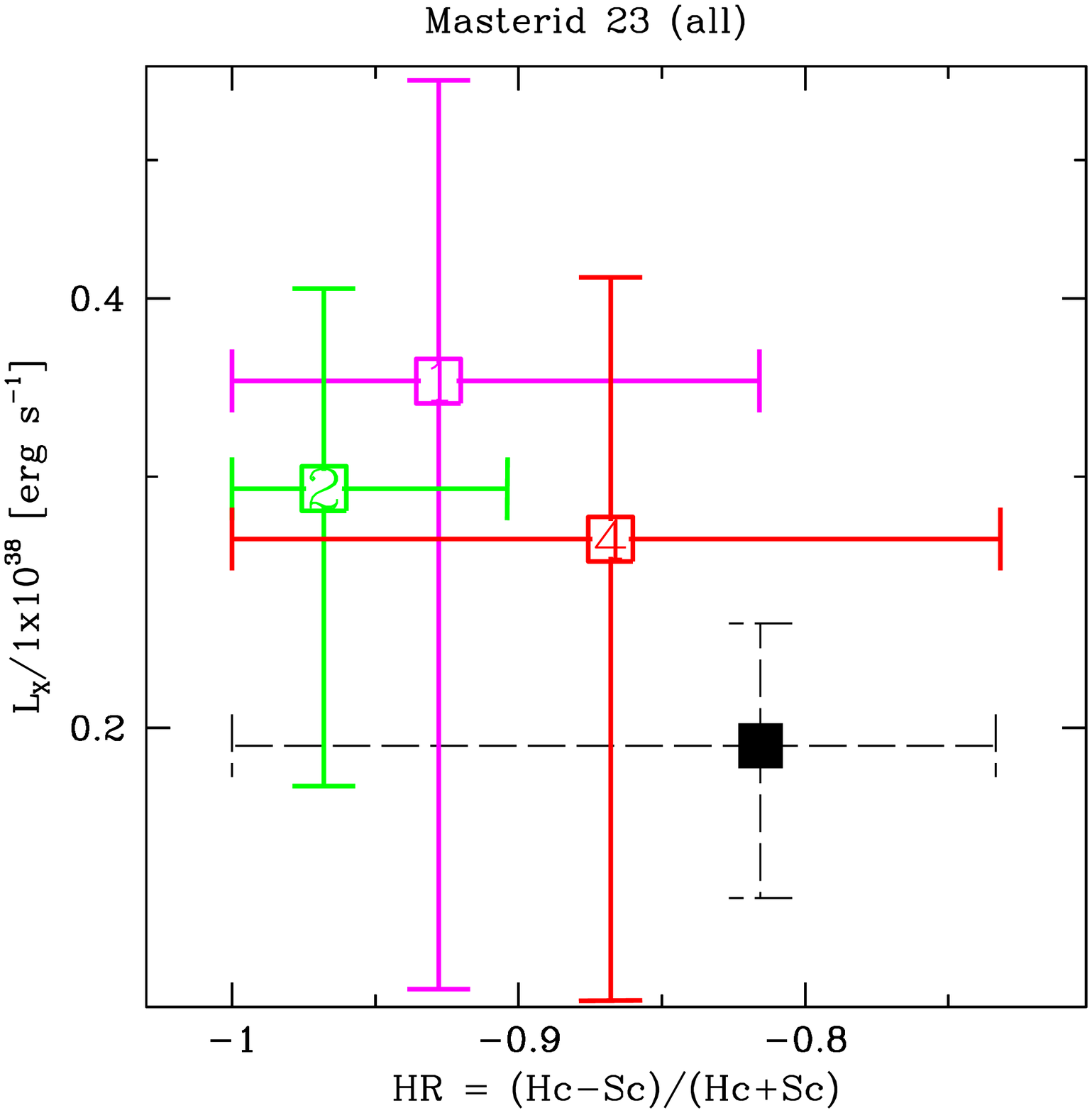}

  \end{minipage}
  \begin{minipage}{0.32\linewidth}
  \centering

    \includegraphics[width=\linewidth]{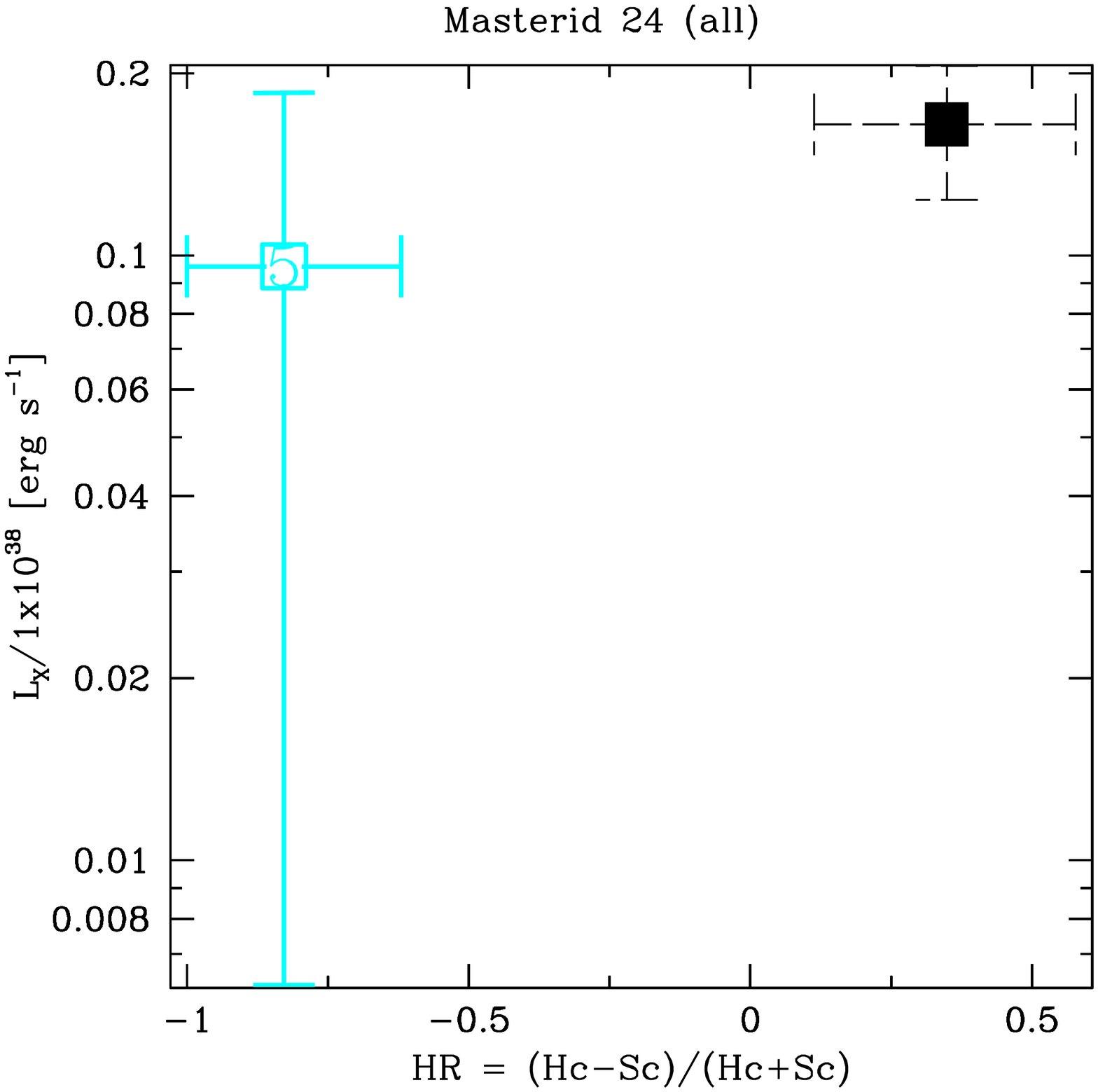}

\end{minipage}
\begin{minipage}{0.32\linewidth}
  \centering

    \includegraphics[width=\linewidth]{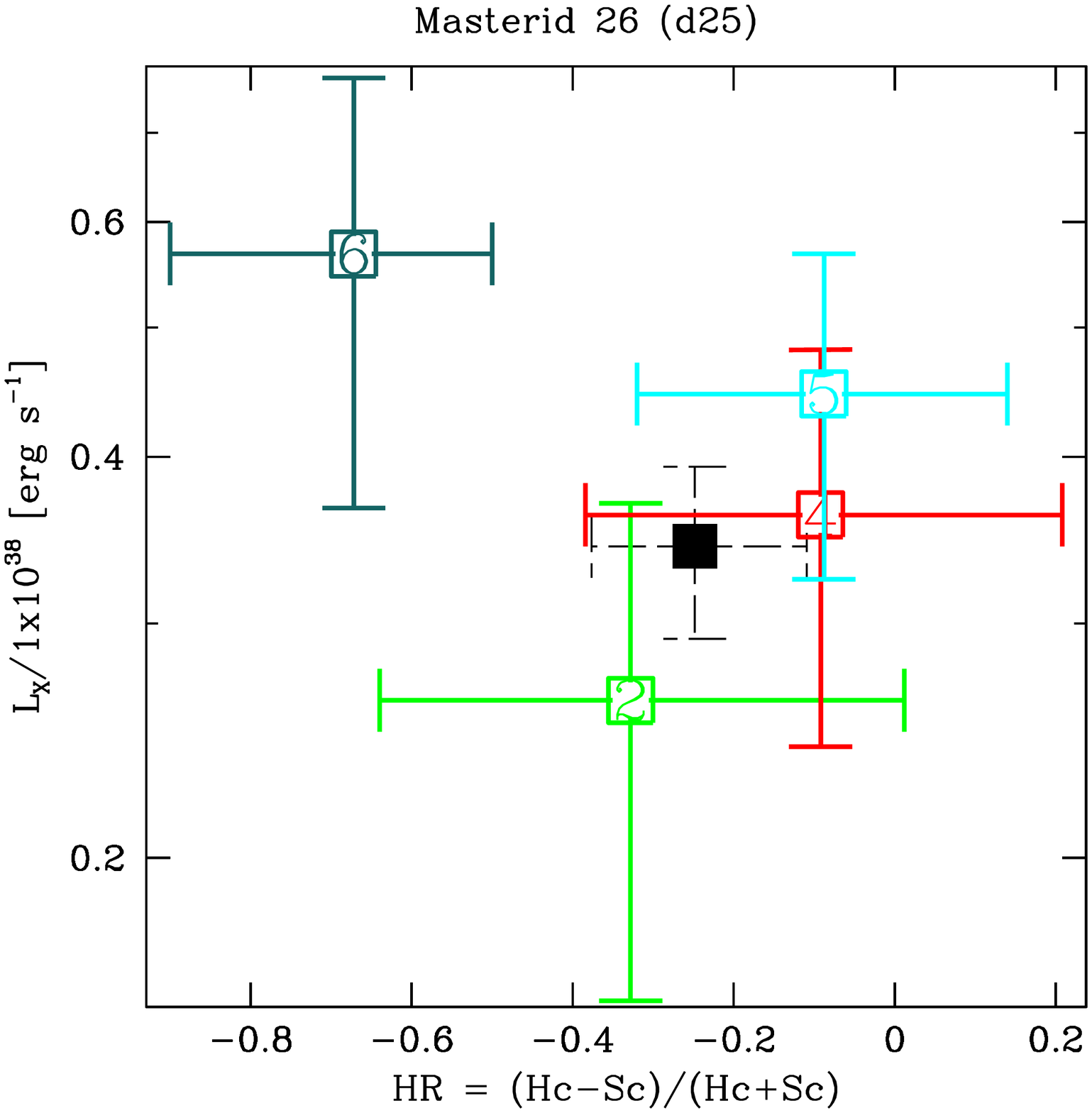}

 \end{minipage}

\begin{minipage}{0.32\linewidth}
  \centering
  
    \includegraphics[width=\linewidth]{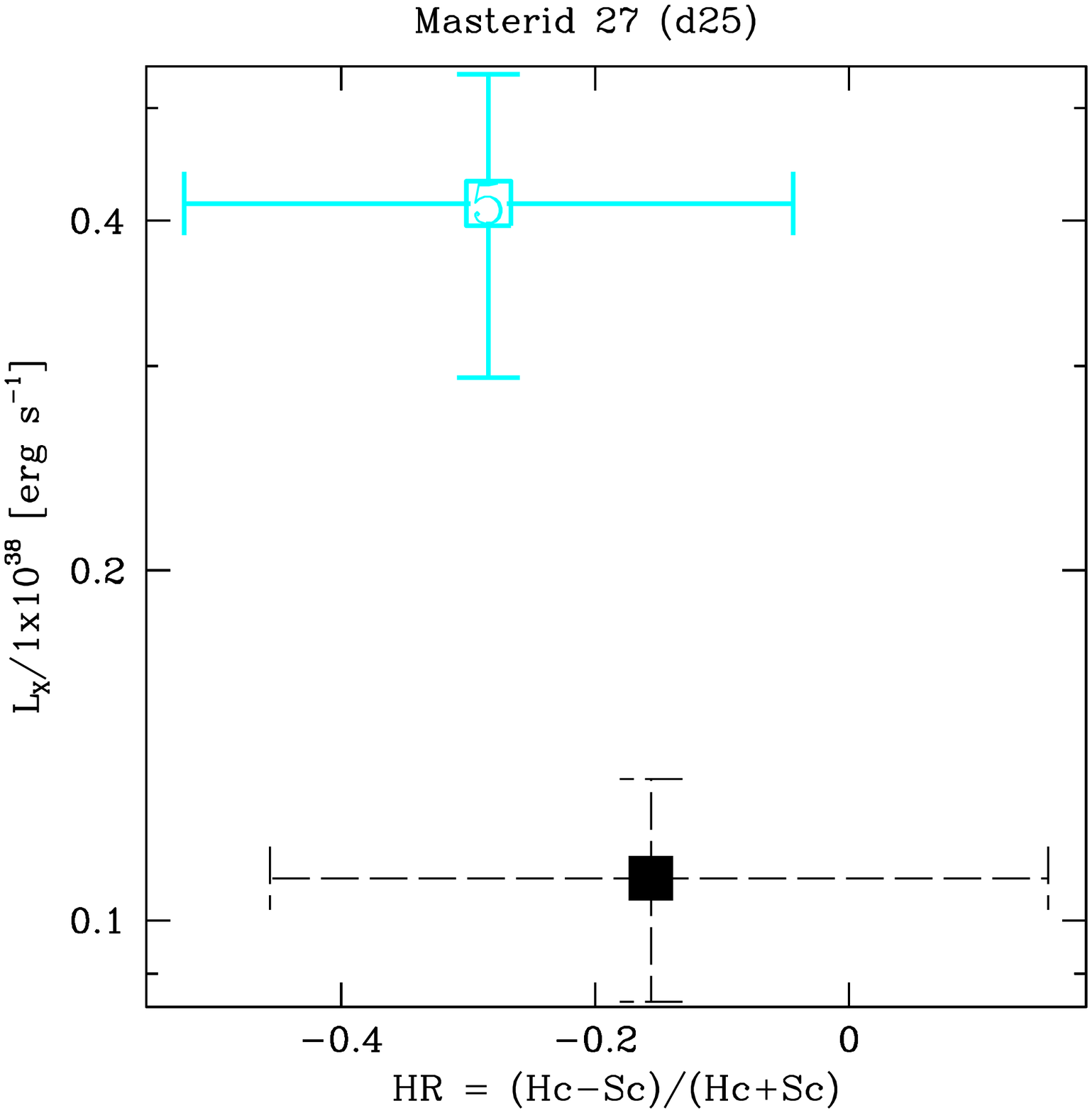}

  \end{minipage}
  \begin{minipage}{0.32\linewidth}
  \centering

    \includegraphics[width=\linewidth]{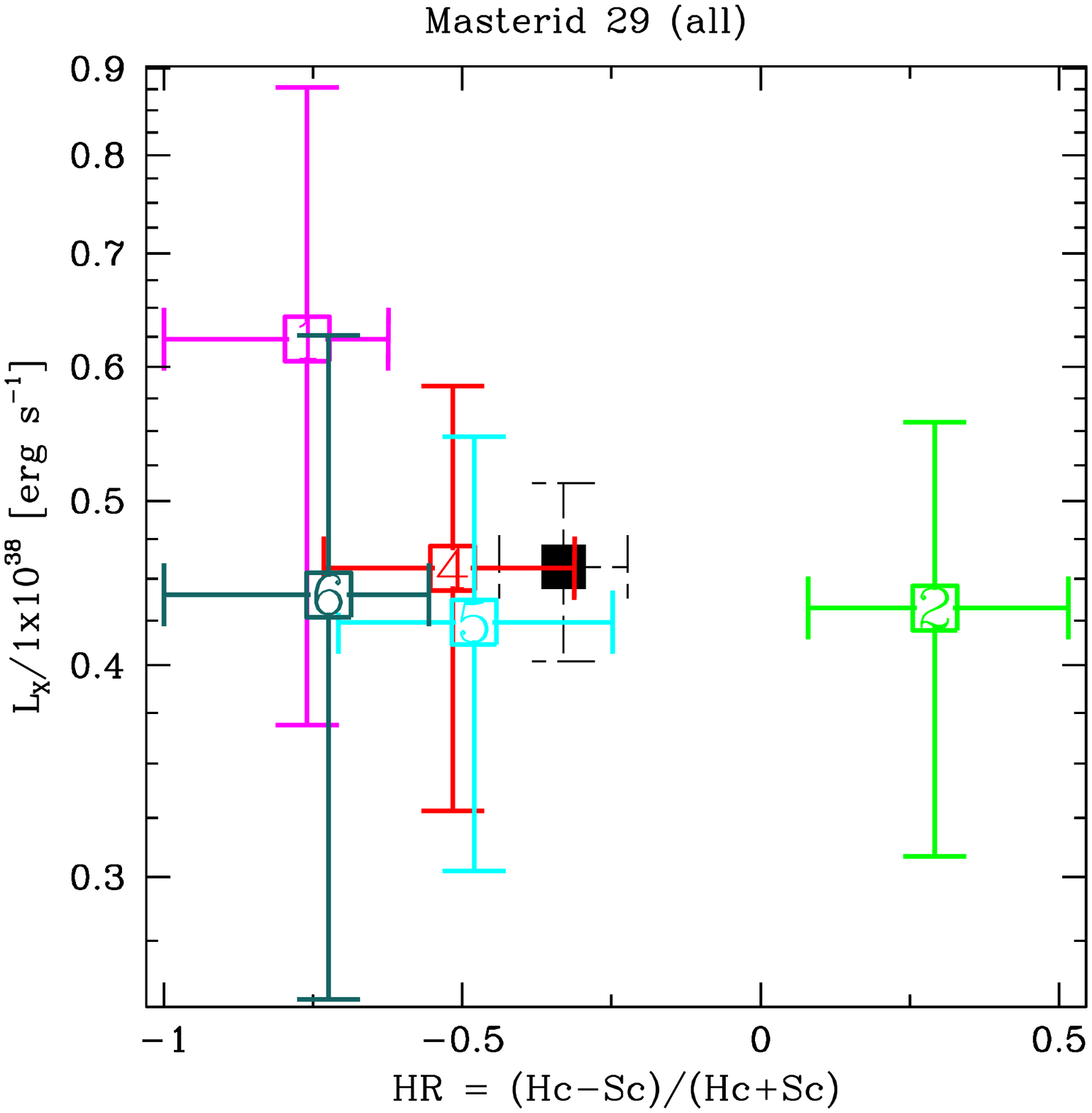}

\end{minipage}
\begin{minipage}{0.32\linewidth}
  \centering

    \includegraphics[width=\linewidth]{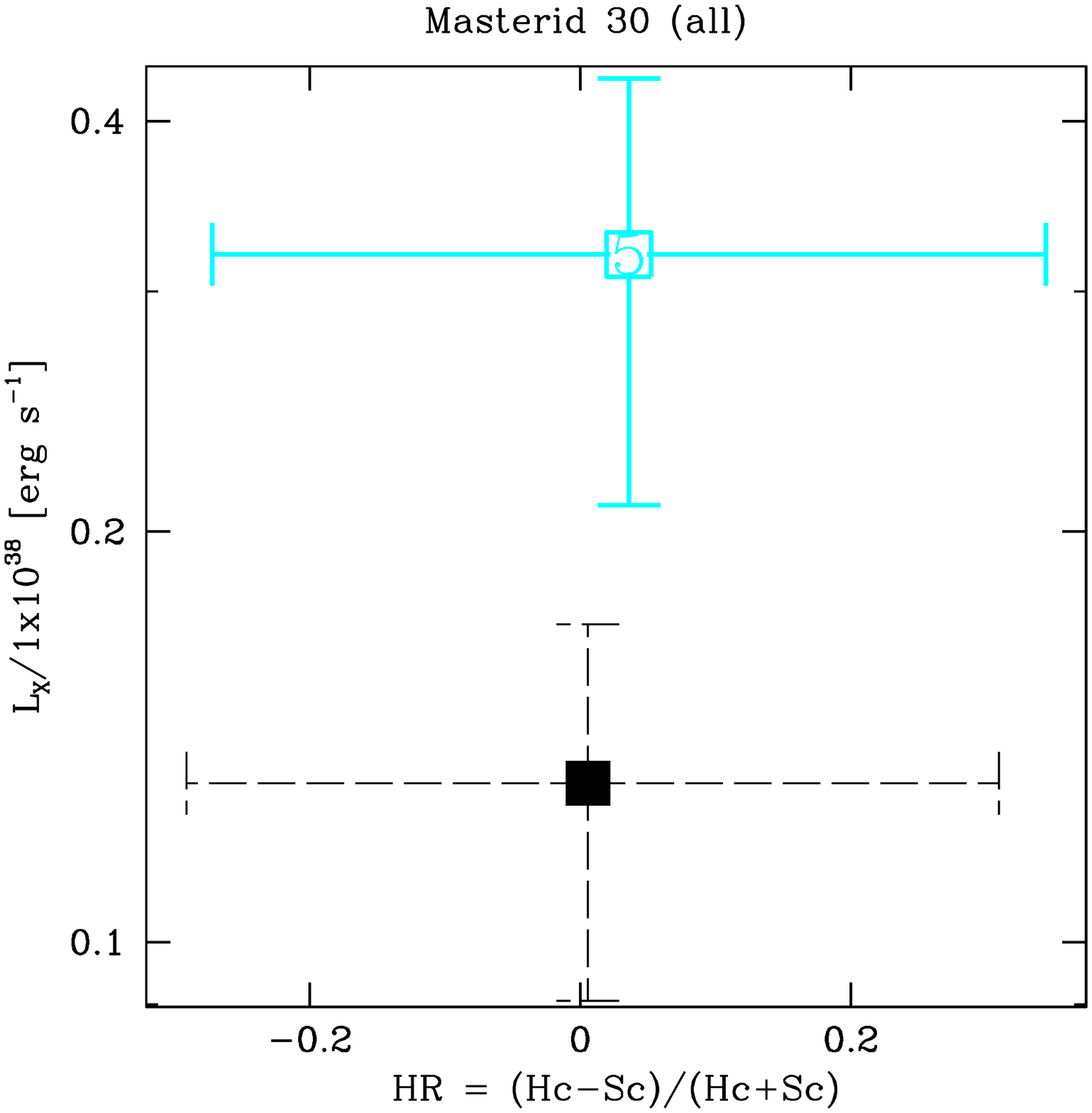}

 \end{minipage}
  
\end{figure}

\begin{figure}
  \begin{minipage}{0.32\linewidth}
  \centering
  
    \includegraphics[width=\linewidth]{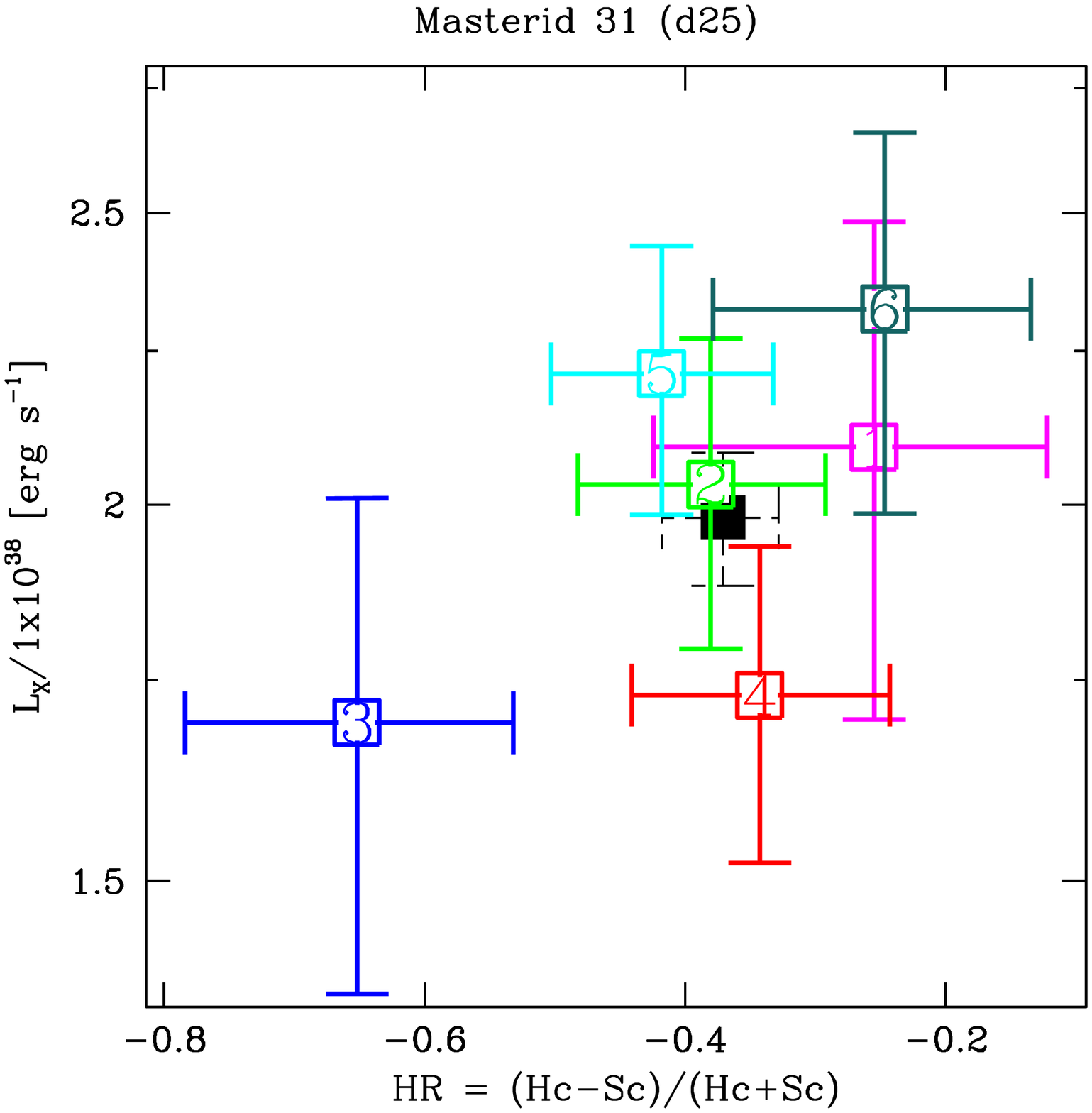}

  \end{minipage}
  \begin{minipage}{0.32\linewidth}
  \centering

    \includegraphics[width=\linewidth]{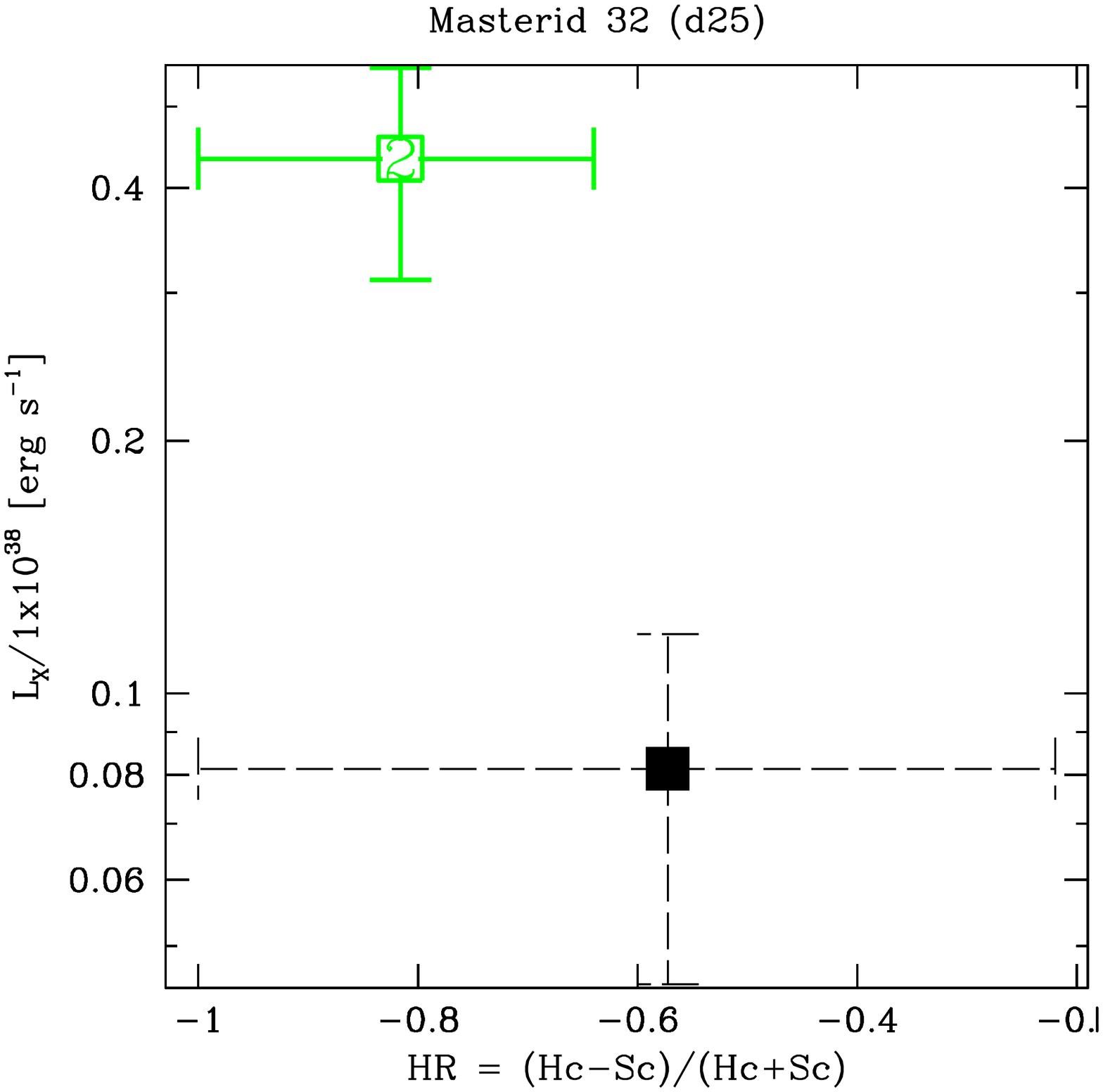}

\end{minipage}
\begin{minipage}{0.32\linewidth}
  \centering

    \includegraphics[width=\linewidth]{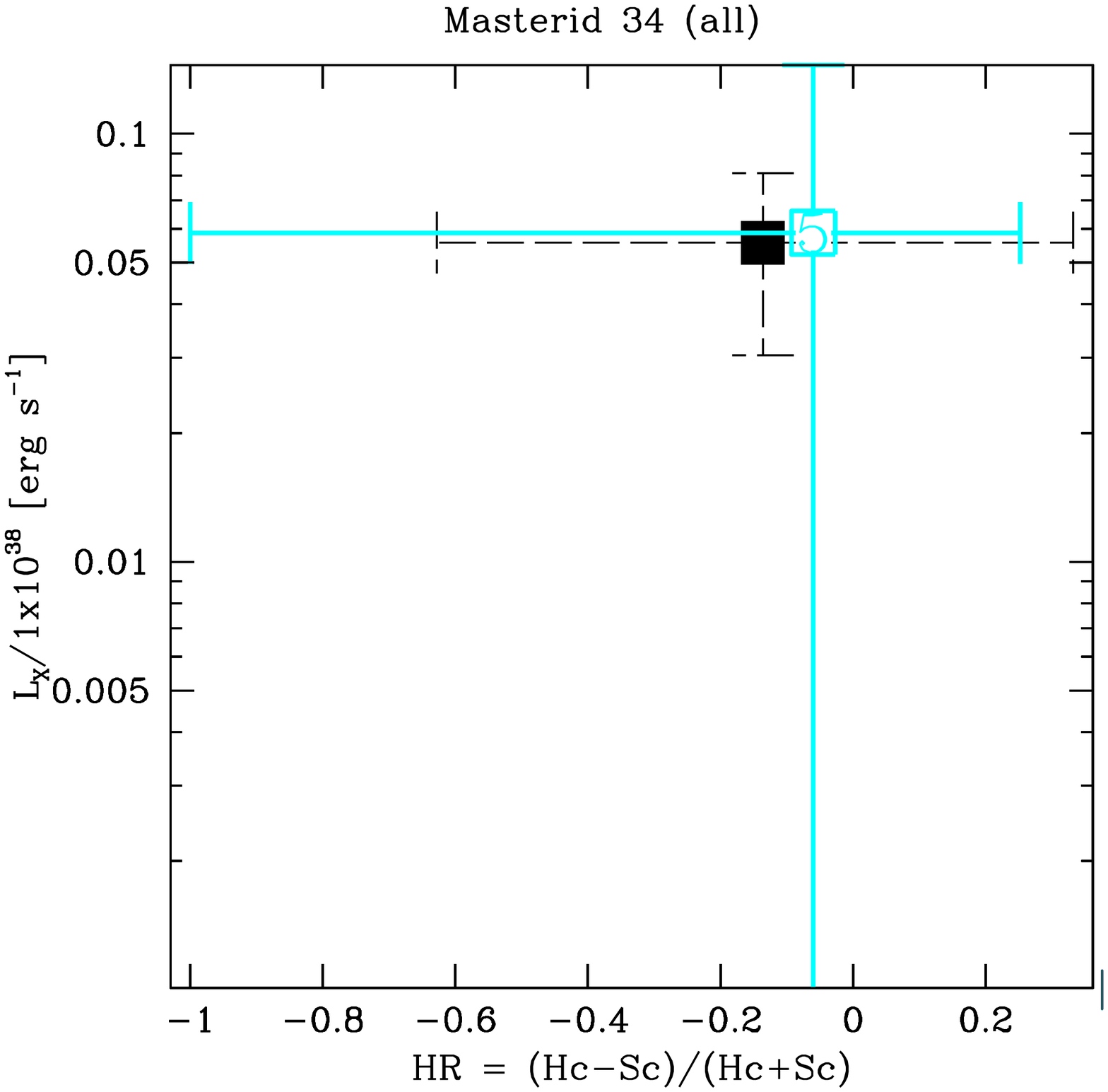}

 \end{minipage}

\begin{minipage}{0.32\linewidth}
  \centering
  
    \includegraphics[width=\linewidth]{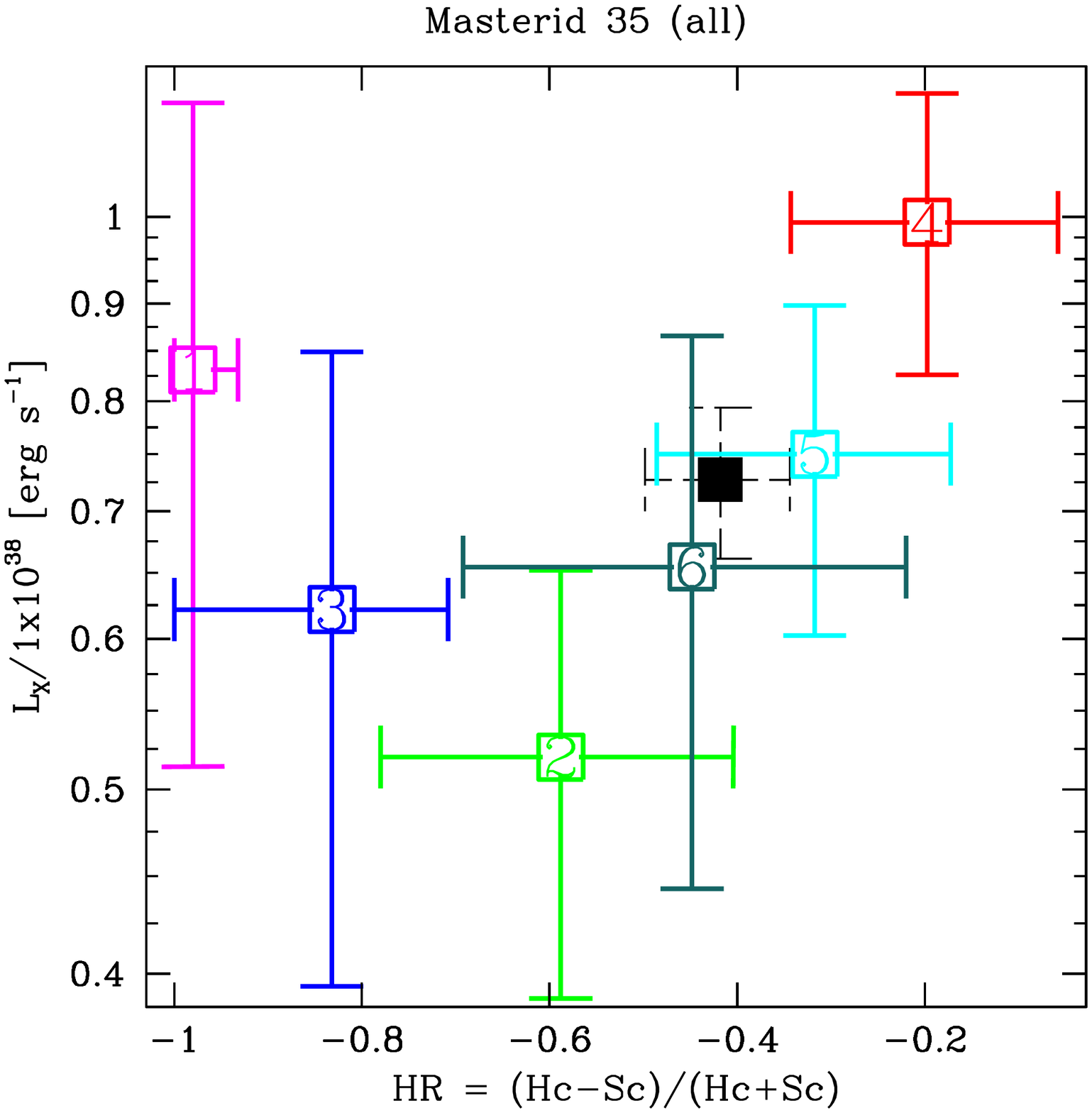}

  \end{minipage}
  \begin{minipage}{0.32\linewidth}
  \centering

    \includegraphics[width=\linewidth]{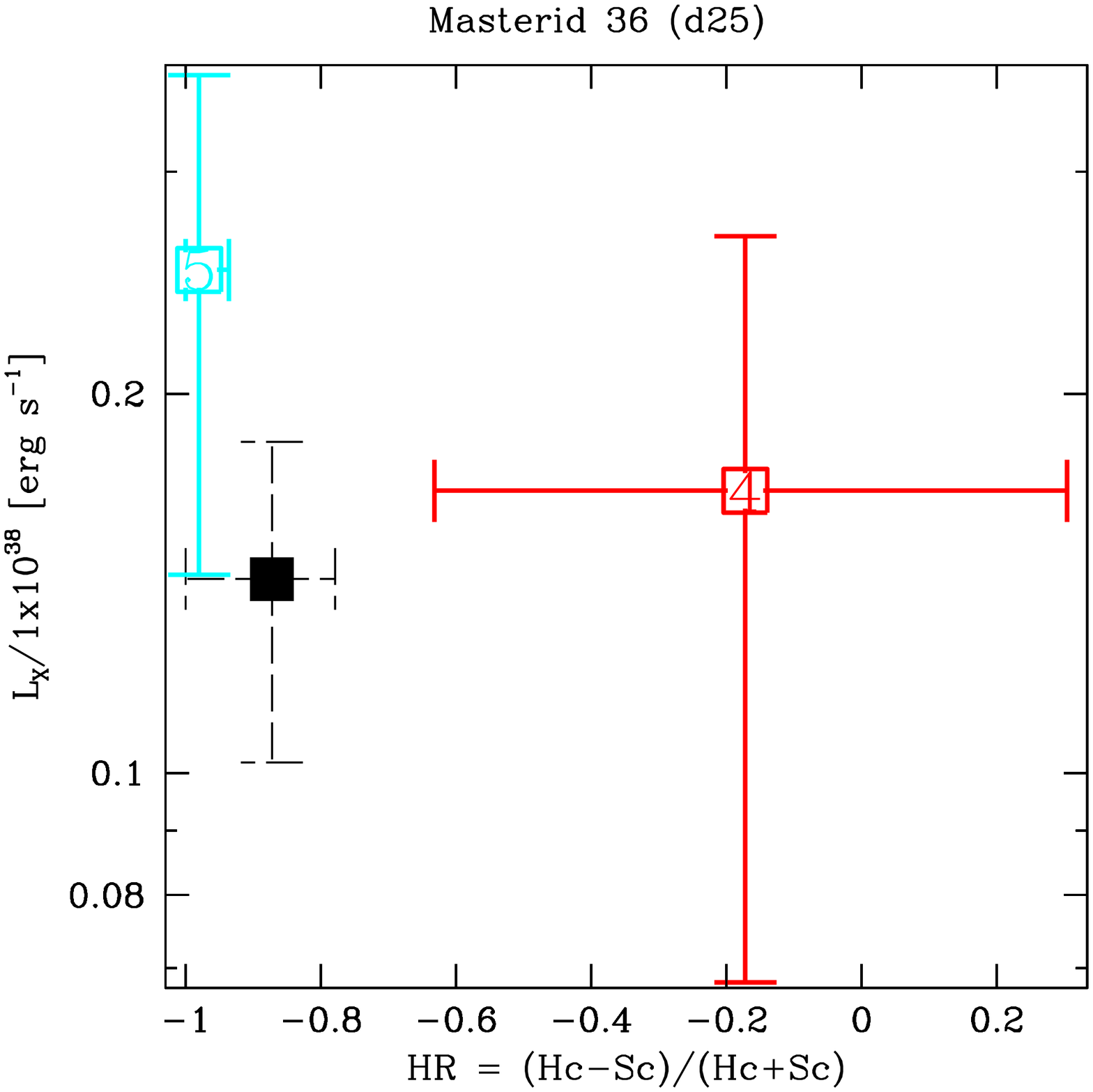}

\end{minipage}
\begin{minipage}{0.32\linewidth}
  \centering

    \includegraphics[width=\linewidth]{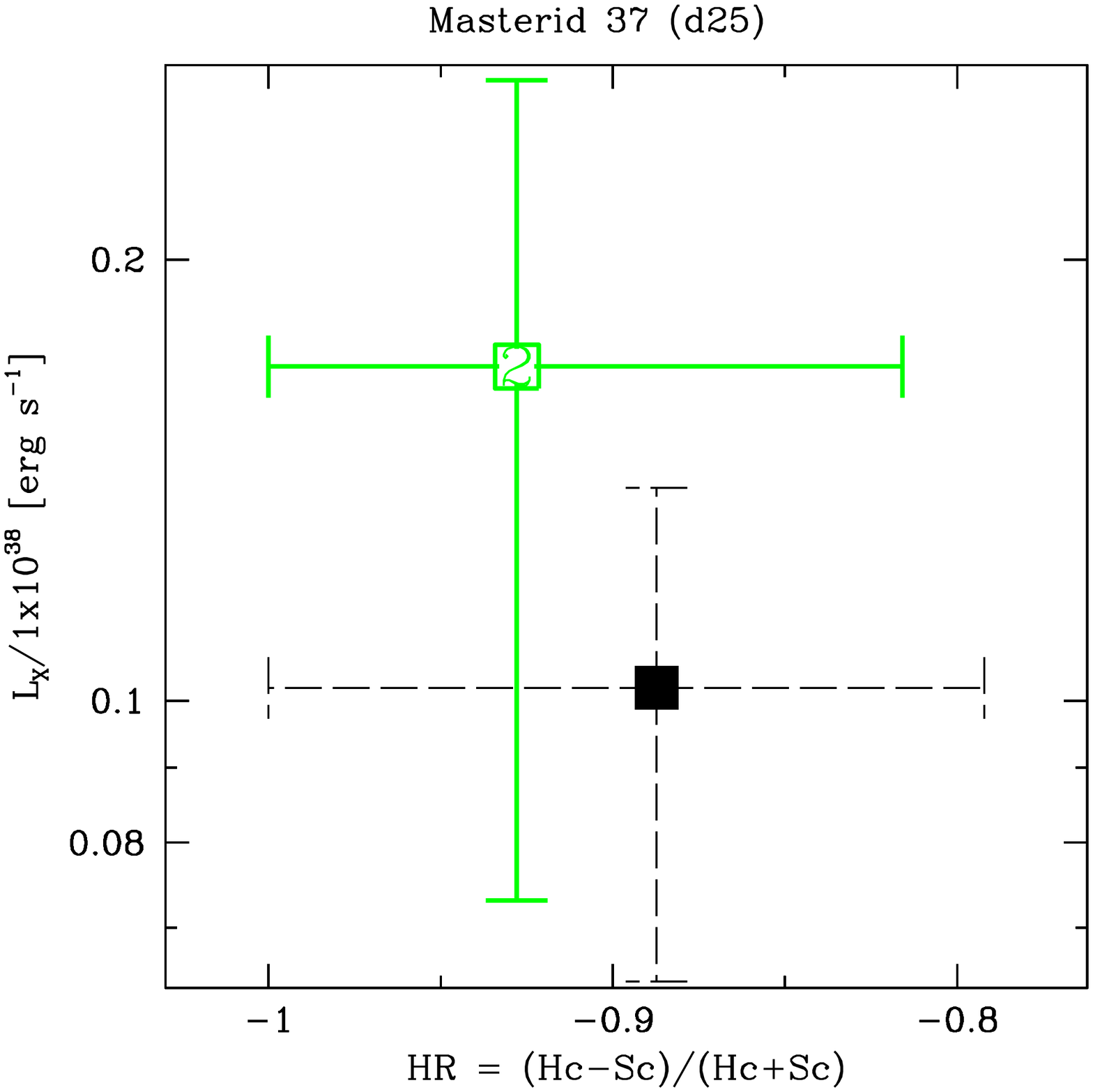}

 \end{minipage}

  \begin{minipage}{0.32\linewidth}
  \centering
  
    \includegraphics[width=\linewidth]{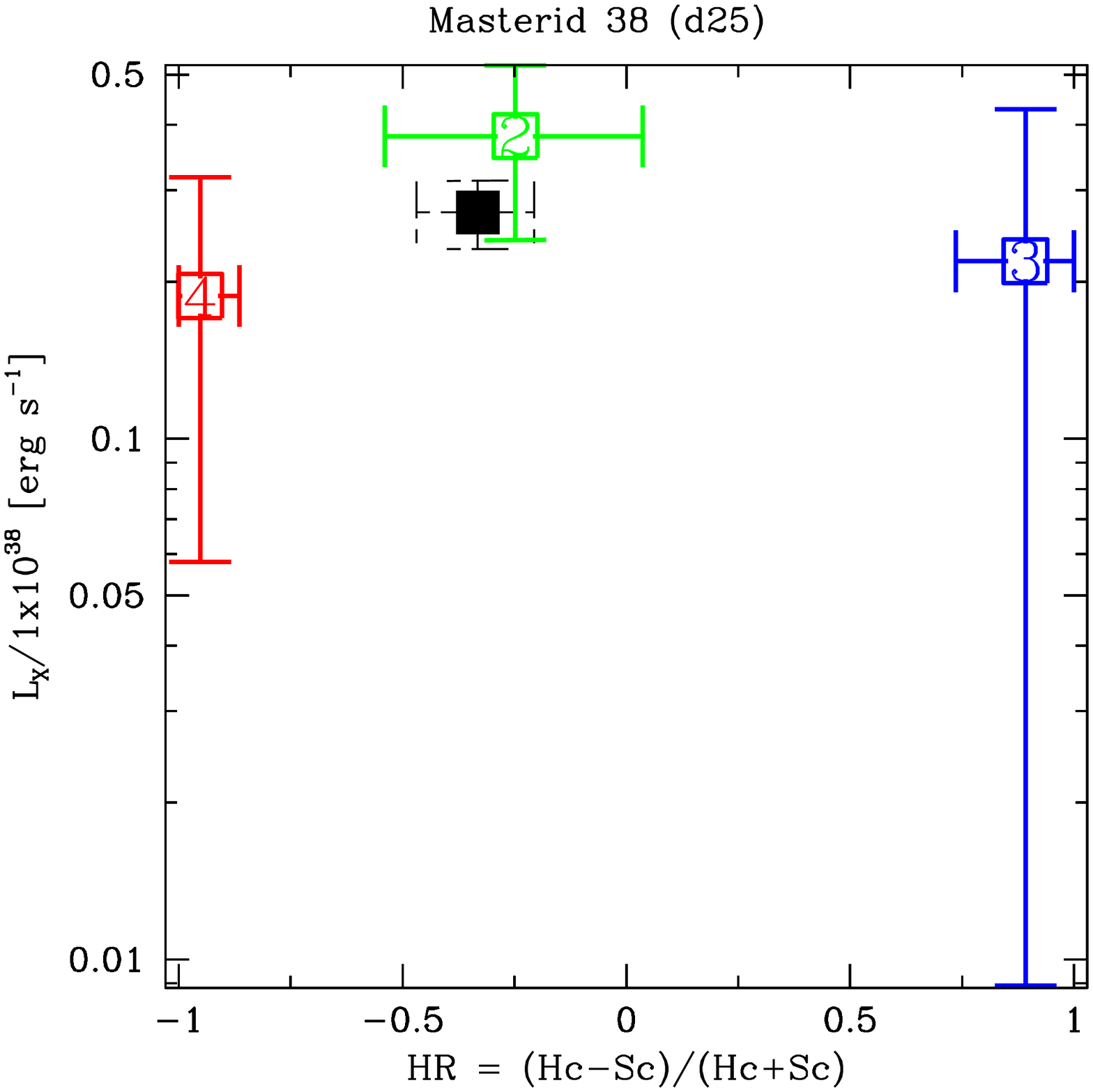}

  \end{minipage}
  \begin{minipage}{0.32\linewidth}
  \centering

    \includegraphics[width=\linewidth]{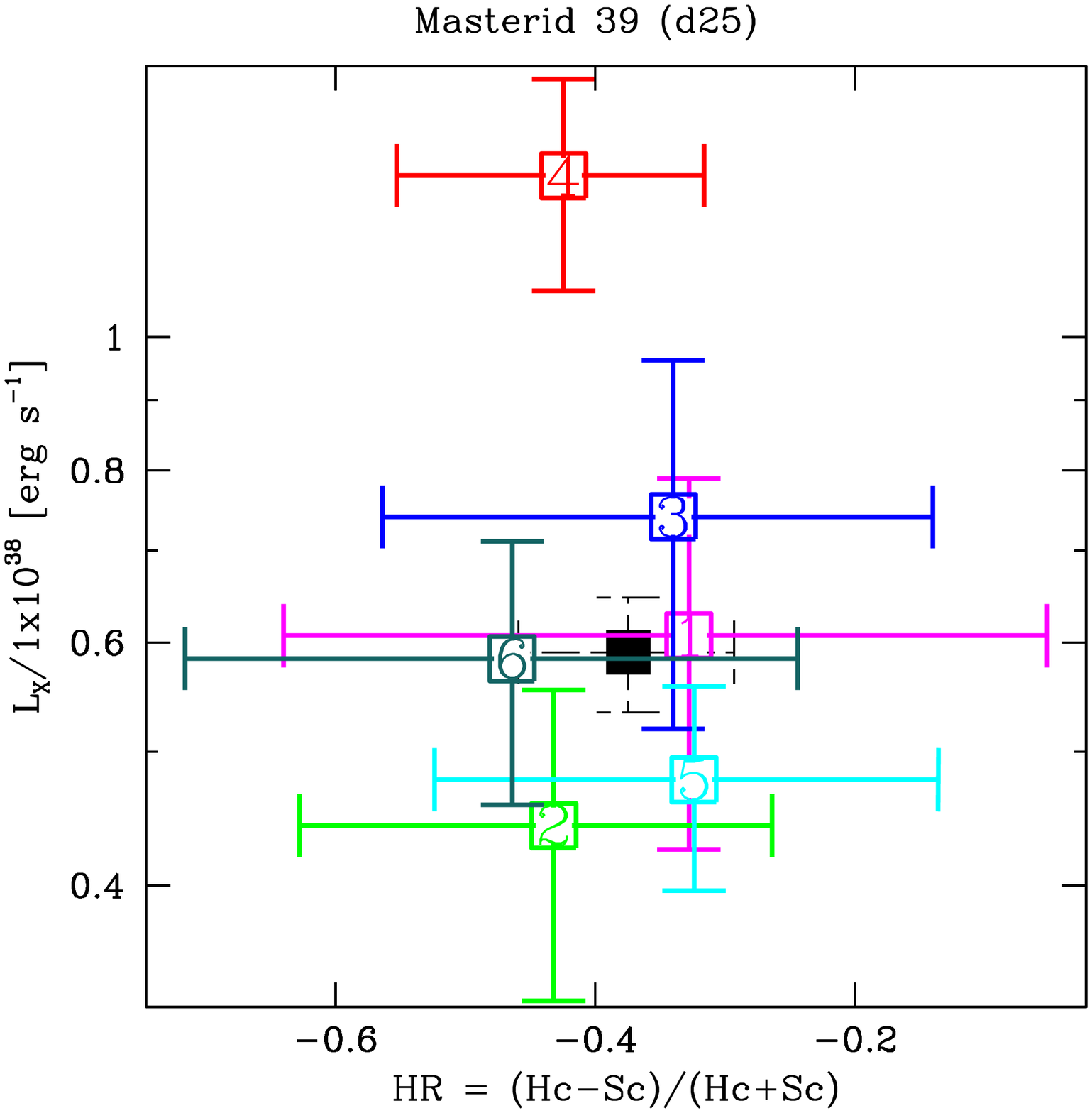}

\end{minipage}
\begin{minipage}{0.32\linewidth}
  \centering

    \includegraphics[width=\linewidth]{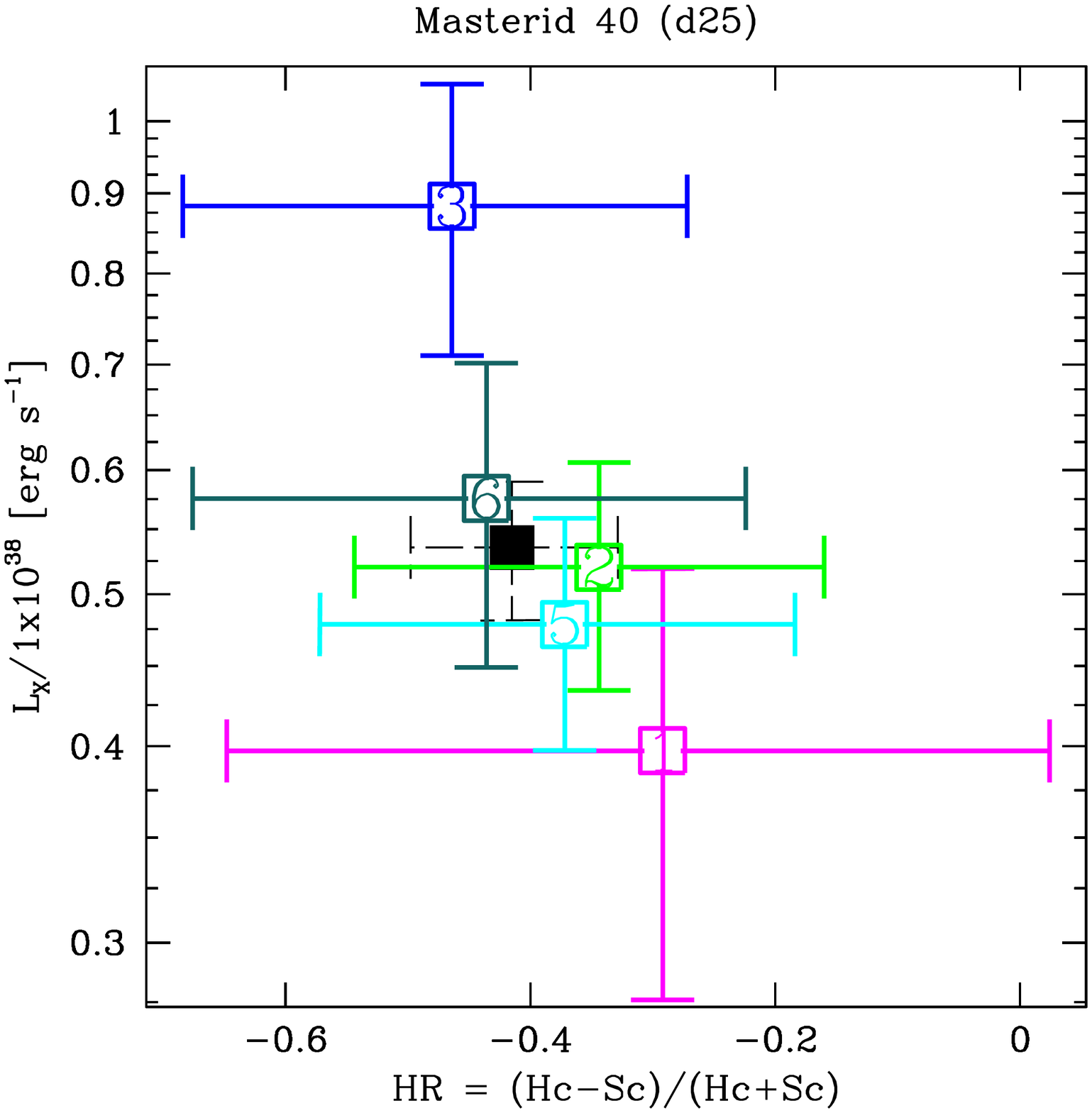}

 \end{minipage}

\begin{minipage}{0.32\linewidth}
  \centering
  
    \includegraphics[width=\linewidth]{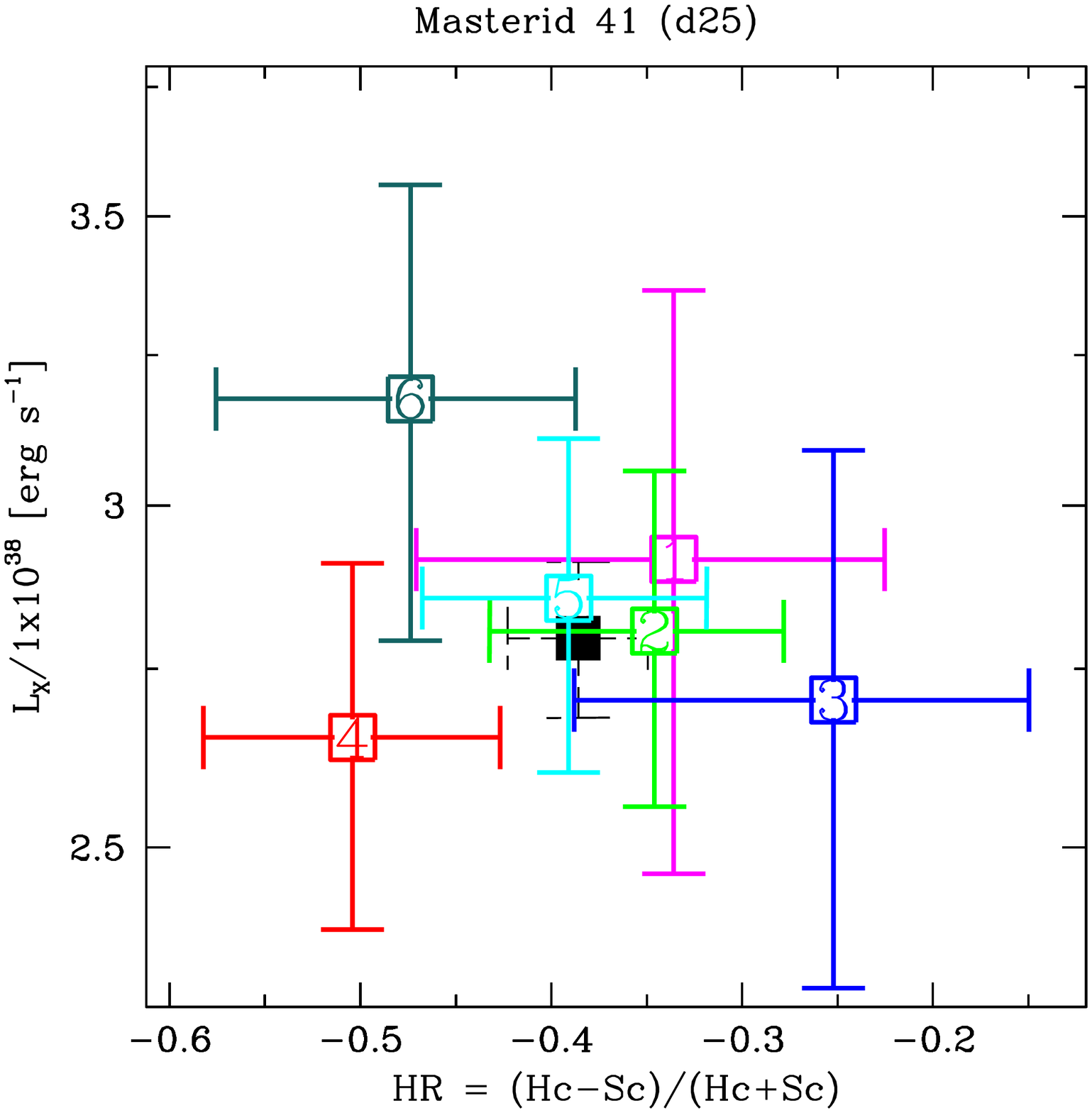}

  \end{minipage}
  \begin{minipage}{0.32\linewidth}
  \centering

    \includegraphics[width=\linewidth]{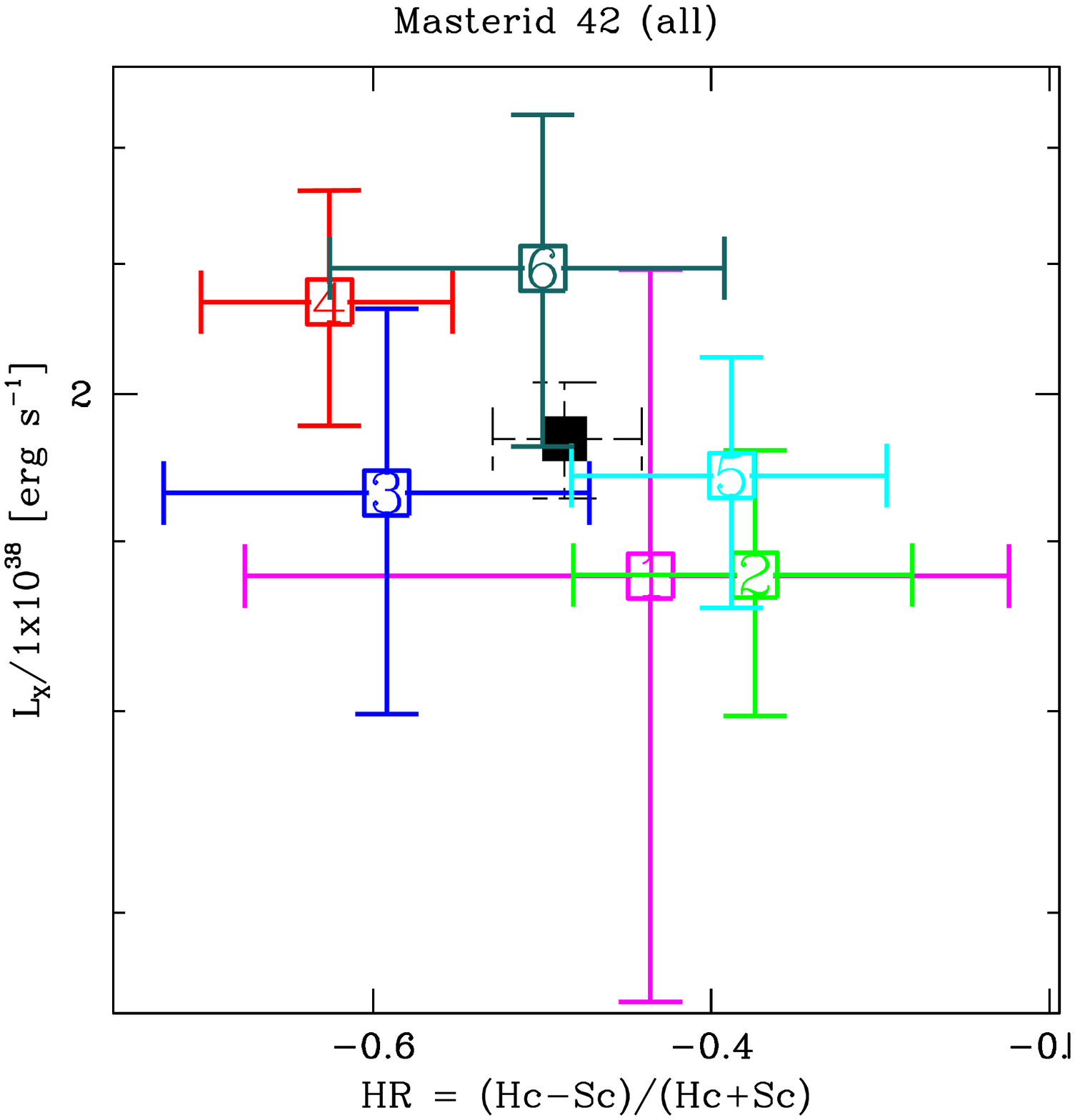}

\end{minipage}
\begin{minipage}{0.32\linewidth}
  \centering

    \includegraphics[width=\linewidth]{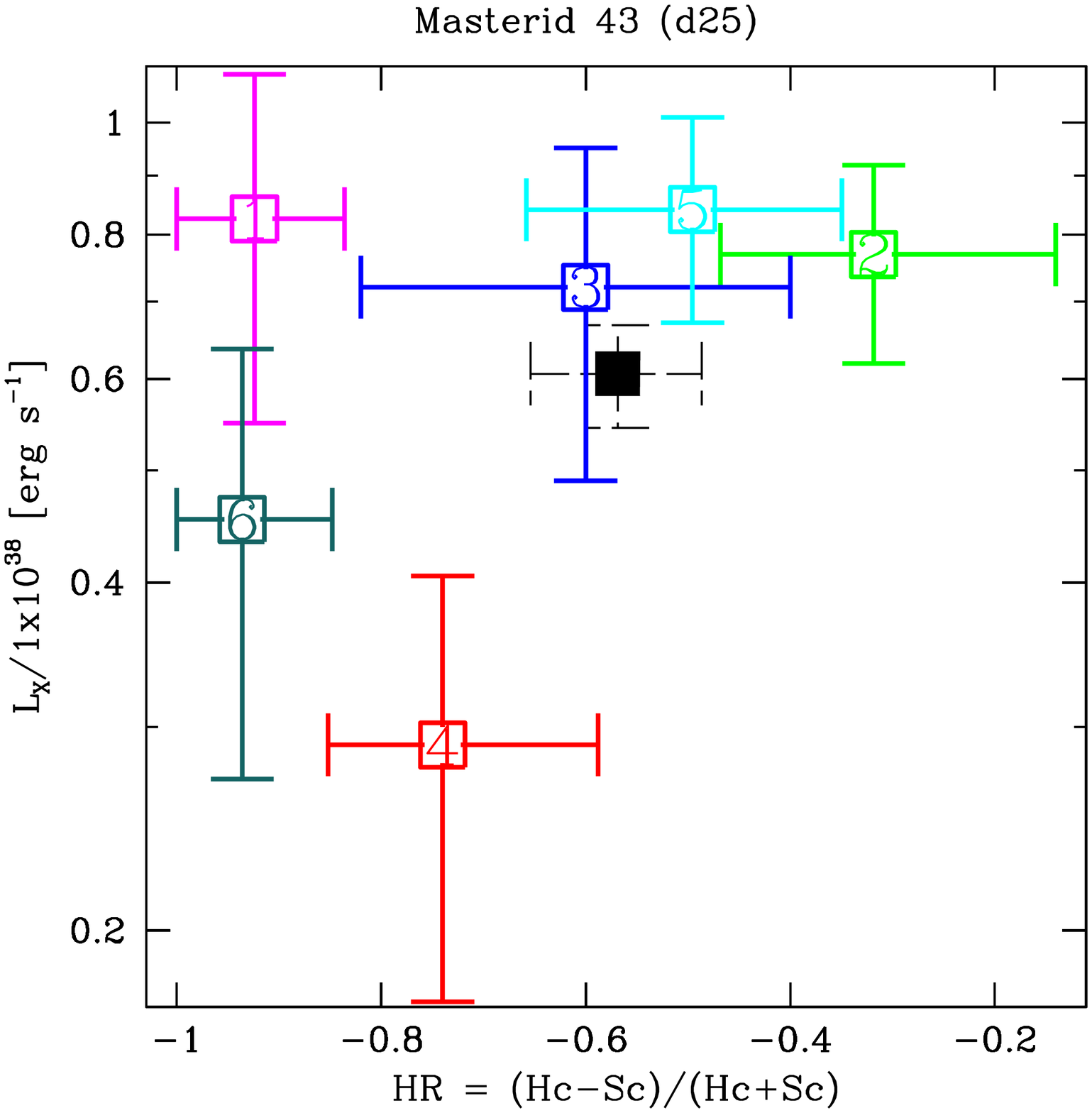}

 \end{minipage}
  
\end{figure}

\begin{figure}
  \begin{minipage}{0.32\linewidth}
  \centering
  
    \includegraphics[width=\linewidth]{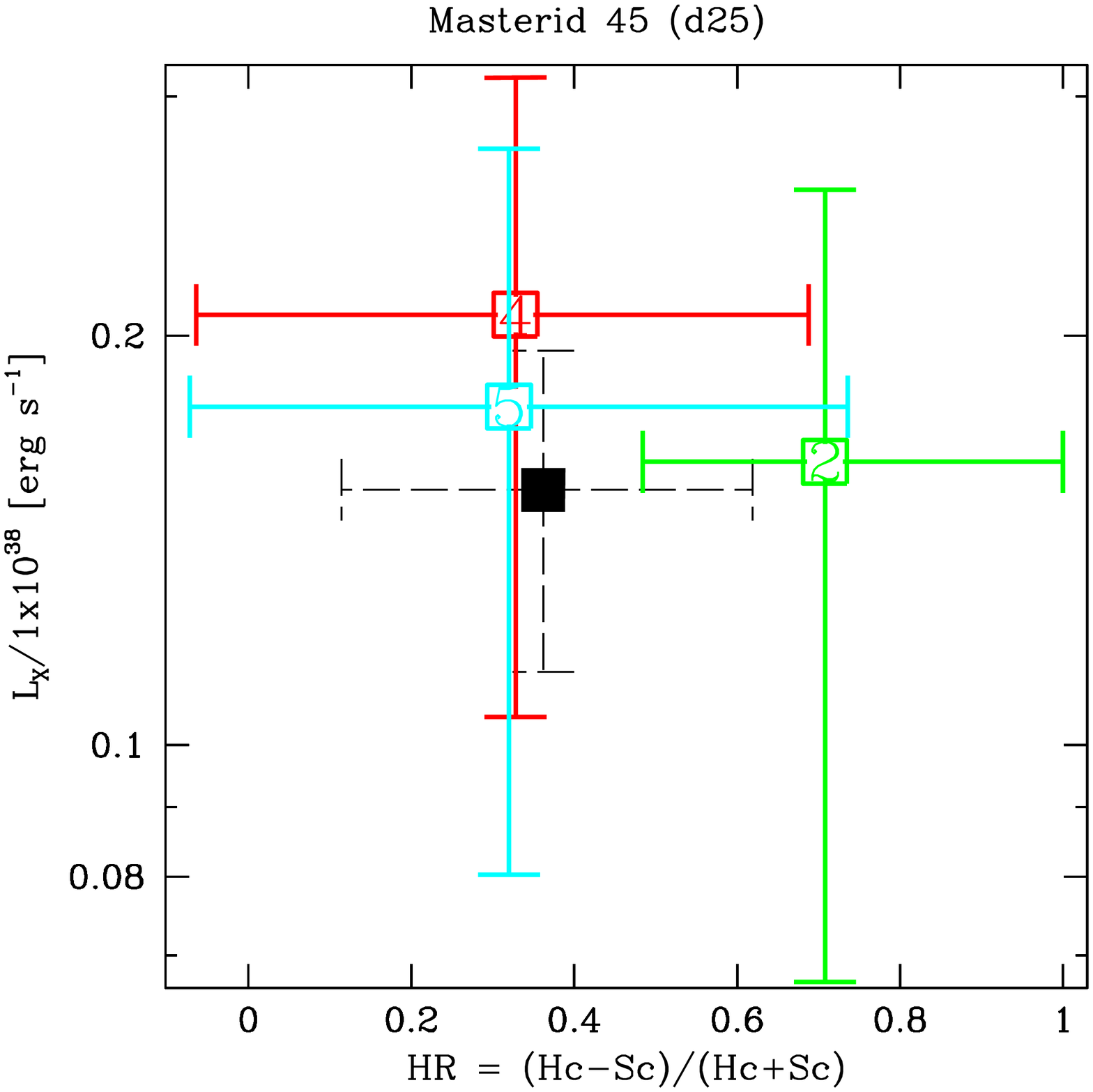}

  \end{minipage}
  \begin{minipage}{0.32\linewidth}
  \centering

    \includegraphics[width=\linewidth]{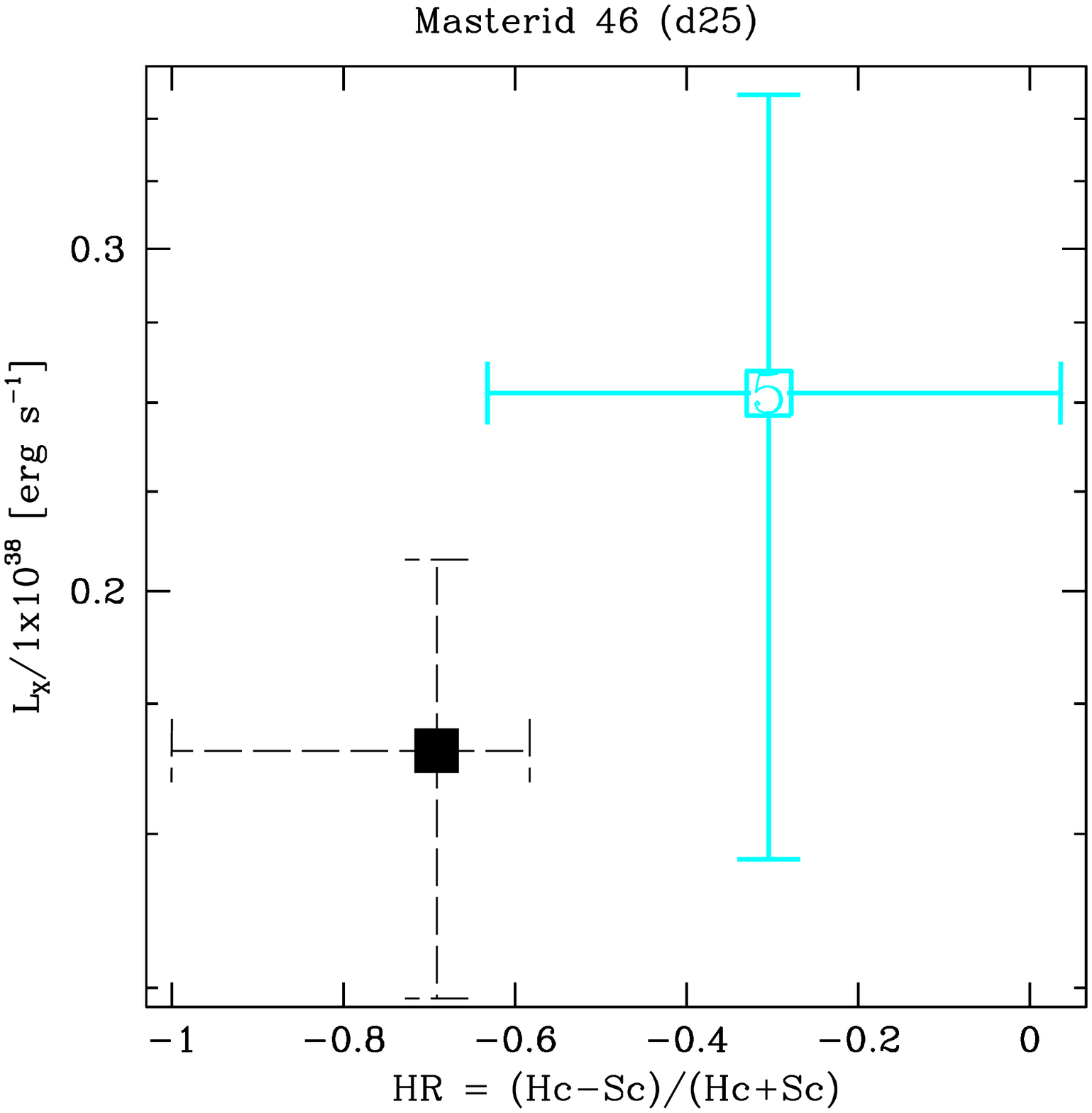}

\end{minipage}
\begin{minipage}{0.32\linewidth}
  \centering

    \includegraphics[width=\linewidth]{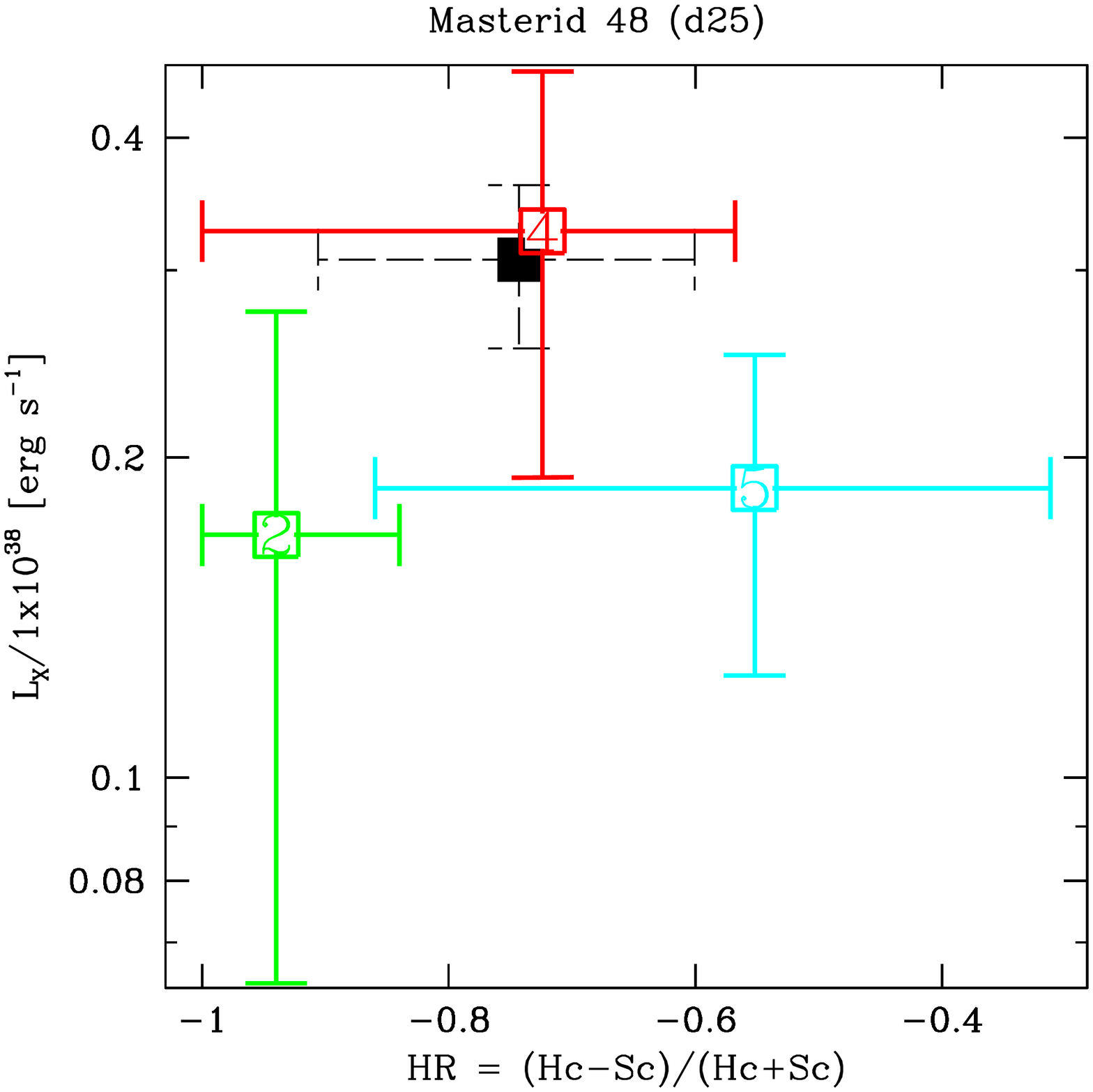}

 \end{minipage}

\begin{minipage}{0.32\linewidth}
  \centering
  
    \includegraphics[width=\linewidth]{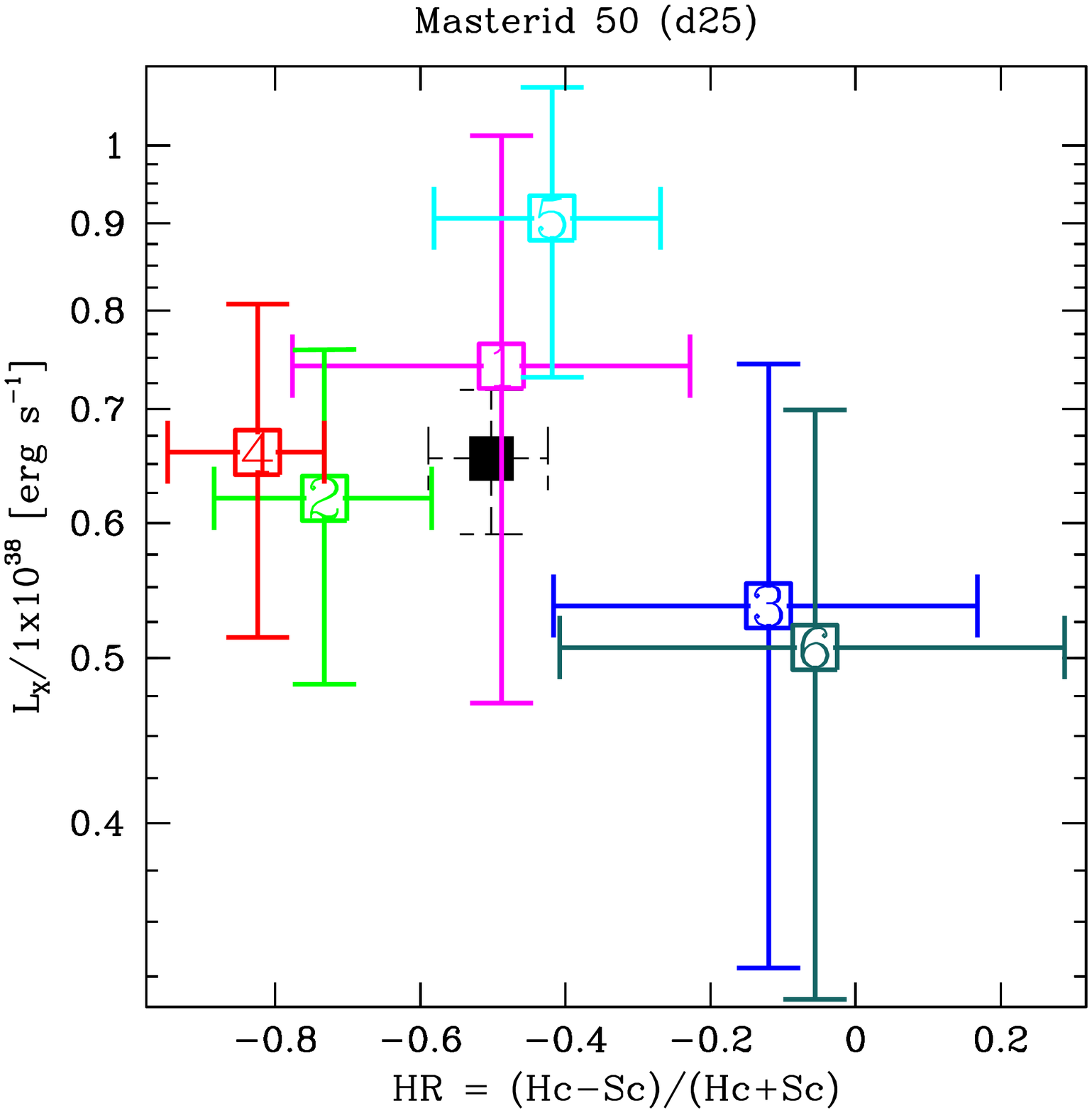}

  \end{minipage}
  \begin{minipage}{0.32\linewidth}
  \centering

    \includegraphics[width=\linewidth]{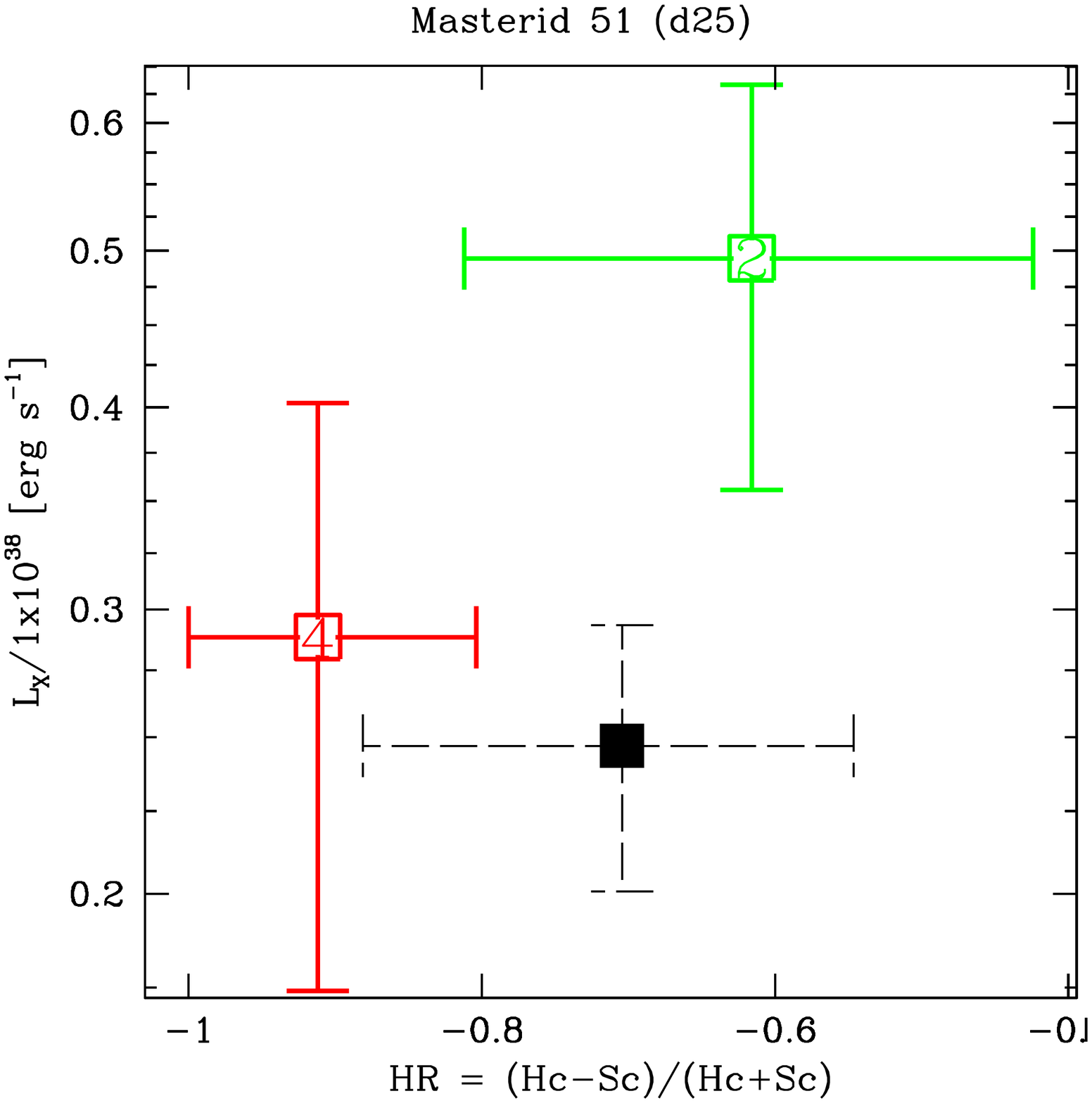}

\end{minipage}
\begin{minipage}{0.32\linewidth}
  \centering

    \includegraphics[width=\linewidth]{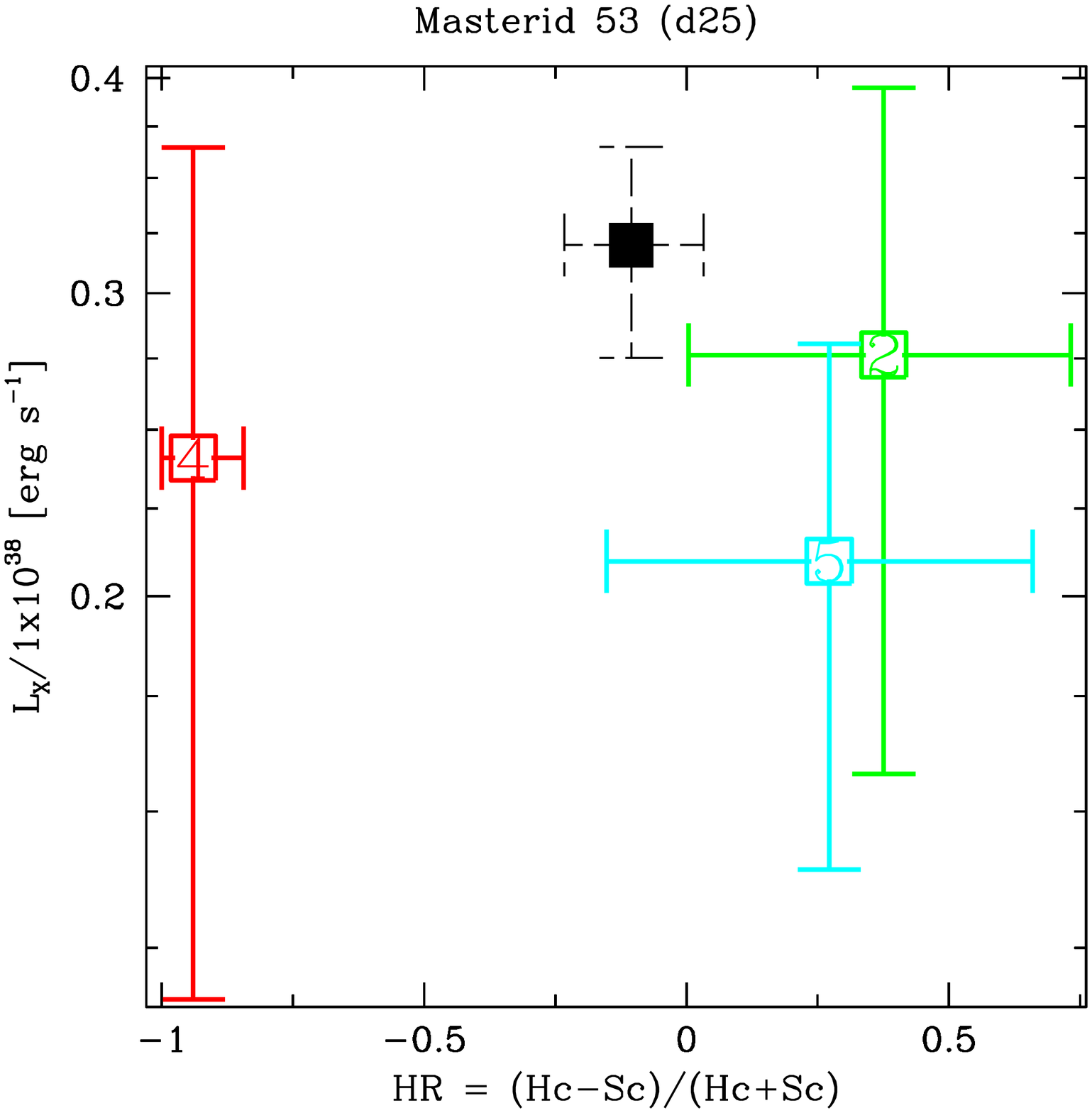}

 \end{minipage}

  \begin{minipage}{0.32\linewidth}
  \centering
  
    \includegraphics[width=\linewidth]{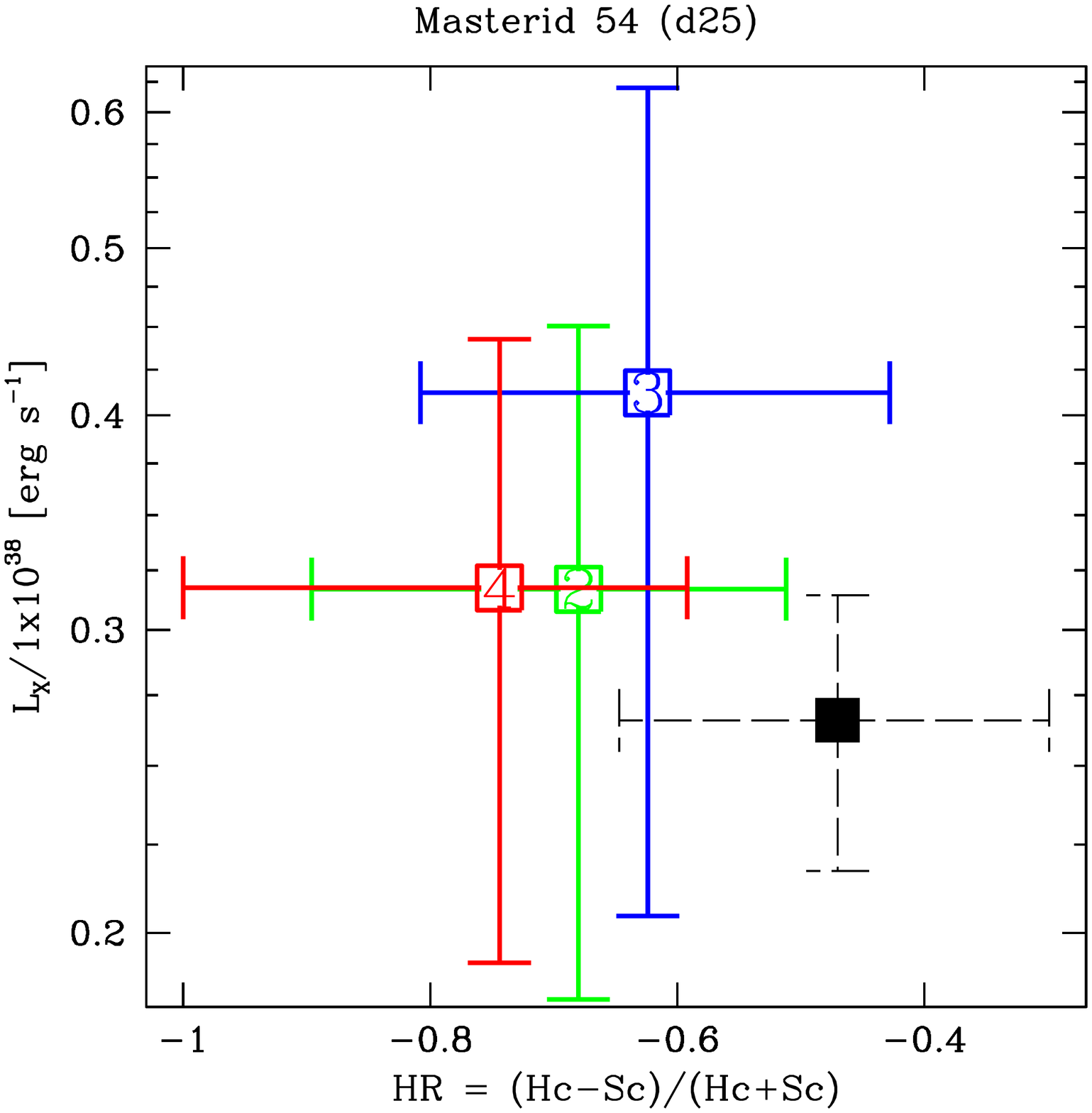}

  \end{minipage}
  \begin{minipage}{0.32\linewidth}
  \centering

    \includegraphics[width=\linewidth]{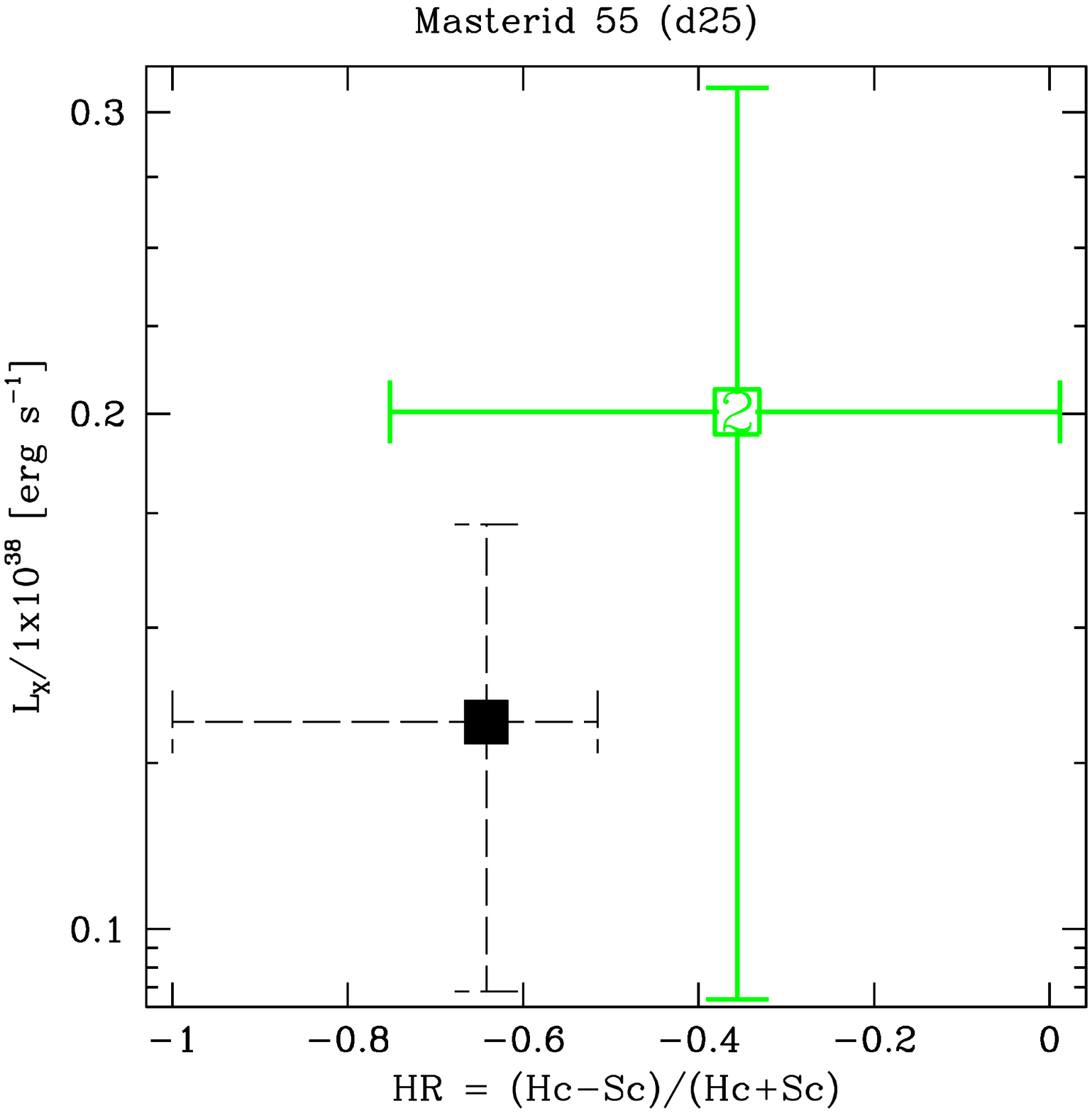}

\end{minipage}
\begin{minipage}{0.32\linewidth}
  \centering

    \includegraphics[width=\linewidth]{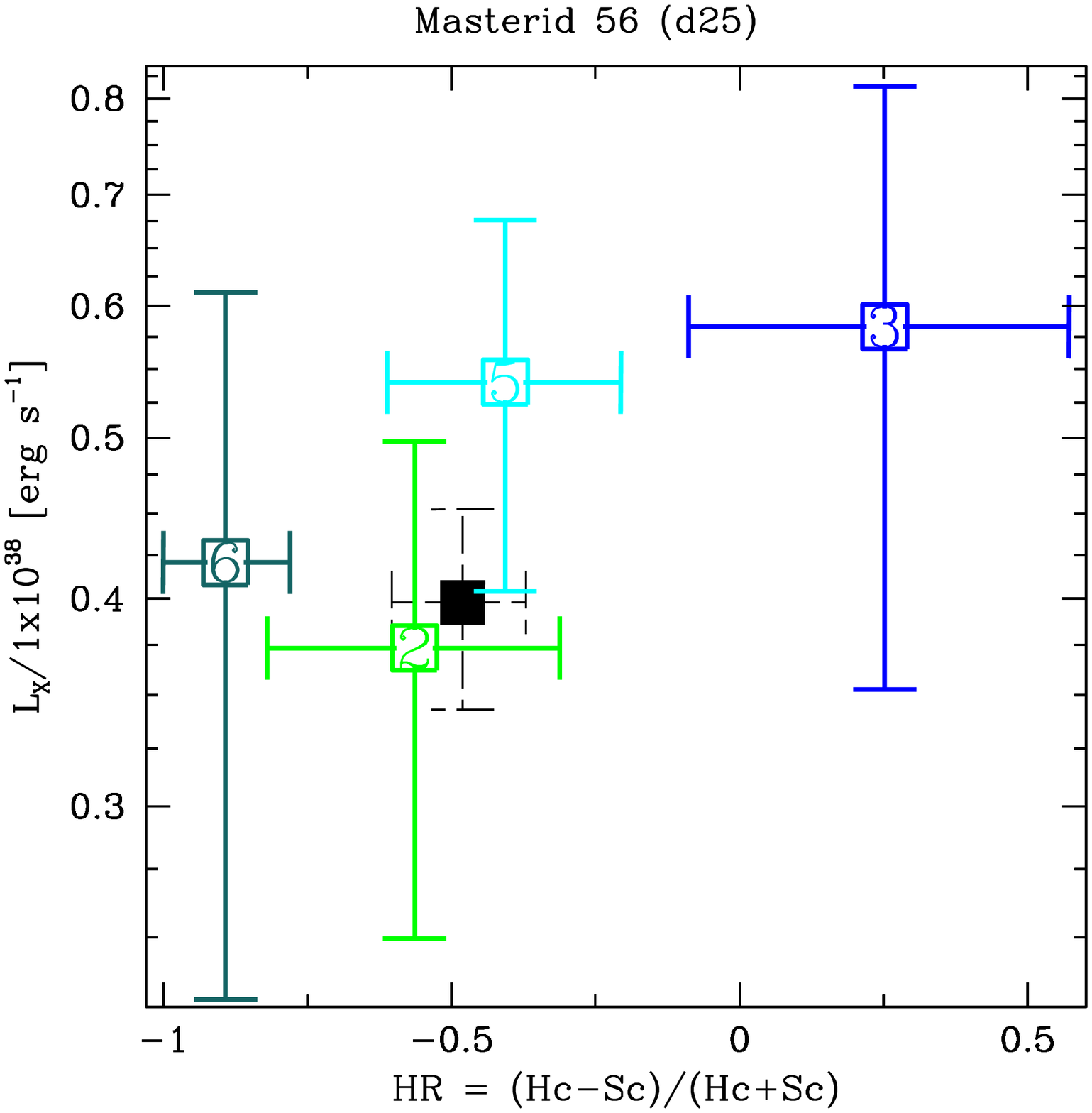}

 \end{minipage}

\begin{minipage}{0.32\linewidth}
  \centering
  
    \includegraphics[width=\linewidth]{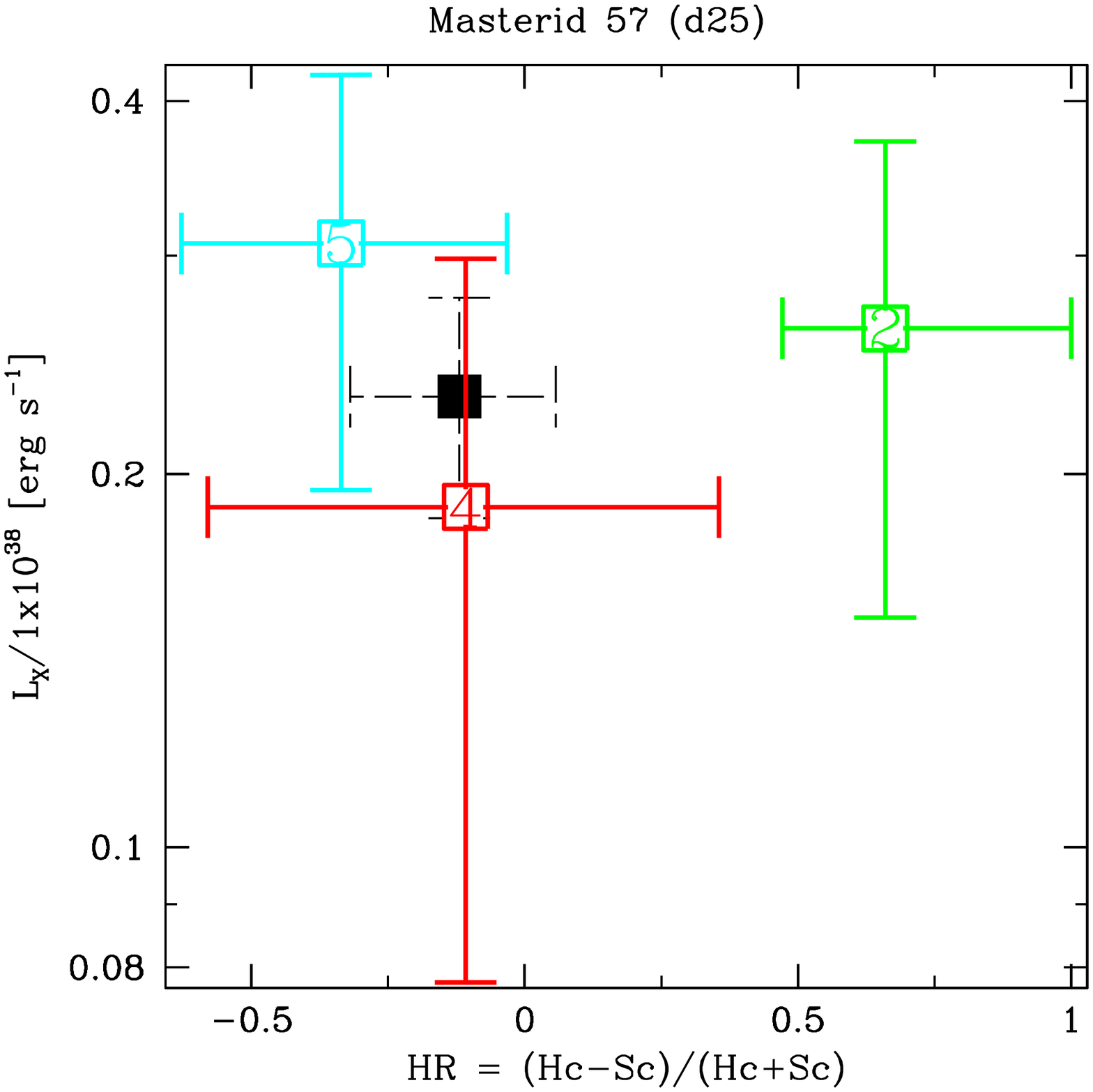}

  \end{minipage}
  \begin{minipage}{0.32\linewidth}
  \centering

    \includegraphics[width=\linewidth]{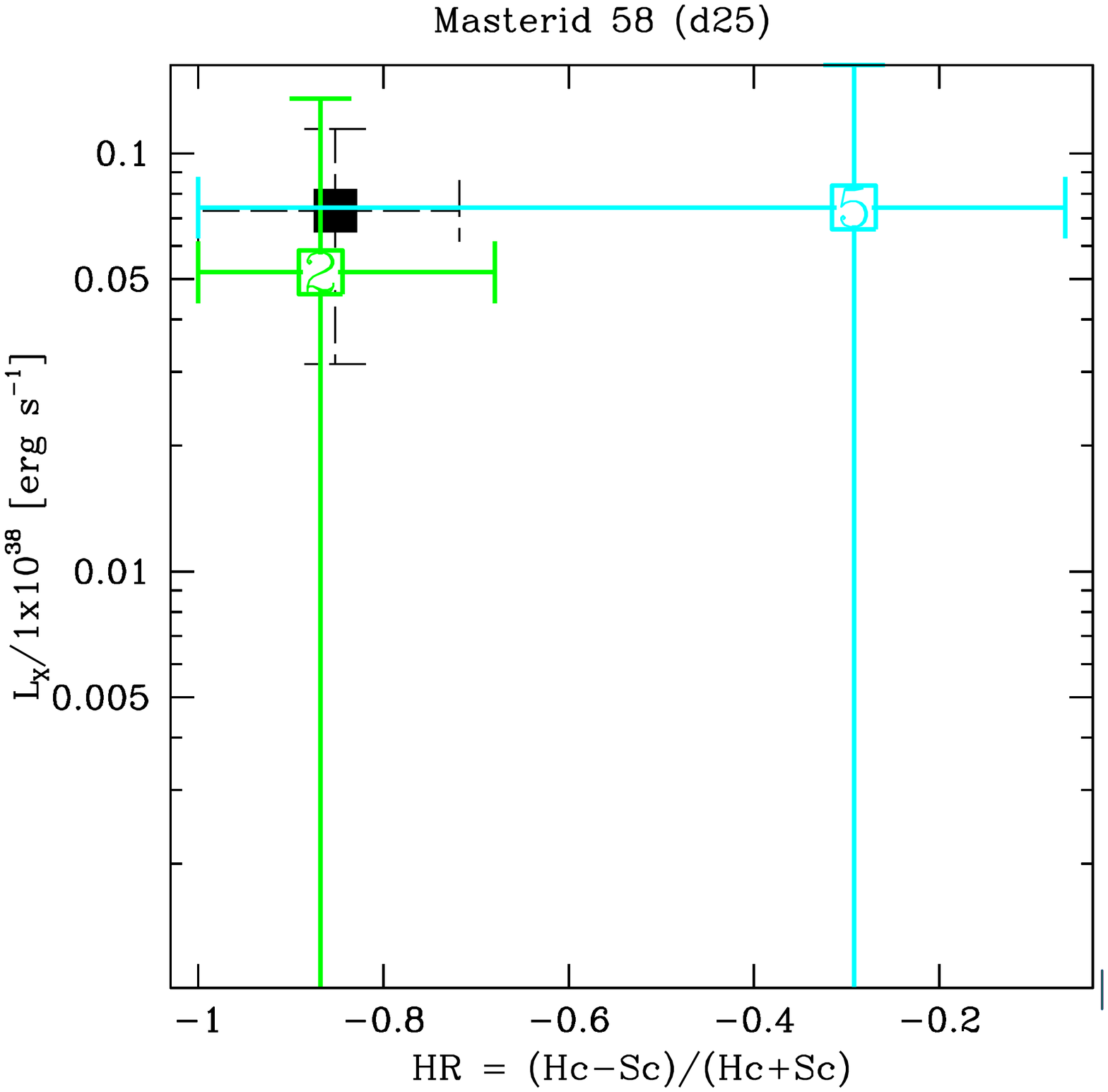}

\end{minipage}
\begin{minipage}{0.32\linewidth}
  \centering

    \includegraphics[width=\linewidth]{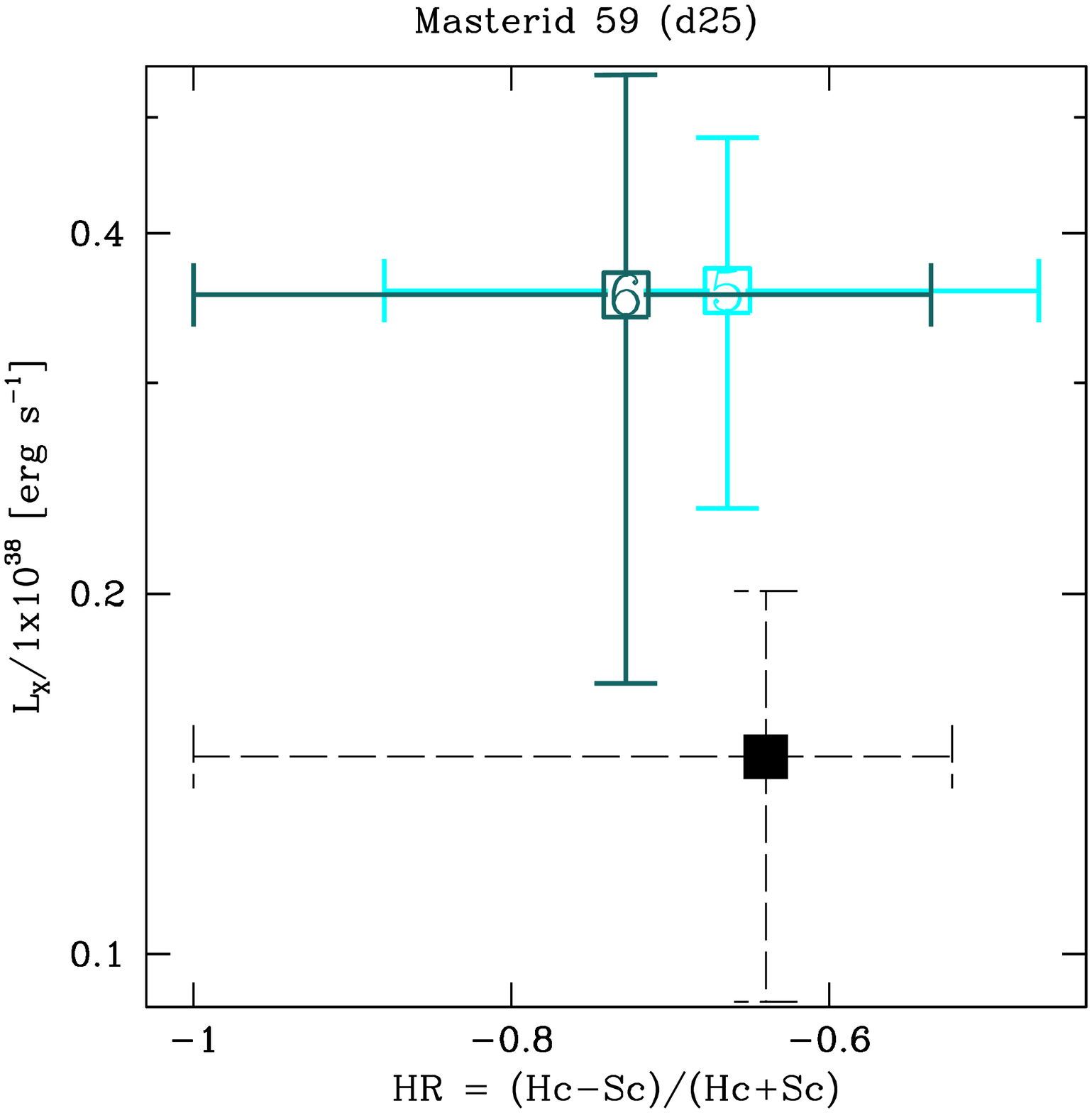}

 \end{minipage}
  
\end{figure}

\begin{figure}
  \begin{minipage}{0.32\linewidth}
  \centering
  
    \includegraphics[width=\linewidth]{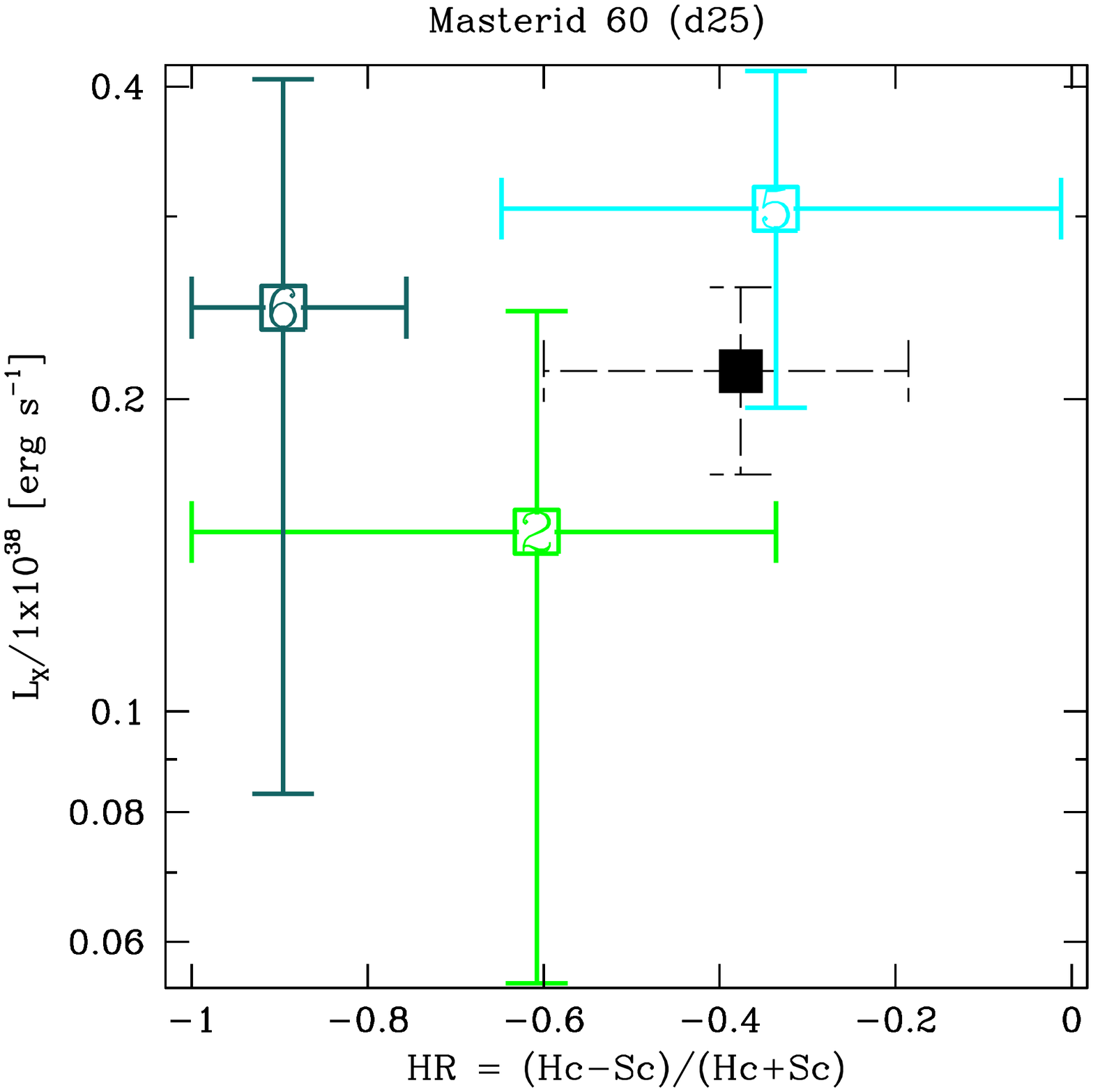}

  \end{minipage}
  \begin{minipage}{0.32\linewidth}
  \centering

    \includegraphics[width=\linewidth]{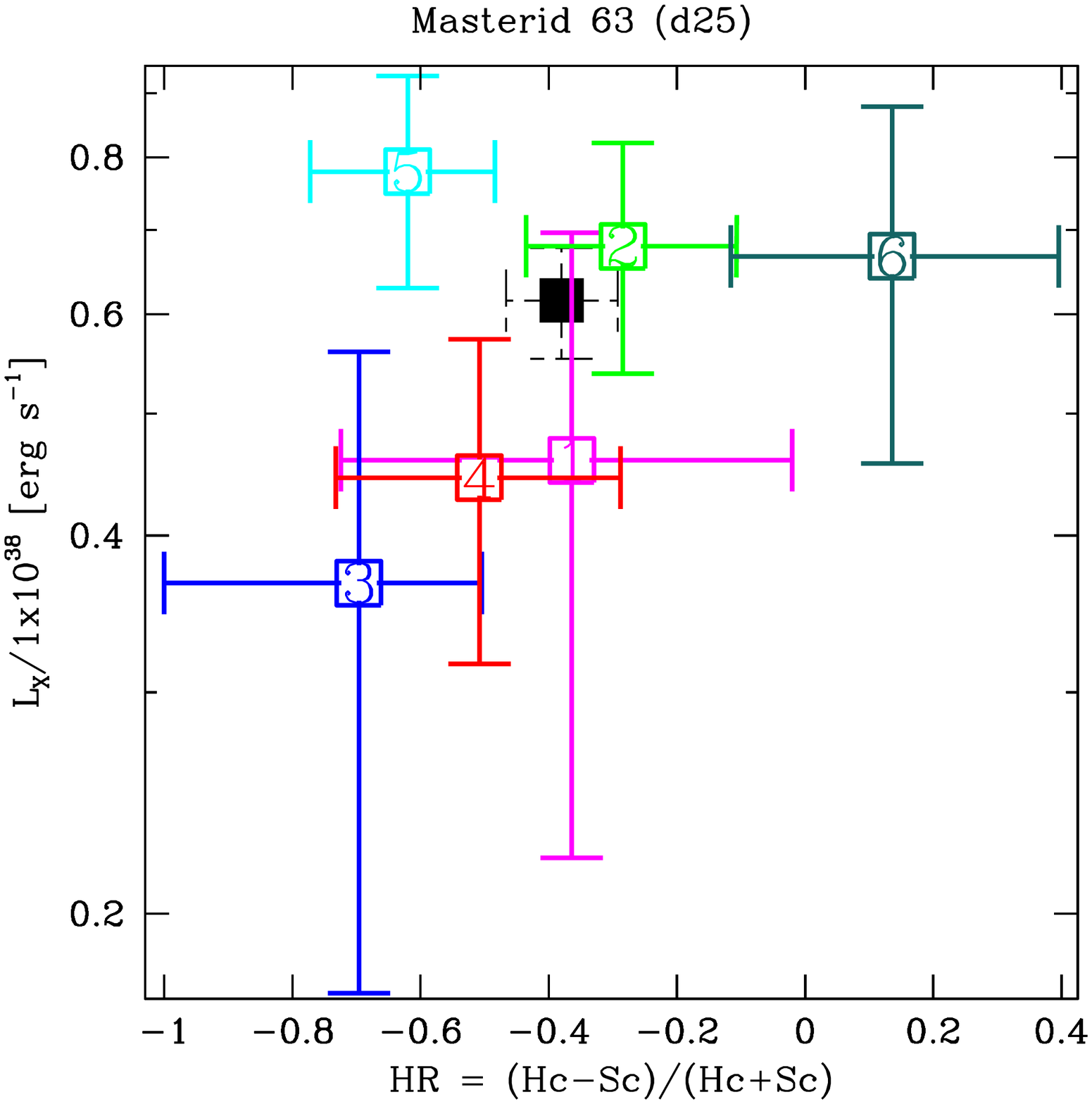}

\end{minipage}
\begin{minipage}{0.32\linewidth}
  \centering

    \includegraphics[width=\linewidth]{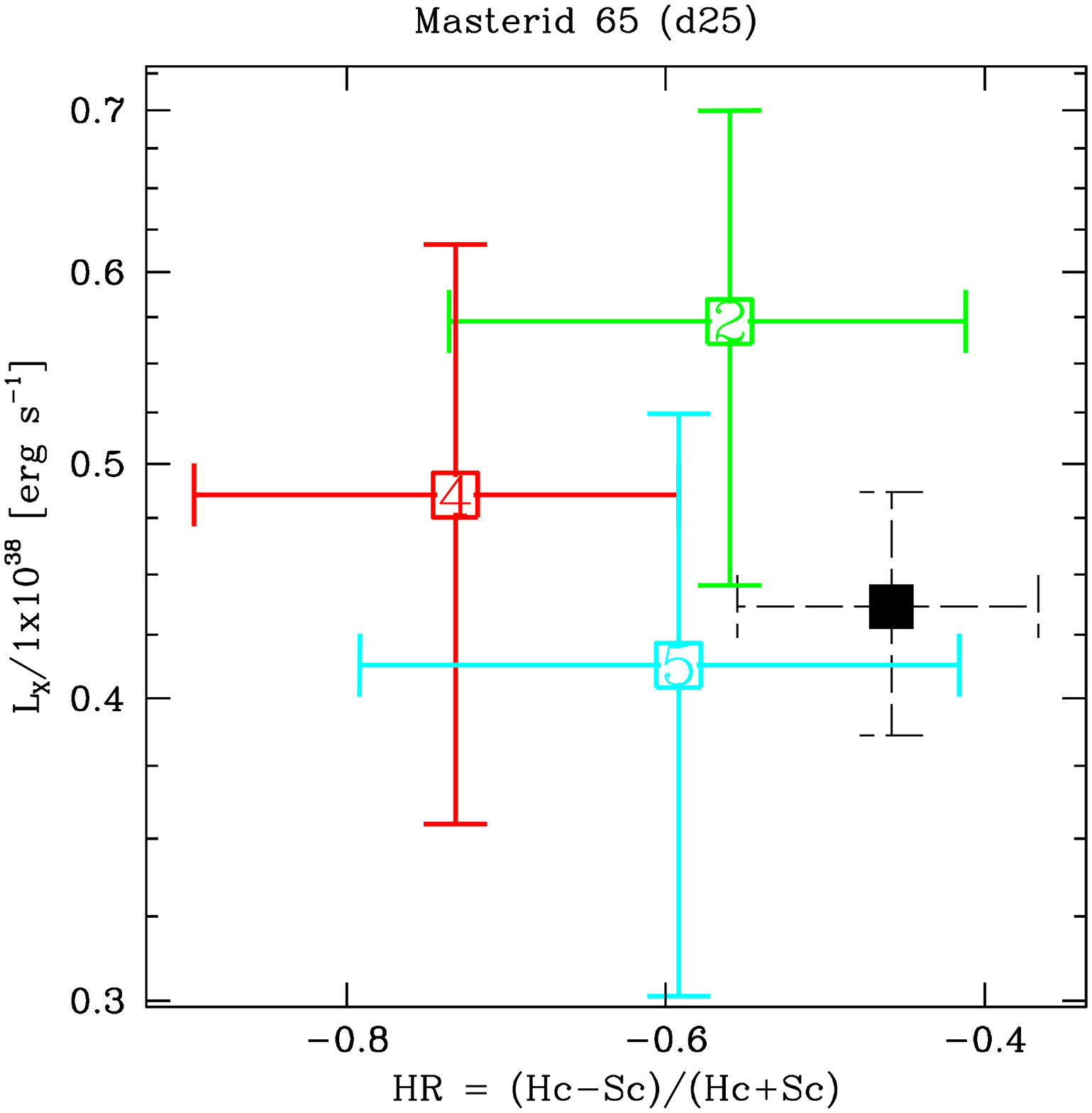}

 \end{minipage}

\begin{minipage}{0.32\linewidth}
  \centering
  
    \includegraphics[width=\linewidth]{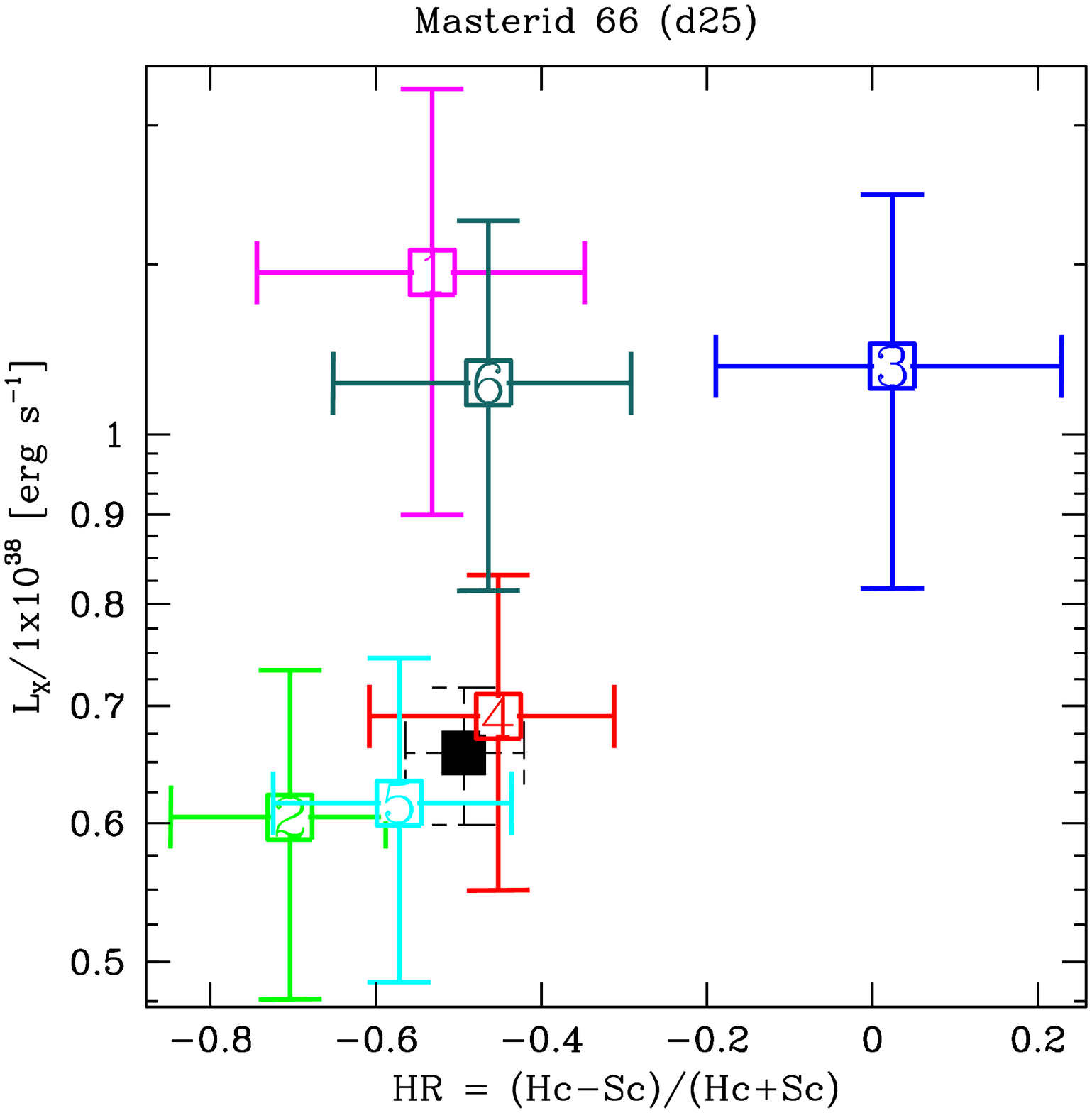}

  \end{minipage}
  \begin{minipage}{0.32\linewidth}
  \centering

    \includegraphics[width=\linewidth]{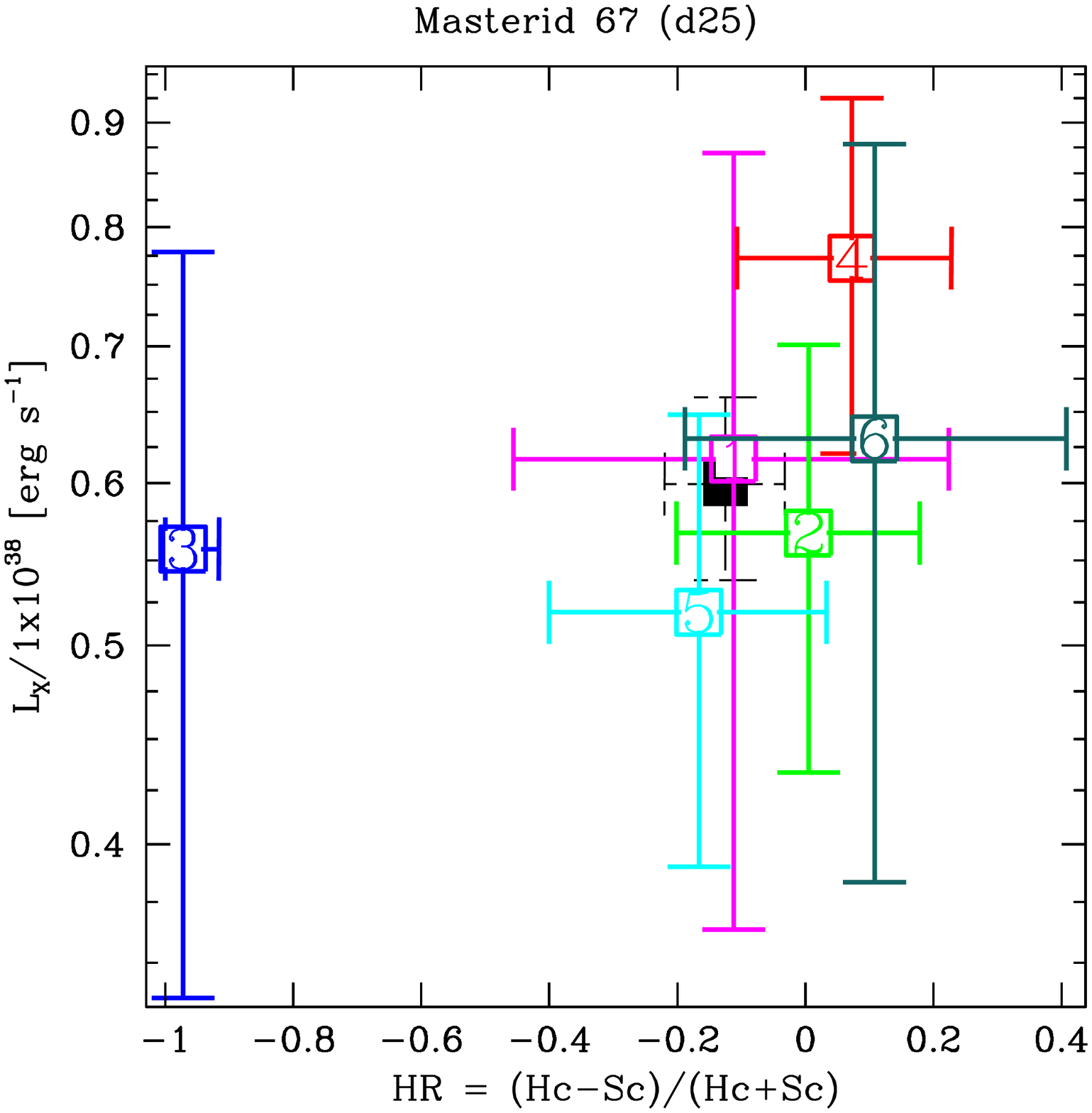}

\end{minipage}
\begin{minipage}{0.32\linewidth}
  \centering

    \includegraphics[width=\linewidth]{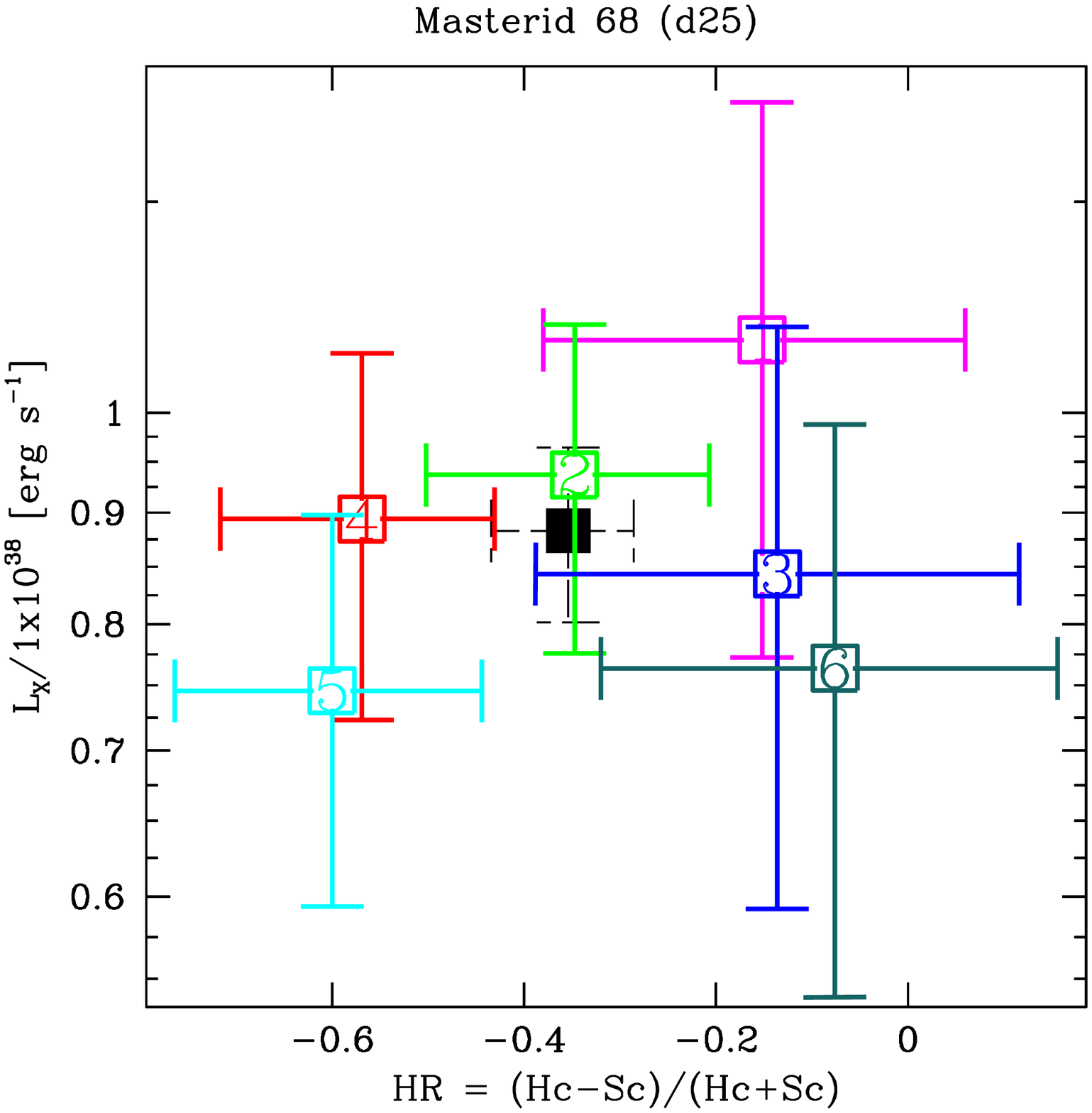}

 \end{minipage}

  \begin{minipage}{0.32\linewidth}
  \centering
  
    \includegraphics[width=\linewidth]{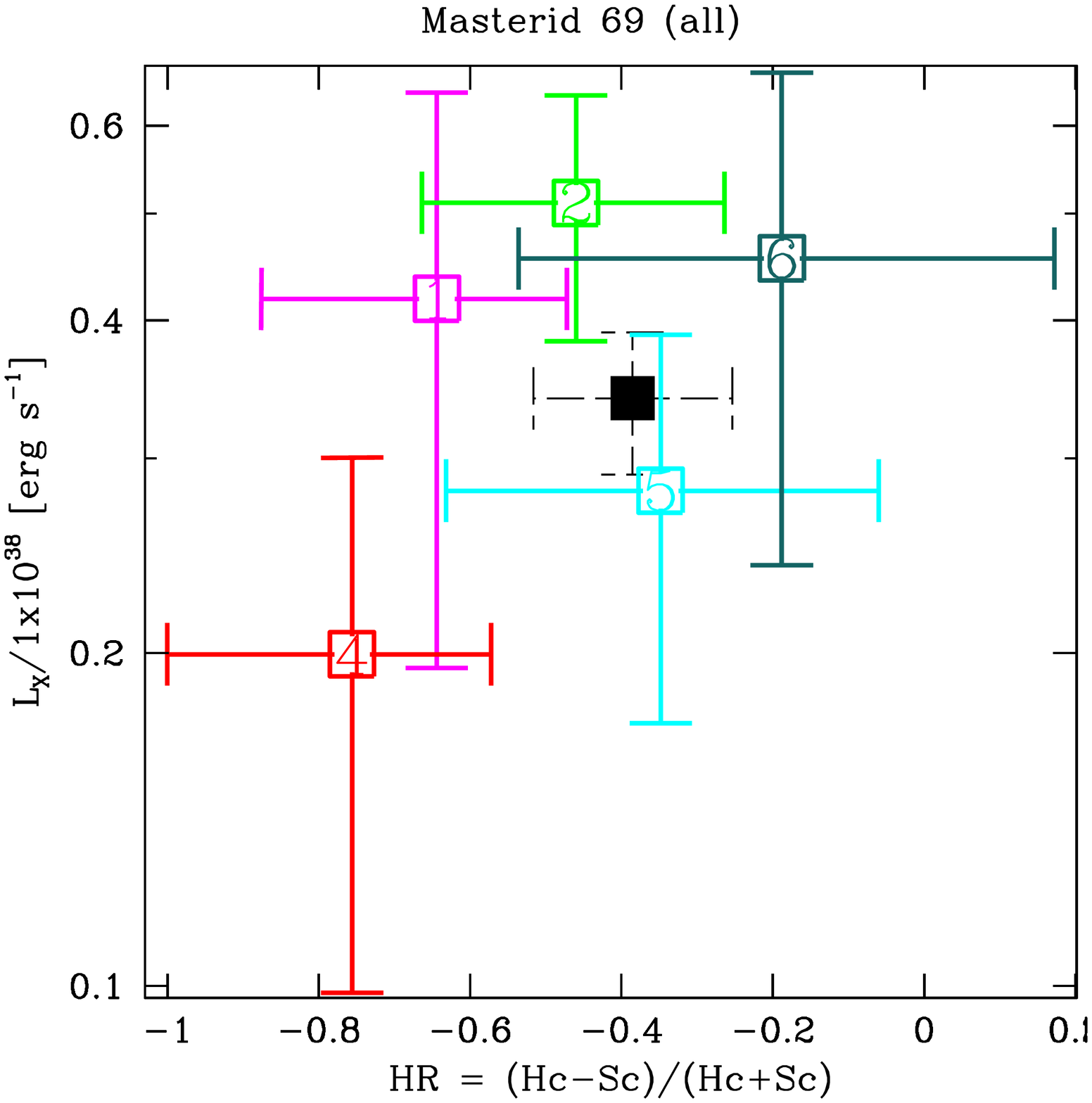}

  \end{minipage}
  \begin{minipage}{0.32\linewidth}
  \centering

    \includegraphics[width=\linewidth]{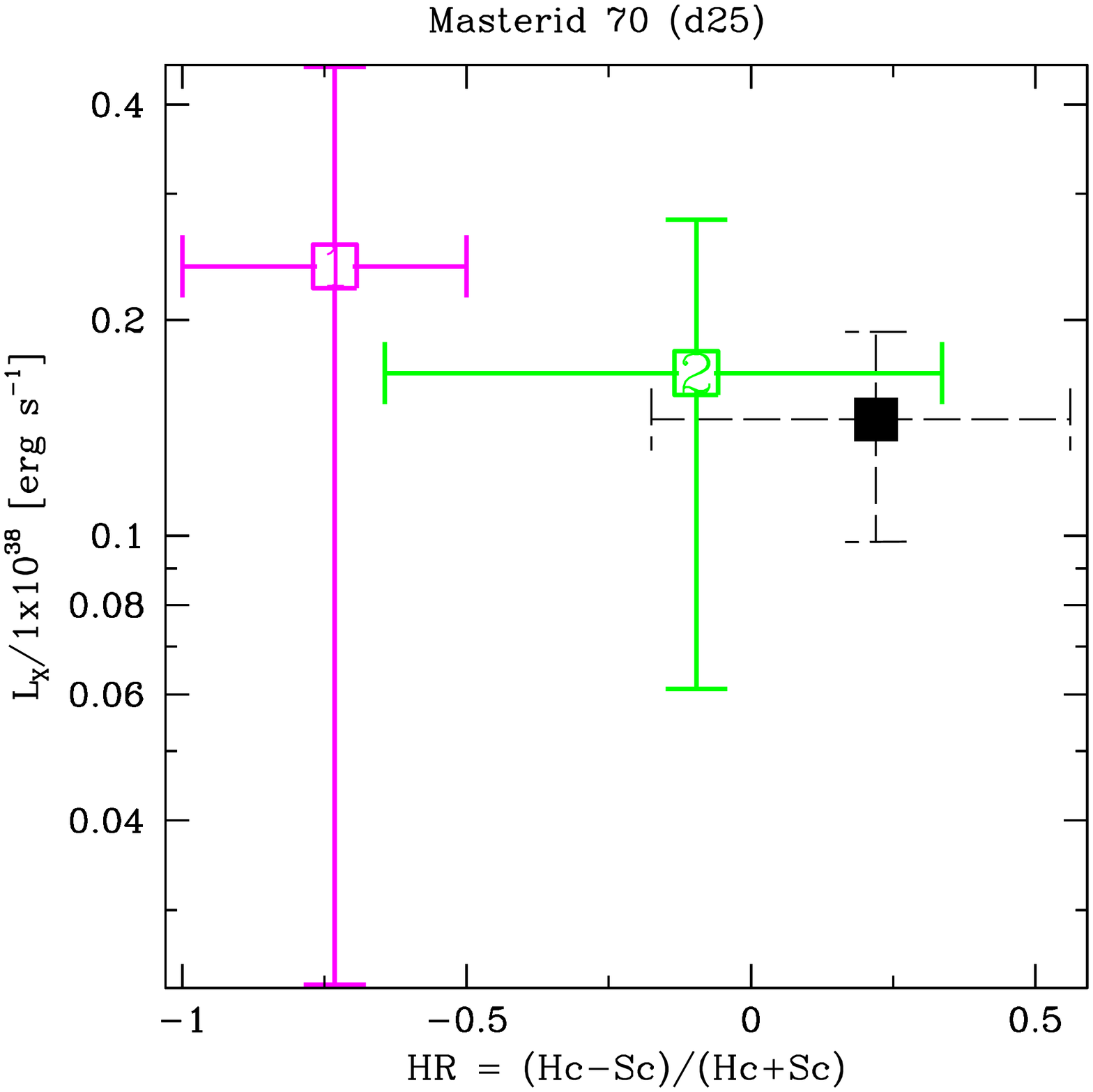}

\end{minipage}
\begin{minipage}{0.32\linewidth}
  \centering

    \includegraphics[width=\linewidth]{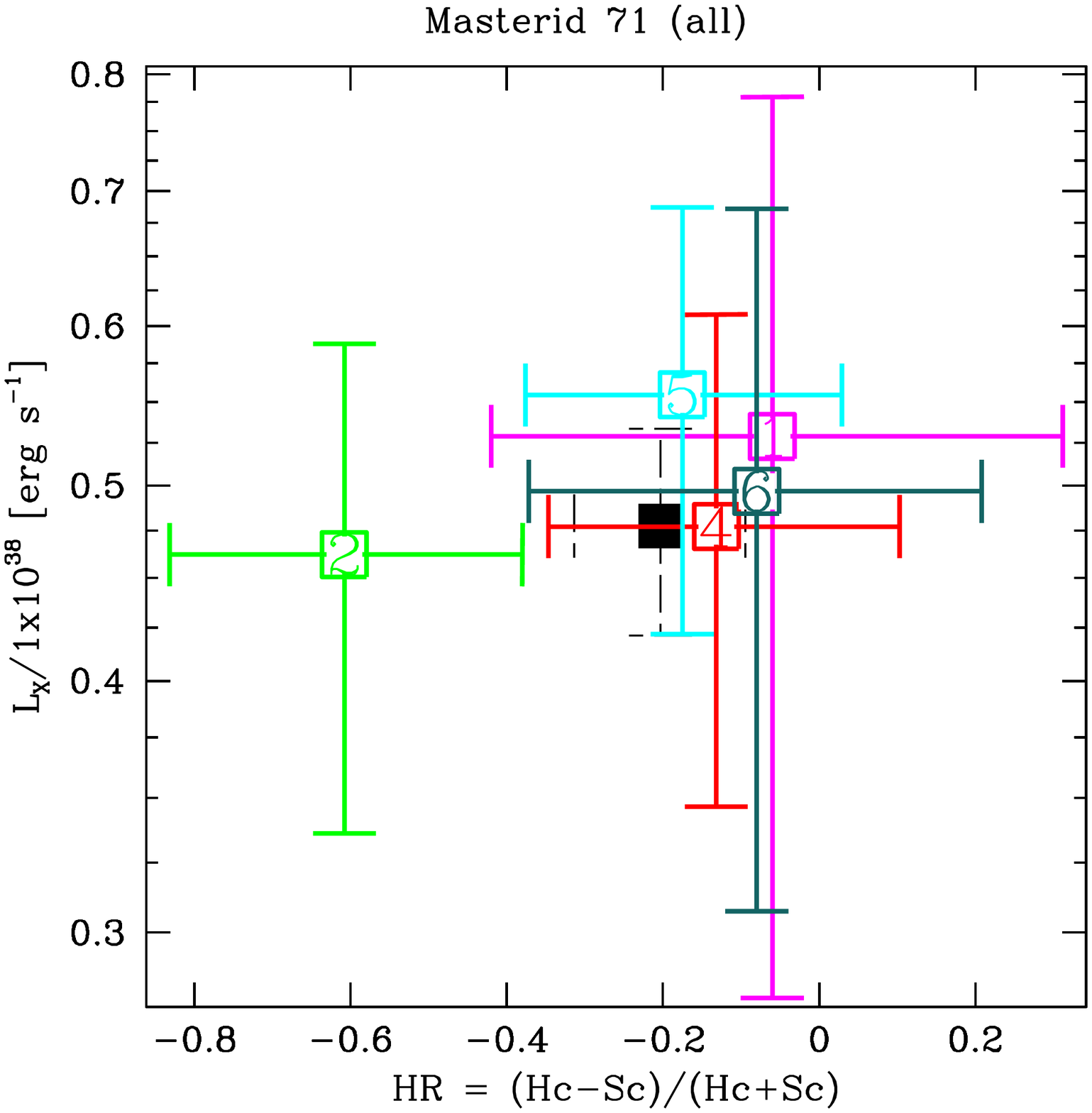}

 \end{minipage}

\begin{minipage}{0.32\linewidth}
  \centering
  
    \includegraphics[width=\linewidth]{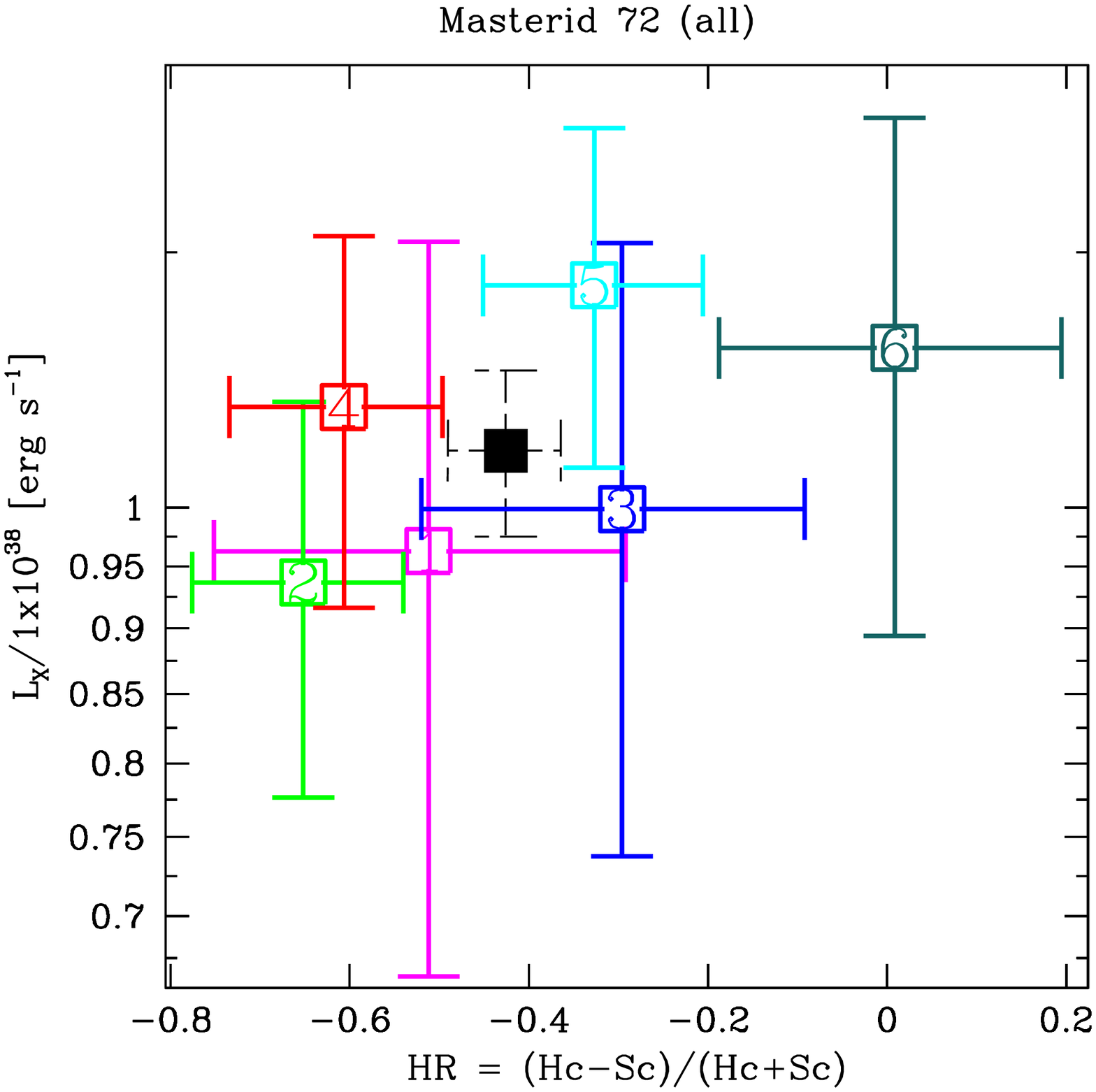}

  \end{minipage}
  \begin{minipage}{0.32\linewidth}
  \centering

    \includegraphics[width=\linewidth]{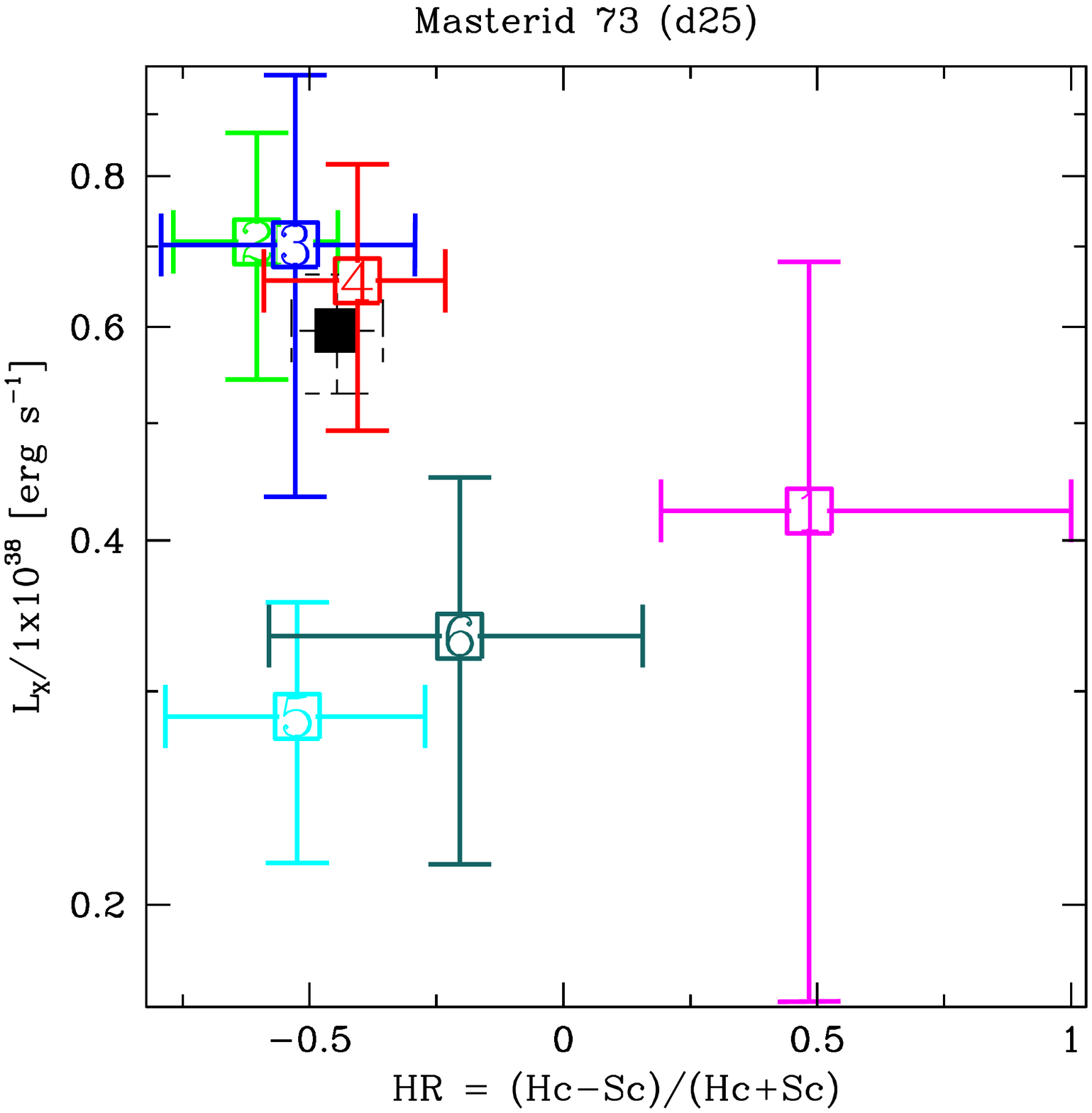}

\end{minipage}
\begin{minipage}{0.32\linewidth}
  \centering

    \includegraphics[width=\linewidth]{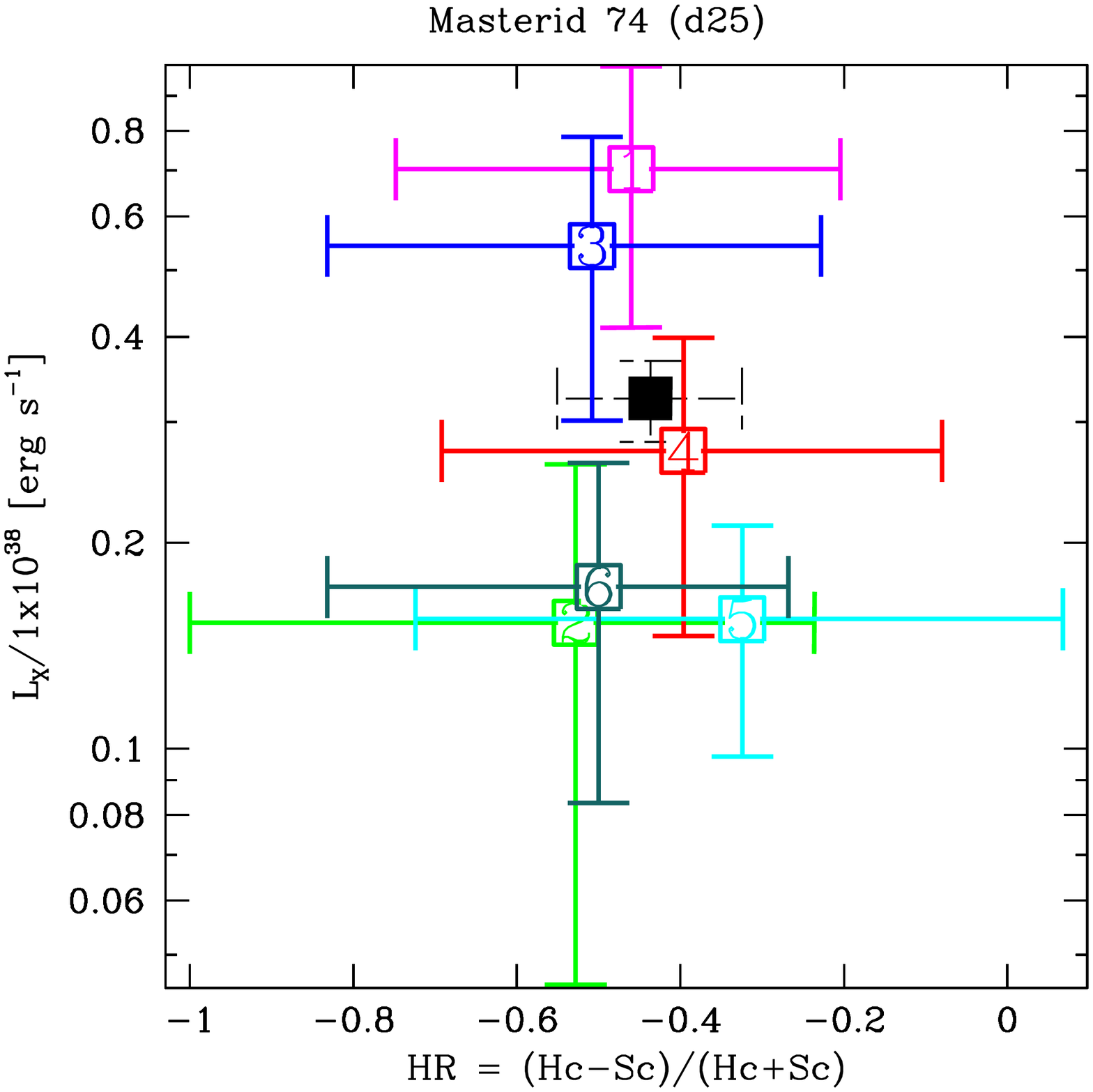}

 \end{minipage}
  
\end{figure}

\begin{figure}
  \begin{minipage}{0.32\linewidth}
  \centering
  
    \includegraphics[width=\linewidth]{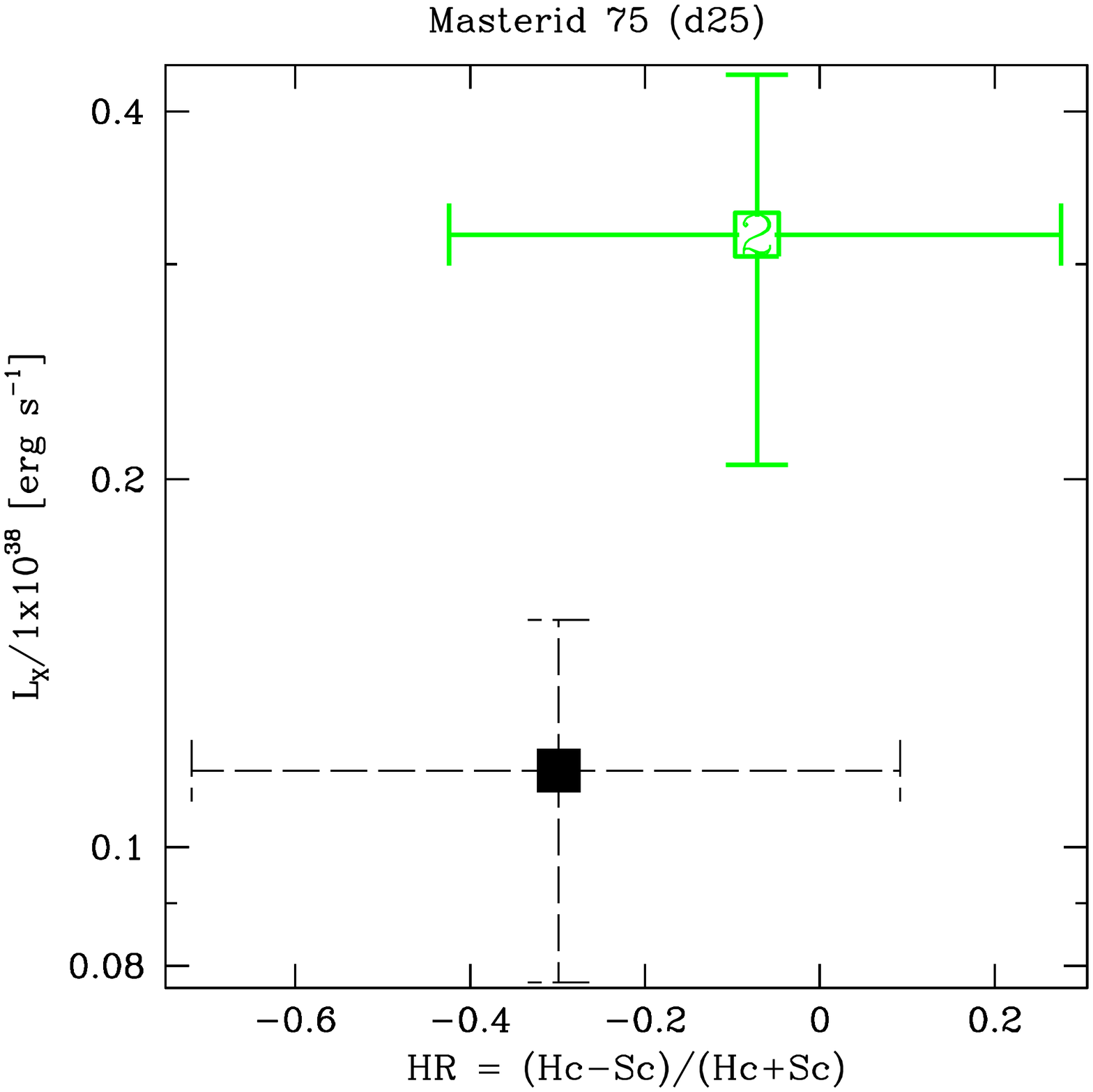}

  \end{minipage}
  \begin{minipage}{0.32\linewidth}
  \centering

    \includegraphics[width=\linewidth]{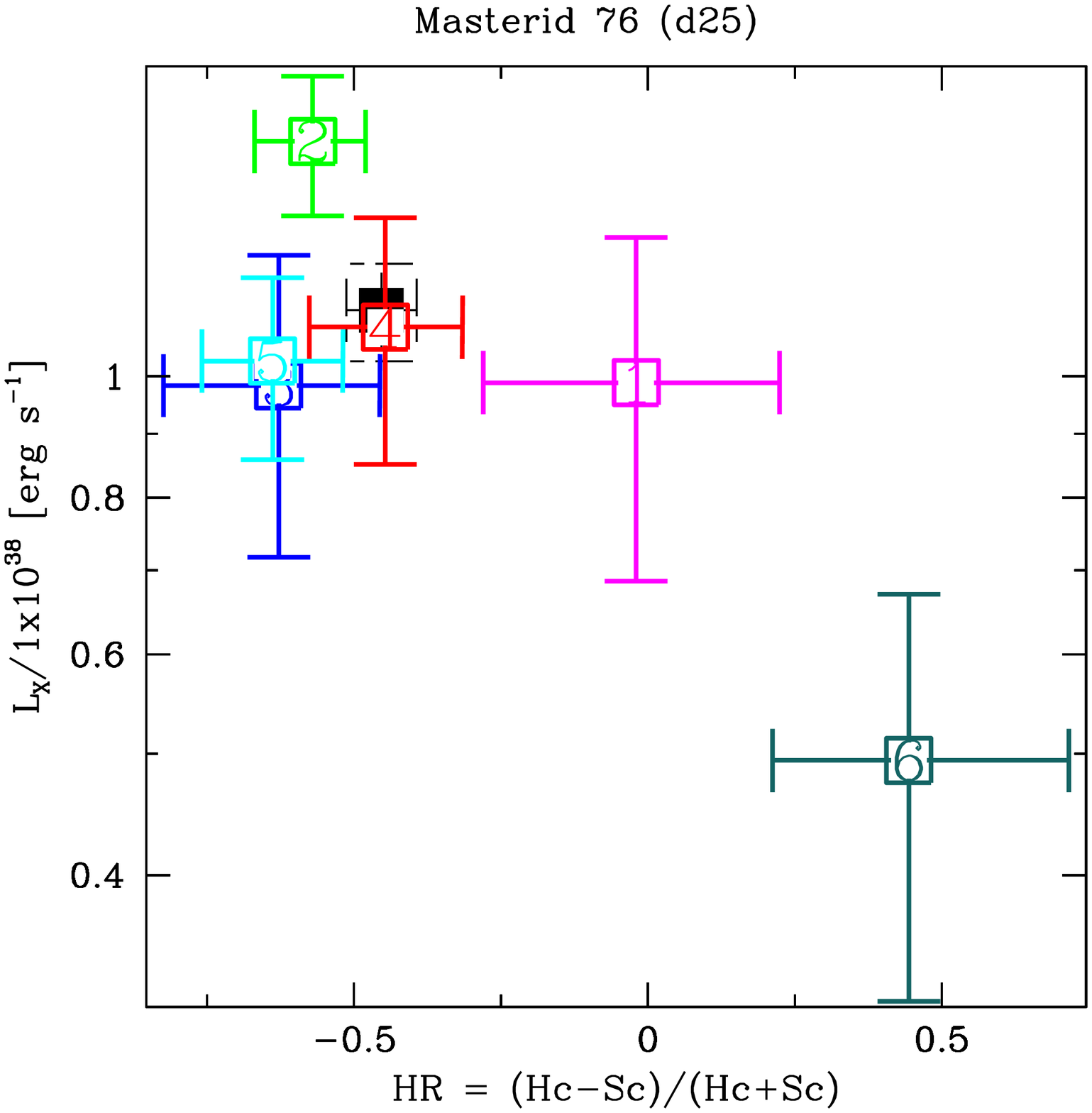}

\end{minipage}
\begin{minipage}{0.32\linewidth}
  \centering

    \includegraphics[width=\linewidth]{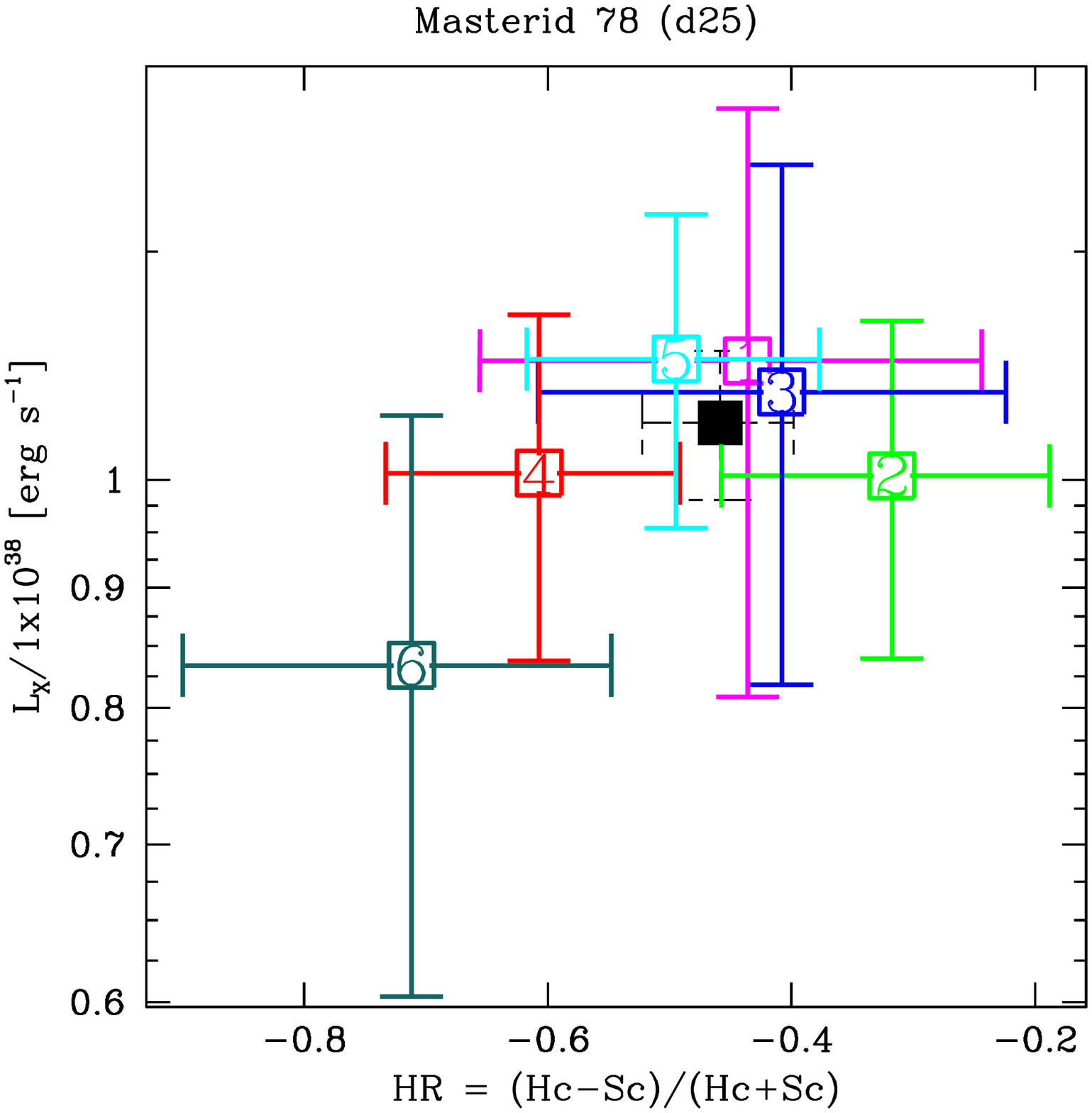}

 \end{minipage}

\begin{minipage}{0.32\linewidth}
  \centering
  
    \includegraphics[width=\linewidth]{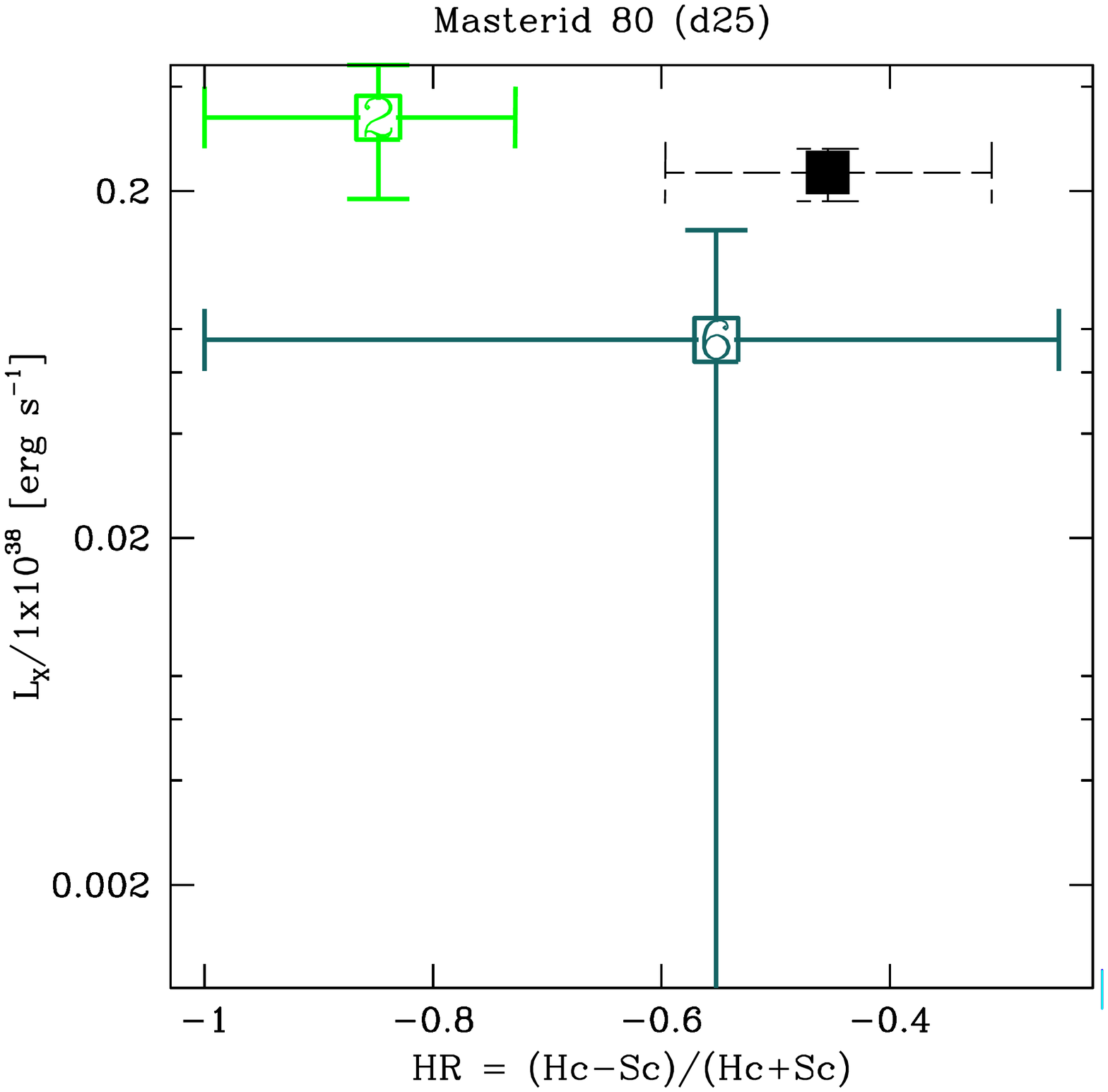}

  \end{minipage}
  \begin{minipage}{0.32\linewidth}
  \centering

    \includegraphics[width=\linewidth]{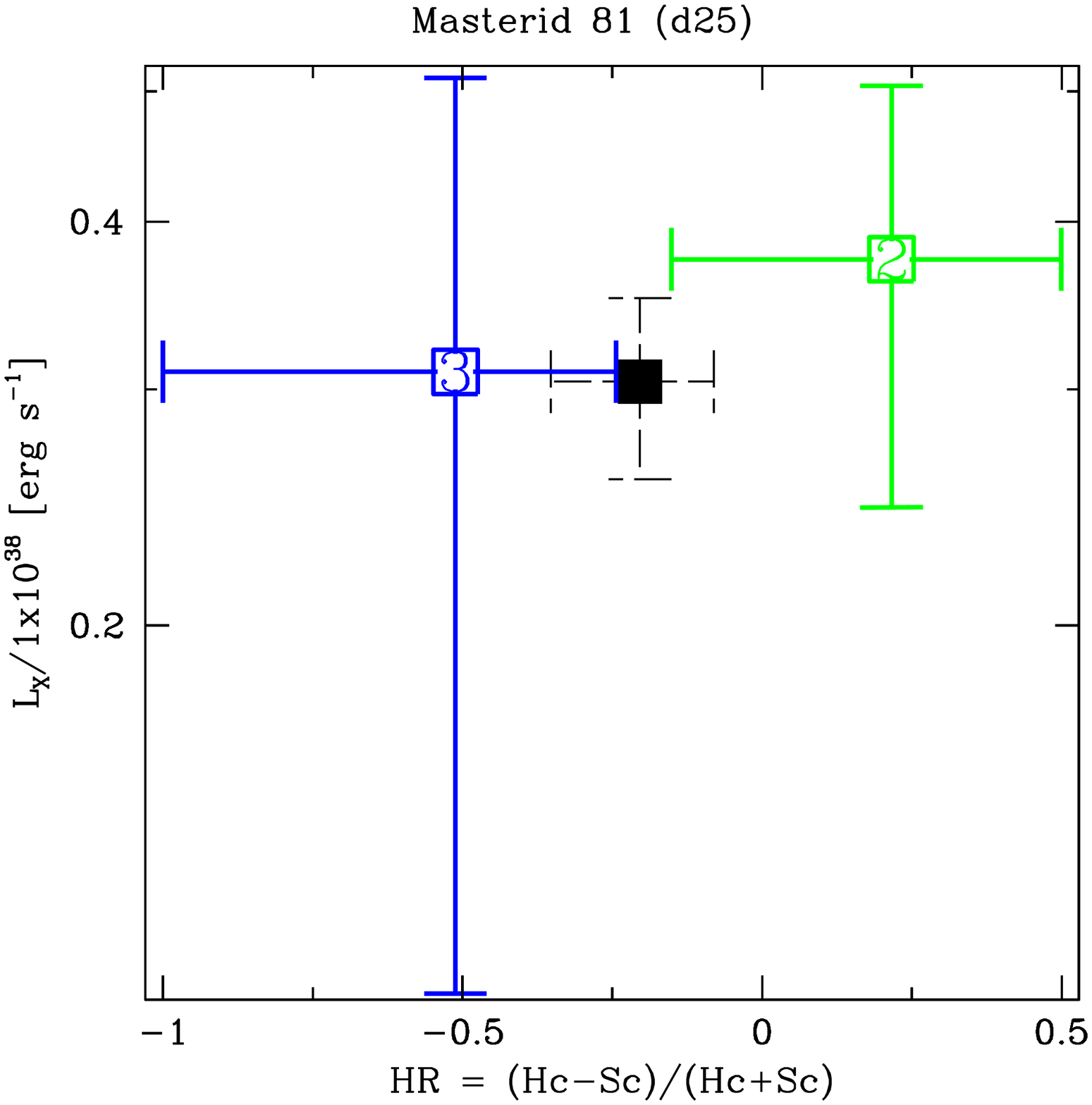}

\end{minipage}
\begin{minipage}{0.32\linewidth}
  \centering

    \includegraphics[width=\linewidth]{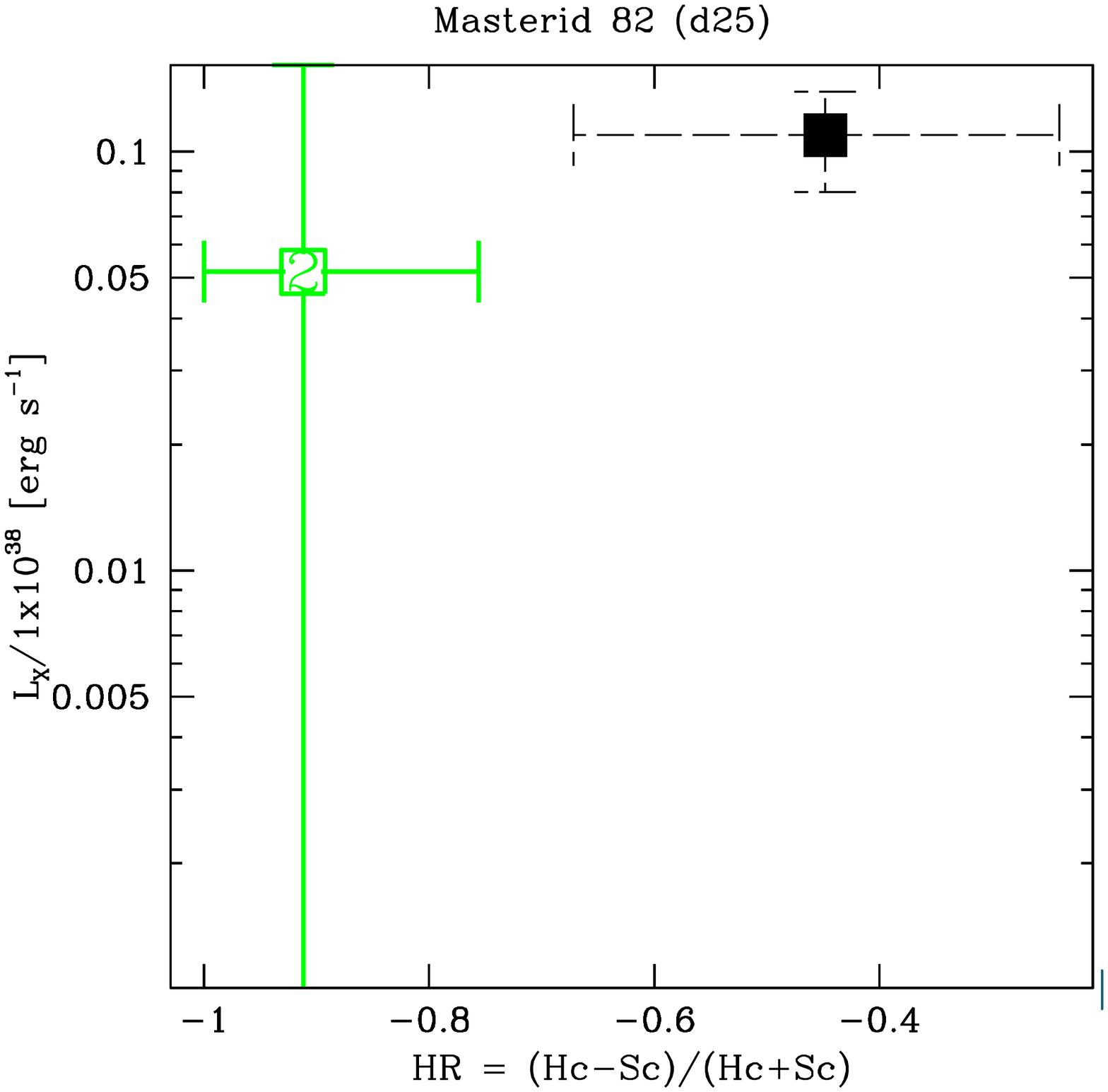}

 \end{minipage}

  \begin{minipage}{0.32\linewidth}
  \centering
  
    \includegraphics[width=\linewidth]{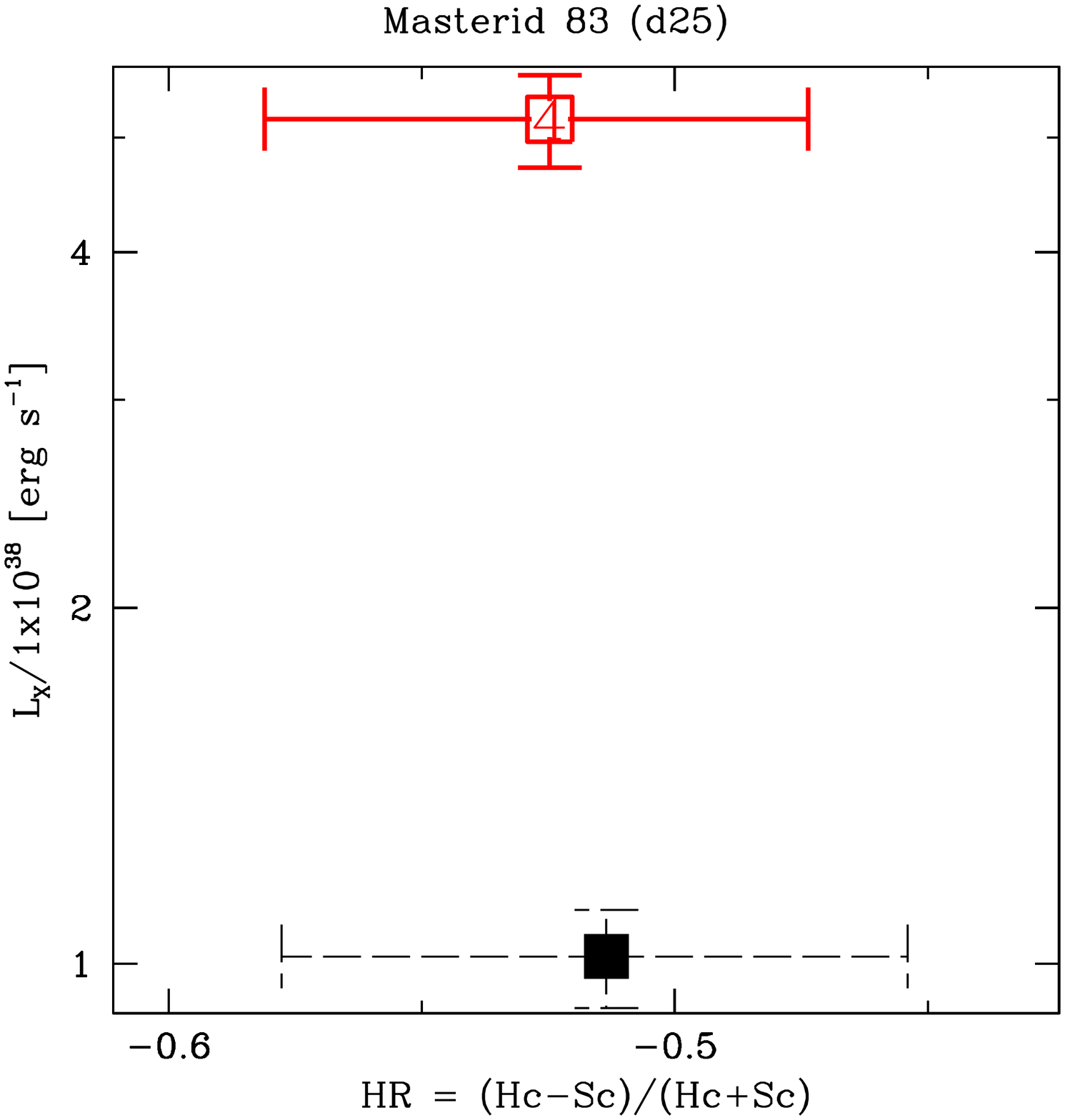}

  \end{minipage}
  \begin{minipage}{0.32\linewidth}
  \centering

    \includegraphics[width=\linewidth]{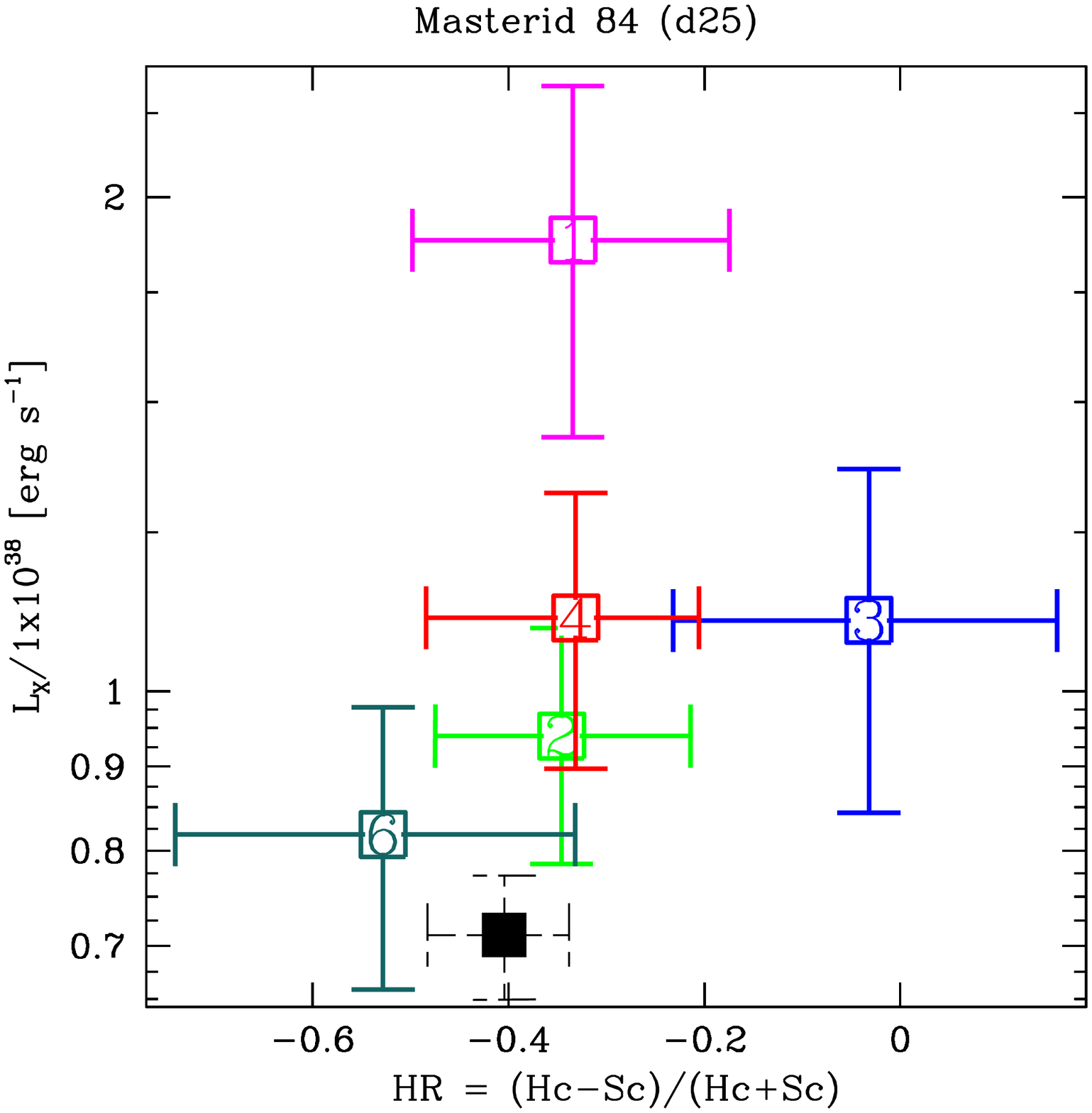}

\end{minipage}
\begin{minipage}{0.32\linewidth}
  \centering

    \includegraphics[width=\linewidth]{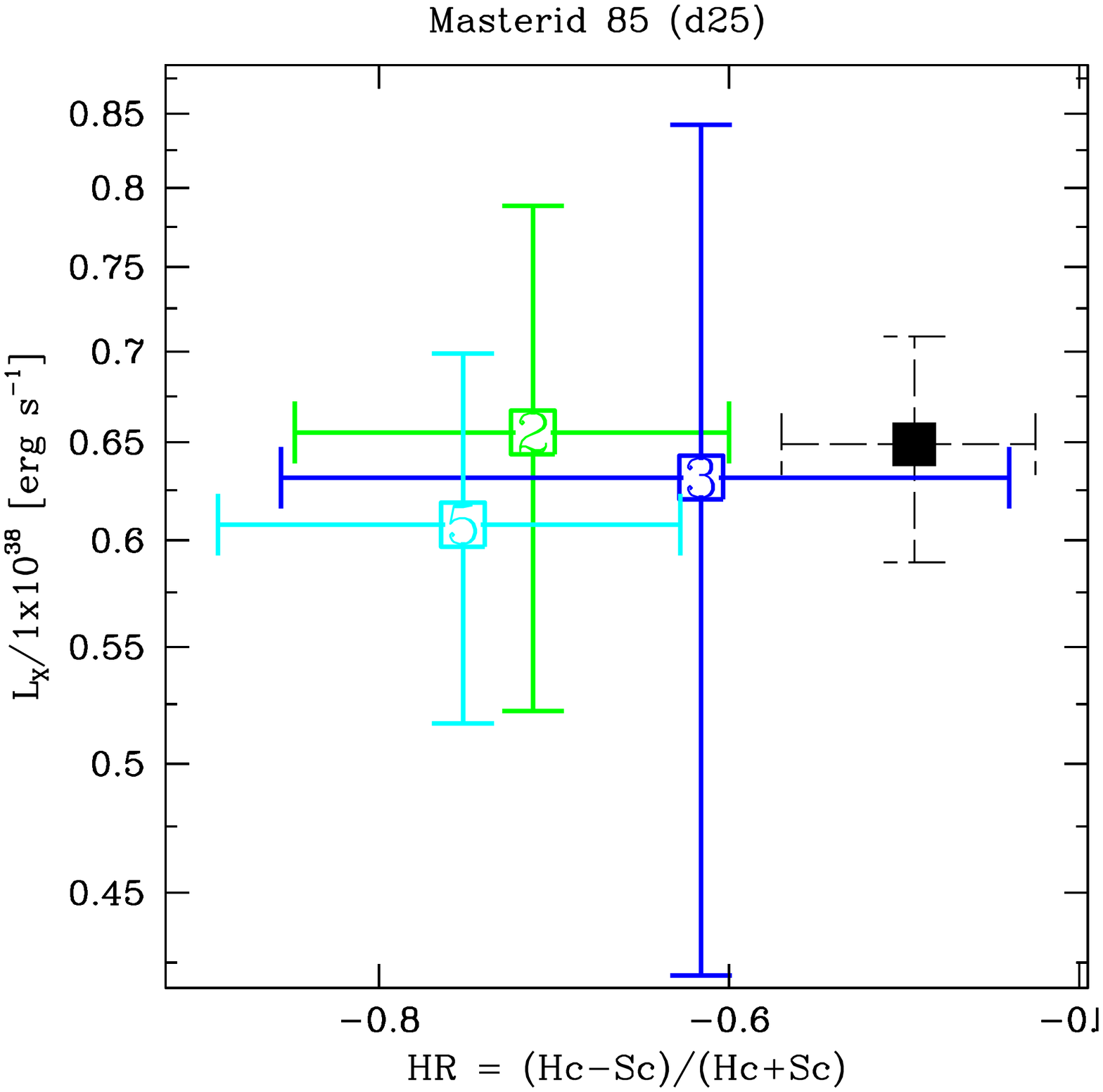}

 \end{minipage}

\begin{minipage}{0.32\linewidth}
  \centering
  
    \includegraphics[width=\linewidth]{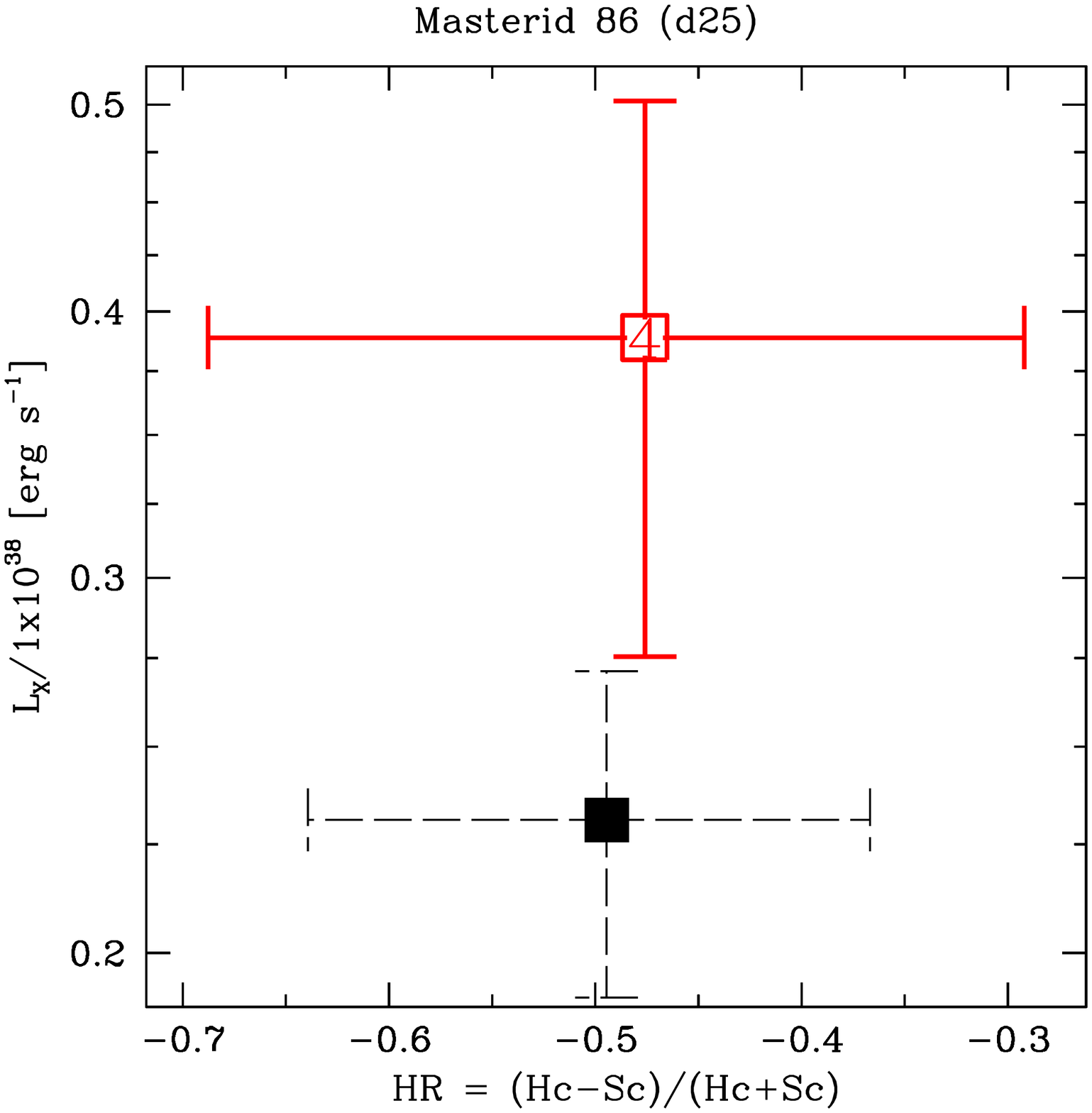}

  \end{minipage}
  \begin{minipage}{0.32\linewidth}
  \centering

    \includegraphics[width=\linewidth]{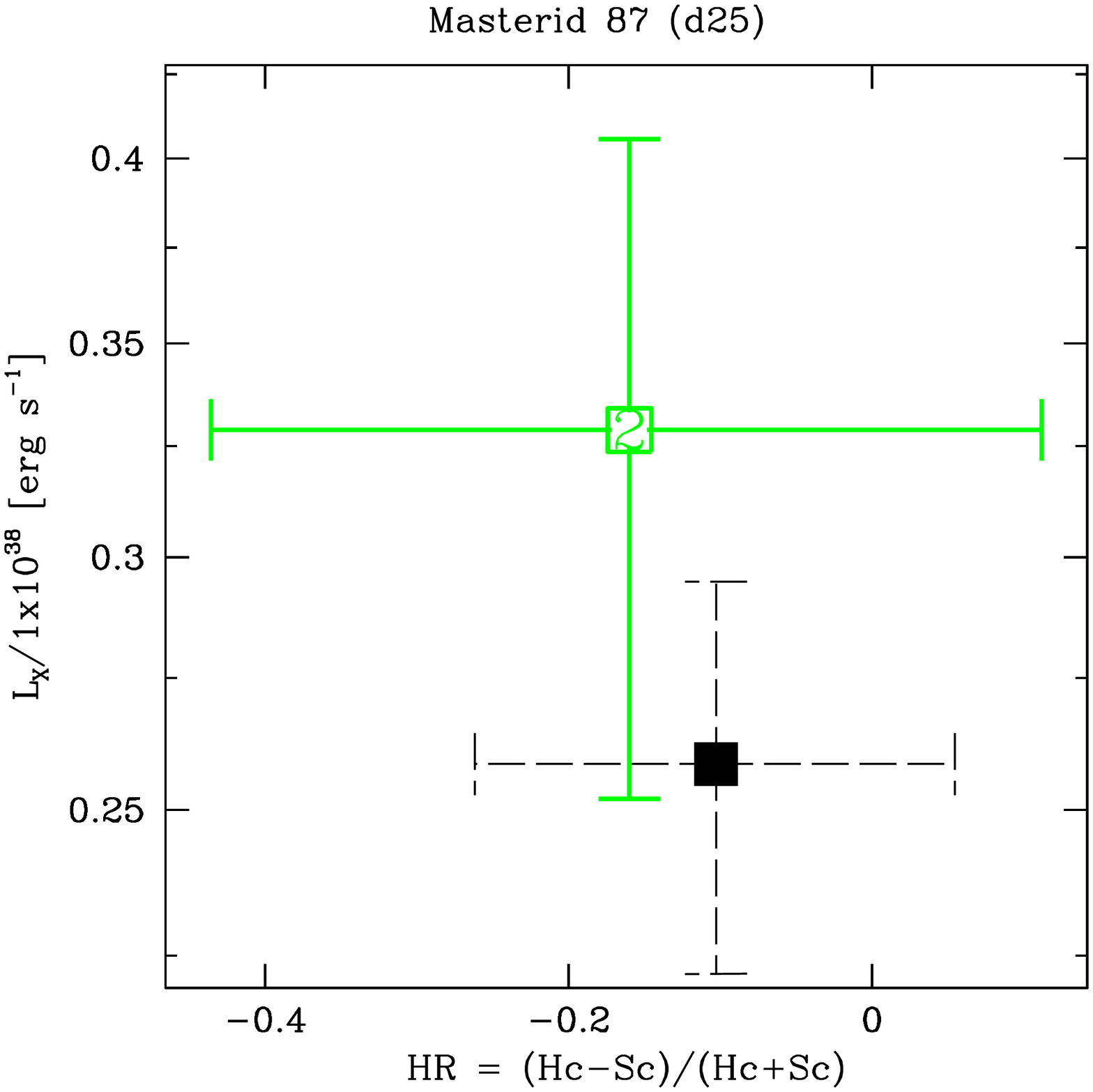}

\end{minipage}
\begin{minipage}{0.32\linewidth}
  \centering

    \includegraphics[width=\linewidth]{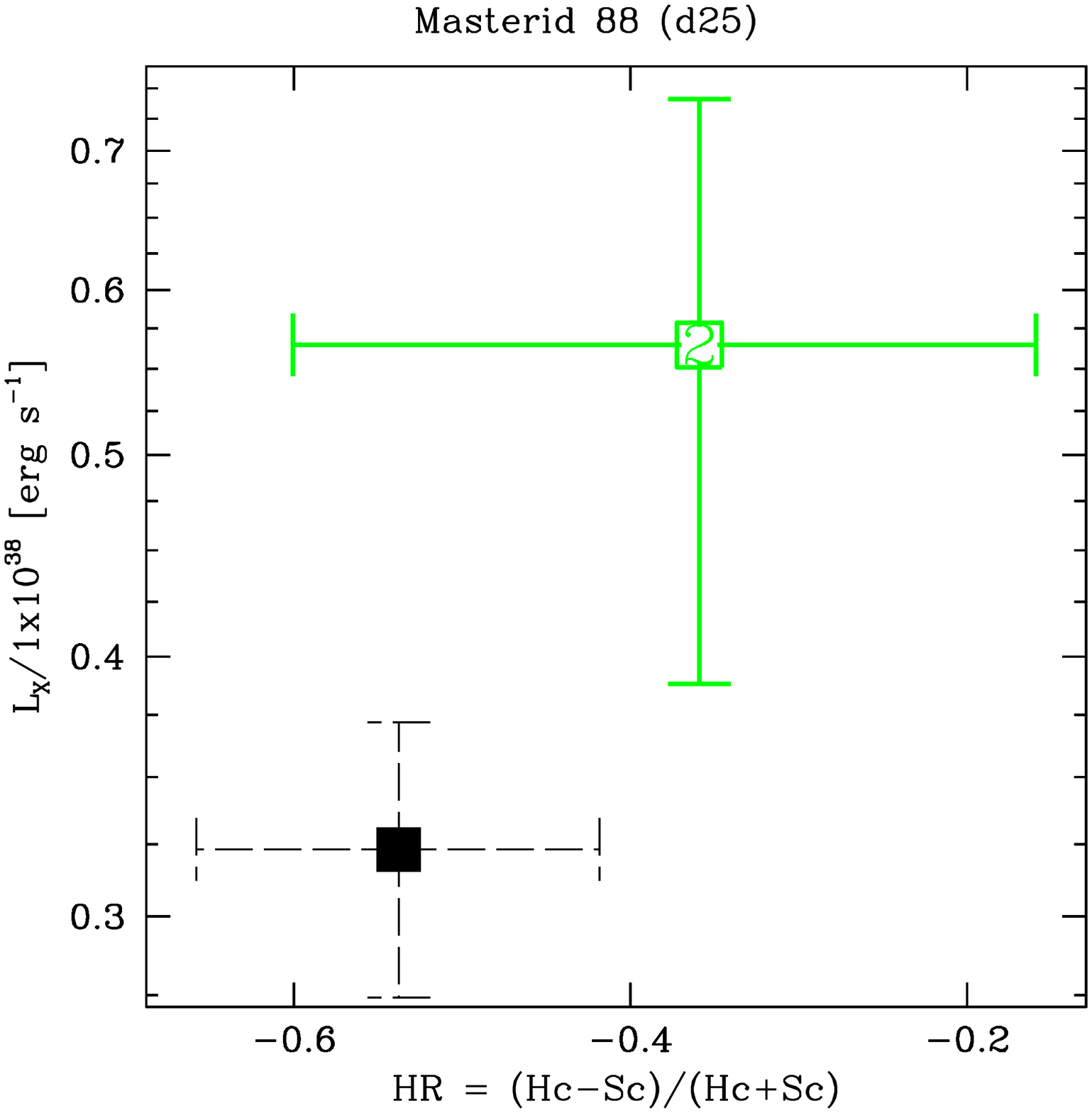}

 \end{minipage}
  
\end{figure}

\begin{figure}
  \begin{minipage}{0.32\linewidth}
  \centering
  
    \includegraphics[width=\linewidth]{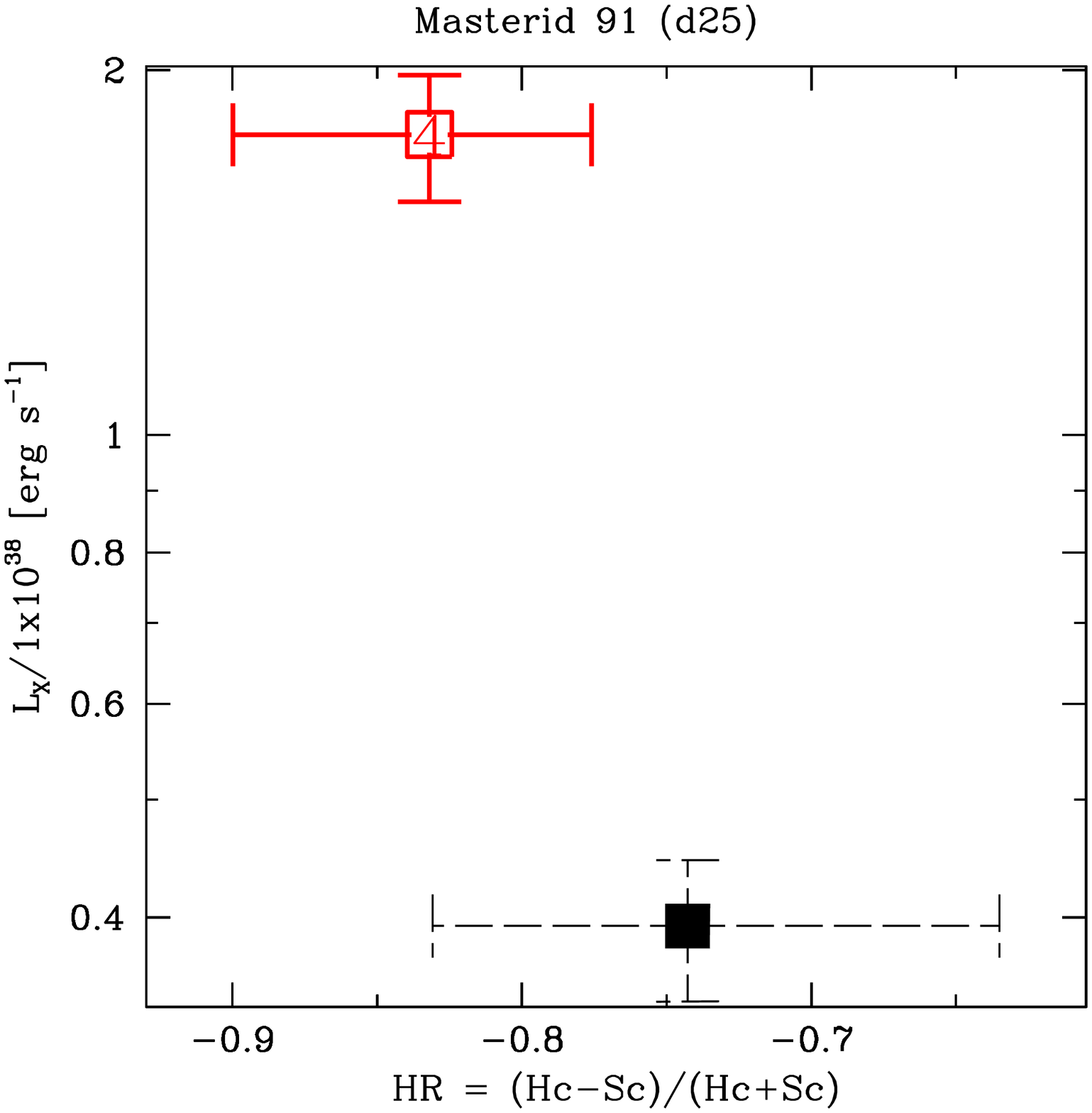}

  \end{minipage}
  \begin{minipage}{0.32\linewidth}
  \centering

    \includegraphics[width=\linewidth]{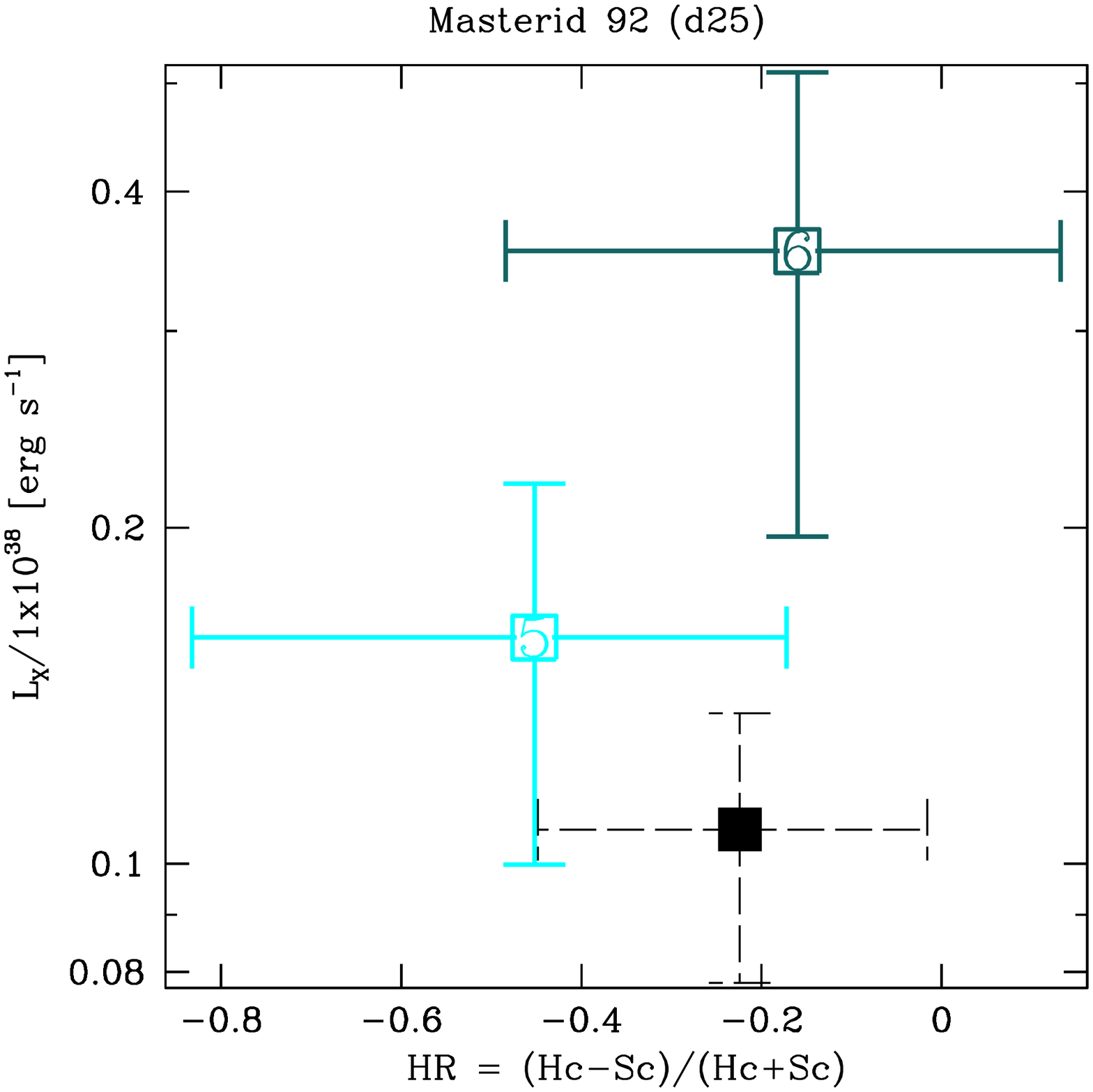}

\end{minipage}
\begin{minipage}{0.32\linewidth}
  \centering

    \includegraphics[width=\linewidth]{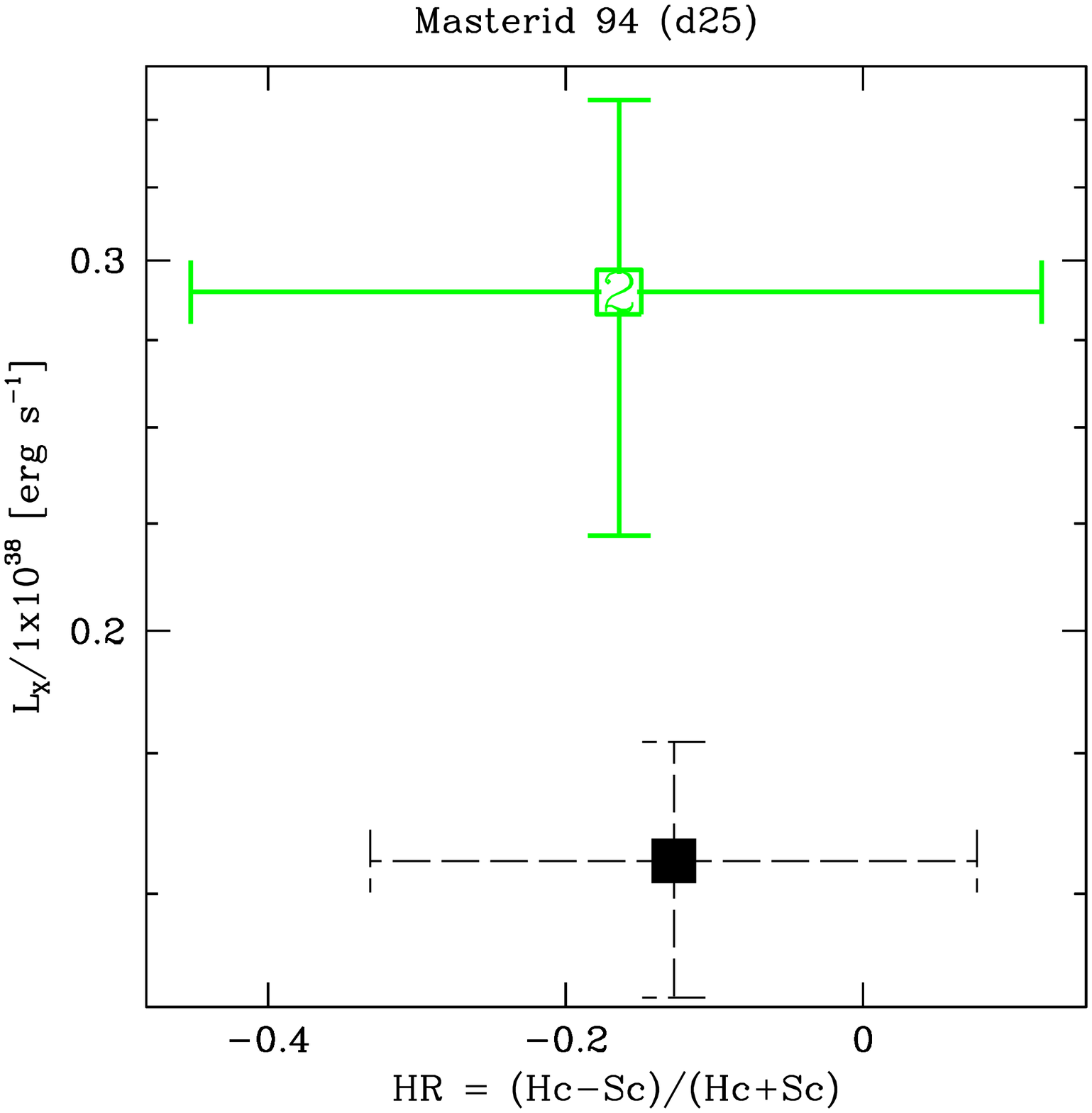}

 \end{minipage}

\begin{minipage}{0.32\linewidth}
  \centering
  
    \includegraphics[width=\linewidth]{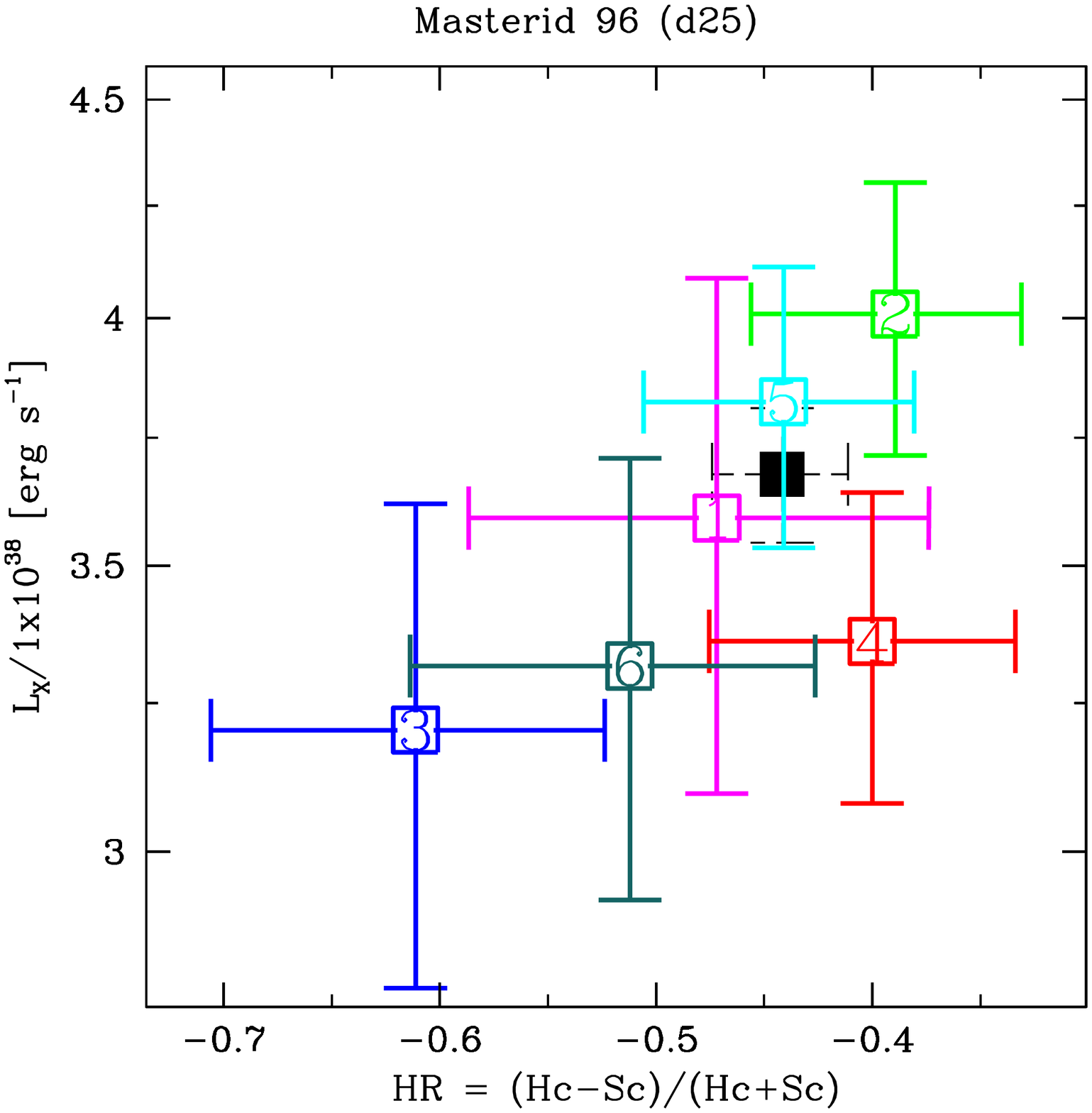}

  \end{minipage}
  \begin{minipage}{0.32\linewidth}
  \centering

    \includegraphics[width=\linewidth]{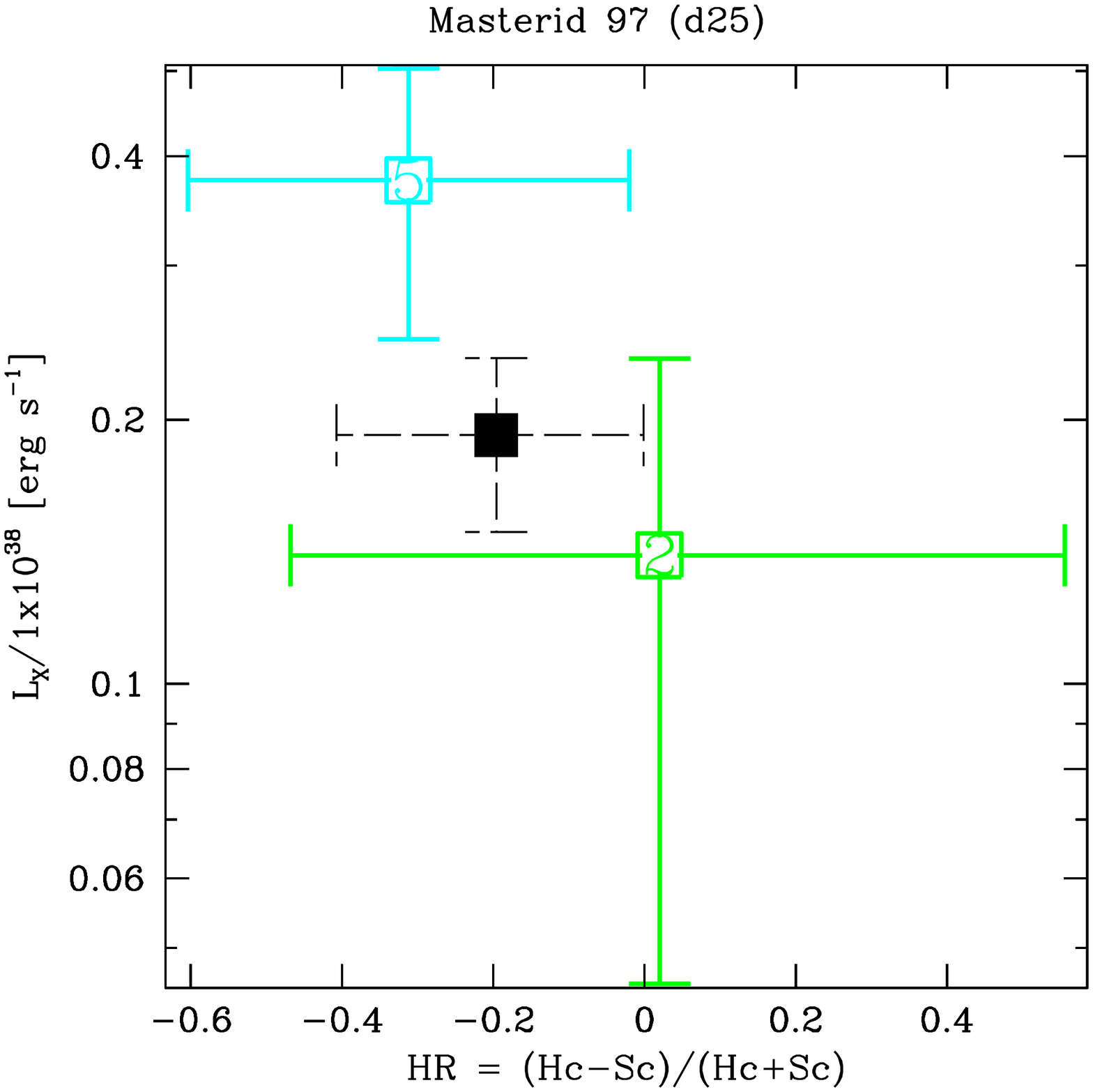}

\end{minipage}
\begin{minipage}{0.32\linewidth}
  \centering

    \includegraphics[width=\linewidth]{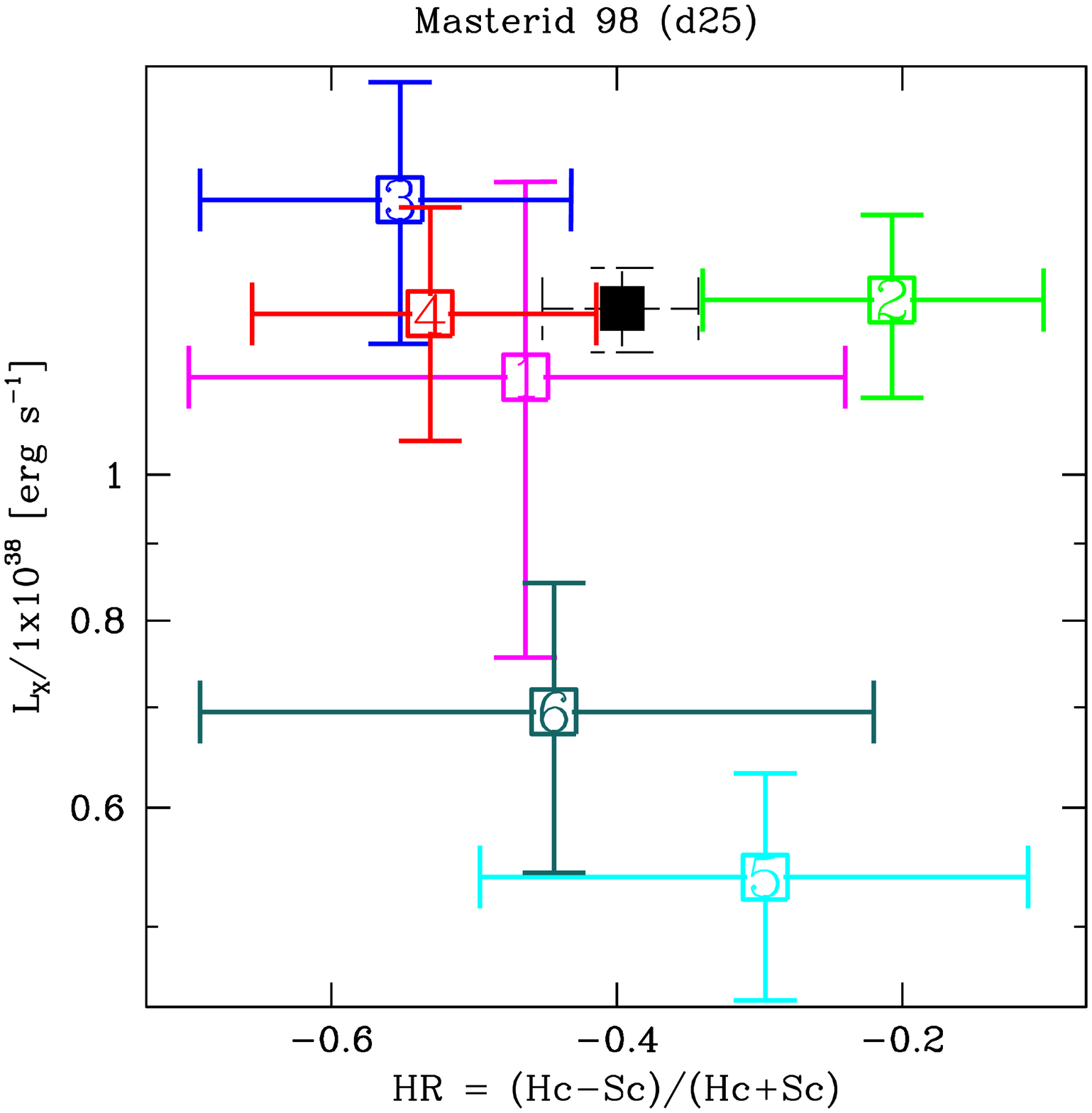}

 \end{minipage}

  \begin{minipage}{0.32\linewidth}
  \centering
  
    \includegraphics[width=\linewidth]{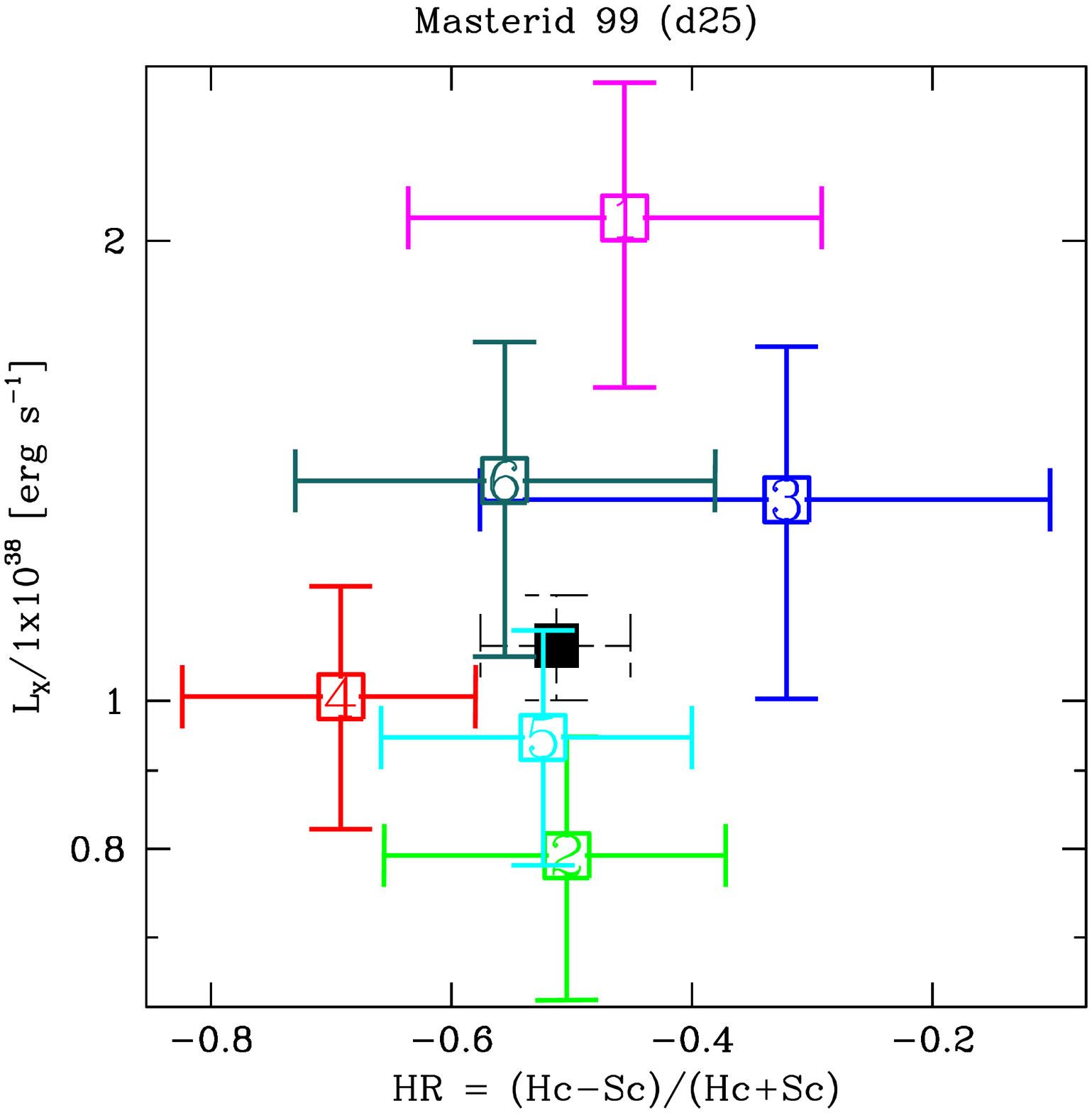}

  \end{minipage}
  \begin{minipage}{0.32\linewidth}
  \centering

    \includegraphics[width=\linewidth]{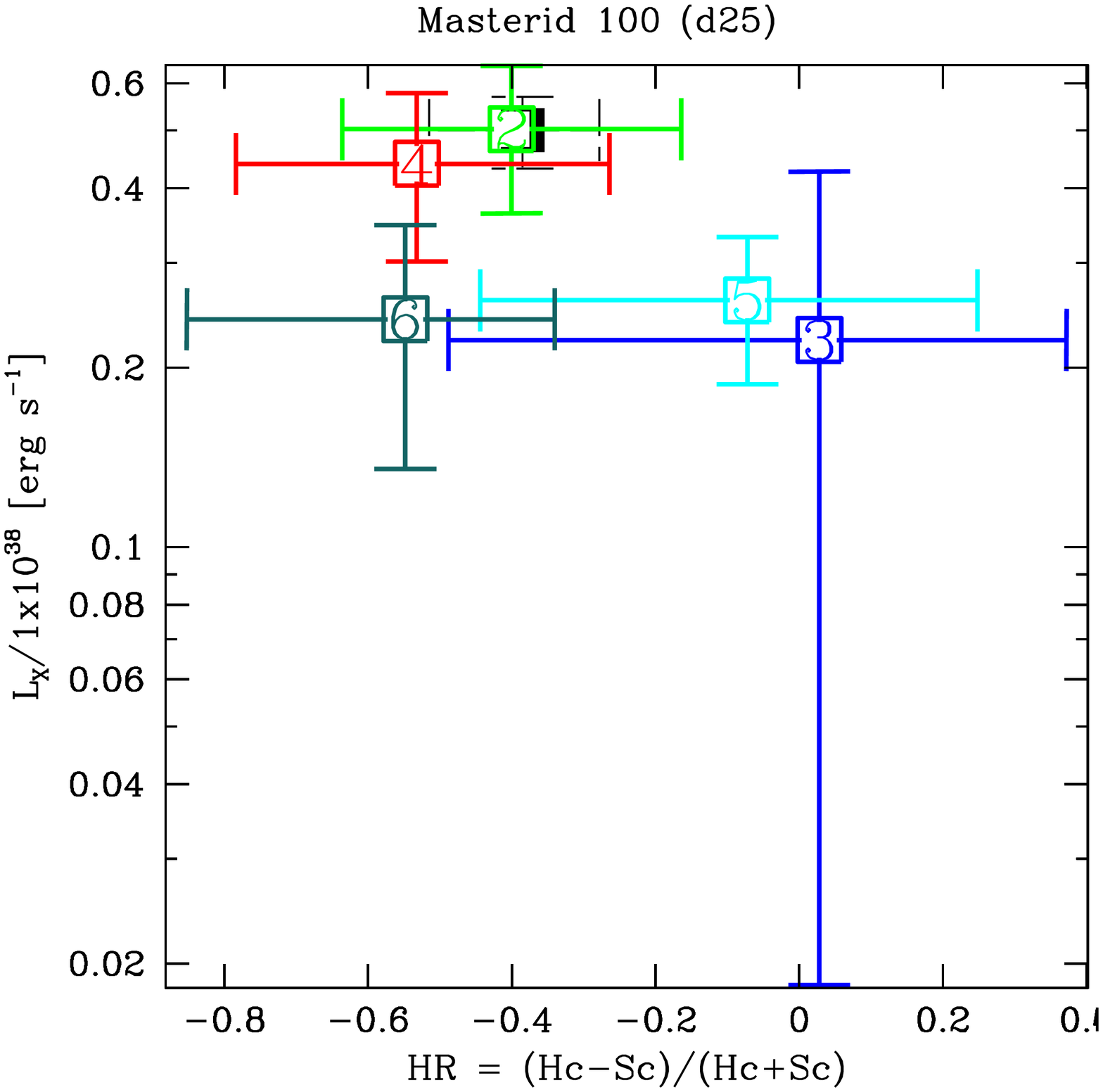}

\end{minipage}
\begin{minipage}{0.32\linewidth}
  \centering

    \includegraphics[width=\linewidth]{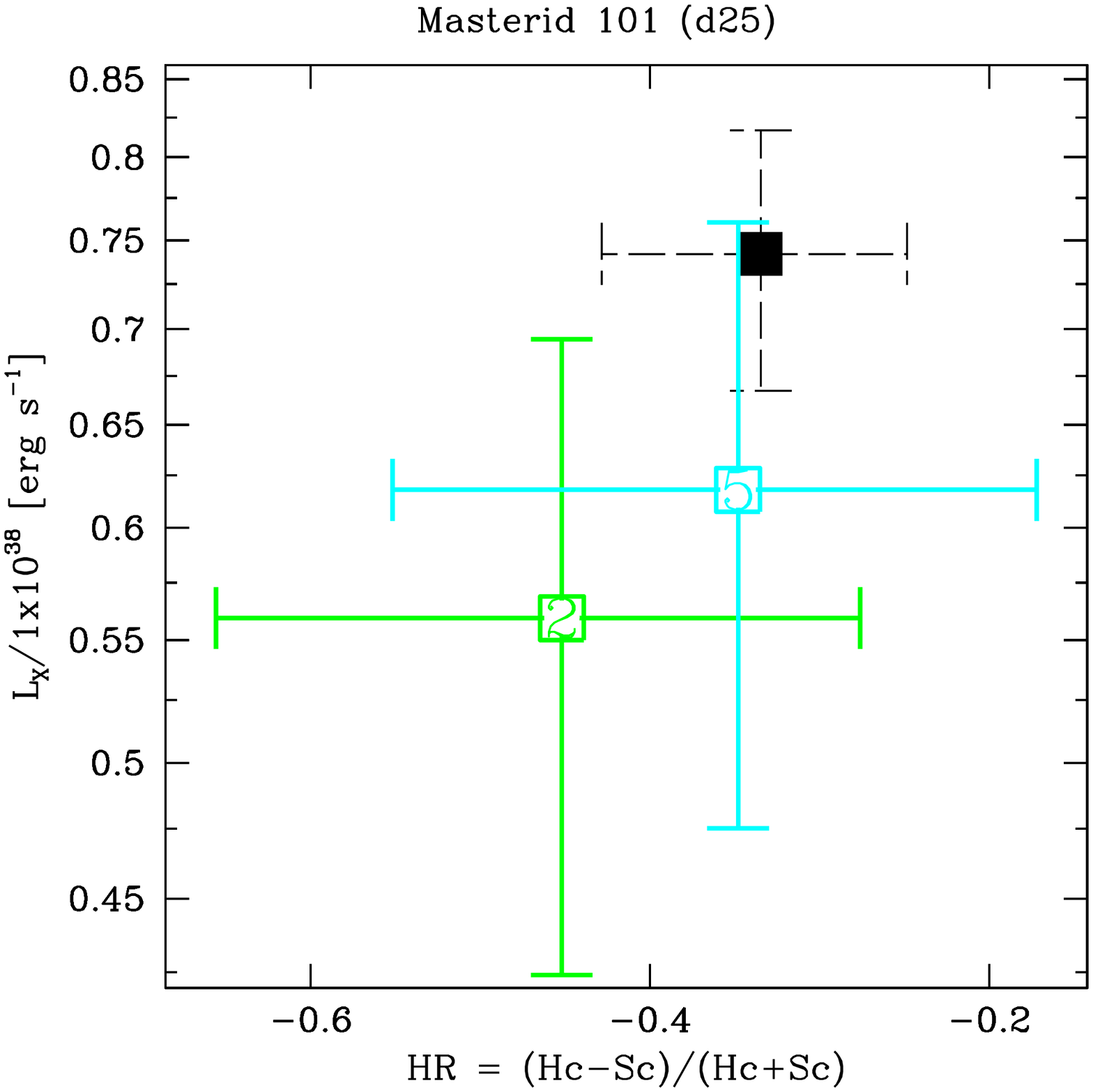}

\end{minipage}

\begin{minipage}{0.32\linewidth}
  \centering
  
    \includegraphics[width=\linewidth]{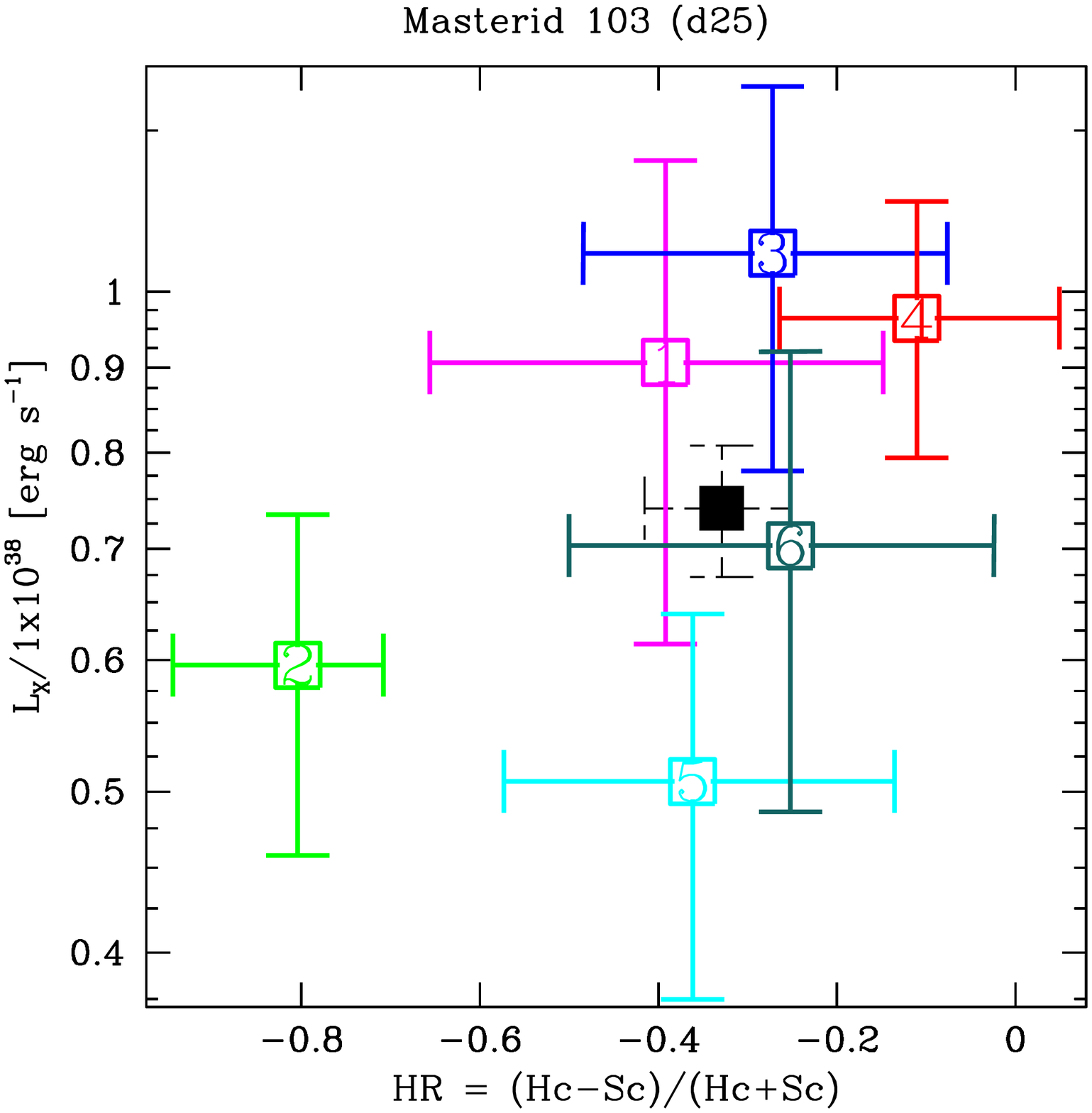}

  \end{minipage}
  \begin{minipage}{0.32\linewidth}
  \centering

    \includegraphics[width=\linewidth]{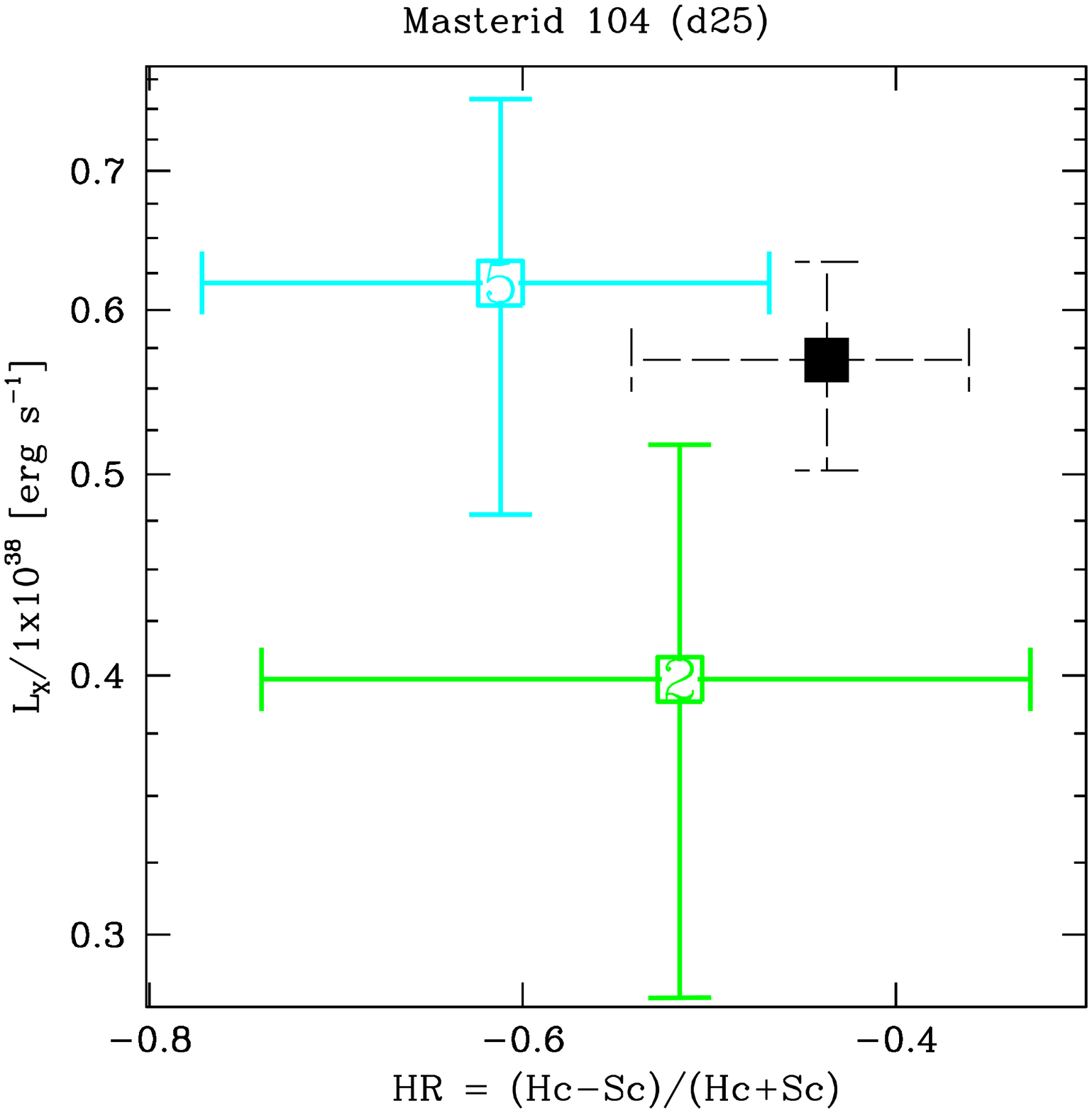}

\end{minipage}
\begin{minipage}{0.32\linewidth}
  \centering

    \includegraphics[width=\linewidth]{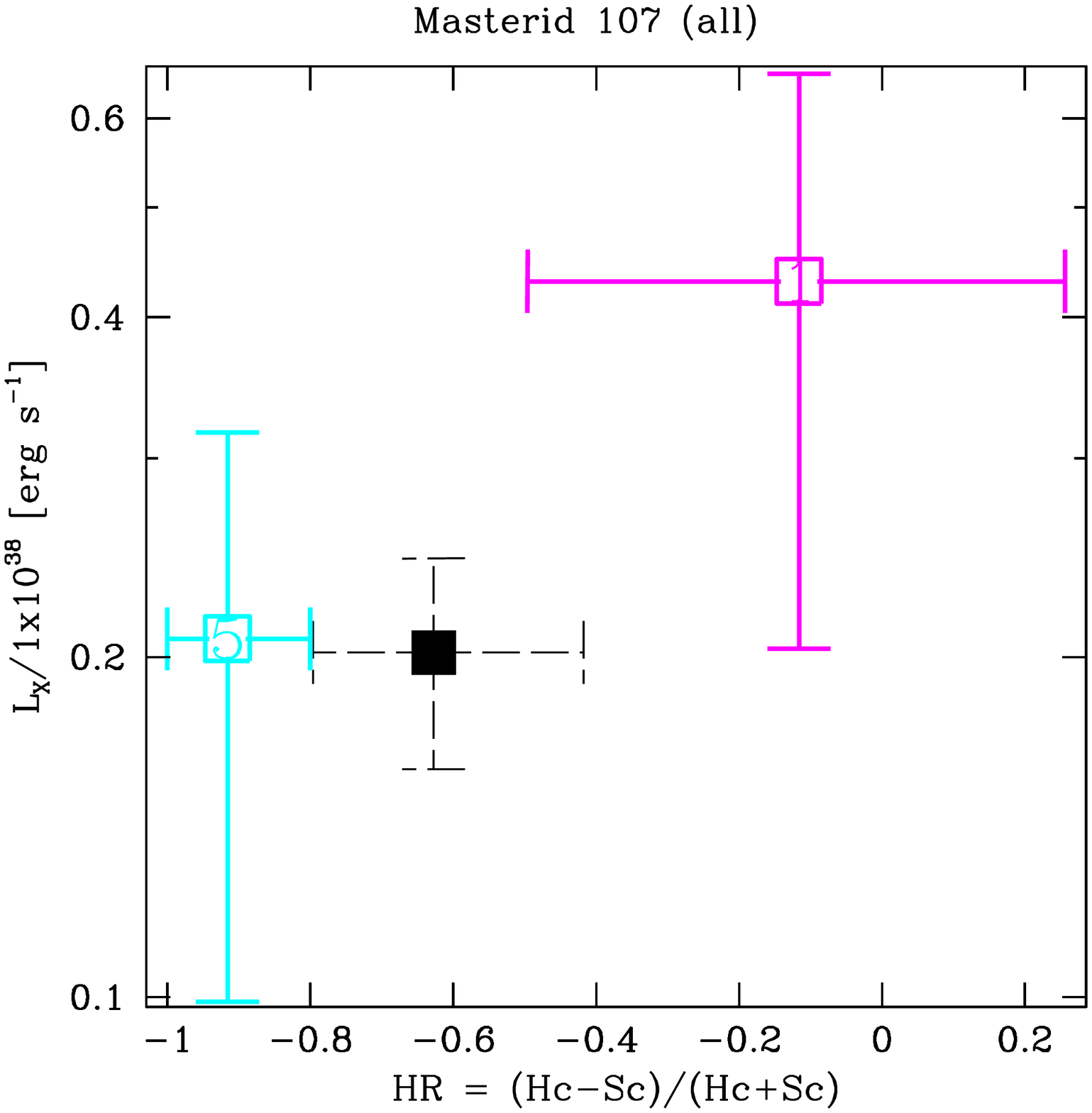}

\end{minipage}
\end{figure}

\begin{figure}
  \begin{minipage}{0.32\linewidth}
  \centering
  
    \includegraphics[width=\linewidth]{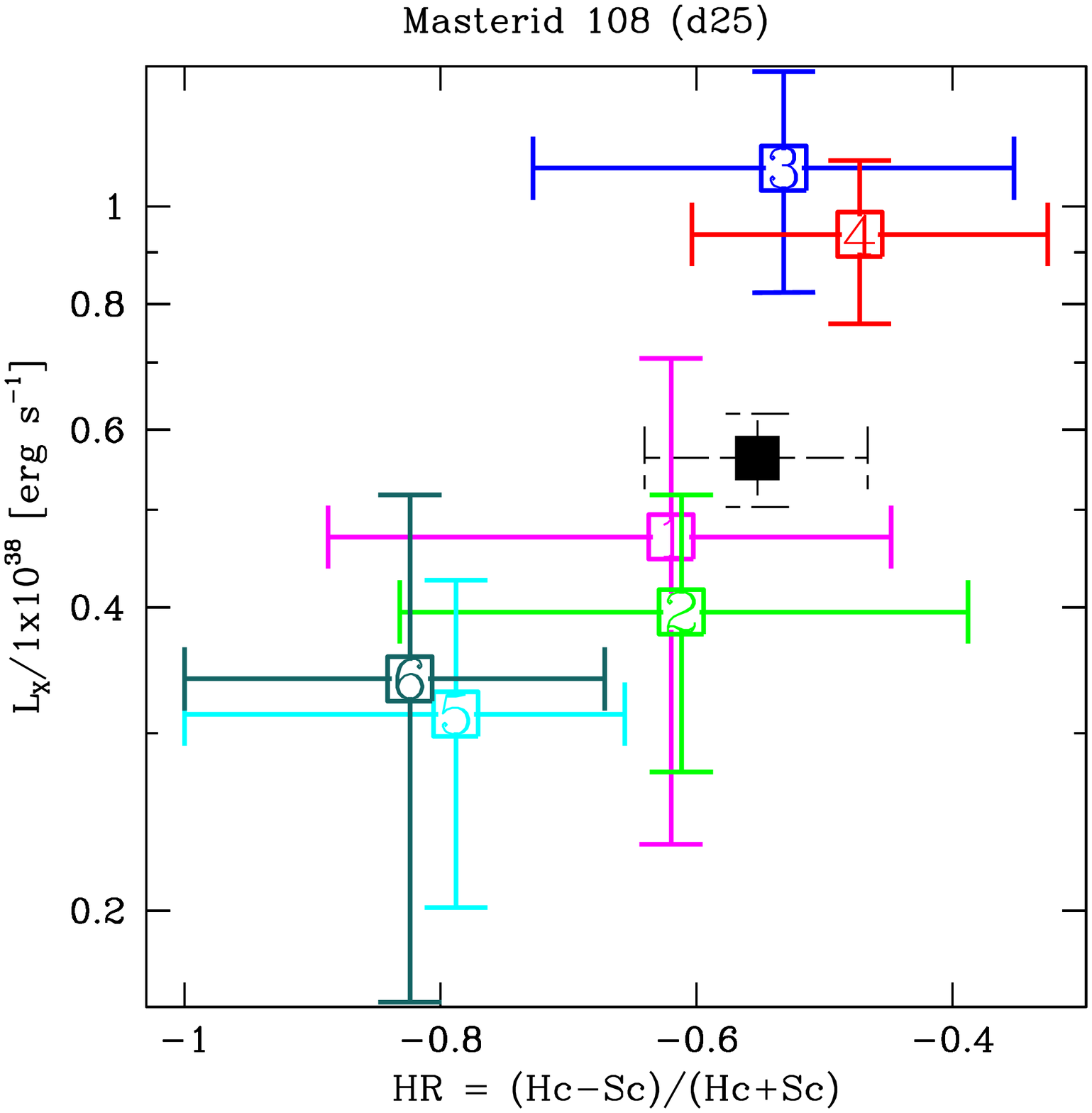}

  \end{minipage}
  \begin{minipage}{0.32\linewidth}
  \centering

    \includegraphics[width=\linewidth]{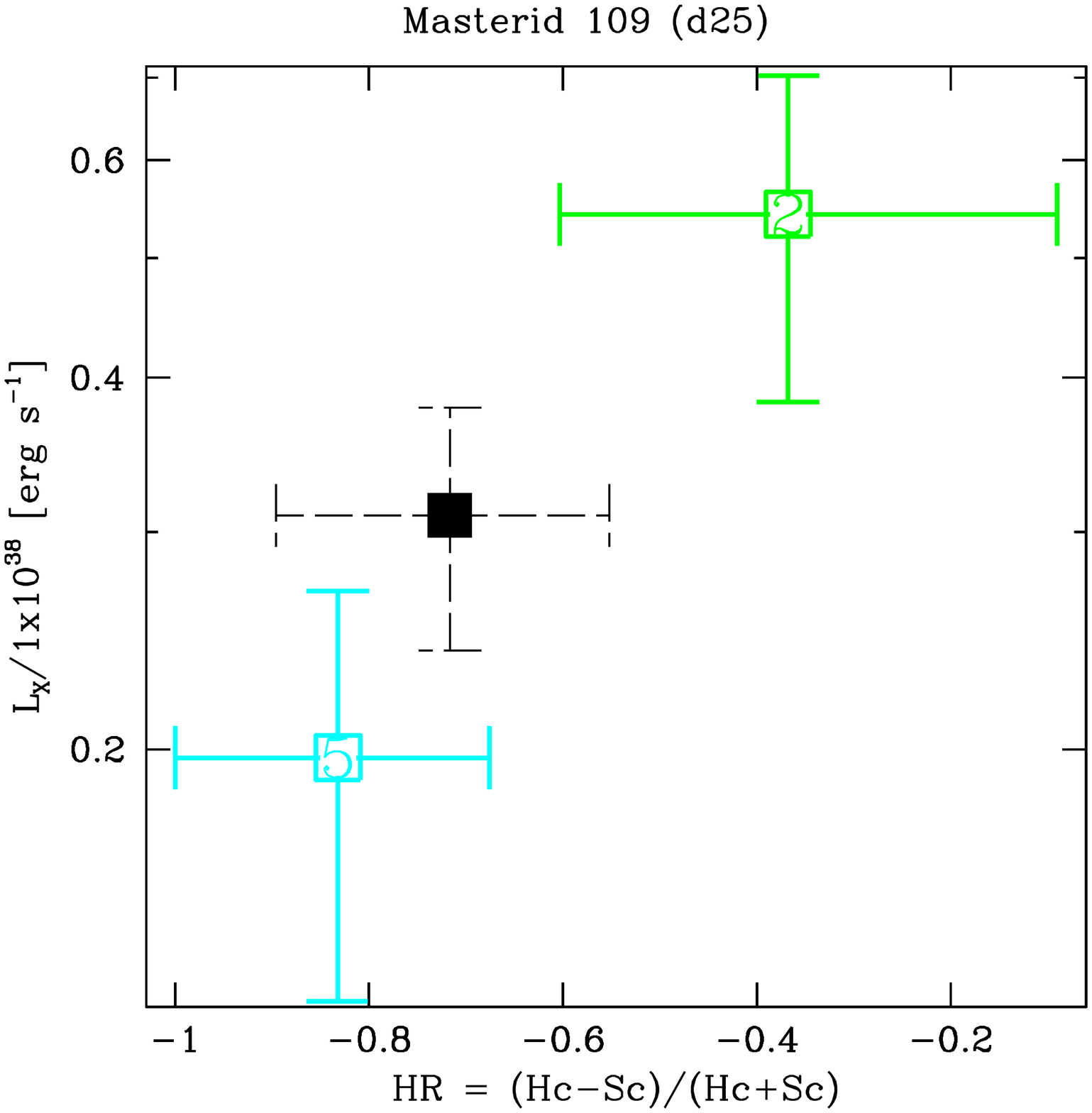}

\end{minipage}
\begin{minipage}{0.32\linewidth}
  \centering

    \includegraphics[width=\linewidth]{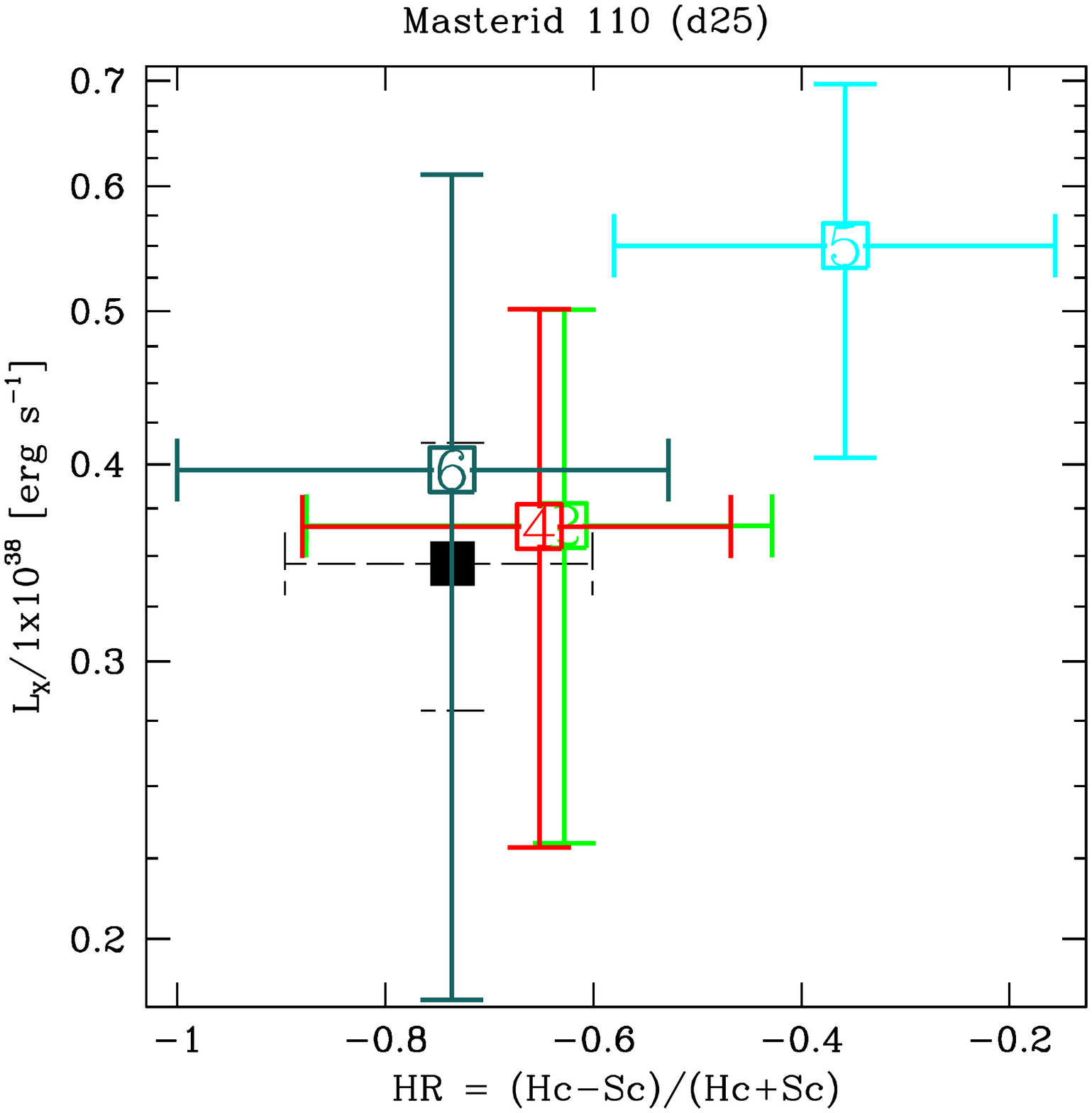}

 \end{minipage}

\begin{minipage}{0.32\linewidth}
  \centering
  
    \includegraphics[width=\linewidth]{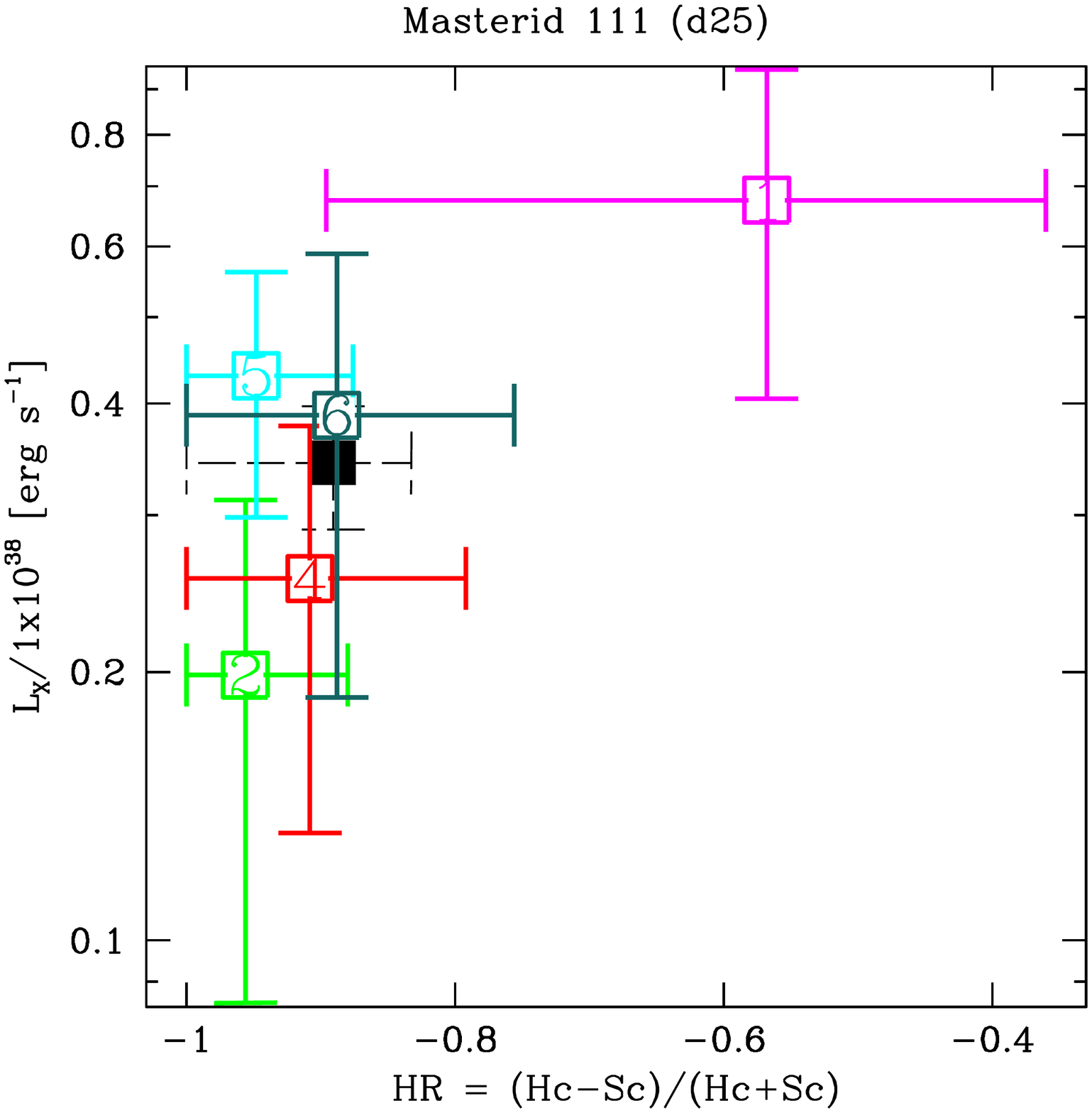}

  \end{minipage}
  \begin{minipage}{0.32\linewidth}
  \centering

    \includegraphics[width=\linewidth]{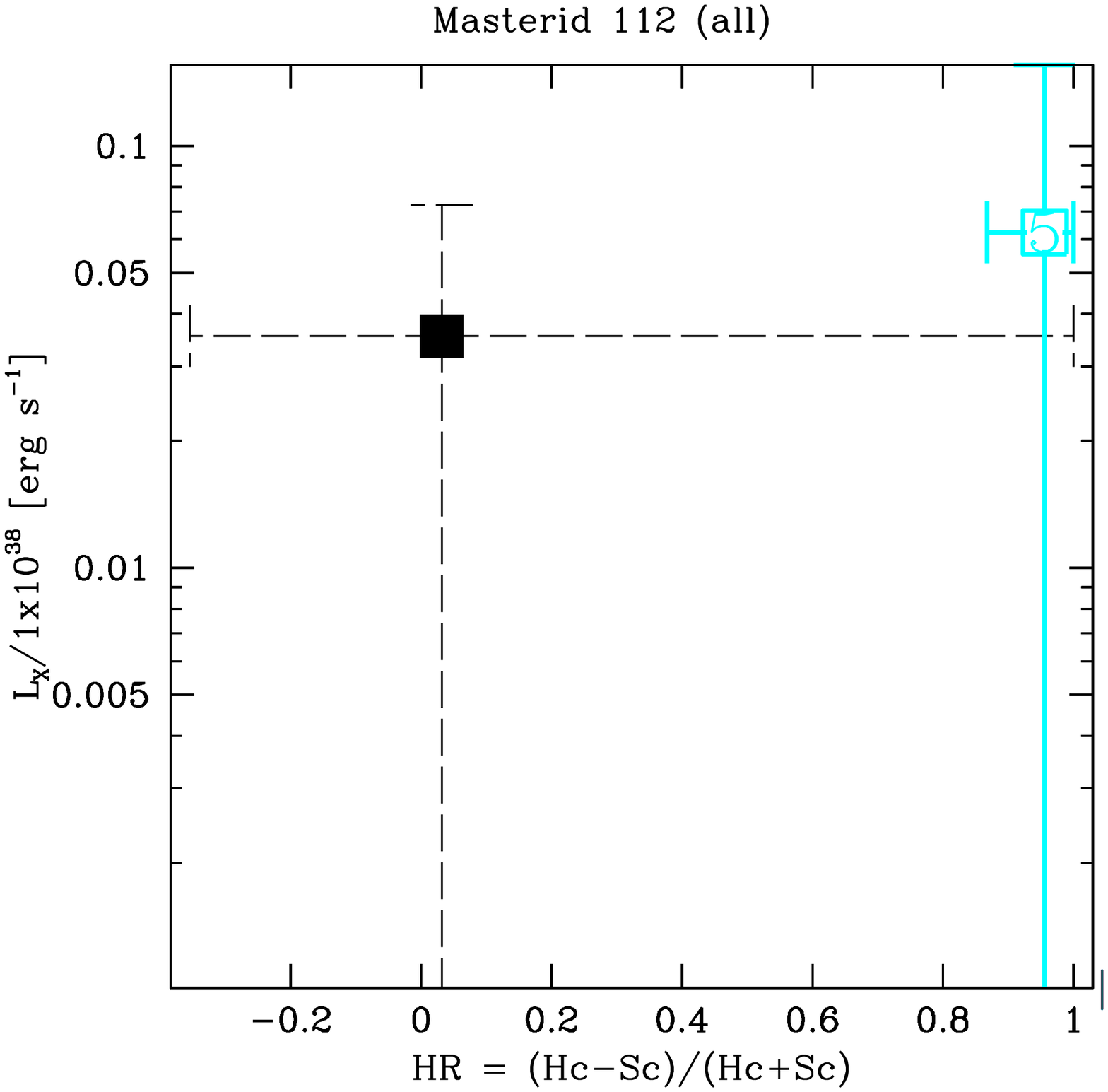}

\end{minipage}
\begin{minipage}{0.32\linewidth}
  \centering

    \includegraphics[width=\linewidth]{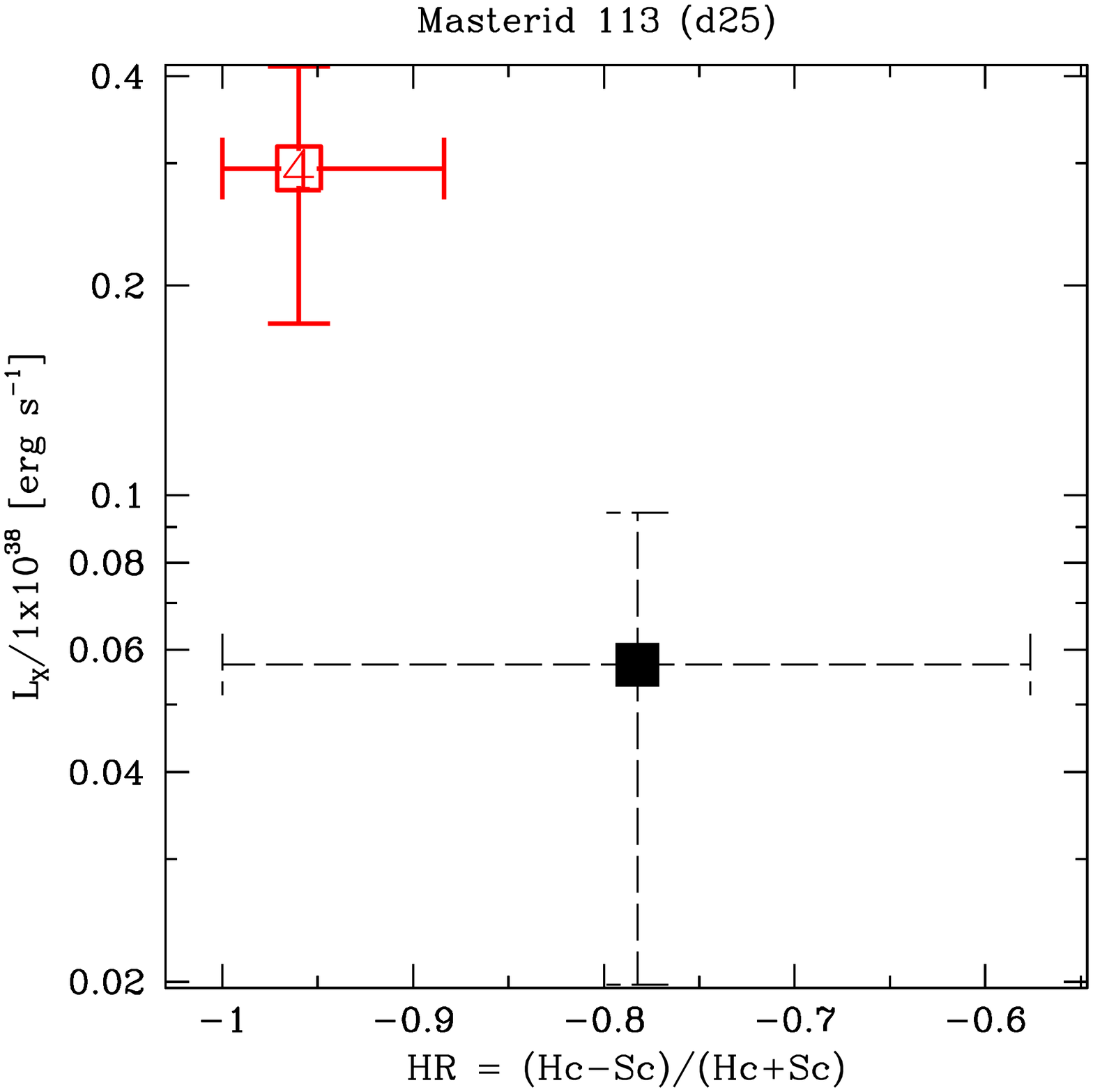}

 \end{minipage}

  \begin{minipage}{0.32\linewidth}
  \centering
  
    \includegraphics[width=\linewidth]{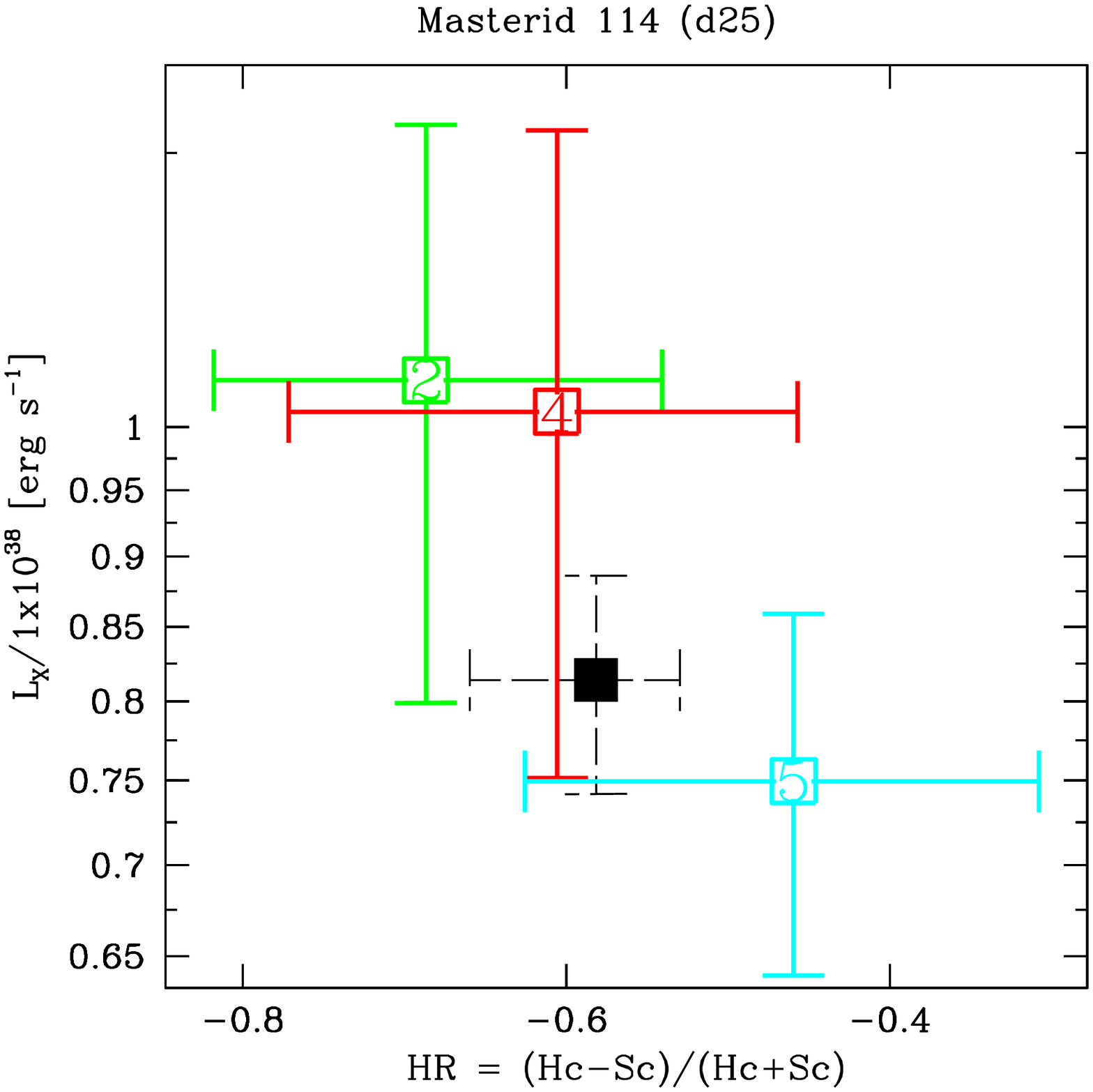}

  \end{minipage}
  \begin{minipage}{0.32\linewidth}
  \centering

    \includegraphics[width=\linewidth]{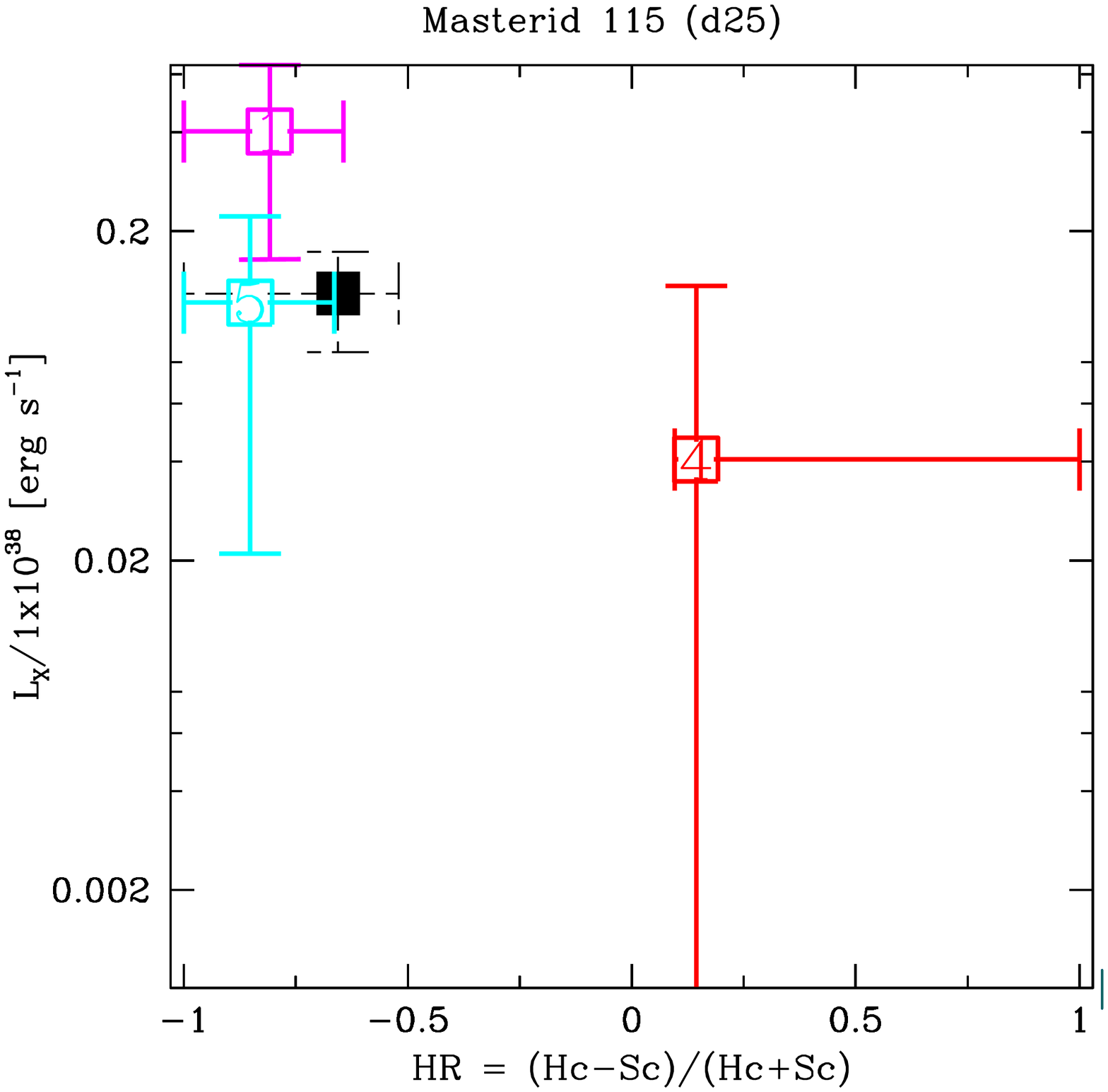}

\end{minipage}
\begin{minipage}{0.32\linewidth}
  \centering

    \includegraphics[width=\linewidth]{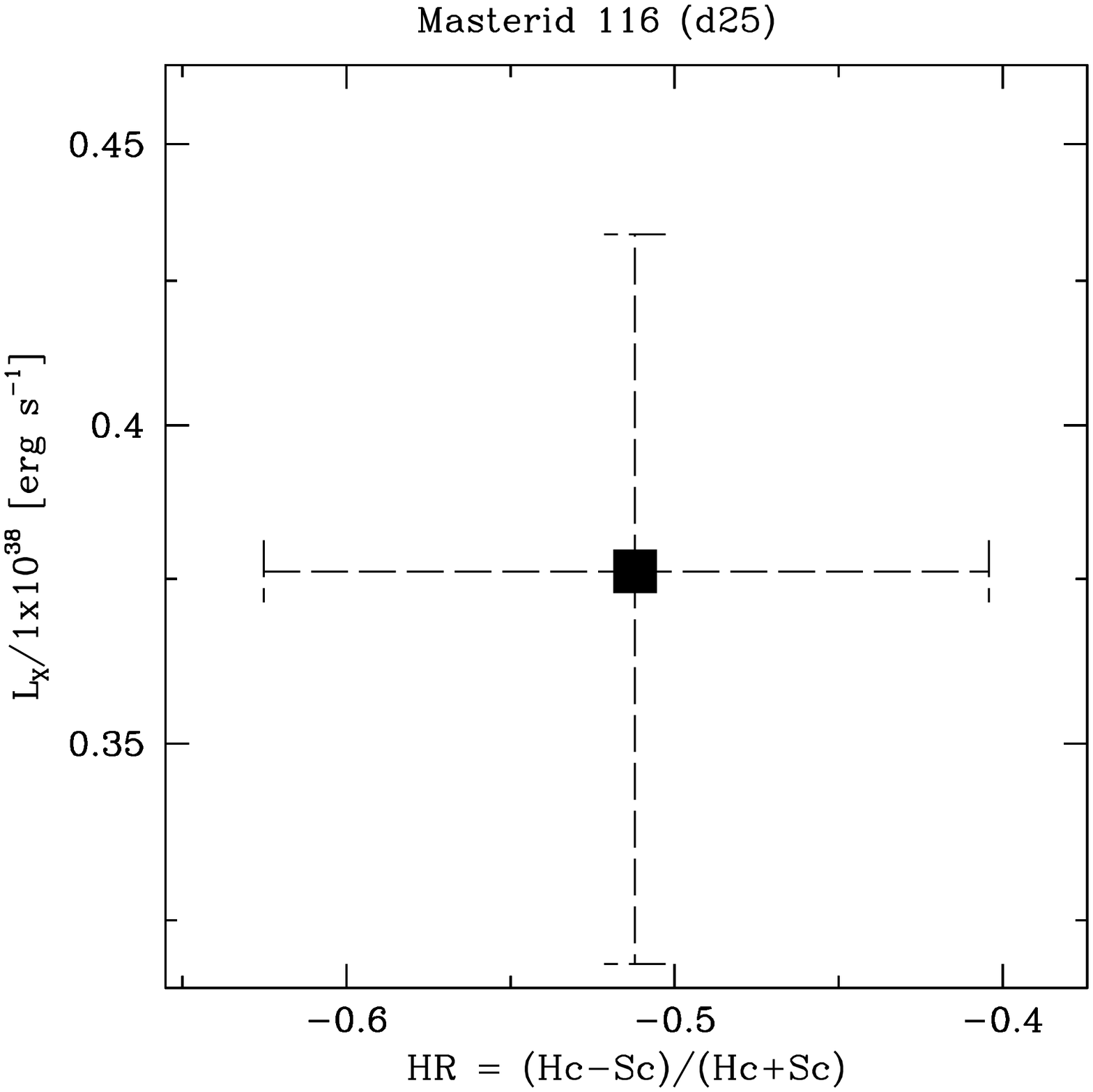}

\end{minipage}

\begin{minipage}{0.32\linewidth}
  \centering
  
    \includegraphics[width=\linewidth]{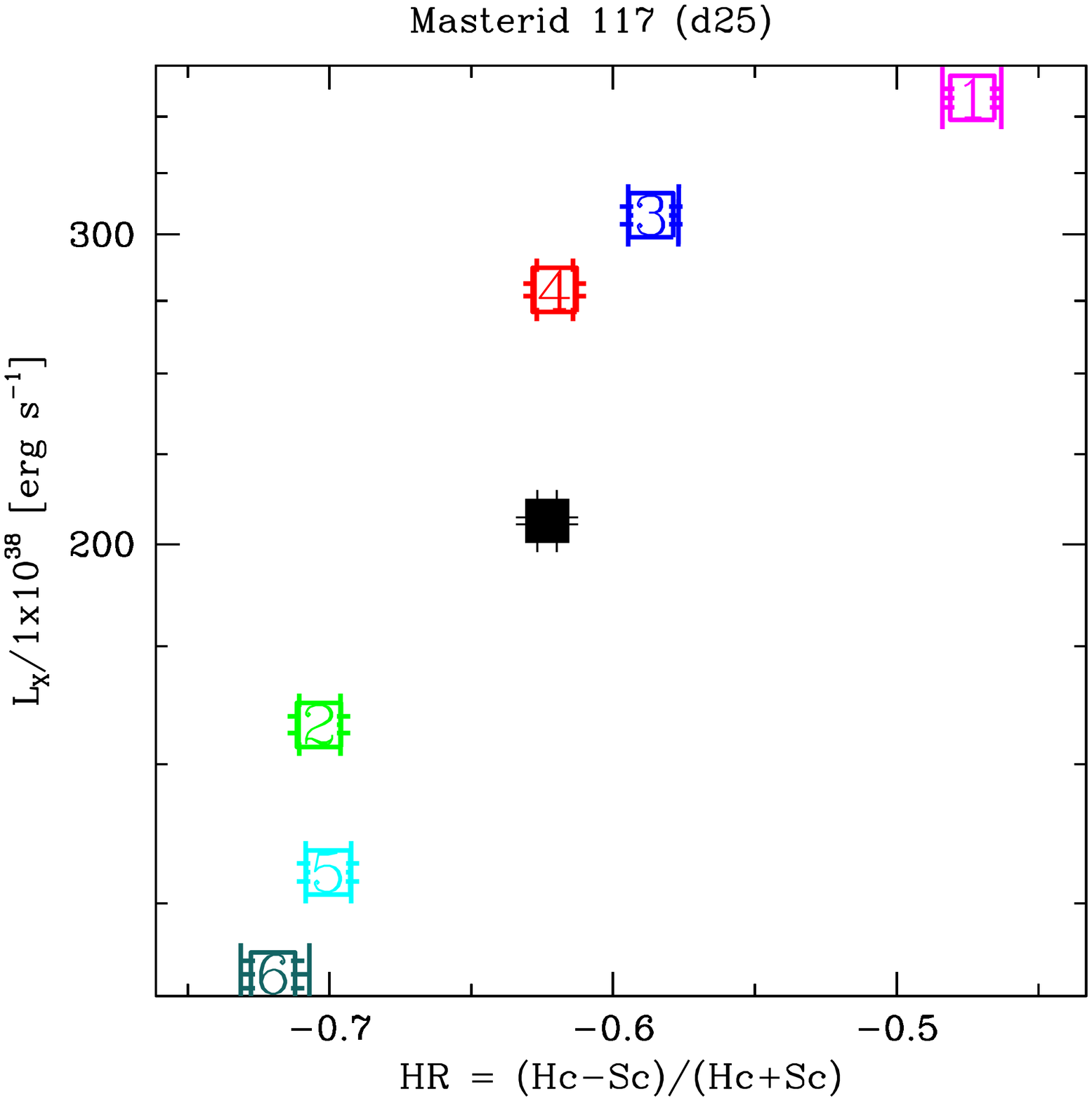}

  \end{minipage}
  \begin{minipage}{0.32\linewidth}
  \centering

    \includegraphics[width=\linewidth]{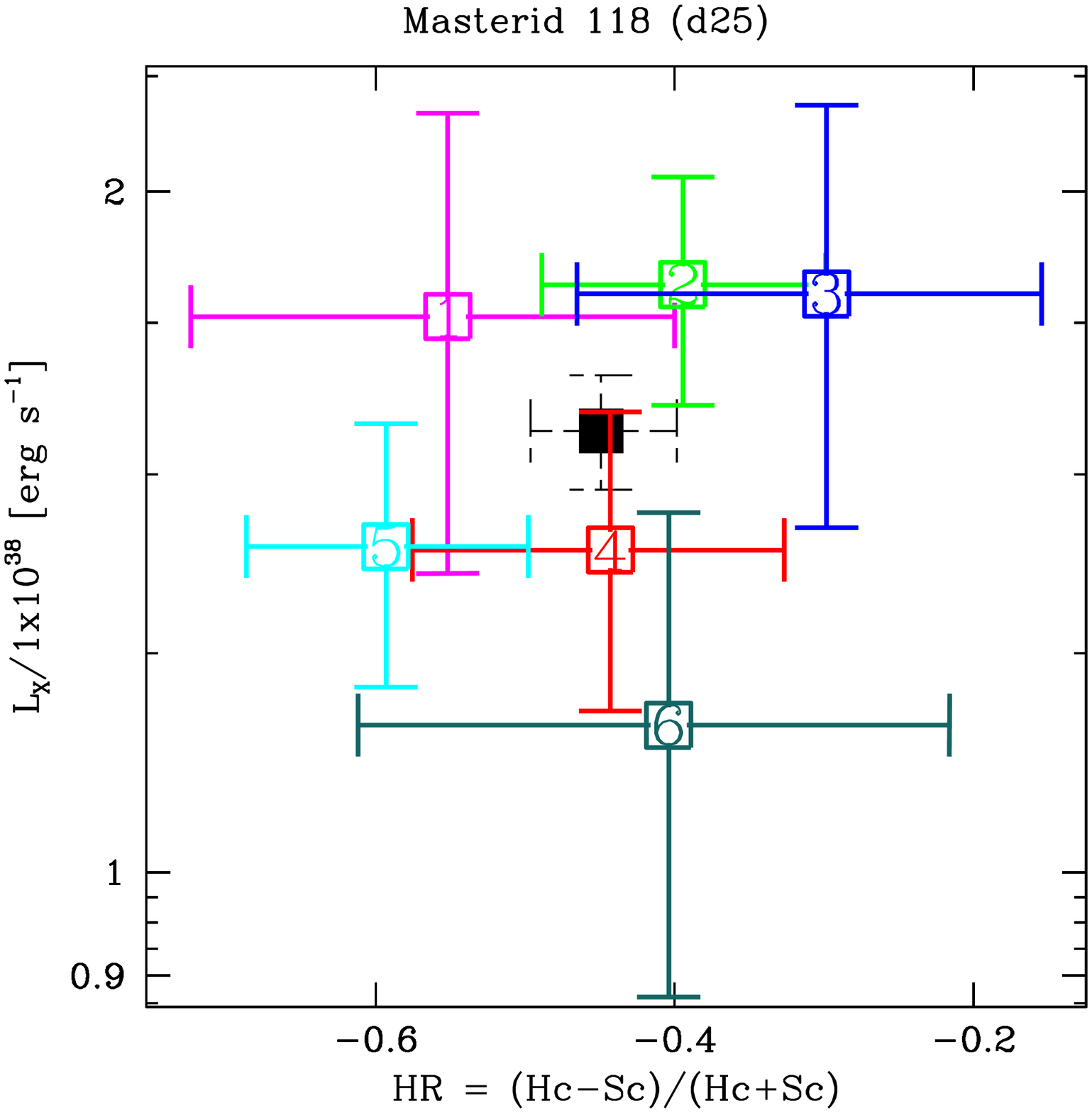}

\end{minipage}
\begin{minipage}{0.32\linewidth}
  \centering

    \includegraphics[width=\linewidth]{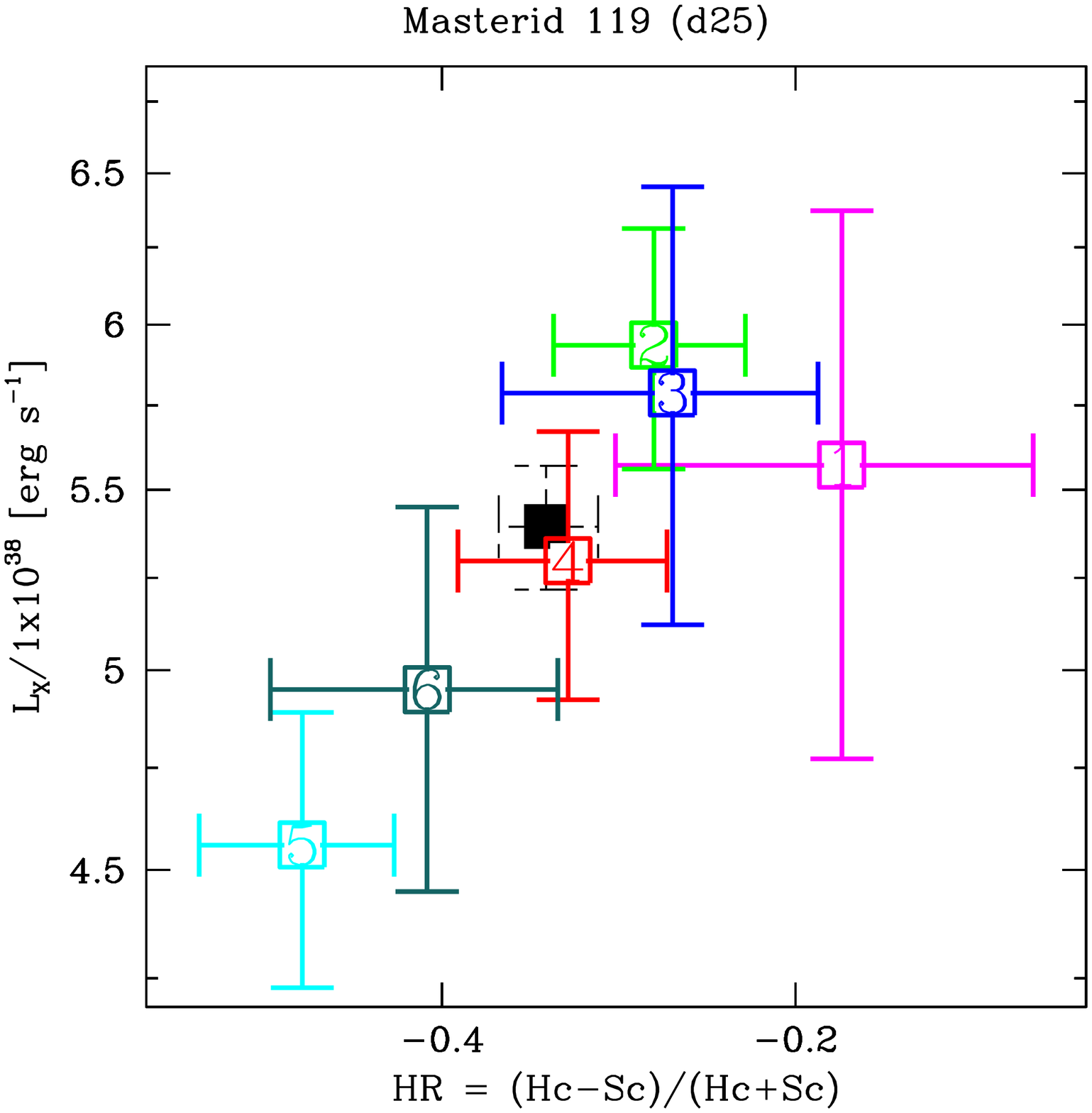}

\end{minipage}
\end{figure}

\begin{figure}
  \begin{minipage}{0.32\linewidth}
  \centering
  
    \includegraphics[width=\linewidth]{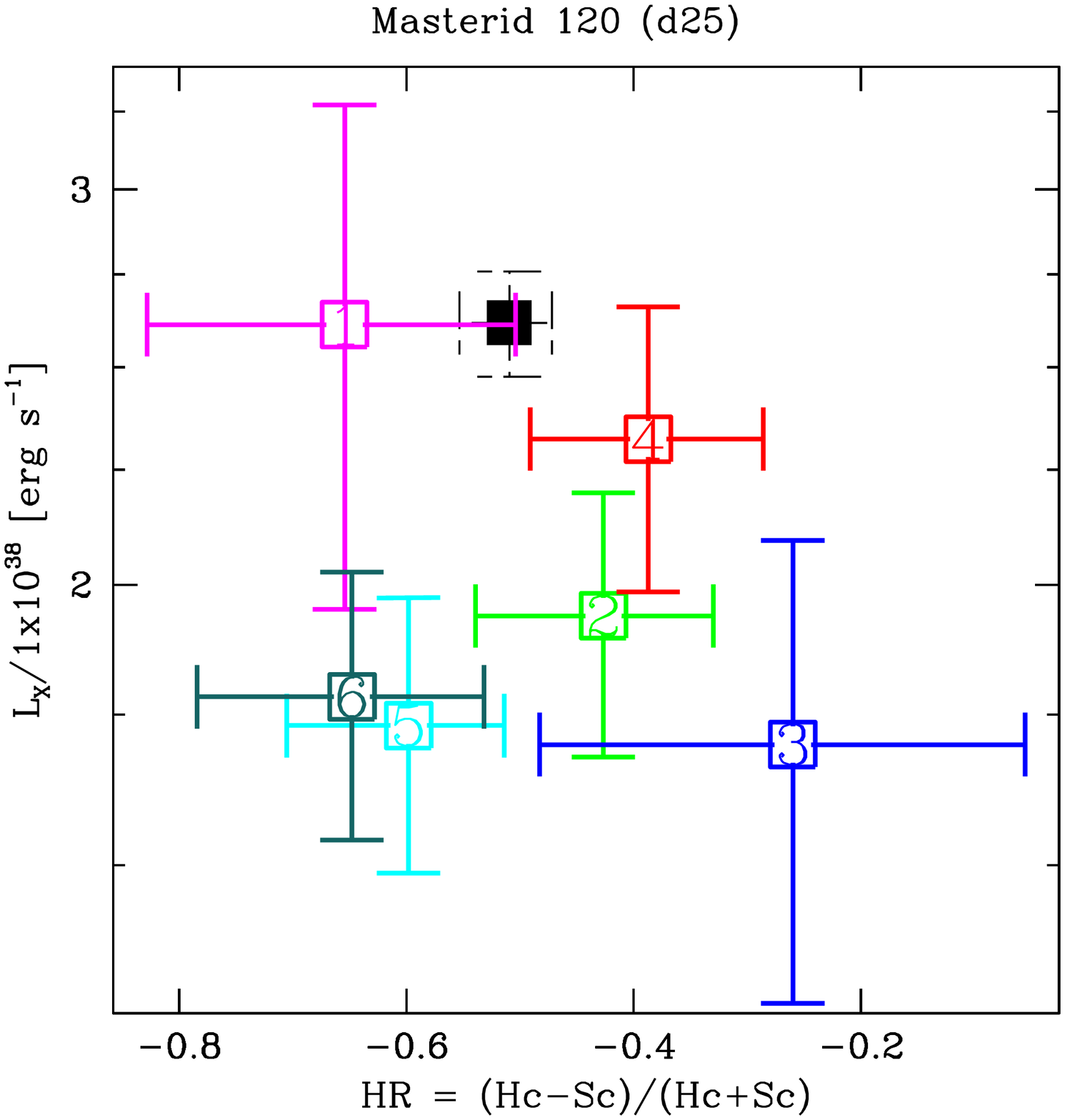}

  \end{minipage}
  \begin{minipage}{0.32\linewidth}
  \centering

    \includegraphics[width=\linewidth]{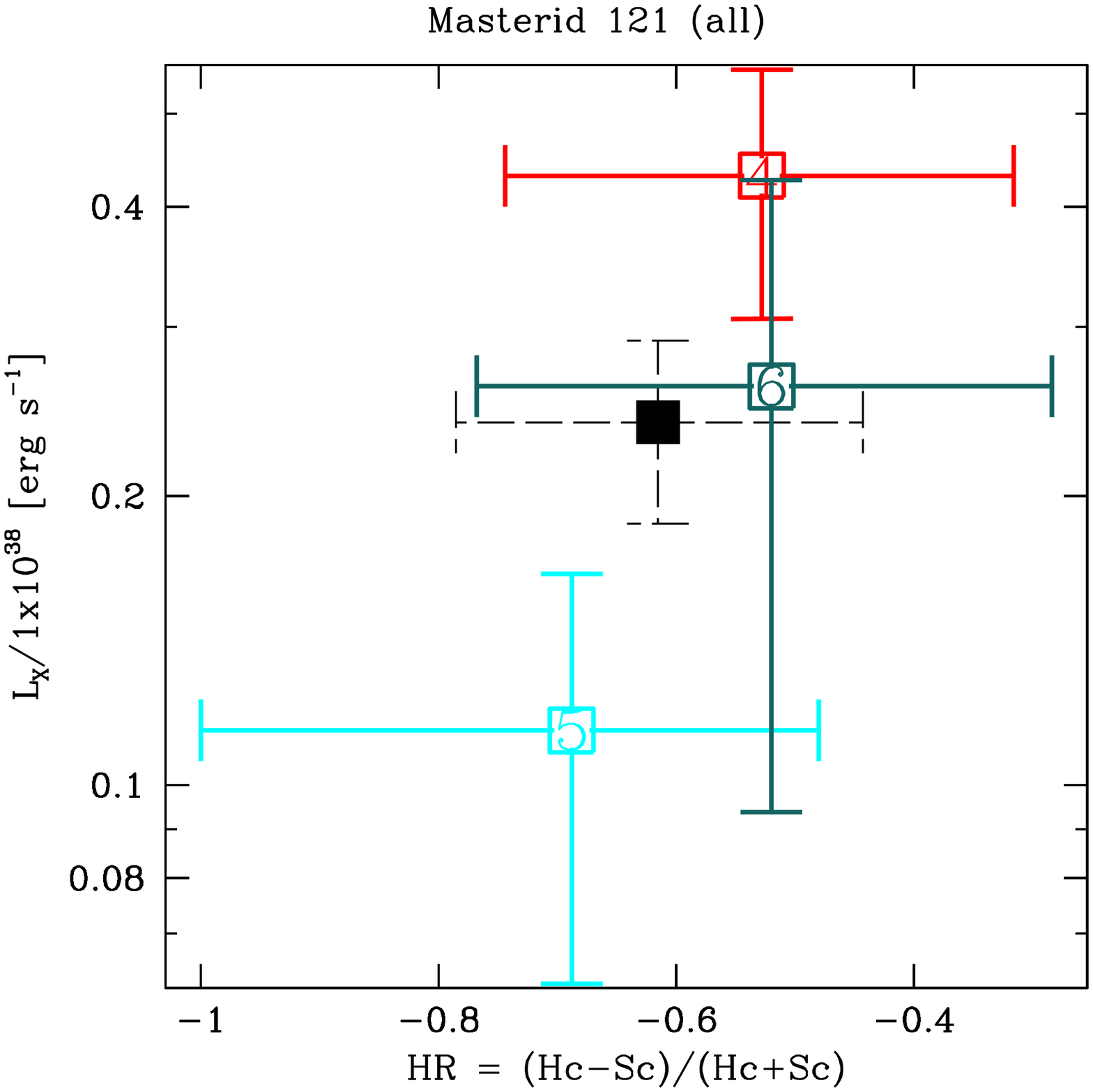}

\end{minipage}
\begin{minipage}{0.32\linewidth}
  \centering

    \includegraphics[width=\linewidth]{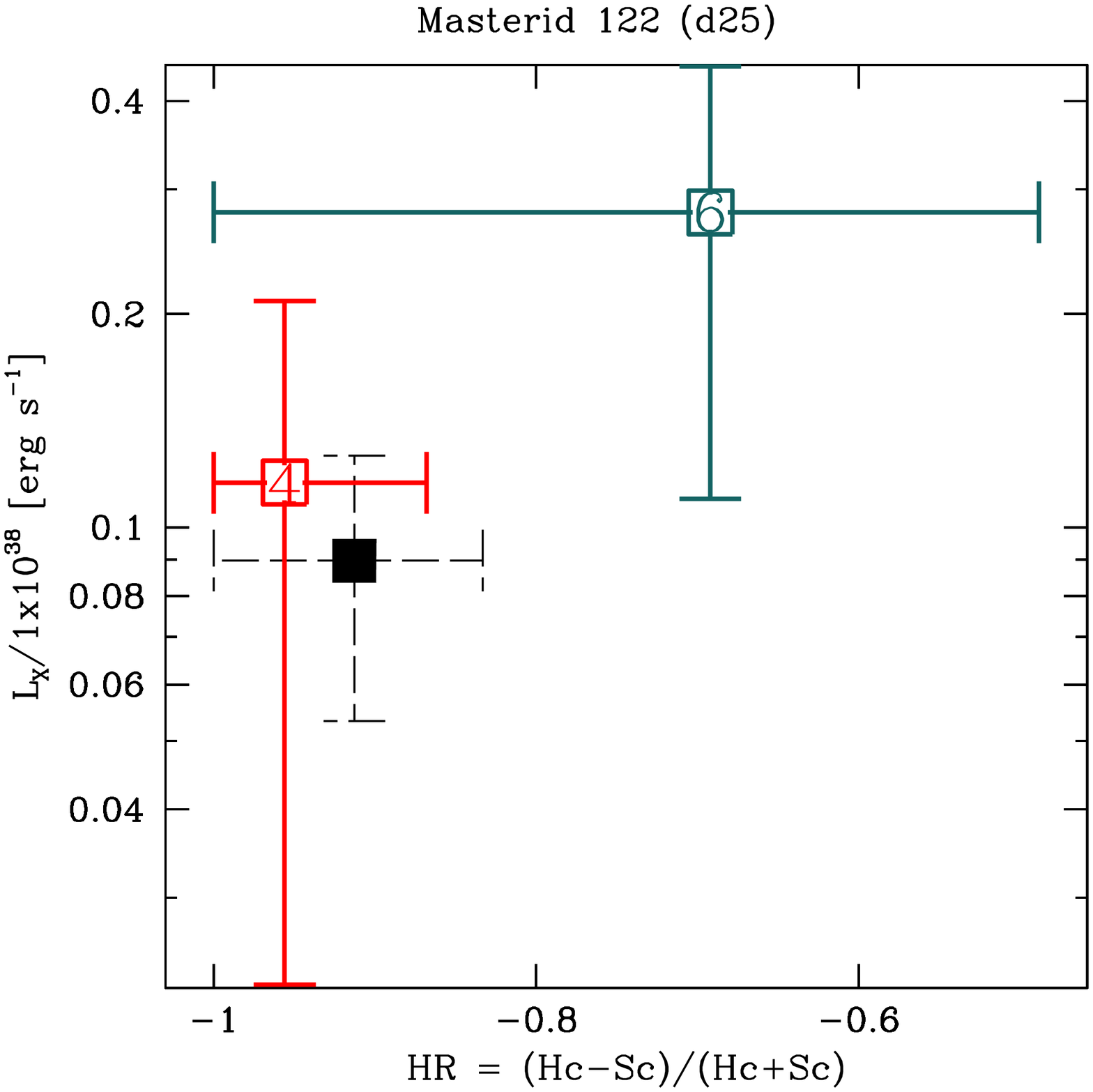}

 \end{minipage}

\begin{minipage}{0.32\linewidth}
  \centering
  
    \includegraphics[width=\linewidth]{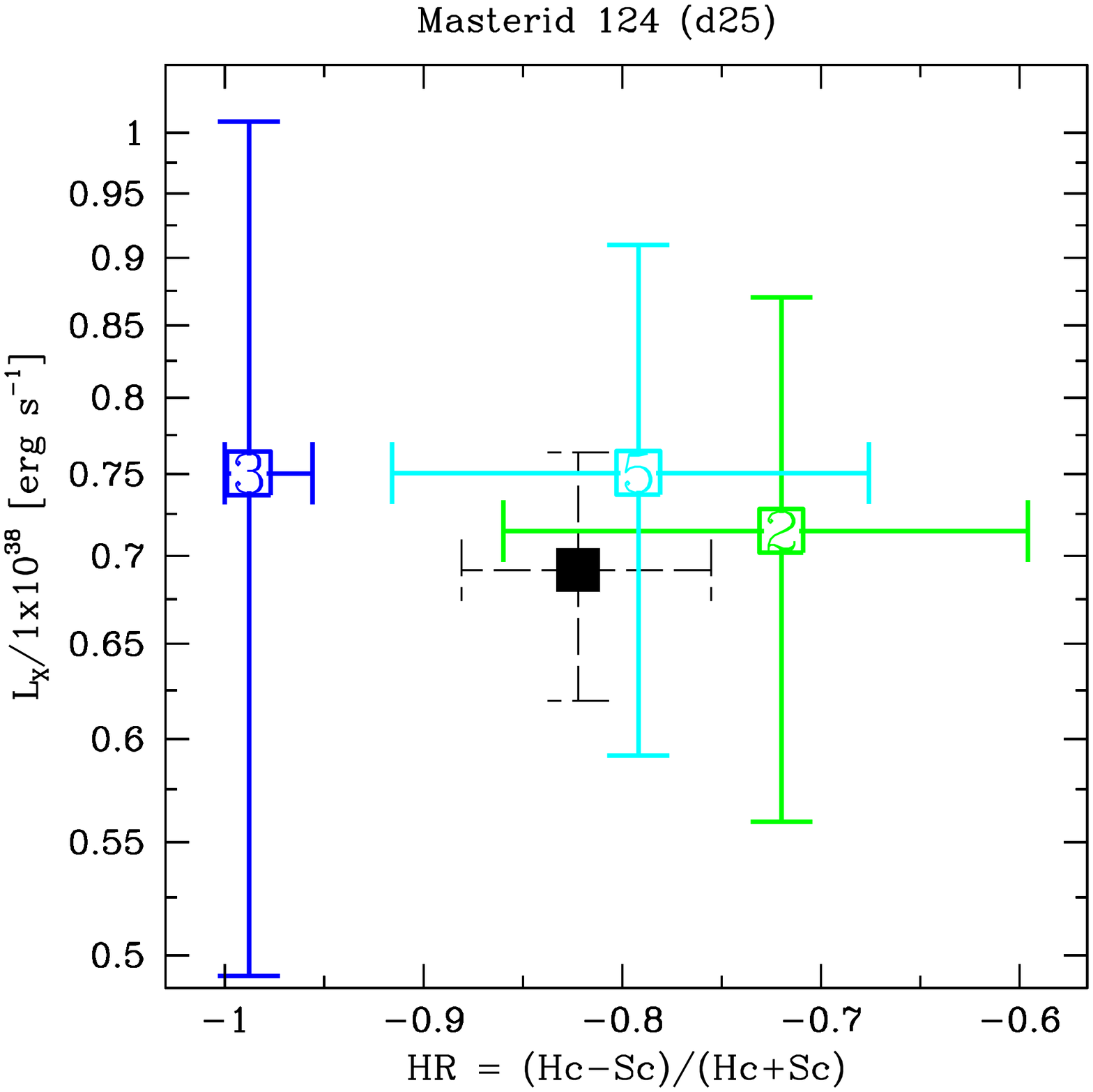}

  \end{minipage}
  \begin{minipage}{0.32\linewidth}
  \centering

    \includegraphics[width=\linewidth]{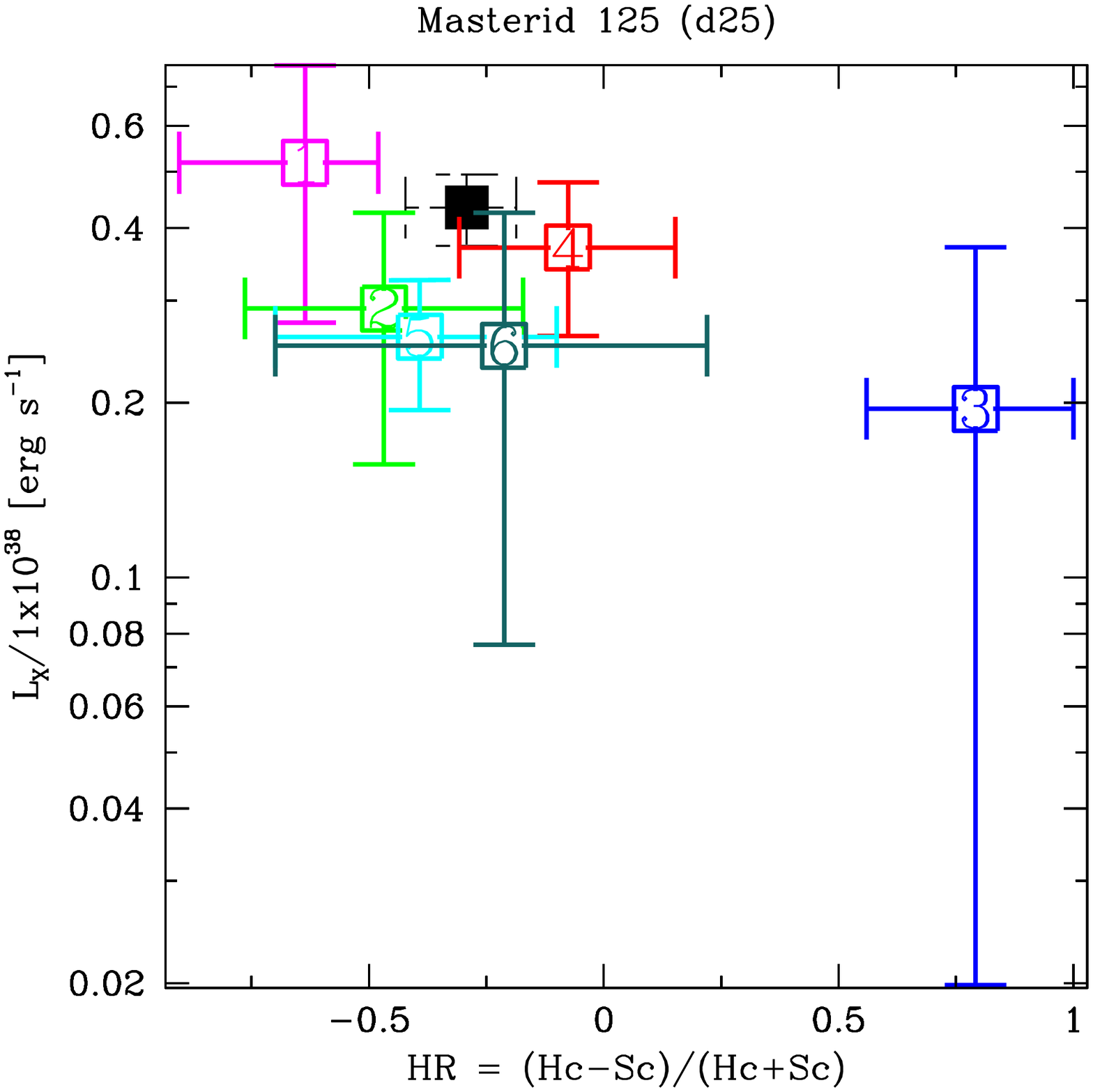}

\end{minipage}
\begin{minipage}{0.32\linewidth}
  \centering

    \includegraphics[width=\linewidth]{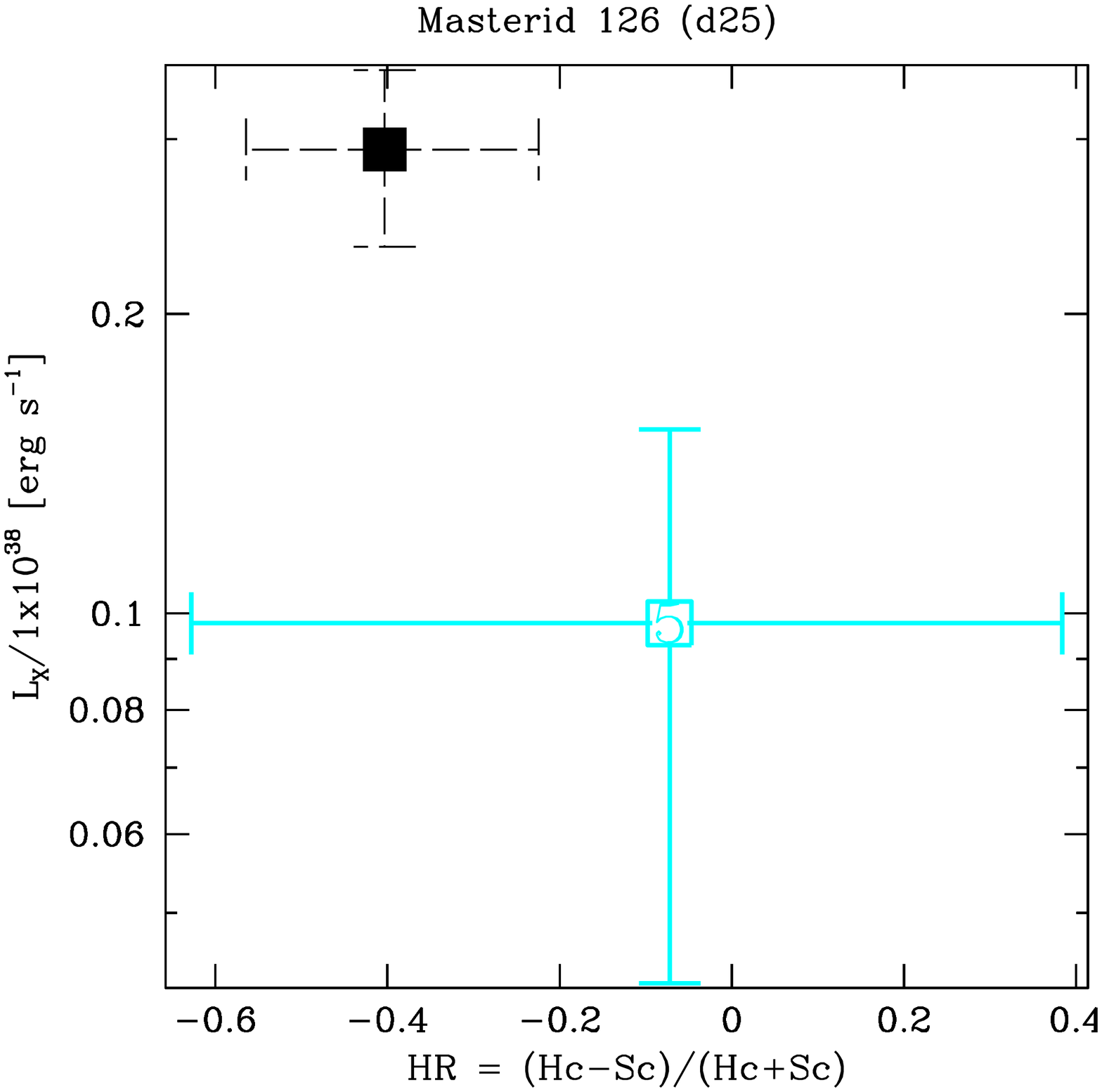}

 \end{minipage}

  \begin{minipage}{0.32\linewidth}
  \centering
  
    \includegraphics[width=\linewidth]{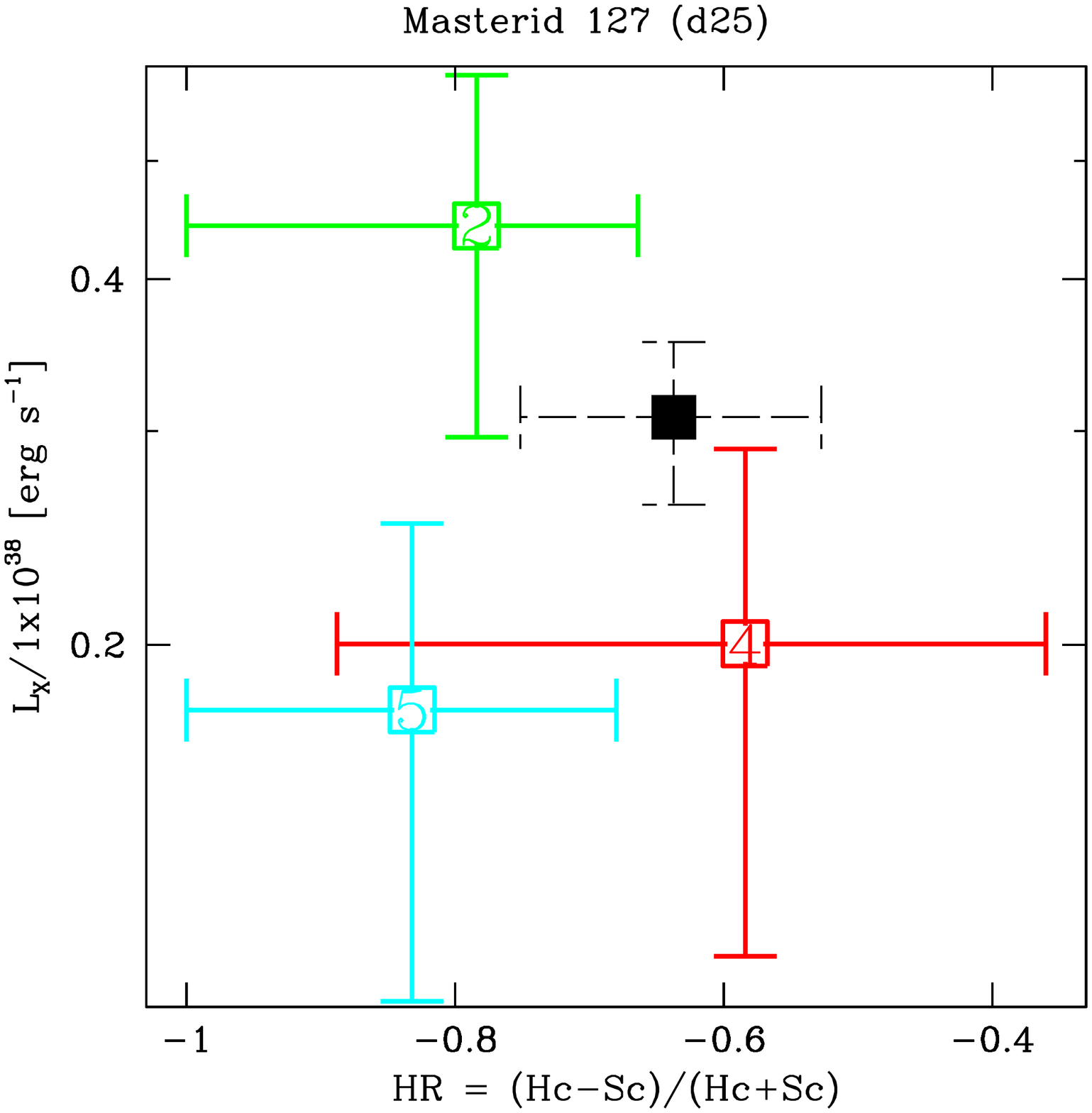}

  \end{minipage}
  \begin{minipage}{0.32\linewidth}
  \centering

    \includegraphics[width=\linewidth]{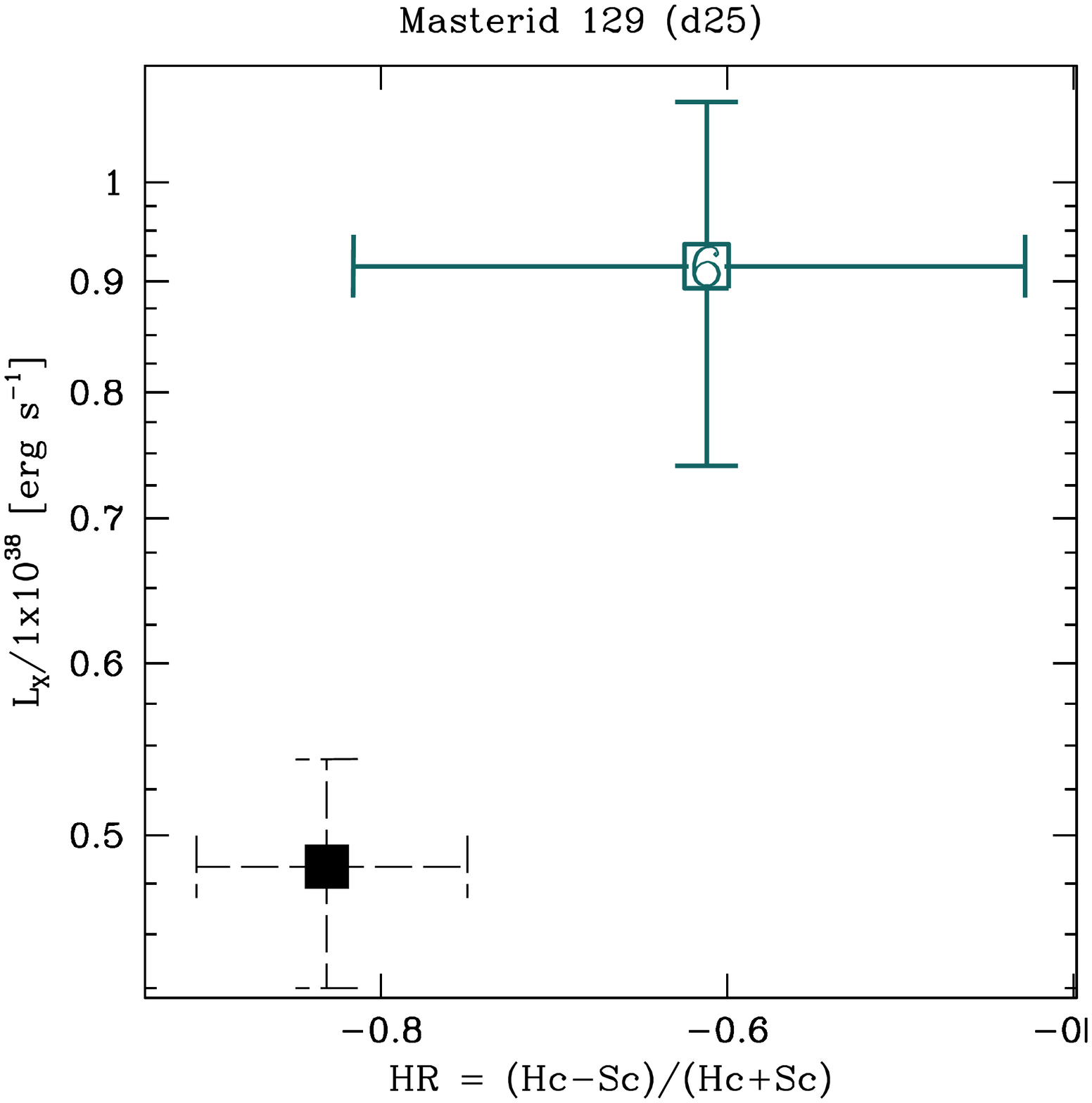}

\end{minipage}
\begin{minipage}{0.32\linewidth}
  \centering

    \includegraphics[width=\linewidth]{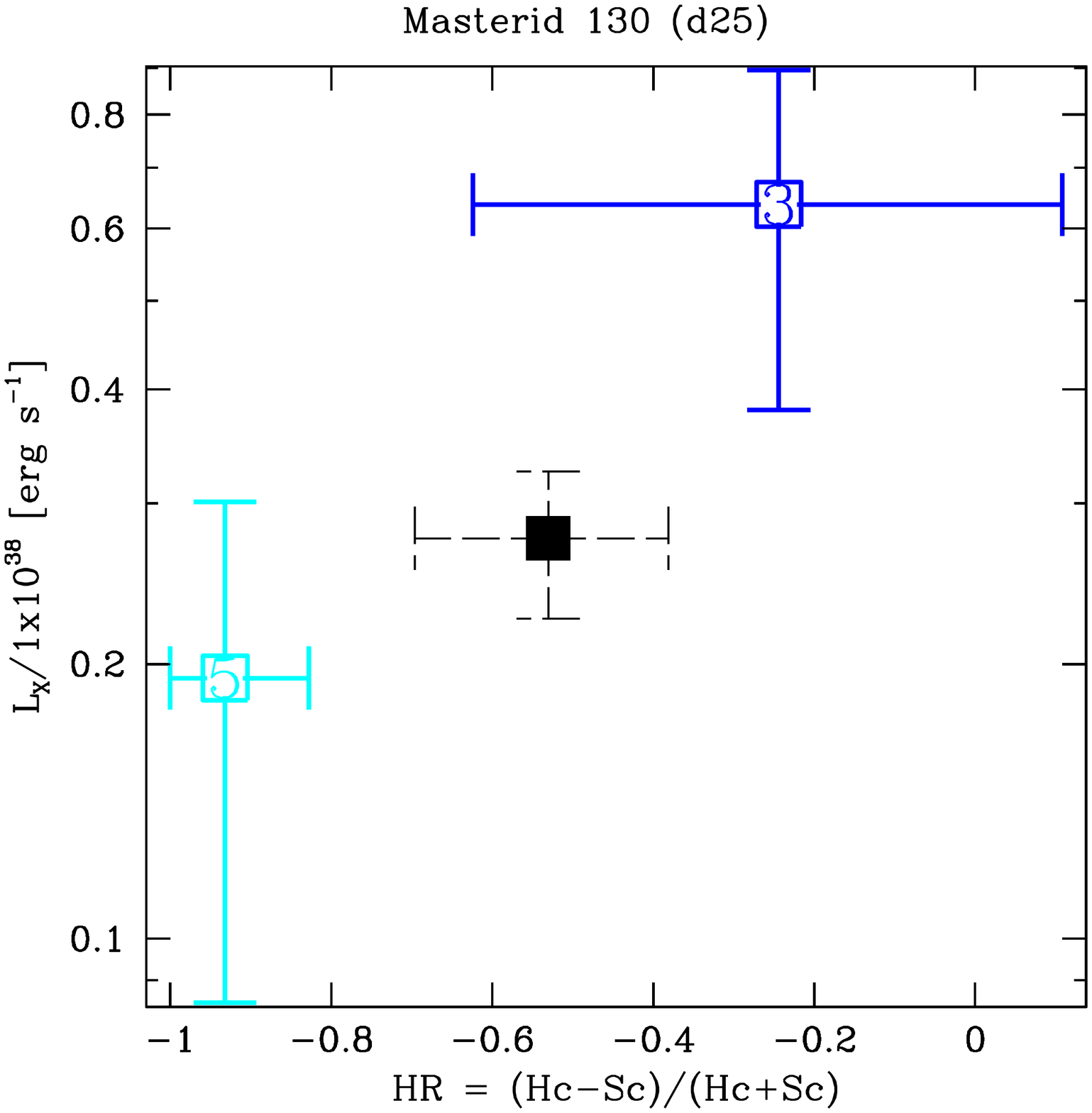}

\end{minipage}

\begin{minipage}{0.32\linewidth}
  \centering
  
    \includegraphics[width=\linewidth]{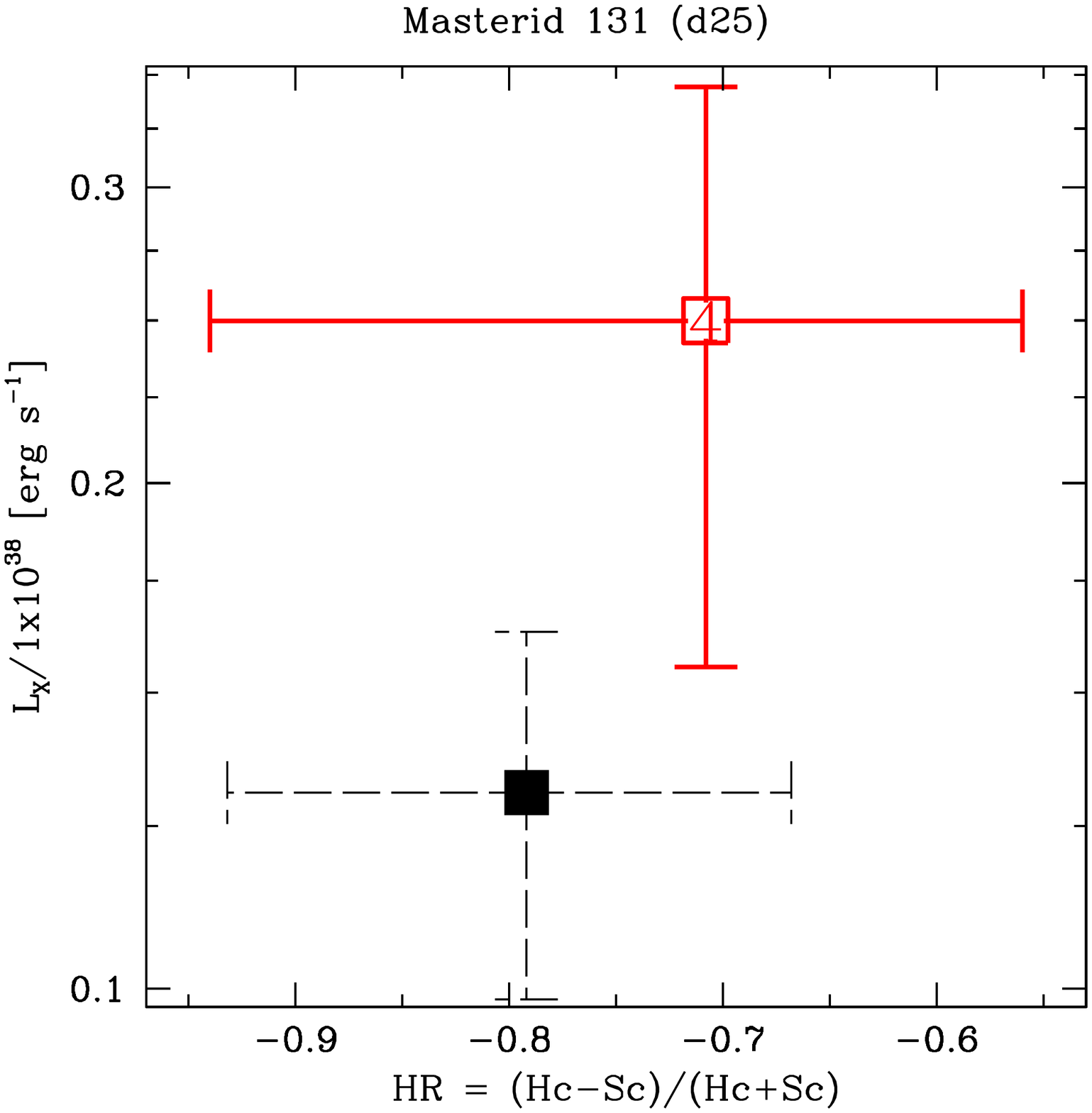}

  \end{minipage}
  \begin{minipage}{0.32\linewidth}
  \centering

    \includegraphics[width=\linewidth]{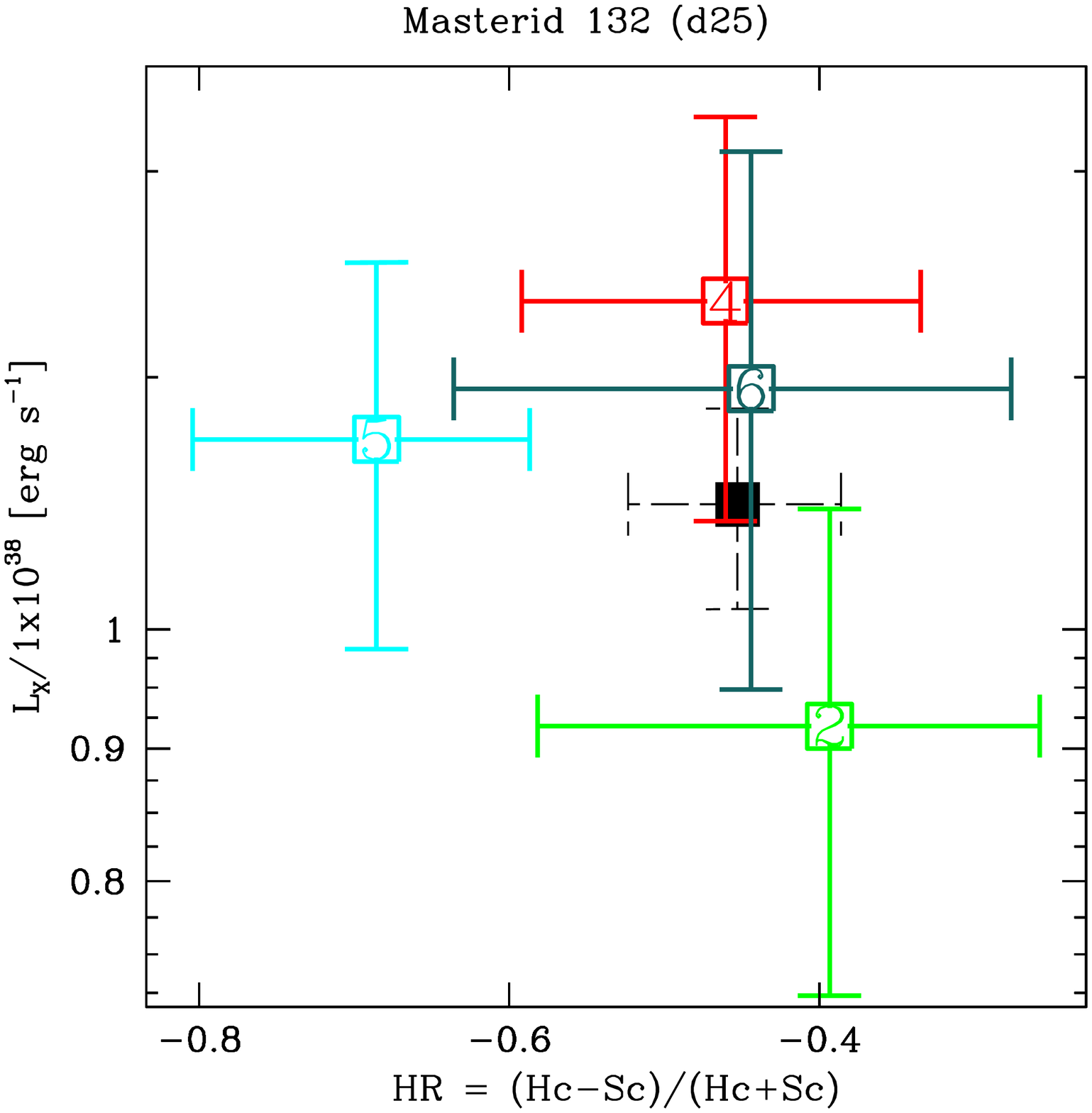}

\end{minipage}
\begin{minipage}{0.32\linewidth}
  \centering

    \includegraphics[width=\linewidth]{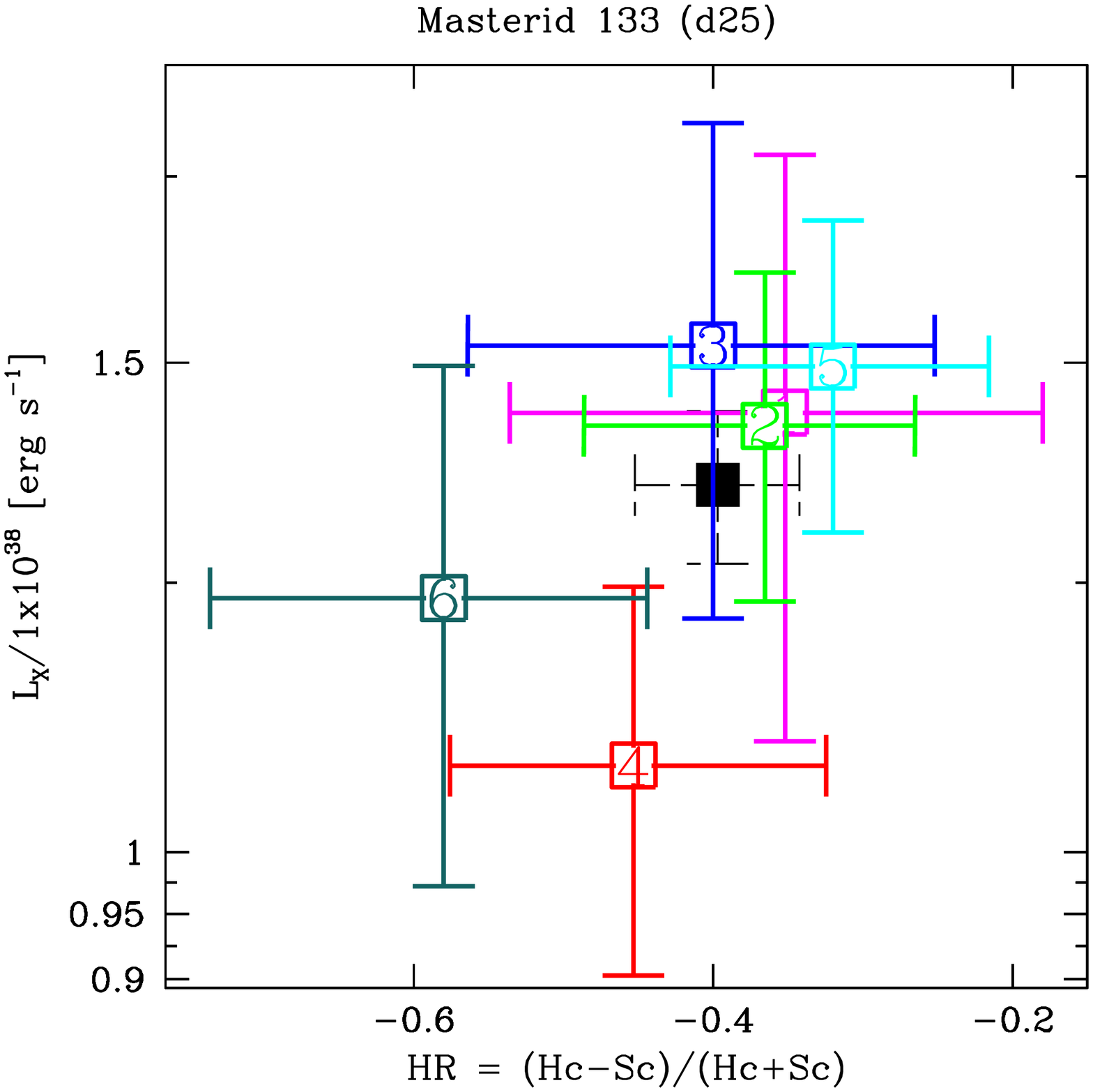}

\end{minipage}
\end{figure}

\begin{figure}
  \begin{minipage}{0.32\linewidth}
  \centering
  
    \includegraphics[width=\linewidth]{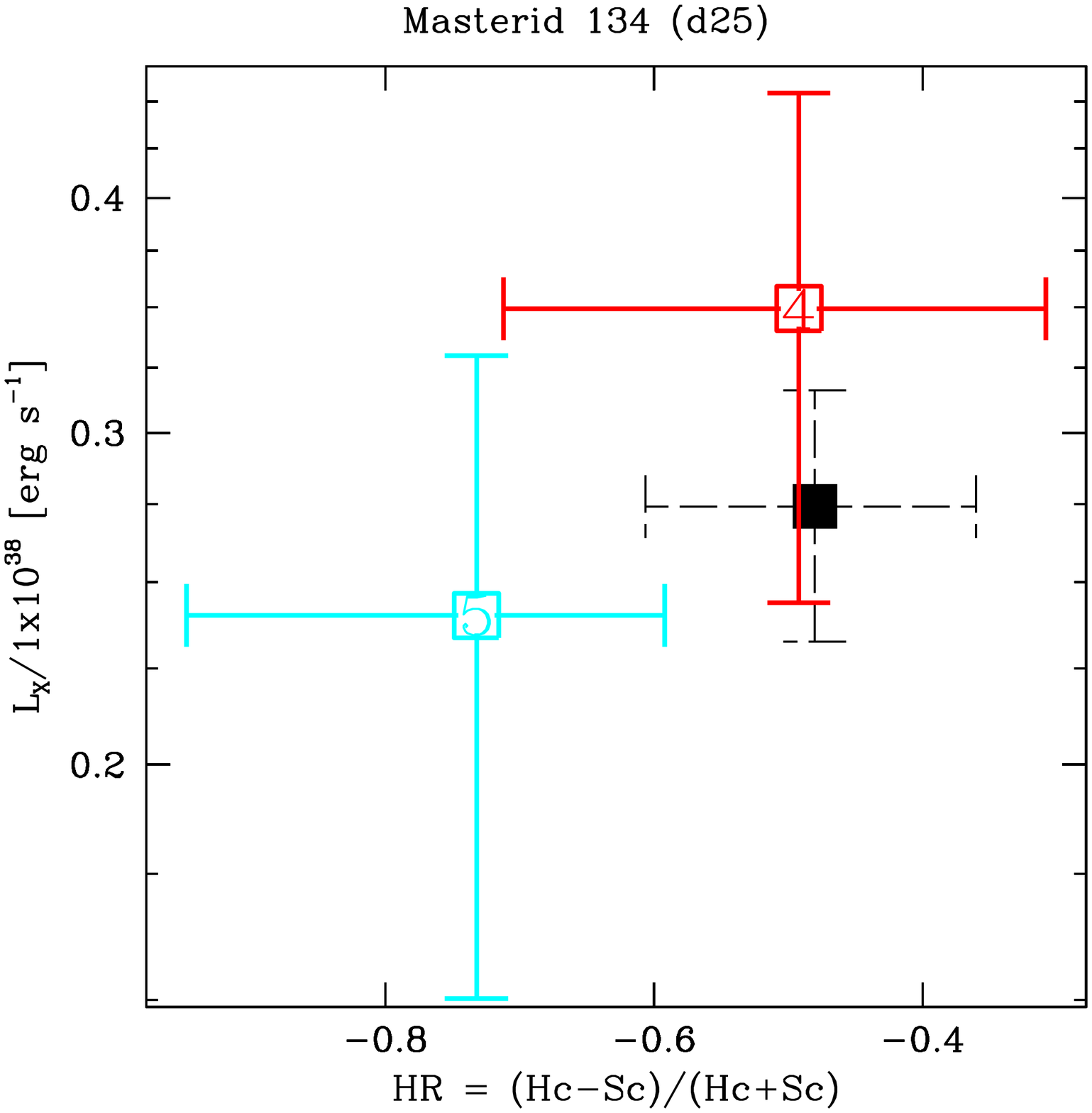}

  \end{minipage}
  \begin{minipage}{0.32\linewidth}
  \centering

    \includegraphics[width=\linewidth]{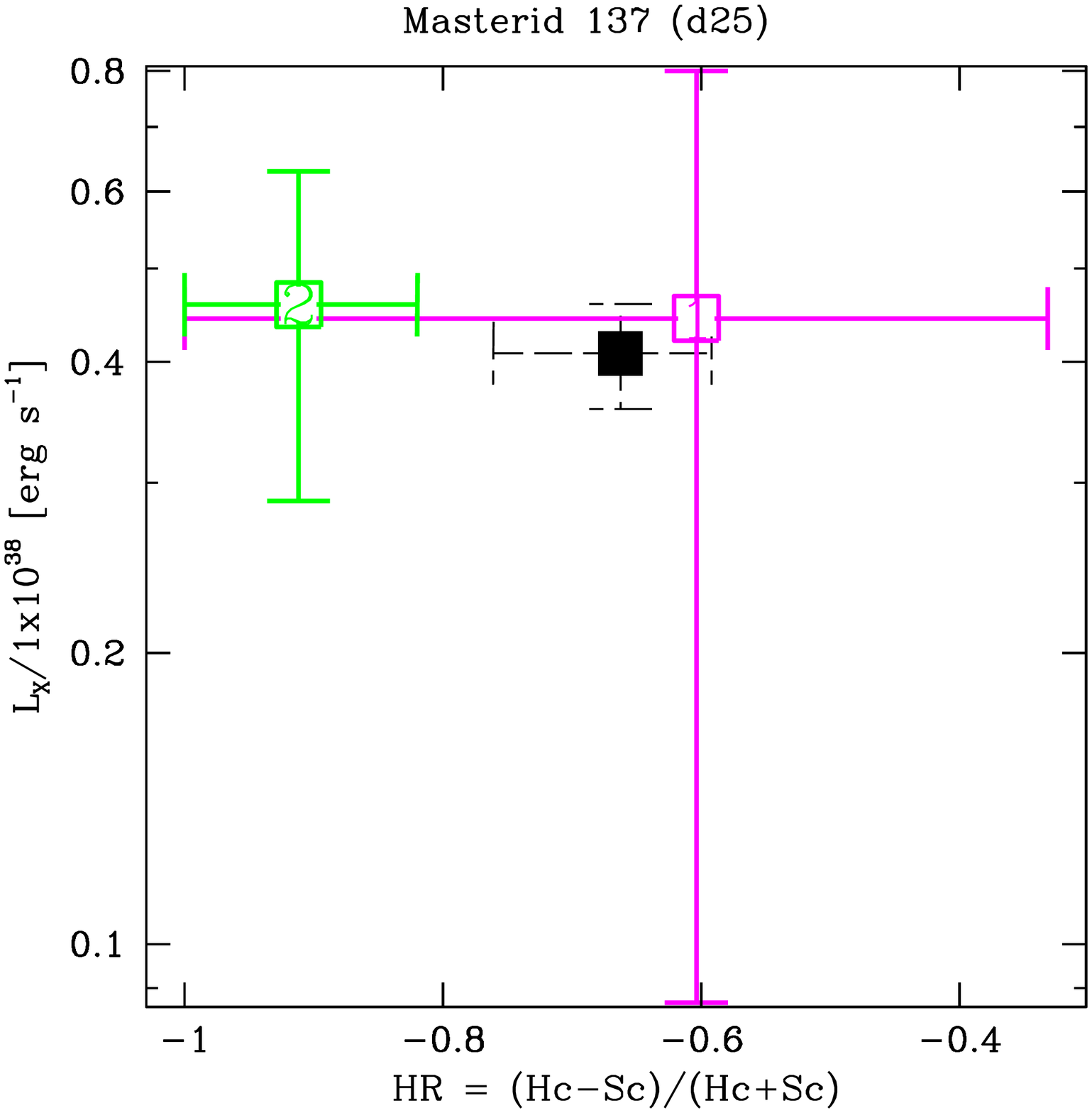}

\end{minipage}
\begin{minipage}{0.32\linewidth}
  \centering

    \includegraphics[width=\linewidth]{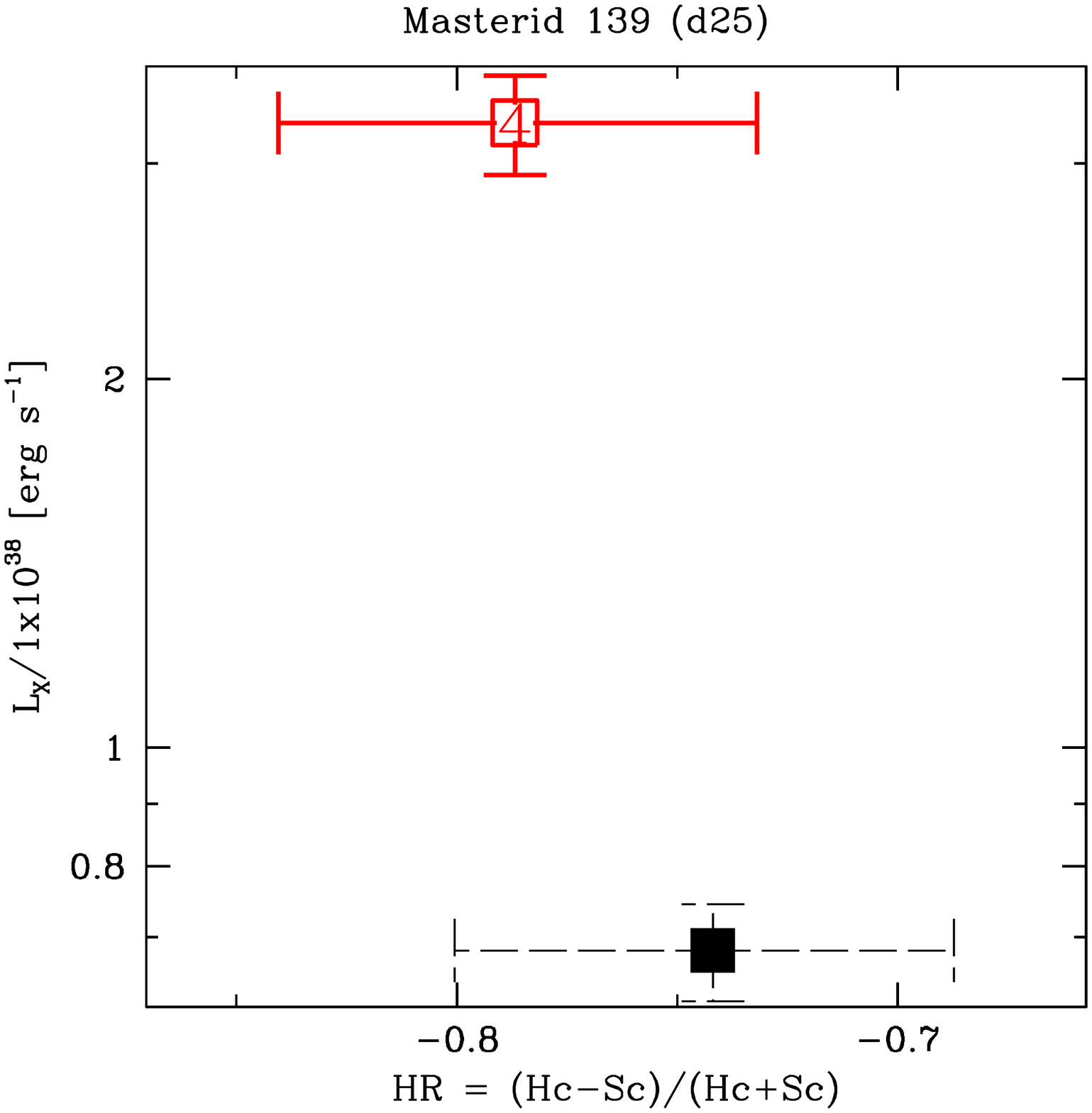}

 \end{minipage}

\begin{minipage}{0.32\linewidth}
  \centering
  
    \includegraphics[width=\linewidth]{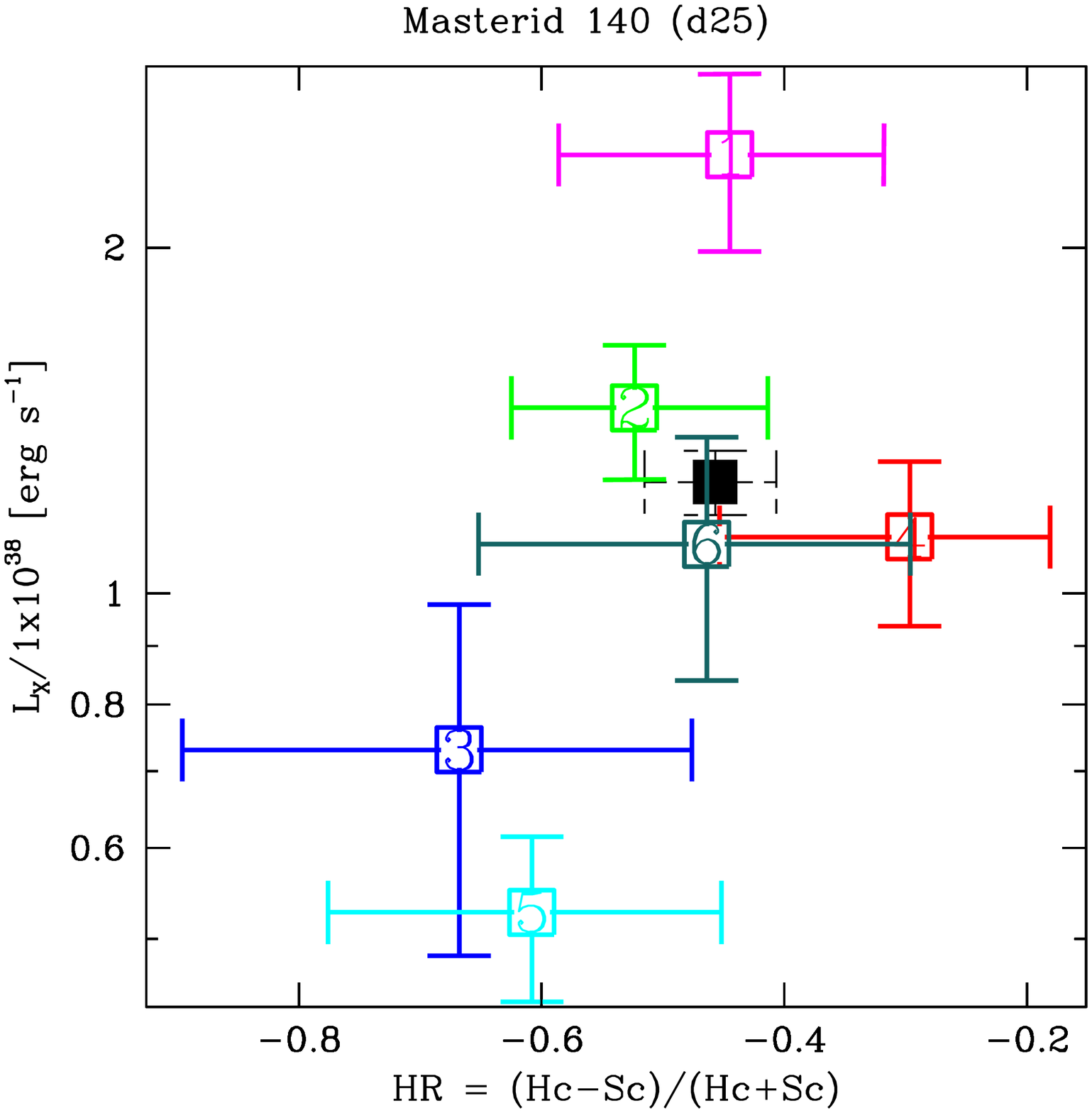}

  \end{minipage}
  \begin{minipage}{0.32\linewidth}
  \centering

    \includegraphics[width=\linewidth]{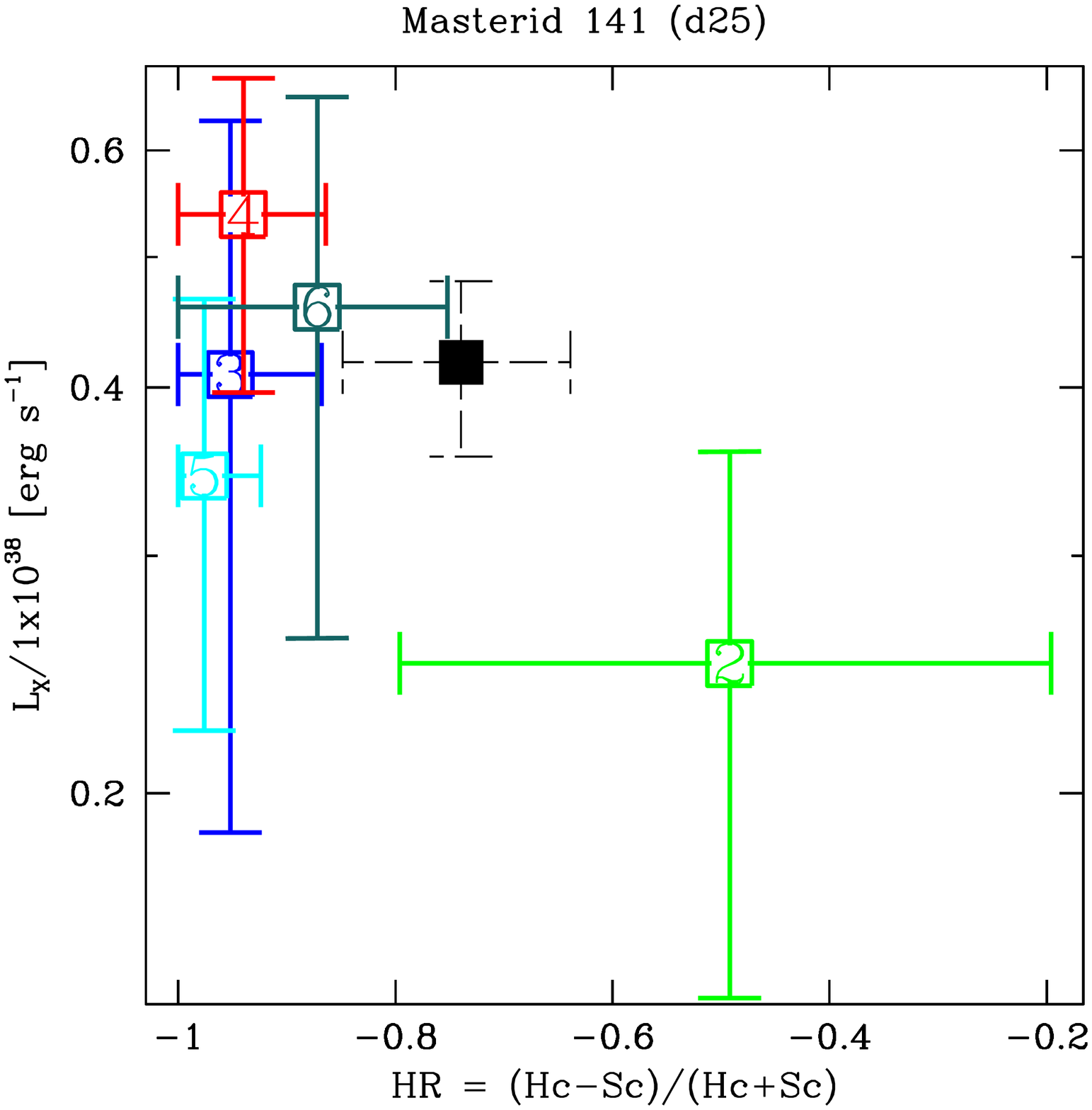}

\end{minipage}
\begin{minipage}{0.32\linewidth}
  \centering

    \includegraphics[width=\linewidth]{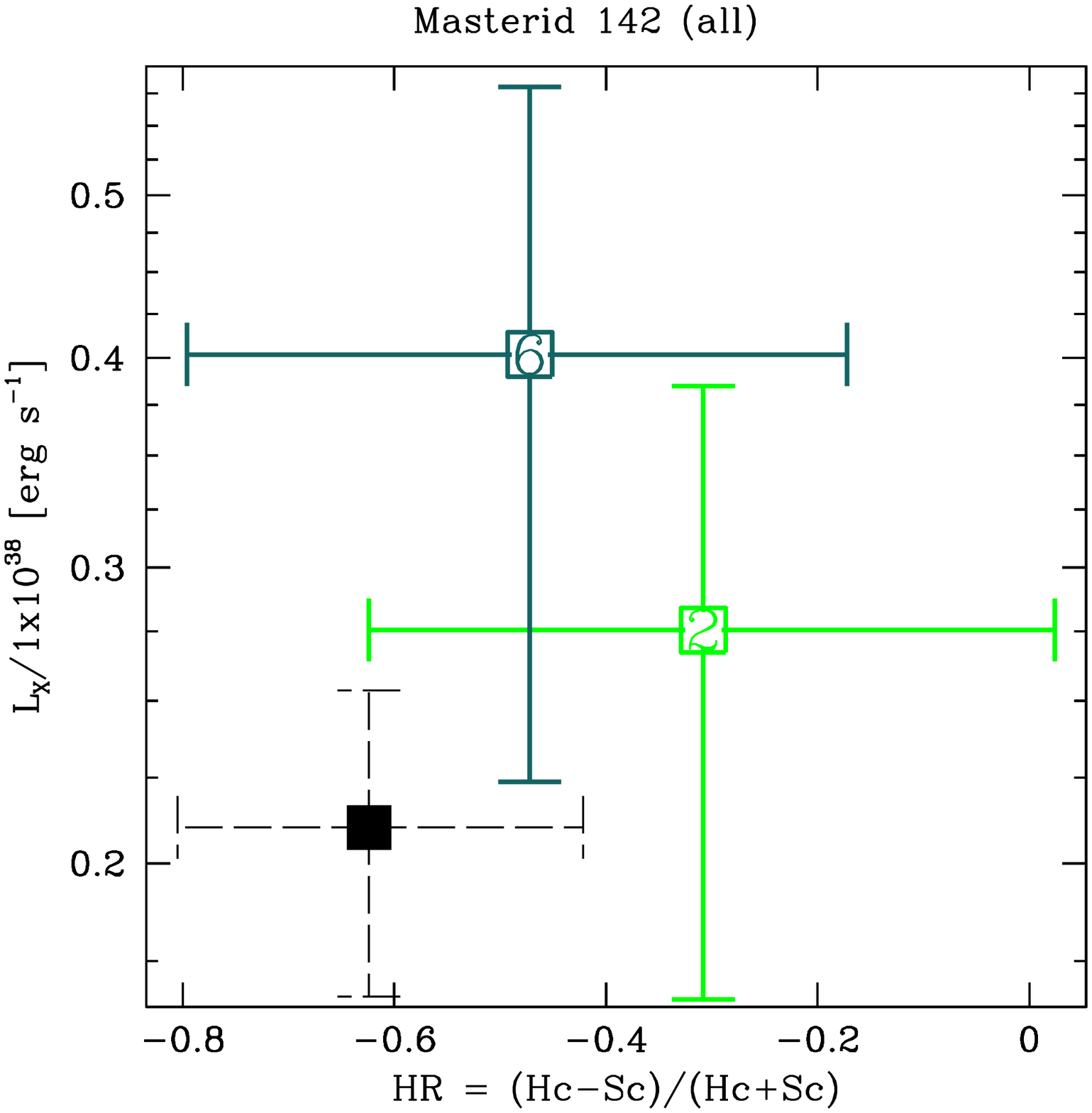}

 \end{minipage}

  \begin{minipage}{0.32\linewidth}
  \centering
  
    \includegraphics[width=\linewidth]{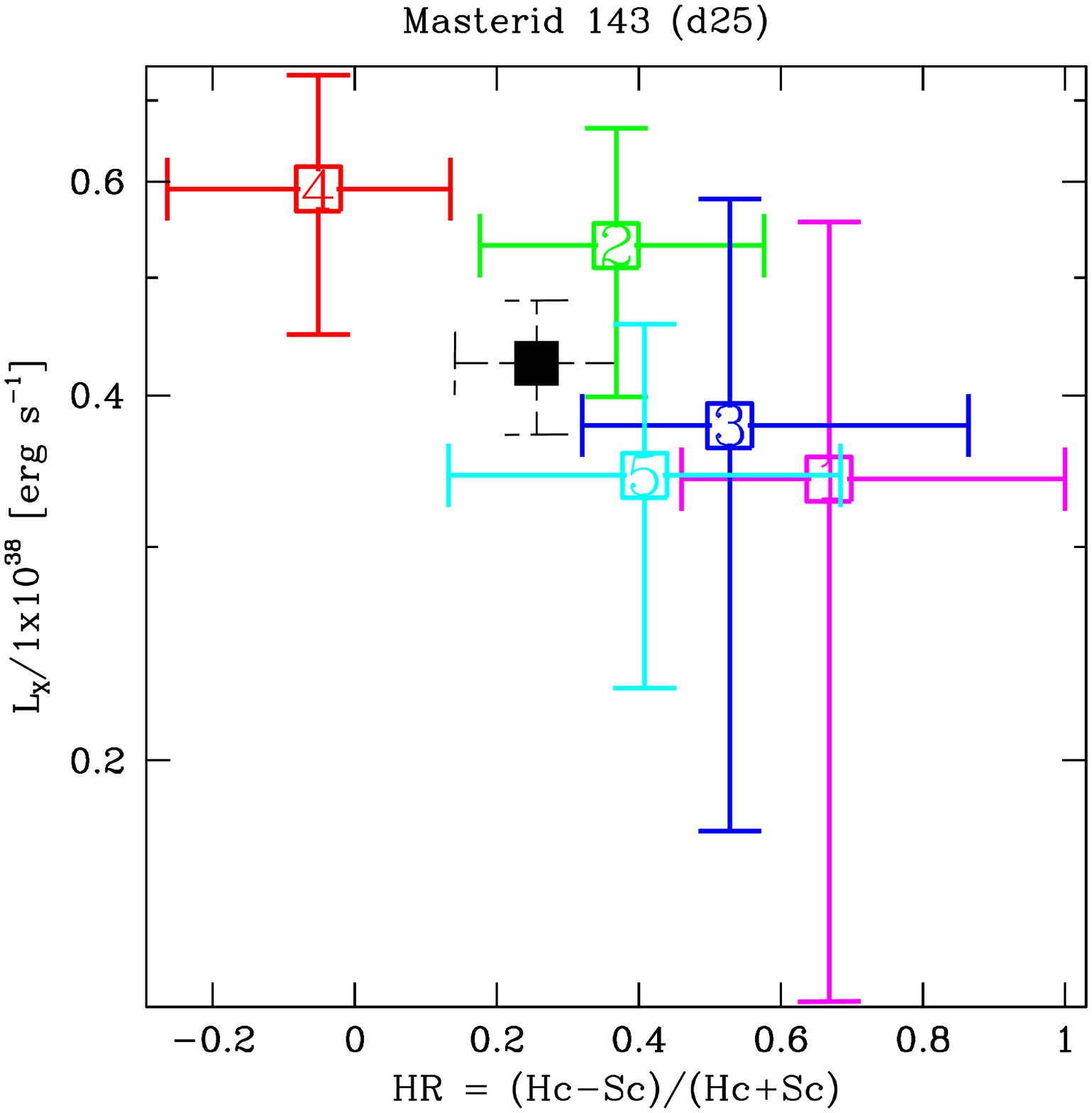}

  \end{minipage}
  \begin{minipage}{0.32\linewidth}
  \centering

    \includegraphics[width=\linewidth]{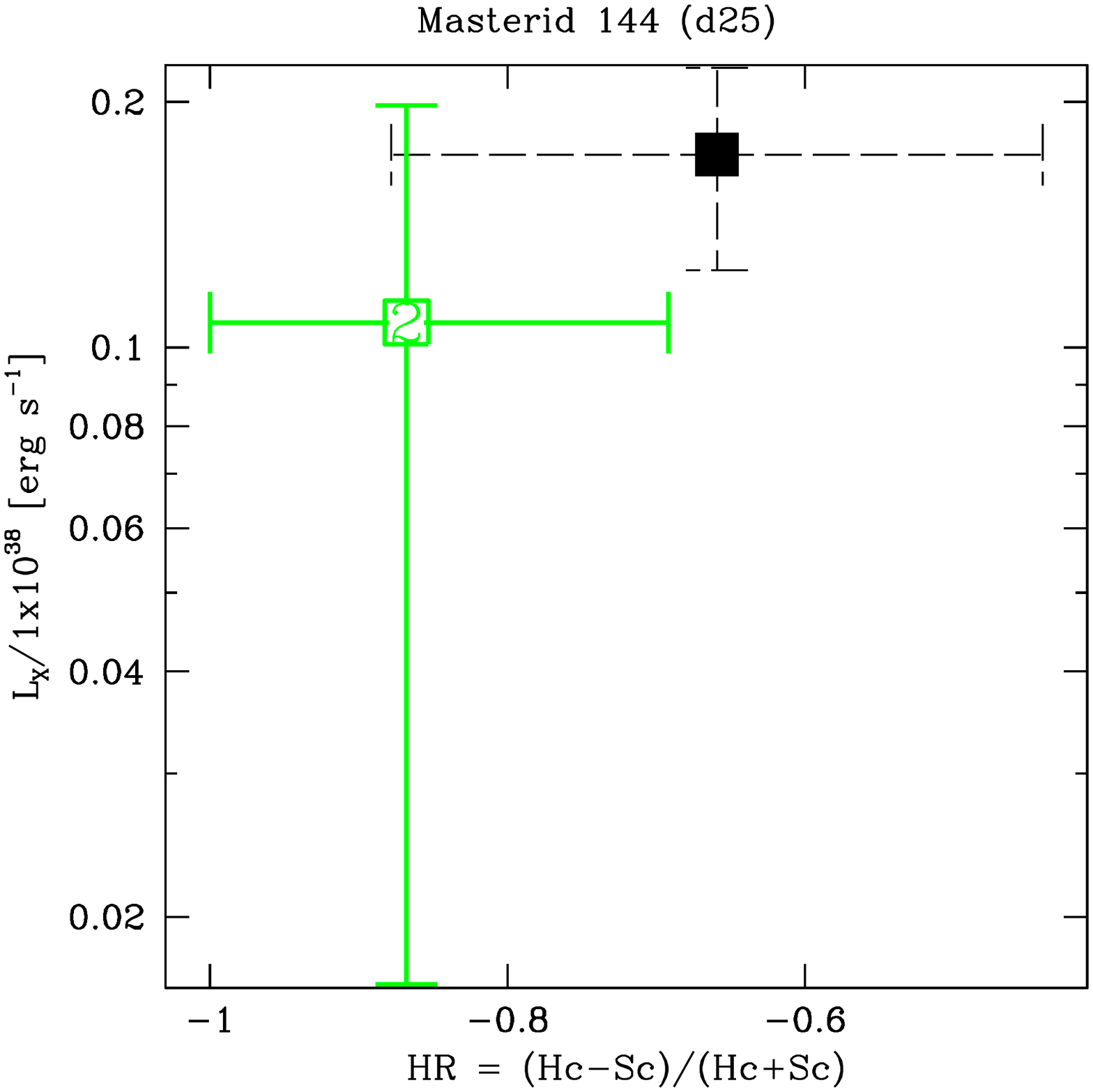}

\end{minipage}
\begin{minipage}{0.32\linewidth}
  \centering

    \includegraphics[width=\linewidth]{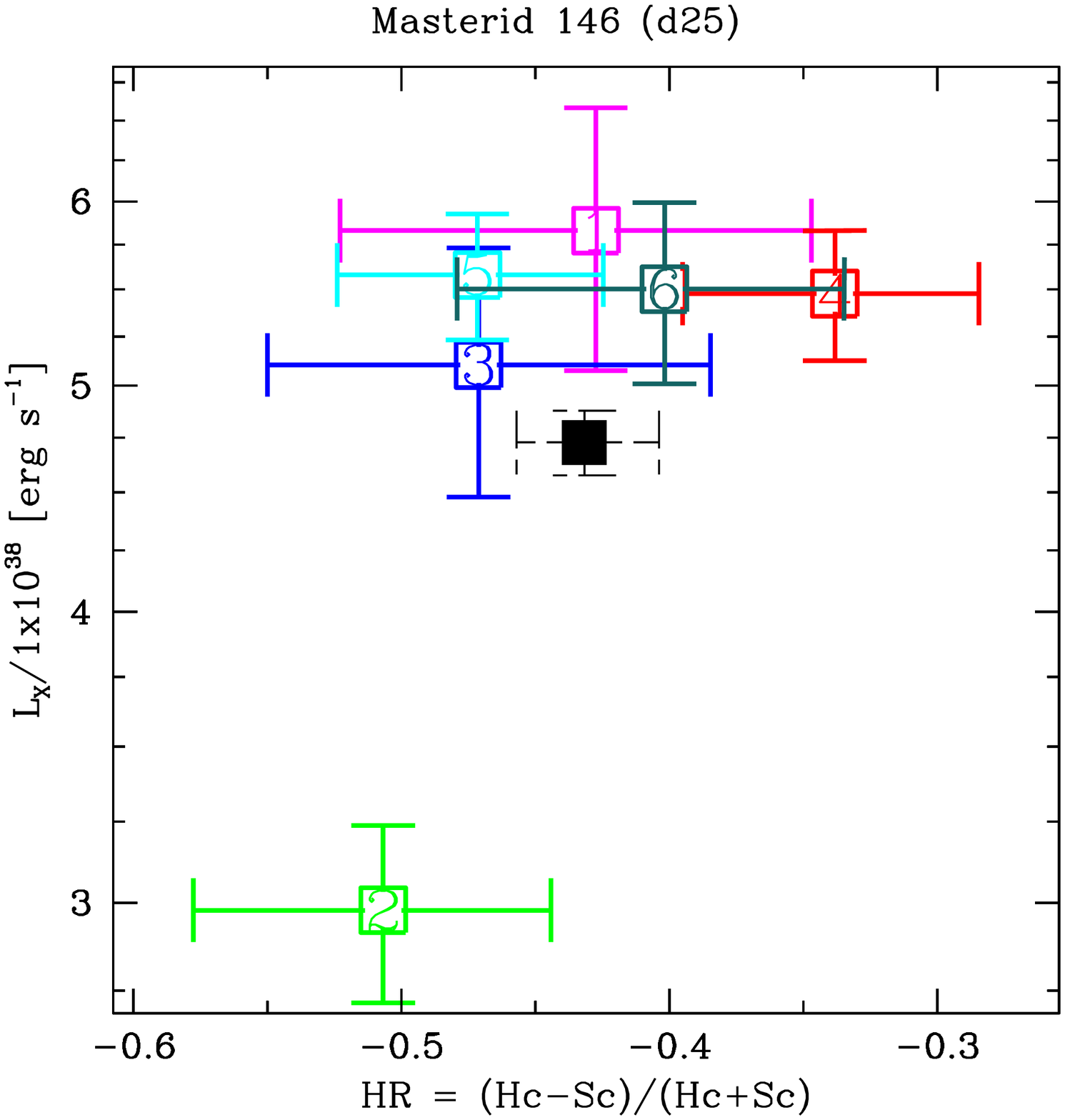}

\end{minipage}

\begin{minipage}{0.32\linewidth}
  \centering
  
    \includegraphics[width=\linewidth]{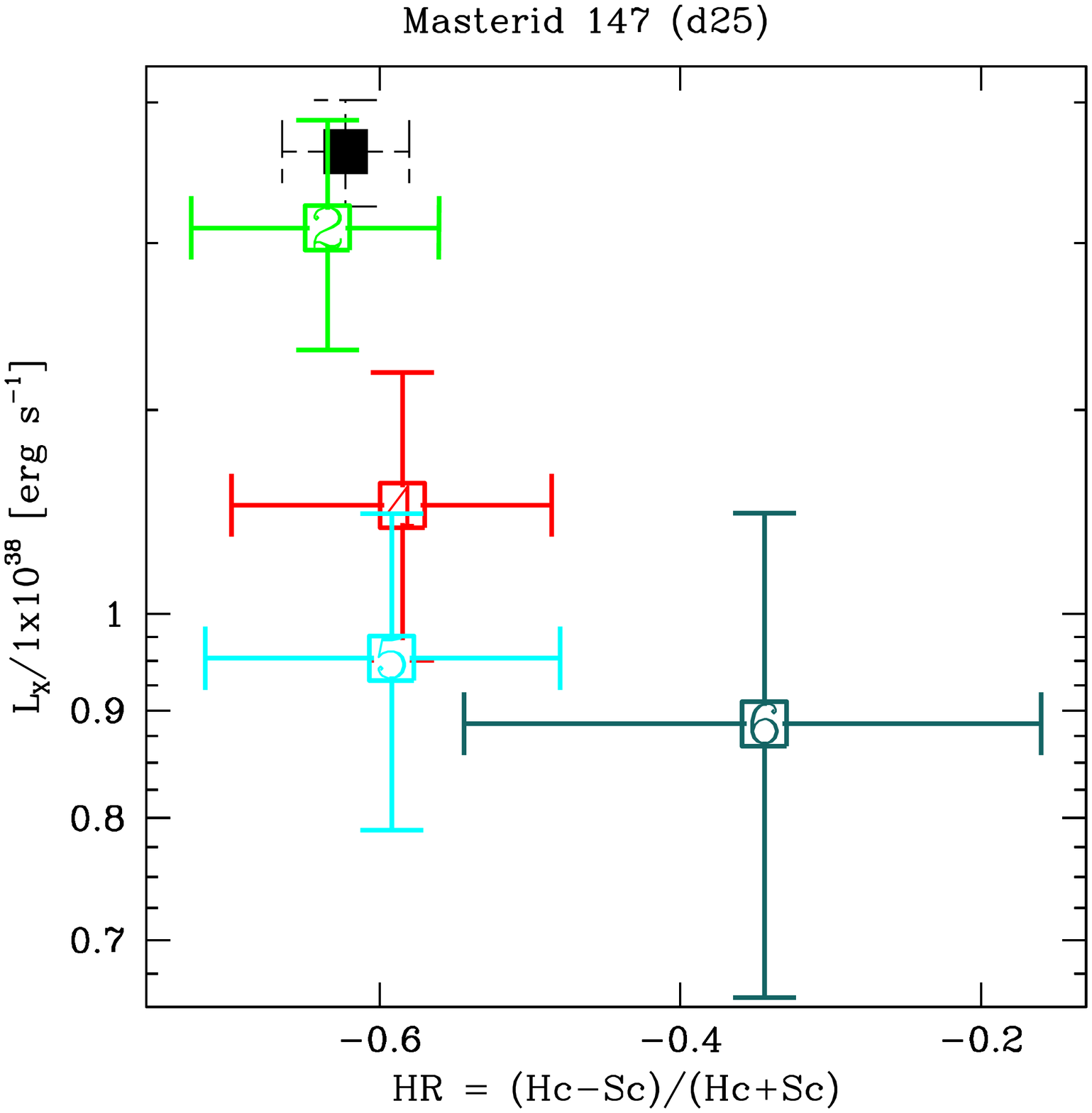}

  \end{minipage}
  \begin{minipage}{0.32\linewidth}
  \centering

    \includegraphics[width=\linewidth]{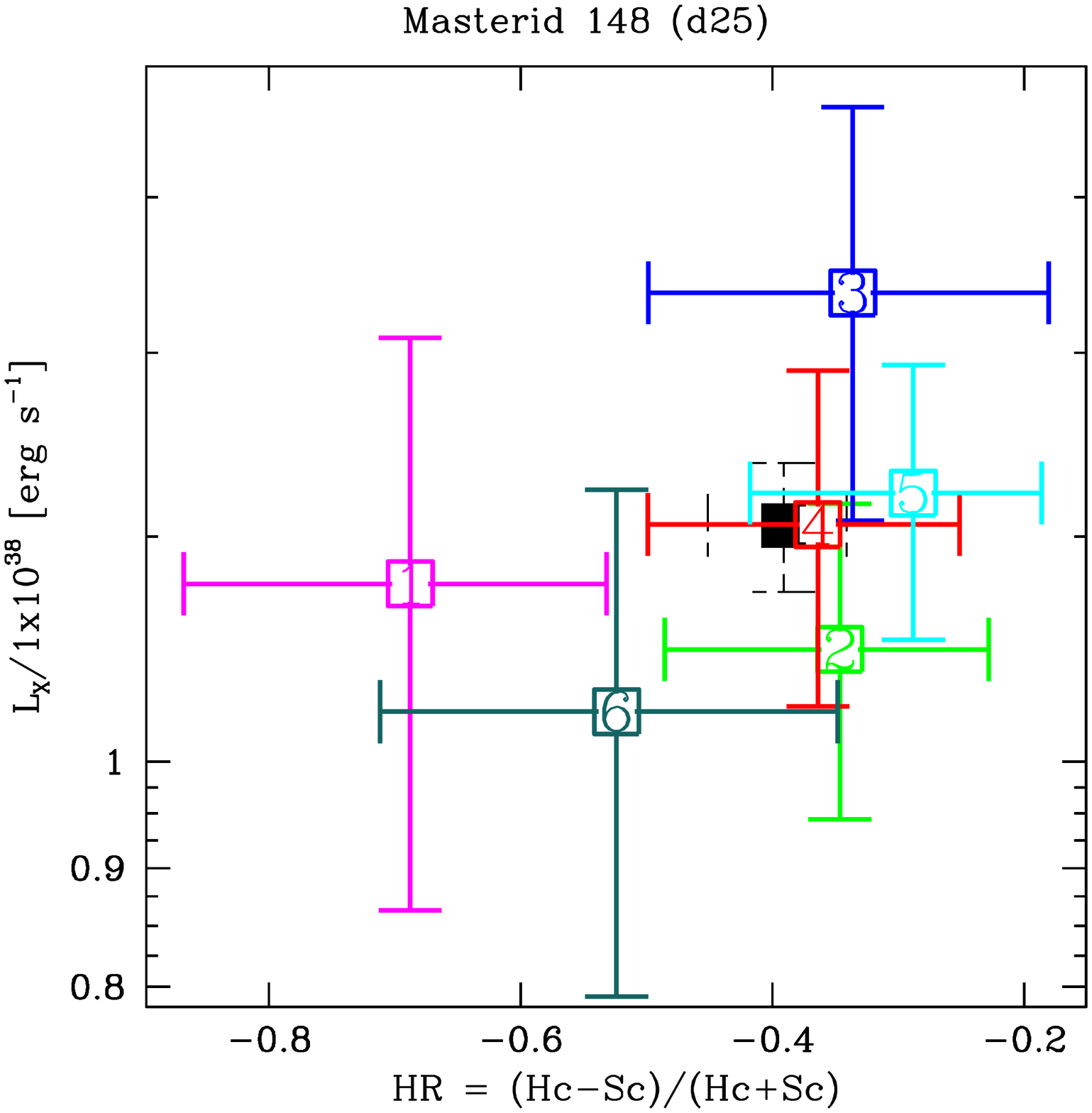}

\end{minipage}
\begin{minipage}{0.32\linewidth}
  \centering

    \includegraphics[width=\linewidth]{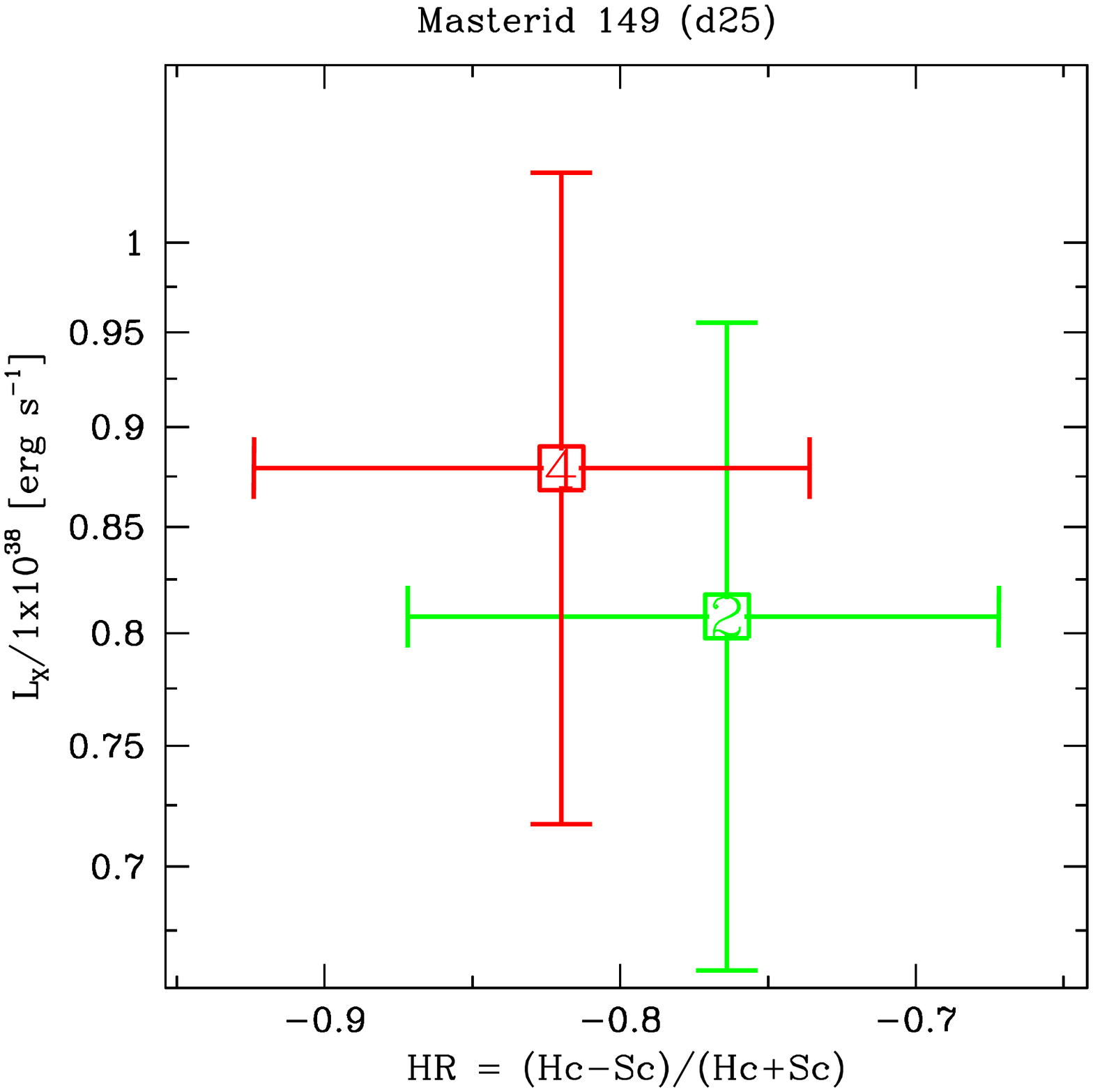}

\end{minipage}
\end{figure}

\begin{figure}
  \begin{minipage}{0.32\linewidth}
  \centering
  
    \includegraphics[width=\linewidth]{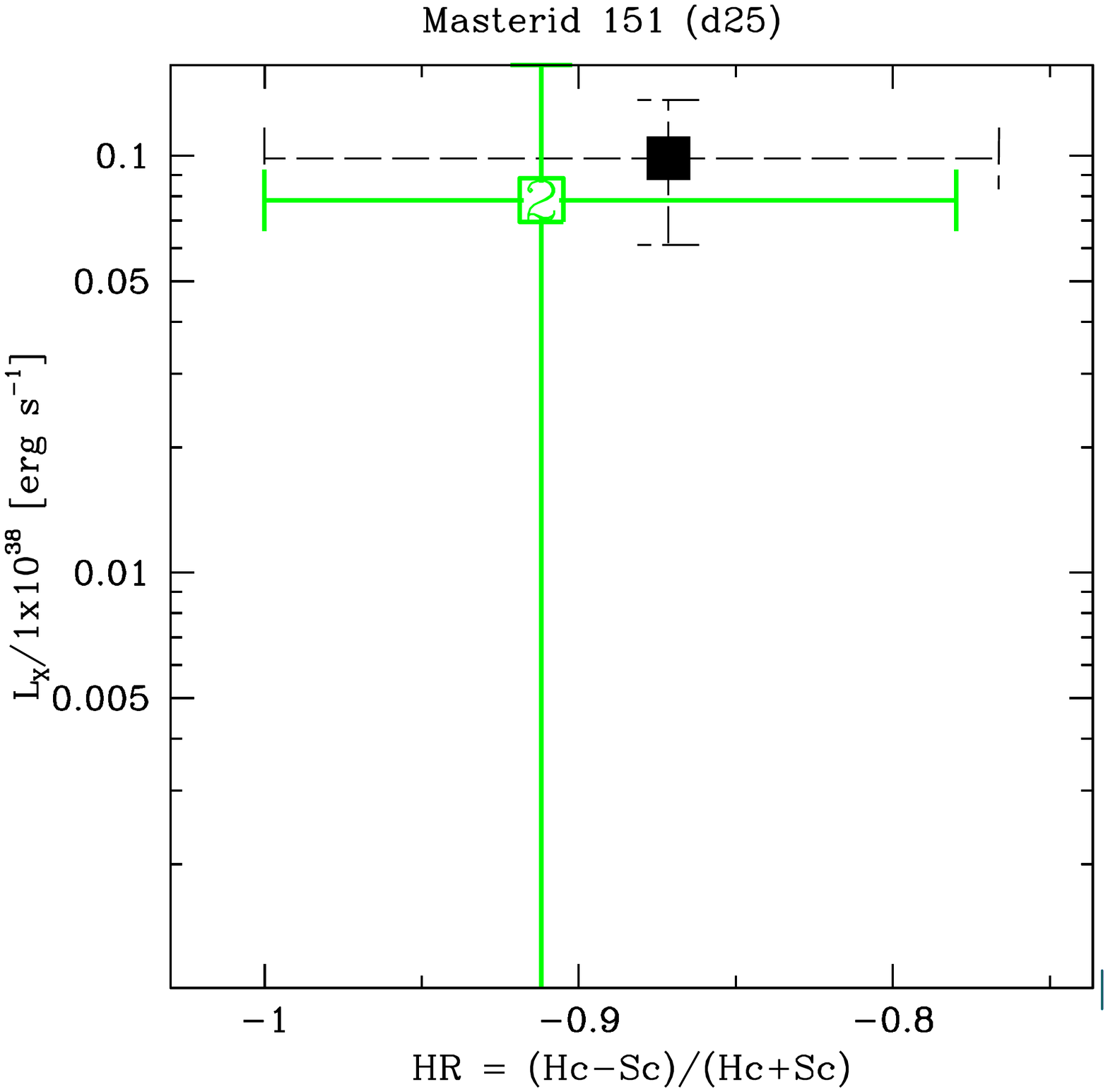}

  \end{minipage}
  \begin{minipage}{0.32\linewidth}
  \centering

    \includegraphics[width=\linewidth]{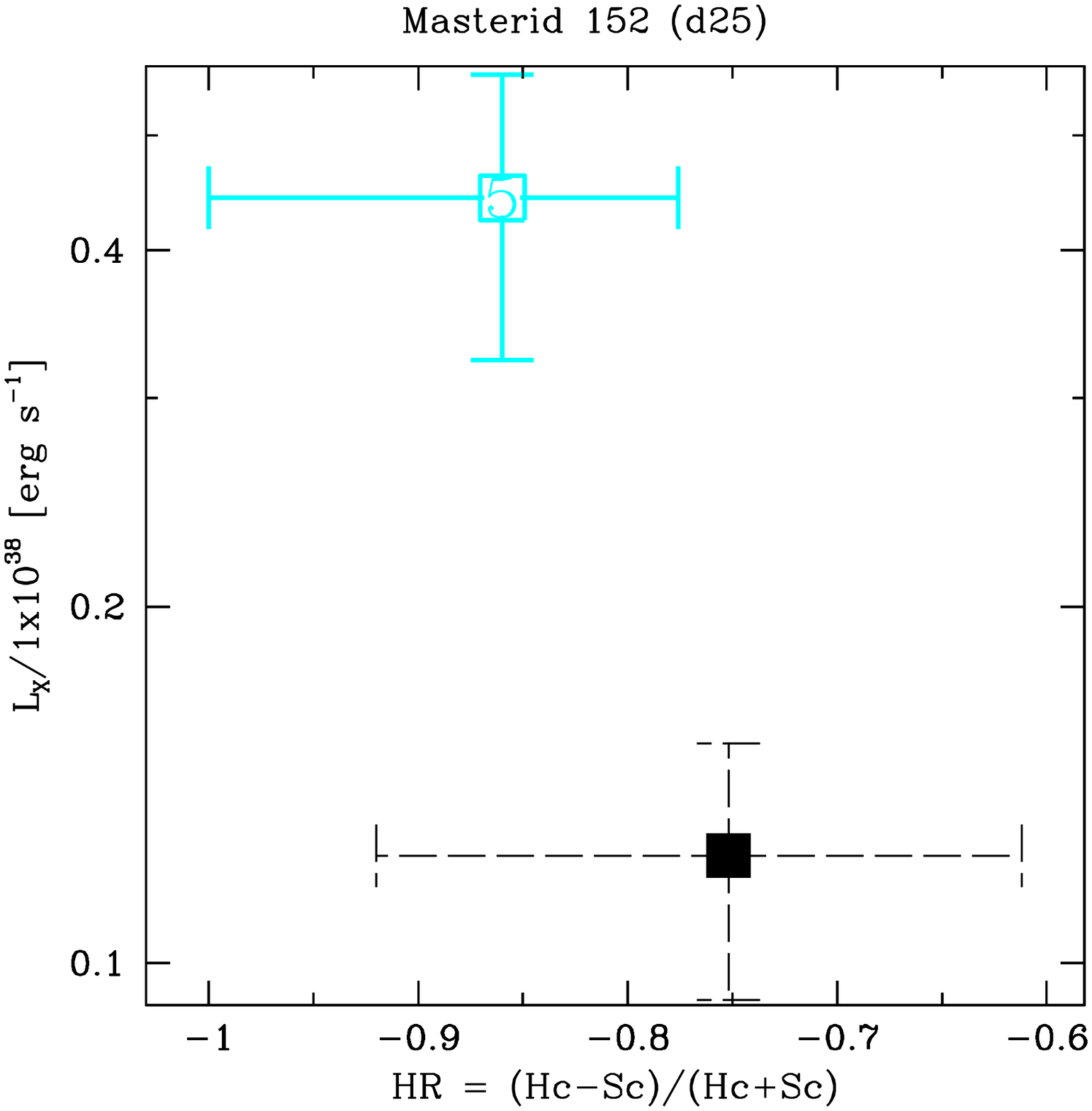}

\end{minipage}
\begin{minipage}{0.32\linewidth}
  \centering

    \includegraphics[width=\linewidth]{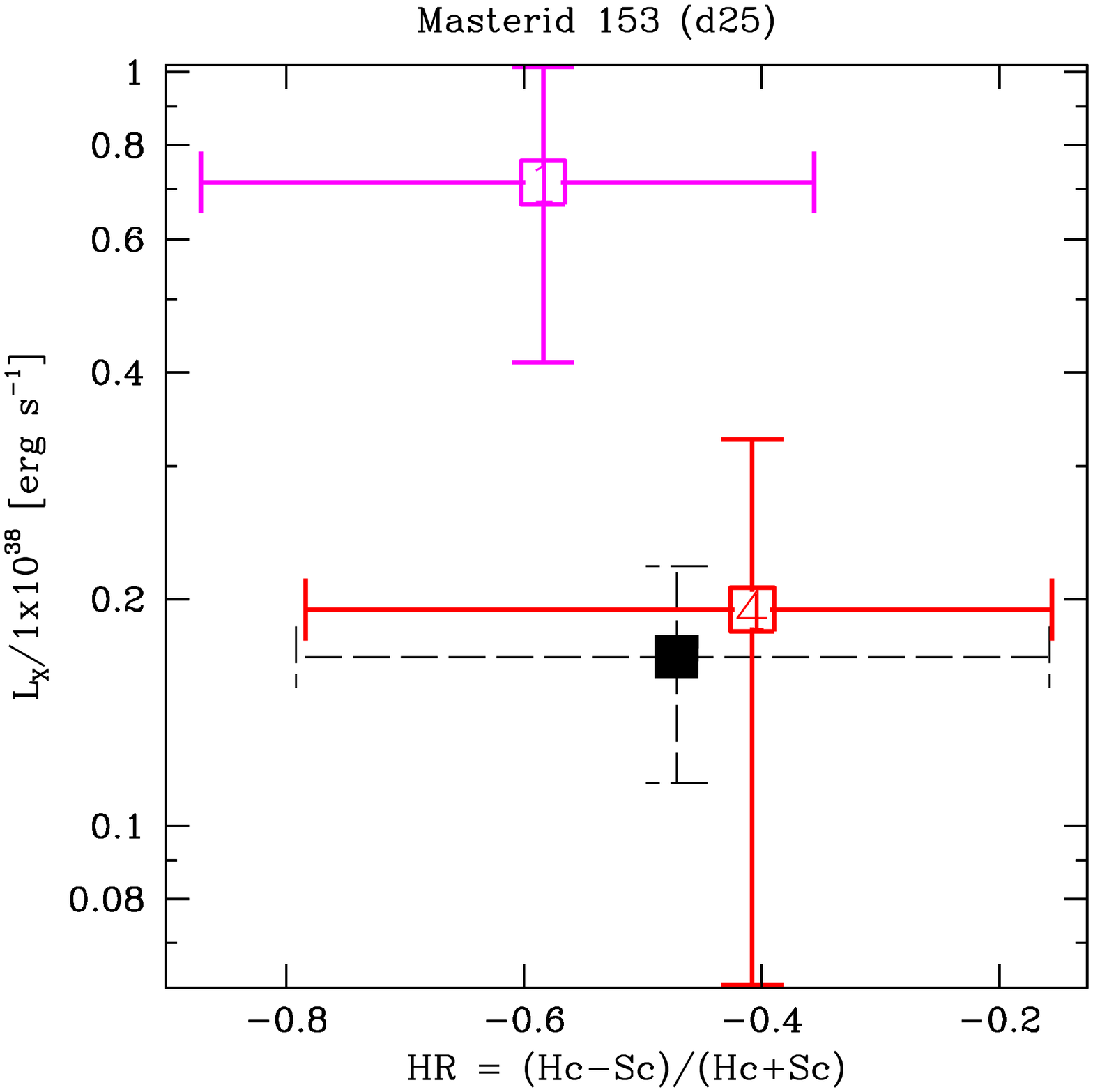}

 \end{minipage}

\begin{minipage}{0.32\linewidth}
  \centering
  
    \includegraphics[width=\linewidth]{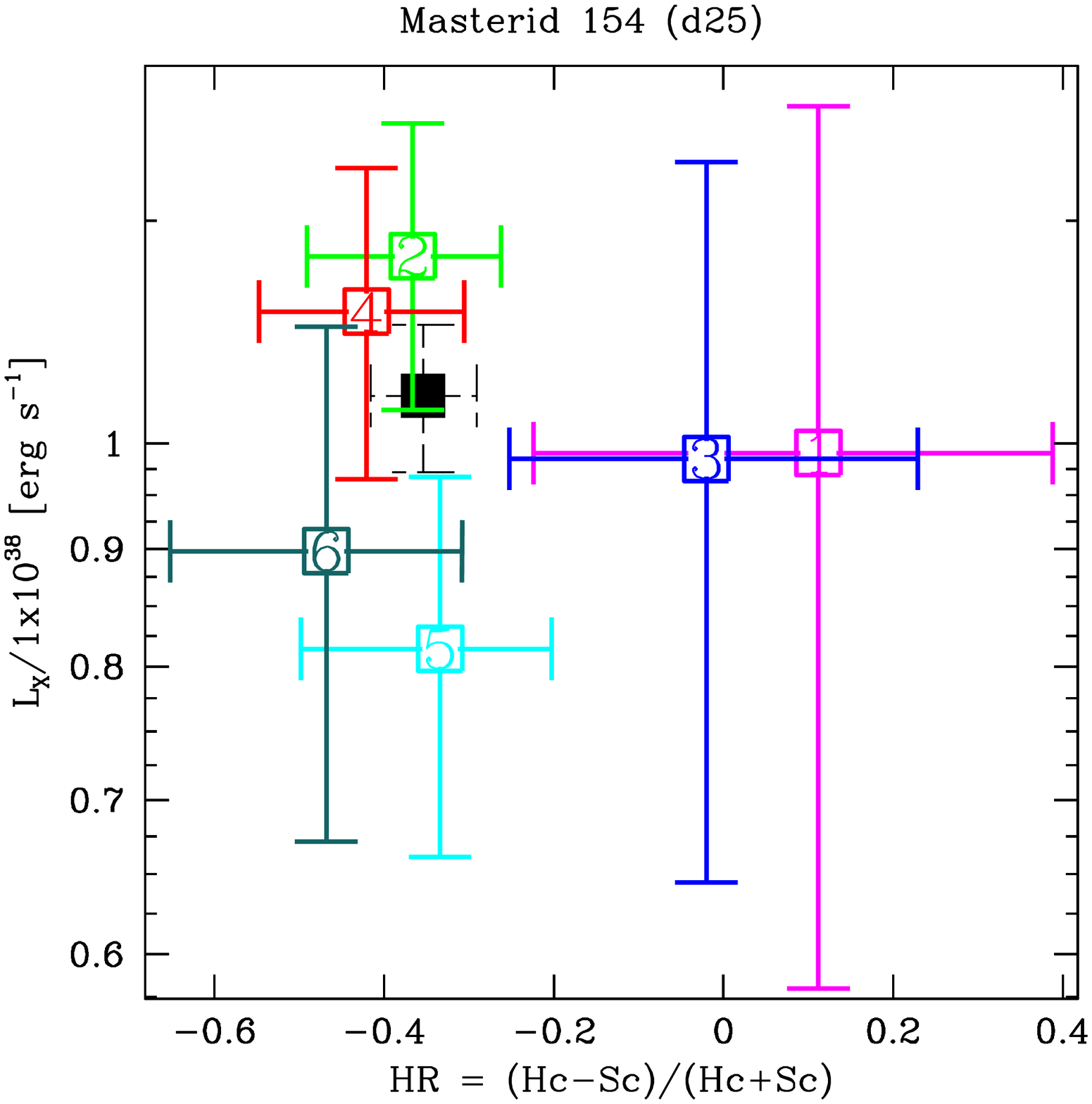}

  \end{minipage}
  \begin{minipage}{0.32\linewidth}
  \centering

    \includegraphics[width=\linewidth]{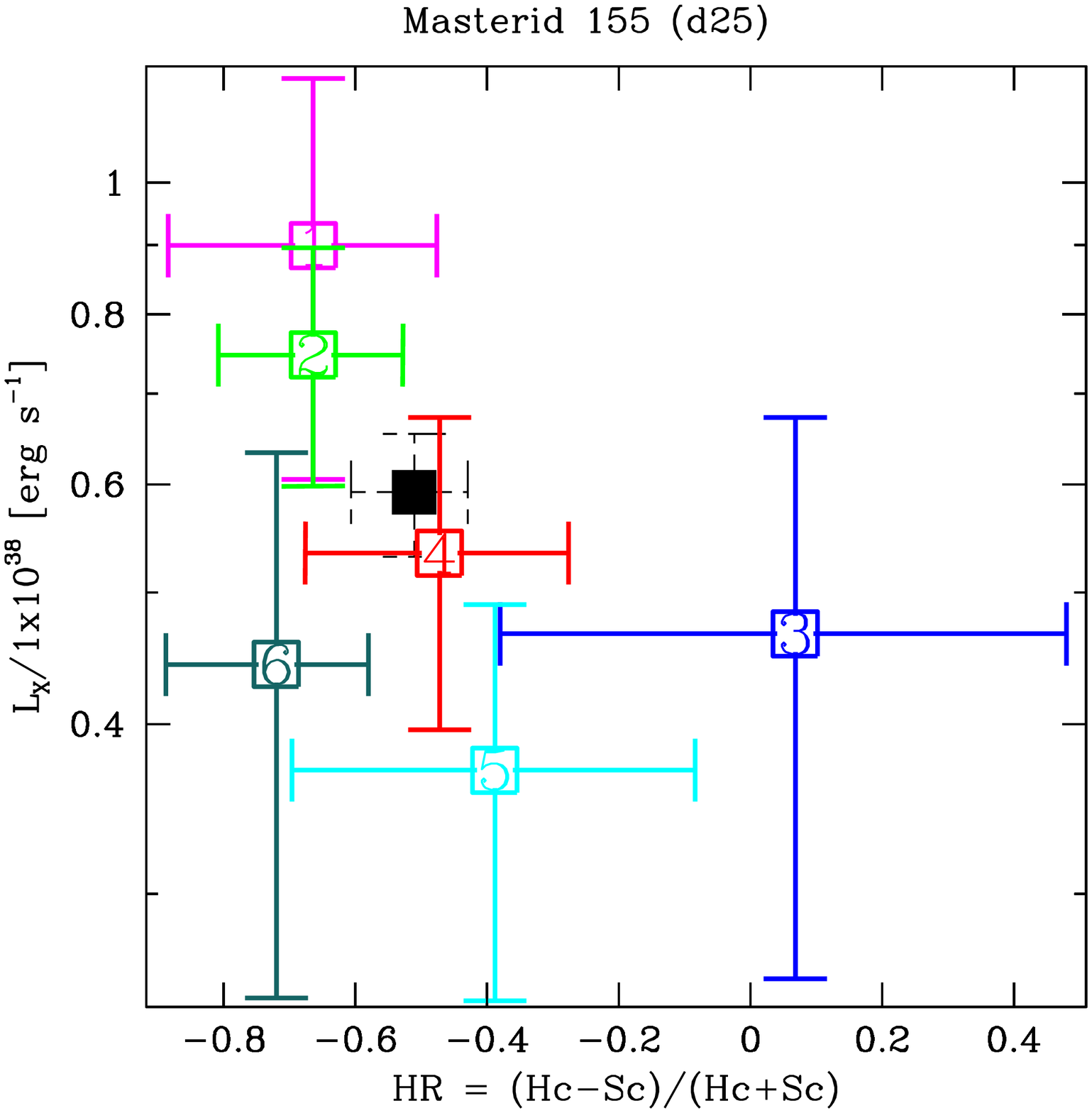}

\end{minipage}
\begin{minipage}{0.32\linewidth}
  \centering

    \includegraphics[width=\linewidth]{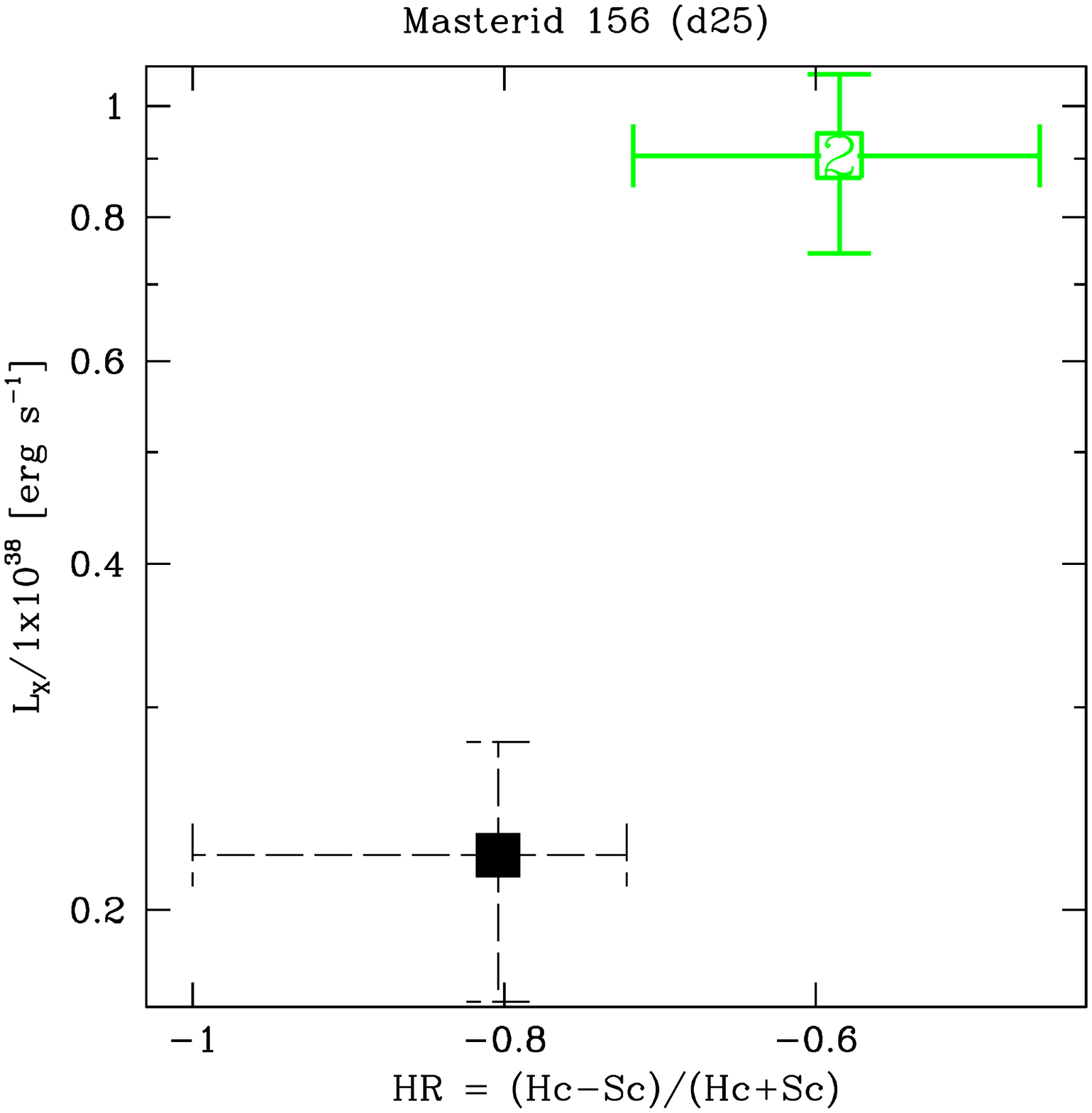}

 \end{minipage}

  \begin{minipage}{0.32\linewidth}
  \centering
  
    \includegraphics[width=\linewidth]{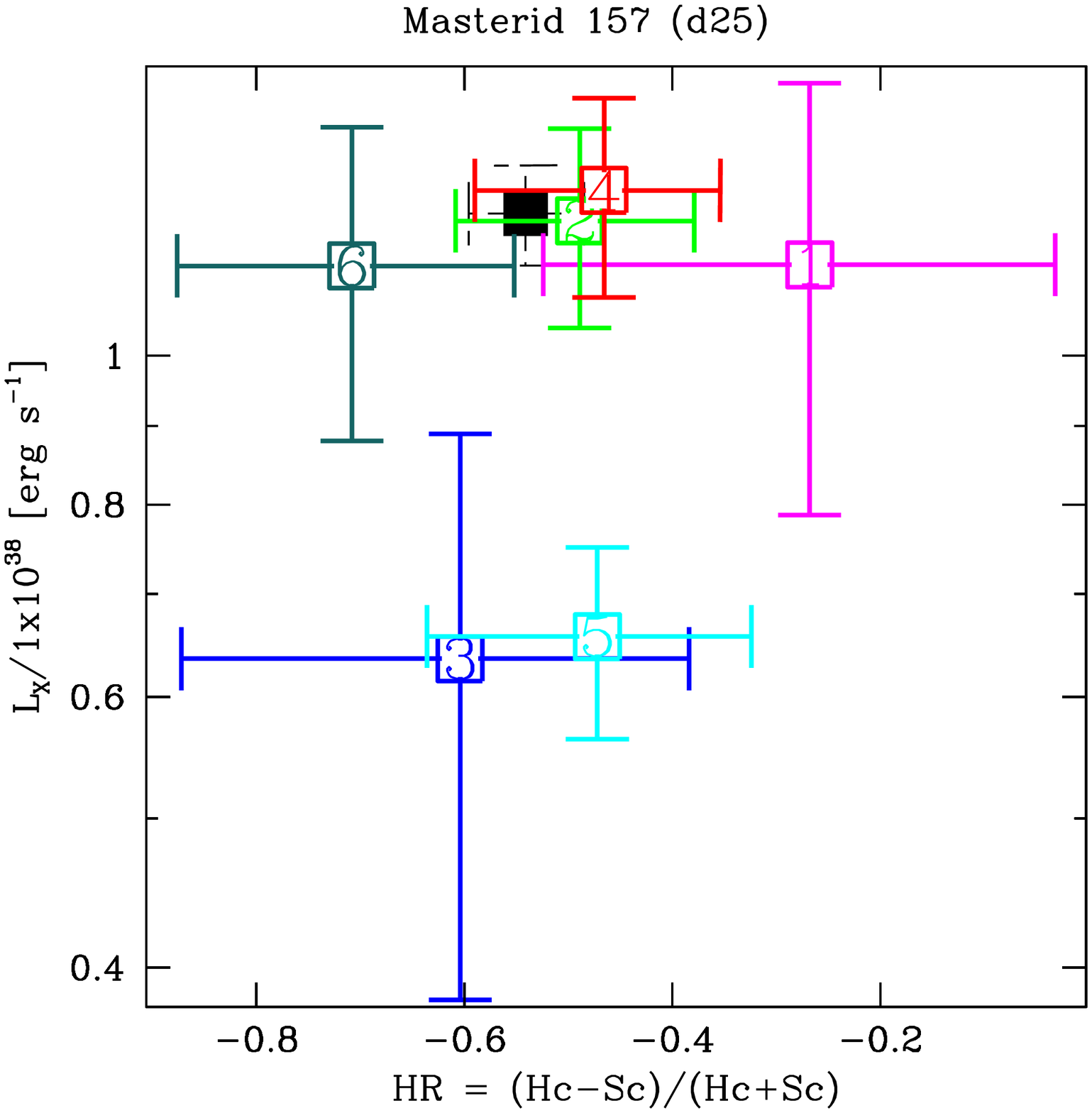}

  \end{minipage}
  \begin{minipage}{0.32\linewidth}
  \centering

    \includegraphics[width=\linewidth]{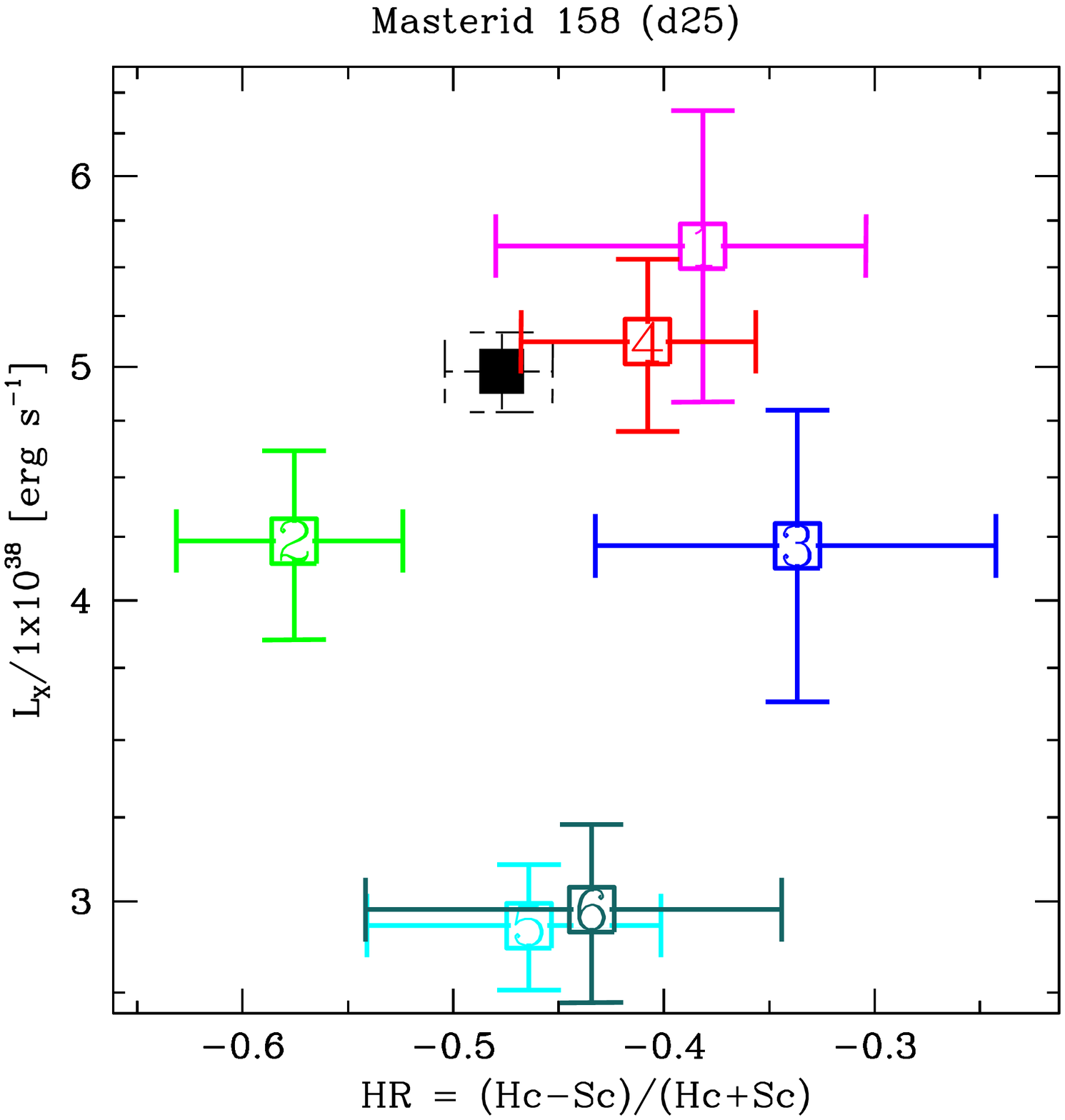}

\end{minipage}
\begin{minipage}{0.32\linewidth}
  \centering

    \includegraphics[width=\linewidth]{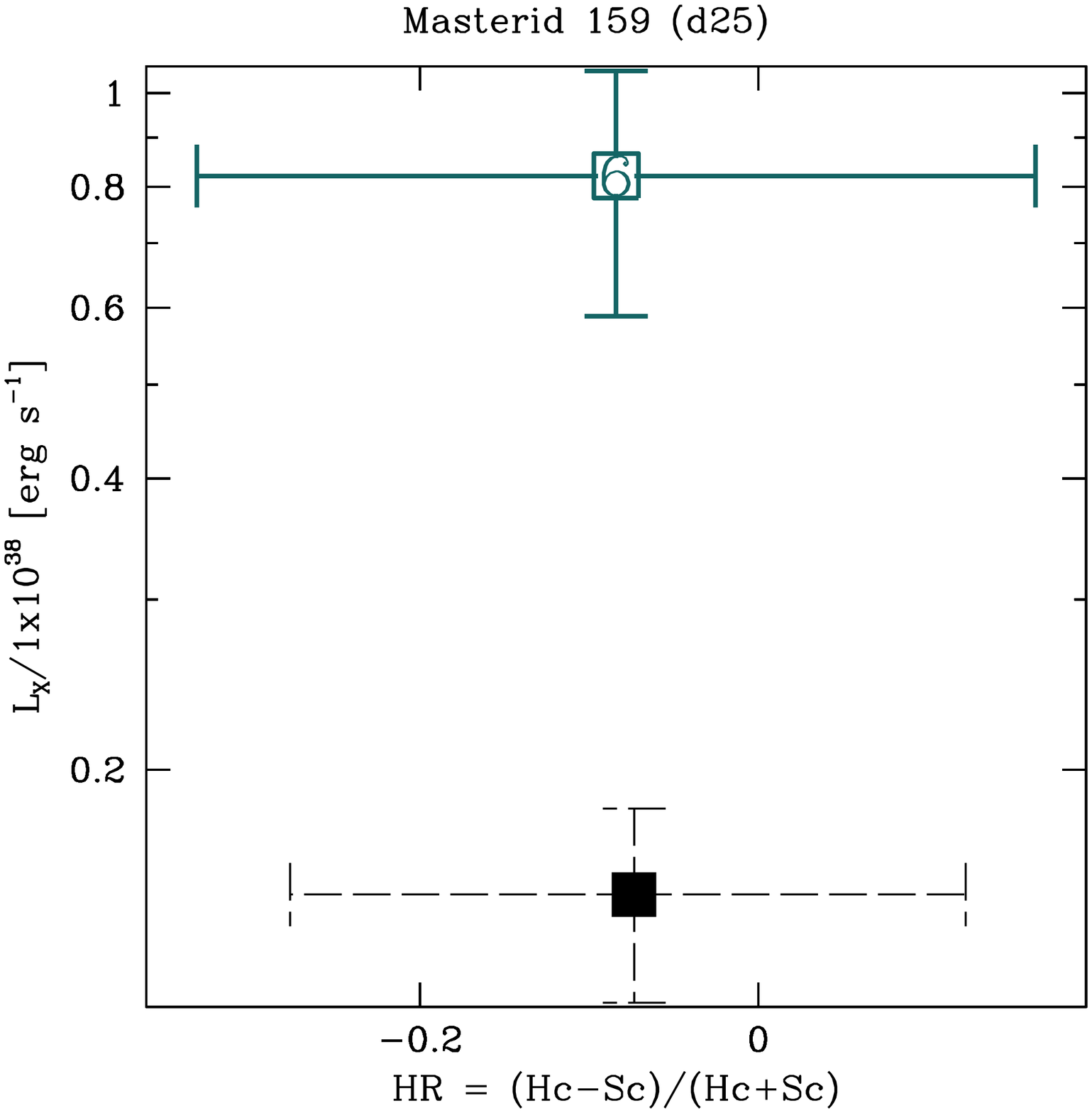}

\end{minipage}

\begin{minipage}{0.32\linewidth}
  \centering
  
    \includegraphics[width=\linewidth]{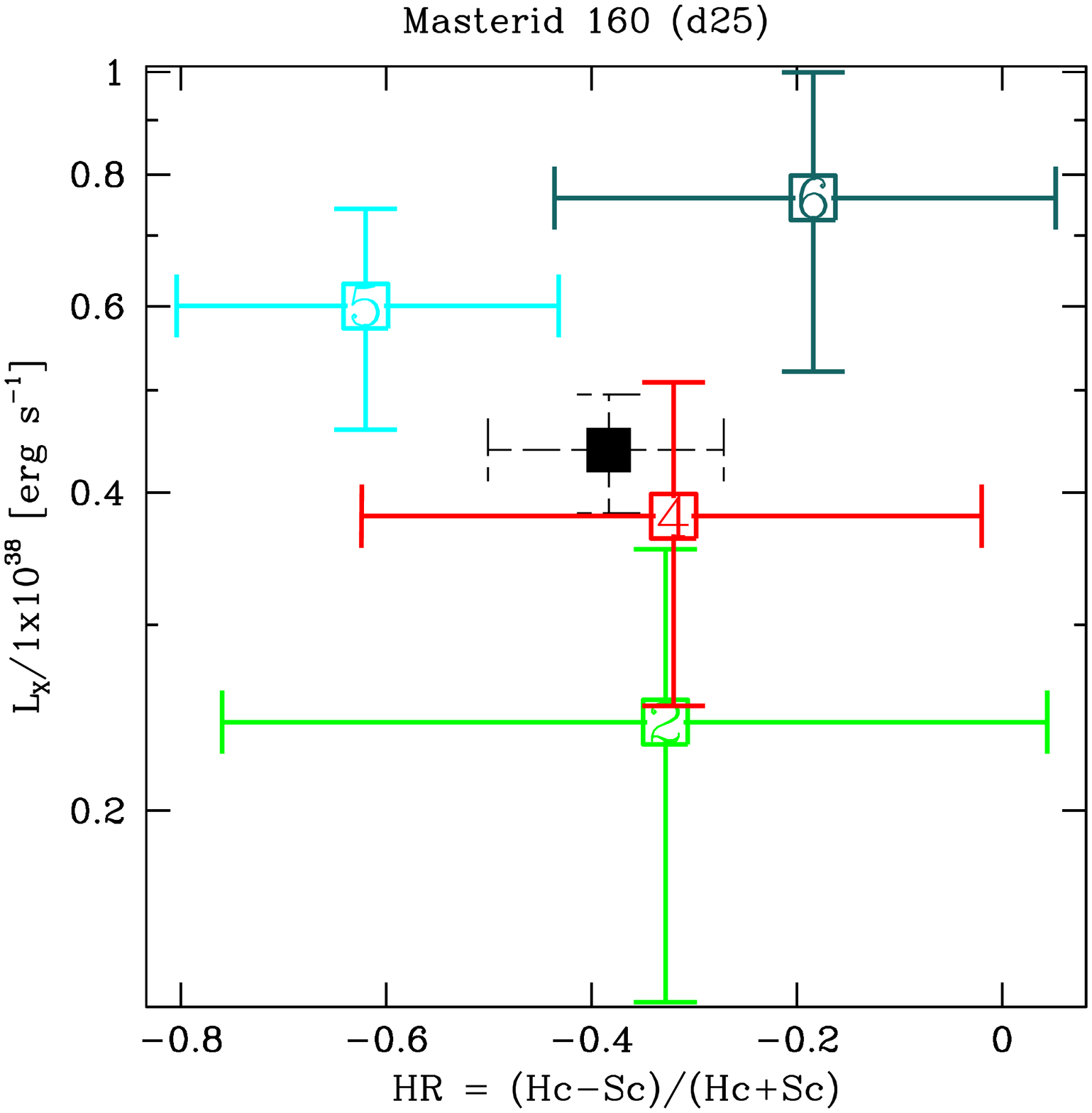}

  \end{minipage}
  \begin{minipage}{0.32\linewidth}
  \centering

    \includegraphics[width=\linewidth]{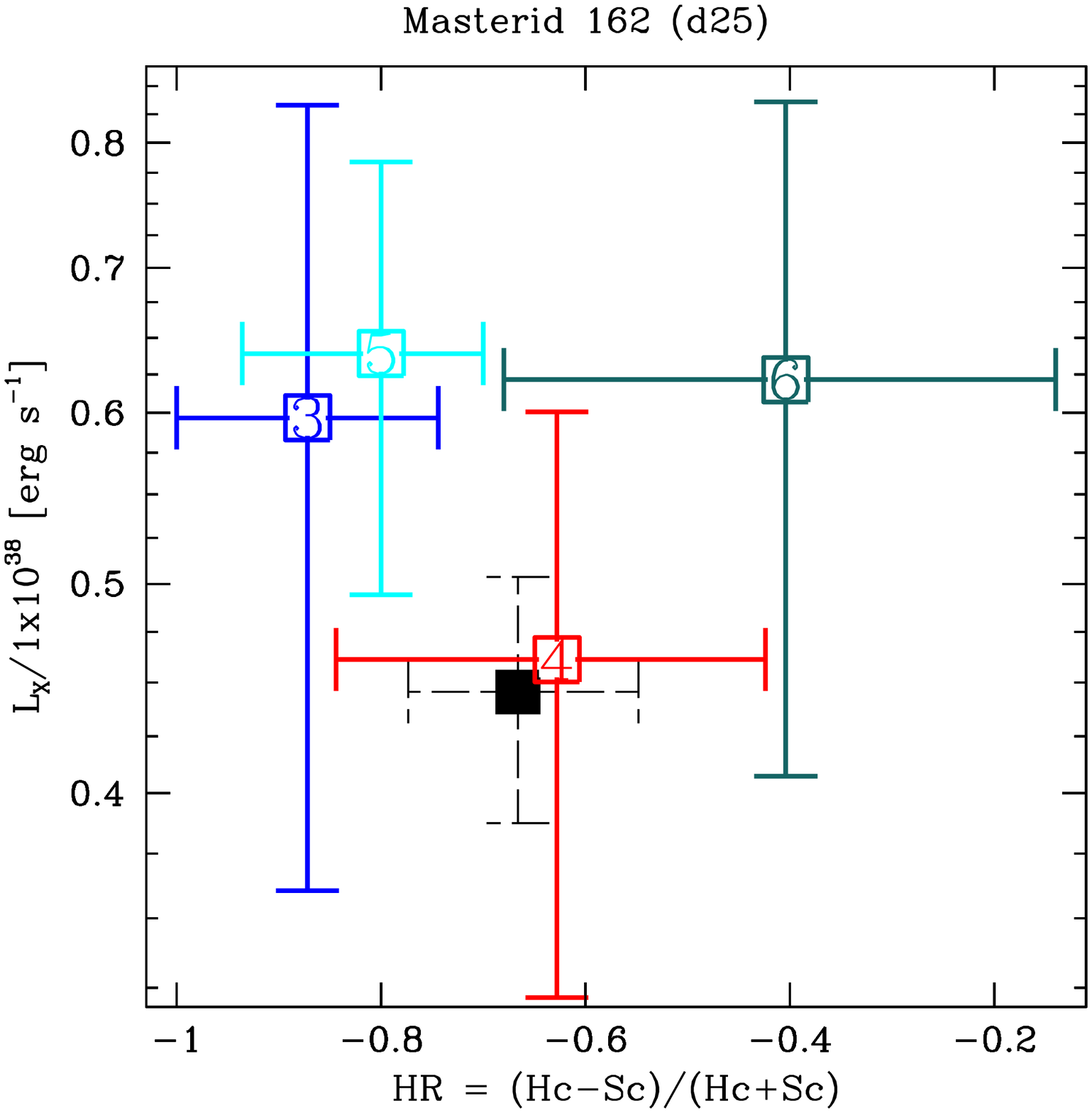}

\end{minipage}
\begin{minipage}{0.32\linewidth}
  \centering

    \includegraphics[width=\linewidth]{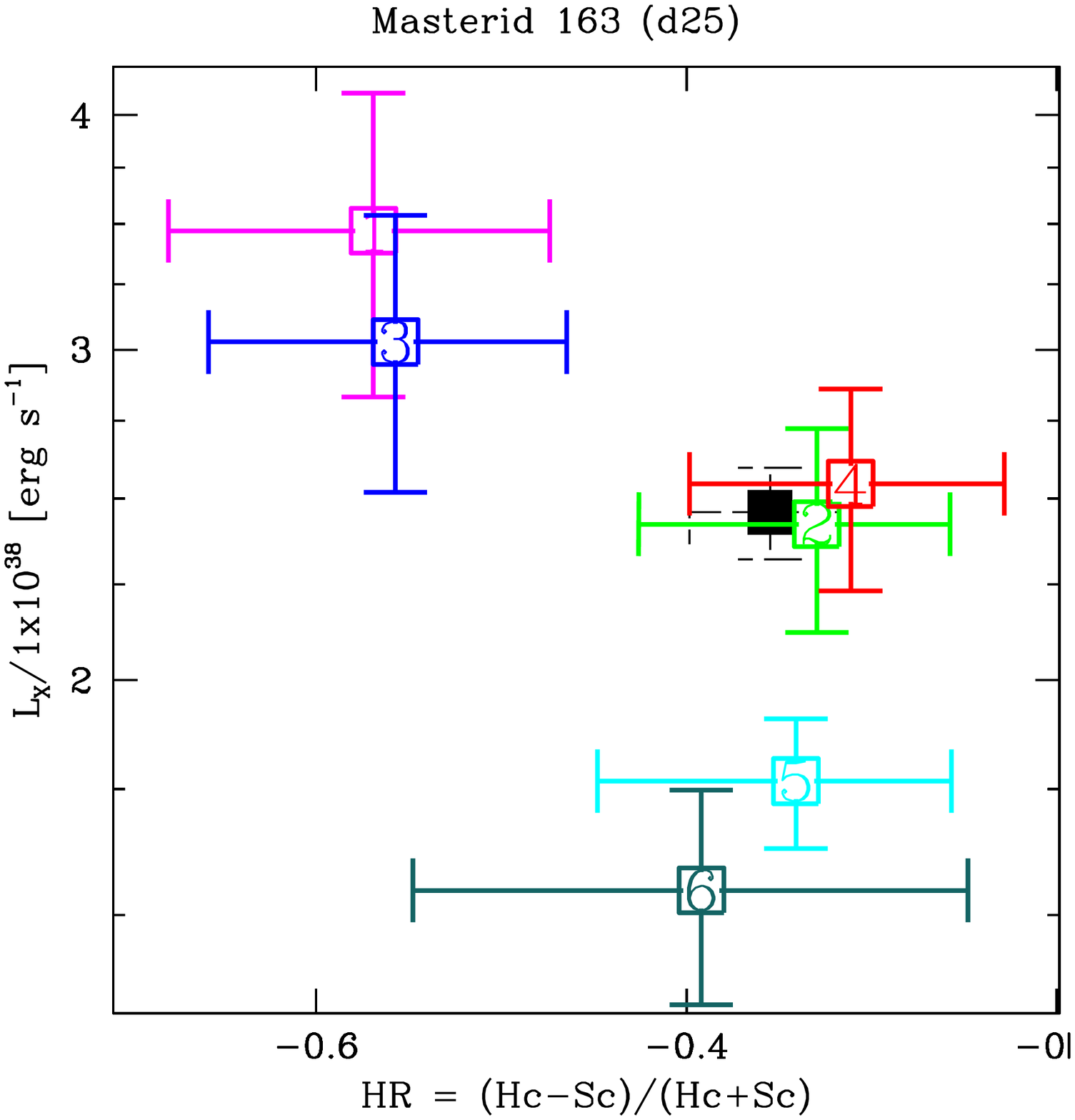}

\end{minipage}
\end{figure}

\begin{figure}
  \begin{minipage}{0.32\linewidth}
  \centering
  
    \includegraphics[width=\linewidth]{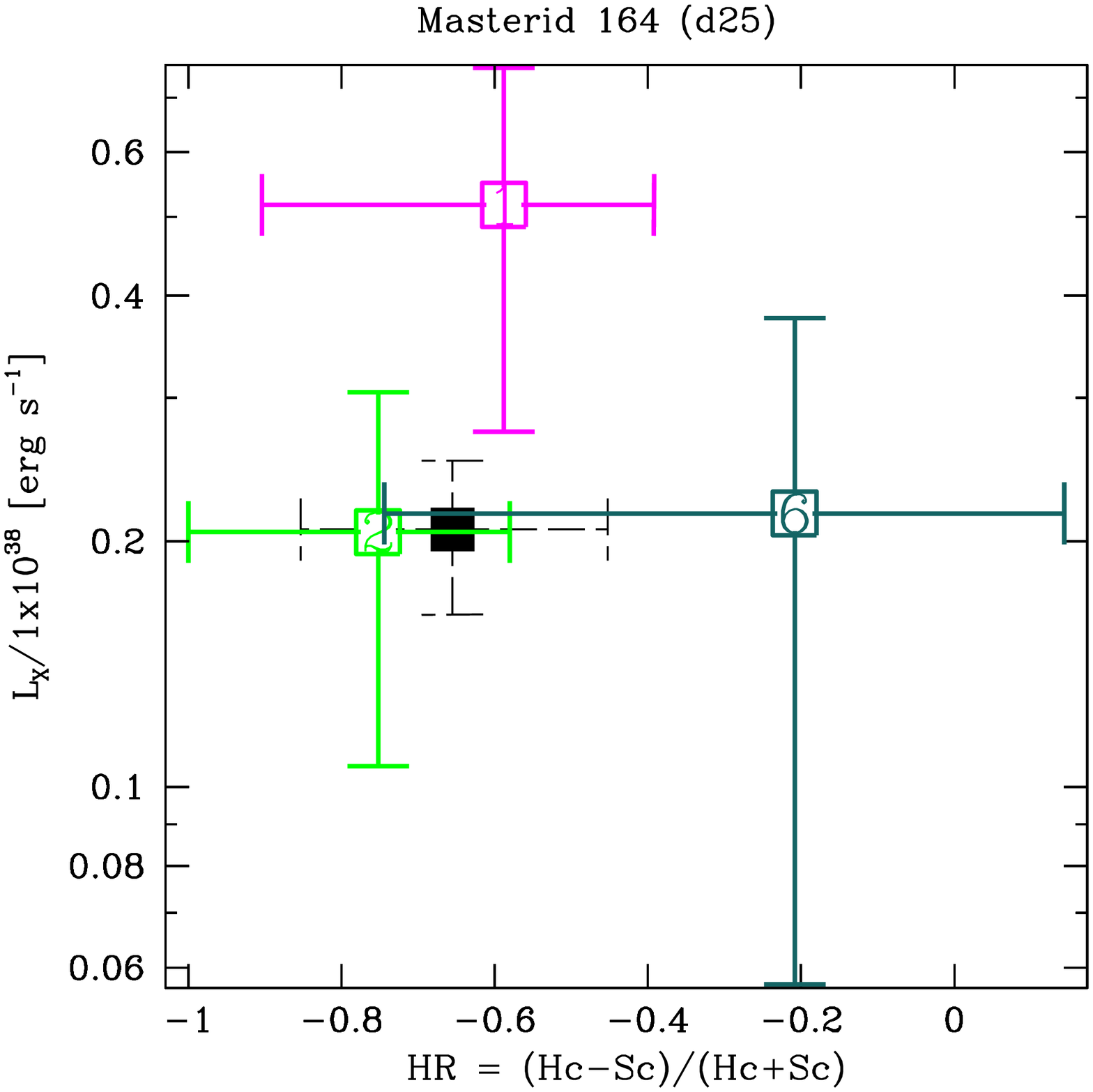}

  \end{minipage}
  \begin{minipage}{0.32\linewidth}
  \centering

    \includegraphics[width=\linewidth]{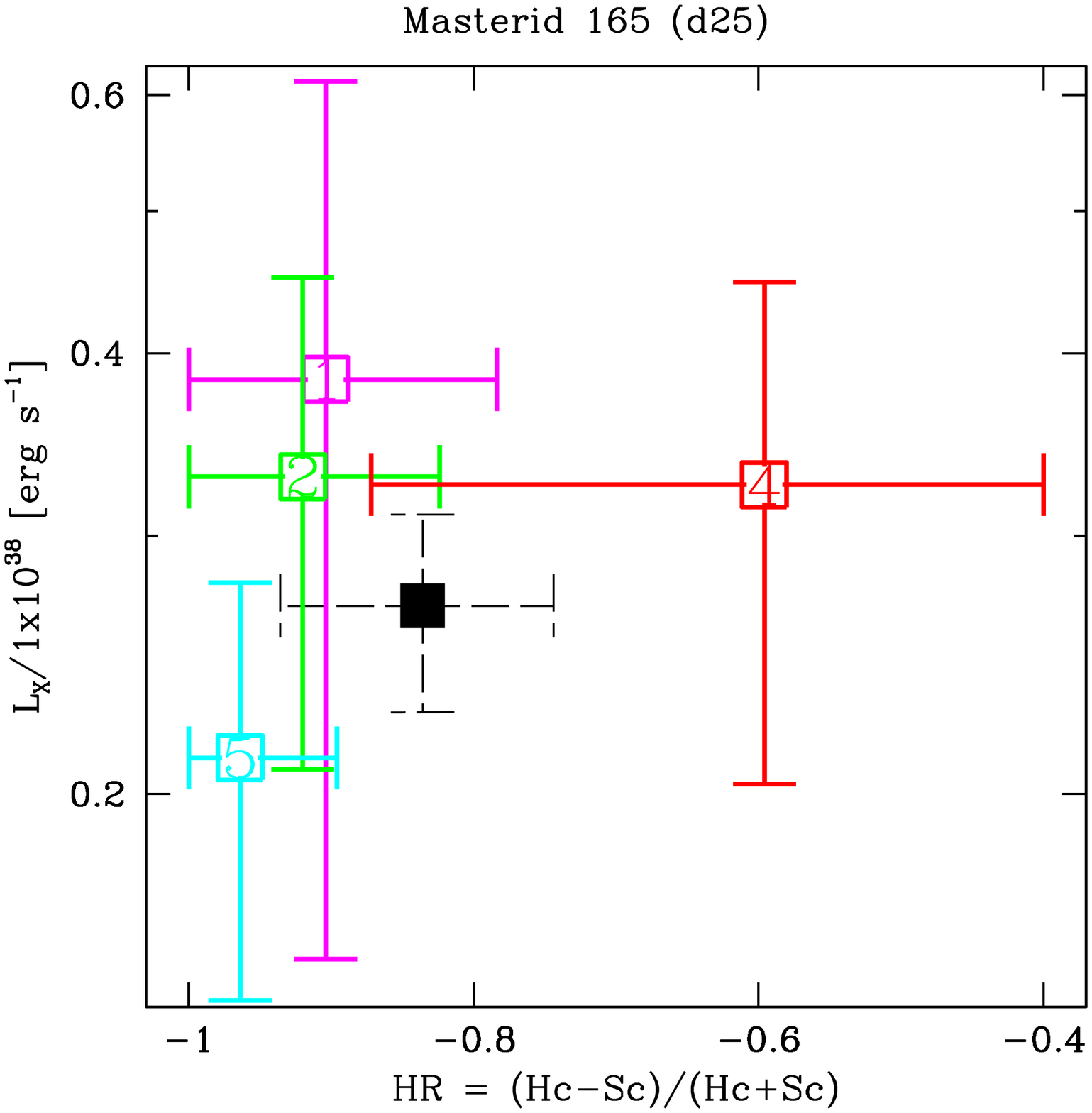}

\end{minipage}
\begin{minipage}{0.32\linewidth}
  \centering

    \includegraphics[width=\linewidth]{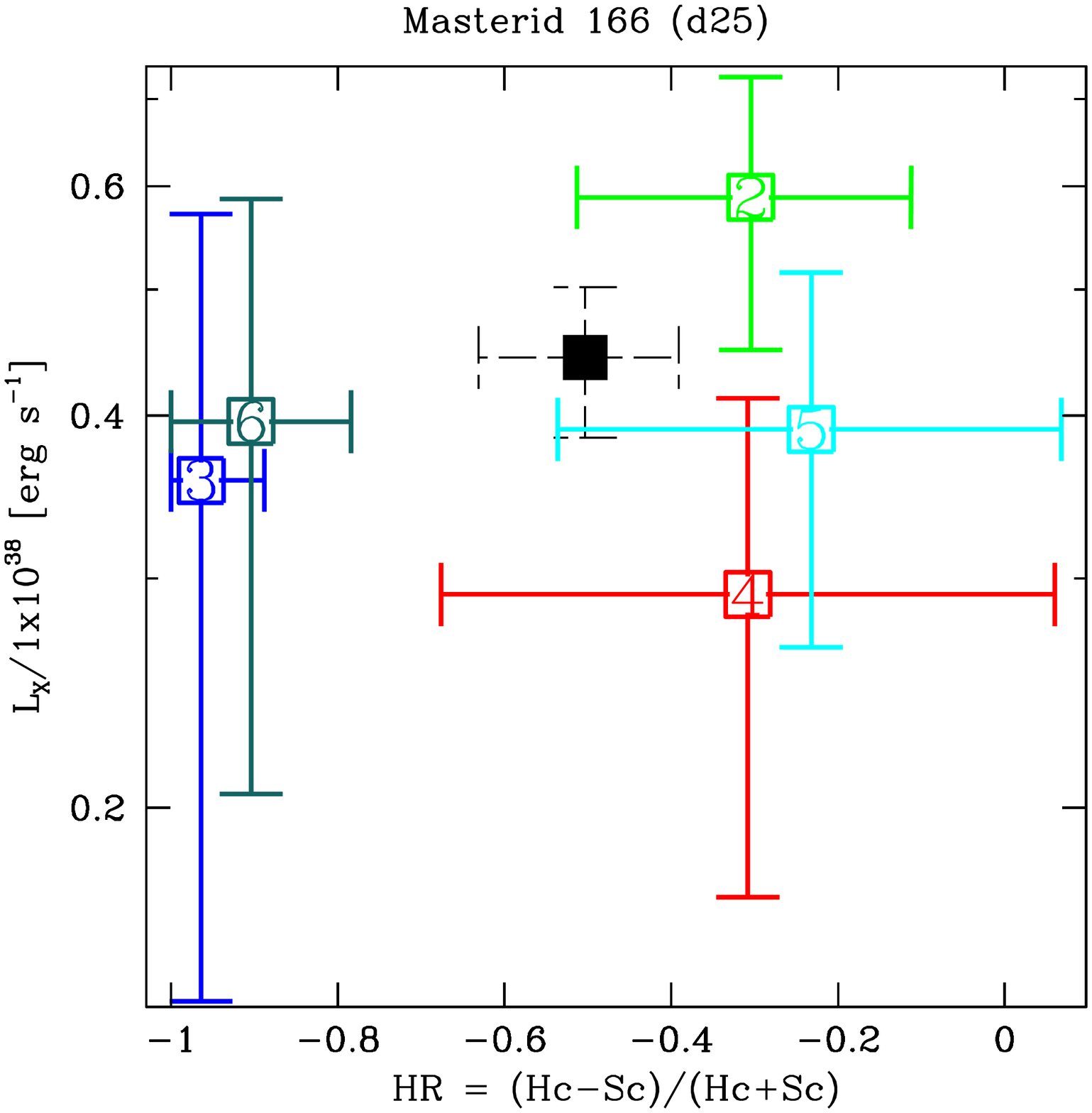}

 \end{minipage}

\begin{minipage}{0.32\linewidth}
  \centering
  
    \includegraphics[width=\linewidth]{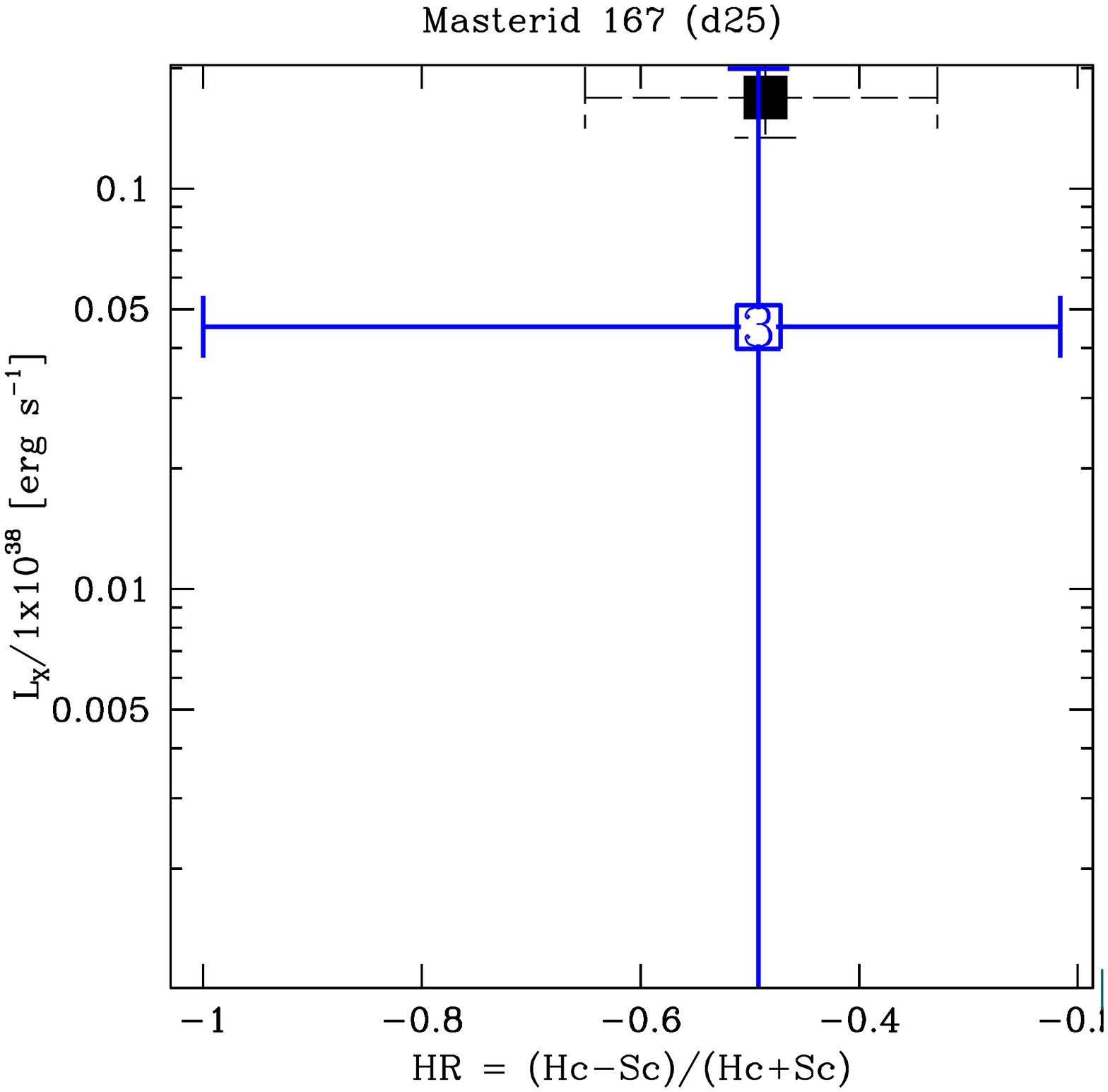}

  \end{minipage}
  \begin{minipage}{0.32\linewidth}
  \centering

    \includegraphics[width=\linewidth]{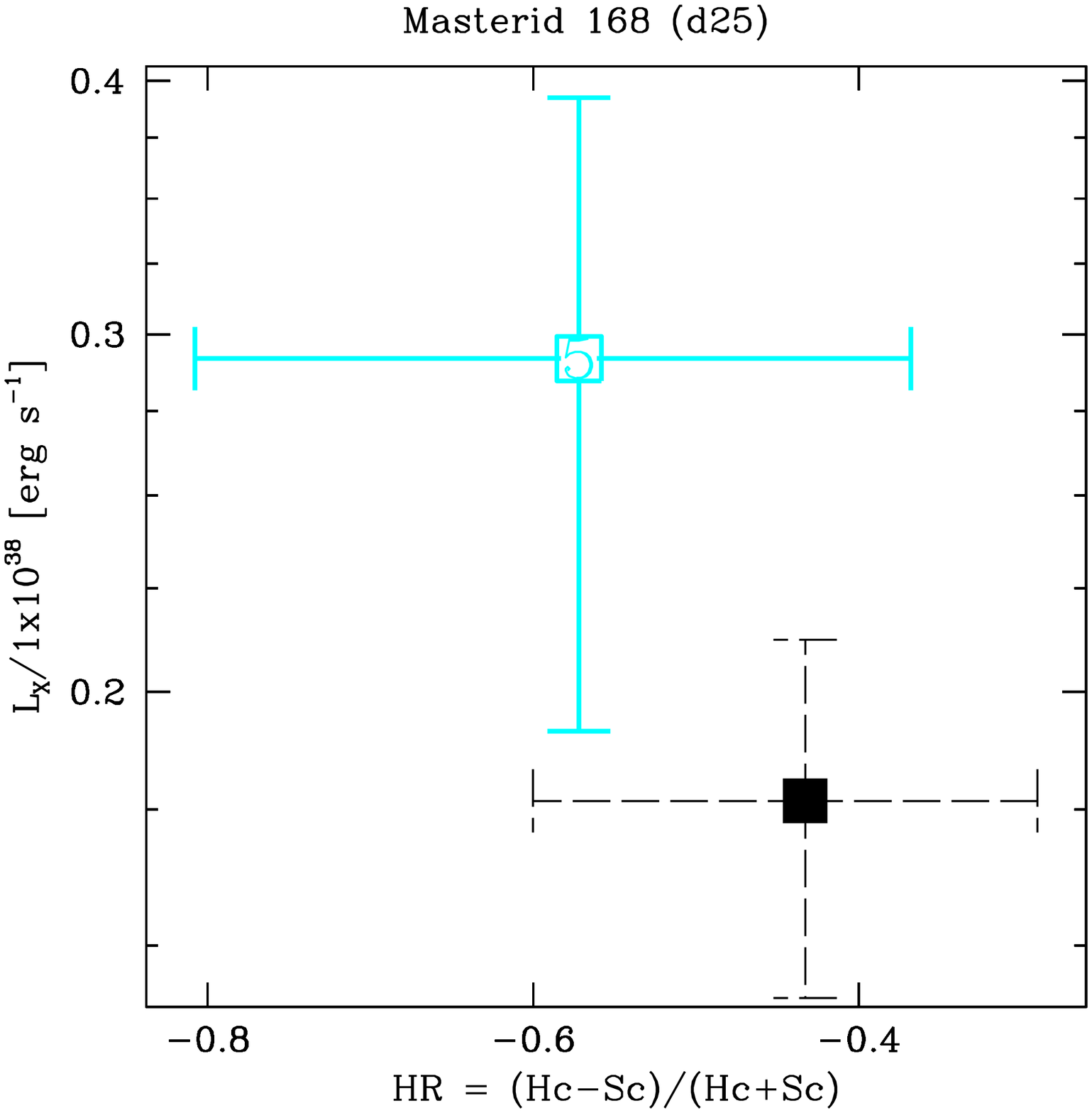}

\end{minipage}
\begin{minipage}{0.32\linewidth}
  \centering

    \includegraphics[width=\linewidth]{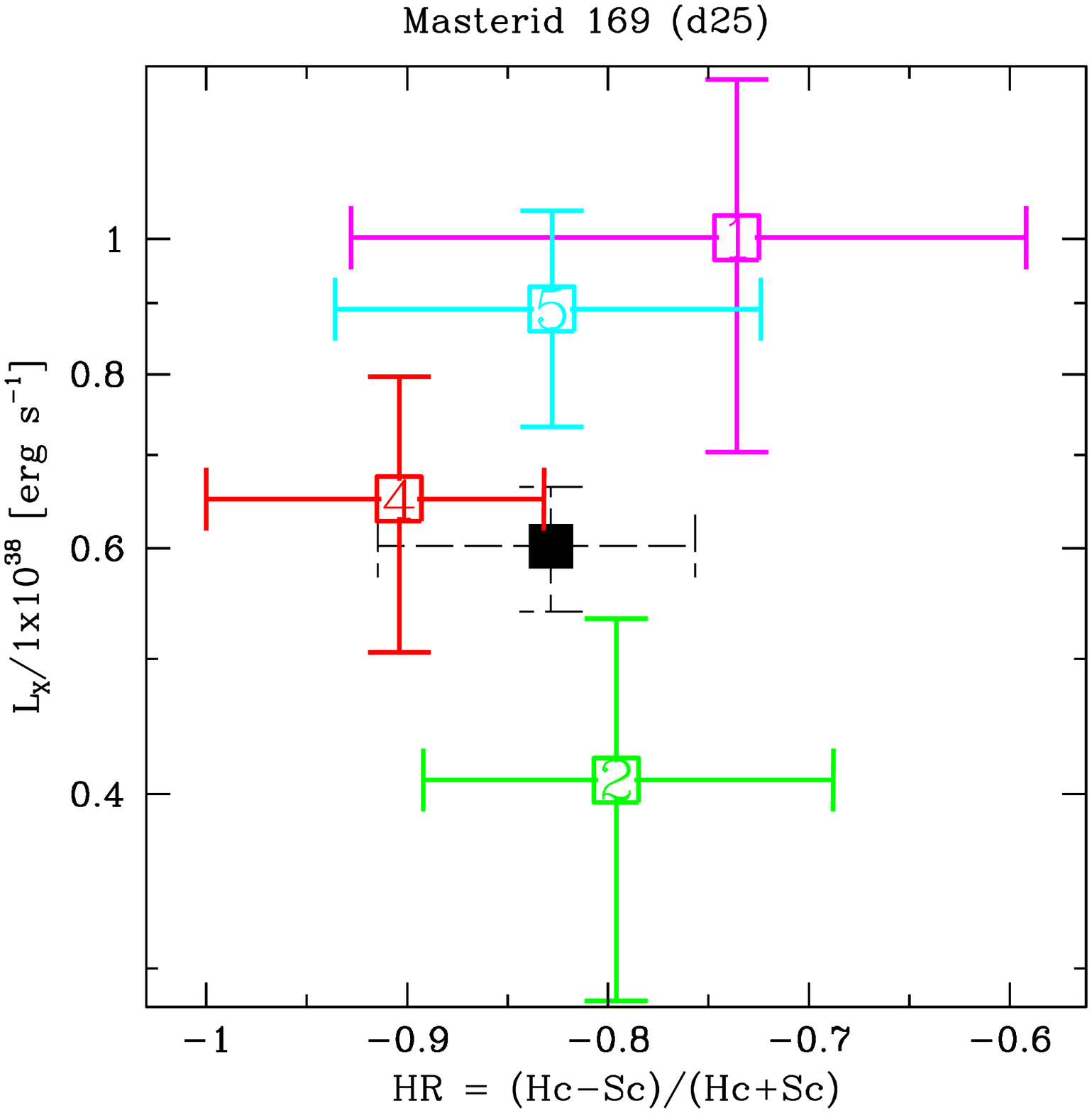}

 \end{minipage}

  \begin{minipage}{0.32\linewidth}
  \centering
  
    \includegraphics[width=\linewidth]{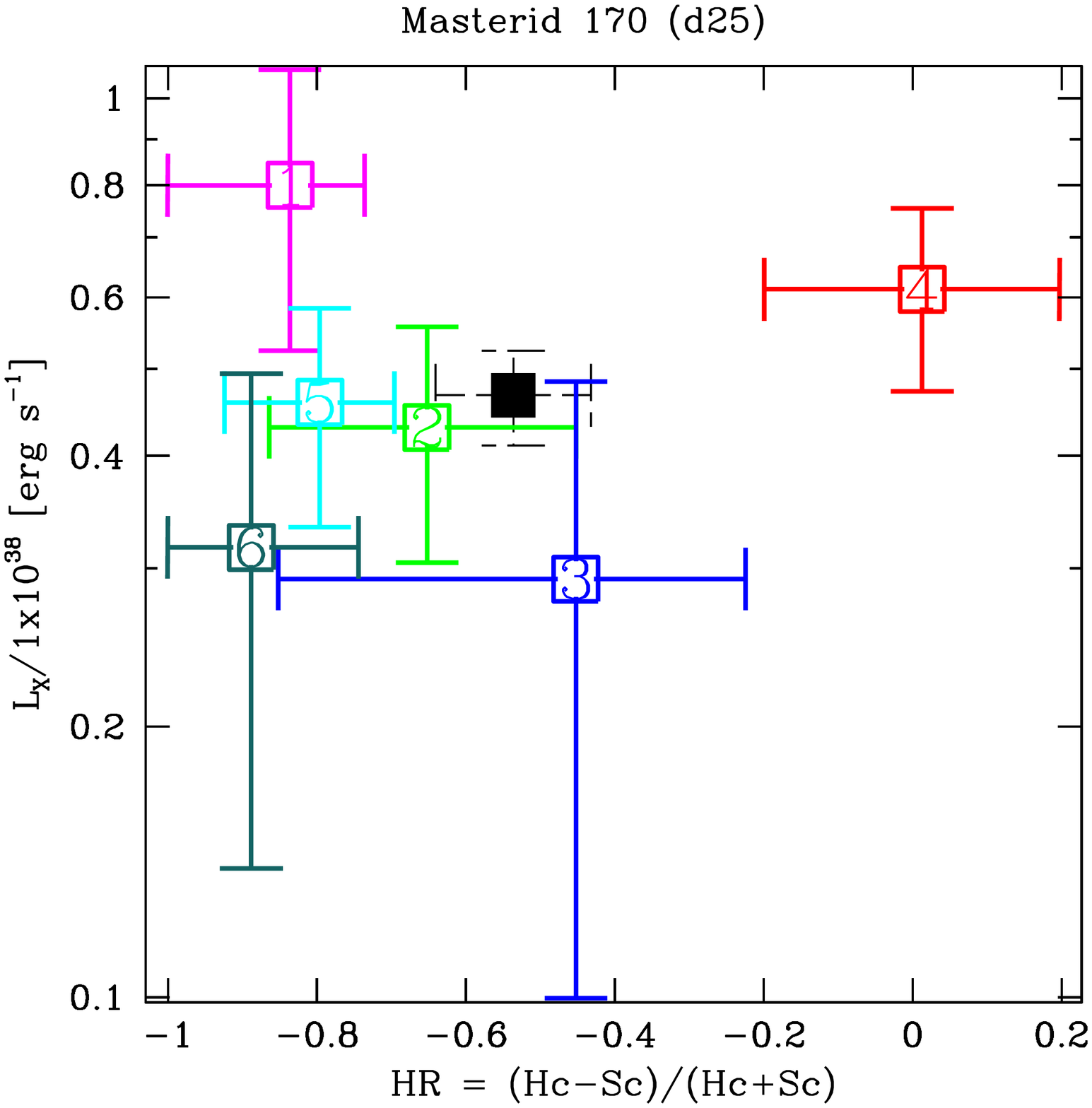}

  \end{minipage}
  \begin{minipage}{0.32\linewidth}
  \centering

    \includegraphics[width=\linewidth]{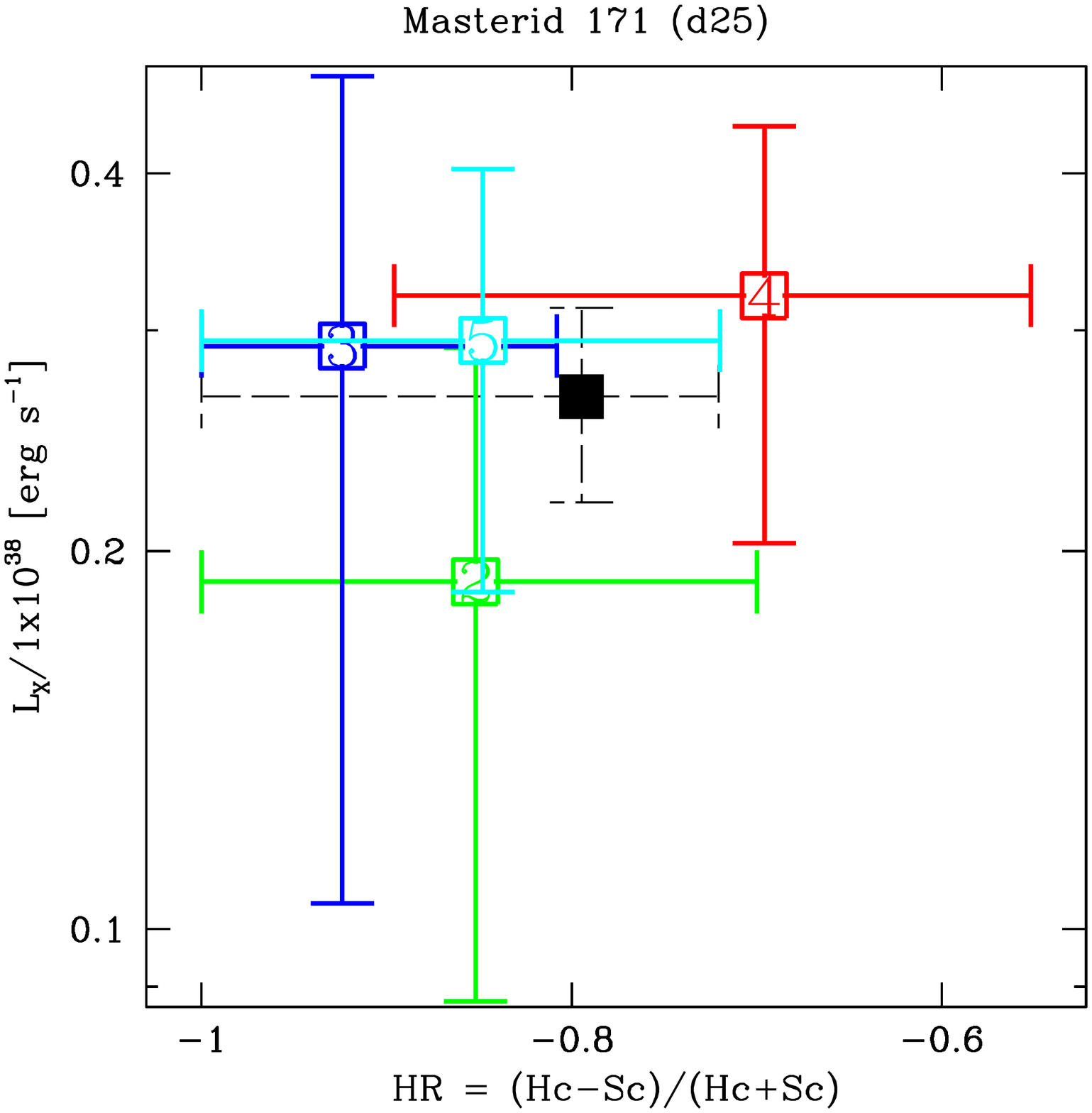}

\end{minipage}
\begin{minipage}{0.32\linewidth}
  \centering

    \includegraphics[width=\linewidth]{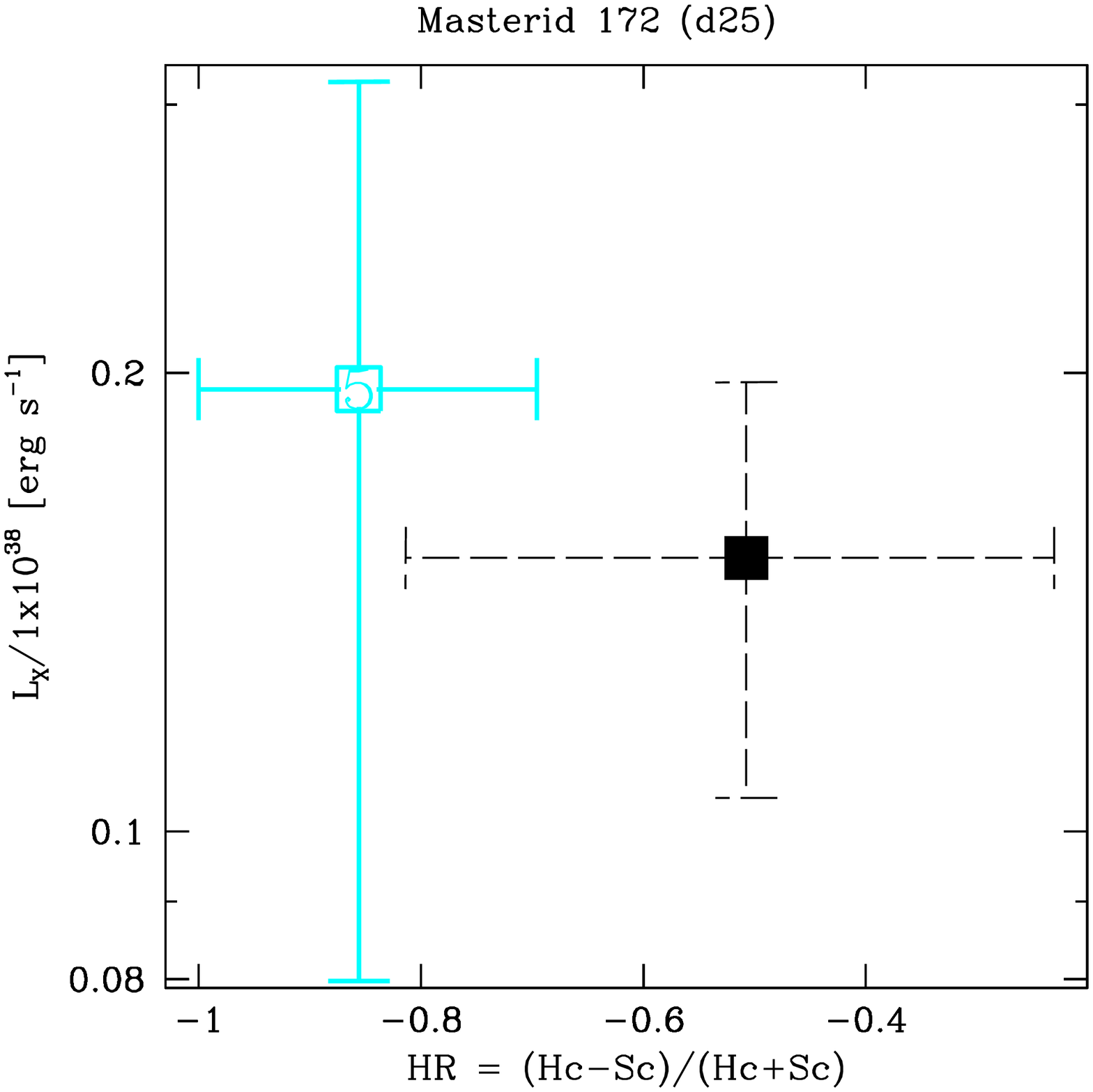}

\end{minipage}

\begin{minipage}{0.32\linewidth}
  \centering
  
    \includegraphics[width=\linewidth]{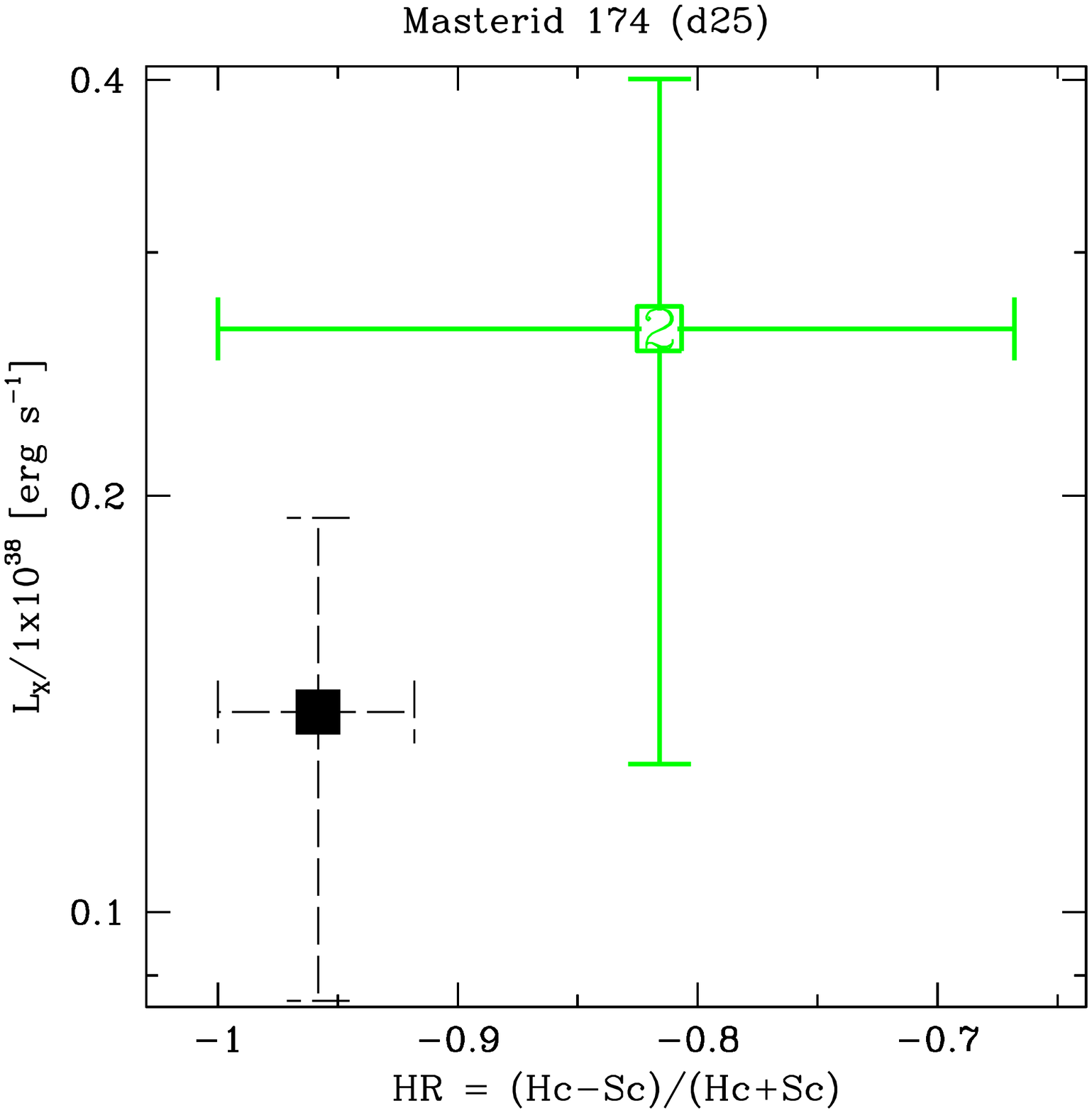}

  \end{minipage}
  \begin{minipage}{0.32\linewidth}
  \centering

    \includegraphics[width=\linewidth]{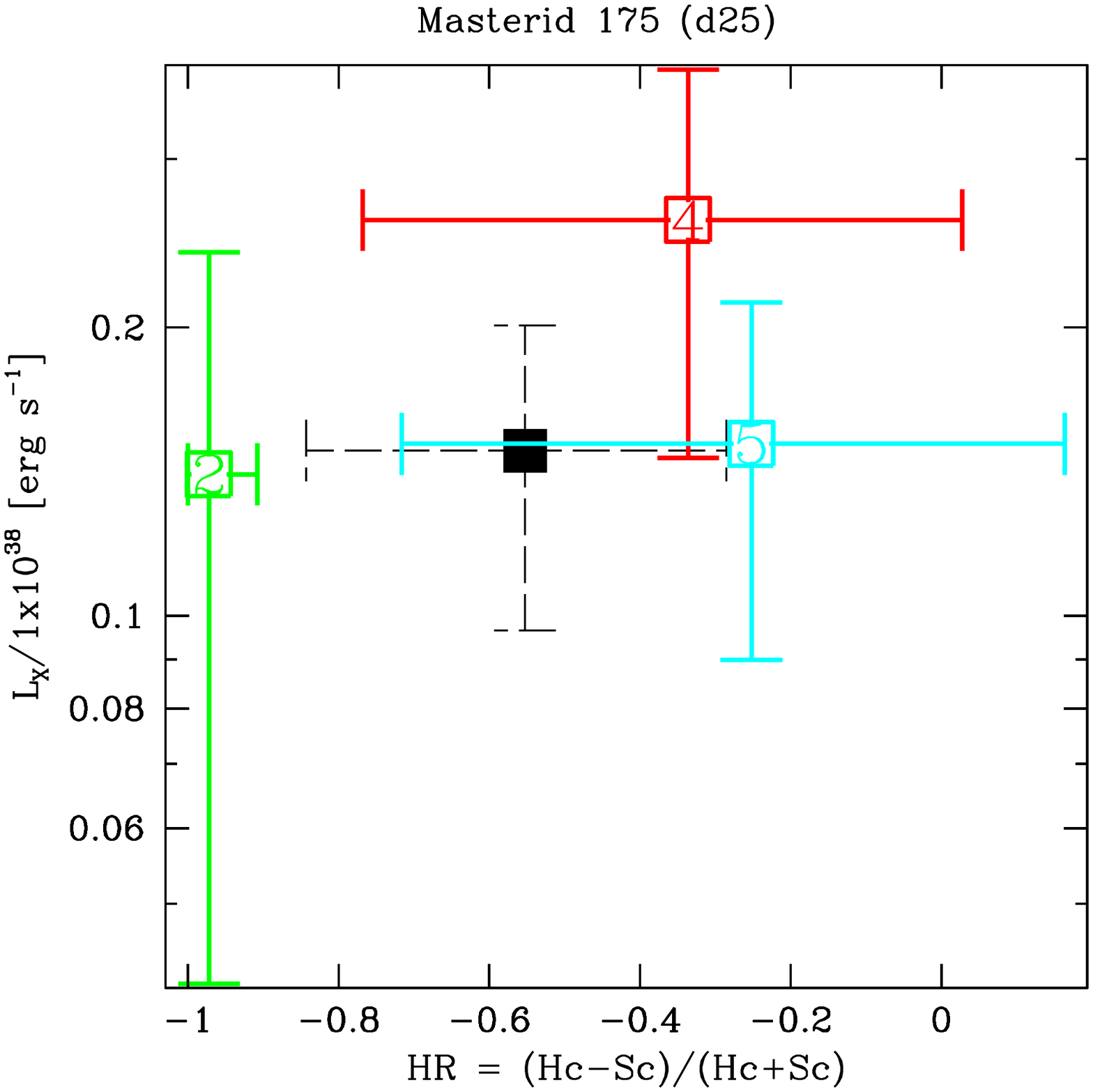}

\end{minipage}
\begin{minipage}{0.32\linewidth}
  \centering

    \includegraphics[width=\linewidth]{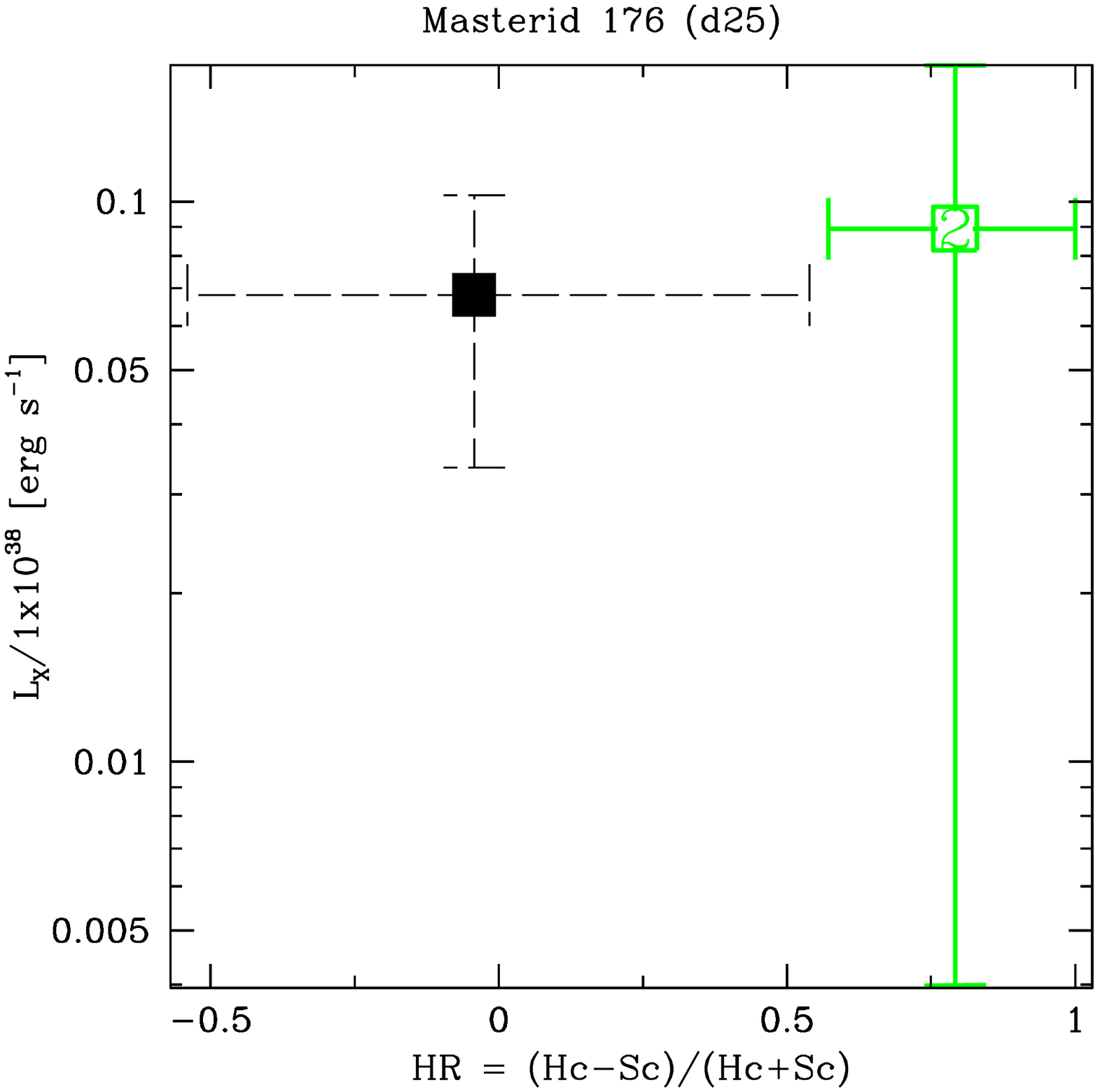}

\end{minipage}
\end{figure}

\begin{figure}
  \begin{minipage}{0.32\linewidth}
  \centering
  
    \includegraphics[width=\linewidth]{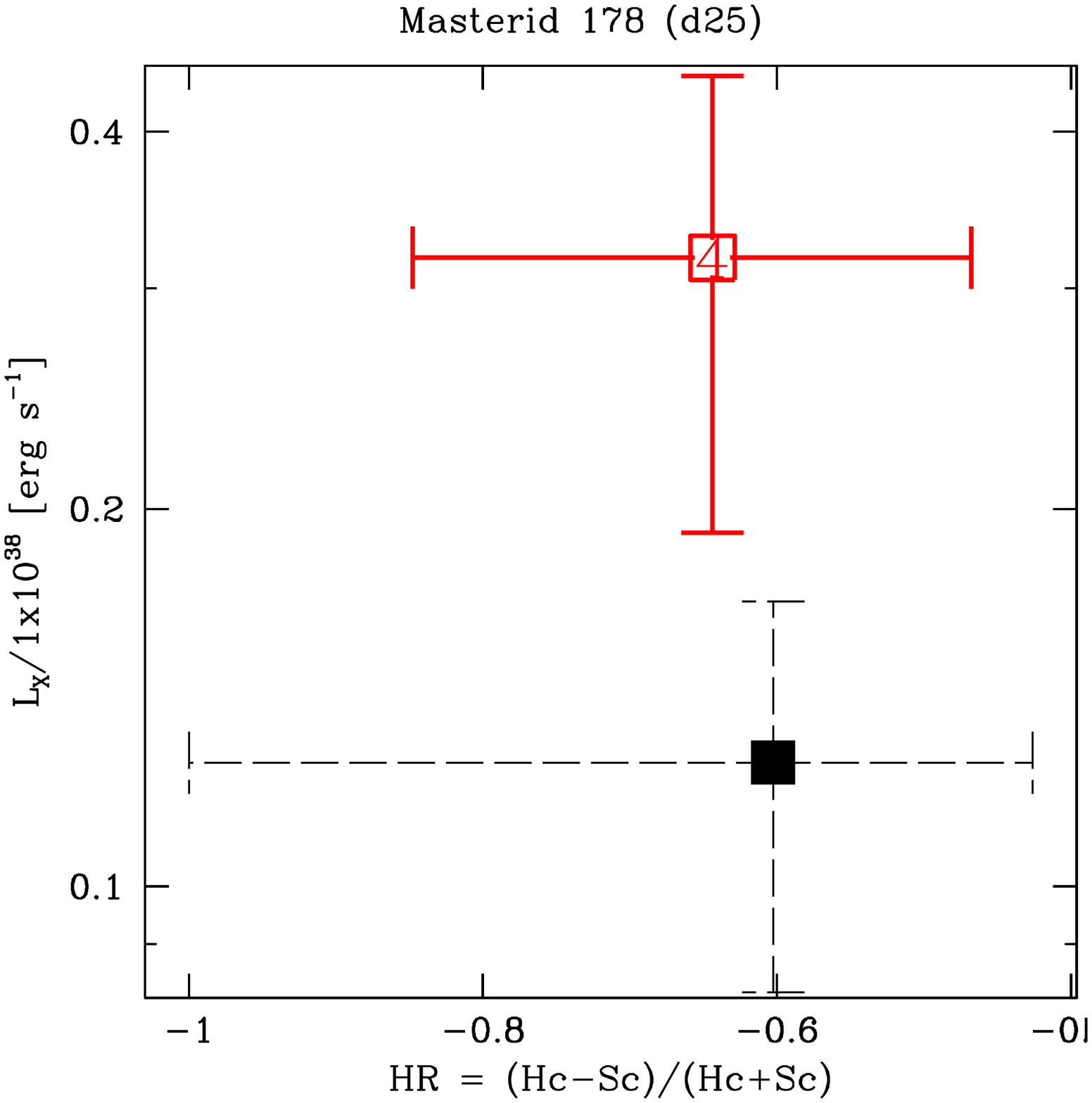}

  \end{minipage}
  \begin{minipage}{0.32\linewidth}
  \centering

    \includegraphics[width=\linewidth]{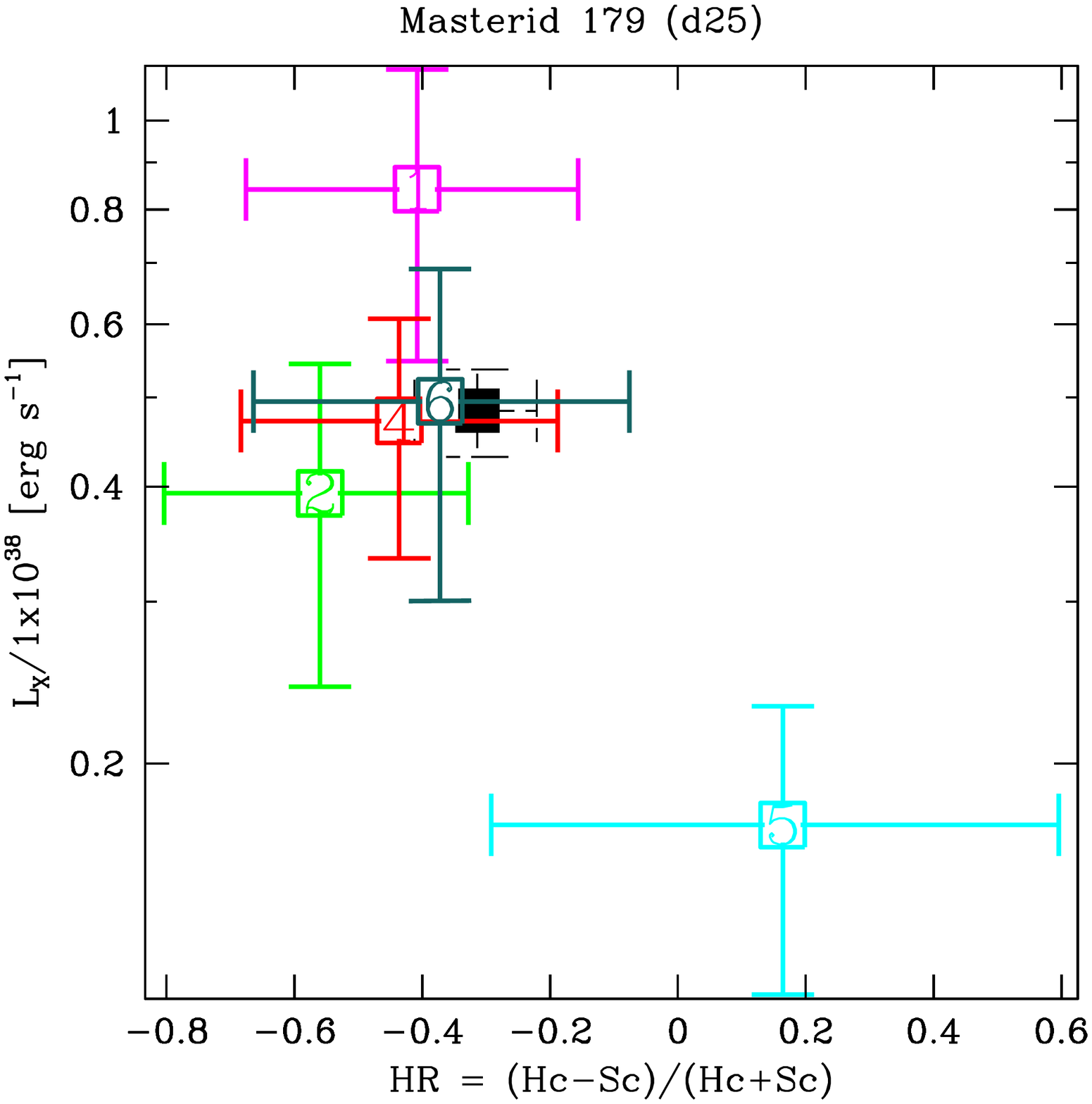}

\end{minipage}
\begin{minipage}{0.32\linewidth}
  \centering

    \includegraphics[width=\linewidth]{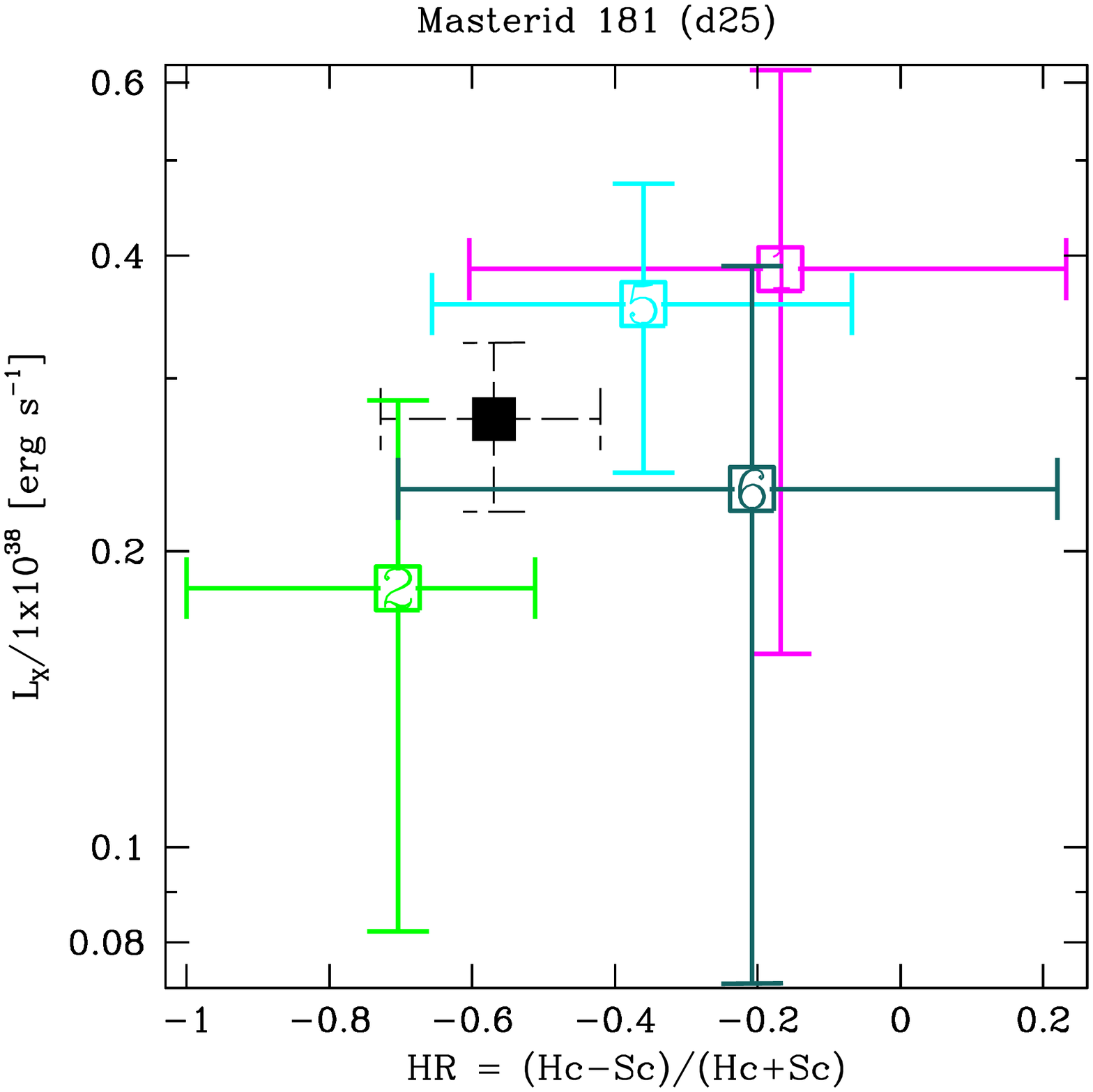}

 \end{minipage}

\begin{minipage}{0.32\linewidth}
  \centering
  
    \includegraphics[width=\linewidth]{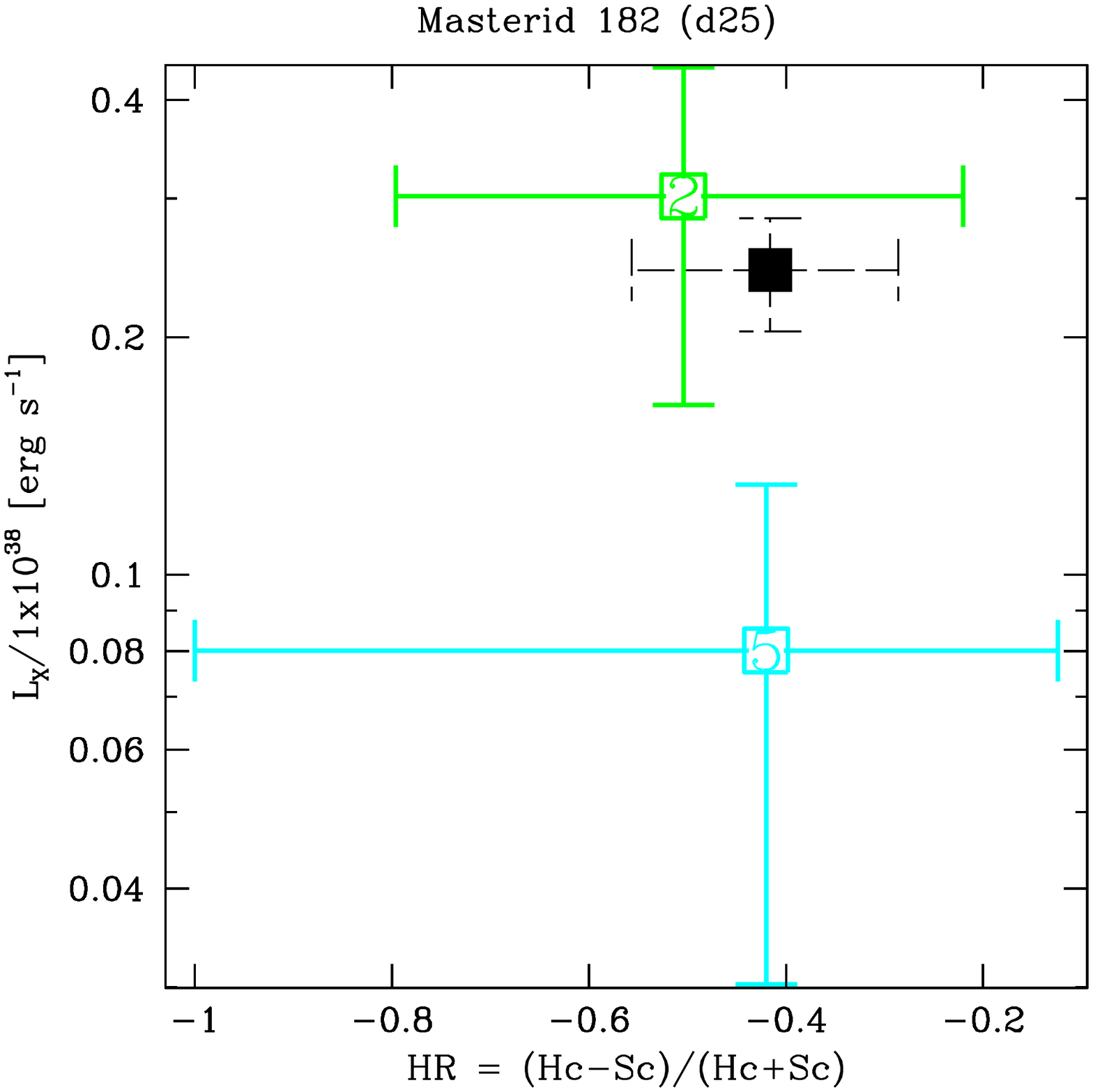}

  \end{minipage}
  \begin{minipage}{0.32\linewidth}
  \centering

    \includegraphics[width=\linewidth]{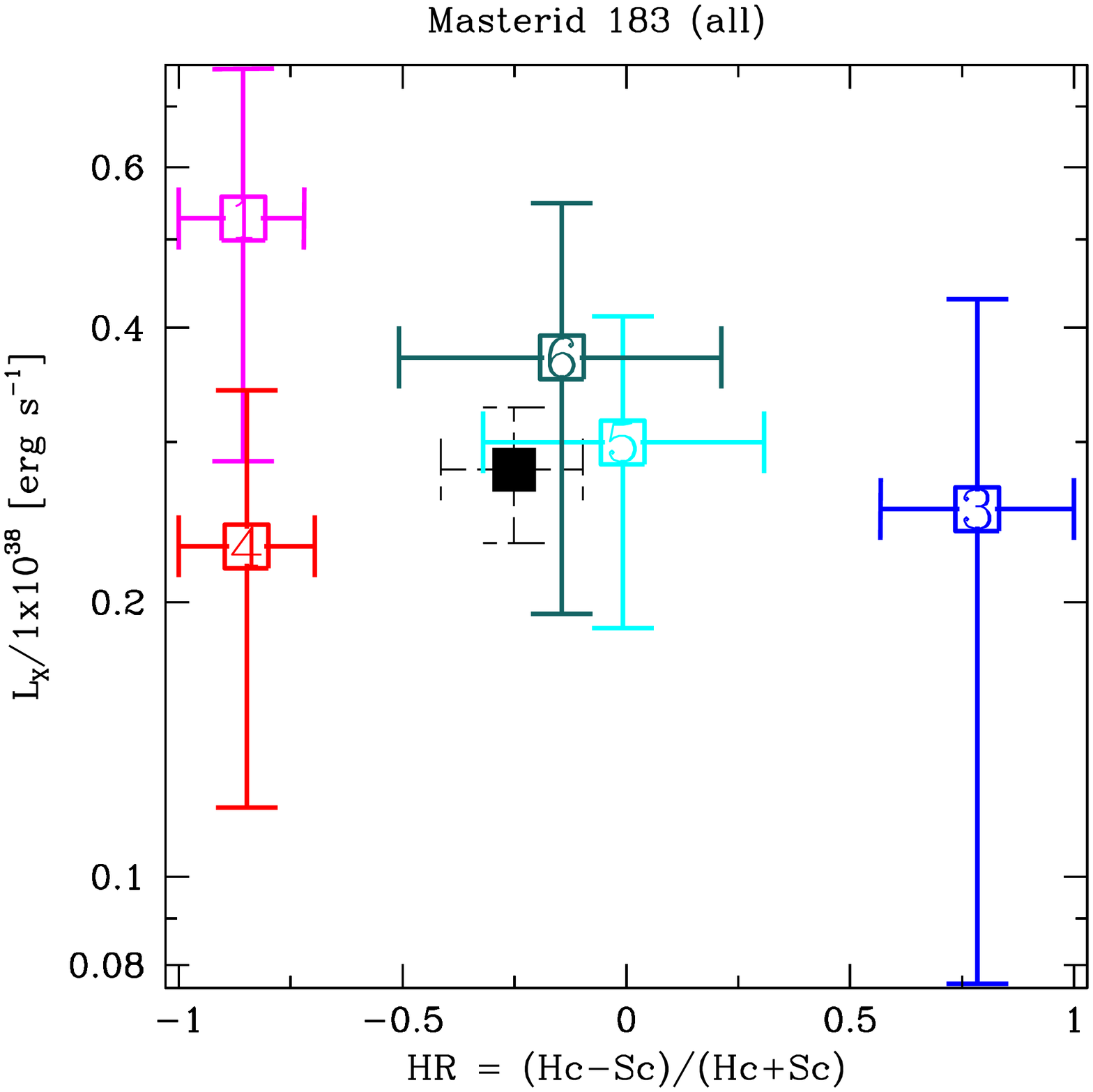}

\end{minipage}
\begin{minipage}{0.32\linewidth}
  \centering

    \includegraphics[width=\linewidth]{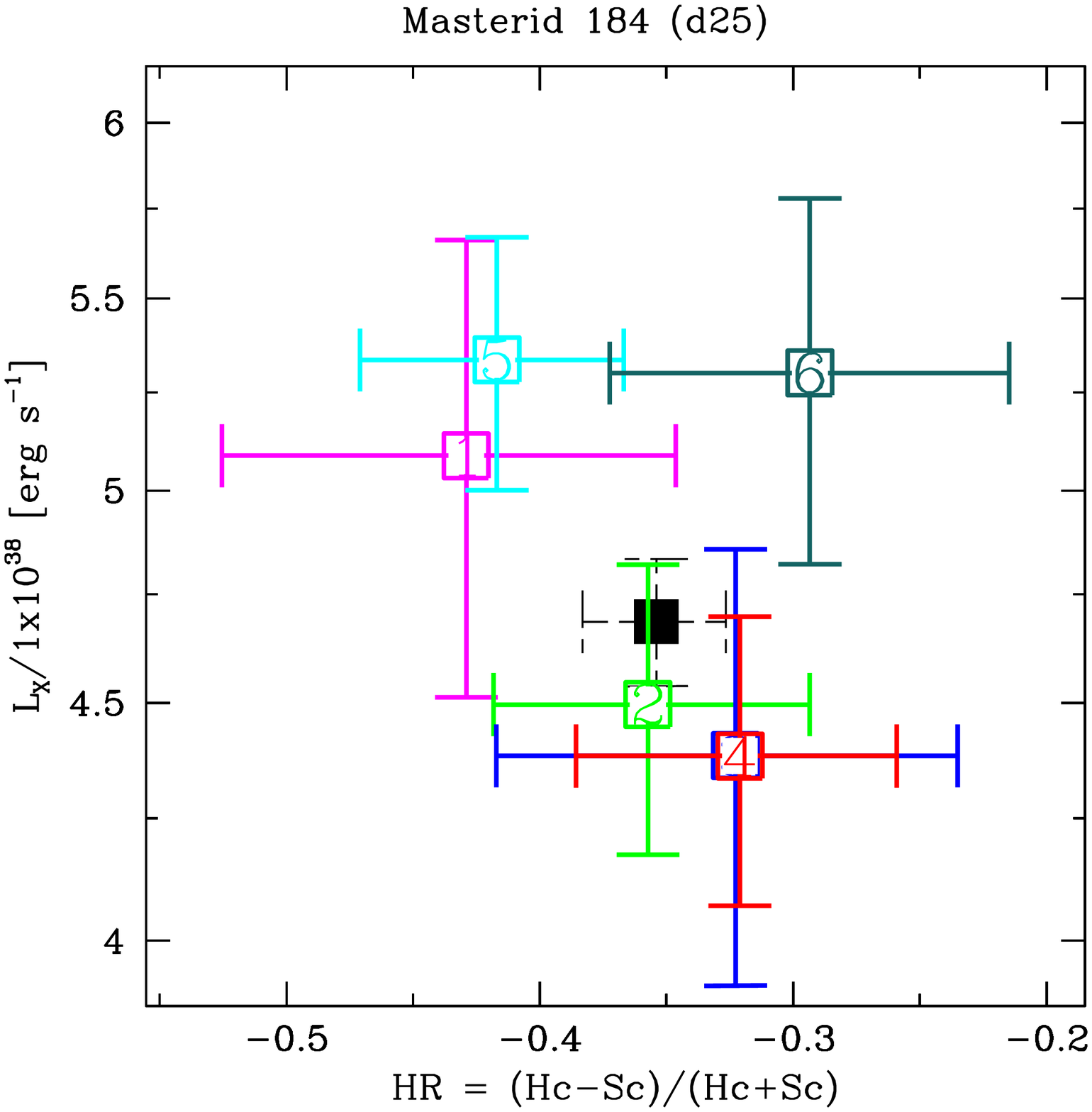}

 \end{minipage}

  \begin{minipage}{0.32\linewidth}
  \centering
  
    \includegraphics[width=\linewidth]{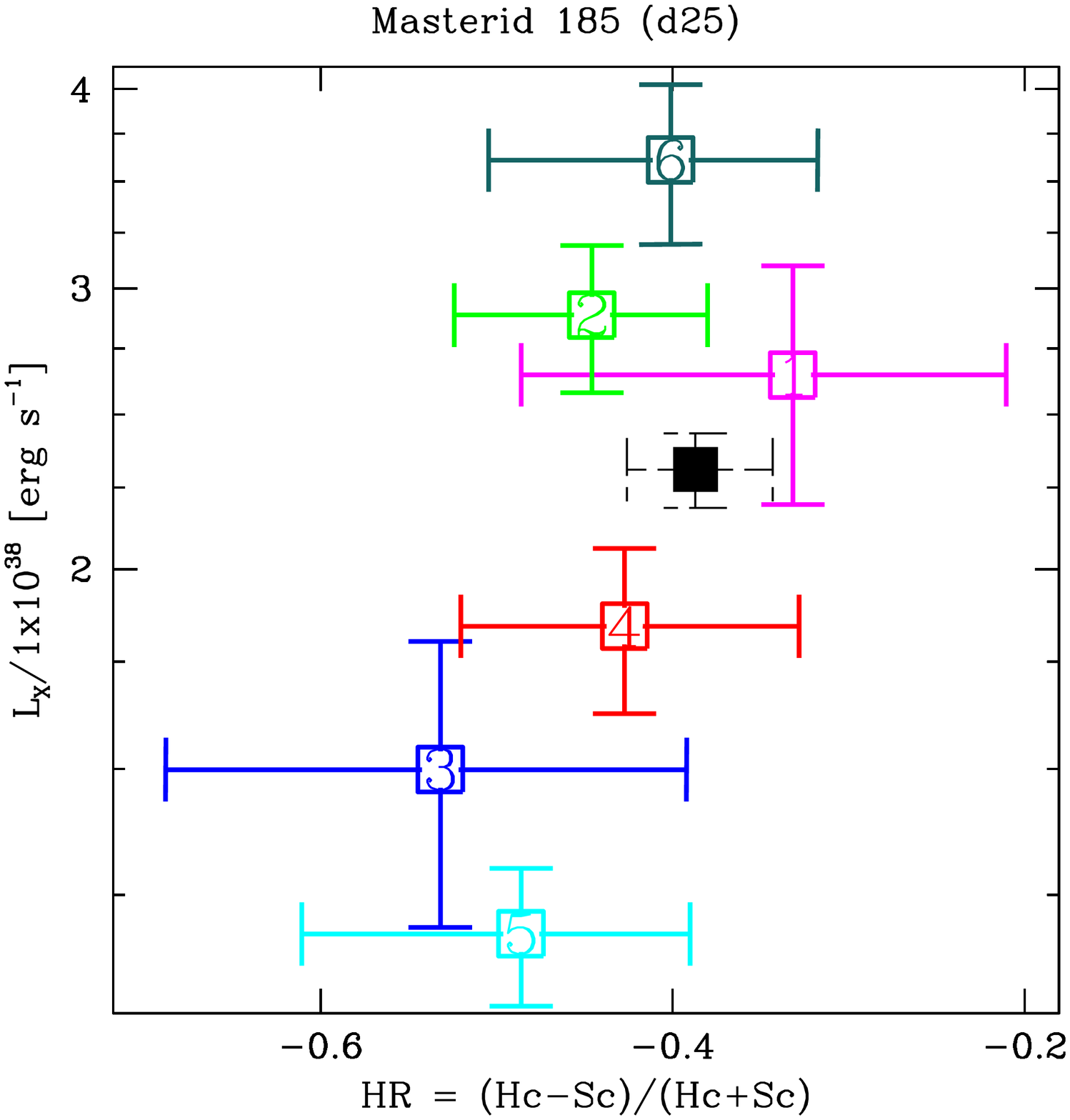}

  \end{minipage}
  \begin{minipage}{0.32\linewidth}
  \centering

    \includegraphics[width=\linewidth]{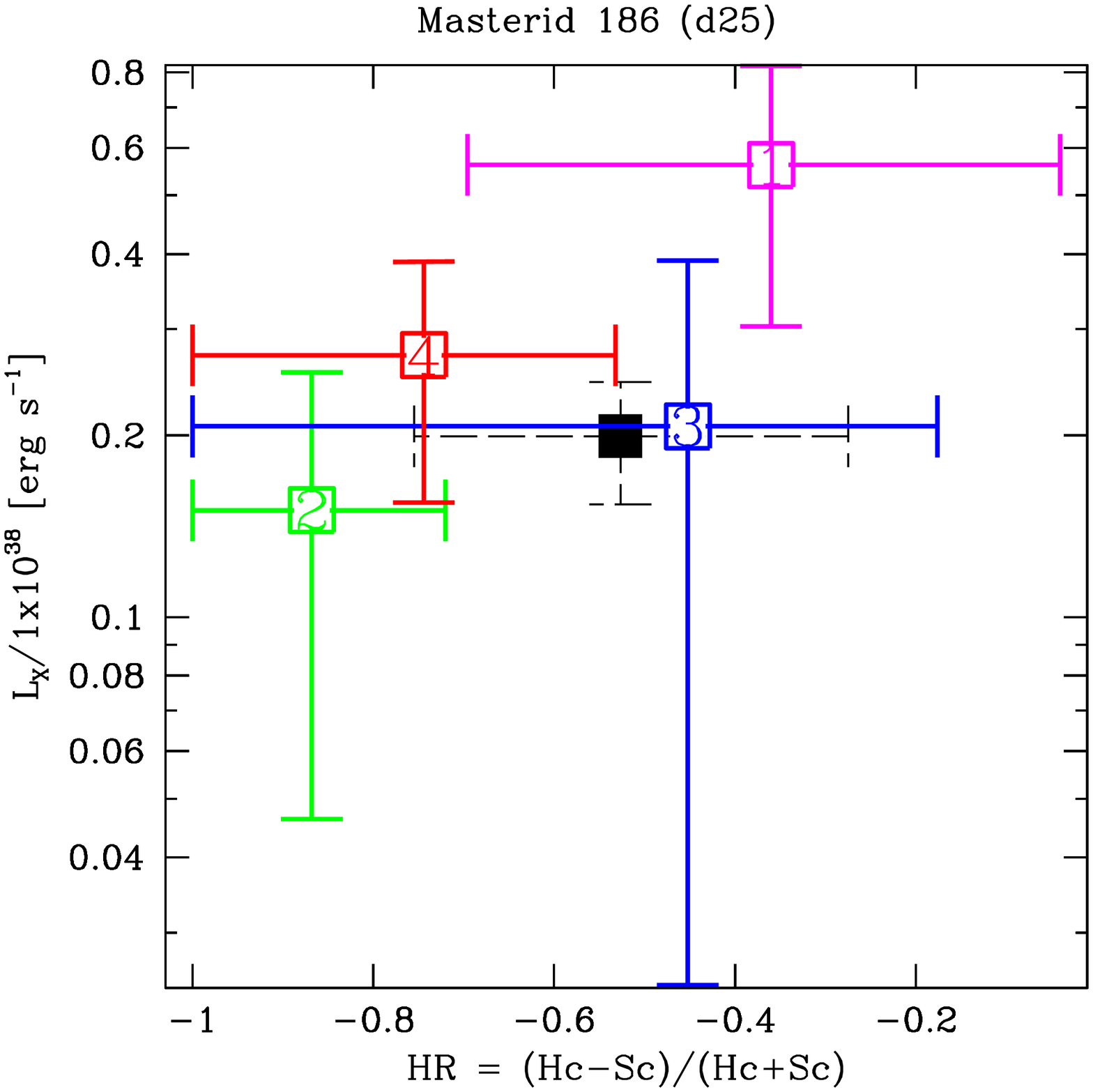}

\end{minipage}
\begin{minipage}{0.32\linewidth}
  \centering

    \includegraphics[width=\linewidth]{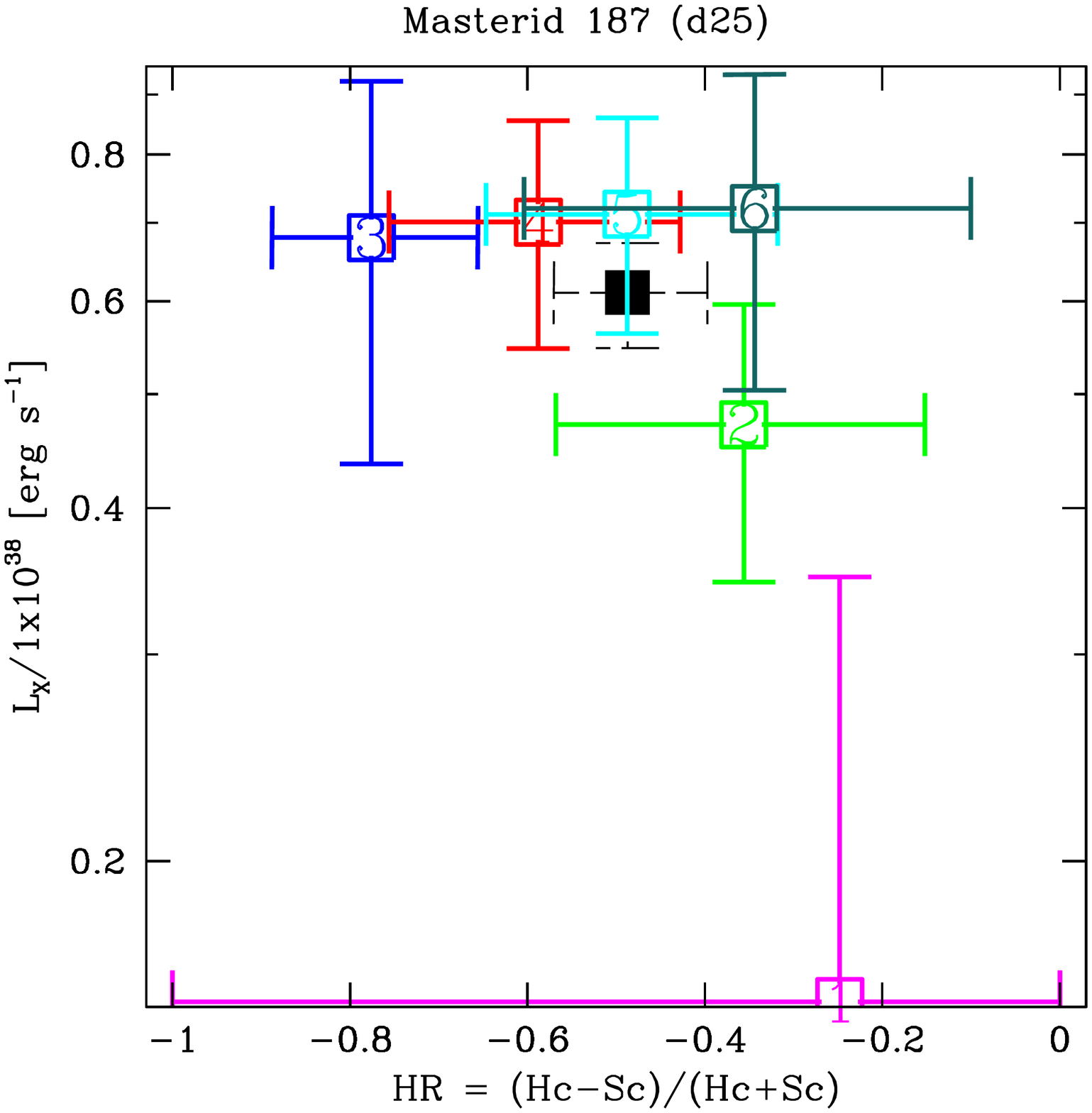}

\end{minipage}

\begin{minipage}{0.32\linewidth}
  \centering
  
    \includegraphics[width=\linewidth]{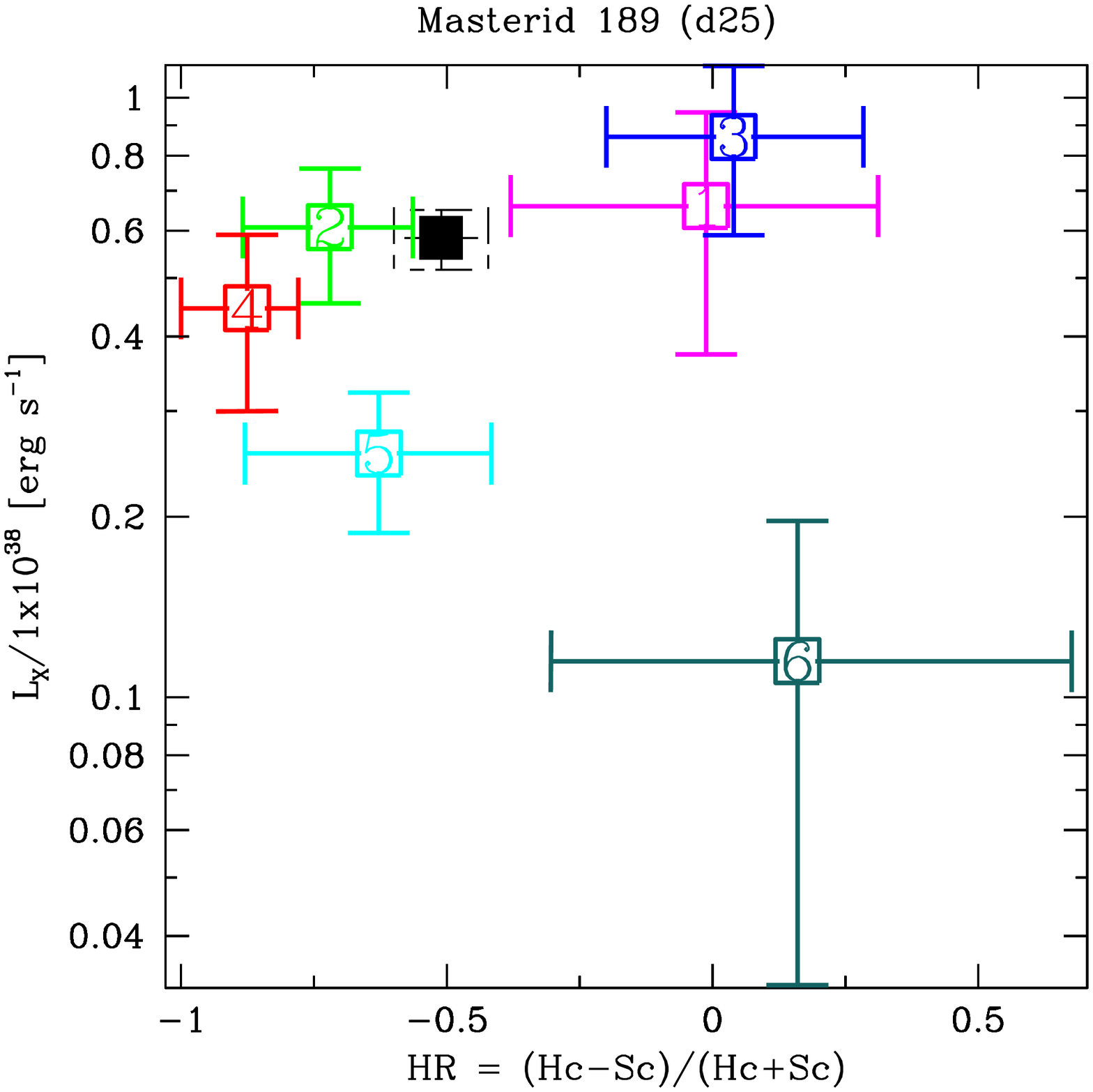}

  \end{minipage}
  \begin{minipage}{0.32\linewidth}
  \centering

    \includegraphics[width=\linewidth]{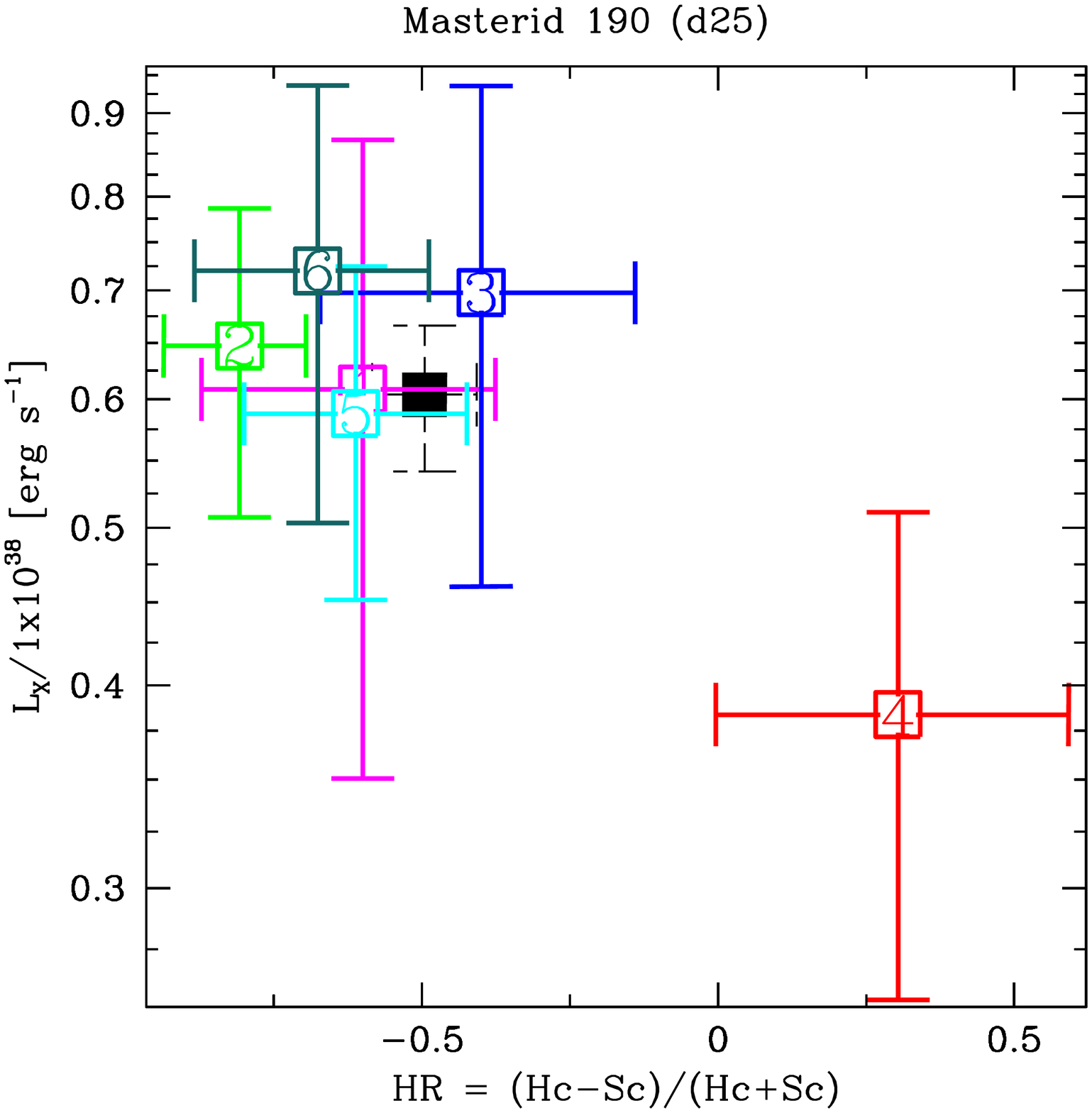}

\end{minipage}
\begin{minipage}{0.32\linewidth}
  \centering

    \includegraphics[width=\linewidth]{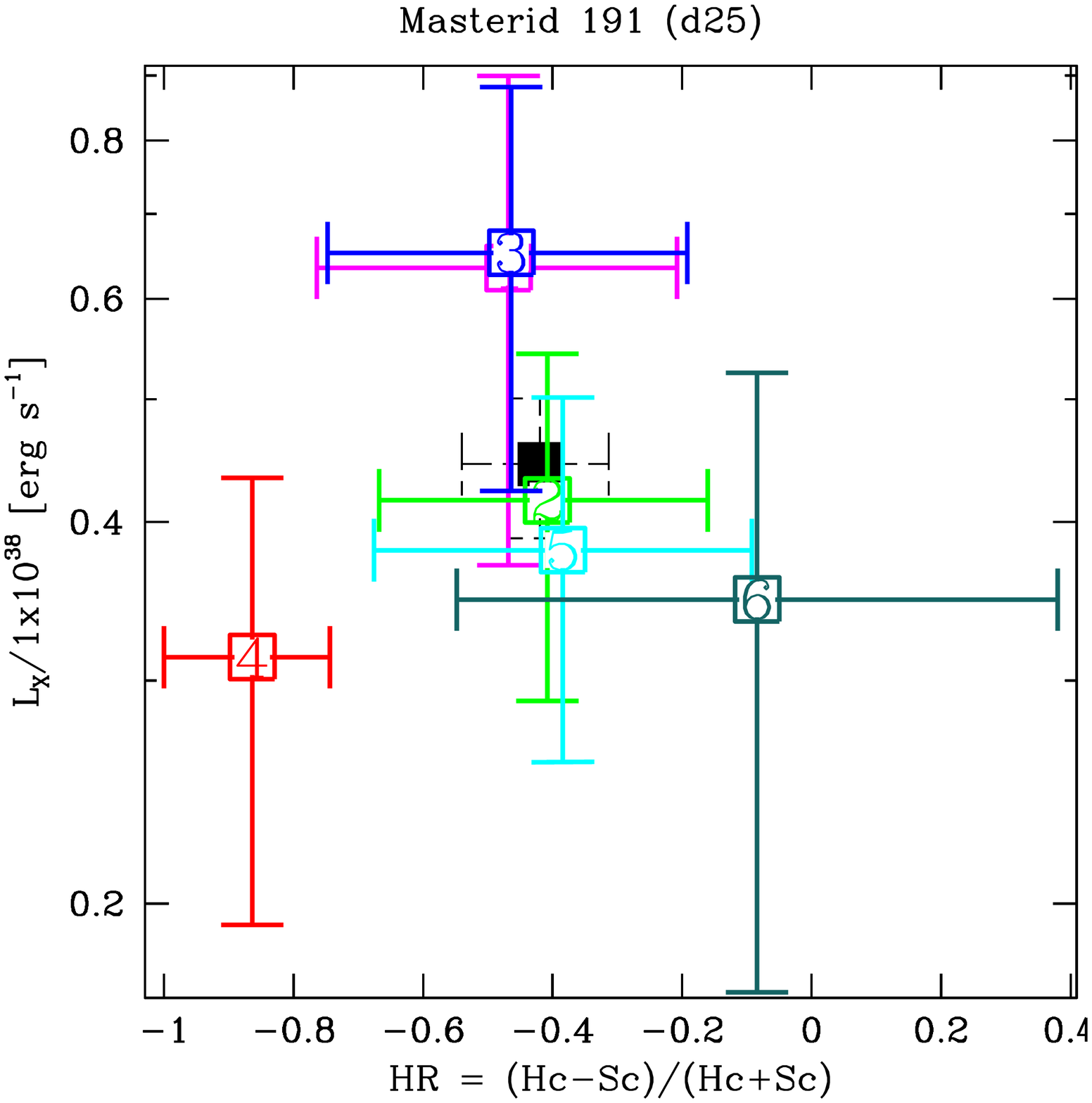}

\end{minipage}
\end{figure}

\begin{figure}
  \begin{minipage}{0.32\linewidth}
  \centering
  
    \includegraphics[width=\linewidth]{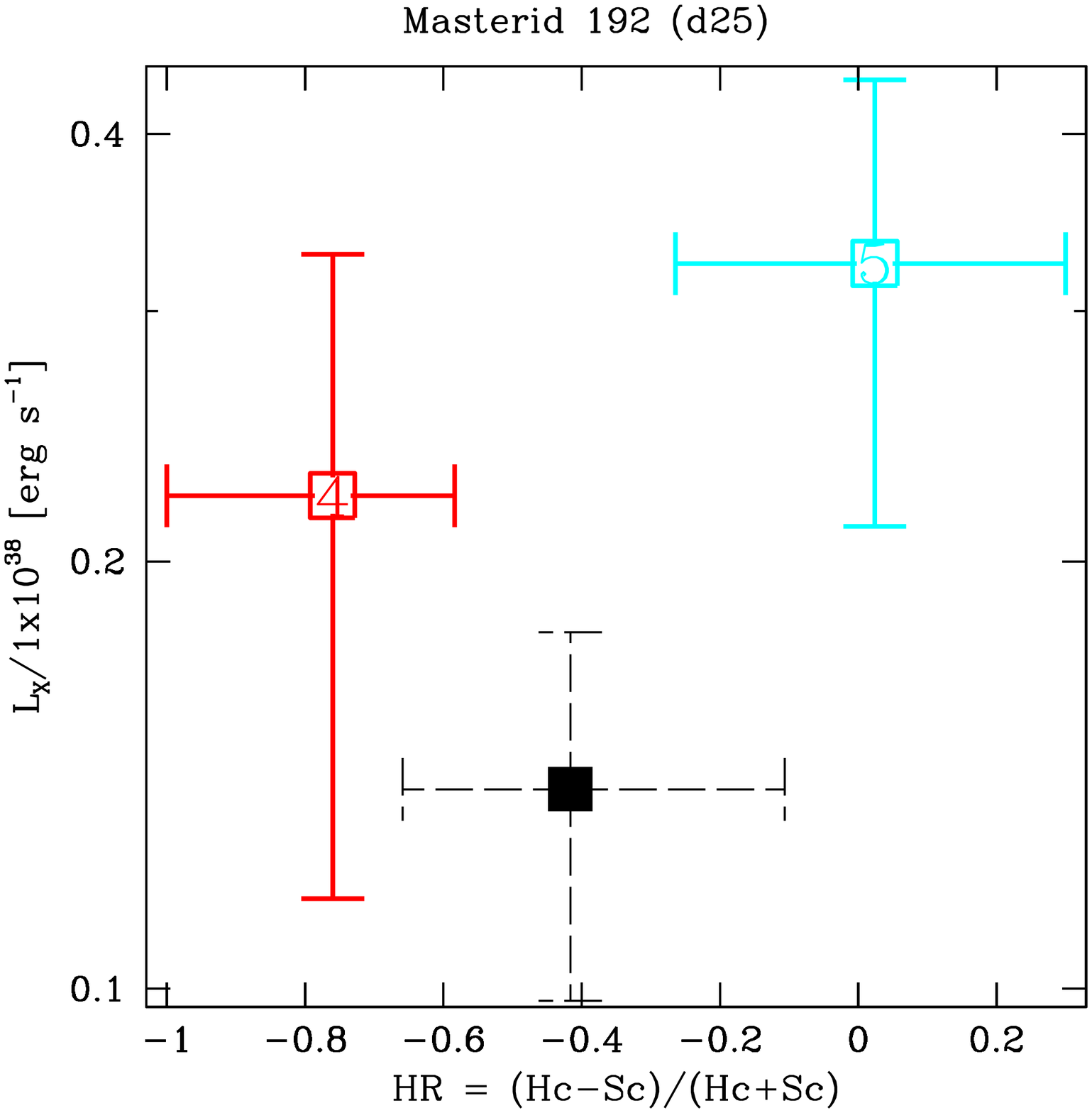}

  \end{minipage}
  \begin{minipage}{0.32\linewidth}
  \centering

    \includegraphics[width=\linewidth]{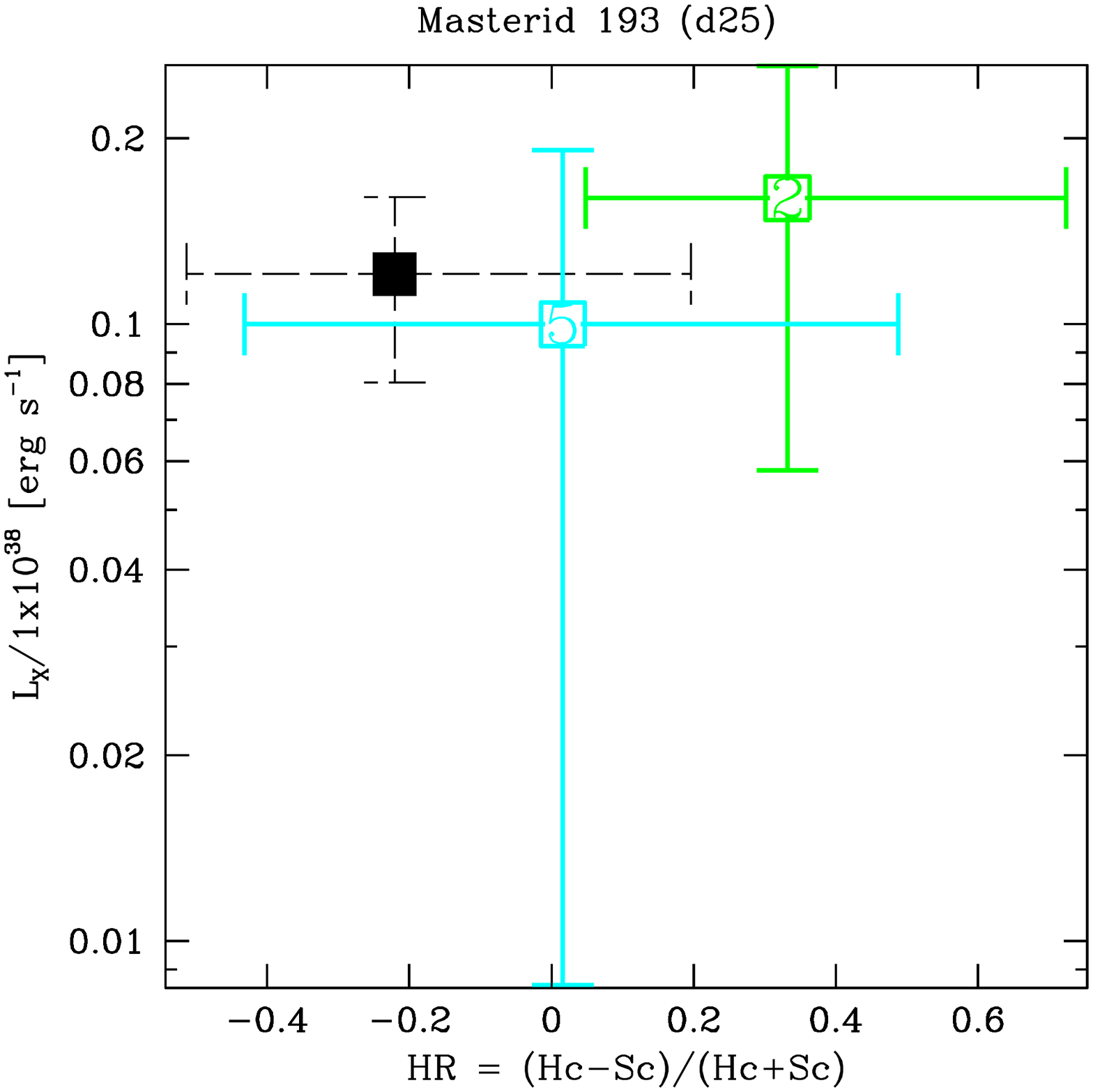}

\end{minipage}
\begin{minipage}{0.32\linewidth}
  \centering

    \includegraphics[width=\linewidth]{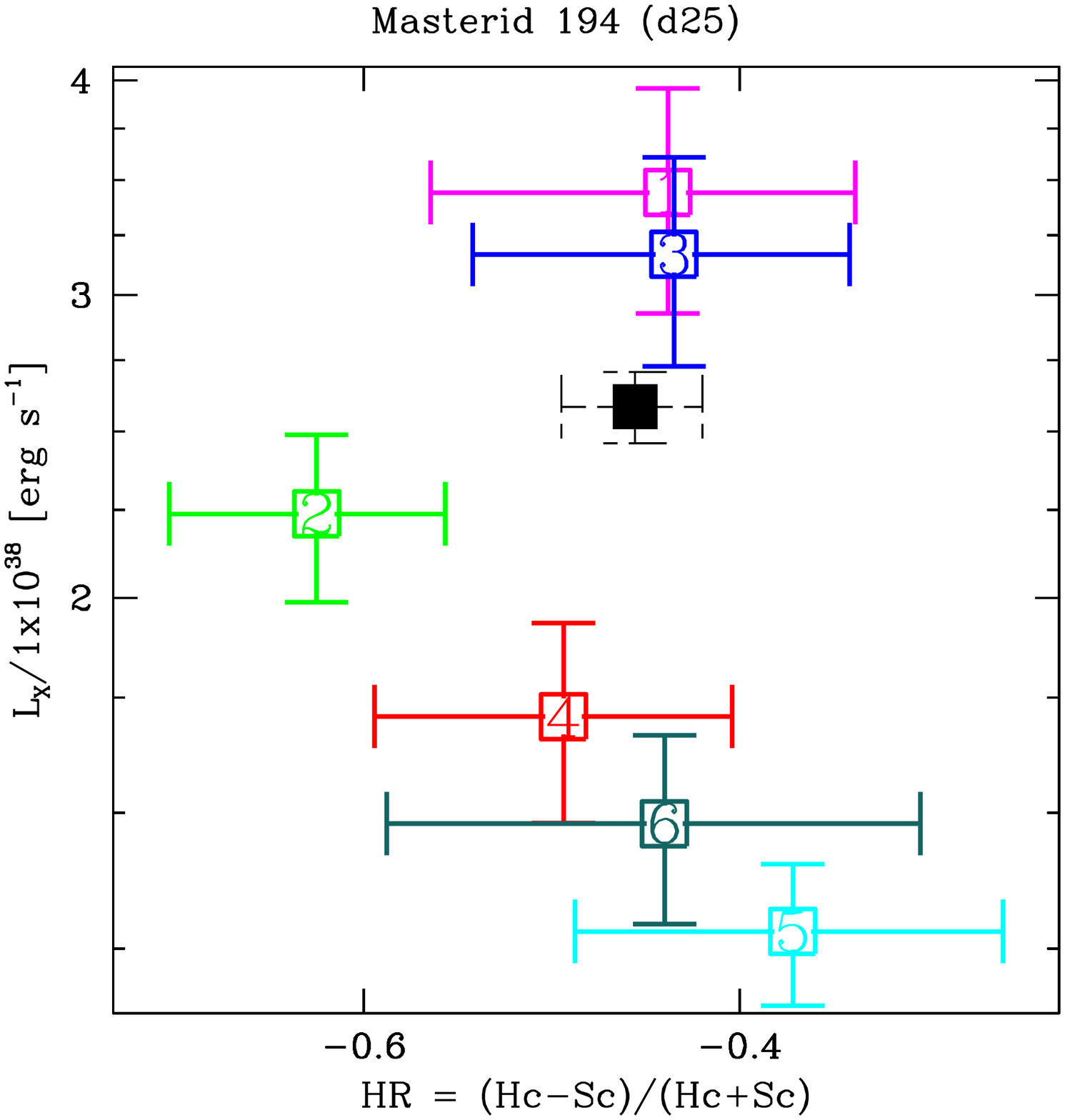}

 \end{minipage}

\begin{minipage}{0.32\linewidth}
  \centering
  
    \includegraphics[width=\linewidth]{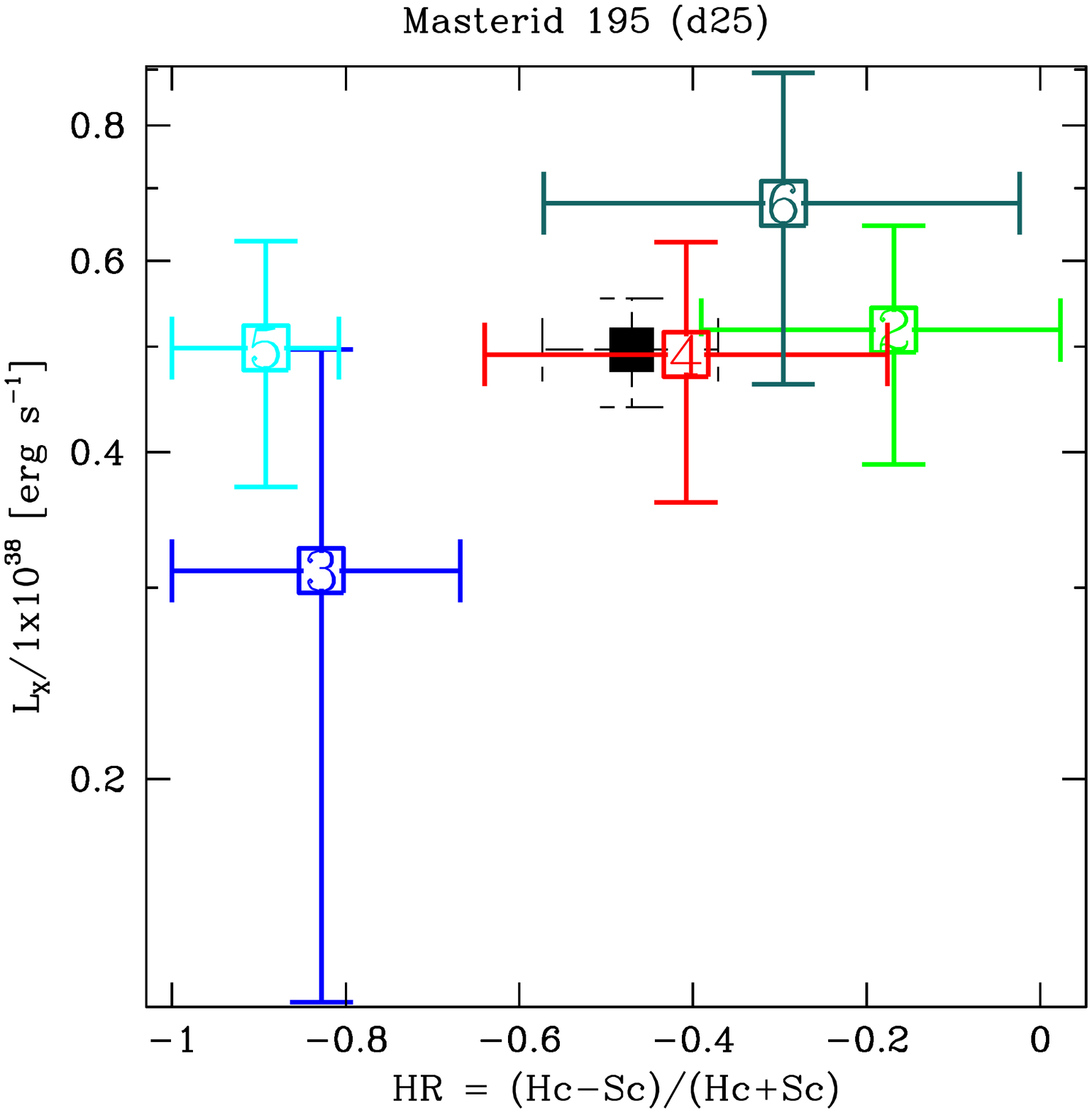}

  \end{minipage}
  \begin{minipage}{0.32\linewidth}
  \centering

    \includegraphics[width=\linewidth]{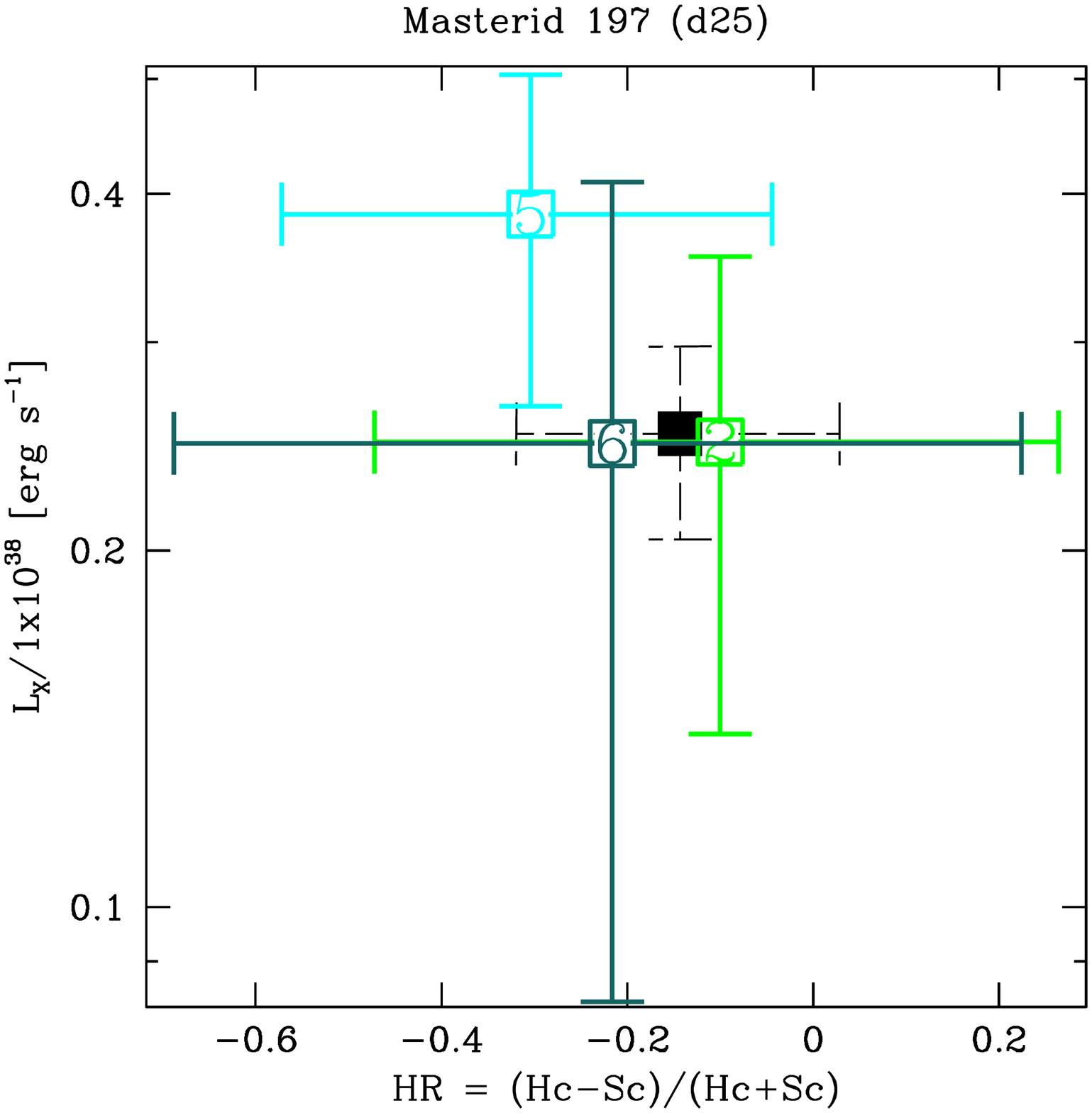}

\end{minipage}
\begin{minipage}{0.32\linewidth}
  \centering

    \includegraphics[width=\linewidth]{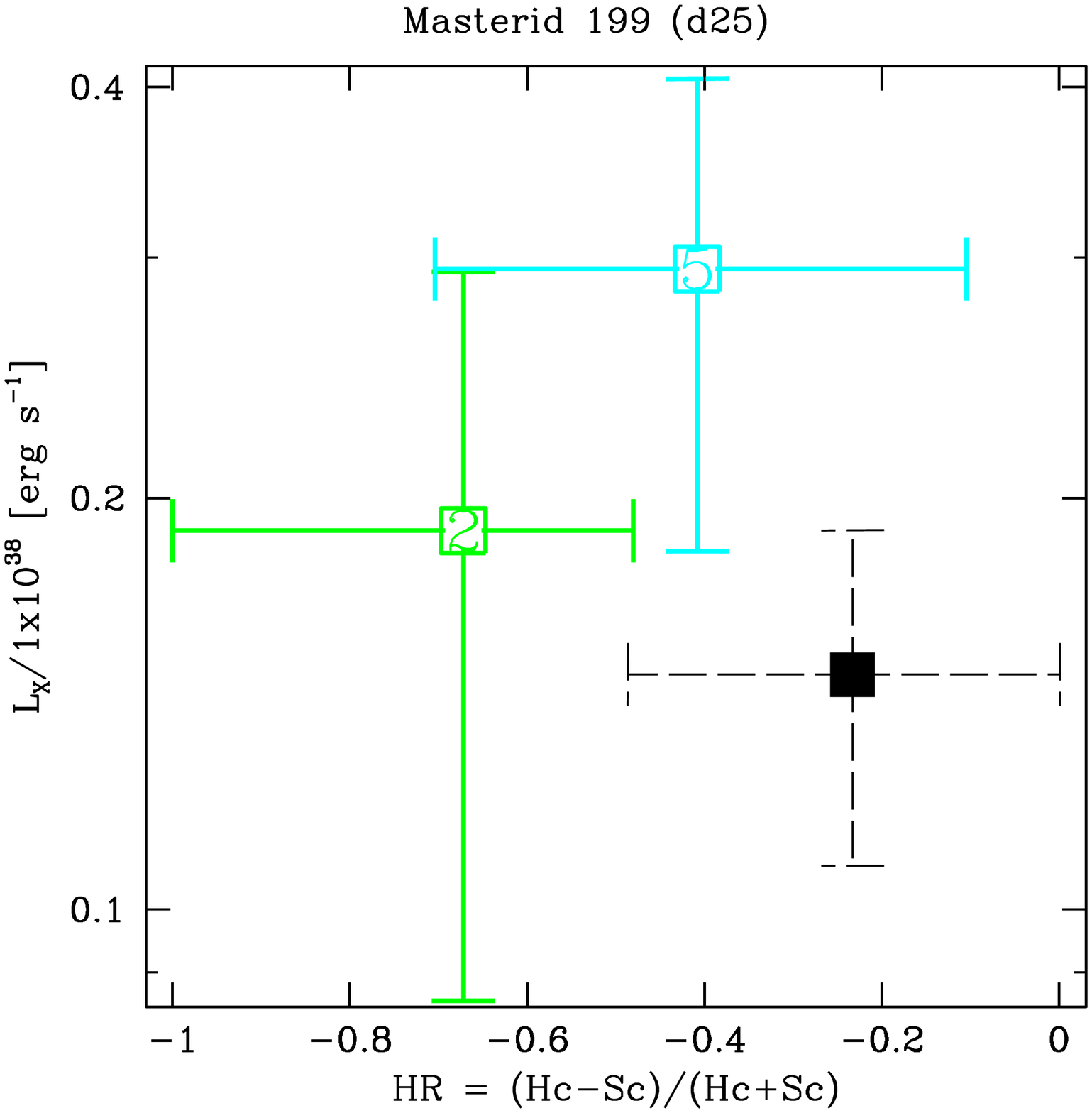}

 \end{minipage}

  \begin{minipage}{0.32\linewidth}
  \centering
  
    \includegraphics[width=\linewidth]{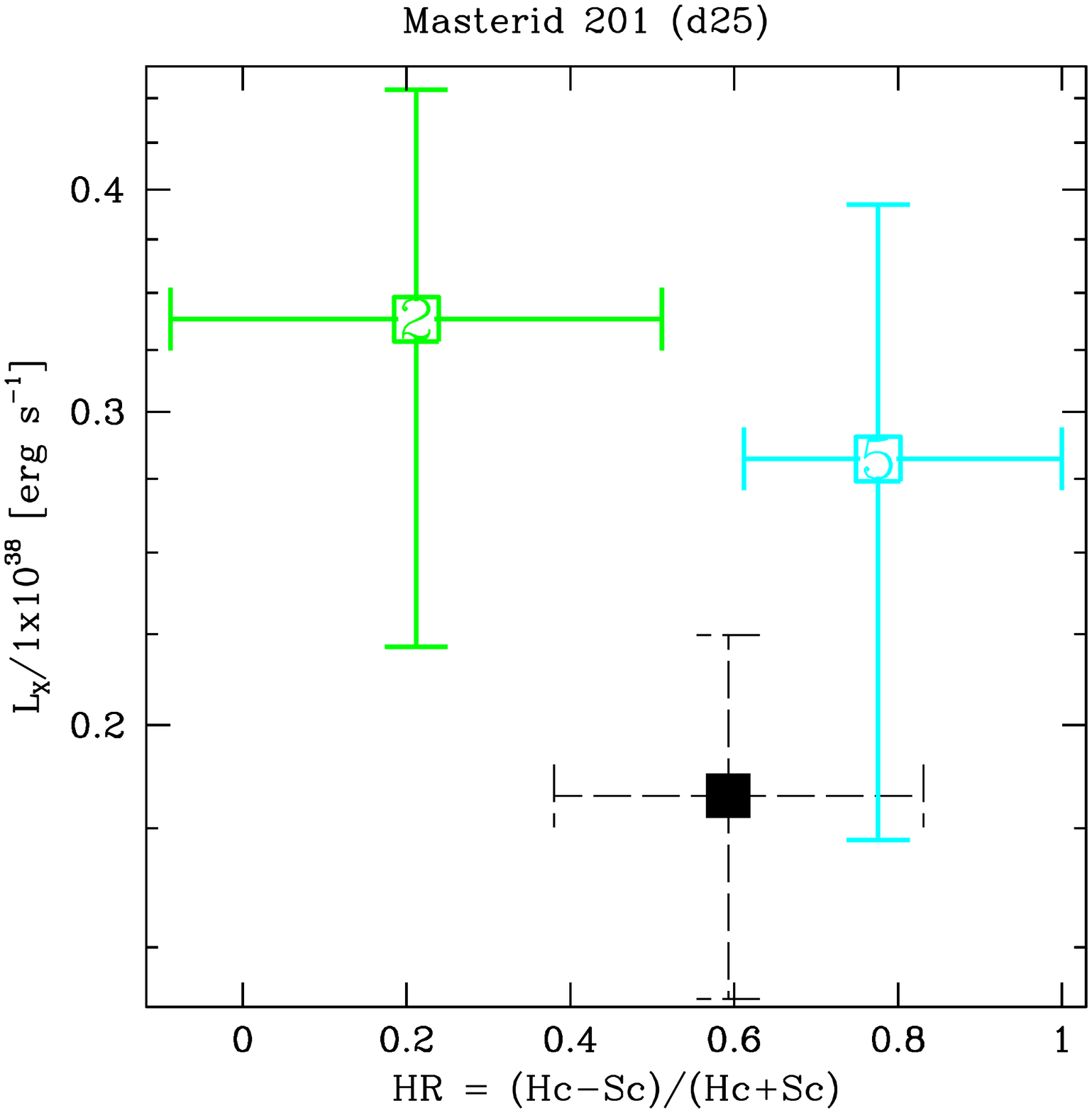}

  \end{minipage}
  \begin{minipage}{0.32\linewidth}
  \centering

    \includegraphics[width=\linewidth]{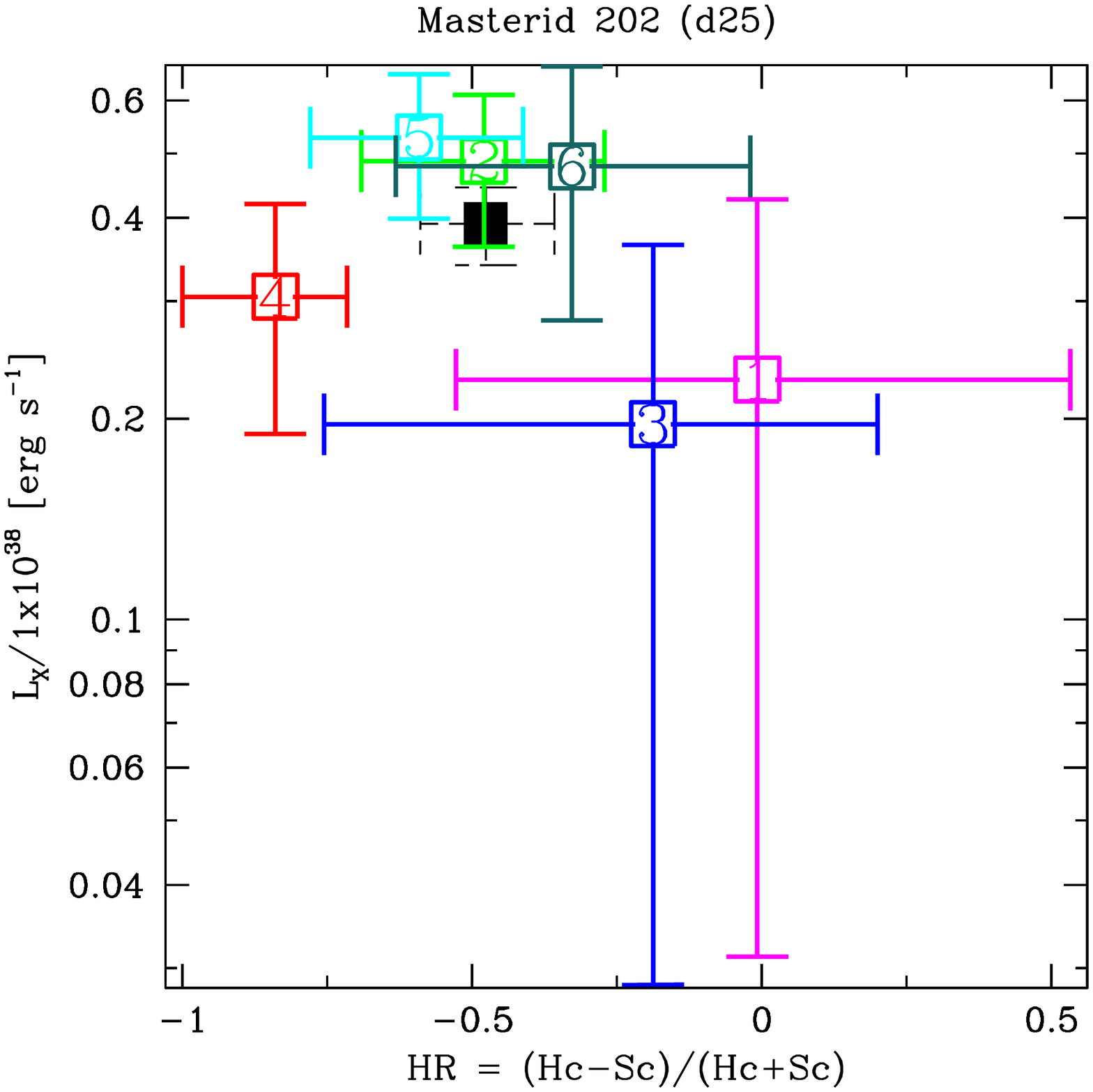}

\end{minipage}
\begin{minipage}{0.32\linewidth}
  \centering

    \includegraphics[width=\linewidth]{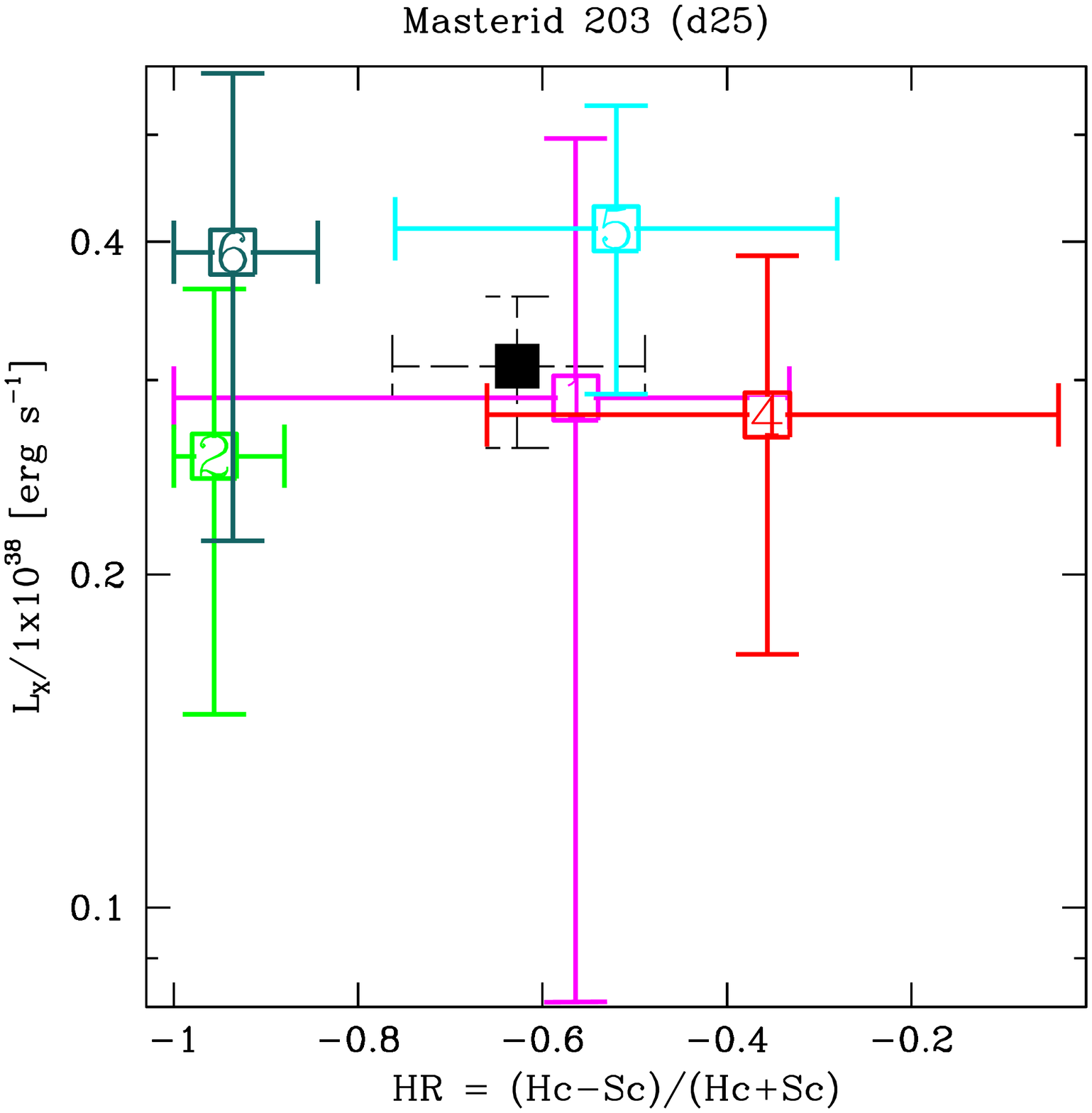}

\end{minipage}

\begin{minipage}{0.32\linewidth}
  \centering
  
    \includegraphics[width=\linewidth]{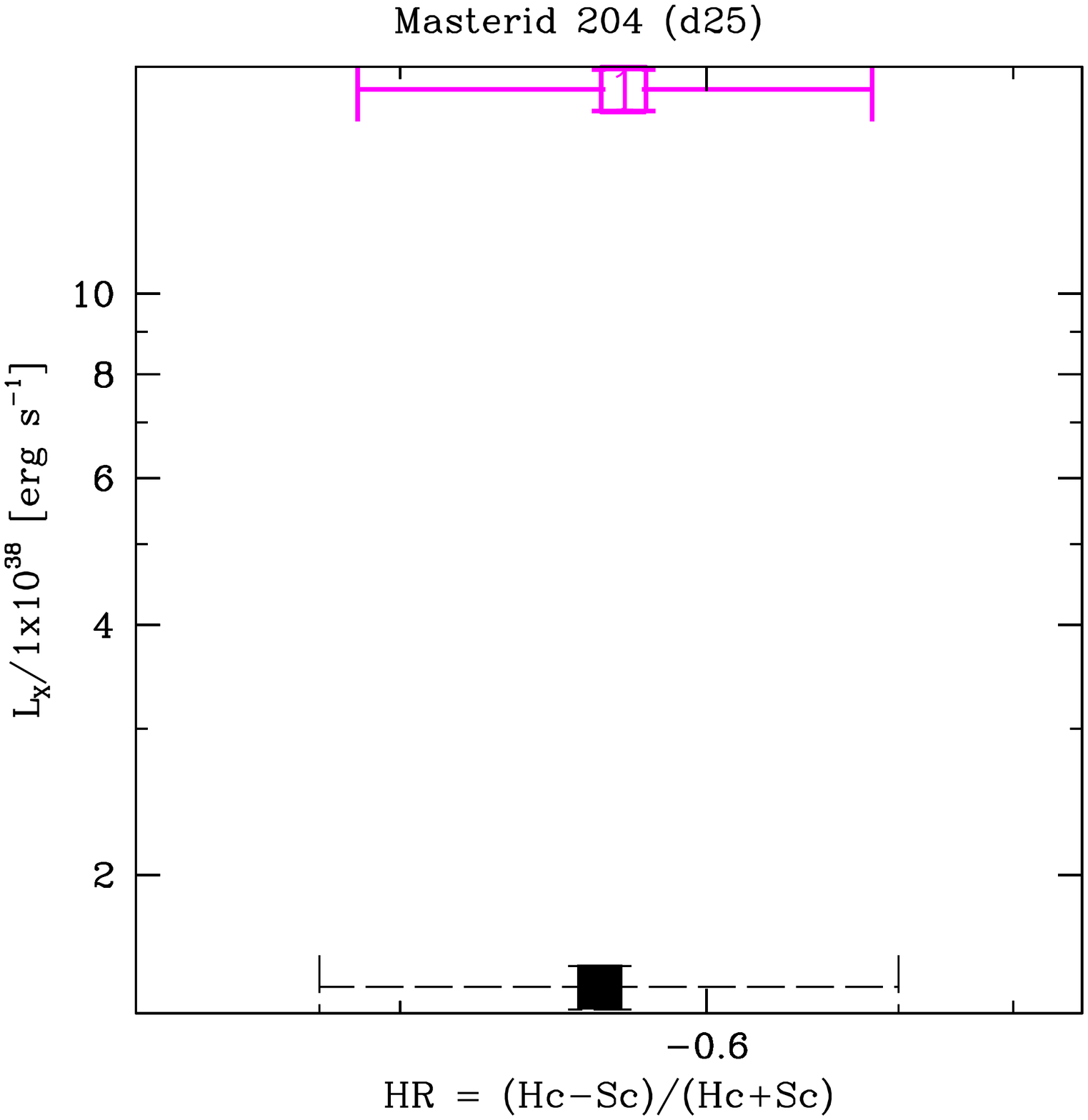}

  \end{minipage}
  \begin{minipage}{0.32\linewidth}
  \centering

    \includegraphics[width=\linewidth]{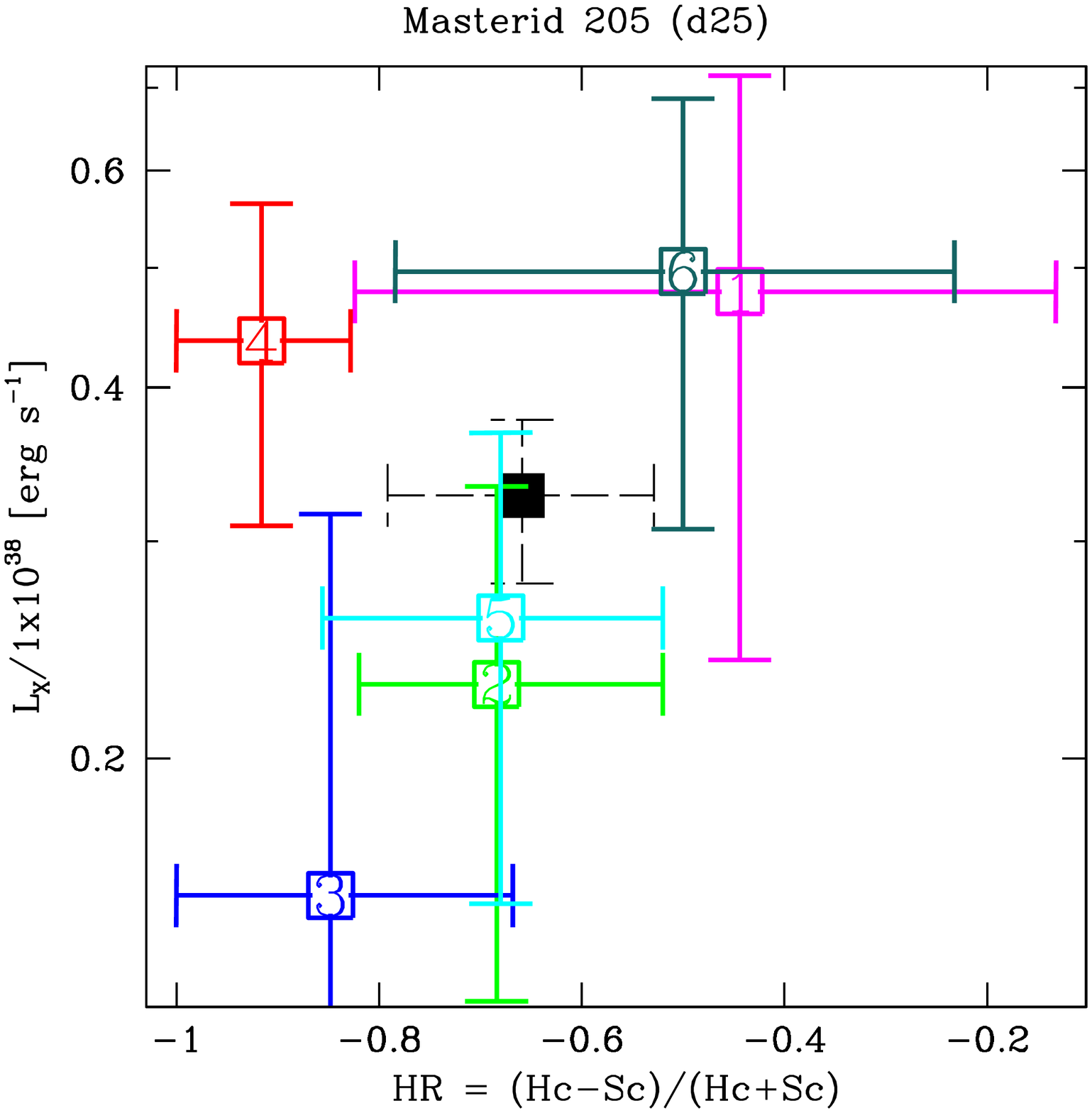}

\end{minipage}
\begin{minipage}{0.32\linewidth}
  \centering

    \includegraphics[width=\linewidth]{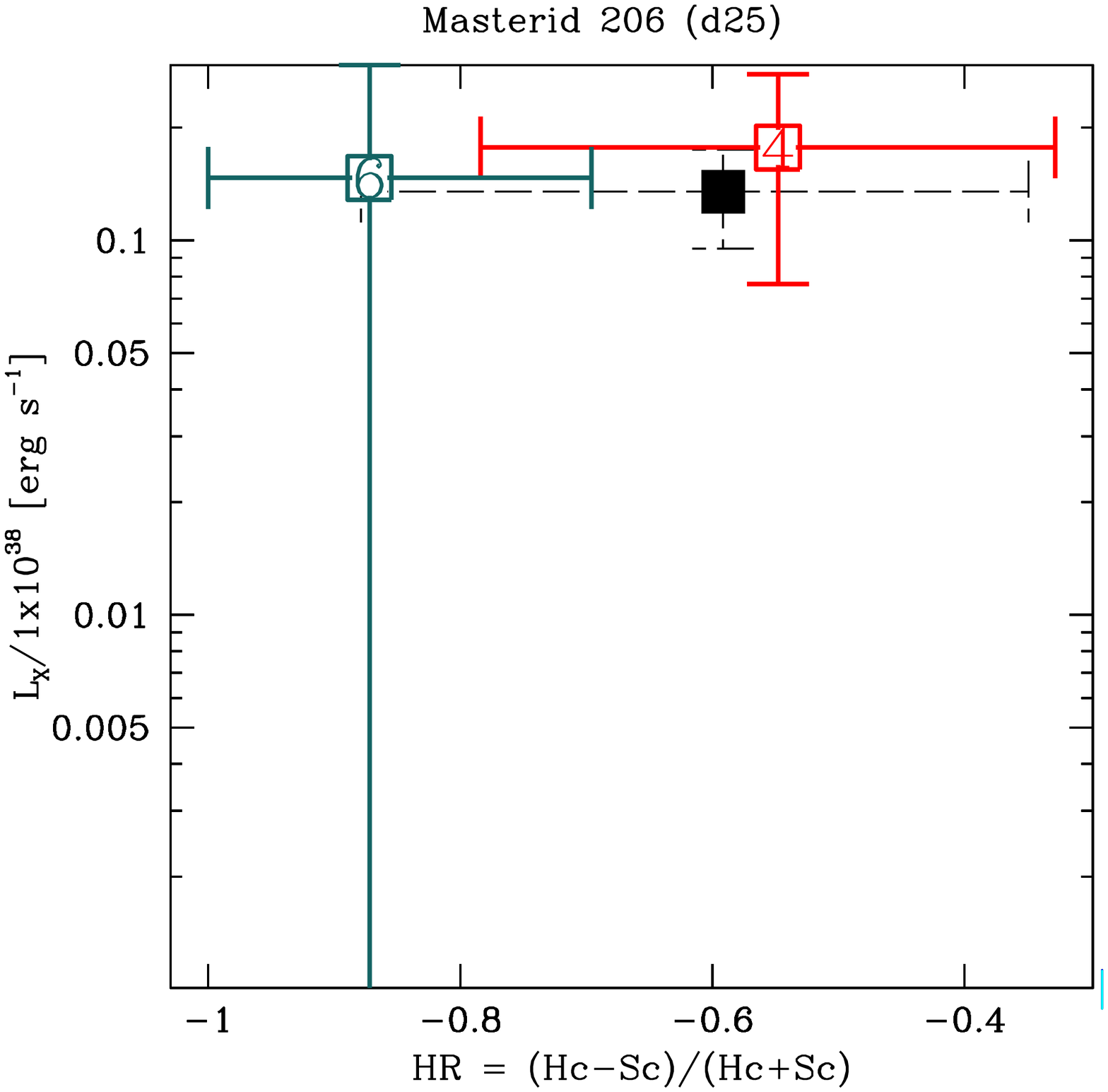}

\end{minipage}
\end{figure}

\begin{figure}
  \begin{minipage}{0.32\linewidth}
  \centering
  
    \includegraphics[width=\linewidth]{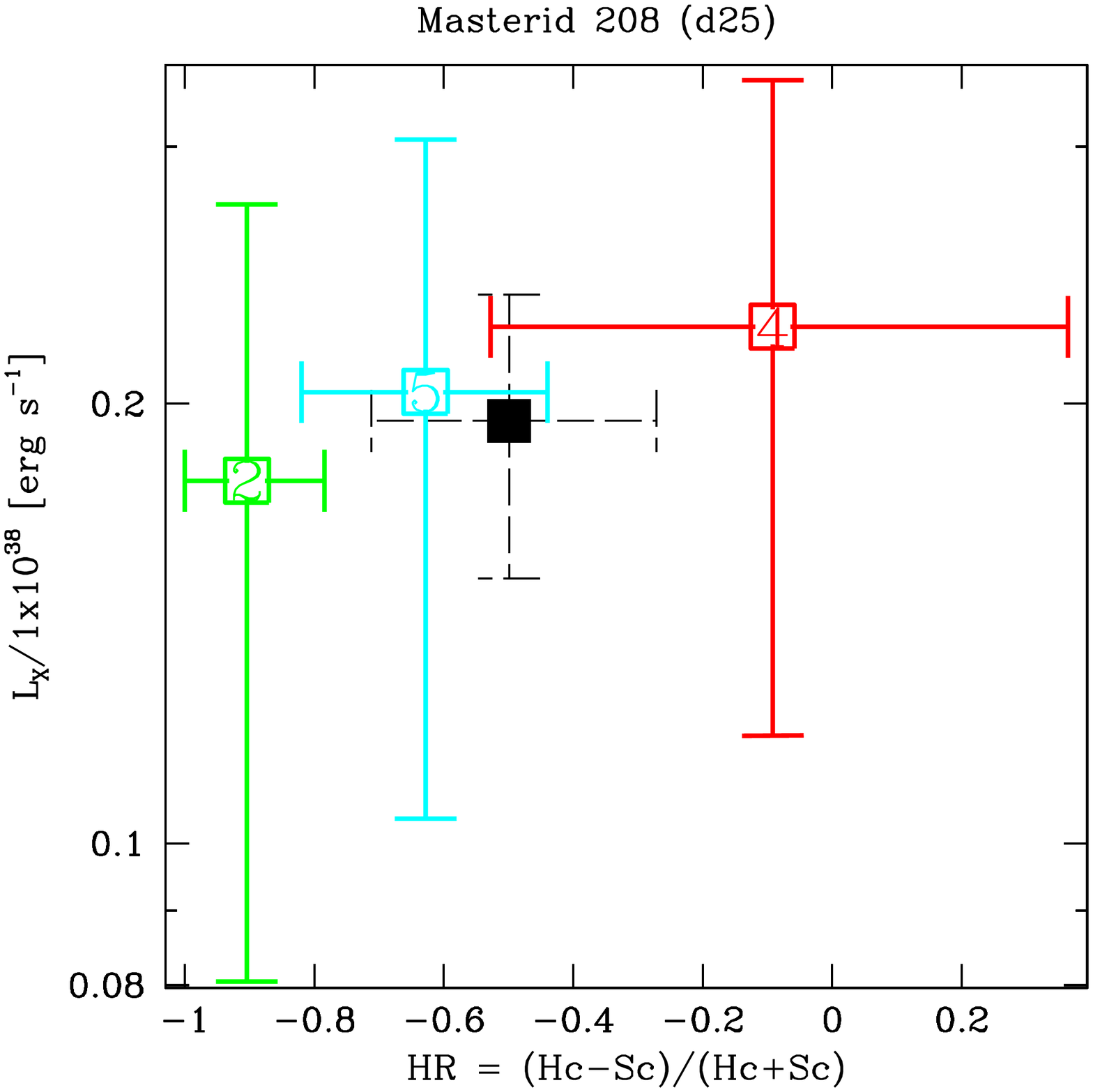}

  \end{minipage}
  \begin{minipage}{0.32\linewidth}
  \centering

    \includegraphics[width=\linewidth]{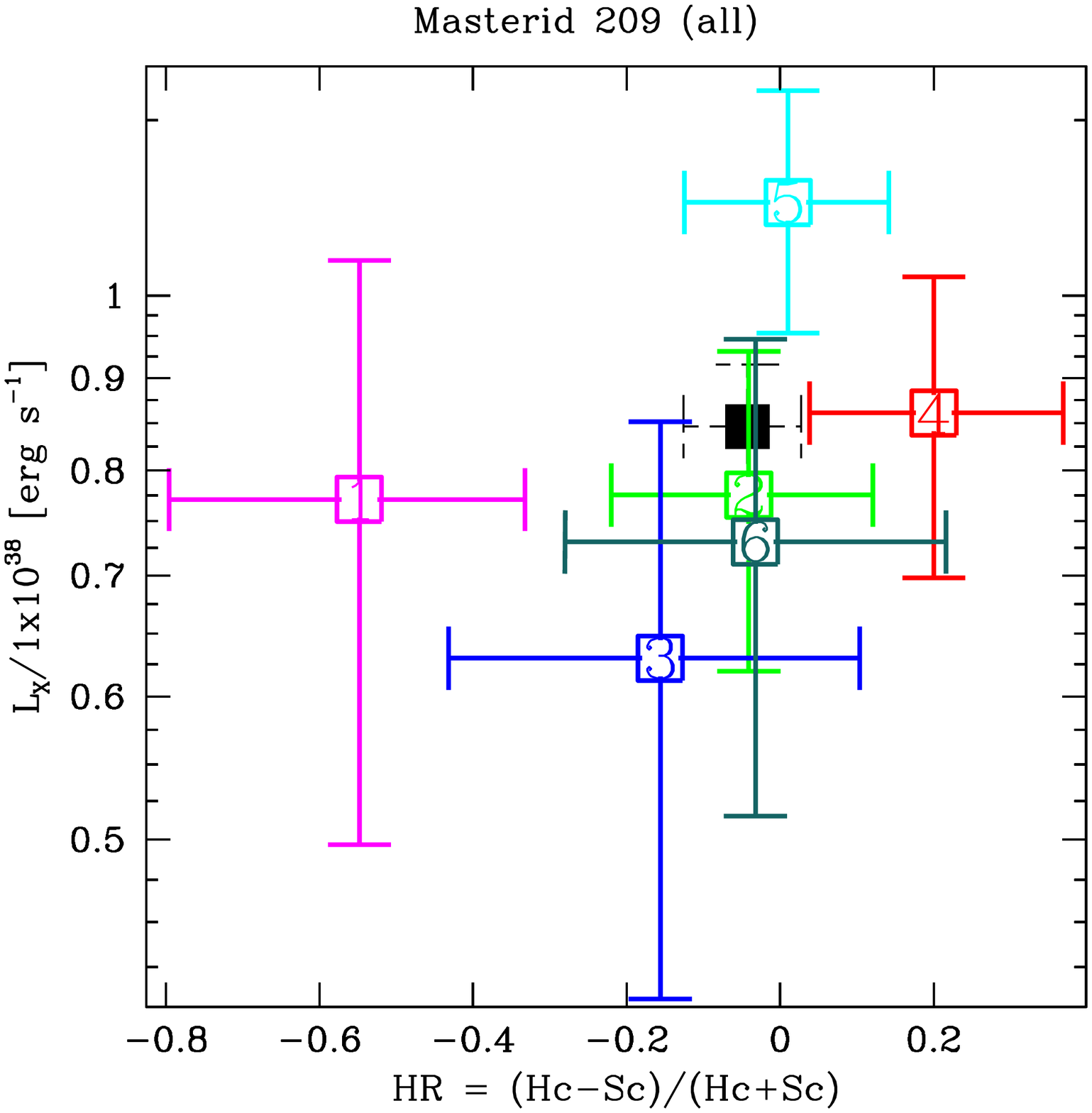}

\end{minipage}
\begin{minipage}{0.32\linewidth}
  \centering

    \includegraphics[width=\linewidth]{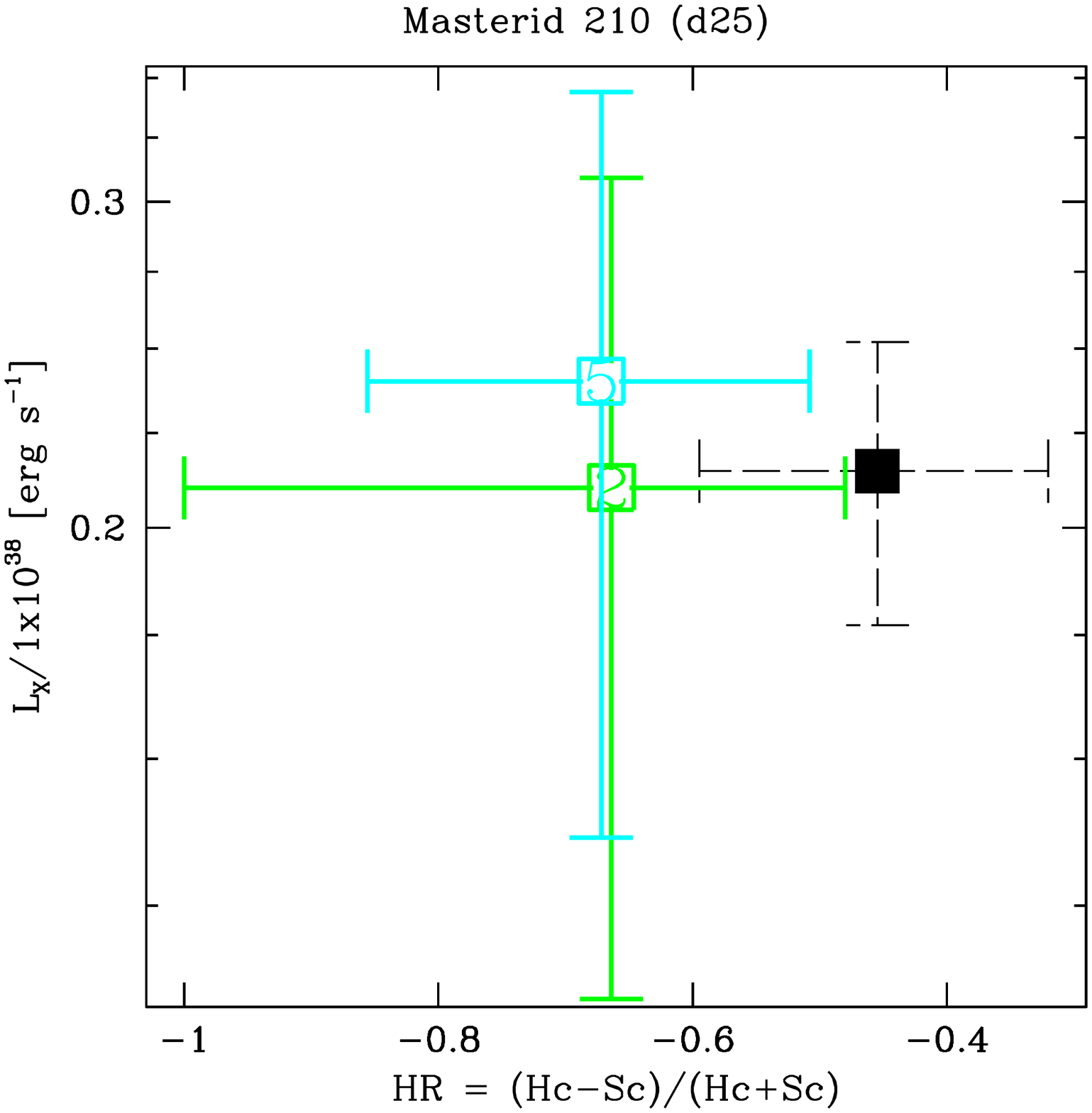}

 \end{minipage}

\begin{minipage}{0.32\linewidth}
  \centering
  
    \includegraphics[width=\linewidth]{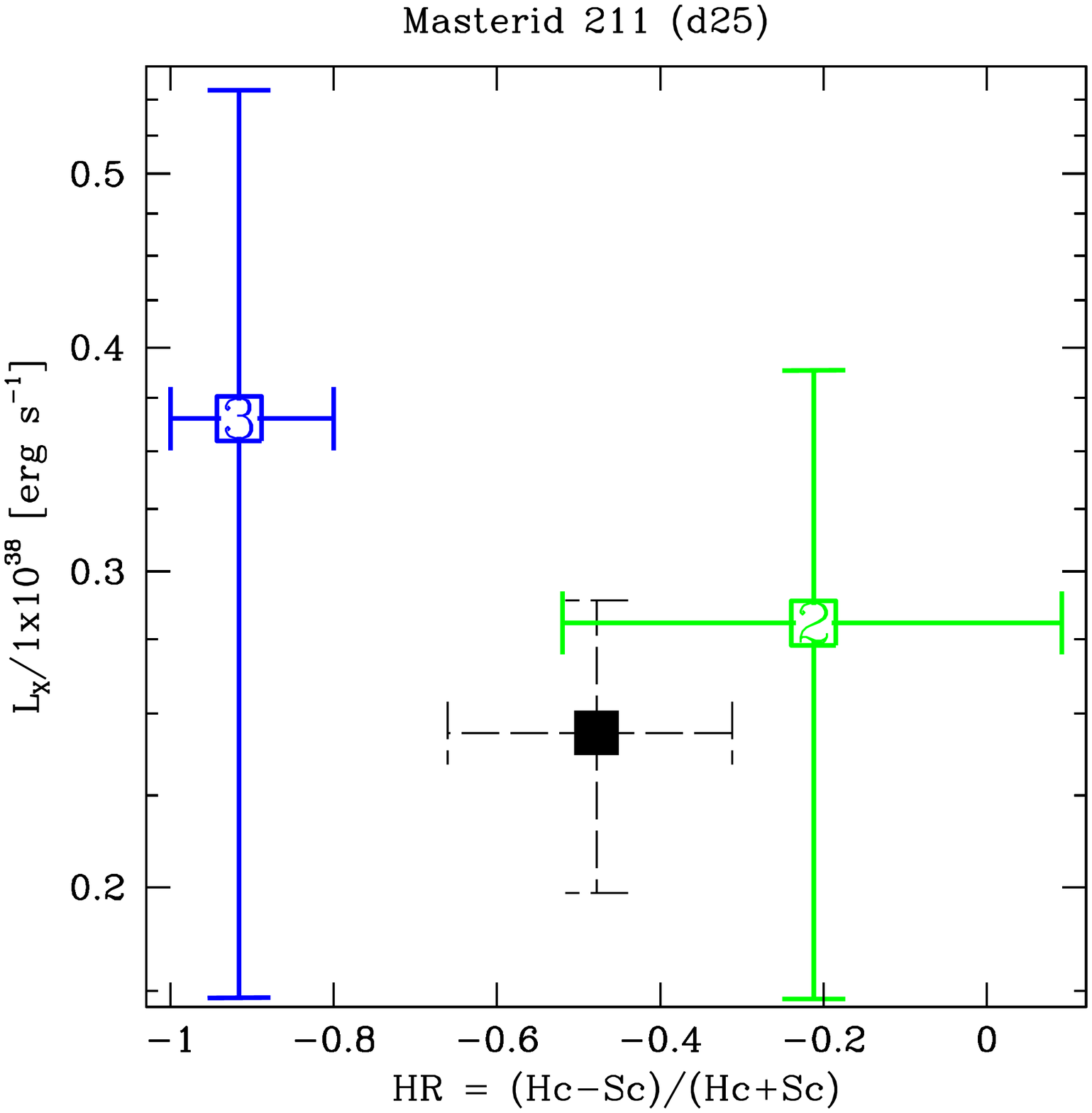}

  \end{minipage}
  \begin{minipage}{0.32\linewidth}
  \centering

    \includegraphics[width=\linewidth]{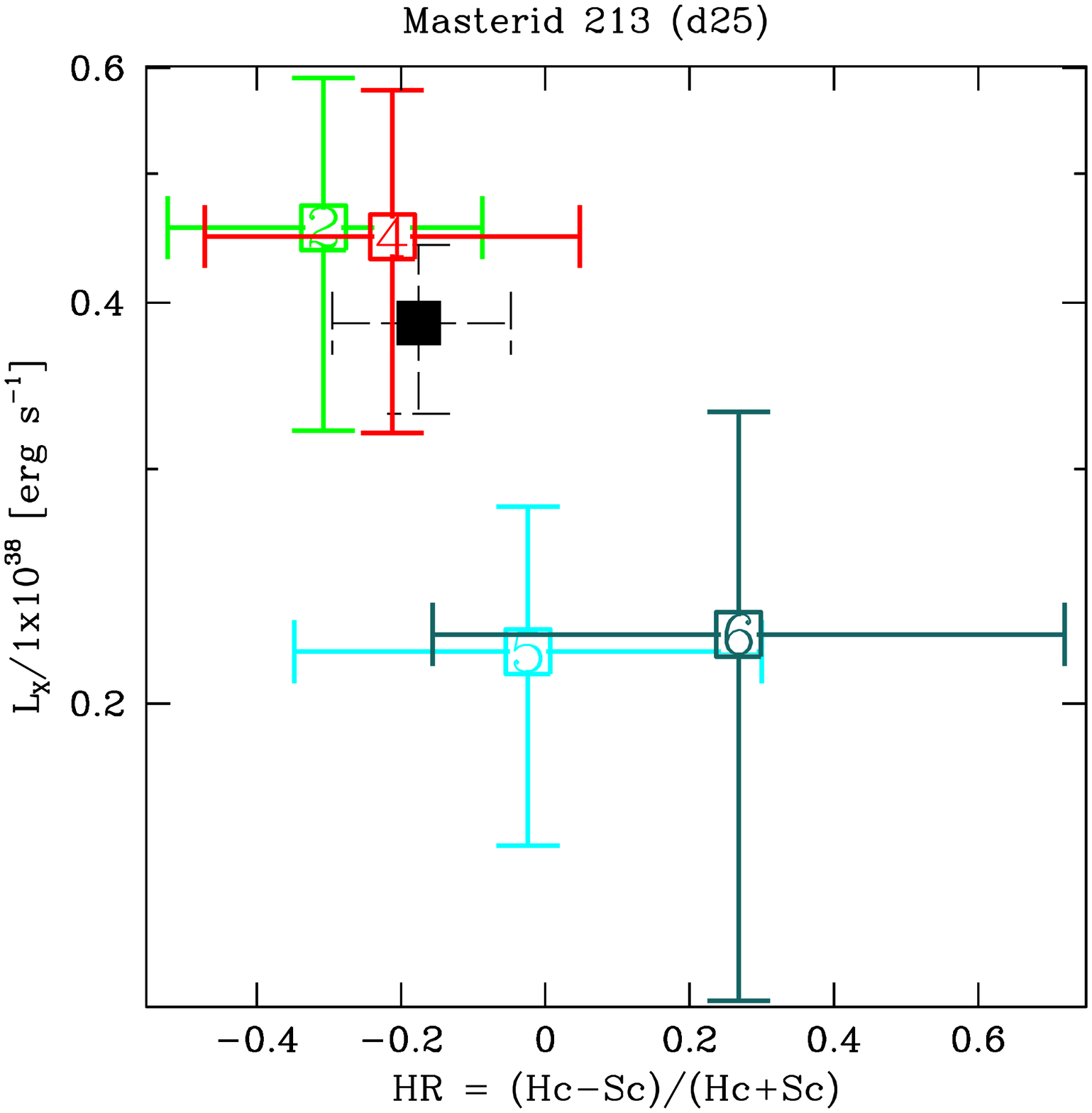}

\end{minipage}
\begin{minipage}{0.32\linewidth}
  \centering

    \includegraphics[width=\linewidth]{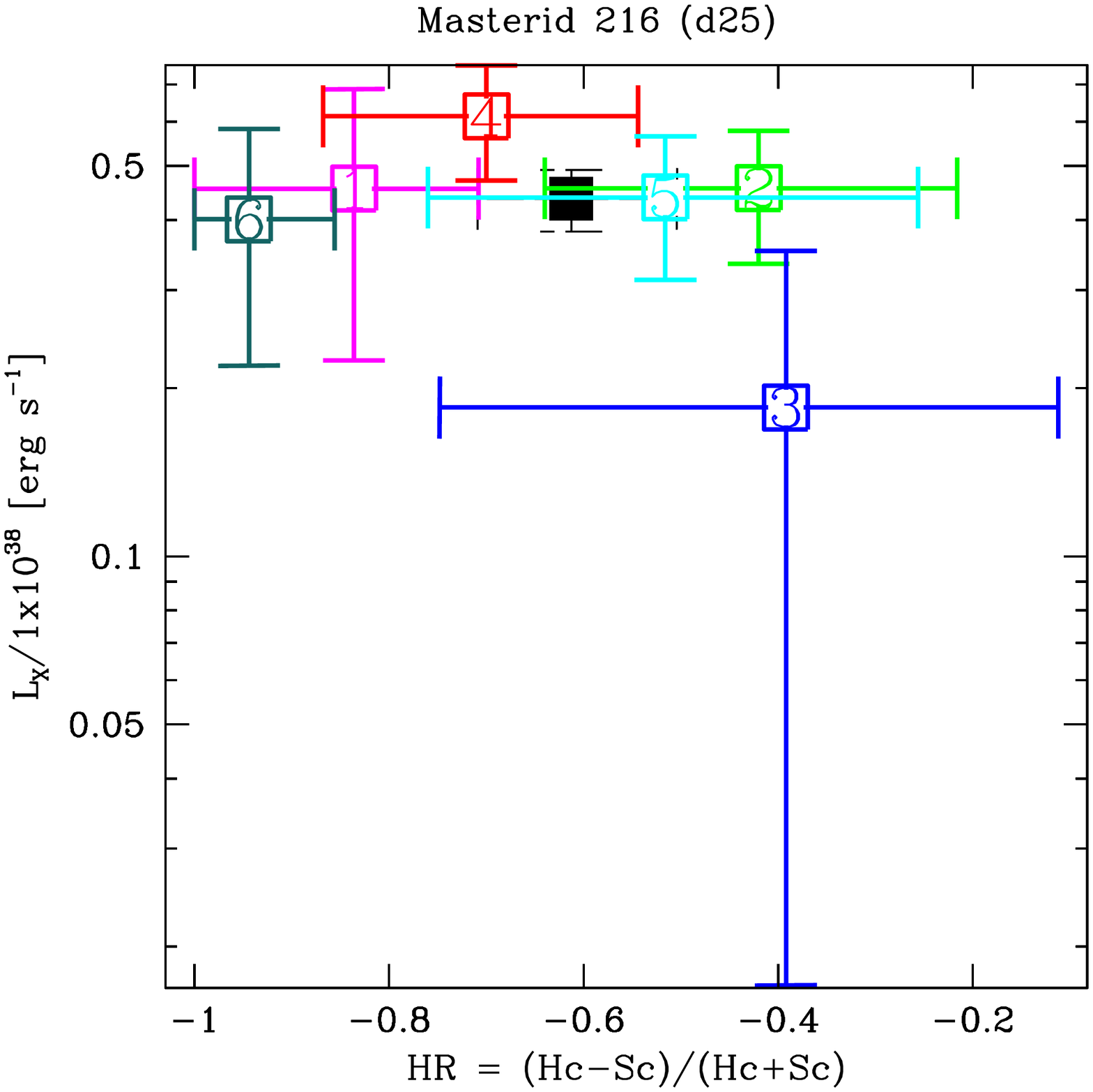}

 \end{minipage}

  \begin{minipage}{0.32\linewidth}
  \centering
  
    \includegraphics[width=\linewidth]{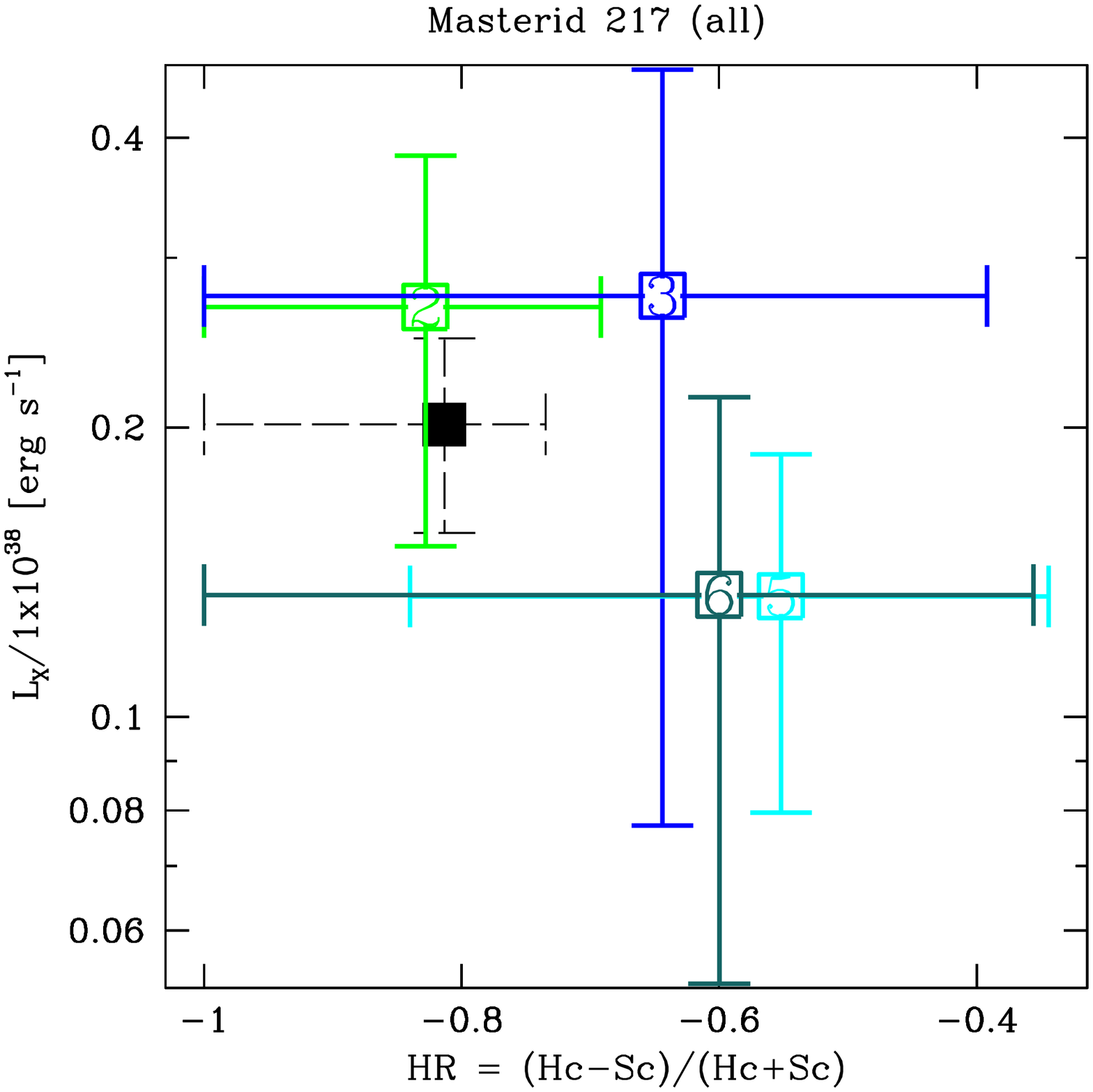}

  \end{minipage}
  \begin{minipage}{0.32\linewidth}
  \centering

    \includegraphics[width=\linewidth]{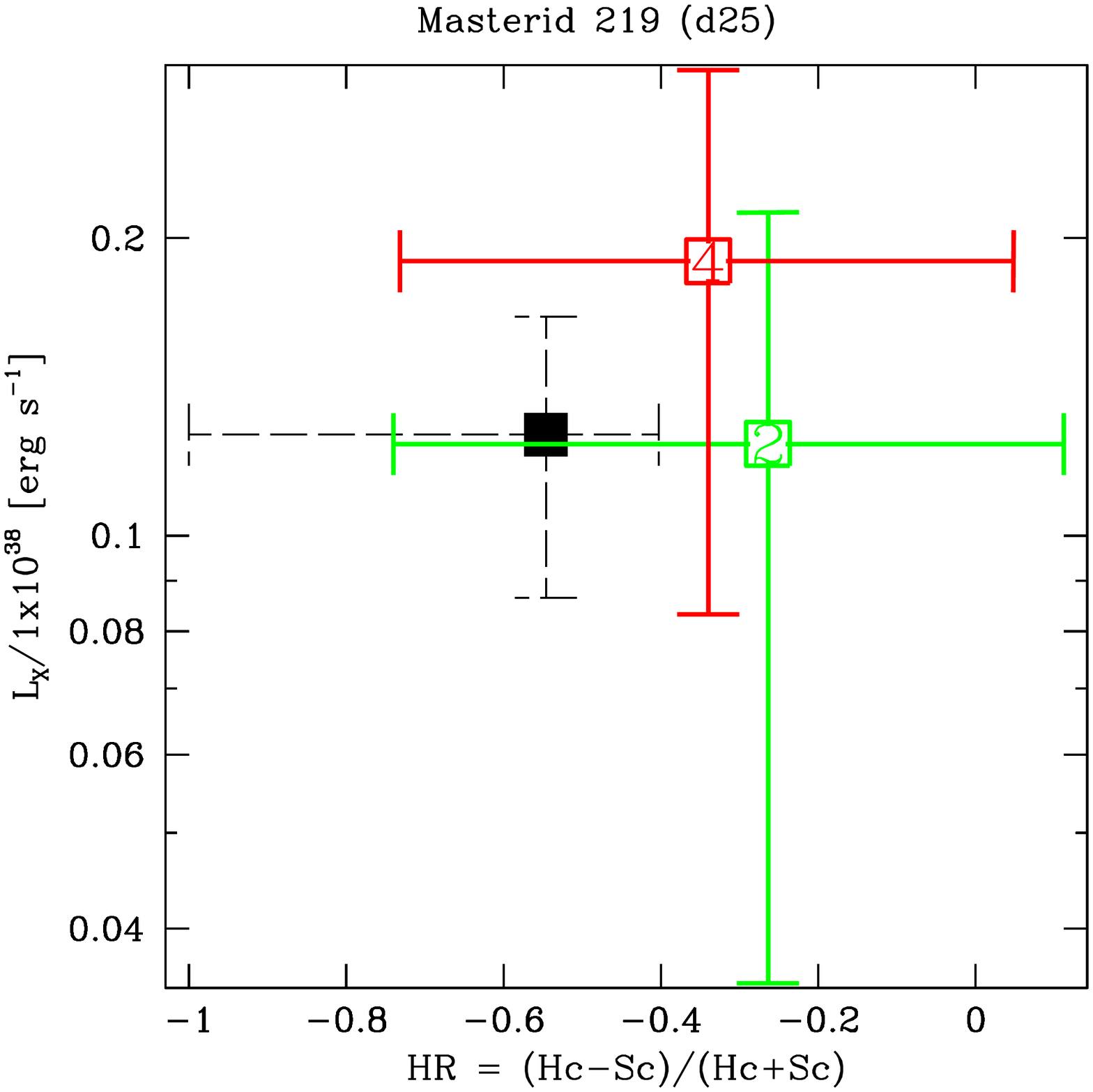}

\end{minipage}
\begin{minipage}{0.32\linewidth}
  \centering

    \includegraphics[width=\linewidth]{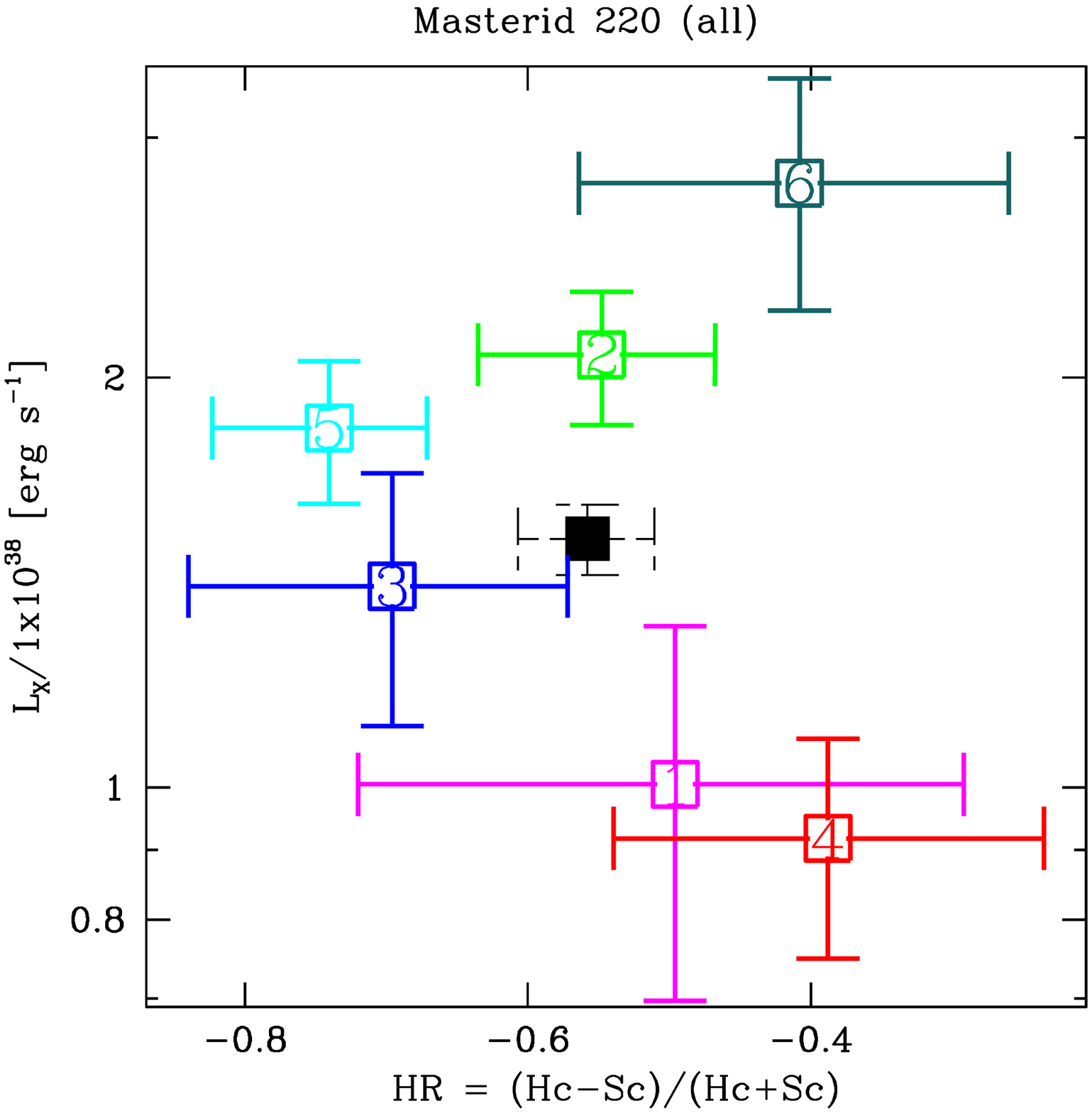}

\end{minipage}

\begin{minipage}{0.32\linewidth}
  \centering
  
    \includegraphics[width=\linewidth]{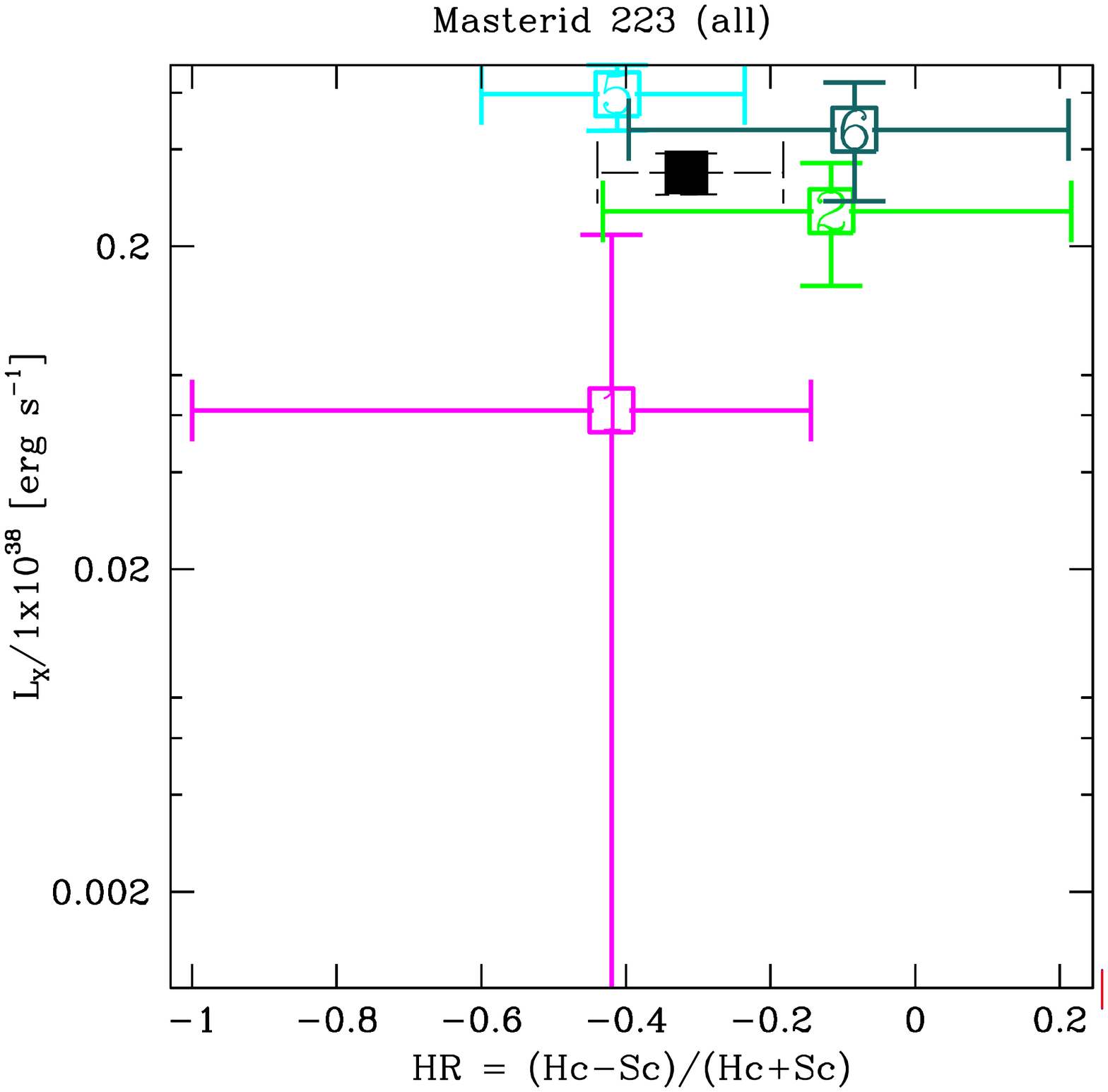}

  \end{minipage}
  \begin{minipage}{0.32\linewidth}
  \centering

    \includegraphics[width=\linewidth]{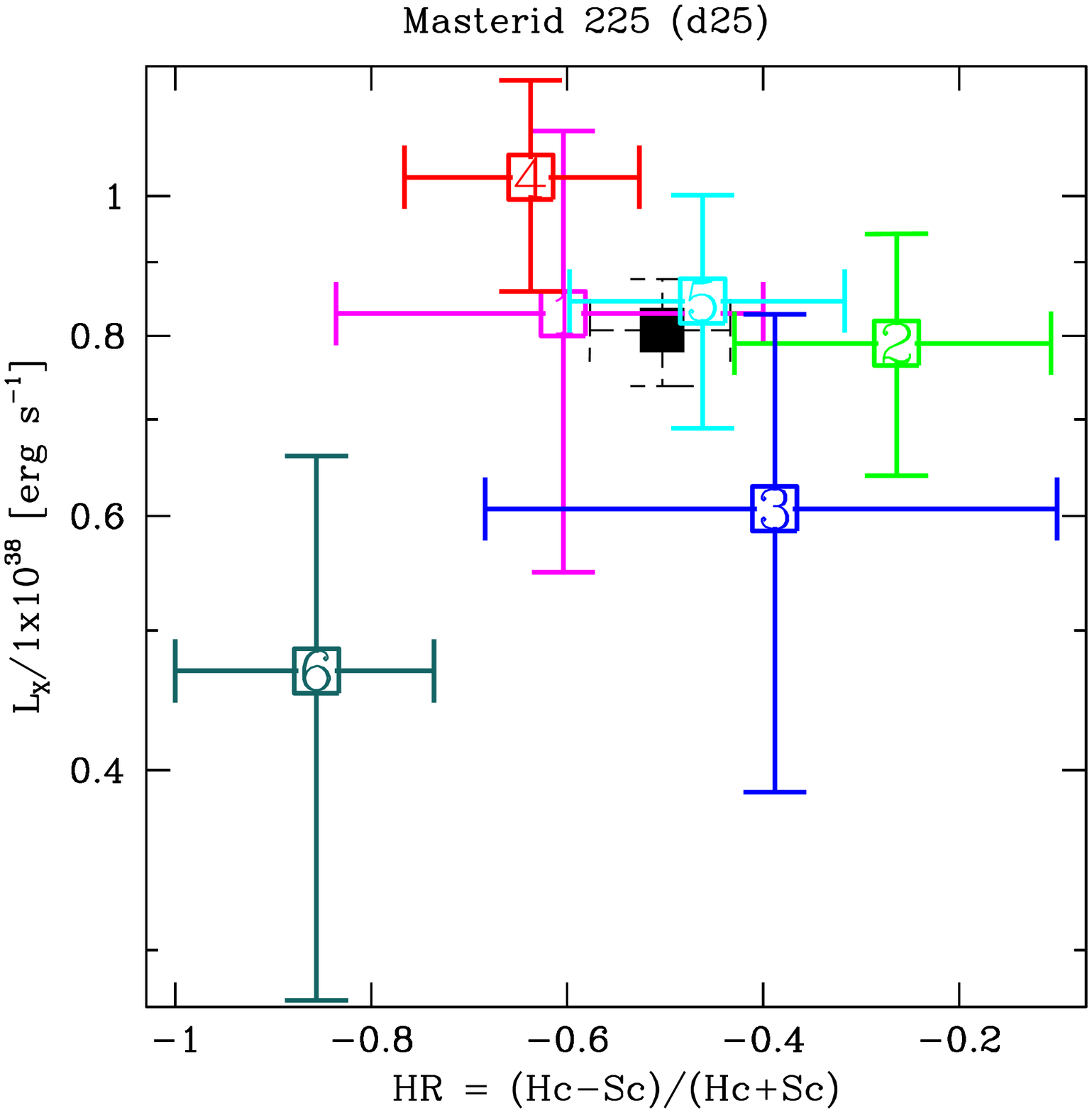}

\end{minipage}
\begin{minipage}{0.32\linewidth}
  \centering

    \includegraphics[width=\linewidth]{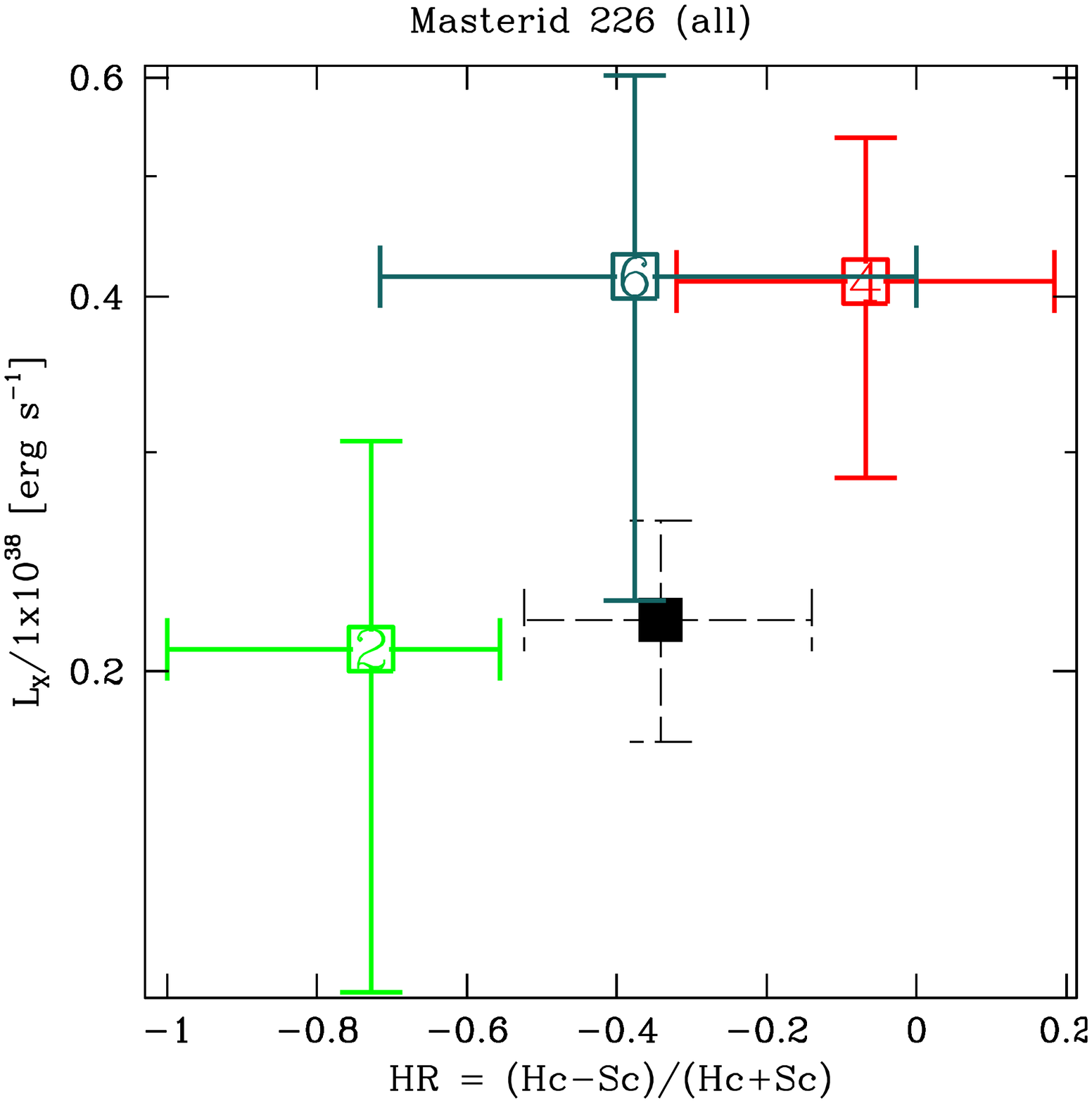}

\end{minipage}
\end{figure}

\clearpage

\begin{figure}
  \begin{minipage}{0.32\linewidth}
  \centering
  
    \includegraphics[width=\linewidth]{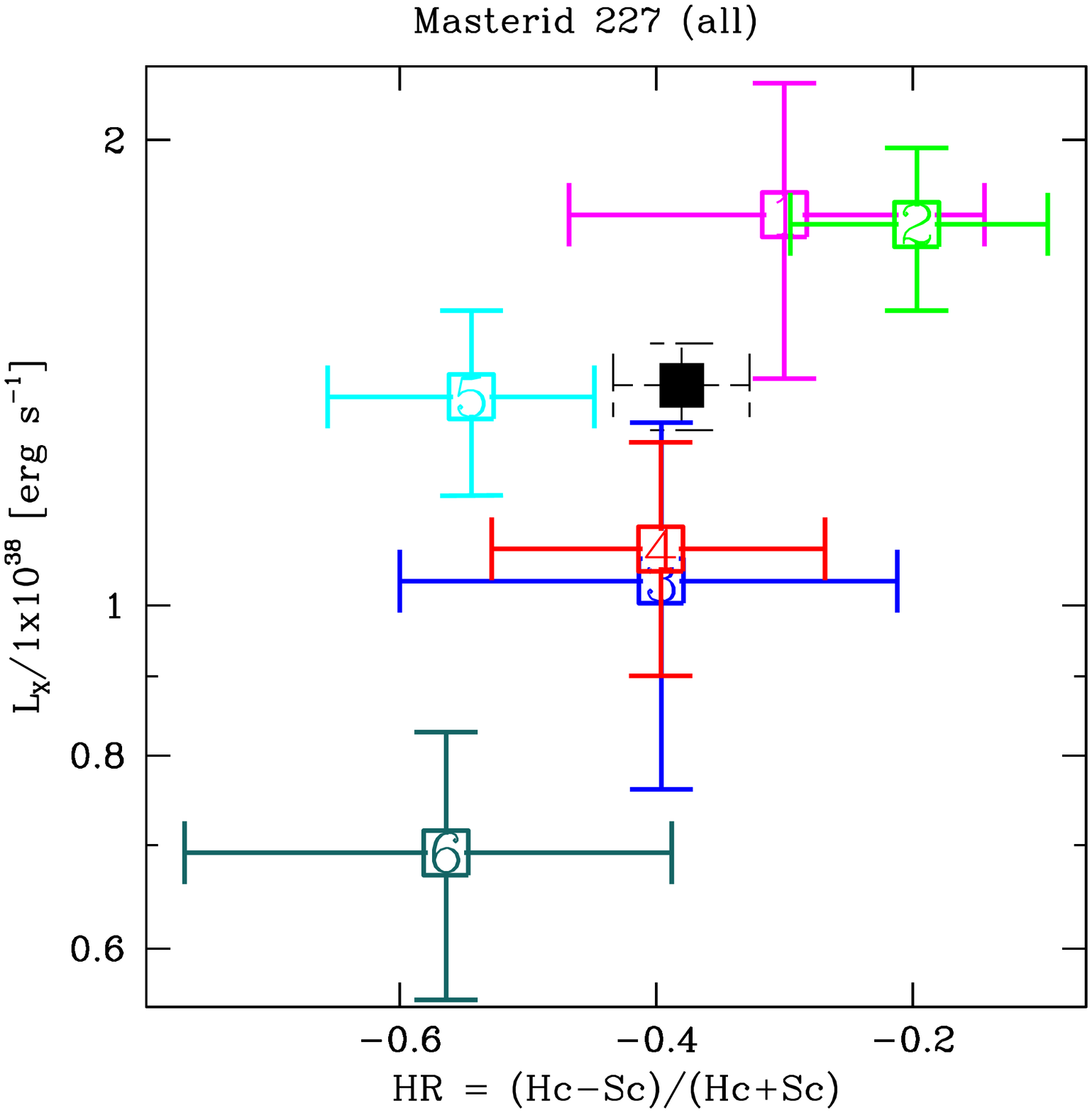}

  \end{minipage}
  \begin{minipage}{0.32\linewidth}
  \centering

    \includegraphics[width=\linewidth]{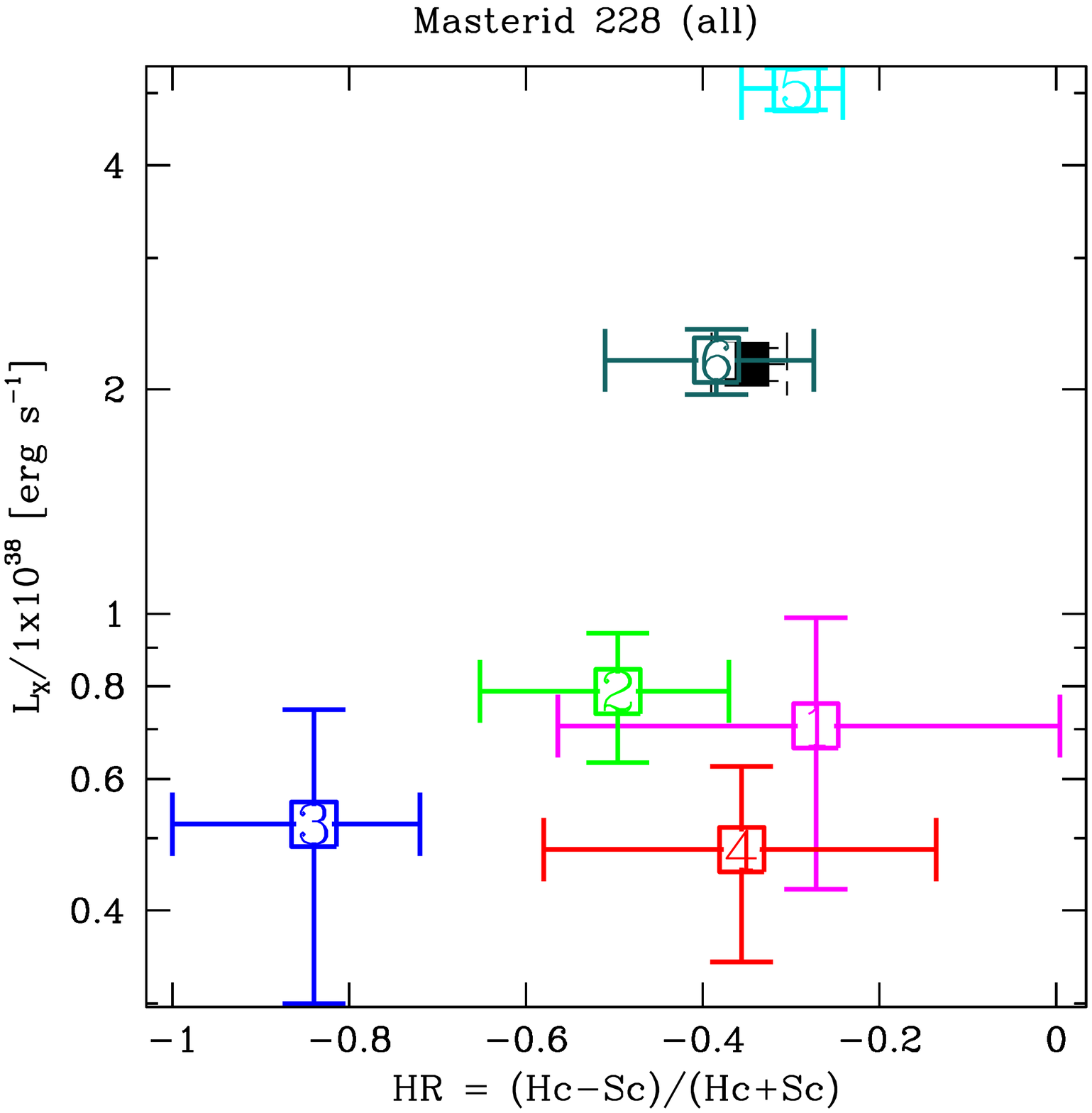}

\end{minipage}
\begin{minipage}{0.32\linewidth}
  \centering

    \includegraphics[width=\linewidth]{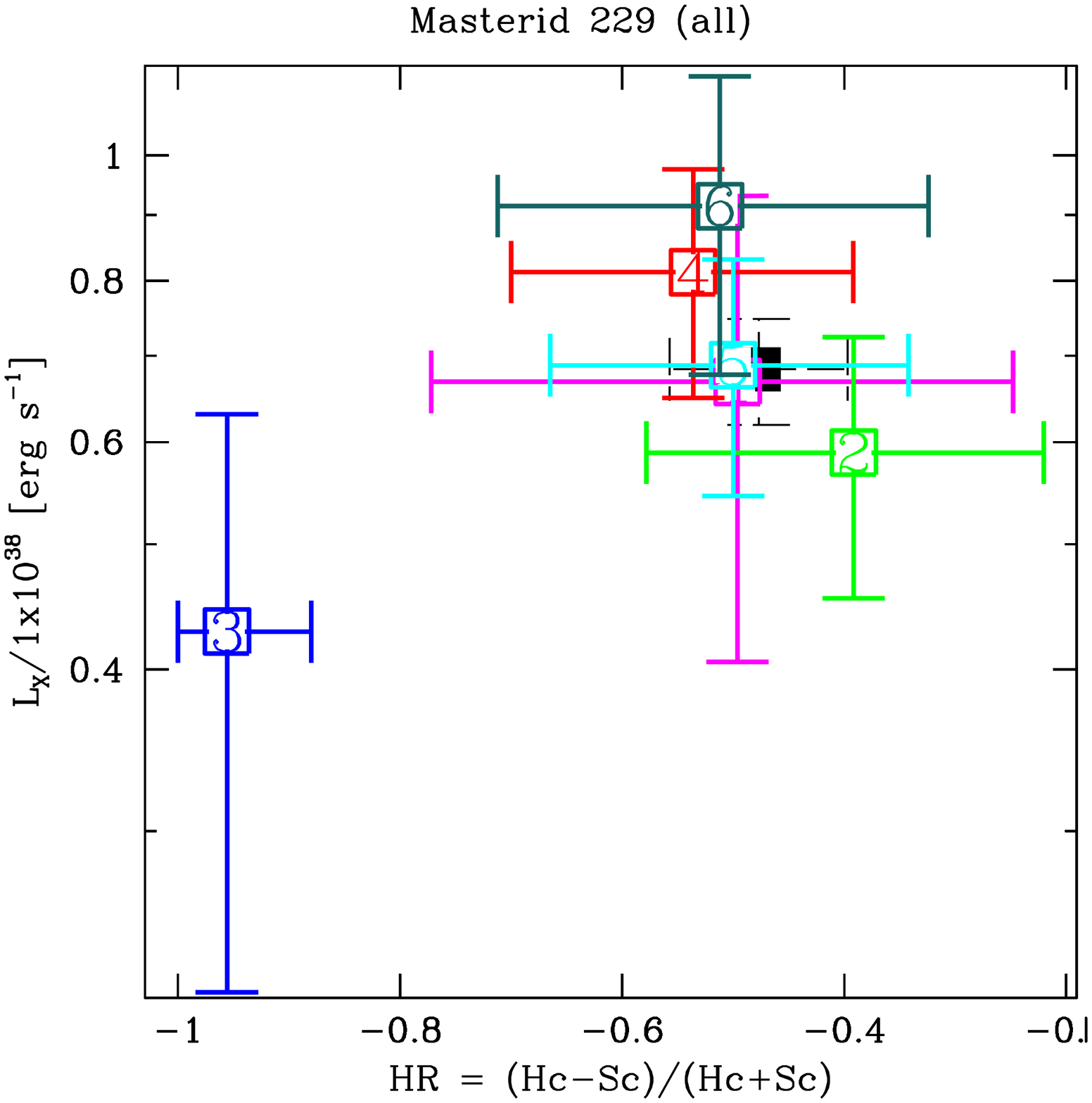}

 \end{minipage}

\begin{minipage}{0.32\linewidth}
  \centering
  
    \includegraphics[width=\linewidth]{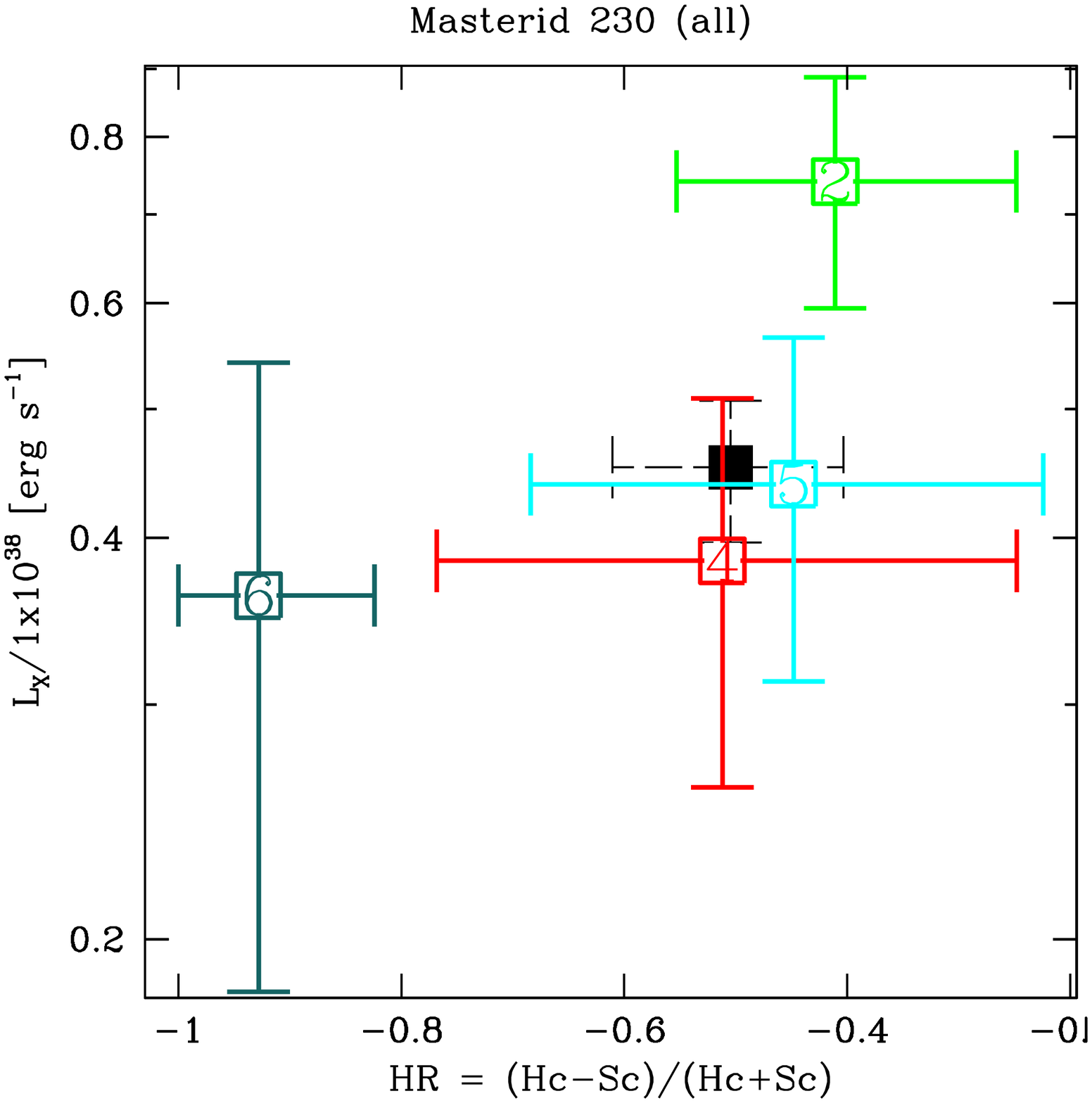}

  \end{minipage}
  \begin{minipage}{0.32\linewidth}
  \centering

    \includegraphics[width=\linewidth]{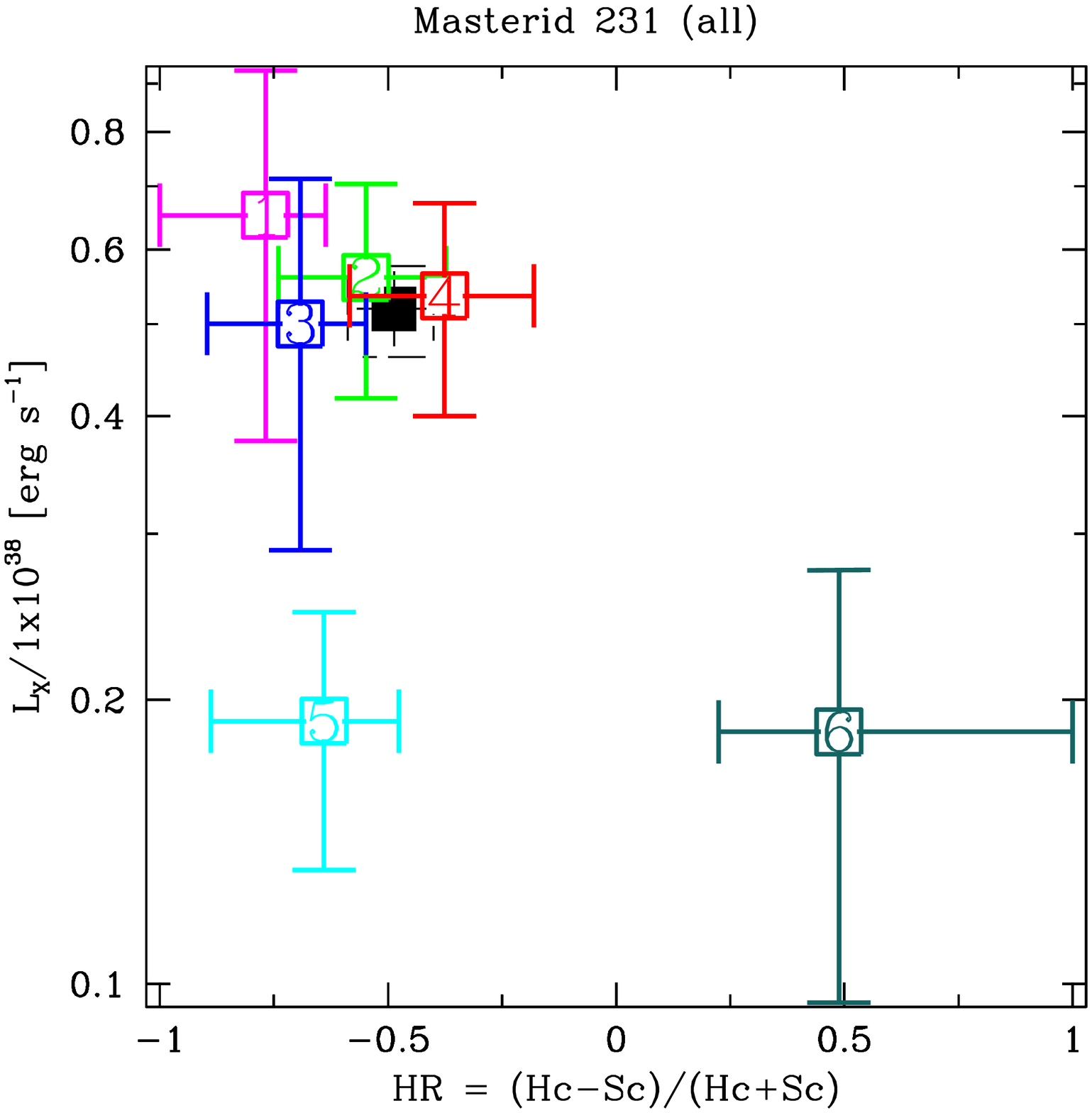}

\end{minipage}
\begin{minipage}{0.32\linewidth}
  \centering

    \includegraphics[width=\linewidth]{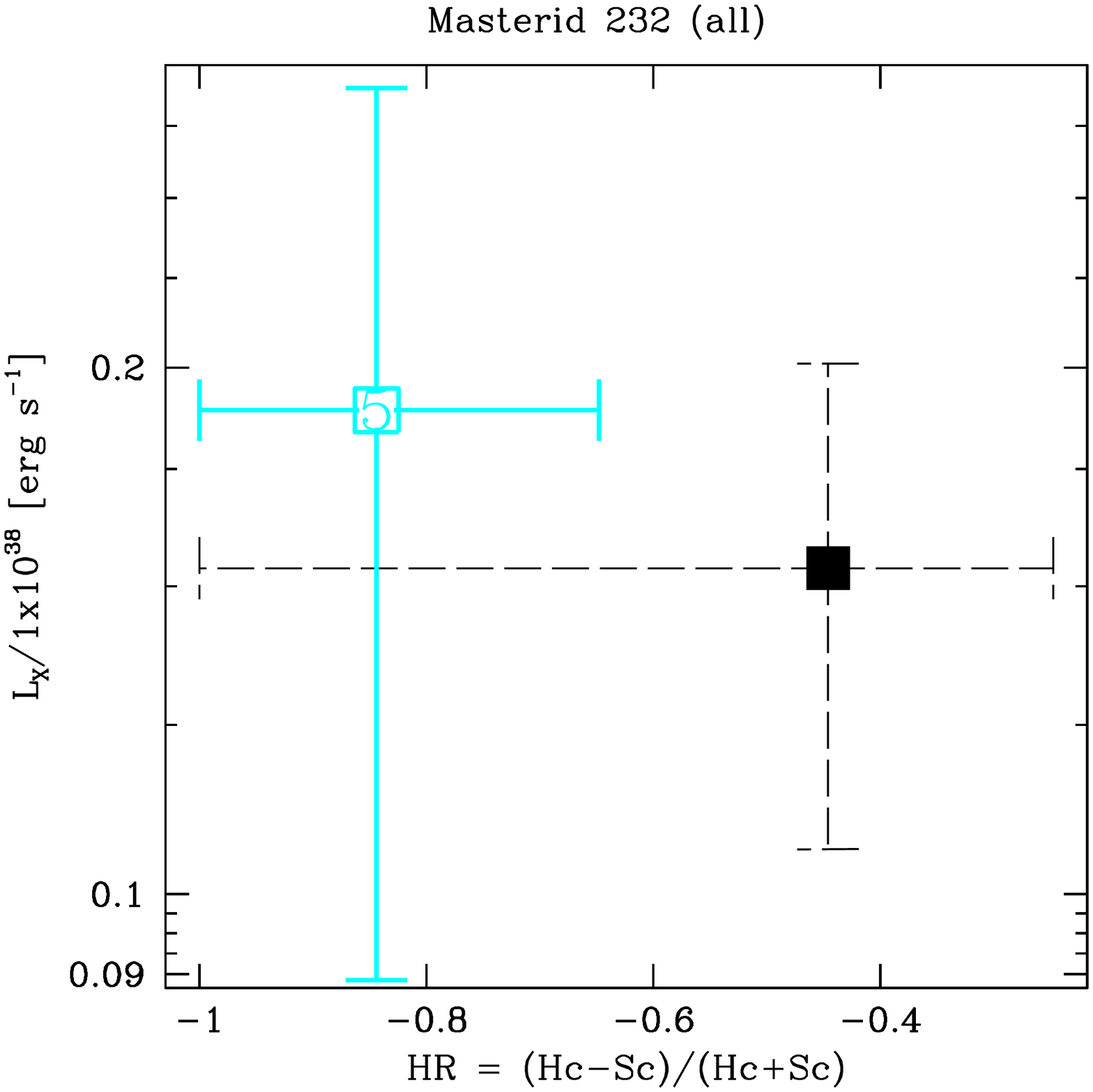}

 \end{minipage}

  \begin{minipage}{0.32\linewidth}
  \centering
  
    \includegraphics[width=\linewidth]{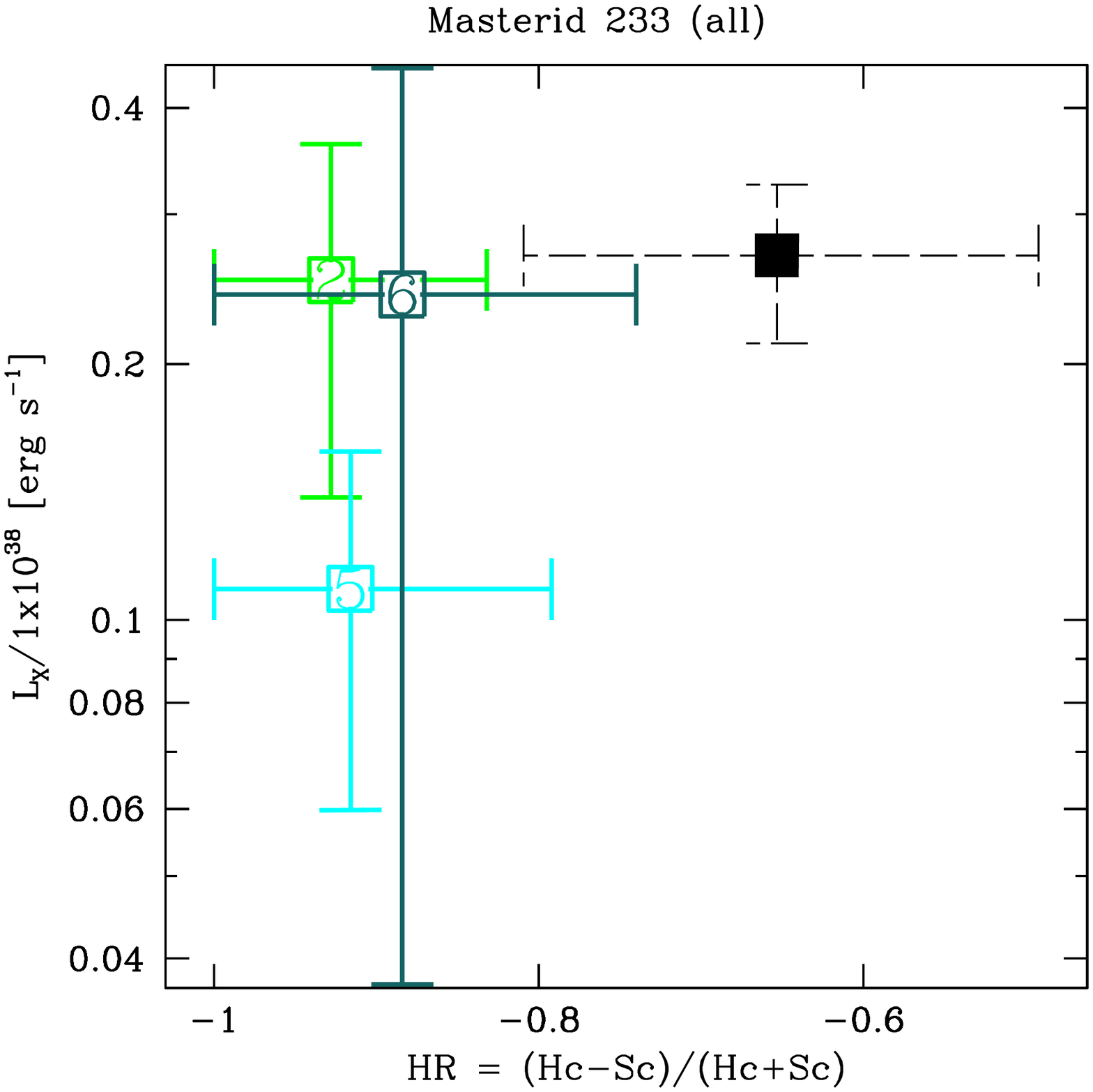}

  \end{minipage}
  \begin{minipage}{0.32\linewidth}
  \centering

    \includegraphics[width=\linewidth]{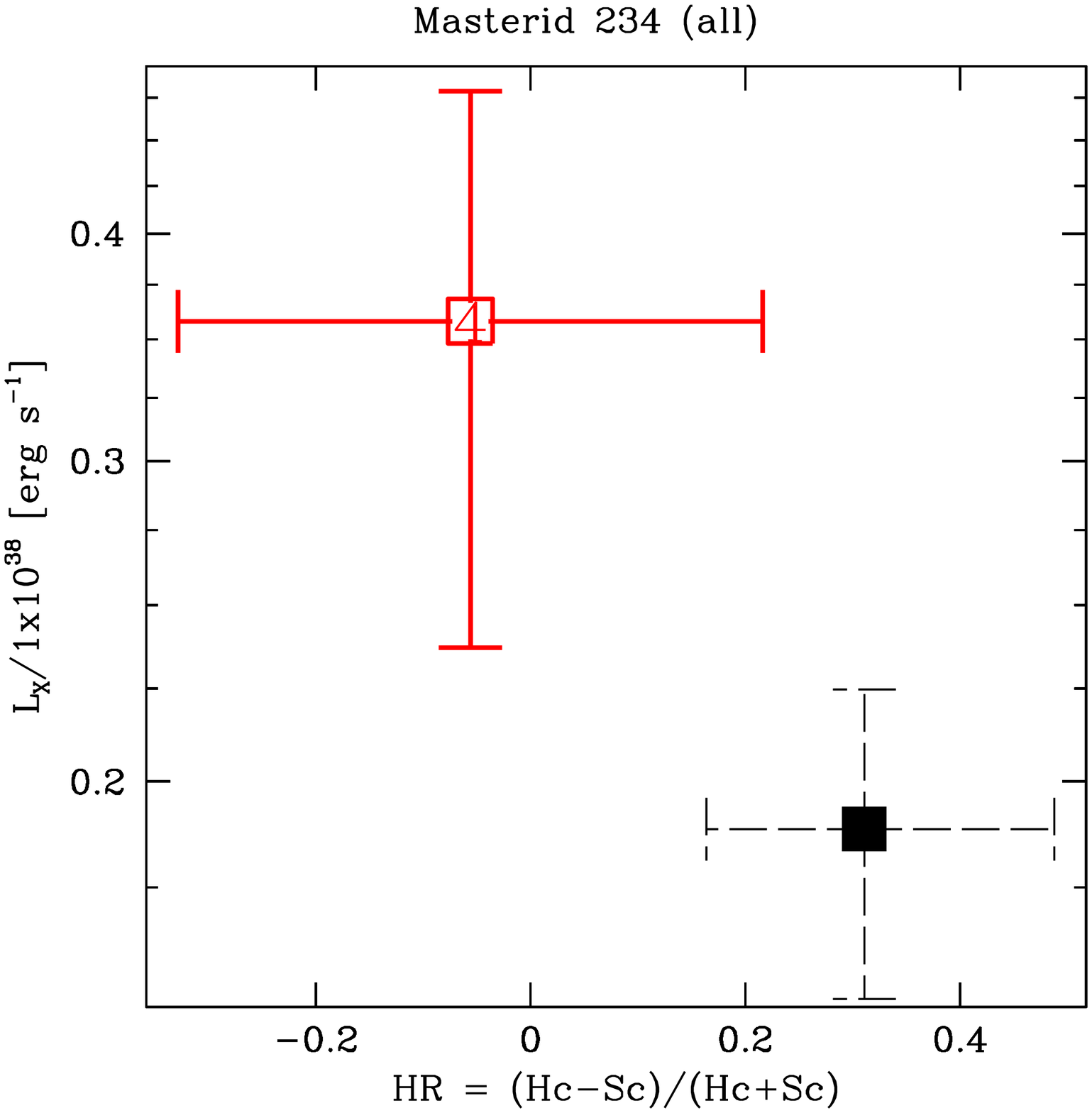}

\end{minipage}
 \begin{minipage}{0.32\linewidth}
  \centering

    \includegraphics[width=\linewidth]{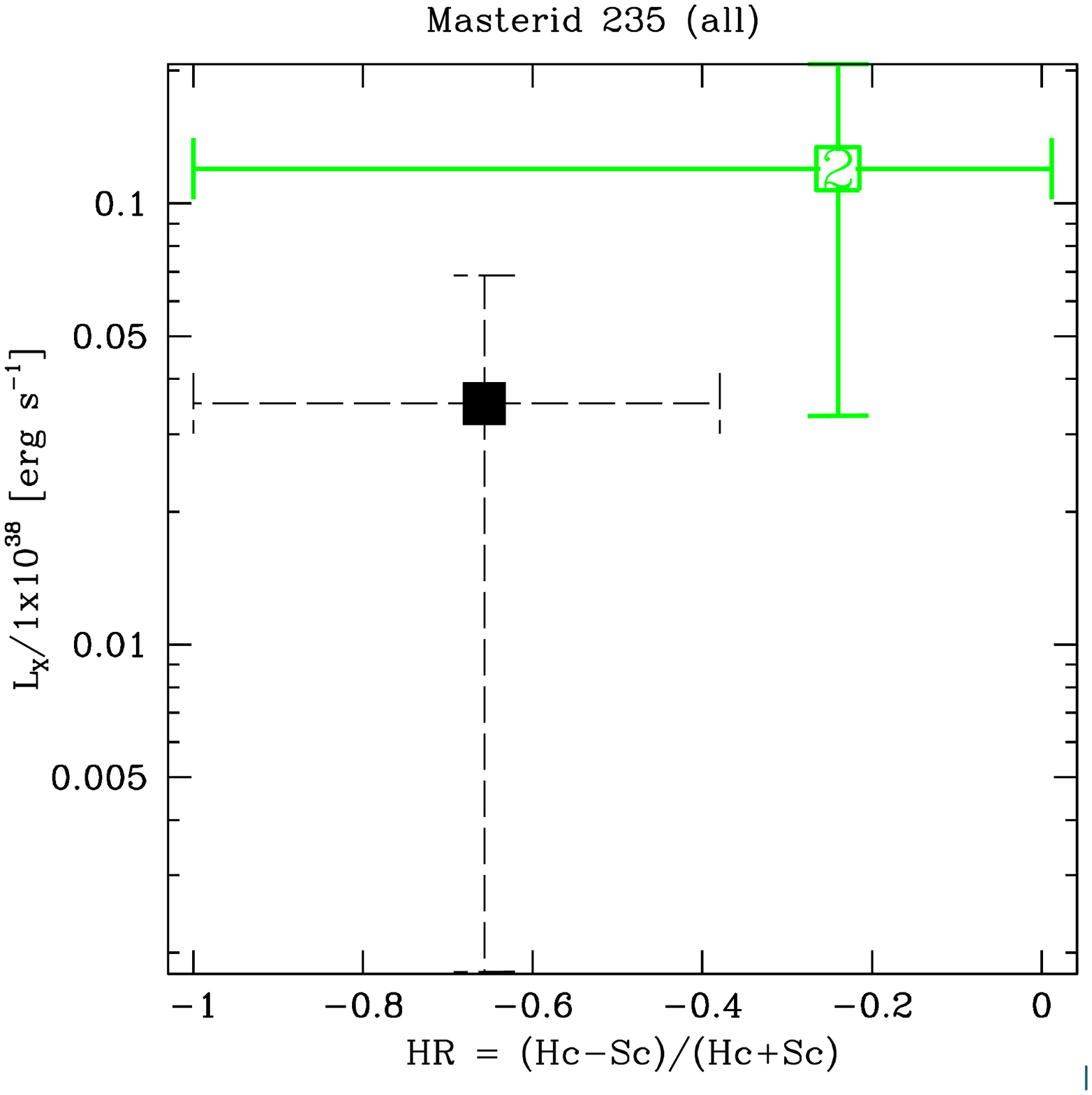}

\end{minipage}
\end{figure}

\begin{figure}
  \begin{minipage}{0.32\linewidth}
  \centering
  
    \includegraphics[width=\linewidth]{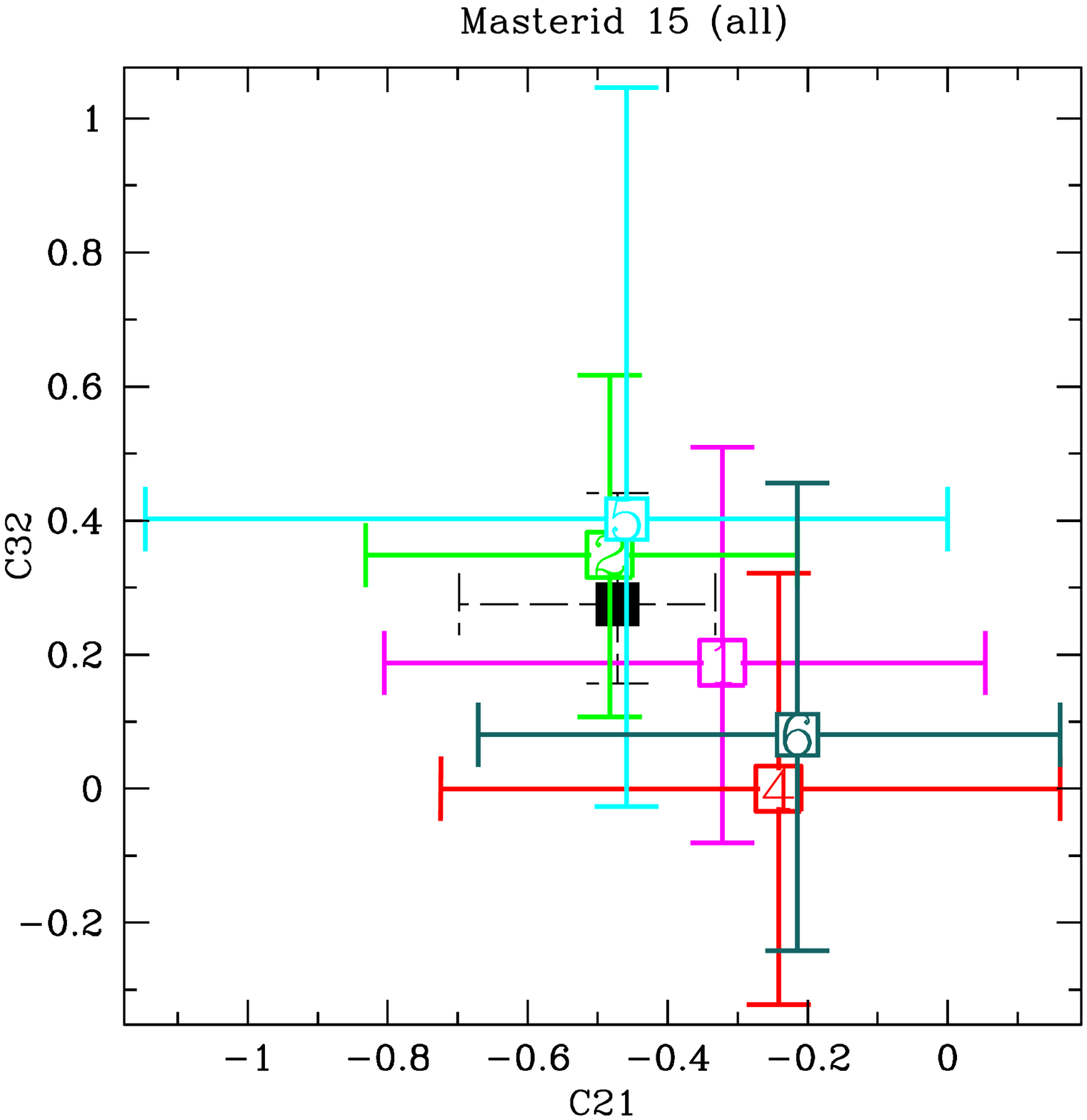}

  \end{minipage}
  \begin{minipage}{0.32\linewidth}
  \centering

    \includegraphics[width=\linewidth]{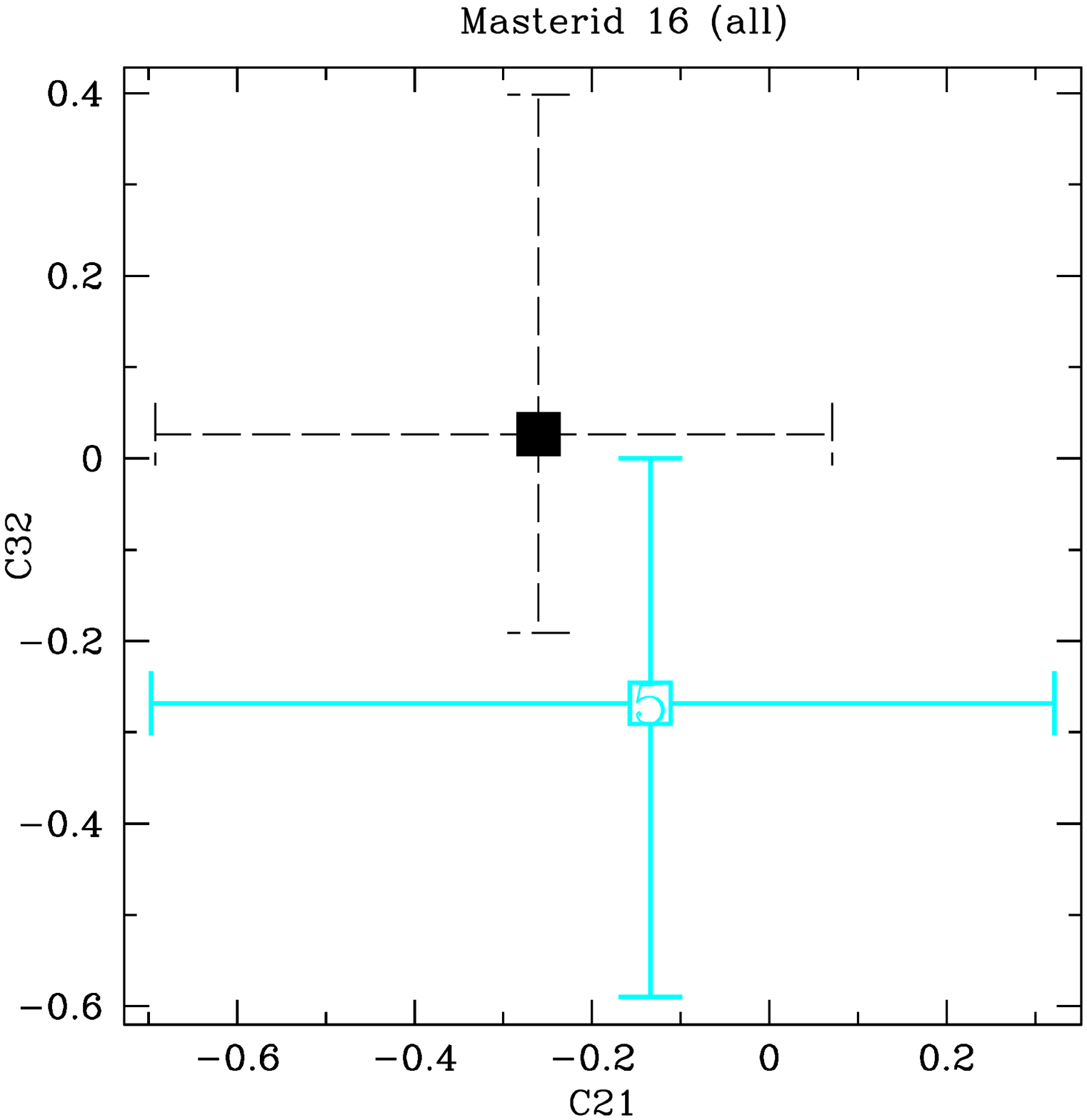}

\end{minipage}
\begin{minipage}{0.32\linewidth}
  \centering

    \includegraphics[width=\linewidth]{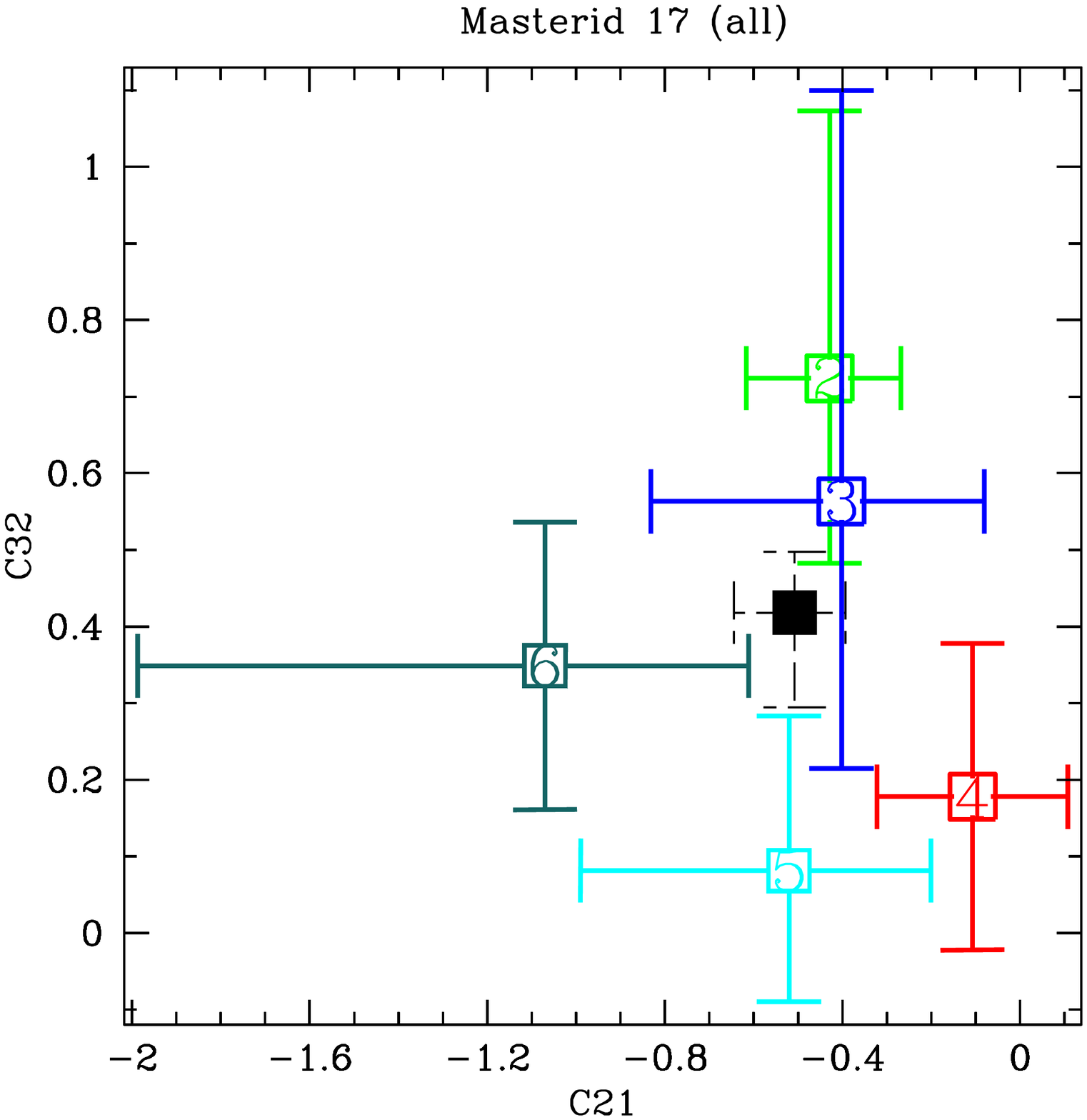}

 \end{minipage}

\begin{minipage}{0.32\linewidth}
  \centering
  
    \includegraphics[width=\linewidth]{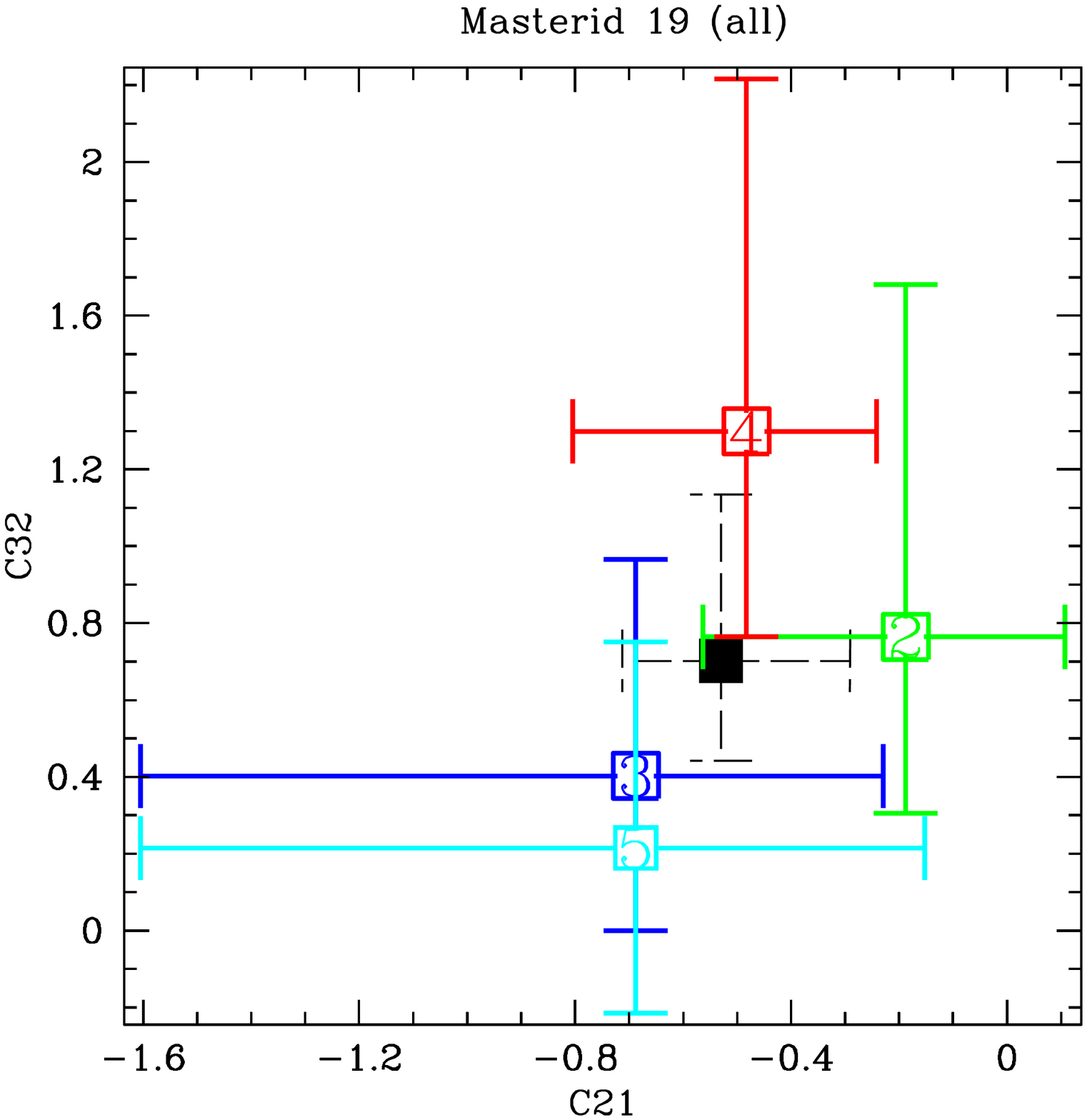}

  \end{minipage}
  \begin{minipage}{0.32\linewidth}
  \centering

    \includegraphics[width=\linewidth]{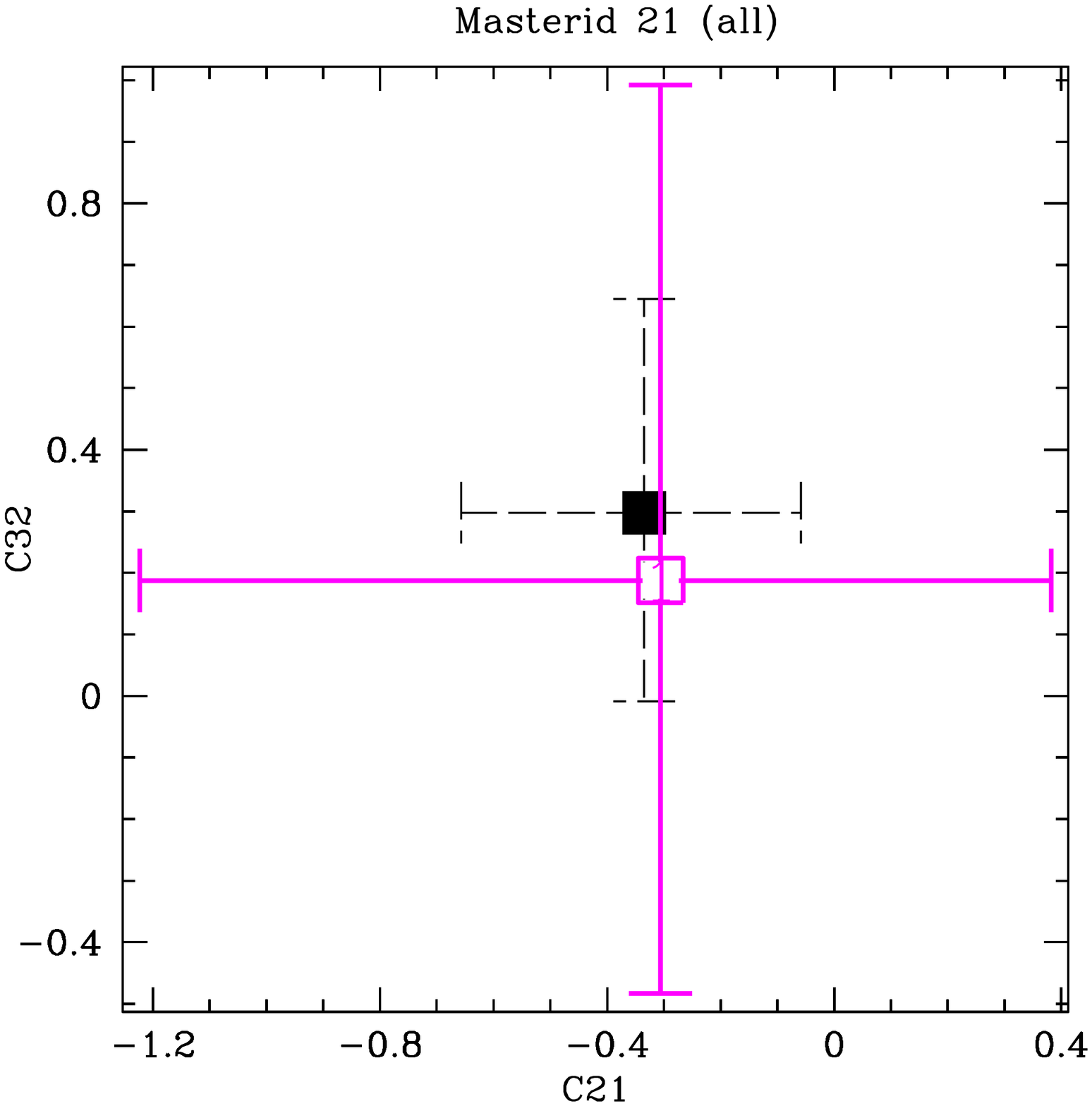}

\end{minipage}
\begin{minipage}{0.32\linewidth}
  \centering

    \includegraphics[width=\linewidth]{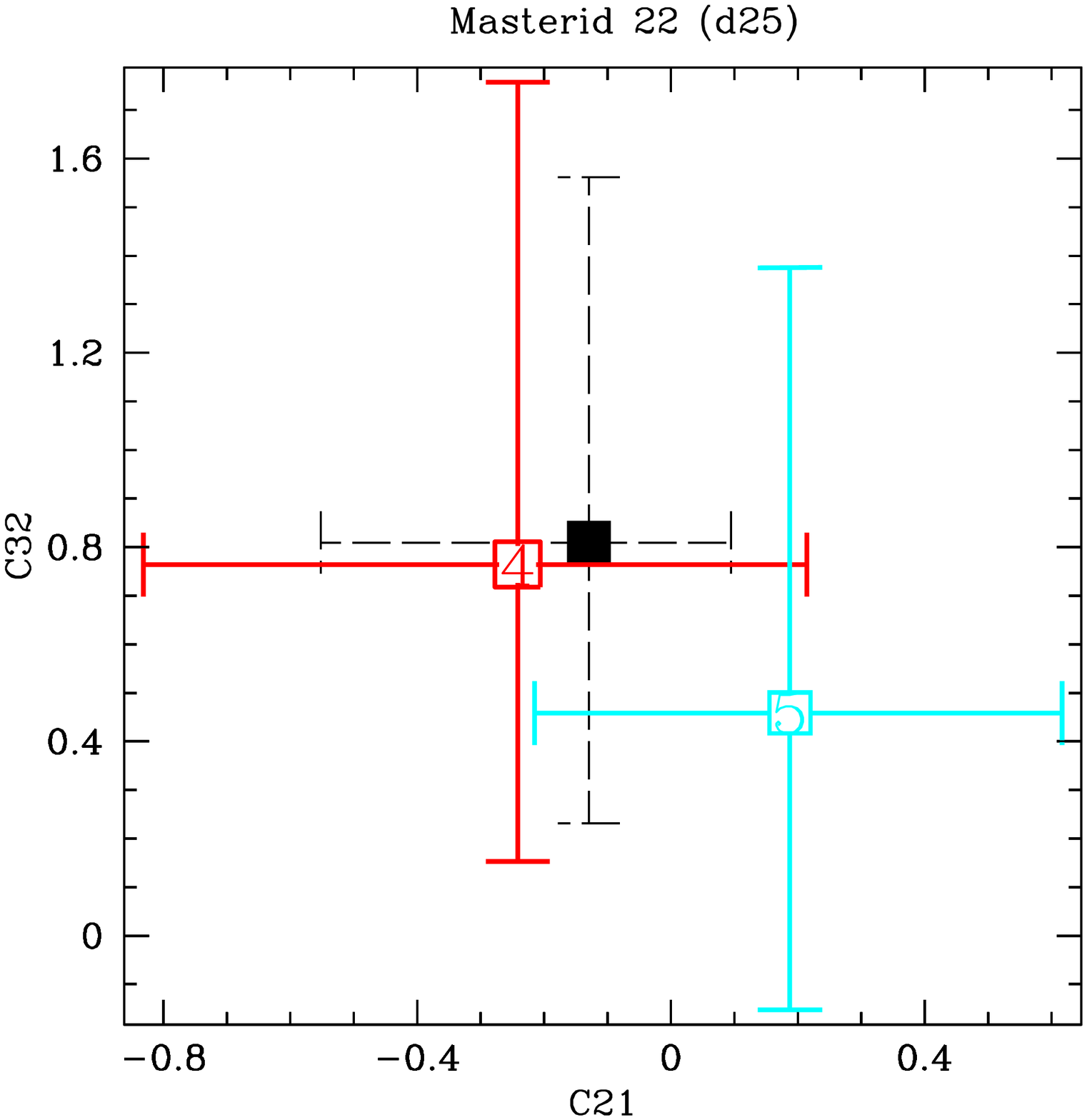}

 \end{minipage}

  \begin{minipage}{0.32\linewidth}
  \centering
  
    \includegraphics[width=\linewidth]{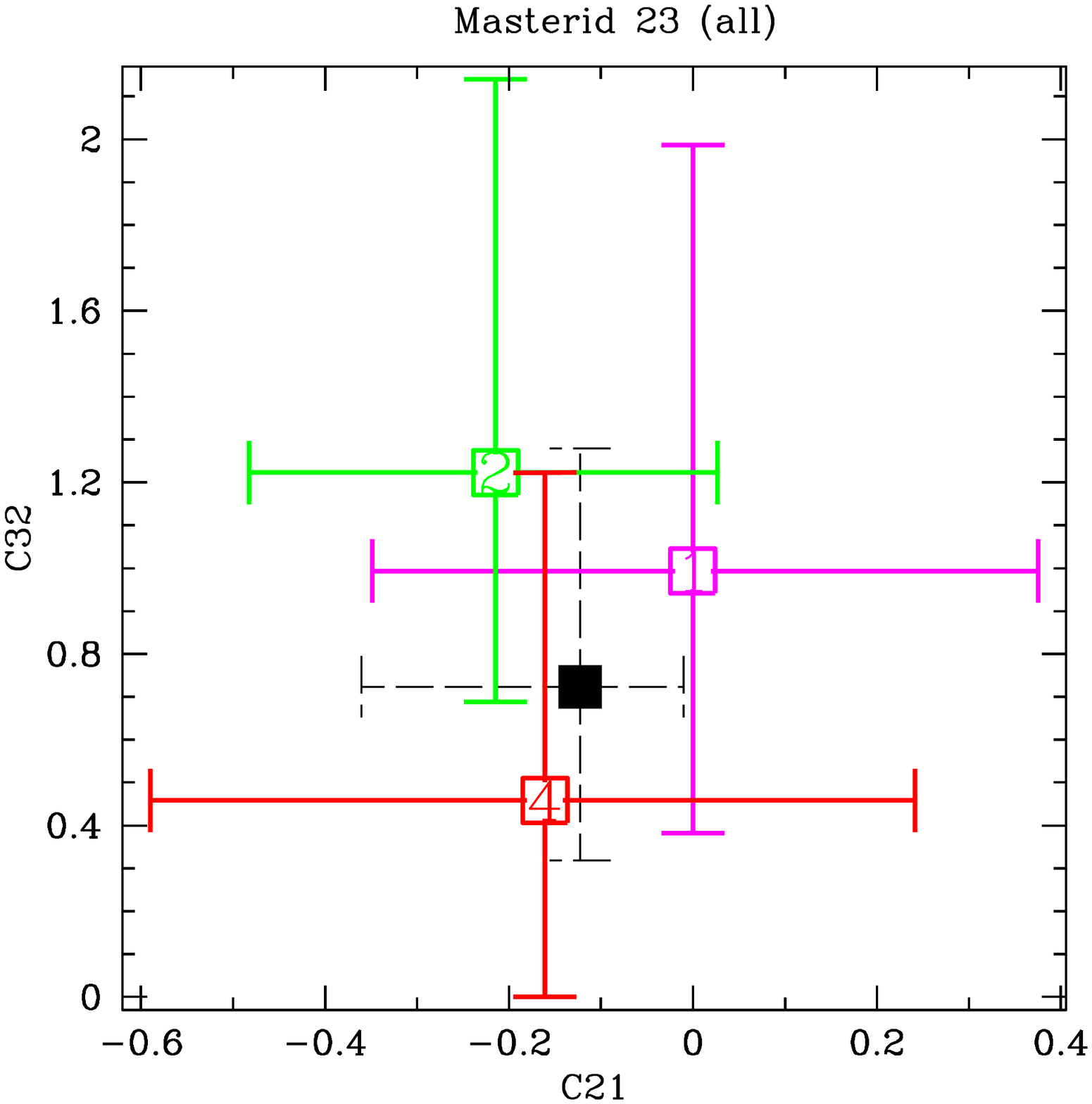}

  \end{minipage}
  \begin{minipage}{0.32\linewidth}
  \centering

    \includegraphics[width=\linewidth]{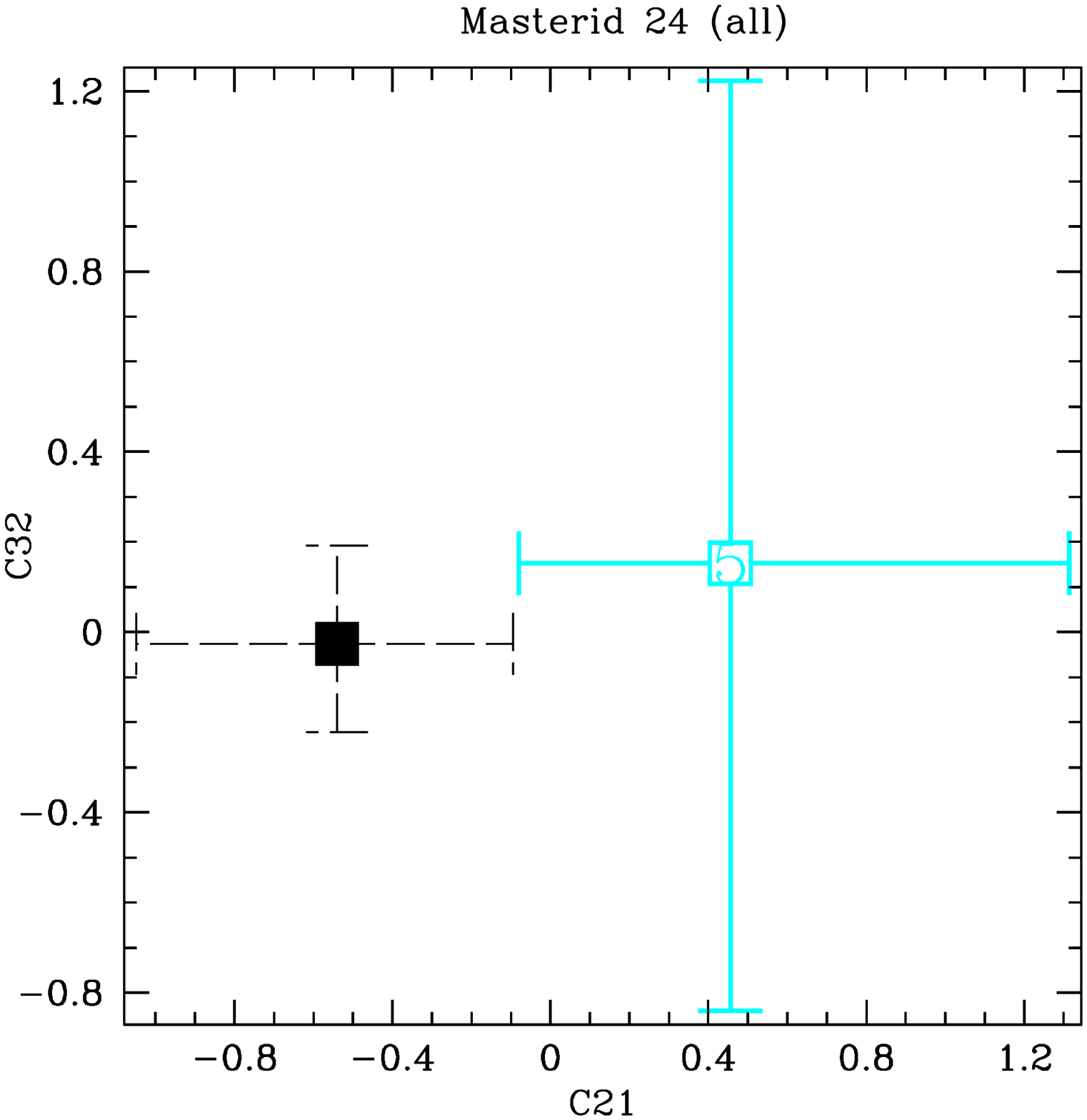}

\end{minipage}
\begin{minipage}{0.32\linewidth}
  \centering

    \includegraphics[width=\linewidth]{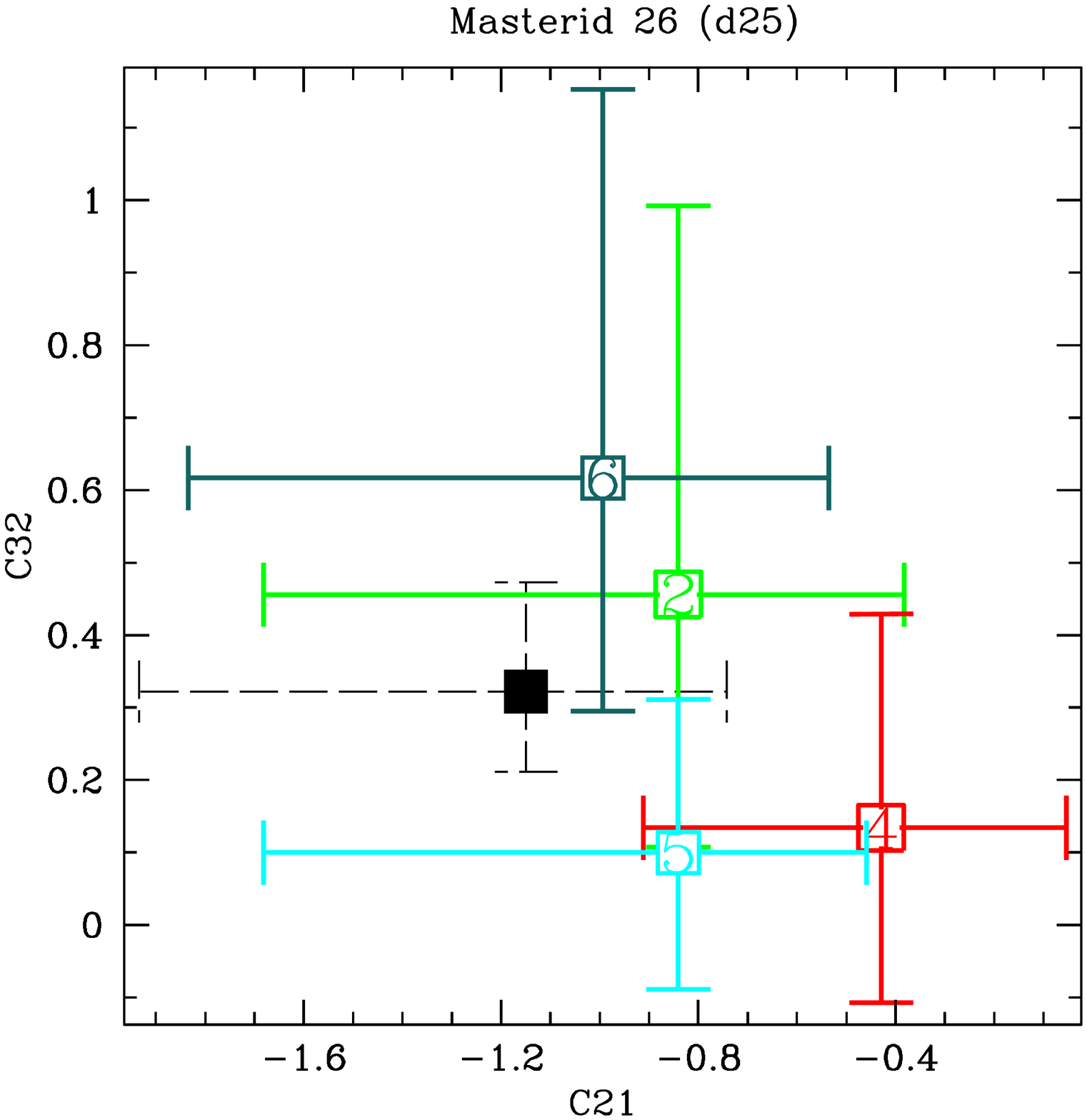}

 \end{minipage}

\begin{minipage}{0.32\linewidth}
  \centering
  
    \includegraphics[width=\linewidth]{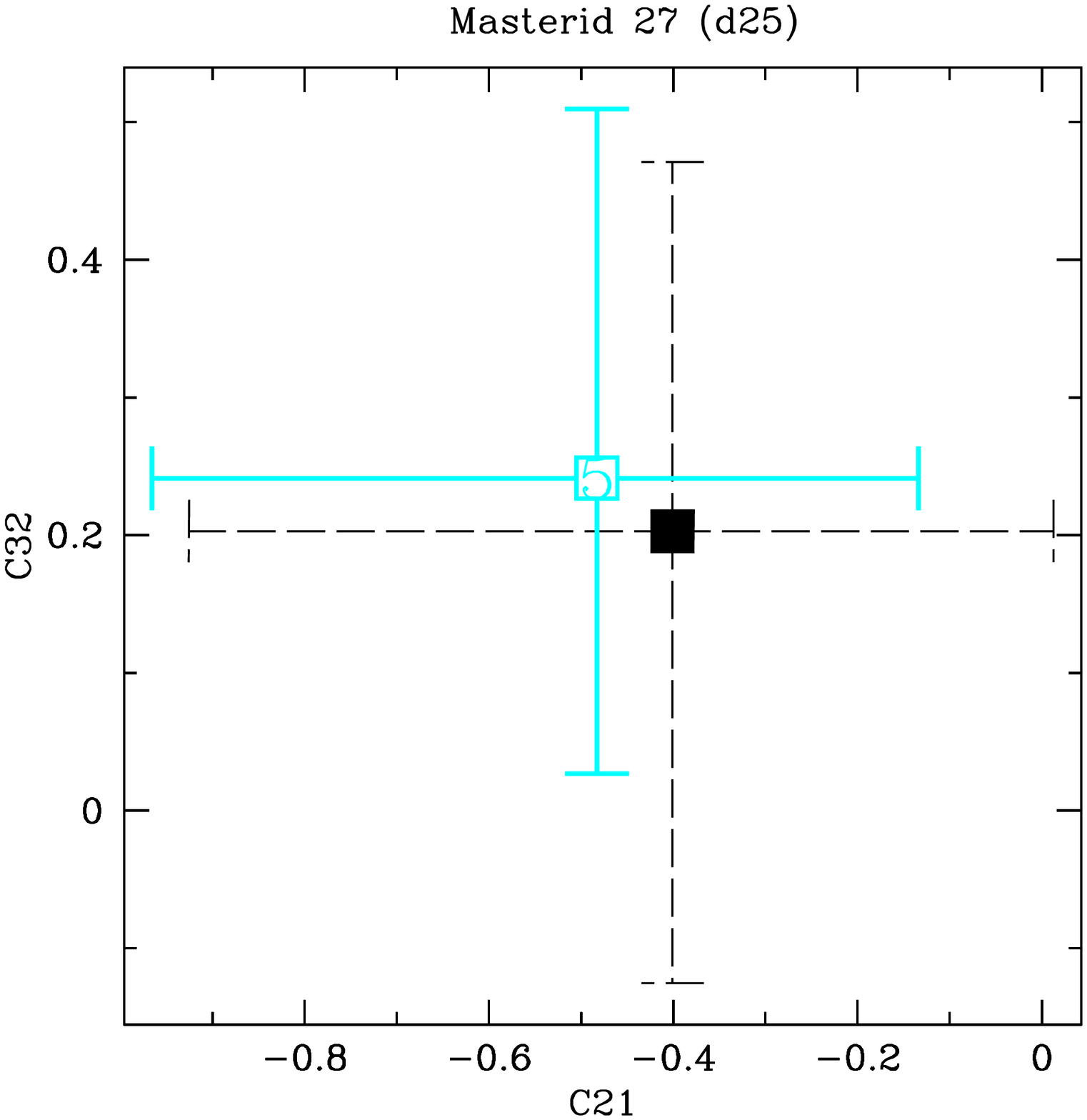}

  \end{minipage}
  \begin{minipage}{0.32\linewidth}
  \centering

    \includegraphics[width=\linewidth]{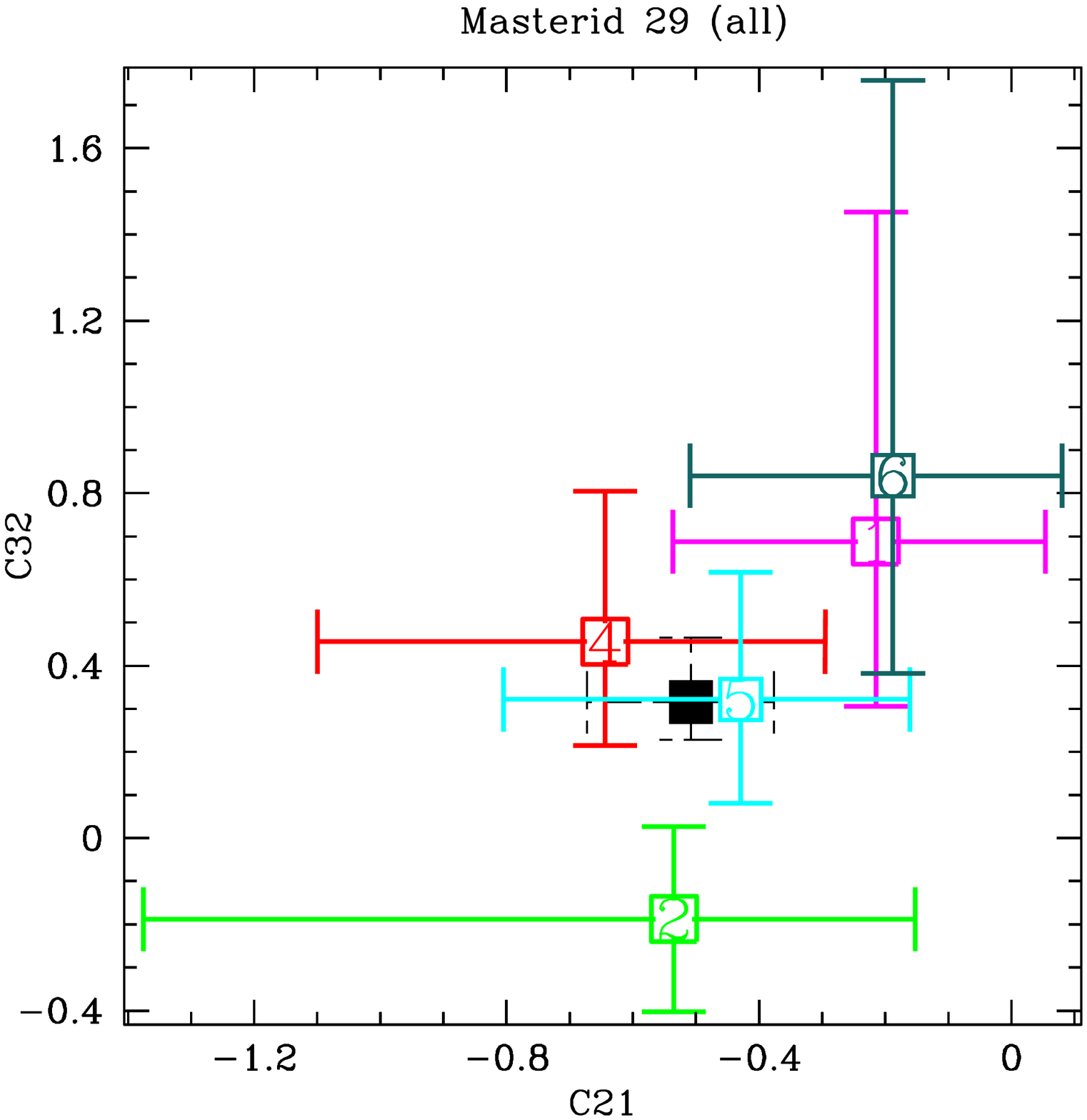}

\end{minipage}
\begin{minipage}{0.32\linewidth}
  \centering

    \includegraphics[width=\linewidth]{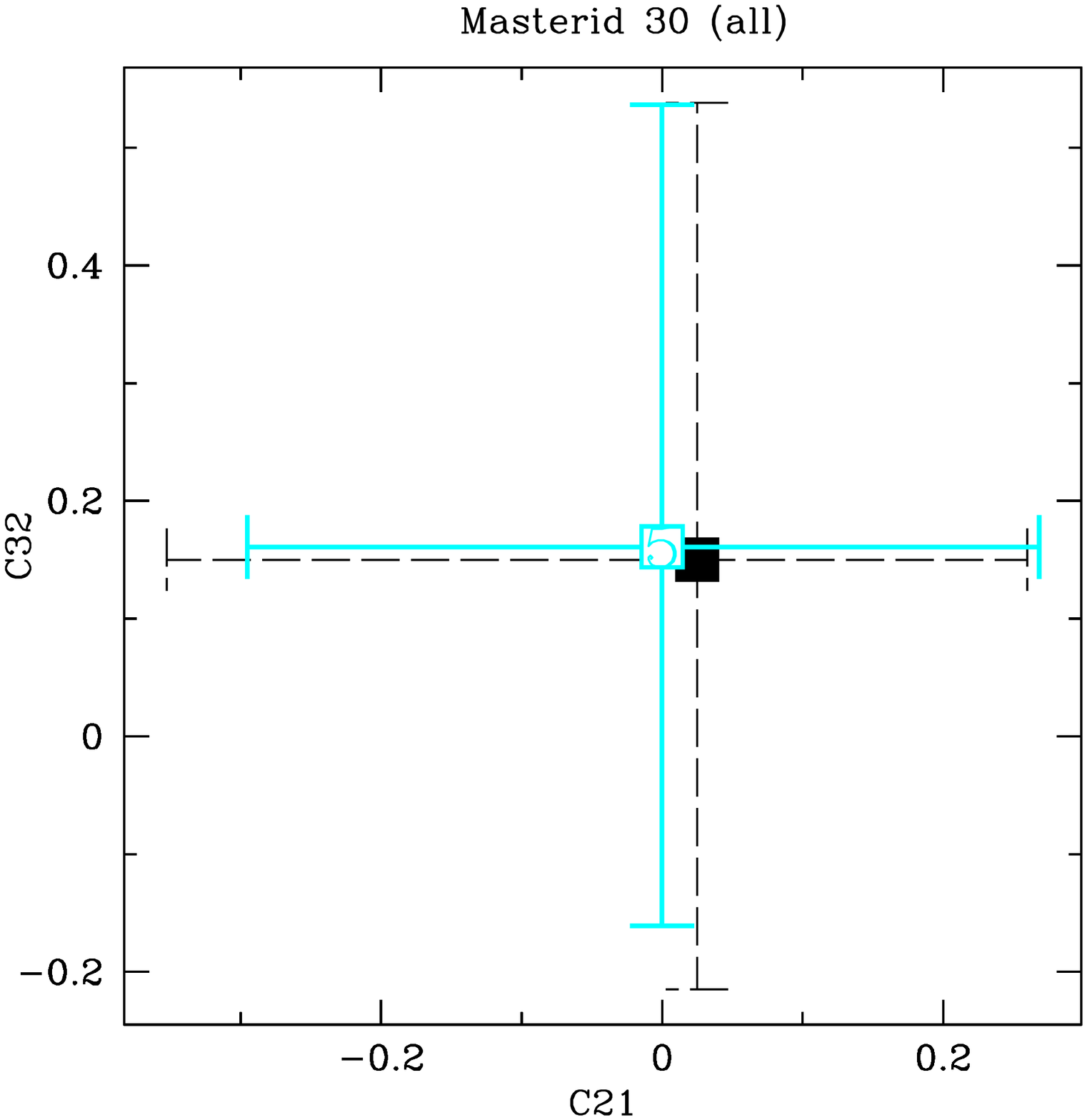}

 \end{minipage}
  
\end{figure}

\begin{figure}
  \begin{minipage}{0.32\linewidth}
  \centering
  
    \includegraphics[width=\linewidth]{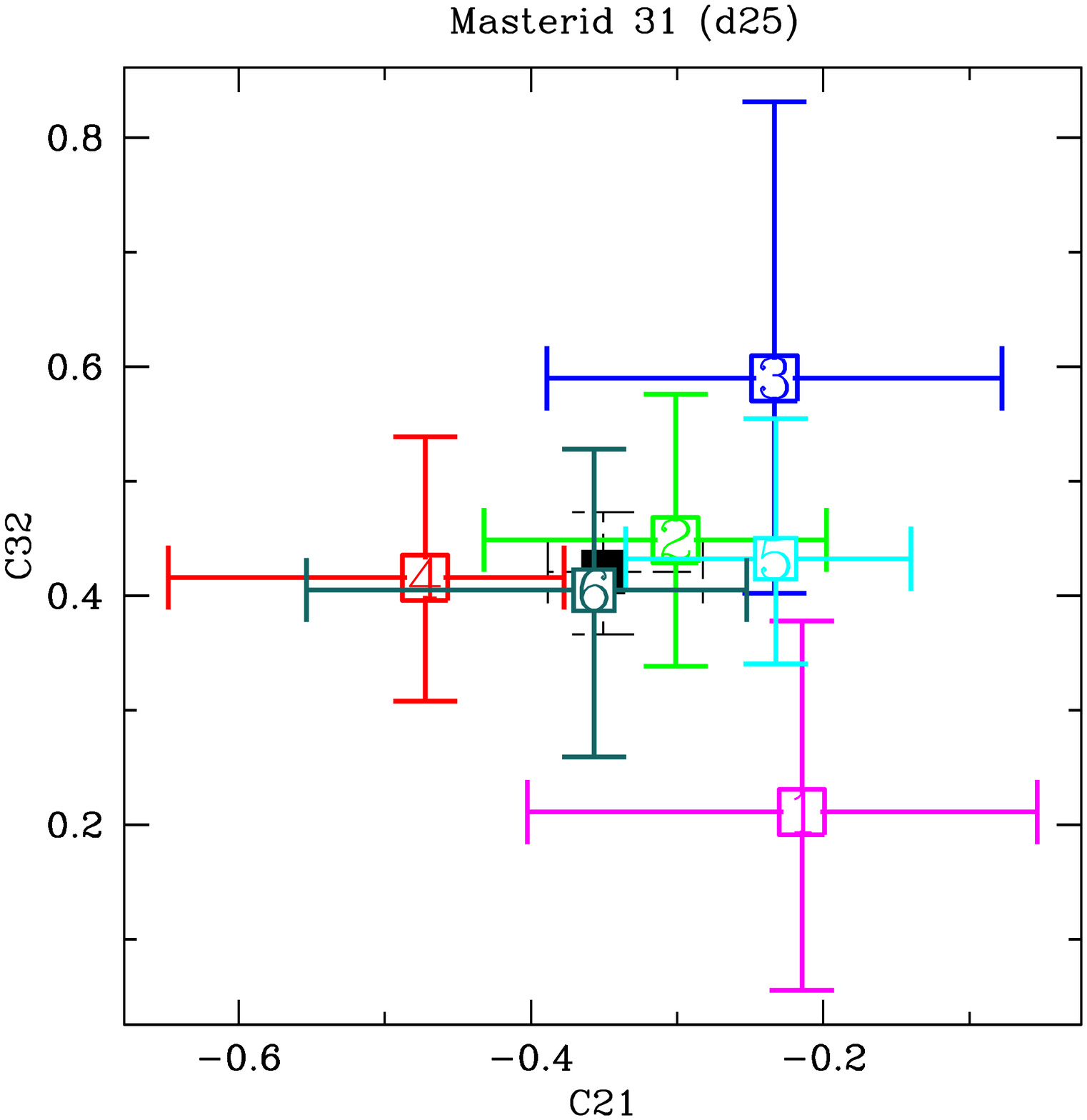}

  \end{minipage}
  \begin{minipage}{0.32\linewidth}
  \centering

    \includegraphics[width=\linewidth]{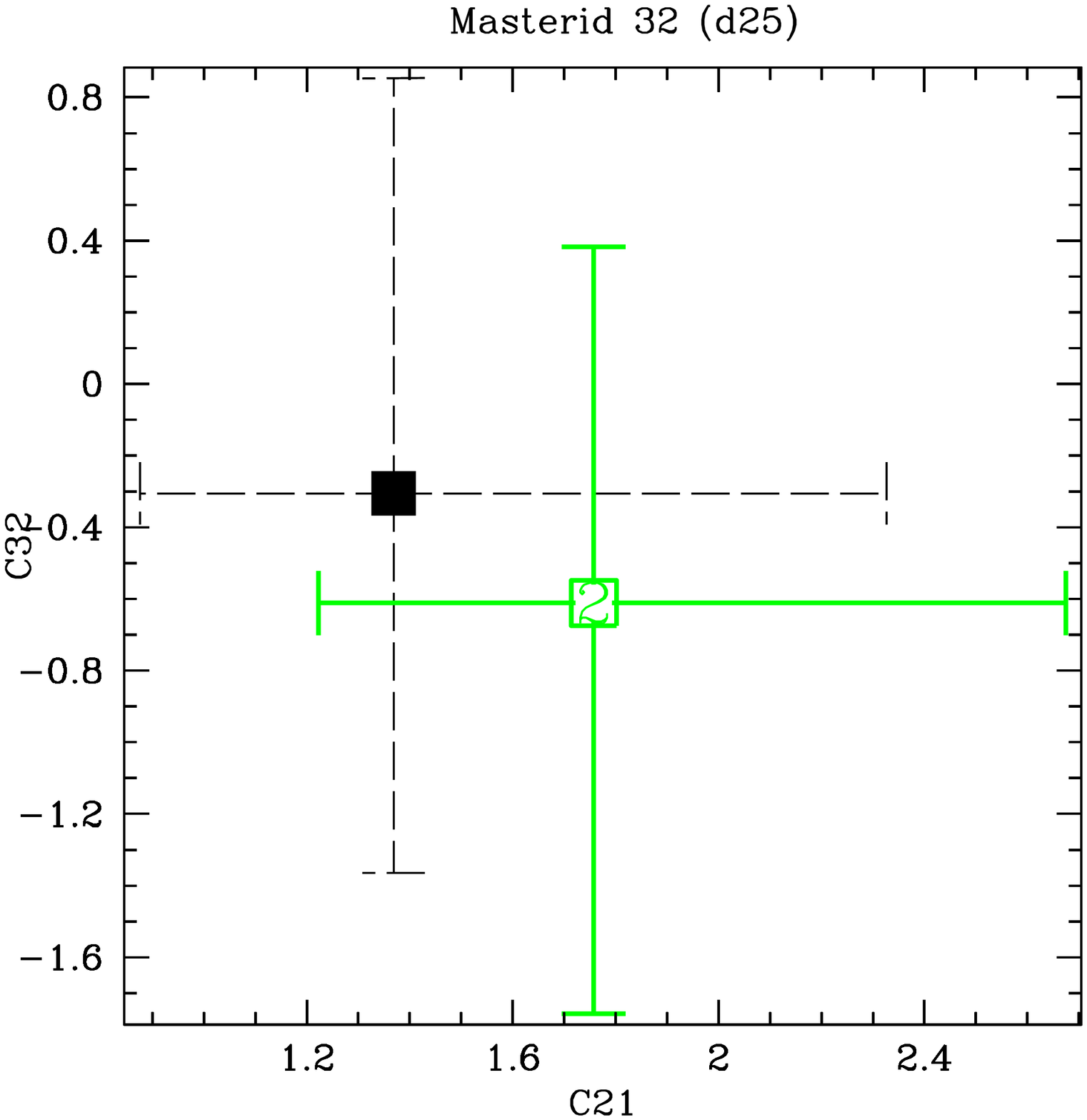}

\end{minipage}
\begin{minipage}{0.32\linewidth}
  \centering

    \includegraphics[width=\linewidth]{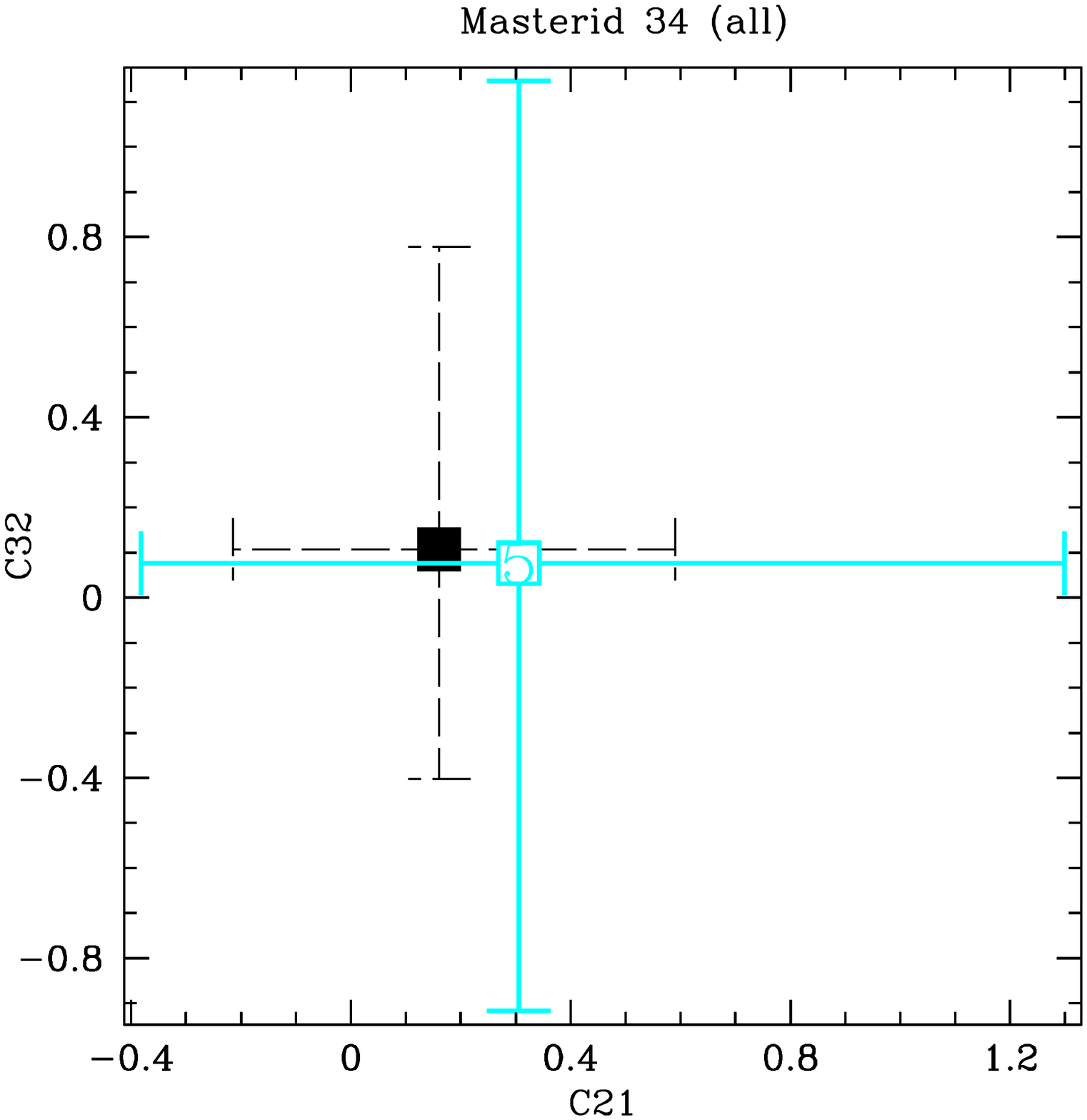}

 \end{minipage}

\begin{minipage}{0.32\linewidth}
  \centering
  
    \includegraphics[width=\linewidth]{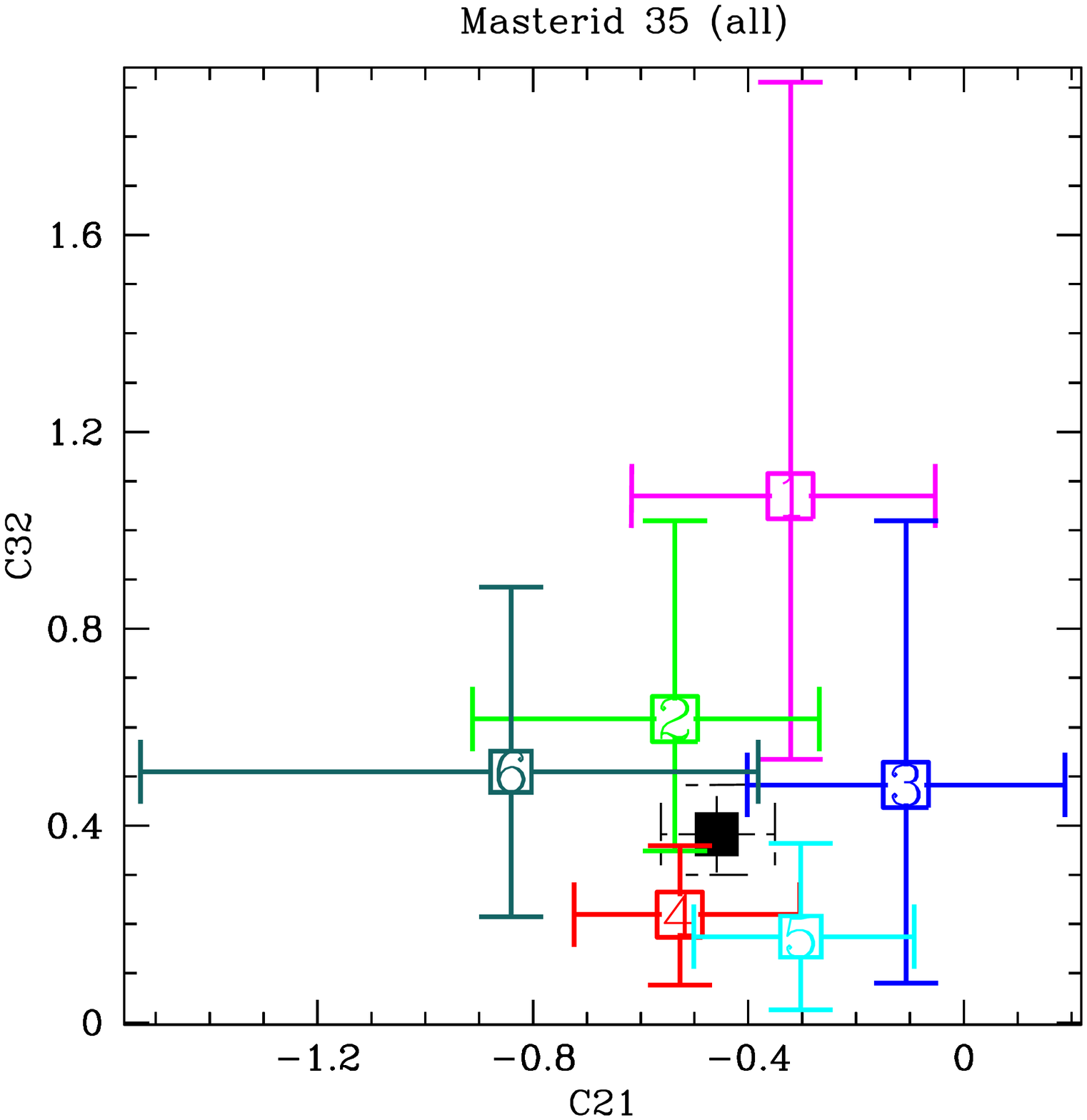}

  \end{minipage}
  \begin{minipage}{0.32\linewidth}
  \centering

    \includegraphics[width=\linewidth]{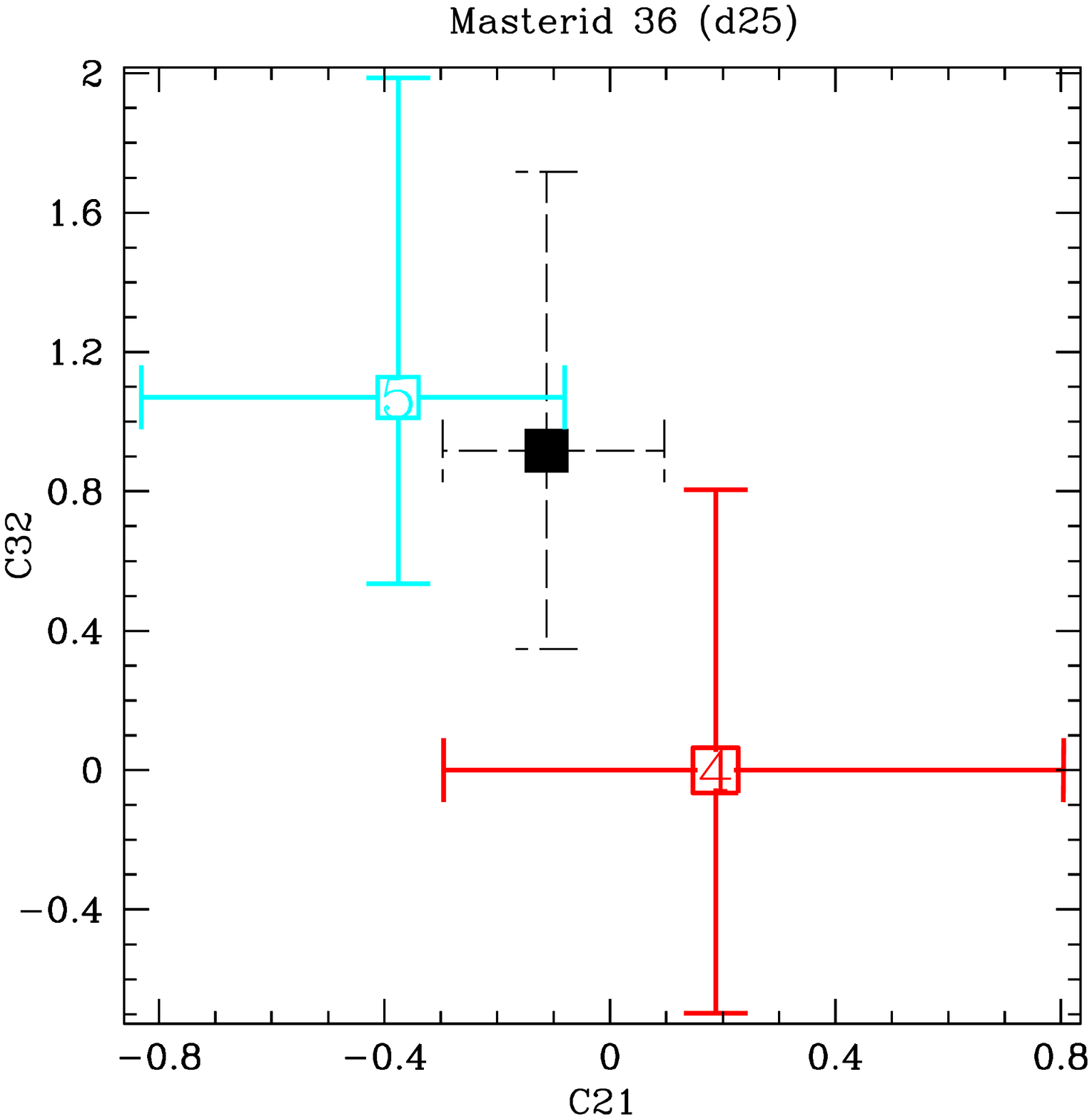}

\end{minipage}
\begin{minipage}{0.32\linewidth}
  \centering

    \includegraphics[width=\linewidth]{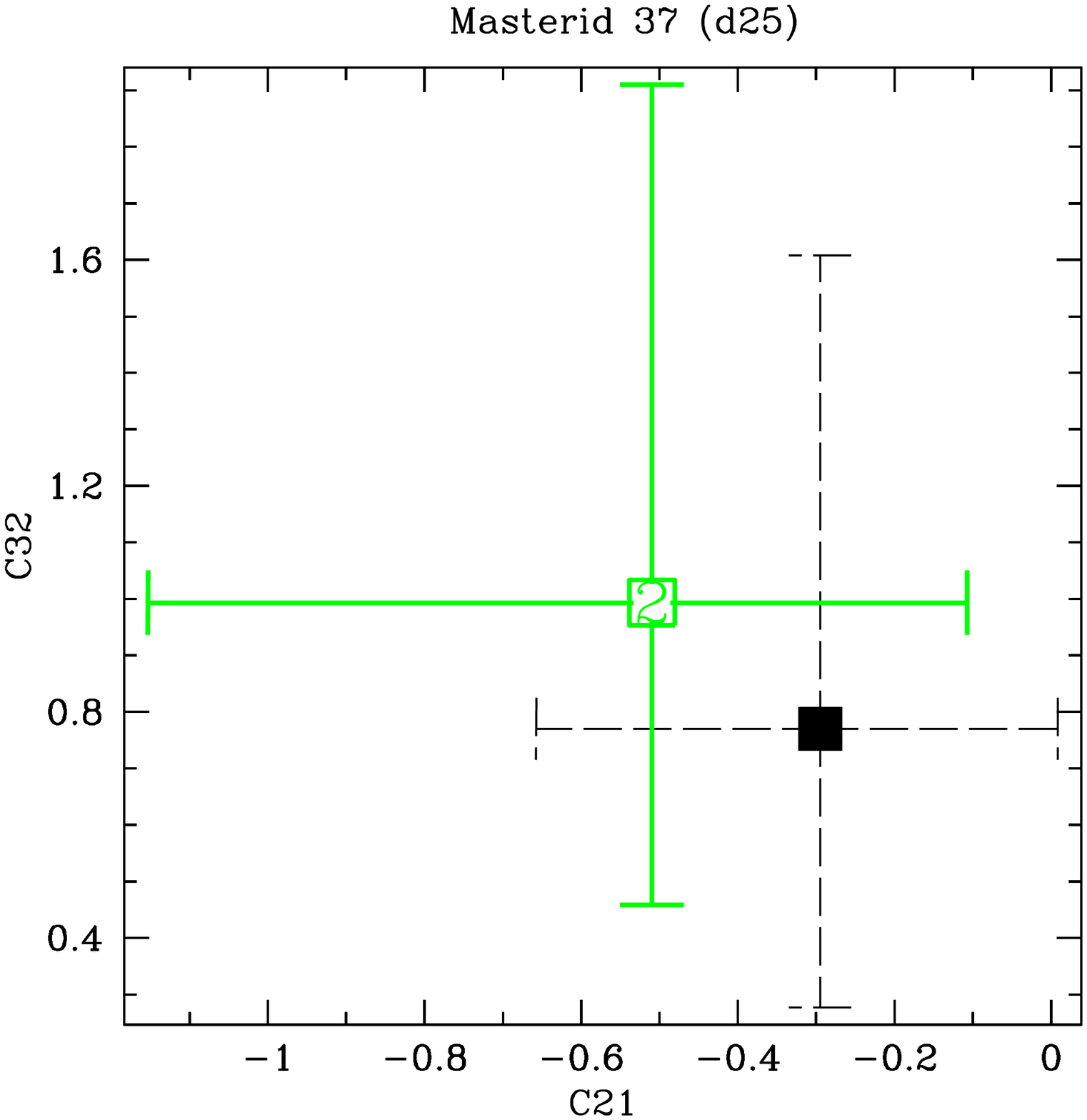}

 \end{minipage}

  \begin{minipage}{0.32\linewidth}
  \centering
  
    \includegraphics[width=\linewidth]{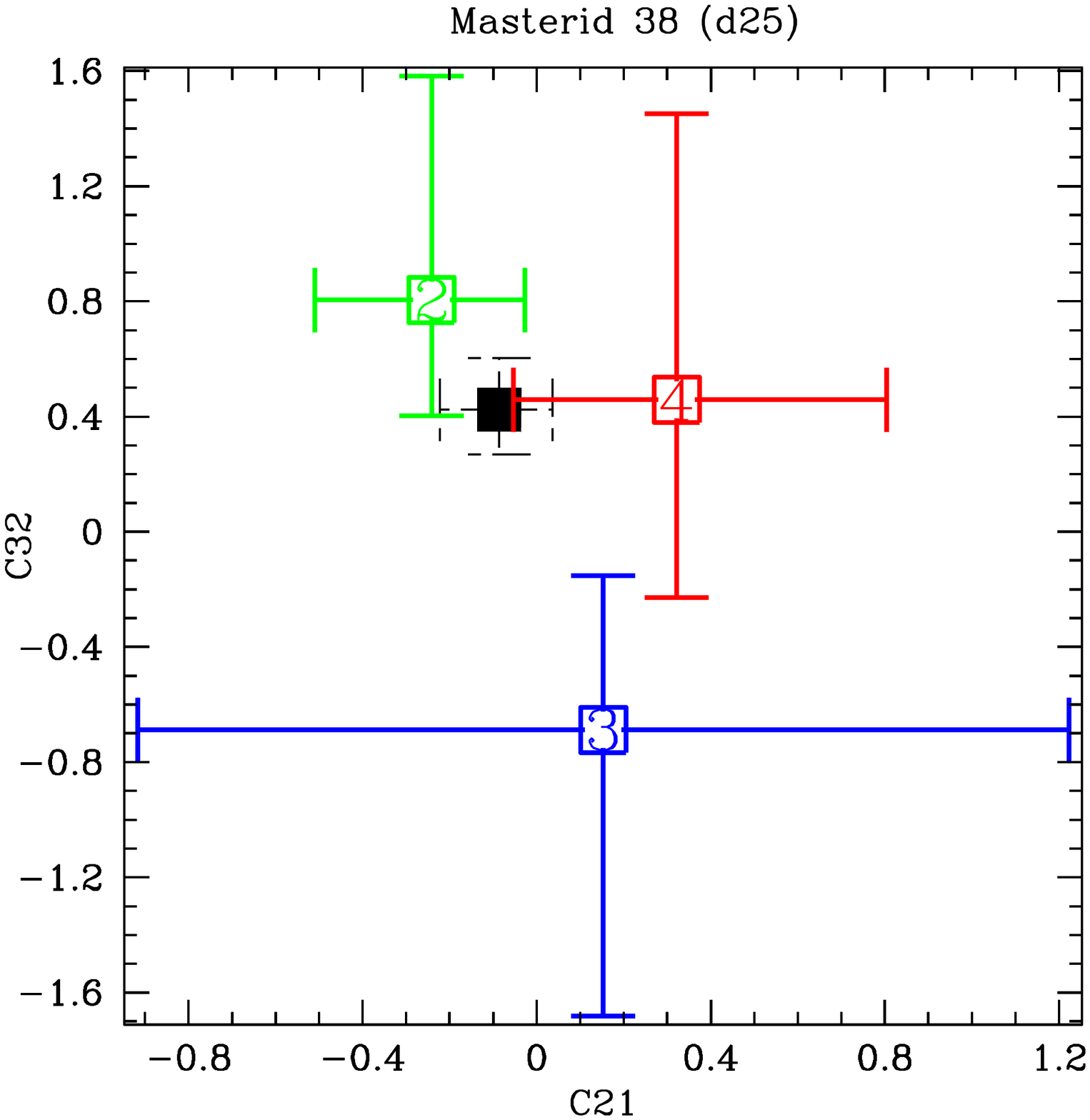}

  \end{minipage}
  \begin{minipage}{0.32\linewidth}
  \centering

    \includegraphics[width=\linewidth]{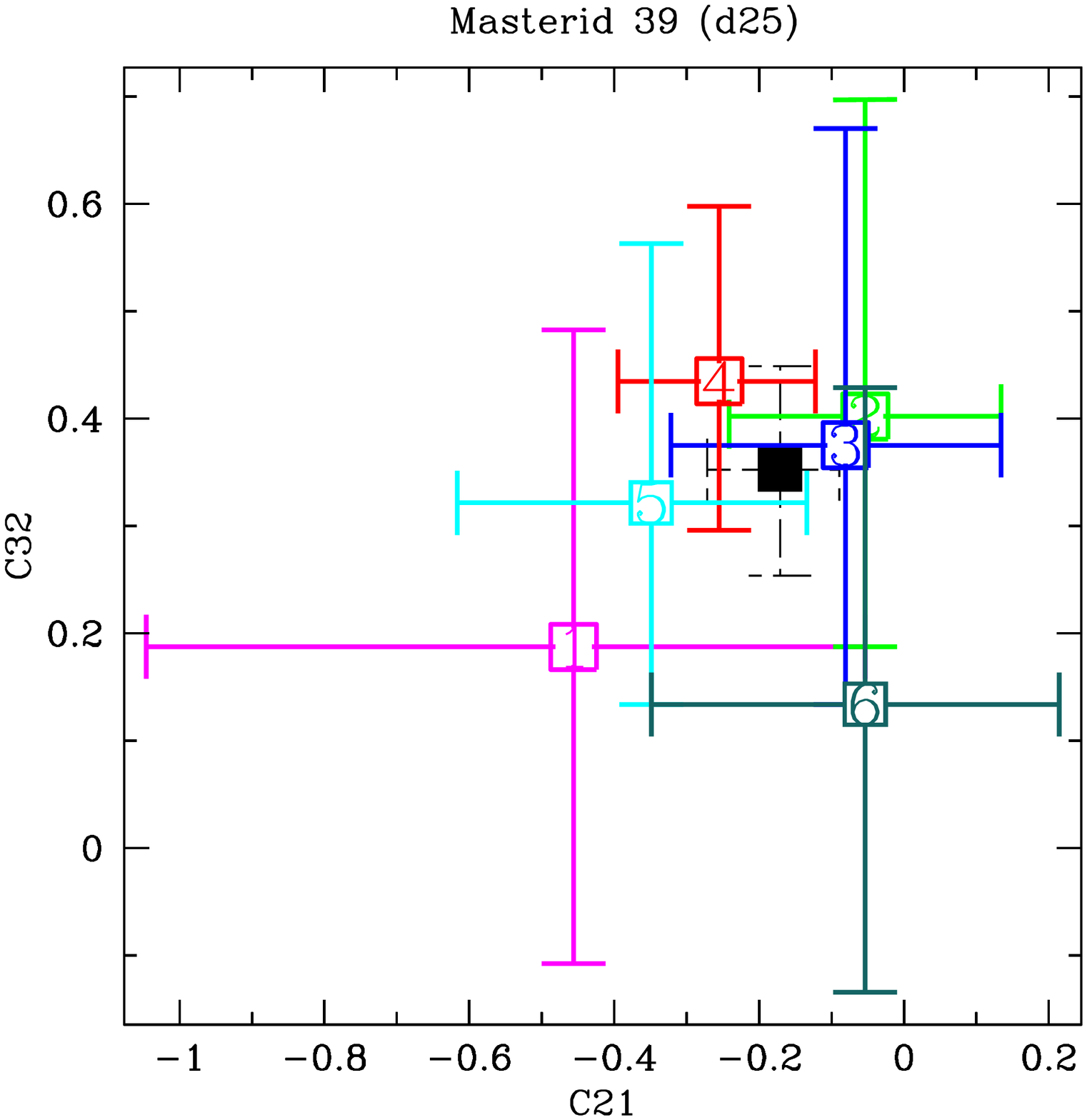}

\end{minipage}
\begin{minipage}{0.32\linewidth}
  \centering

    \includegraphics[width=\linewidth]{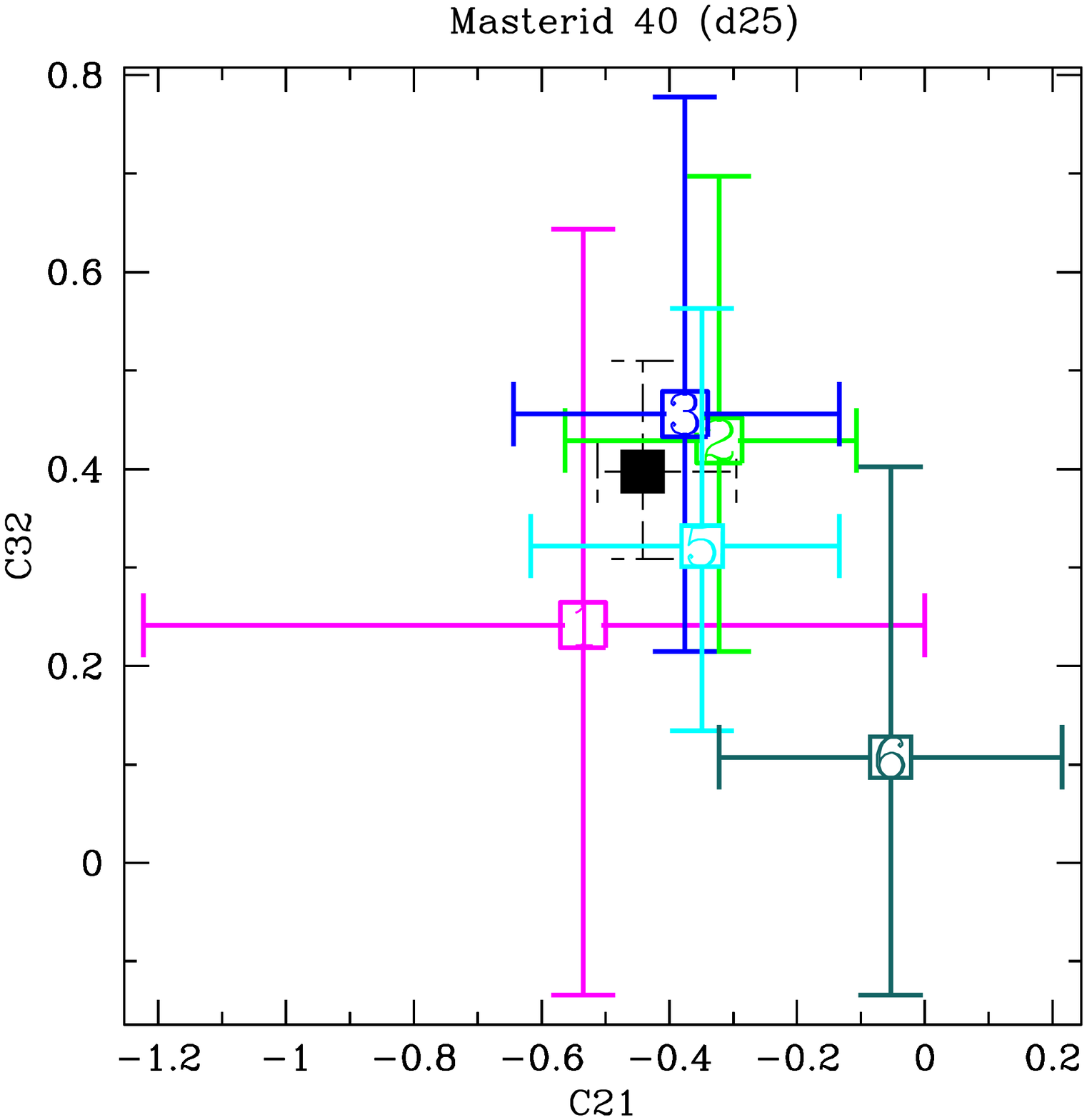}

 \end{minipage}

\begin{minipage}{0.32\linewidth}
  \centering
  
    \includegraphics[width=\linewidth]{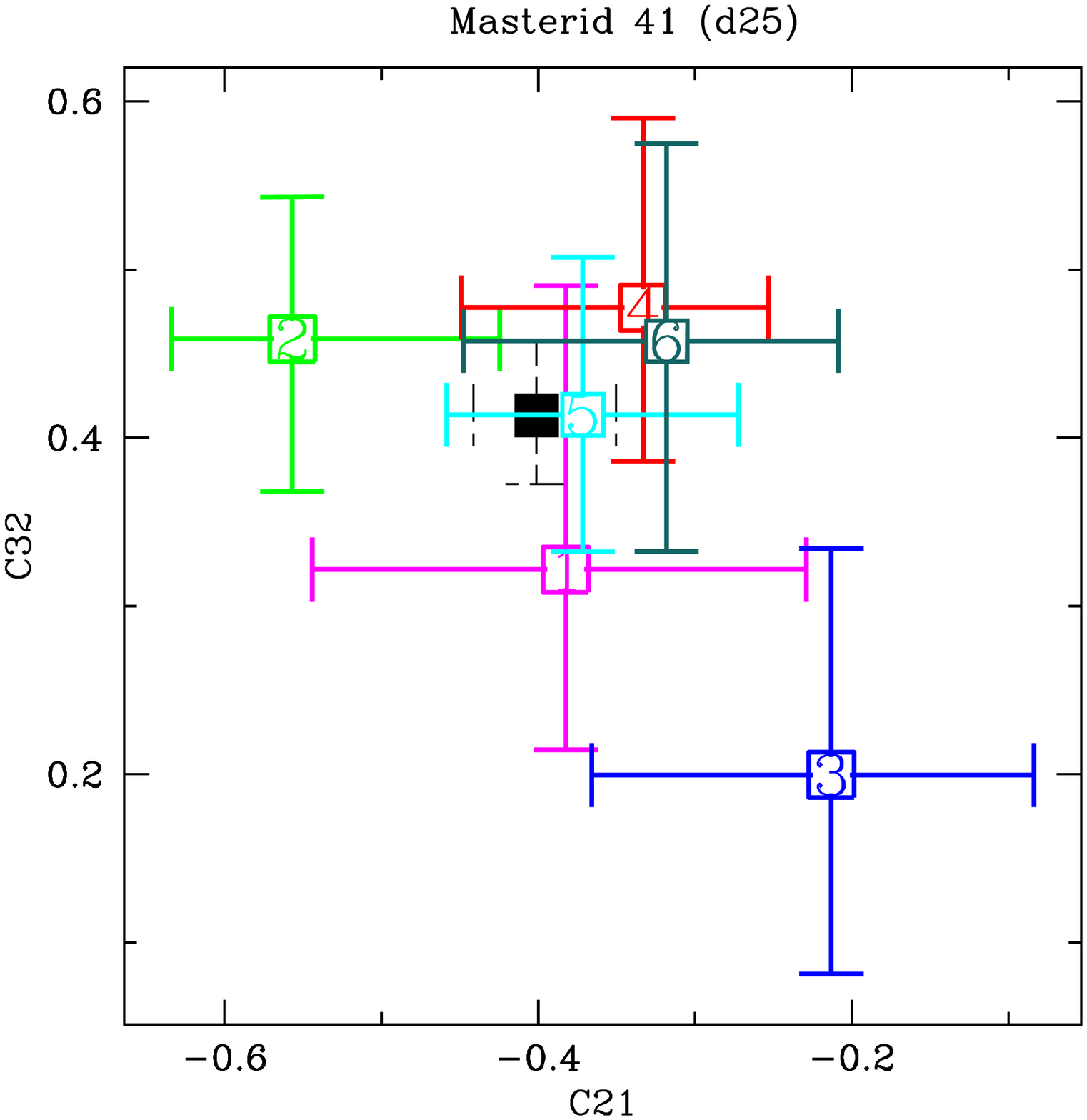}

  \end{minipage}
  \begin{minipage}{0.32\linewidth}
  \centering

    \includegraphics[width=\linewidth]{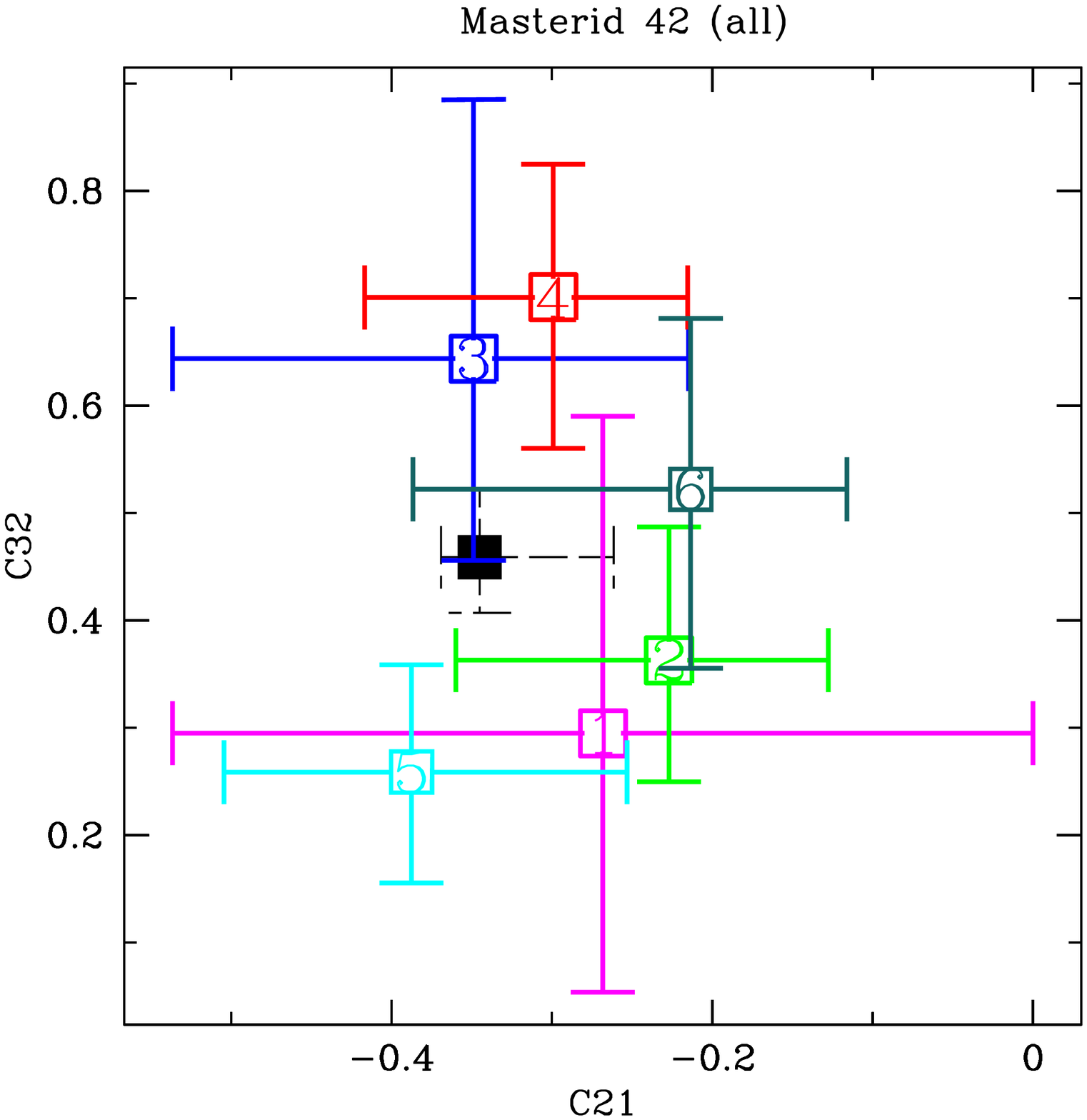}

\end{minipage}
\begin{minipage}{0.32\linewidth}
  \centering

    \includegraphics[width=\linewidth]{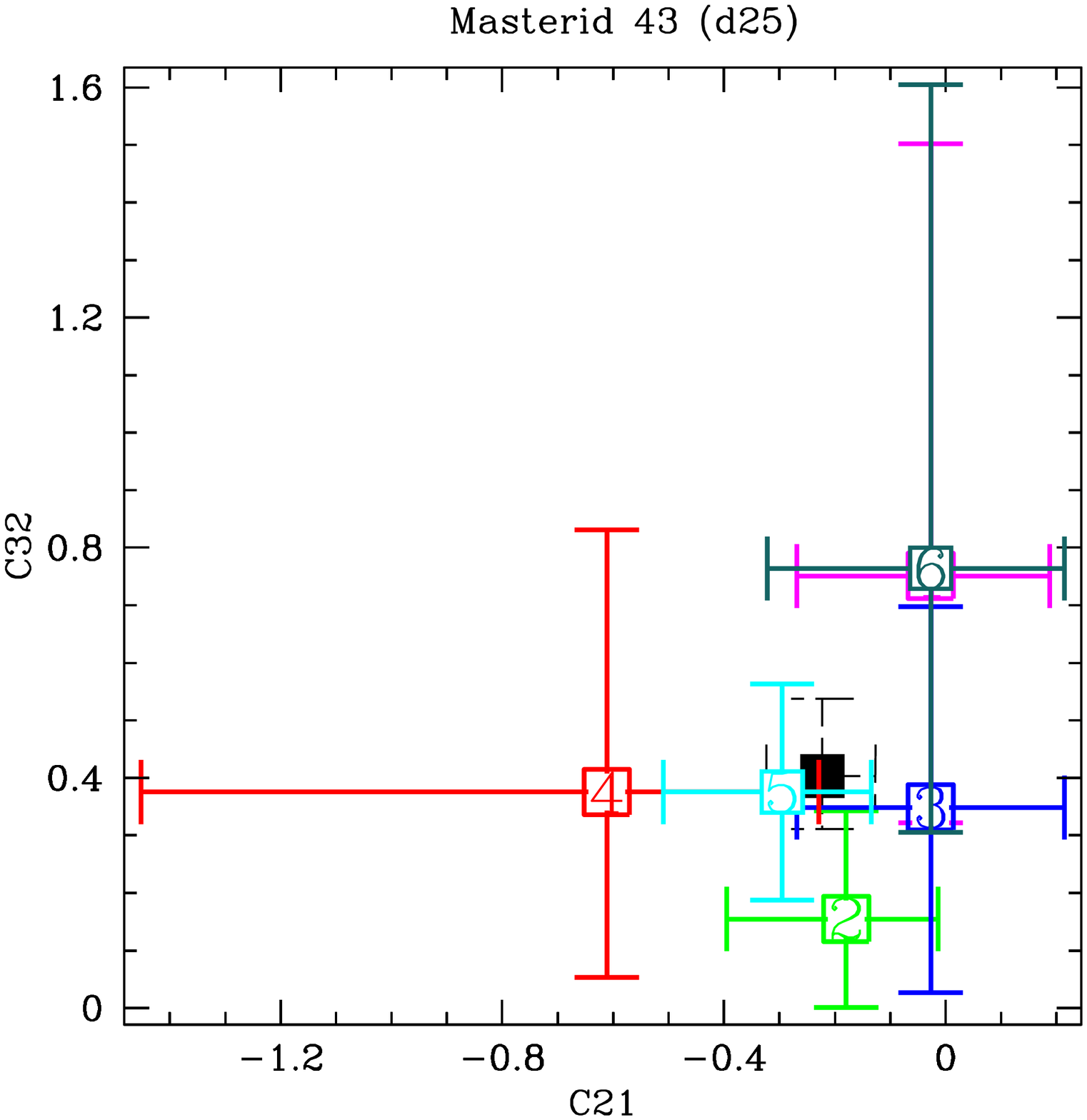}

 \end{minipage}
  
\end{figure}

\begin{figure}
  \begin{minipage}{0.32\linewidth}
  \centering
  
    \includegraphics[width=\linewidth]{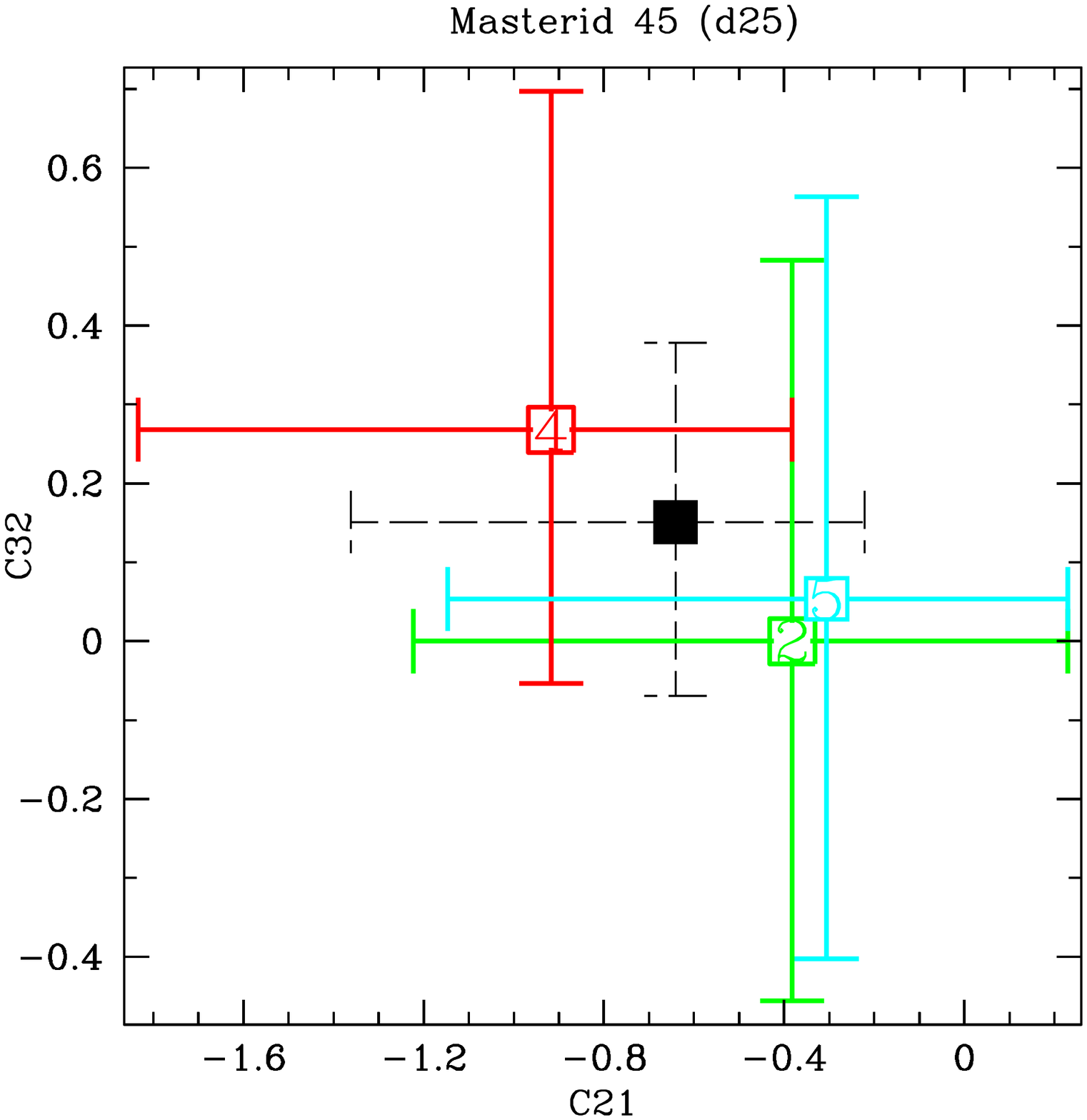}

  \end{minipage}
  \begin{minipage}{0.32\linewidth}
  \centering

    \includegraphics[width=\linewidth]{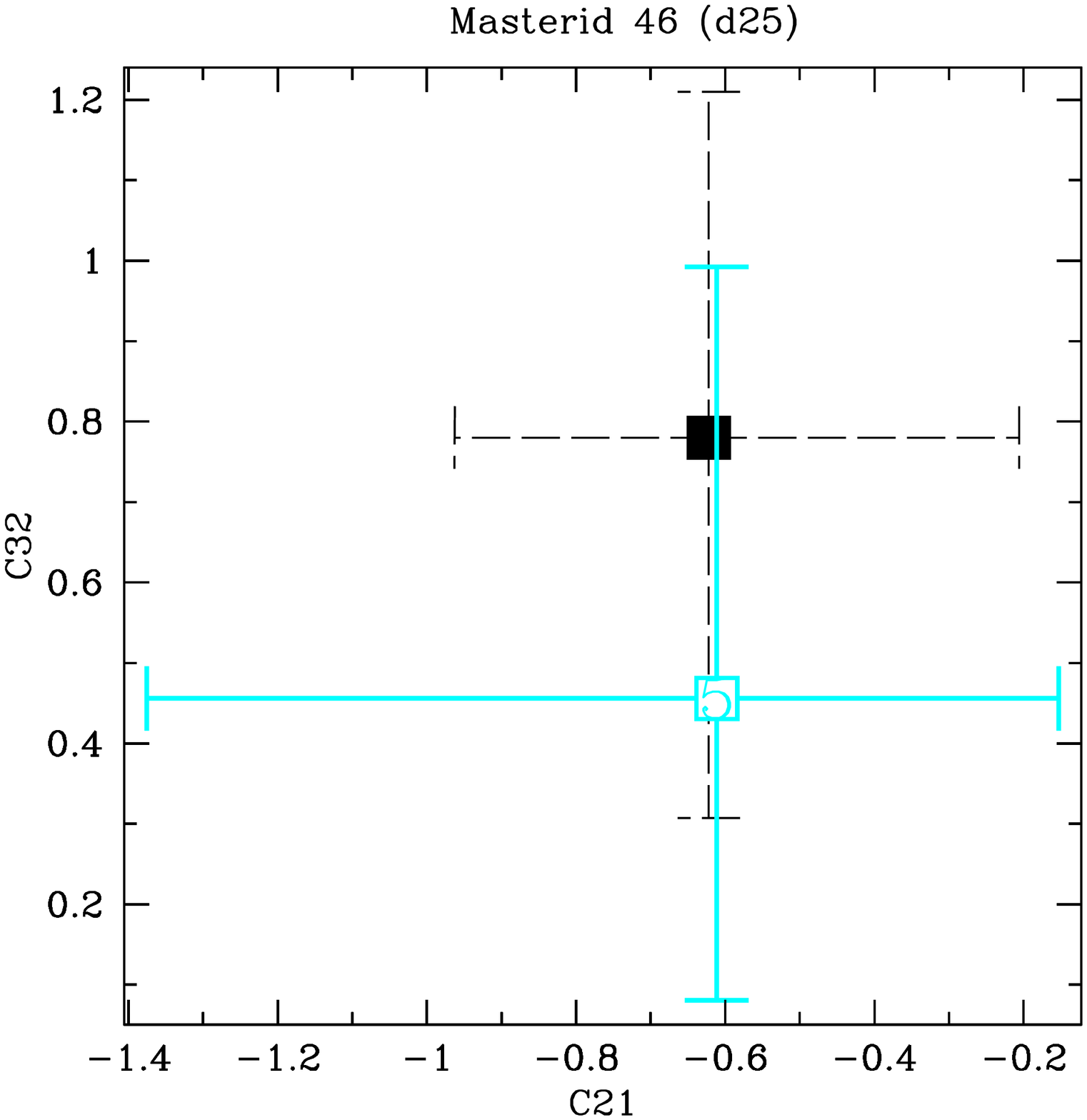}

\end{minipage}
\begin{minipage}{0.32\linewidth}
  \centering

    \includegraphics[width=\linewidth]{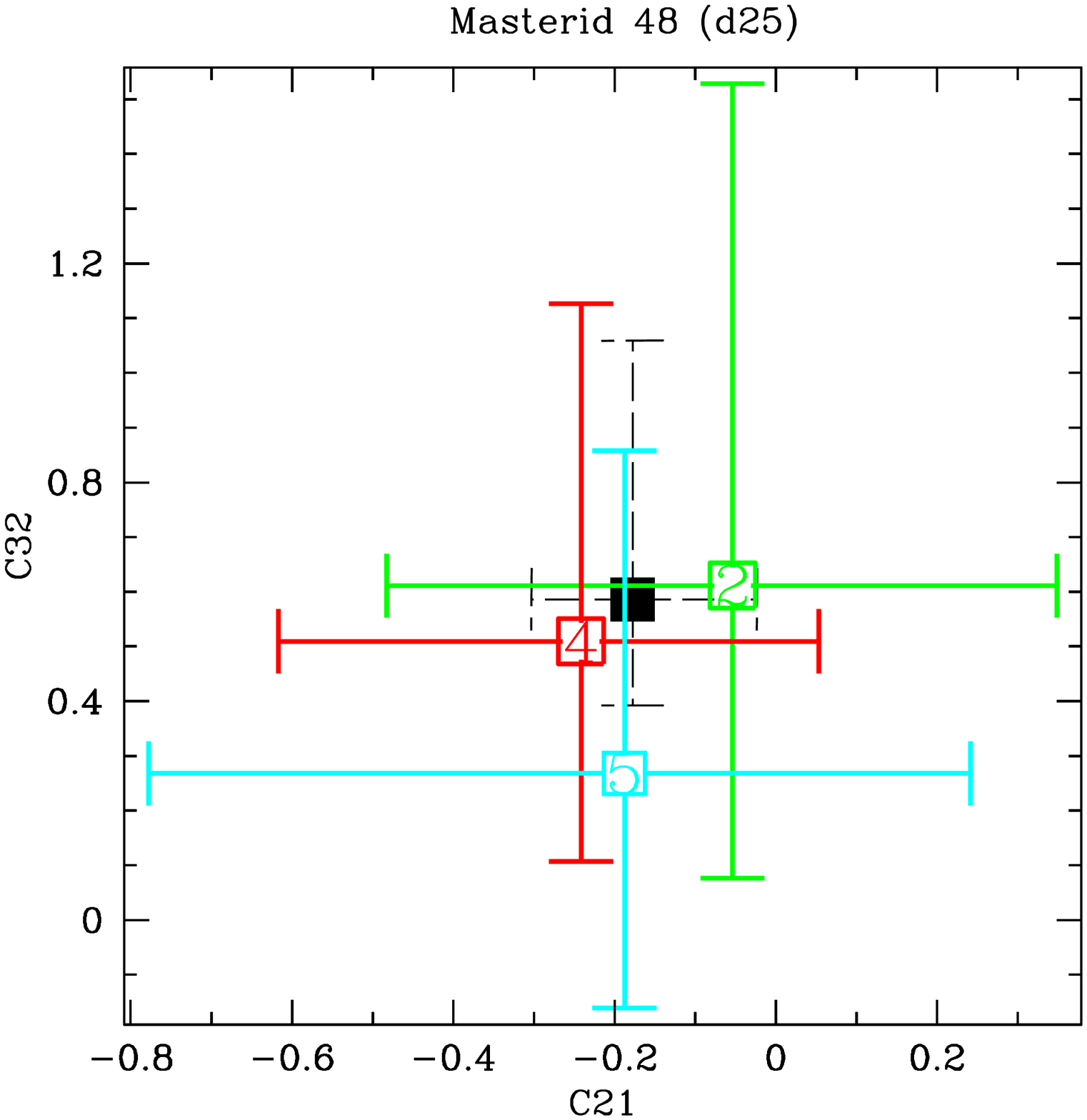}

 \end{minipage}

\begin{minipage}{0.32\linewidth}
  \centering
  
    \includegraphics[width=\linewidth]{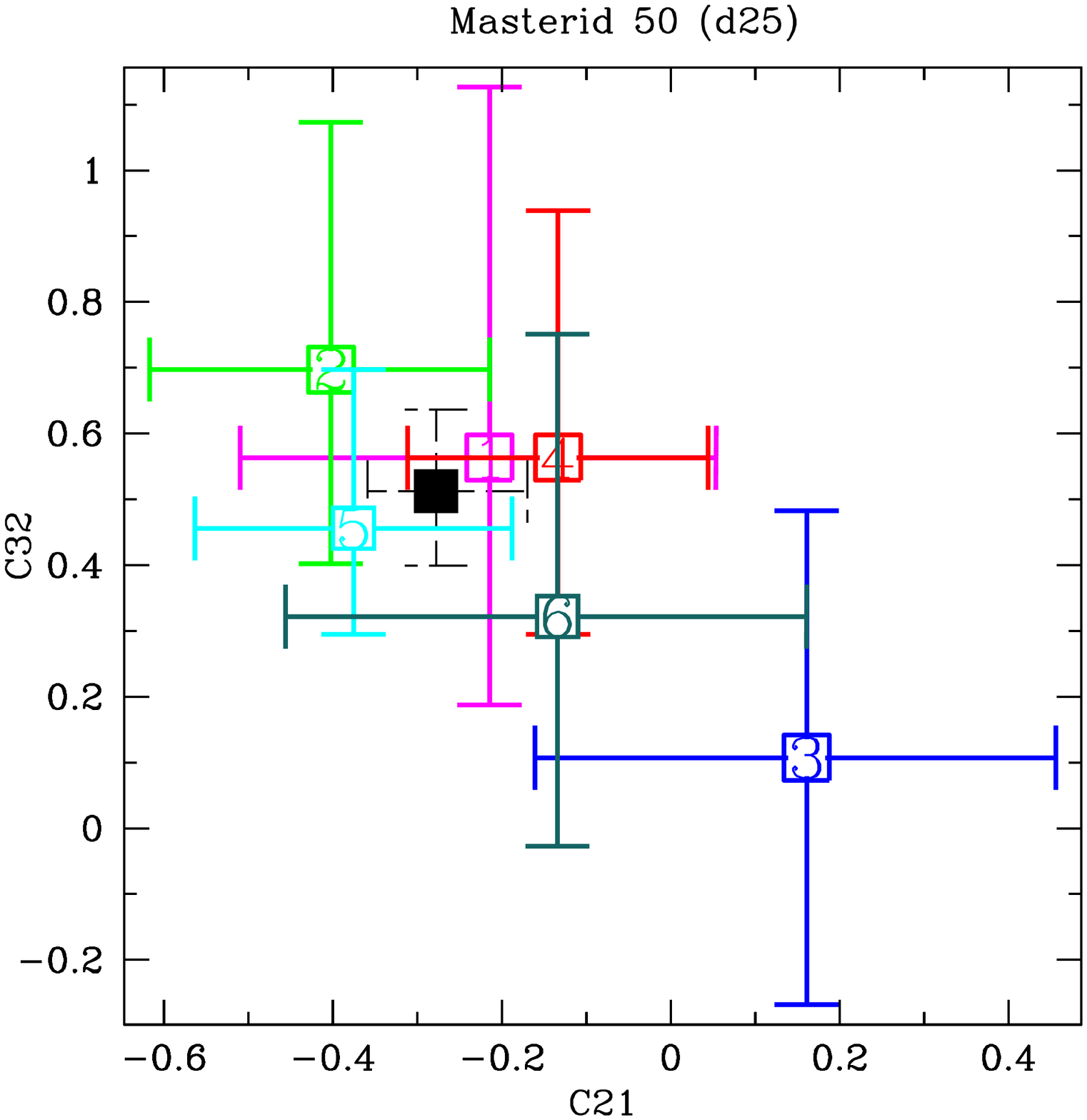}

  \end{minipage}
  \begin{minipage}{0.32\linewidth}
  \centering

    \includegraphics[width=\linewidth]{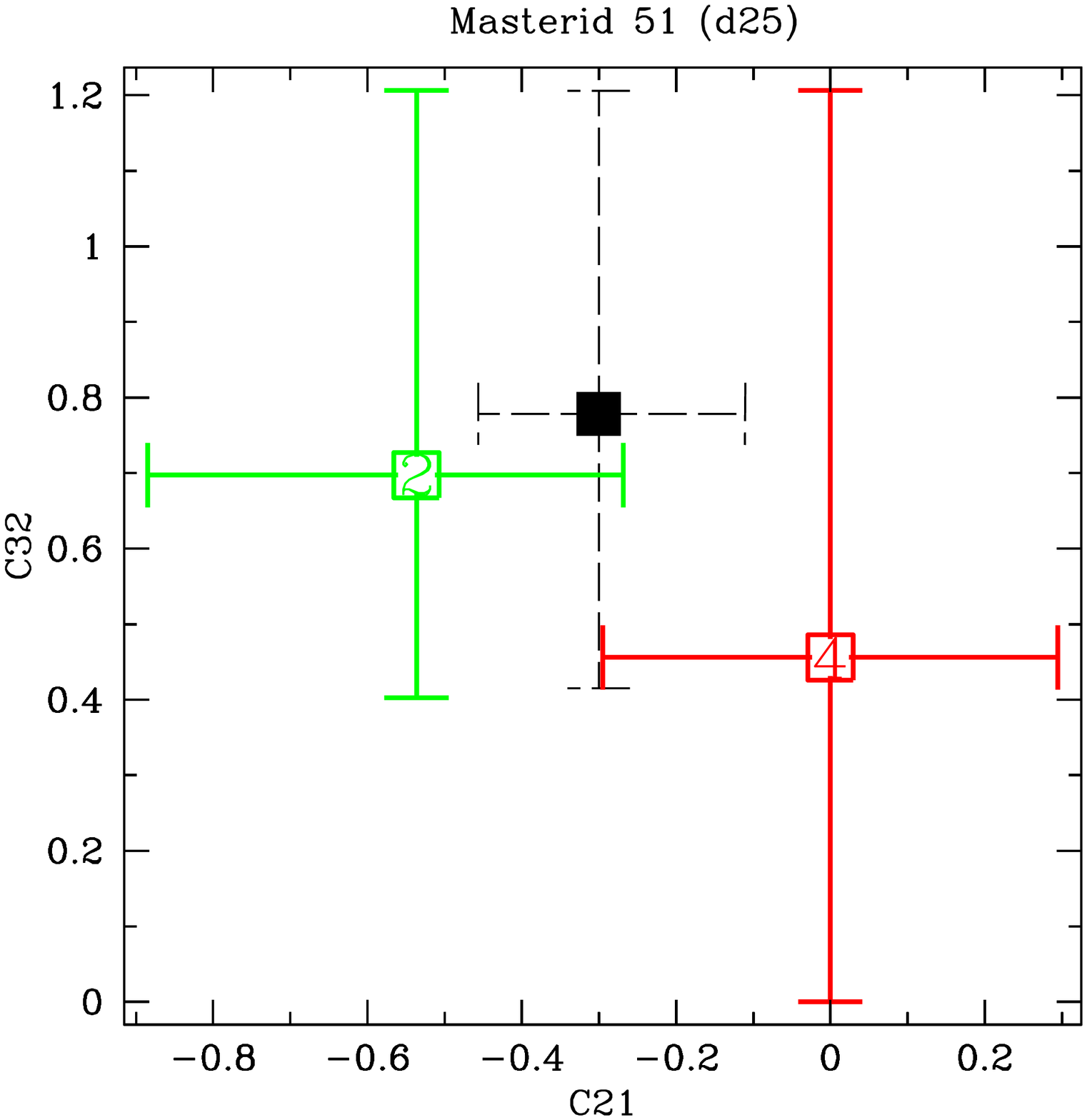}

\end{minipage}
\begin{minipage}{0.32\linewidth}
  \centering

    \includegraphics[width=\linewidth]{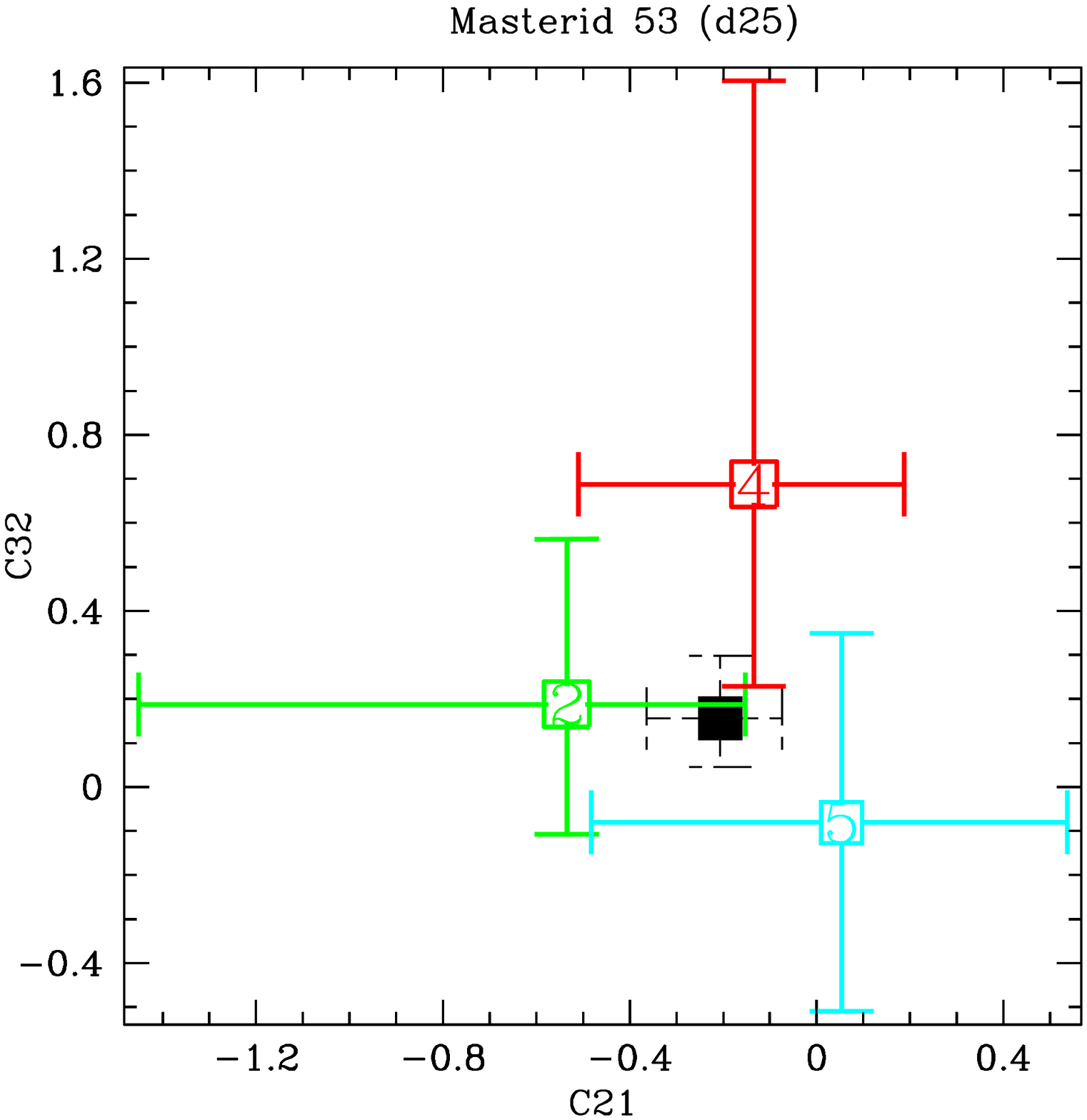}

 \end{minipage}

  \begin{minipage}{0.32\linewidth}
  \centering
  
    \includegraphics[width=\linewidth]{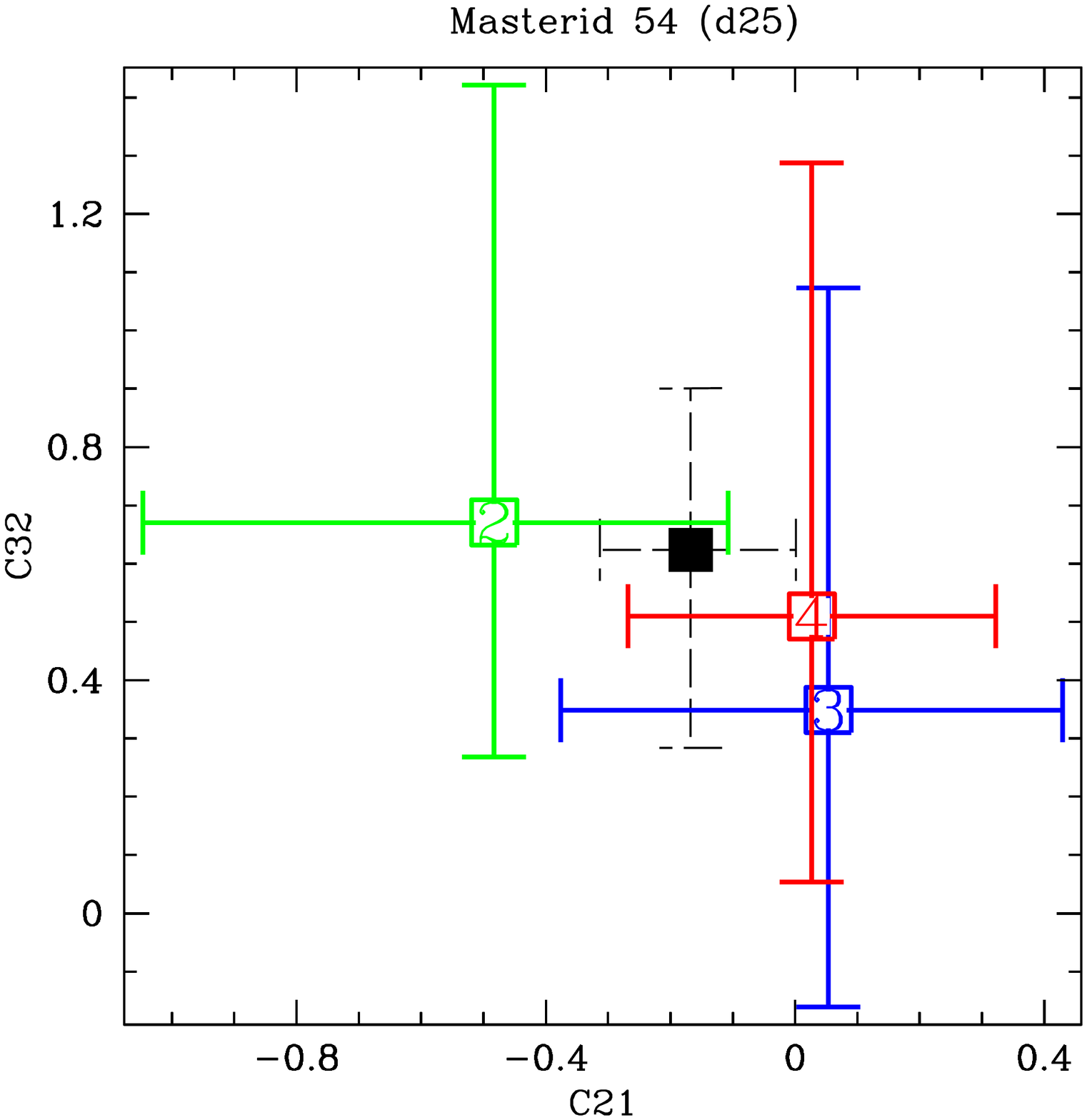}

  \end{minipage}
  \begin{minipage}{0.32\linewidth}
  \centering

    \includegraphics[width=\linewidth]{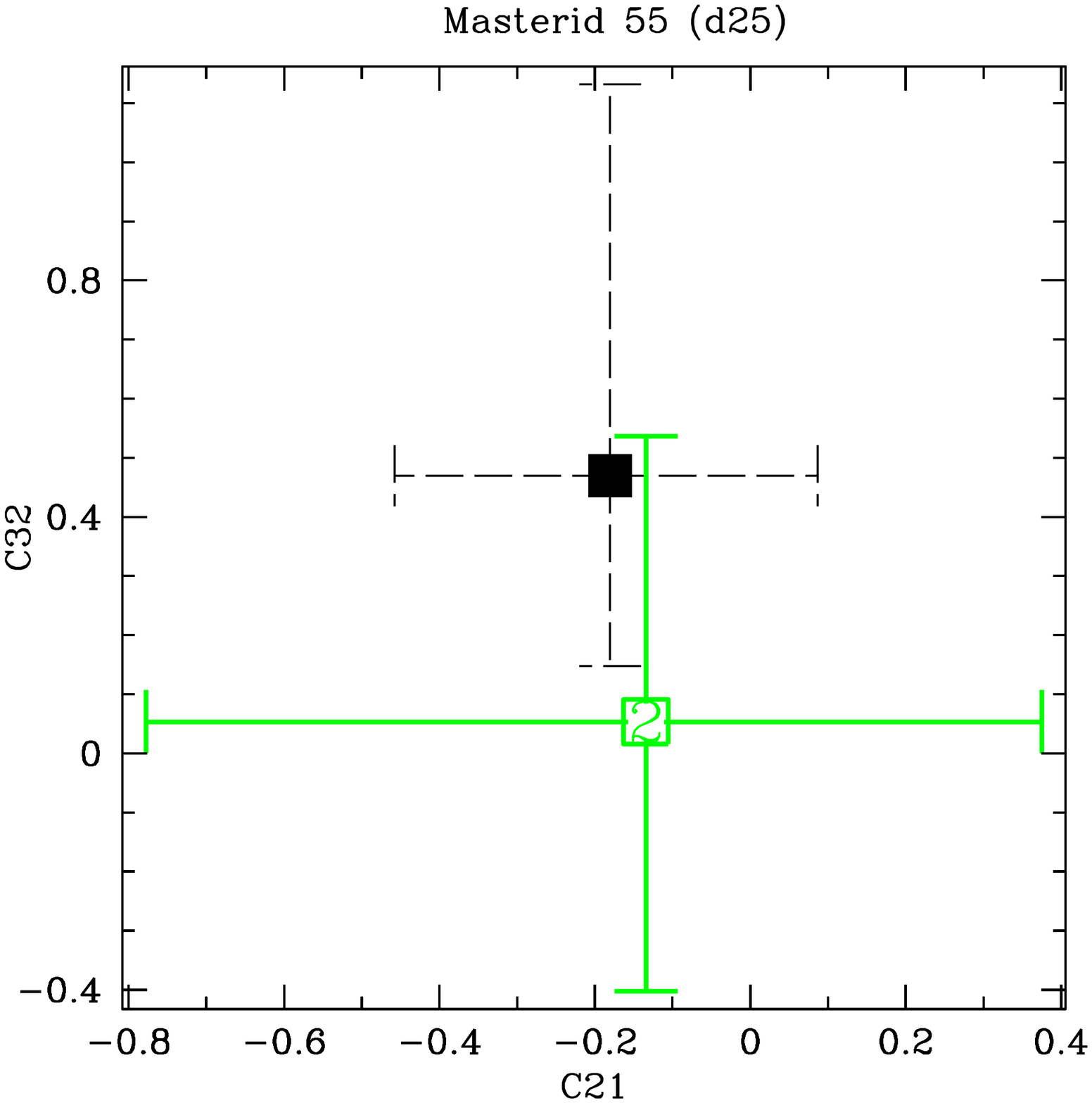}

\end{minipage}
\begin{minipage}{0.32\linewidth}
  \centering

    \includegraphics[width=\linewidth]{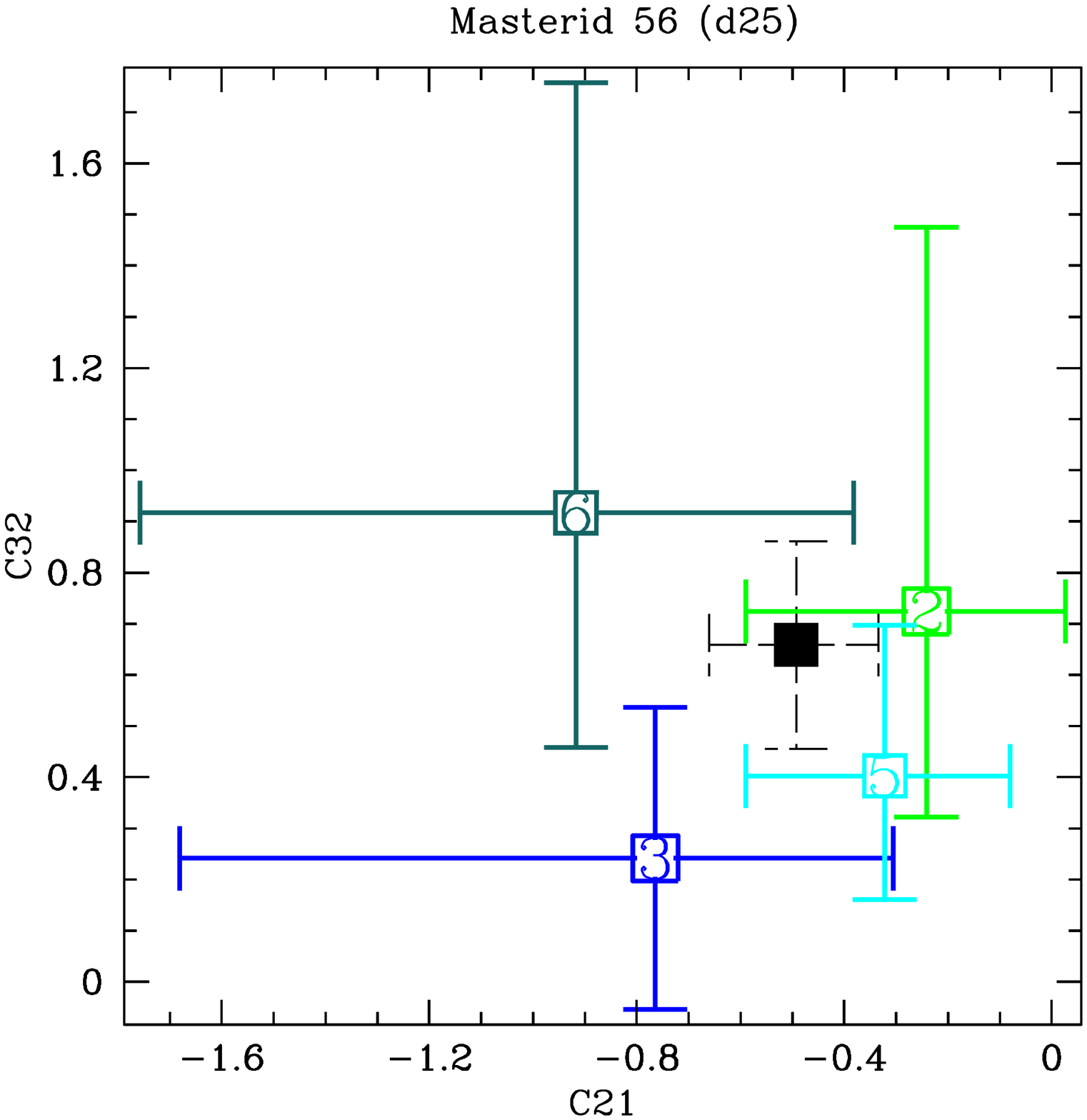}

 \end{minipage}

\begin{minipage}{0.32\linewidth}
  \centering
  
    \includegraphics[width=\linewidth]{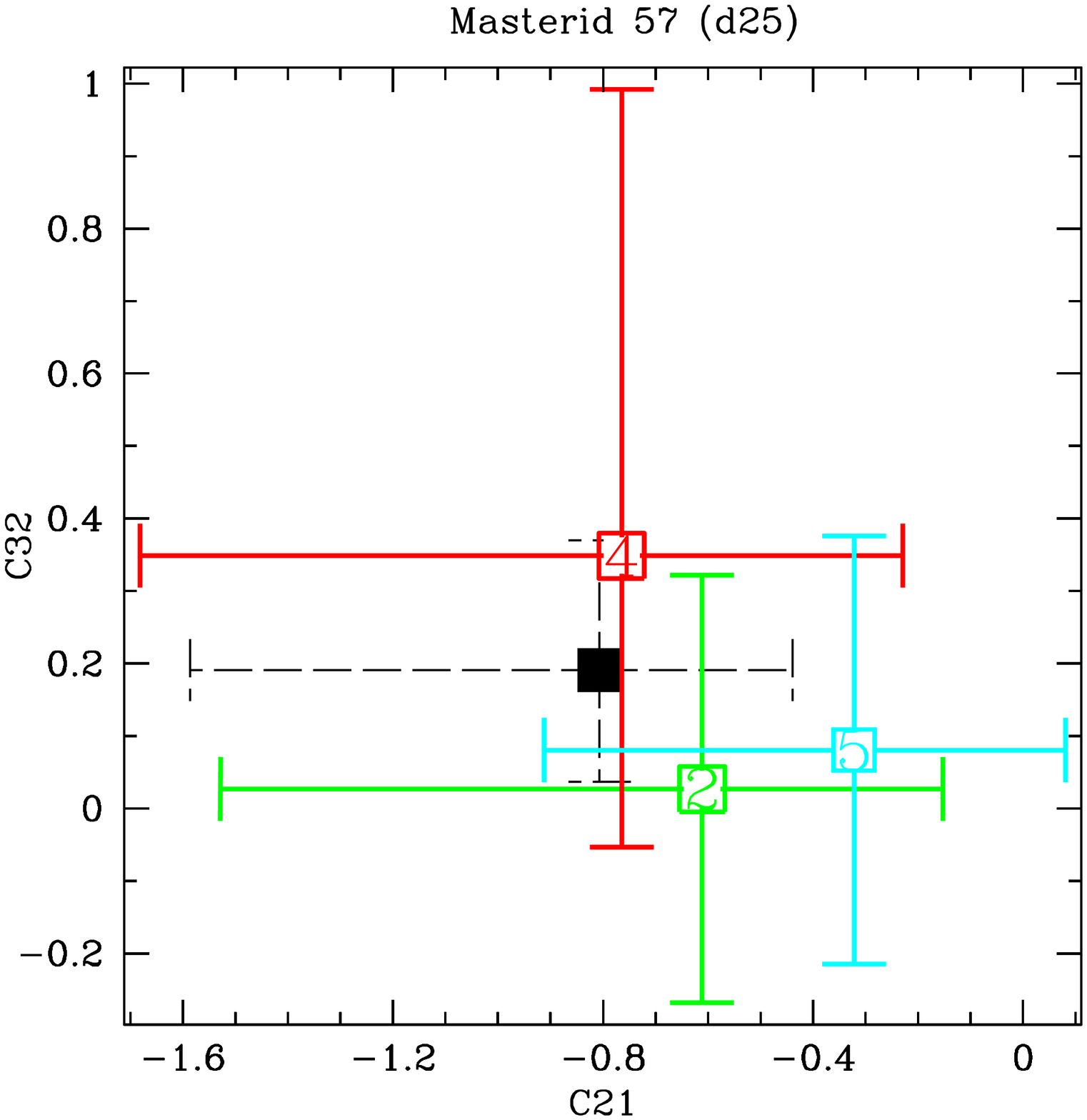}

  \end{minipage}
  \begin{minipage}{0.32\linewidth}
  \centering

    \includegraphics[width=\linewidth]{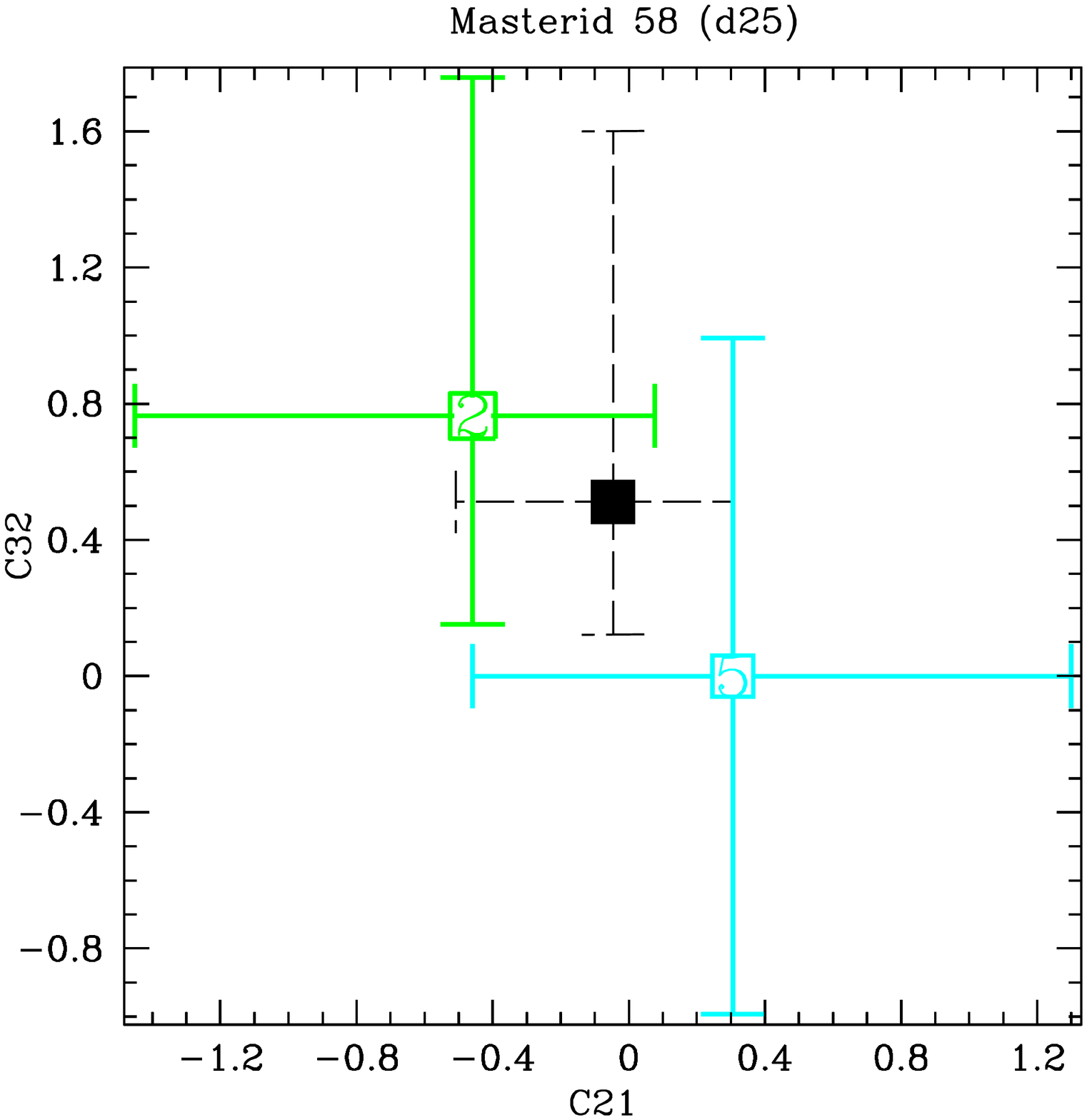}

\end{minipage}
\begin{minipage}{0.32\linewidth}
  \centering

    \includegraphics[width=\linewidth]{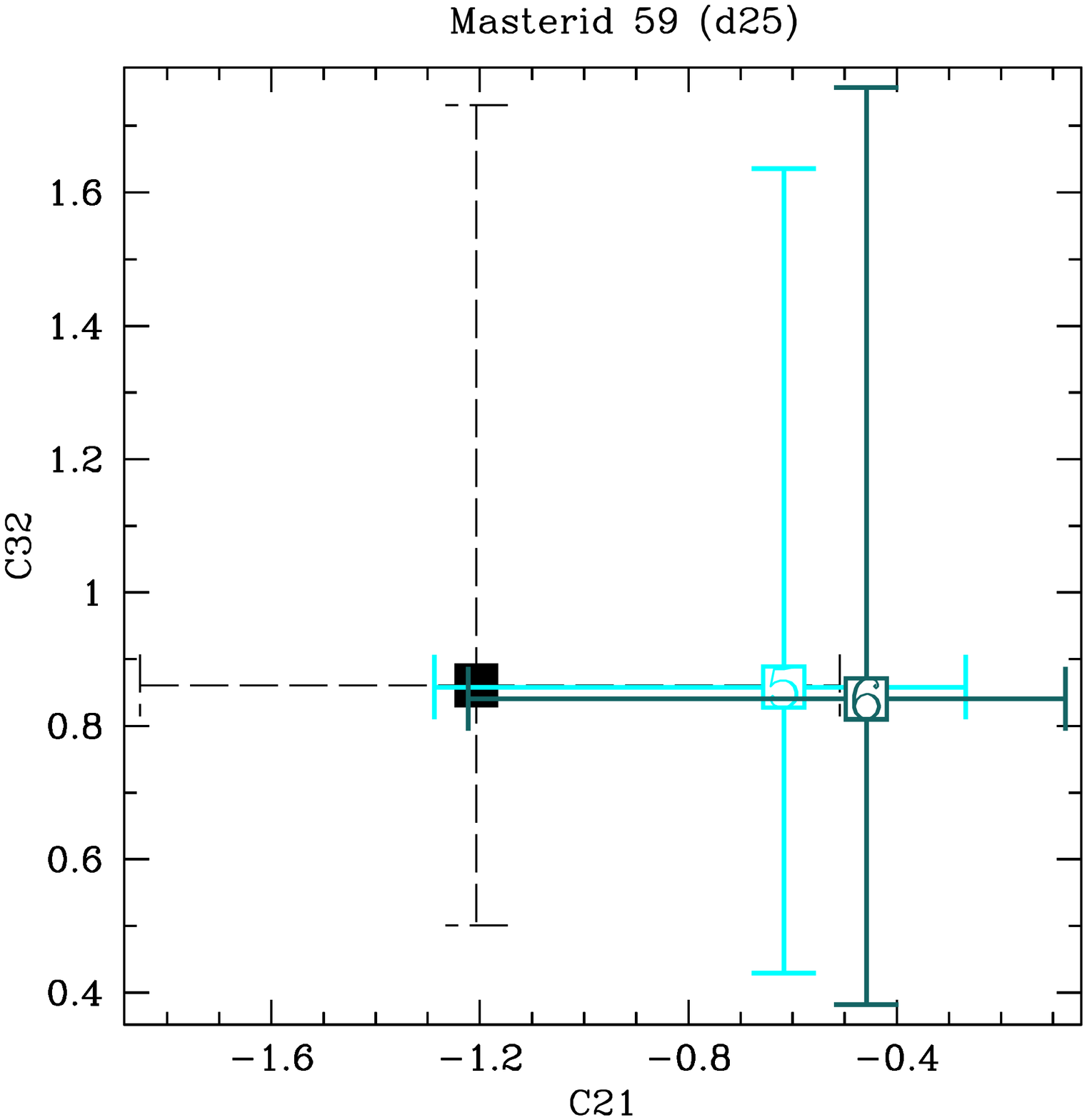}

 \end{minipage}
  
\end{figure}

\begin{figure}
  \begin{minipage}{0.32\linewidth}
  \centering
  
    \includegraphics[width=\linewidth]{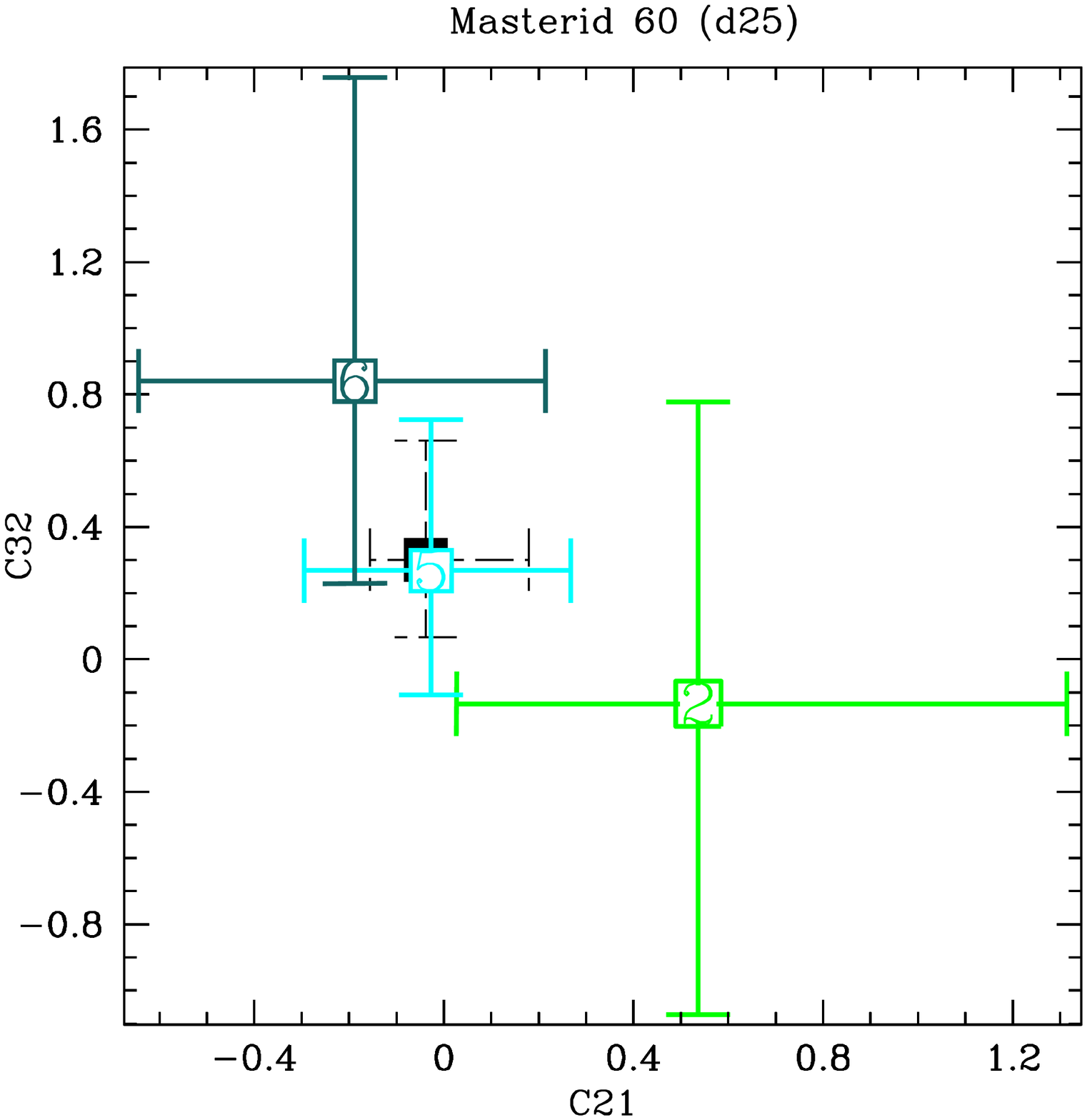}

  \end{minipage}
  \begin{minipage}{0.32\linewidth}
  \centering

    \includegraphics[width=\linewidth]{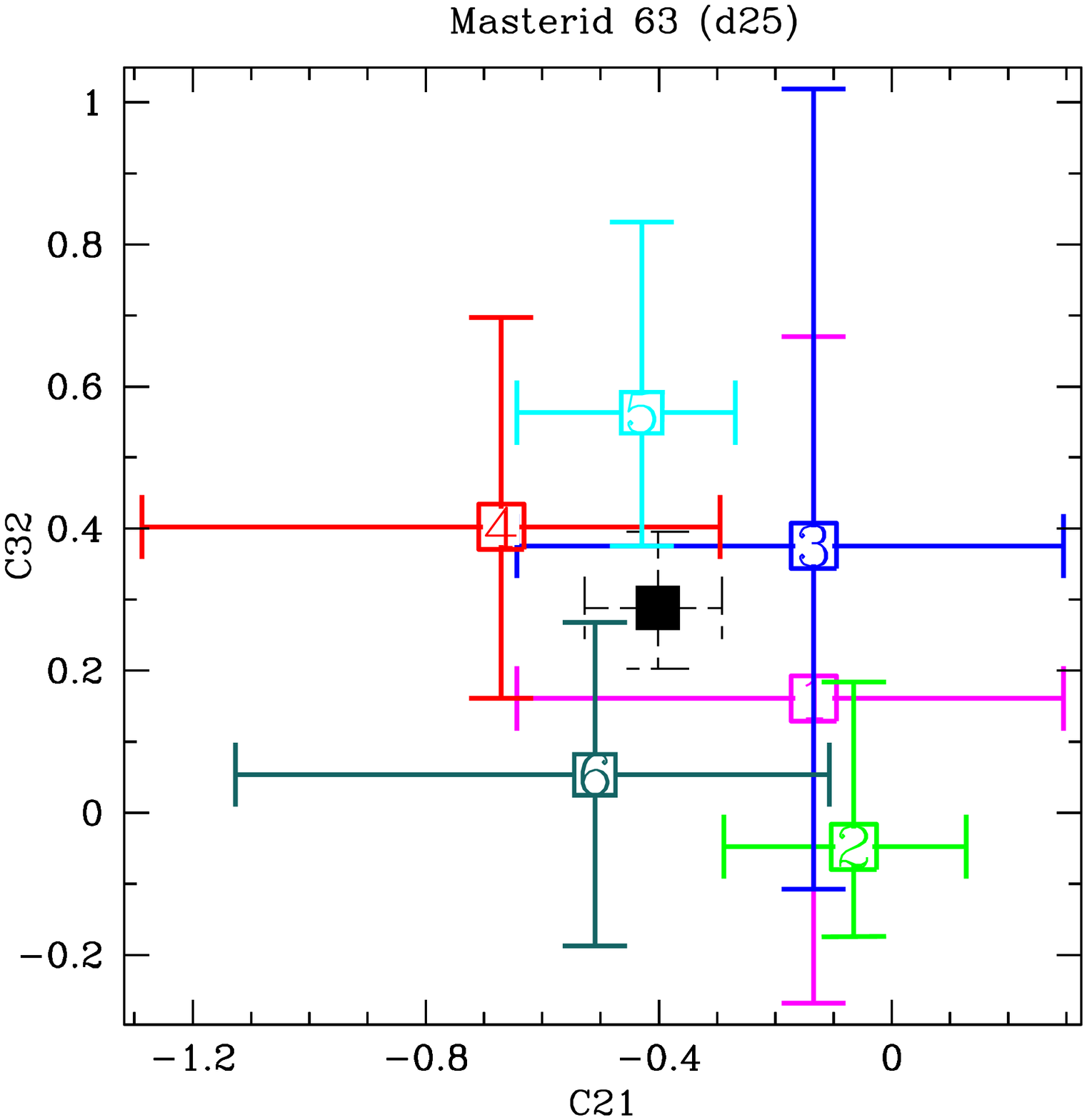}

\end{minipage}
\begin{minipage}{0.32\linewidth}
  \centering

    \includegraphics[width=\linewidth]{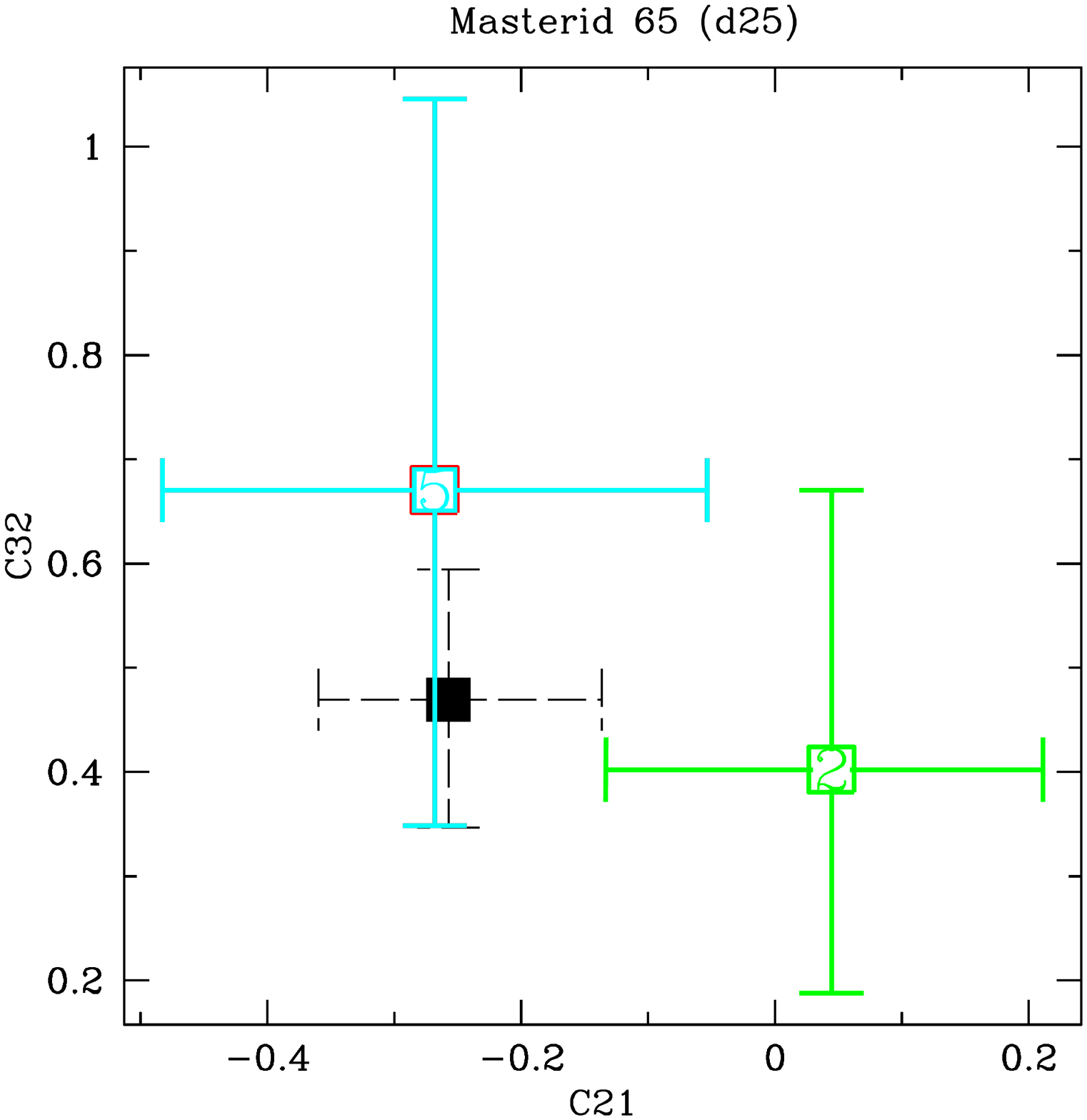}

 \end{minipage}

\begin{minipage}{0.32\linewidth}
  \centering
  
    \includegraphics[width=\linewidth]{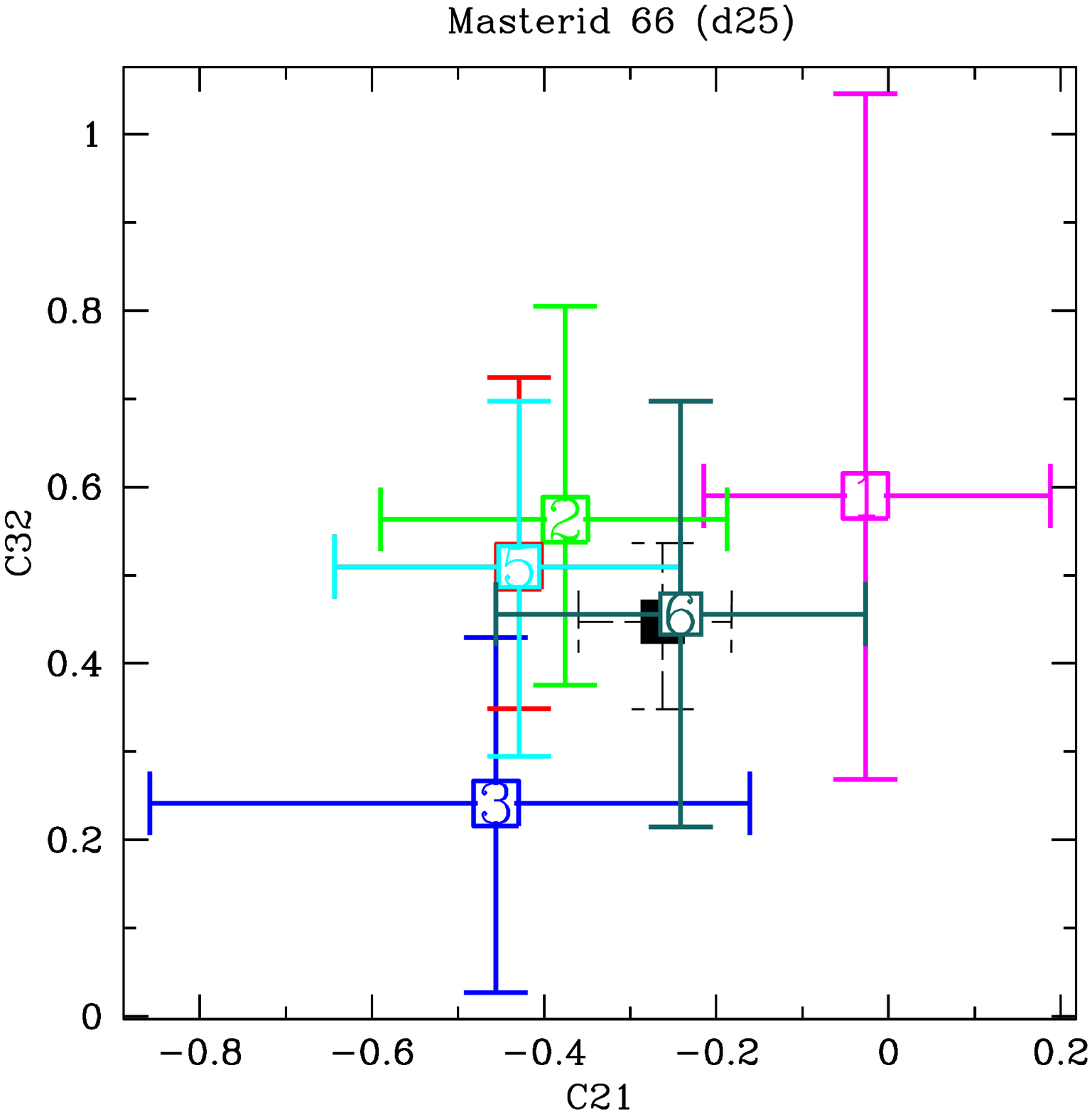}

  \end{minipage}
  \begin{minipage}{0.32\linewidth}
  \centering

    \includegraphics[width=\linewidth]{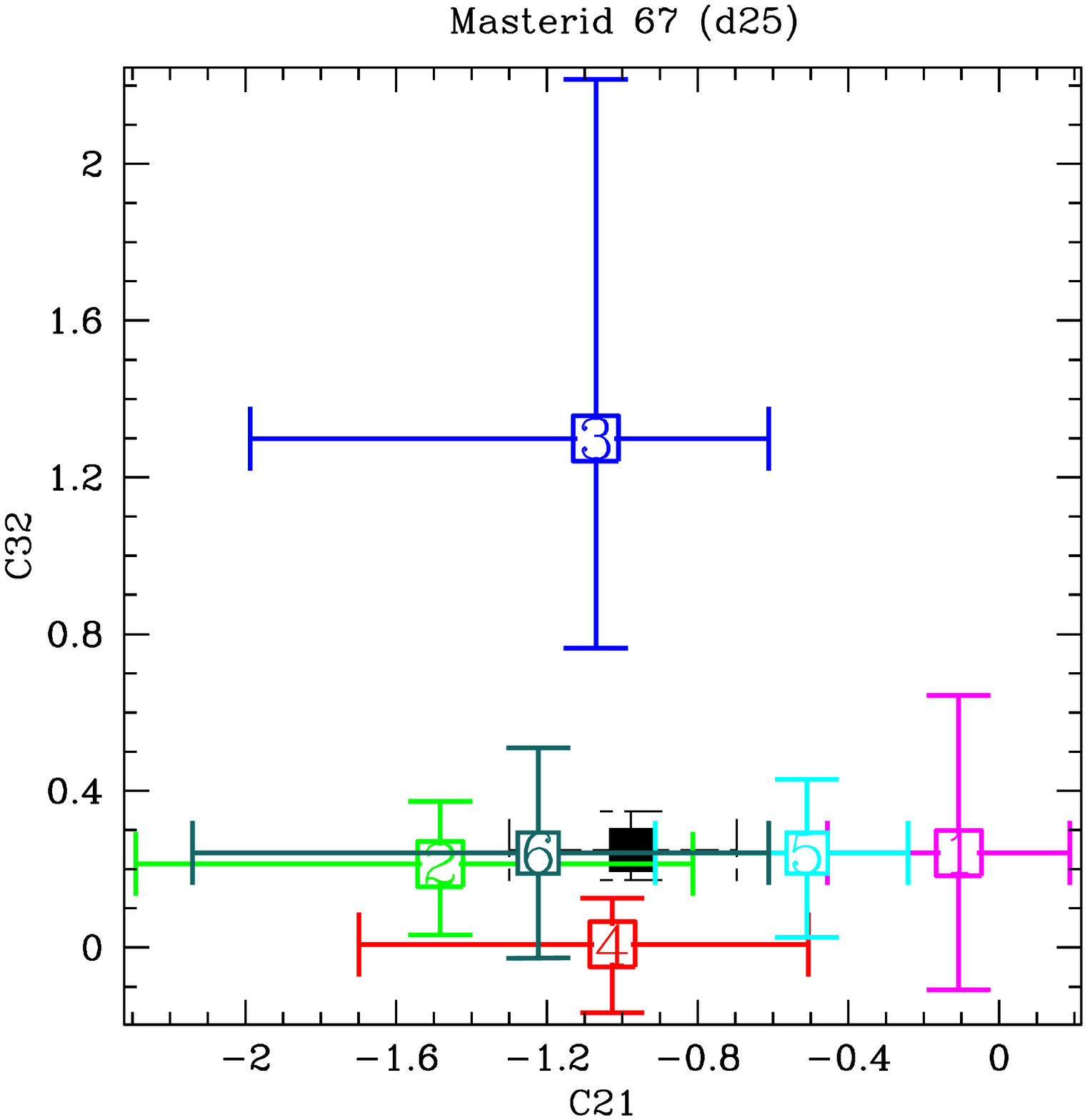}

\end{minipage}
\begin{minipage}{0.32\linewidth}
  \centering

    \includegraphics[width=\linewidth]{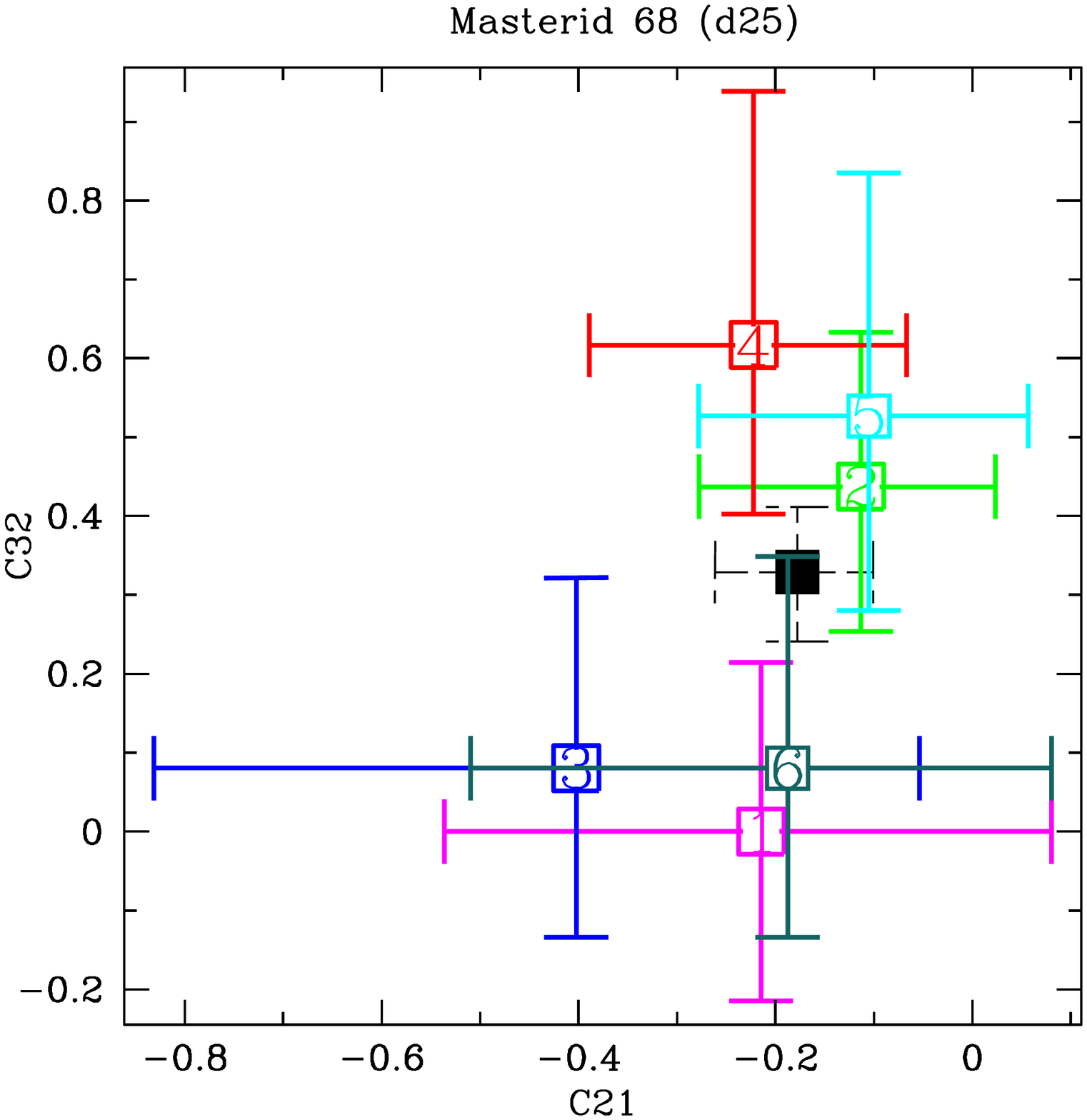}

 \end{minipage}

  \begin{minipage}{0.32\linewidth}
  \centering
  
    \includegraphics[width=\linewidth]{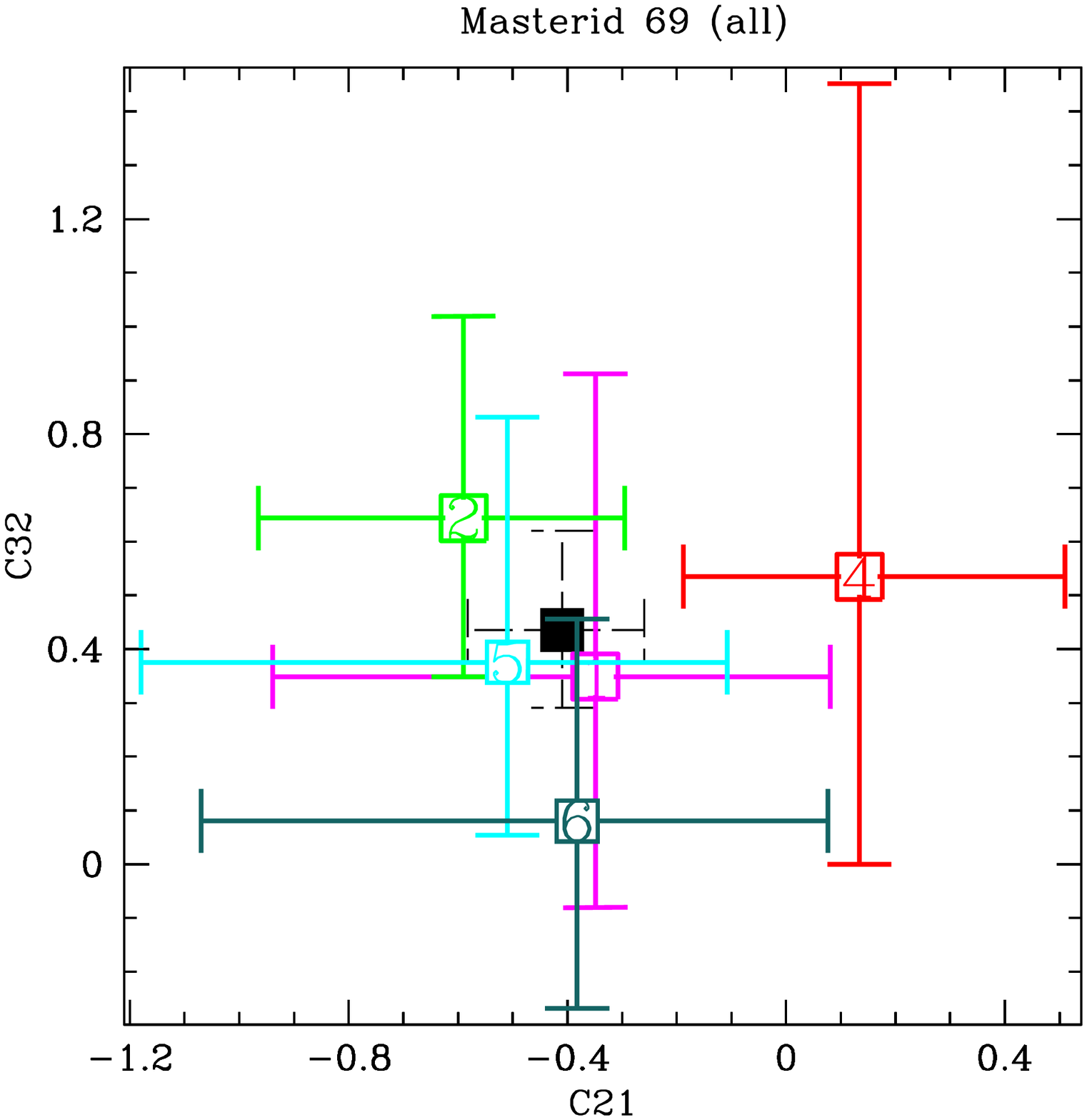}

  \end{minipage}
  \begin{minipage}{0.32\linewidth}
  \centering

    \includegraphics[width=\linewidth]{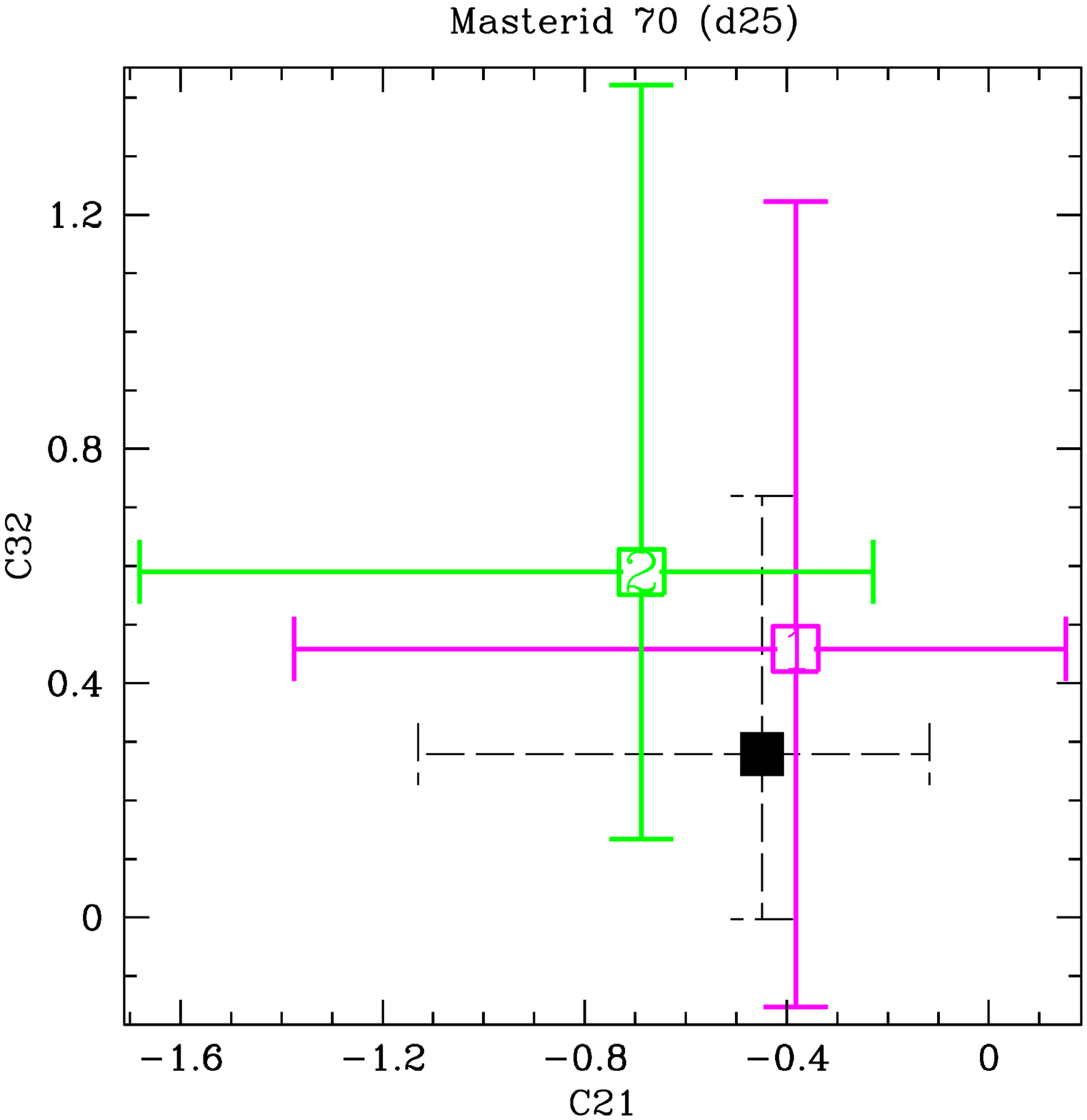}

\end{minipage}
\begin{minipage}{0.32\linewidth}
  \centering

    \includegraphics[width=\linewidth]{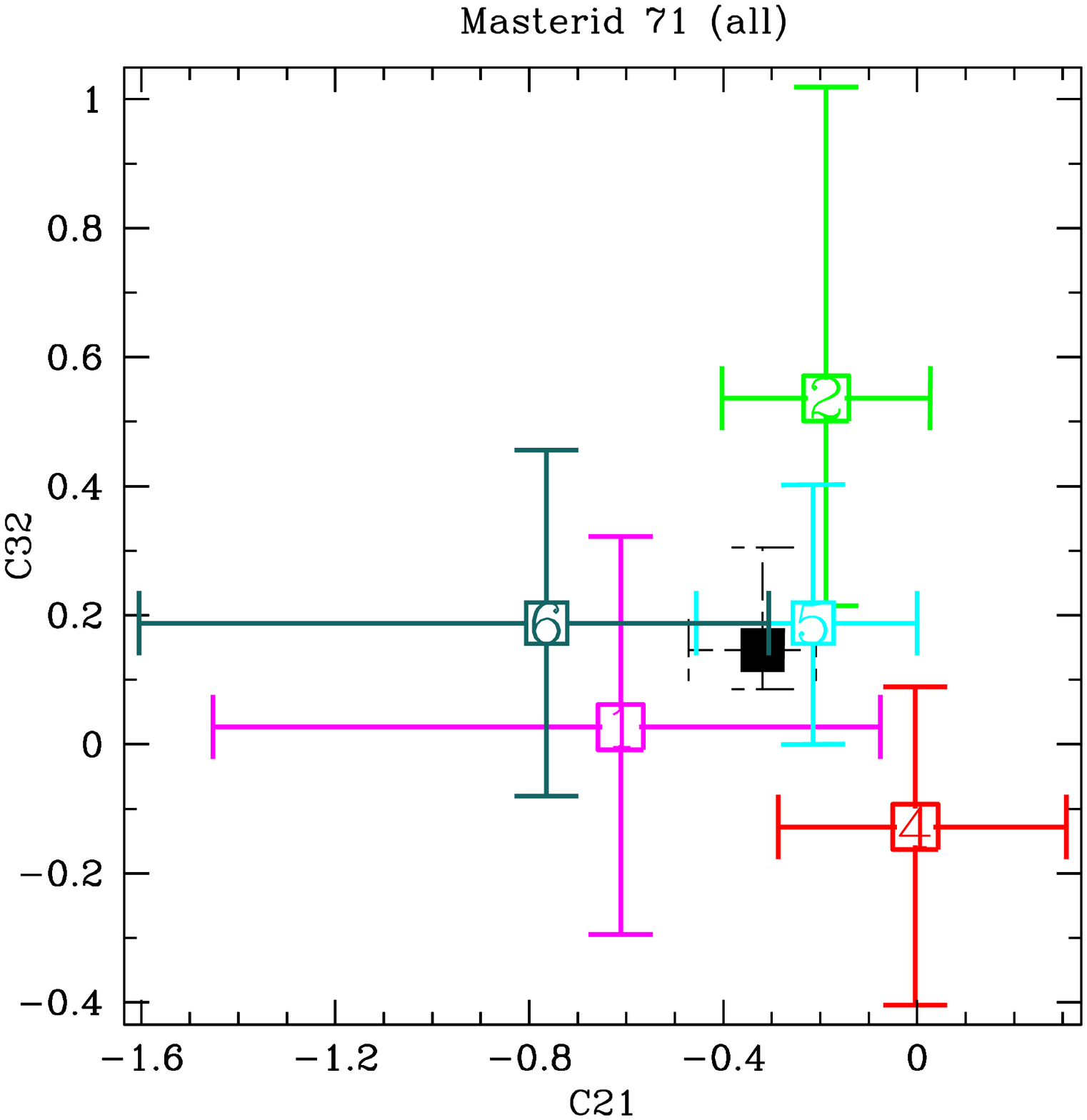}

 \end{minipage}

\begin{minipage}{0.32\linewidth}
  \centering
  
    \includegraphics[width=\linewidth]{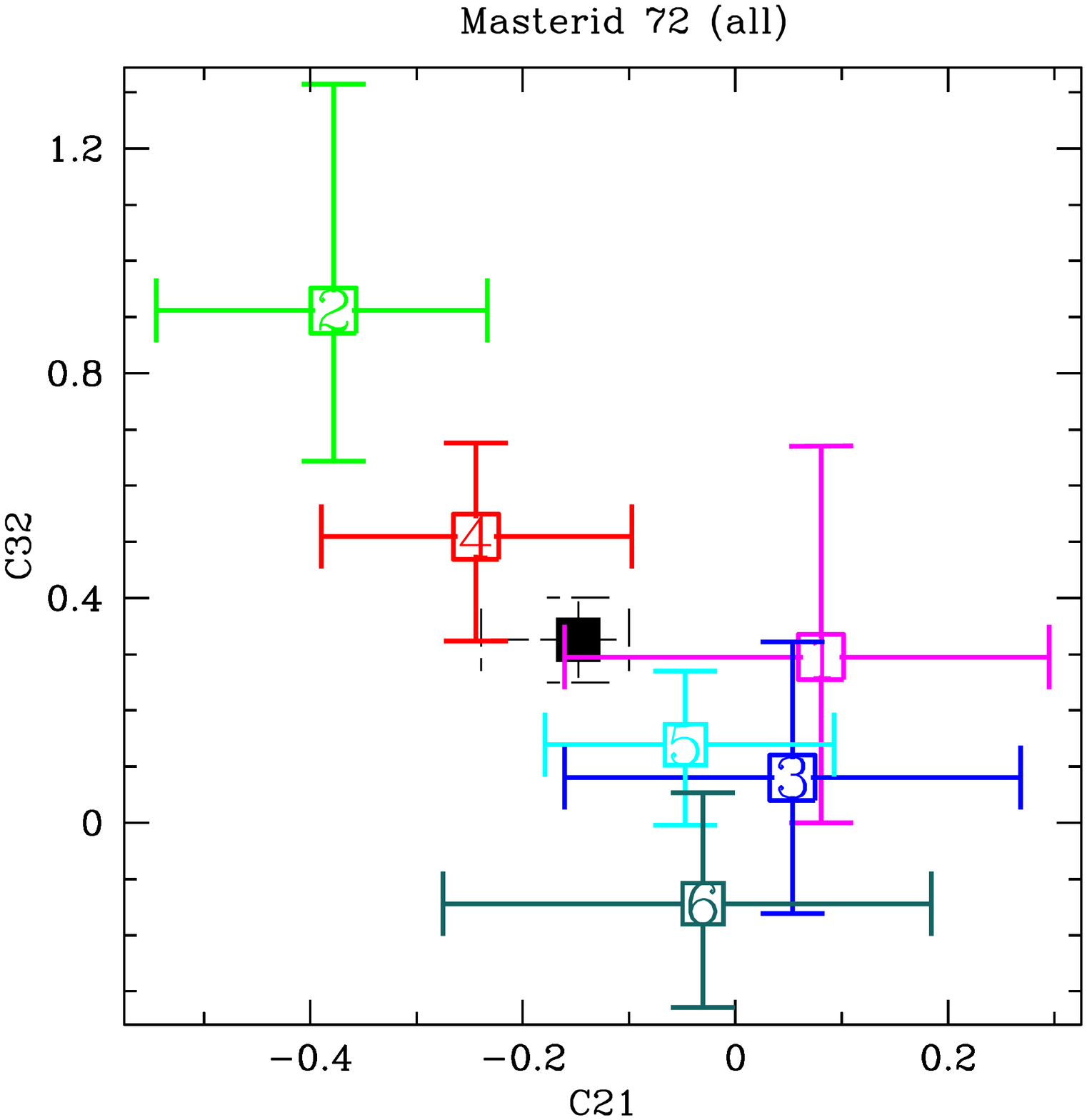}

  \end{minipage}
  \begin{minipage}{0.32\linewidth}
  \centering

    \includegraphics[width=\linewidth]{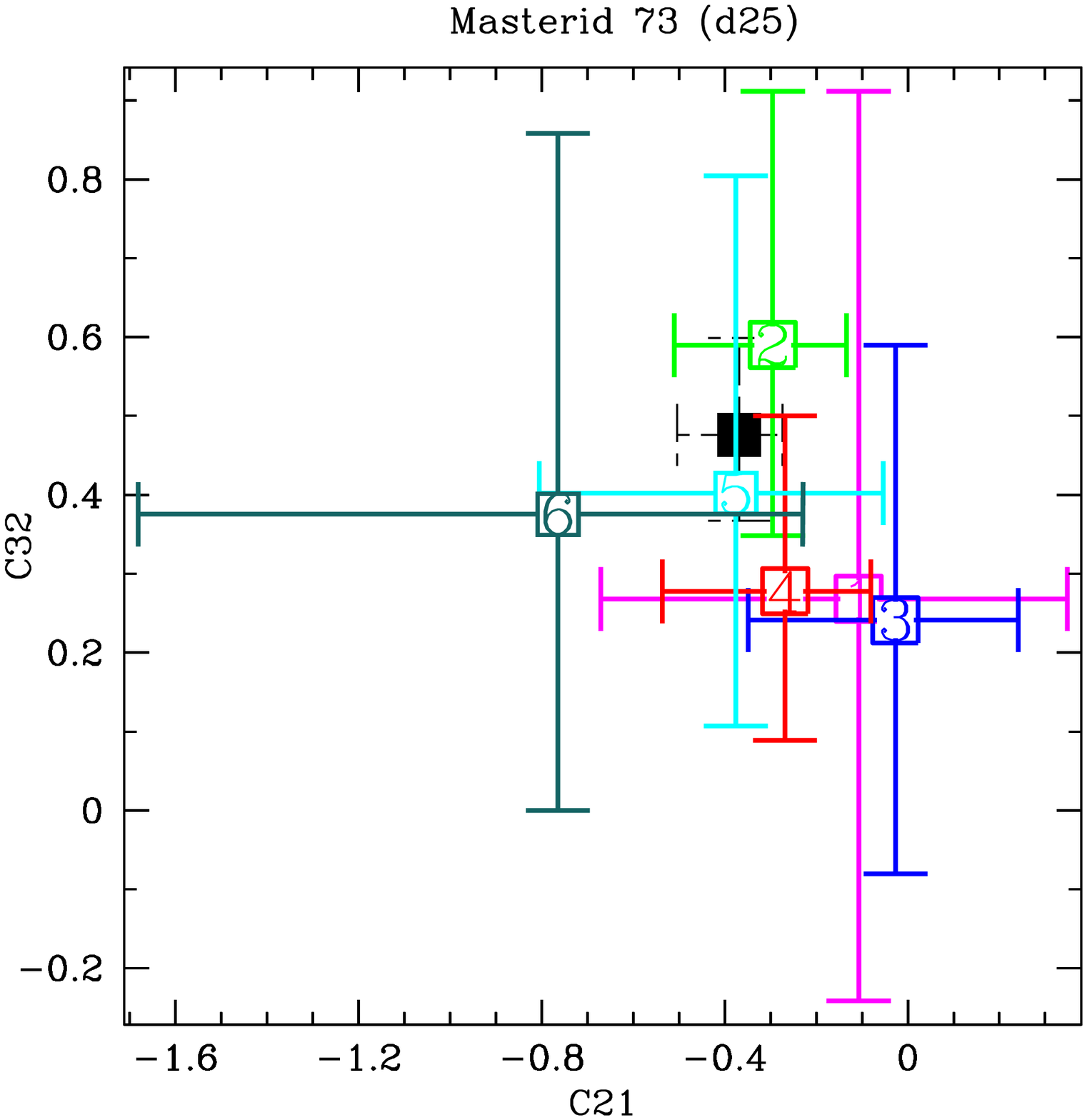}

\end{minipage}
\begin{minipage}{0.32\linewidth}
  \centering

    \includegraphics[width=\linewidth]{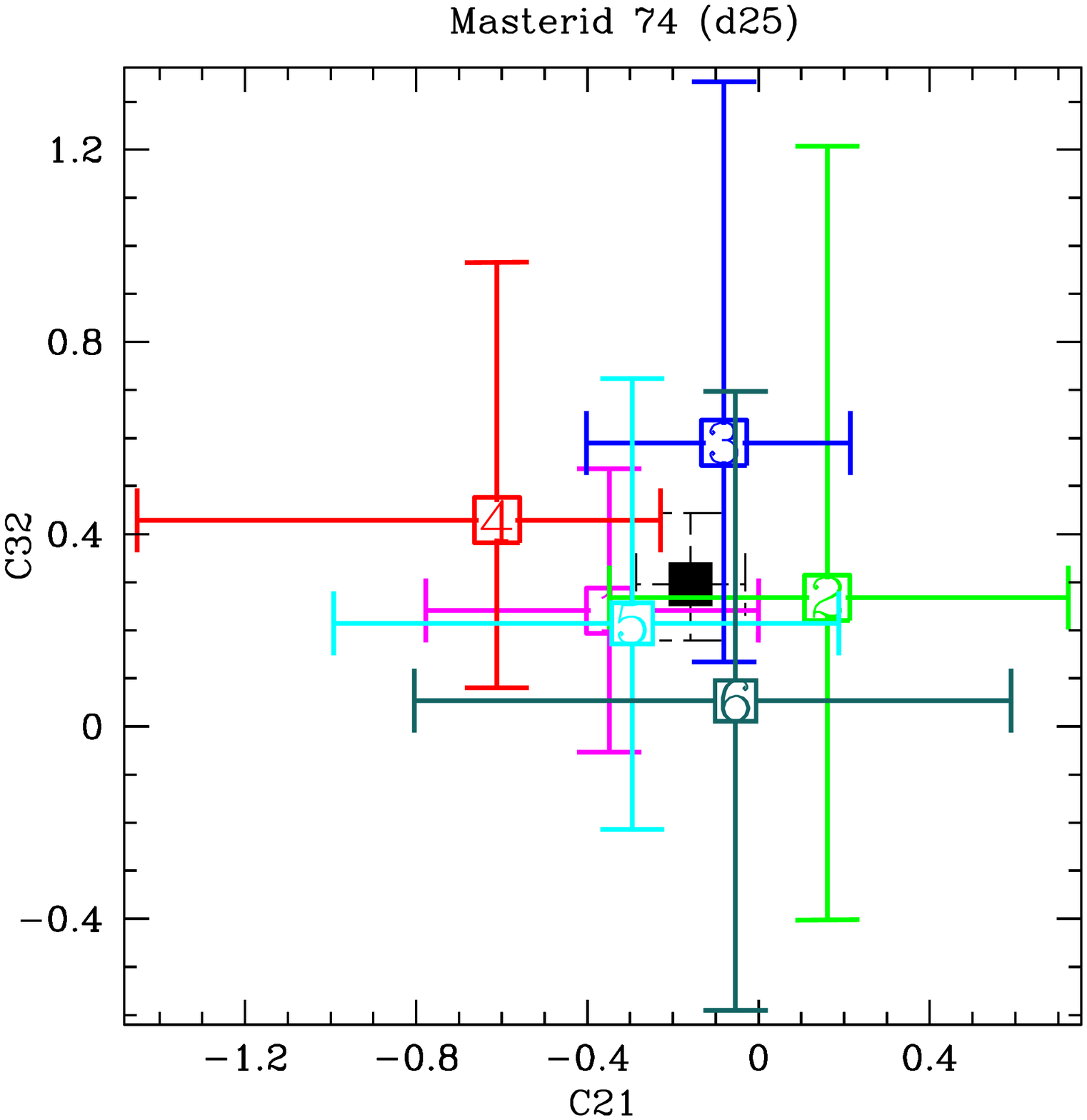}

 \end{minipage}
  
\end{figure}
\clearpage

\begin{figure}
  \begin{minipage}{0.32\linewidth}
  \centering
  
    \includegraphics[width=\linewidth]{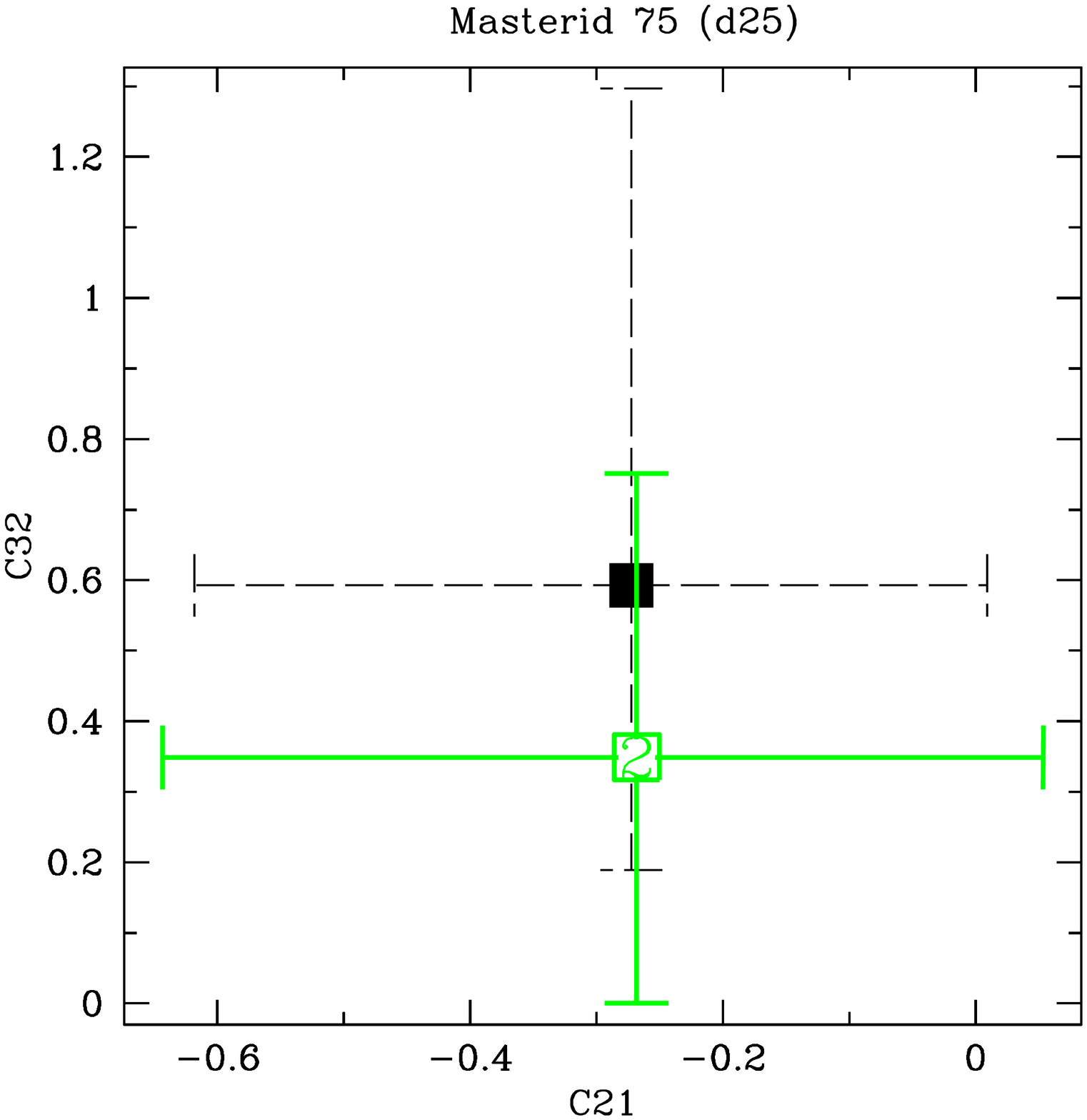}

  \end{minipage}
  \begin{minipage}{0.32\linewidth}
  \centering

    \includegraphics[width=\linewidth]{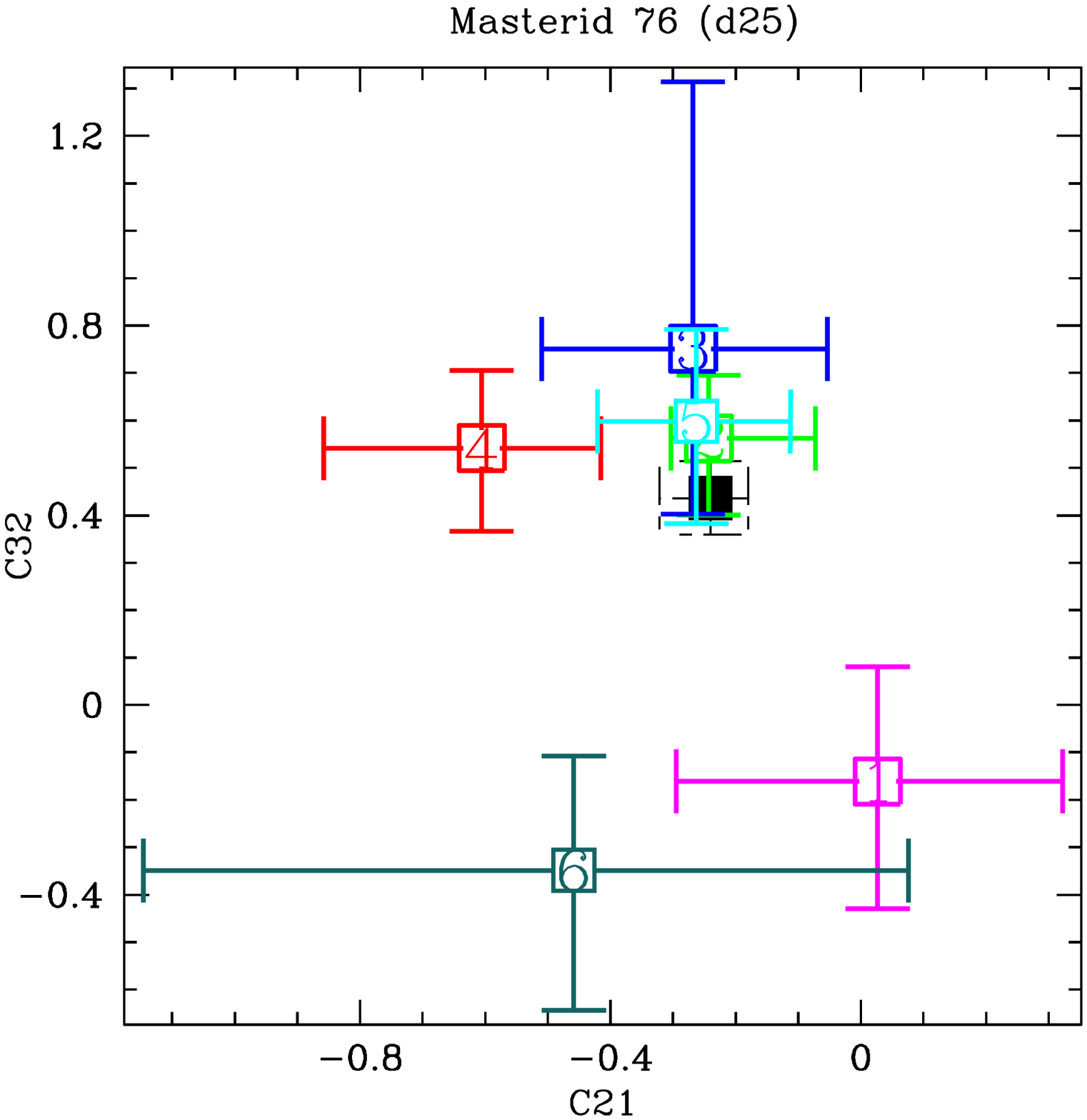}

\end{minipage}
\begin{minipage}{0.32\linewidth}
  \centering

    \includegraphics[width=\linewidth]{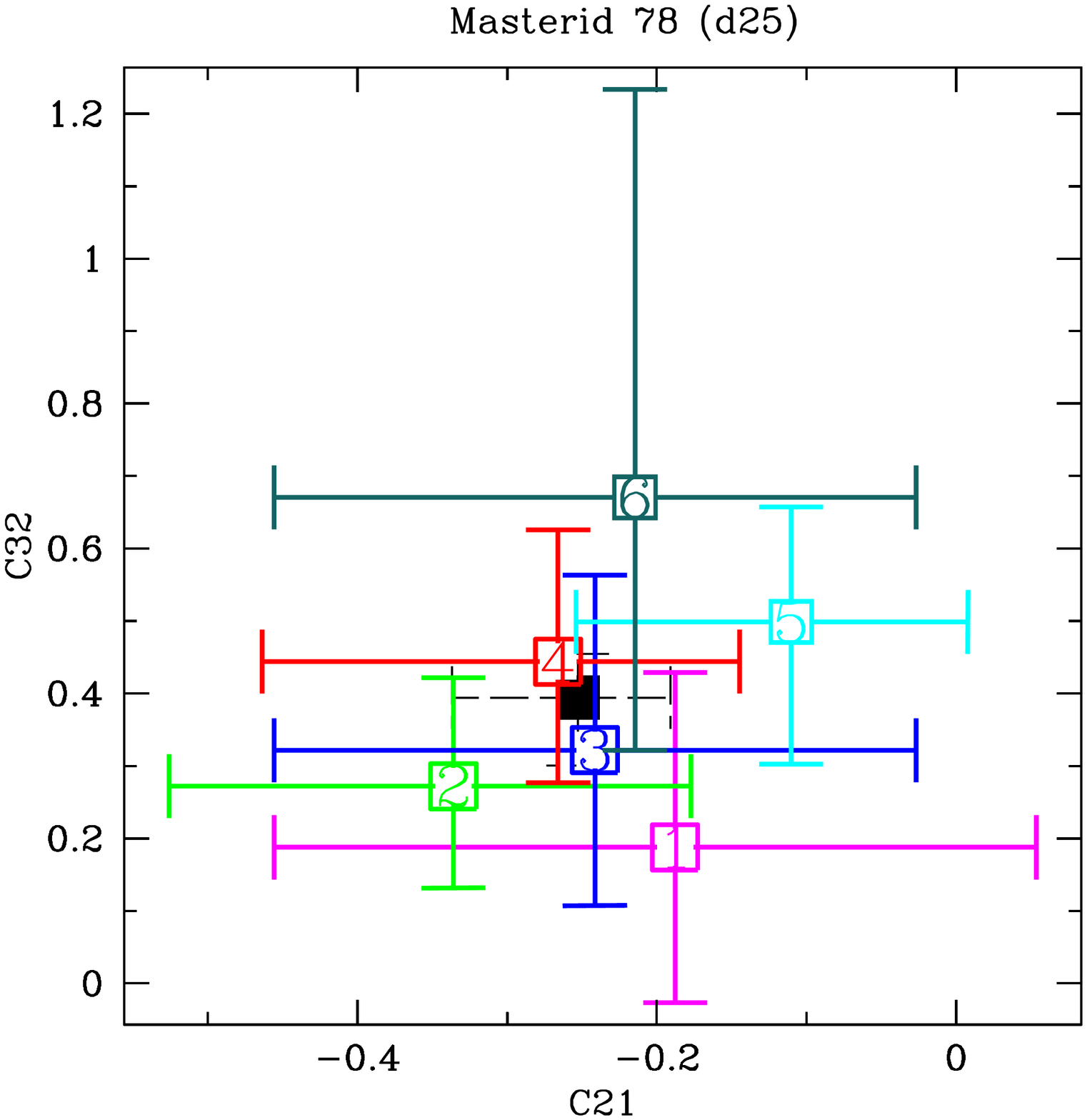}

 \end{minipage}

\begin{minipage}{0.32\linewidth}
  \centering
  
    \includegraphics[width=\linewidth]{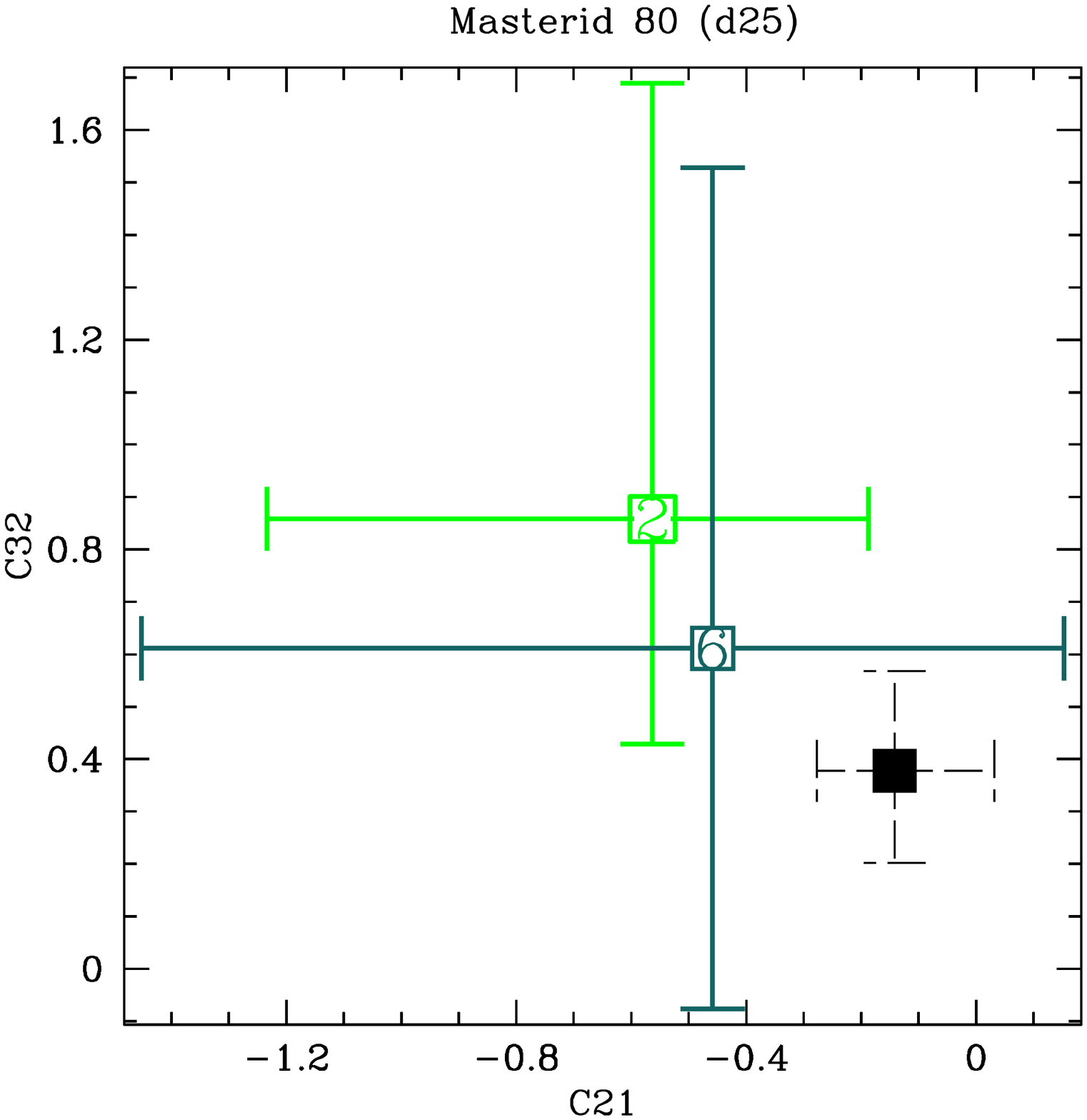}

  \end{minipage}
  \begin{minipage}{0.32\linewidth}
  \centering

    \includegraphics[width=\linewidth]{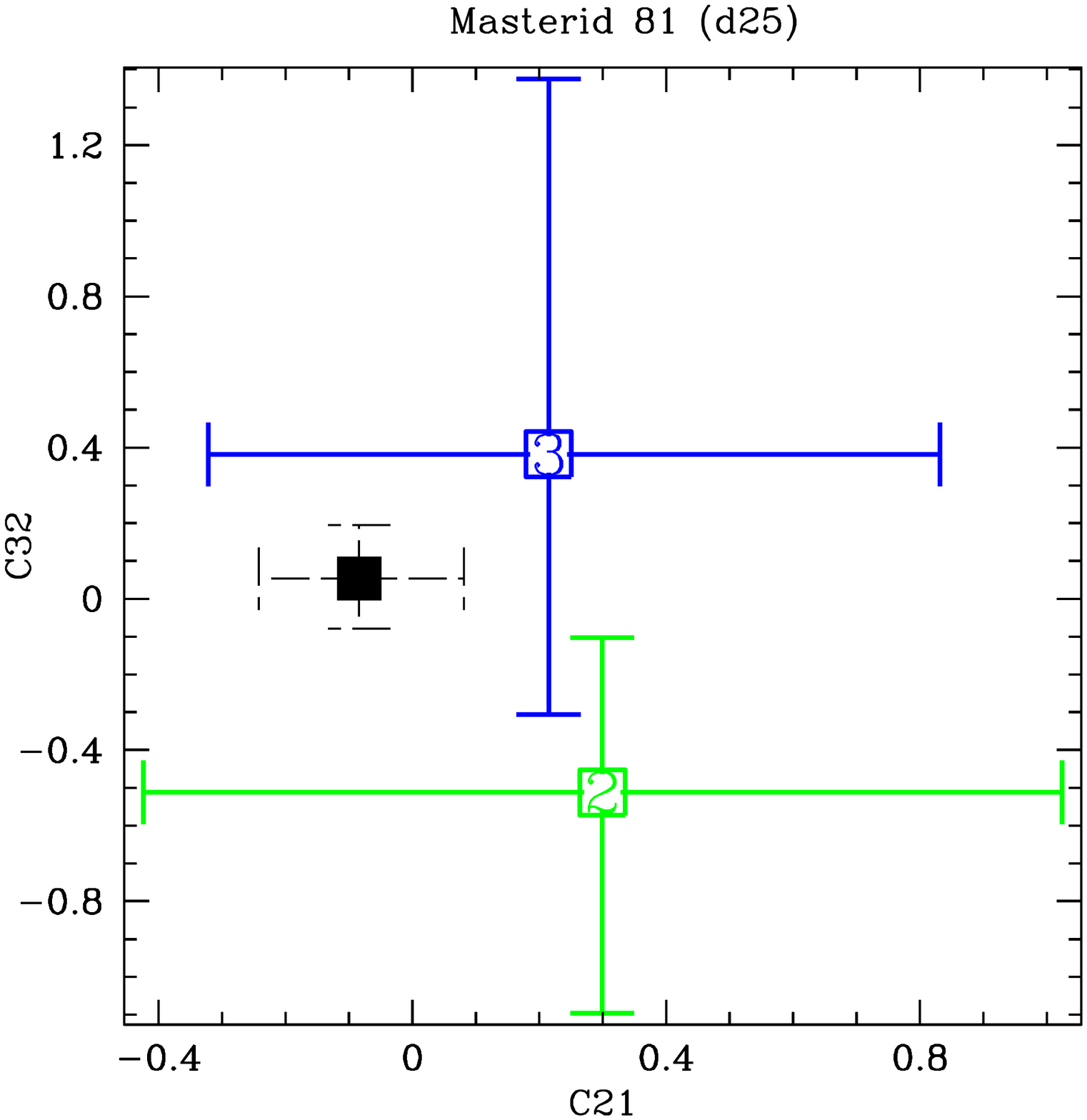}

\end{minipage}
\begin{minipage}{0.32\linewidth}
  \centering

    \includegraphics[width=\linewidth]{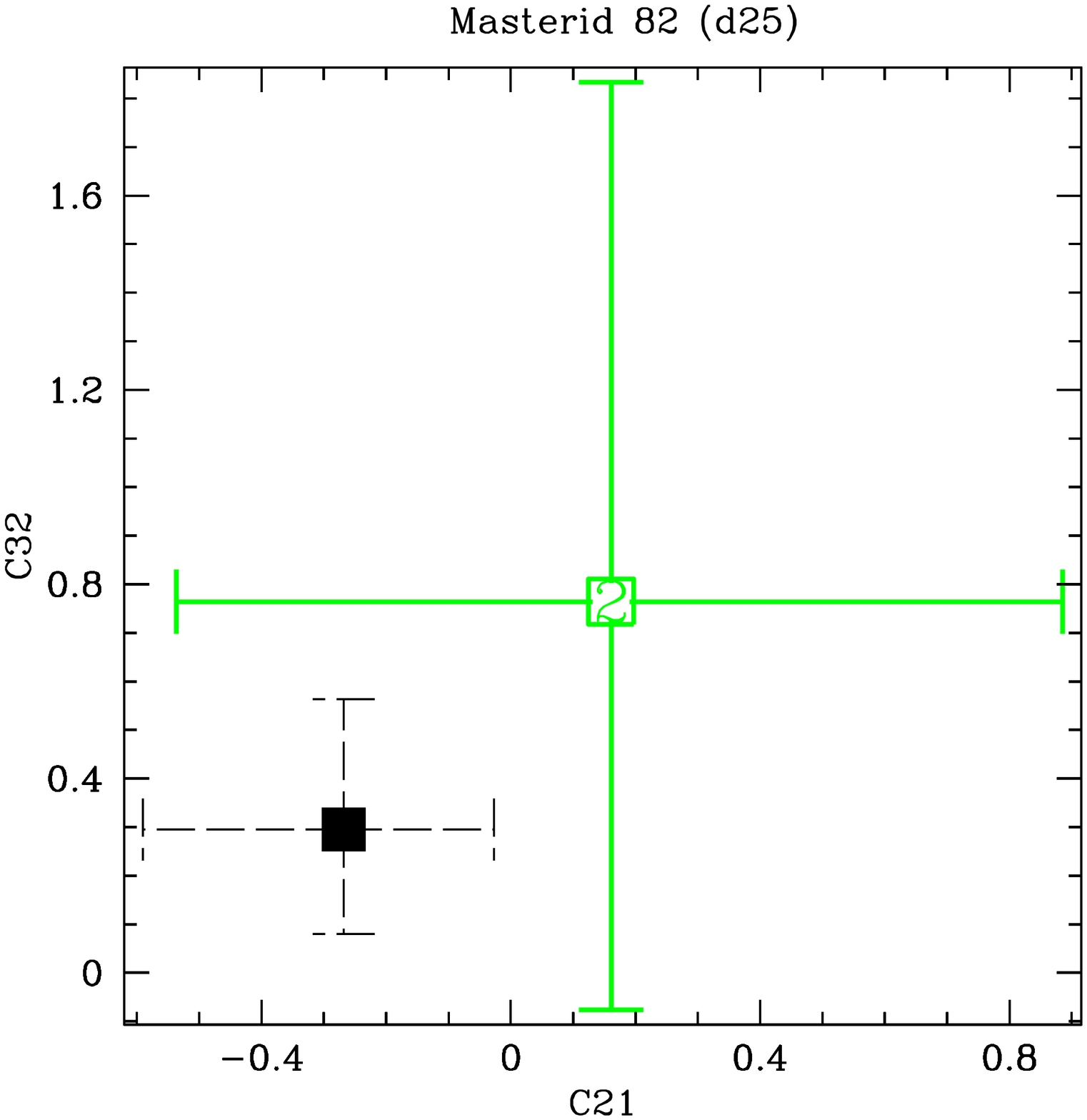}

 \end{minipage}

  \begin{minipage}{0.32\linewidth}
  \centering
  
    \includegraphics[width=\linewidth]{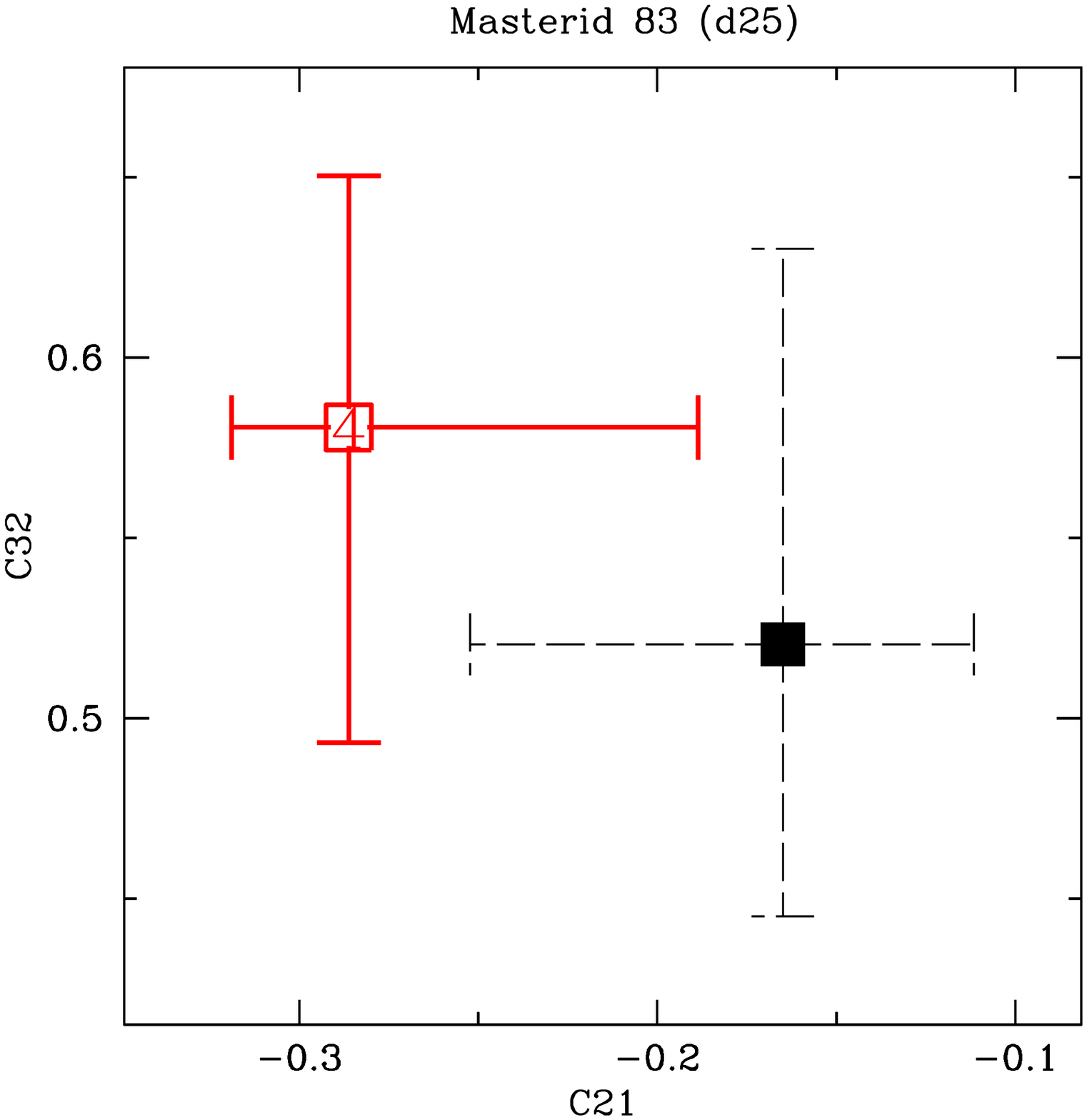}

  \end{minipage}
  \begin{minipage}{0.32\linewidth}
  \centering

    \includegraphics[width=\linewidth]{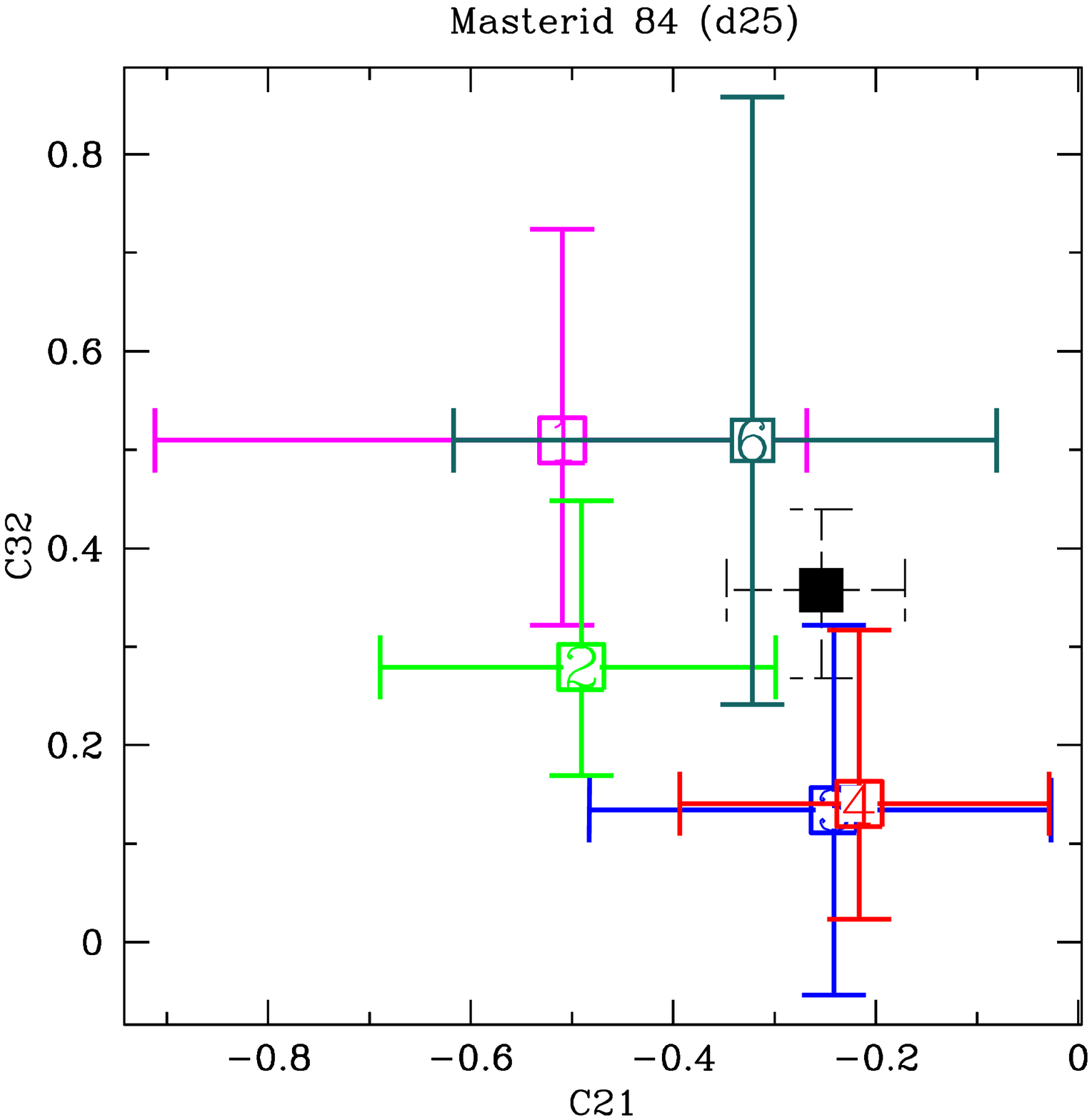}

\end{minipage}
\begin{minipage}{0.32\linewidth}
  \centering

    \includegraphics[width=\linewidth]{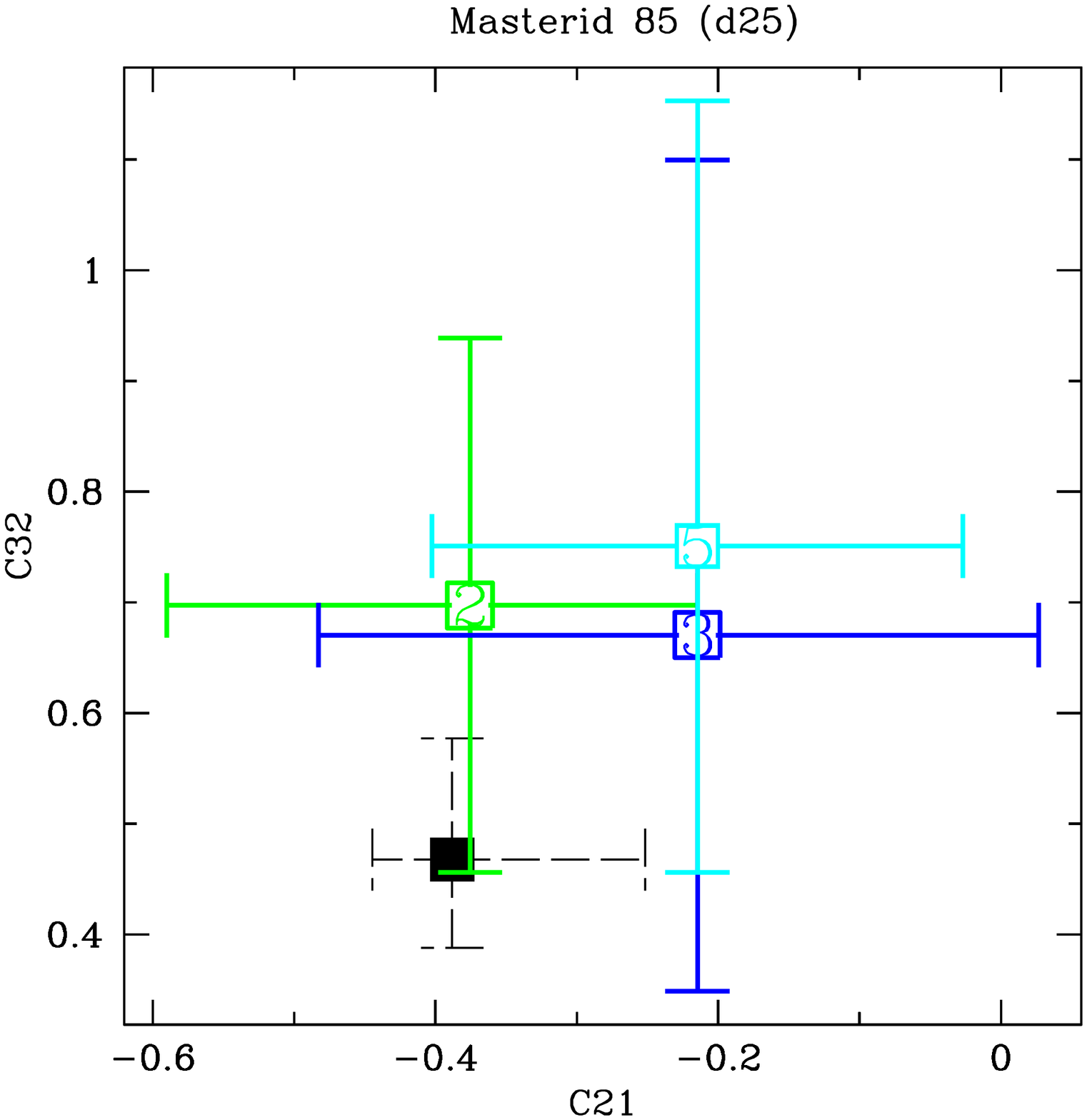}

 \end{minipage}

\begin{minipage}{0.32\linewidth}
  \centering
  
    \includegraphics[width=\linewidth]{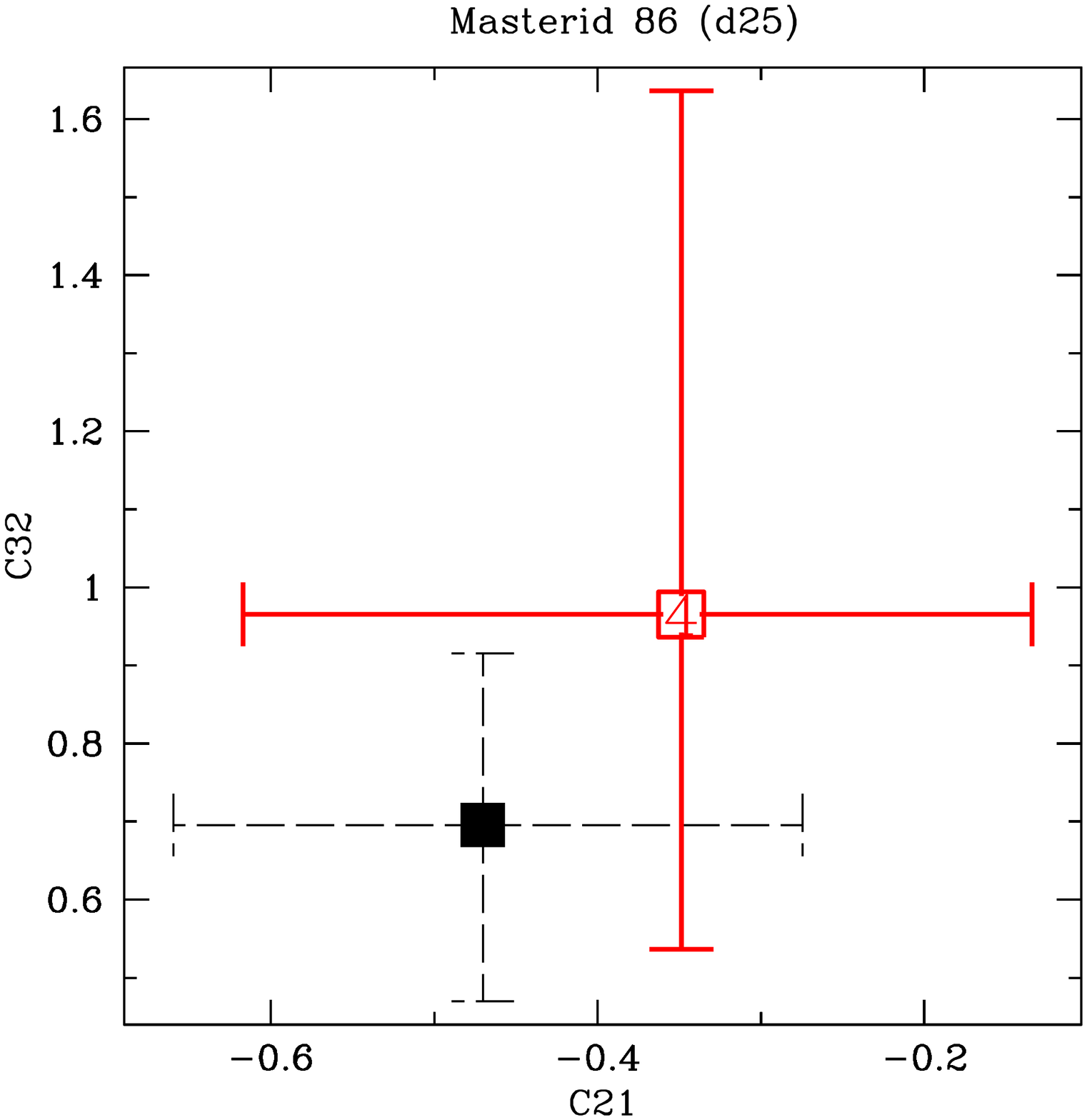}

  \end{minipage}
  \begin{minipage}{0.32\linewidth}
  \centering

    \includegraphics[width=\linewidth]{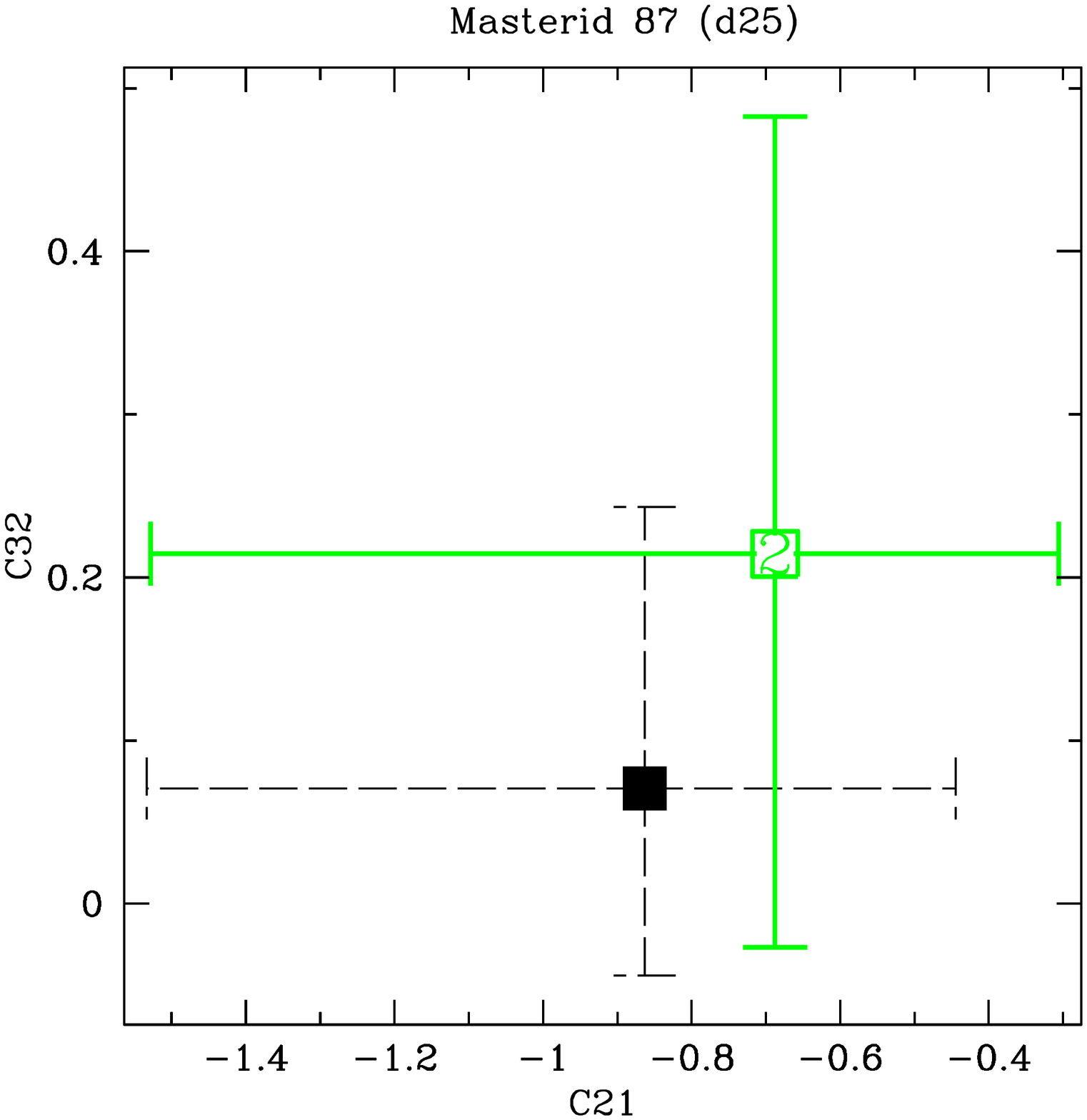}

\end{minipage}
\begin{minipage}{0.32\linewidth}
  \centering

    \includegraphics[width=\linewidth]{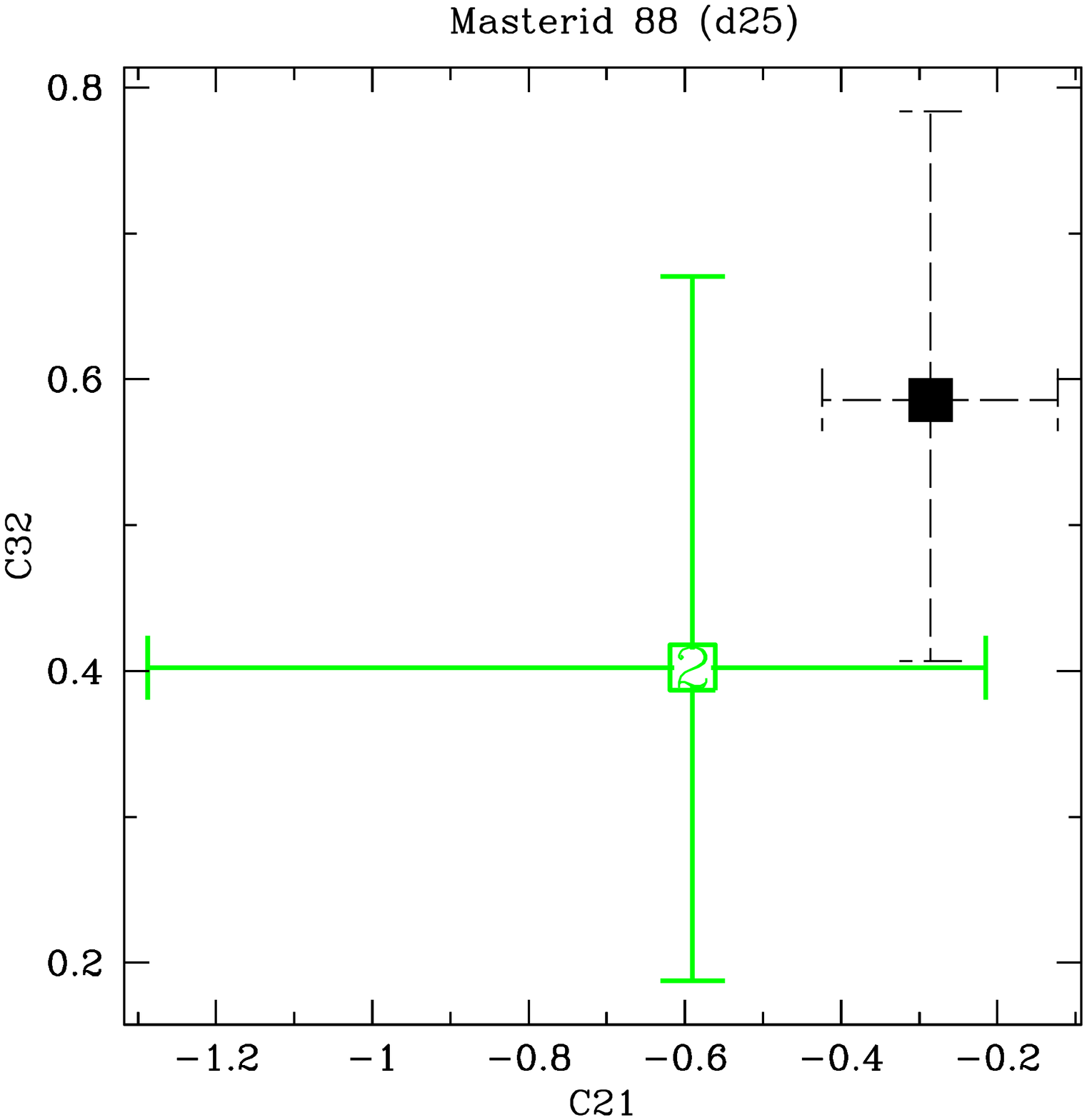}

 \end{minipage}
  
\end{figure}

\begin{figure}
  \begin{minipage}{0.32\linewidth}
  \centering
  
    \includegraphics[width=\linewidth]{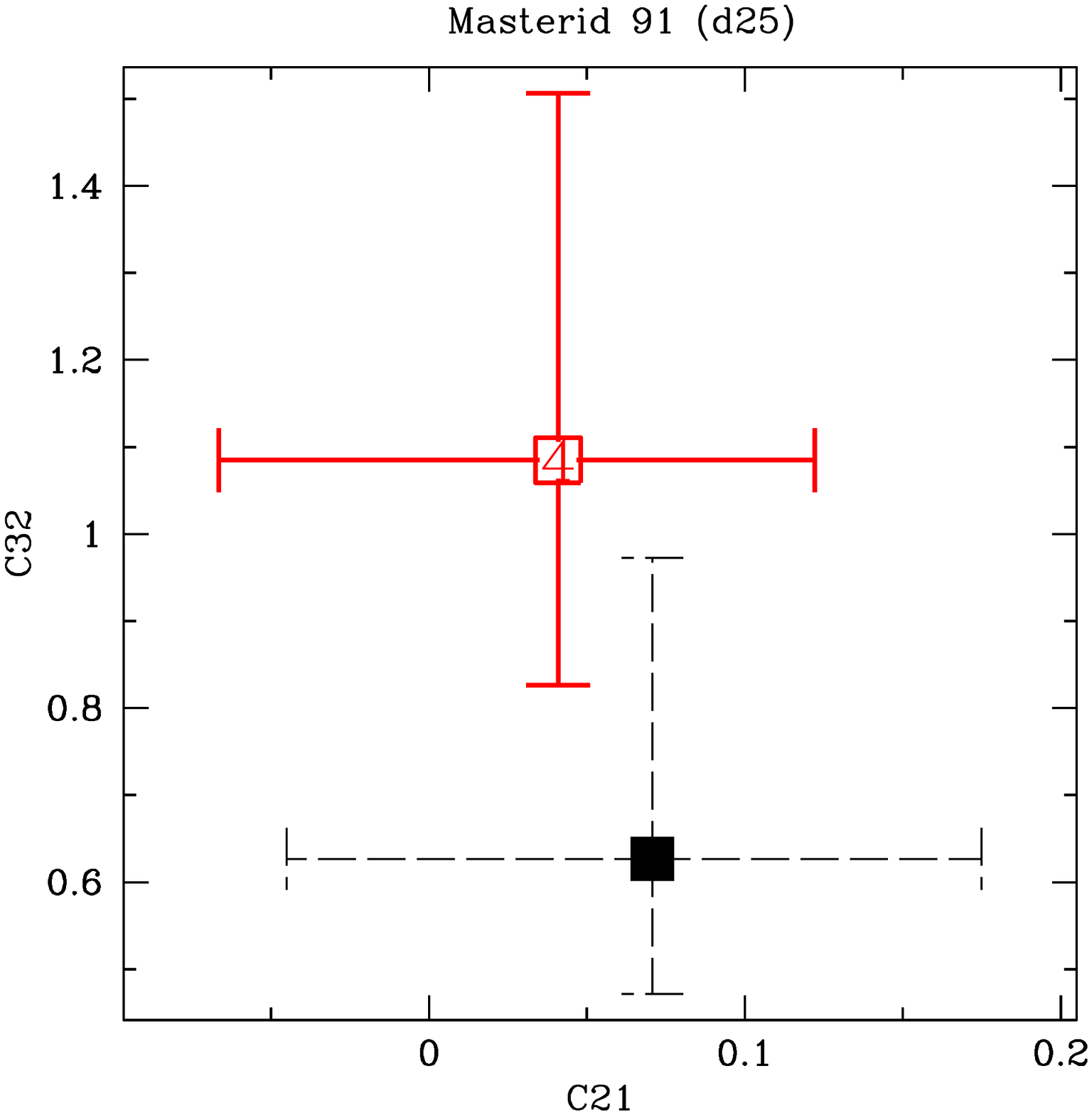}

  \end{minipage}
  \begin{minipage}{0.32\linewidth}
  \centering

    \includegraphics[width=\linewidth]{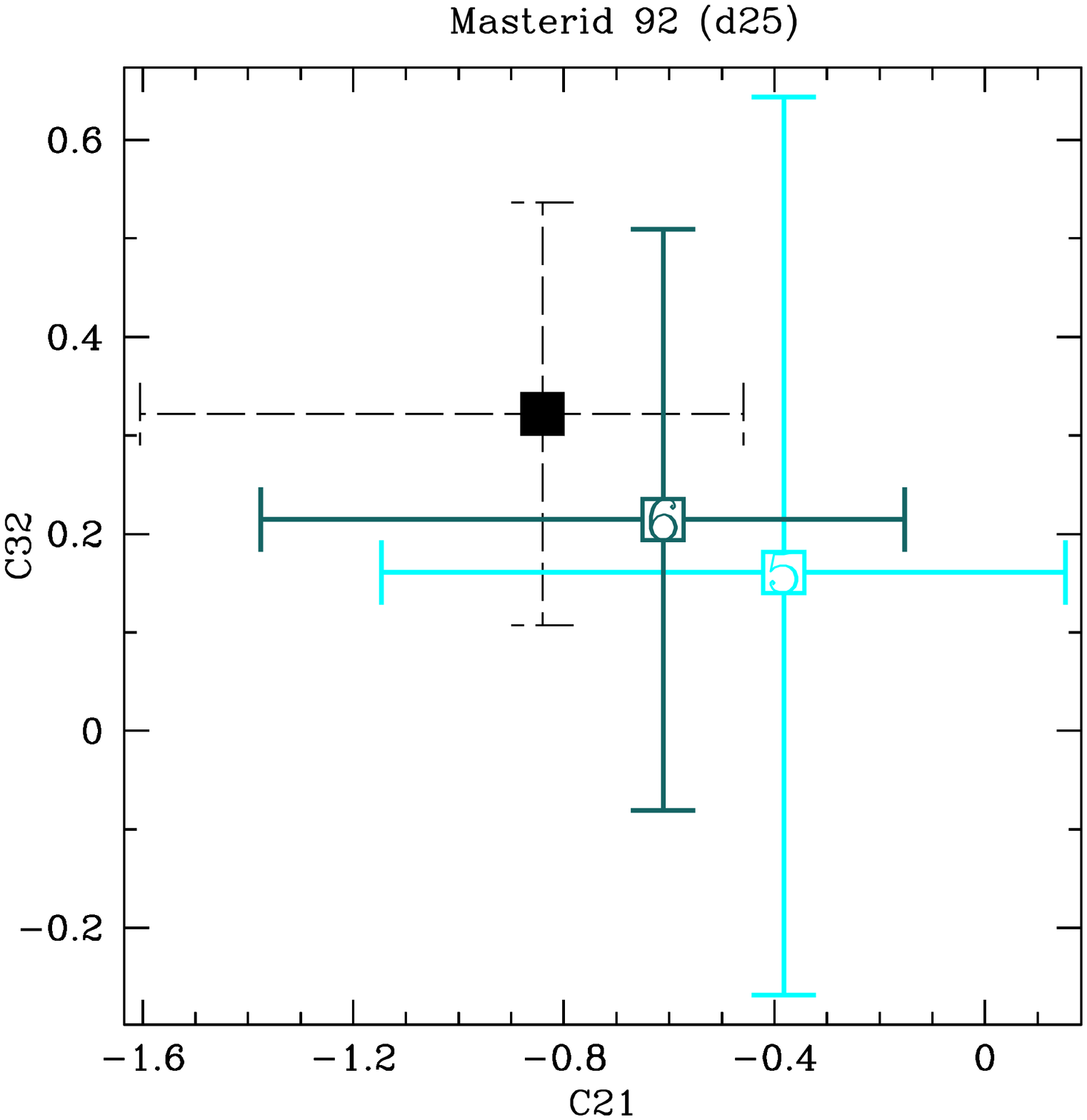}

\end{minipage}
\begin{minipage}{0.32\linewidth}
  \centering

    \includegraphics[width=\linewidth]{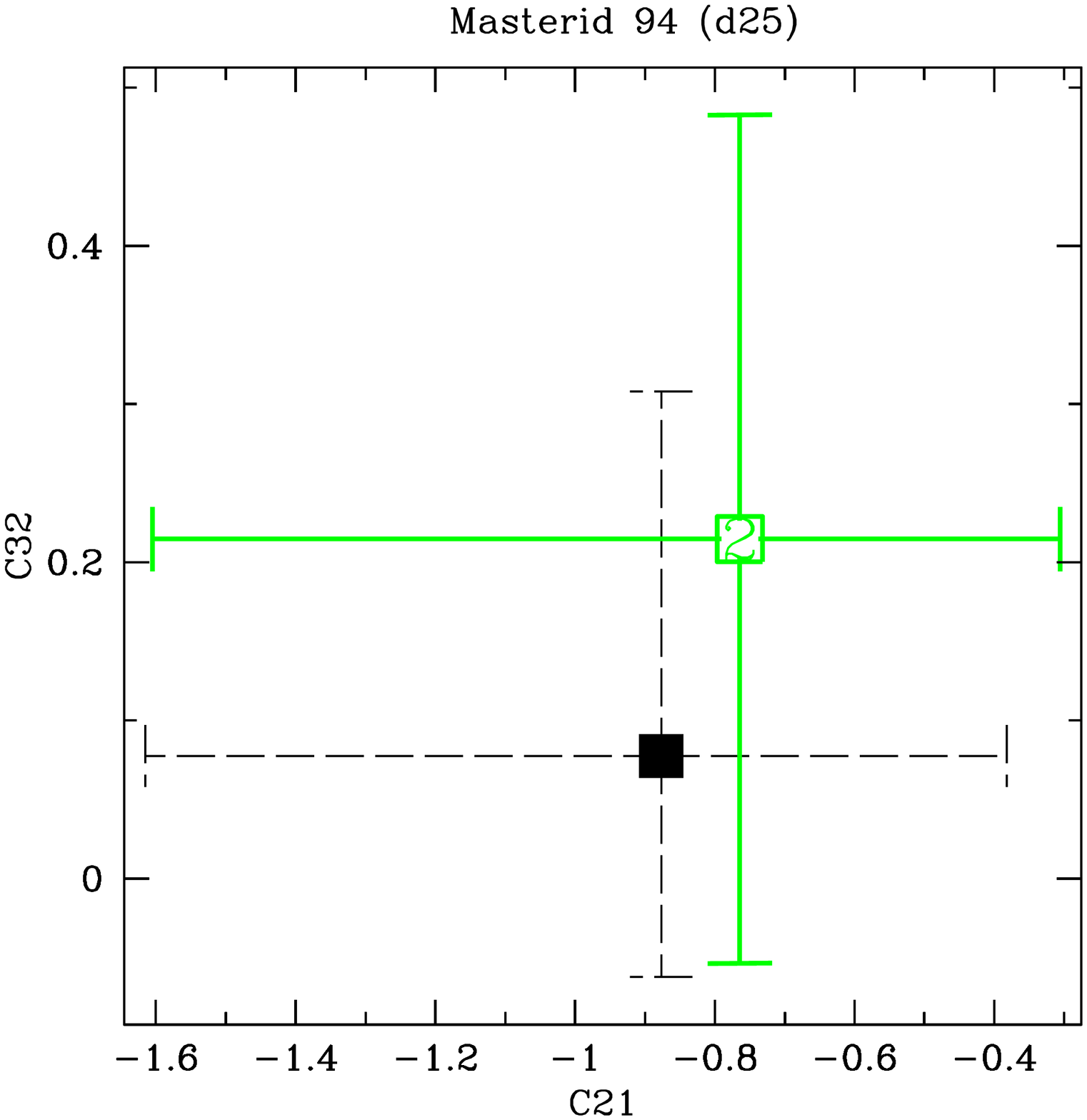}

 \end{minipage}

\begin{minipage}{0.32\linewidth}
  \centering
  
    \includegraphics[width=\linewidth]{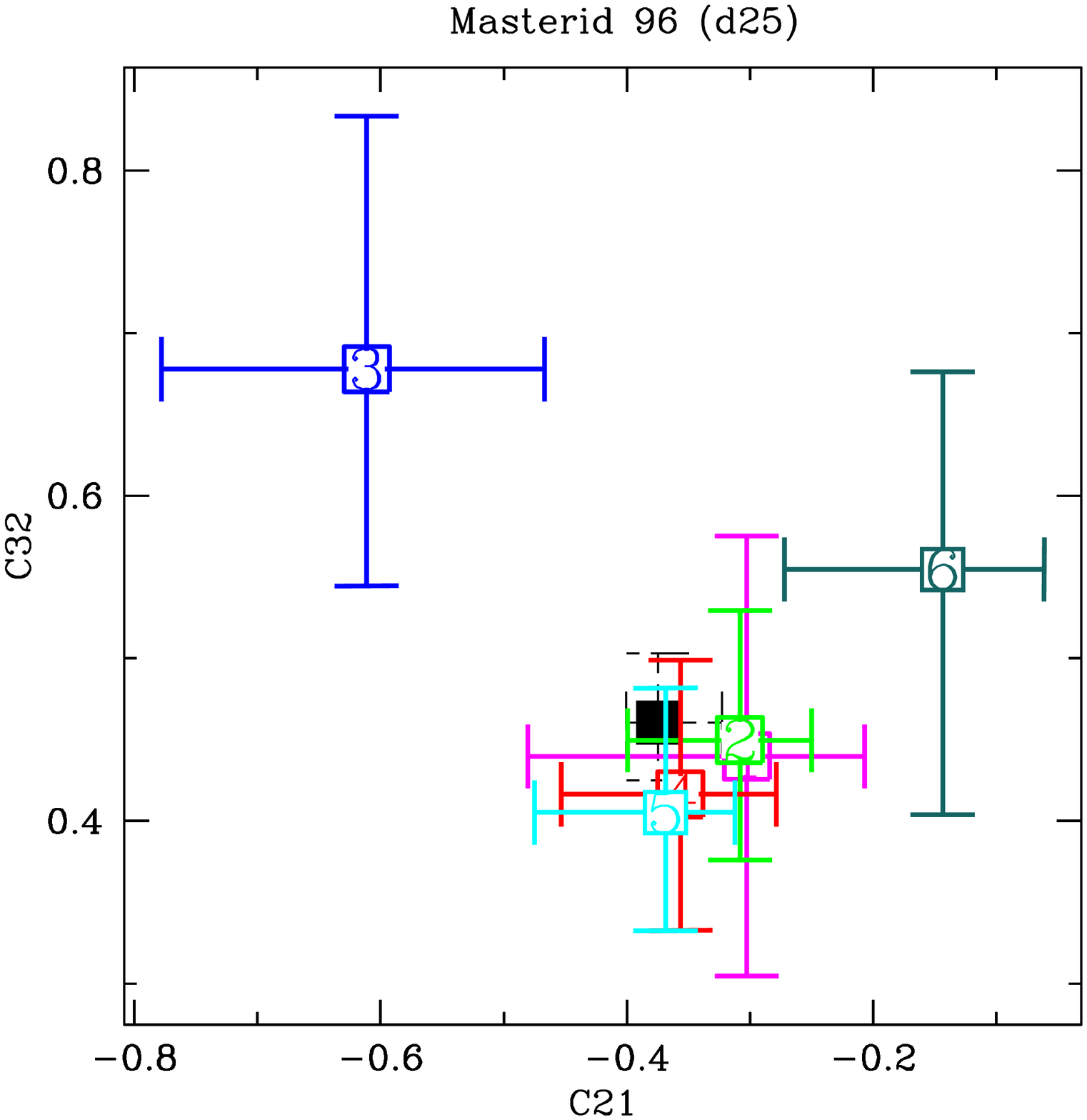}

  \end{minipage}
  \begin{minipage}{0.32\linewidth}
  \centering

    \includegraphics[width=\linewidth]{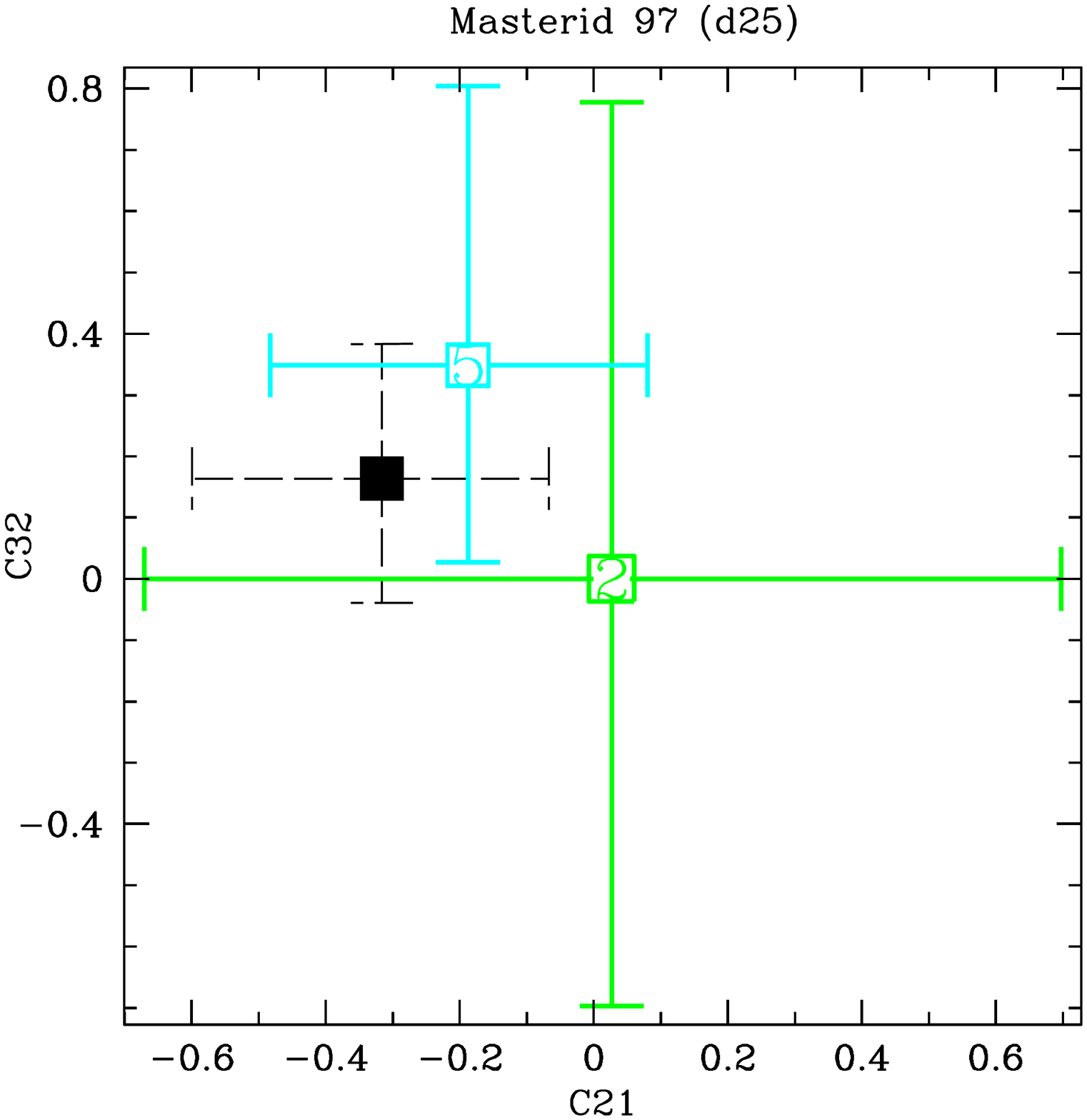}

\end{minipage}
\begin{minipage}{0.32\linewidth}
  \centering

    \includegraphics[width=\linewidth]{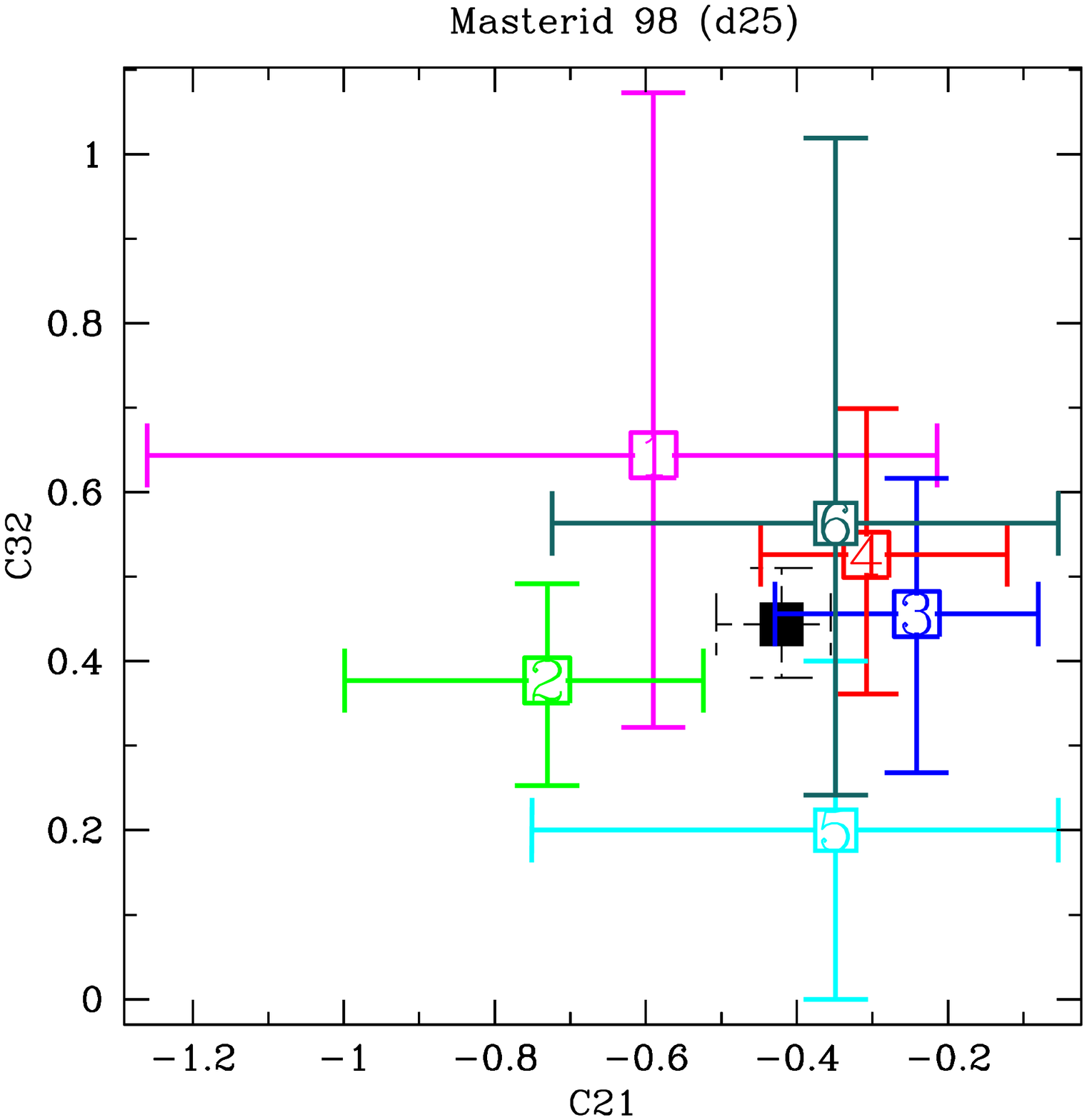}

 \end{minipage}

  \begin{minipage}{0.32\linewidth}
  \centering
  
    \includegraphics[width=\linewidth]{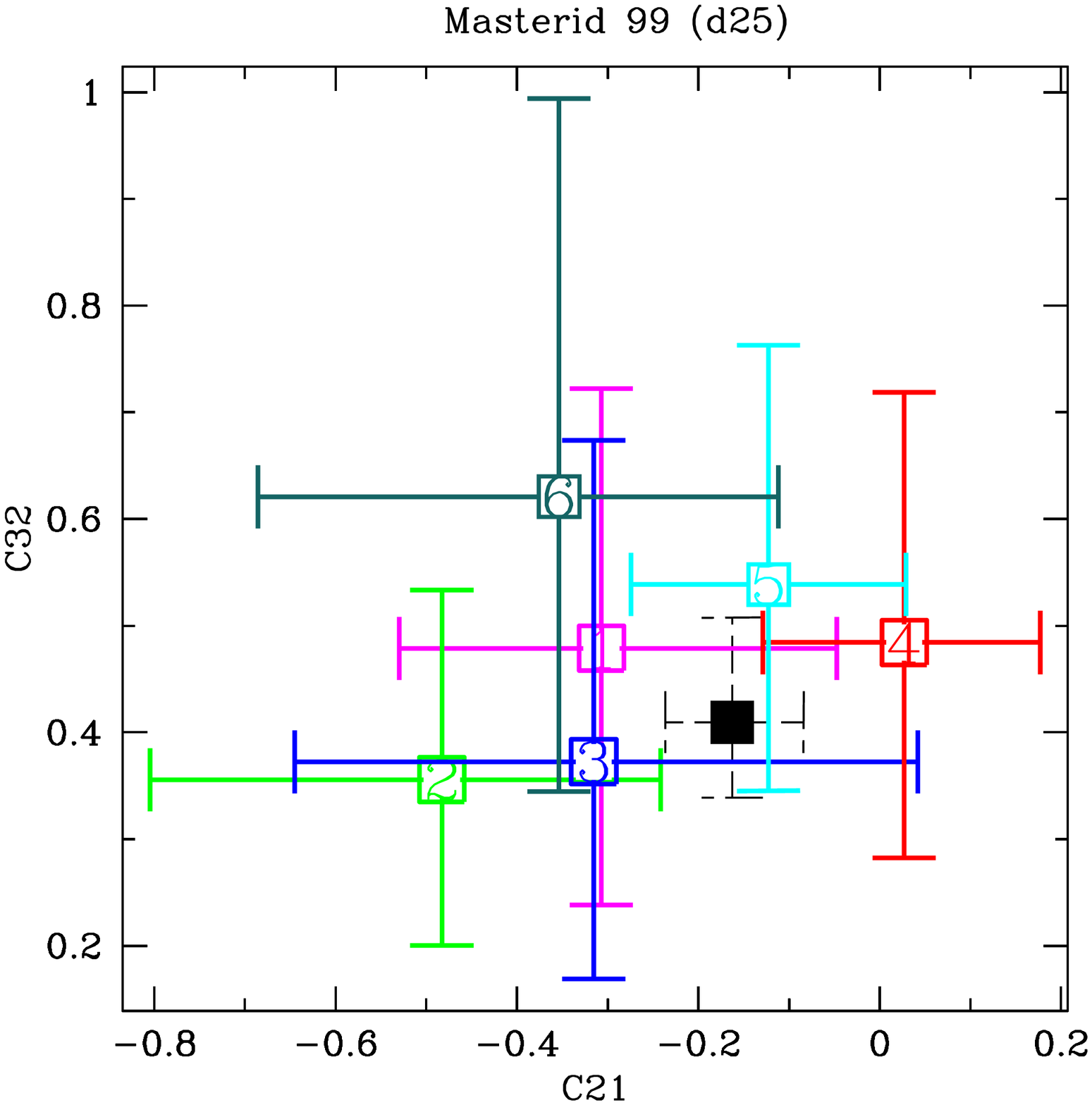}

  \end{minipage}
  \begin{minipage}{0.32\linewidth}
  \centering

    \includegraphics[width=\linewidth]{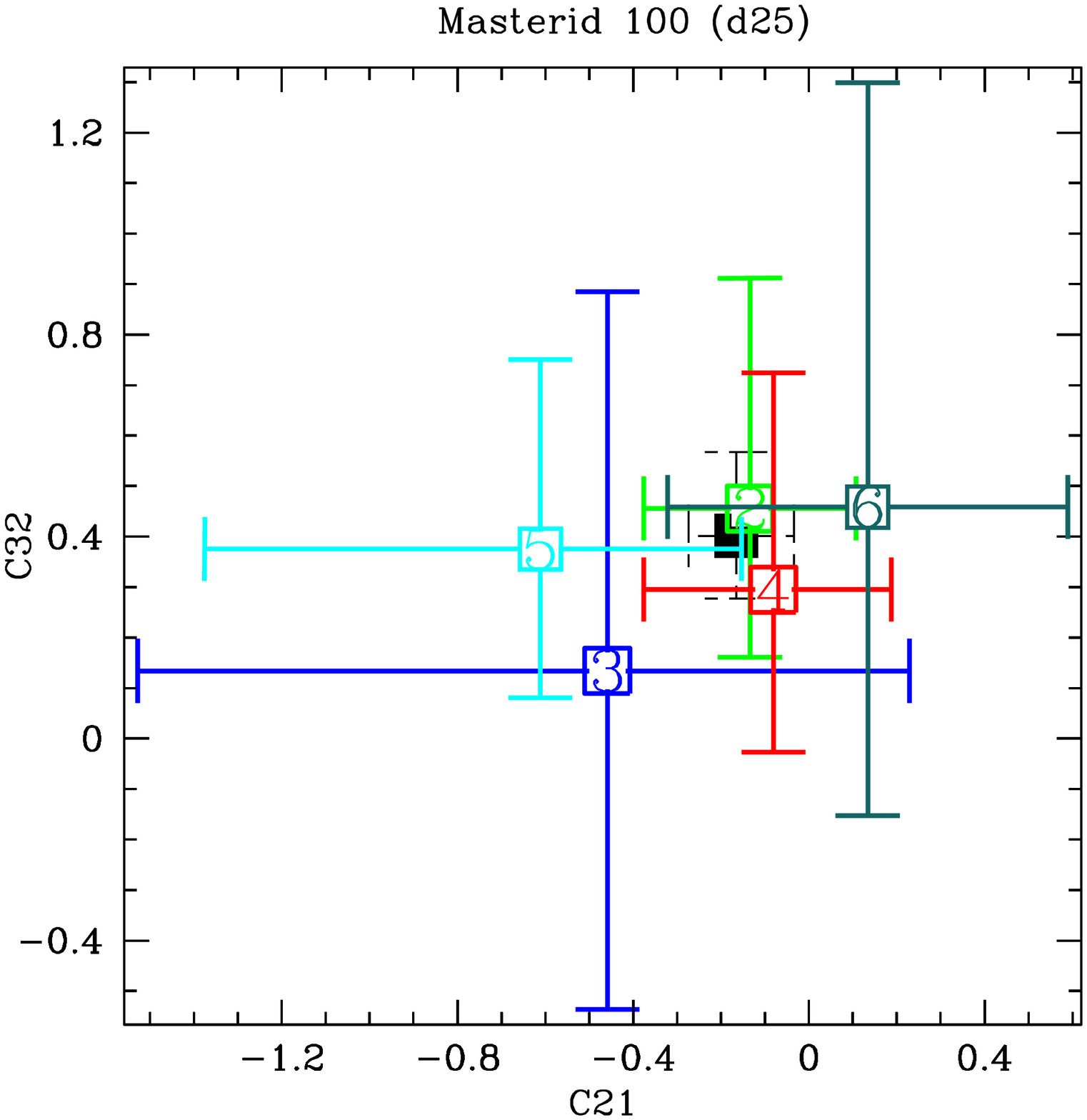}

\end{minipage}
\begin{minipage}{0.32\linewidth}
  \centering

    \includegraphics[width=\linewidth]{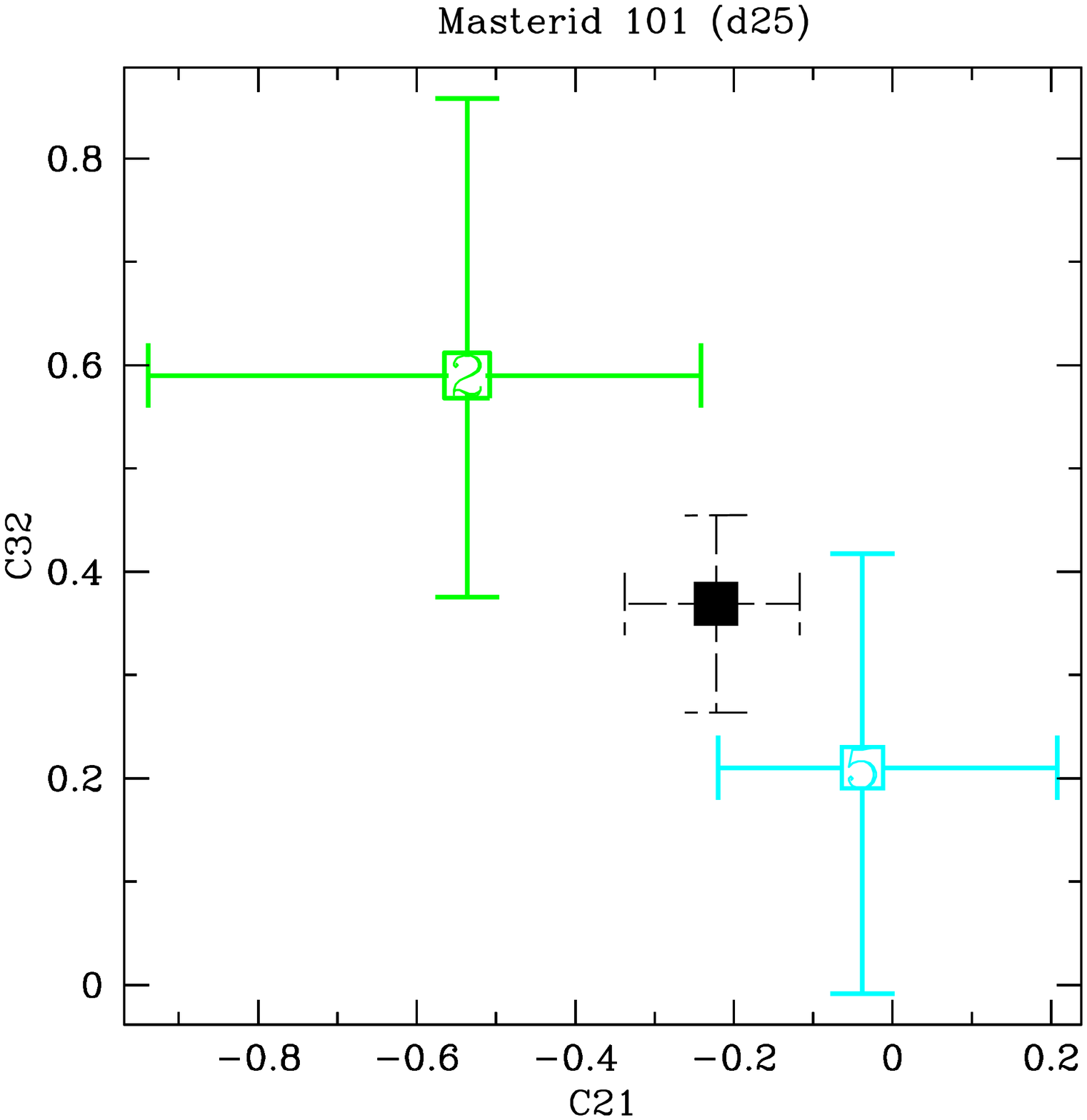}

\end{minipage}

\begin{minipage}{0.32\linewidth}
  \centering
  
    \includegraphics[width=\linewidth]{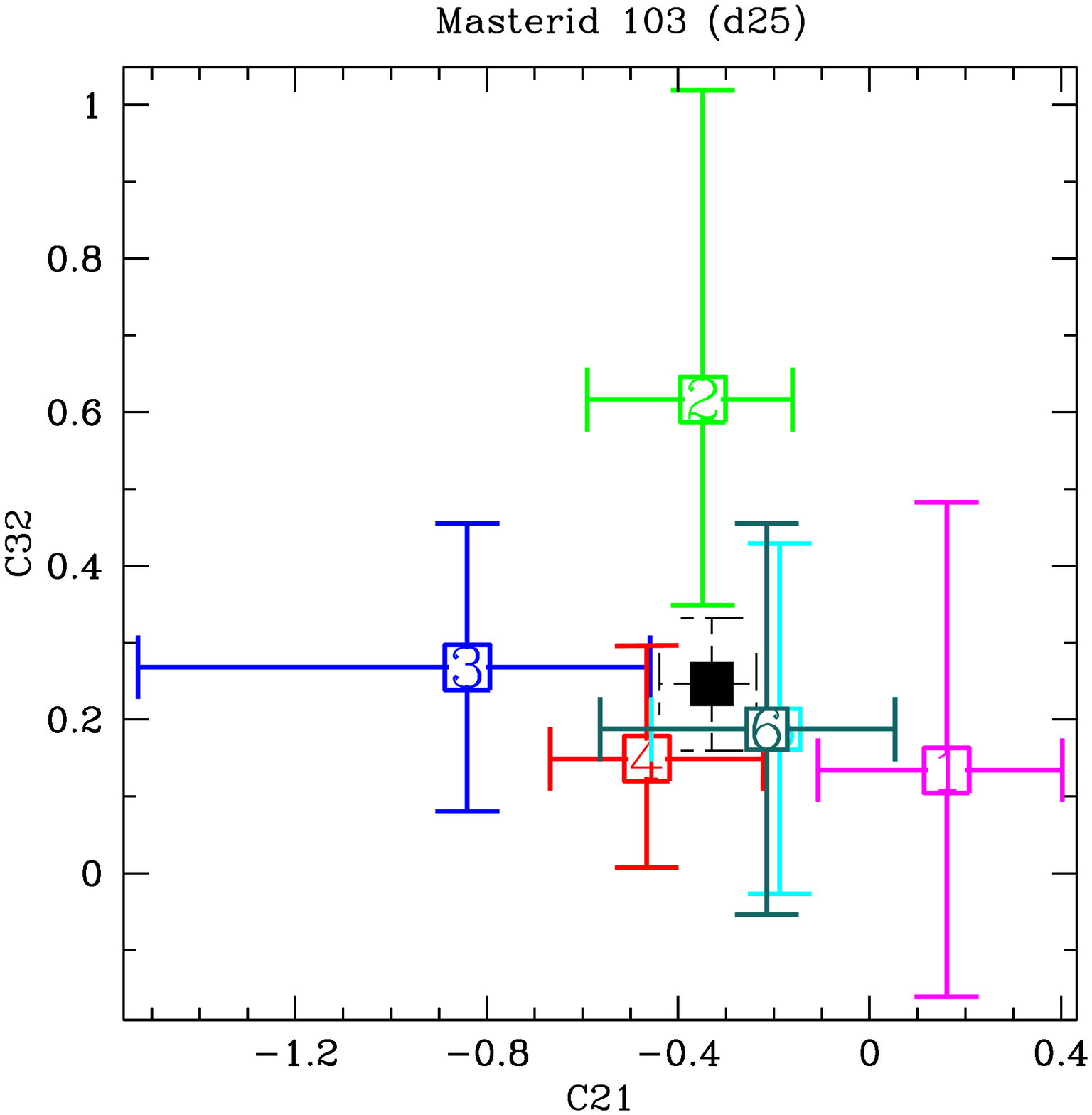}

  \end{minipage}
  \begin{minipage}{0.32\linewidth}
  \centering

    \includegraphics[width=\linewidth]{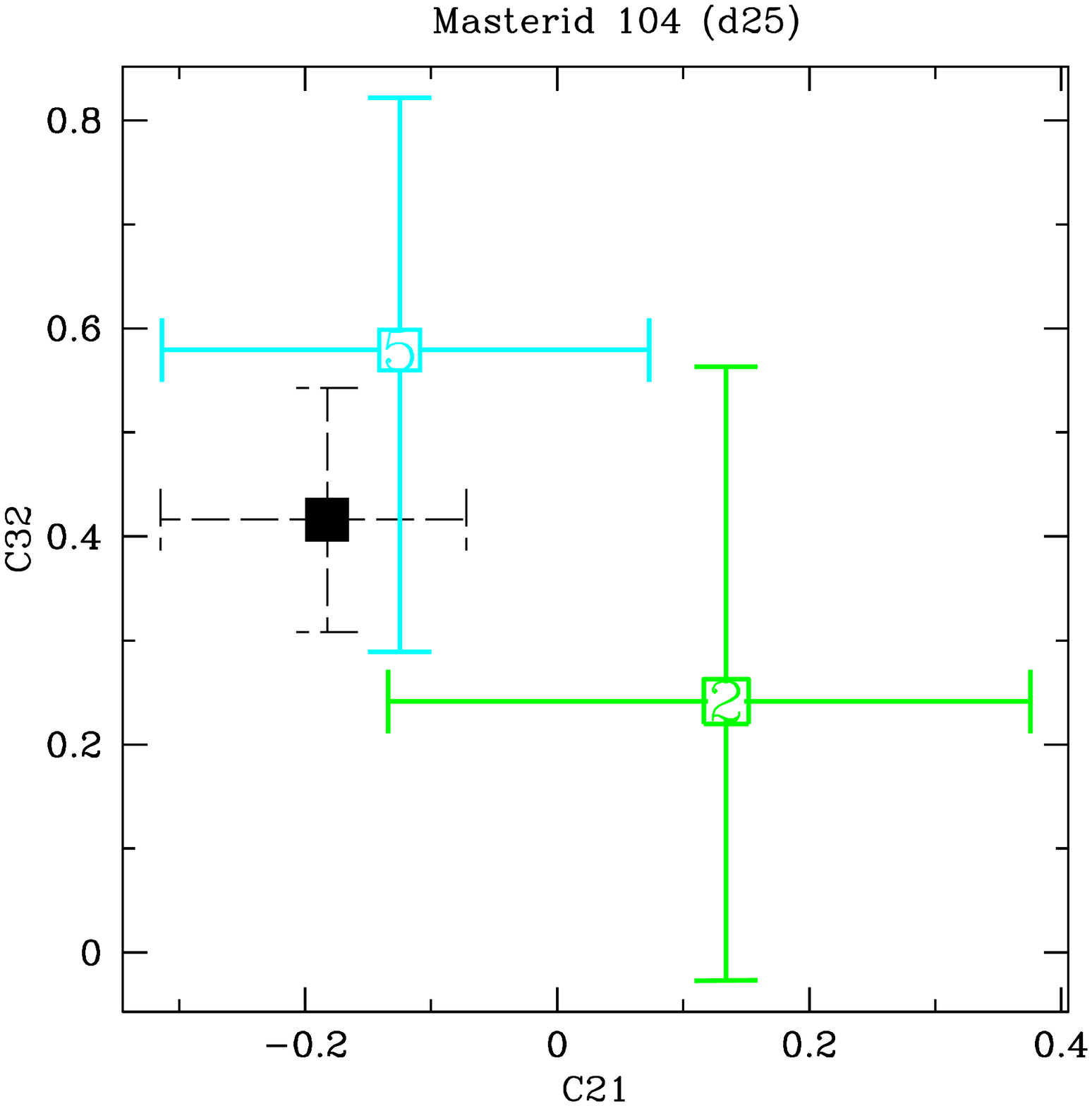}

\end{minipage}
\begin{minipage}{0.32\linewidth}
  \centering

    \includegraphics[width=\linewidth]{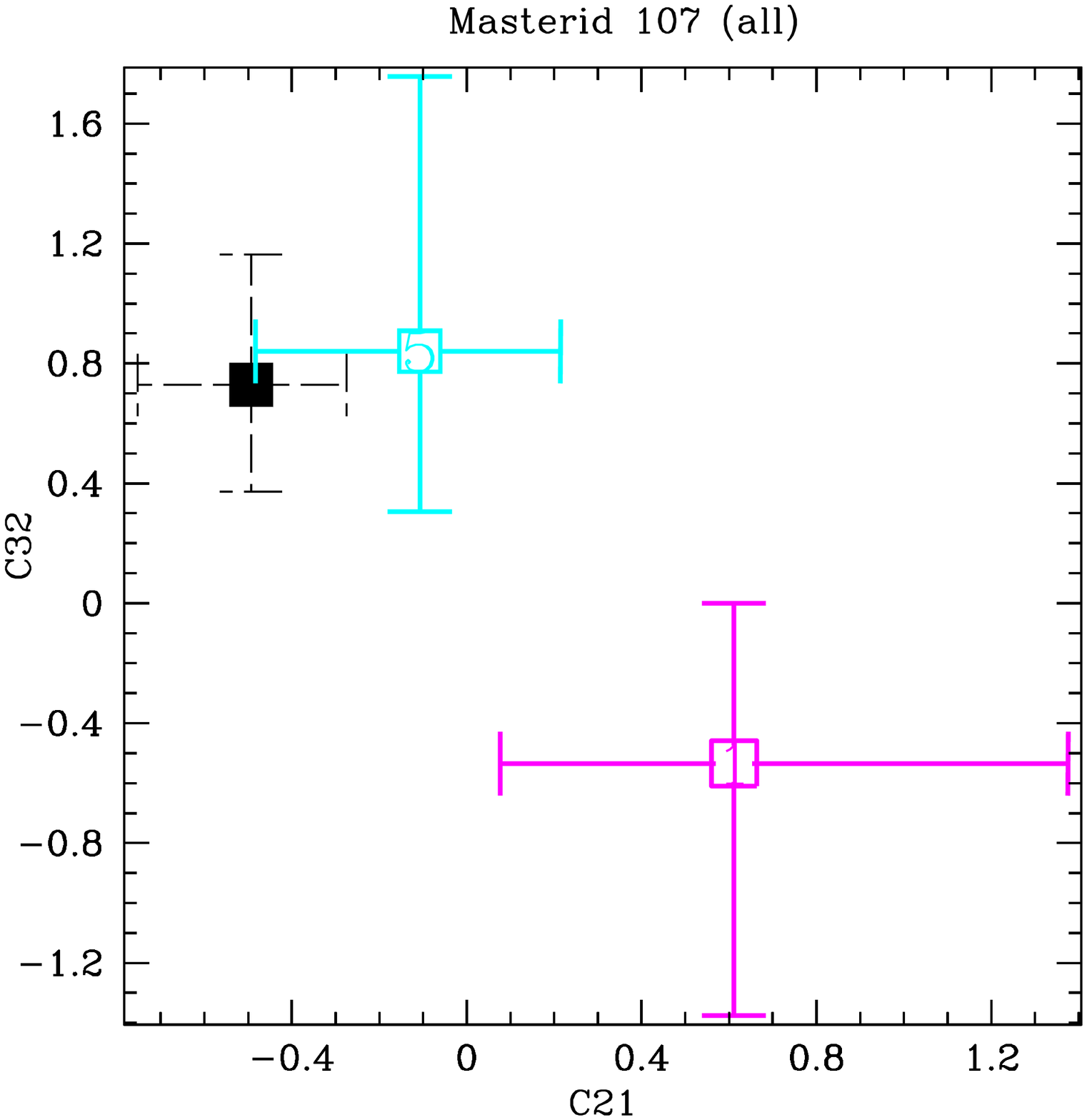}

\end{minipage}
\end{figure}

\begin{figure}
  \begin{minipage}{0.32\linewidth}
  \centering
  
    \includegraphics[width=\linewidth]{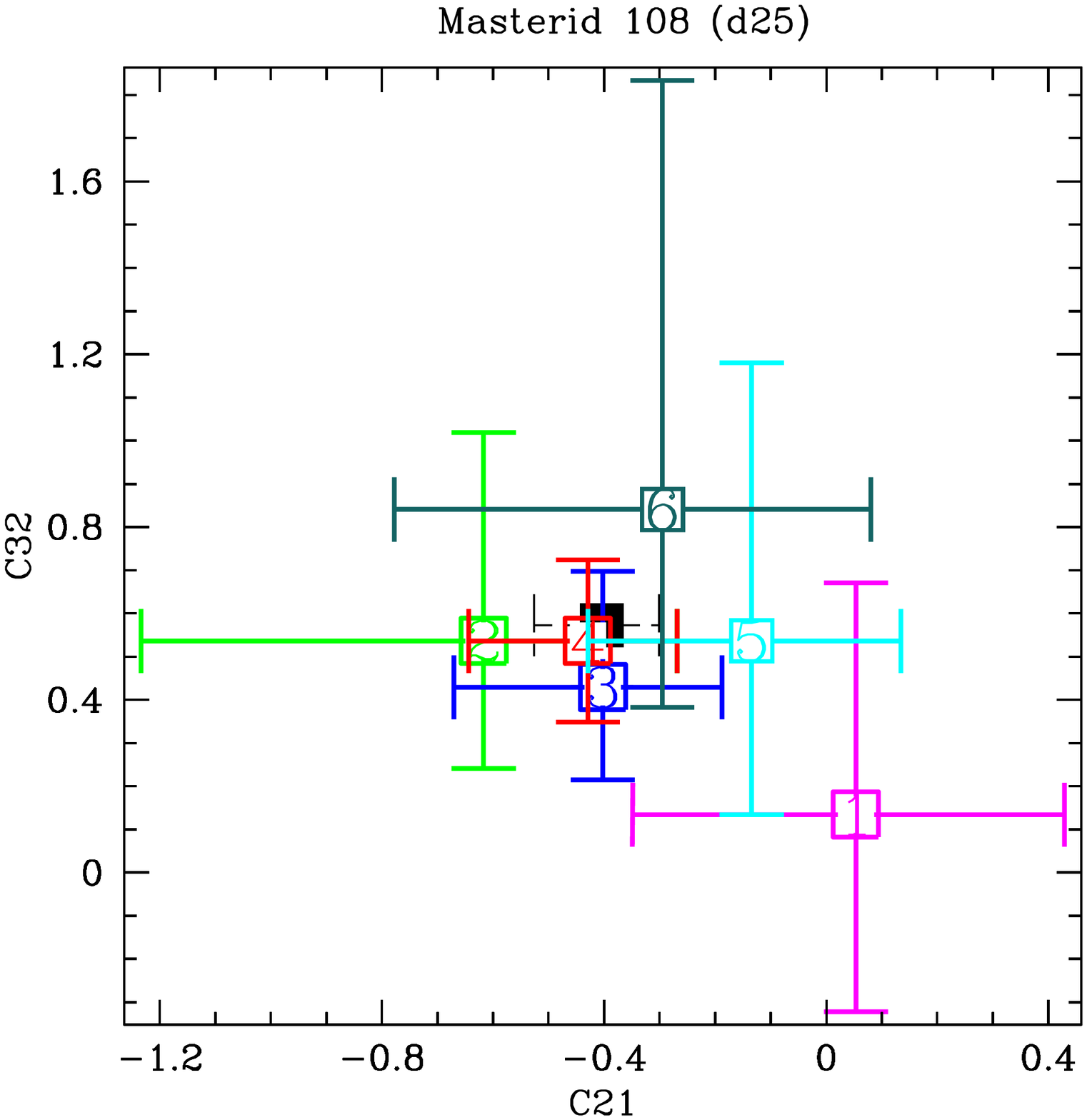}

  \end{minipage}
  \begin{minipage}{0.32\linewidth}
  \centering

    \includegraphics[width=\linewidth]{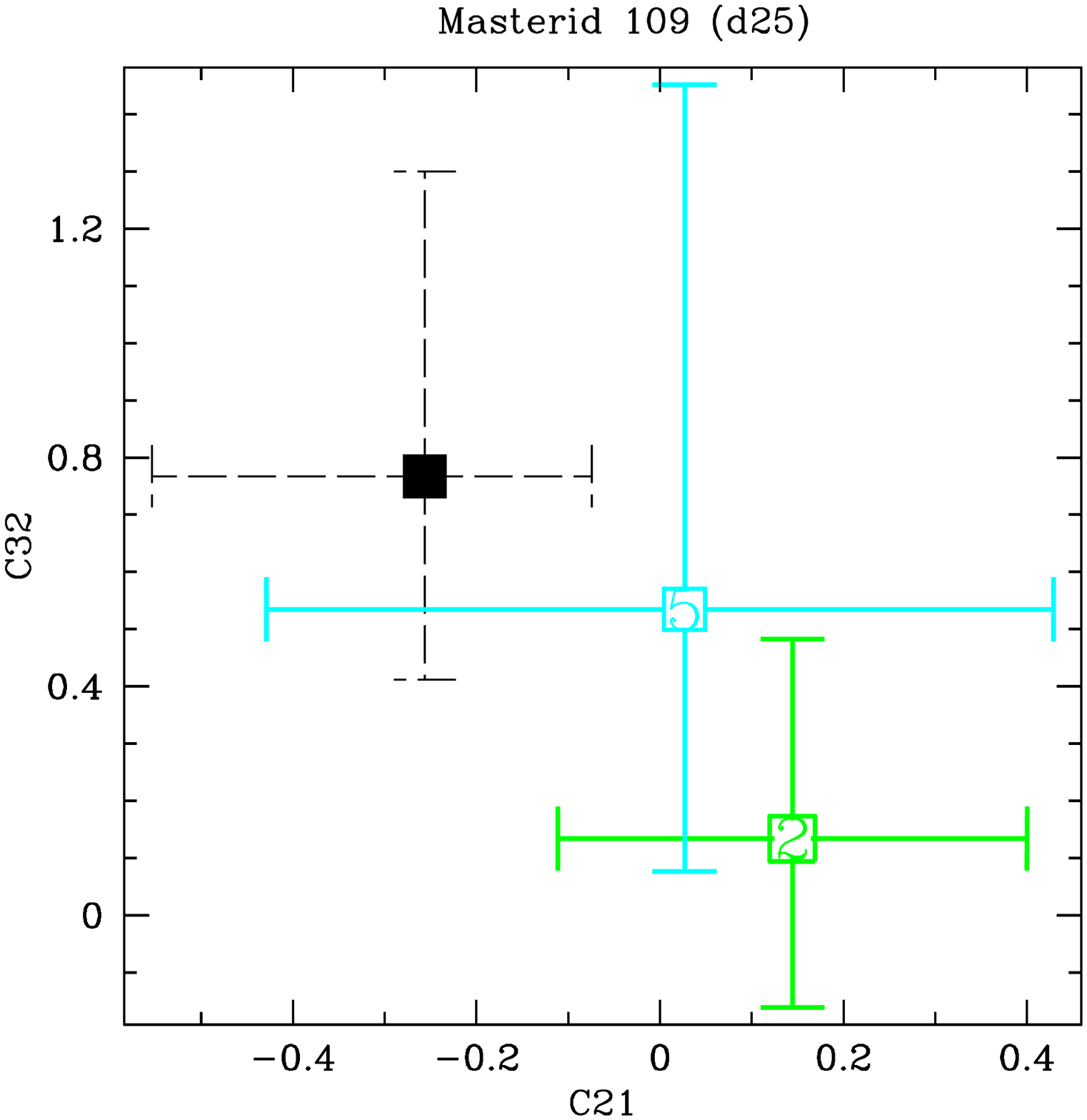}

\end{minipage}
\begin{minipage}{0.32\linewidth}
  \centering

    \includegraphics[width=\linewidth]{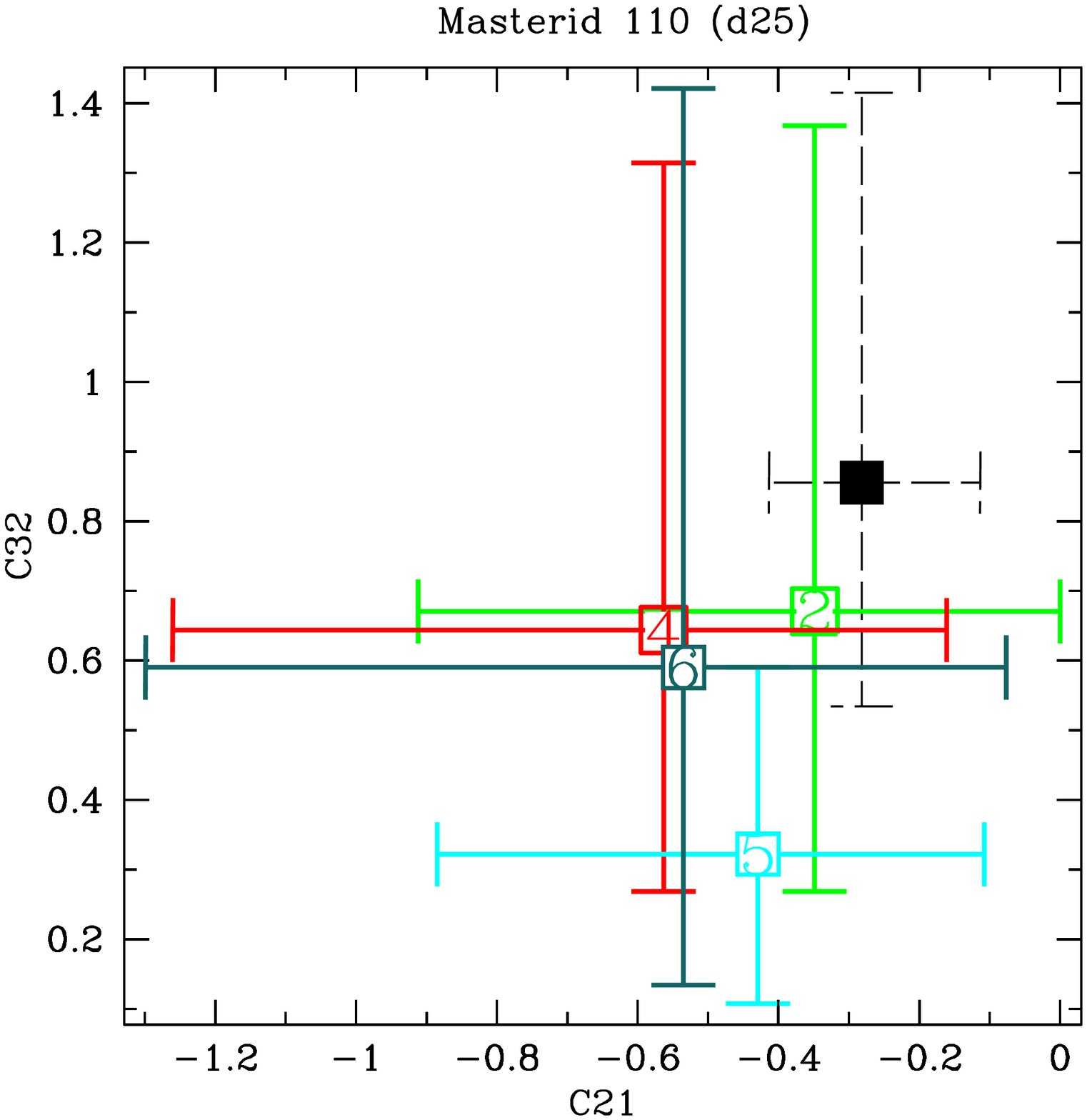}

 \end{minipage}

\begin{minipage}{0.32\linewidth}
  \centering
  
    \includegraphics[width=\linewidth]{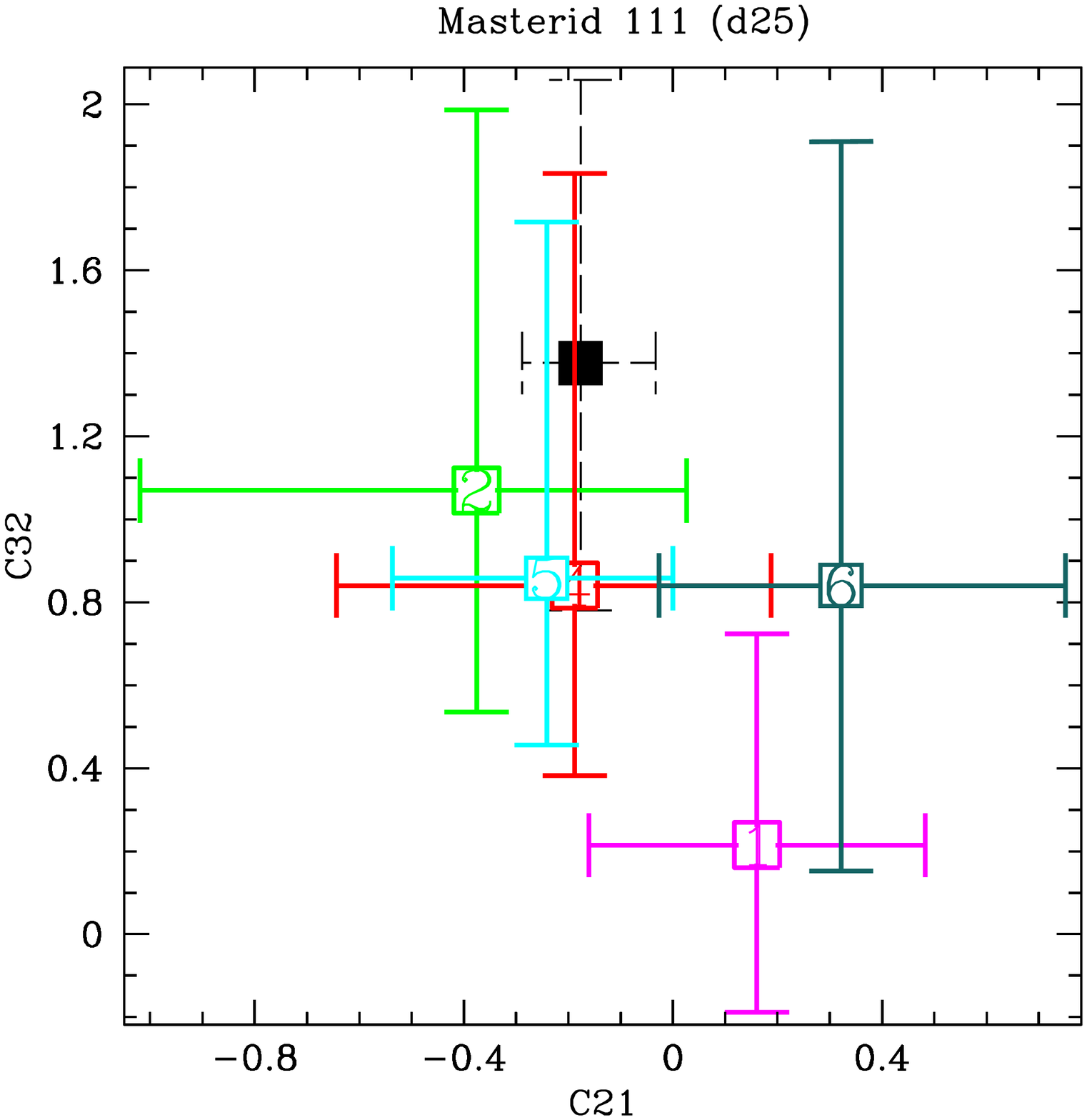}

  \end{minipage}
  \begin{minipage}{0.32\linewidth}
  \centering

    \includegraphics[width=\linewidth]{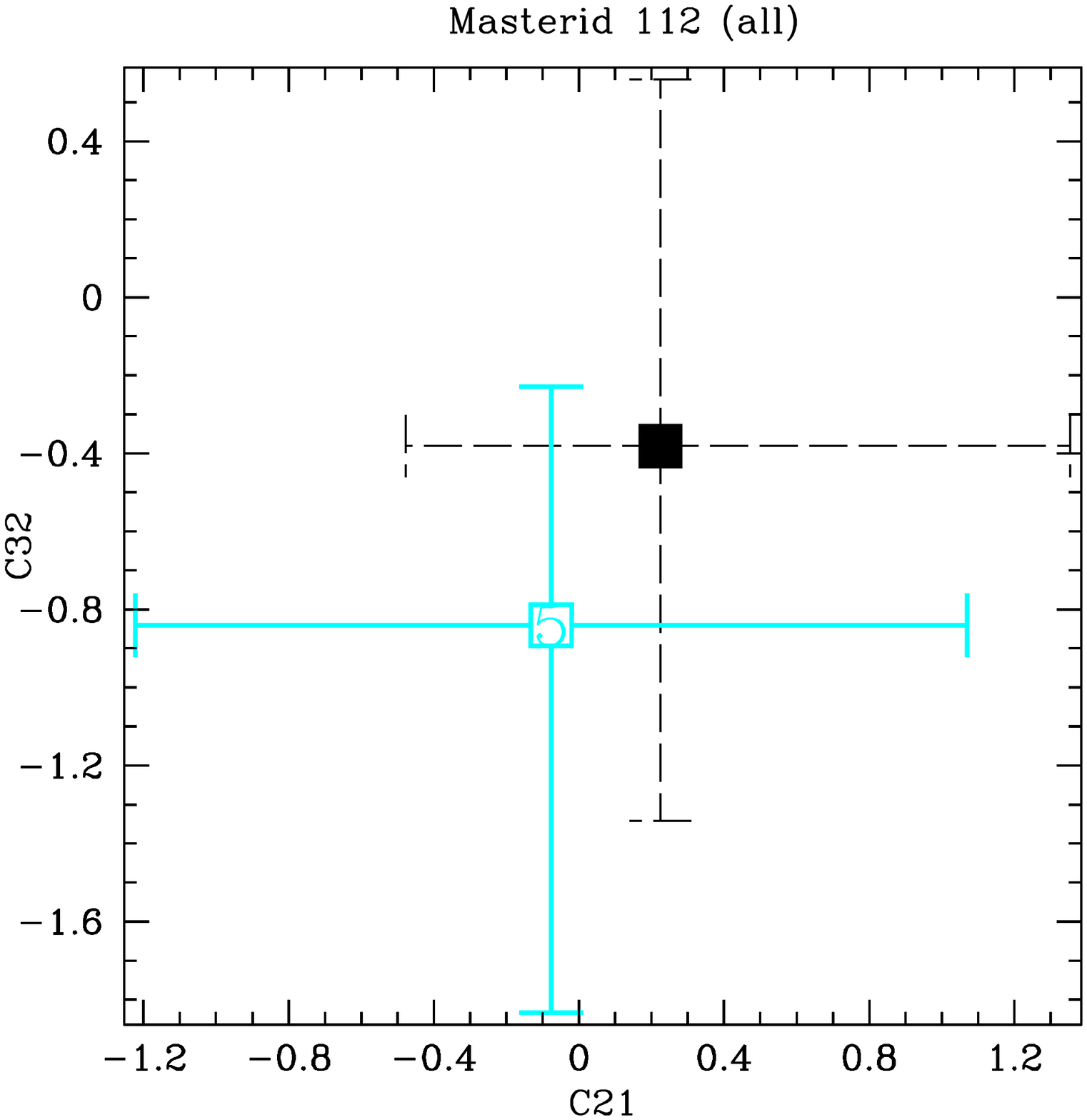}

\end{minipage}
\begin{minipage}{0.32\linewidth}
  \centering

    \includegraphics[width=\linewidth]{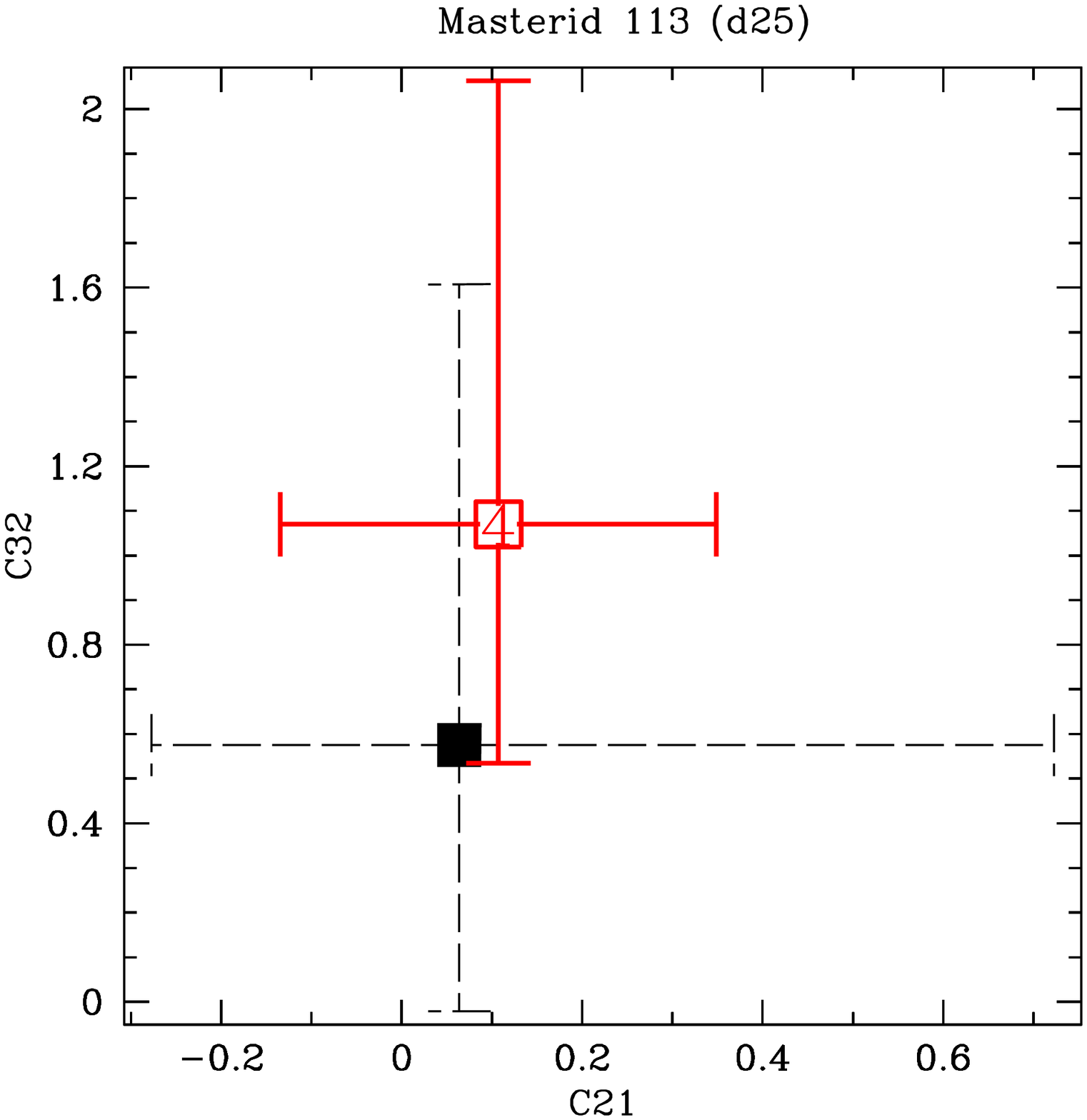}

 \end{minipage}

  \begin{minipage}{0.32\linewidth}
  \centering
  
    \includegraphics[width=\linewidth]{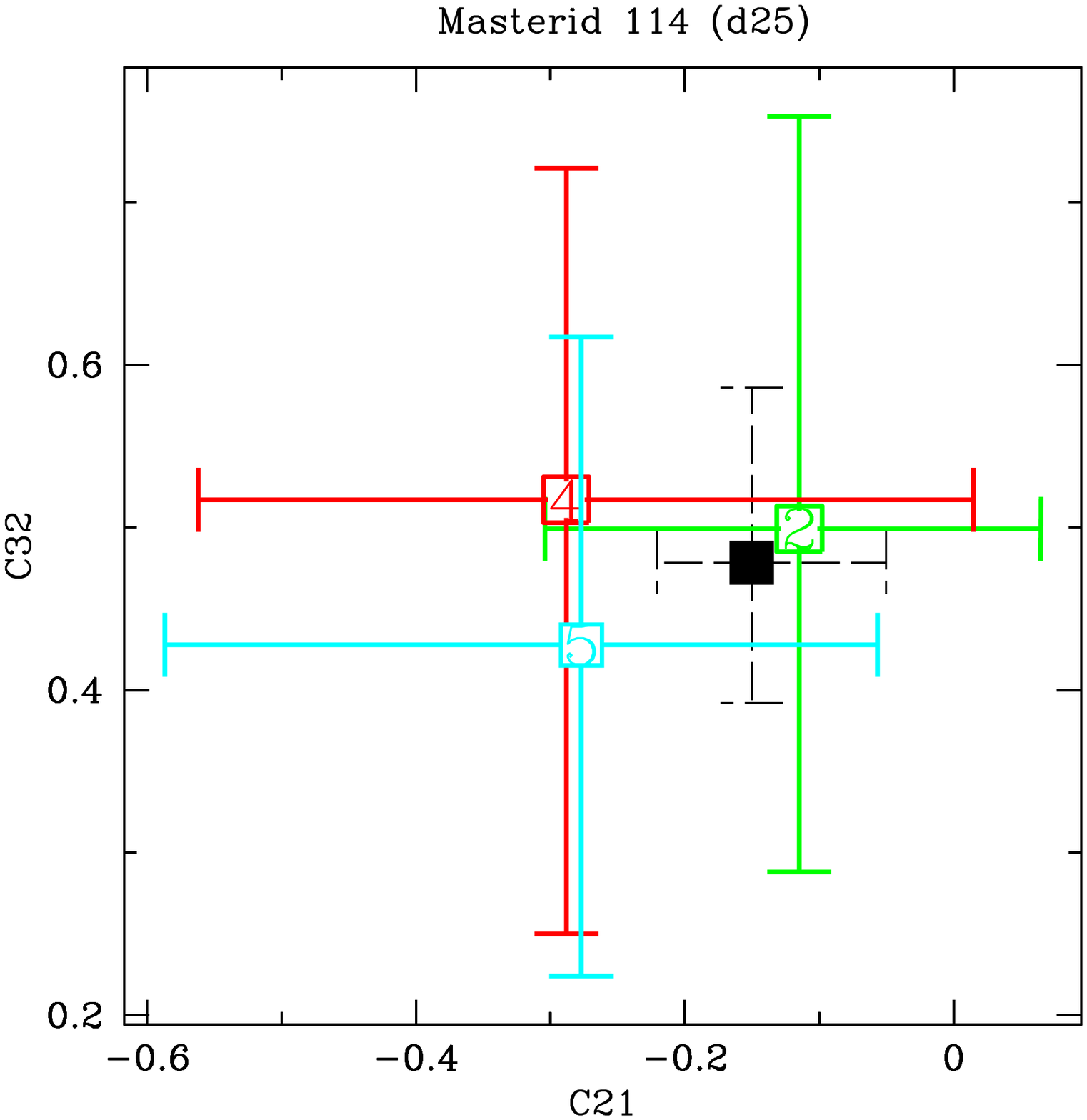}

  \end{minipage}
  \begin{minipage}{0.32\linewidth}
  \centering

    \includegraphics[width=\linewidth]{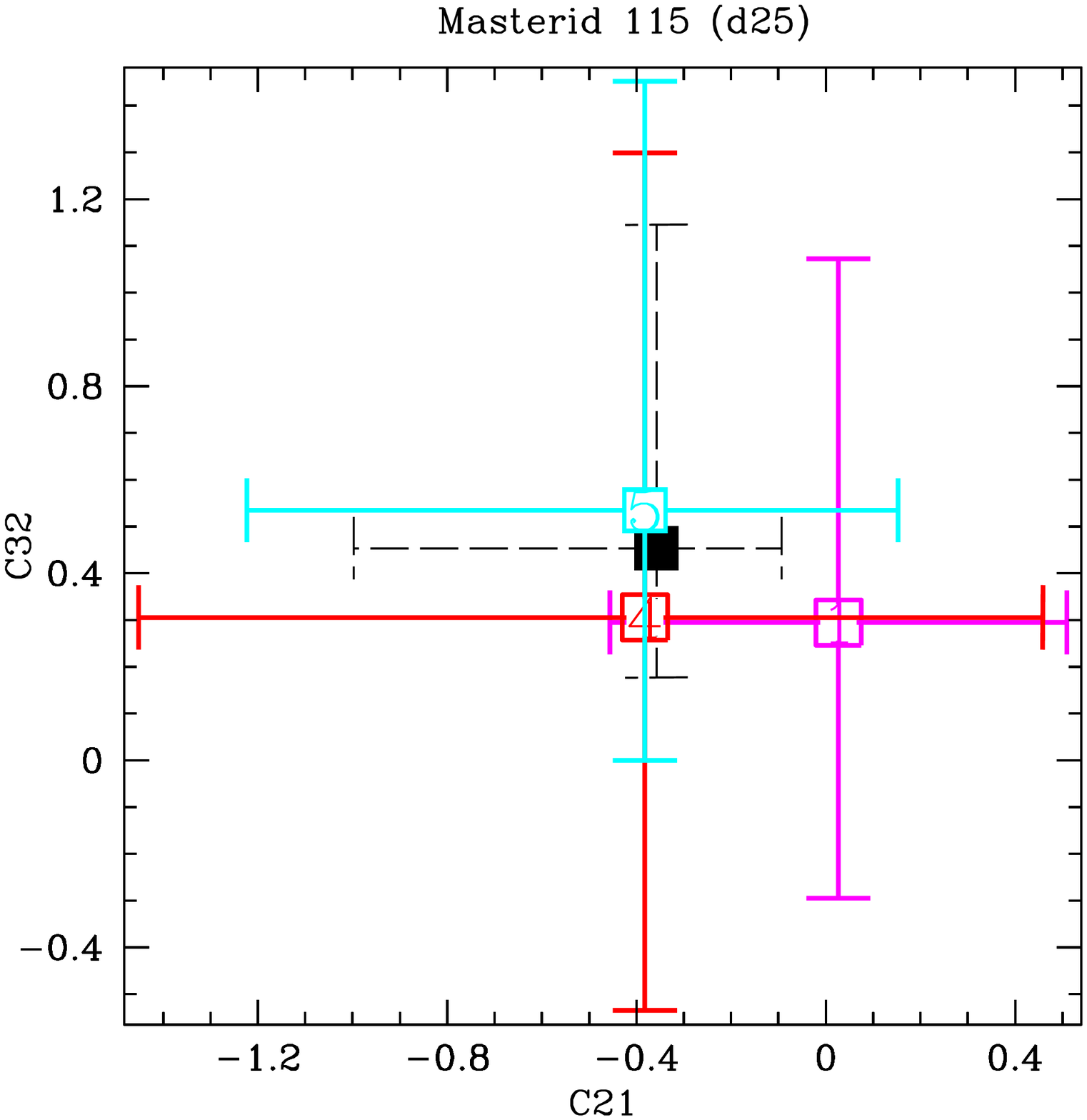}

\end{minipage}
\begin{minipage}{0.32\linewidth}
  \centering

    \includegraphics[width=\linewidth]{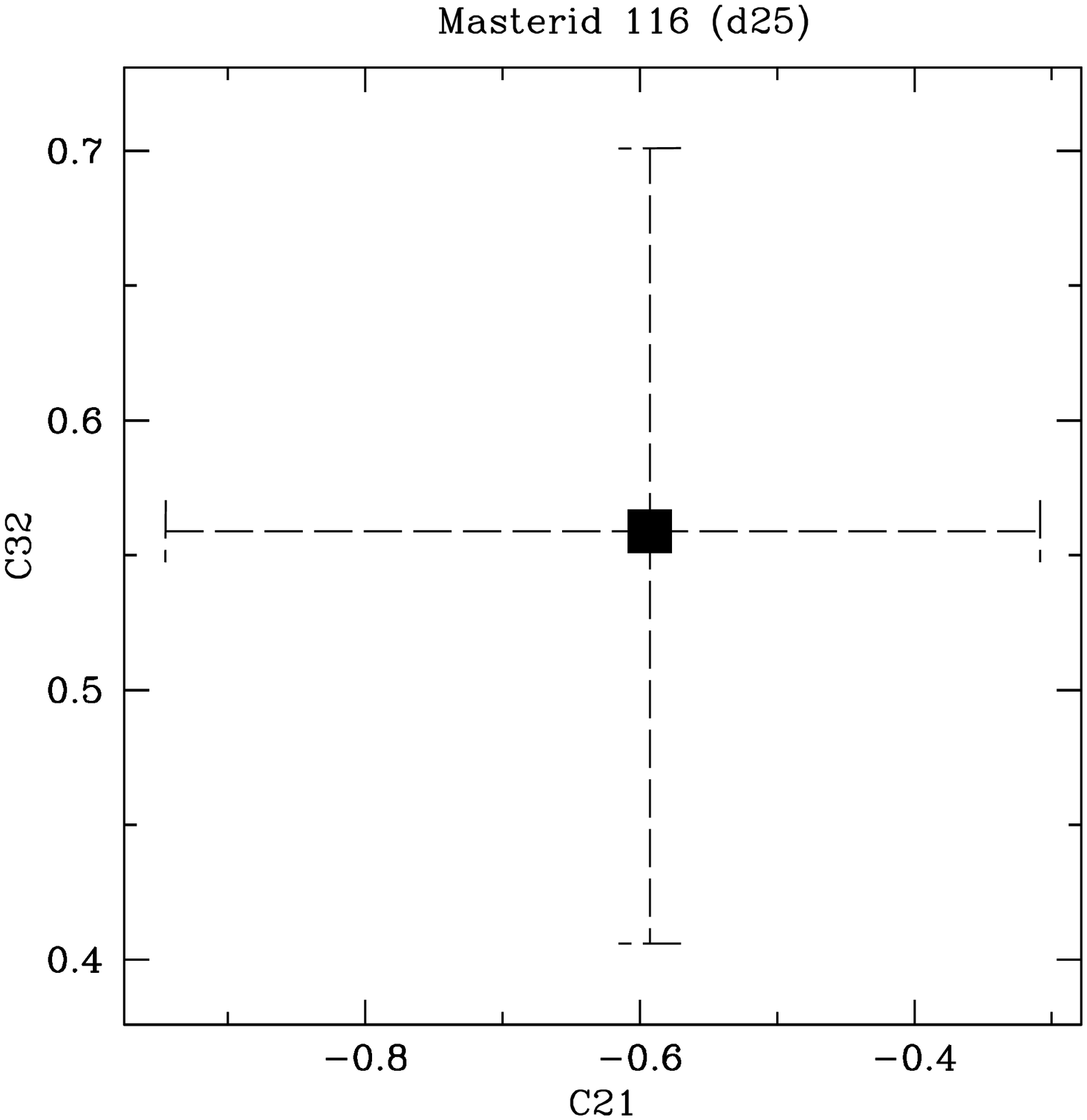}

\end{minipage}

\begin{minipage}{0.32\linewidth}
  \centering
  
    \includegraphics[width=\linewidth]{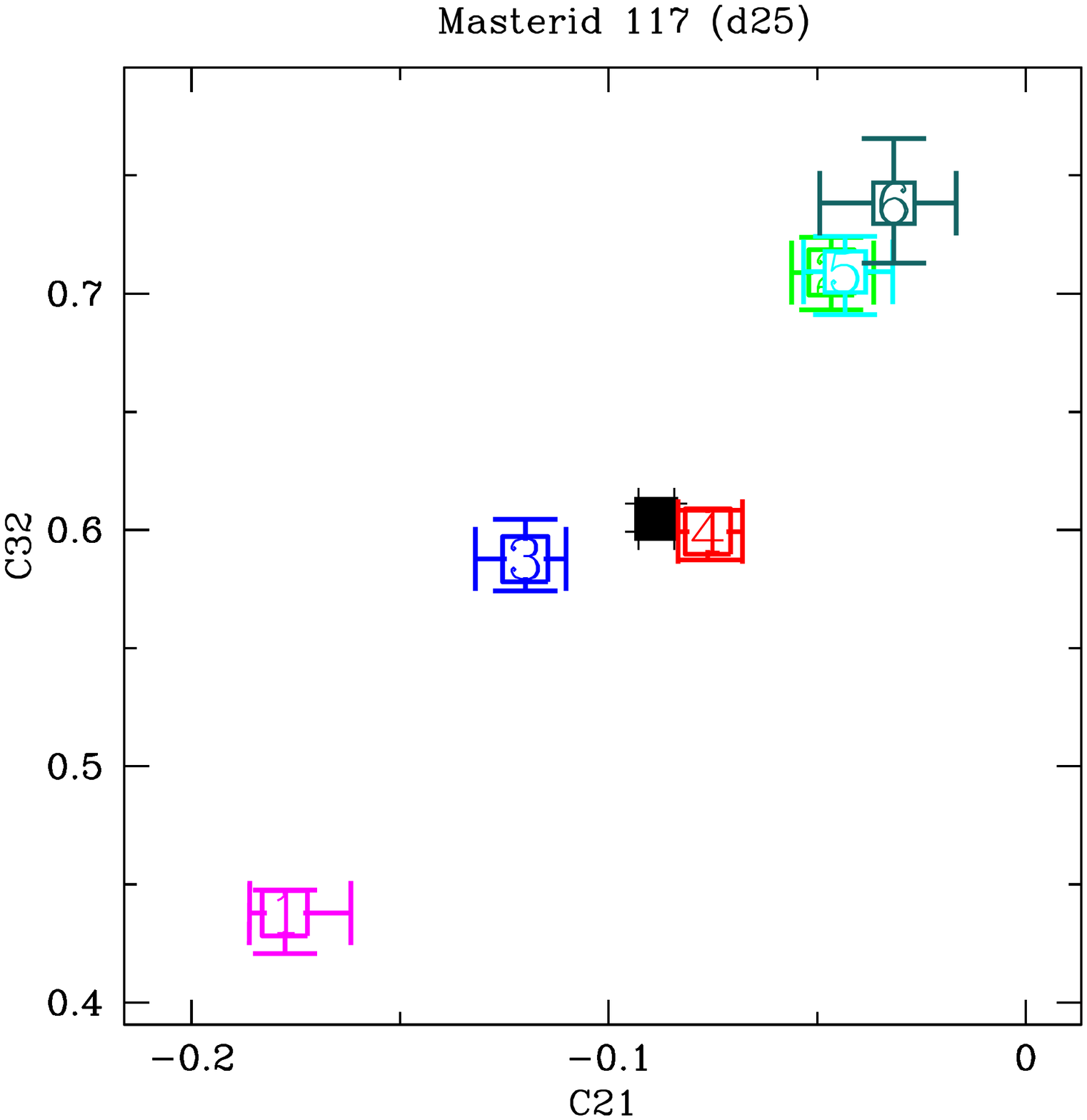}

  \end{minipage}
  \begin{minipage}{0.32\linewidth}
  \centering

    \includegraphics[width=\linewidth]{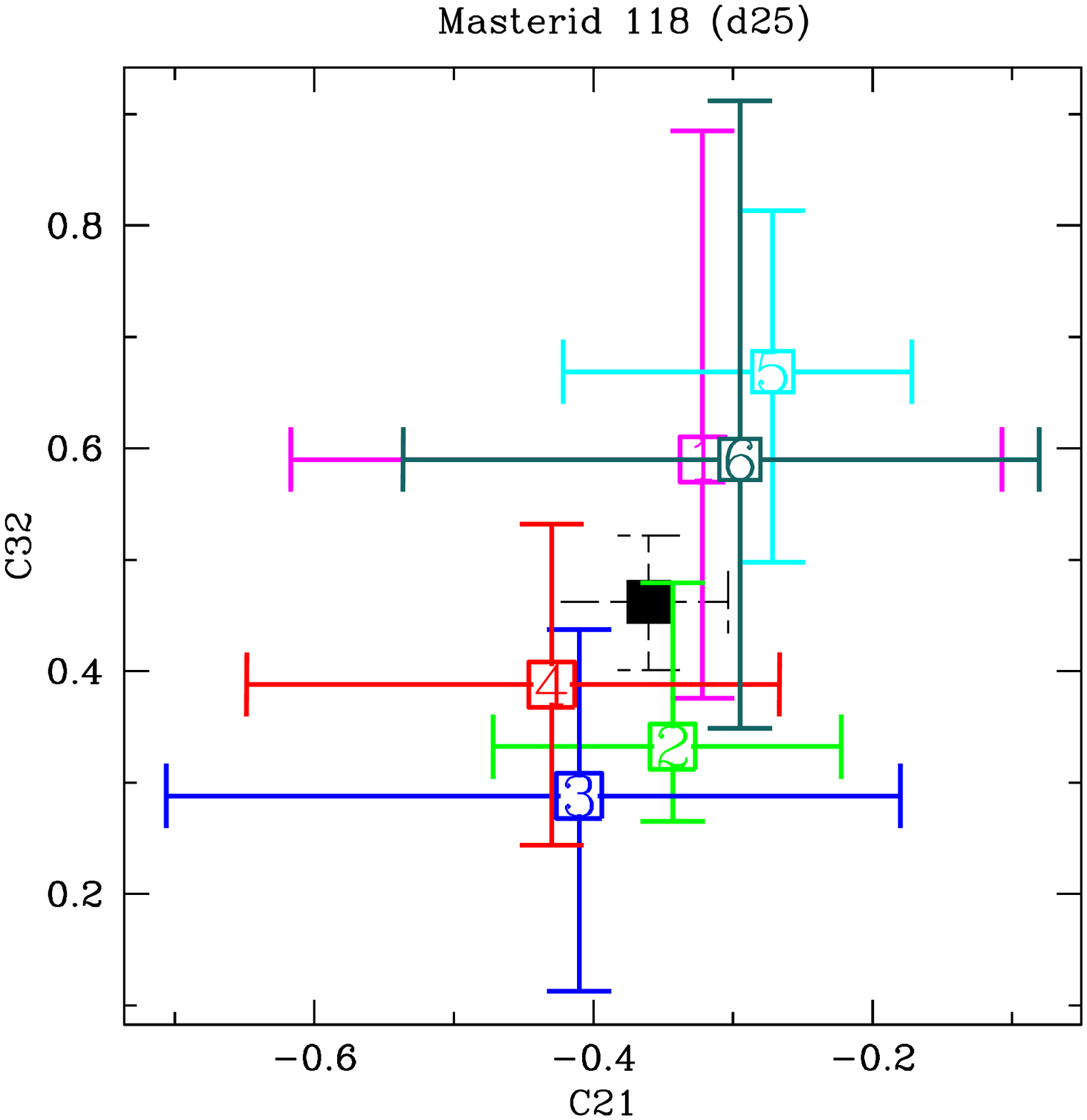}

\end{minipage}
\begin{minipage}{0.32\linewidth}
  \centering

    \includegraphics[width=\linewidth]{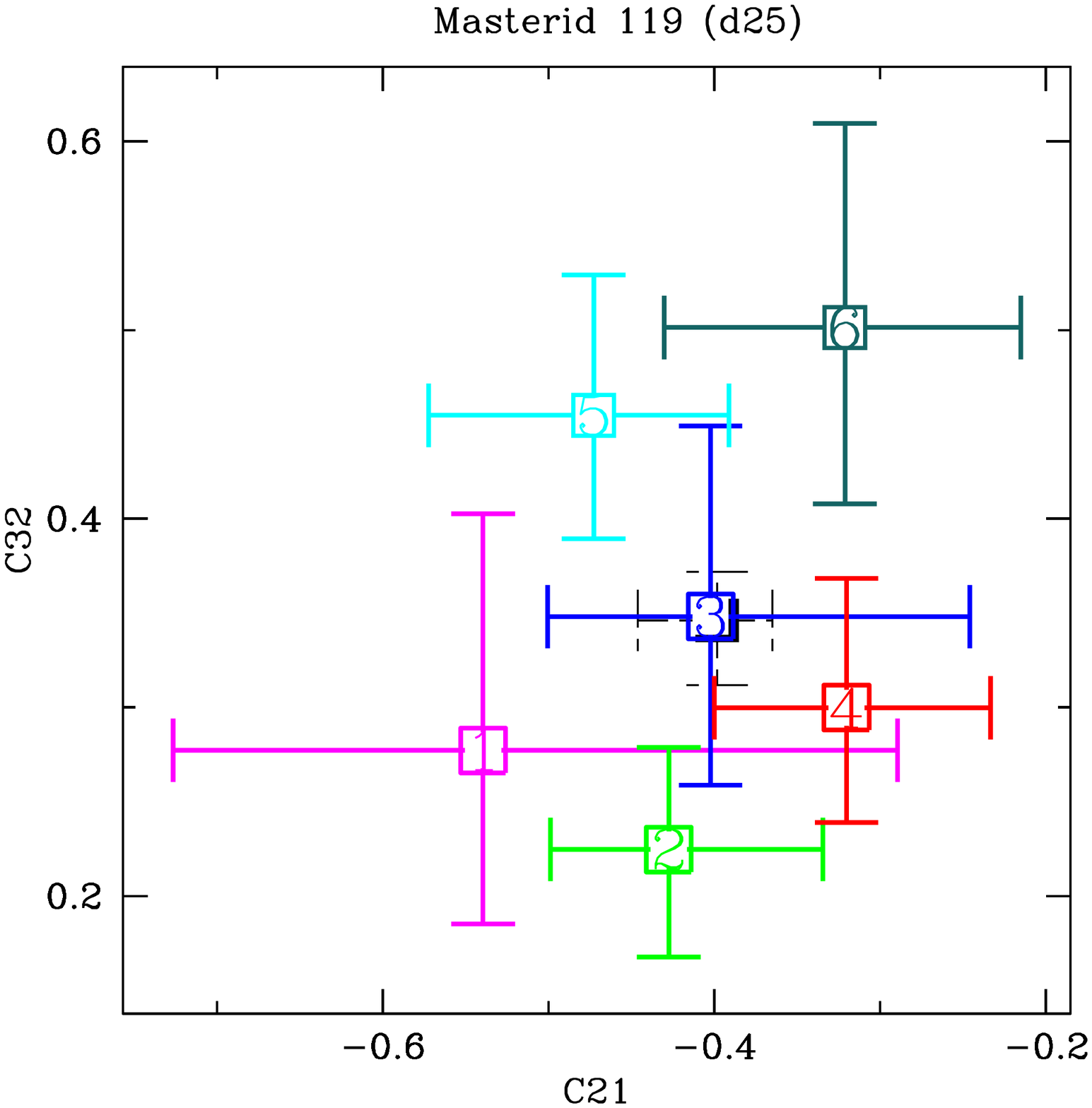}

\end{minipage}
\end{figure}

\begin{figure}
  \begin{minipage}{0.32\linewidth}
  \centering
  
    \includegraphics[width=\linewidth]{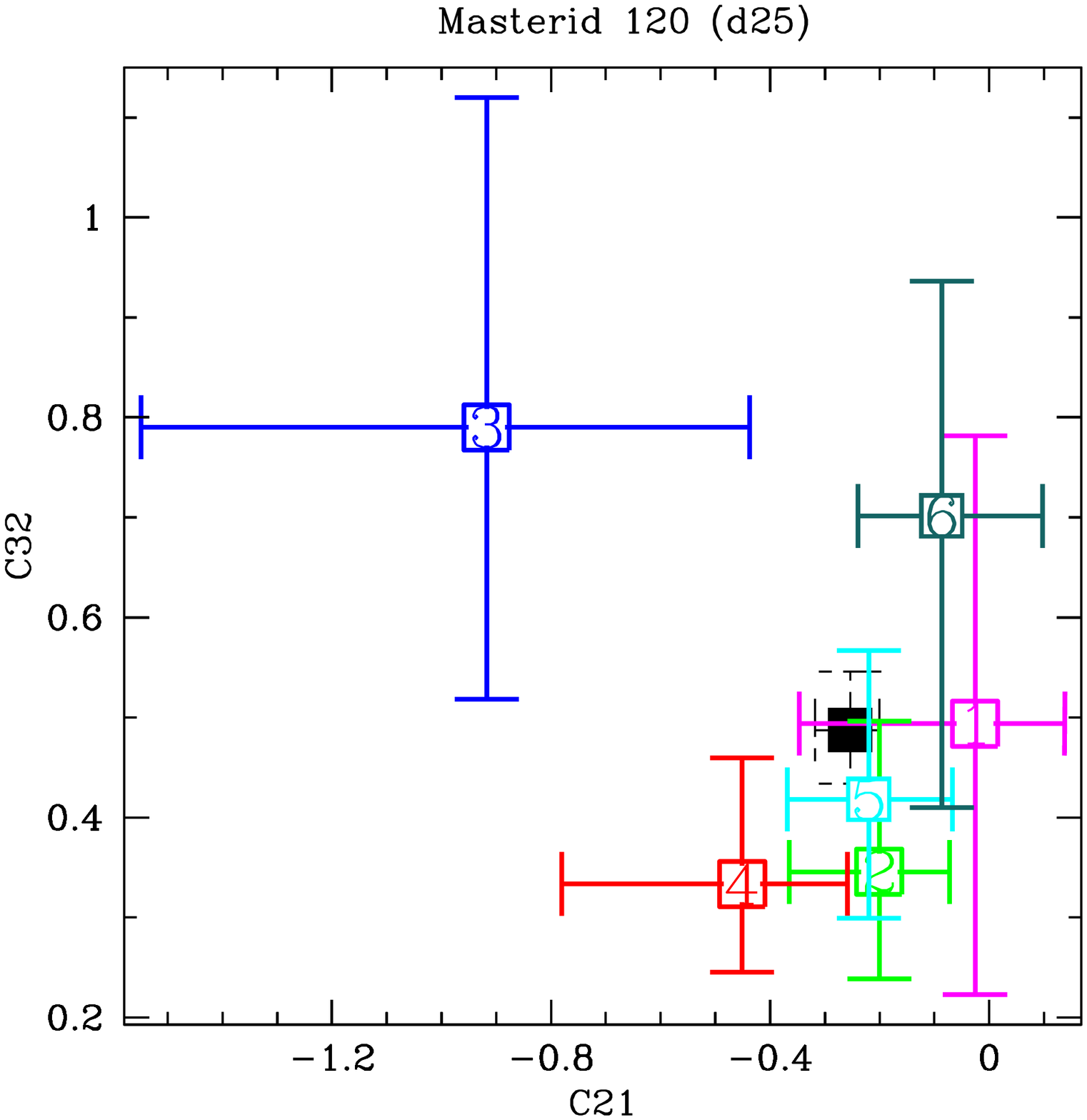}

  \end{minipage}
  \begin{minipage}{0.32\linewidth}
  \centering

    \includegraphics[width=\linewidth]{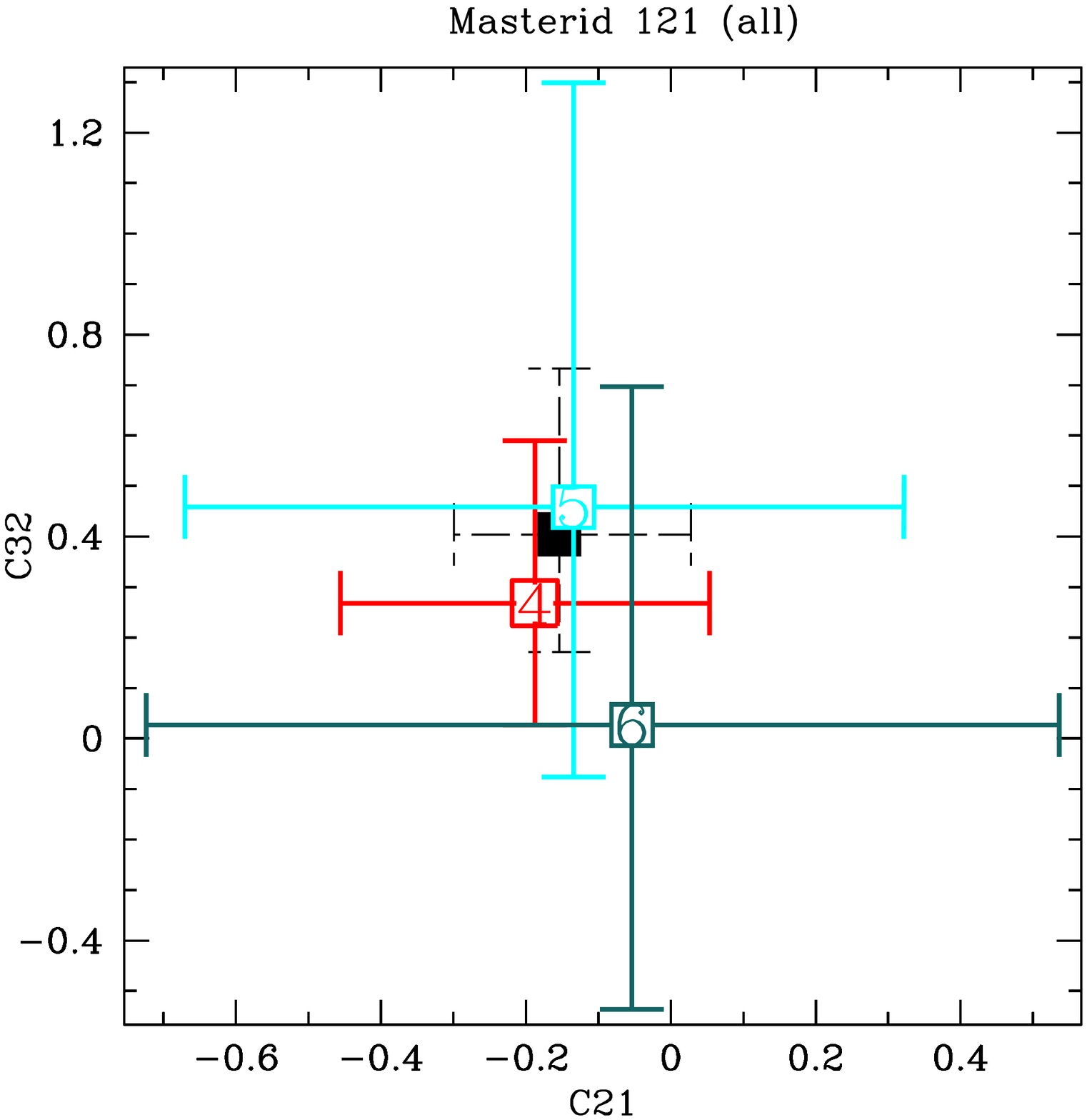}

\end{minipage}
\begin{minipage}{0.32\linewidth}
  \centering

    \includegraphics[width=\linewidth]{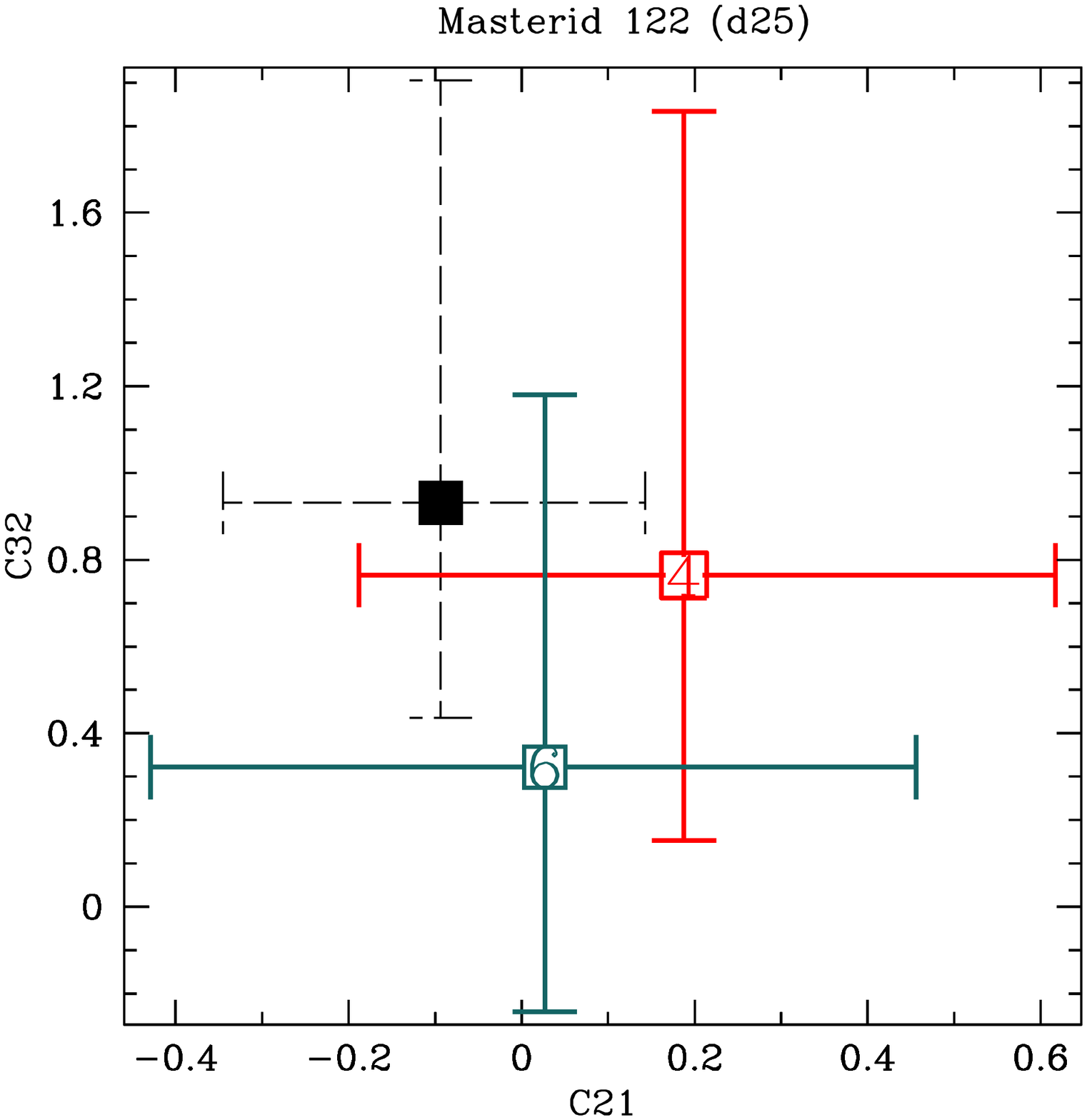}

 \end{minipage}

\begin{minipage}{0.32\linewidth}
  \centering
  
    \includegraphics[width=\linewidth]{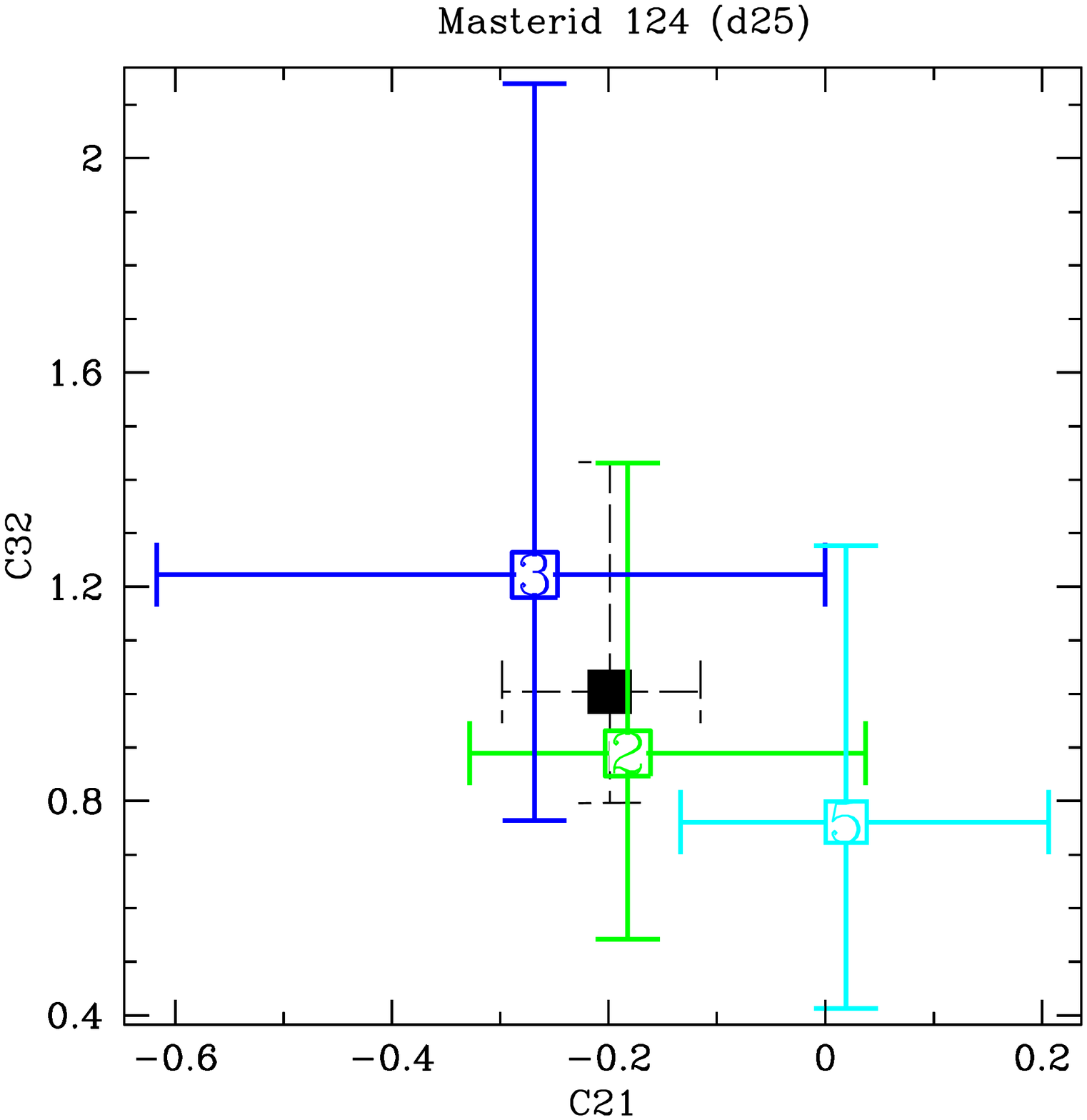}

  \end{minipage}
  \begin{minipage}{0.32\linewidth}
  \centering

    \includegraphics[width=\linewidth]{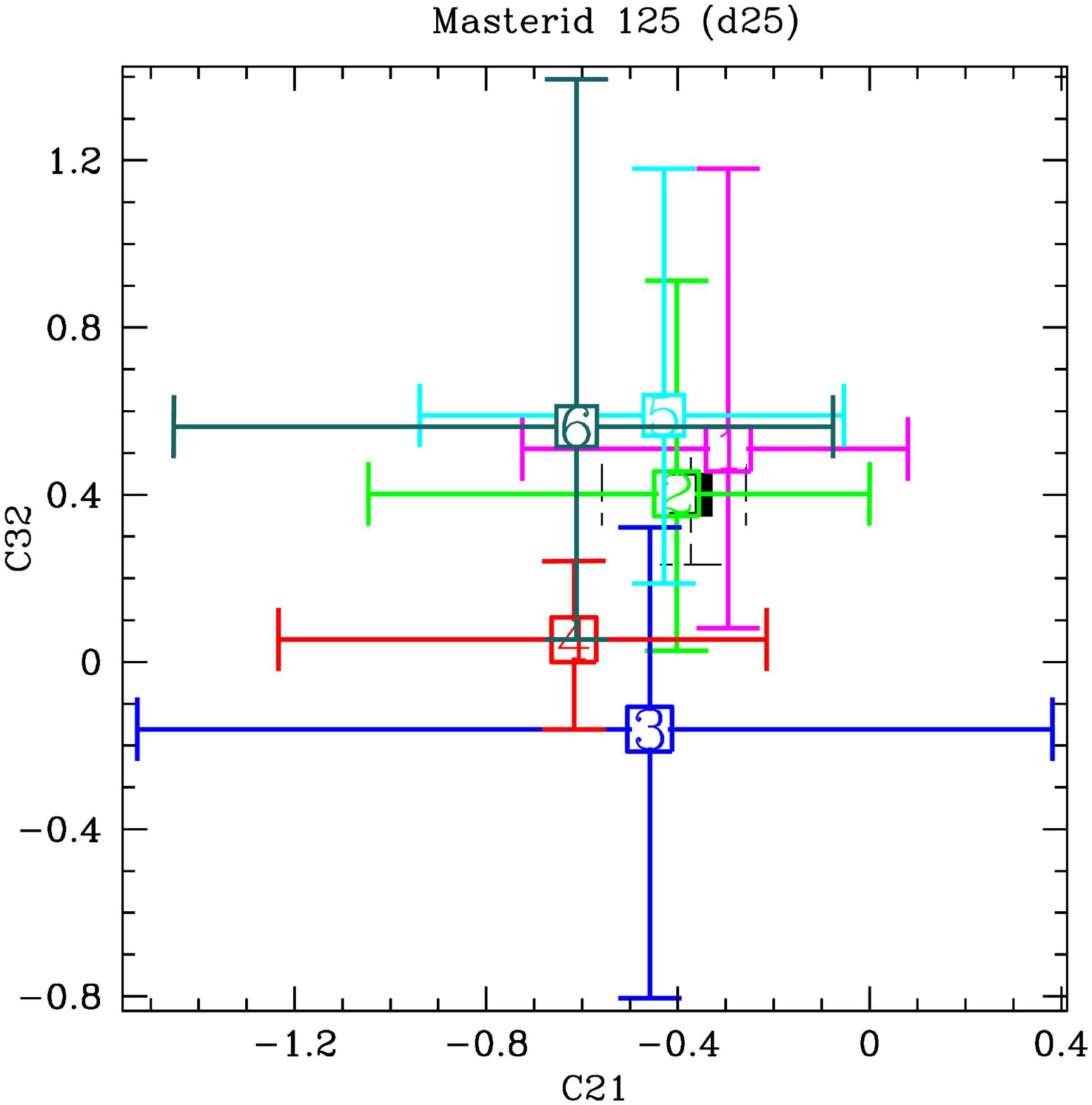}

\end{minipage}
\begin{minipage}{0.32\linewidth}
  \centering

    \includegraphics[width=\linewidth]{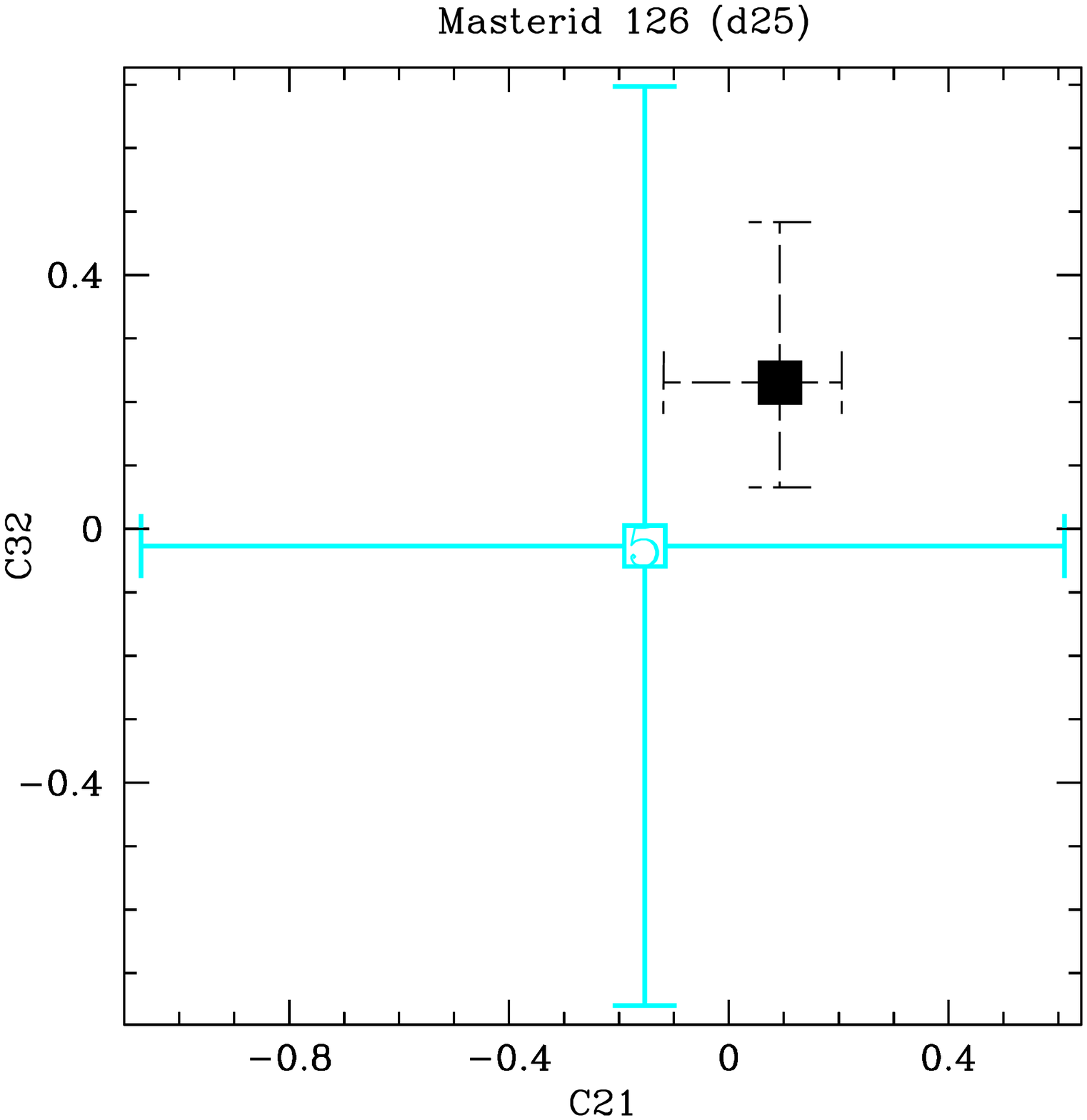}

 \end{minipage}

  \begin{minipage}{0.32\linewidth}
  \centering
  
    \includegraphics[width=\linewidth]{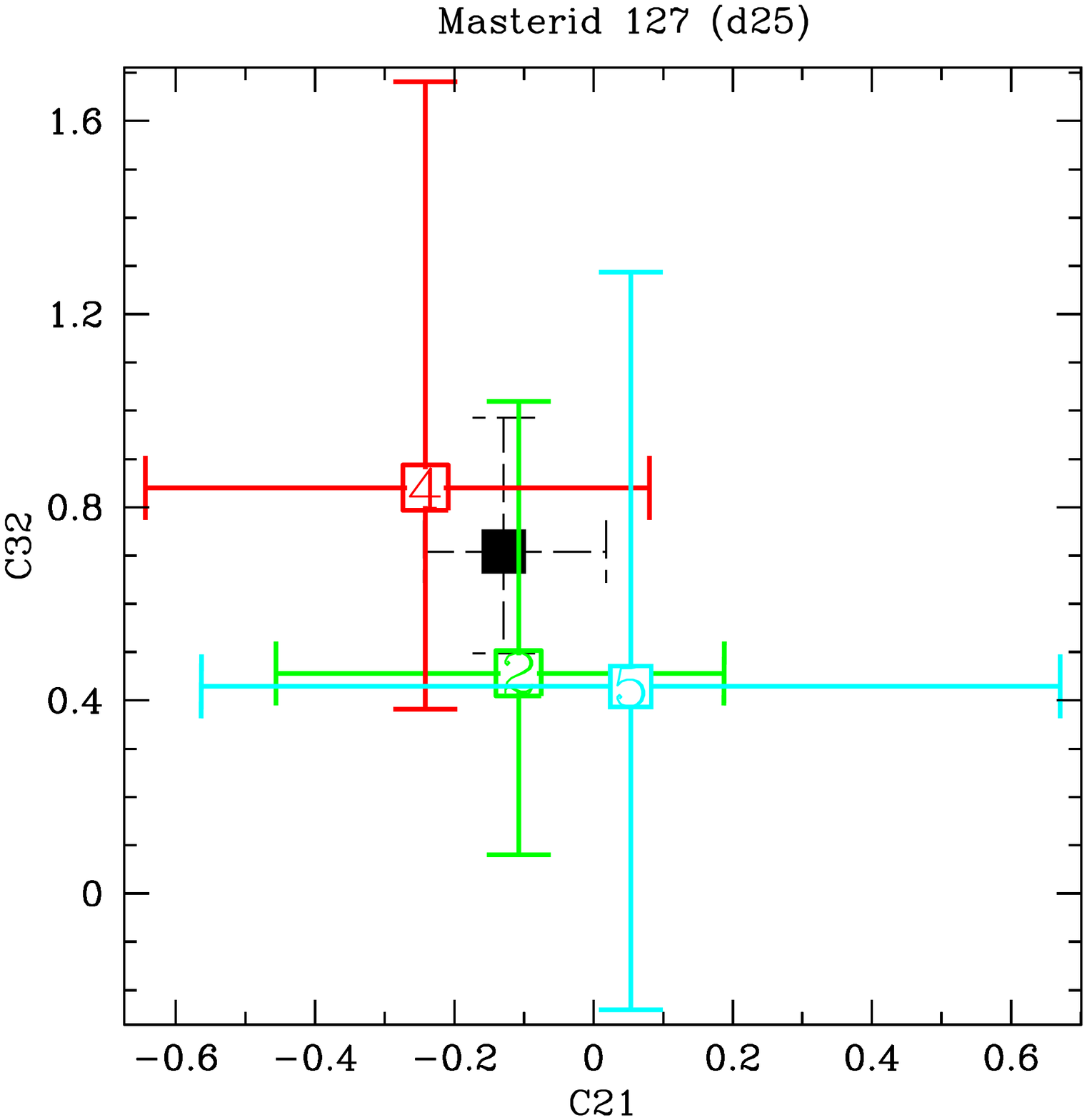}

  \end{minipage}
  \begin{minipage}{0.32\linewidth}
  \centering

    \includegraphics[width=\linewidth]{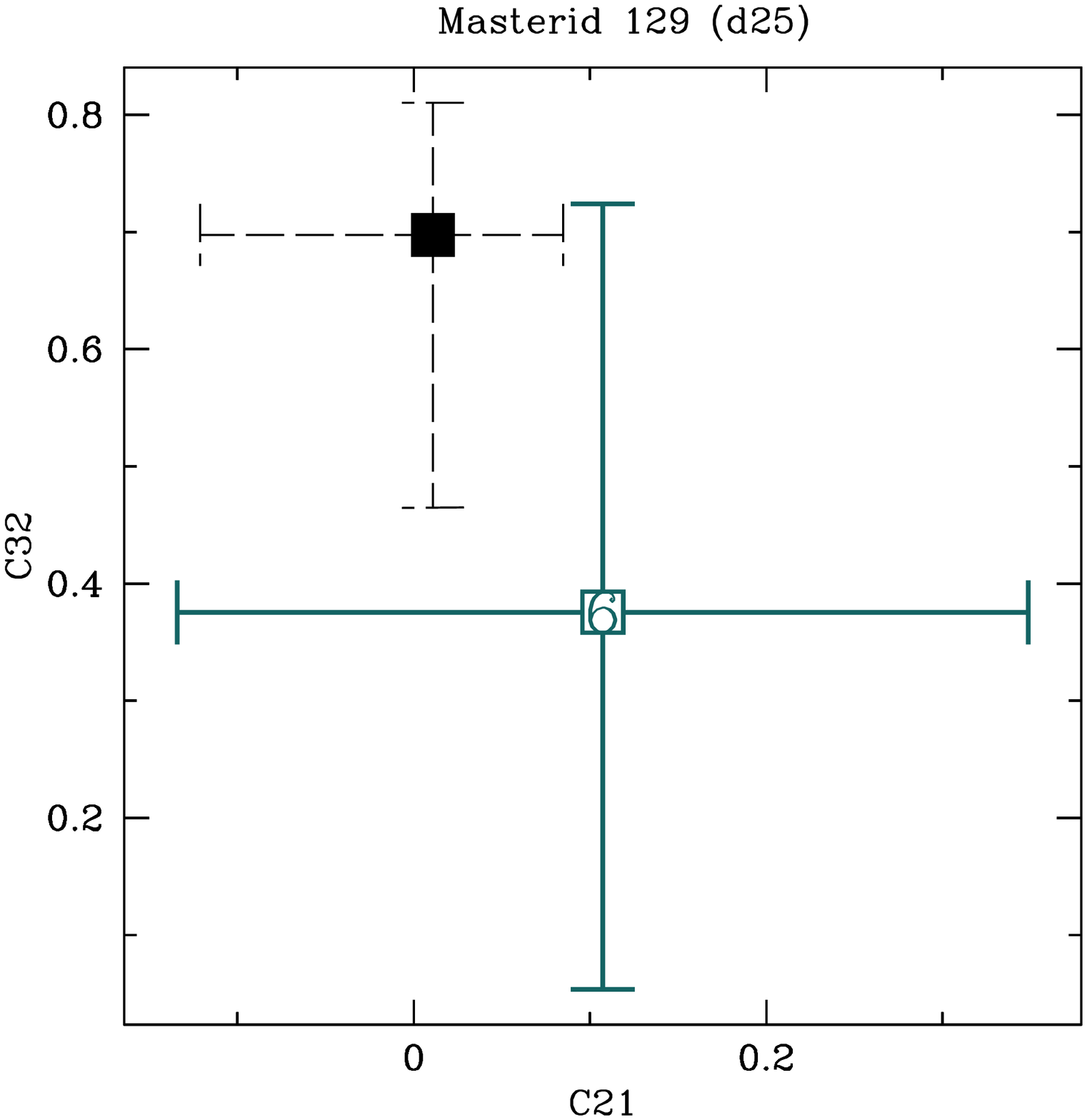}

\end{minipage}
\begin{minipage}{0.32\linewidth}
  \centering

    \includegraphics[width=\linewidth]{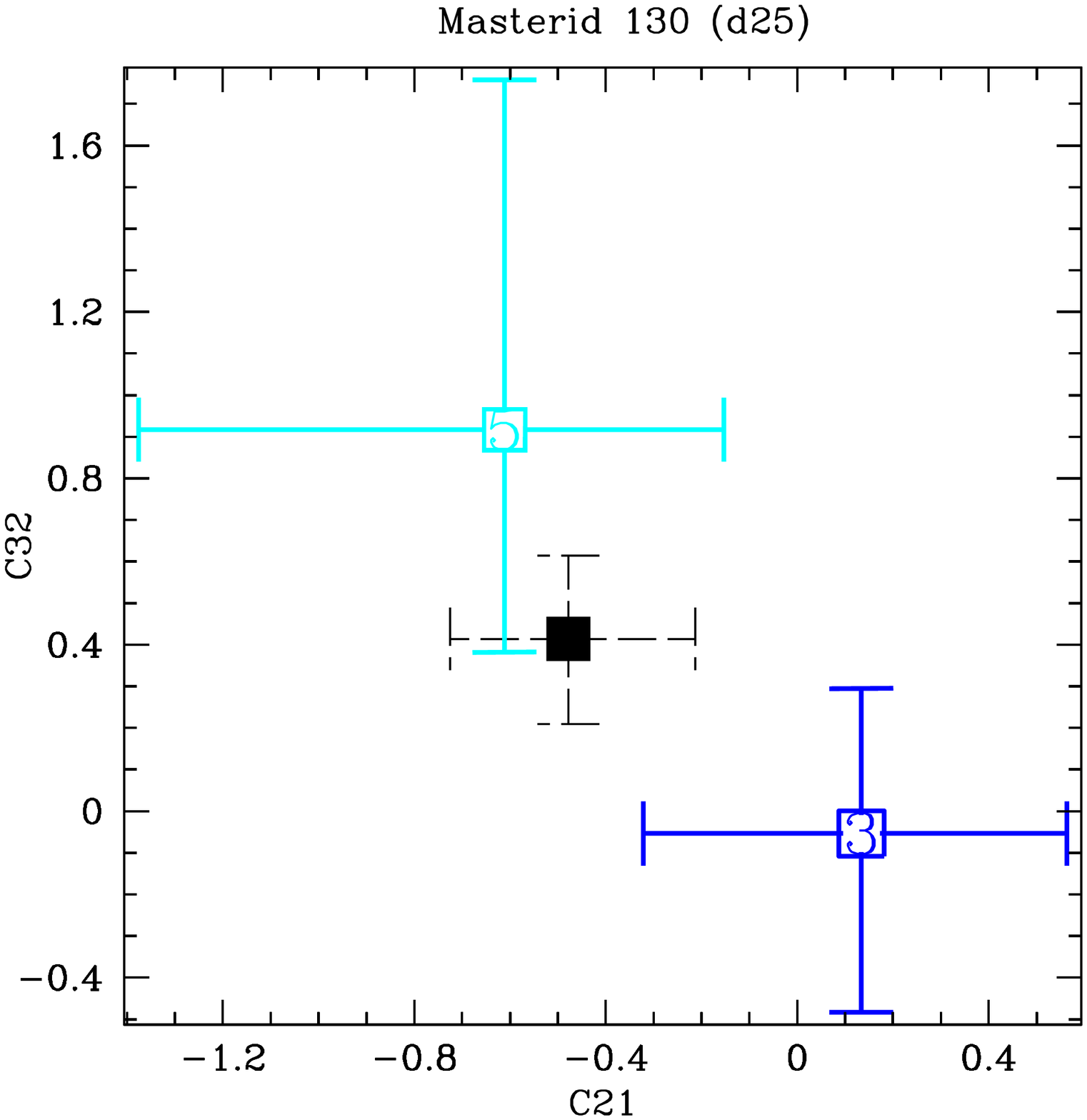}

\end{minipage}

\begin{minipage}{0.32\linewidth}
  \centering
  
    \includegraphics[width=\linewidth]{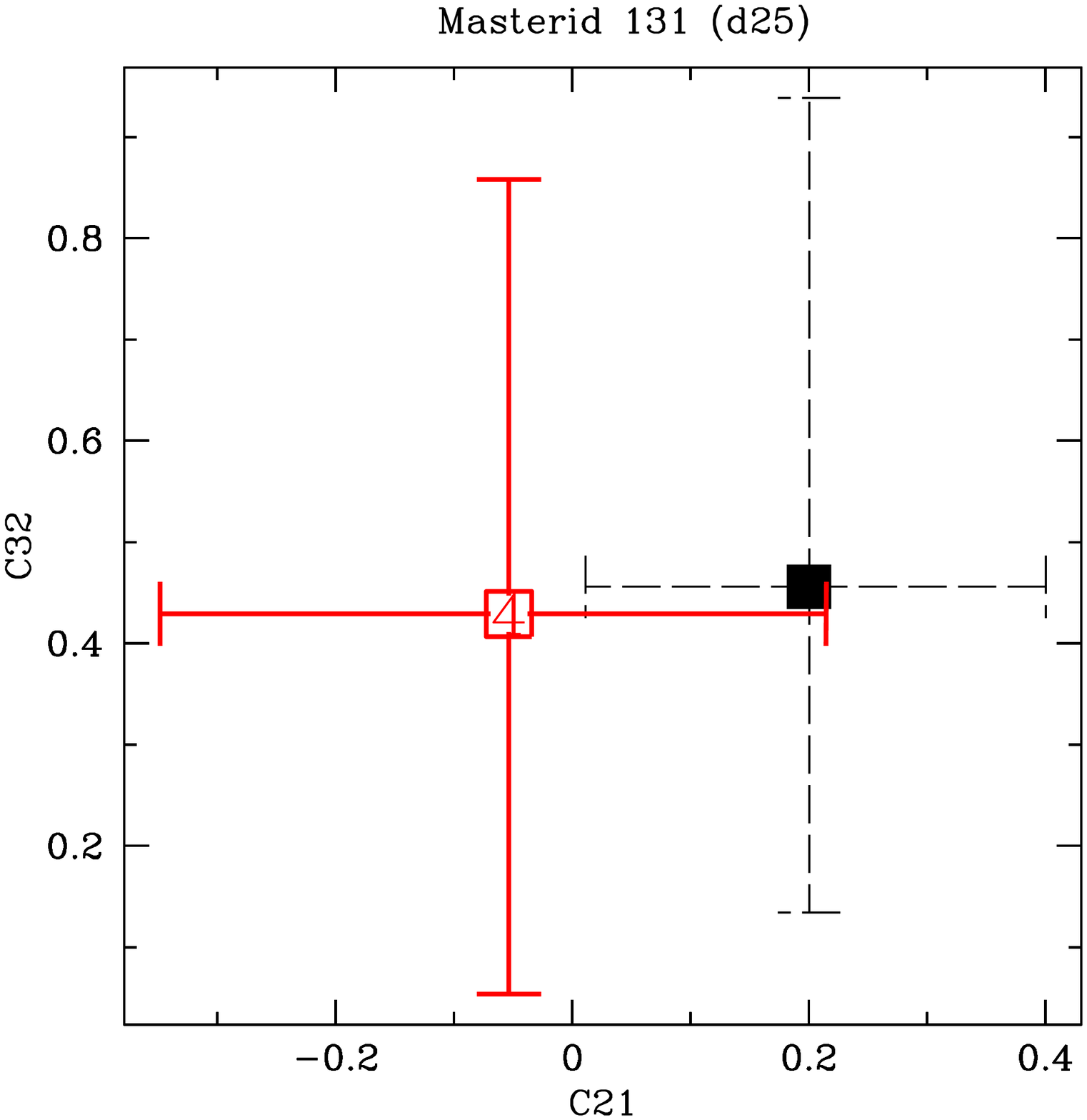}

  \end{minipage}
  \begin{minipage}{0.32\linewidth}
  \centering

    \includegraphics[width=\linewidth]{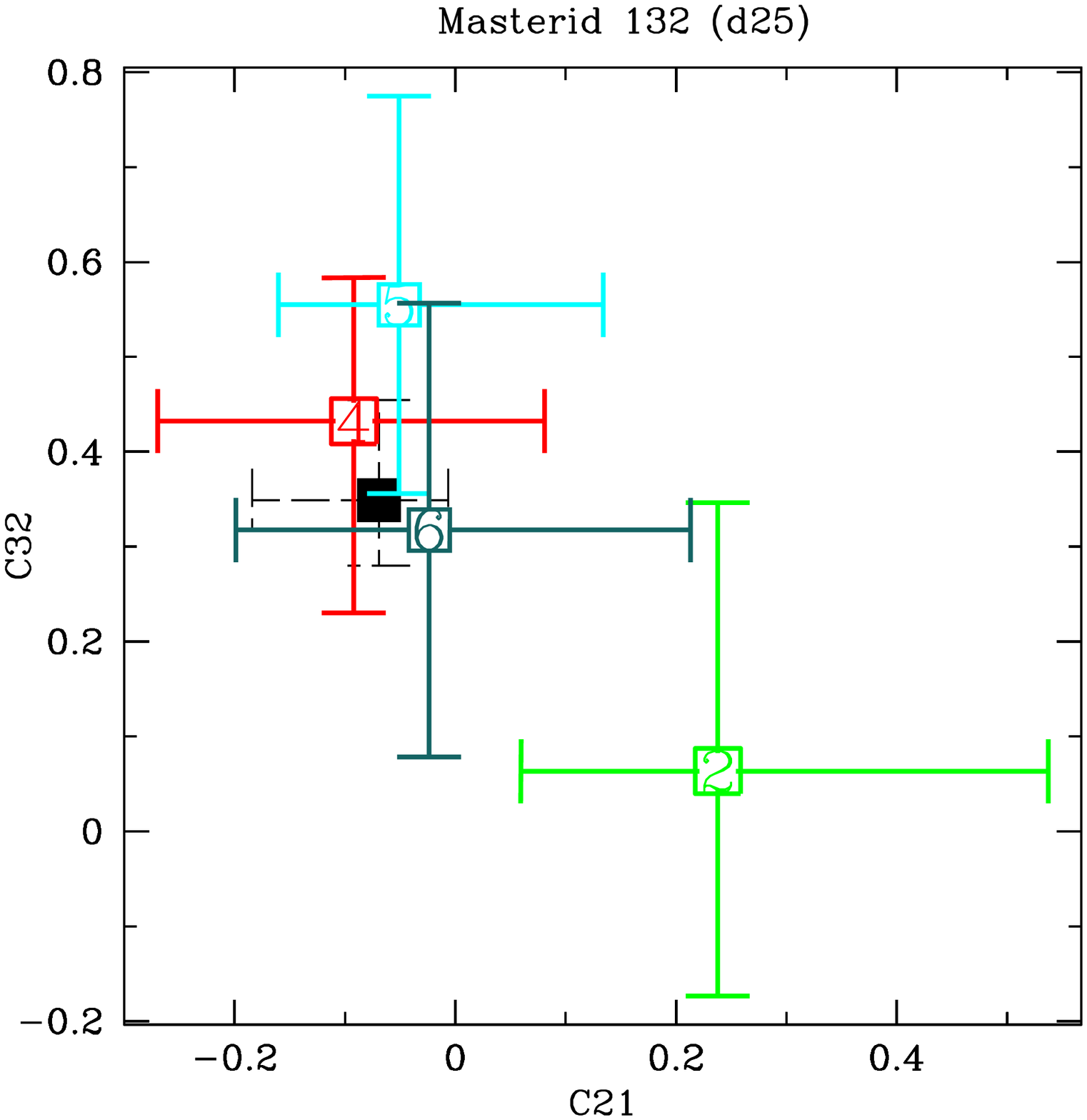}

\end{minipage}
\begin{minipage}{0.32\linewidth}
  \centering

    \includegraphics[width=\linewidth]{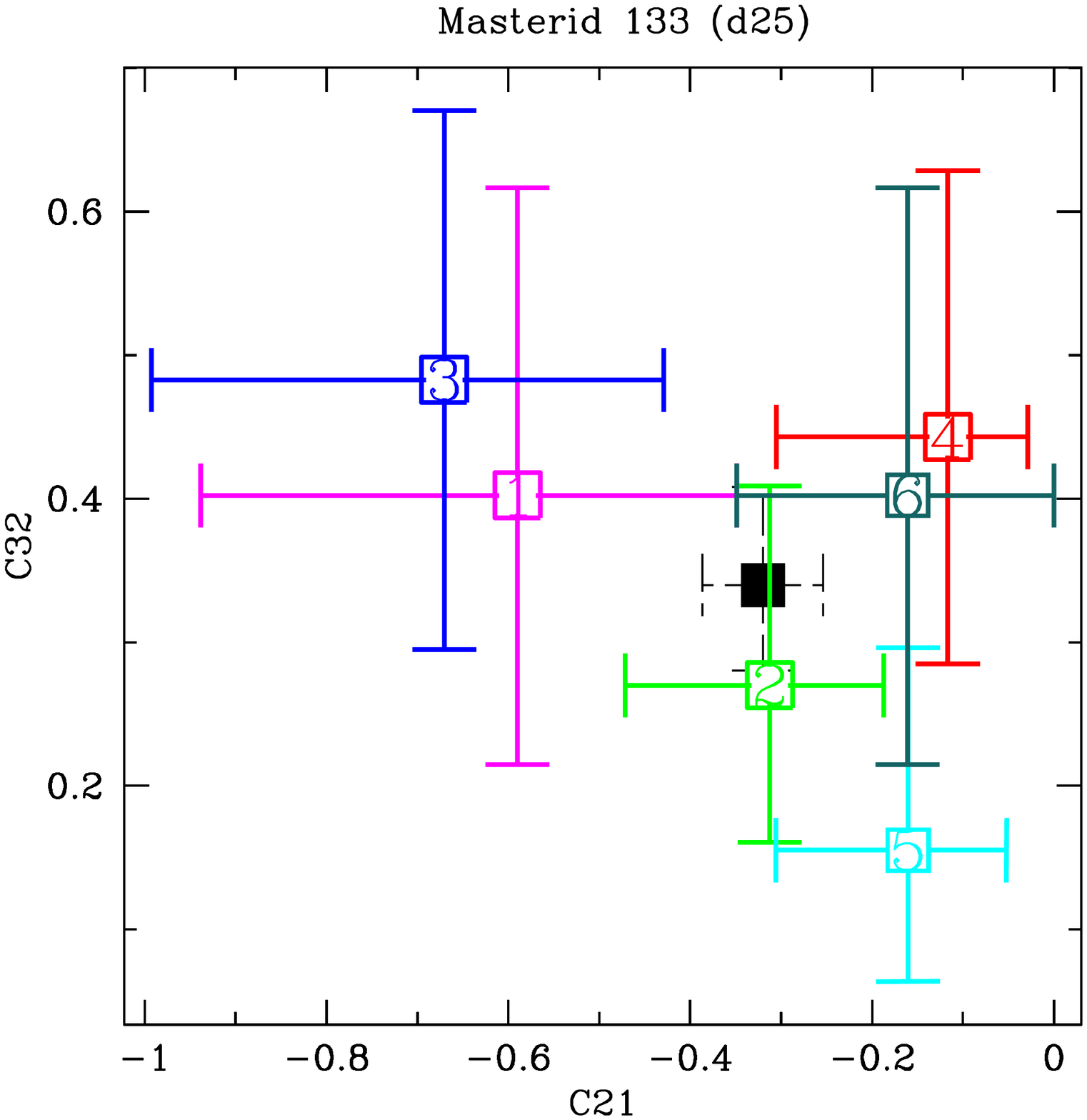}

\end{minipage}
\end{figure}

\begin{figure}
  \begin{minipage}{0.32\linewidth}
  \centering
  
    \includegraphics[width=\linewidth]{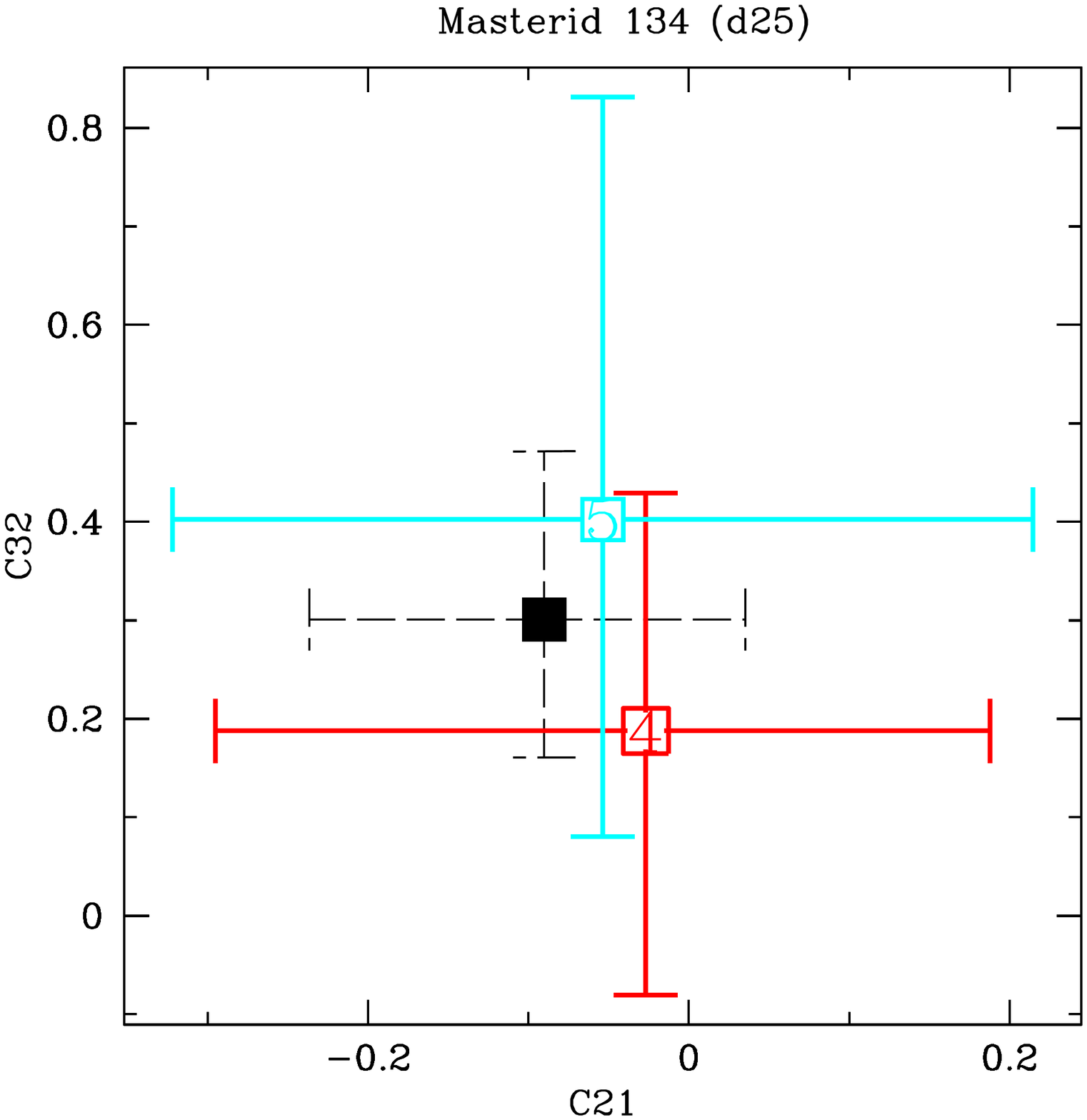}

  \end{minipage}
  \begin{minipage}{0.32\linewidth}
  \centering

    \includegraphics[width=\linewidth]{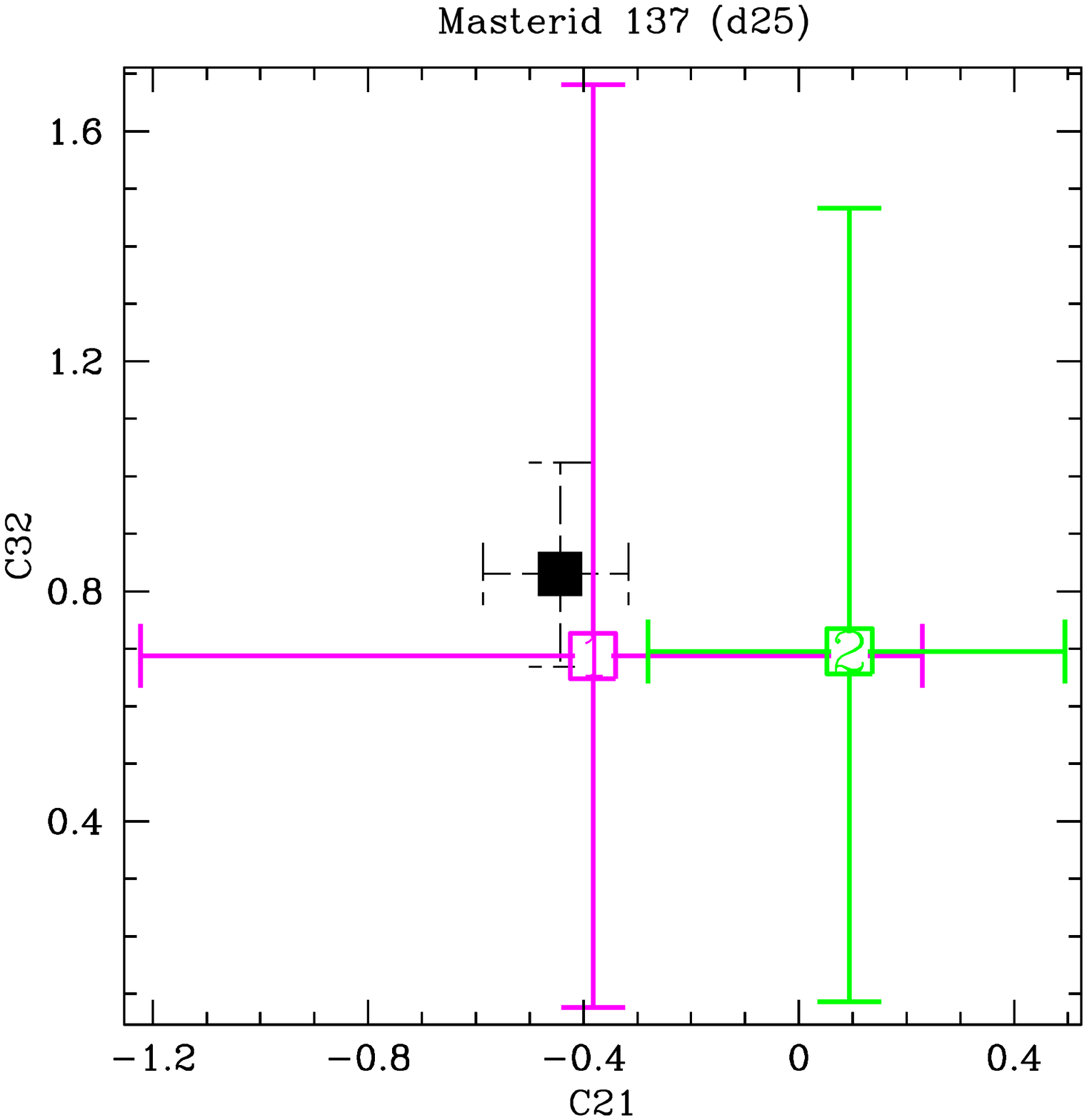}

\end{minipage}
\begin{minipage}{0.32\linewidth}
  \centering

    \includegraphics[width=\linewidth]{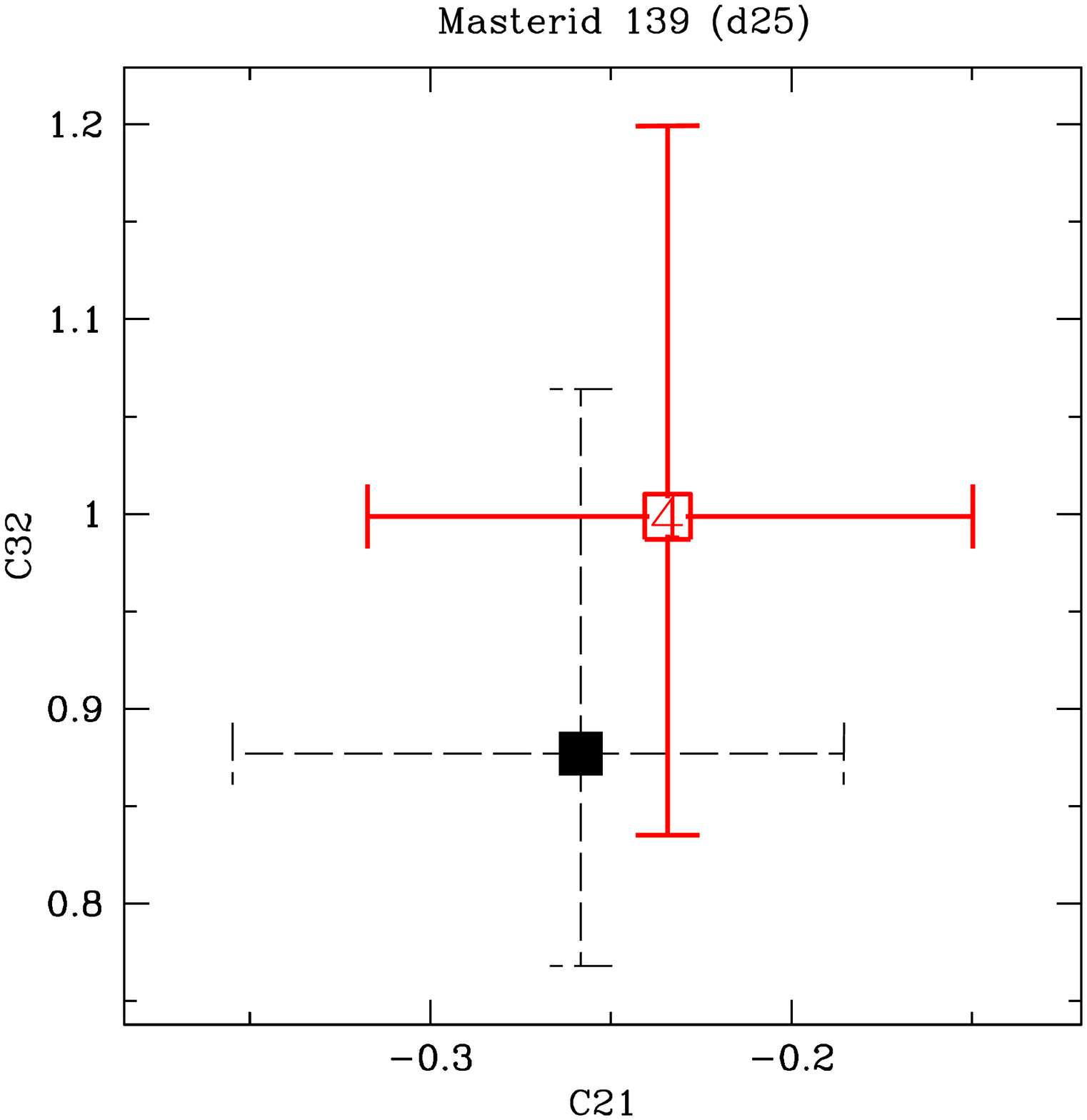}

 \end{minipage}

\begin{minipage}{0.32\linewidth}
  \centering
  
    \includegraphics[width=\linewidth]{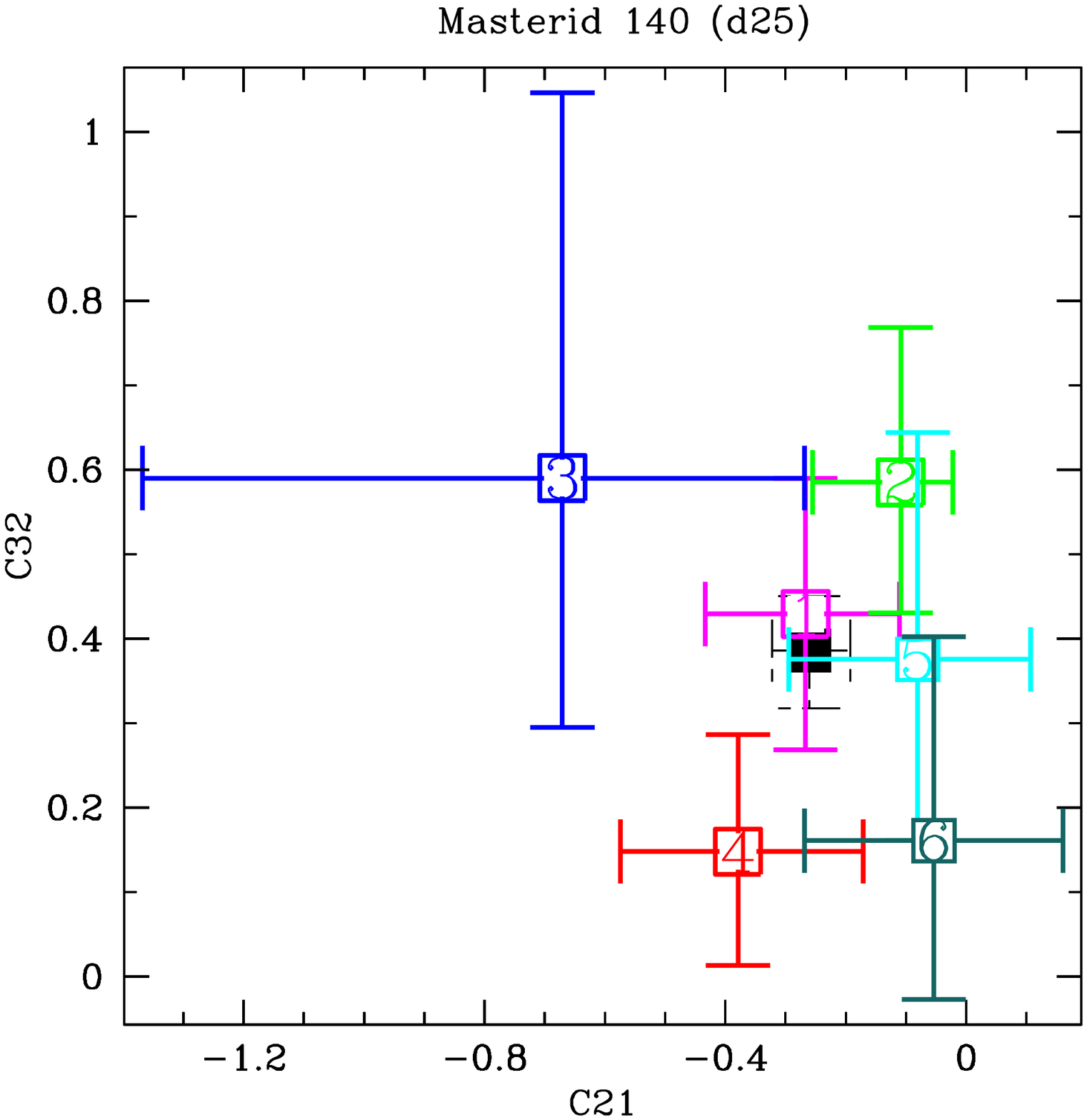}

  \end{minipage}
  \begin{minipage}{0.32\linewidth}
  \centering

    \includegraphics[width=\linewidth]{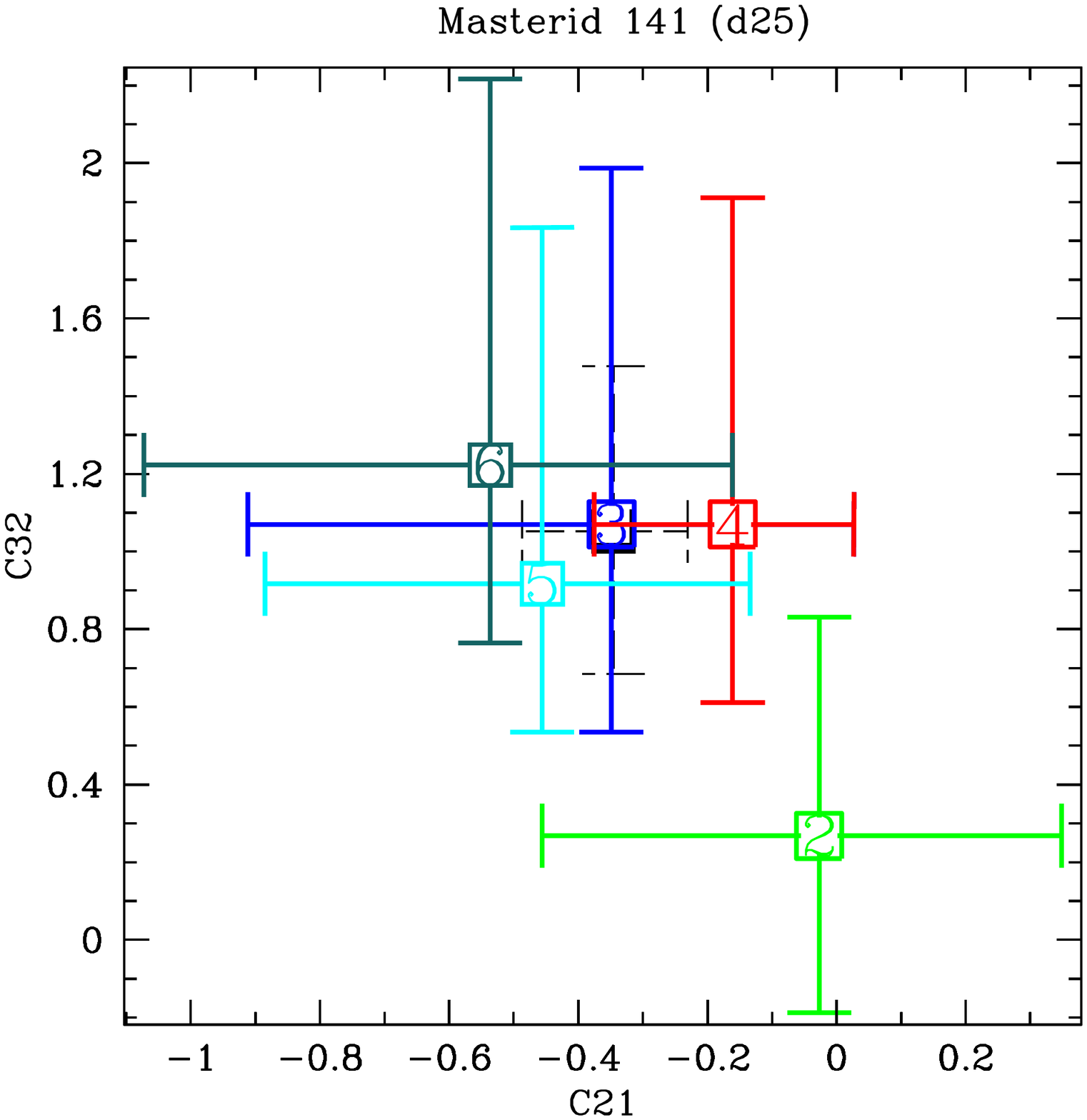}

\end{minipage}
\begin{minipage}{0.32\linewidth}
  \centering

    \includegraphics[width=\linewidth]{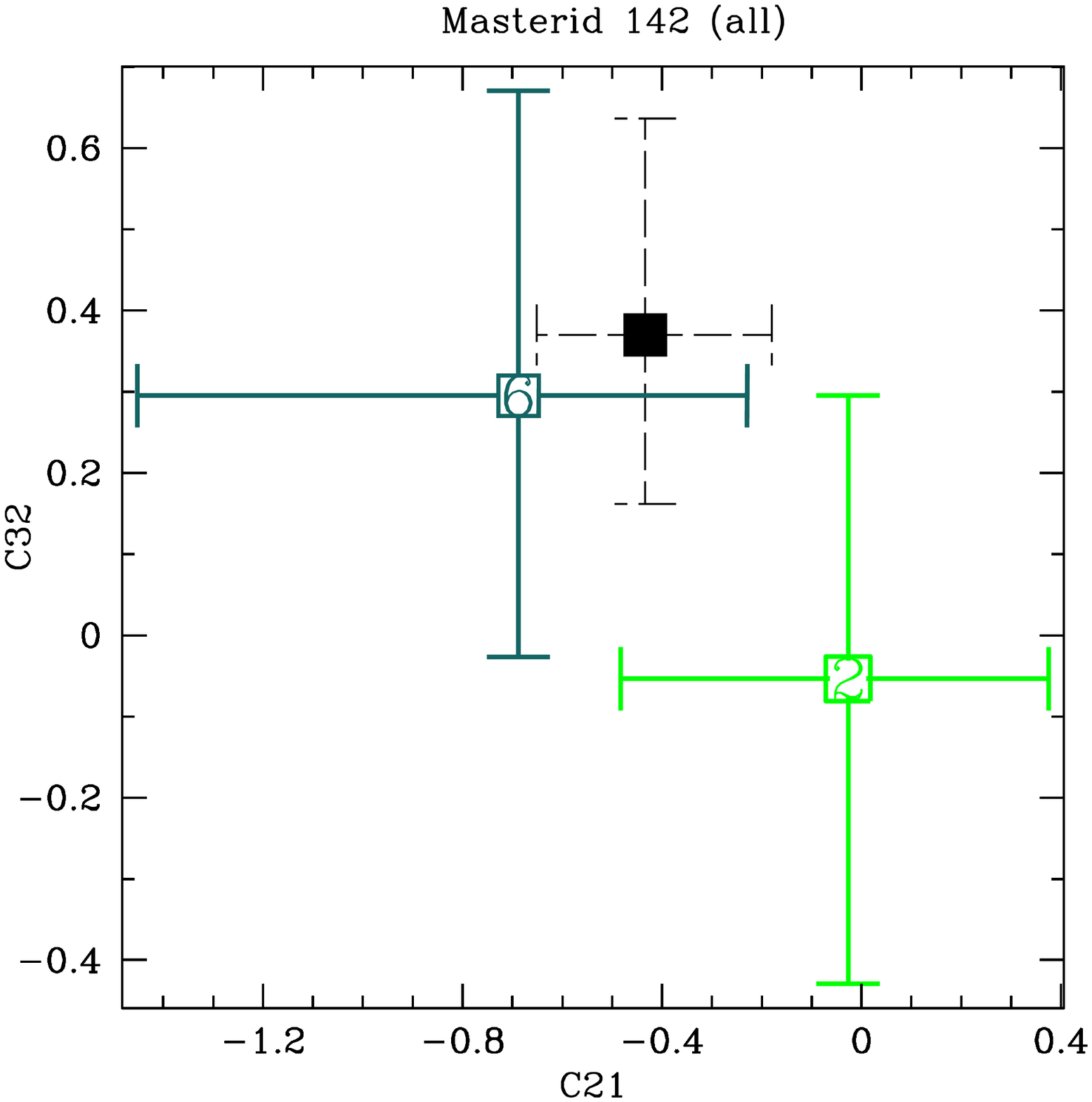}

 \end{minipage}

  \begin{minipage}{0.32\linewidth}
  \centering
  
    \includegraphics[width=\linewidth]{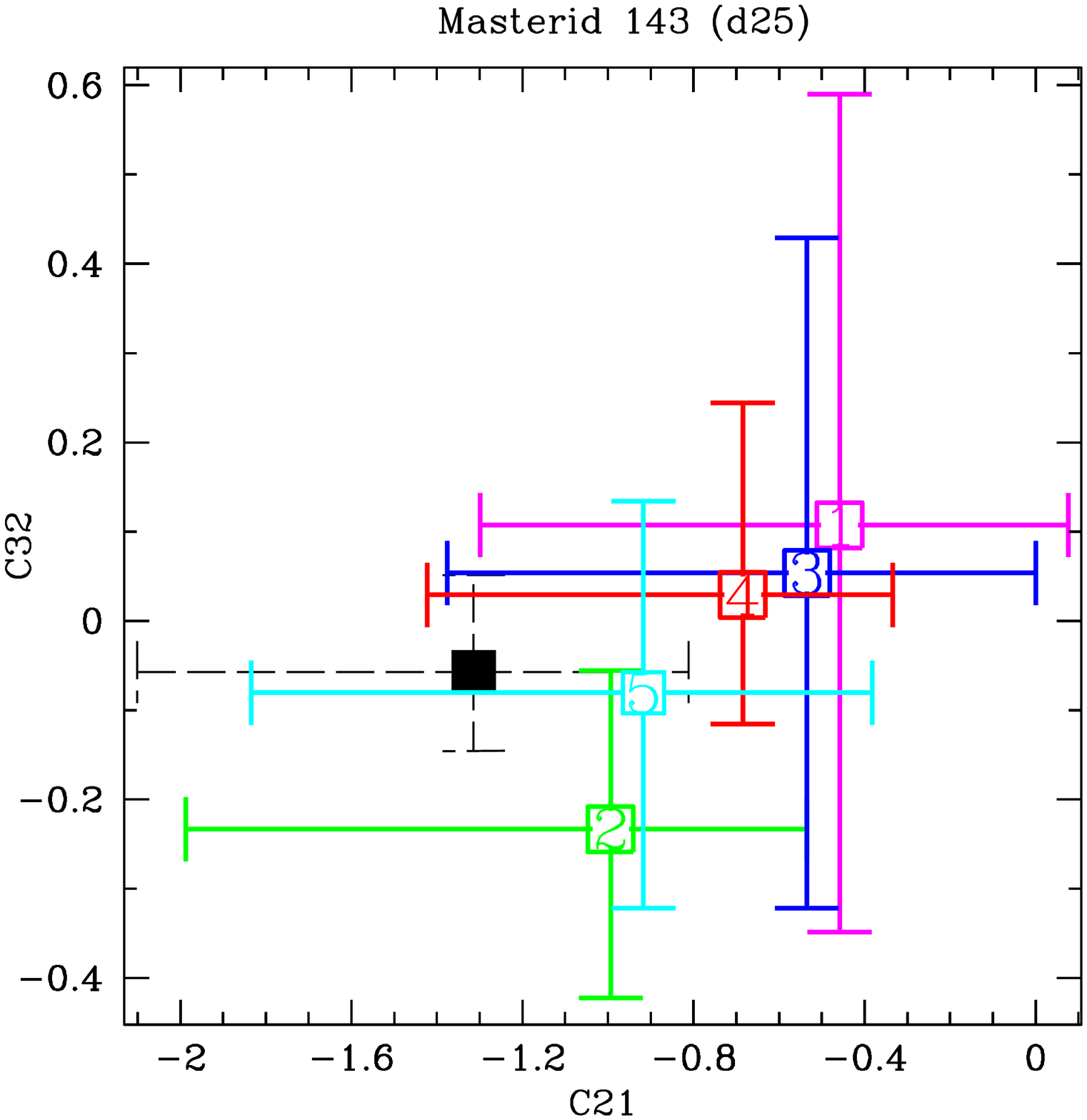}

  \end{minipage}
  \begin{minipage}{0.32\linewidth}
  \centering

    \includegraphics[width=\linewidth]{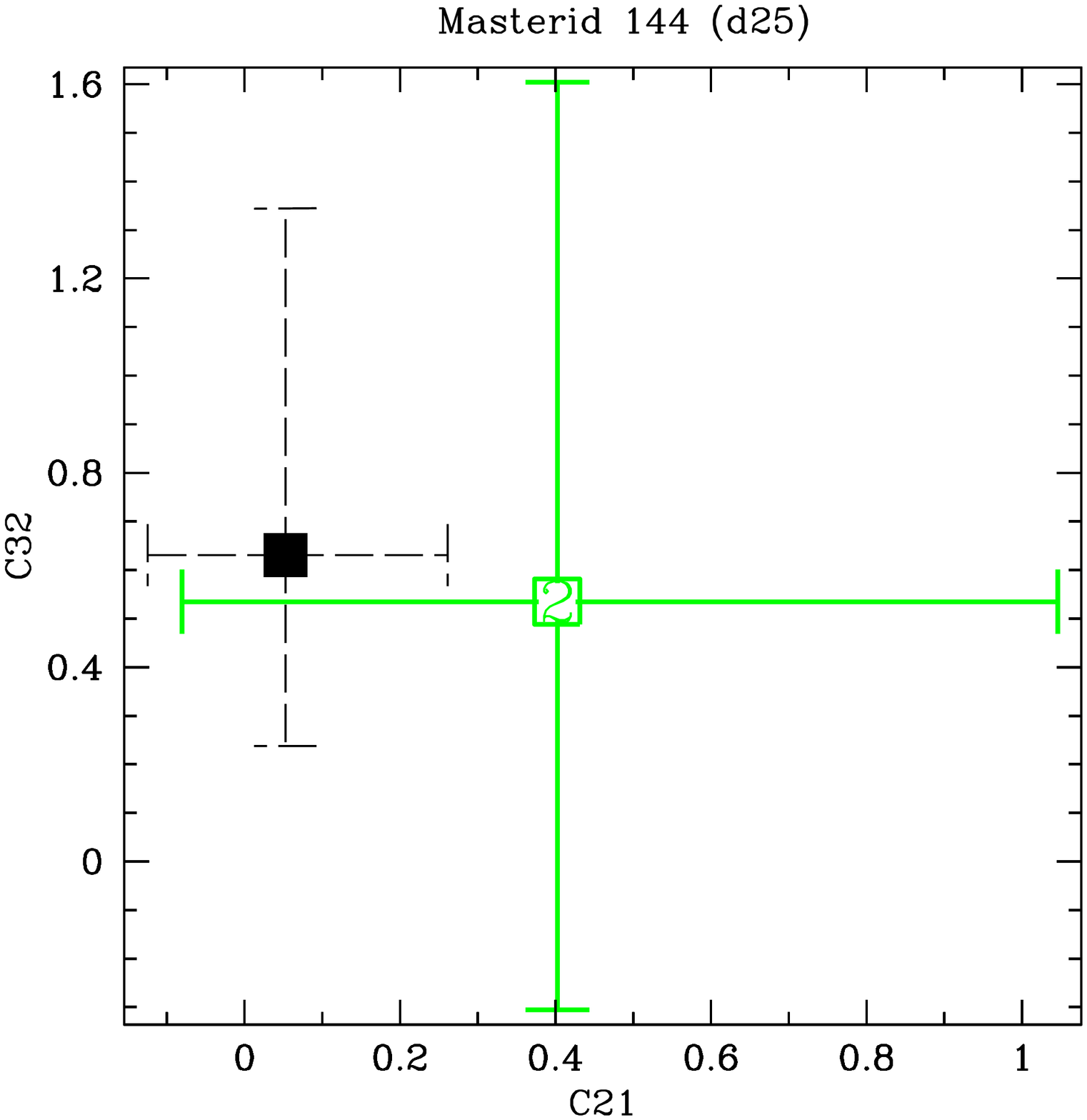}

\end{minipage}
\begin{minipage}{0.32\linewidth}
  \centering

    \includegraphics[width=\linewidth]{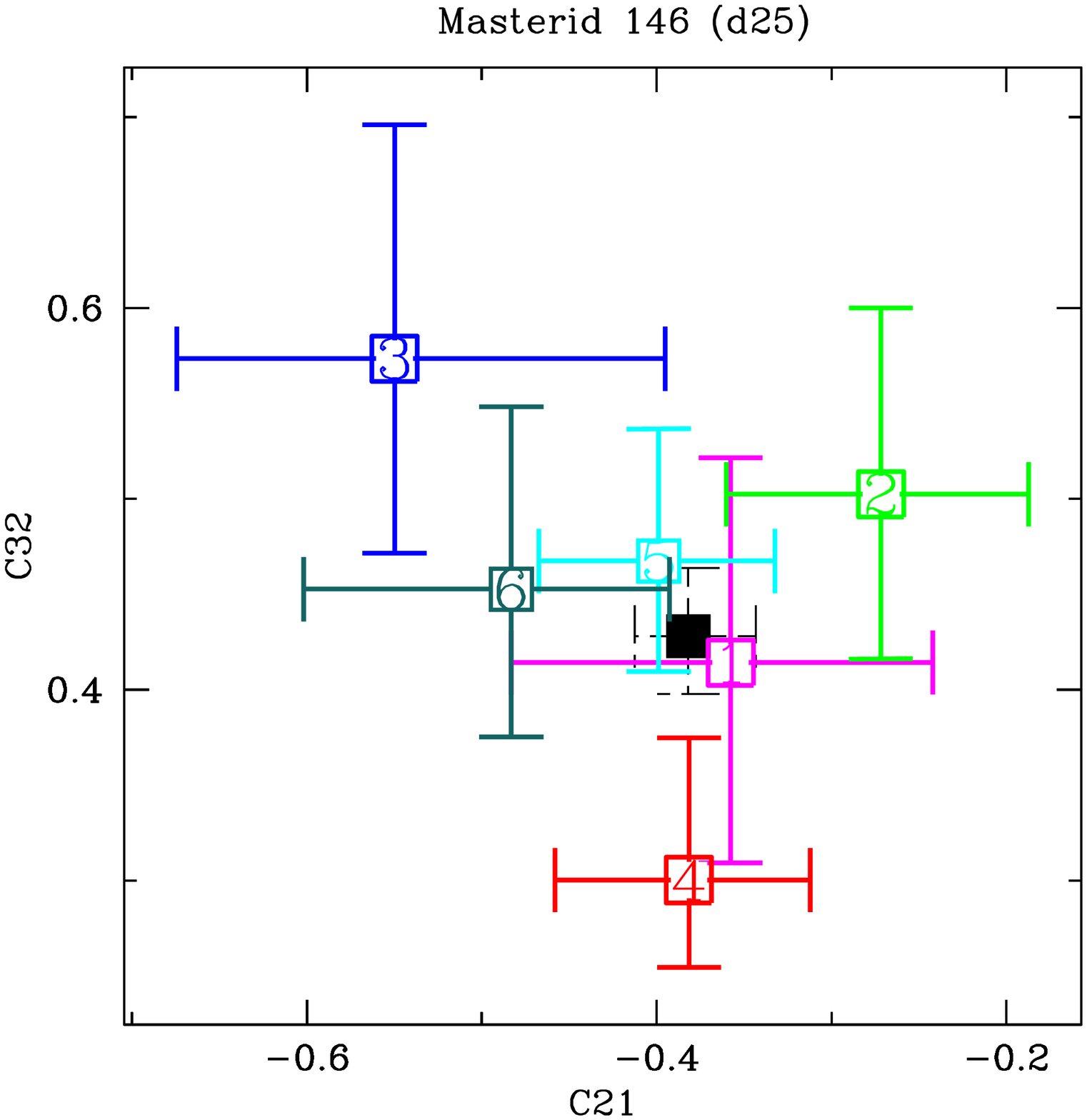}

\end{minipage}

\begin{minipage}{0.32\linewidth}
  \centering
  
    \includegraphics[width=\linewidth]{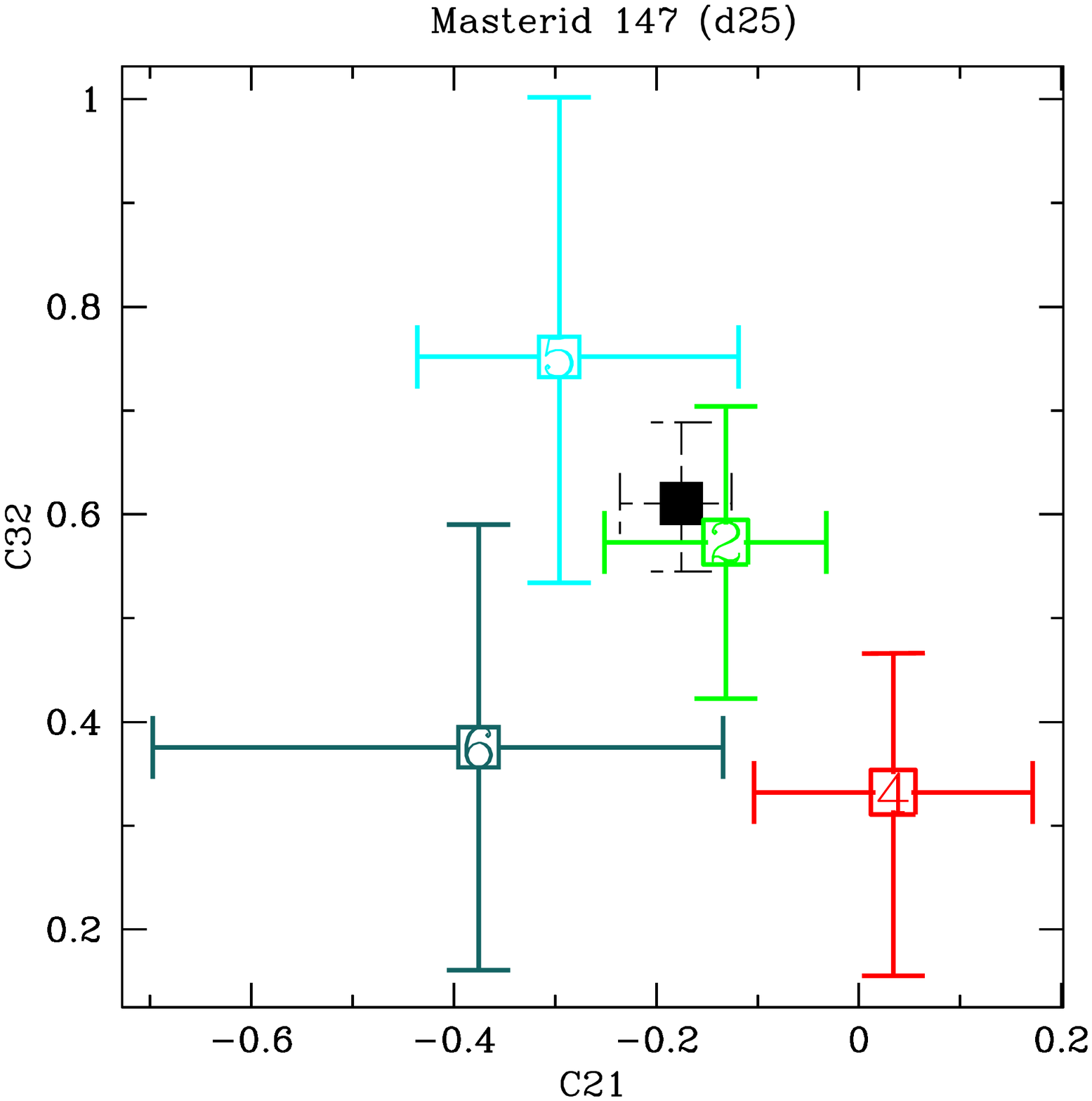}

  \end{minipage}
  \begin{minipage}{0.32\linewidth}
  \centering

    \includegraphics[width=\linewidth]{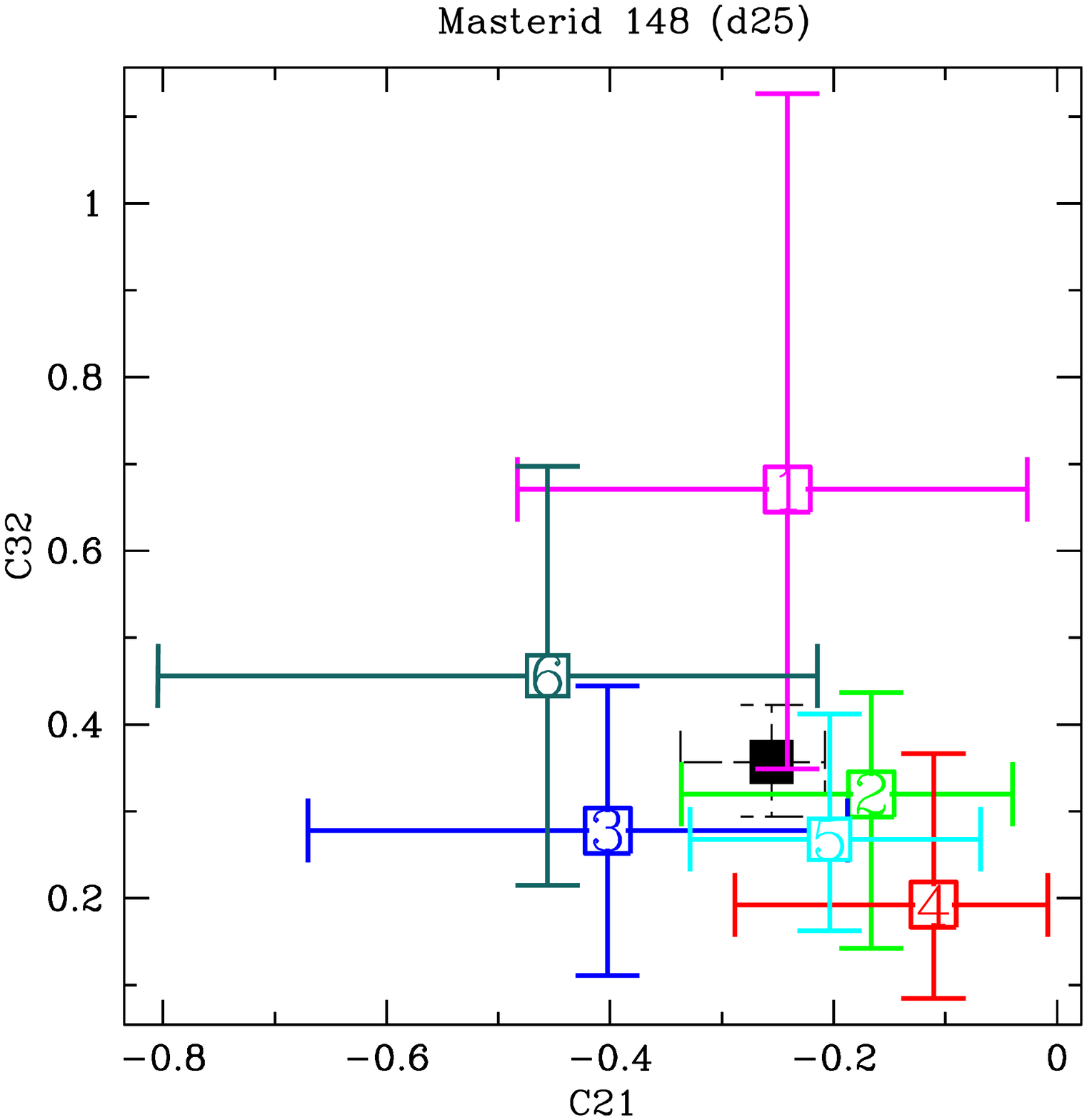}

\end{minipage}
\begin{minipage}{0.32\linewidth}
  \centering

    \includegraphics[width=\linewidth]{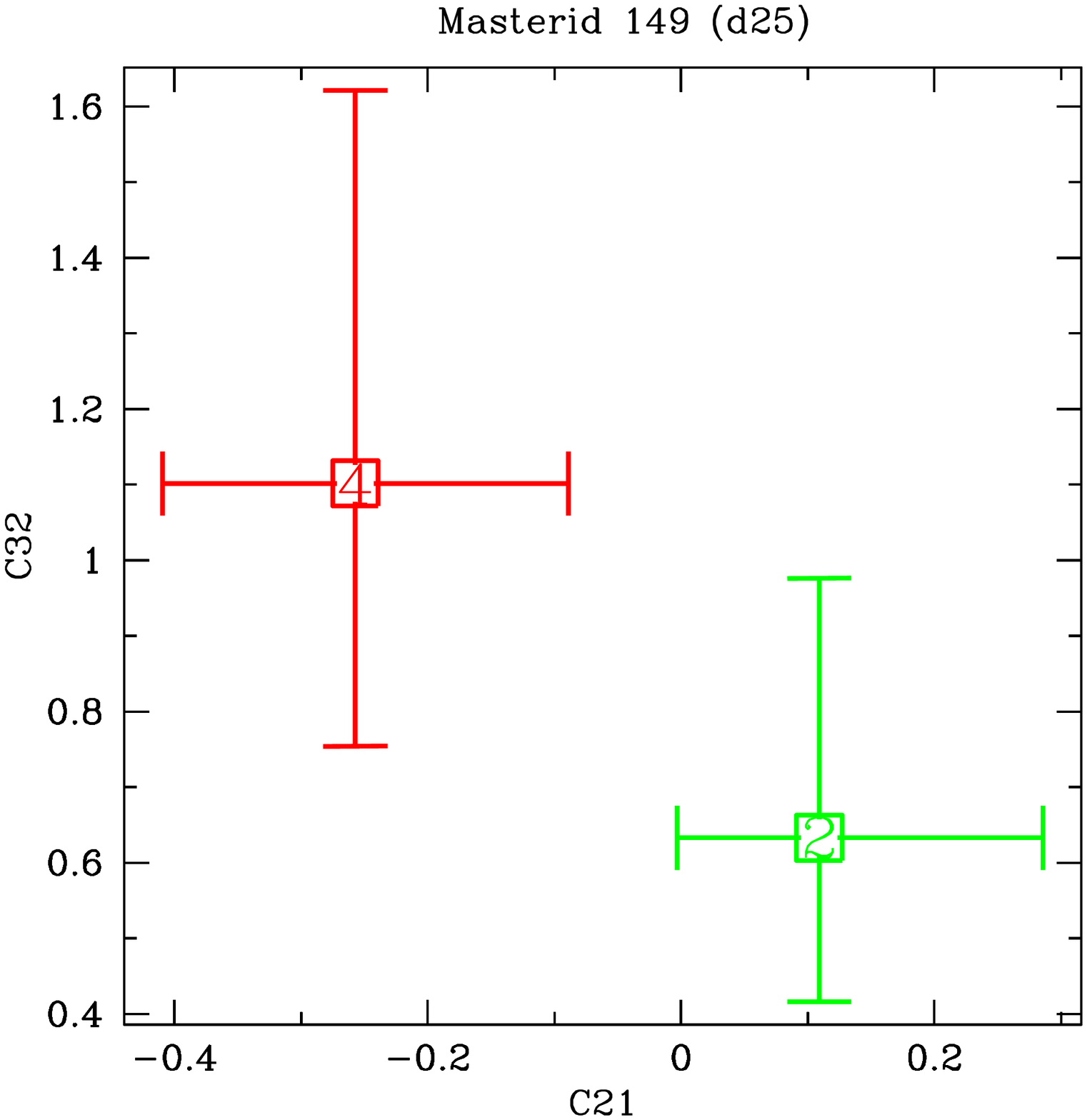}

\end{minipage}
\end{figure}

\clearpage

\begin{figure}
  \begin{minipage}{0.32\linewidth}
  \centering
  
    \includegraphics[width=\linewidth]{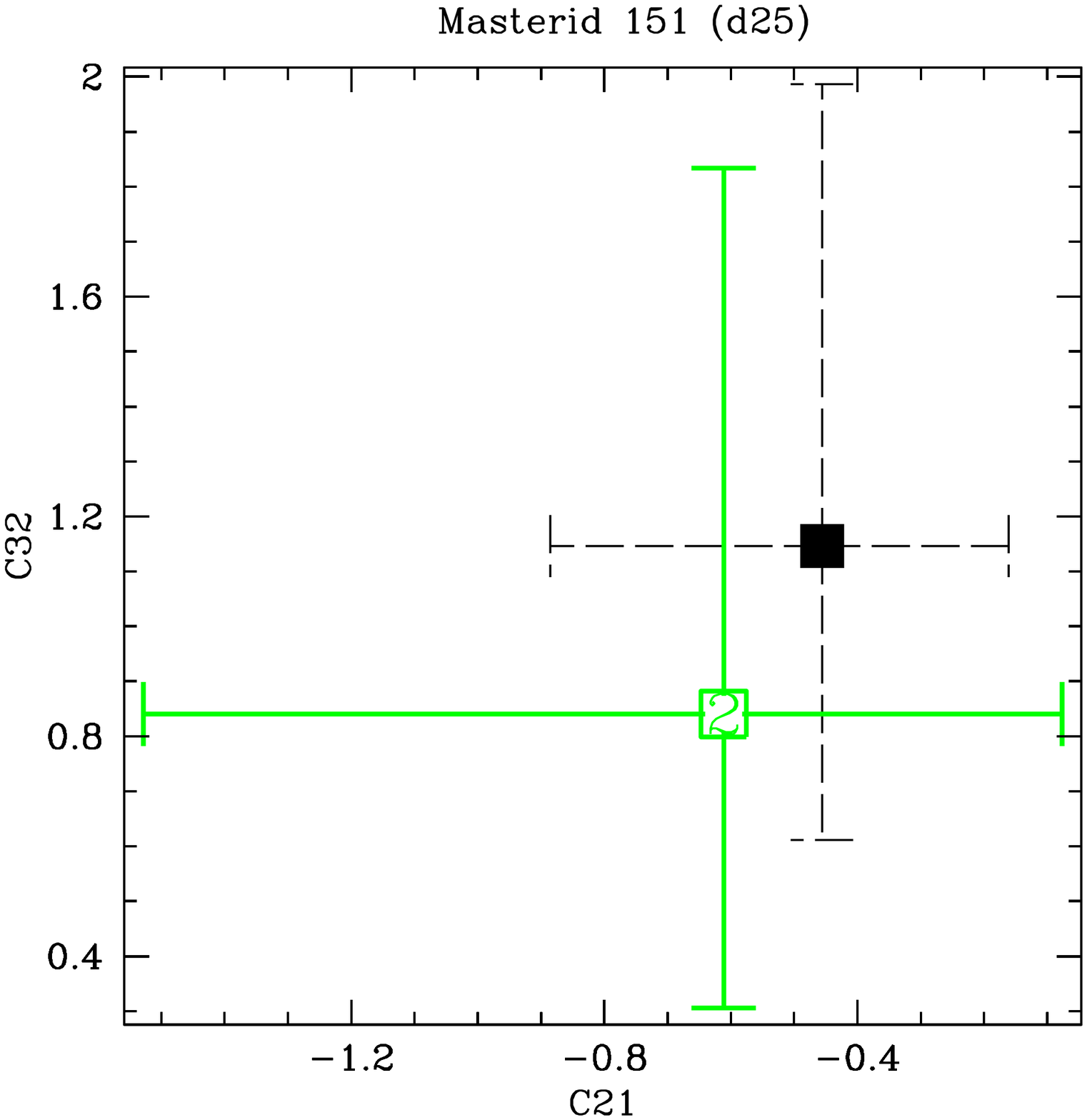}

  \end{minipage}
  \begin{minipage}{0.32\linewidth}
  \centering

    \includegraphics[width=\linewidth]{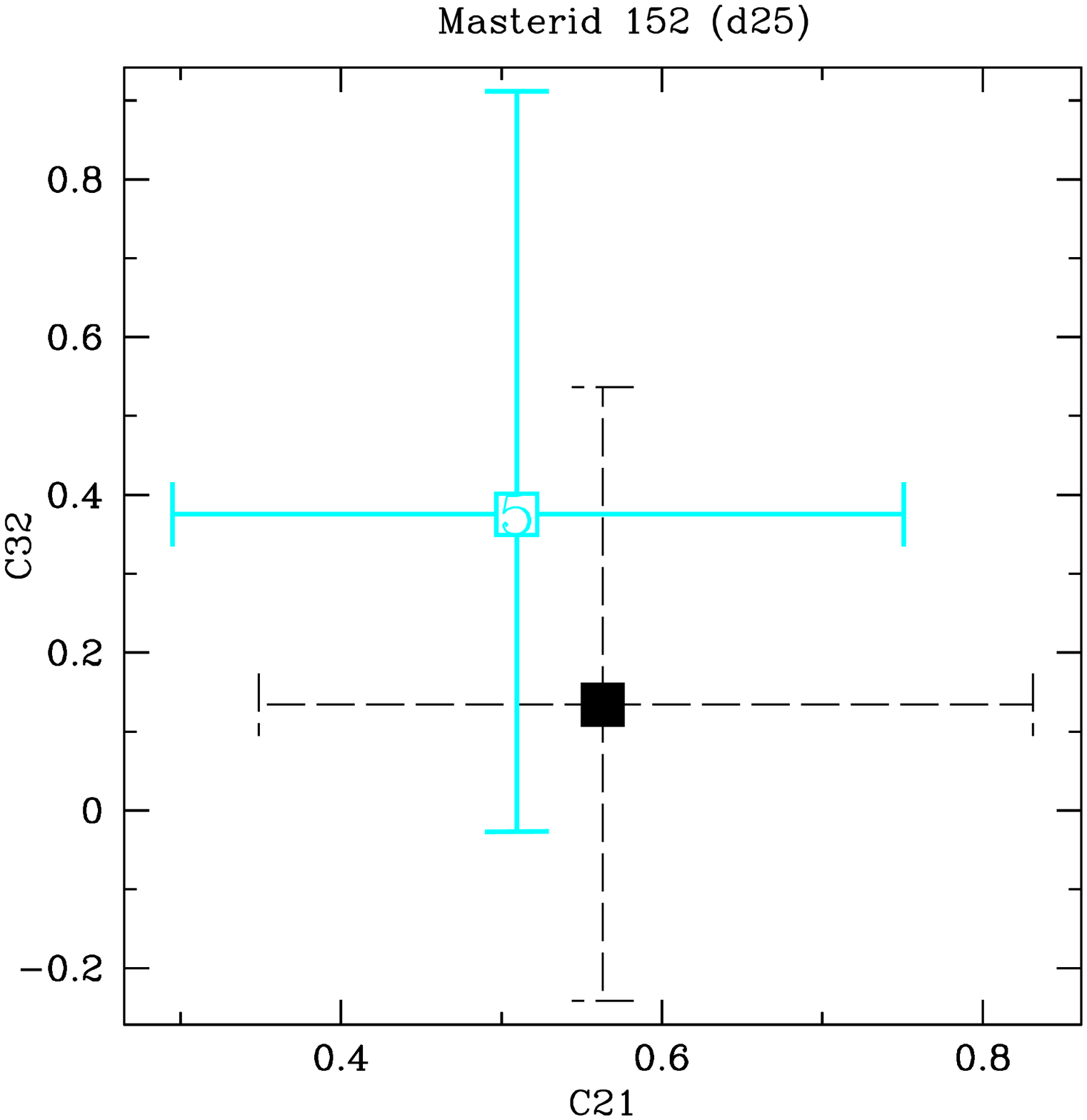}

\end{minipage}
\begin{minipage}{0.32\linewidth}
  \centering

    \includegraphics[width=\linewidth]{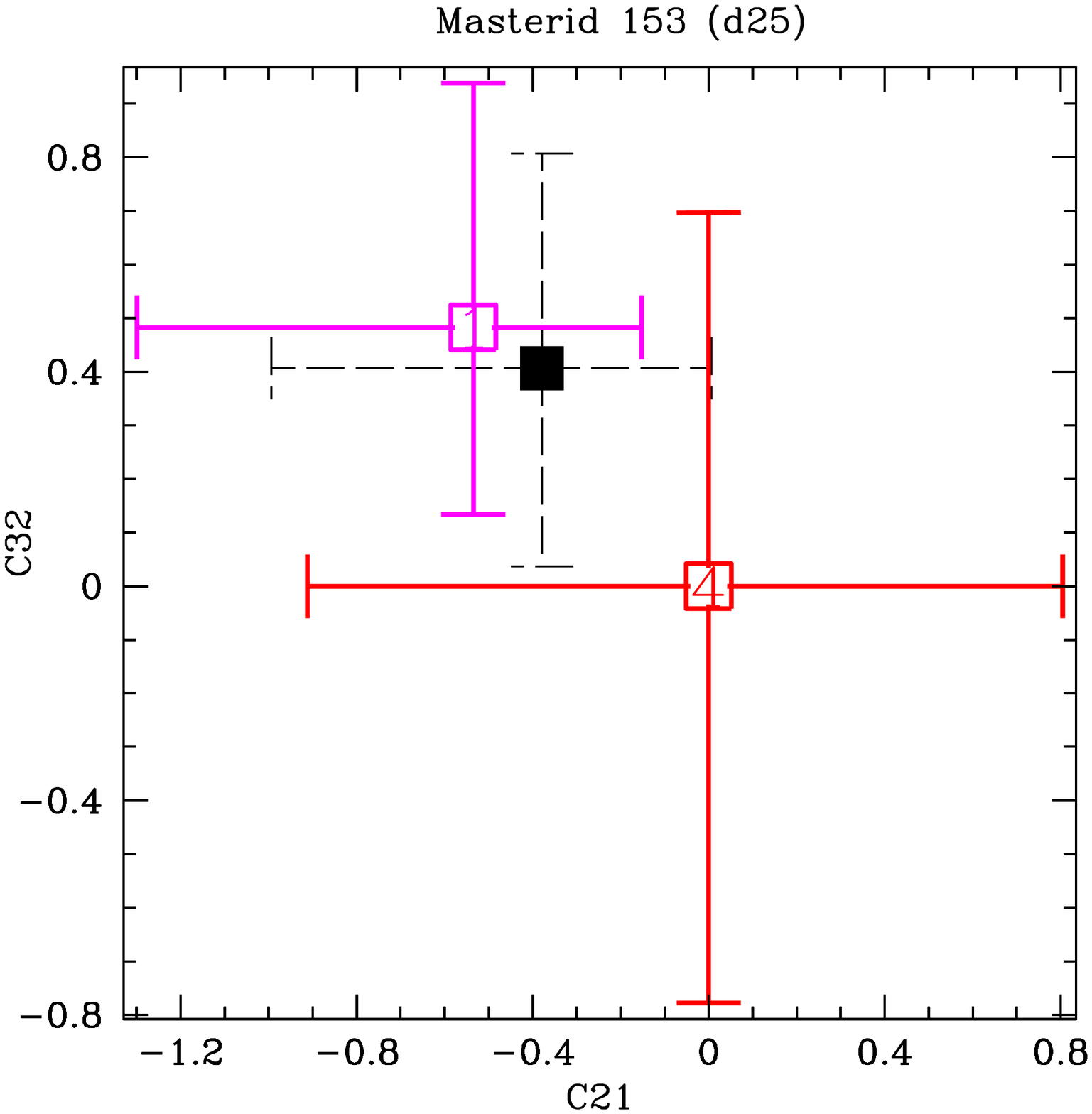}

 \end{minipage}

\begin{minipage}{0.32\linewidth}
  \centering
  
    \includegraphics[width=\linewidth]{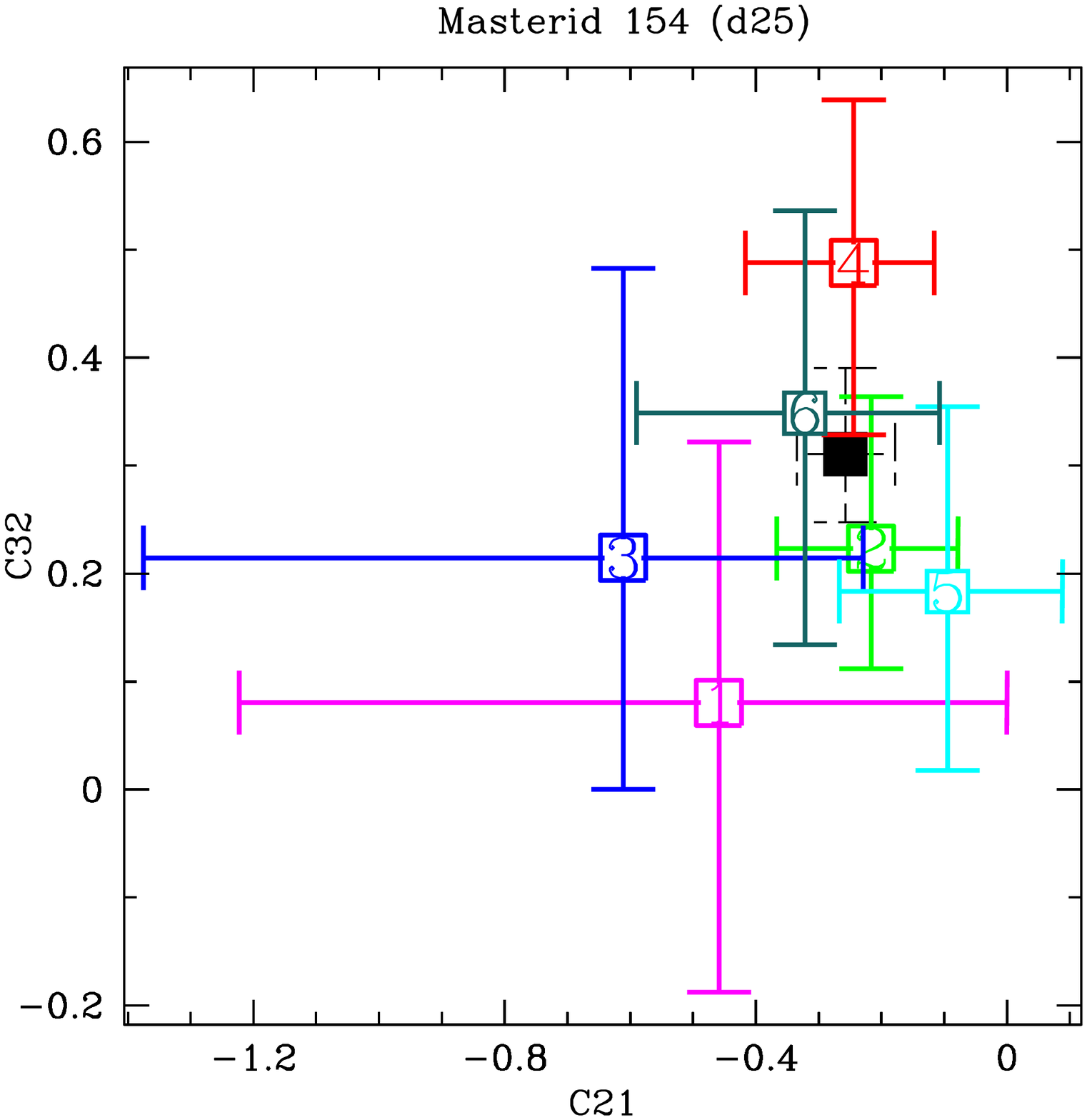}

  \end{minipage}
  \begin{minipage}{0.32\linewidth}
  \centering

    \includegraphics[width=\linewidth]{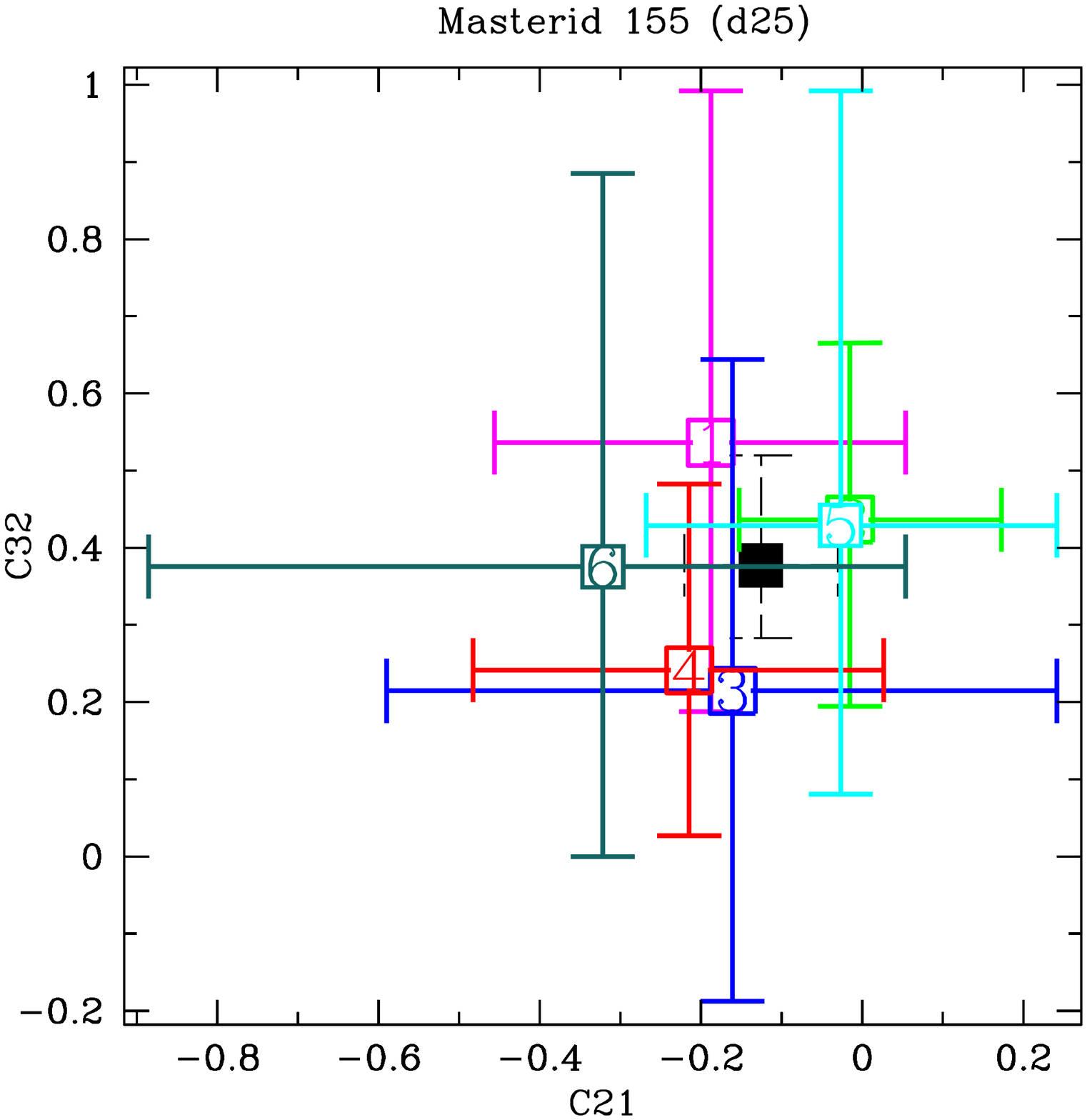}

\end{minipage}
\begin{minipage}{0.32\linewidth}
  \centering

    \includegraphics[width=\linewidth]{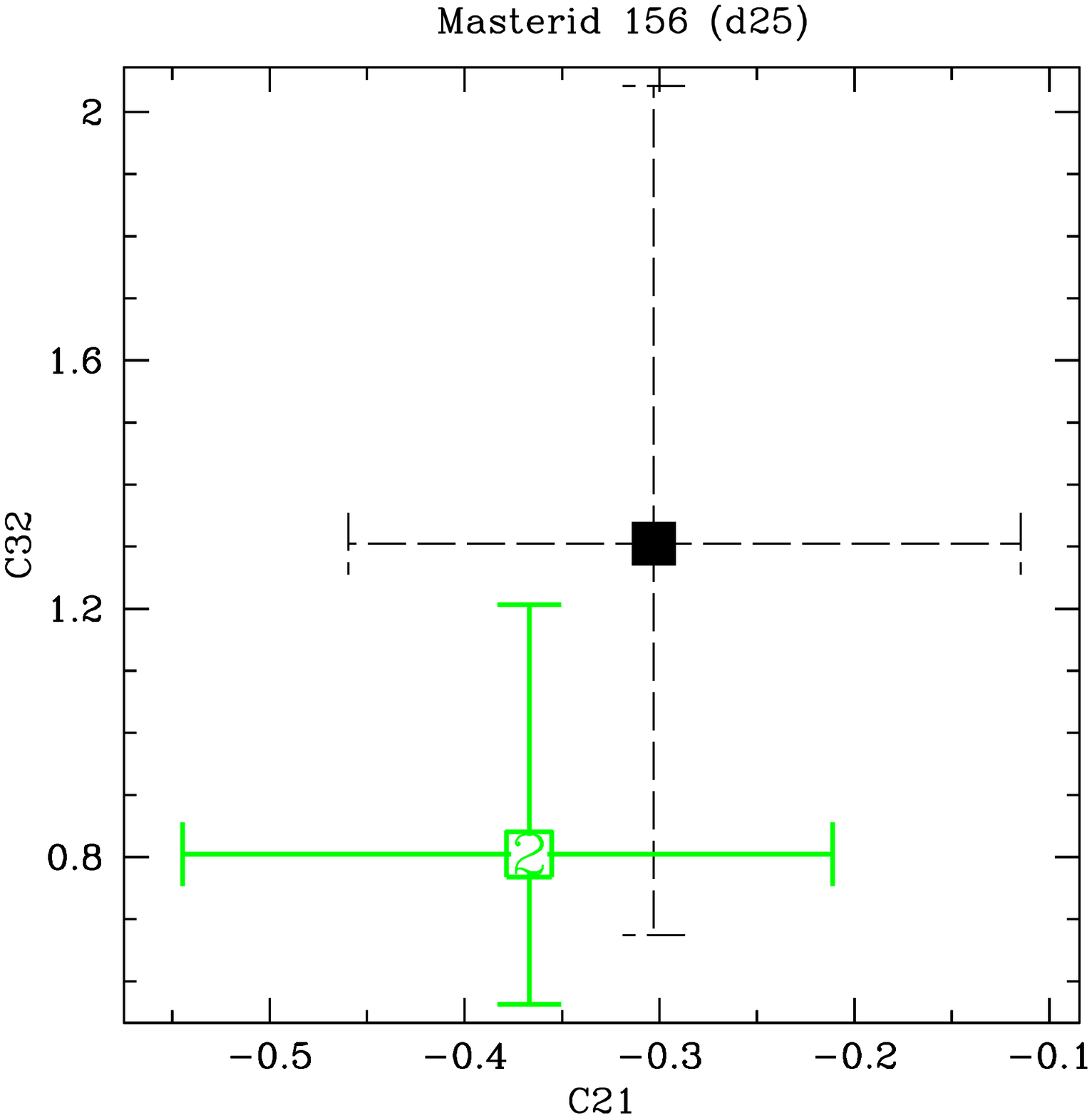}

 \end{minipage}

  \begin{minipage}{0.32\linewidth}
  \centering
  
    \includegraphics[width=\linewidth]{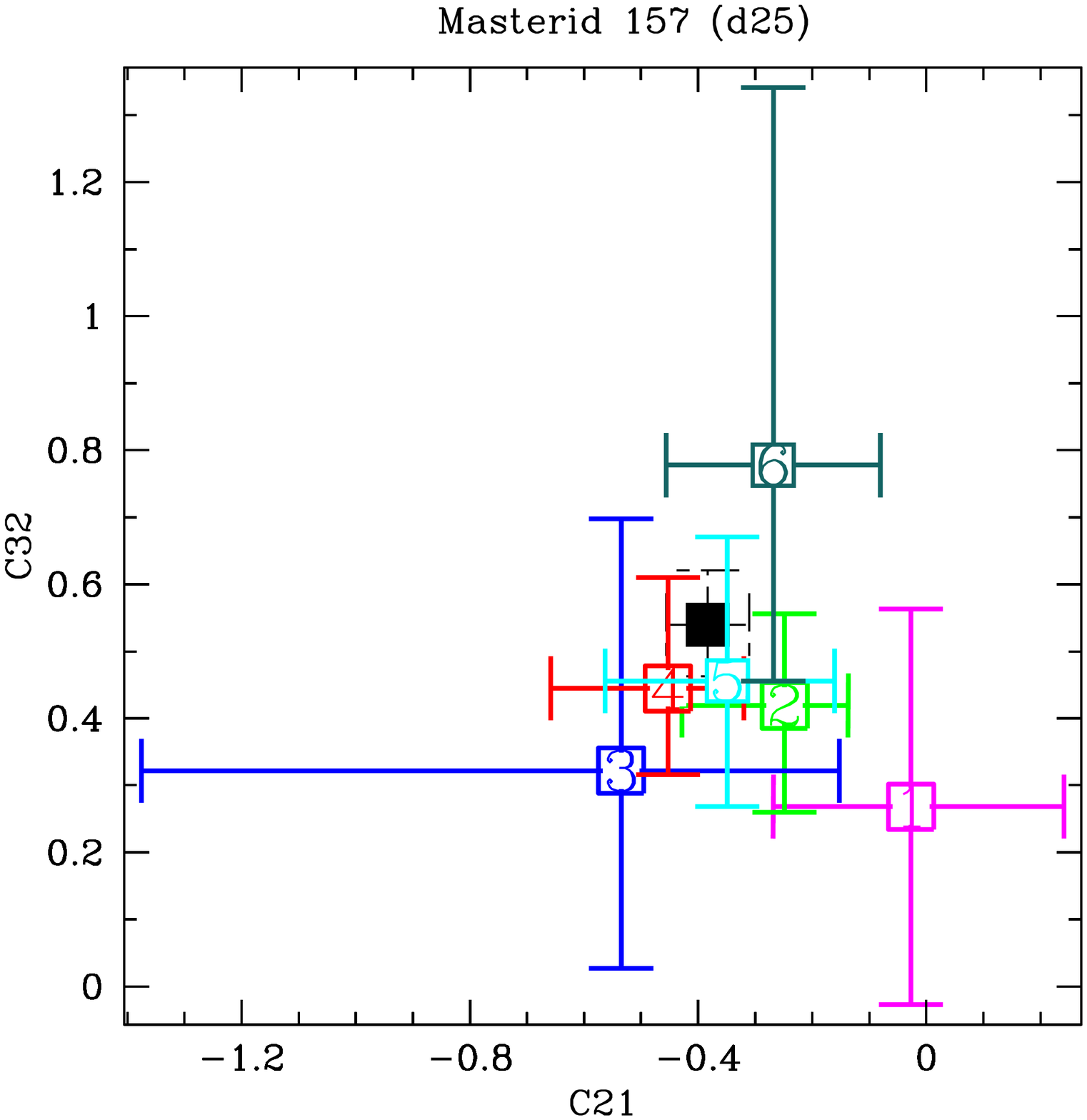}

  \end{minipage}
  \begin{minipage}{0.32\linewidth}
  \centering

    \includegraphics[width=\linewidth]{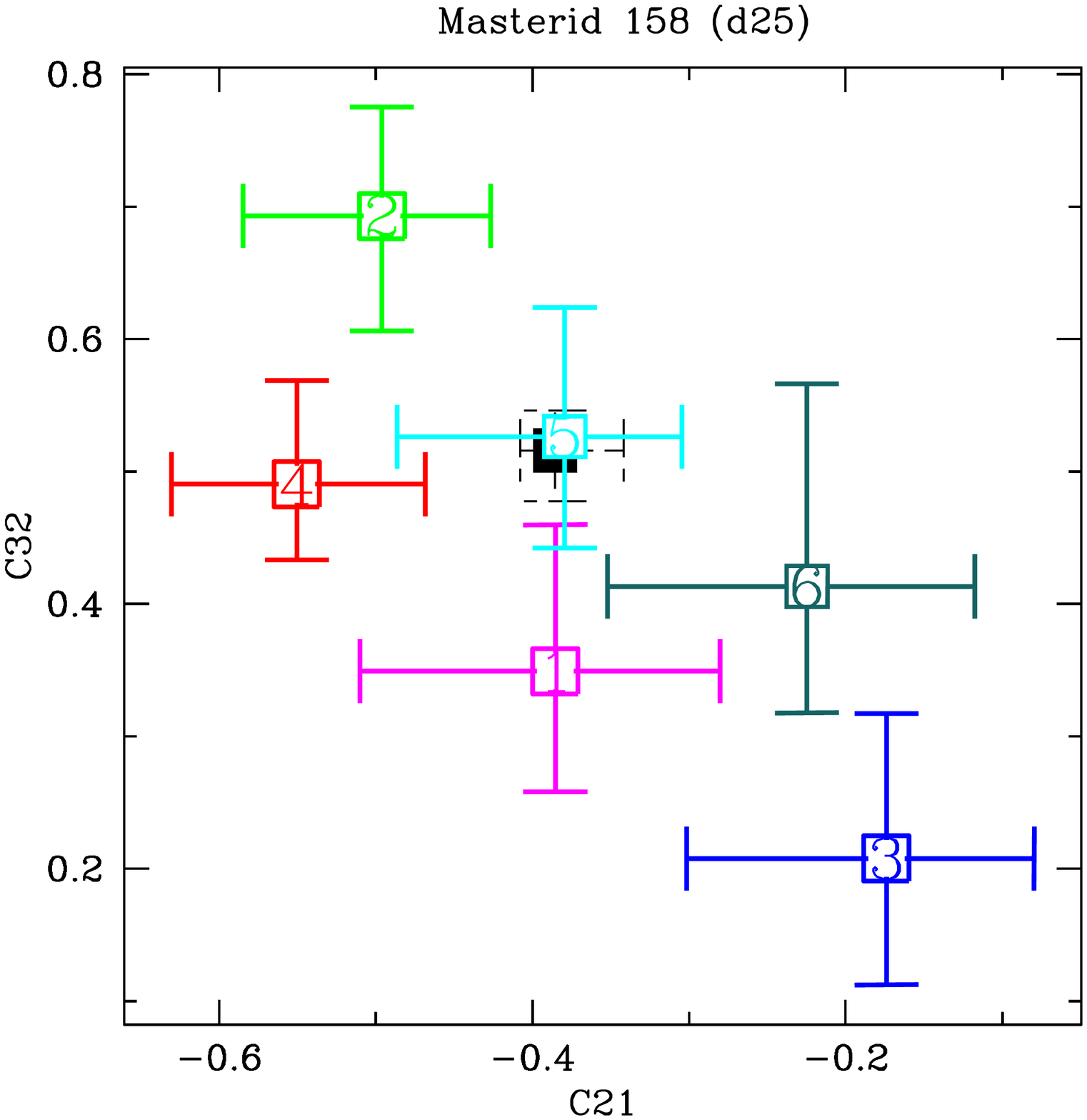}

\end{minipage}
\begin{minipage}{0.32\linewidth}
  \centering

    \includegraphics[width=\linewidth]{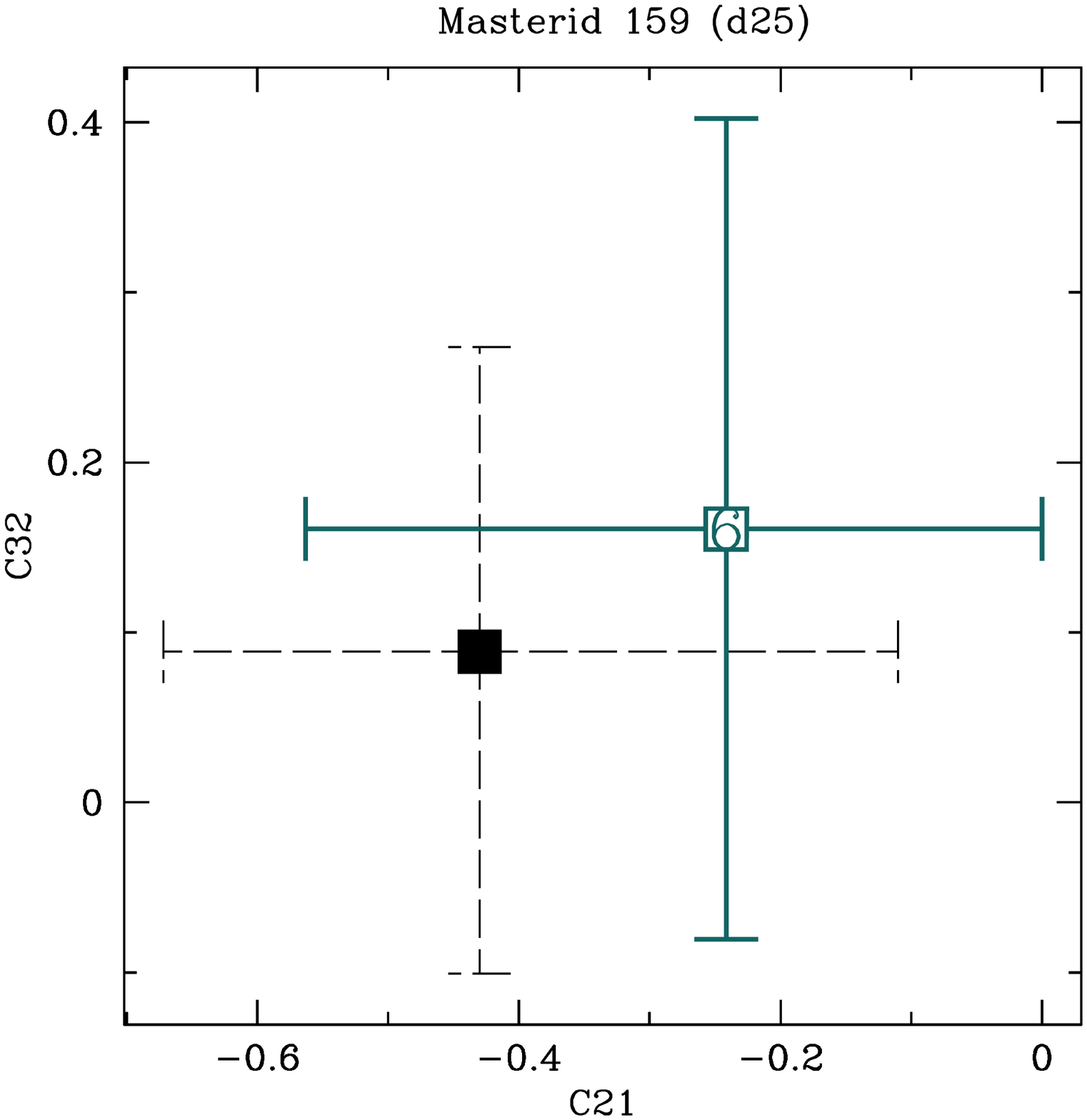}

\end{minipage}

\begin{minipage}{0.32\linewidth}
  \centering
  
    \includegraphics[width=\linewidth]{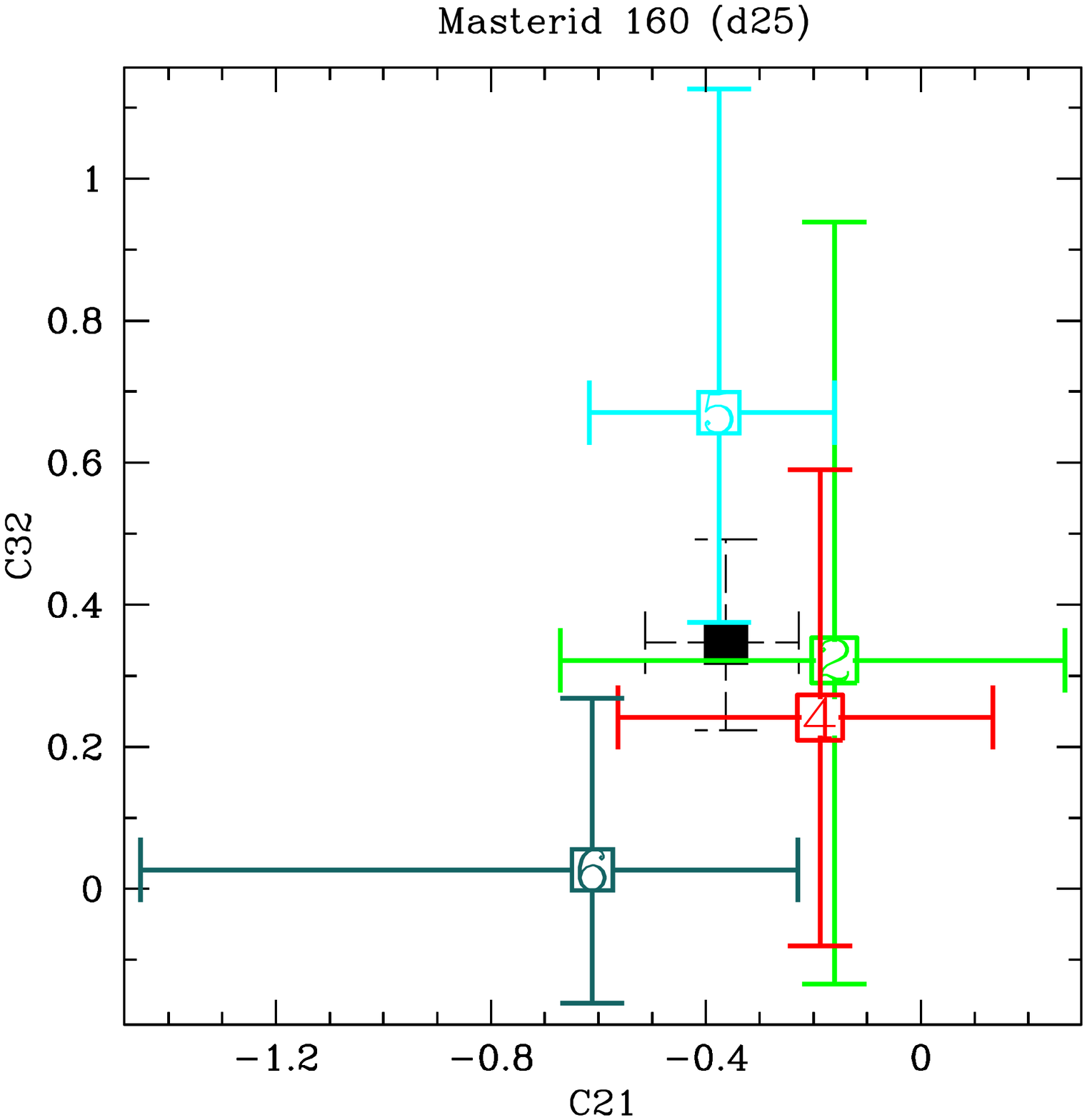}

  \end{minipage}
  \begin{minipage}{0.32\linewidth}
  \centering

    \includegraphics[width=\linewidth]{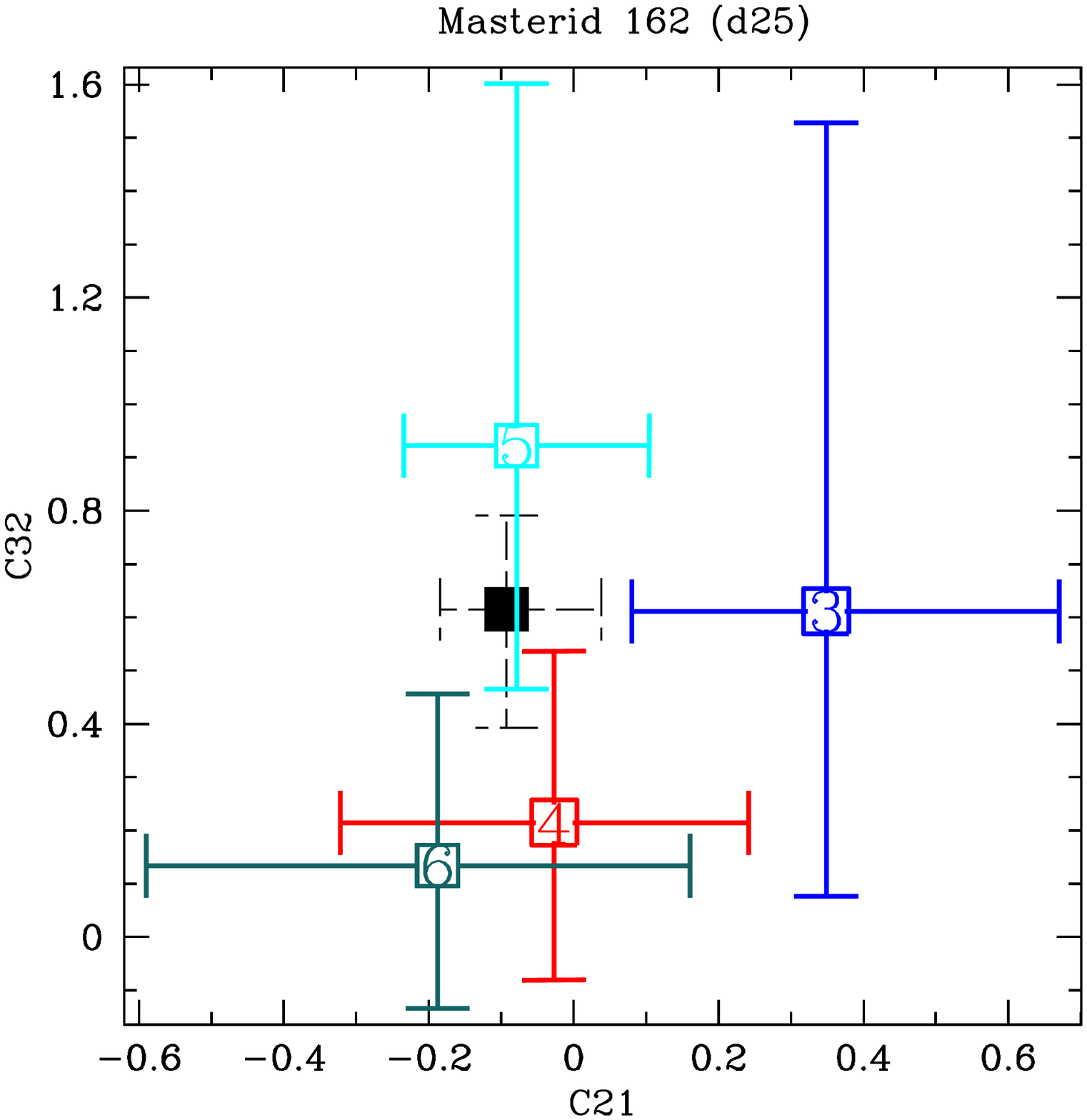}

\end{minipage}
\begin{minipage}{0.32\linewidth}
  \centering

    \includegraphics[width=\linewidth]{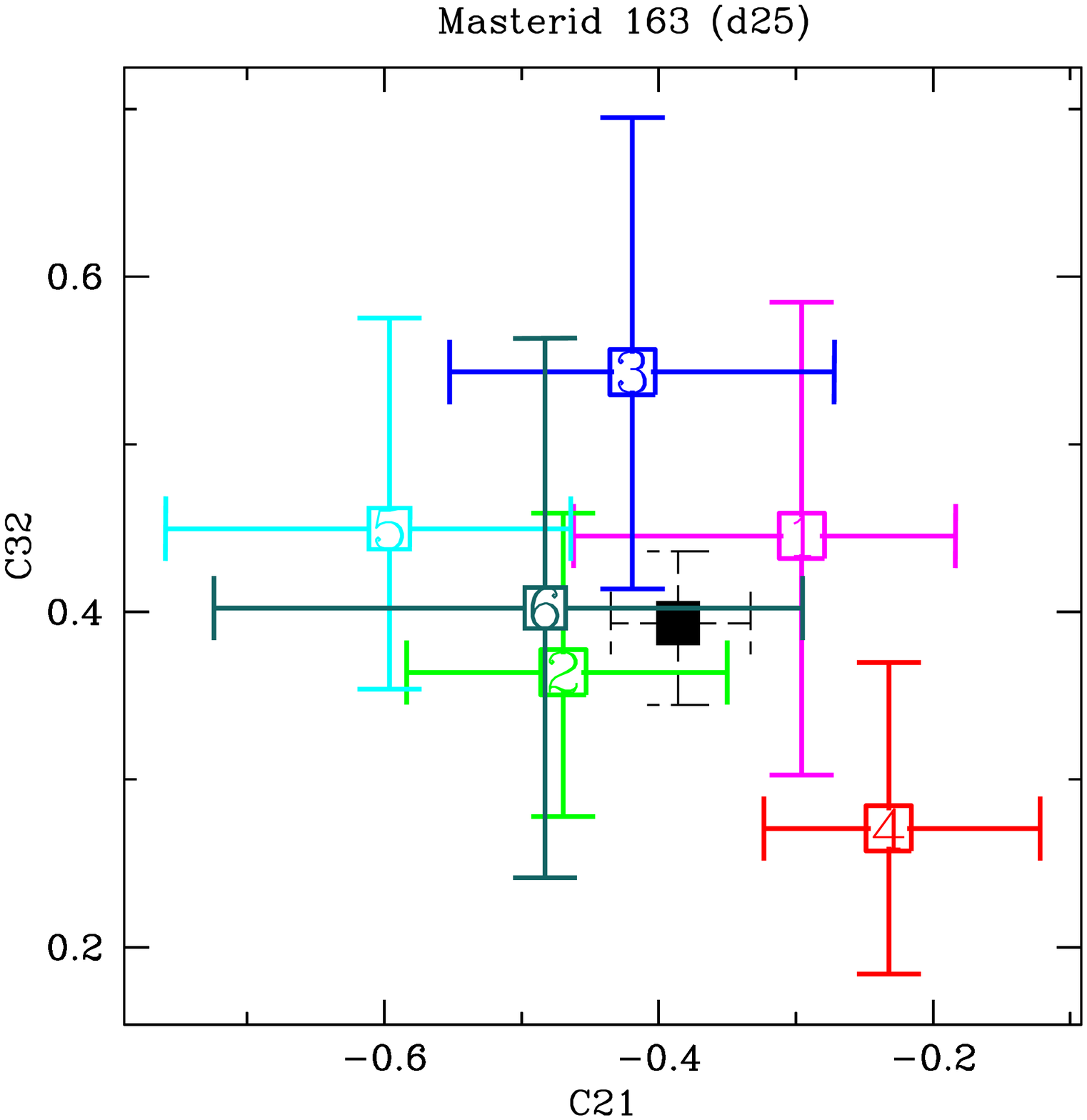}

\end{minipage}
\end{figure}

\begin{figure}
  \begin{minipage}{0.32\linewidth}
  \centering
  
    \includegraphics[width=\linewidth]{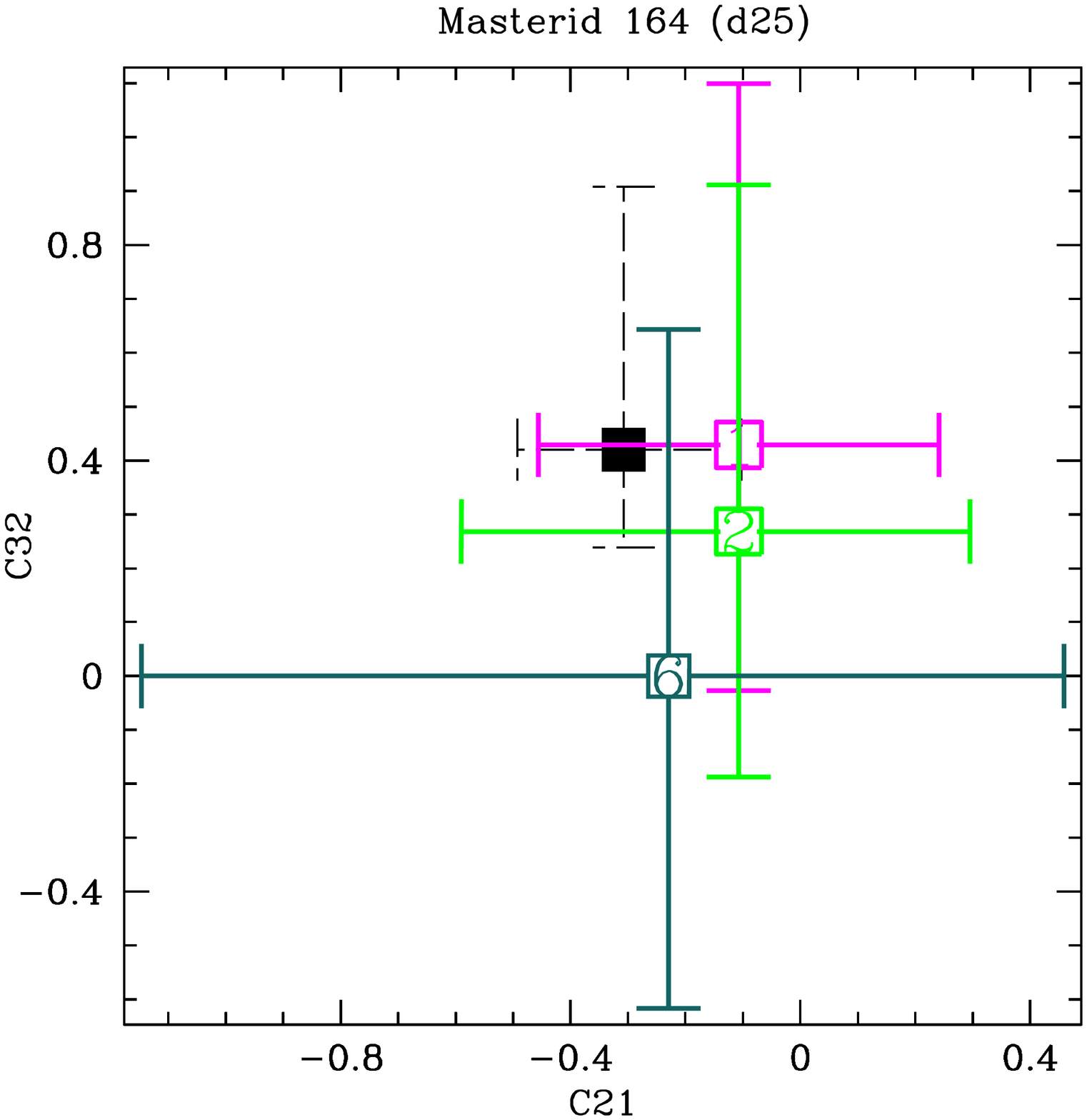}

  \end{minipage}
  \begin{minipage}{0.32\linewidth}
  \centering

    \includegraphics[width=\linewidth]{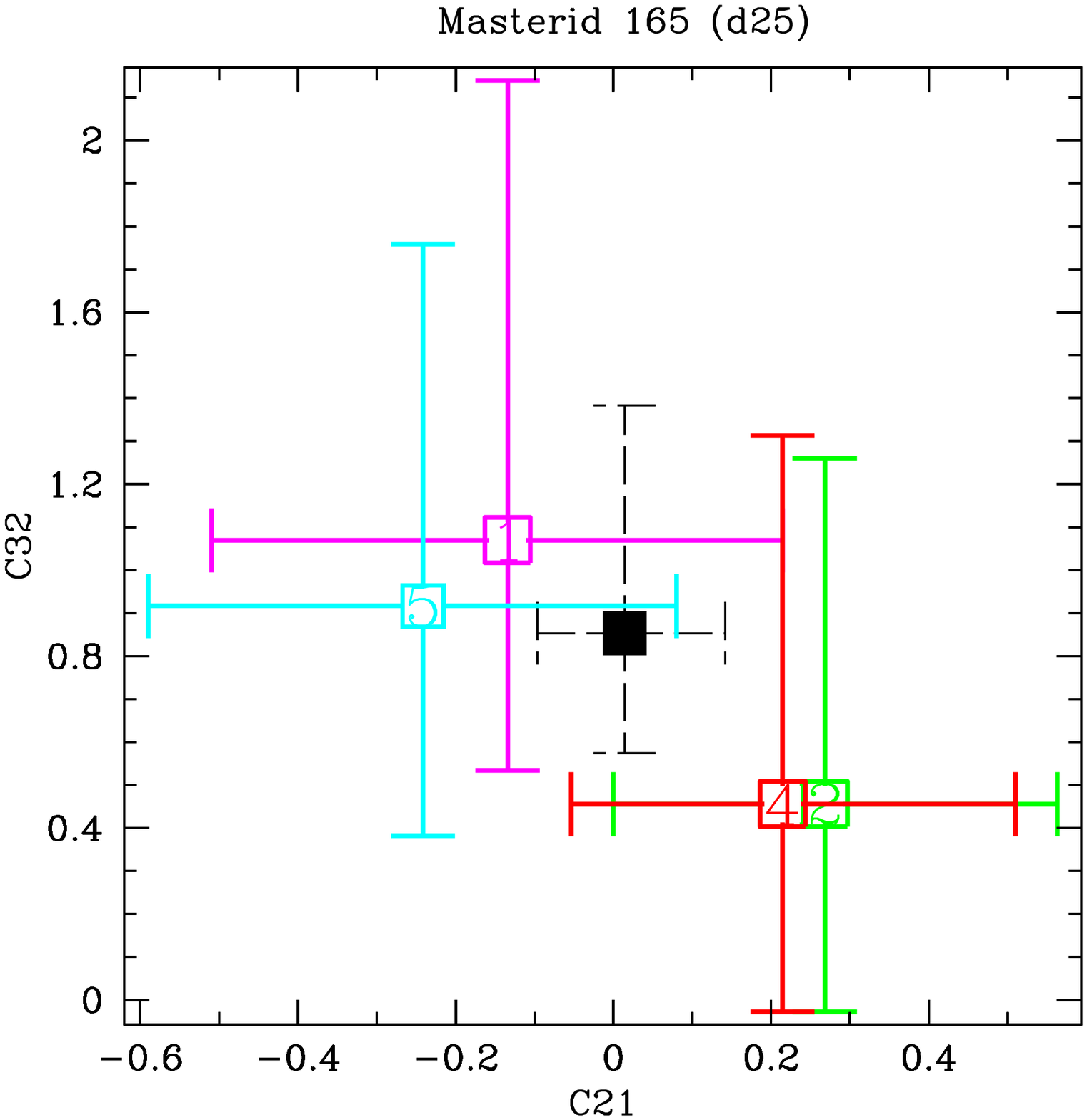}

\end{minipage}
\begin{minipage}{0.32\linewidth}
  \centering

    \includegraphics[width=\linewidth]{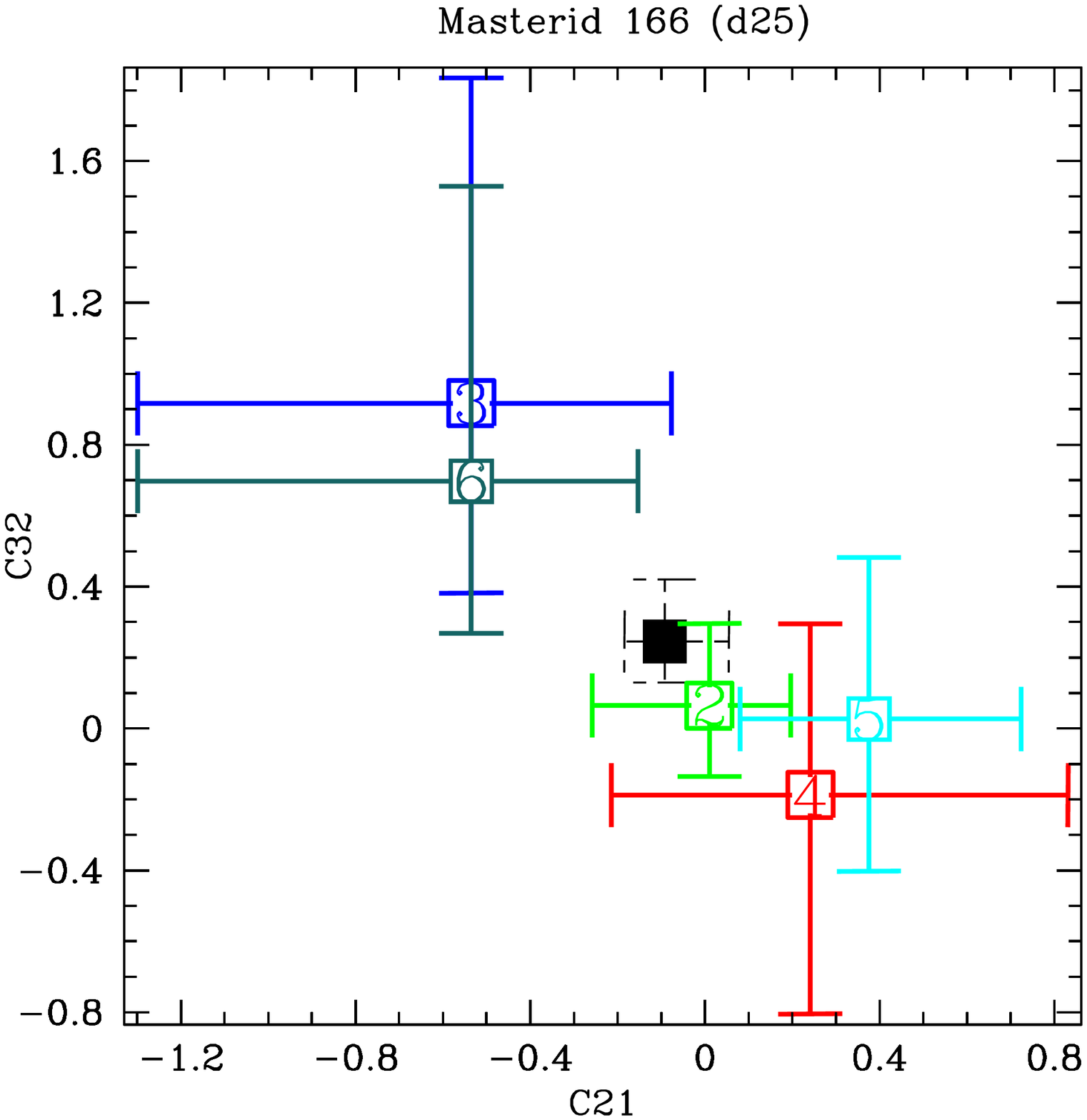}

 \end{minipage}

\begin{minipage}{0.32\linewidth}
  \centering
  
    \includegraphics[width=\linewidth]{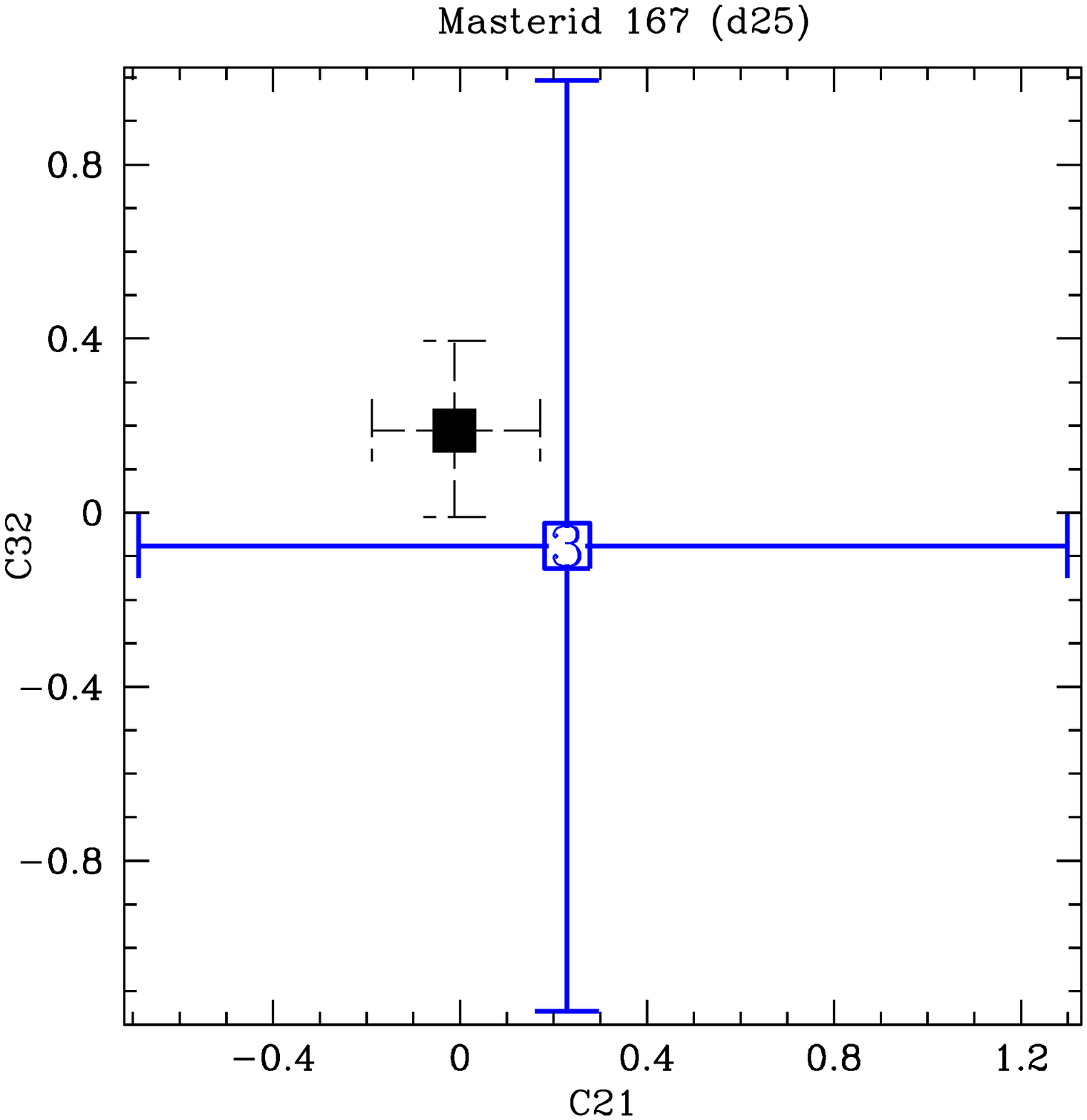}

  \end{minipage}
  \begin{minipage}{0.32\linewidth}
  \centering

    \includegraphics[width=\linewidth]{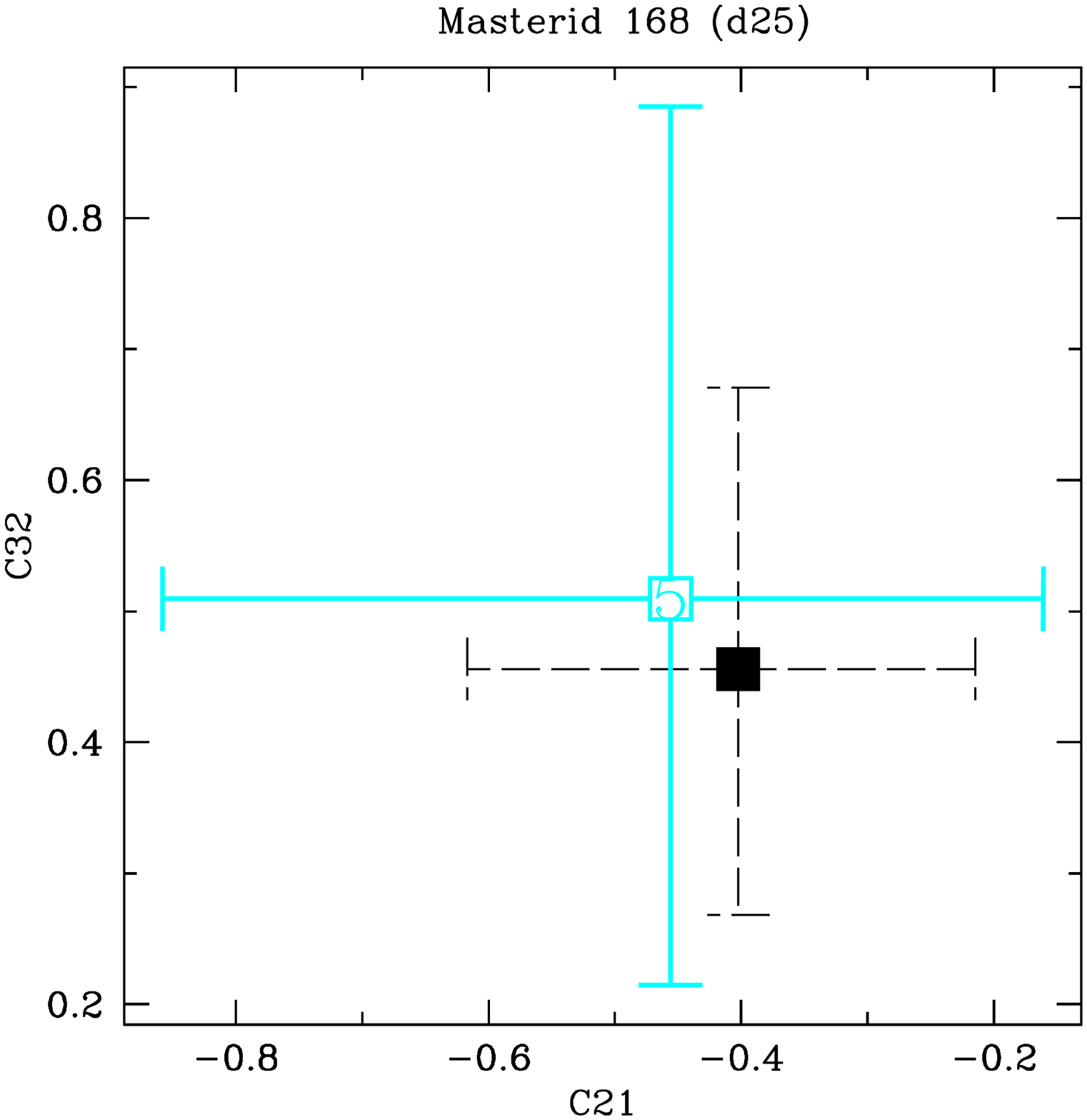}

\end{minipage}
\begin{minipage}{0.32\linewidth}
  \centering

    \includegraphics[width=\linewidth]{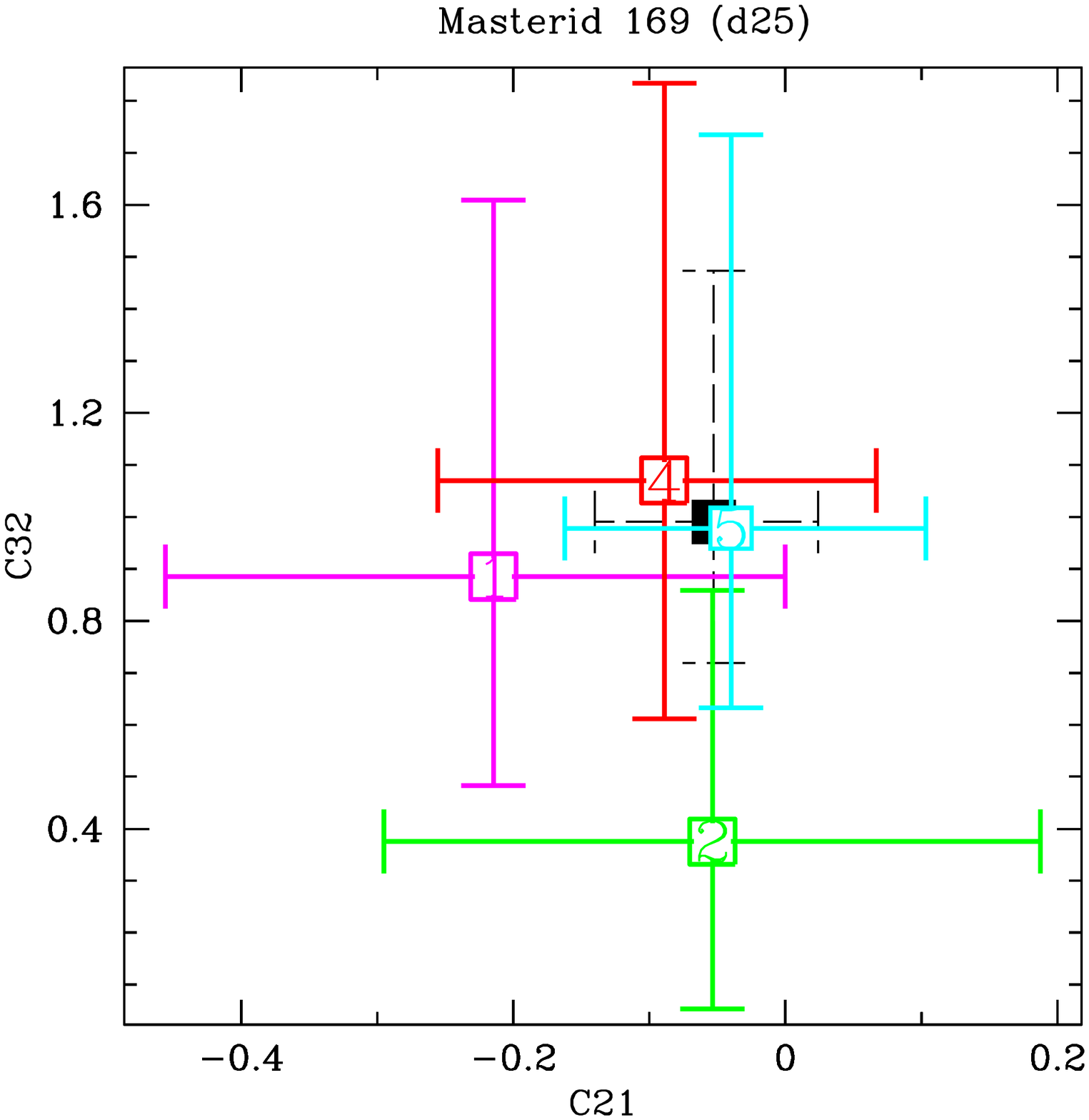}

 \end{minipage}

  \begin{minipage}{0.32\linewidth}
  \centering
  
    \includegraphics[width=\linewidth]{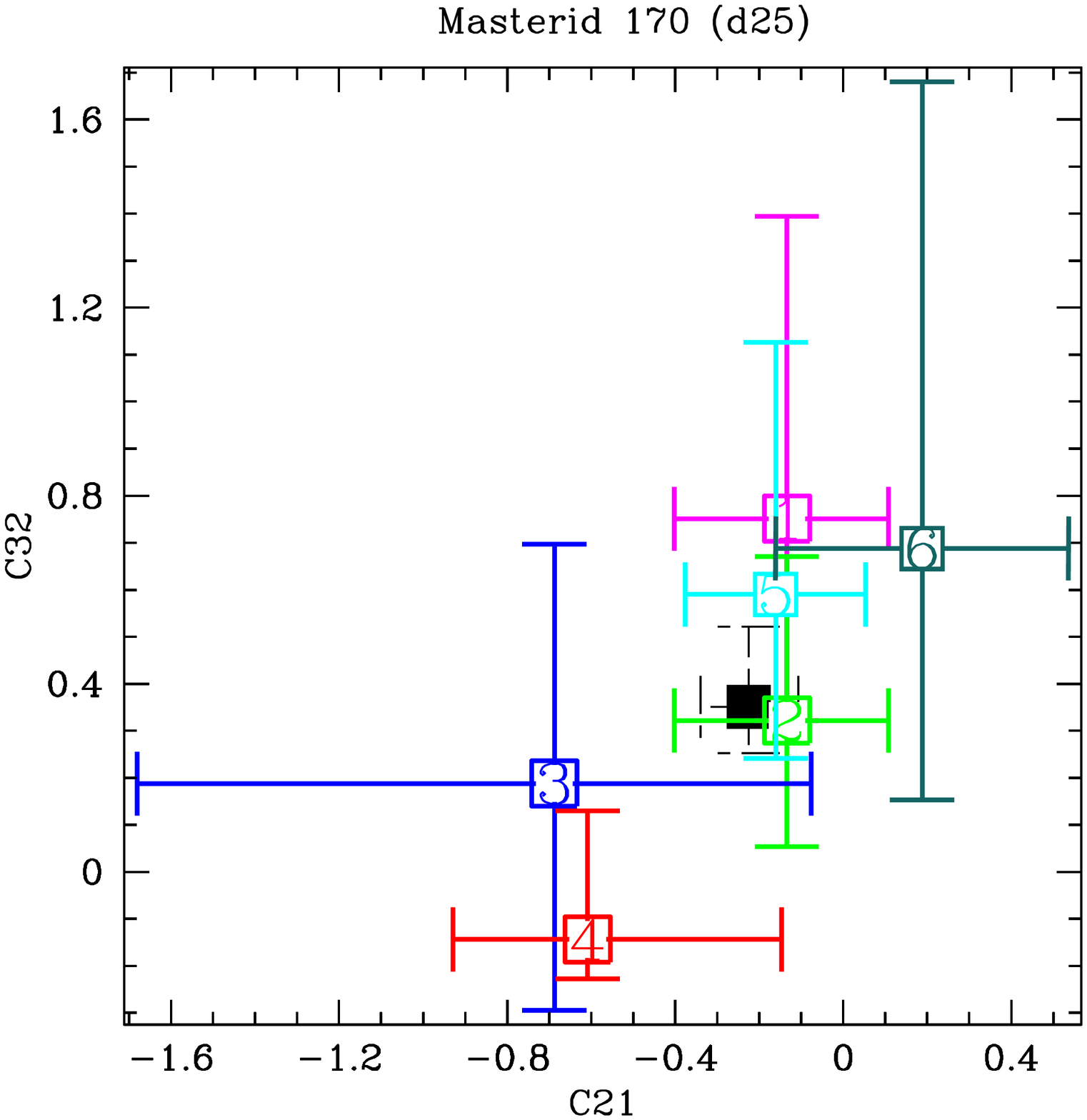}

  \end{minipage}
  \begin{minipage}{0.32\linewidth}
  \centering

    \includegraphics[width=\linewidth]{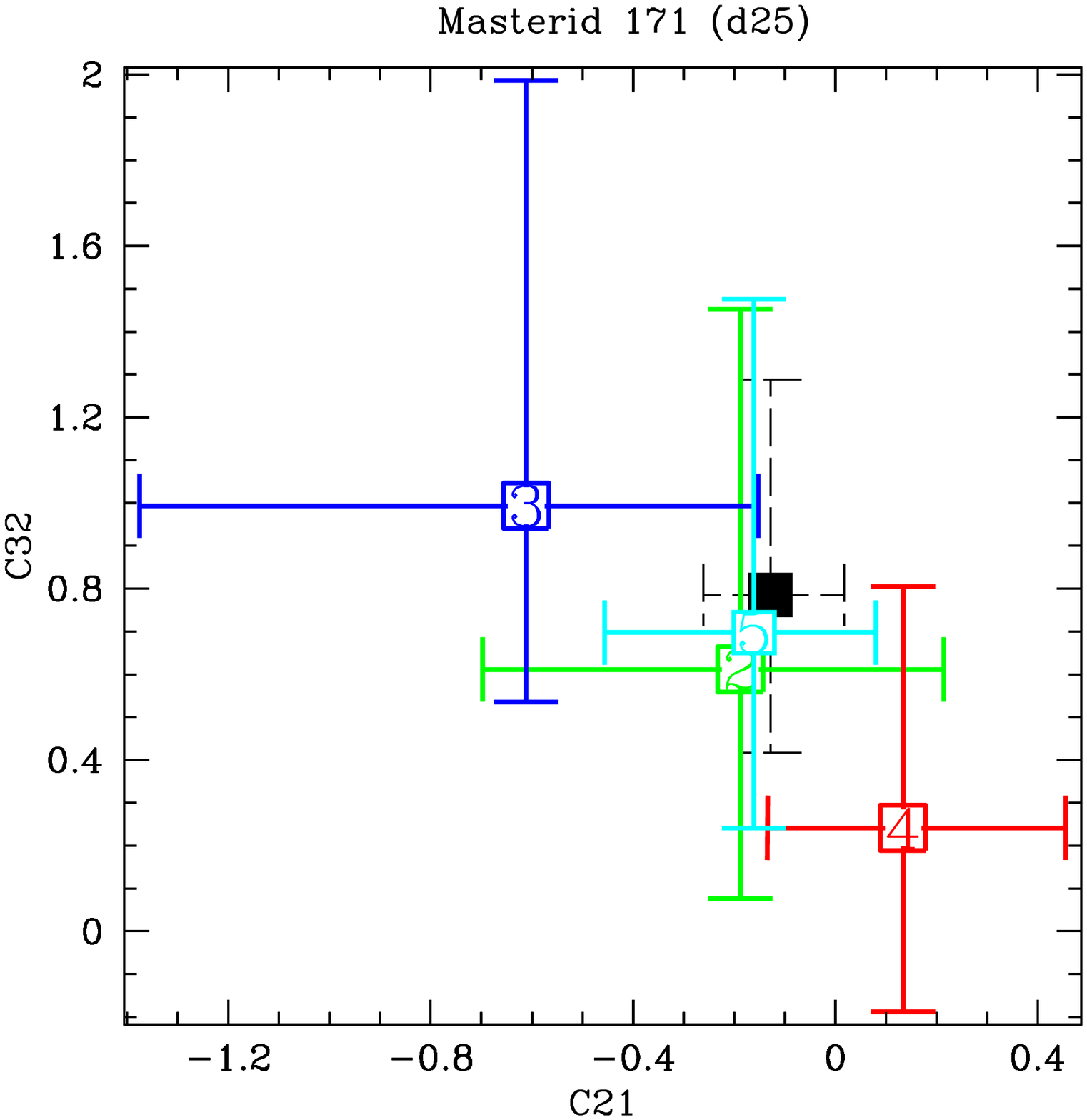}

\end{minipage}
\begin{minipage}{0.32\linewidth}
  \centering

    \includegraphics[width=\linewidth]{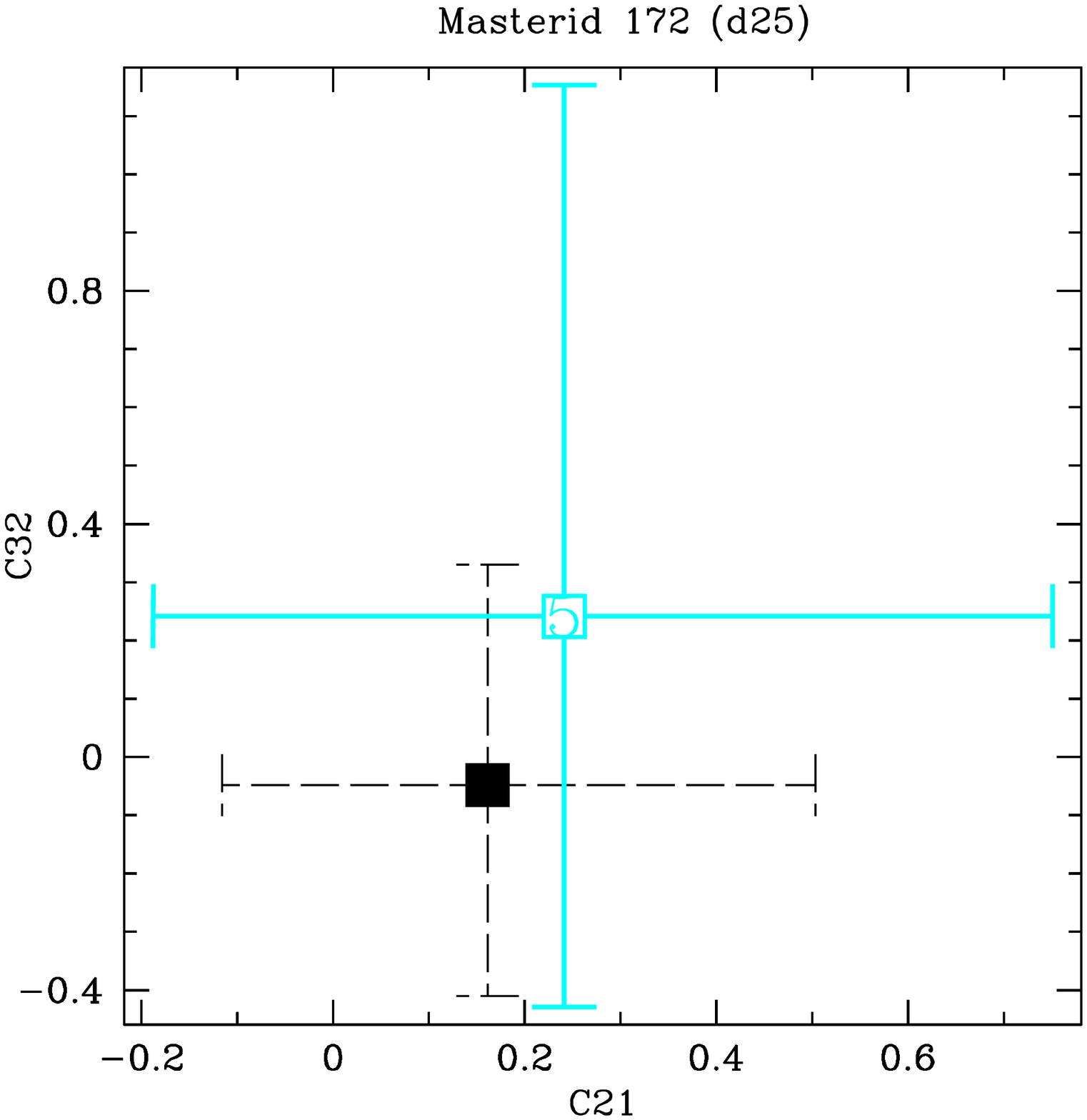}

\end{minipage}

\begin{minipage}{0.32\linewidth}
  \centering
  
    \includegraphics[width=\linewidth]{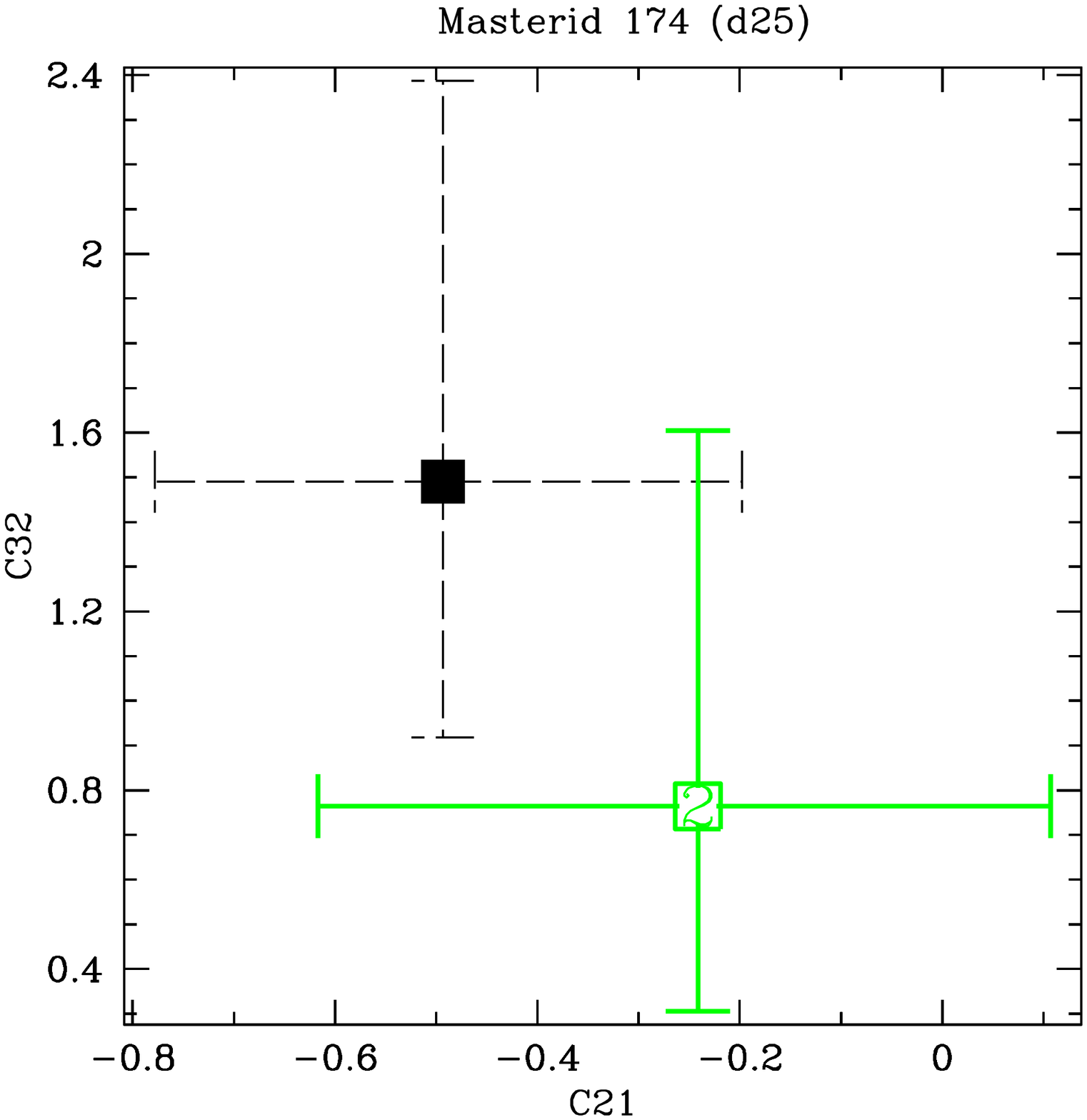}

  \end{minipage}
  \begin{minipage}{0.32\linewidth}
  \centering

    \includegraphics[width=\linewidth]{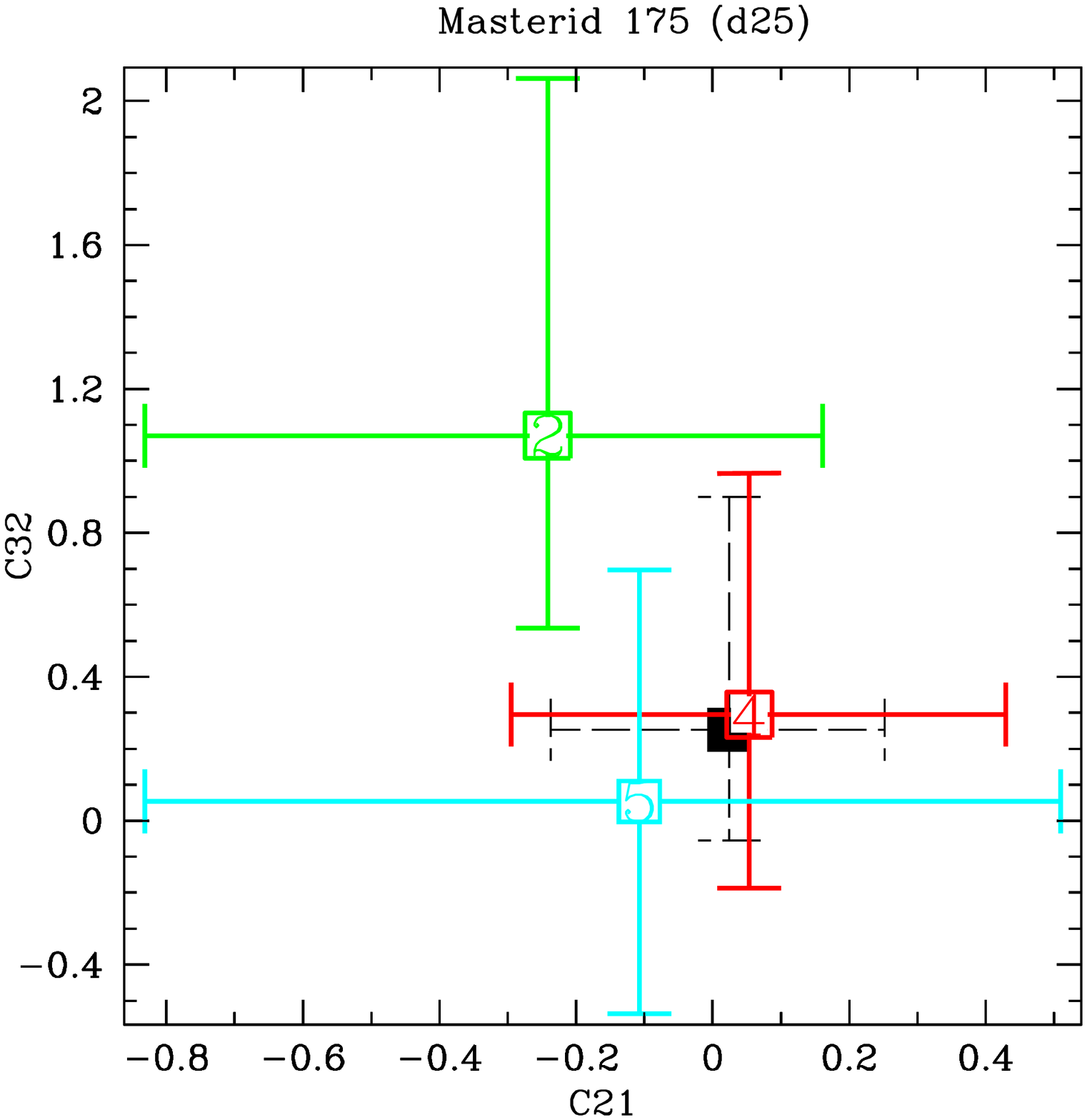}

\end{minipage}
\begin{minipage}{0.32\linewidth}
  \centering

    \includegraphics[width=\linewidth]{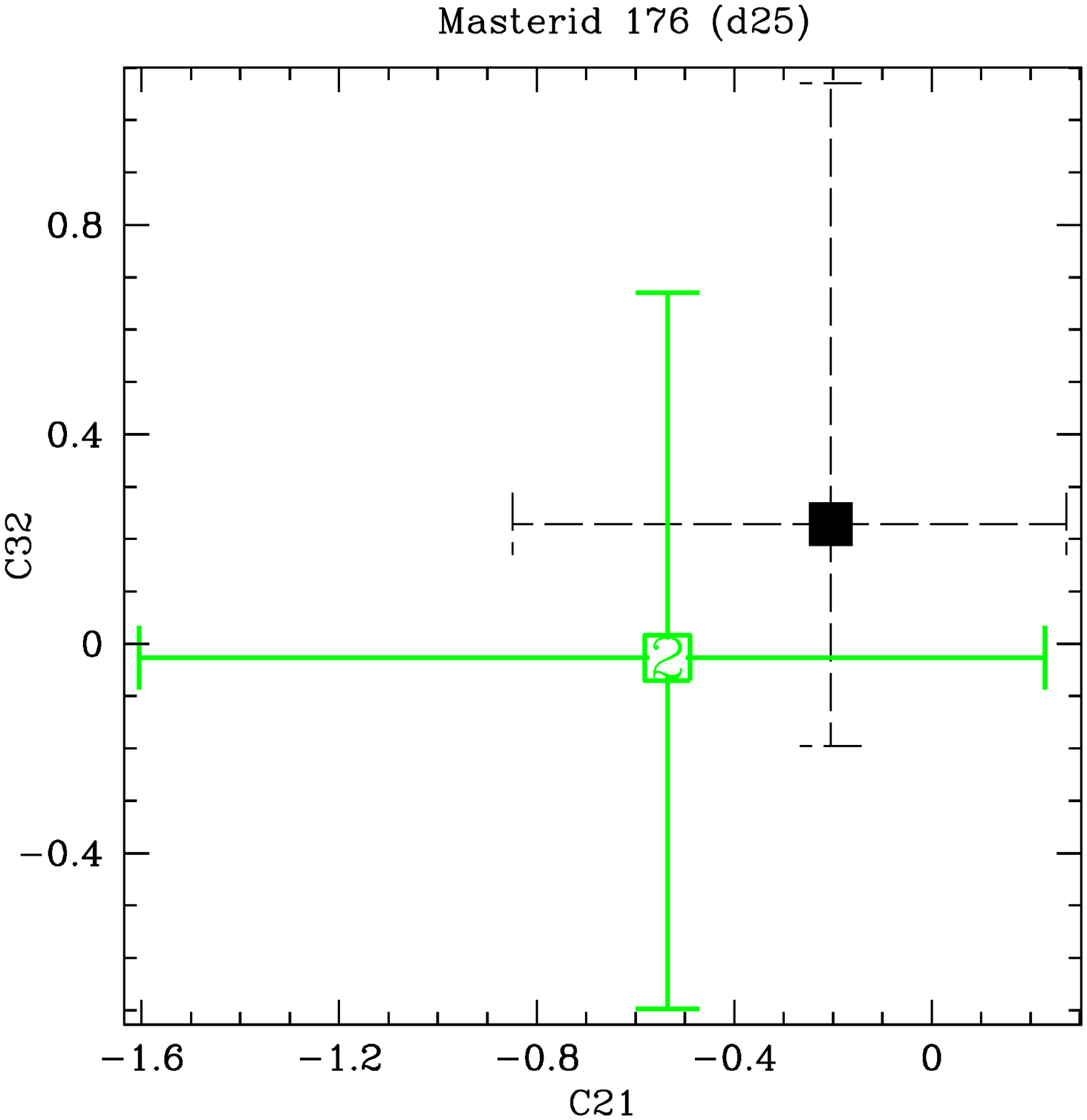}

\end{minipage}
\end{figure}

\begin{figure}
  \begin{minipage}{0.32\linewidth}
  \centering
  
    \includegraphics[width=\linewidth]{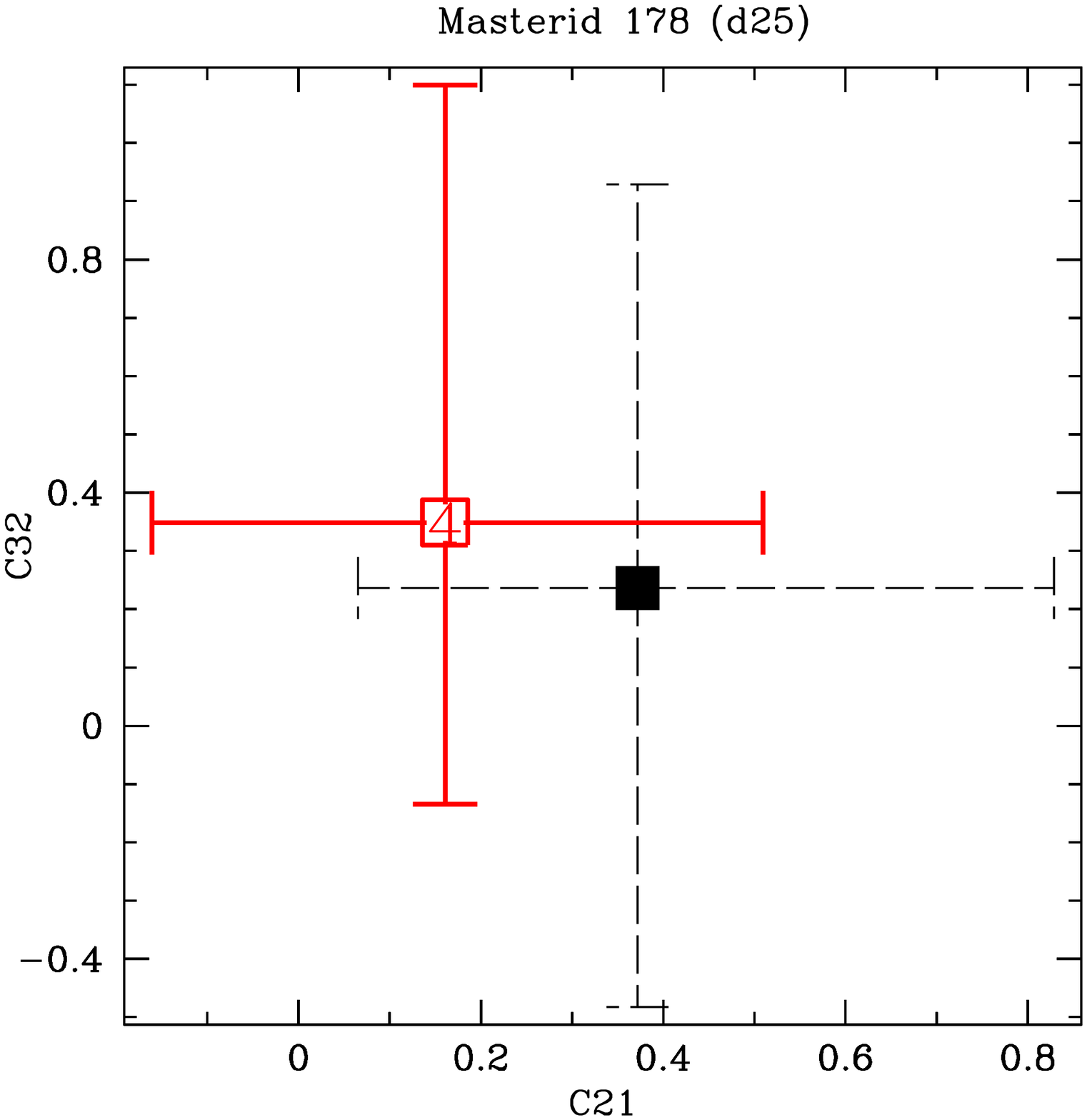}

  \end{minipage}
  \begin{minipage}{0.32\linewidth}
  \centering

    \includegraphics[width=\linewidth]{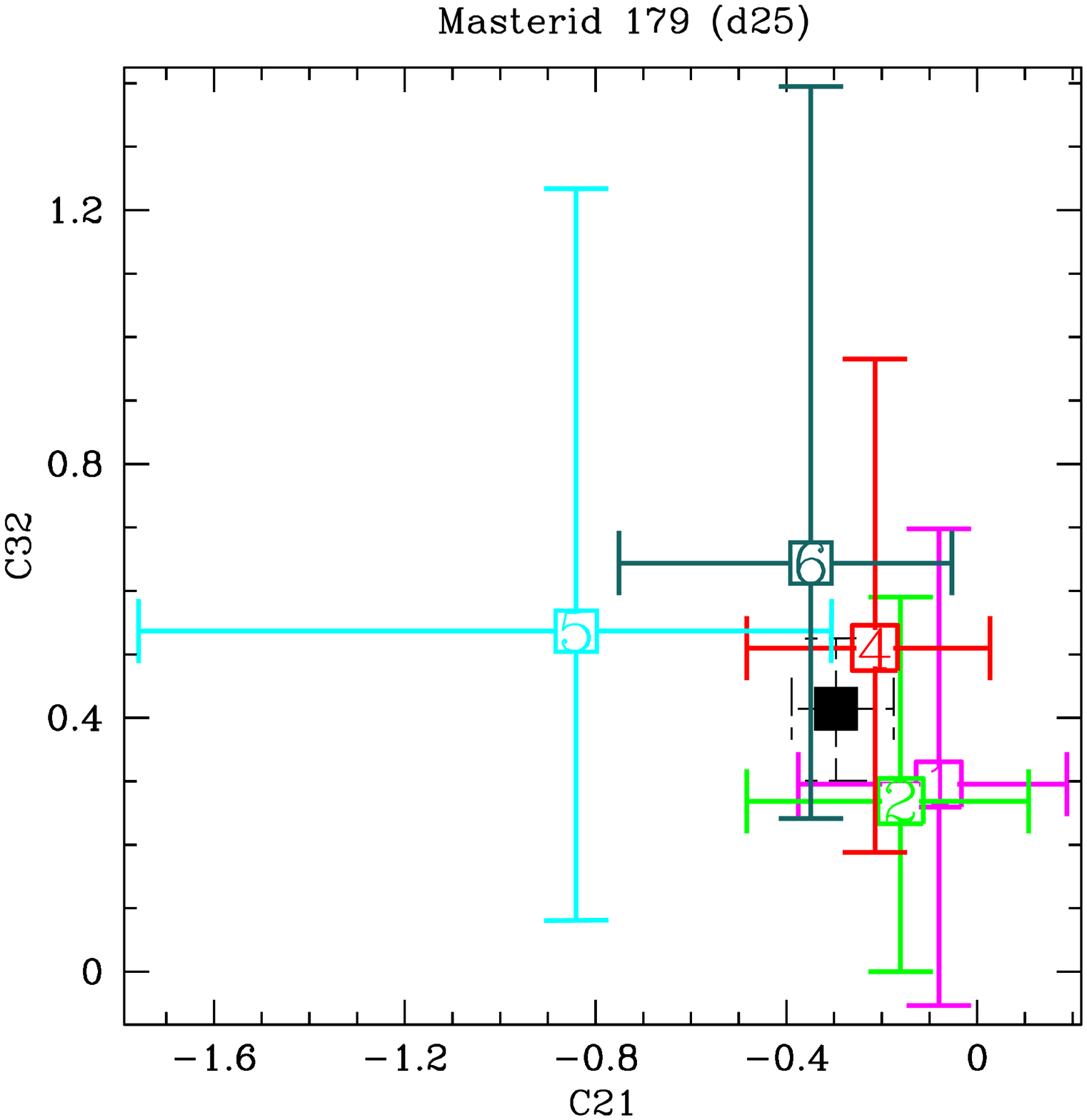}

\end{minipage}
\begin{minipage}{0.32\linewidth}
  \centering

    \includegraphics[width=\linewidth]{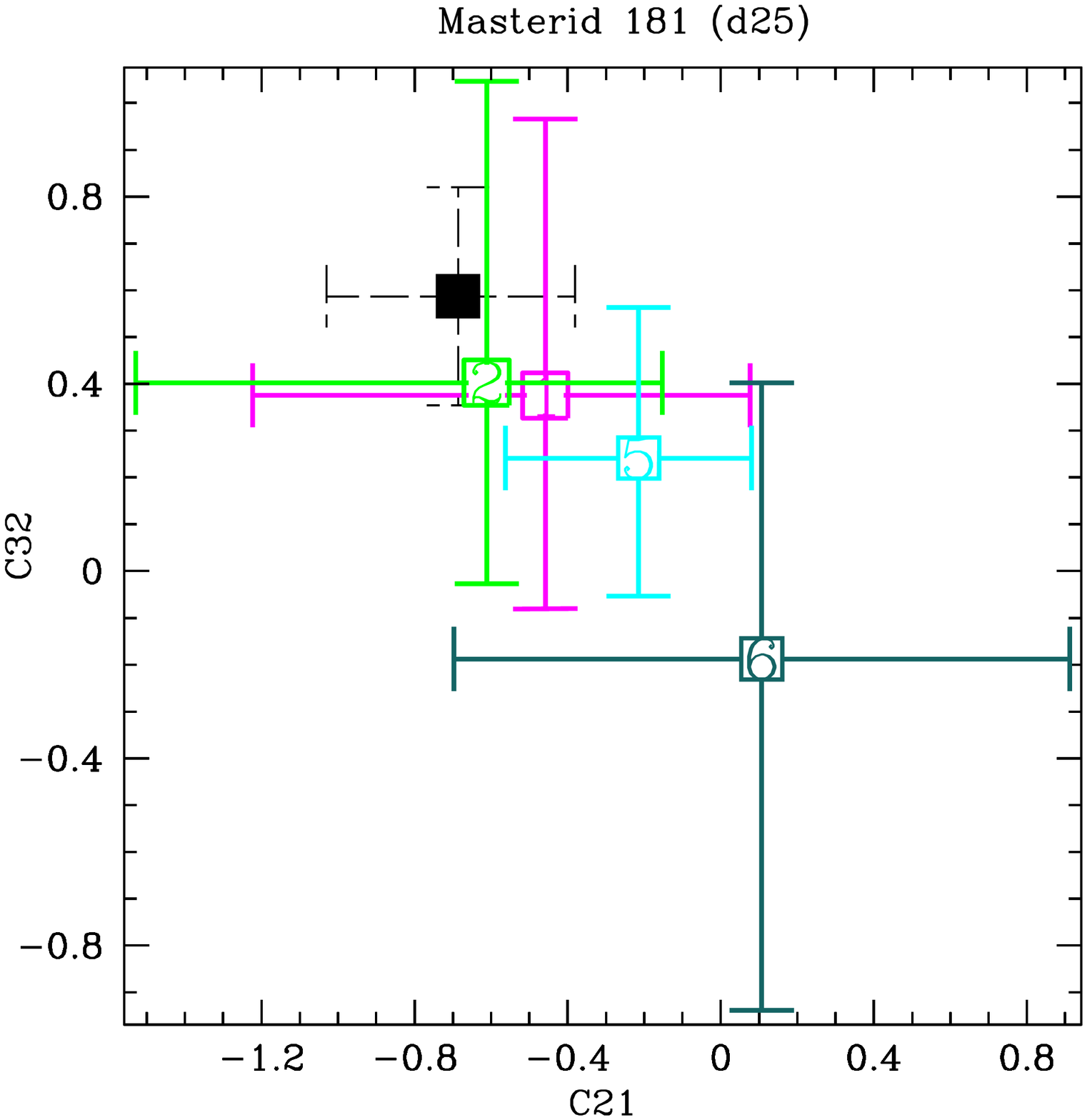}

 \end{minipage}

\begin{minipage}{0.32\linewidth}
  \centering
  
    \includegraphics[width=\linewidth]{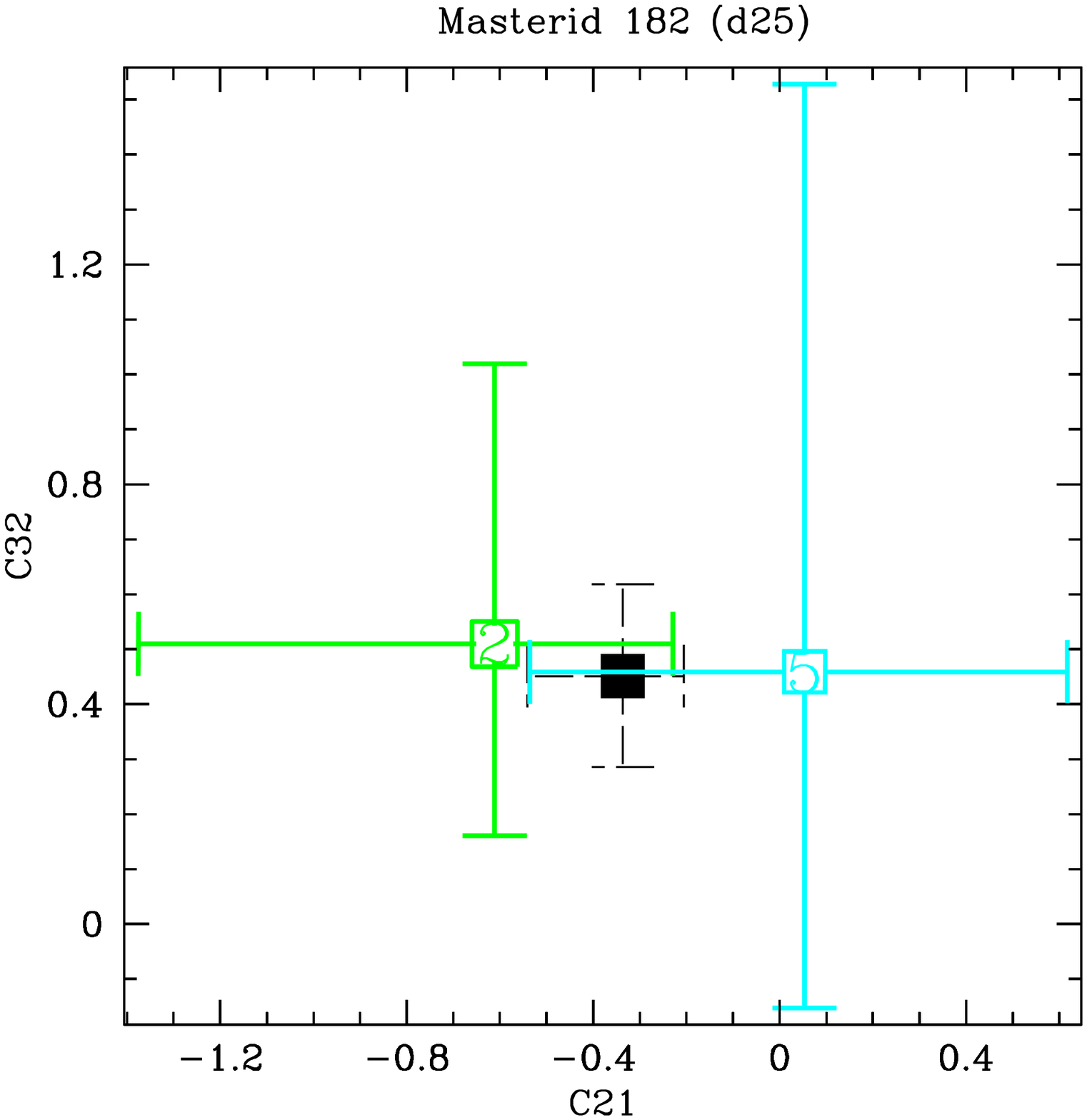}

  \end{minipage}
  \begin{minipage}{0.32\linewidth}
  \centering

    \includegraphics[width=\linewidth]{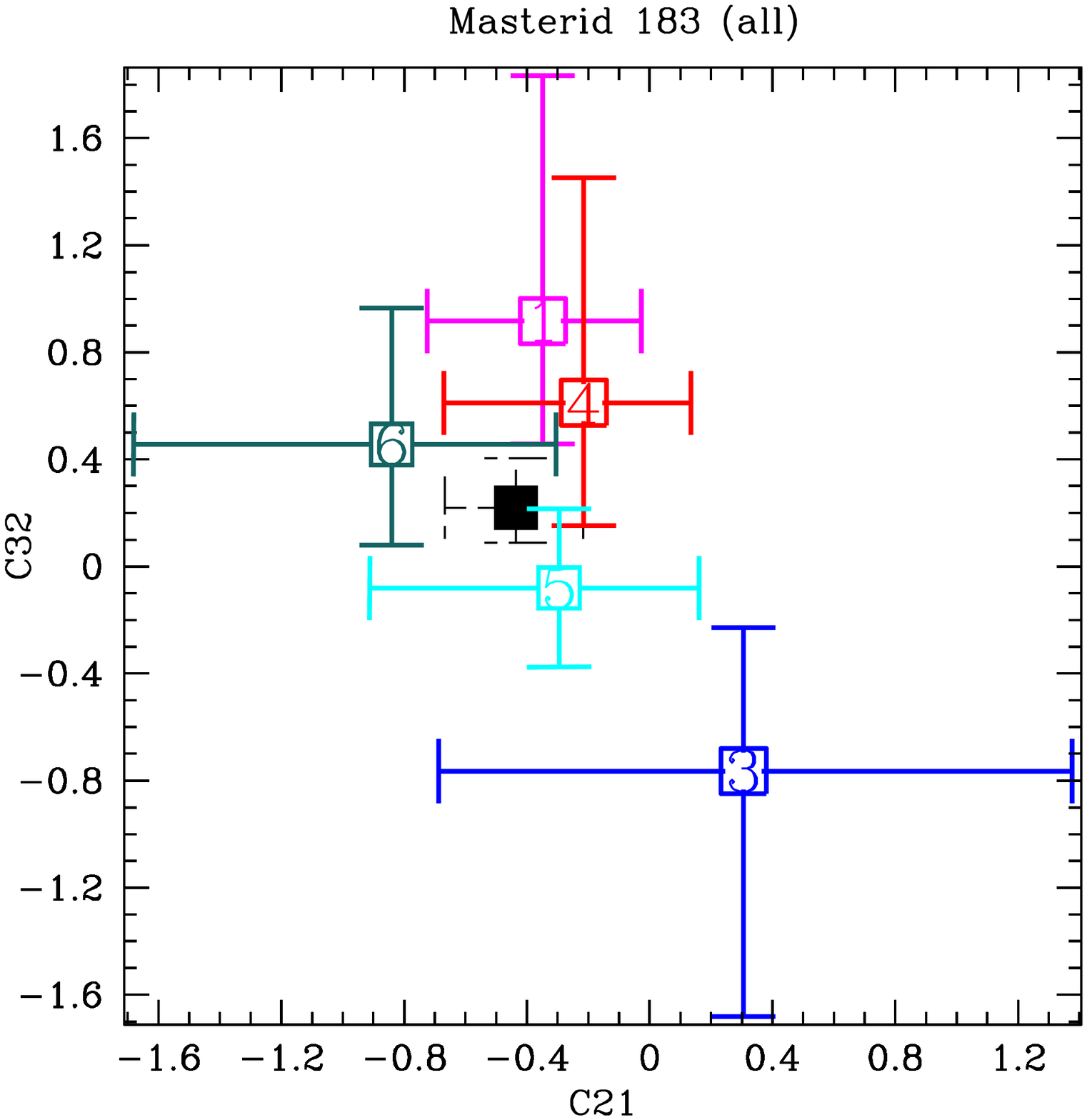}

\end{minipage}
\begin{minipage}{0.32\linewidth}
  \centering

    \includegraphics[width=\linewidth]{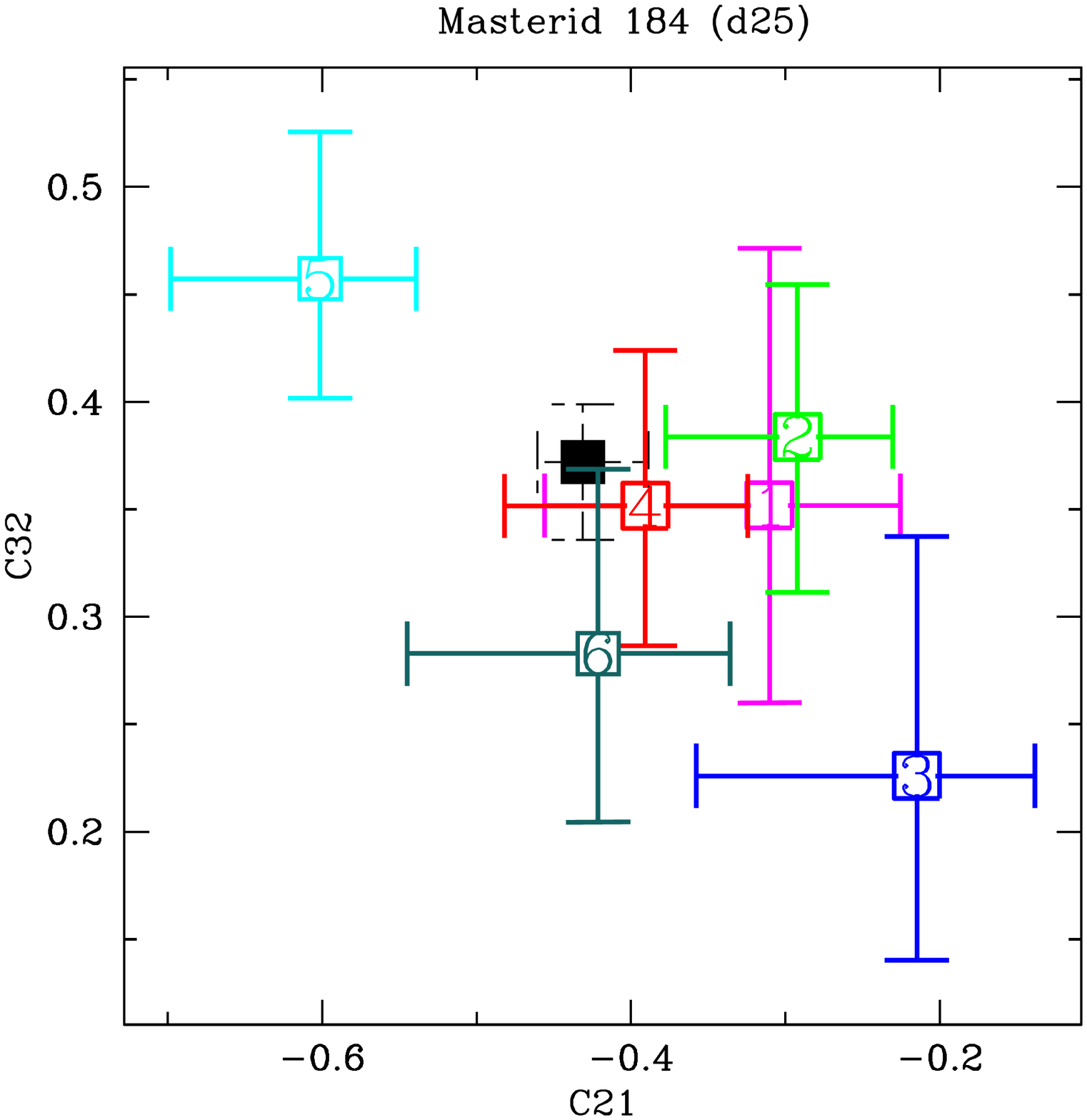}

 \end{minipage}

  \begin{minipage}{0.32\linewidth}
  \centering
  
    \includegraphics[width=\linewidth]{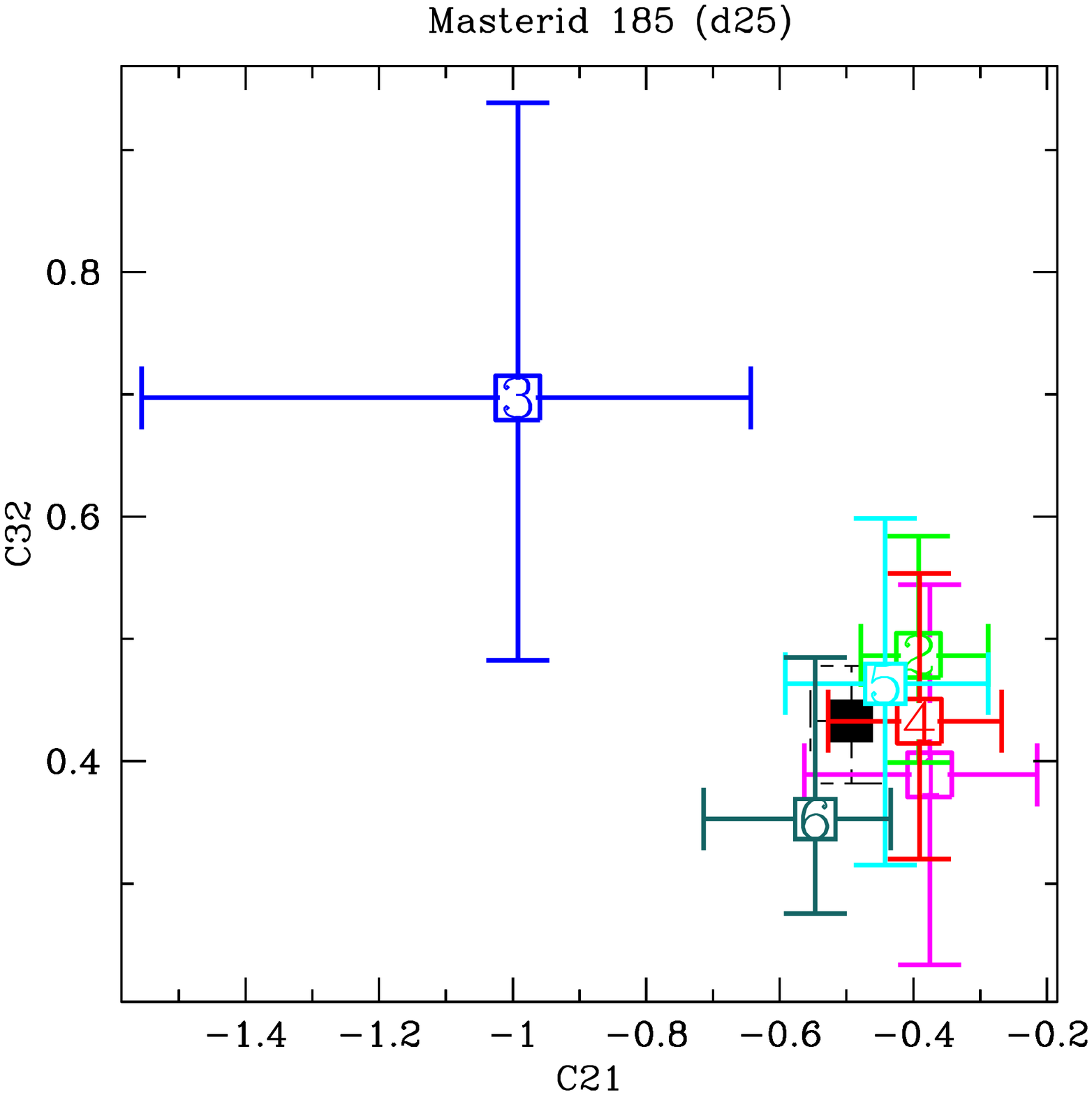}

  \end{minipage}
  \begin{minipage}{0.32\linewidth}
  \centering

    \includegraphics[width=\linewidth]{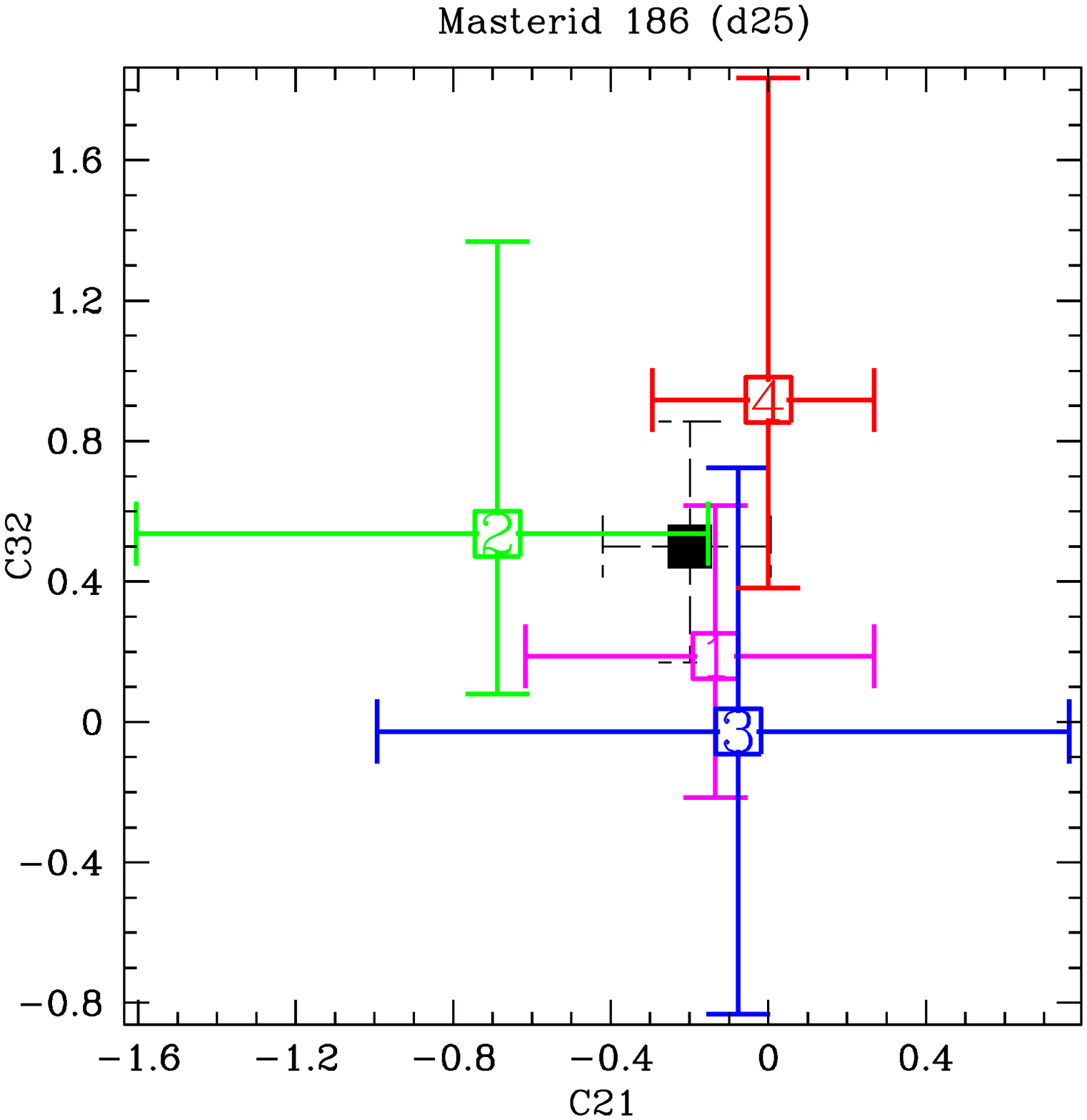}

\end{minipage}
\begin{minipage}{0.32\linewidth}
  \centering

    \includegraphics[width=\linewidth]{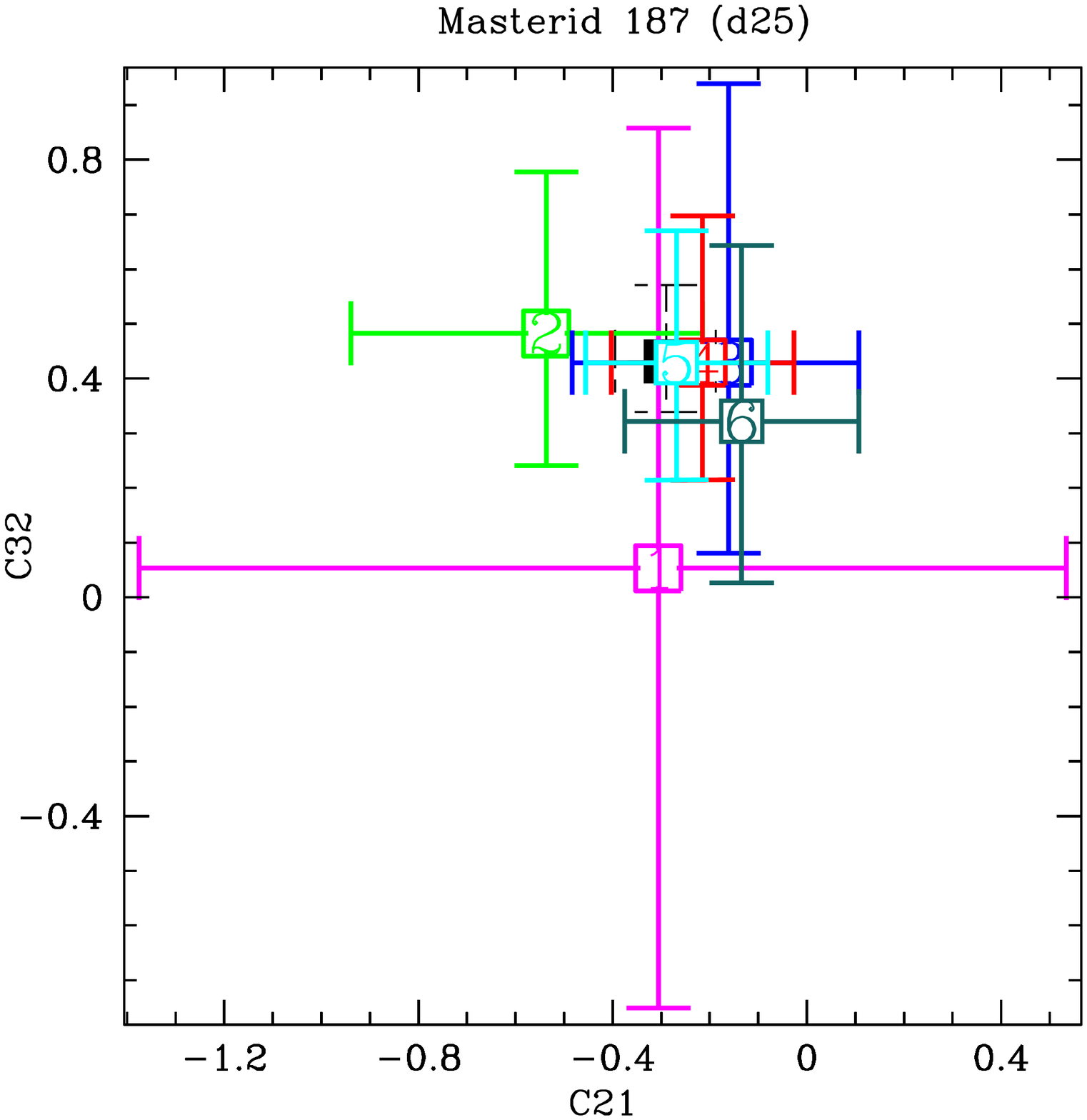}

\end{minipage}

\begin{minipage}{0.32\linewidth}
  \centering
  
    \includegraphics[width=\linewidth]{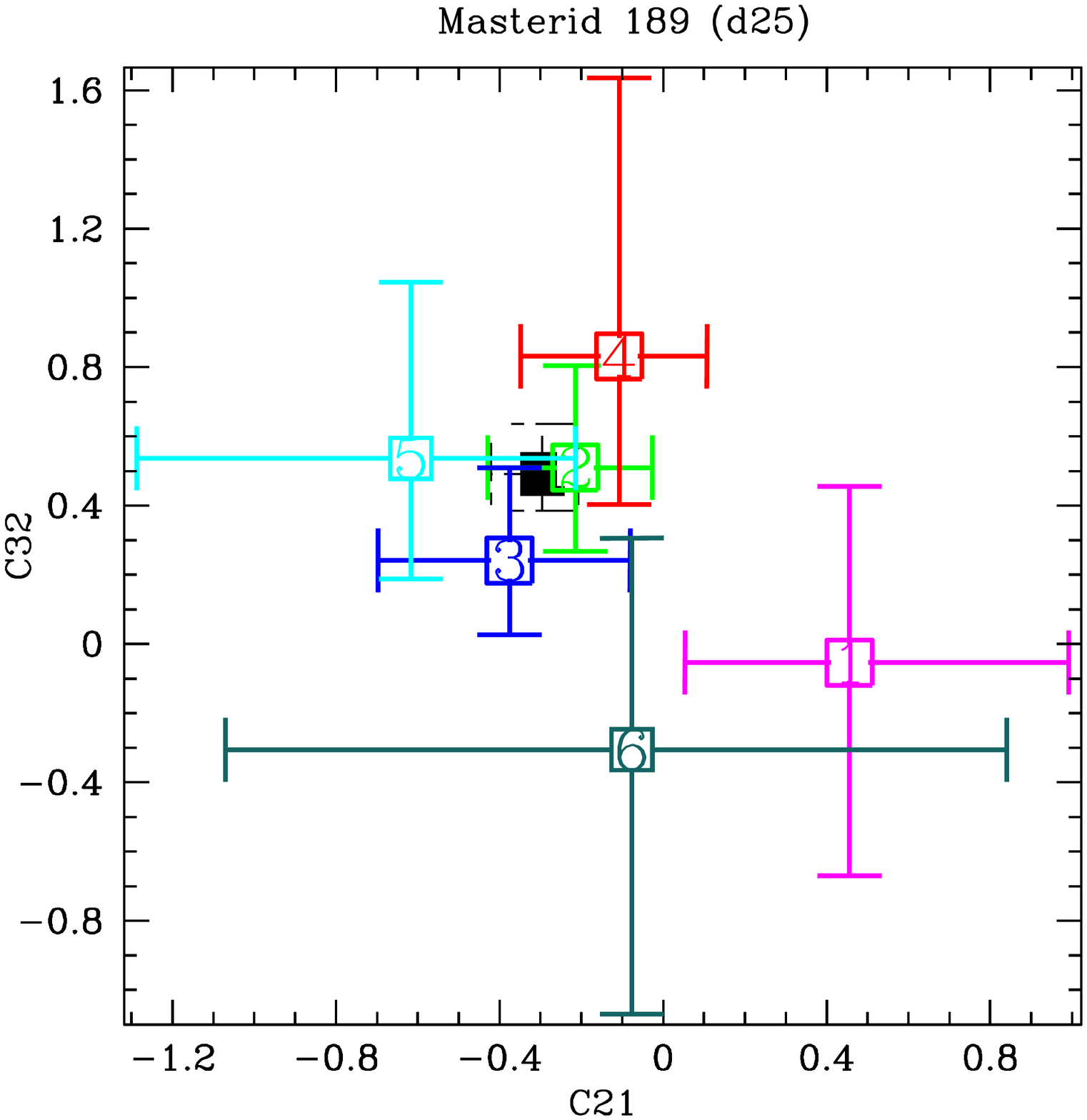}

  \end{minipage}
  \begin{minipage}{0.32\linewidth}
  \centering

    \includegraphics[width=\linewidth]{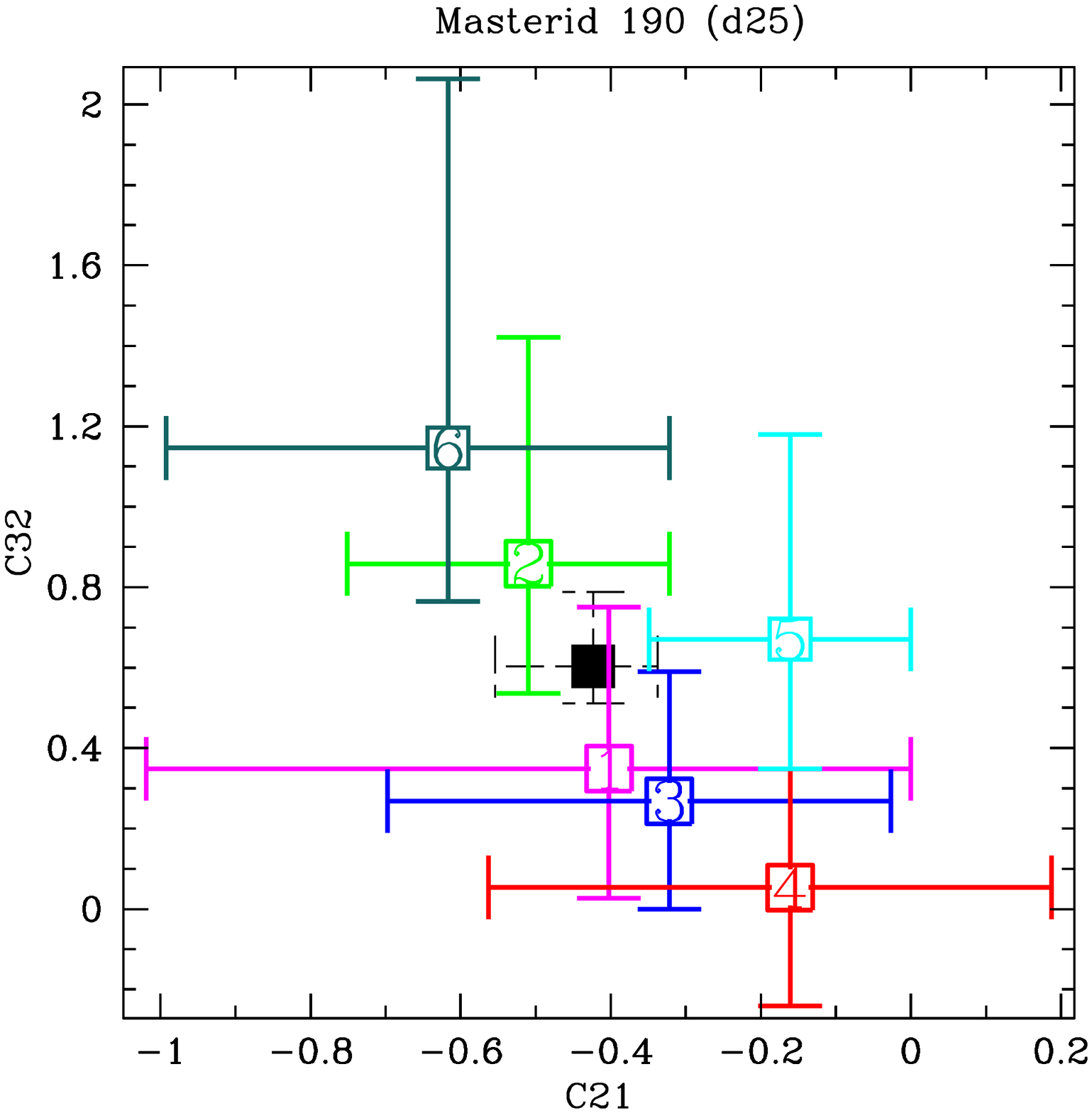}

\end{minipage}
\begin{minipage}{0.32\linewidth}
  \centering

    \includegraphics[width=\linewidth]{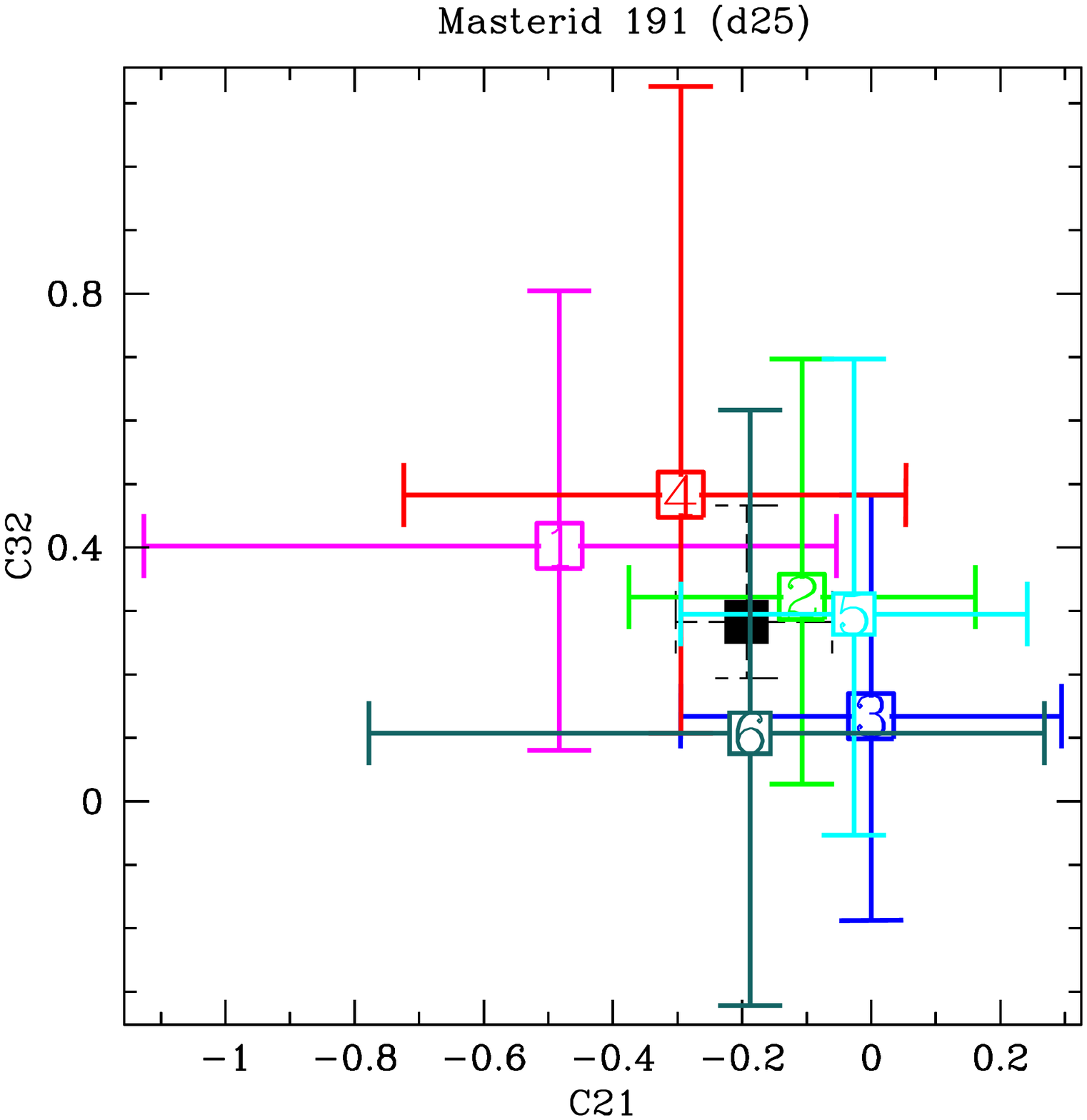}

\end{minipage}
\end{figure}

\begin{figure}
  \begin{minipage}{0.32\linewidth}
  \centering
  
    \includegraphics[width=\linewidth]{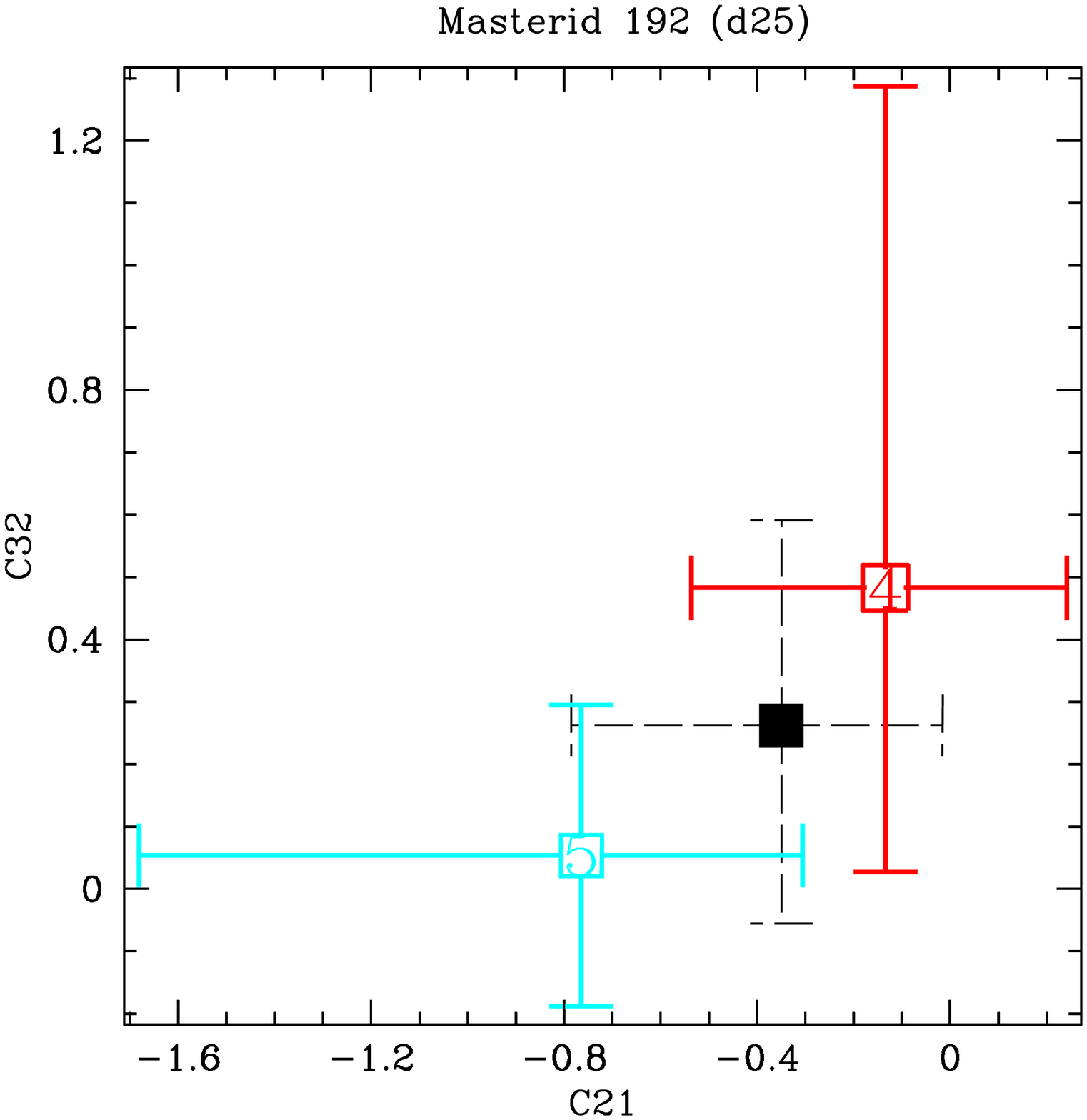}

  \end{minipage}
  \begin{minipage}{0.32\linewidth}
  \centering

    \includegraphics[width=\linewidth]{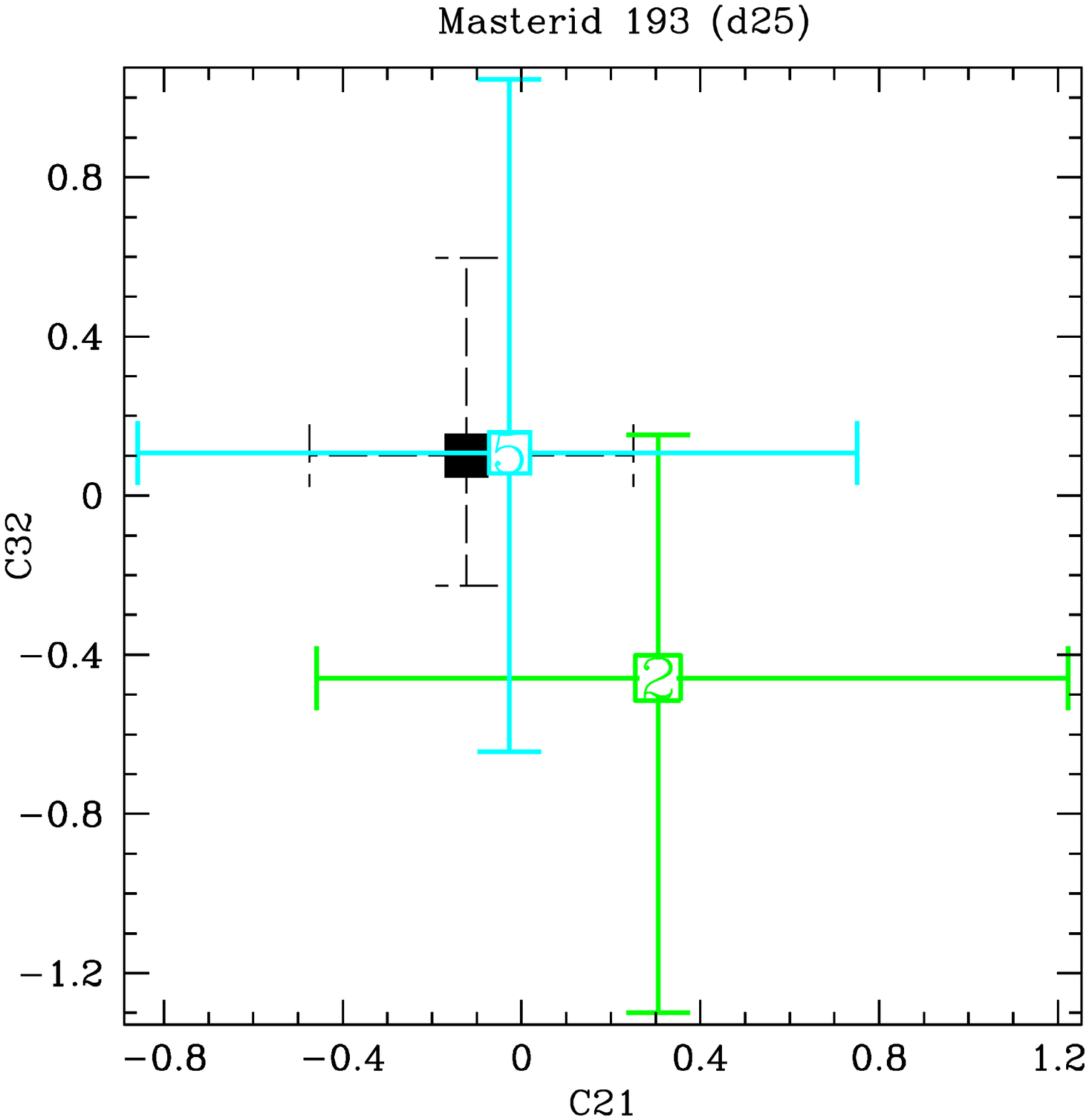}

\end{minipage}
\begin{minipage}{0.32\linewidth}
  \centering

    \includegraphics[width=\linewidth]{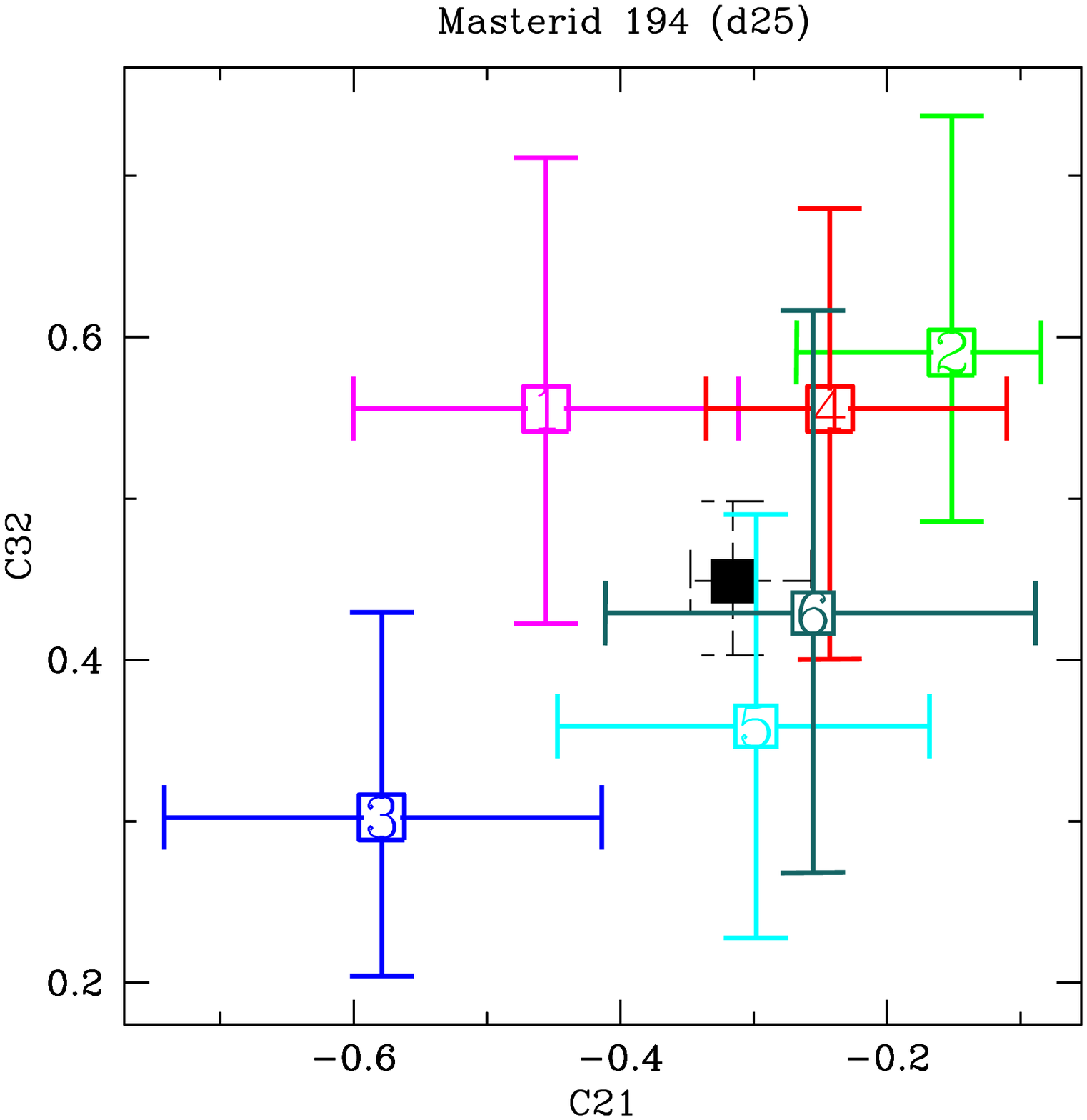}

 \end{minipage}

\begin{minipage}{0.32\linewidth}
  \centering
  
    \includegraphics[width=\linewidth]{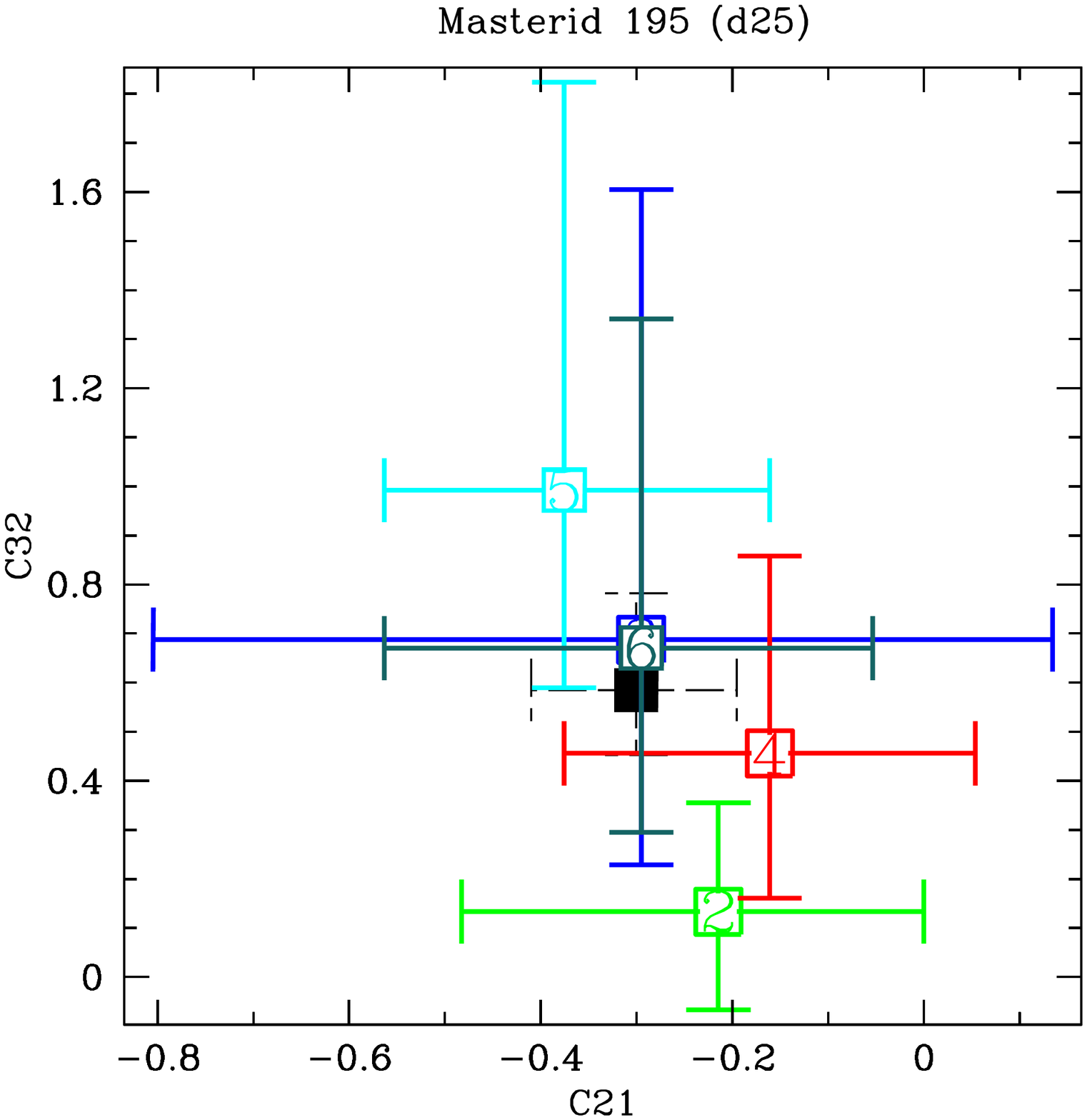}

  \end{minipage}
  \begin{minipage}{0.32\linewidth}
  \centering

    \includegraphics[width=\linewidth]{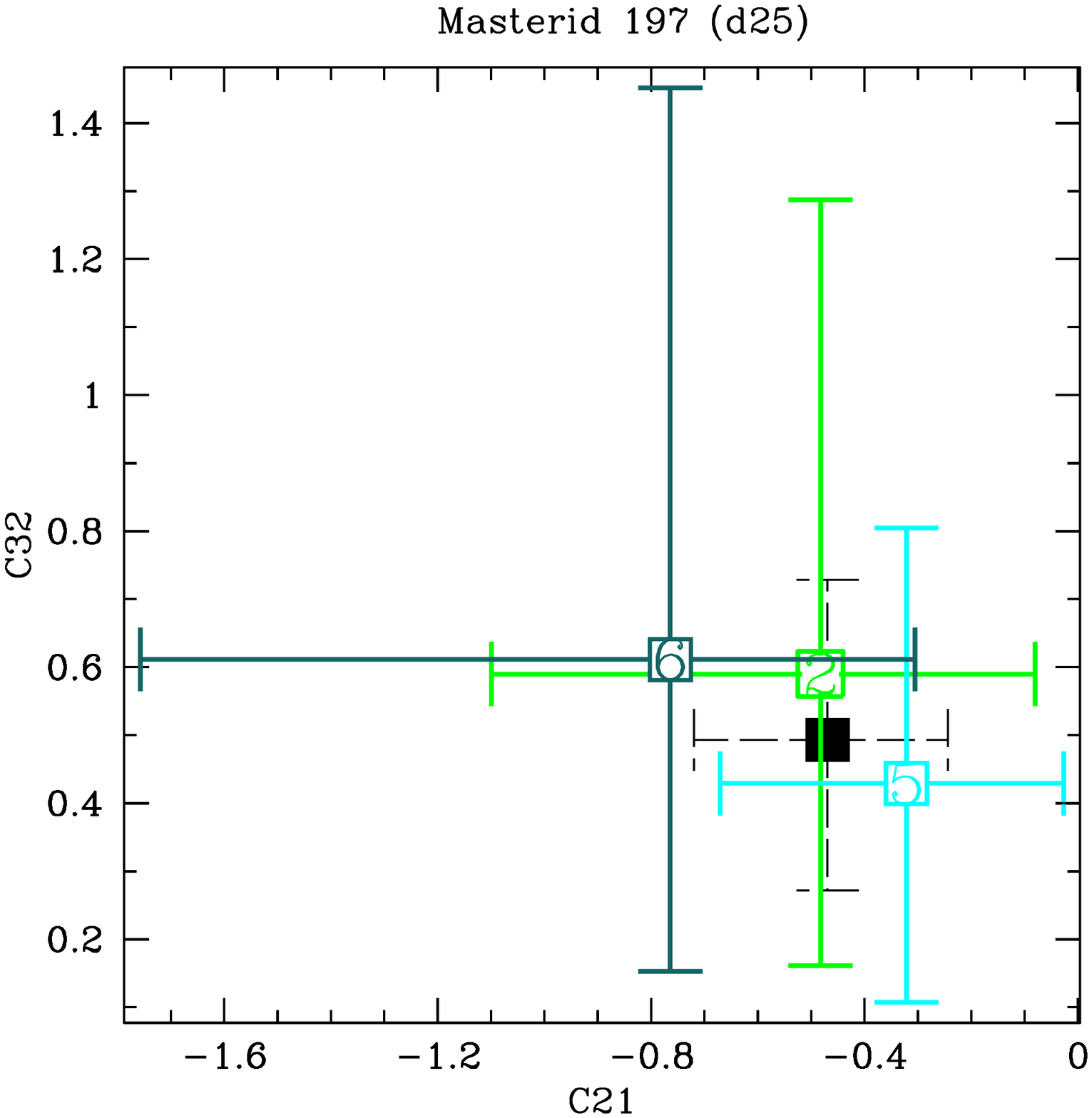}

\end{minipage}
\begin{minipage}{0.32\linewidth}
  \centering

    \includegraphics[width=\linewidth]{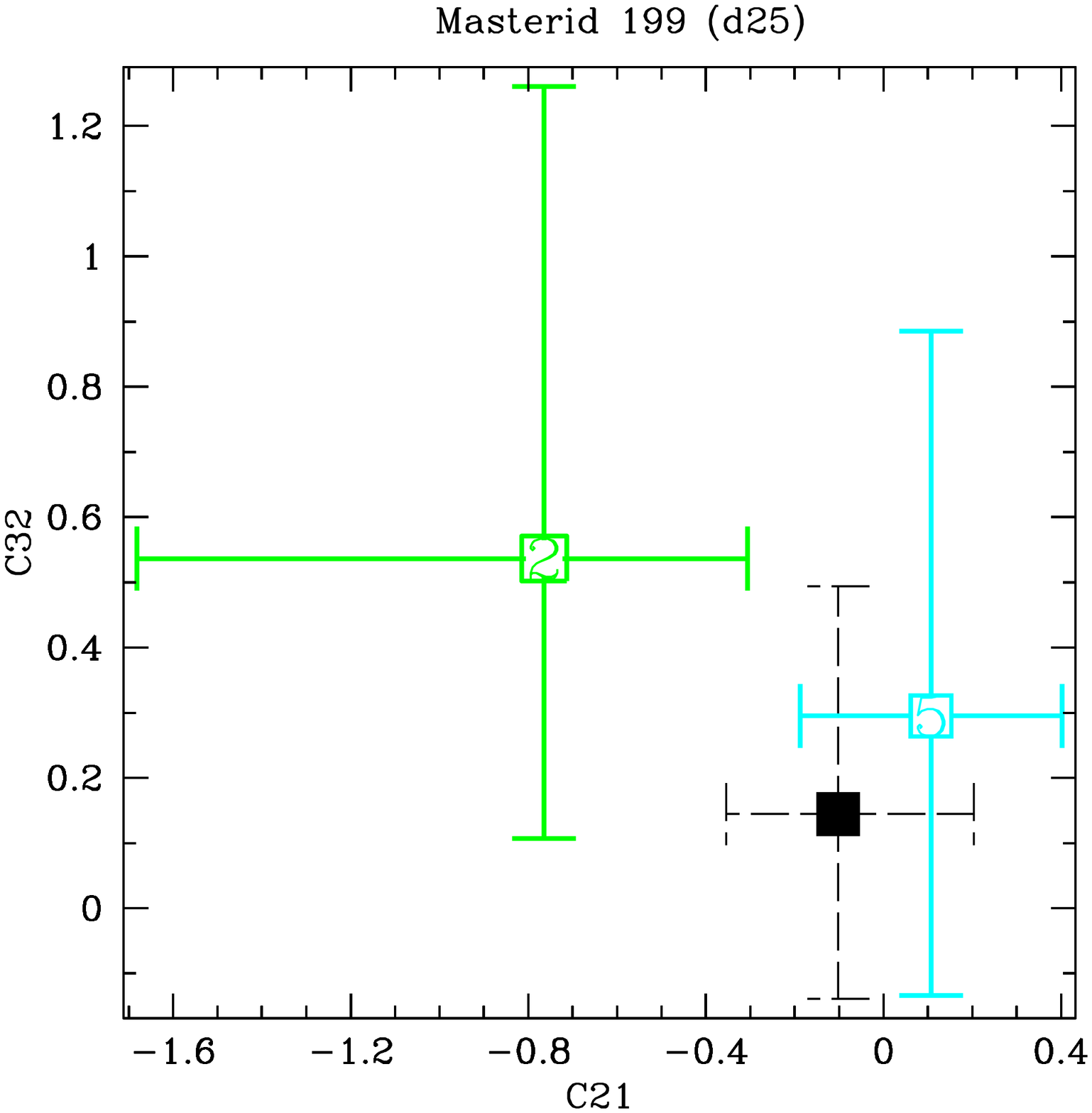}

 \end{minipage}

  \begin{minipage}{0.32\linewidth}
  \centering
  
    \includegraphics[width=\linewidth]{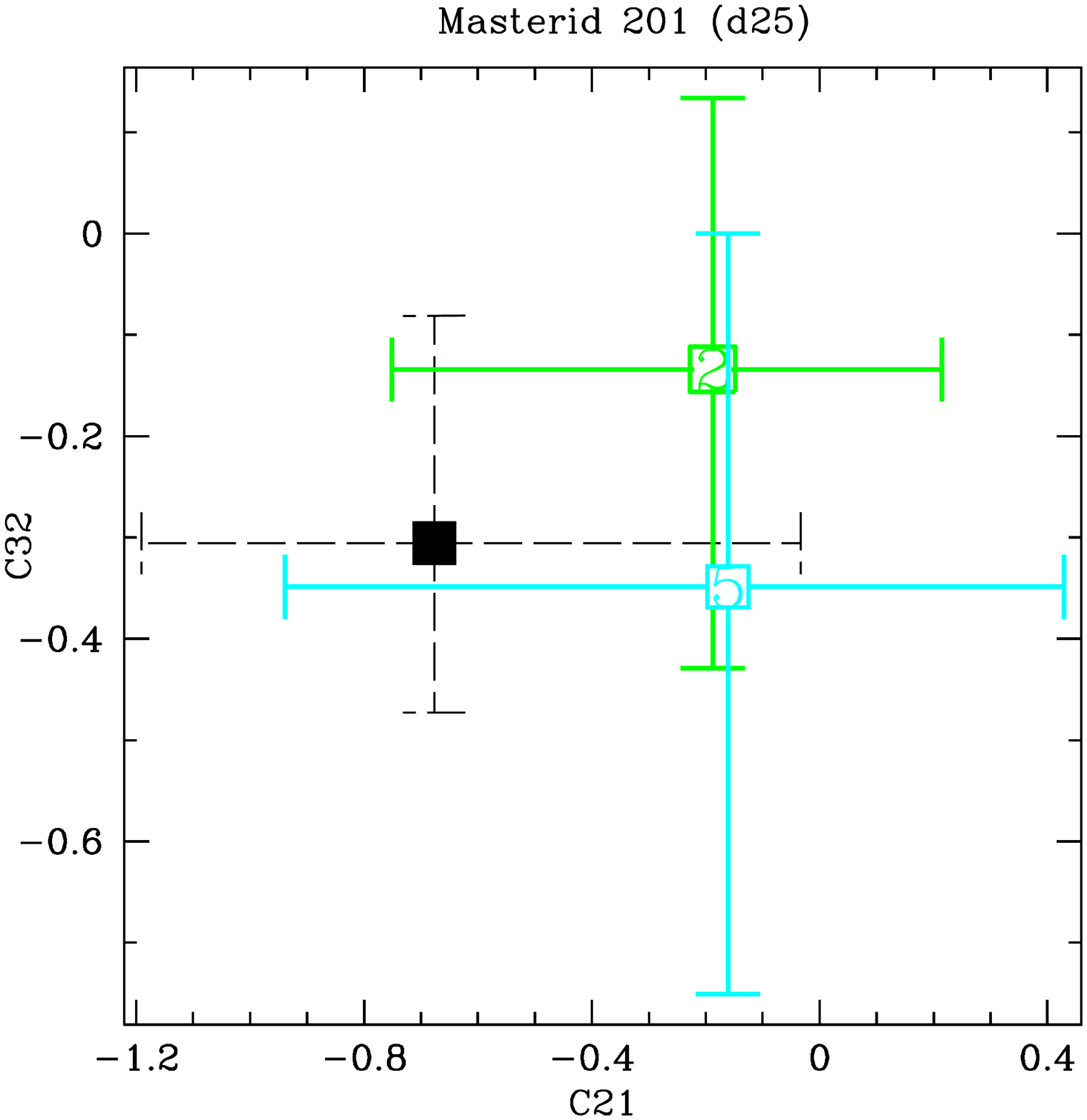}

  \end{minipage}
  \begin{minipage}{0.32\linewidth}
  \centering

    \includegraphics[width=\linewidth]{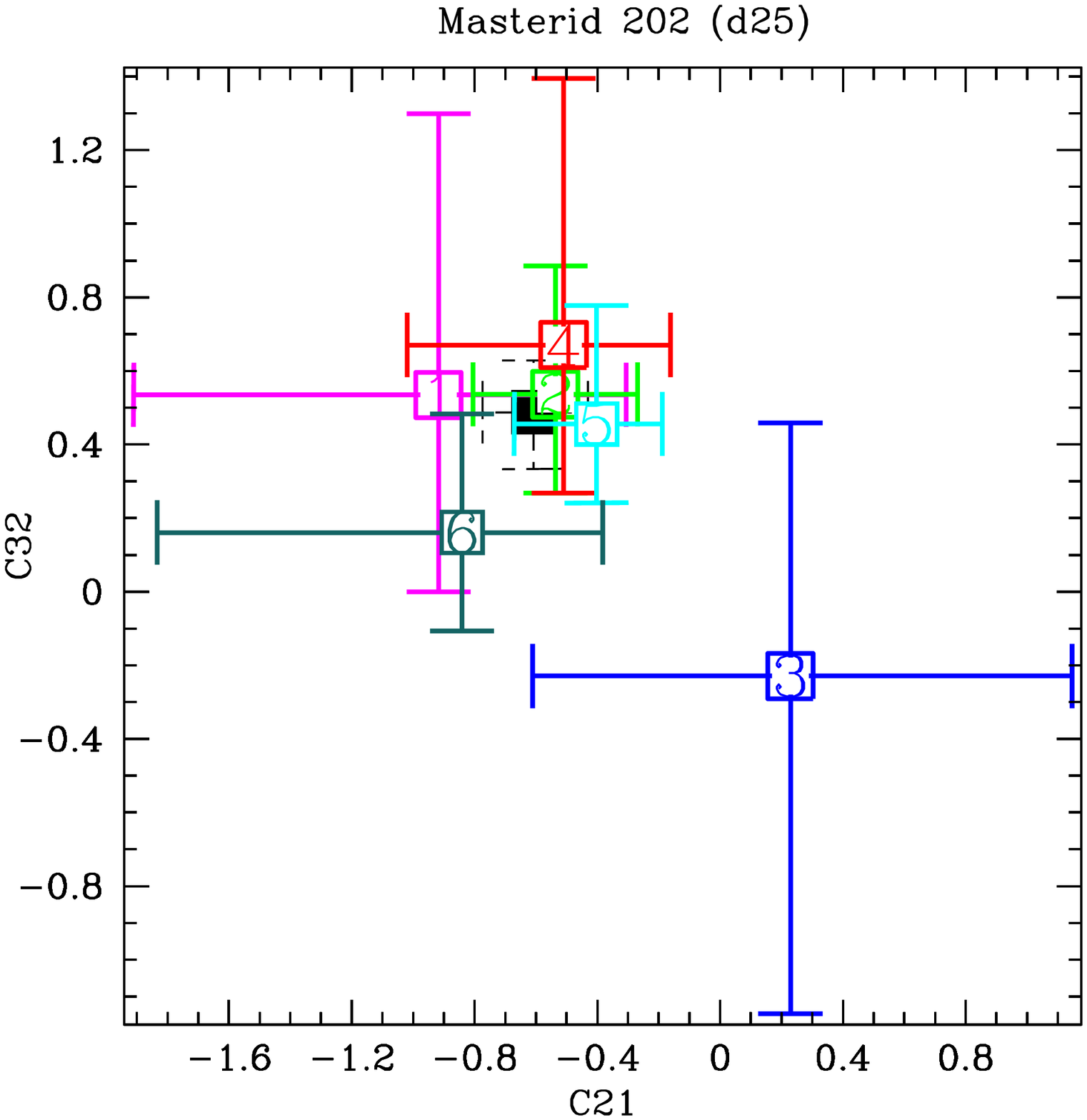}

\end{minipage}
\begin{minipage}{0.32\linewidth}
  \centering

    \includegraphics[width=\linewidth]{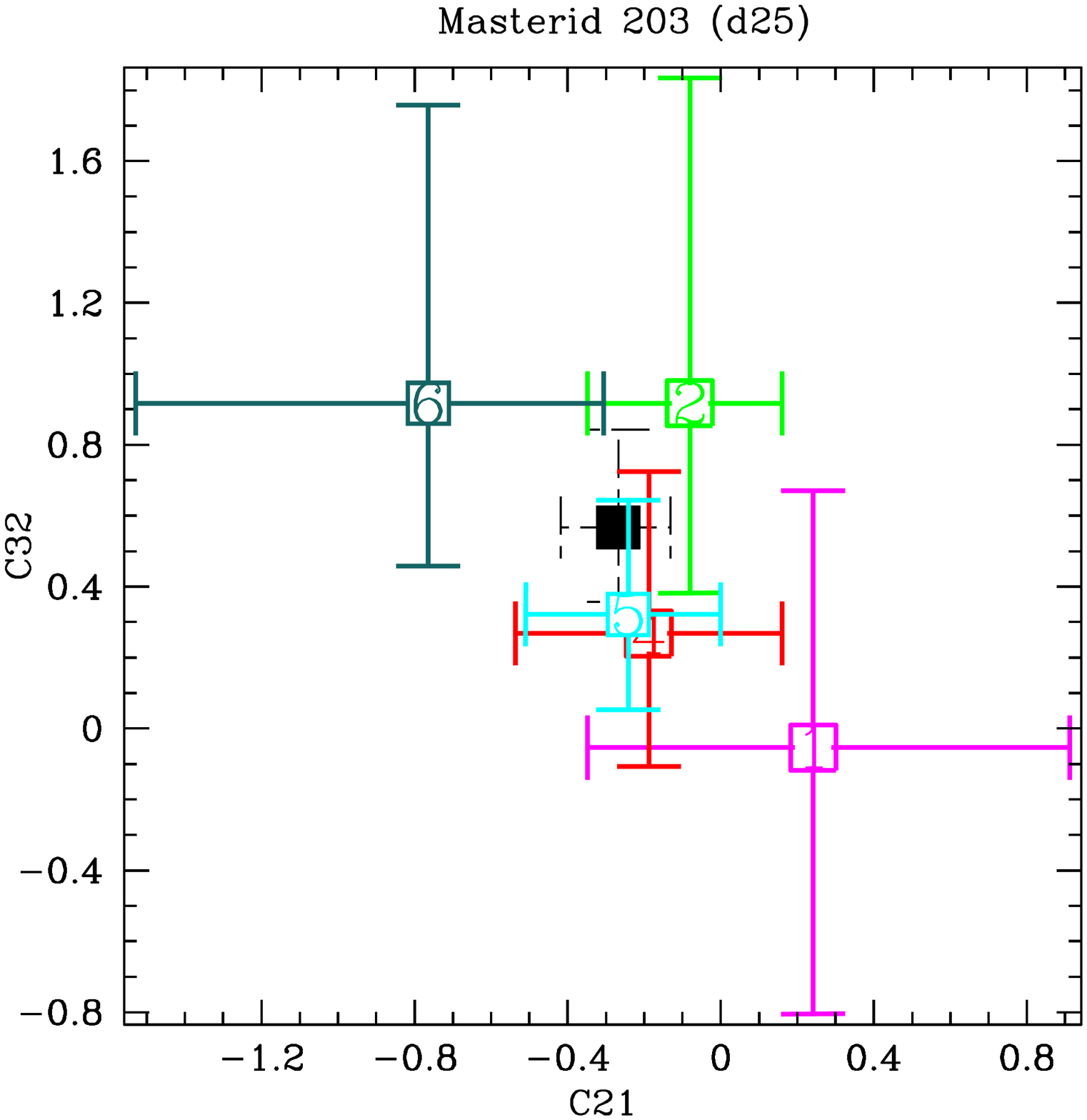}

\end{minipage}

\begin{minipage}{0.32\linewidth}
  \centering
  
    \includegraphics[width=\linewidth]{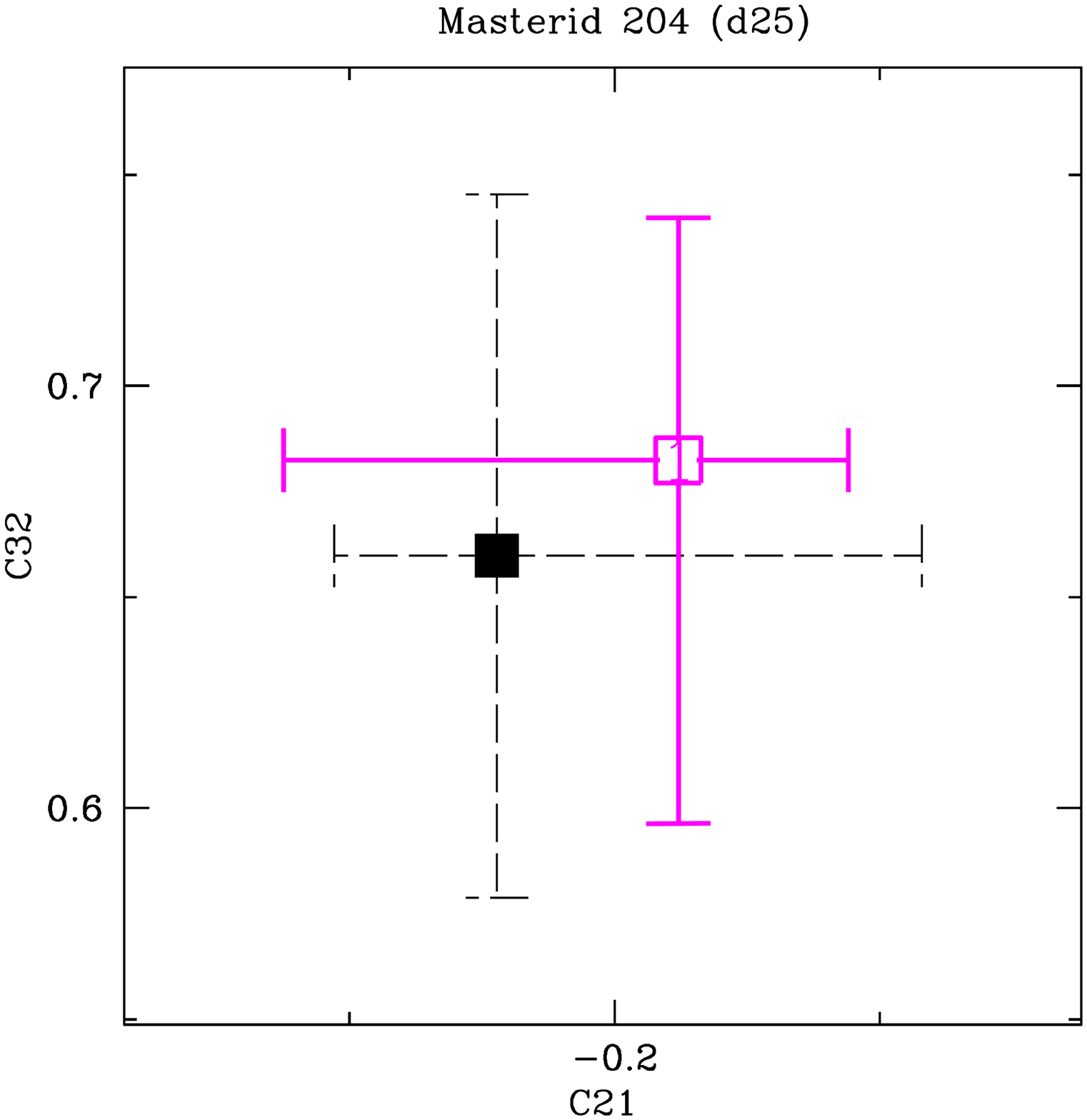}

  \end{minipage}
  \begin{minipage}{0.32\linewidth}
  \centering

    \includegraphics[width=\linewidth]{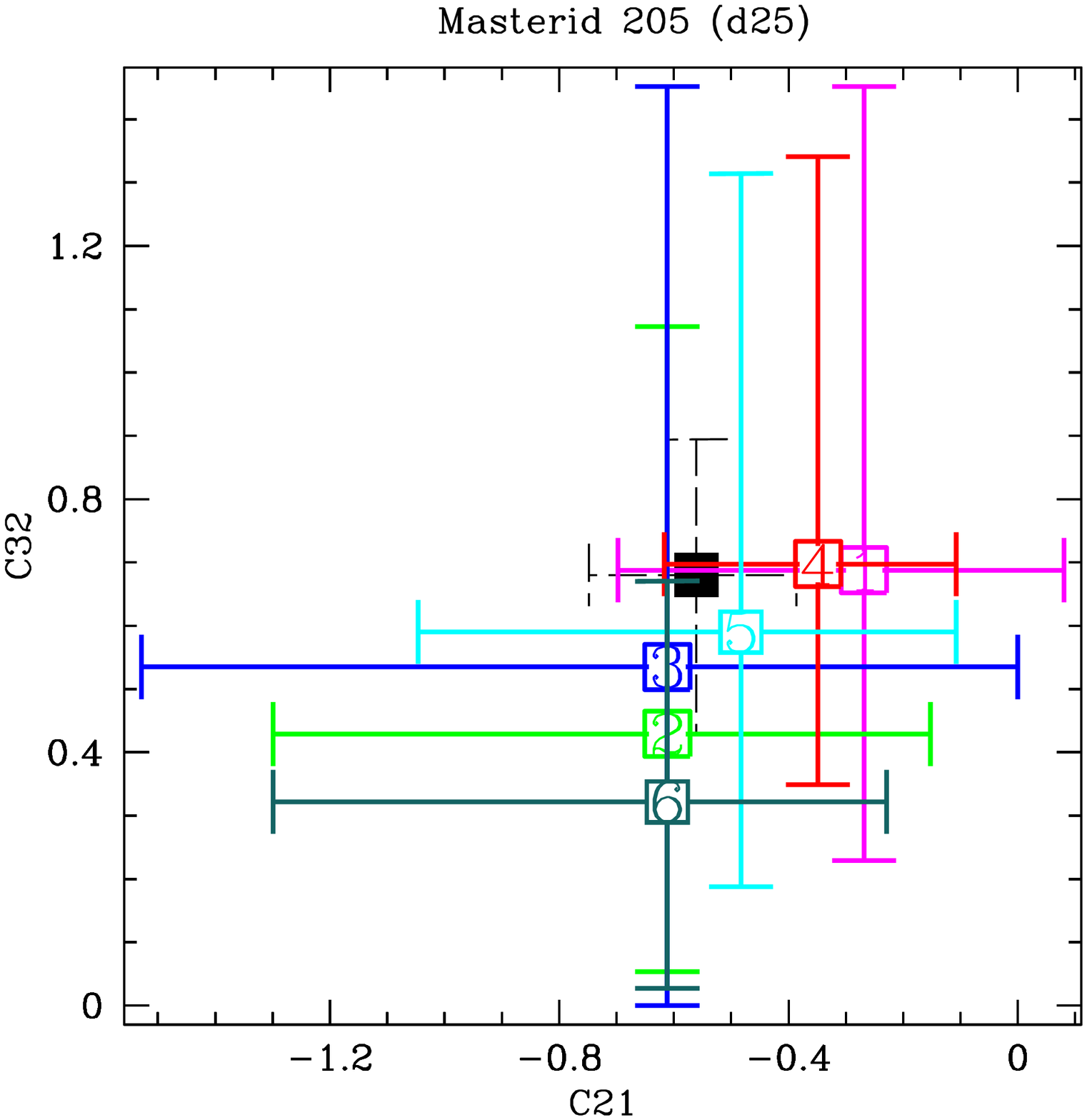}

\end{minipage}
\begin{minipage}{0.32\linewidth}
  \centering

    \includegraphics[width=\linewidth]{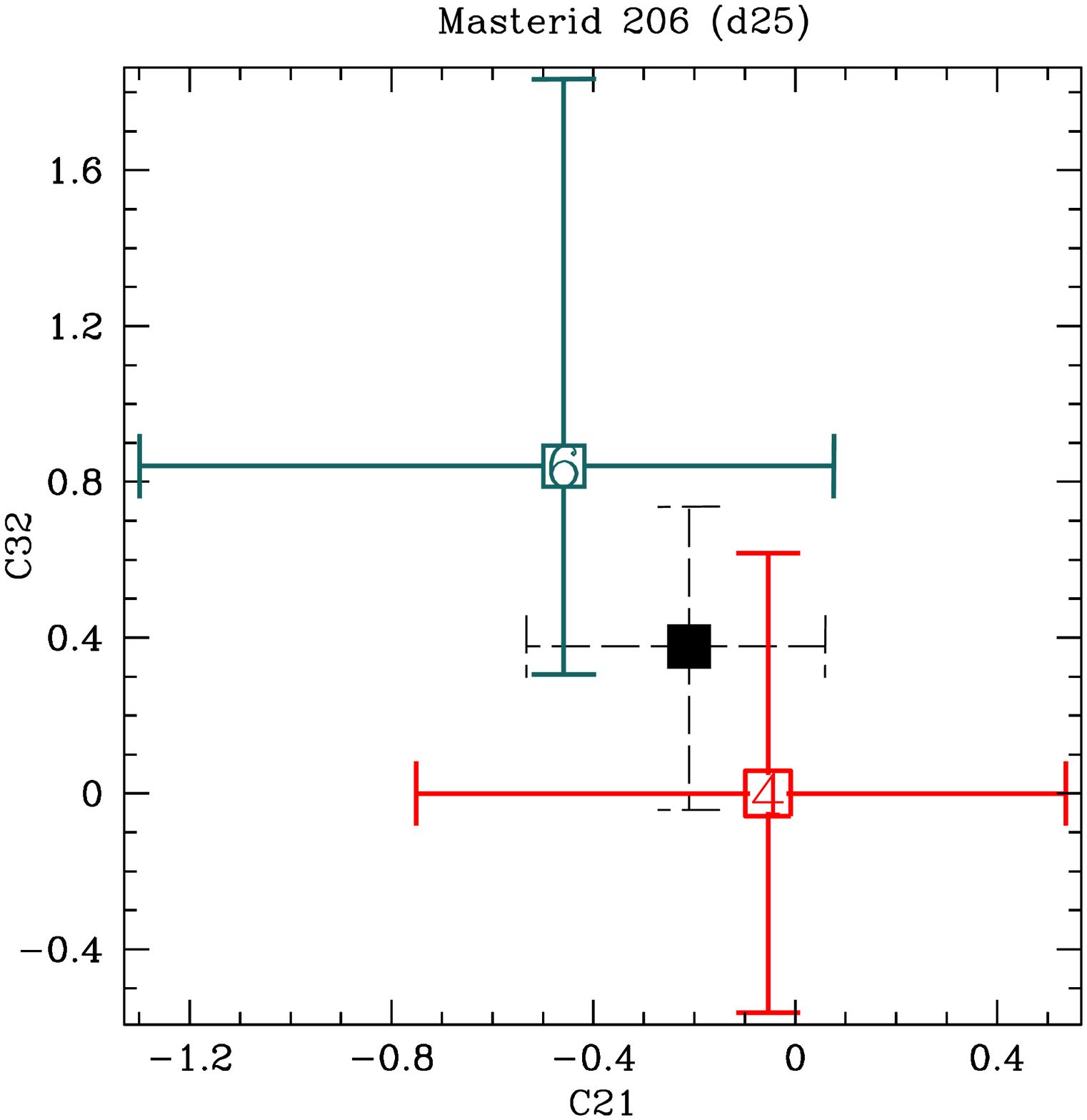}

\end{minipage}
\end{figure}

\begin{figure}
  \begin{minipage}{0.32\linewidth}
  \centering
  
    \includegraphics[width=\linewidth]{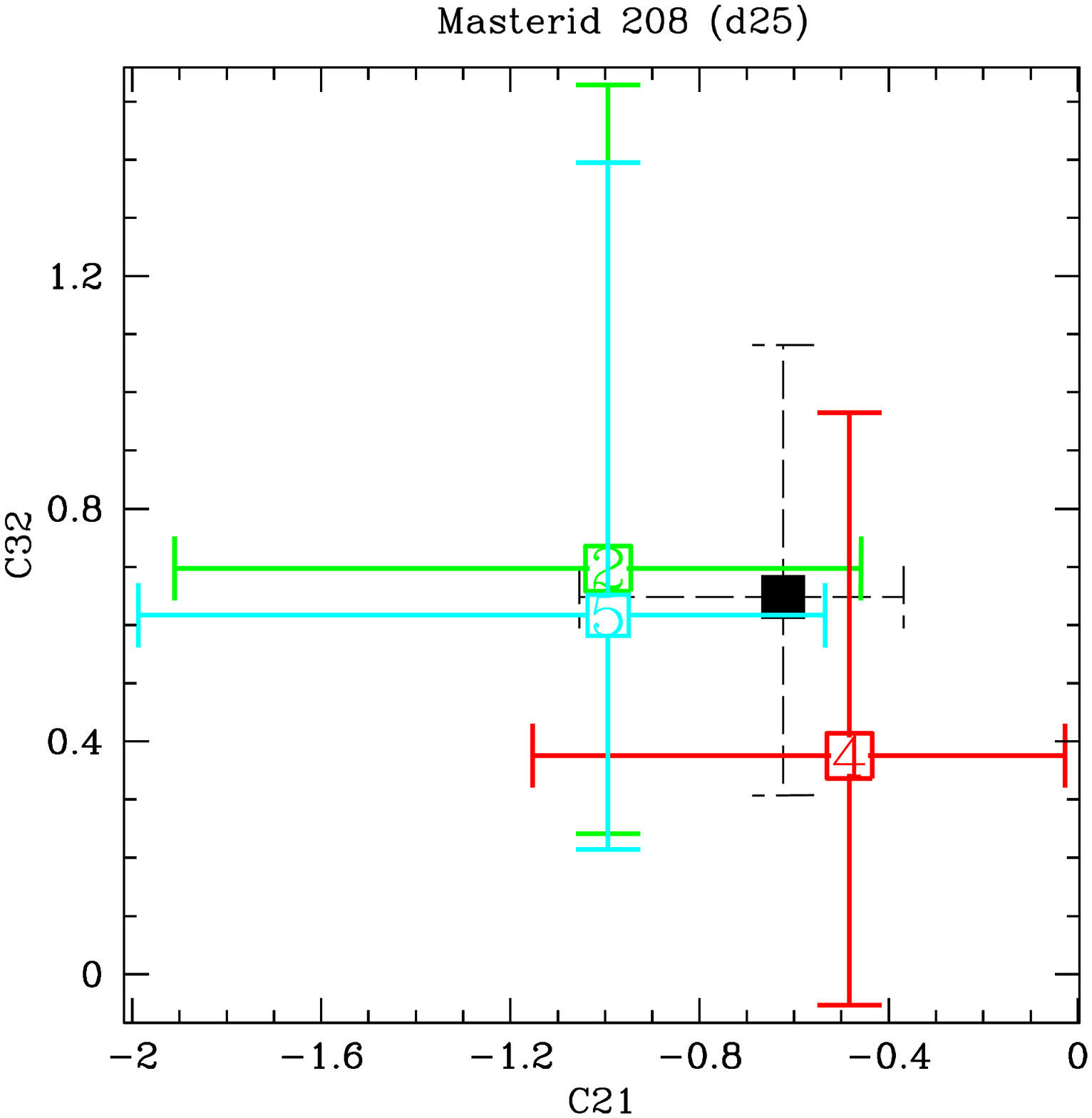}

  \end{minipage}
  \begin{minipage}{0.32\linewidth}
  \centering

    \includegraphics[width=\linewidth]{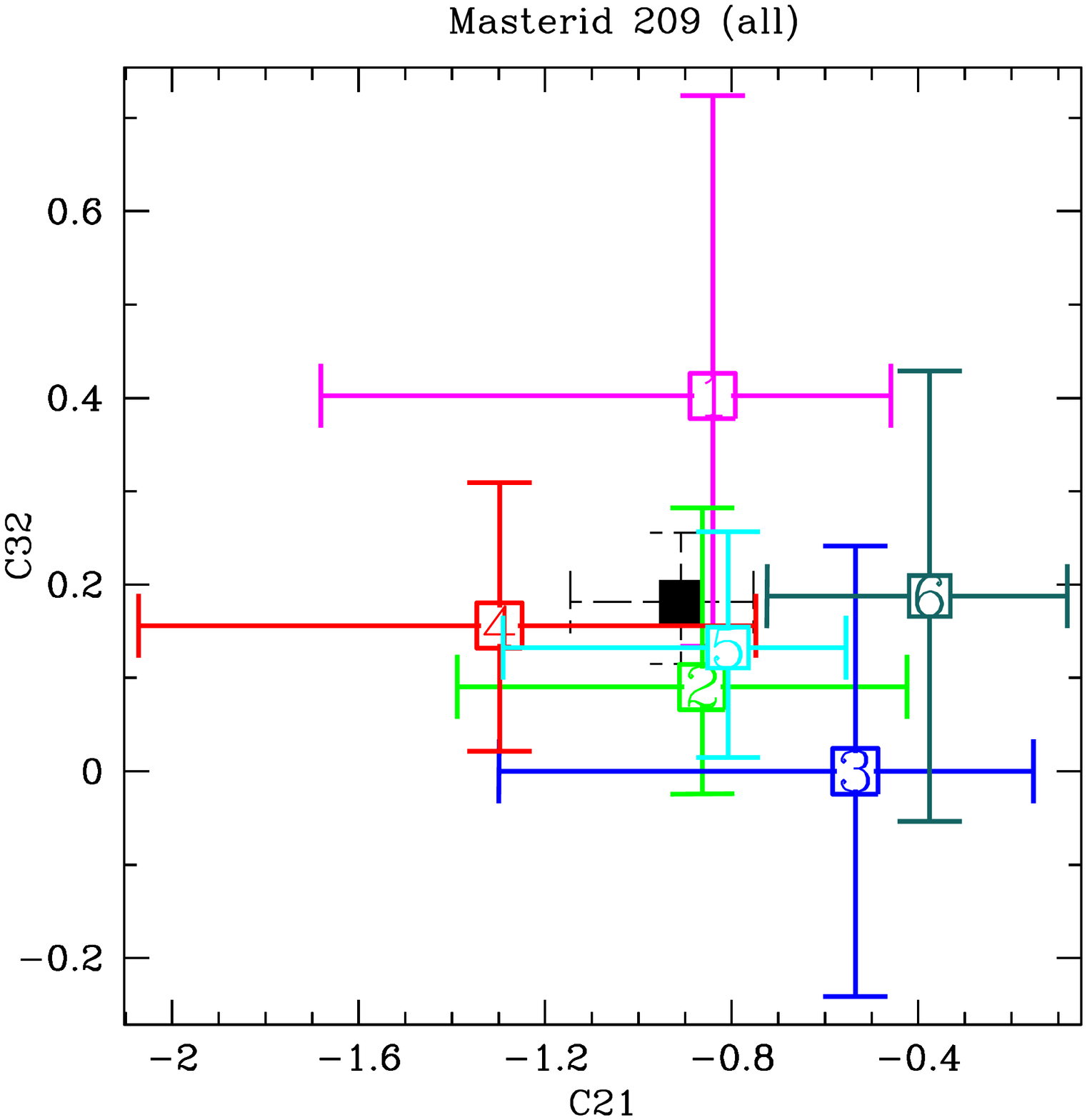}

\end{minipage}
\begin{minipage}{0.32\linewidth}
  \centering

    \includegraphics[width=\linewidth]{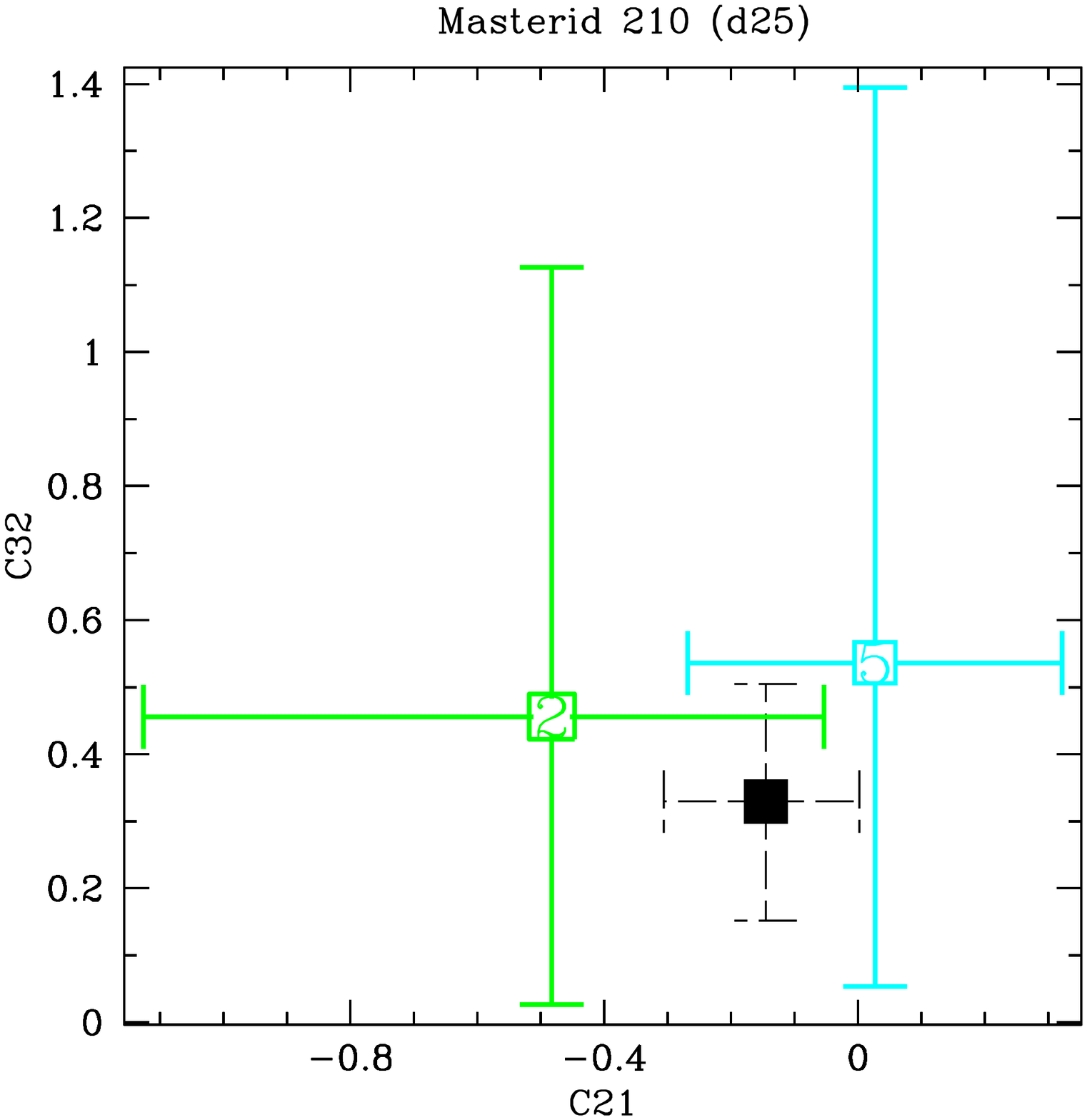}

 \end{minipage}

\begin{minipage}{0.32\linewidth}
  \centering
  
    \includegraphics[width=\linewidth]{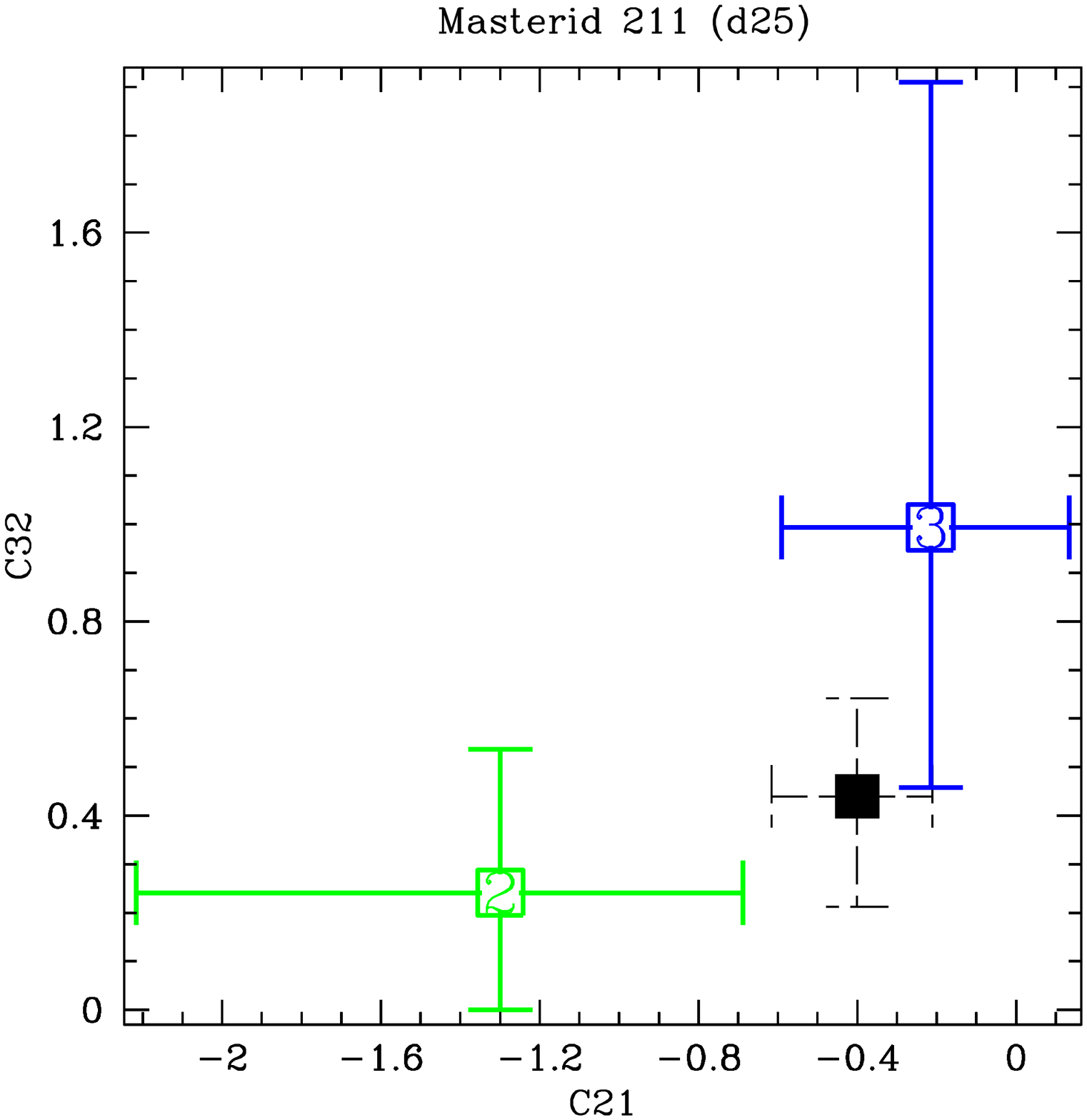}

  \end{minipage}
  \begin{minipage}{0.32\linewidth}
  \centering

    \includegraphics[width=\linewidth]{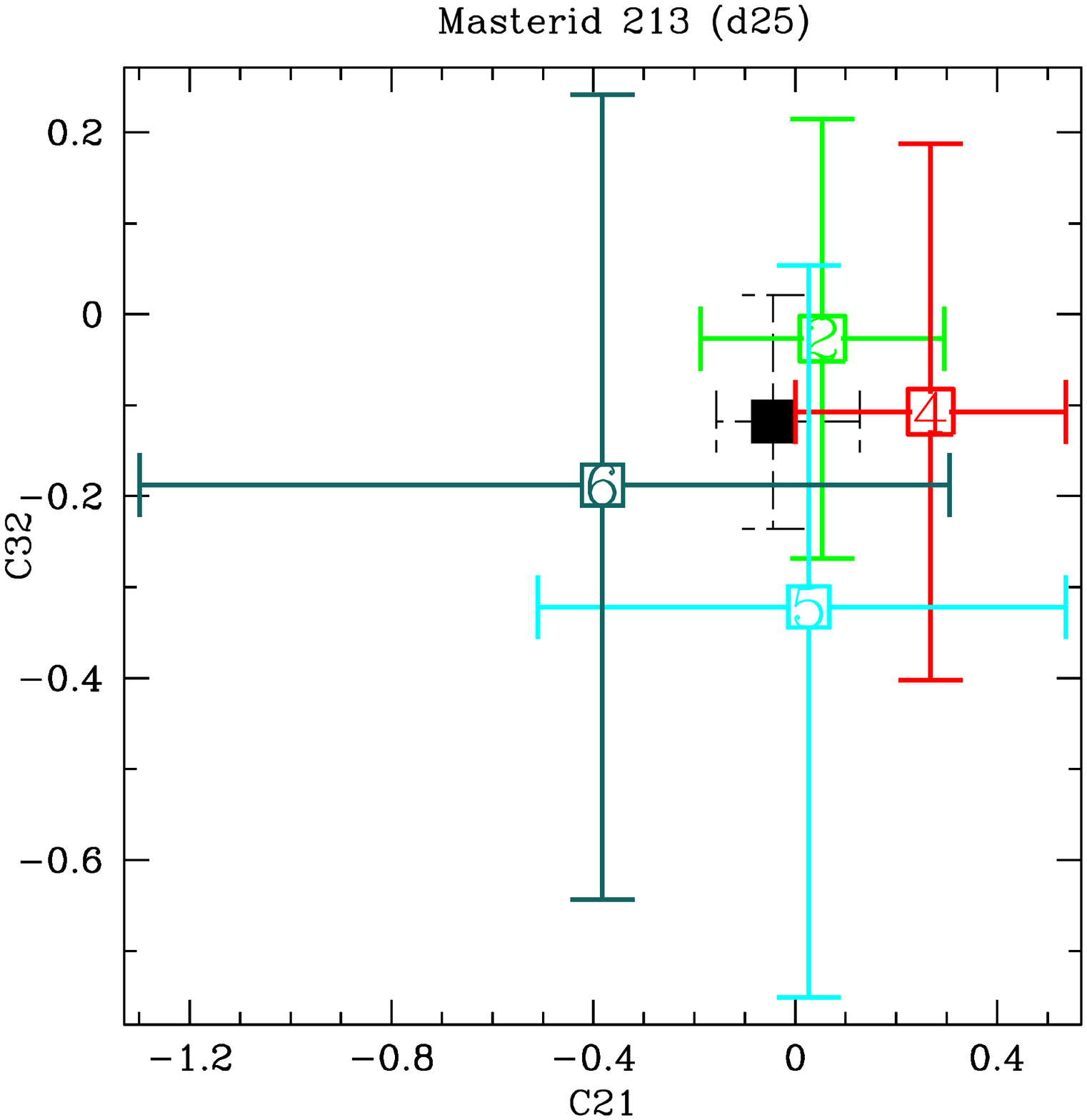}

\end{minipage}
\begin{minipage}{0.32\linewidth}
  \centering

    \includegraphics[width=\linewidth]{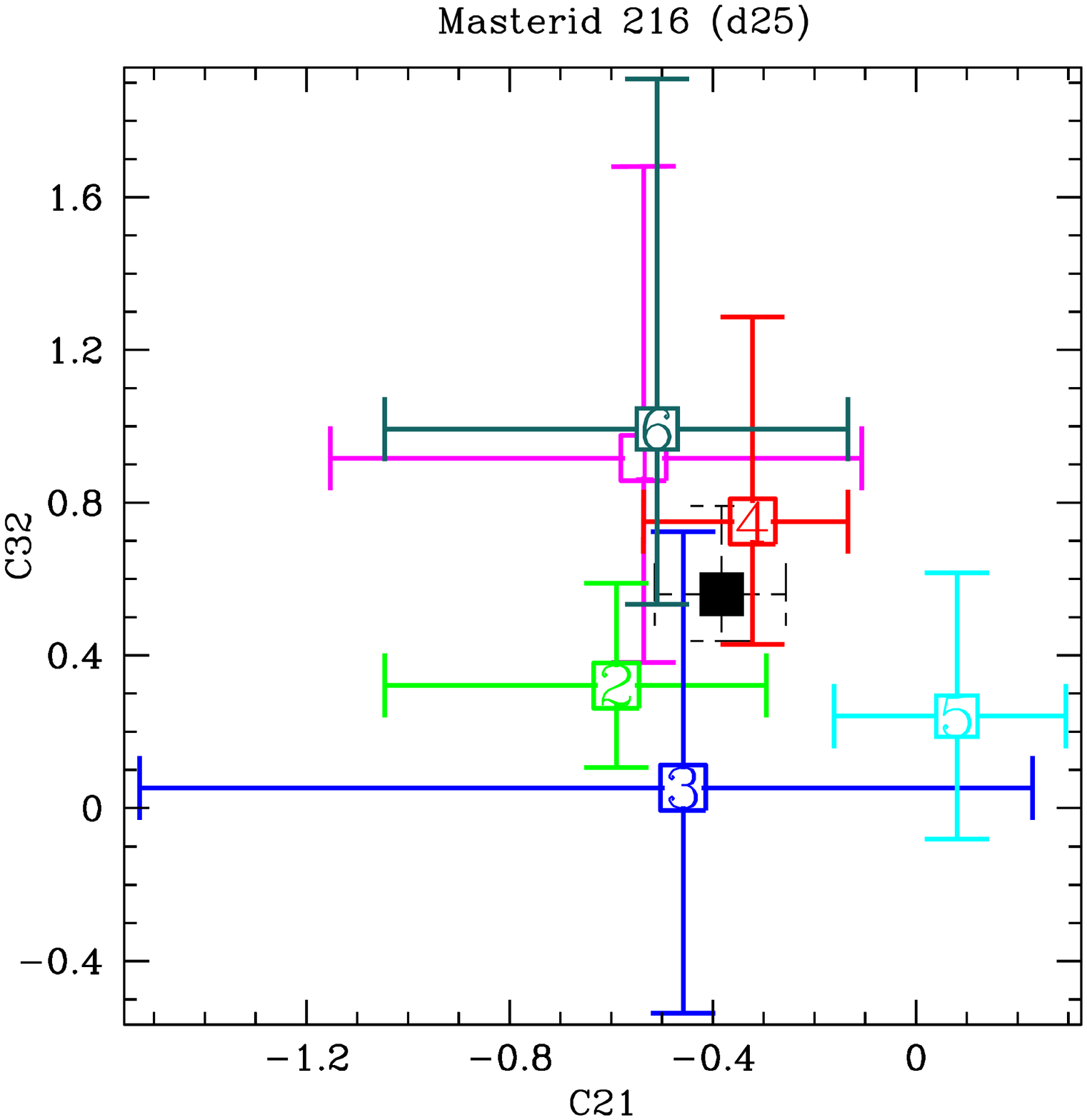}

 \end{minipage}

  \begin{minipage}{0.32\linewidth}
  \centering
  
    \includegraphics[width=\linewidth]{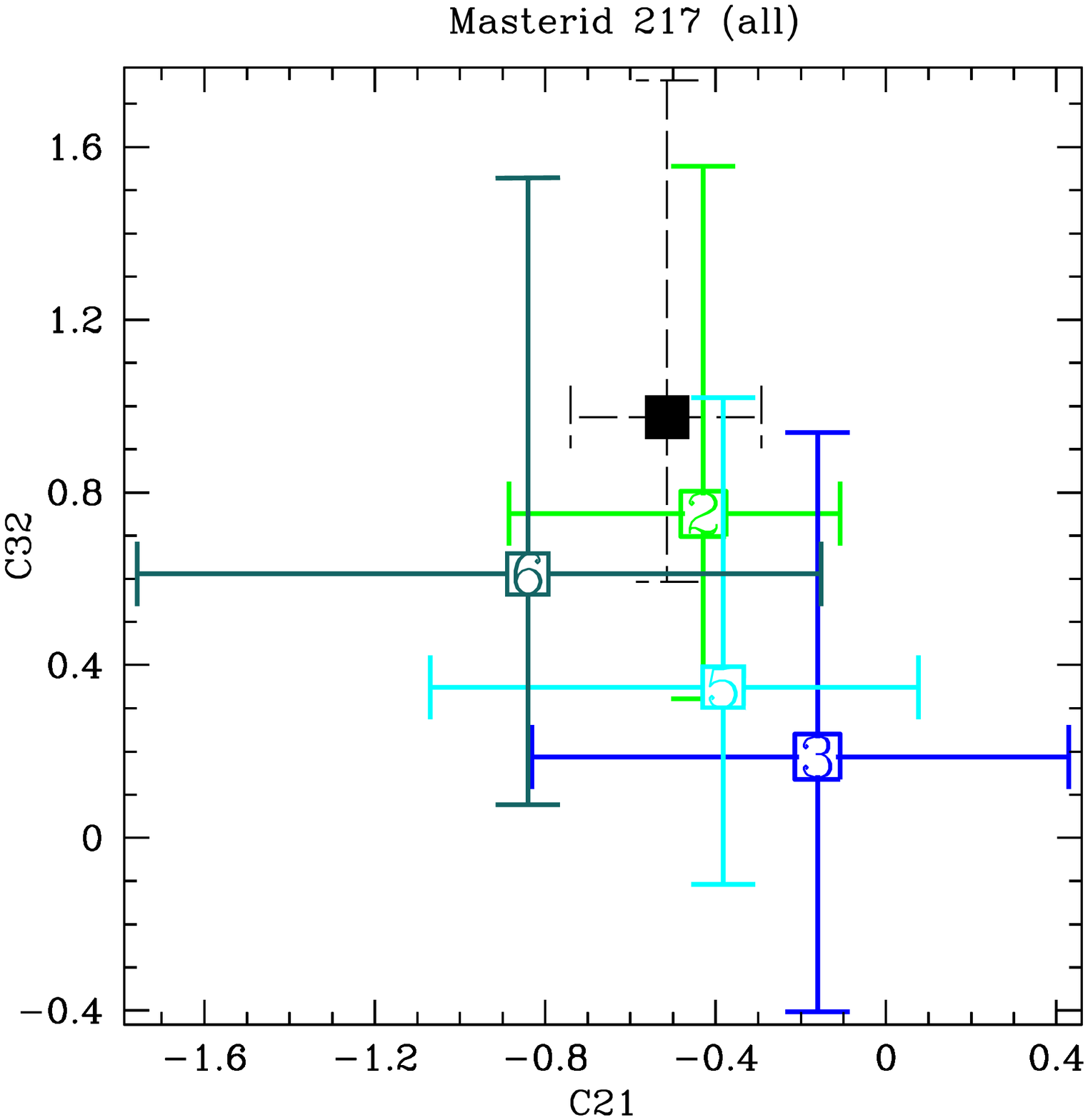}

  \end{minipage}
  \begin{minipage}{0.32\linewidth}
  \centering

    \includegraphics[width=\linewidth]{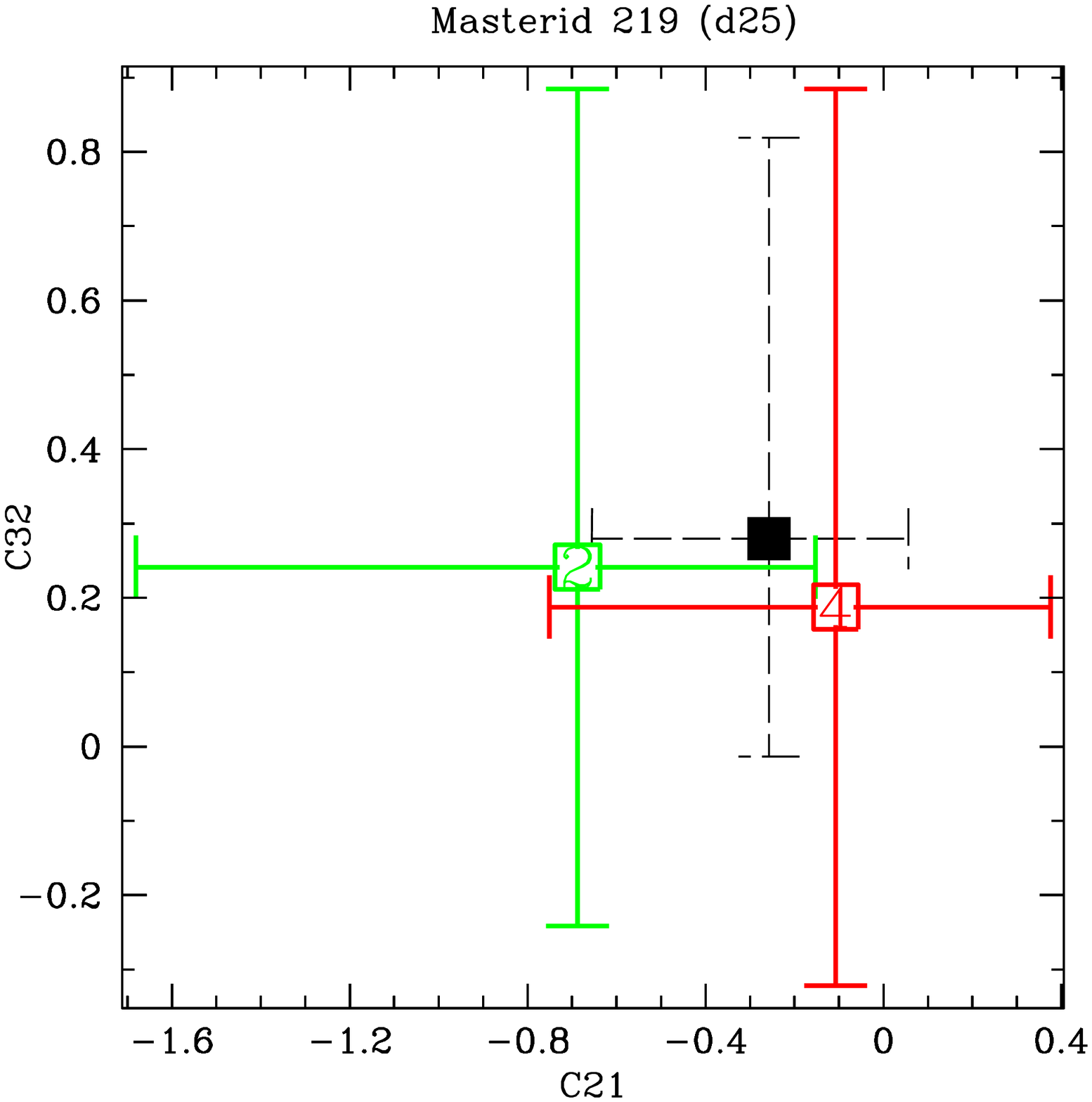}

\end{minipage}
\begin{minipage}{0.32\linewidth}
  \centering

    \includegraphics[width=\linewidth]{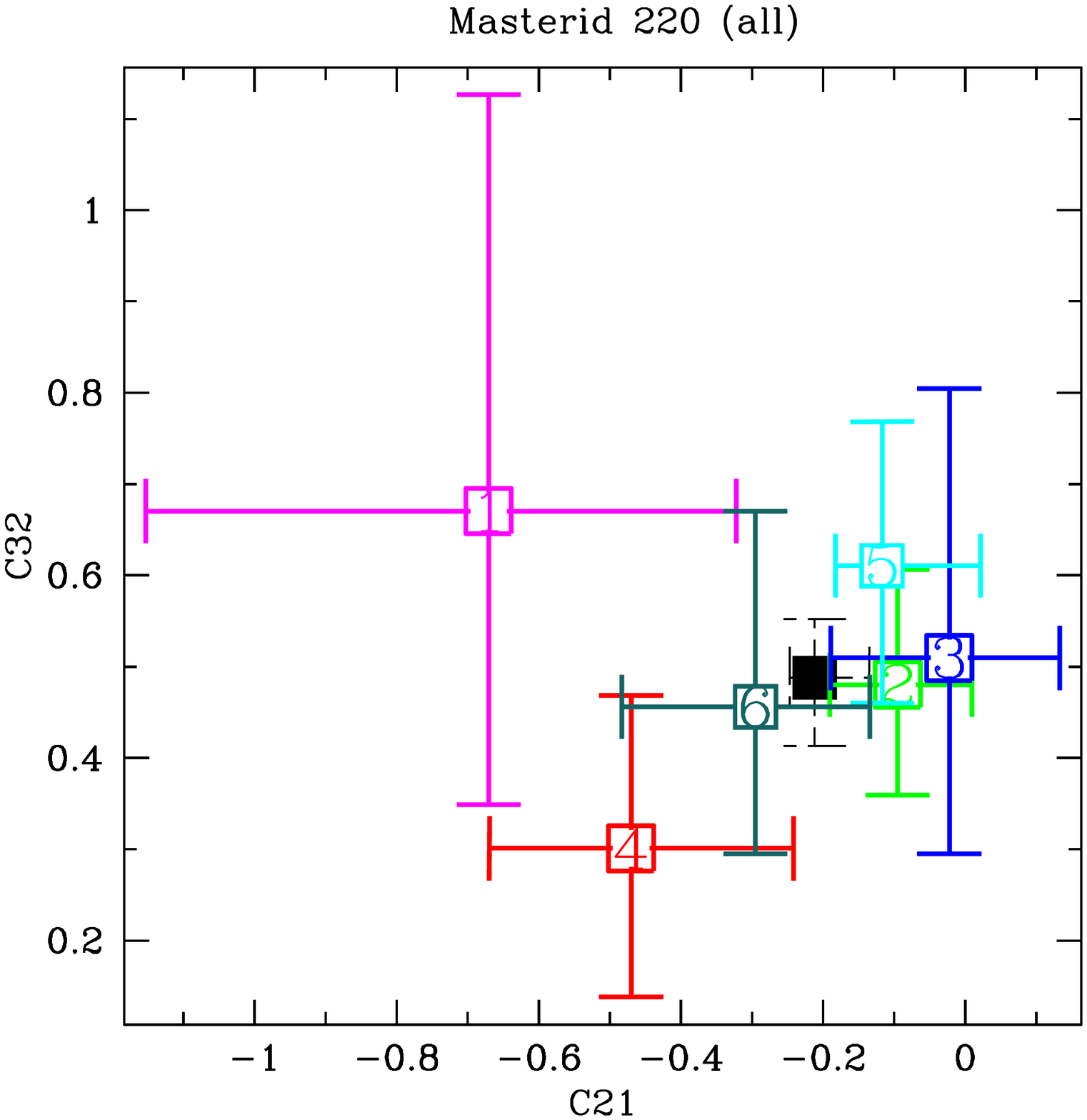}

\end{minipage}

\begin{minipage}{0.32\linewidth}
  \centering
  
    \includegraphics[width=\linewidth]{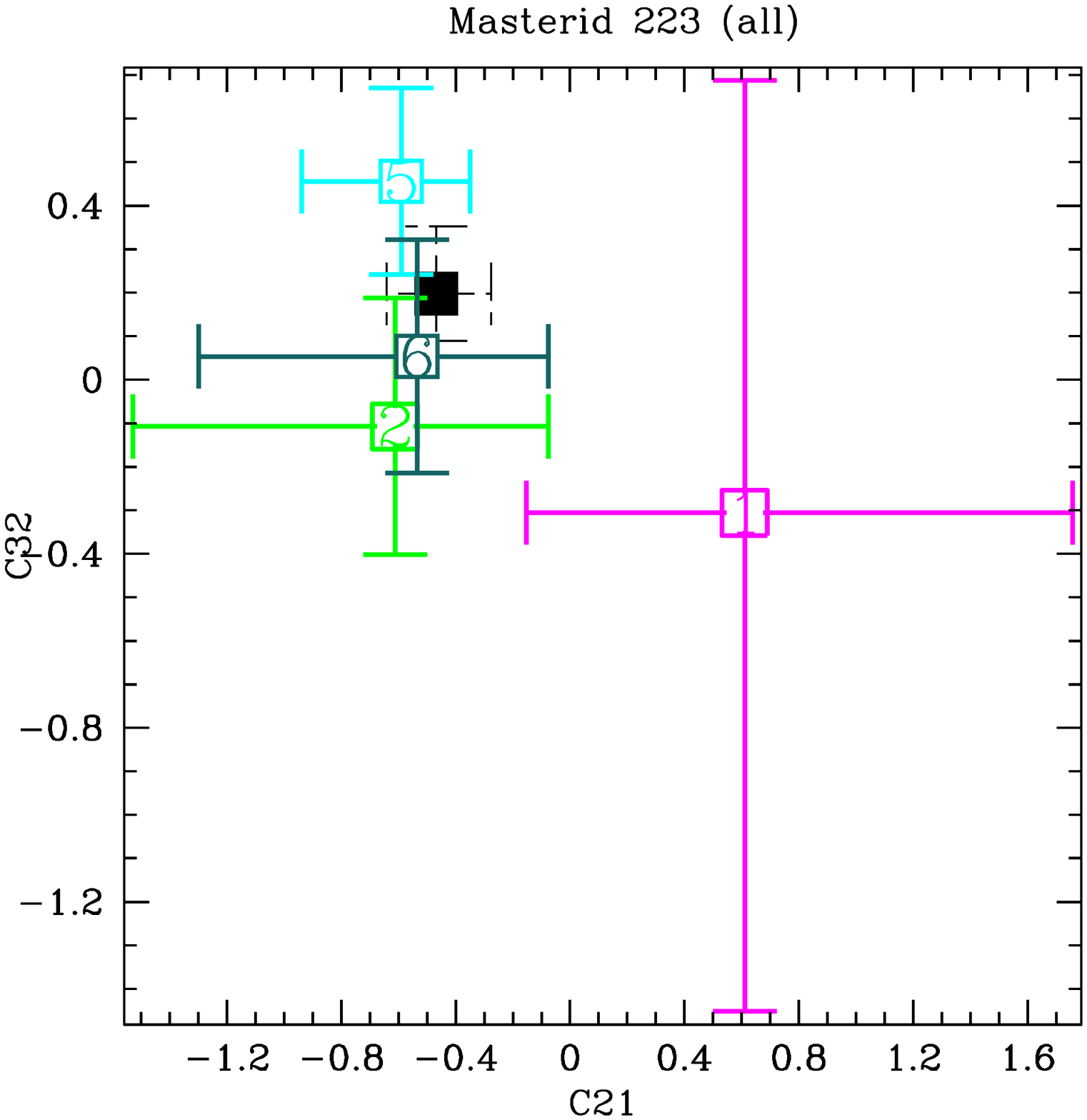}

  \end{minipage}
  \begin{minipage}{0.32\linewidth}
  \centering

    \includegraphics[width=\linewidth]{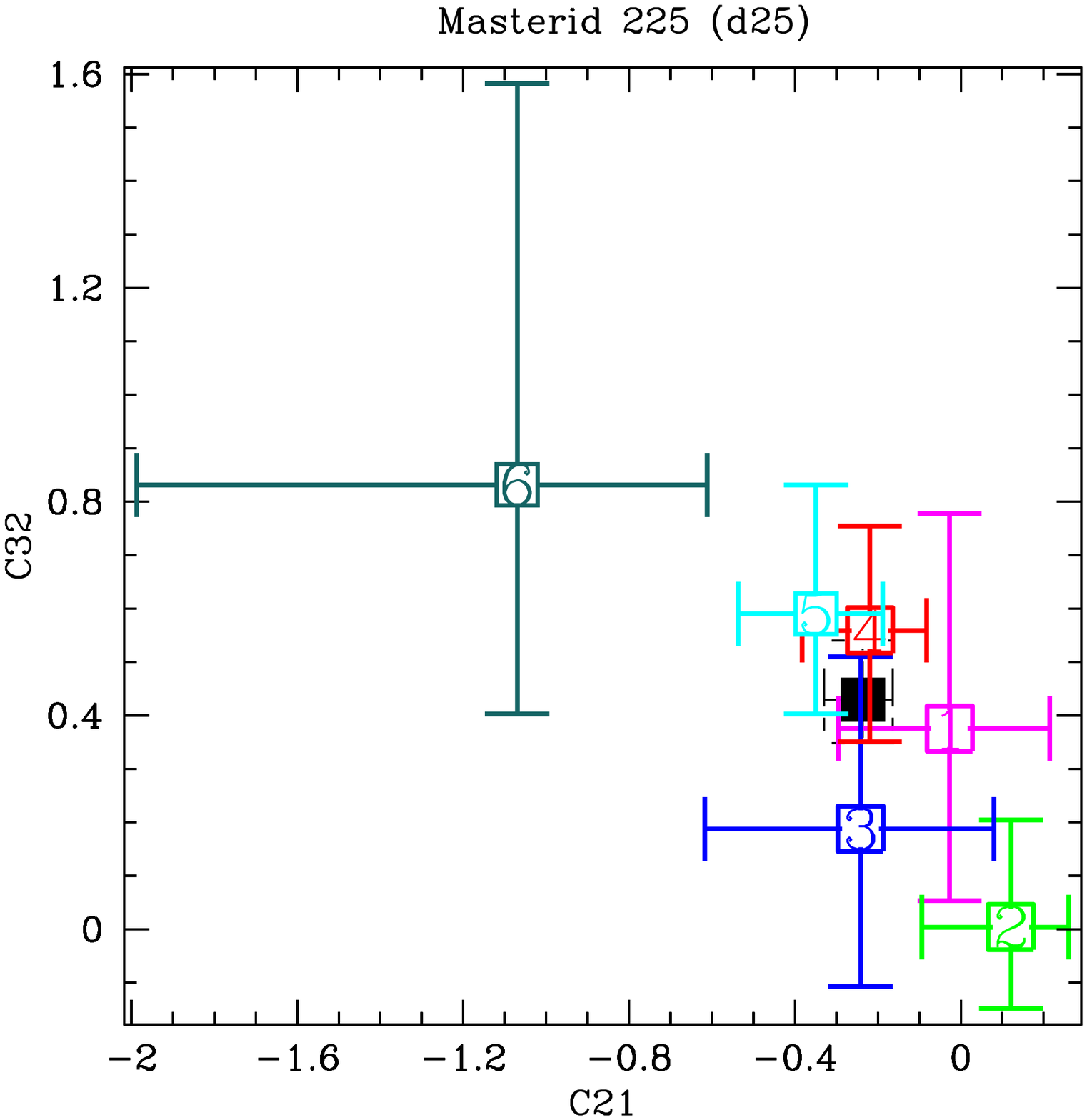}

\end{minipage}
\begin{minipage}{0.32\linewidth}
  \centering

    \includegraphics[width=\linewidth]{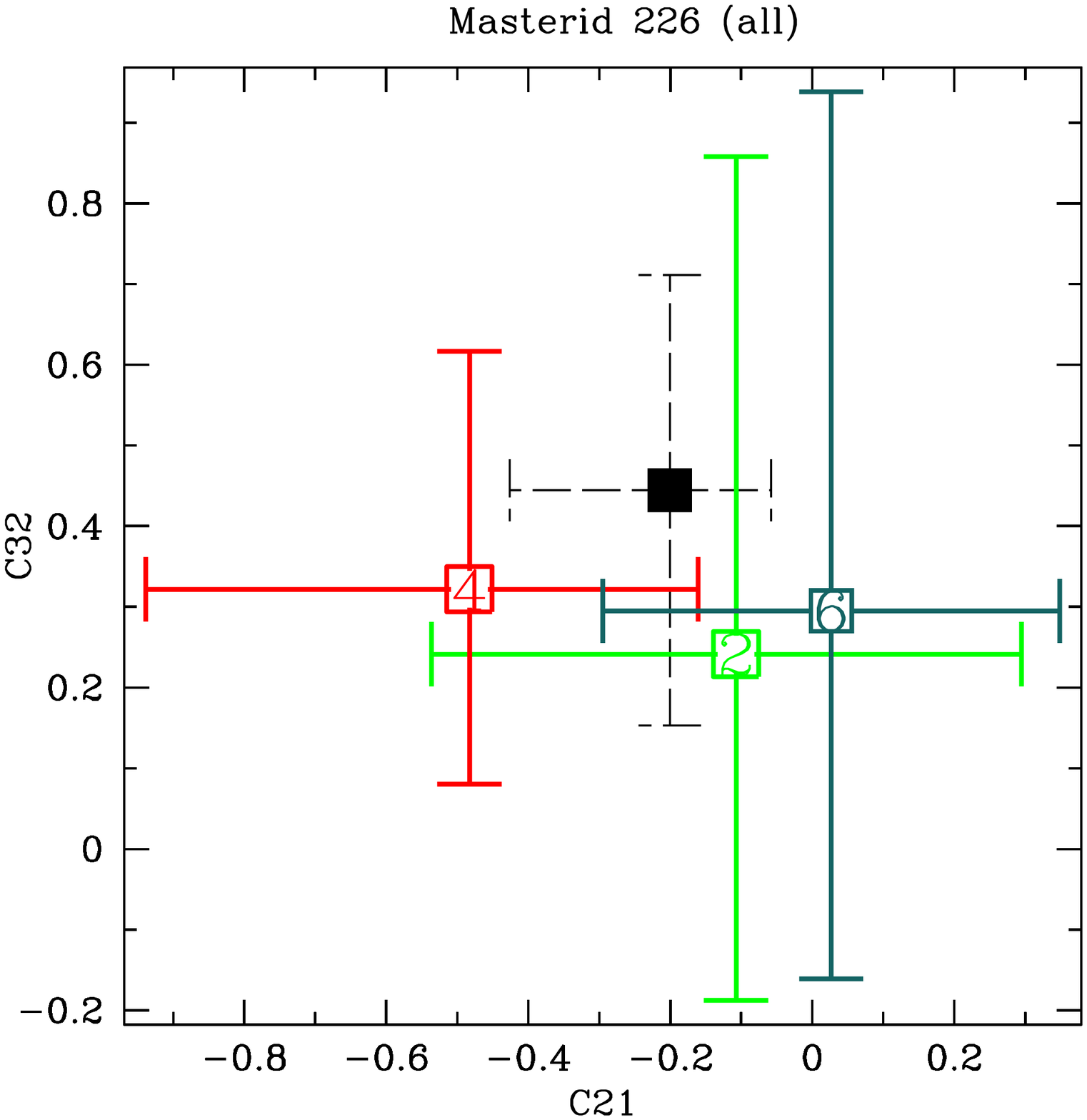}

\end{minipage}
\end{figure}

\begin{figure}
  \begin{minipage}{0.32\linewidth}
  \centering
  
    \includegraphics[width=\linewidth]{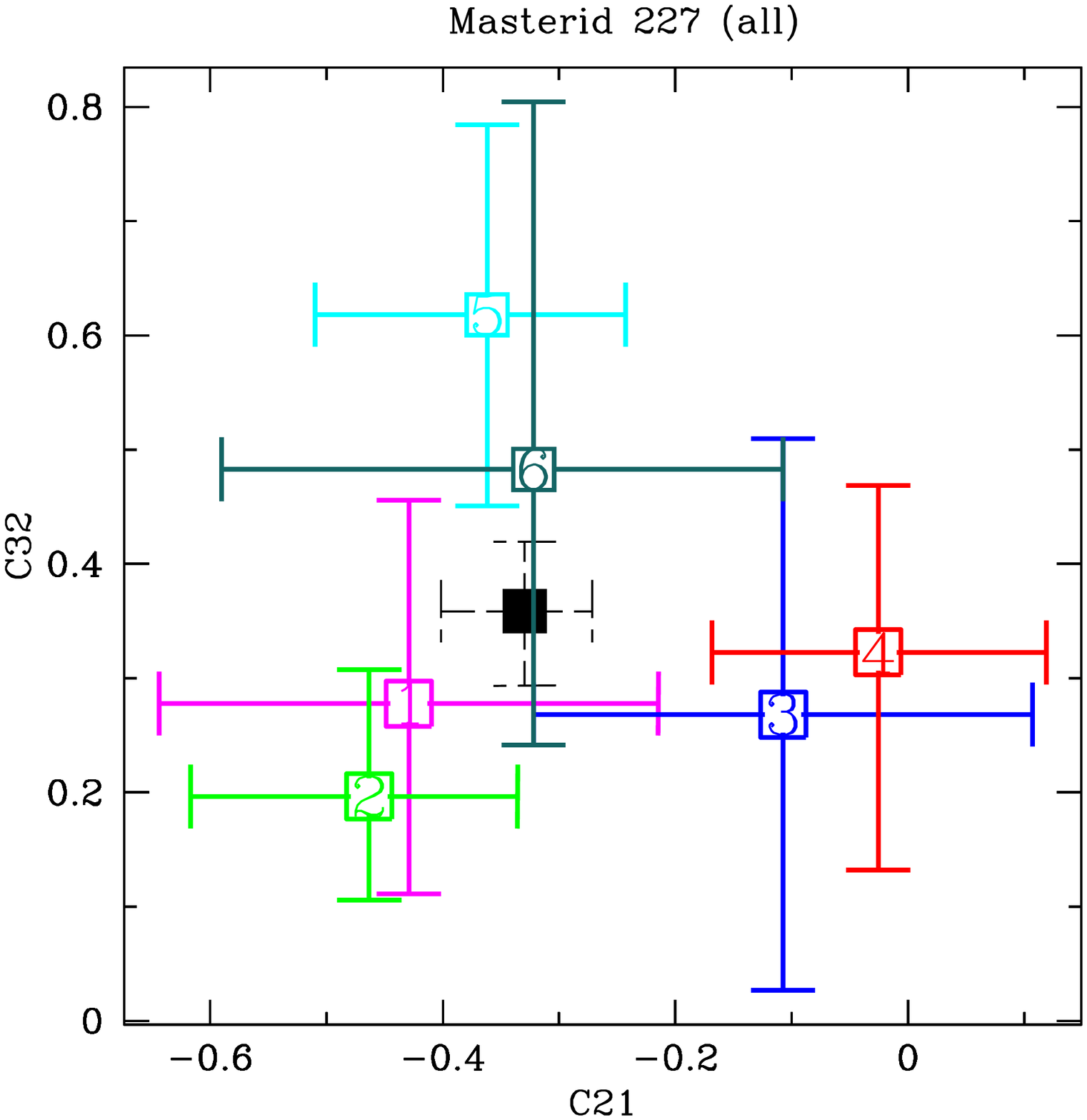}

  \end{minipage}
  \begin{minipage}{0.32\linewidth}
  \centering

    \includegraphics[width=\linewidth]{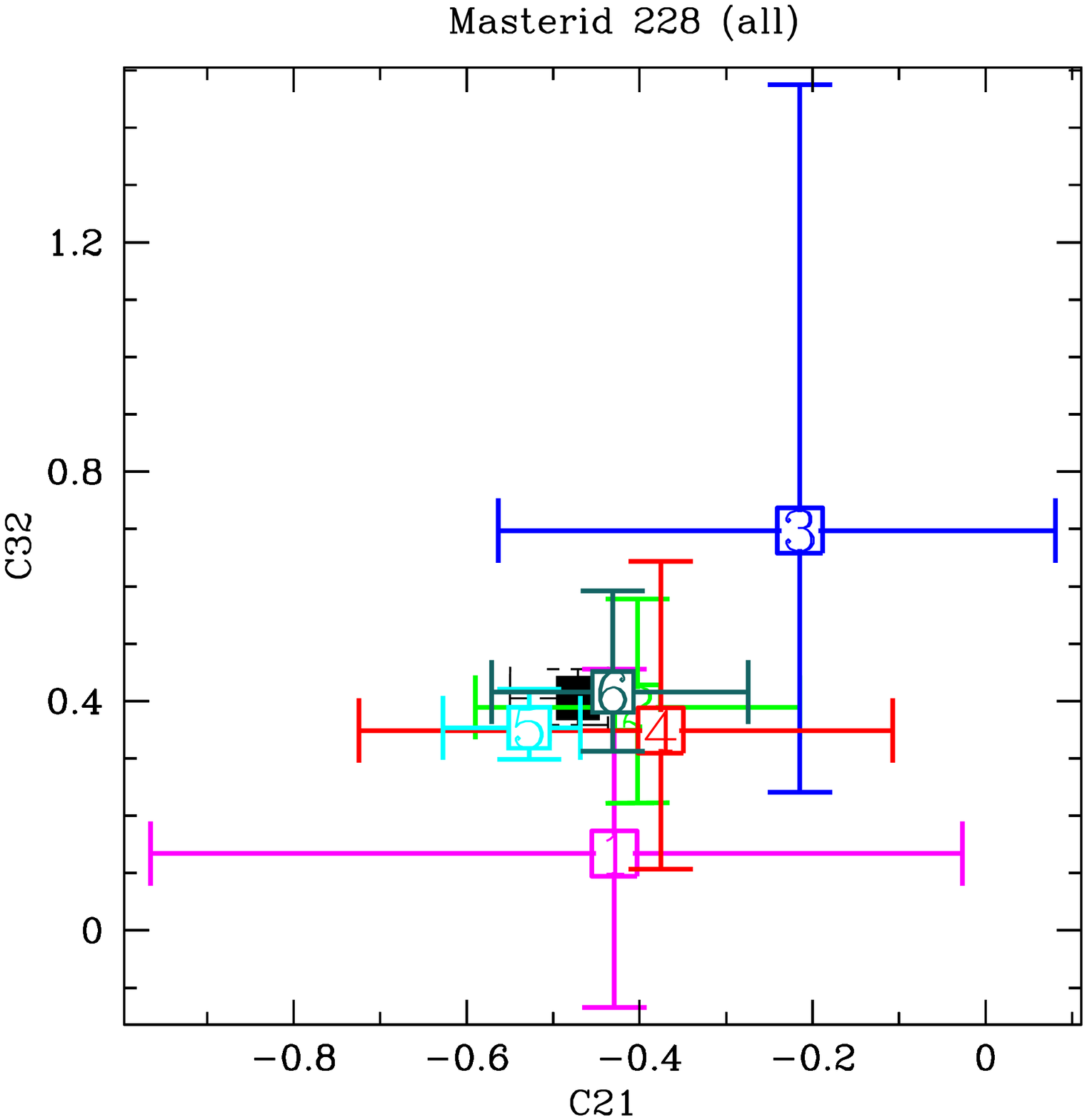}

\end{minipage}
\begin{minipage}{0.32\linewidth}
  \centering

    \includegraphics[width=\linewidth]{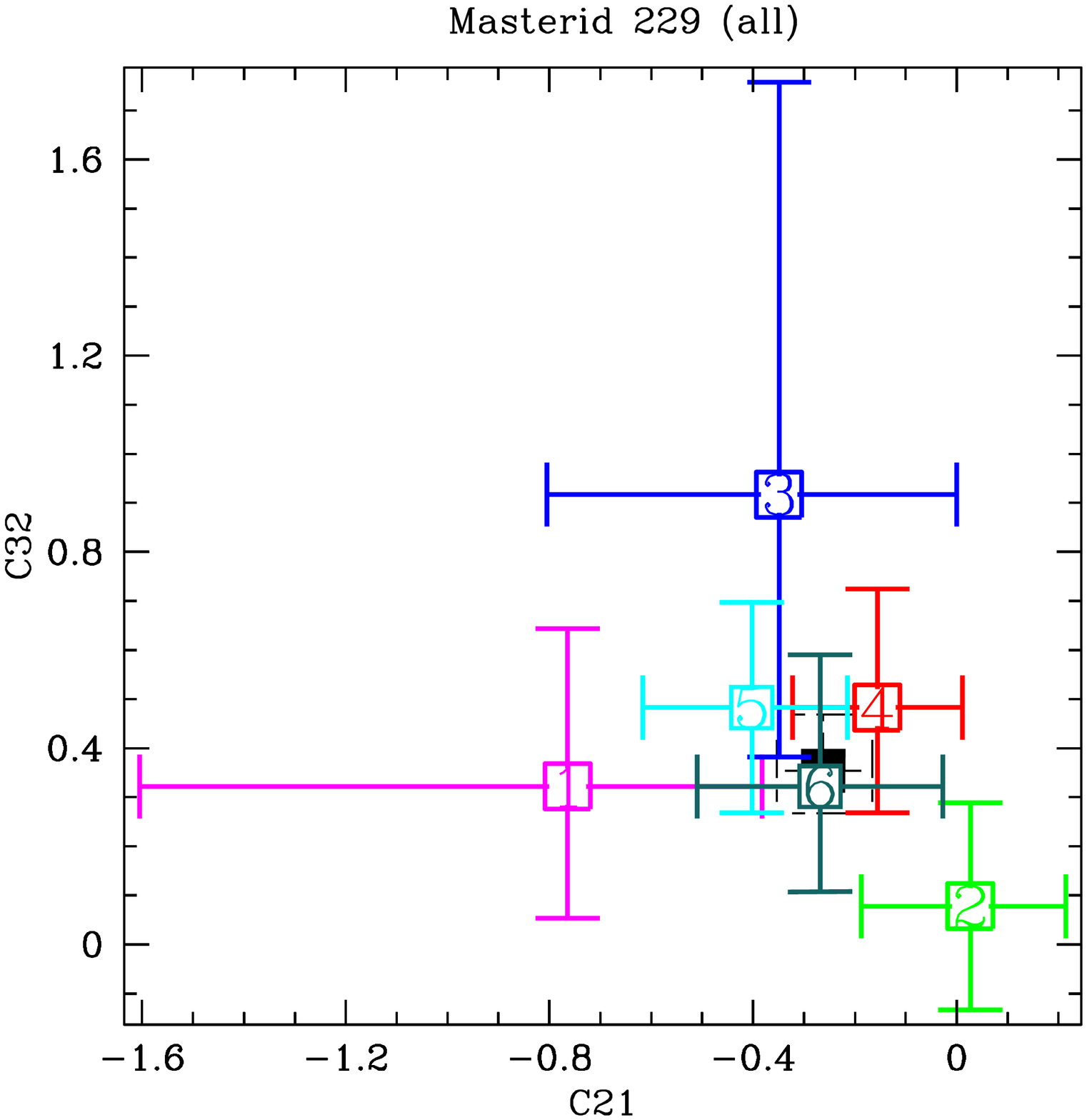}

 \end{minipage}

\begin{minipage}{0.32\linewidth}
  \centering
  
    \includegraphics[width=\linewidth]{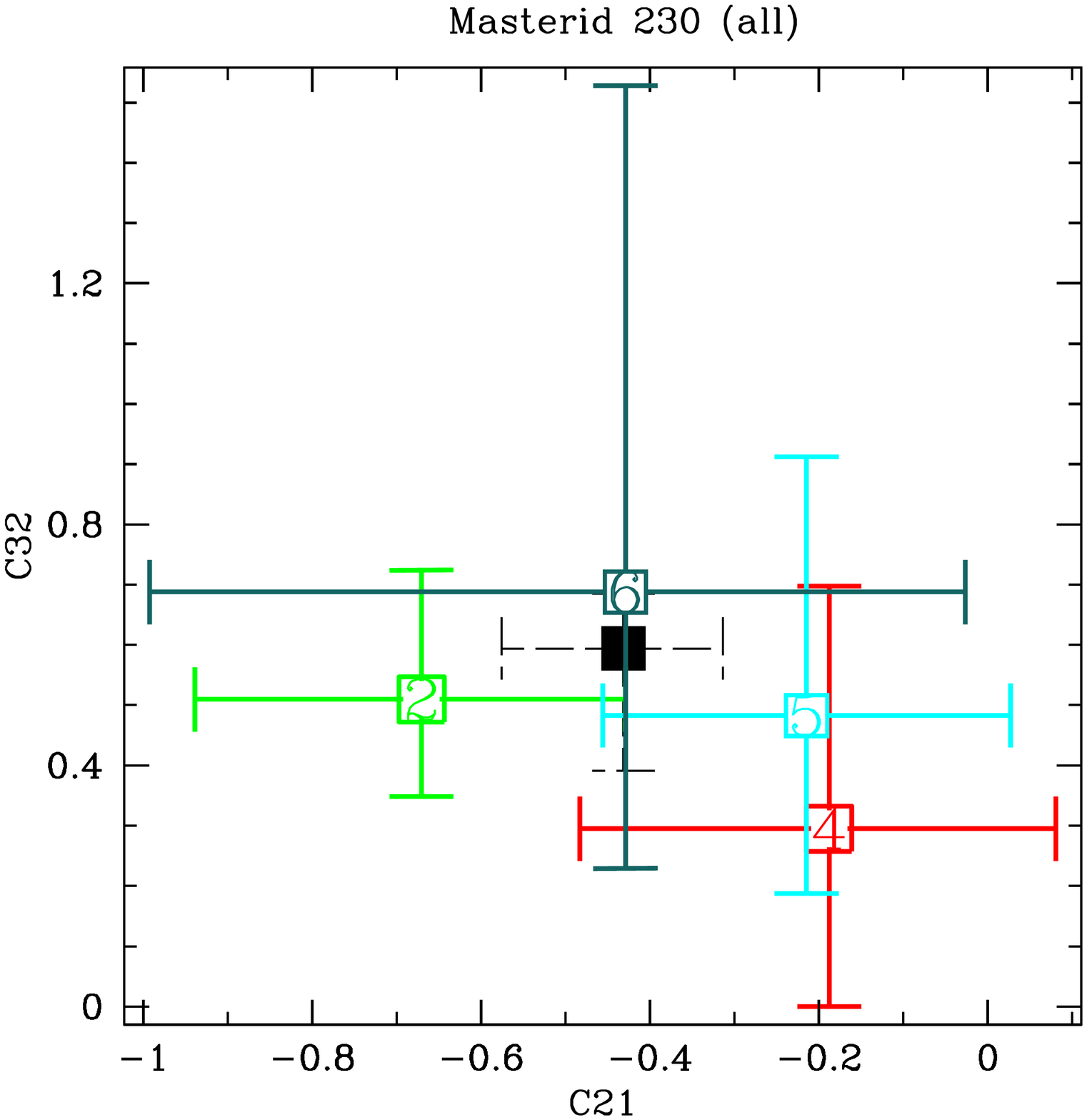}

  \end{minipage}
  \begin{minipage}{0.32\linewidth}
  \centering

    \includegraphics[width=\linewidth]{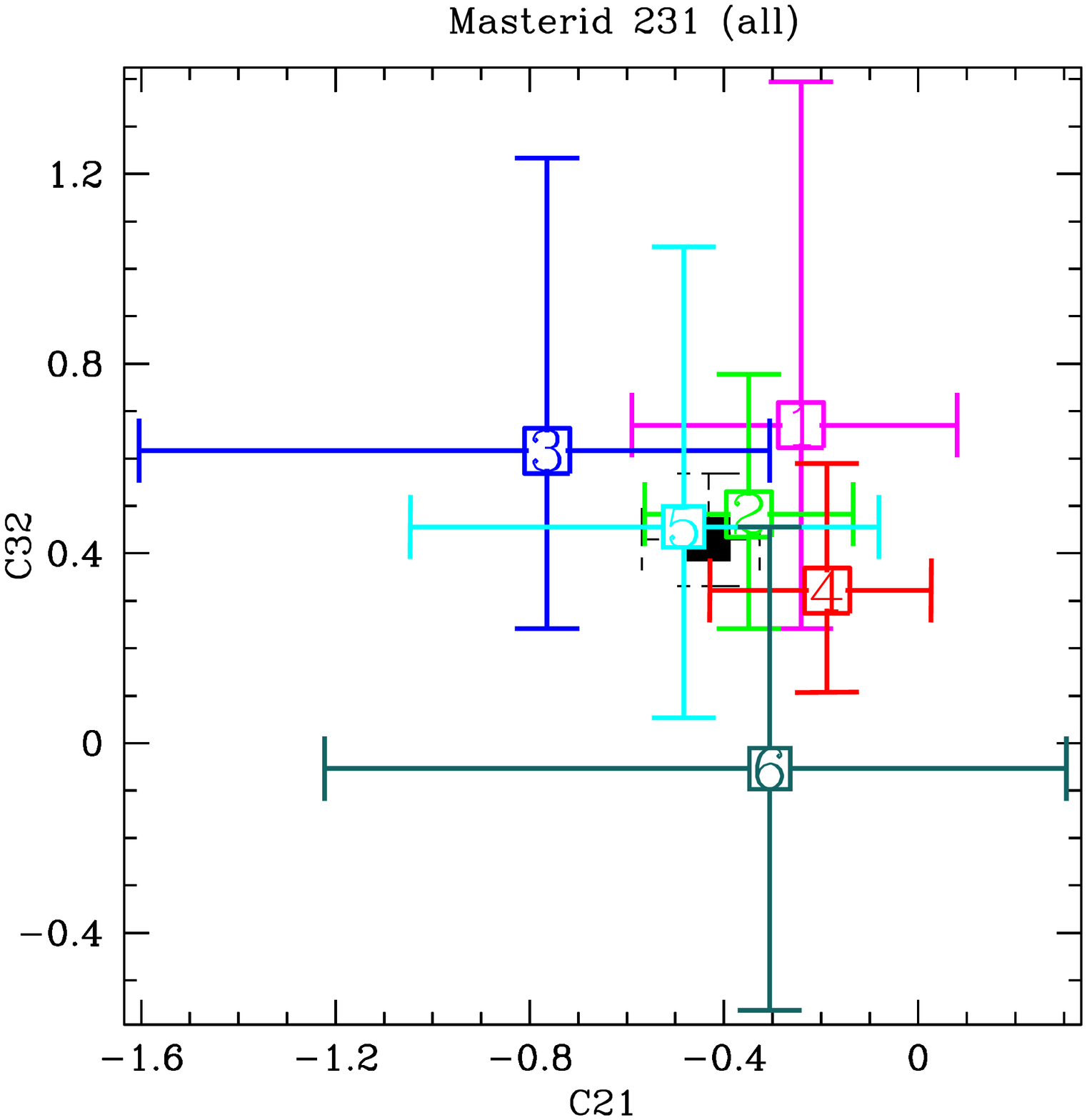}

\end{minipage}
\begin{minipage}{0.32\linewidth}
  \centering

    \includegraphics[width=\linewidth]{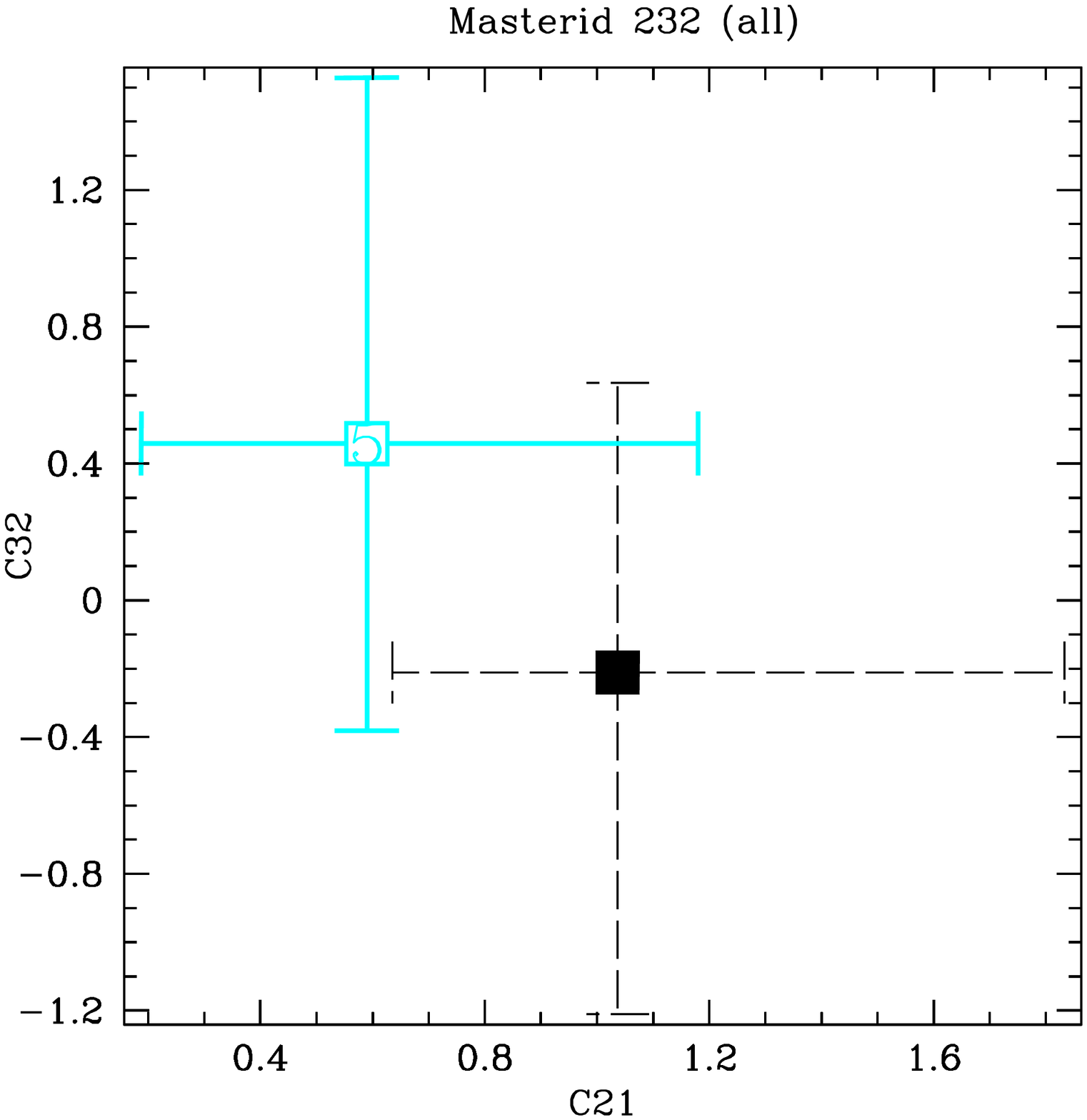}

 \end{minipage}

  \begin{minipage}{0.32\linewidth}
  \centering
  
    \includegraphics[width=\linewidth]{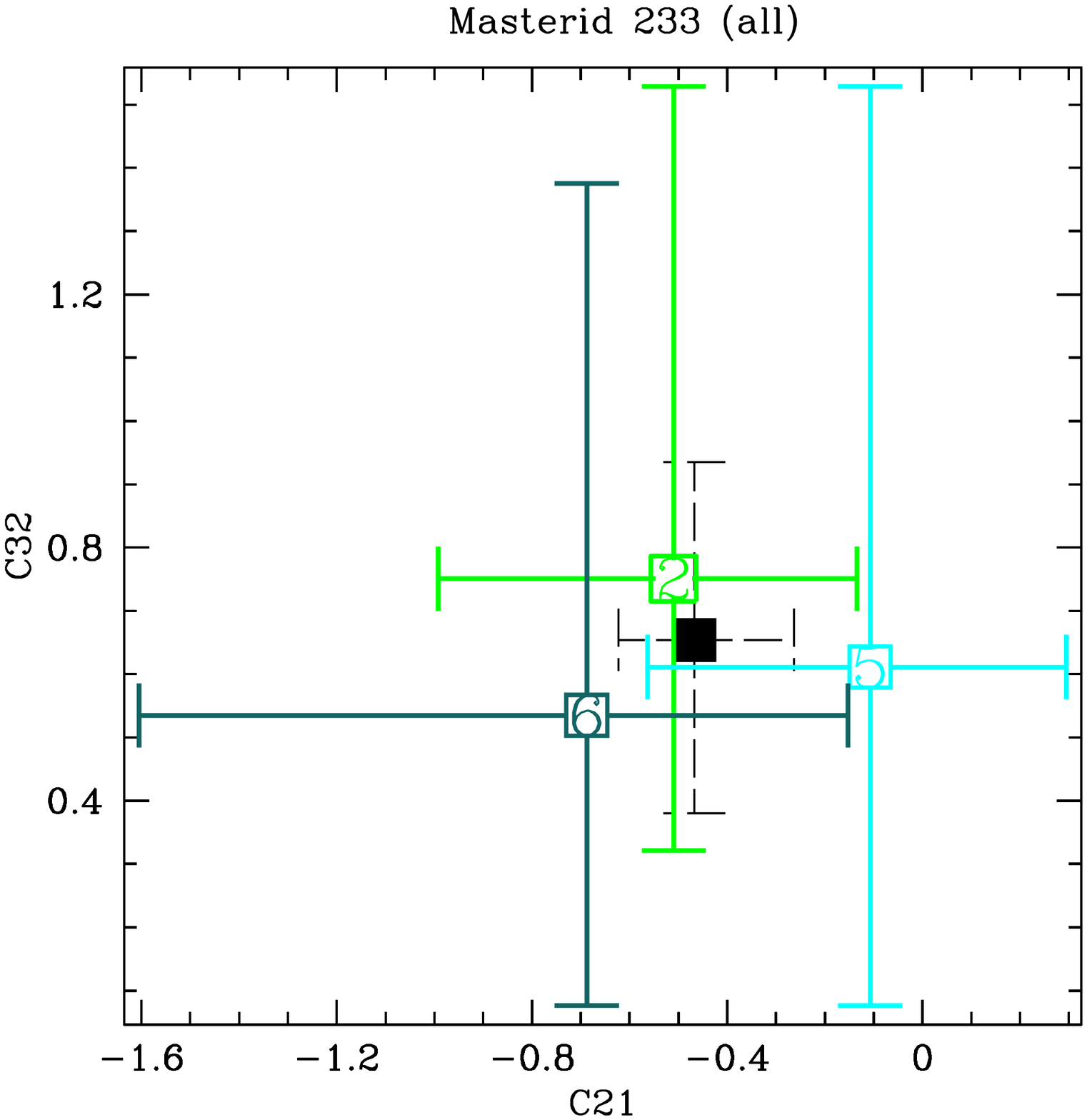}

  \end{minipage}
  \begin{minipage}{0.32\linewidth}
  \centering

    \includegraphics[width=\linewidth]{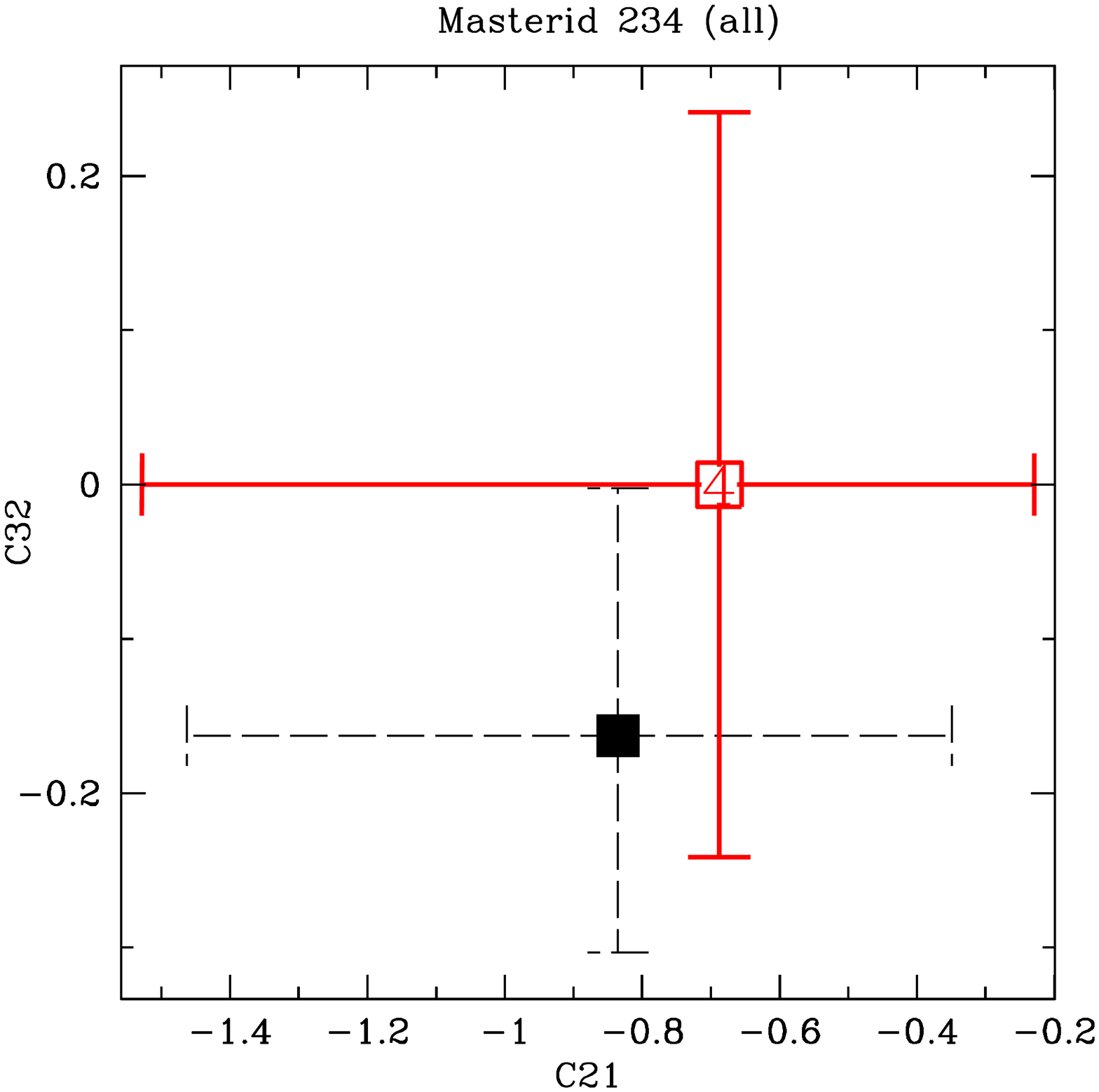}

\end{minipage}
\begin{minipage}{0.32\linewidth}
  \centering

    \includegraphics[width=\linewidth]{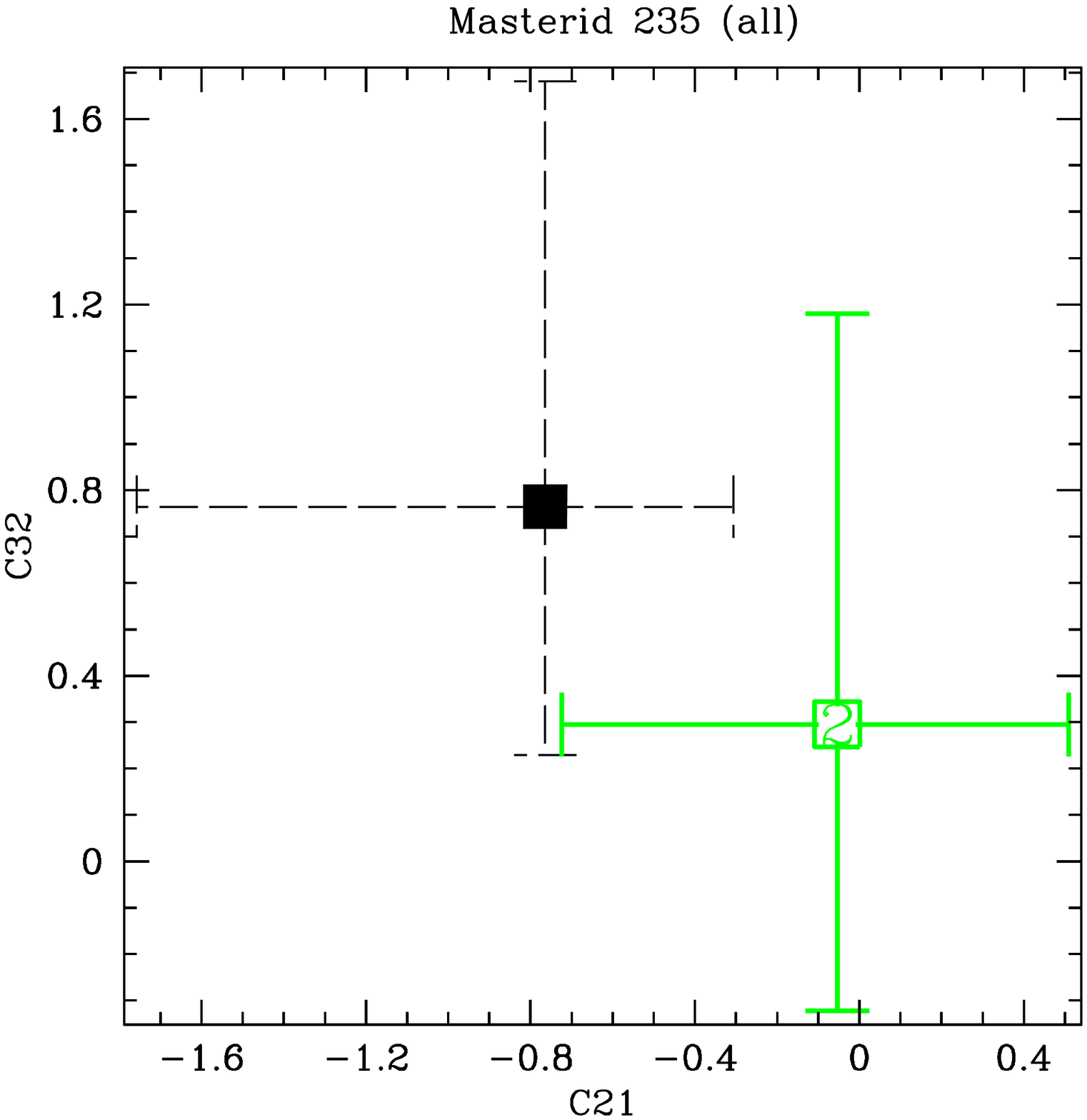}

\end{minipage}
\end{figure}

\begin{thebibliography}{}

\bibitem[Bildsten \& Deloye. 2004]{} Bildsten, L. \& Deloye, C.~J.
2004, ApJ, 607, L119

\bibitem[Brassington \etal\ 2008]{} Brassington, N., J., \etal\ 2008, ApJS, 179, 142

\bibitem[Cappellari \etal\ 2006]{} Cappellari, M., \etal\ 2006, MNRAS, 366, 1126

\bibitem[Fabbiano. 2006]{} Fabbiano, G. 2006, ARAA, 44, 323

\bibitem[Feng 2006]{}Feng, H. \& Kaaret, P. 2006 ApJL, 650, L75 

\bibitem[Freeman. 2002]{}Freeman, P.~E., Kashyap, V., Rosner, R. \& Lamb D.~Q., 2002, ApJS, 138, 185

\bibitem[Gehrels. 1986]{}Gehrels, N. 1986, ApJ, 303, 336

\bibitem[Giacconi. 1974]{}Giacconi, R. 1974, in X-ray Astronomy, eds. R. Giacconi \& H. Gursky, p.155, Dordrecht: Reidel\

\bibitem[Gilfanov. 2004]{}Gilfanov, M. 2004, MNRAS, 349, 146

\bibitem[Grindlay. 1984]{}Grindlay, J. E.  1984, Adv. Space Res., 3, 19

\bibitem[Harris. 1991]{}Harris, W. E. 1991, ARAA, 29, 543

\bibitem[Ho. 1997]{}Ho, L. C., Filippenko, A. V., Sargent, W. L. W. \& Peng, C. Y. 1997, ApJS, 112, 391 

\bibitem[Irwin. 2003]{}Irwin, J. A., Athey, A. E. \& Bregman, J. N. 2003, ApJ, 587, 356

\bibitem[Irwin. 2004]{}Irwin, J. A., Bregman, J. N. \& Athey, A. E. 2004, ApJ, 601, L143

\bibitem[Irwin. 2005]{}Irwin, J. 2005, ApJ, 631, 511

\bibitem[Irwin. 2006]{}Irwin, J. 2006, MNRAS, 371, 1903

\bibitem[Kenter. 2005]{}Kenter, A., \etal\ 2005, ApJS, 161, 9

\bibitem[Kim \etal\ 2004]{}Kim, D.-W., Cameron, R. A., Drake, J. J., \etal\ 2004a, ApJS, 150, 19

\bibitem[Kim \etal\ 2004]{}Kim, D.-W., Wilkes, B., Green, P., \etal\
 2004b, ApJ, 600, 59

\bibitem[Kim \etal\ 2006]{}Kim, D.-W., \etal\ 2006, ApJ, 652, 1090

\bibitem[Kim \etal\ 2007a]{}Kim, M., \etal\ 2007a, ApJS, 169, 401

\bibitem[Kim \etal\ 2007b]{}Kim, M., \etal\ 2007b, ApJ, 659, 29

\bibitem[King 2002]{}King, A. R. 2002, MNRAS, 335, L13

\bibitem[Kraft \etal\ 2001]{}Kraft, R. P., \etal\ 2001, ApJ, 560, 675

\bibitem[Kundu \& Whitmore]{}Kundu, A., Whitmore, B. C. 2001, ApJ, 121, 2950

\bibitem[McClintock \& Remillard 2006]{} McClintock, J. E., \& Remillard, R. A. 2006, in Compact Stellar X-Ray Sources, ed. W. H. G. Lewin \& M. van der Klis (Cambridge: Cambridge Univ. Press), in press

\bibitem[Park et al 2006]{} Park, T., Kashyap, V., Siemiginowska, A.,
van Dyk, D., Zezas, A., Heinke, C., Wargelin, B. 2006, ApJ, 652, 610

\bibitem[Piro \& Bilsten. 2002]{}Piro, A. L. \& Bildsten, L. 2002,
ApJ, 571, L103

\bibitem[Scargle 1998]{}Scargle, J.D. 1998, ApJ, 504, 405 

\bibitem[Sivakoff 2008]{}Sivakoff, G.~R., Jord\'{a}n, A., Juett,
A.~M., Sarazin, C.~L., \& Irwin, J.~A., 2008, astro-ph/0806.0627

\bibitem[Terlevich \& Forbes. 2002]{}Terlevich, A. I. \& Forbes, D. A. 2002, MNRAS, 330, 547

\bibitem[Tonry \etal\ 2001]{}Tonry J., \etal\ 2001, ApJ, 546, 681

\bibitem[van Dyk \etal\ 2001]{} van Dyk, D., \etal\ 2001, ApJ, 548, 224

\bibitem[Verbunt \& van den Heuvel. 1995]{}Verbunt, F. \& van den Heuvel, E. P. J. 1995, in X-ray Binaries, eds. Lewin, W. H. G., van Paradijs, J., van den Heuvel, E. P. J., Cambridge, UK: CUP, p.457

\bibitem[Voss \& Gilfanov 2006]{}Voss, R. \& Gilfanov, M. 2006, A\&A, 447, 71

\bibitem[Weisskopf 2000]{}Weisskopf, M. C., Tananbaum, H. D., Van Speybroeck, L. P. \& O'Dell, S. L. 2000, Proc. SPIE, 4012, 2

\bibitem[Williams \etal\ 2008]{}Williams, B.~F., \etal\ 2008, ApJ, 680, 1120

\bibitem[Xu \etal\ 2005]{}Xu, Y., Xu, H., Zhang, Z., Kundu, A., Wang, Y. \& Wu, X.-P., 2005, 631, 809

\bibitem[Zezas \etal\ 2002]{}Zezas, A, Fabbiano, G., Rots, A. H.,
Murray, S. S. 2002, ApJ, 577, 710

\end{thebibliography}
\end{document}